\documentclass[onecolumn]{aastex61}
\pdfoutput=1
\usepackage[colorinlistoftodos]{todonotes}
\usepackage{float}
\usepackage{chngcntr}
\counterwithin*{footnote}{section}
\usepackage[english]{babel}
\usepackage[toc,page]{appendix}
\usepackage[section]{placeins}
\usepackage{graphicx}
\usepackage{amsmath}
\usepackage[colorinlistoftodos]{todonotes}
\usepackage{float}
\usepackage{natbib}
\usepackage{gensymb}
\usepackage[section]{placeins}
\usepackage{epstopdf}
\usepackage[colorinlistoftodos]{todonotes}
\usepackage{float}
\usepackage[english]{babel}
\usepackage[utf8x]{inputenc}

\begin{document}

\title{The VLA Nascent Disk And Multiplicity Survey of Perseus Protostars (VANDAM).\\
 IV. Free-Free Emission from Protostars: Links to Infrared Properties, Outflow Tracers, and Protostellar Disk Masses.}
 
\author{Łukasz Tychoniec}
\affiliation{Leiden Observatory, Leiden University, P.O. Box 9513, NL-2300RA Leiden, The Netherlands} 
\affiliation{Astronomical Observatory, Faculty of Physics, Adam Mickiewicz University, Słoneczna 36, PL-60268 Poznań, Poland}

\author{John J. Tobin}
\affiliation{Leiden Observatory, Leiden University, P.O. Box 9513, NL-2300RA Leiden, The Netherlands} 
\affiliation{Homer L. Dodge Department of Physics and Astronomy, University of Oklahoma, 440 W. Brooks Street, Norman, OK 73019, USA}

\author{Agata Karska}
\affiliation{Centre for Astronomy, Nicolaus Copernicus University in Toruń, Faculty of Physics, Astronomy and Informatics, Grudziadzka 5, PL-87100 Toruń, Poland} 

\author{Claire Chandler}
\affiliation{National Radio Astronomy Observatory, P.O. Box O, 1003 Lopezville Road, Socorro, NM 87801-0387, USA}

\author{Michael M. Dunham}
\affiliation{Harvard-Smithsonian Center for Astrophysics, 60 Garden St., Cambridge, MA, USA}
\affiliation{Department of Physics, State University of New York Fredonia, Fredonia, NY 14063, USA}

\author{Robert J. Harris}
\affiliation{Department of Astronomy, University of Illinois, Urbana, IL 61801, USA}

\author{Kaitlin M. Kratter}
\affiliation{Department of Astronomy and Steward Observatory, University of Arizona, 933 N Cherry Ave, Tucson, AZ 85721, USA}

\author{Zhi-Yun Li}
\affiliation{Department of Astronomy, University of Virginia, Charlottesville, VA 22903, USA}

\author{Leslie W. Looney} 
\affiliation{Department of Astronomy, University of Illinois, Urbana, IL 61801, USA}

\author{Carl Melis}
\affiliation{Center for Astrophysics and Space Sciences, University of California, San Diego, CA 92093, USA}

\author{Laura M. P\'erez}
\affiliation{Center for Astrophysics and Space Sciences, University of California, San Diego, CA 92093, USA}
\affiliation{Universidad de Chile, Departamento de Astronoma, Camino El Observatorio 1515, Las Condes, Santiago, Chile}

\author{Sarah I. Sadavoy}
\affiliation{Harvard-Smithsonian Center for Astrophysics, 60 Garden St., Cambridge, MA, USA}
\affiliation{Max-Planck-Institut f{\"u}r Astronomie, K{\"o}nigstuhl 17, D-69117 Heidelberg, Germany}

\author{Dominique Segura-Cox}
\affiliation{Department of Astronomy, University of Illinois, Urbana, IL 61801, USA}

\author{Ewine F. van Dishoeck}
\affiliation{Leiden Observatory, Leiden University, P.O. Box 9513, NL-2300RA Leiden, The Netherlands} 
\affiliation{Max-Planck Institut f{\"u}r Extraterrestrische Physik, Giessenbachstrasse 1, 85748 Garching, Germany}

\correspondingauthor{Łukasz Tychoniec}
\email{tychoniec@strw.leidenuniv.nl}

\begin{abstract}
Emission from protostars at centimeter radio wavelengths has been shown to trace the free-free emission arising from ionizing shocks as a result of jets and outflows driven by protostars. 
Therefore, measuring properties of protostars at radio frequencies can provide valuable insights into the nature of their outflows and jets.
We present a C-band (4.1 cm and 6.4 cm) survey of all known protostars (Class 0 and Class I) in Perseus as part of the VLA Nascent Disk and Multiplicity (VANDAM) Survey. 
We examine the known correlations
between radio flux density and protostellar parameters such as bolometric luminosity and outflow force, for our sample. We also investigate the relationship between radio 
flux density and far-infrared line luminosities from \textit{Herschel}. We show that free-free emission originates most likely from J-type shocks; however, the large scatter indicates that those two types of emission probe different time and spatial scales. Using C-band fluxes, we removed  an estimation of free-free contamination from the corresponding Ka-band (9 mm) flux densities that primarily probe dust emission from embedded disks. We find that the compact ($<$~1\arcsec) dust emission is lower for Class I sources (median dust mass 96 M$_{\oplus}$)  relative to Class 0 (248 M$_{\oplus}$), but several times higher than in Class II (5-15 M$_{\oplus}$). If this compact dust emission is tracing primarily the embedded disk, as is likely for many sources, this result provides evidence for decreasing disk masses with protostellar evolution, with sufficient mass for forming giant planet cores primarily at early times.
\end{abstract}

\section{Introduction}
Stars are born through a collapse of cold cores of dust and gas, usually within molecular clouds. A significant fraction of the parental core material is, however, dispersed by powerful outflows and jets rather than incorporated into the protostar \citep[e.g.,][]{Arce2006, Offner2014}. Both outflows and jets are key features observed in star-forming regions toward most young stellar objects \citep{Frank2014}. Outflow properties are expected to reflect the age and activity of the embedded protostar. For example, studies have shown that outflows decrease in force with protostellar evolution \citep[e.g.,][]{Bontemps1996, Yildiz2015} and outflow ejection rates correlate with accretion onto the central protostar \citep[e.g.,][]{Shu1994, Mottram2017}. Those characteristics suggest that the earliest stages of star formation are essential to investigate because this is the period where stars accumulate most of their mass and are interacting most vigorously with the core and cloud by means of outflows.

Ejecta from the protostar can have different forms. Fast, supersonic jets are well collimated and they interact
with cold gas around the protostar in shock events. While likely consisting of atomic gas, it is observed that they can also be composed of high-velocity molecular gas, especially in very young sources \citep[e.g.,][]{Bachiller1990, Tafalla2004, Hirano2010}. Molecules, however, are most frequently observed in the much wider, and slower outflow, which contains more mass than a jet. The relationship between the outflow and the jet is still strongly debated, but
there is a growing body of evidence, both from observations  \citep[e.g.,][]{Nisini2015, Dionatos2017} and simulations \citep[e.g.,][]{Machida2014} suggesting that the collimated jet is also powering the wide molecular outflow.

Radio continuum emission from protostars is a unique tracer of the ionized component of the protostellar jet. Radio emission from protostars often appears as an unresolved and compact counterpart to the infrared and submillimeter detections. With high-resolution observations, extended radio emission is often elongated along the direction of the large-scale jets \citep[e.g.,][]{Curiel1989, Anglada1995}, suggesting it is tracing the base of the collimated jet. The radio jets from protostars are most often found toward those in the intermediate and high-mass regime \citep[e.g.,][]{Rodriguez1989a, Curiel1993, Girart2002}, but examples of low-mass protostars with radio jets are known as well \citep[e.g.,][]{Rodriguez1997, Tychoniec2018}.

Emission at centimeter wavelengths can track various processes in the protostellar environment. The radio spectral index ($\alpha$; where $F_\nu \sim \nu ^\alpha$) can be used to distinguish between different types of emission.
Thermal dust emission usually has a steep spectrum with  $\alpha = 2 + \beta$ where $\beta \lesssim 1$ 
for dense disks with large grains \citep{Kwon2009,Testi2014}. Dust emission is still detectable at $\sim$ 1 cm, but is not expected to contribute significantly at C-band. The free-free emission from ionized gas has a spectral 
index with typical values from -0.1 to 2.0 \citep{Panagia1975, Rodriguez2003}. Spectral indices below -0.1 are indicative
 of non-thermal emission generally associated with synchrotron emission resulting from high-velocity electrons interacting with magnetic fields \citep[e.g.,][] {Rybicki1979}. 
This mechanism has been verified as a possibility since polarization in a protostellar radio jet with a negative spectral index has been detected 
\citep{Carrasco-Gonzalez2010}. More evolved pre-main sequence stars can exhibit the negative spectral indices due to the gyrosynchrotron emission from the stellar coronae \citep[e.g.,][]{Dzib2013}.

Understanding the contribution of different mechanisms of emission at radio wavelengths is essential not only to analyze ionized jets but  also to analyze the dust emission at radio wavelengths. The free-free emission can significantly contribute to the continuum at shorter wavelengths thereby increasing the measured flux densities. Any free-free contamination must be removed to obtain accurate measurements of dust properties and masses of the youngest protostellar disks.

To date, numerous studies have examined radio emission from protostars. Several authors have compiled existing observations and identified general trends between radio emission and protostellar properties \citep[e.g.,][]{Anglada1995, Furuya2003, Shirley2007, Wu2004}, while others conducted surveys of molecular clouds. However, the surveys so far lacked sensitivity, resolution and/or sample size \citep[e.g.,][]{Reipurth2004, Scaife2011, Dzib2013, Pech2016}.

The VLA Nascent Disk and Multiplicity Survey (VANDAM) \citep{Tobin2015a} is able to overcome previous limitations by targeting the largest homogeneous sample of protostars at 0.8, 1.0, 4.1, and 6.4 cm observing wavelengths. The VANDAM survey targeted all known Class 0 and Class I protostars in the Perseus molecular cloud, providing unbiased observations of the radio jets from those sources.
Perseus is a natural choice for this survey, hosting not only the greatest number of young stellar objects among the nearby clouds but also the largest fraction of Class 0 and Class I protostars \citep{Evans2009}. The distance to Perseus (235 pc; \citealt{Hirota2011}) guarantees high spatial resolution observations.

In this paper, we present C-band observations (4.1 and 6.4 cm) from the NSF's Karl G. Jansky Very Large Array
of all known protostars in the Perseus molecular cloud, including flux densities and derived spectral indices. We also calculate masses of compact dust emission at 9 mm from Ka-band observations, taking into account the free-free contributions based on the C-band data. Furthermore, we compare those parameters with protostellar properties such as bolometric luminosity and temperature, molecular and atomic far-infrared line luminosities, and outflow force.

\subsection{The Sample}
A total of 95 protostars were targeted by the VANDAM survey in C-band, summarized in Table \ref{tab:table1}. The sample was selected using \textit{Spitzer}, 
\textit{Herschel}, and Bolocam observations \citep{Enoch2009, Evans2009, Sadavoy2014}. The sources have bolometric luminosities between 0.1 L$_{\odot}$  and 33 L$_{\odot}$, spanning the low-mass regime.
 For a detailed description of the source sample selection, see \cite{Tobin2016}. The non-detection of three Class II sources in Ka-band: EDJ2009-161, EDJ2009-333, and EDJ2009-268 resulted in them being excluded from the C-band observations. On the other hand, serendipitous Ka-band detections of the Class II sources: EDJ2009-233, EDJ2009-173, EDJ2009-235, and the pre-main sequence binary system SVS3, are included in the C-band sample. 

\section{Observations and Analysis}
We conducted C-band observations with the VLA in A-configuration between 2014 February 28 and 2014 April 12. The C-band data (4.1 and 6.4 cm) were taken in 8-bit mode, yielding 2 GHz of bandwidth divided into sixteen
128 MHz sub-bands with 2 MHz channels and full polarization products.

We centered 1 GHz basebands at 4.7 and 7.4 GHz avoiding some persistent radio frequency interference in these bands. The observations in two different frequencies allow the measurement of the spectral index which is crucial in the characterization of the sources and discriminating between protostars and extragalactic sources.
The radio source 3C48 was both the absolute flux density and bandpass calibrator and J0336+3218 was the complex gain calibrator. The estimated absolute flux calibration uncertainty is  $\sim$ 5$\%$
 and is not included in the reported flux density errors. This error will not influence the spectral index, as it is obtained from observations at the two ends of the same band, and thus limited only by the 
uncertainty of the flux calibrator model \citep[$\sim$2\%;][]{Perley2017}.
 Further details of the calibration and data reduction of the C-band observations are described in the previous VANDAM papers \citep{Tobin2015a,Tychoniec2018}

The large primary beam of the C-band observations - 5\arcmin~and 7.2\arcmin~FWHM  for 4.1 and 6.4 cm, respectively - means that fewer pointings are necessary, as compared to Ka-band observations and 38 fields were observed in total.  Due to the overlap of the fields, some sources have multiple detections. In those cases, the detection with the lowest distance to the primary beam center was used in the analysis. The typical size of the synthesized beam was 0\farcs3-0\farcs4  with a typical RMS noise of 4-6 $\mu$Jy. Separate characteristics of each field are provided in Table \ref{tab:table2}.
We used the AEGEAN source finder version r903 \citep{Hancock2012} to identify sources in all the fields with a specific \textit{seed} threshold, defining the lowest peak value for the source to be claimed real, set to 6$\sigma$. With the CASA \citep[version 4.2.2;][]{McMullin2007} \textit{imstat} procedure we obtained RMS over the whole image and we used it as an input in the source finder code. Field C15, C16, and C21, have prominent radio galaxies that created artifacts in the maps. For these fields, we measured the noise value manually in an area unaffected by the bright sources.  Frames were also cross-checked manually for the protostars not detected by the source finder code and detections over 3$\sigma$ at protostellar positions were added to the sample.

Based on the method described above, the list of objects was created and we performed 2D Gaussian fitting with the CASA task \textit{imfit} to measure flux densities and corresponding errors. Unresolved sources with relatively faint emission (below 15$\sigma$) were fit using Gaussians with position angle and sizes that matched the synthesized beam to avoid unrealistic fit parameters.
For sources with extended emission, the source finder code provided multiple peaks of emission that were subsequently used in the \textit{imfit} task as the Gaussian peaks.  For these sources, the resulting flux density is the sum of all components.  Finally, we corrected fluxes for the primary beam attenuation.

In this work, we explore correlations between measured flux densities and protostellar properties. Due to a large number of non-detections of known protostars, proper accounting of upper limits enables us to derive more accurate correlations from the data. For correlations, we use The Space Telescope Data Analysis System (STSDAS) {\it statistics} package, that allows one to analyze datasets with upper and/or lower limits. To estimate the correlation strengths, we use Spearman's rank correlation coefficient ($\rho$), obtained with STSDAS {\it spearman} procedure which also provides the probability of no correlation ($P$). The Expectation-Maximization algorithm (EM) is used to obtain parameters of the best linear fit to the data with the procedure {\it emmethod}. For equations and implementation of the data censoring, see \cite{Isobe1986}. To determine if two sets of values are statistically different, we use log-rank test and a Kaplan-Meier (KM) estimator to produce cumulative distribution functions. Both procedures are implemented within the LIFELINES package for Python \citep{DavidsonPilon2017} which takes upper limits into account.

\section{Results}

\subsection{Detections}
From the targeted protostars in Table \ref{tab:table1}, we report detections in C-band for 60 out of 95 systems (63\%) in either 4.1 or 6.4 cm.  Specifically, 31 out of 46 Class 0 (67\%) and 21 out of 37 Class I  (56\%) protostars were detected. We detect 9 of 12 (75\%) of targeted Class II systems, but this sample is smaller and biased towards more embedded sources. Out of all systems, 23 have multiple stellar components (21 binary and 2 triple systems) as identified by \cite{Tobin2016}; three of those are unresolved in C-band, which results in 117 targeted individual protostars. We detect 11 components of multiple systems (6 Class 0, 3 Class I, and 2 Class II). Thus, the total number of protostars with measured flux in at least one of the wavelengths in C-band is 71, making a detection rate of 61\% with 37/57 (65\%) Class 0, 24/45 (53\%) Class I, and  10/15 (75\%) Class II protostars. For known protostars that were not detected, we used 3$\sigma$ upper limits based on the RMS of the field, corrected for the primary beam attenuation.

For binary systems, we additionally calculated the combined flux of all components together for comparison with parameters that were obtained for unresolved systems. For example, when comparing with outflow force, it is not possible to determine which of the close
companions is the outflow driving source, and the same applies to the bolometric luminosity. Far-infrared observations have lower resolution than available with interferometry, so one obtains the luminosity of both components. However, when comparing with bolometric temperature, we compare the flux densities separately for each component of the multiple system, assuming that both companions are at the same evolutionary stage, which is generally a good assumption \citep{Murillo2016}.
The summary of measured flux densities and spectral indices is presented in Table \ref{tab:table3}.

Apart from the targeted protostars, we serendipitously detected a plethora of radio sources within the large
 C-band primary beam. All of them were compared with the SIMBAD catalog. Some of them were detected previously and 17 sources from this sample were marked by various authors as YSO candidates. Due to their tentative classification, they are not considered in the further analysis. However, we note that 8 of them have positive radio spectral indices in C-band as expected for protostars. The  more evolved pre-main sequence stars may exhibit negative indices \citep[e.g.,][]{Dzib2013}, and distinguishing them from extragalactic sources is difficult by means of spectral index, thus making cross-matched catalogs important. The summary of the sources with possible protostellar nature is presented in Table \ref{tab:table4}.

In Table \ref{tab:table5} we present 59 previously detected sources of various nature, including 16 stars (2 T-Tauri stars), 27 radio, 8 X-ray, 4 infrared unclassified radio sources, 1 brown dwarf, and 3 associated with starless cores. Negative spectral indices prevail in this sample, indicating non-thermal processes. For stars, the non-thermal emission is probably related to coronal activity, while for unclassified sources it would point to their extragalactic nature. For 12 sources, \cite{Pech2016} reported new detections, and we list them in Table \ref{tab:table6}.

Across the entire sample we detect 490 new sources. Table \ref{tab:table7} lists these new detections. We assume that most of them are extragalactic. To test this, we estimate the expected amount of extragalactic sources based on the equations from \cite{Anglada1998} (see their Appendix) derived from number counts of radio sources \citep{Condon1984, Rodriguez1989c}.
For a detection threshold $F_{\lambda}$, the expected number of extragalactic sources per primary beam is:
\begin{equation}
\textrm{N}_{6.4} = 1.15\  F_{6.4}^{-0.75}
\end{equation}
\begin{equation}
\textrm{N}_{4.1} = 0.40\  F_{4.1}^{-0.75}
\end{equation}
With the 6 $\sigma$ threshold used in the source finder we obtain values of:

\begin{equation}
\textrm{N}_{\ 6.4} \sim 16\textrm{, for}\ F_{6.4} \geq 30\ \mu \textrm{Jy},\
\end{equation}
\begin{equation}
\textrm{N}_{\ 4.1} \sim 7\textrm{, for} \ F_{4.1} \geq 24\ \mu \textrm{Jy},\
\end{equation}

For the new detections, we find average numbers of  $\textrm{N}_{\ 6.4}\sim15$ and 
$\textrm{N}_{\ 4.1}\sim11$ per field. These average values are broadly consistent with the expected number of extragalactic sources, although 4.1 cm value is a bit high. This estimate depends on an assumed spectral index of the extragalactic sources ($\alpha = -0.7$). If some of the sources have flatter indices, we would expect even more of them to be detected at 4.1 cm than predicted.

\subsection{Flux densities from protostars}

\begin{figure}[H]
\centering
\includegraphics[width=0.4\linewidth]{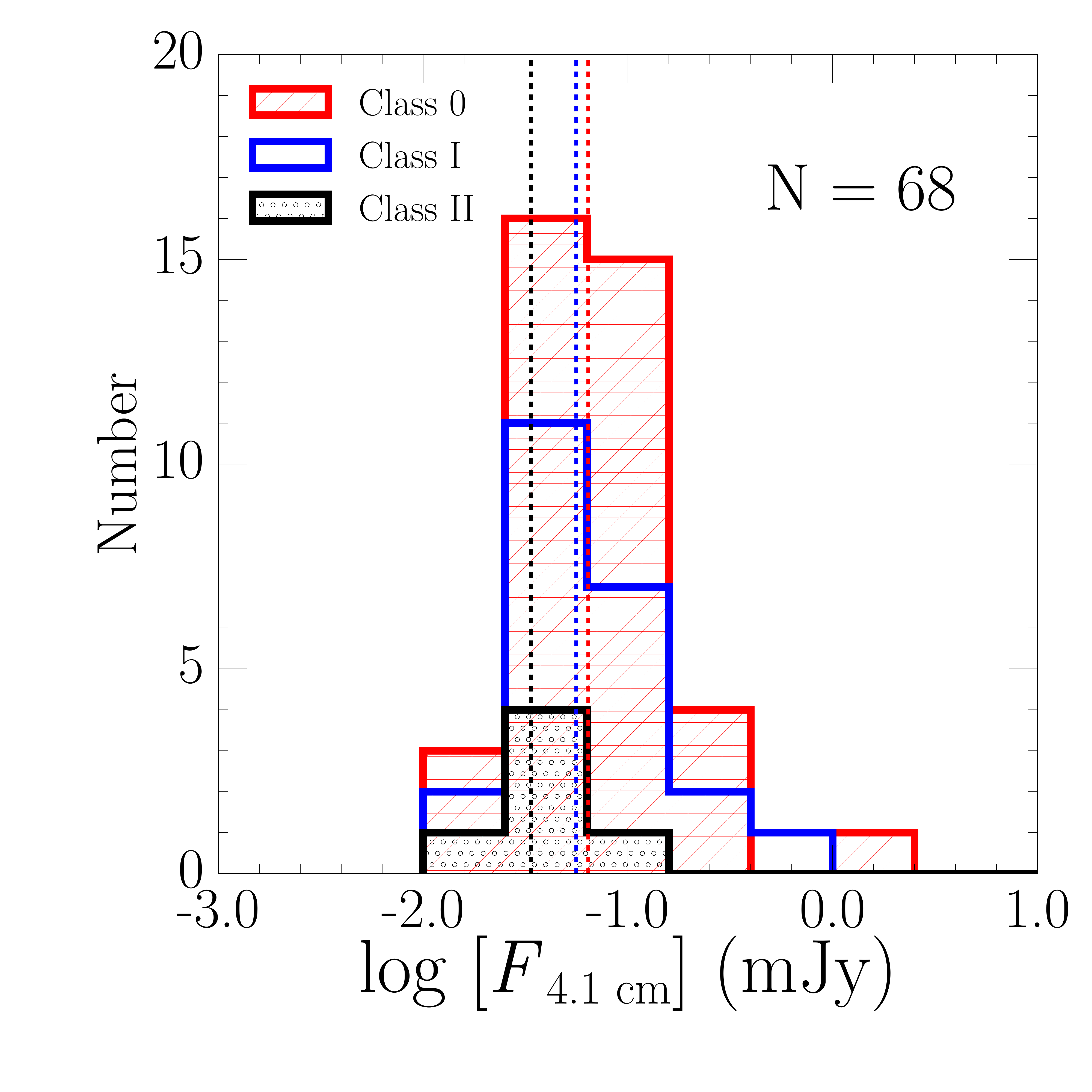}
\includegraphics[width=0.4\linewidth]{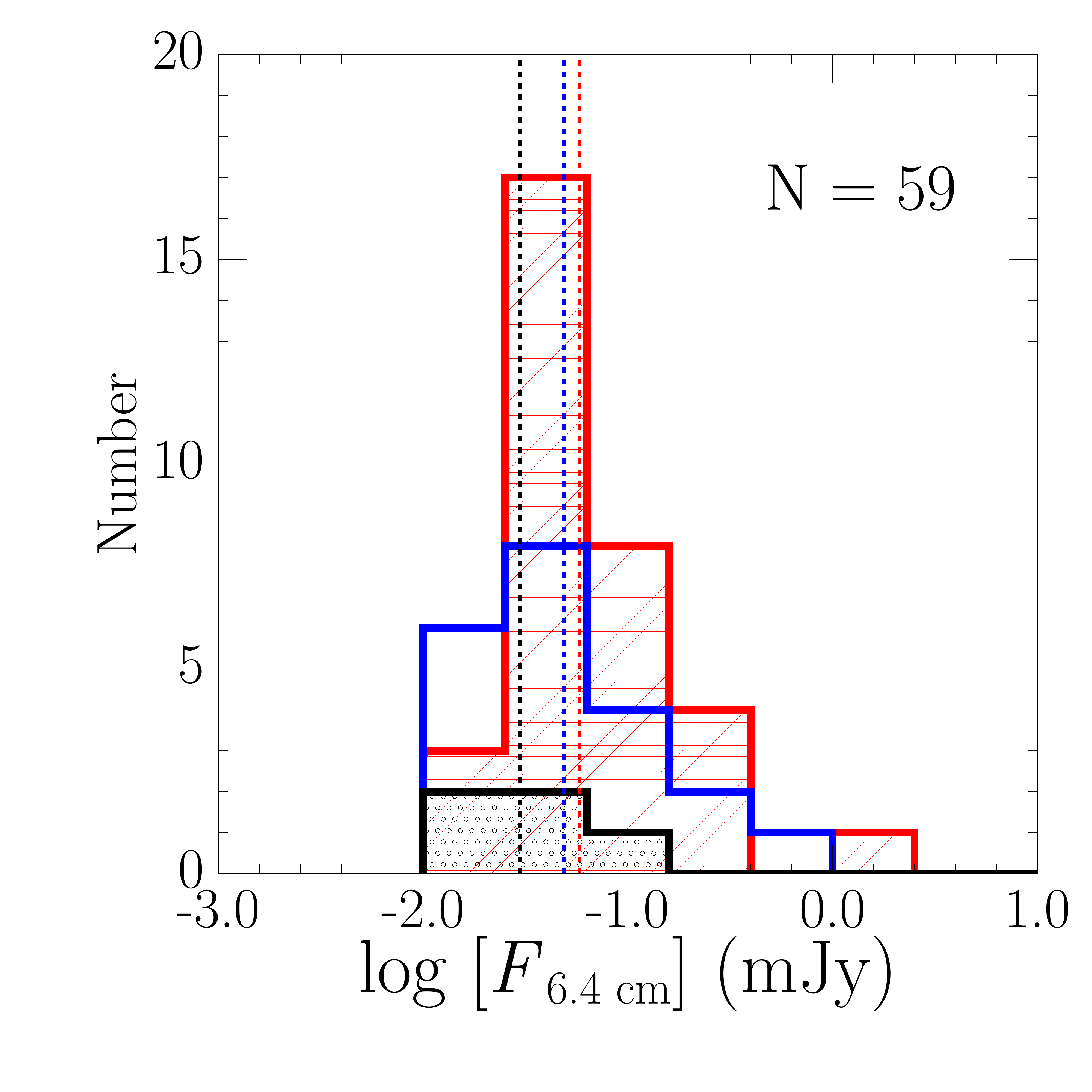}

\caption{Distribution of flux densities for 4.1 cm (left) and 6.4 cm (right). Dashed lines show the median for each evolutionary class. The median values for 4.1 cm flux are 0.064 mJy, 0.056 mJy, and 0.034 mJy, for Class 0, Class I, and Class II. The median values for 6.4 cm flux are 0.058 mJy, 0.048 mJy, and 0.033 mJy for Class 0, Class I, and Class II}
  \label{fig:fluxhistogram}
\end{figure}

Figure \ref{fig:fluxhistogram} shows histograms of flux densities at 4.1 and 6.4 cm from the known protostars. We use the log-rank test to estimate probabilities 
for Class 0 and Class I fluxes to be drawn from the same sample. We obtain high probabilities of 64\% and 54\% for 4.1 cm and 6.4 cm respectively, 
consistent with no difference between the two samples. This result, combined with no significant difference between the fraction of detected protostars (65\% for Class 0 and 53\% for Class I) indicates that the radio emission mechanism 
should not differ between the two evolutionary classes. This result might indicate that the thermal radio jets are not driven by the release of accretion energy, 
which is expected to decrease from Class 0 to Class I \citep{Fischer2017}.
This is in agreement with \cite{Pech2016}, who show for a smaller sample of protostars consistent
fluxes between Class 0 and Class I. However, other sample-limited studies suggest that the radio emission mechanisms could be different for Class 0 and Class I protostars \citep{Scaife2011}.

\begin{figure}[H]
\centering
\includegraphics[width=0.8\linewidth]{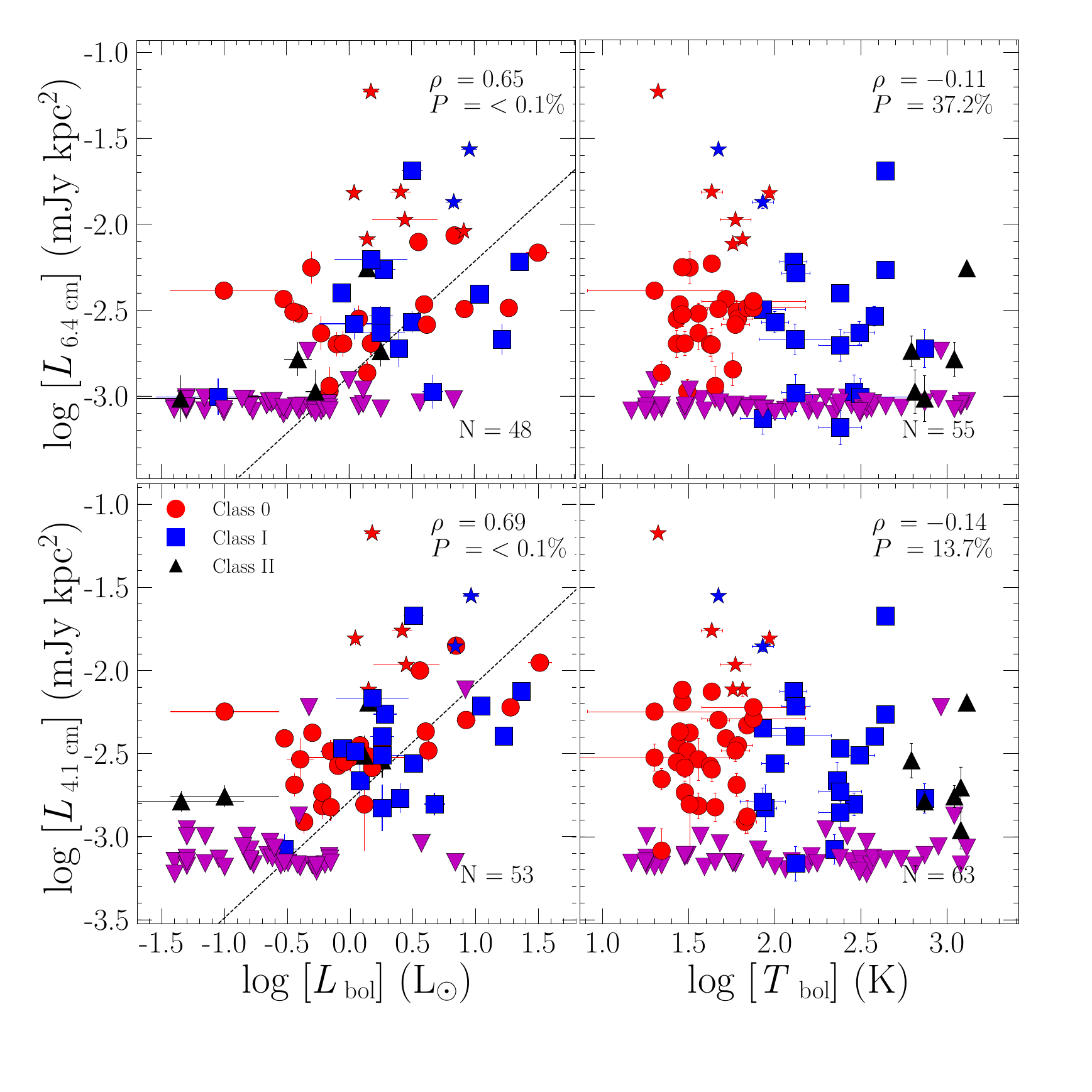}
\caption{Luminosity at 4.1 cm (bottom) and 6.4 cm (top) compared with bolometric luminosity (left) and temperature (right). 
 Spearman's rank correlation coefficient and probability of no correlation is shown in the top-right corner.
 Sources with resolved radio jets are marked as stars and upper limits as magenta triangles.}
\label{fig:fig2}
\end{figure}

Figure \ref{fig:fig2} compares the C-band flux densities corrected for distance (radio luminosities: $L_{\lambda} = F_{\lambda} \times D^2$) with the 
bolometric luminosity and temperature of protostars. The bolometric luminosity is a marker of the protostellar mass and the current accretion rate, 
and the bolometric temperature is often used to infer protostellar evolutionary status. The values used here are taken from multiple works analyzing spectral energy
 distribution of protostars in Perseus \citep{Enoch2009, Sadavoy2014, Young2015, Murillo2016}. We find no correlation with the bolometric temperature, suggesting that the radio emission is
independent of the evolutionary class. Previous studies \citep{Dzib2013, Dzib2015, Pech2016} are consistent with this result at least for the Class 0 to Class II regime.
On the other hand, the radio luminosity shows a weak correlation with the bolometric luminosity.  The EM algorithm provides following fitting parameters:

\begin{equation} \label{eq:5}
\log (L_{\textrm{4.1 cm}})=(-2.78\pm0.07) + (0.70\pm0.10)\ \log(L_{\textrm{bol}}), \rho=0.69
\end{equation}
\begin{equation} \label{eq:6}
\log (L_{\textrm{6.4 cm}})=(-2.89\pm0.06) + (0.67\pm0.10)\ \log(L_{\textrm{bol}}), \rho=0.65
\end{equation}

\subsection{Spectral indices}
With the two C-band fluxes, we calculate the radio spectral index, which is a reliable tool to discriminate between thermal and non-thermal emission processes. We measure the spectral index following:
\begin{equation}
\label{eq:spectral_index}
\alpha=\frac{\ln (F_{\nu_1}/ F_{\nu_2})}{\ln(\nu_1/\nu_2)}
\end{equation}
To calculate the spectral index errors we use a standard propagation of error \citep{Chiang2012}. 

\begin{figure}[H]
\centering
\includegraphics[width=0.4\linewidth]{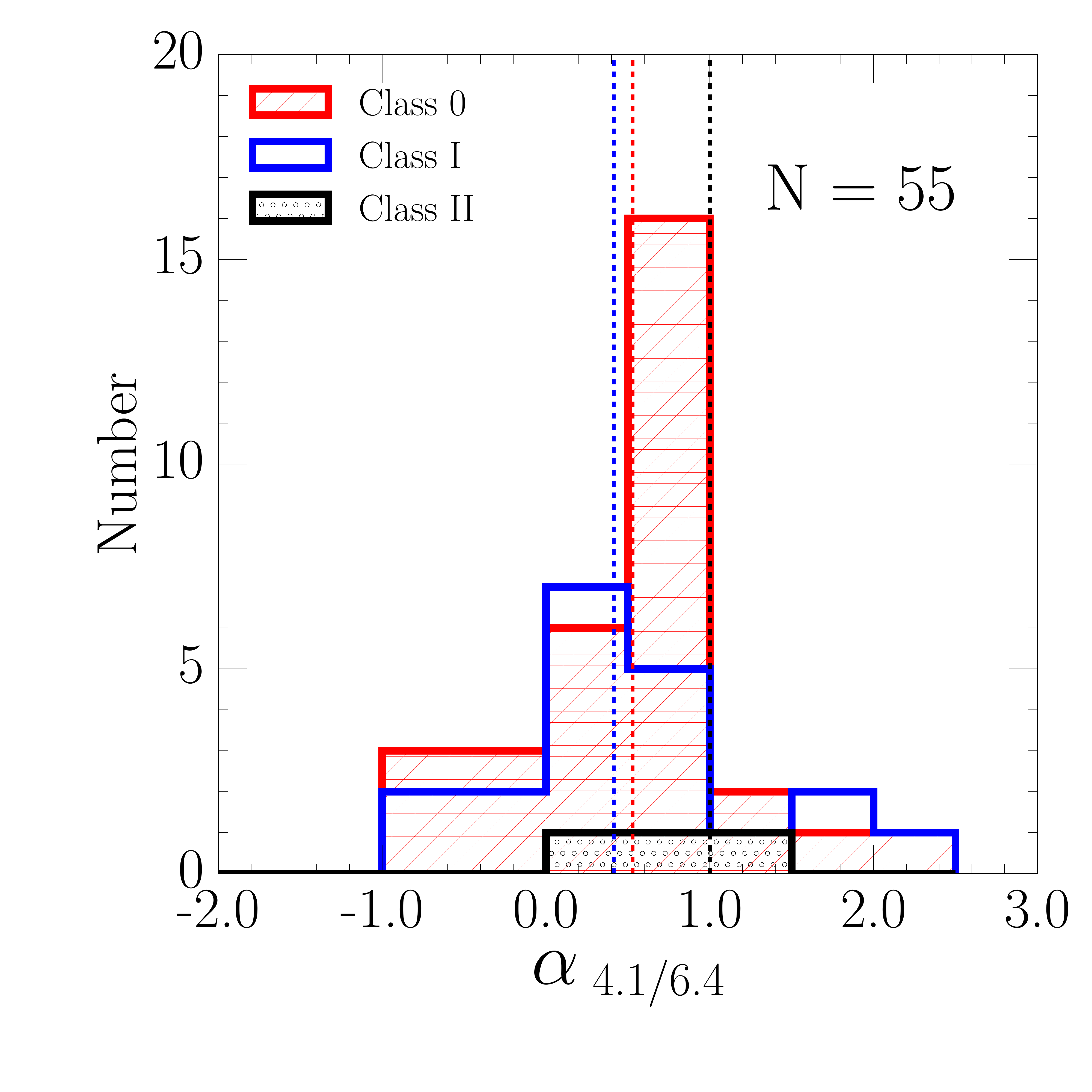}
\caption{ Distribution of spectral indices. Dashed lines show the median values for each evolutionary class. Median values are 0.52, 0.41, 0.99, 0.51 for Class 0, Class I, Class II, and total sample respectively. The statistical probability of Class 0 and Class I spectral indices to be drawn from the same sample is 58\%.}
\label{fig:index_hist}
\end{figure}

Figure \ref{fig:index_hist} shows histograms of the spectral indices for each evolutionary stage. The median values for each distribution are 0.52 for Class 0, 0.41 for Class I, and 0.99 for Class II; the overall median is 0.52. The result from log-rank test for Class 0 and Class I is a 58\% probability of these two being drawn from the same sample, thus there is no evidence for evolutionary trend in radio spectral indices.
The median value for the total sample is in very good agreement with \cite{Shirley2007} who analyzed a sample of sources with wider range of luminosities, and obtained a median index of 0.5.
The median value is also similar to the expected spectral index of $\sim 0.6$ from an unresolved collimated wind \citep{Reynolds1986}. The spectral index is also consistent with the value of 0.6 obtained for spherical winds of stars \cite{Panagia1975}. Thus, with a median value of $\alpha =  0.52$ we cannot determine the origin of the radio emission from the spectral index alone.
Nevertheless, we can rule out some mechanisms from the radio emission. \cite{Rodriguez1993}  conclude that highly negative spectral indices like $\alpha < -0.1$ are explained solely by synchrotron emission and cannot arise from free-free emission. Thus, it is important to list those protostellar sources which fall below the free-free regime. The sources with highly negative spectral indices are Per-emb-9 ($-0.92\pm 0.63$), and Per-emb-19 ($-0.91\pm 0.49$); they are Class 0  objects with low bolometric luminosity 
($L_{\textrm{bol}}$ $<$ 0.6 L$_{\odot}$). The emission from these protostars is compact, but as their signal to noise ratio is low, indicated by the high error of the spectral index measurement, 
they remain consistent with $\alpha>-0.1$ within 2$\sigma$ uncertainty.

\begin{figure}[H]
\centering
\includegraphics[width=0.45\linewidth]{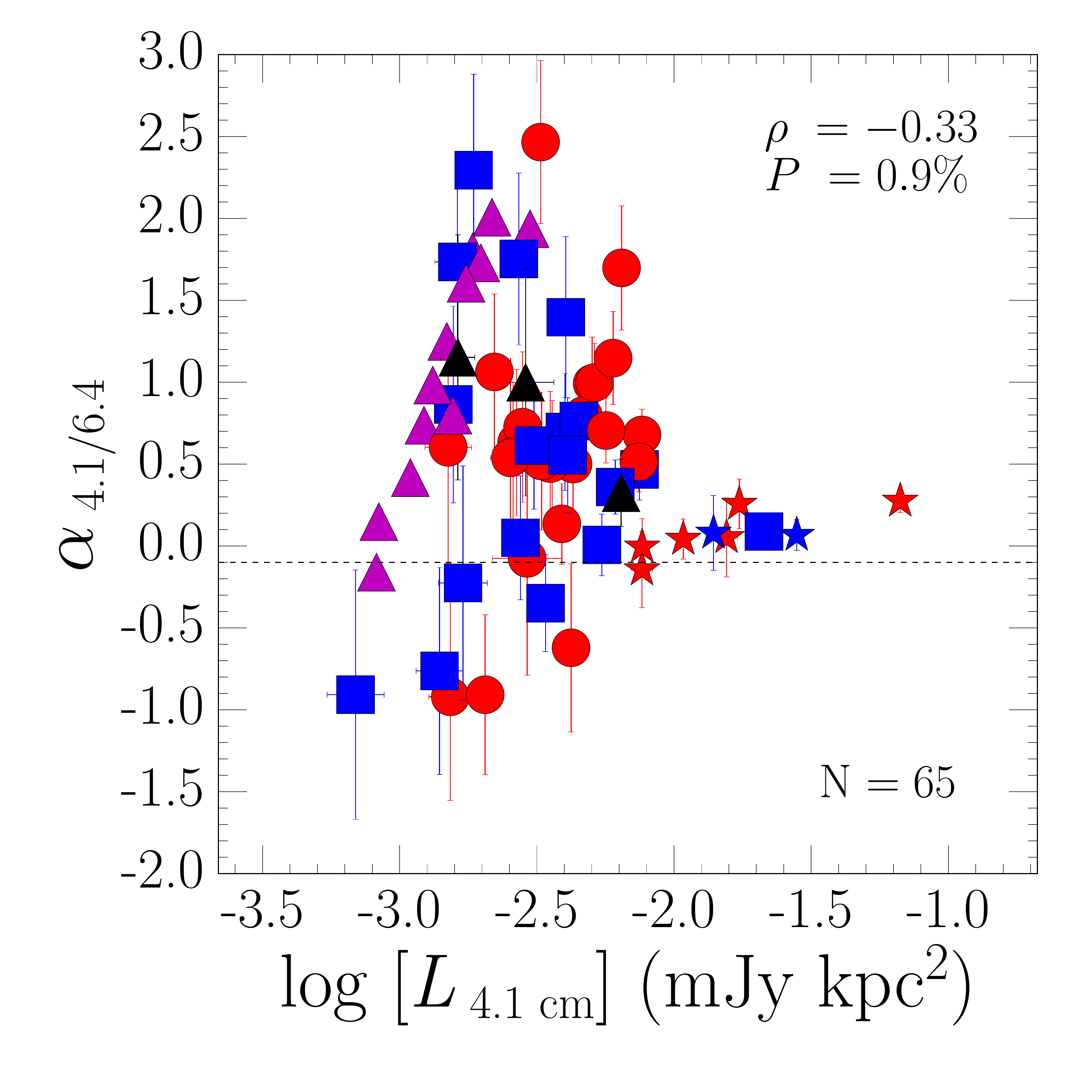}
\includegraphics[width=0.45\linewidth]{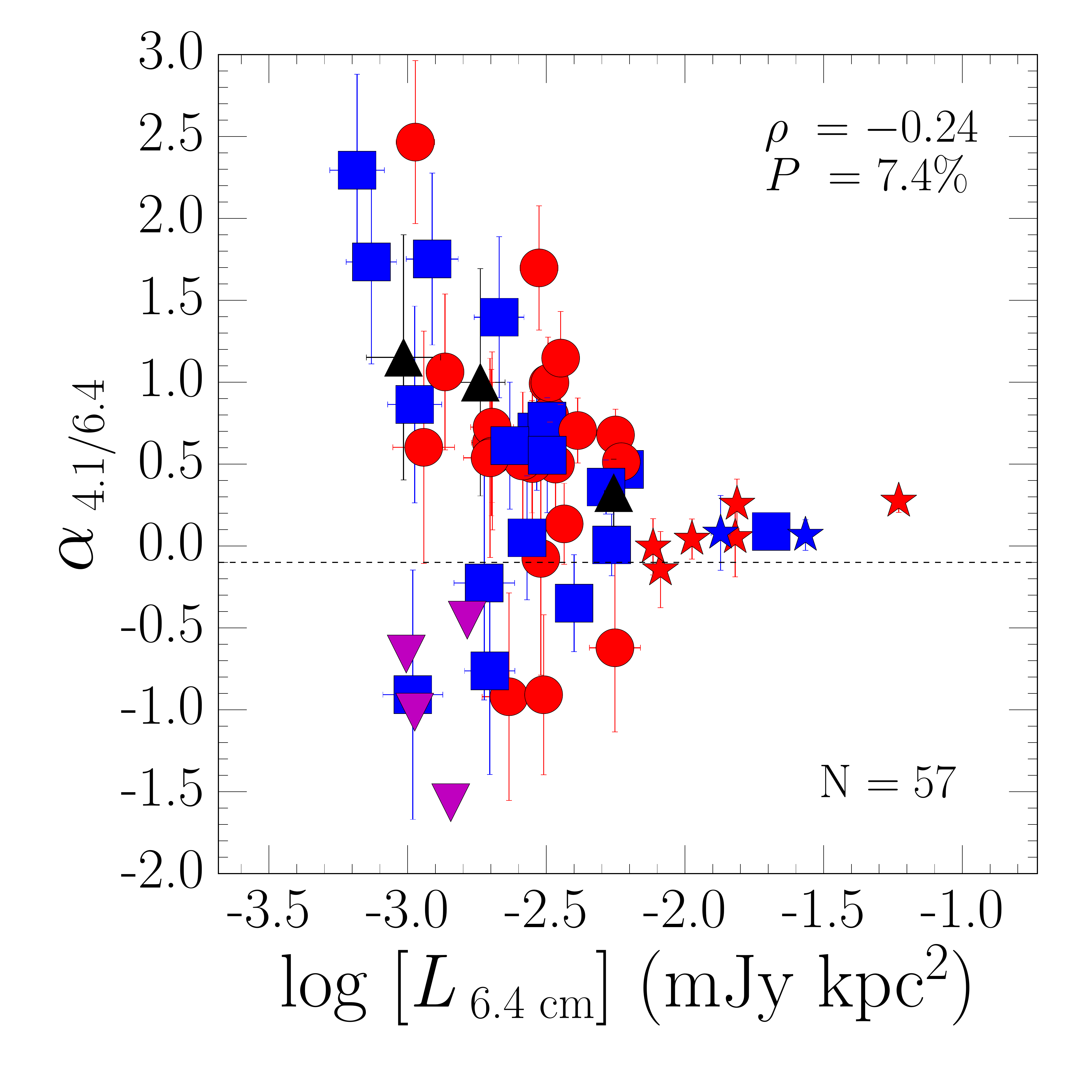}
\includegraphics[width=0.45\linewidth]{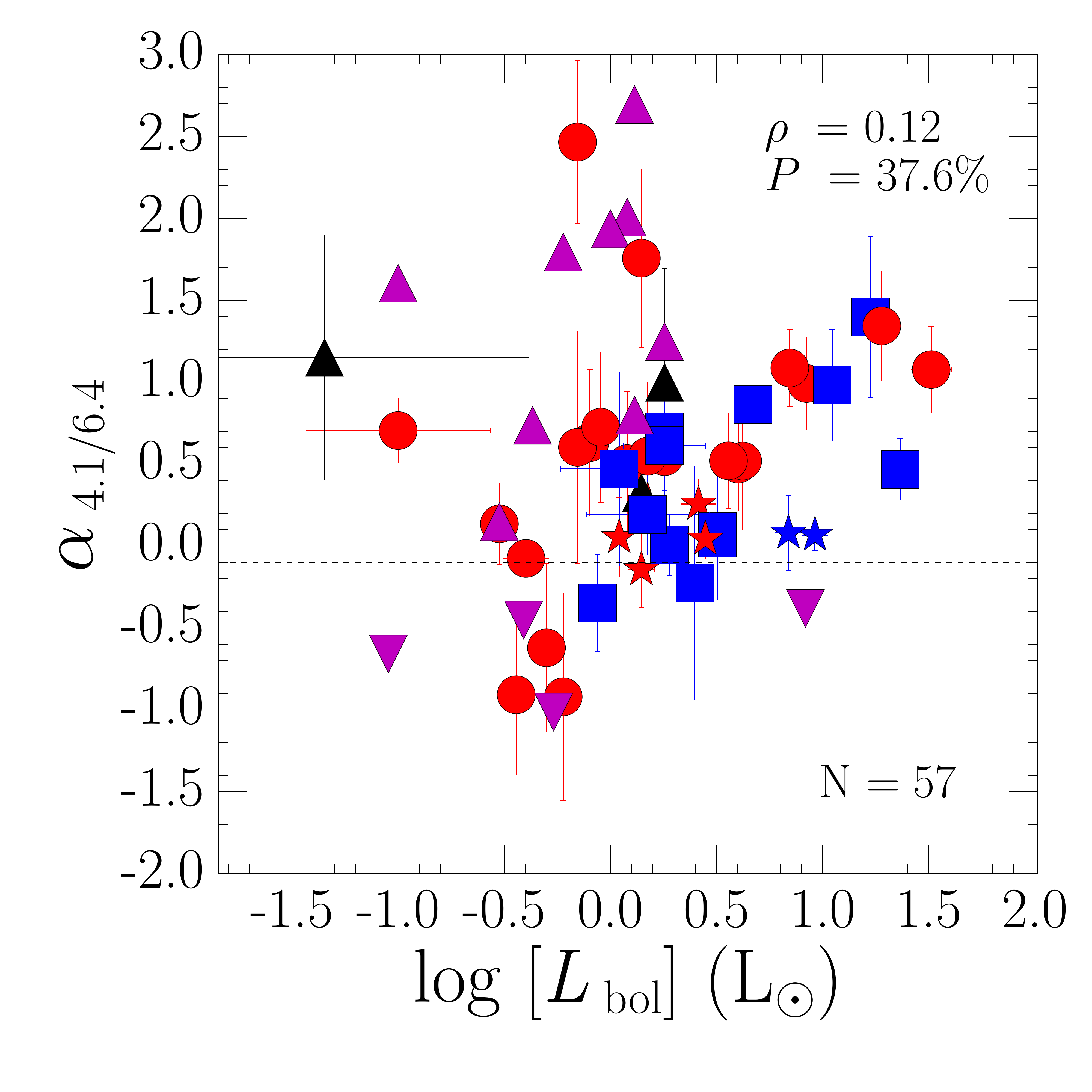}
\includegraphics[width=0.45\linewidth]{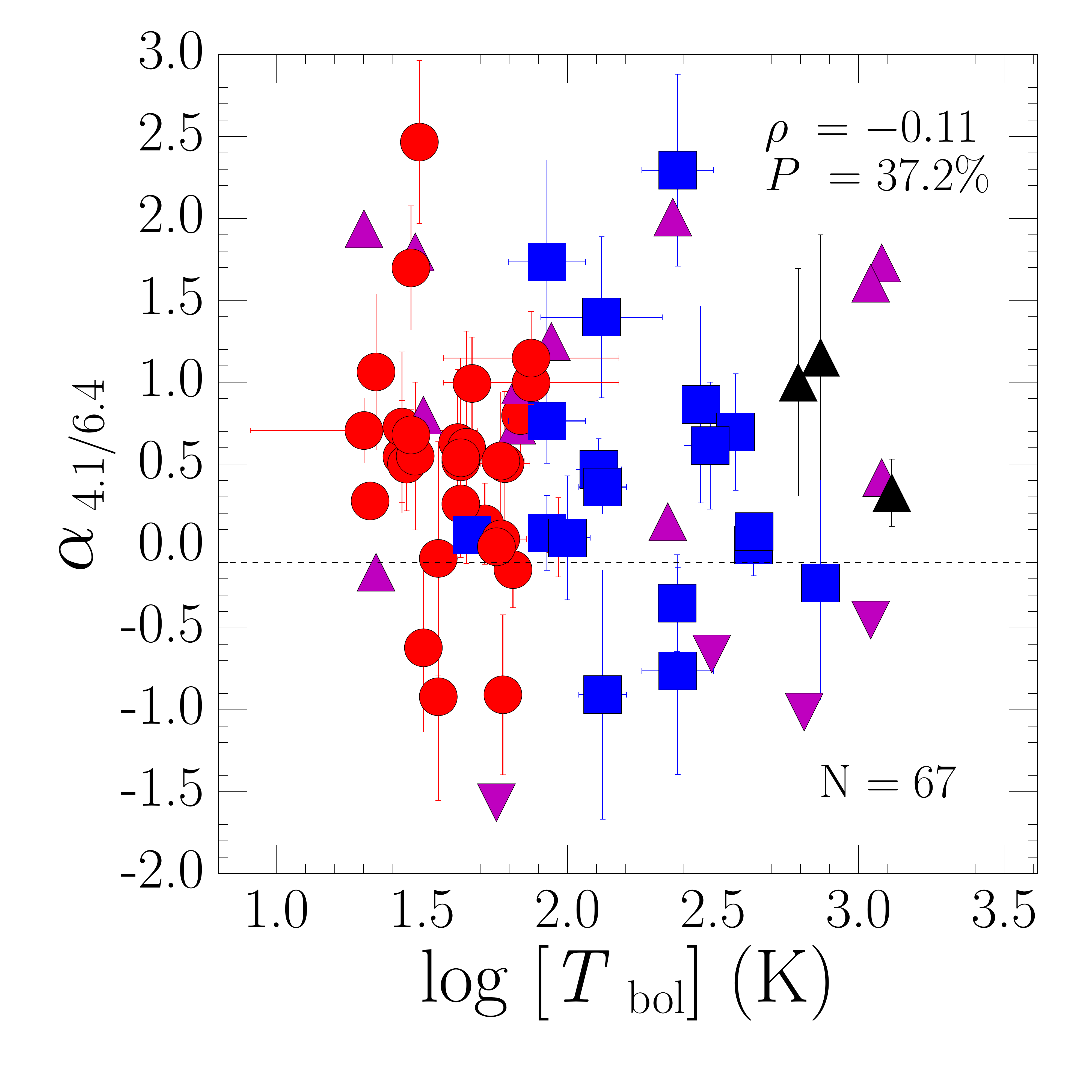}
\caption{ Spectral indices between 4.1 cm and 6.4 cm compared with luminosity at 4.1 cm (top left) and 6.4 cm (top right) and with
 bolometric luminosity (bottom left) and temperature (bottom right). The dashed line indicates minimum value of the spectral index for the free-free emission ($\alpha$ = -0.1).
Sources with resolved radio jets are marked as stars, upper limits as magenta triangles facing down, and lower limits as magenta triangles facing up.}
\label{fig:index}
\end{figure}

Figure \ref{fig:index} compares the observed spectral index with the radio luminosity for the known protostars in our sample. 
It is important to note that the most luminous radio sources ($> 0.01 $ mJy kpc$^2$) have spectral indices below the median for the whole sample,
 near the optically thin limit for the free-free emission which is -0.1. We conclude that it is caused by the emission from optically thin regions of a jet.
 Interestingly, most of those sources exhibit resolved radio jets \citep{Tychoniec2018} so lower spectral indices come most likely from the outflow positions 
where the emission is optically thin or non-thermal emission might contribute. Lower spectral indices from resolved jets were theoretically predicted by \cite{Reynolds1986}. 
The most luminous sources exhibit significantly less scatter than the lower luminosity sources. This can be explained by shock ionization dominating the emission of the bright sources,
 while other, less prominent processes can contribute at low radio luminosities.

\clearpage

We also show the spectral index compared with bolometric luminosity and temperature, in Figure  \ref{fig:index}. We find no correlation between bolometric temperature and spectral index, which suggests that the radio spectral index does not change systematically with protostellar evolution. We found a similar result as in Figure \ref{fig:index_hist}. A trend in spectral indices with increasing bolometric luminosity can be noted by eye. Removing the four outliers and ignoring upper and lower limits seems to give more hints for correlation ($\rho$ =0.49, $P$=0.2\%; see Figure {\ref{fig:lbolindex_out}} in the Appendix \ref{apA}). 
On the other hand including upper and lower limits in the statistical analysis casts doubt on any relation between the two values ($\rho$ =0.12, $P$=50\%).
 This relation was also investigated by \cite{Shirley2007} with the conclusion that the optical depth of the emission is not dependent on the source luminosity. 
Their sample of sources with obtained spectral indices included only three sources with $L_{\rm bol}\ >\ 100\ L_\odot$. Even if the relation is  unclear, we suggest this requires further study. The enhanced capabilities of VLA demonstrated in this work, can be used in a more massive cloud, 
where protostars with wider range of bolometric luminosity are present. This could show if the free-free emission becomes optically thick for sources with more ionizing radiation.

\subsection{Multiple systems}

The VANDAM survey detected a large number of multiple systems in the Perseus molecular cloud. Due to the superior Ka-band resolution, a detailed analysis of
multiplicity was performed with the 8 mm and 1 cm VLA observations \citep{Tobin2016}. A total of 13 new systems with separations below 500 au were detected. Here we examine the emission at longer wavelengths toward these close multiples.

\subsubsection{Comments on systems below 30 au:}
The VLA Ka-band data showed multiplicity on $\sim$ 30 au scales toward 3 sources:  Per-emb-2, Per-emb-5, and Per-emb-18. C-band observations offer lower resolution than Ka-band, which makes detection of the closest binaries impossible. We describe the C-band emission properties of those sources below.

Per-emb-2 appears slightly extended along the direction of the binary at 4.1 cm. The 6.4 cm map, however, is unresolved and peaks at the position of the Per-emb-2-B source. The spectral index map shows steeper values toward the Per-emb-2-A source similar to the Ka-band resolved maps. \cite{Tobin2016} found a similarly steeper spectral index toward Per-emb-2-A from Ka-band data, and suggested that 2-B source is more affected by free-free emission. While unresolved, it appears that most of the C-band flux is aligned with 2-B source but the S/N is low. Per-emb-5 is clearly detected only at 4.1 cm. Its emission is centered on the position of Per-emb-5-B, and its C-band spectral index is consistent with the flat values obtained in the Ka-band.

Per-emb-18  has a steep spectral index in the Ka-band, suggesting that the free-free emission is significantly contributing to the flux at the source position. This source has been identified as a resolved radio jet by VLA C-band observations with a position angle consistent with a large-scale H$_2$ outflow \citep{Davis2008} and perpendicular to the position angle of the binary system \cite{Tychoniec2018}. The extended dust structure to the east of Per-emb-18 is seen only in the low-resolution Ka-band image as noted by \cite{Tobin2016} and it is not detected in C-band, further suggesting that this clump is not hosting a protostar nor powering a strong outflow.

\subsubsection{Comments on possible close multiples from VANDAM:}
\cite{Tobin2016} reported four sources with marginally resolved structures, but not significant enough to report a new detection. 
The Ka-band maps for EDJ2009-183 from \cite{Tobin2016} shows extended emission that could be attributed to a protostellar component. 
This emission is marginally detected in the 4.1 cm map, indicating that it might a be faint thermal jet which is also supported 
by the C-band flat spectral index (0.05 $\pm$ 0.38). EDJ2009-156-B is completely unresolved in C-band, but the spectral index suggests
 a significant contribution of free-free emission to the Ka-band. Per-emb-25 is slightly extended in 4.1 cm map. Interestingly it is peaked at the position of the possible companion, not at the well-detected primary source, making it a strong candidate for a binary. A steep spectral slope in the C-band does not indicate a large contribution from free-free emission Per-emb-52 is a non-detection, preventing further interpretation of the Ka-band data.

\subsubsection{Systems with separation $>$~30 - 500 au:}
\begin{figure}[H]
\centering
\includegraphics[width=0.45\linewidth]{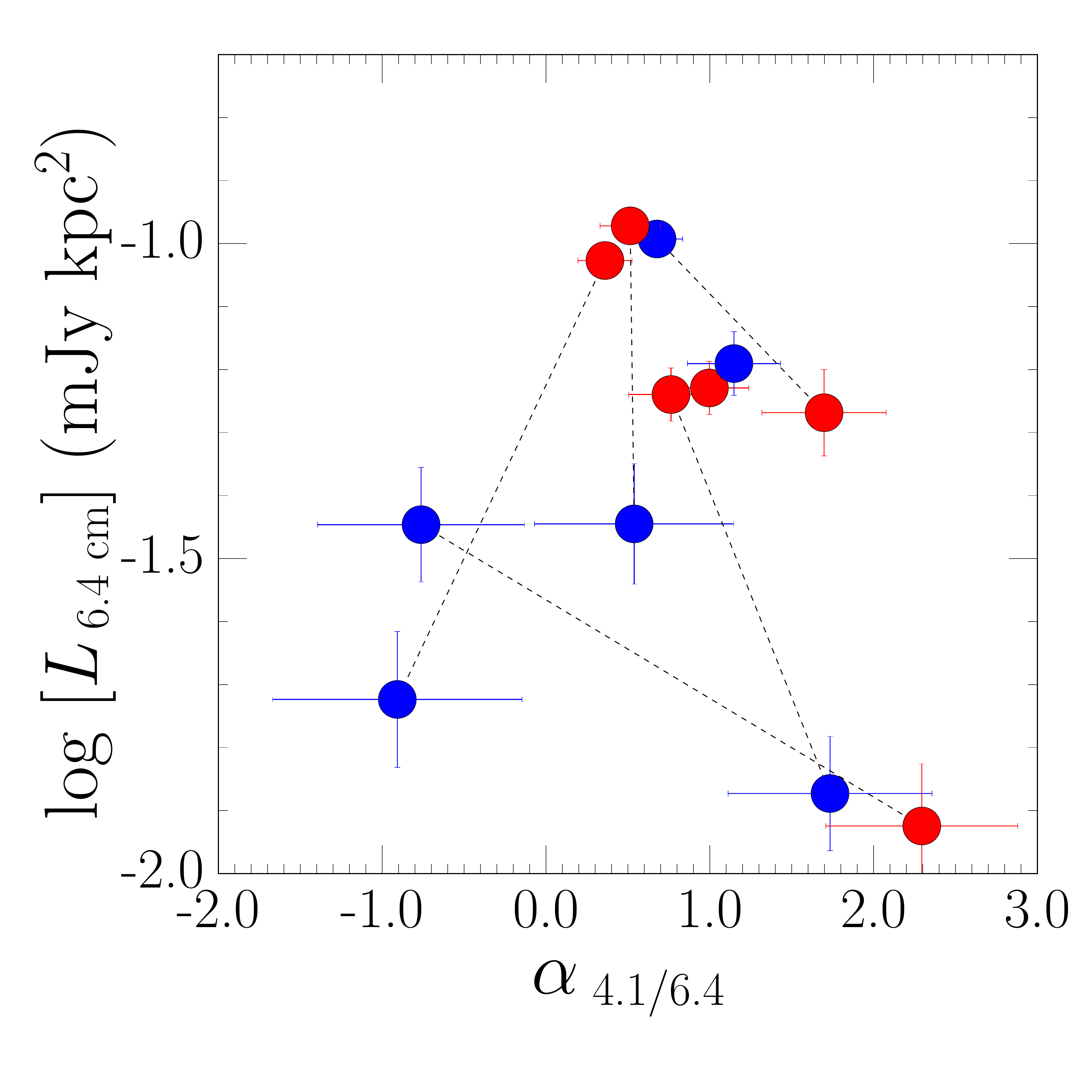}
\hspace{1 cm}
\includegraphics[width=0.45\linewidth]{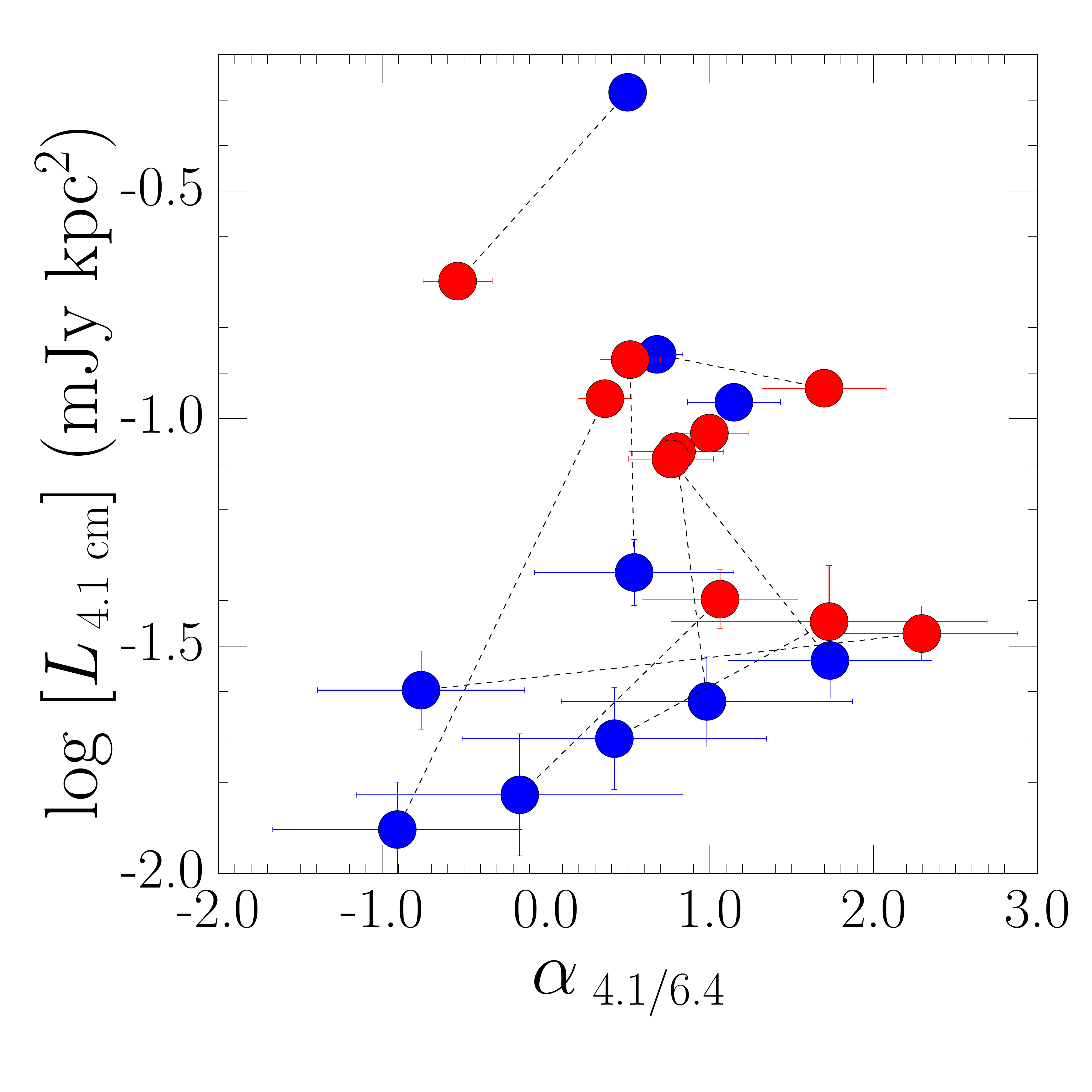}
\caption{Plots showing 4.1 cm (left) and 6.4 cm (right) luminosity of the binary systems compared with the spectral index. Red bullet represents the more luminous component of the binary in Ka-band observations \cite{Tobin2016}. Dashed lines are connecting components of the same system.}
\label{fig:multiples}
\end{figure}

\cite{Tobin2016} found 19 systems with sources separated by 30 au to 500 au. We detect 10 (50\%) of these systems in at least one of the C-band sub-bands. We also identify an additional source in SVS3 that was not detected in Ka-band. A comparison of their fluxes, spectral indices and dust masses is presented in Table  \ref{tab:table8}. Among detected multiples, some of them have very similar fluxes while for others one of the companions dominates the radio emission. There is no dependence between flux differences and separation. We also find variations in spectral index between the companions.
 While most of the compact dust differences are moderate, there is the notable example of Per-emb-12 where the A component has a mass $\sim 17$ times greater than the B component. In the case of Per-emb-12, the B source has greater flux in C-band while in Ka-band the A companion is an order of magnitude brighter. Figure \ref{fig:multiples} illustrates the differences in flux densities and spectral index between the multiple systems.

\subsection{Non-detections}

Radio emission coincident with protostars is well established as a common phenomenon. In this section, we investigate the nature of protostellar sources where we note the absence of the emission at C-band. The most natural explanation for the non-detection arise from the sensitivity of our observations. Even though our sensitivity is quite good $\sim$ 5 $\mu$Jy RMS, we still may miss the lowest luminosity protostars. The correlation between radio and bolometric luminosity shows that sources with low bolometric luminosities should have lower C-band fluxes \citep{Anglada1995,Shirley2007}. Indeed, most of our non-detections (except Per-emb-29 and Per-emb-21) have bolometric luminosities below 0.7 L$_\odot$. On the other hand, many of the sources below that threshold have significant radio flux.
All the First Hydrostatic Starless Core (FHSC) candidates and Very Low Luminosity Objects (VeLLOs): B1-bN \citep{Hirano1999, Pezzuto2012, Gerin2015}, Per-bolo-58 \citep{Enoch2010}, L1451-MMS \citep{Pineda2011}, Per-bolo-45 \citep{Schnee2012}, and L1448IRS2E \citep{Chen2010}, were not detected, probably due to their low luminosity. In contrast, Per-emb-29 and Per-emb-21 are not detected in our C-band observations. Per-emb-21 has L$_\odot = 6.9$ and Per-emb-29 L$_\odot$=3.7 and we would expect them to have a significant radio flux. It is possible that moderate long-term variability of the free-free emission is tightly connected to the episodic nature of the outflow/accretion events.

\subsection{Updating radio and bolometric luminosity correlations}
Radio emission from low-mass protostars cannot be explained by photoionization because the ionizing flux from the stars is too low \citep{Rodriguez1989b,Cabrit1992,Anglada1995}. Instead radio emission is attributed to shocks from the jets, which is supported by similar position angles between radio and molecular emission from the outflows \citep[and references therein]{Anglada1995}. Correlation of the radio flux and the bolometric luminosity also supports this hypothesis, as more luminous sources are expected to power more energetic outflows \citep{Bontemps1996, Wu2004}, therefore producing stronger ionizing shocks.

The most up-to-date and complete comparison of the radio flux and bolometric luminosity was provided by \cite{Shirley2007}, who compiled data from various works \citep{Anglada1995,Anglada1998,Furuya2003,Eiroa2005}.
We are able to improve upon this characterization using both the VANDAM sample alone, and by combining it with \cite{Shirley2007} data. The VANDAM observations include lower luminosity protostars than those used in \cite{Shirley2007}, hence we can extend the analysis of the bolometric and radio luminosity correlation.

We updated the distances and scaled the bolometric luminosities from the \cite{Shirley2007} 
consisting of 45 sources at 3.6 cm and 34 at 6 cm. We merged the samples with the 4.1 cm and 6.4 cm sources
 from VANDAM which resulted in a sample size of 98 and 82 for each wavelength respectively (detections only).
For merged VANDAM and Shirley sample we found stronger correlations, with the following linear fitting parameters:
\begin{equation}
\log (L_{\textrm{4.1 cm}})=(-2.66\pm0.06) + (0.91\pm0.06)\ \log(L_{\textrm{bol}}), \rho = 0.82
\end{equation}
\begin{equation}
\log (L_{\textrm{6.4 cm}})=(-2.80\pm0.07) + (1.00\pm0.07)\ \log(L_{\textrm{bol}}), \rho = \boldsymbol{0.79}
\end{equation}

\begin{figure}[H]
\centering
\includegraphics[width=0.4\linewidth]{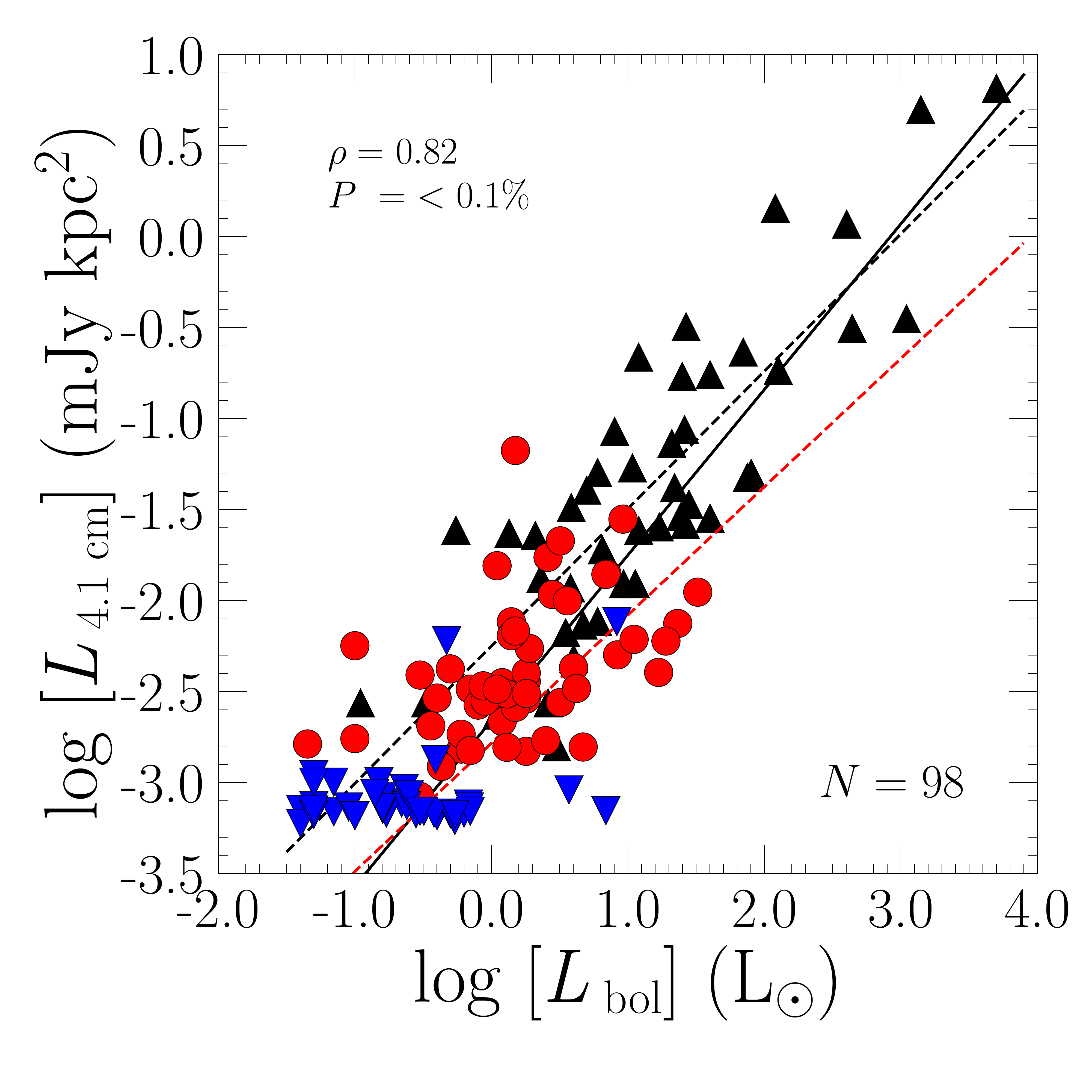}
\hspace{1 cm}
\includegraphics[width=0.4\linewidth]{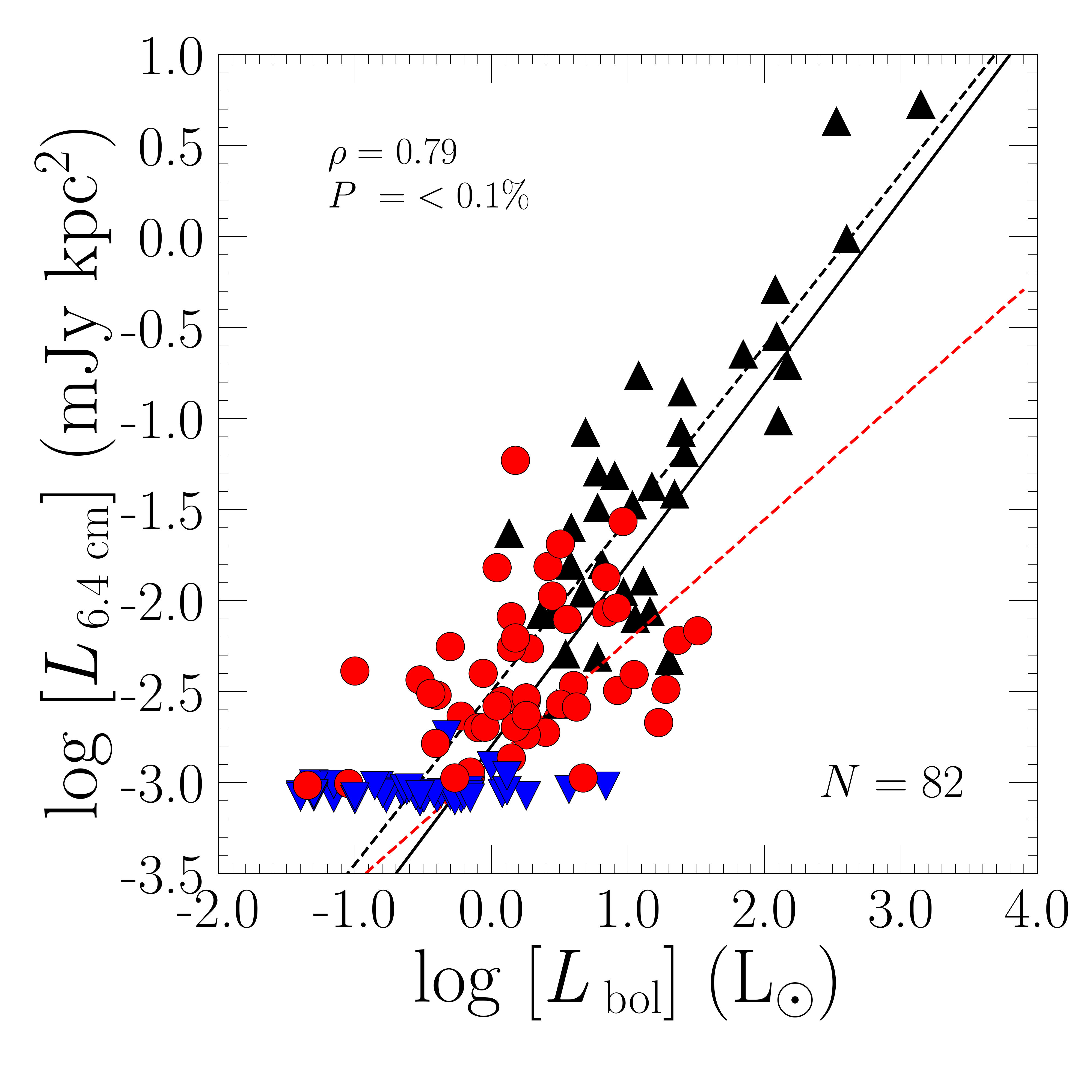}
\caption{Radio luminosities plotted against bolometric luminosities of the sources. Red circles represent VANDAM sources, black triangles are the sources from \cite{Shirley2007}, and blue triangles are upper limits of the VANDAM data. Red and black dashed lines show the linear fits to the VANDAM and \citep{Shirley2007} samples, respectively. 
The solid line represents fit to the merged sample. Spearman’s rank correlation coefficient and the probability of no correlation for the merged sample is shown in the left top corner.}
\label{fig:lbol_sh}
\end{figure}

The correlation for the merged sample appears robust and does not differ significantly from the correlation from \cite{Shirley2007}. On the other hand, the linear fit parameters to the VANDAM
data are different than for the merged sample, even considering the errors. The somewhat weak correlation in the VANDAM sample alone (see Equations \ref{eq:5} and \ref{eq:6}) results from the scatter within the sample that can be explained 
by the variable nature of free-free emission. Moreover, a small contribution from the synchrotron emission can cause additional scatter \citep[e.g.,][]{Tychoniec2018}.
 Only by analyzing protostars spanning several orders of magnitude in luminosity can one derive a robust trend. For example, extended thermal jets can give a
 temporal rise to the flux. The Perseus results fill out the low-luminosity end of the overall distribution significantly better than before. \cite{Morata2015}
 analyzed a sample of proto-brown dwarfs showing that they have radio fluxes higher than expected from their bolometric luminosities. This possibly suggests that  
correlation is flatter at the very low luminosities, but it is not evident with our data.

\section{Correlations with molecular outflow tracers}
\subsection{Far-infrared line emission}

To characterize the relationship between radio emission and outflows, we use tracers of jets and outflows from observations of far-infrared molecular and atomic lines. The far-infrared regime is crucial to understand the cooling processes of gas in star-forming clouds; since it predominantly traces warm gas, emission at these wavelengths is expected to probe the currently shocked material \citep[e.g.,][]{Nisini2002, Karska2013, Manoj2013, Manoj2016}.  Thus, we expect to observe a correlation between far-infrared line luminosities and radio luminosity which is likely tracing the shock-ionized gas.

We compare the VANDAM observations with data obtained by The Photoconductor Array Camera and Spectrometer (PACS) instrument \citep{Poglitsch2010} onboard the \textit{Herschel Space Observatory} \citep{Pilbratt2010}.  The data come from two \textit{Herschel} key programs: WISH \citep{vanDishoeck2011} and DIGIT \citep{Green2013}, as well as from an open time program WILL \citep{Mottram2017}. The PACS spectrometer is an Integral Field Unit (IFU) instrument with 25 spatial pixels (so-called spaxels) a field of view of $\sim$ 50\arcsec; each spaxel is  9\farcs4 x 9\farcs4, corresponding to a physical resolution of about 2200 au at the distance to Perseus. The wavelength coverage of the PACS instrument (55 - 210 $\mu$m) allows one to study some of the key far-IR cooling agents of the shocked gas e.g., CO, H$_2$O, OH, [O I]. Almost half of the sources analyzed within the sample shows extended emission on the scales of $\sim\ 10^4$  au, most commonly in [O I] \citep{Karska2018}. By contrast, VLA observations in C-band primarily trace the emission from the inner 60 au.  Comparing such different scales as represented by radio and infrared observations can be challenging. PACS observations trace the outflow history averaged over the past $10^2-10^3$ yr while the VLA gives insight on timescales as short as a few years \citep[e.g.,][]{Hull2016}. We can then analyze how the nature of the outflow varies in time.

\begin{figure}[H]
 \centering
  \includegraphics[width=0.75\linewidth]{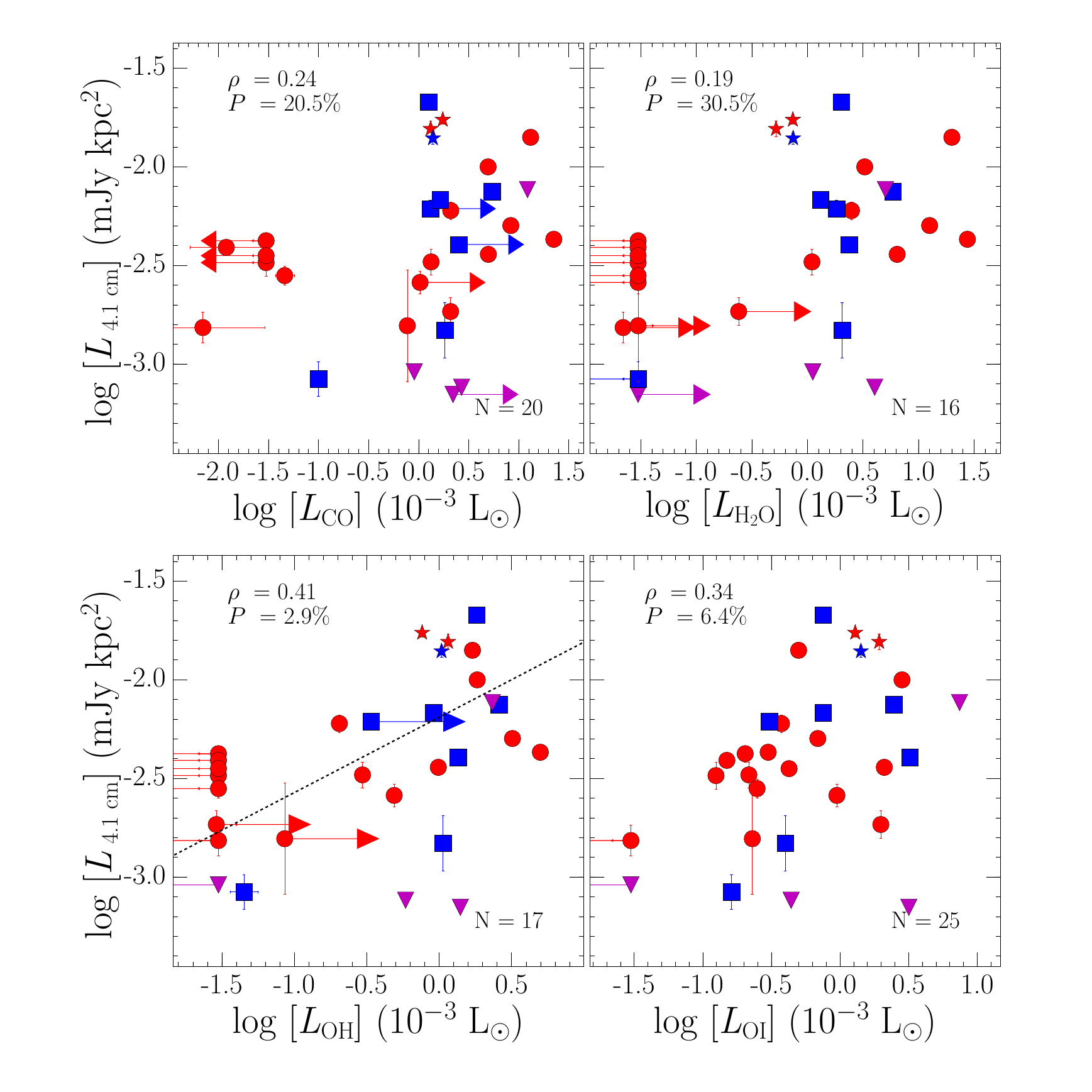}
\caption{Luminosity at 4.1 cm compared with CO (top left), H\raisebox{-.4ex}{\scriptsize 2}O (top right), [O I] (bottom left) and OH (bottom right) far-IR line luminosity. Upper limits for radio luminosities are plotted as magenta triangles, and lower or upper limits for {\it Herschel} line luminosities are indicated with arrows. Spearman’s rank correlation coefficient and the probability of no correlation is shown in the right top corner (for a combined sample of Class 0 and Class I protostars). }
  \label{fig:fig7}
\end{figure}

In Figure \ref{fig:fig7} we compare the radio luminosity at 4.1 cm with far-infrared luminosities of carbon monoxide (CO; J$_{\rm up}>$14), water vapor (H\raisebox{-.4ex}{\scriptsize 2}O), oxygen [O I] and hydroxyl radical (OH). Similar figures with 6.4 cm luminosities are given in the Appendix \ref{apA} (Fig \ref{fig:fig16}). The line luminosities are calculated by co-adding fluxes of the lines detected within the PACS wavelength range, and scaled with distance. We generally see very weak correlations or no evidence of correlations between radio luminosity and far-IR line luminosities. Nevertheless, we explore possible relations. The radio luminosity at 4.1 cm is weakly correlated with OH ($\rho$ = 0.41, $P$= 2.9\%), with a stronger relation for Class I ($\rho$ = 0.64, $P$ = 7.0\%); and with [O I] ($\rho$ = 0.34, $P$ = 6.4\%), also showing a stronger dependence for Class I ($\rho$ = 0.52, $P$= 13.9\%). For 6.4 cm we can only see a weak correlation with OH ($\rho$ = 0.43, $P$= 2.1\%), and [O I] ($\rho$ = 0.33, $P$= 8.0\%). No correlation with $\rho\ >$  0.4 is observed for H\raisebox{-.4ex}{\scriptsize 2}O and CO line luminosities and radio luminosity. Correlation coefficients are summarized in Table \ref{tab:table10}. 

The correlation between radio luminosity and the far-IR line luminosities may be linked to the correlations of those quantities with bolometric luminosity. \cite{Karska2013} show that the correlation of bolometric luminosity and far-IR lines are relatively weak (e.g., r = 0.63 for CO, r = 0.53 for [O I]); the extension over many orders of magnitude in source luminosity shows that the correlation is significant \citep[r$>$0.92 for CO][]{SanJose-Garcia2013}. Accordingly, on the scale of one cloud, and with a narrow range of protostellar luminosities, many other phenomena, such as long-term variability of both radio and far-IR emission can result in a large scatter.

Moderate correlation of radio luminosity with OH and [O I], together with none for CO and H\raisebox{-.4ex}{\scriptsize 2}O is interesting, as it informs us about the physical origin of the emission. As discussed above, ionization that produces free-free emission is expected to come from shocks. Shocks are divided into two main types: J-type (jump) shocks, with a sharp jump in conditions between pre- and post-shock gas and C-type (continuous) shocks where the change in temperature and density is less dramatic and occurs in a continuous manner \citep[e.g.,][]{Draine1983, Neufeld1989, Hollenbach1989}.

Observations of OH and [O I] with \textit{Herschel} are interpreted as arising in dissociative J-type shocks \citep{vanKempen2010, Wampfler2013}; up to 50\% of CO emission may result from them as well, and less than 10\% of the H\raisebox{-.4ex}{\scriptsize 2}O (\citealt{Karska2014a}, \citealt{Mottram2014}).
Comparing this to our results, we can infer that ionization that results in free - free emission
is likely caused by J - type shocks. Alternatively, UV radiation from accretion shocks or central protostar
can explain some of the ionization. In that case, C - type shocks with significant UV contribution could cause the observed ionization.


The observed scatter and weak correlations between far-infrared line and radio continuum fluxes suggest that the ionized collimated jet close to the protostar is not directly related to the large-scale outflow. This is most likely related to the different physical scales compared here - far-IR lines observed with {\it Herschel} trace material excited in multiple ejection events, while the free-free emission probed by the 
VLA corresponds only to the most recent ejection. This could potentially be related to the accretion activity, however, a correlation of radio emission and accretion bursts observed through infrared variability has not yet been established \citep{Galvan-Madrid2015}.

\subsection{Molecular outflow force}
The discovery of correlations between the outflow force and the radio luminosity was crucial for linking the free-free emission from the protostars to the jet/outflow \citep[e.g.,][]{Cabrit1992, Anglada1995}.  We examine this relation for the protostars in Perseus, and we add this subset to the sample of known protostellar radio sources with calculated outflow forces to solidify the correlation.

\begin{figure}[H]
\centering

  \includegraphics[width=0.4\linewidth]{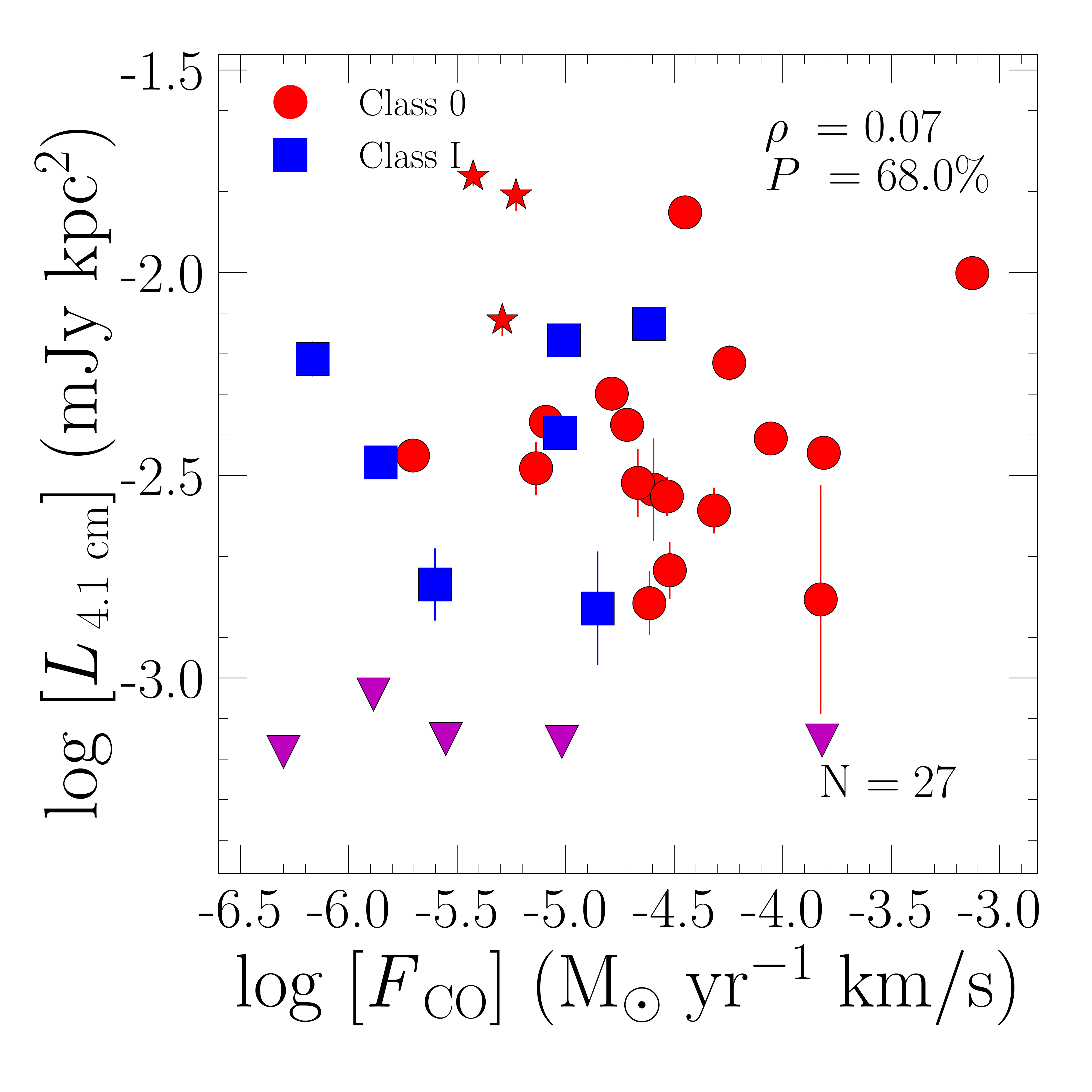}
  \includegraphics[width=0.4\linewidth]{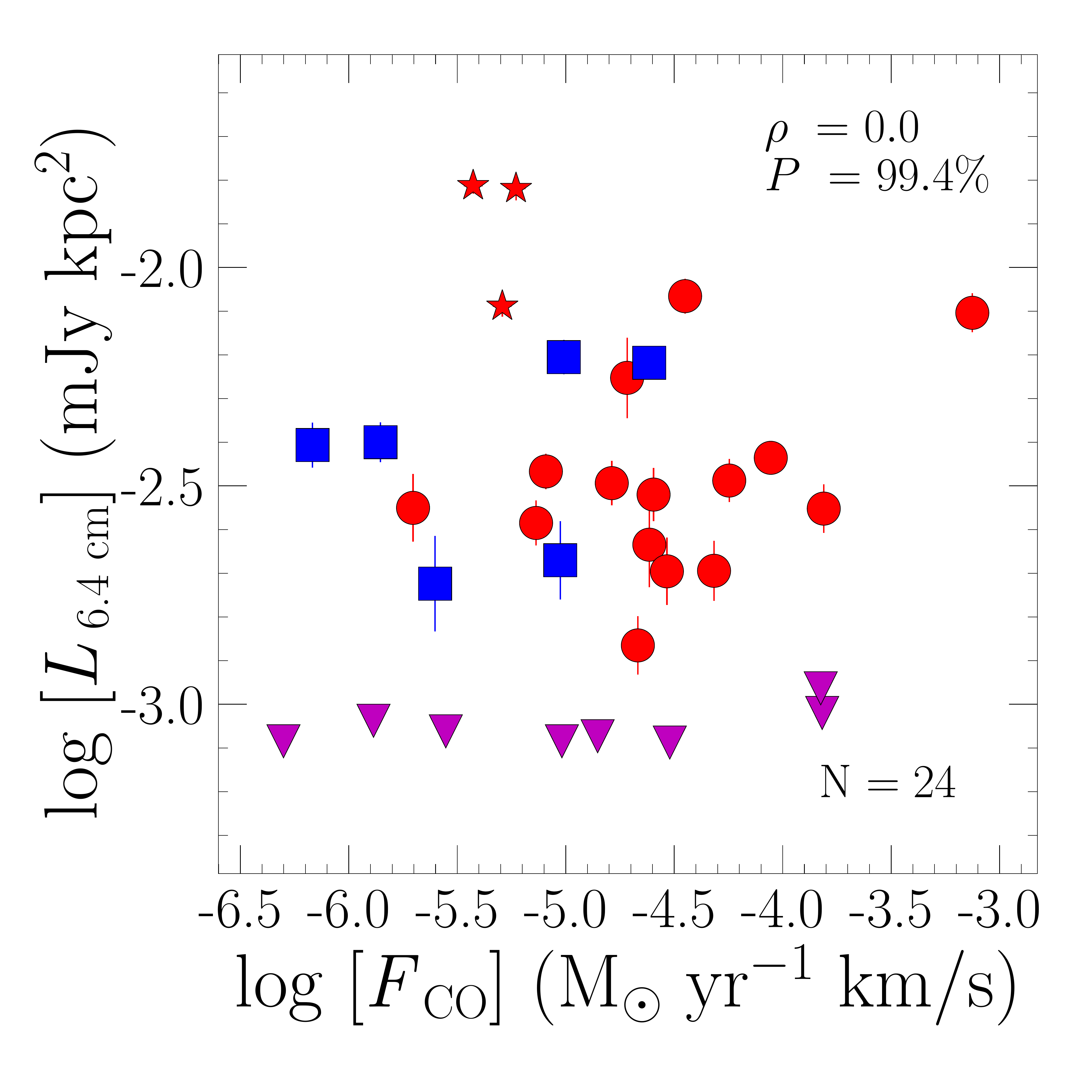}

\caption{Radio luminosity at 4.1 cm (left) and 6.4 cm (right) compared with outflow force from various observations of CO. Upper limits are marked as magenta triangles. The Spearman's rank correlation coefficient and the probability of no correlation are shown in the top-right corner.}
\label{fig:fout}
\end{figure}

Outflow forces for Perseus protostars were taken from \cite{Mottram2017} and \cite{Hatchell2007a} 
who used CO 3-2 James Clerk Maxwell Telescope (JCMT) observations to measure them.
 We present a comparison of the radio luminosity and outflow force in Figure \ref{fig:fout}. No significant correlation is observed in these comparisons. When using different observations for outflow forces there is a caveat of introducing additional error through different scales observed and different methods used. This issue can introduce even an order of magnitude errors \citep{Marel2013}.

The lack of correlations of radio luminosity with outflow force/momentum differs with a number of other studies \citep[e.g.,][]{Cabrit1992,Anglada1995, Shirley2007} 
but all those works used a much wider range of protostellar luminosities in order to derive their correlations. It is important to keep in mind that the molecular outflow force is probed over much greater scales than radio emission, as noted above. It means that while radio emission probes very recent ejection activity, the molecular outflow is averaged over much longer timescales.

\begin{figure}[H]
\centering
\begin{minipage}{.45\textwidth}
  \centering
  \includegraphics[width=1\linewidth]{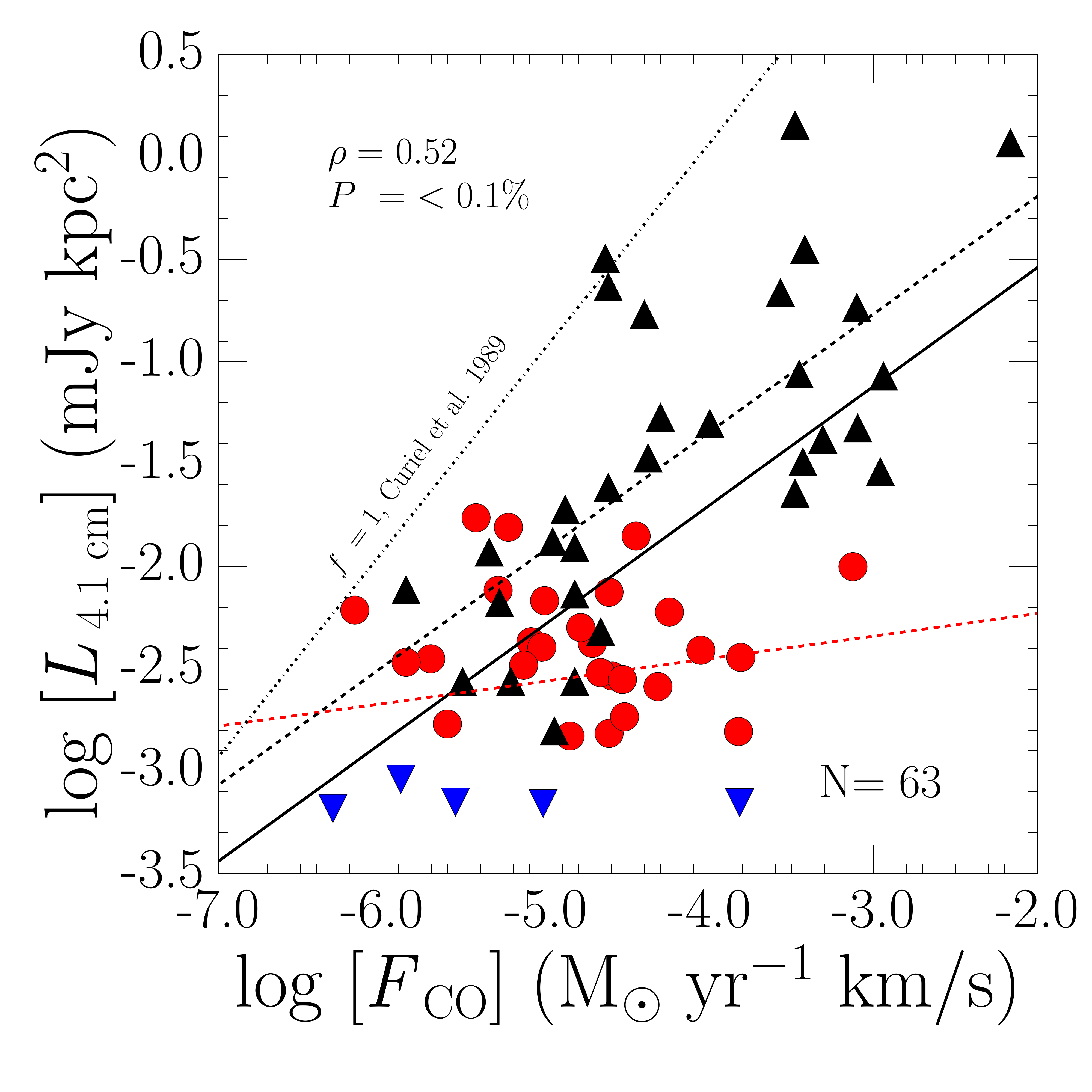}
  \label{fig:colow}
\end{minipage}
\begin{minipage}{.45\textwidth}
  \centering
  \includegraphics[width=1\linewidth]{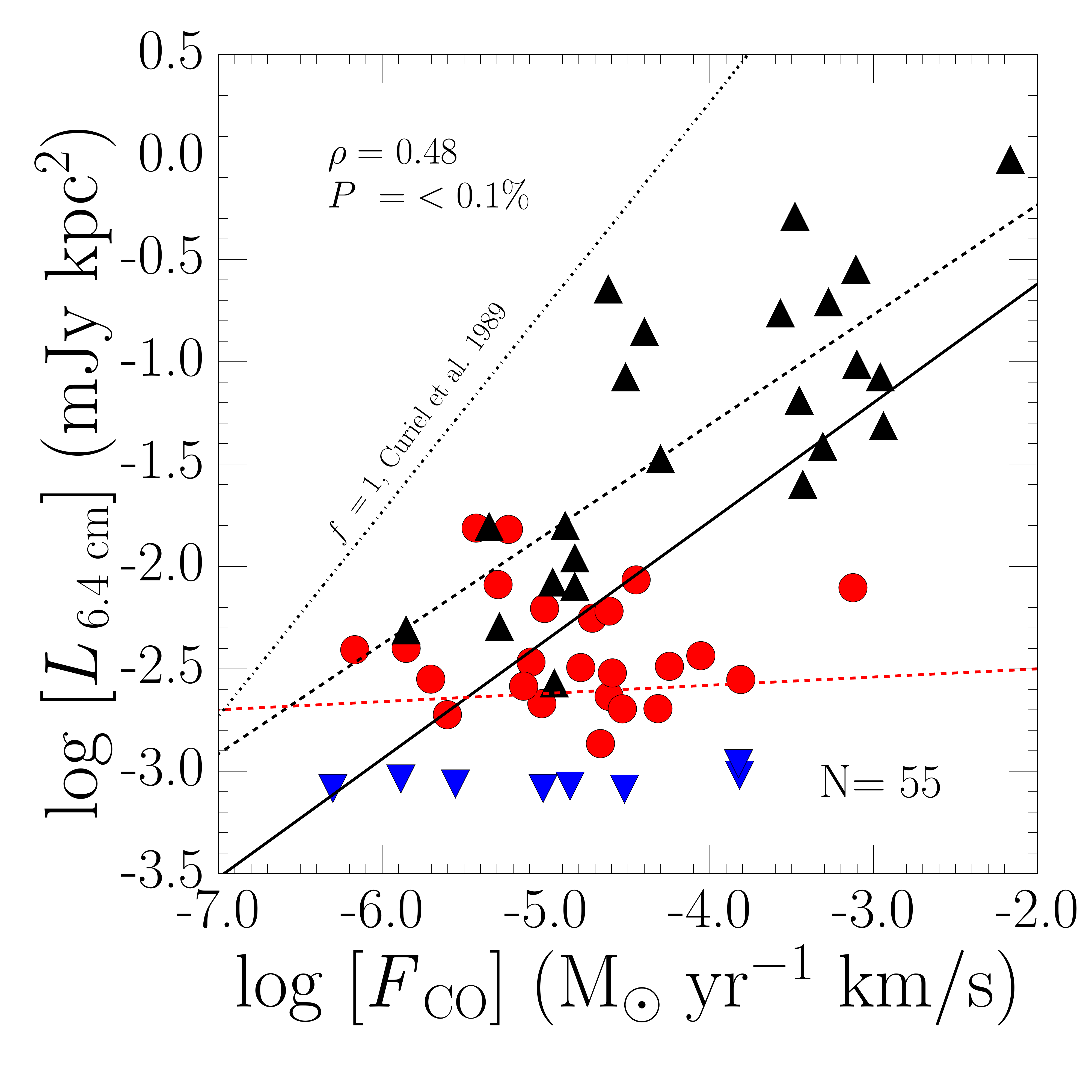}
  \label{fig:oilow}
\end{minipage}%

\caption{ Radio luminosity at 4.1 cm (left) and 6.4 cm (right) plotted against outflow force from CO observations. Red and black dashed lines shows linear fits to the VANDAM and \cite{Shirley2007} samples respectively. Solid lines represent fits to the merged sample. Black dash-dot line represent the expected C-band fluxes from the outflow force alone, assuming 100\% efficiency following \cite{Curiel1989}. This line correspond to the maximum C-band fluxes that can be produced from the CO outflows.}
\label{fig:fout_scaife}
\end{figure}

To determine if the relation remains valid for a wider range of luminosities, we combine the VANDAM sample with data collected by \cite{Shirley2007}, and plot them together in Figure \ref{fig:fout_scaife}. We updated distances to the sources included in the sample based on the most recent observations. We again find that the merged sample produces a correlation consistent with that of \cite{Shirley2007}. As we noted  for bolometric luminosity, the correlations are more clear when spanning more orders of magnitude in source luminosity.
For the merged VANDAM and \cite{Shirley2007} sample we fit linear functions with the EM algorithm:

\begin{equation}
\log (L_{\textrm{4.1 cm}})=(0.62\pm0.45) + (0.58\pm0.09)\ \log(F_{\textrm{CO}}), \rho = 0.52
\end{equation}
\begin{equation}
\log (L_{\textrm{6.4 cm}})=(0.54\pm0.49) + (0.58\pm0.11)\ \log(F_{\textrm{CO}}), \rho = 0.48
\end{equation}

The \cite{Scaife2011, Scaife2012} observed a weaker correlation between the 1.8 cm radio luminosity and the outflow force. Those authors checked if
the outflow force is sufficient to produce observed radio flux by calculating the minimum outflow force needed for ionization based on an equation from \cite{Curiel1989}:
\begin{equation}\label{eq:fout}
\log L_\nu = 4.24 +\log [F_{\textrm{out}}f(5\textrm{GHz}/\nu)]
\end{equation}
where $f$ is the ionization efficiency factor.
The \cite{Scaife2011} concluded that their sample had outflow forces that were too small to produce the observed radio flux,  although the emission at 1.8 cm is likely to have contributions from dust. Here we perform a similar analysis, 
and the minimum outflow force necessary to produce the observed C-band fluxes is plotted in Figure \ref{fig:fout_scaife}.
 The f=1 case is shown by the dotted line. This case represents the upper limit of the expected C-band fluxes based on 100\% outflow efficiency.
 Thus, we find that the outflow force can easily produce the observed C-band radio emission for both the VANDAM and the \cite{Shirley2007} samples. 
We note that the energy produced by the outflow is enough to generate the observed radio flux for all the sources, both from Perseus as well as the \cite{Shirley2007} sample.

\section{Mass of the protostellar disks} \label{diskmasses}

\subsection{Calculating the mass}

Some of the key questions in star formation are (1) how early do disks form and (2) how do they evolve and form planets. 
The properties of the youngest disks are still not very well defined. The VANDAM survey in the Ka-band with unprecedented resolution (15 au) found several resolved disk candidates
 \citep{Tobin2016, Segura-Cox2016}, but follow-up kinematic data are needed to determine whether or not these structures are rotationally-supported disks.
 The 9 mm Ka-band emission comes from $< 0\farcs5$ scales and likely originates from a disk or compact inner envelope. The observations of the most point-like disk candidates are not consistent with envelope profiles (D. Segura-Cox et al. 2018, in preparation). Therefore compact dust emission at 9 mm is likely tracing genuine disks. Calculating disk masses for an unbiased sample of very young protostars can provide important insights on the early stages of their evolution.

Disk mass can be estimated from the thermal dust emission, assuming the dust is optically thin. Ka-band observations are sensitive to radiation coming from cold and large grains in regions with high densities, which is most likely a direct progenitor of the disk, if not the disk itself. However, continuum emission in the Ka-band may also include 
a substantial thermal free - free component which can contribute to the emission even at wavelengths shorter than those measured by Ka-band \citep[e.g.,][]{Choi2009}. Thus, to accurately estimate the disk mass one needs to remove any free-free contamination from the Ka-band fluxes. For the VANDAM survey we expect free-free emission to contribute significantly to the Ka-band emission for many sources, because the median spectral indices for the sample between 8 mm and 1 cm are below 2 \citep{Tobin2016}. These values are lower than the typical spectral indices expected for dust, $\alpha = 2 + \beta$, where $\beta < 1$ is expected for dense disks
\citep[e.g.,][]{Draine2006, Kwon2009, Testi2014}. In this section, we assess the contribution of free-free emission on the Ka-band flux to subtract it and hence derive dust-only flux densities to calculate the masses of the embedded disks.

\begin{figure}[H]
\centering
\includegraphics[width=0.3\linewidth]{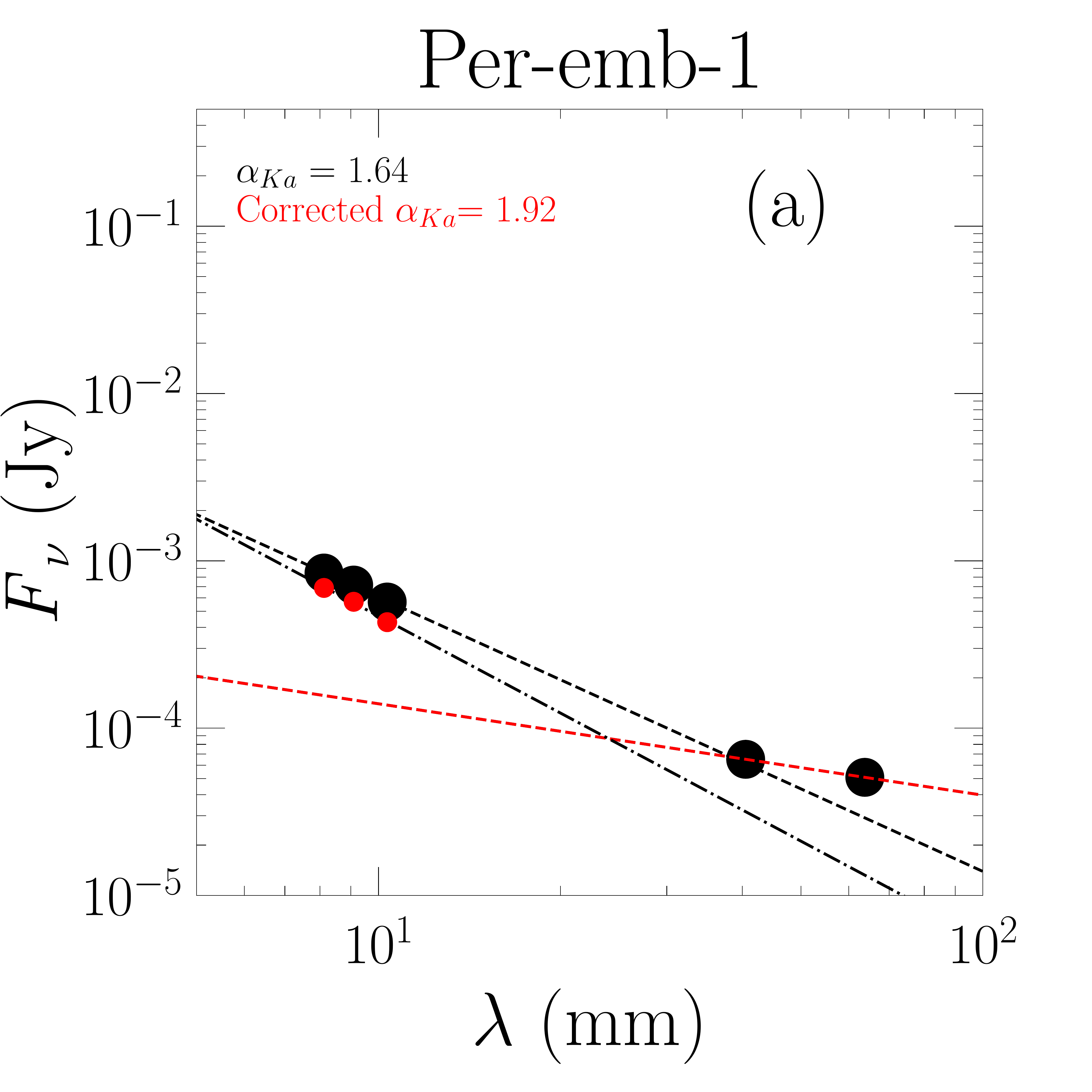}
\includegraphics[width=0.3\linewidth]{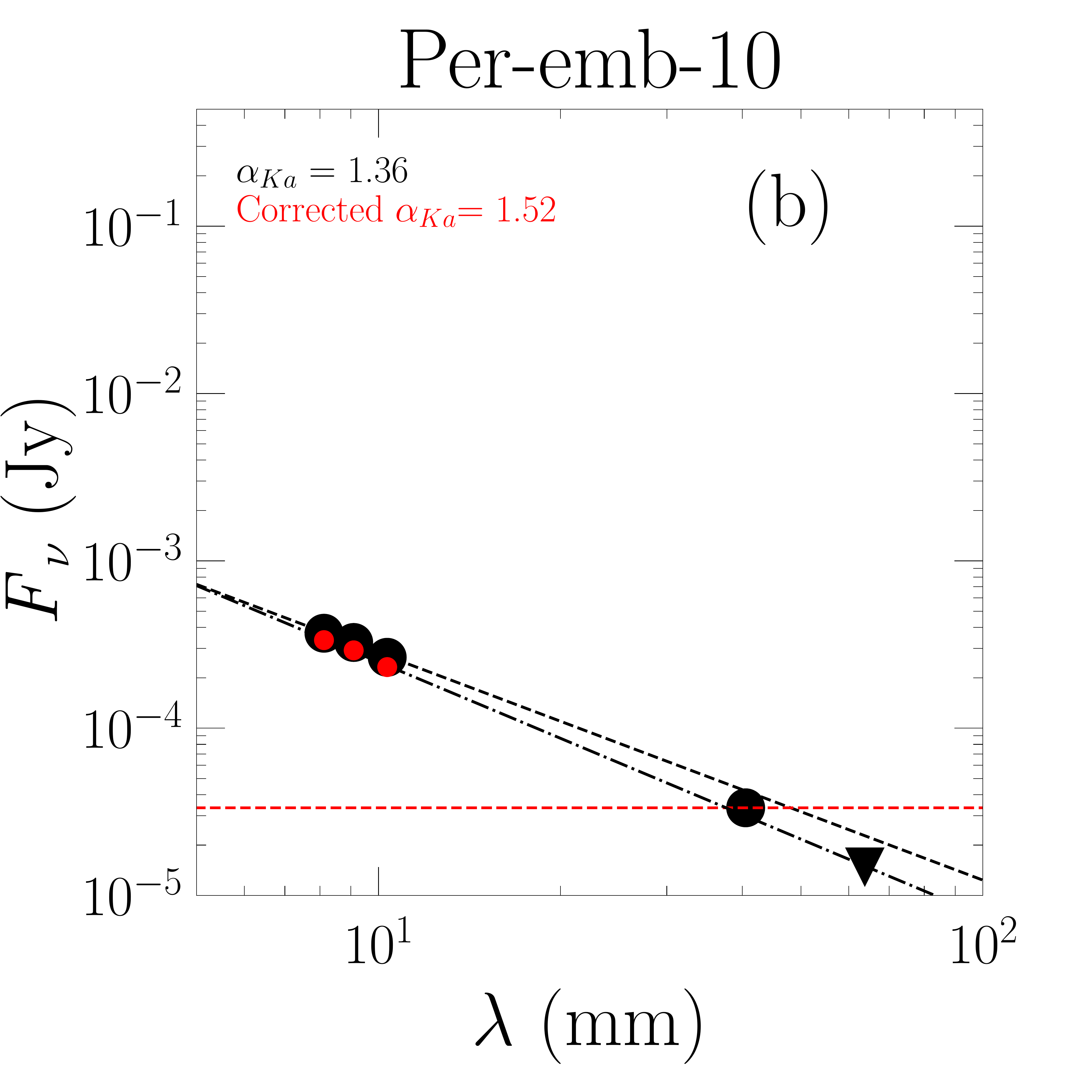}
\includegraphics[width=0.3\linewidth]{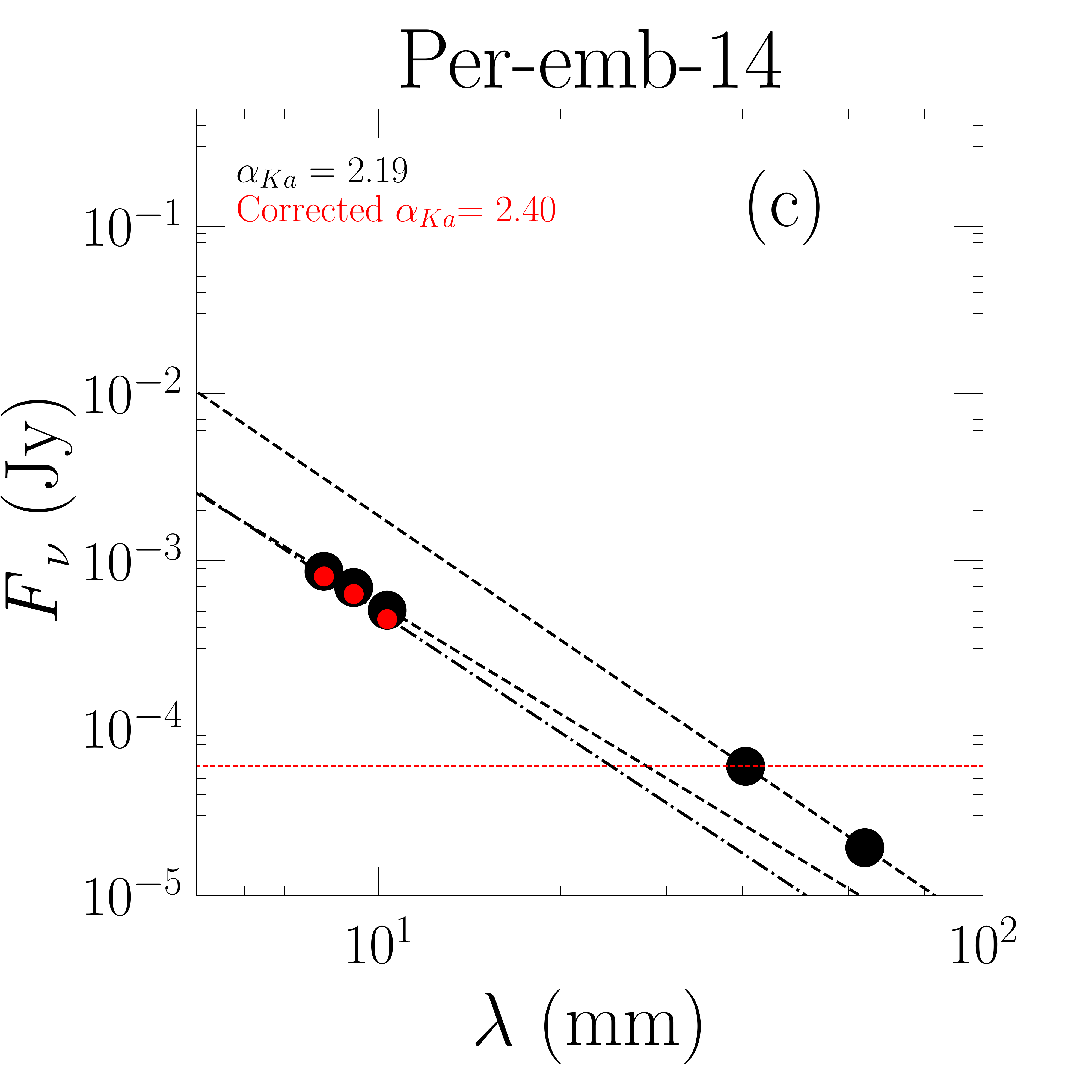}
\includegraphics[width=0.3\linewidth]{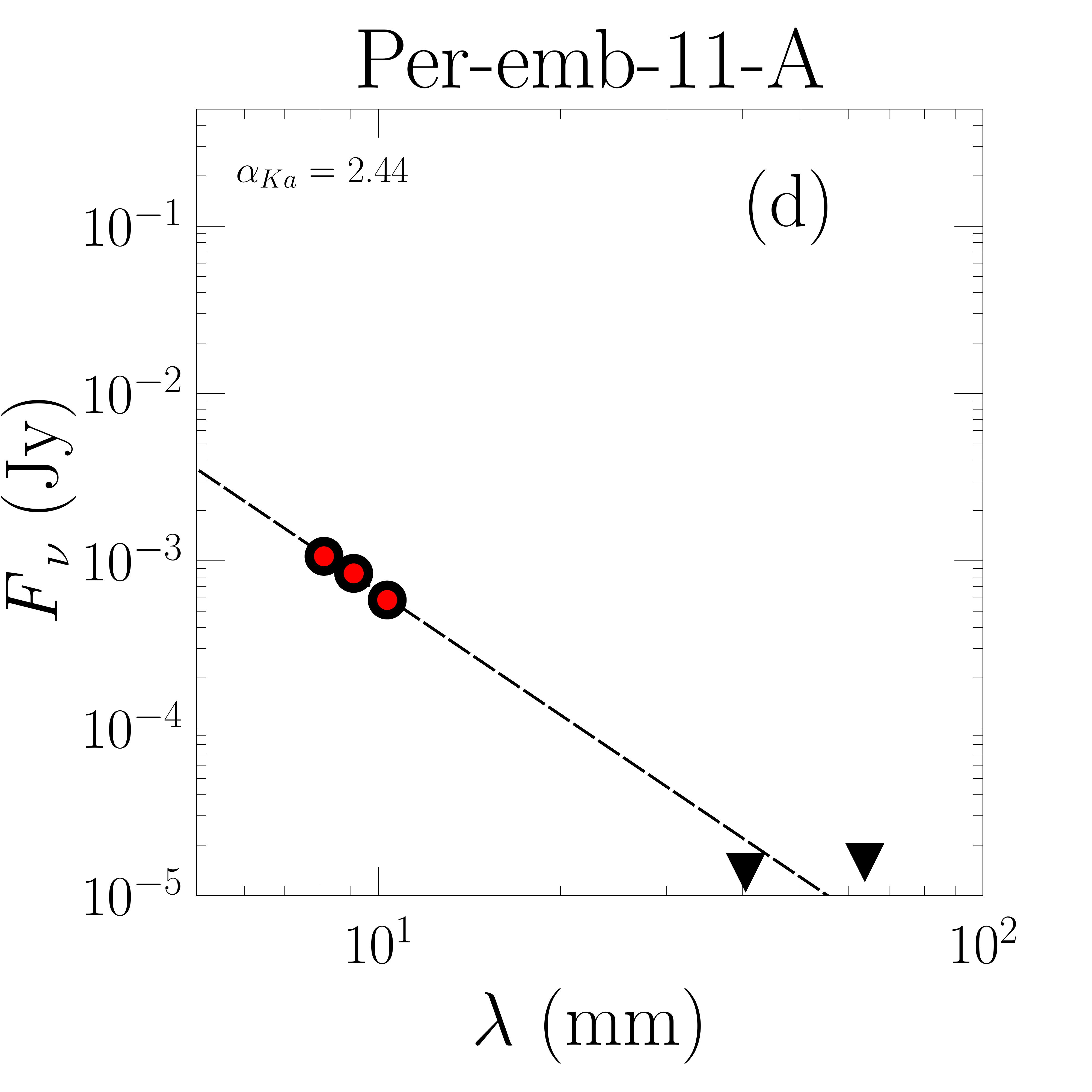}
\includegraphics[width=0.3\linewidth]{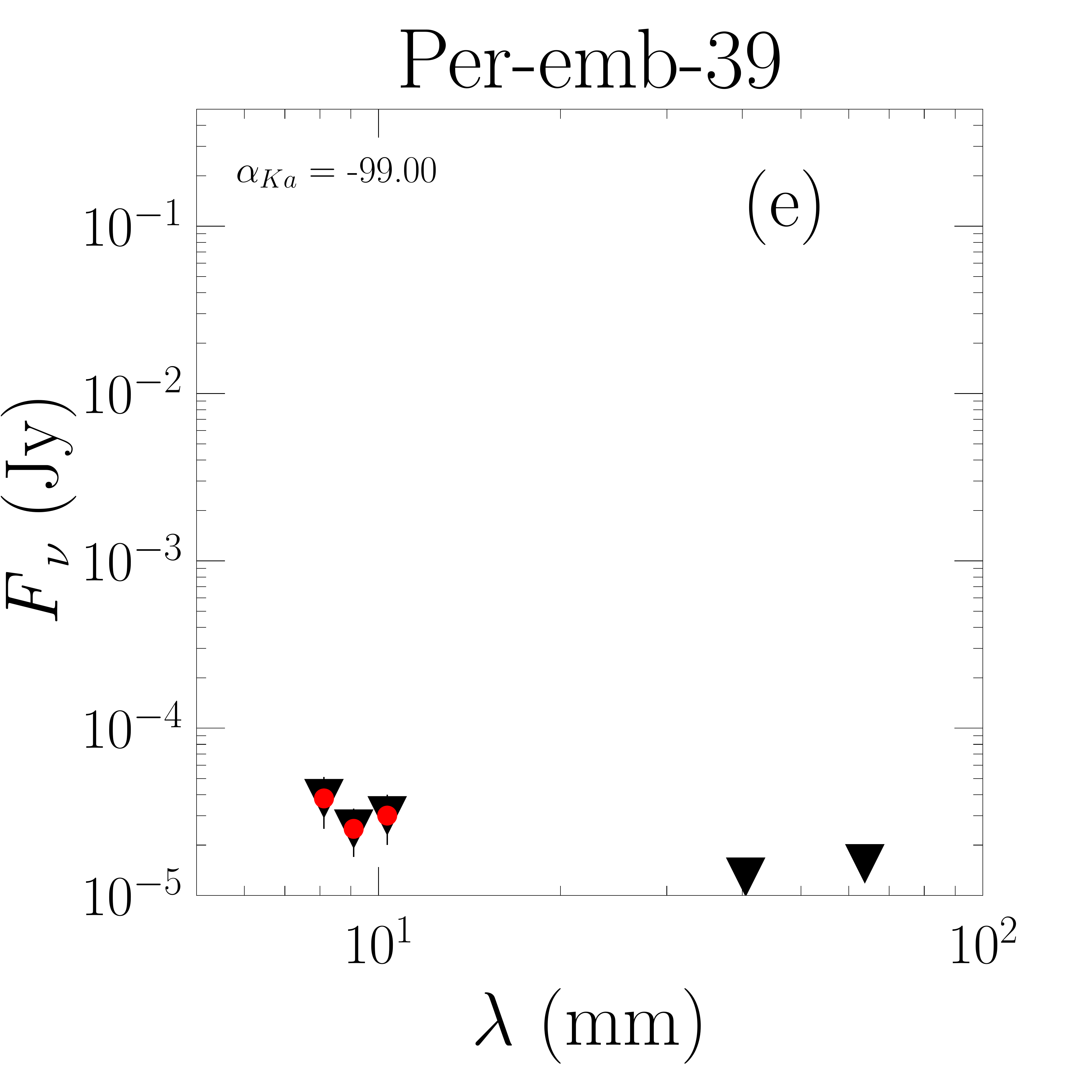}
\includegraphics[width=0.3\linewidth]{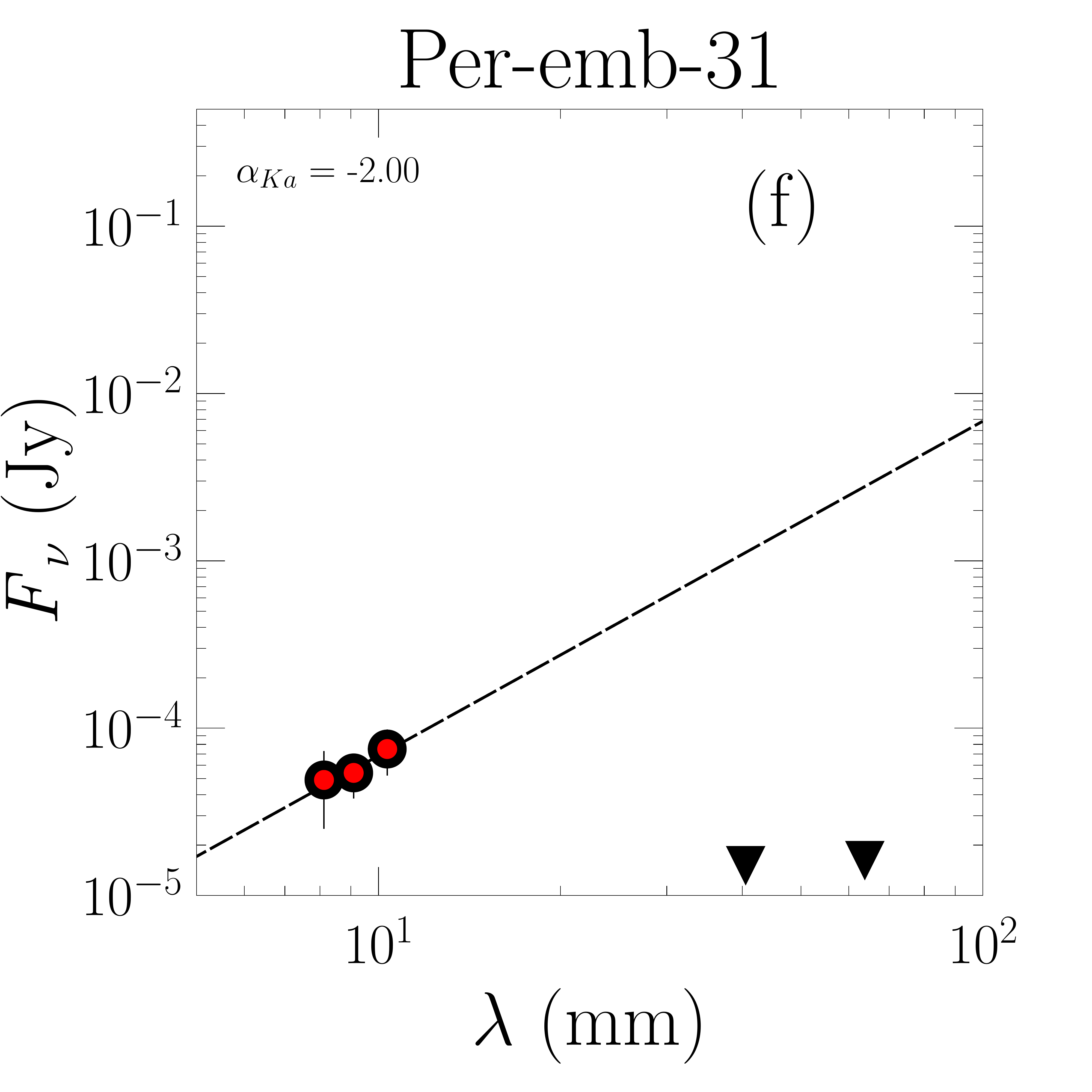}

\caption{Example radio spectral energy distributions for our disk candidates. Each panel shows a different case in how we corrected the Ka-band data for free-free contamination. See text for details. Black bullets represent Ka-band and C-band flux densities, and triangles are upper limits. Red bullets mark the corrected Ka-band flux densities. Dotted lines are linear fits to the original data, and the red line represents the function from which the free-free contribution was estimated. Dash-dot line marks fit to the corrected Ka-band flux densities.}
\label{fig:flagslopes}

\end{figure}

We fit a linear function to C-band logarithmic fluxes and then assumed that the value of this function at 9 mm is the free - free contribution to the total 9 mm flux. We use the Ka-band 9 mm flux density taken in the B configuration, because the beam size is comparable to that of C-band observations taken in the A configuration. 
Figure \ref{fig:flagslopes} represents each of the cases in our sample. 
In case (a), both C-band fluxes are well-detected and we determined the free-free contribution in the Ka-bands 
from the C-band spectral index; in case (b), the source is detected at one C-band wavelength. To calculate the free-free contribution in case (b) we use the detected C-band flux and assume a free-free spectral index of 0. For case (c), we find a steeper slope for the C-band fluxes than the Ka-band fluxes,
 which can arise if the free-free emission is optically thick 
\citep{Ghavamian1998}. Since we expect any free-free emission at Ka-band to be optically thin,  we use the 4.1 cm fluxes and an assumed spectral index of zero. In case (d)
 neither of the C-band fluxes are detected and we assume there is no free-free contamination at Ka-band for these sources. In case (e) we have non-detections in both C-band and Ka-band and we calculate upper limits of these disk masses assuming no free-free contamination. For case (f), we obtain a negative or flat spectral index in Ka-band, which suggests the radio emission is not tracing dust even at 9 mm, and we provide an upper limit.

We consider the disk masses from these sources as upper limits and remove no free-free emission. Radio spectra for all of the sources are presented in the Appendix \ref{sec:apC}. For close binaries, unresolved by C-band (Per-emb-2, Per-emb-5, Per-emb-18), we assume they share a common disk and we treat them as single protostars. This analysis is subject to many uncertainties. Free-free emission with a positive slope should turnover at wavelengths shorter than 4 cm, which would decrease the amount of the actual contribution. Ka-band and C-band observations were taken at different epochs (8 months later) and variability of the free-free emission can affect the analysis. The contribution of synchrotron emission can also affect the spectral index. Finally, measured disk masses from Ka-band observations should be considered lower limits because the emission at the 8 mm and 1 cm is sensitive to the largest dust grains in the innermost parts of the disk \citep{Segura-Cox2016}. Nevertheless, we can still compare the disk properties across the VANDAM sample and identify trends with evolution, given that they are observed uniformly.

After correcting the Ka-band fluxes for free-free contamination, we calculate the mass of the disk, following the equation from \cite{Hildebrand1983}:
\begin{equation}
\label{eq:dustmass}
M=\frac{D^2F_\lambda}{{\kappa}_\lambda B_\lambda(T_{\rm dust})}
\end{equation}

where $D$ is the distance to the protostar ($\sim$ 235 pc), $F_{\lambda}$ is the flux density from thermal dust emission,
$\kappa _{\lambda}$ is the dust opacity, $B$($T_{\rm dust}$) is the Planck blackbody function for an assumed dust temperature of 30 K, typical temperature assumed for cold dust \citep{Whitney2003}.
The value of $\kappa _{\lambda}$ is based on the \cite{Ossenkopf1994} dust opacity models:
\begin{equation}
\kappa _{\lambda} = 0.00899 \Big(\frac{1.3\  \textrm{mm}}{\lambda}\Big)^\beta \textrm{cm}^2 \textrm{g}^{-1}
\end{equation}
which for $\lambda=9\ \textrm{mm}$ and $\beta= 1$ \citep{Andrews2009} typical for disks, 
assuming a gas to dust mass ratio of 100:1, gives a value: $\kappa_{\ 9\ \textrm{mm}} = 0.00128 \textrm{cm}^2 \textrm{g}^{-1}$. Table \ref{tab:table9} lists the calculated disk masses for the VANDAM sources.

\begin{figure}[H]
\centering
  \includegraphics[width=0.4\linewidth]{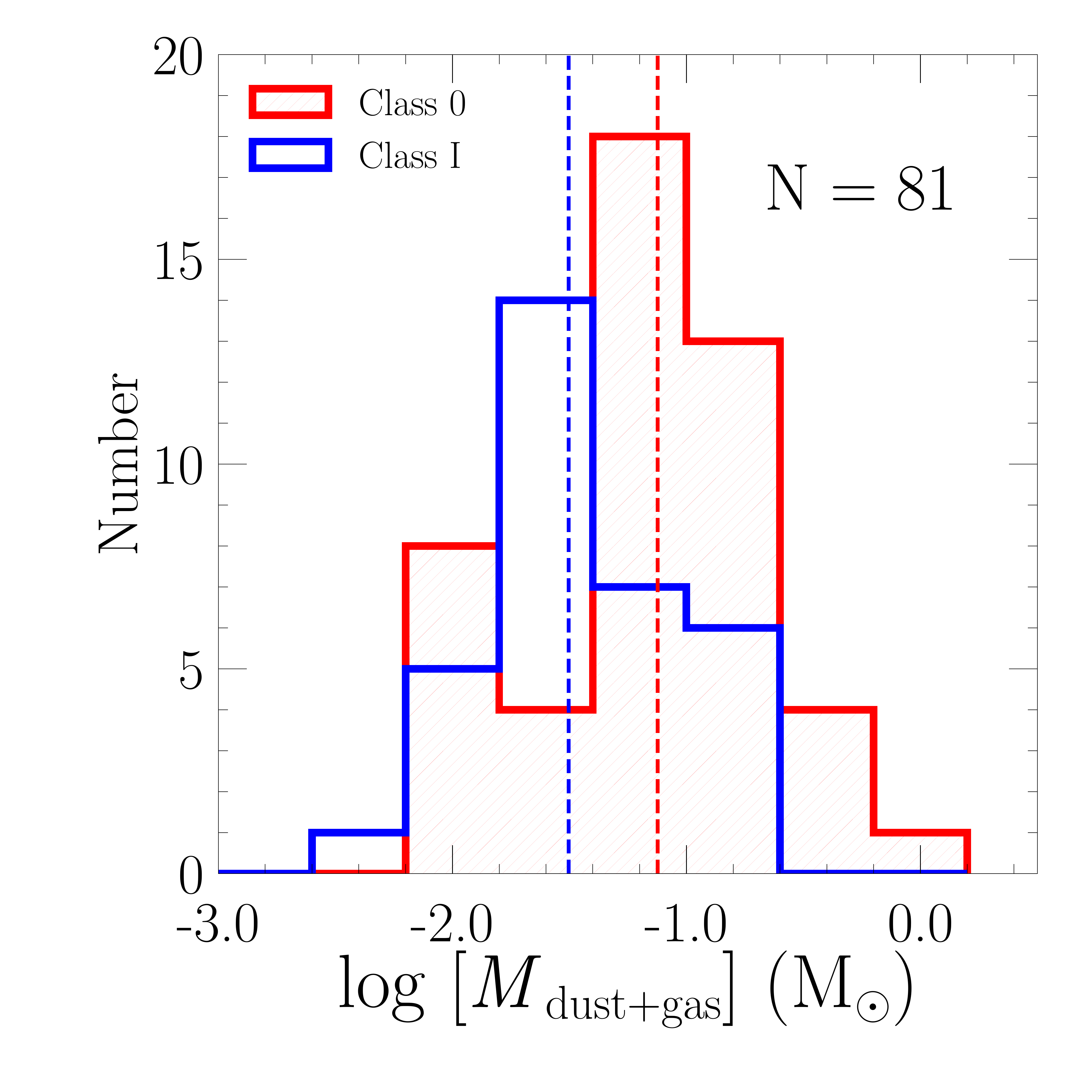}
  \includegraphics[width=0.4\linewidth]{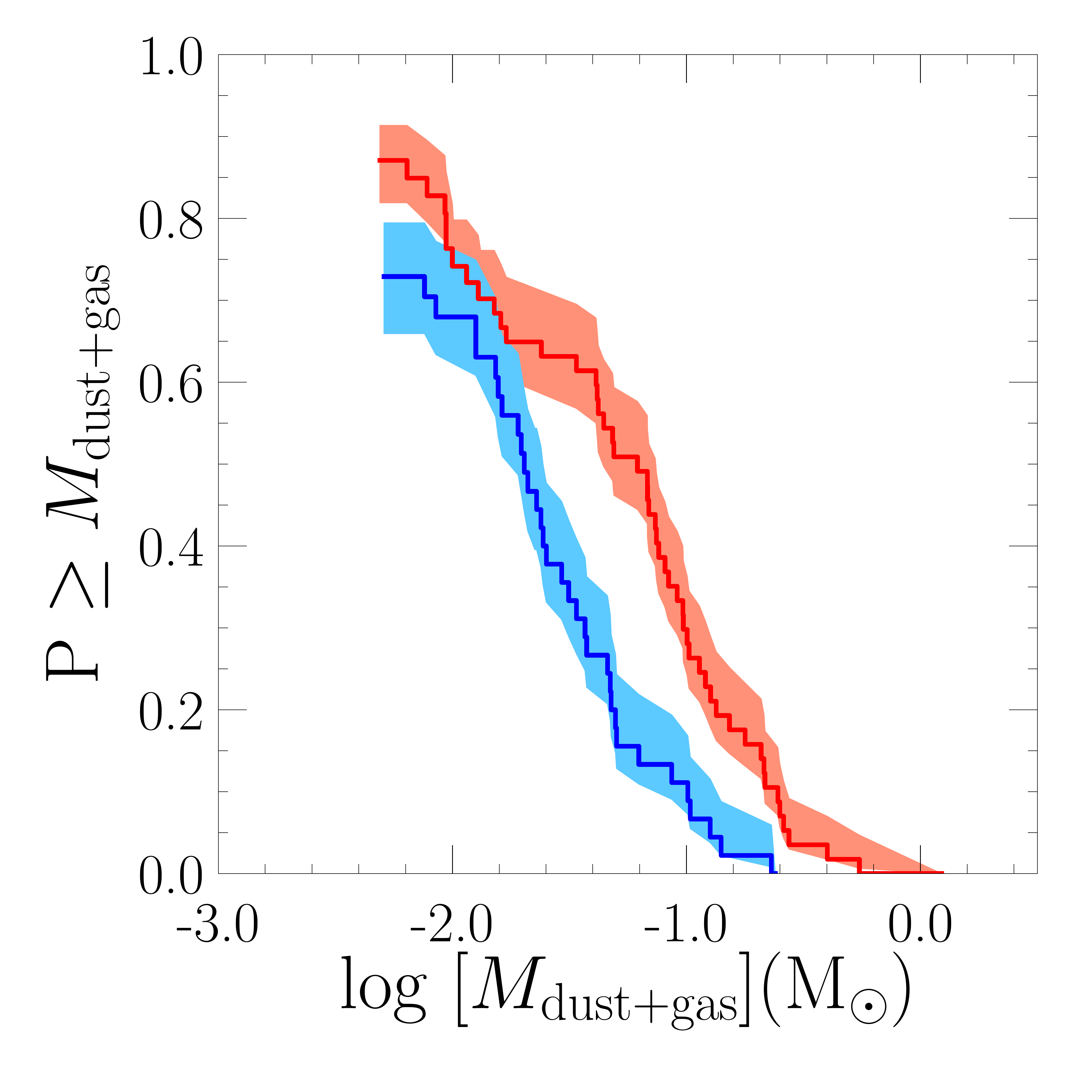}
\caption{Left: Histogram of disk masses for each evolutionary class, obtained with a fixed temperature of dust, T=30 K. Medians are shown with dashed lines, with respective colors. 
Median values are 0.075 M$_\odot$, 0.031 M$_\odot$, and 0.049 M$_\odot$ for Class 0, Class I, and total sample respectively.
The statistical probability of Class 0 and Class I values of the disk mass to be drawn from the same sample is 2.5\%.
Right: Cumulative distribution obtained with K-M method with 1$\sigma$ errors shown.}
\label{fig:masshist}
\end{figure}

\begin{figure}[H]
\centering
  \includegraphics[width=0.4\linewidth]{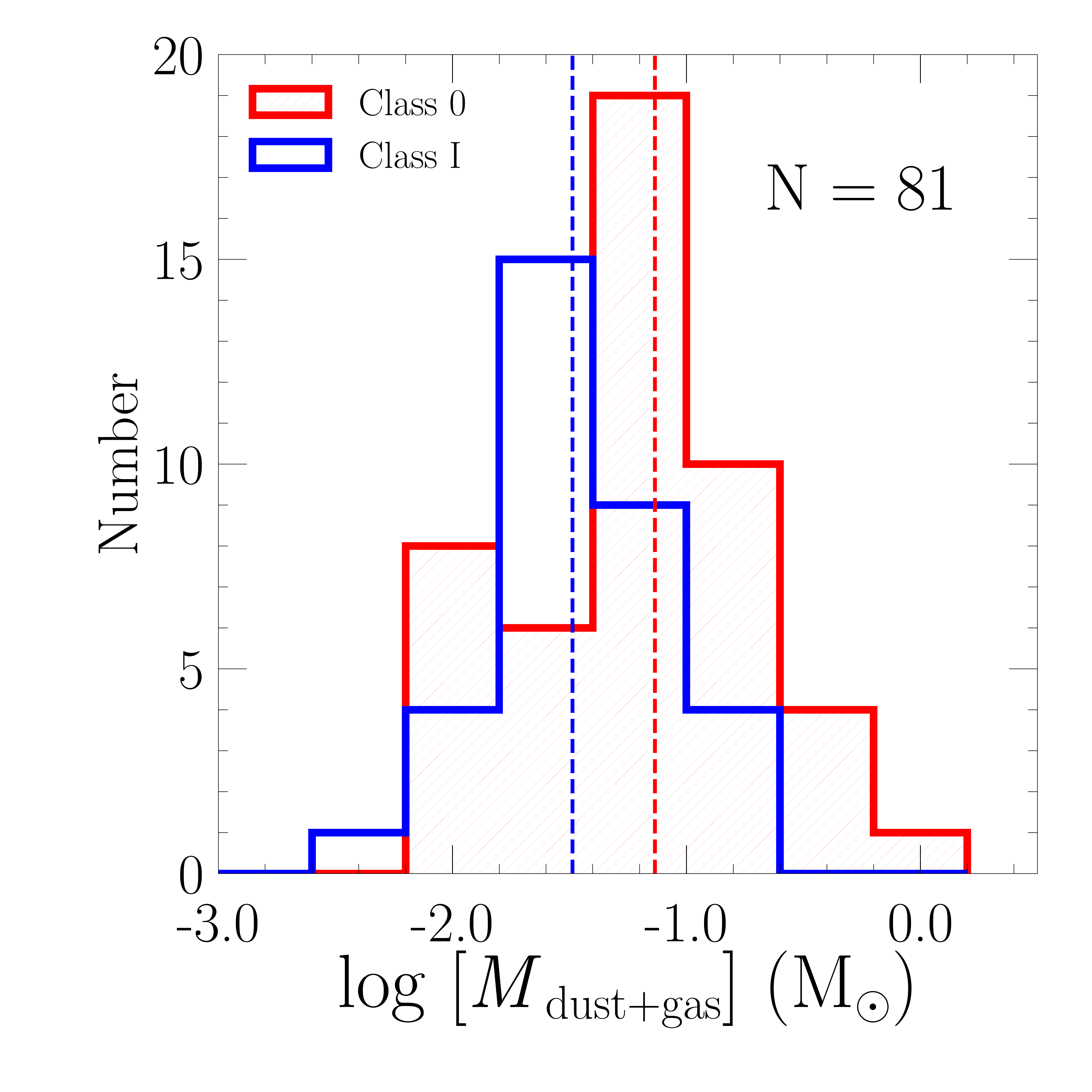}
  \includegraphics[width=0.4\linewidth]{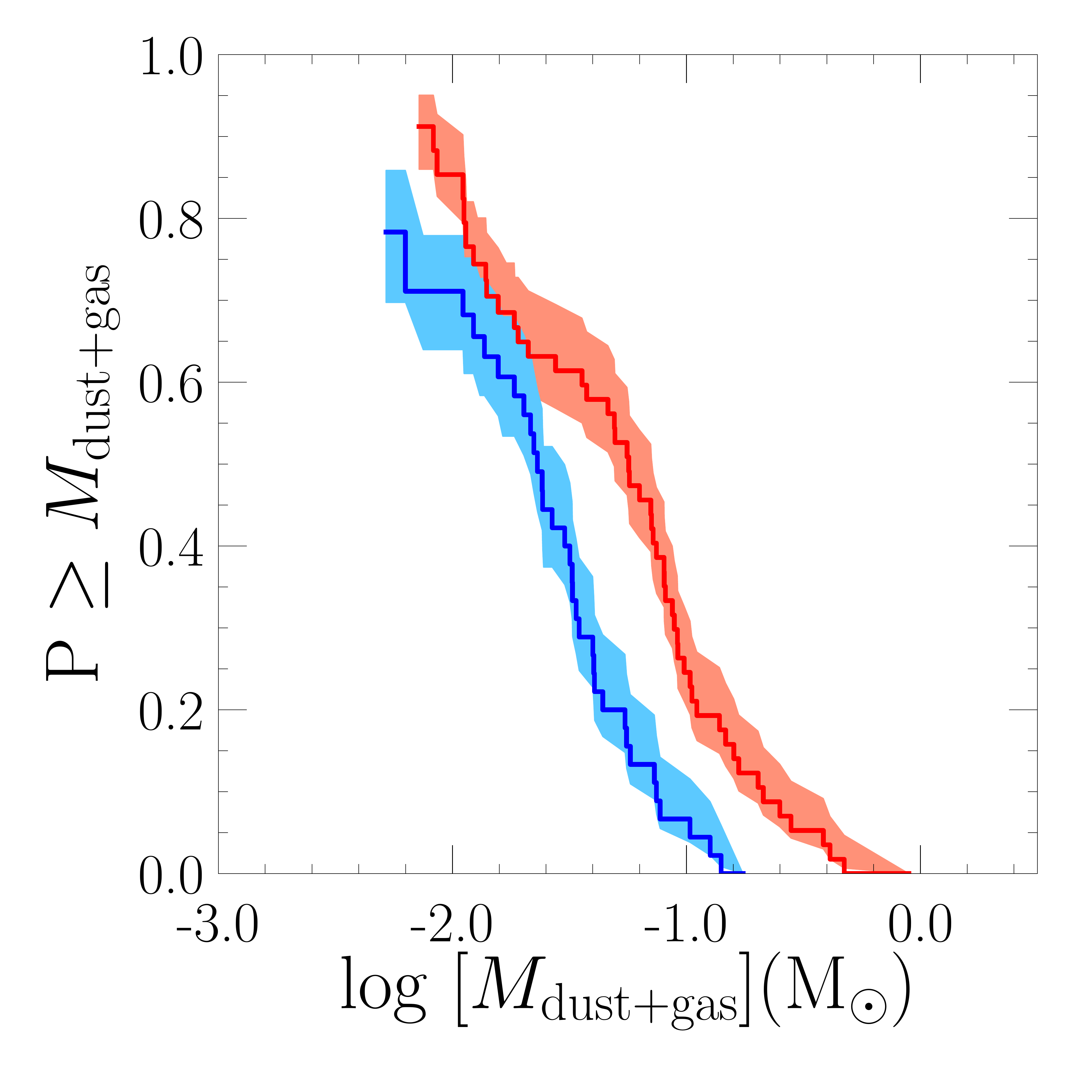}
\caption{Similar to Figure \ref{fig:masshist} but for masses calculated using temperatures determined 
$T_{\textrm{dust}}= 30 [\textrm{K}] \times (L_{\textrm{bol}} /\textrm{L}{\odot})^{1/4}$. Median values are 0.073 M$_\odot$, 0.033 M$_\odot$, 0.055 M$_\odot$ for Class 0, Class I 
and total sample respectively. We find the statistical probability of 1.5\% that the Class 0 and Class I disk masses are drawn from the same sample.}
\label{fig:masshist2}
\end{figure}

\begin{figure}[H]
\centering
\includegraphics[width=0.4\linewidth]{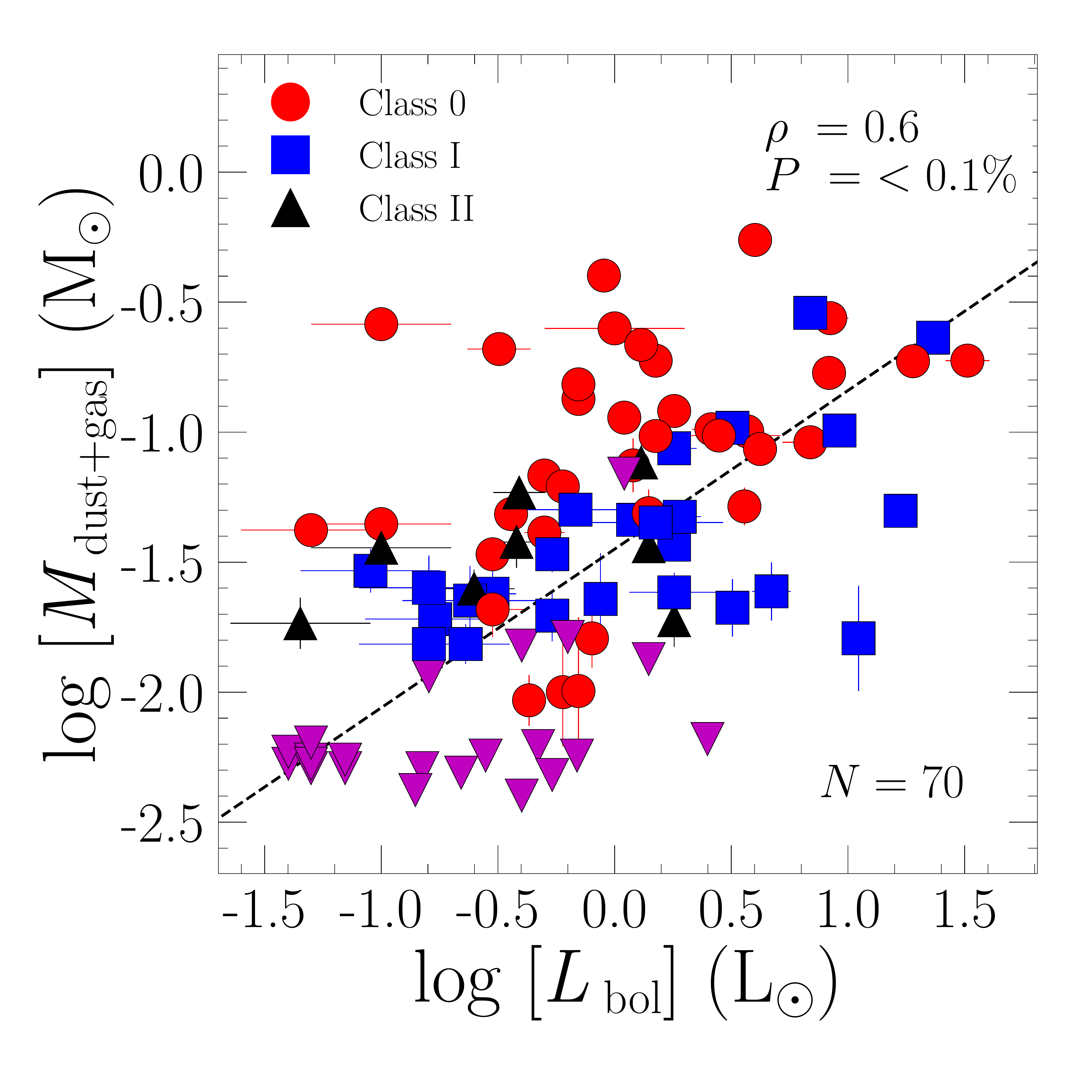}
\includegraphics[width=0.4\linewidth]{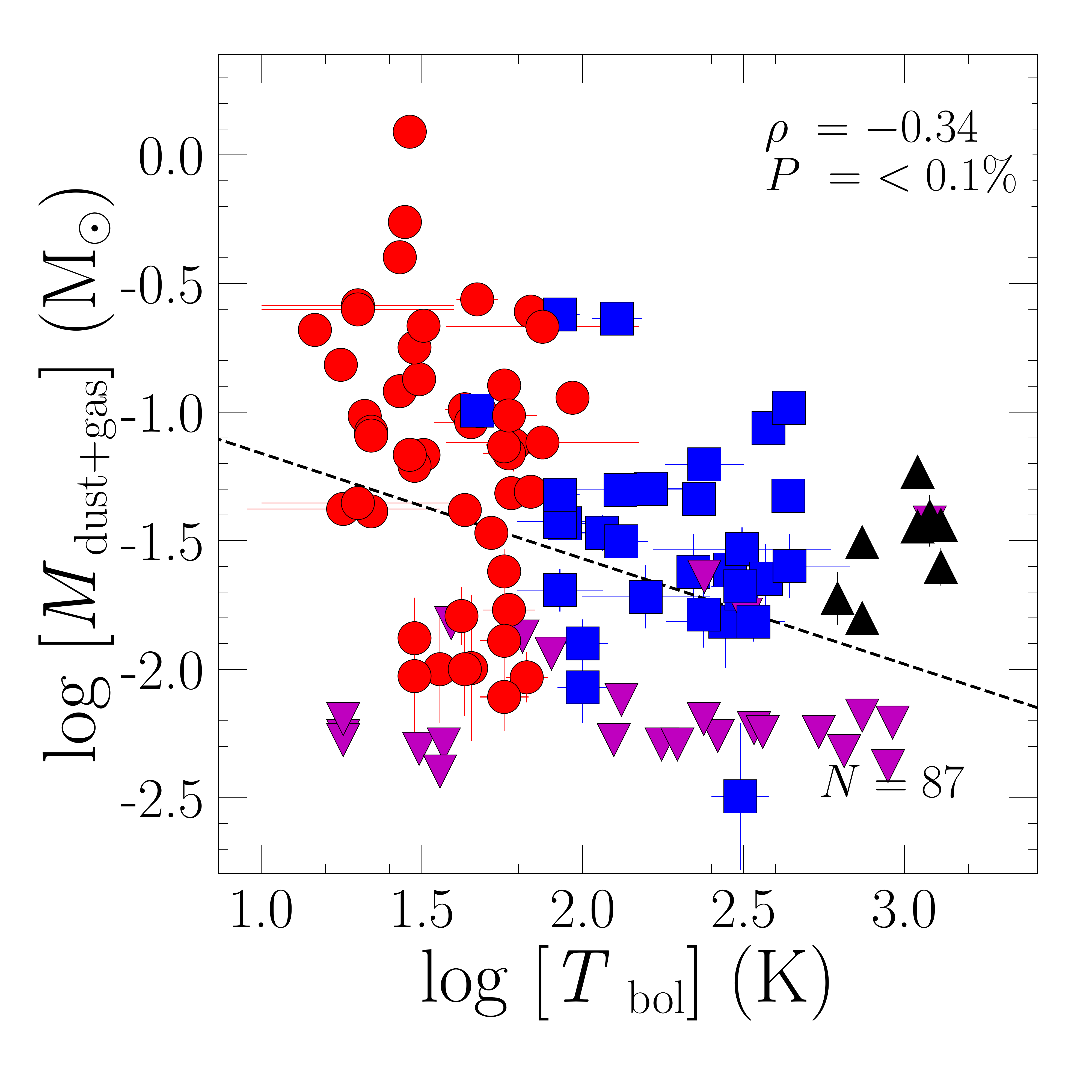}
\caption{Disk mass compared with bolometric luminosity (left) and temperature (right). Upper limits are marked as magenta triangles. 
Spearman's rank correlation coefficient and the probability of no correlation are shown in the top-right corner. Dashed lines represents the 
EM algorithm fit to the data. Note that for bolometric luminosity all multiple systems are combined together, while for bolometric temperature each component 
is considered separately but with the same bolometric temperature}, which results in different sample sizes.
\label{fig:masstbol}
\end{figure}

The calculated masses are consistent with those obtained by \cite{Segura-Cox2016} for seven sources from the VANDAM sample. \cite{Segura-Cox2016} used the same Ka-band data to model the disk structure and removed free-free contamination using a point-source model of free-free emission. For Per-emb-8, however, \cite{Segura-Cox2016} modeled a higher disk mass of 0.12-0.24 M$_\odot$, where we obtained a value of $0.097\pm0.006$ M$_\odot$ with our free-free correction. This source exhibits particularly strong, extended free-free emission with a resolved radio jet \citep{Tychoniec2018}, such that the free-free emission contributes roughly 43\% of the Ka-band continuum. For Per-emb-12-A (IRAS 4A) \cite{Cox2015} obtained 2.3 M$_\odot$ from uncorrected VANDAM data (they used $\beta = 1.3$, which further increases the estimated mass). With our corrected Ka-band fluxes, we find a mass of 1.2 M$_\odot$, which is still remarkably large, but more consistent with the typical masses ($< 1 $M$_\odot$) of the low-mass protostellar disks \citep[e.g.,][]{Jorgensen2009,Enoch2011}.

Figure \ref{fig:masshist} shows the distribution of mass for each evolutionary stage and the cumulative distribution obtained with the Kaplan-Meier estimator, 
for Class 0 and Class I only. We notice a clear decrease in mass between Class 0 and Class I with median values of 0.075 M$_\odot$ and 0.031 M$_\odot$ respectively.
The log-rank test was used to test the probability of drawing Class 0 and Class I datasets from the same sample. We find the probability of only 2.5\%, 
indicating that Class 0 and Class I mass distributions are statistically different. The sample size of Class II sources is too small to draw statistical conclusions
 (the median mass is 0.036 M$_\odot$). The median mass for the Class 0 and I sample together is 0.049 M$_\odot$.

As a constant dust temperature for all the sources is a very simplistic assumption, we also tried to account for the source luminosity by scaling the assumed dust temperature with the bolometric luminosity following: $T_{\textrm{dust}}= 30 [\textrm{K}] \times (L_{\textrm{bol}} /\textrm{L}_{\odot})^{1/4}$. Figure \ref{fig:masshist2} shows the mass distribution in this case. Obtained values are still consistent with an evolutionary decrease of masses, with log-rank test indicating a 1.5 \% chance of Class 0 and Class I distributions being drawn
 from the same sample. Taking into account the inescapable limitations, it is clear that disk mass does not grow between Class 0 and Class I, 
which suggests that disks form early during the star formation process and have the highest masses at an early age.

Figure \ref{fig:masstbol} shows disk masses compared with bolometric temperature and luminosity.
 We observe a weak correlation ($\rho=0.60$, $P$ $<$0.01\%) between the disk mass and the bolometric luminosity (Figure \ref{fig:masstbol}). 
As the latter is used as a proxy of protostellar mass (with many caveats), this result is reminiscent of the correlation between 
the disk mass and stellar mass observed for the more evolved disks \citep{Natta2000, Williams2011, Ansdell2017}. The noticeable 
decrease of disk mass with bolometric temperature is seen, hinting at a dependency between disk mass and evolution ($\rho =\ -0.34$, $P$ $<$0.1\%), 
as already apparent from the distributions of disk masses for Class 0 and Class I discussed above. 

\begin{figure}[H]
\centering
  \includegraphics[width=0.4\linewidth]{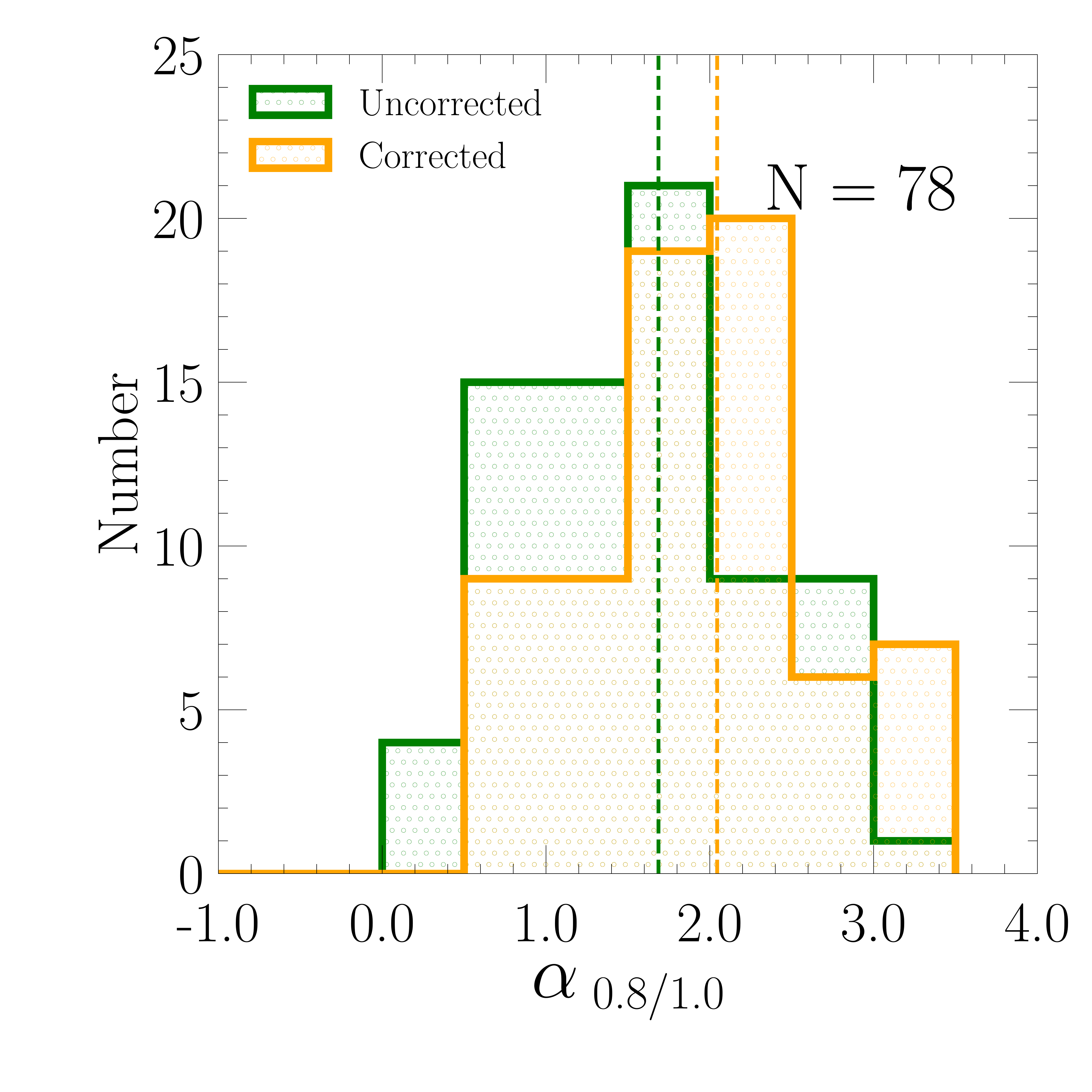}
  \includegraphics[width=0.4\linewidth]{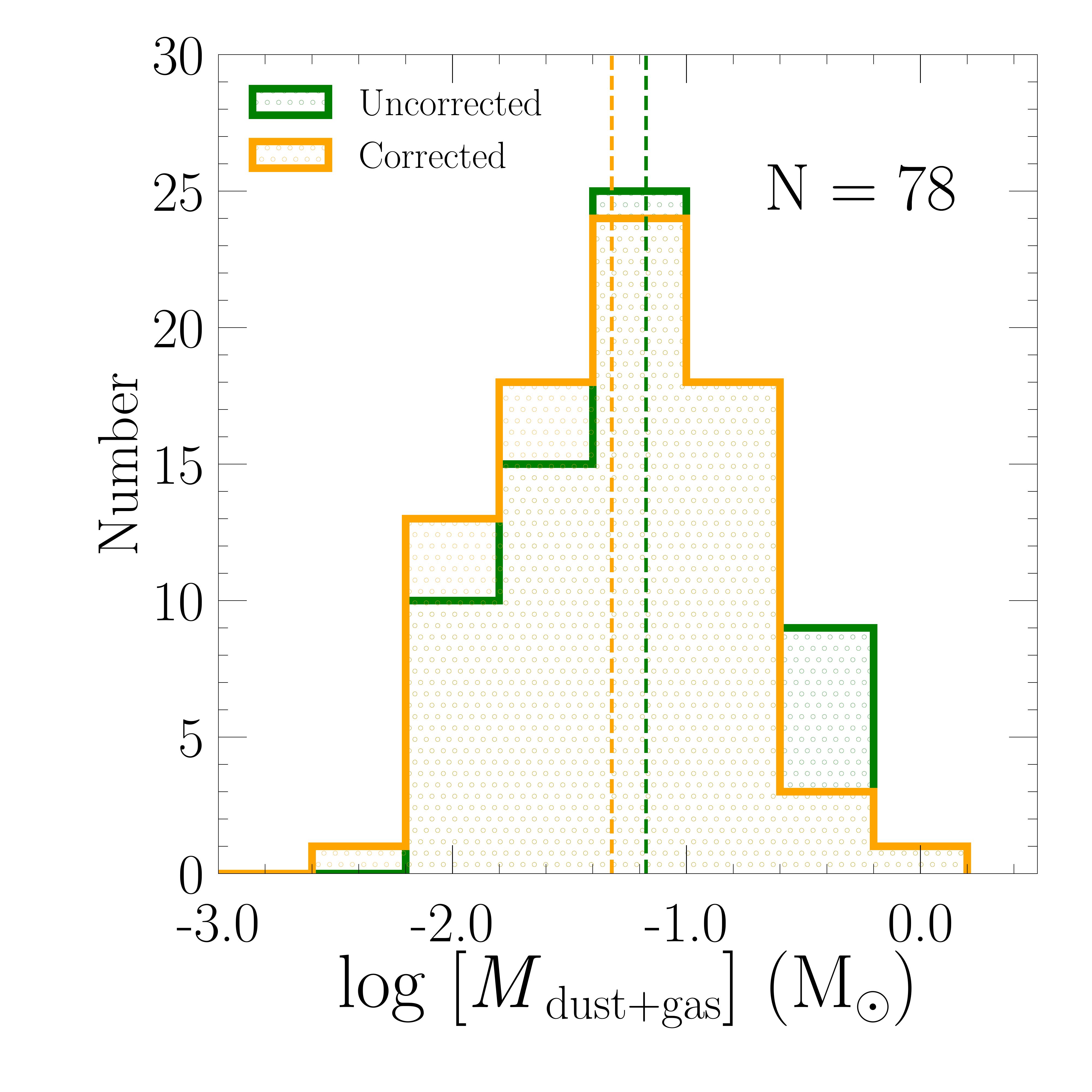}

\caption{Histograms of Ka-band spectral index (left) and disk masses (right). Values not corrected for free-free contribution are shown in green and corrected in yellow.
Median values and log-rank probabilities of being drawn from the same sample are: 1.69 (uncorrected), 2.04 (corrected), 
and 0.6\% for the spectral index distributions and 0.067 M$_\odot$ (uncorrected), 0.048 M$_\odot$ (corrected), and 37\% for the disk masses.}
\label{fig:masshist3}
\end{figure}

Finally, we assess the impact of the free-free emission on the calculated disk masses and spectral indices in Ka-band. 
Figure \ref{fig:masshist3} presents the distributions of masses and spectral indices with and without correction.
 When the correction is not applied, the spectral indices between 8 mm and 1 cm are flatter (median spectral index drops from 2.04 to 1.69) 
and the disk masses increase (from 0.048 M$_\odot$ to 0.067 M$_\odot$). The spectral index change is statistically significant, 
with log-rank: 0.6\% while mass change is less robust (log-rank: 37\%). Thus, the free-free contribution can to some extent explain the shallower 
than expected spectral indices observed in the Ka-band \cite{Tobin2016}, and it seems that without correcting for the free-free contribution the 
masses of the disks are slightly overestimated.

\subsection{Evolutionary trend in dust mass}
The advent of the Atacama Large Millimeter/submillimeter Array (ALMA) has made possible studying the gas and dust content of protoplanetary disks with 
unprecedented sensitivity. Surveys of Class II disks at different mean ages show that the disk dust mass consistently decreases
with age within the Class II population \citep{Ansdell2016, Ansdell2017,Barenfeld2016}.
 The other outstanding conclusion from those pioneering surveys was that there is not enough dust content to form gas giant planet cores even in the youngest 
Class II disks \citep[e.g.,][]{Ansdell2016}. Studying even younger embedded disks can answer important questions: does the decrease of the dust mass with age 
happen as early as from Class 0 to Class I, so in the first 0.5 Myr; is the dust mass in the embedded disks high enough to allow the formation of the cores of gas giants?  

\cite{Tobin2015} find a median mass of $\sim$ 0.05 M$_\odot$ for 9 Class 0 protostars in Perseus.
They note that the value is about an order of magnitude higher than the disk masses for Class II objects by \cite{Andrews2005},
and about 5 times larger than the median mass for Class I disks from \cite{Jorgensen2009}.
With the much greater sample of VANDAM sources, we confirm a decrease in disk mass with evolutionary class.
However, \cite{Jorgensen2009} pointed out that some of the assumptions used
to derive mass, such as constant opacity and temperature could confuse the real picture.
They considered models provided by \cite{Visser2009} which suggest that dust temperature
decreases from Class 0 to Class I due to the systematic decrease in the luminosity between Class 0 and Class I. After applying factors to simulate the evolutionary effects, \cite{Jorgensen2009} find that the apparent trend between Class 0 and Class I masses becomes insignificant. 
\cite{Fischer2017} recently show a decrease in bolometric luminosity for a large sample of protostars in Orion, 
which might occur due to the decrease of the envelope emission, while protostellar luminosity still increases. 
\cite{Dunham2014b} suggest using different temperatures for Class 0 and Class I disks based on pure hydrodynamical simulations. 
However, our attempt to take into account the possible difference in luminosity between the two evolutionary classes still yields a statistically significant difference between disk masses of Class 0 and Class I protostars.

\begin{figure}[H]
\centering
  \includegraphics[width=0.6\linewidth]{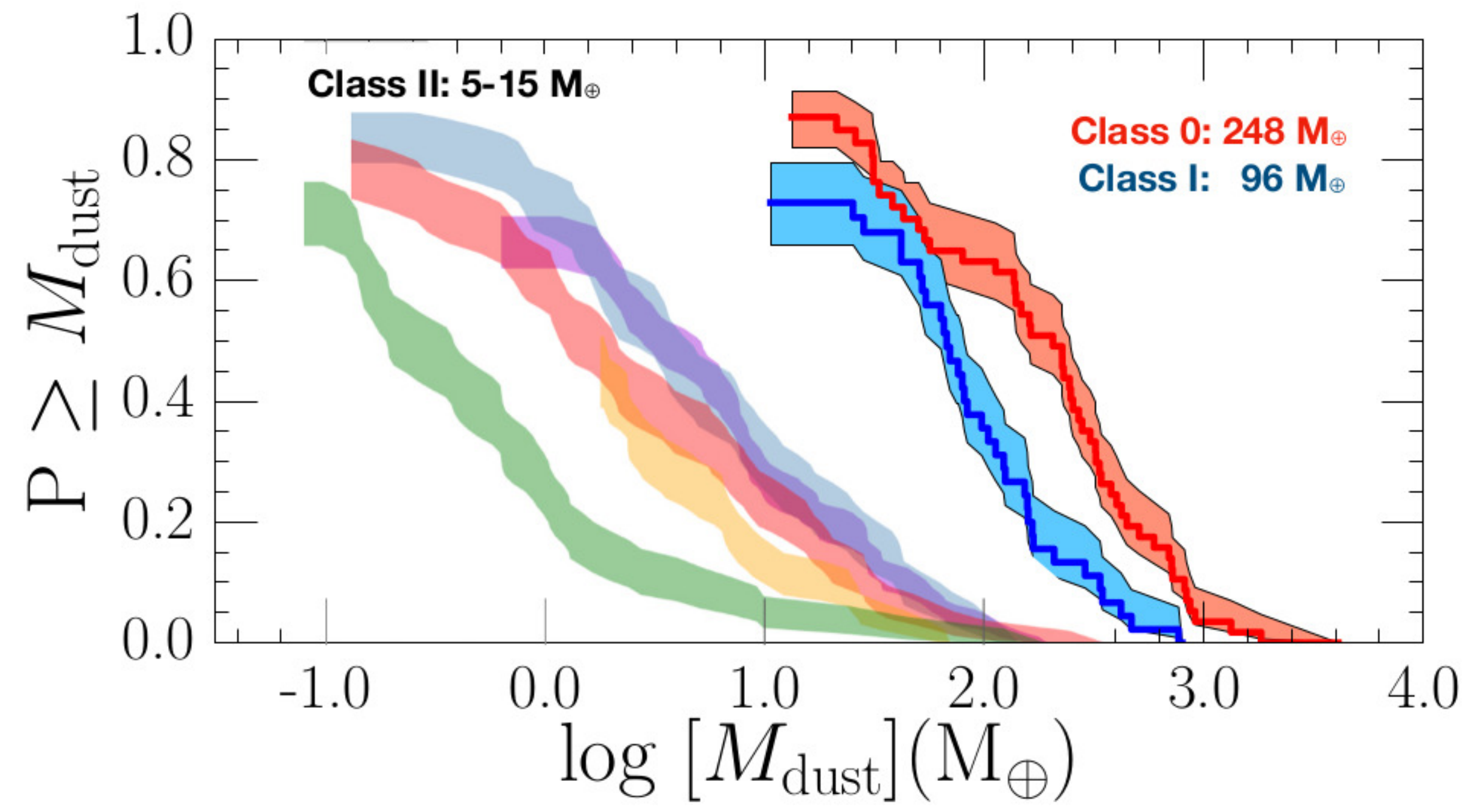}

\caption{Cumulative distributions of disk masses in units of Earth mass. Class II distributions for four regions adapted from \cite{Ansdell2017}. 
Different Class II star forming regions are presented: Taurus \citep[purple,][]{Andrews2013}, Lupus \citep[blue,][]{Ansdell2016}, 
Chamaeleon I \citep[red,][]{Pascucci2016}, $\sigma$ Orionis \citep[yellow,][]{Ansdell2017}, and Upper Sco \citep[green,][]{Barenfeld2016}.}
\label{fig:vandamansdell}
\end{figure}

Figure \ref{fig:vandamansdell} shows a comparison between VANDAM results and Class II surveys presented in \cite{Ansdell2017}. 
The observed decrease of mass between Class 0 and Class I protostars, and further to Class II, shows that a significant fraction of dust is dispersed or incorporated into larger bodies. If the latter scenario is considered, the amount of dust-only mass available for planet formation (248 M$_{\oplus}$\footnote{values converted to Earth masses and without multiplying by 100 to exclude gas mass}) is enough to form solid cores of the giant planets. The further decrease in mass to 96 M$_{\oplus}$ in Class I shows that significant grain growth could occur at those early stages \citep{Miotello2014,Sheehan2017}. Recent ALMA surveys of Class II disks yield masses of 5-15 M$_{\oplus}$ for different star-forming regions \citep[e.g.,][]{Barenfeld2016,Pascucci2016,Ansdell2016}.
 It shows that if the core accretion is considered as a planet formation route, it may begin very early in Class 0, and the physical conditions at those early stages should be considered in planet formation models.

\section{Conclusions}
We observed all known (84) Class 0 and I protostellar systems in the Perseus molecular cloud with the VLA at C-band (4.1 cm and 6.4 cm). The major conclusions of this work are as follows:

\begin{enumerate}

\item The detection rate is 61\% for Class 0 and 53\% for Class I protostars. Both flux densities and spectral indices do not show a significant difference between the two evolutionary stages indicating that strength and nature of the emission is independent of evolution at least through the protostellar phase.

\item The spectral index from 4.1 cm to 6.4 cm for the detected protostars has a median value of $\alpha _{median}=0.51$, consistent with moderately optically thick thermal free - free emission. The C-band spectral index shows no correlation with protostellar bolometric luminosity and temperature. Sources with resolved thermal jets have typically lower spectral indices consistent with optically thin emission from the jet, in addition to being the brightest free-free objects.

\item We detect all components in half of the close ($<$ 500 au)  binary systems present in a sample. Protostellar companions within the same system can have very different flux densities and spectral indices. There are also examples of systems where brighter Ka-band component appears fainter in C-band.

\item We greatly extended  the group of the protostars characterized at centimeter wavelengths especially in the low-luminosity end. 
However, the radio luminosity from the protostars only in Perseus is weakly correlated with bolometric luminosity, 
by combining these data with previous observations spanning a larger range of $L_{\rm bol}$ we obtained a good correlation. The linear fit for the Perseus-only sample shows a flatter relation between radio and bolometric luminosity than for the merged sample.

\item We investigate correlations between the radio luminosity and molecular and atomic far-IR line luminosity from \textit{Herschel}.
 We obtain moderate correlations for OH and [O I]. Comparing this result with shock models we conclude that the ionization observed
as free-free emission is predominantly a result of J - type shocks. Extending this analysis to a sample of high-mass protostars can provide further insight.

\item We update the correlation between the radio luminosity and  outflow force from protostars. Within the range of luminosities in Perseus there is no correlation, 
but inclusion of a greater range of radio luminosities and outflow forces results in a moderate correlation, consistent with previous studies. We find that the molecular outflow forces are sufficient to produce the obtained radio fluxes in our sample. It shows that while generally shock ionization is a viable explanation of free-free emission, molecular outflows have different characteristics than a thermal radio jet, likely due to the different scales probed.

\item We calculate the disk masses around protostars,  using Ka-band (9 mm) flux densities corrected for free-free contribution from the C-band.
 A statistically significant difference is observed between Class 0 and Class I disk masses with the median disk mass being more than a factor of two higher for Class 0 protostars median dust mass 248 M$_{\oplus}$ in Class 0). By the Class II phase, the median disk dust mass has dropped by an order of magnitude. This result suggests that protoplanetary disks have their highest masses at early times - with a dust mass reservoir sufficient to form giant planet cores - and that grains can grow rapidly in the embedded phase.
 The C-band contribution lowers between 8 mm and 1 cm, while the measured disk mass is not significantly changed.

\end{enumerate}

\acknowledgments{ The authors thank the anonymous referee, whose comments enhanced the quality of the manuscript. The authors wish to thank Y. Shirley and J. Mottram for sharing data from their previous works, and G. Anglada for insightful comments on the draft.
ŁT thanks S. van Terwisga and A. Bosman for stimulating discussions. Astrochemistry in Leiden is supported by the Netherlands Research School for Astronomy (NOVA), by a Royal Netherlands Academy of Arts and Sciences
 (KNAW) professor prize, and by the European Union A-ERC grant 291141 CHEMPLAN. ŁT is supported by Leiden/ESA Astrophysics Program for Summer Students (LEAPS). J.J.T. acknowledges support from the Homer L. Dodge endowed chair, and J.J.T. and Ł.T. acknowledge support from grant 639.041.439 from the Netherlands Organisation for Scientific Research (NWO). J.J.T. acknowledges past support provided by NASA through Hubble Fellowship grant \#HST-HF-51300.01-A awarded by the Space Telescope Science Institute, which is operated by the Association of Universities for Research in Astronomy, 
Inc., for NASA, under contract NAS 5-26555. AK acknowledges support from the Polish National Science Center grant 2016/21/D/ST9/01098. ŁT and AK acknowledge support from the HECOLS International Associated Laboratory, 
supported in part by the Polish NCN grant DEC-2013/08/M/ST9/00664. ZYL is supported in part by NASA NNX 14AB38G, NSF AST-1313083 and AST-1716259. 
The National Radio Astronomy Observatory is a facility of the 
National Science Foundation operated under cooperative agreement by Associated Universities, Inc. This research made use of NASA's Astrophysics Data System.
}
\facilities{VLA}

\software{Astropy \citep{Astropy2013}, APLpy  \citep{Robitaille2012}, 
Matplotlib  \citep{Hunter2007}, MPFIT \citep{Markwardt2009},  lifelines \citep{DavidsonPilon2017}, AEGEAN \citep{Hancock2012}, CASA \citep{McMullin2007}, STSDAS, ds9
}

\newpage
\bibliography{my_bib.bib}

\clearpage

\appendix

\section{Additional correlations} \label{apA}

We show a spectral index plotted against bolometric luminosities, but with upper/lower limits excluded and with four outliers  removed: Per-emb-14, EDJ2009-156, SVS13A2, L1448NW.
\begin{figure}[H]
\centering
\includegraphics[width=0.4\linewidth]{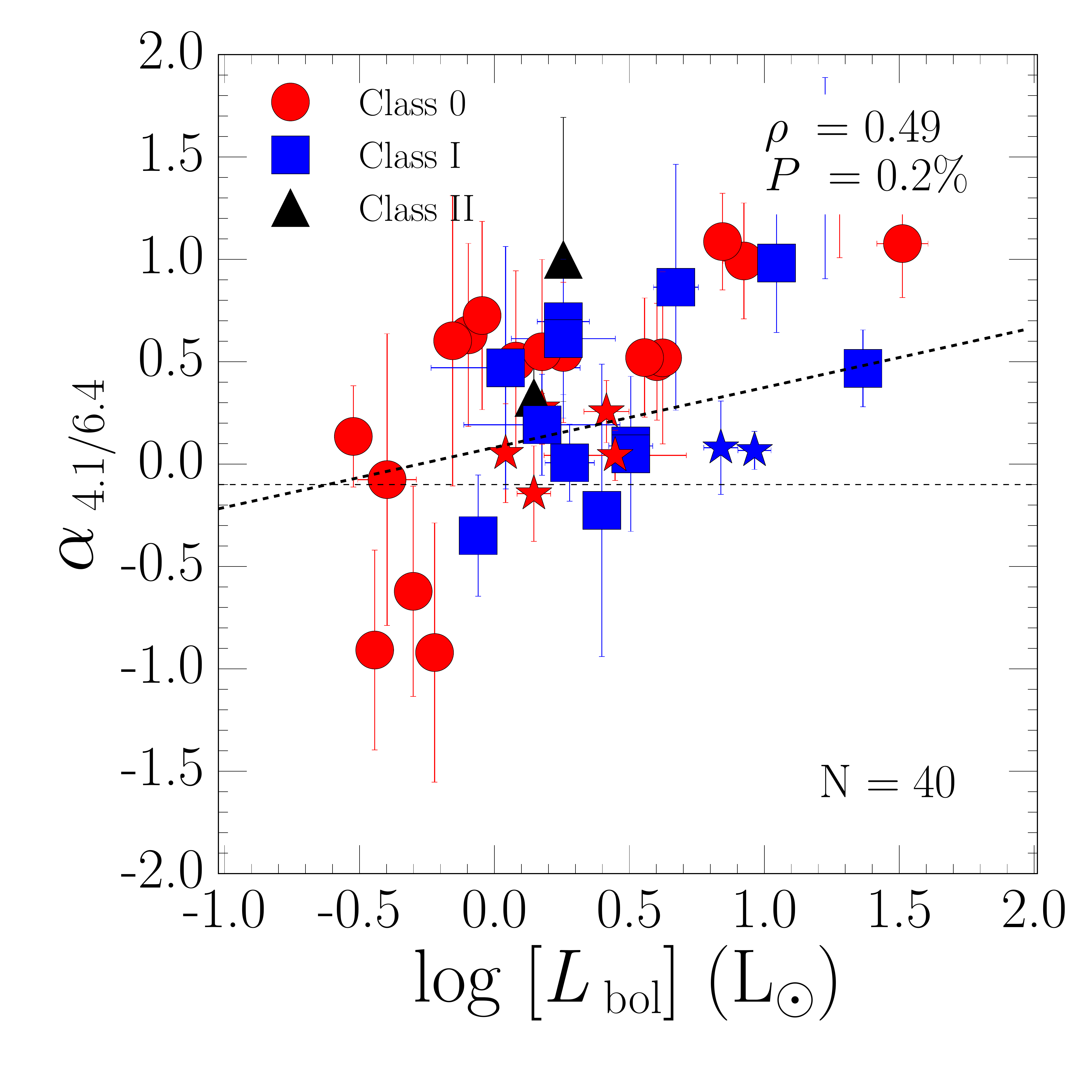}
\caption{Left: Spectral index compared with bolometric luminosity with four outliers removed: Per-emb-14, EDJ2009-156, SVS13A2, L1448NW. }
\label{fig:lbolindex_out}
\end{figure}
We present in Figure \ref{fig:fig16} radio luminosity at 6.4 cm plotted against far-IR line luminosities. The relations at 6.4 cm are similar to those obtained at 4.1 cm.
\begin{figure}[H]
 \centering
  \includegraphics[width=0.8\linewidth]{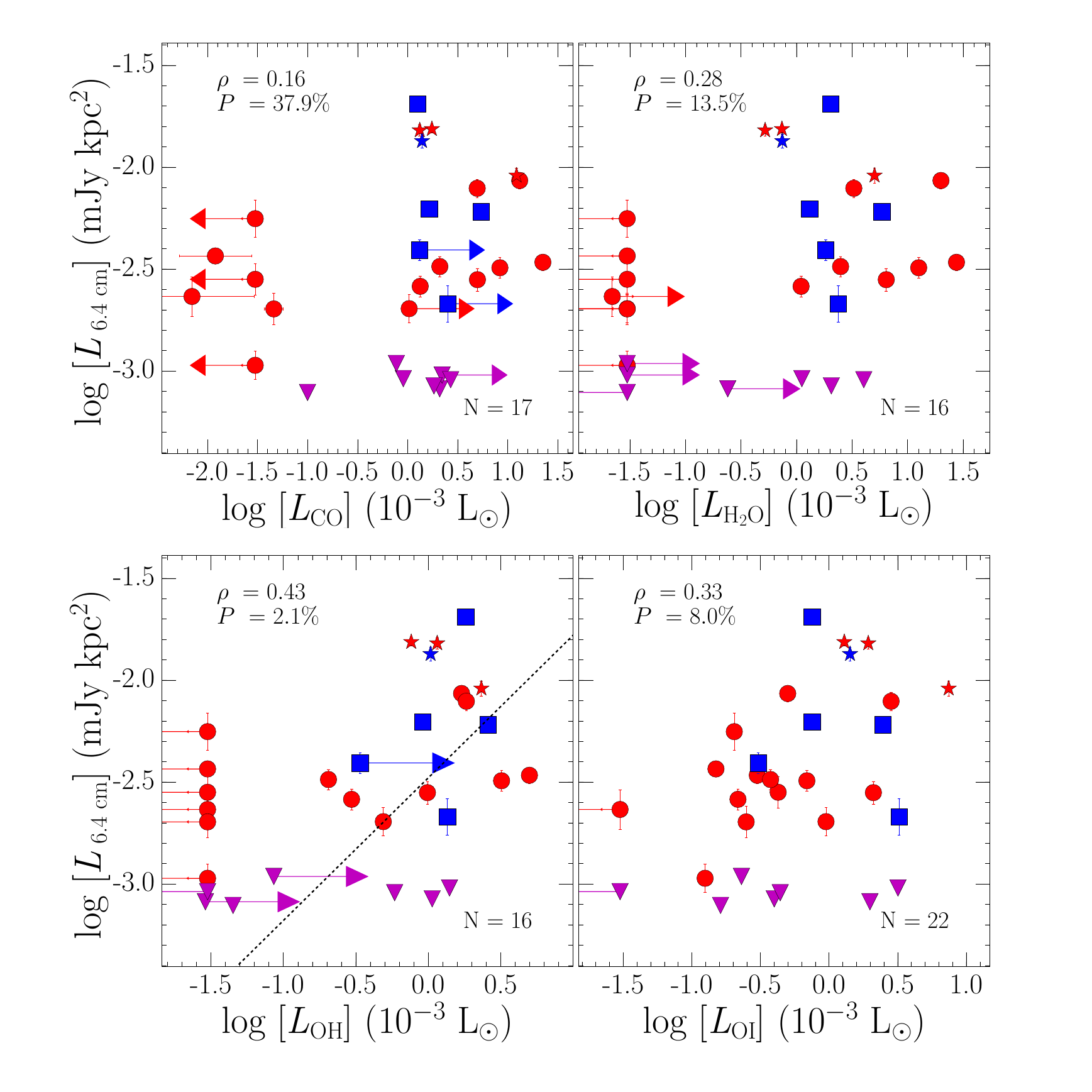}
\caption{Luminosity at 6.4 cm compared with CO (top left), H\raisebox{-.4ex}{\scriptsize 2}O (top right), [O I] (bottom left), and OH (bottom right) line luminosities. Upper limtis for radio luminosities are plotted as magenta triangles, and lower or upper limits for {\it Herschel} line luminosities are indicated with arrows. Spearman's rank correlation coefficient and probability of no correlation is shown in the right top corner (for combined sample of Class 0 and Class I protostars).}
  \label{fig:fig16}
\end{figure}
\section{Interesting sources}
\begin{figure}[H]
  \includegraphics[width=0.29\linewidth]{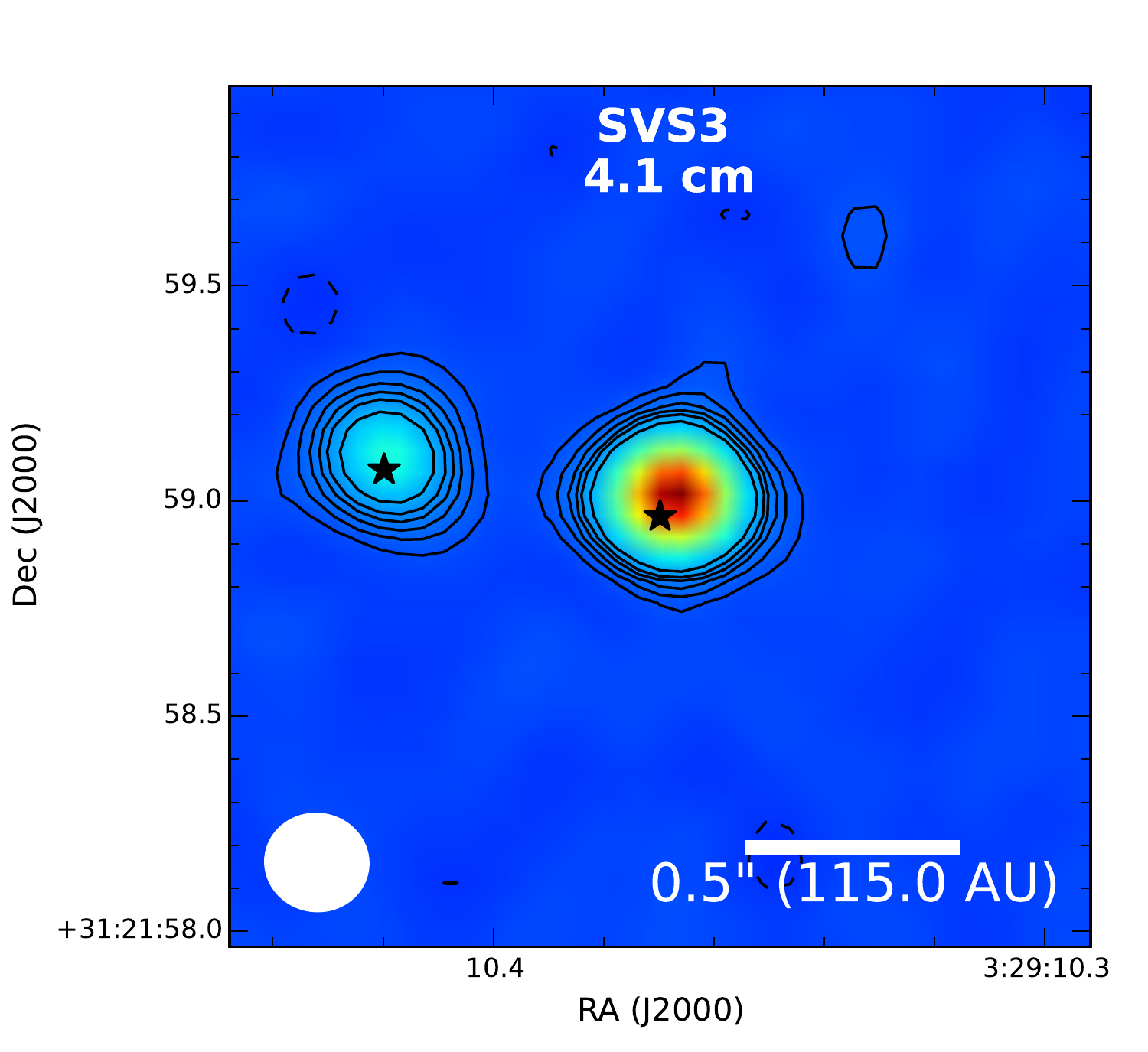}
  \includegraphics[width=0.29\linewidth]{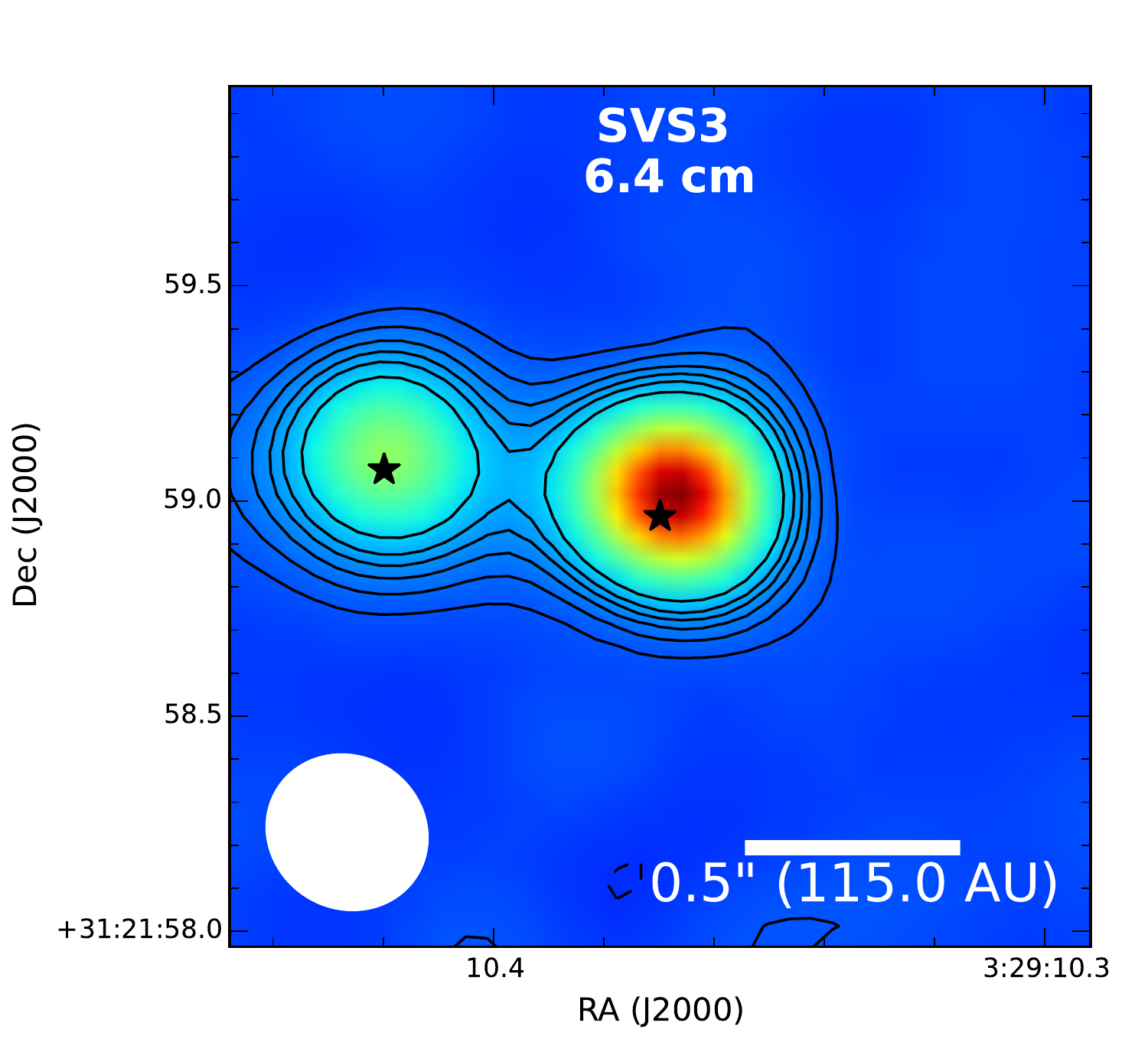}
  \includegraphics[width=0.36\linewidth]{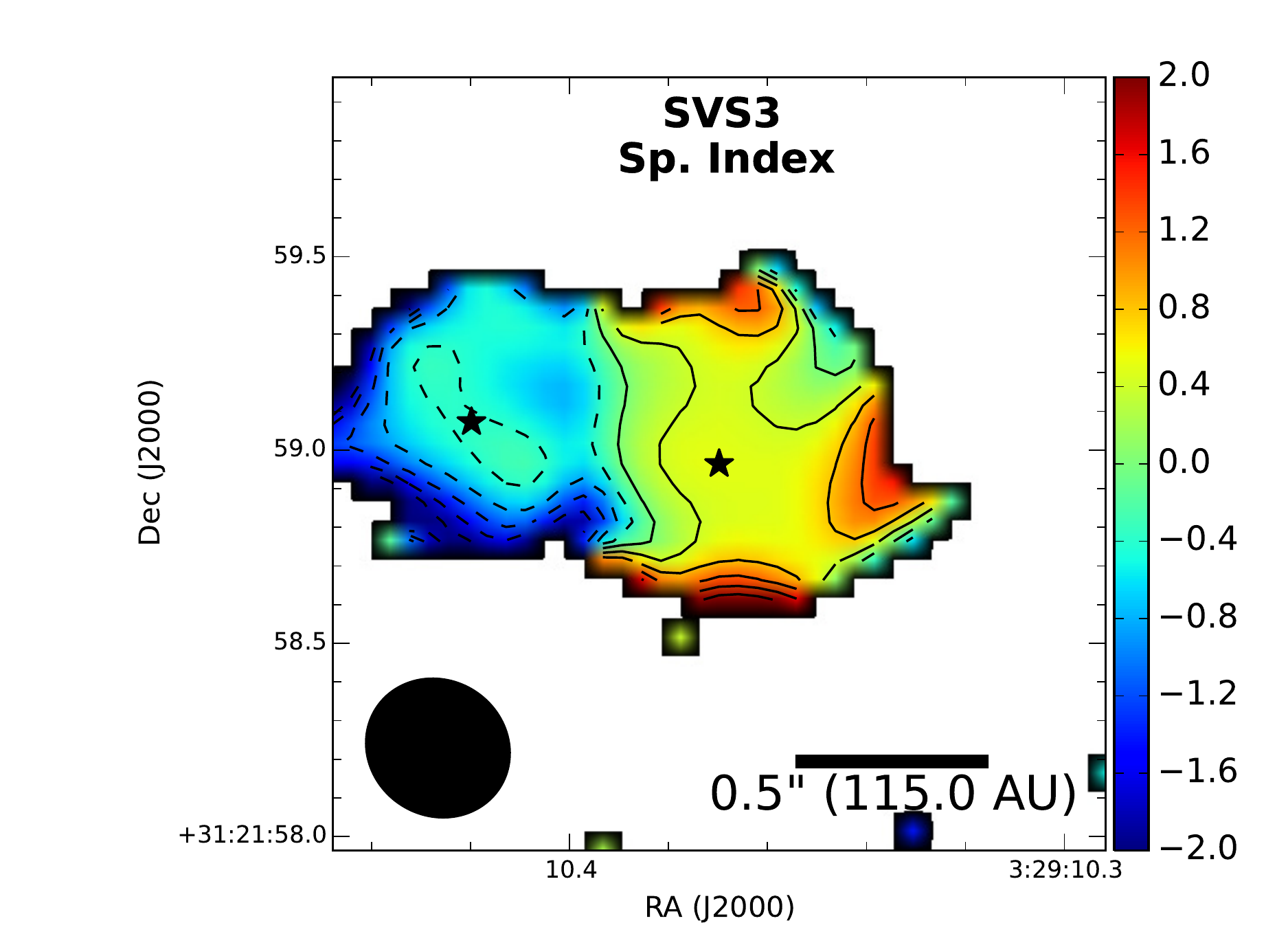}
\caption{Maps of SVS3 system centered on a SVS3-B component. From left to right: 4.1 cm, 6.4 cm and spectral index map.}
\label{fig:svs3}
\end{figure}

We note the serendipitous detection of a few intriguing sources in the C-band observations, although not having protostellar nature.
The source SVS3 - also known as IRAS 03260-3111 - is a visible star illuminating the surrounding cloud. It is actually a binary first reported as such by \citep{Connelley2008a}. In C-band observations we clearly see both components with SVS3-B being the brighter of the two (Figure \ref{fig:svs3}). 
SVS3-A was well detected with Ka-band observations \citep{Tobin2016} with negative spectral index. It appears to have negative spectral index in C-band observations as well. This altogether points to this star probably being a bright synchrotron source originating from coronal activity, characteristic for more evolved sources and is also supported by X-ray detection \citep{Preibisch1997, Getman2002}.  SVS3-B was not detected in the Ka-band observations indicating that there is very little dust present in this component. Free-free emission associated with the source appears spherical and resolved, consistent with ionized wind from the evolved, luminous star.

\begin{figure}[H]
 \centering
  \includegraphics[width=0.30\linewidth]{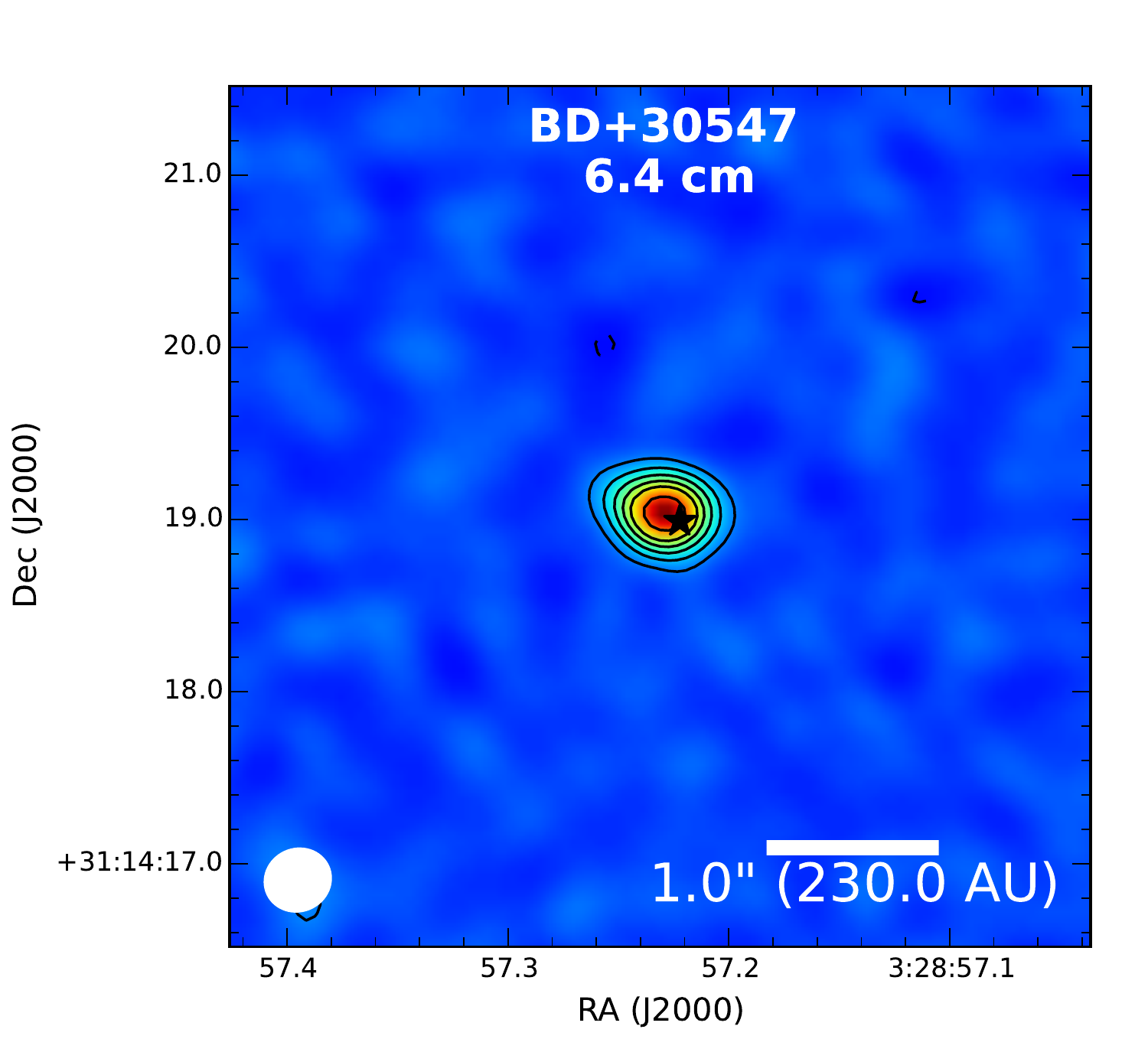}
  \includegraphics[width=0.30\linewidth]{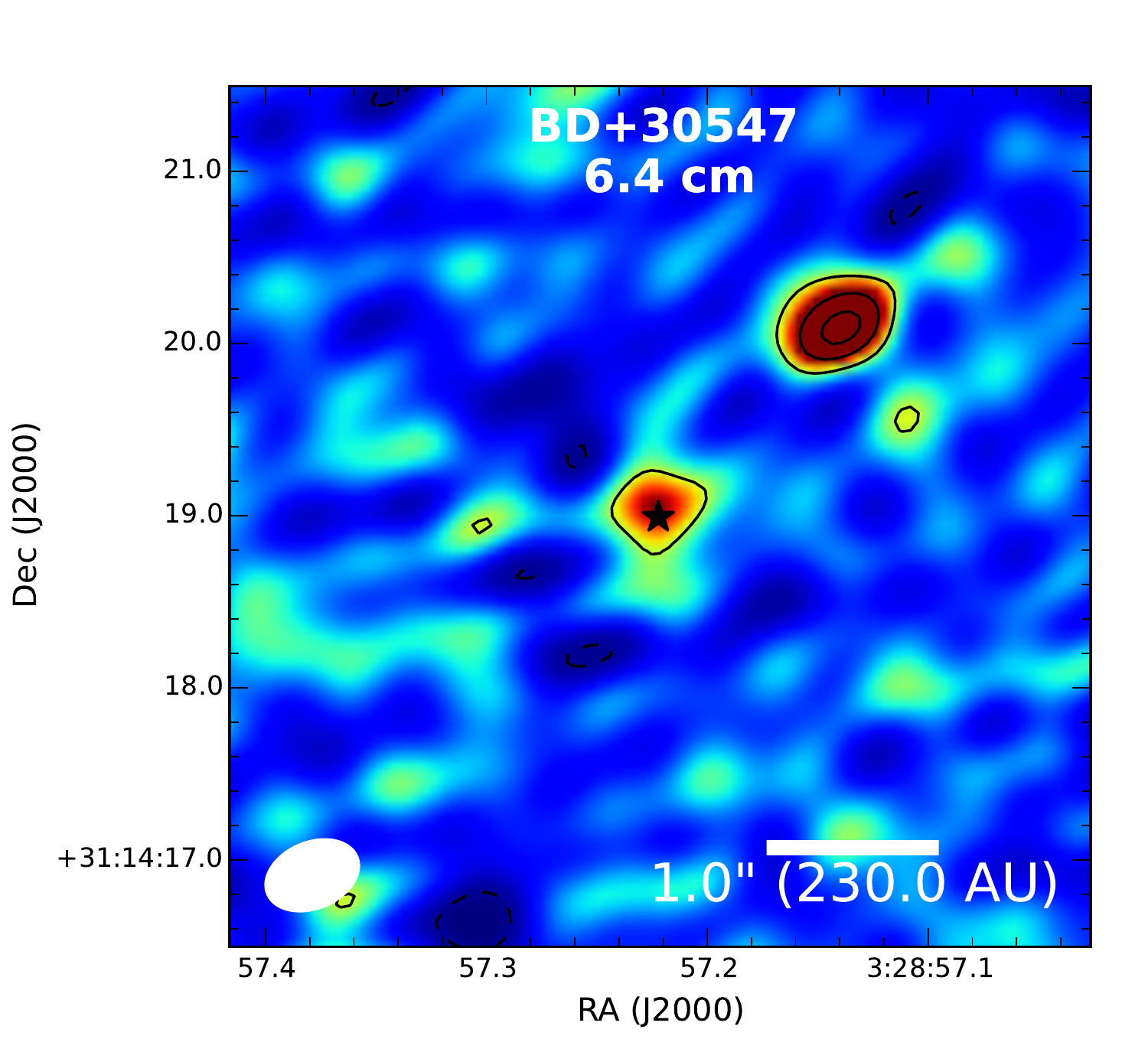}
\caption{BD+30 547 maps at 6.4 cm from two different epochs. Significant decrease in brightness of the central source can be observed, as well as the appearance of the new source of emission north-west of the BD+30 547.}
\label{fig:BD+30547}
\end{figure}

Source BD +30 547 is a visible star with strong variability detected previously \citep{Rodriguez1999}. It also appears variable in our observations. We note a potential transient source appearing about 1" north-west away from BD +30 547 in only one of the maps, from five fields available (Figure \ref{fig:BD+30547}).

We detect source 2MASS J03293053+3127280 reported previously as a brown dwarf \citep{Wilking2004}. 
Extremely low bolometric luminosity of this source 0.001 L$_{\odot}$ makes explanation of the emission difficult. 
 The source has a positive spectral index, excluding possibility of coronal activity, rather pointing at a surprisingly powerful stellar wind from a brown dwarf.
Recently, \cite{Rodriguez2017} report a detection of brown dwarfs with radio fluxes order-of-magnitude more powerful than expected.

\clearpage

\section{Free-free and dust slopes} \label{sec:apC}
\begin{figure}[H]
 \centering

  \includegraphics[width=0.30\linewidth]{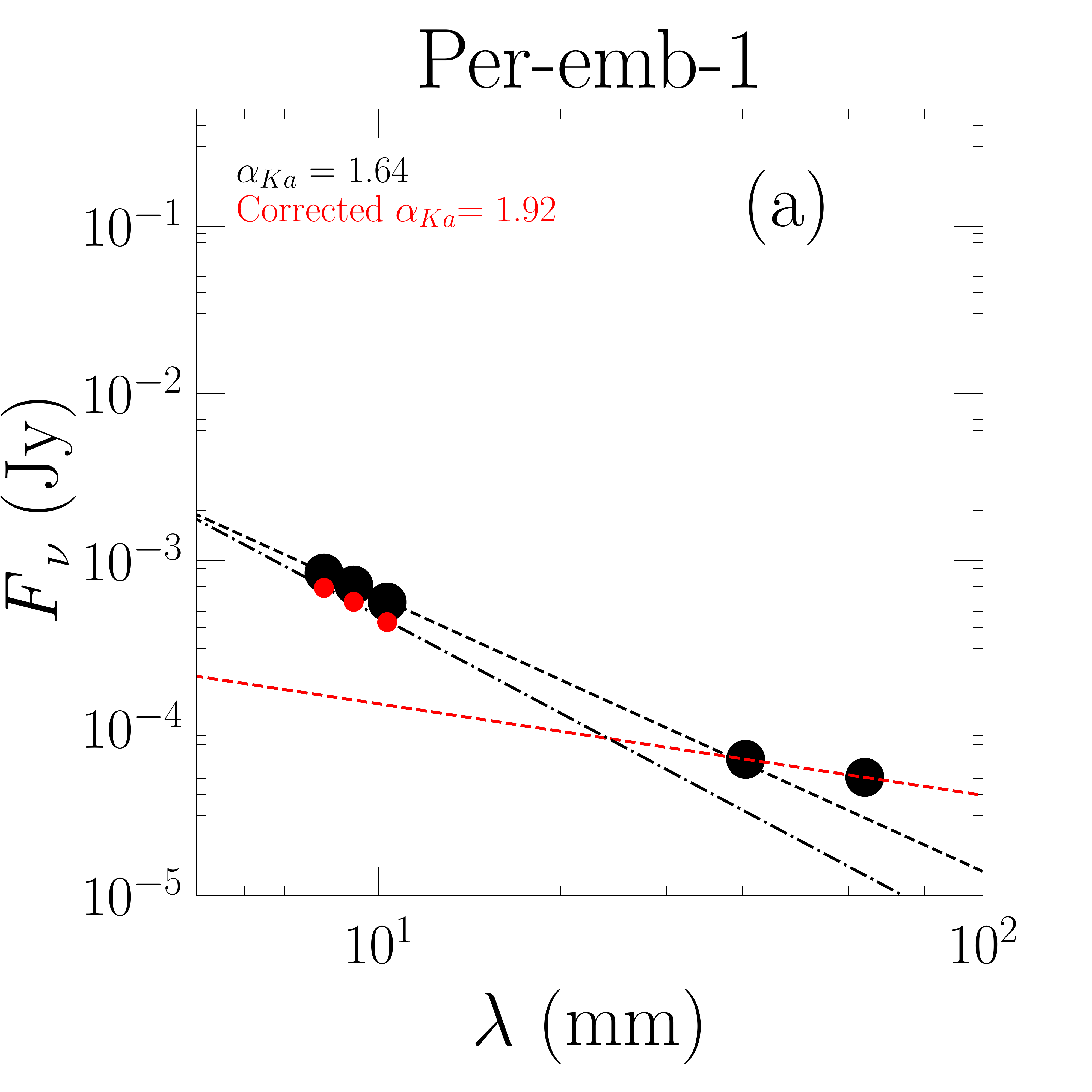}
  \includegraphics[width=0.30\linewidth]{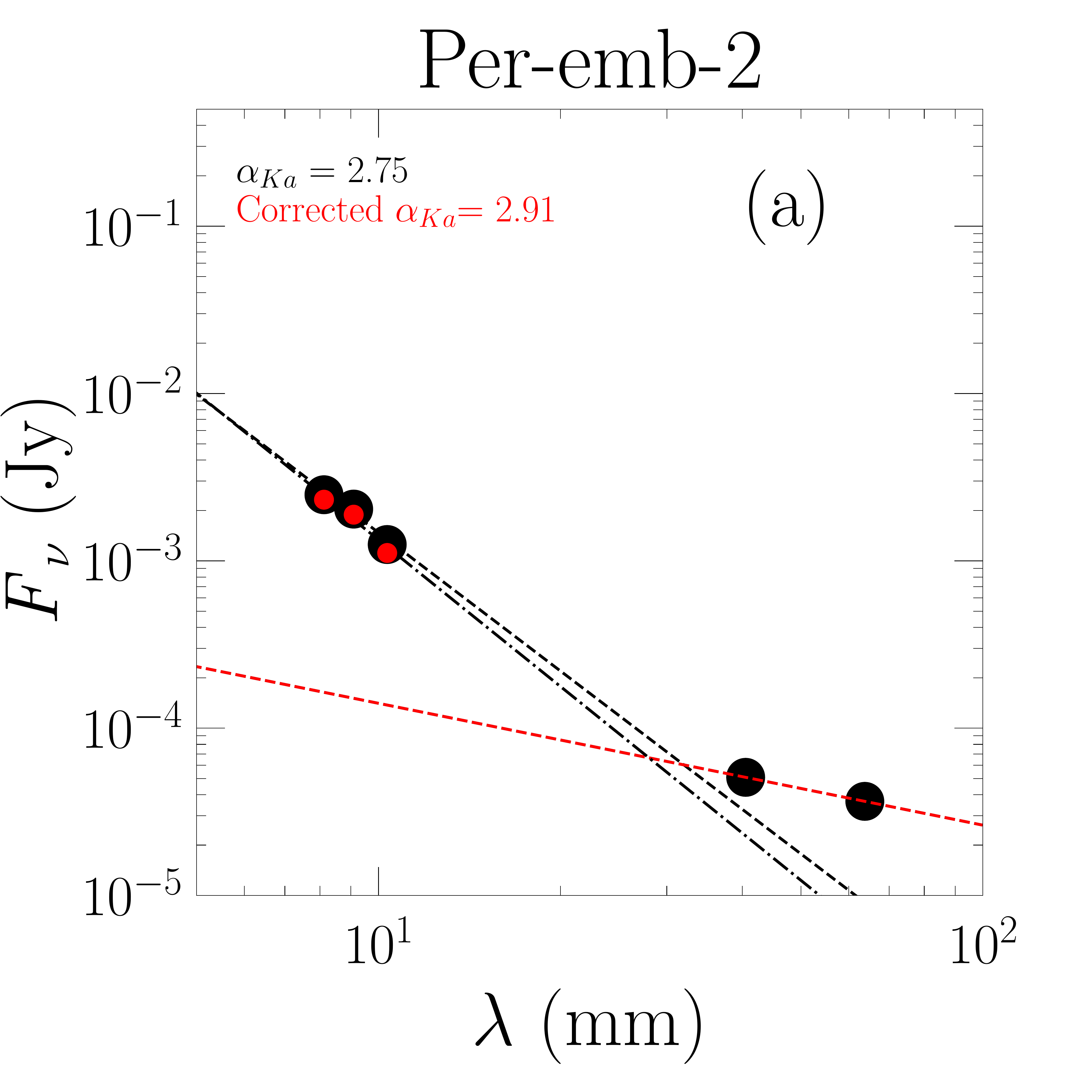}
  \includegraphics[width=0.30\linewidth]{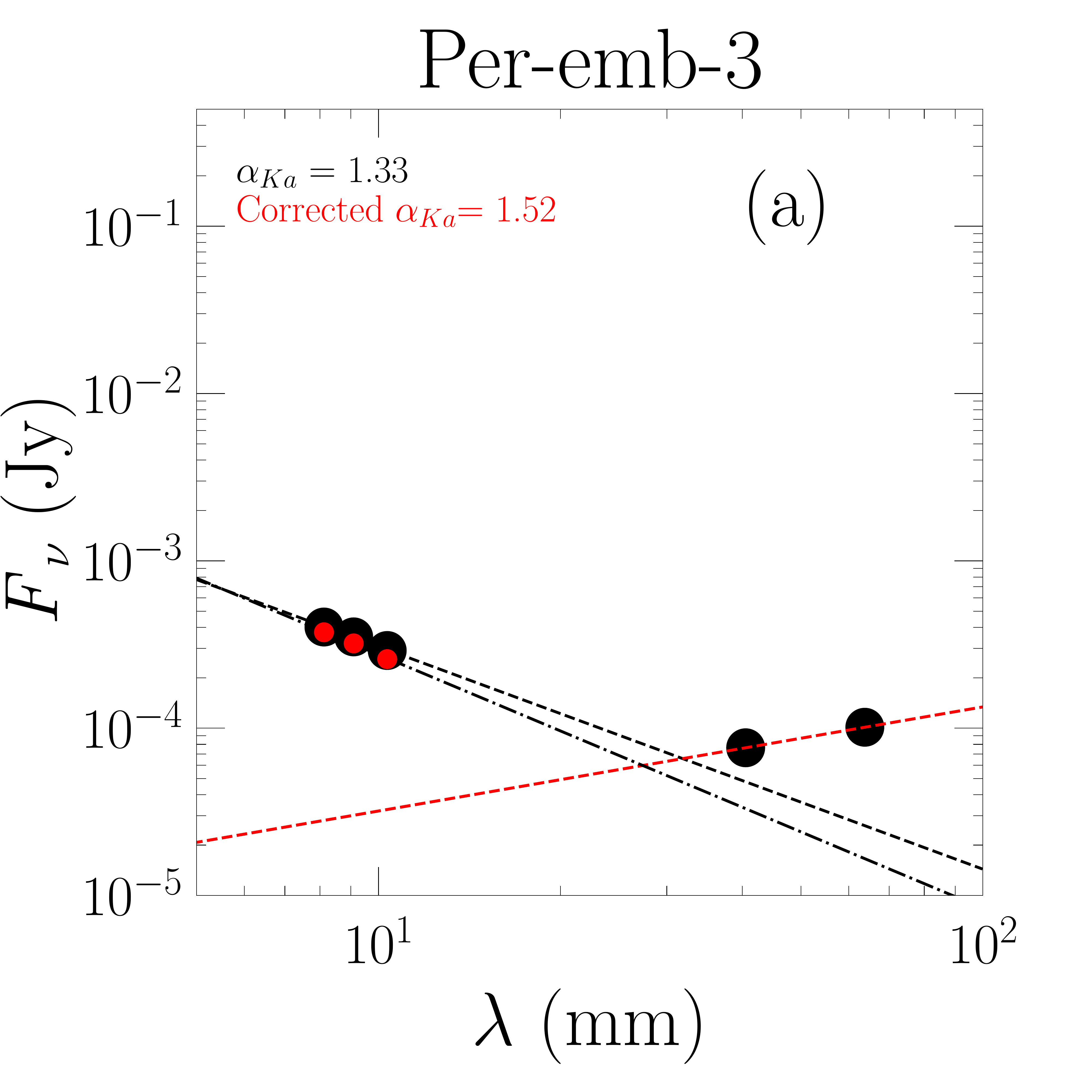}
  \includegraphics[width=0.30\linewidth]{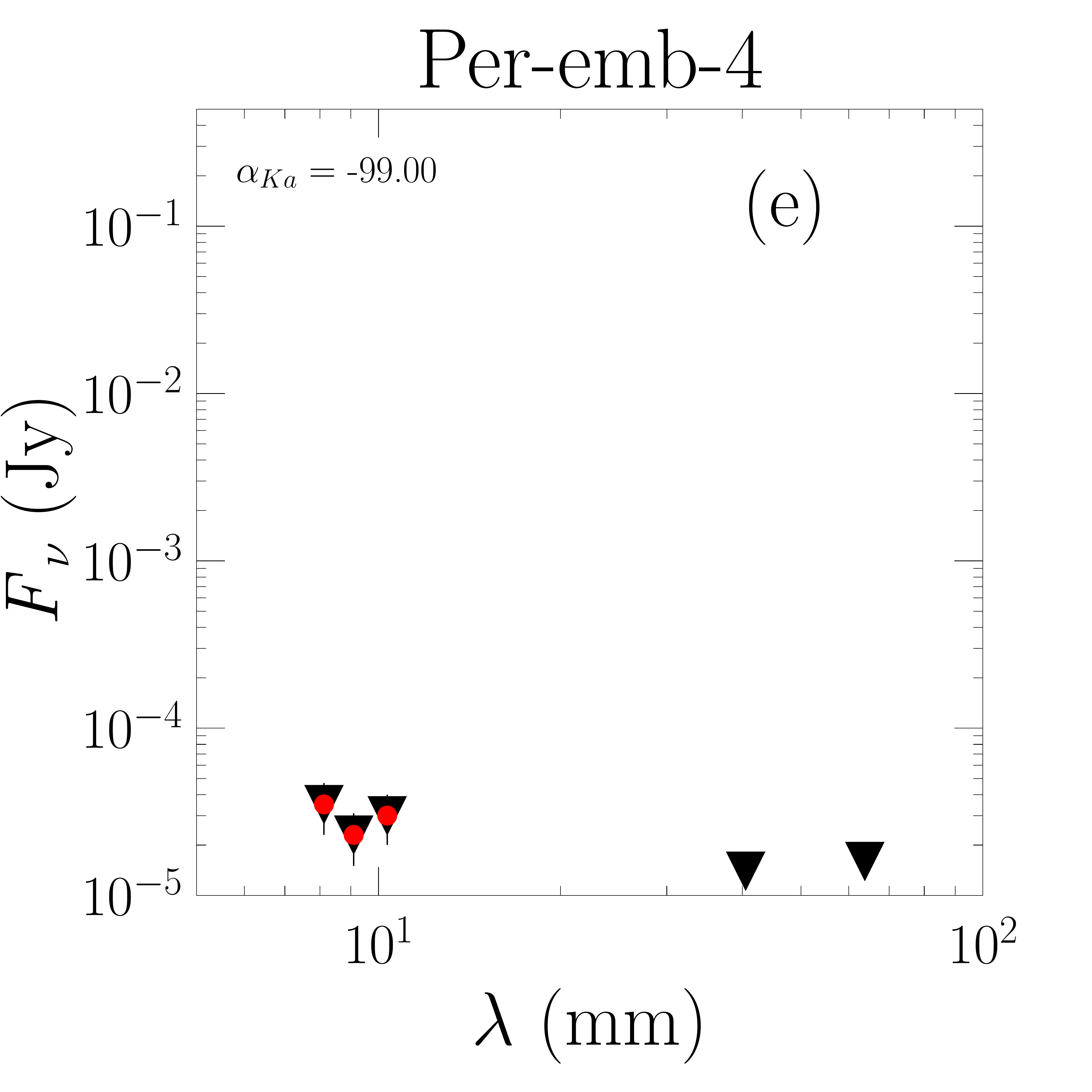}
  \includegraphics[width=0.30\linewidth]{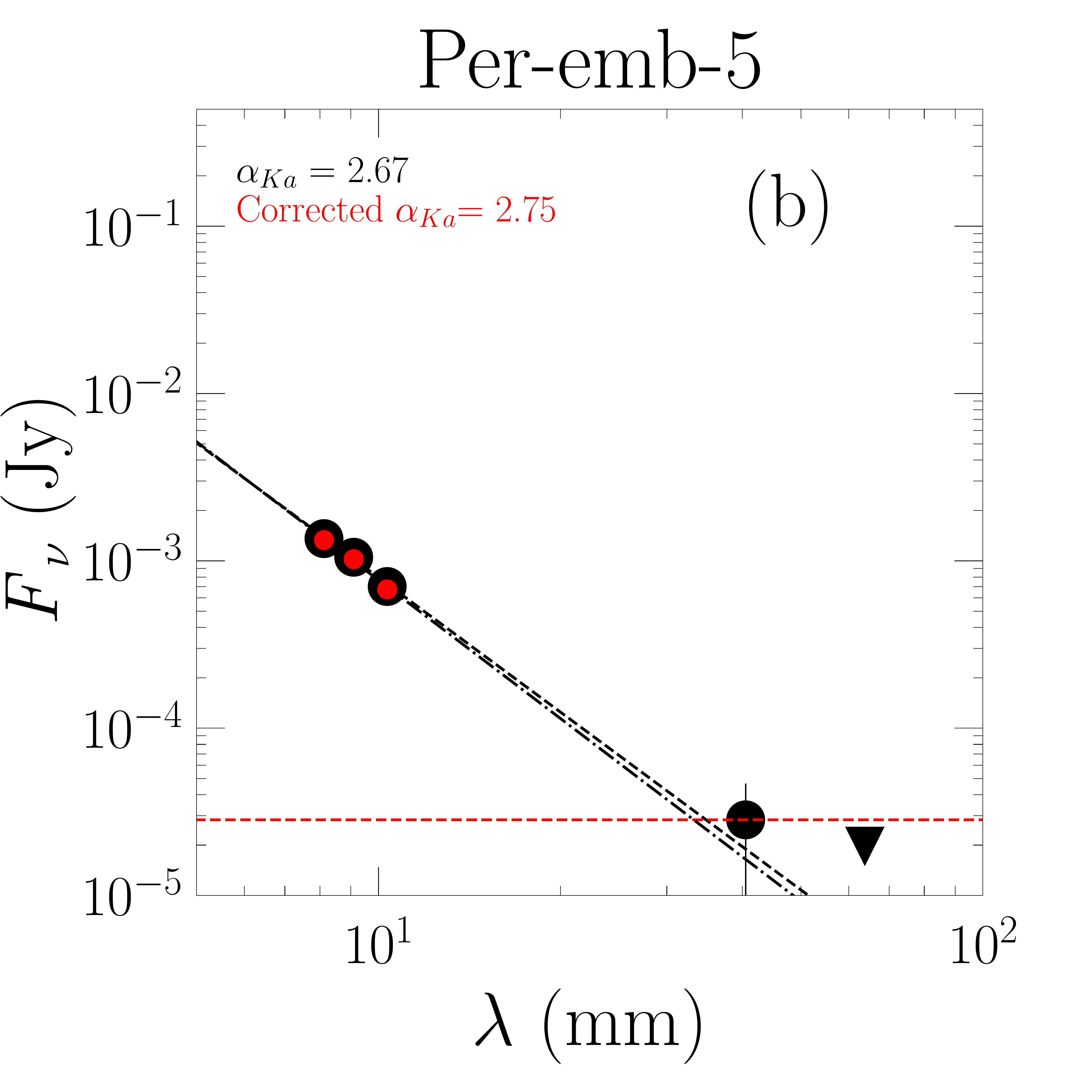}
  \includegraphics[width=0.30\linewidth]{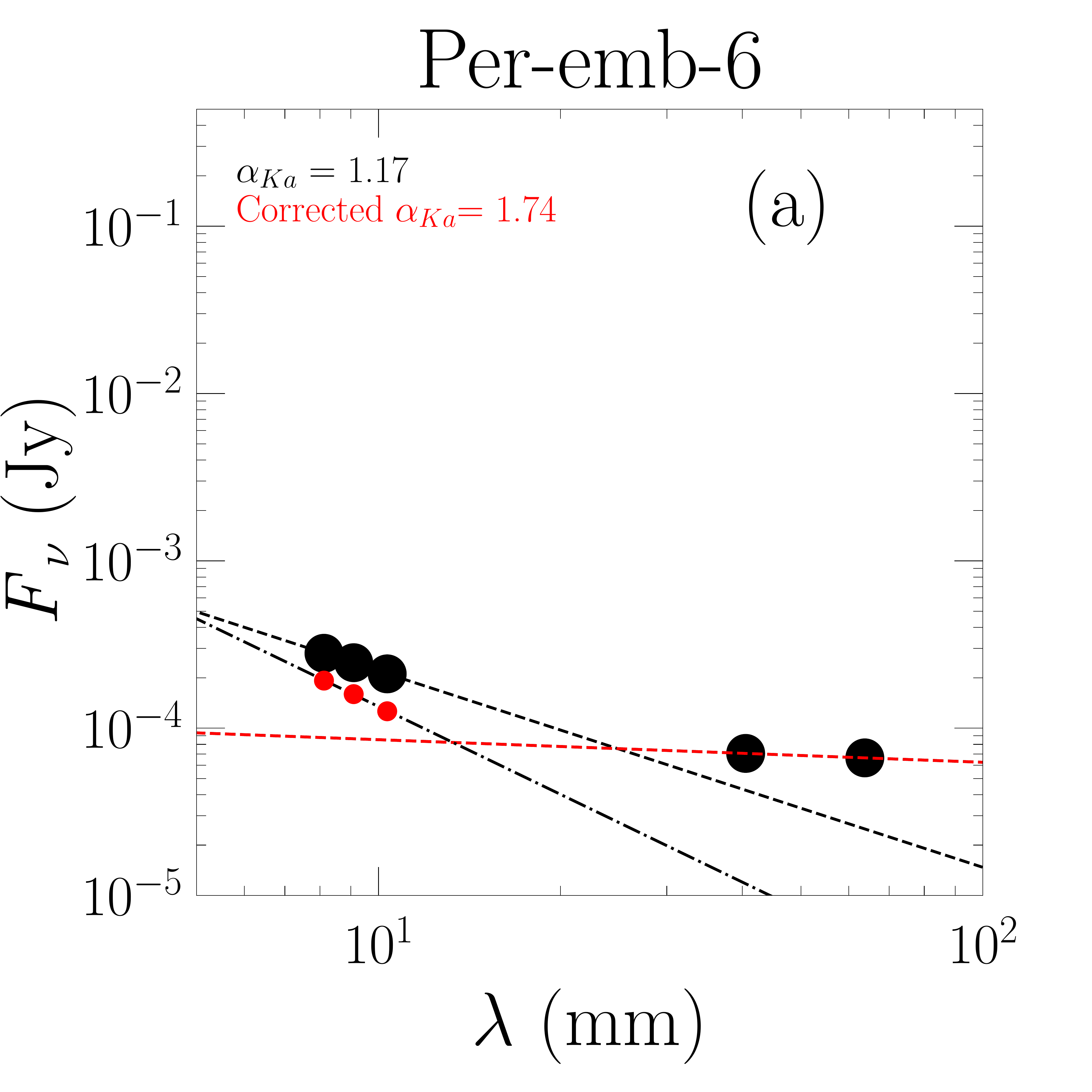}
  \includegraphics[width=0.30\linewidth]{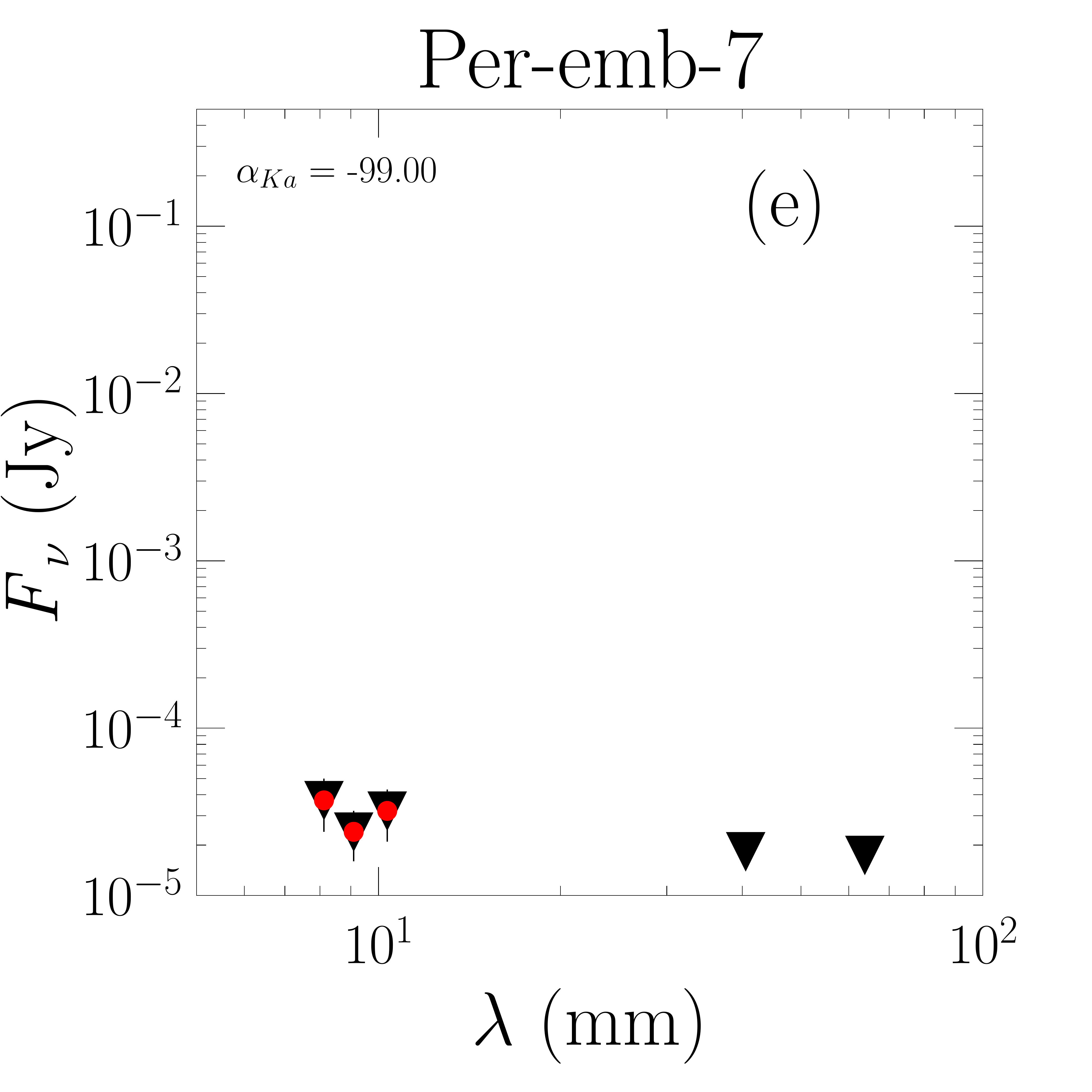}
  \includegraphics[width=0.30\linewidth]{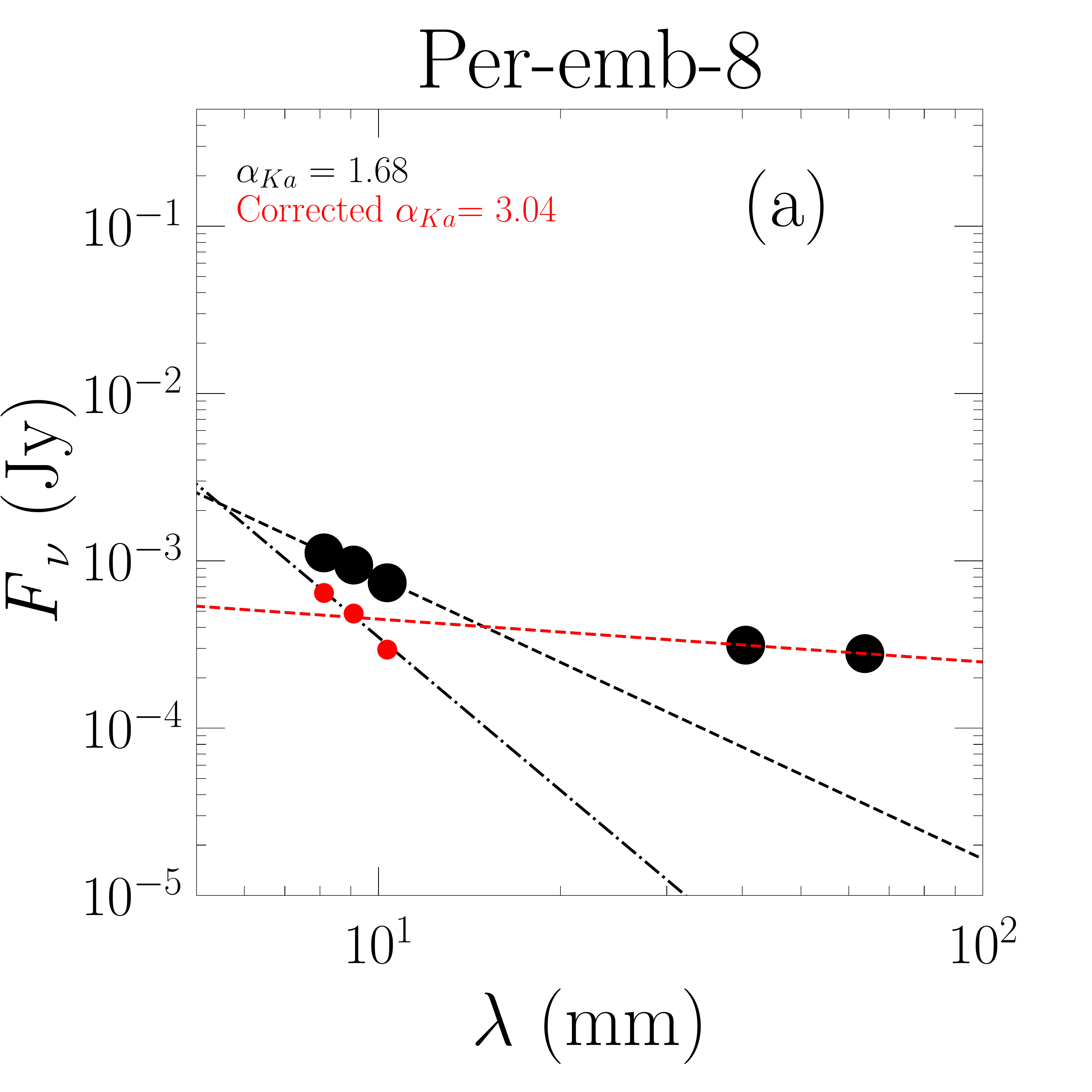}
  \includegraphics[width=0.30\linewidth]{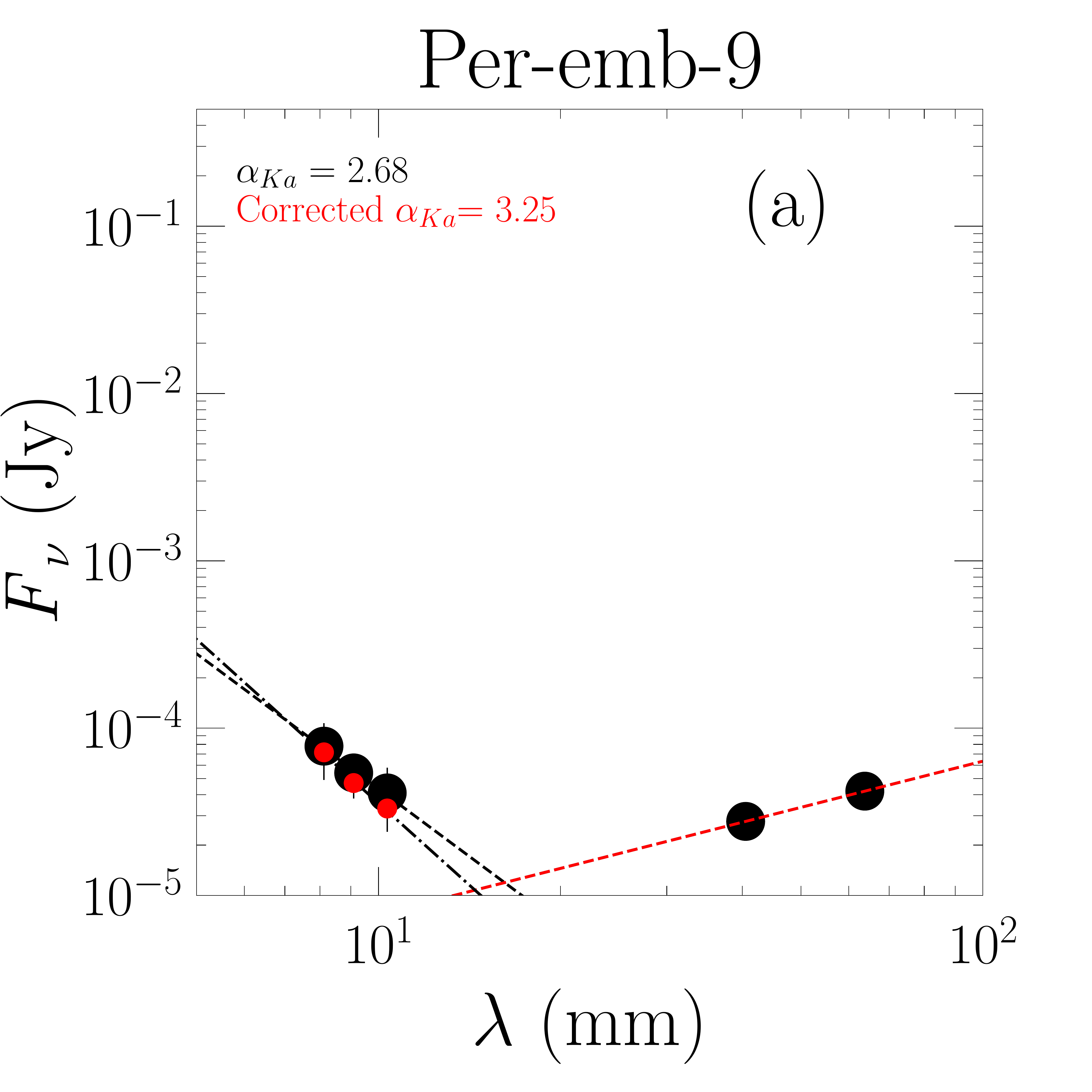}

\caption{Radio spectral energy distributions for all protostars in VANDAM survey. Black bullets represent Ka-band and C-band flux densities, and triangles are upper limits. 
Red bullets mark the corrected Ka-band flux densities. Dotted lines are linear fits to the original data, and the red line represents the function from which the free-free contribution was estimated. 
Dash-dot line marks fit to the corrected Ka-band flux densities. Labels a-e indicate a different case of correcting Ka-band data for free-free contamination.}
\label{fig:massall1}
\end{figure}

 Figure \ref{fig:massall1} shows the radio spectra for all of the sources from the VANDAM survey. Linear fit to the logarithmic fluxes, applied free-free correction to the dust flux, and the flag used when 
considering the protostellar disk mass are showed as described in Section \ref{diskmasses}.

\begin{figure}[H]
\centering
  \includegraphics[width=0.30\linewidth]{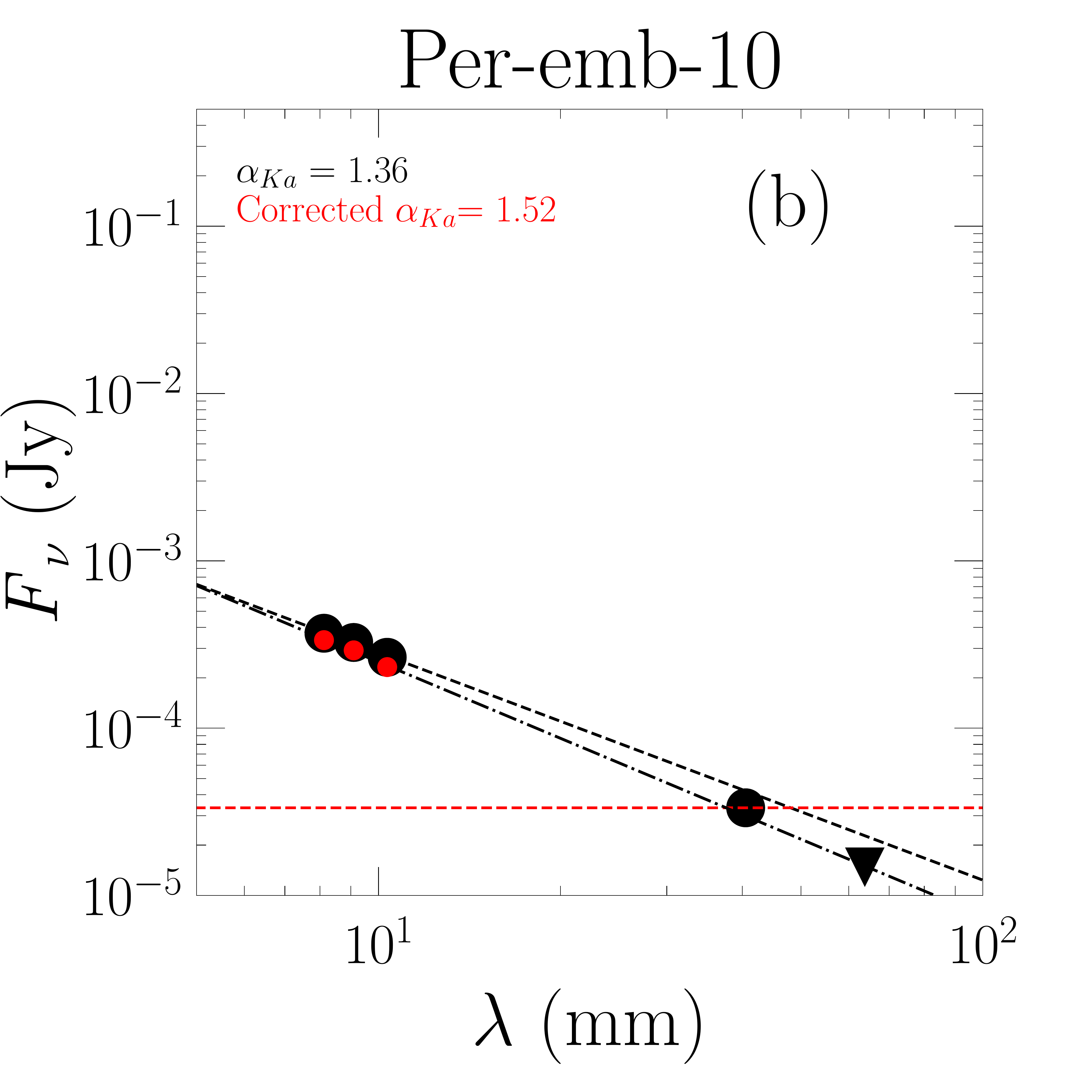}
  \includegraphics[width=0.30\linewidth]{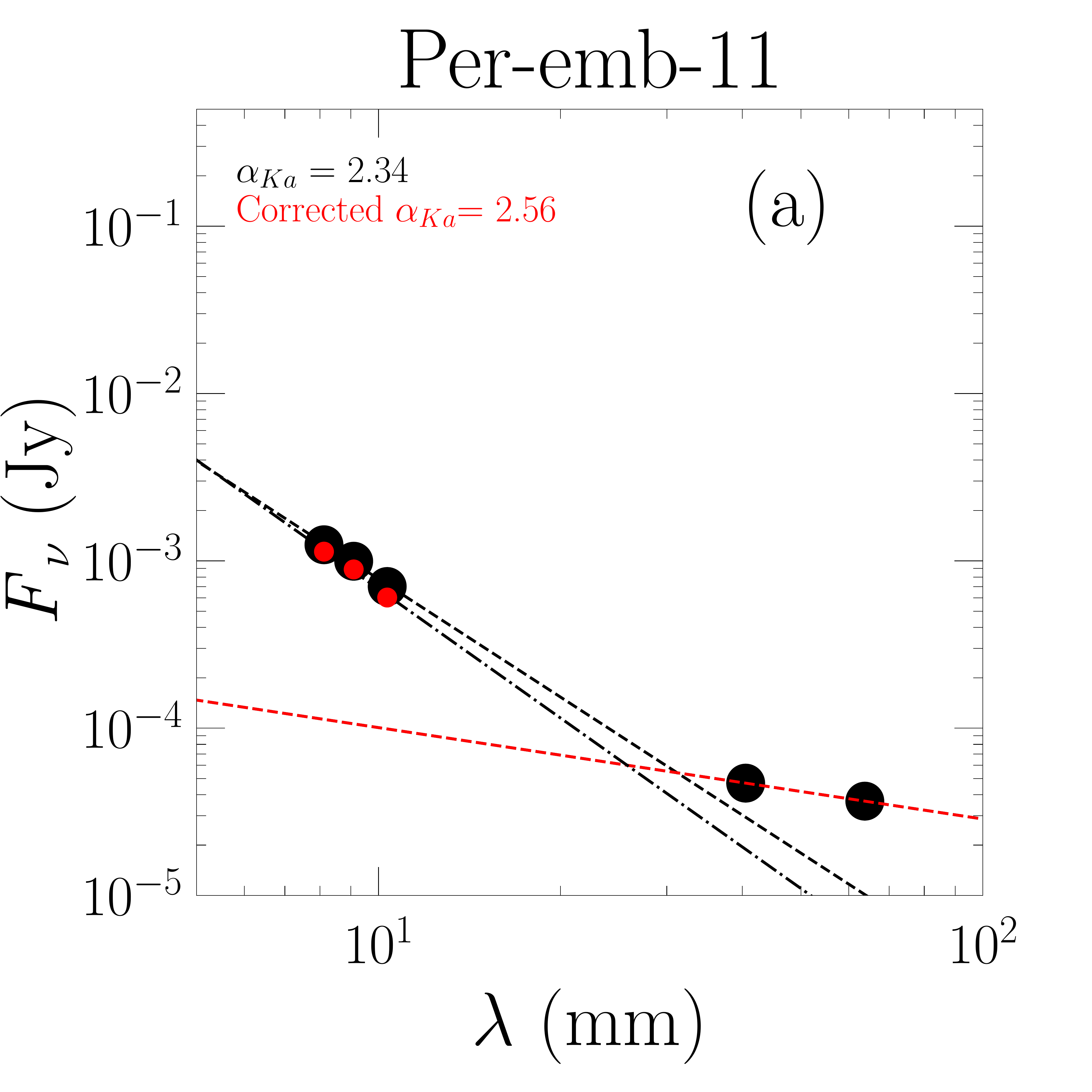}
  \includegraphics[width=0.30\linewidth]{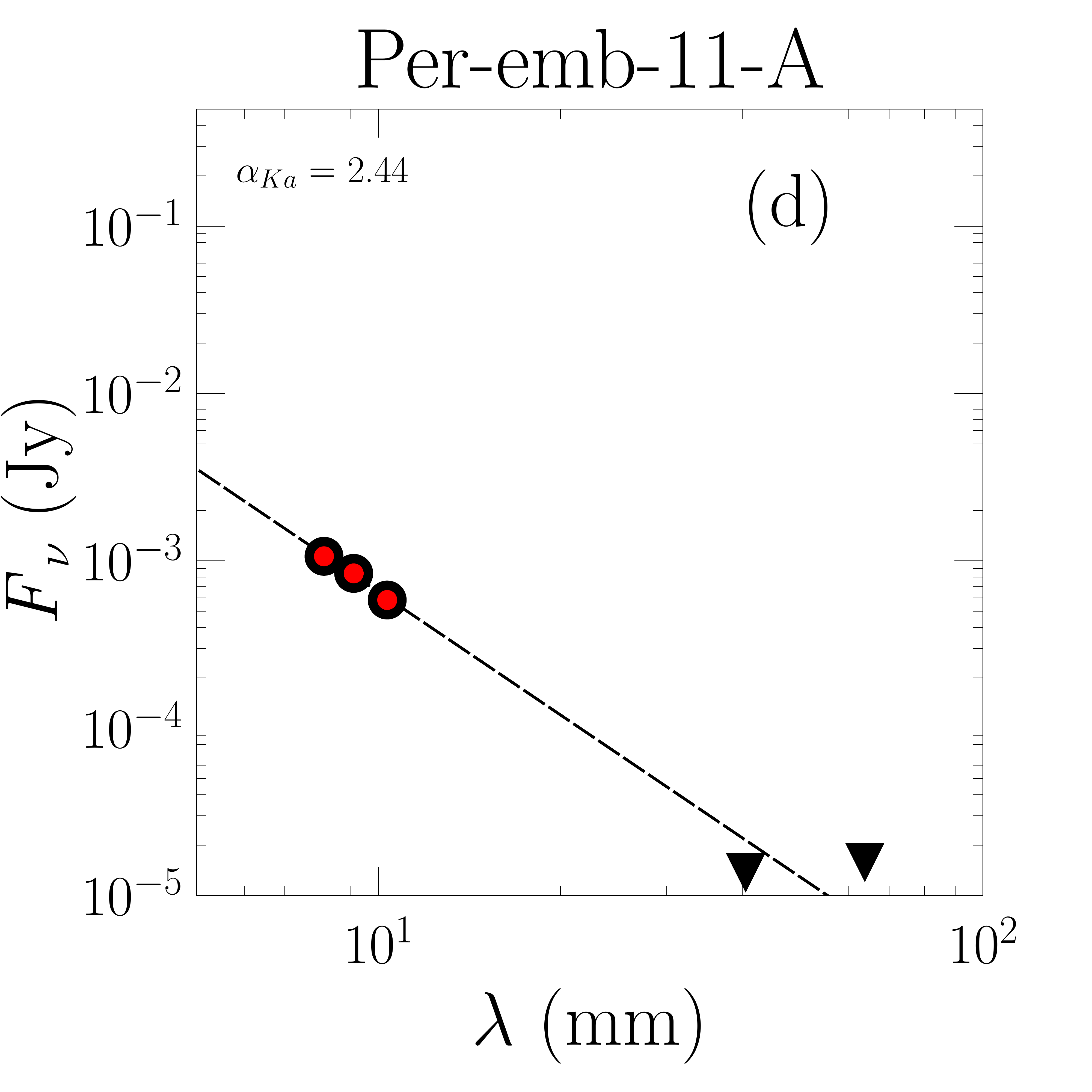}
  \includegraphics[width=0.30\linewidth]{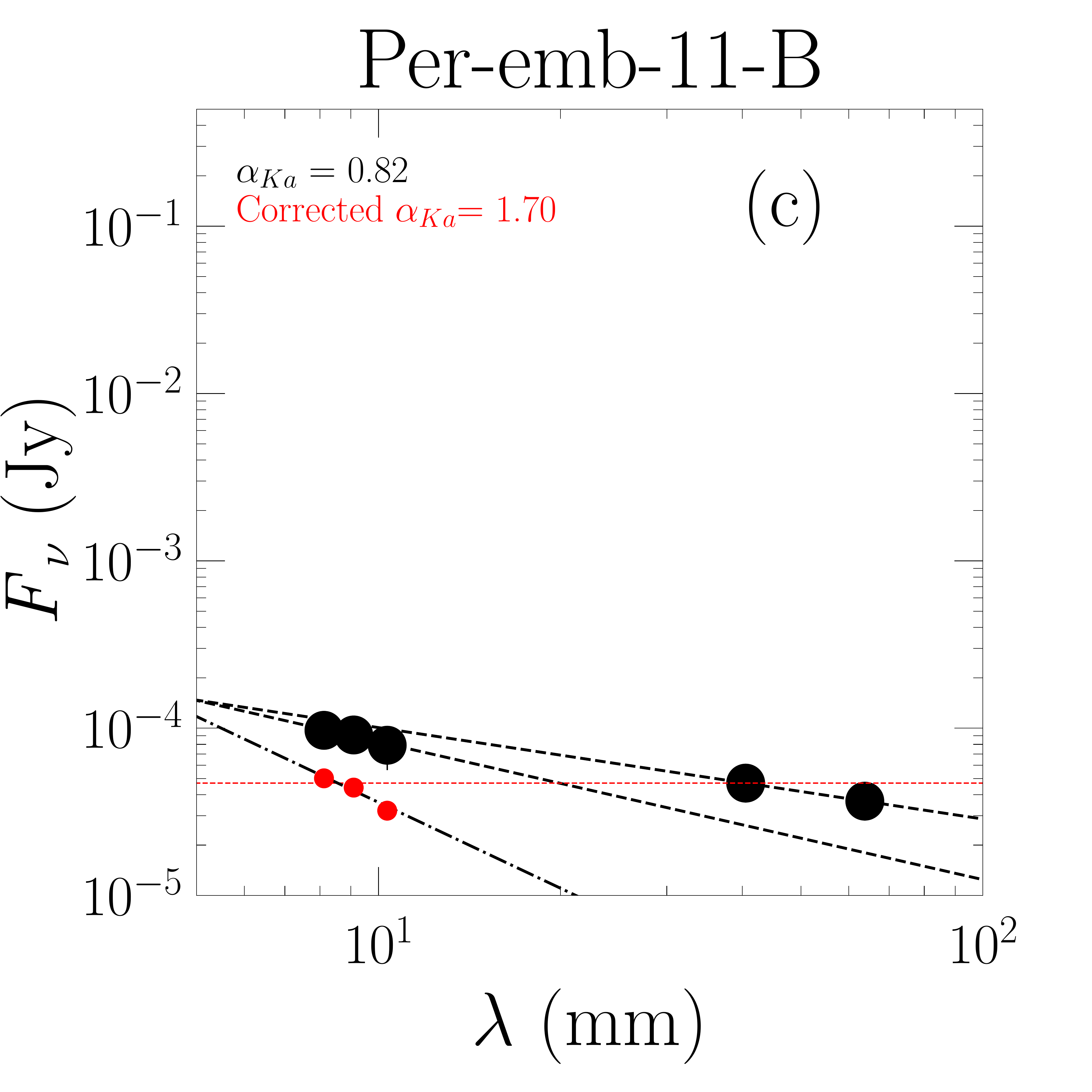}
  \includegraphics[width=0.30\linewidth]{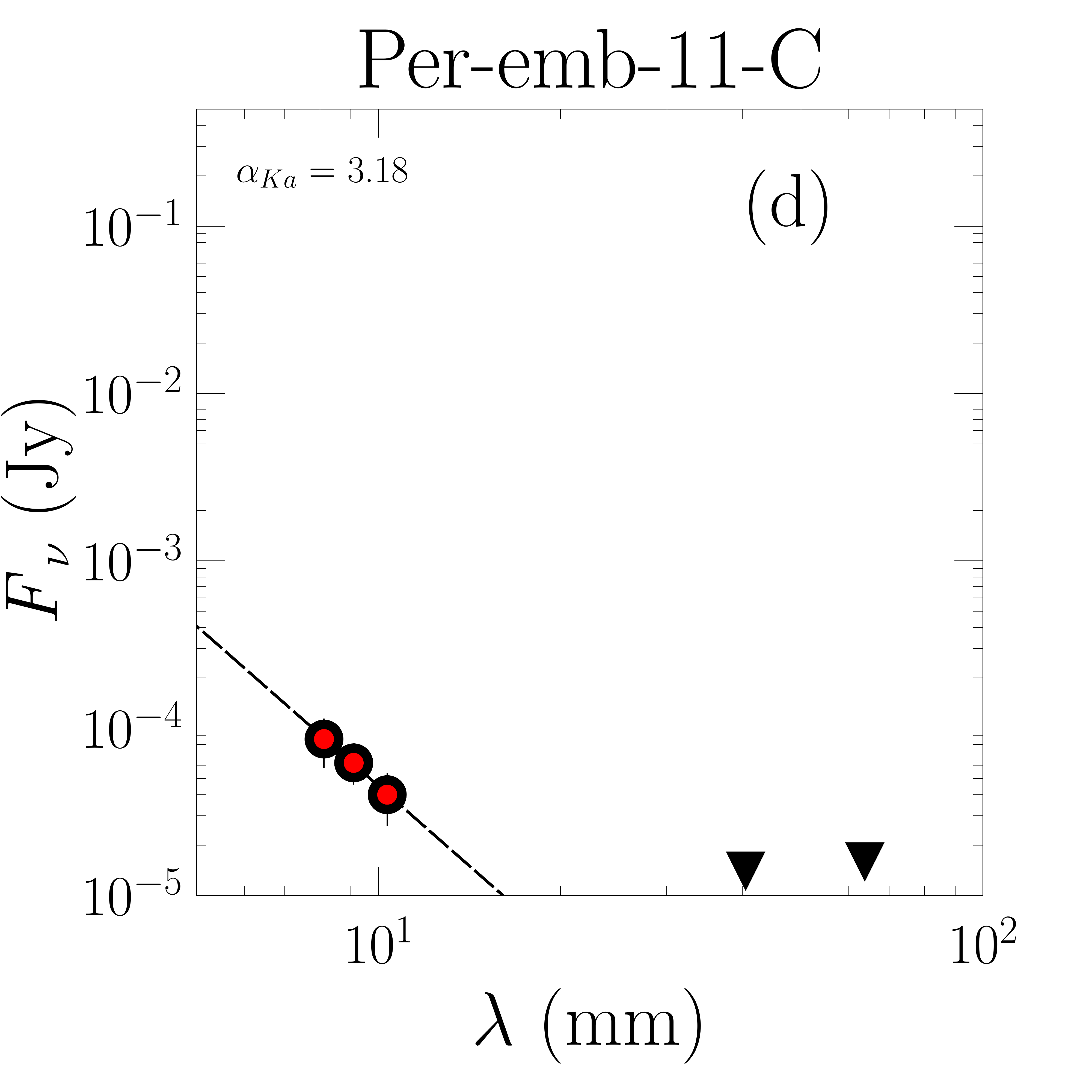}
  \includegraphics[width=0.30\linewidth]{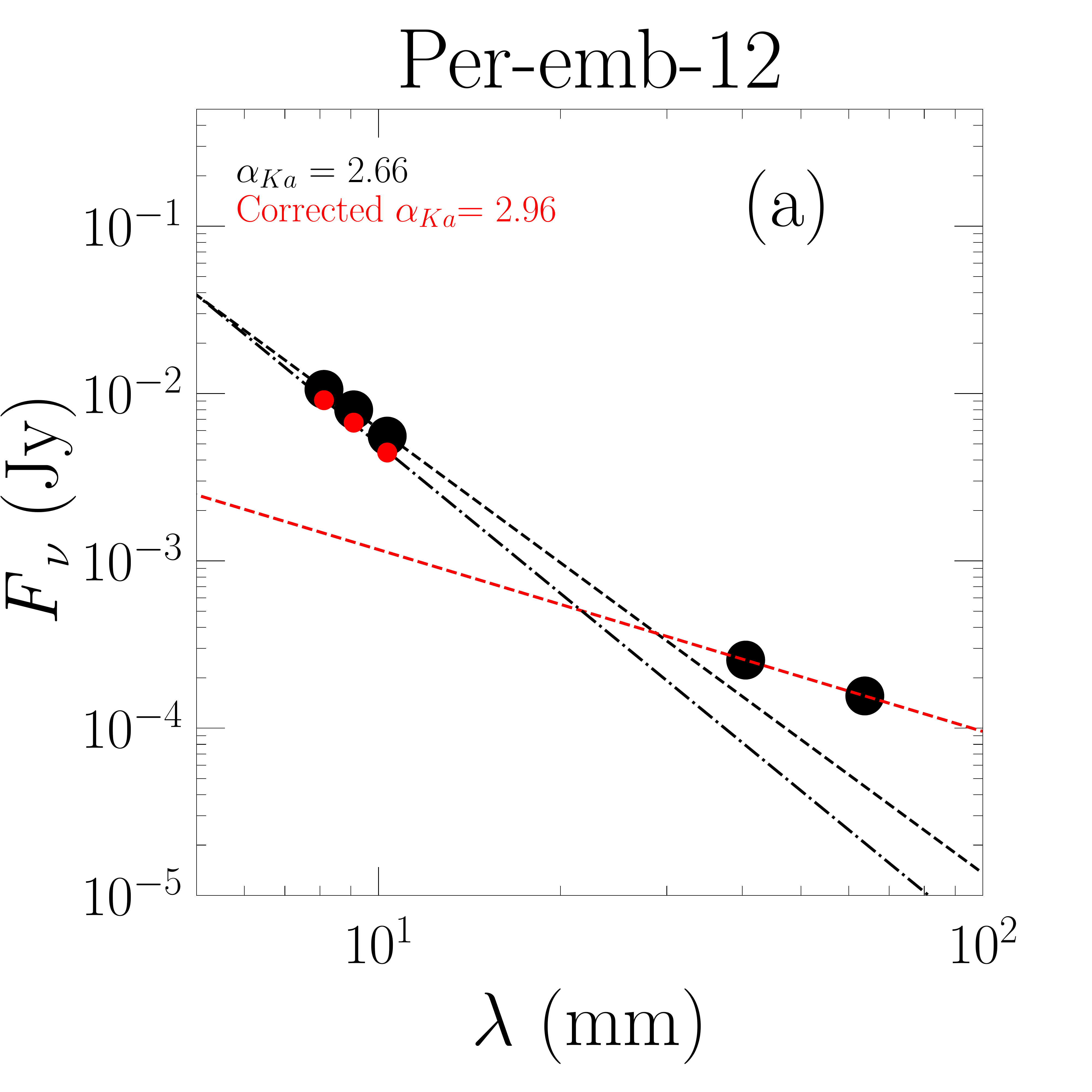}
  \includegraphics[width=0.30\linewidth]{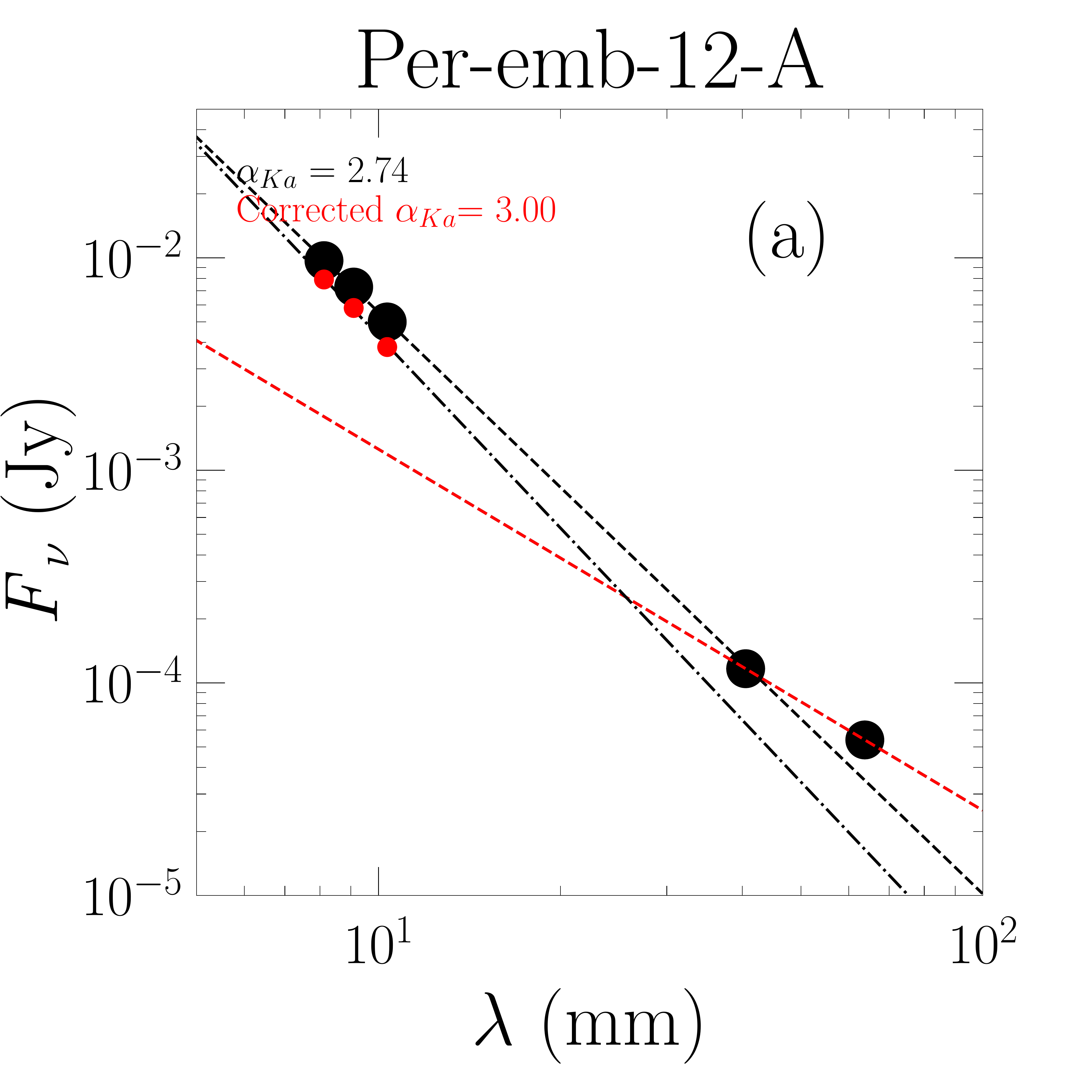}
  \includegraphics[width=0.30\linewidth]{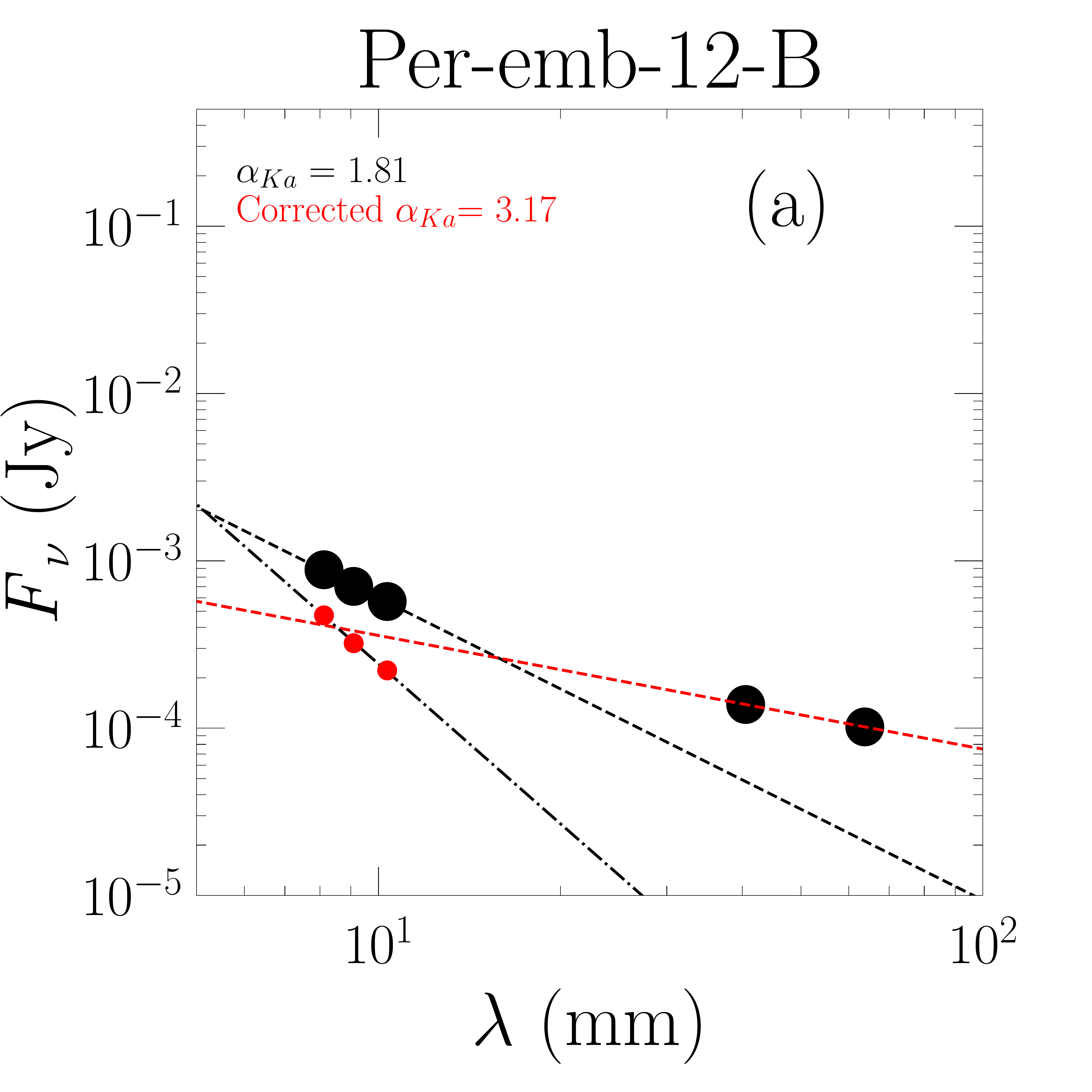}
  \includegraphics[width=0.30\linewidth]{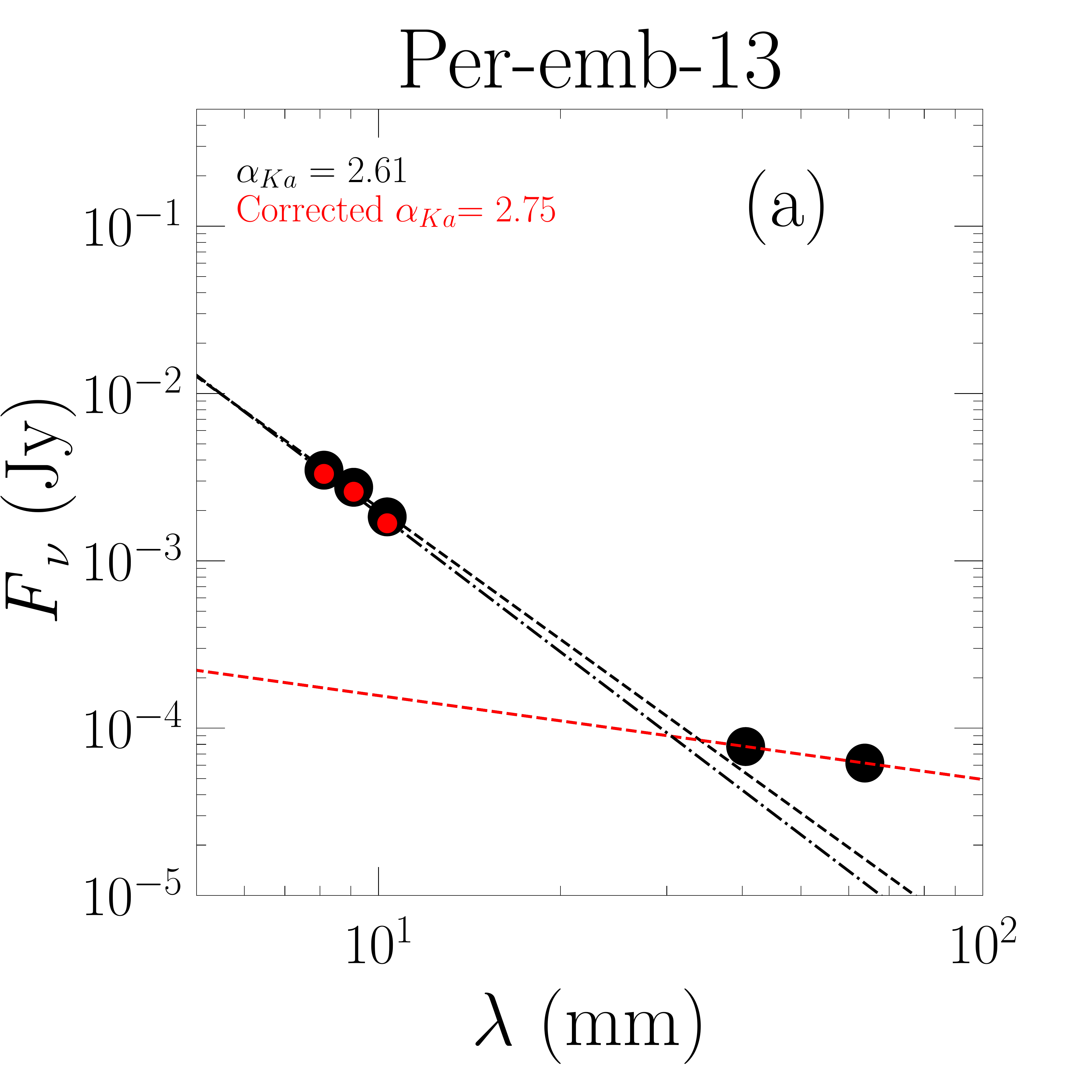}
  \includegraphics[width=0.30\linewidth]{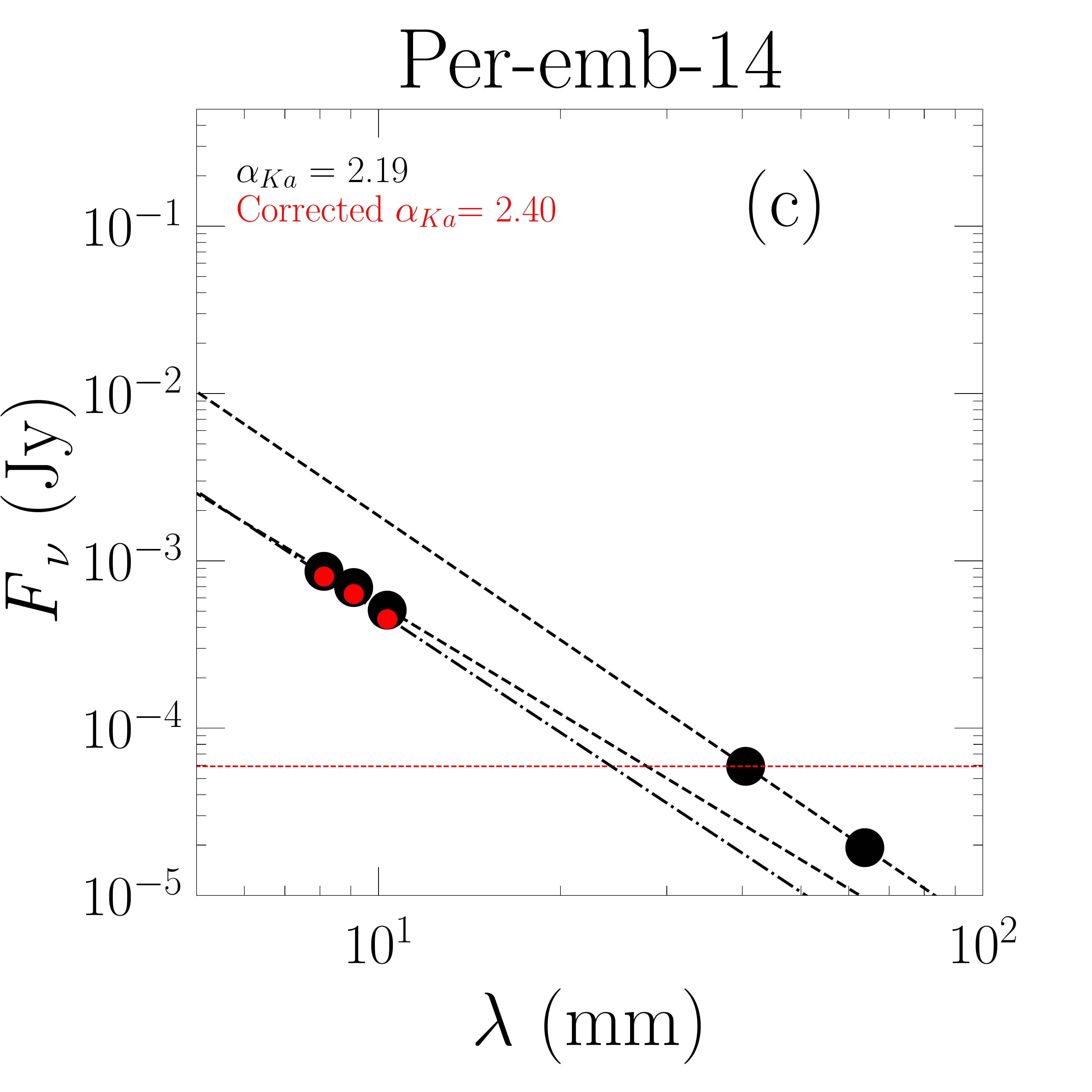}
  \includegraphics[width=0.30\linewidth]{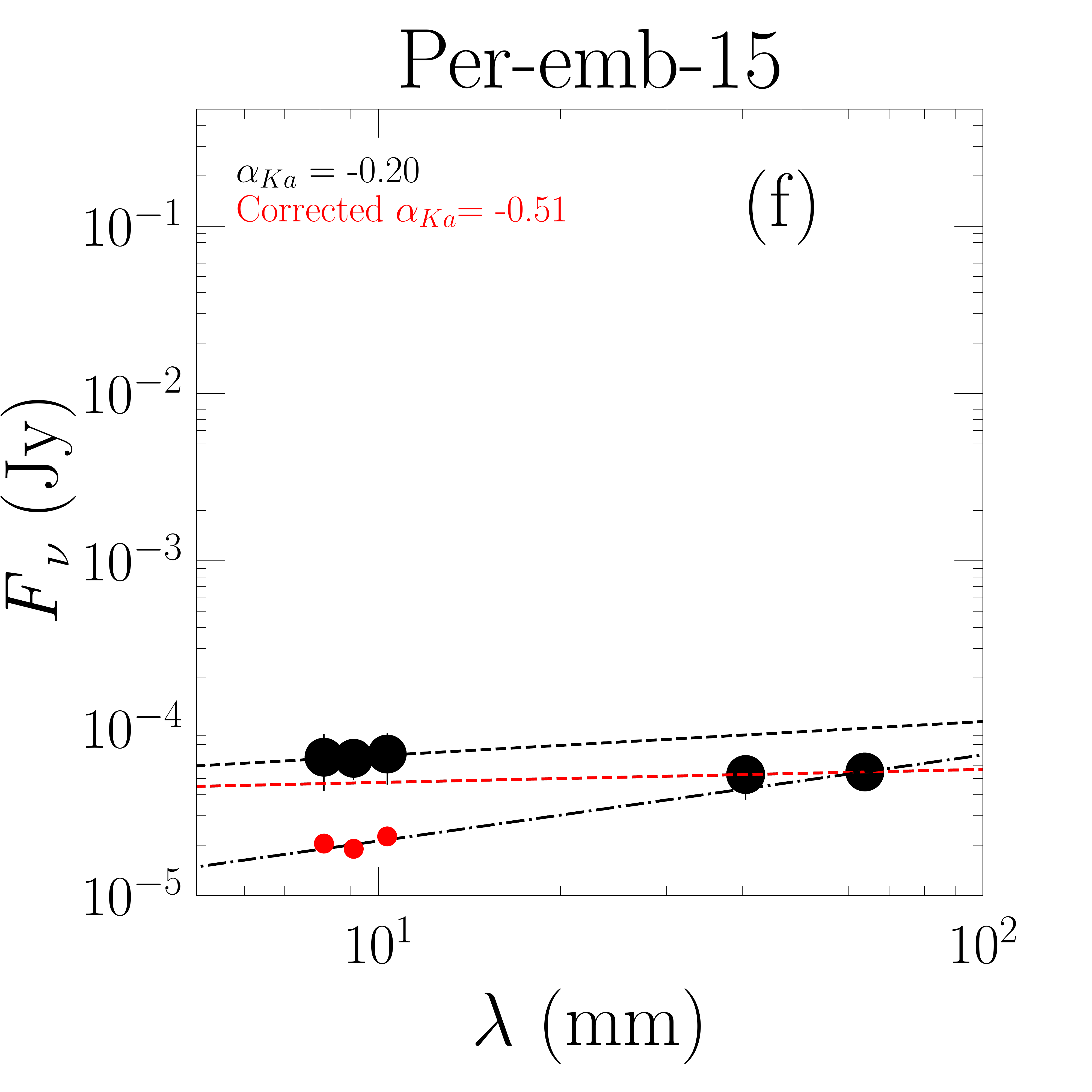}
  \includegraphics[width=0.30\linewidth]{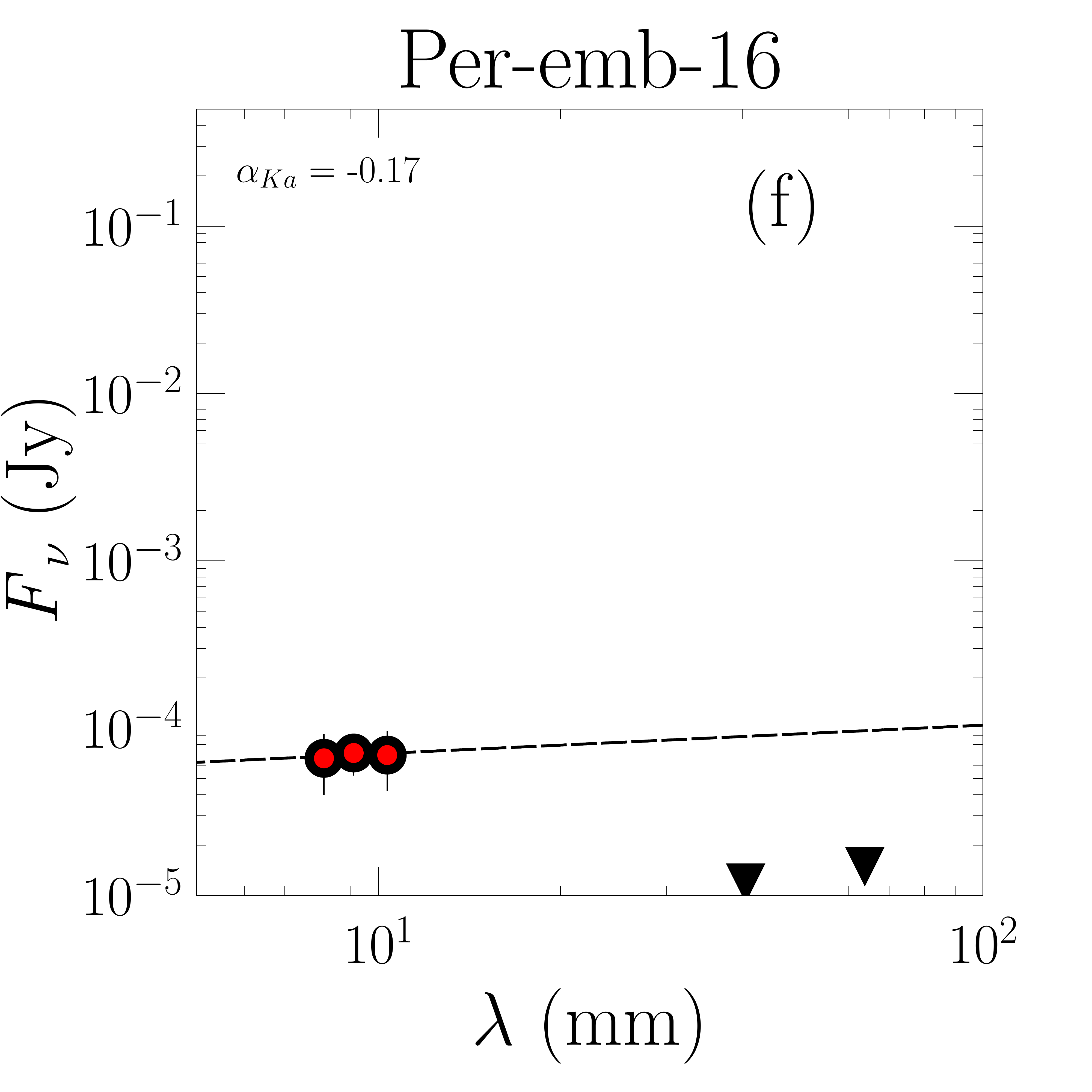}

\end{figure}
\begin{figure}[H]
\centering
  \includegraphics[width=0.30\linewidth]{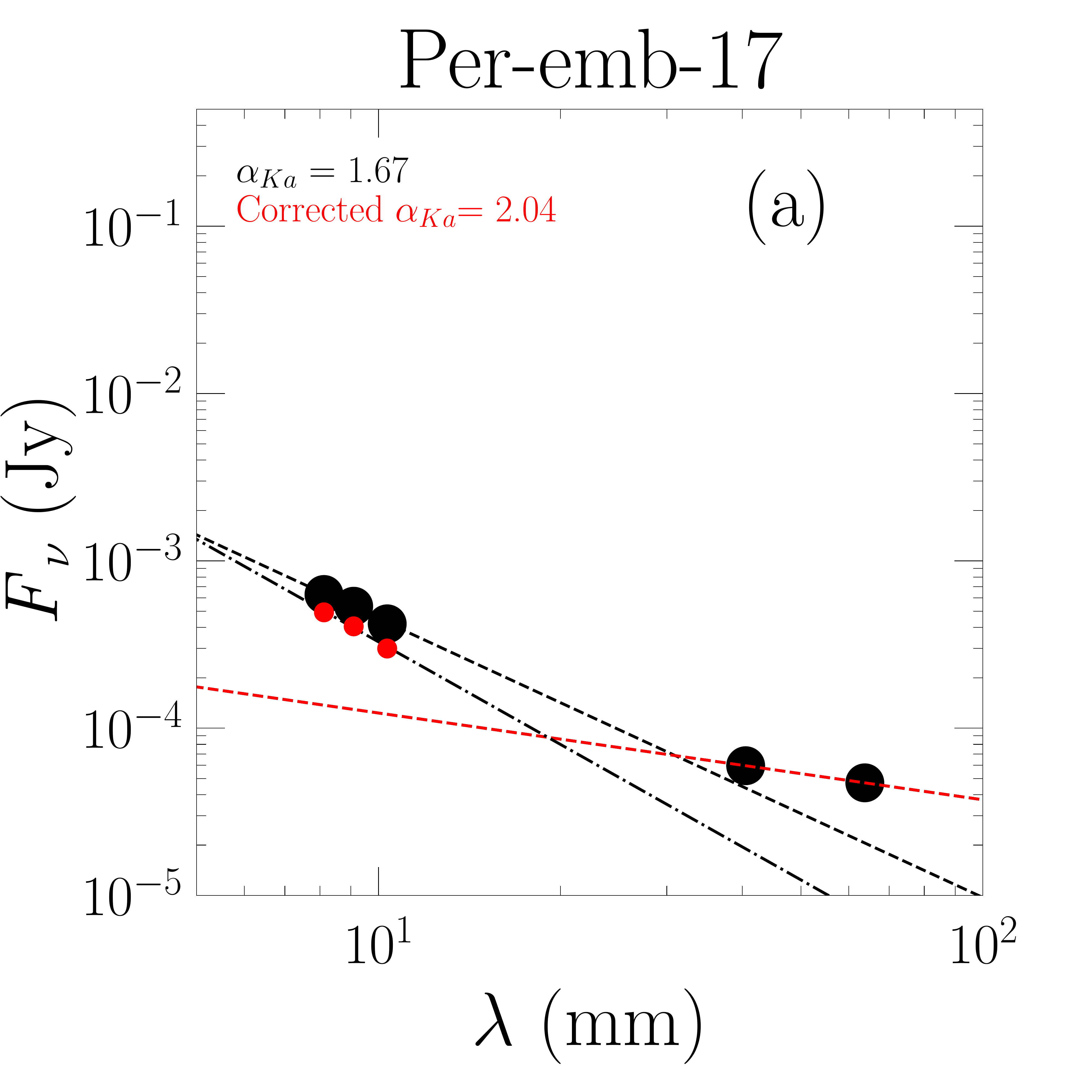}
  \includegraphics[width=0.30\linewidth]{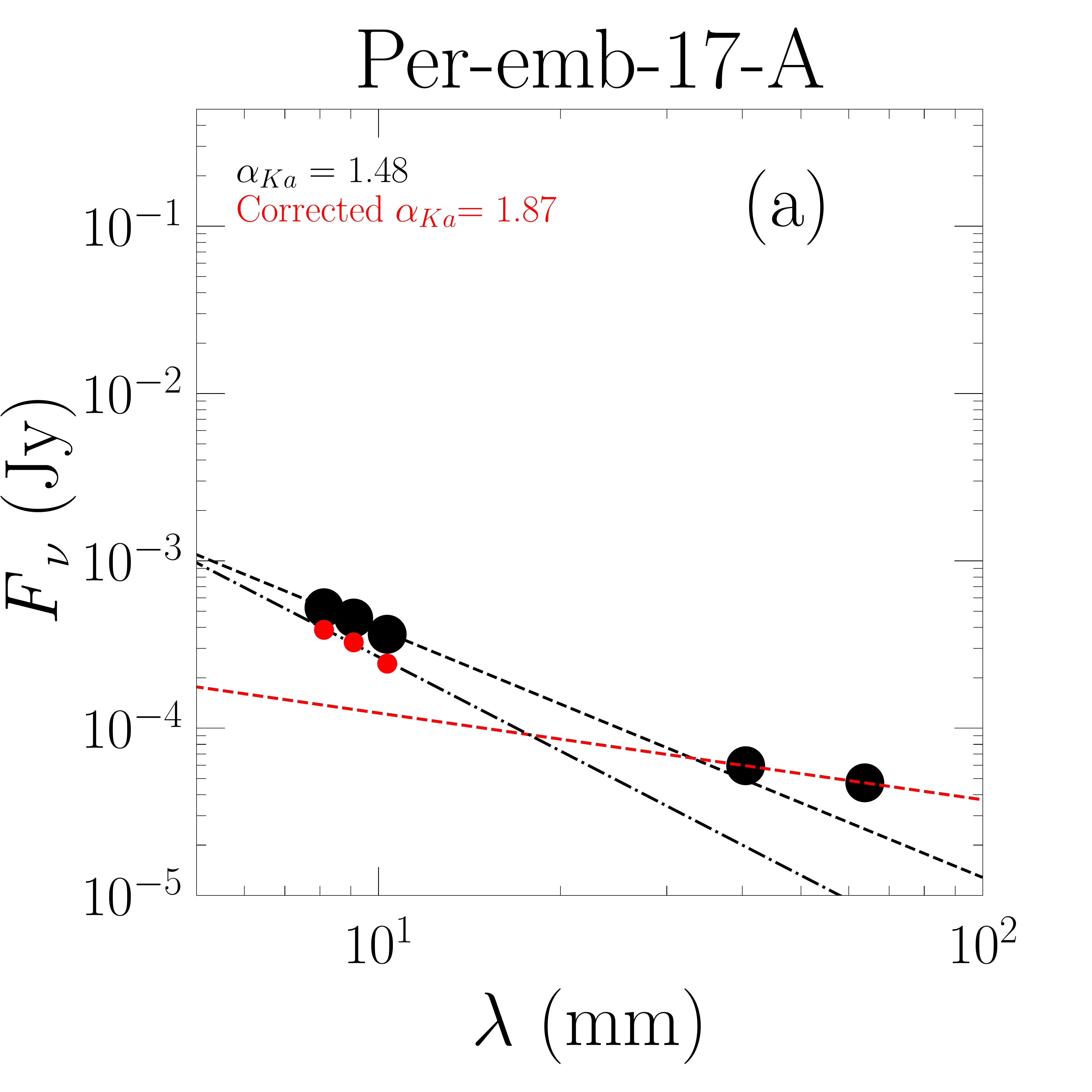}
  \includegraphics[width=0.30\linewidth]{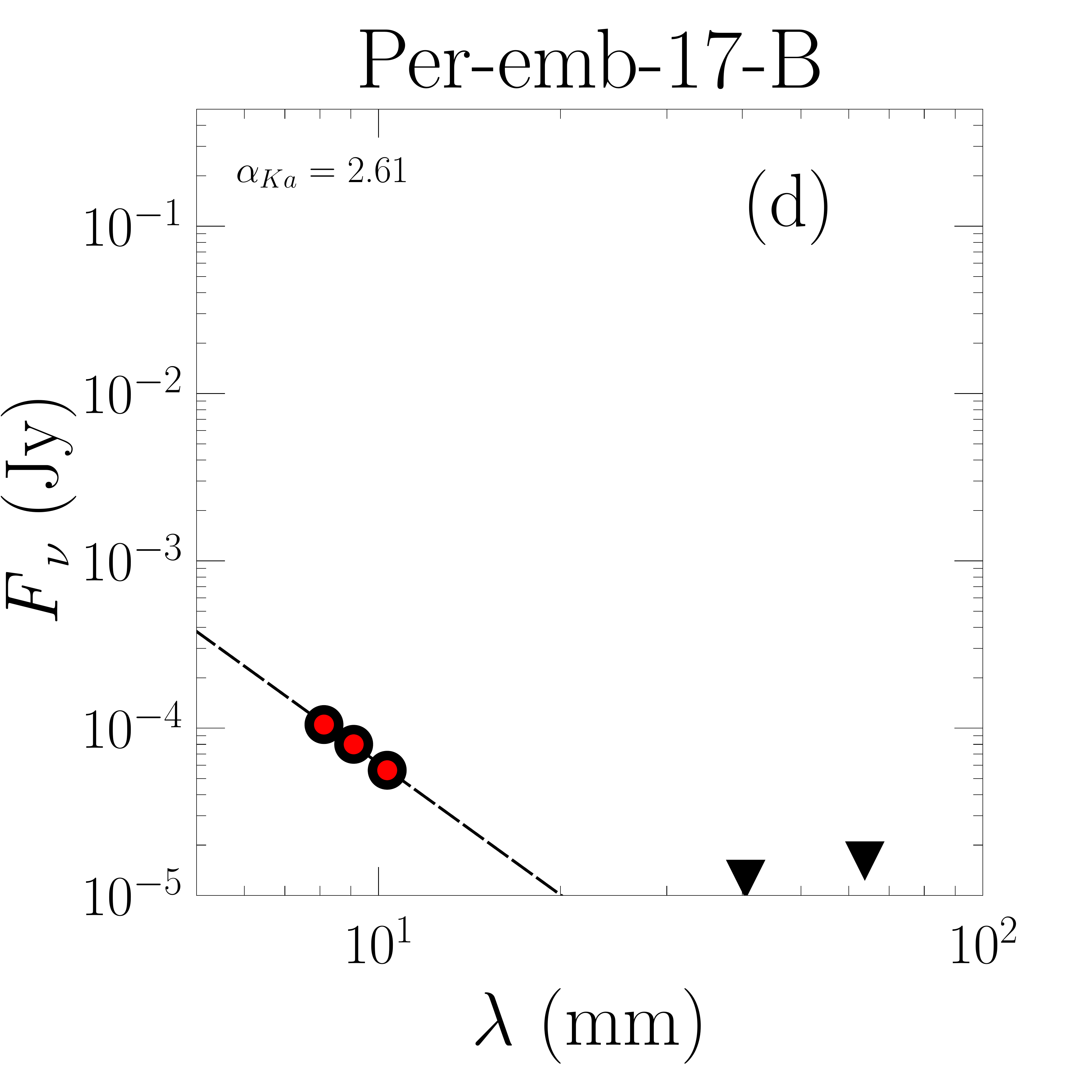}
  \includegraphics[width=0.30\linewidth]{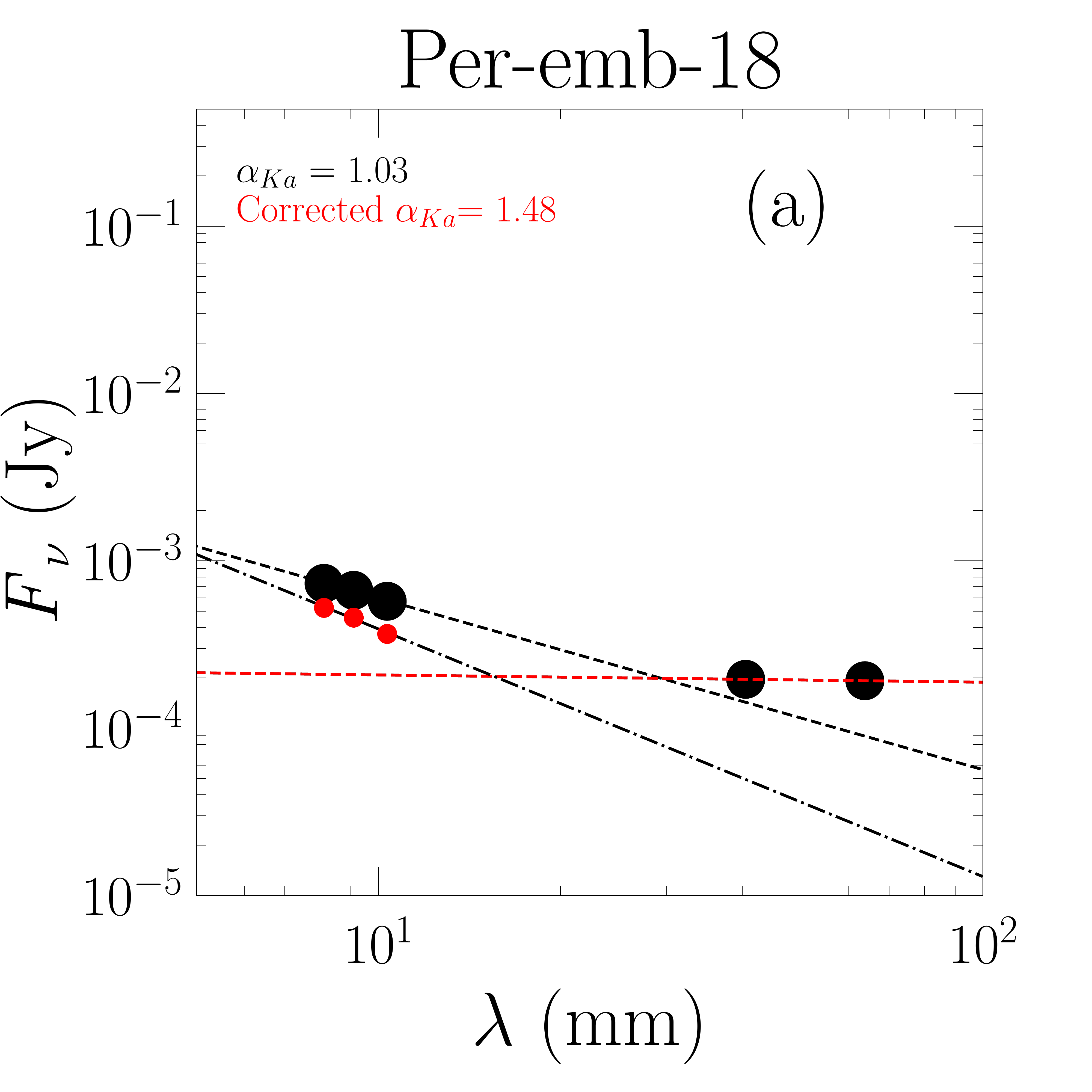}
  \includegraphics[width=0.30\linewidth]{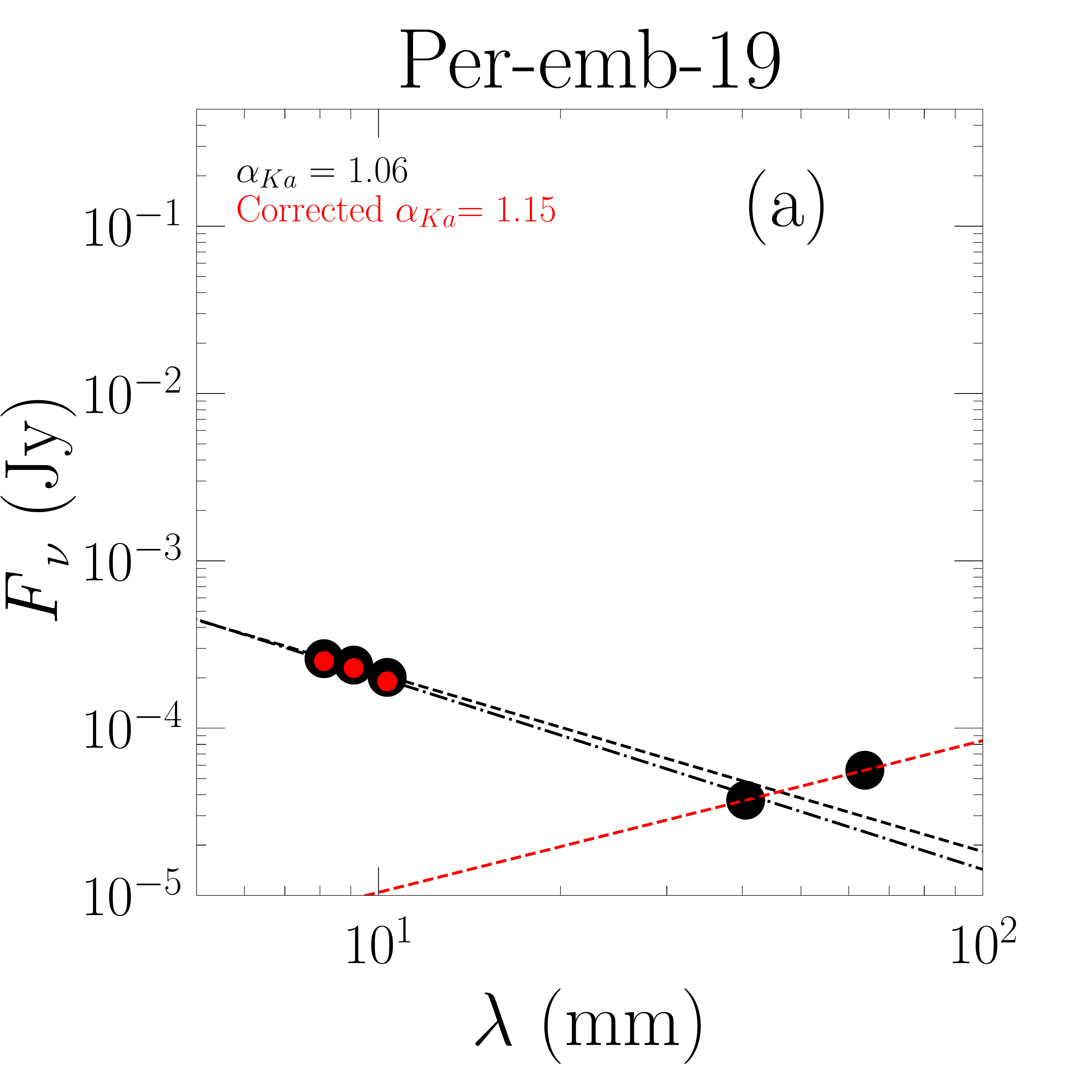}
  \includegraphics[width=0.30\linewidth]{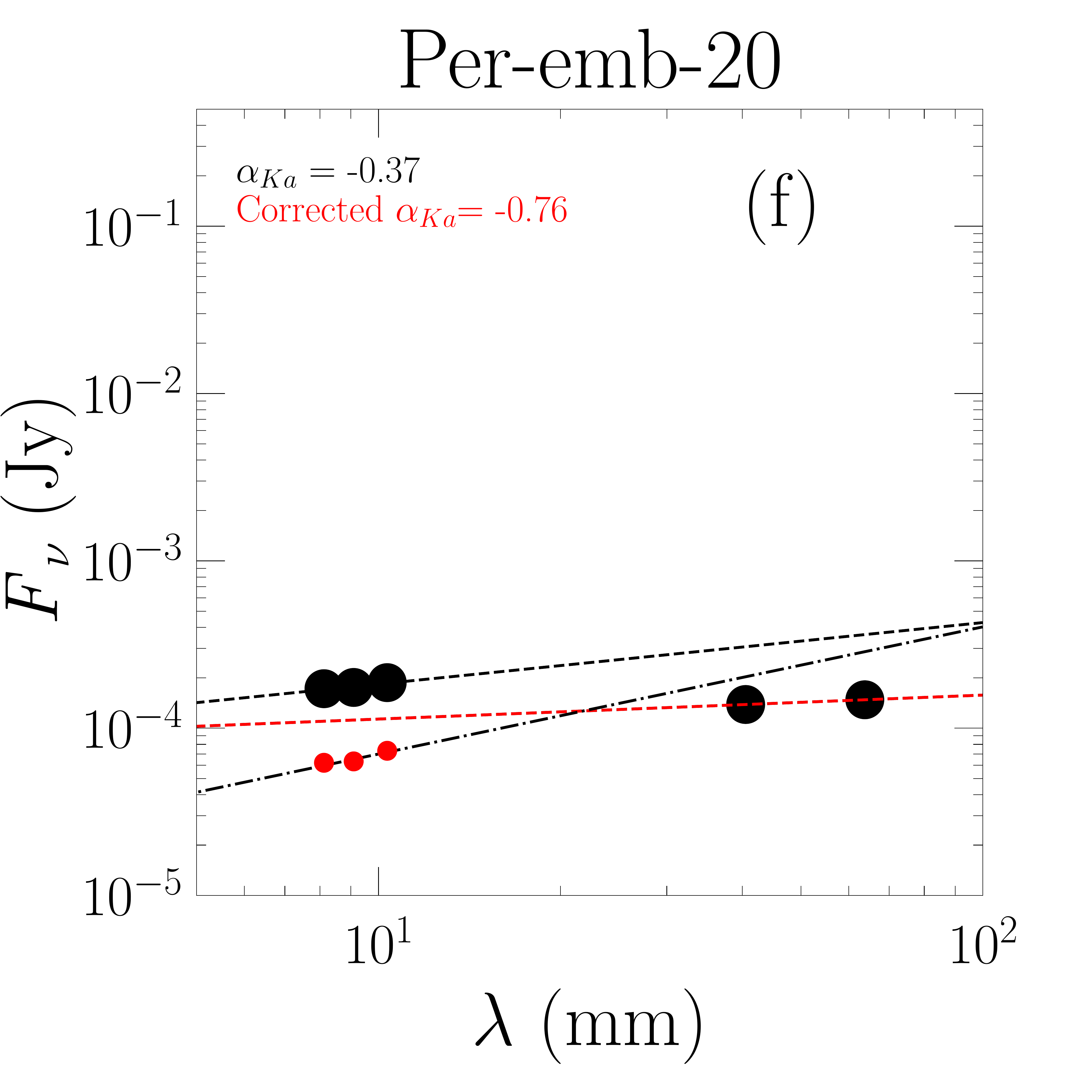}
  \includegraphics[width=0.30\linewidth]{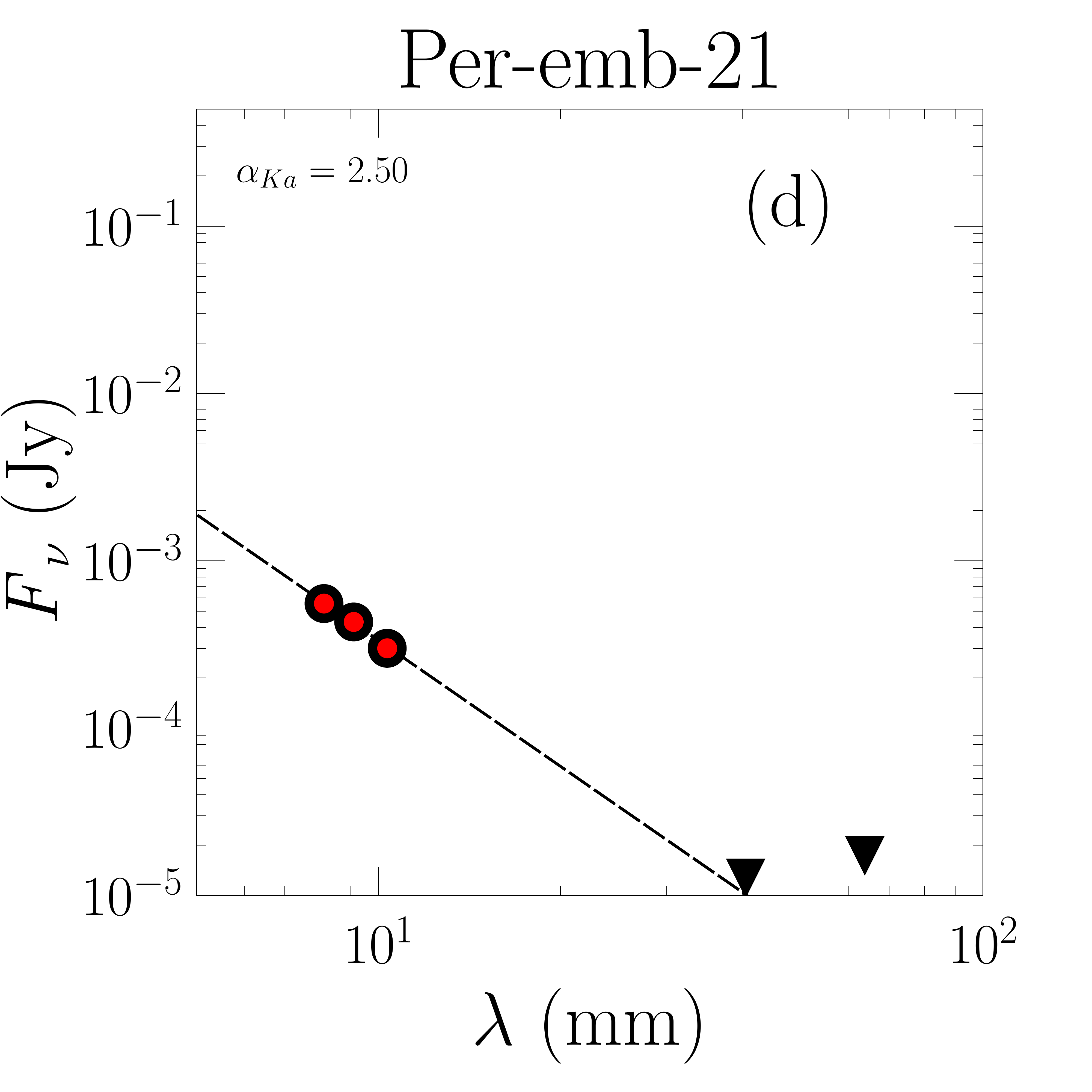}
  \includegraphics[width=0.30\linewidth]{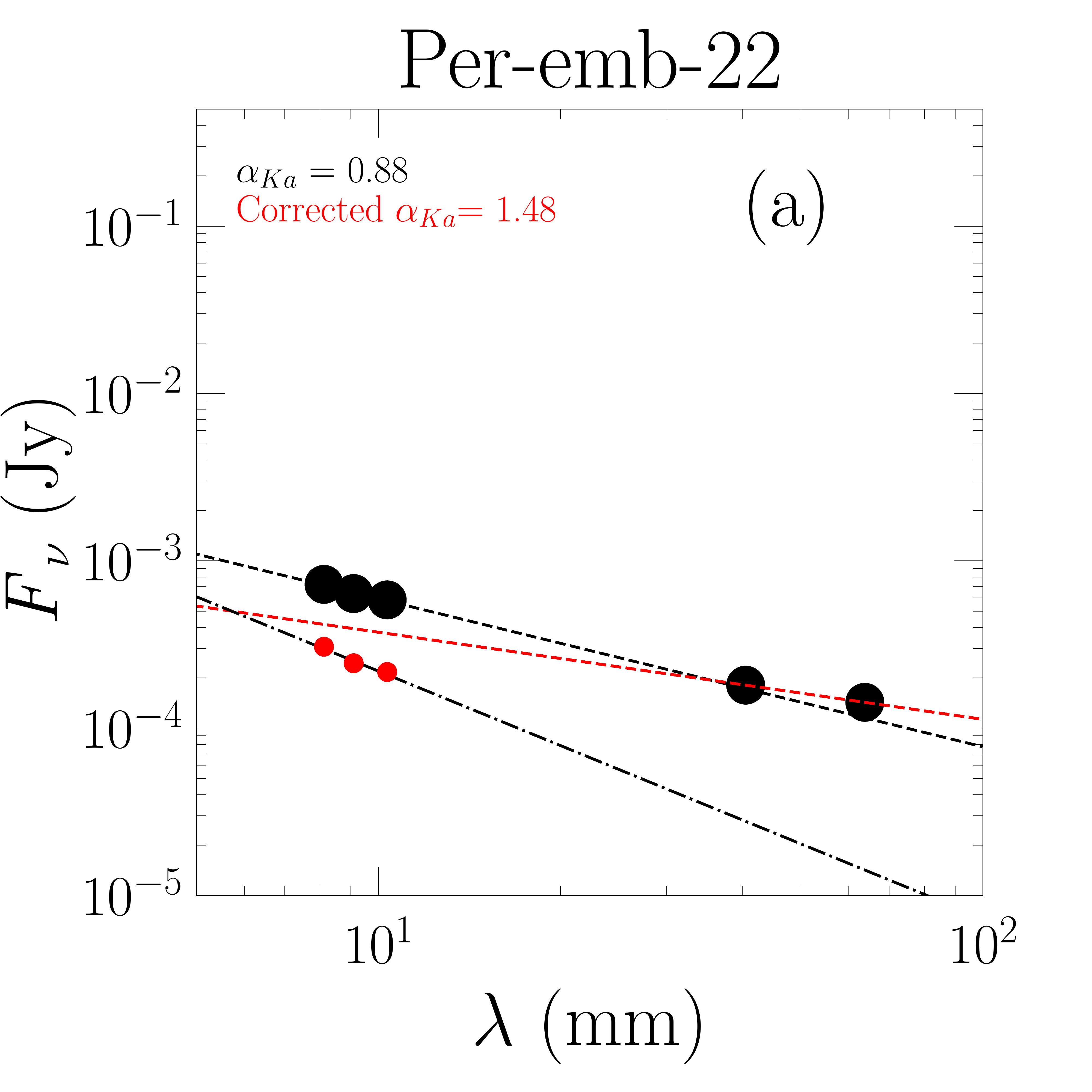}
  \includegraphics[width=0.30\linewidth]{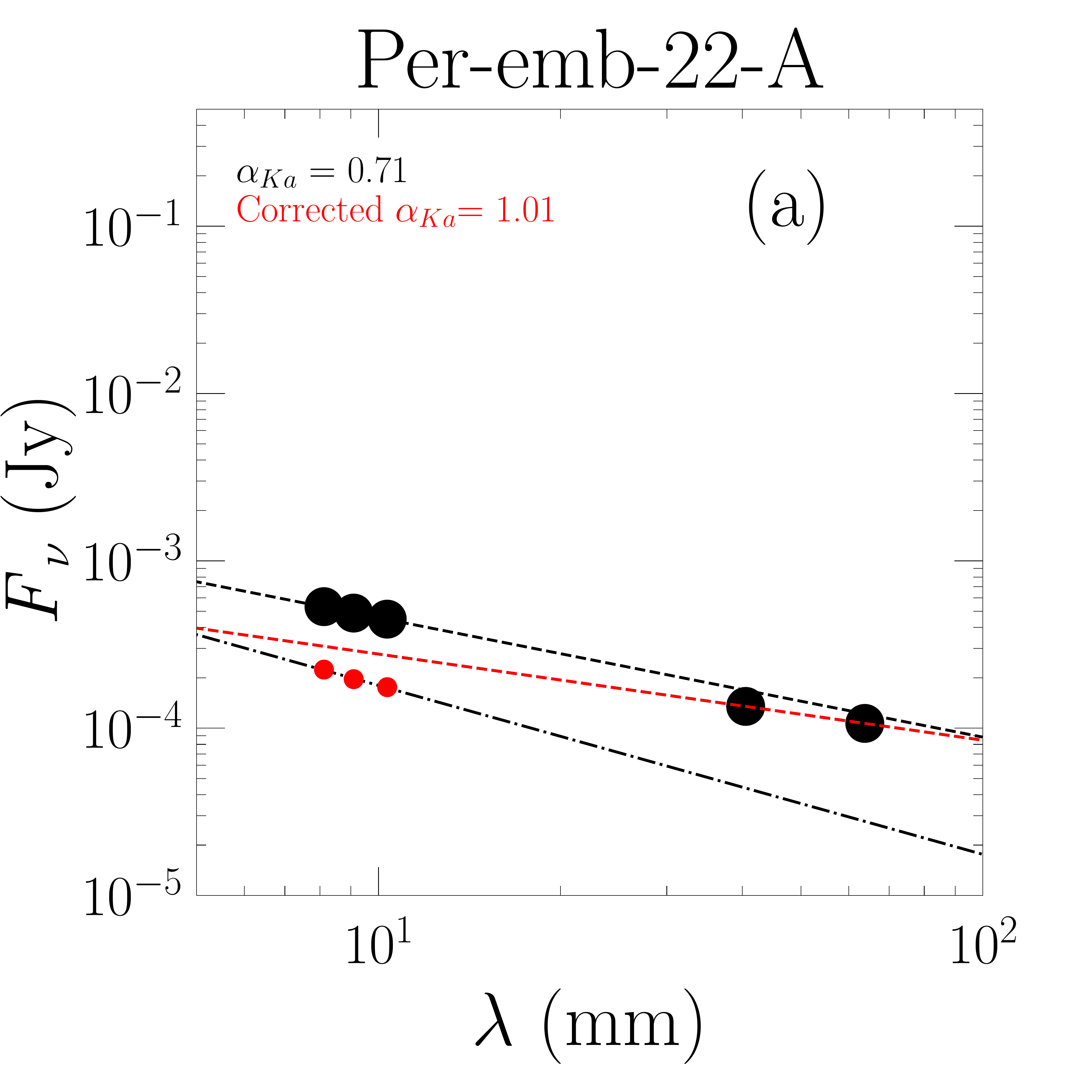}
  \includegraphics[width=0.30\linewidth]{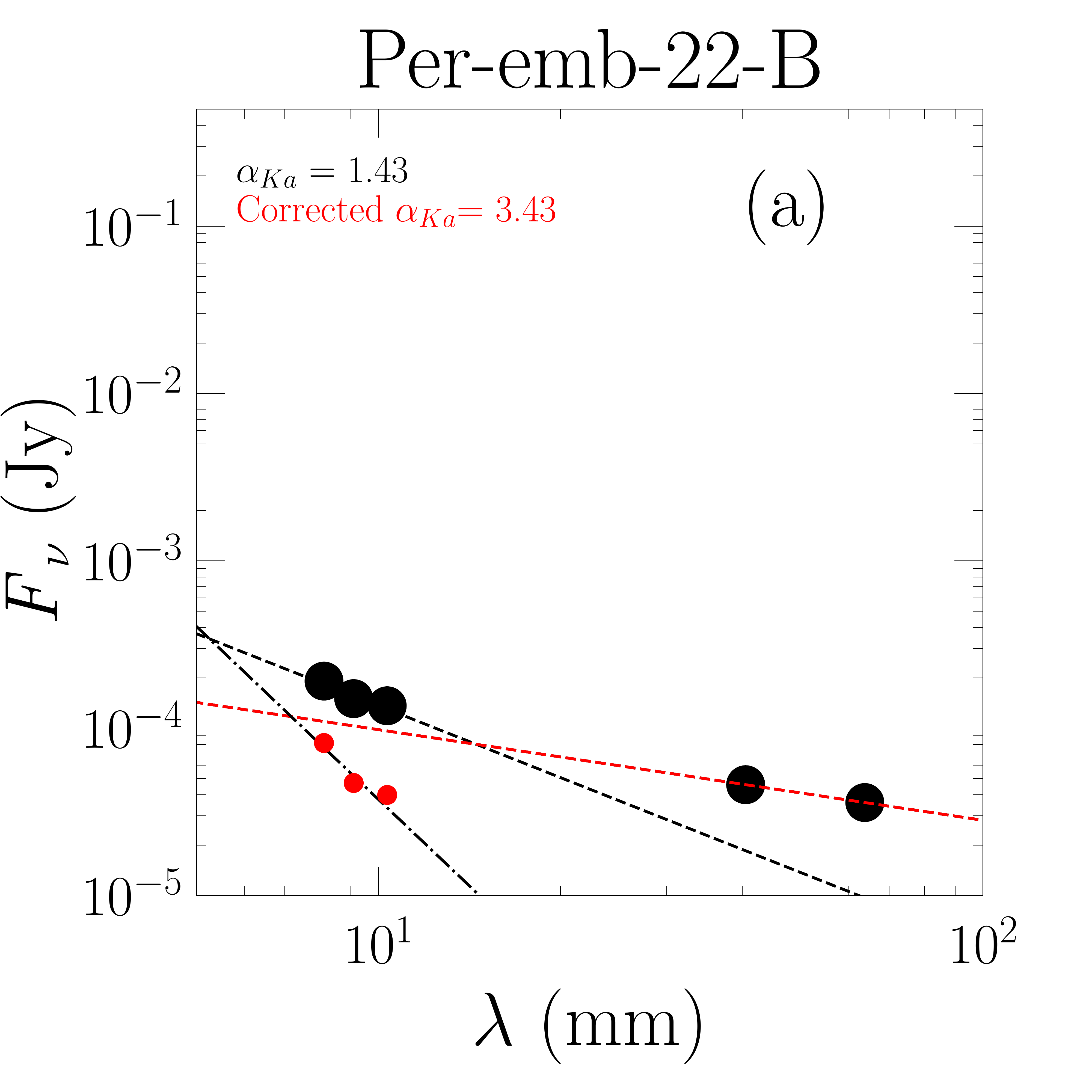}
  \includegraphics[width=0.30\linewidth]{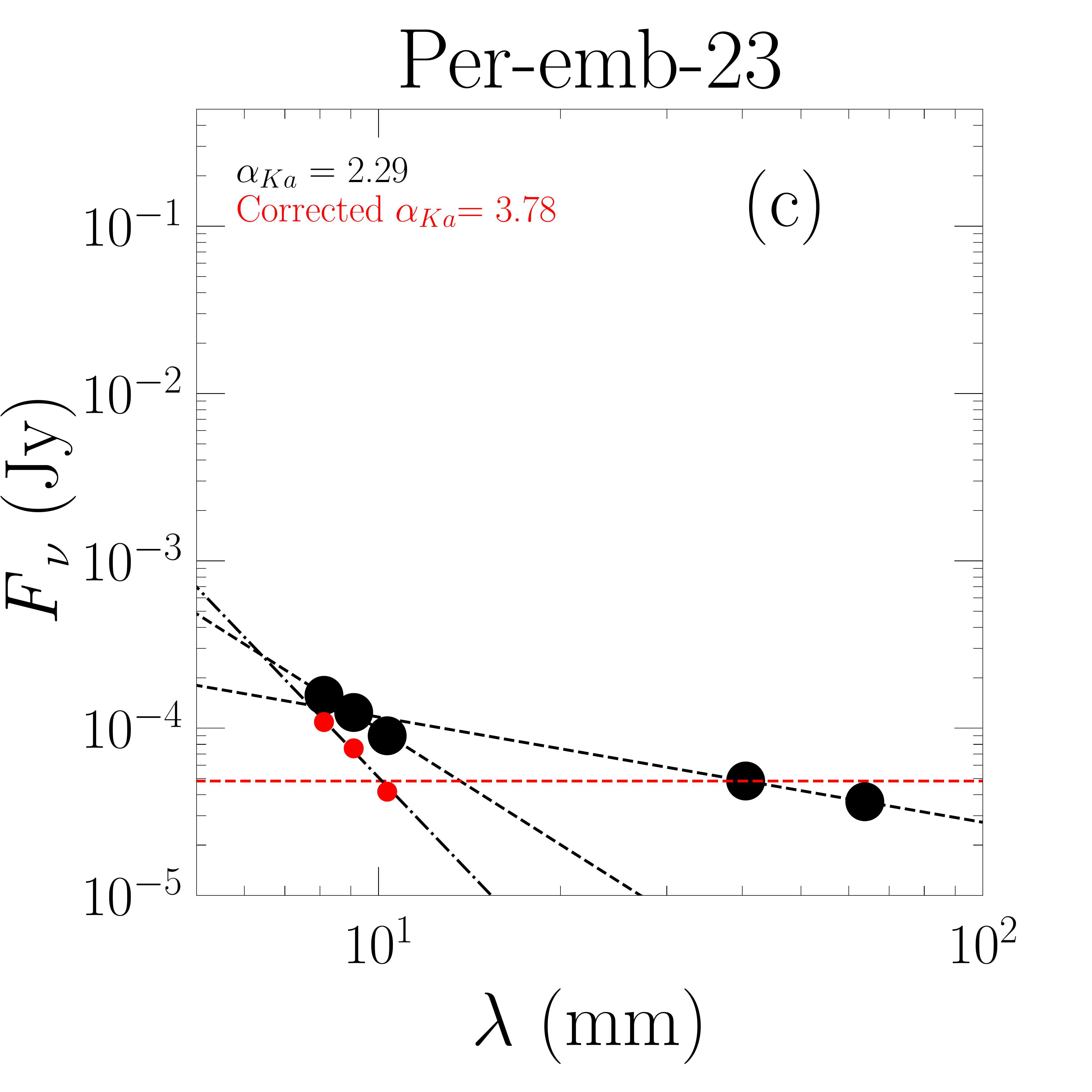}
  \includegraphics[width=0.30\linewidth]{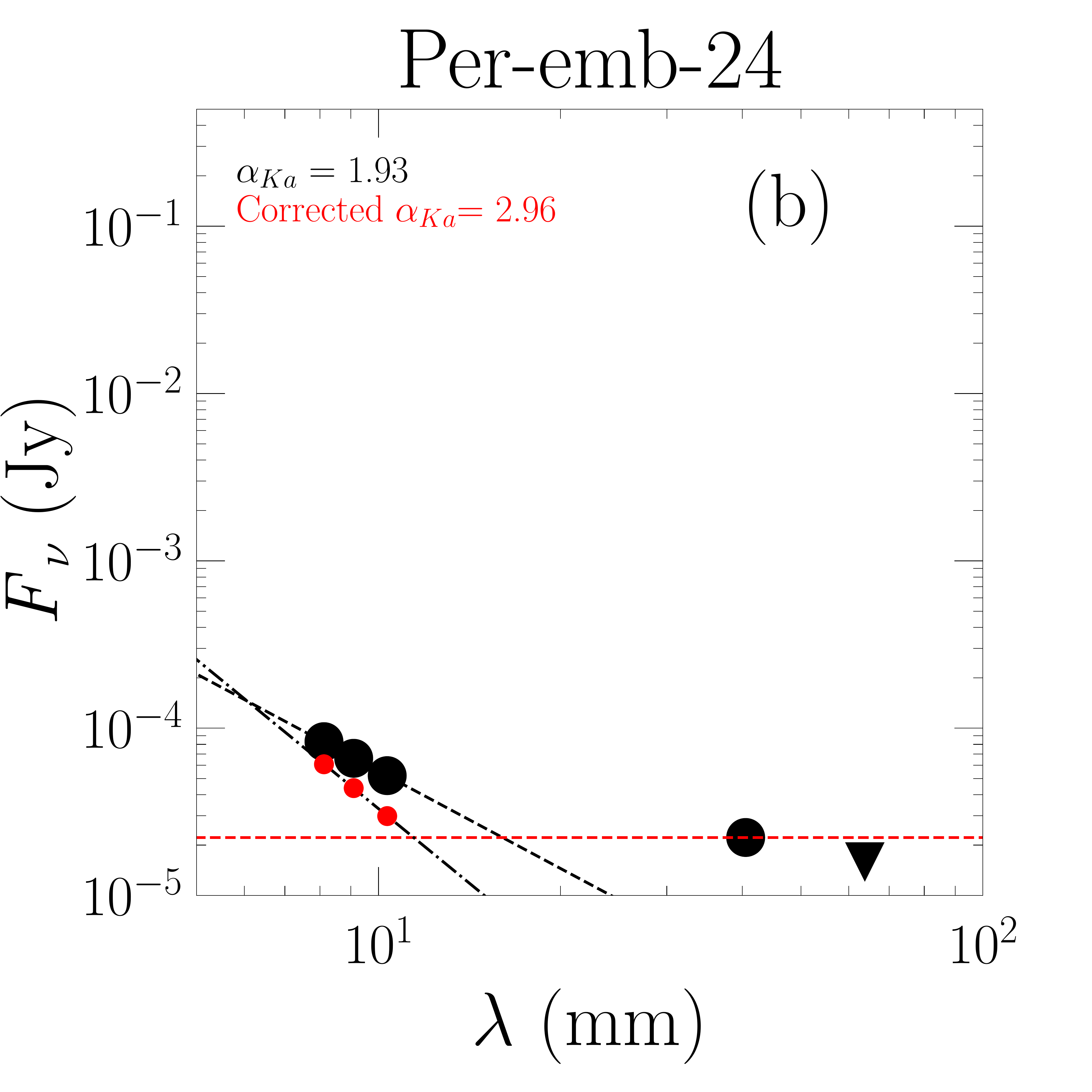}

\end{figure}
\begin{figure}[H]
\centering
  \includegraphics[width=0.30\linewidth]{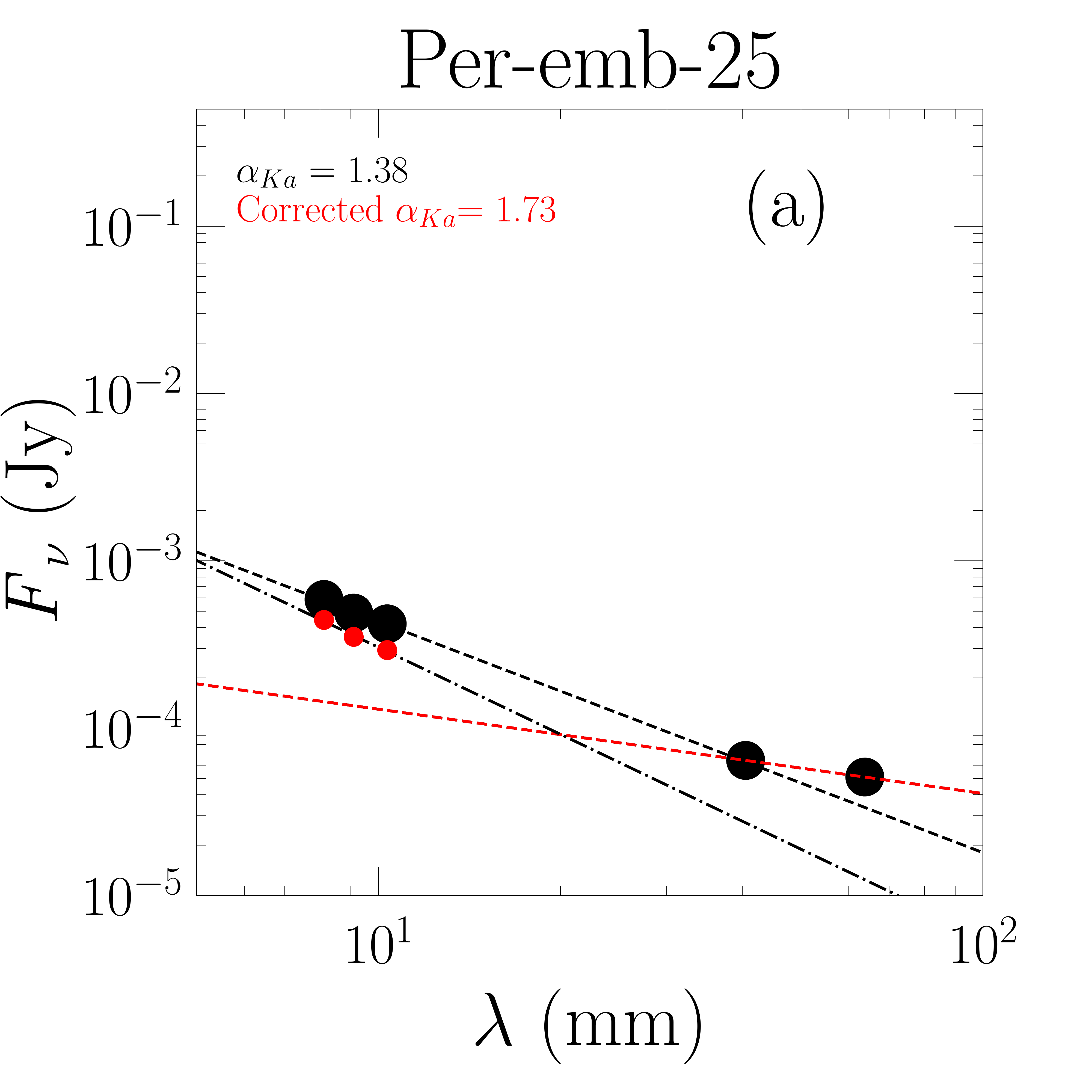}
  \includegraphics[width=0.30\linewidth]{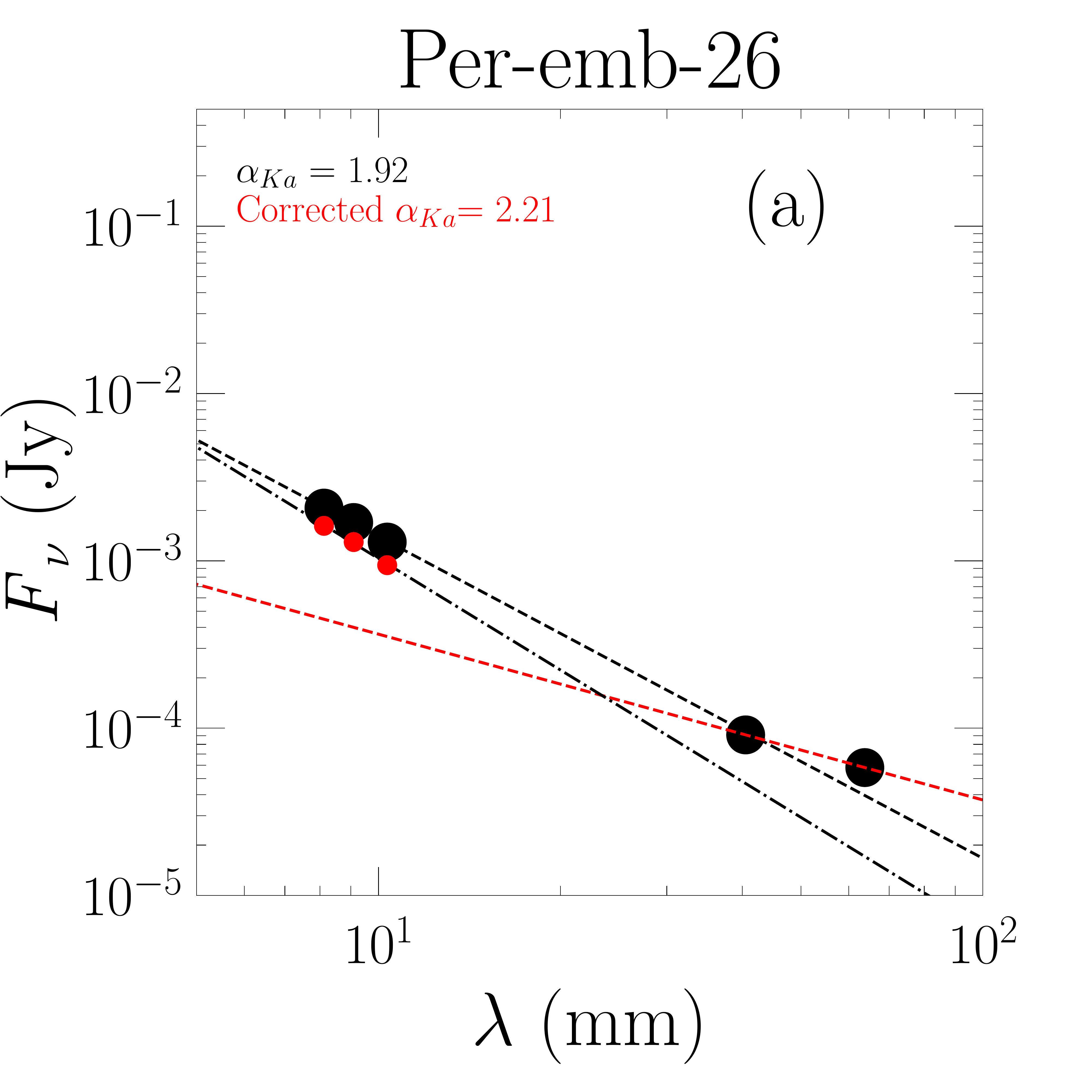}
  \includegraphics[width=0.30\linewidth]{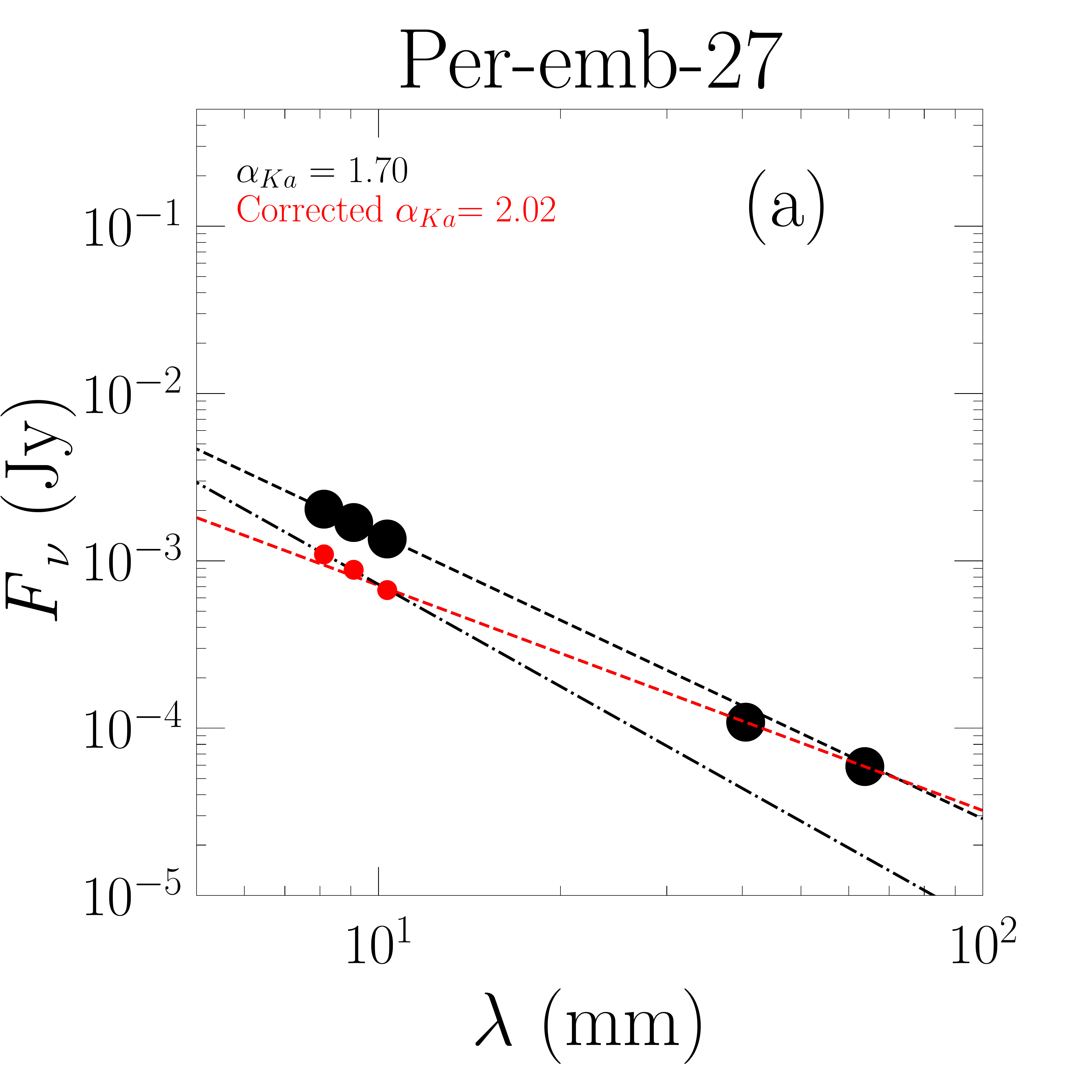}
  \includegraphics[width=0.30\linewidth]{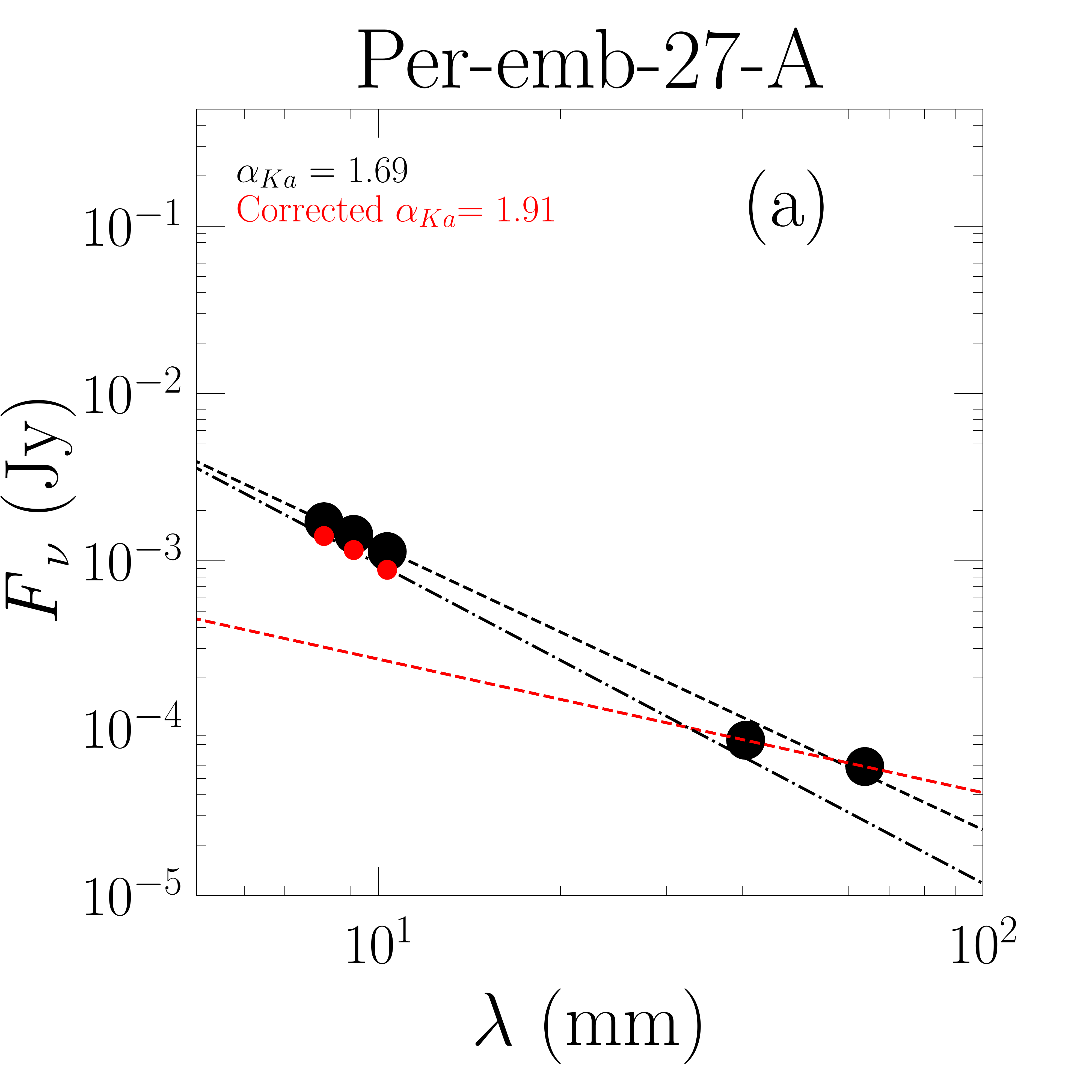}
  \includegraphics[width=0.30\linewidth]{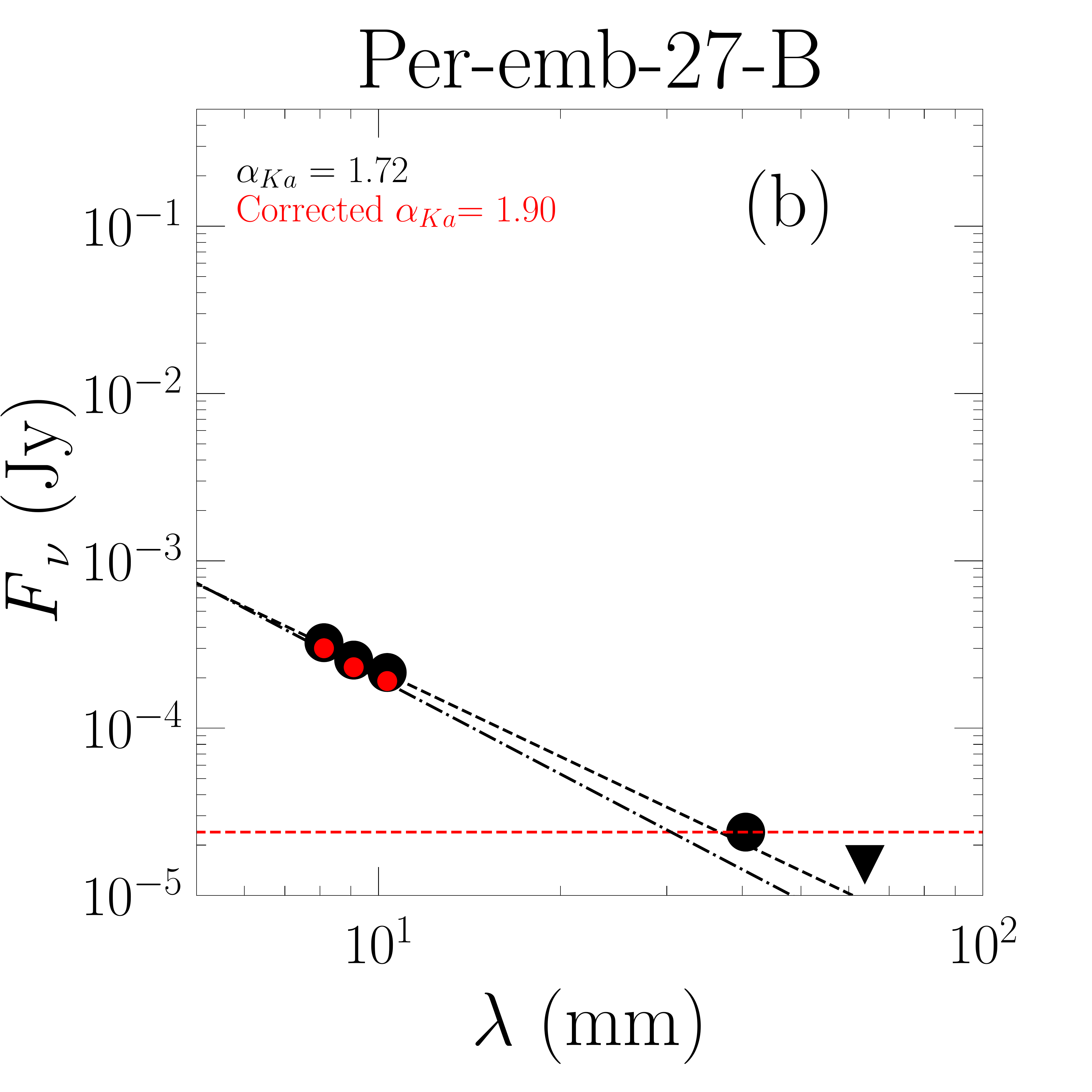}
  \includegraphics[width=0.30\linewidth]{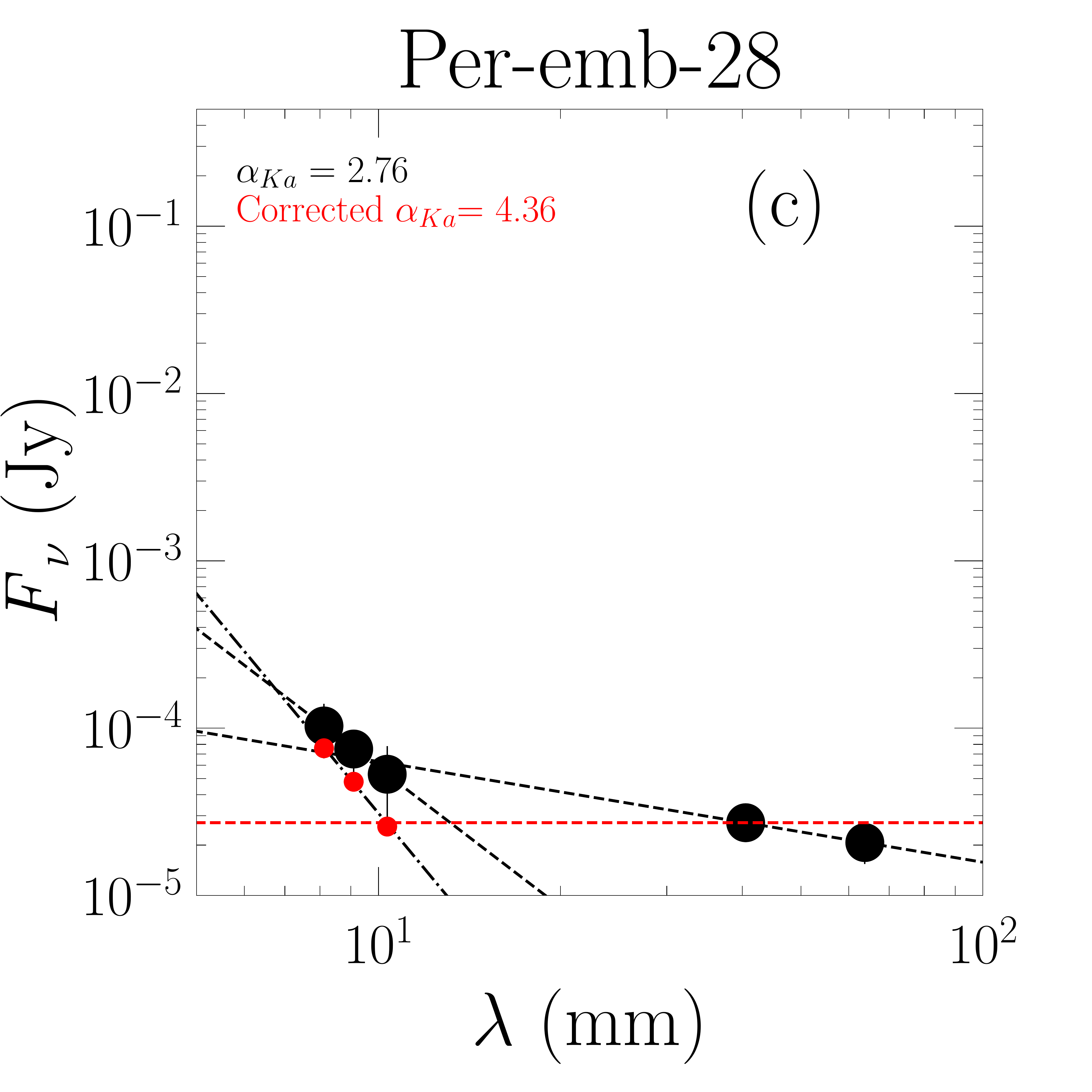}
  \includegraphics[width=0.30\linewidth]{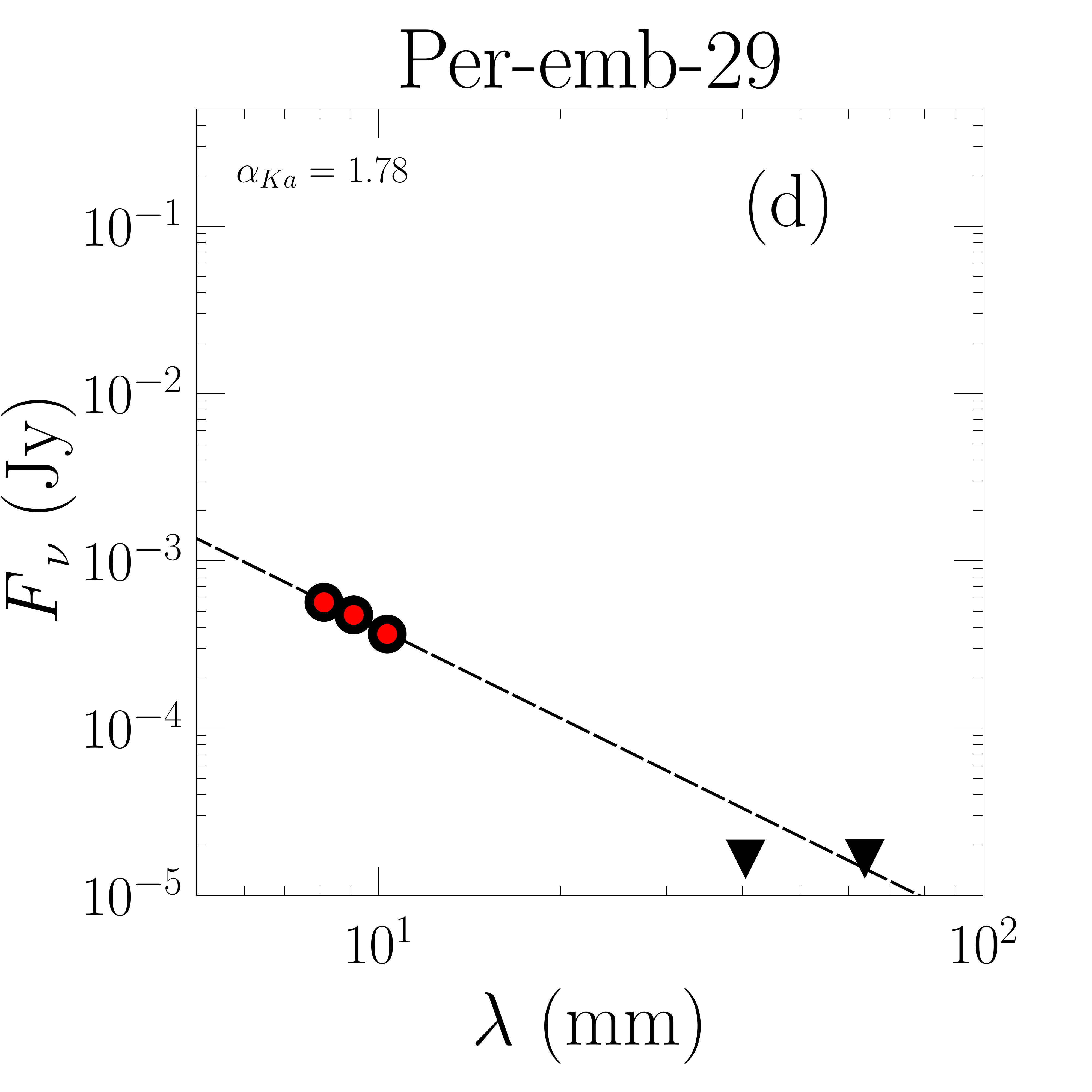}
  \includegraphics[width=0.30\linewidth]{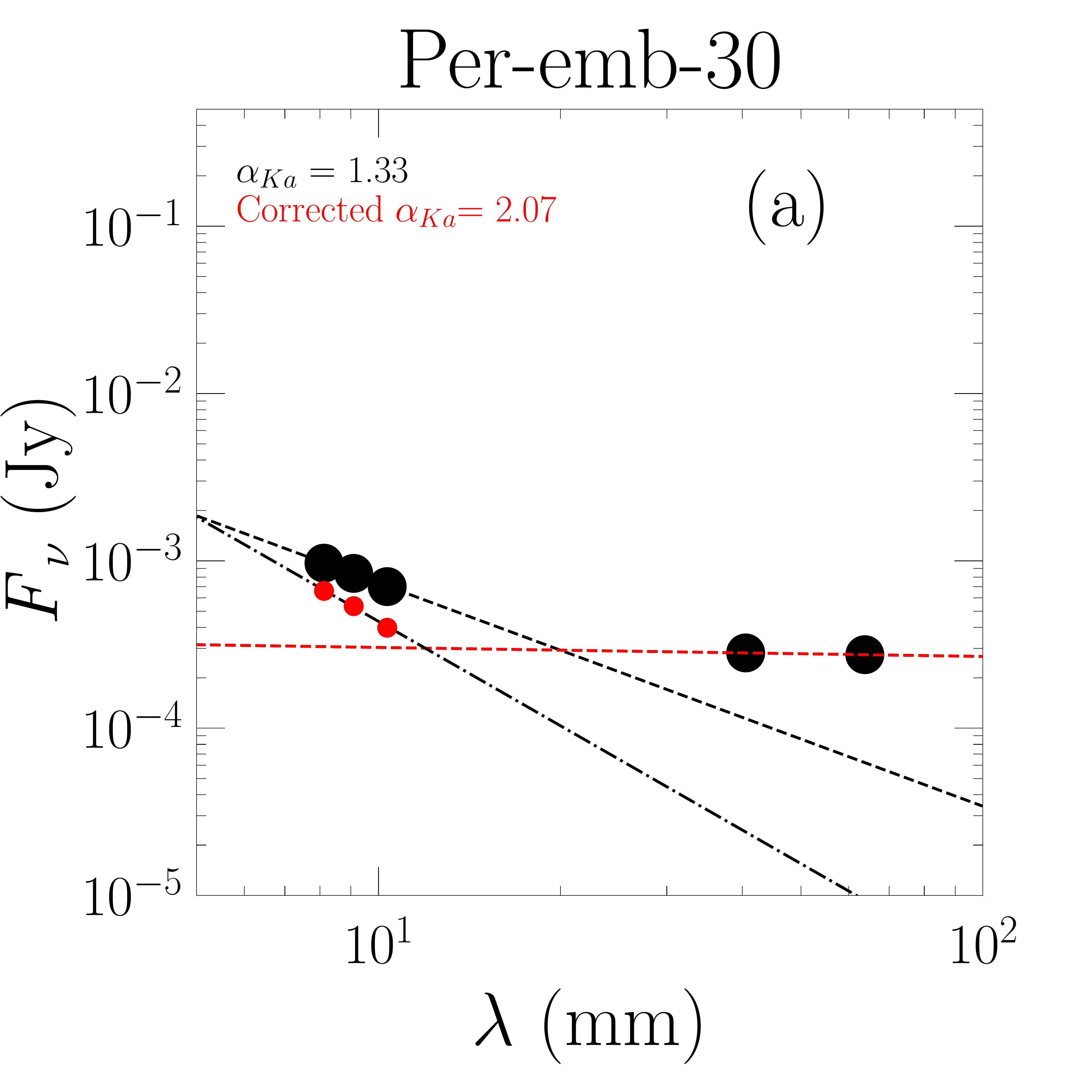}
  \includegraphics[width=0.30\linewidth]{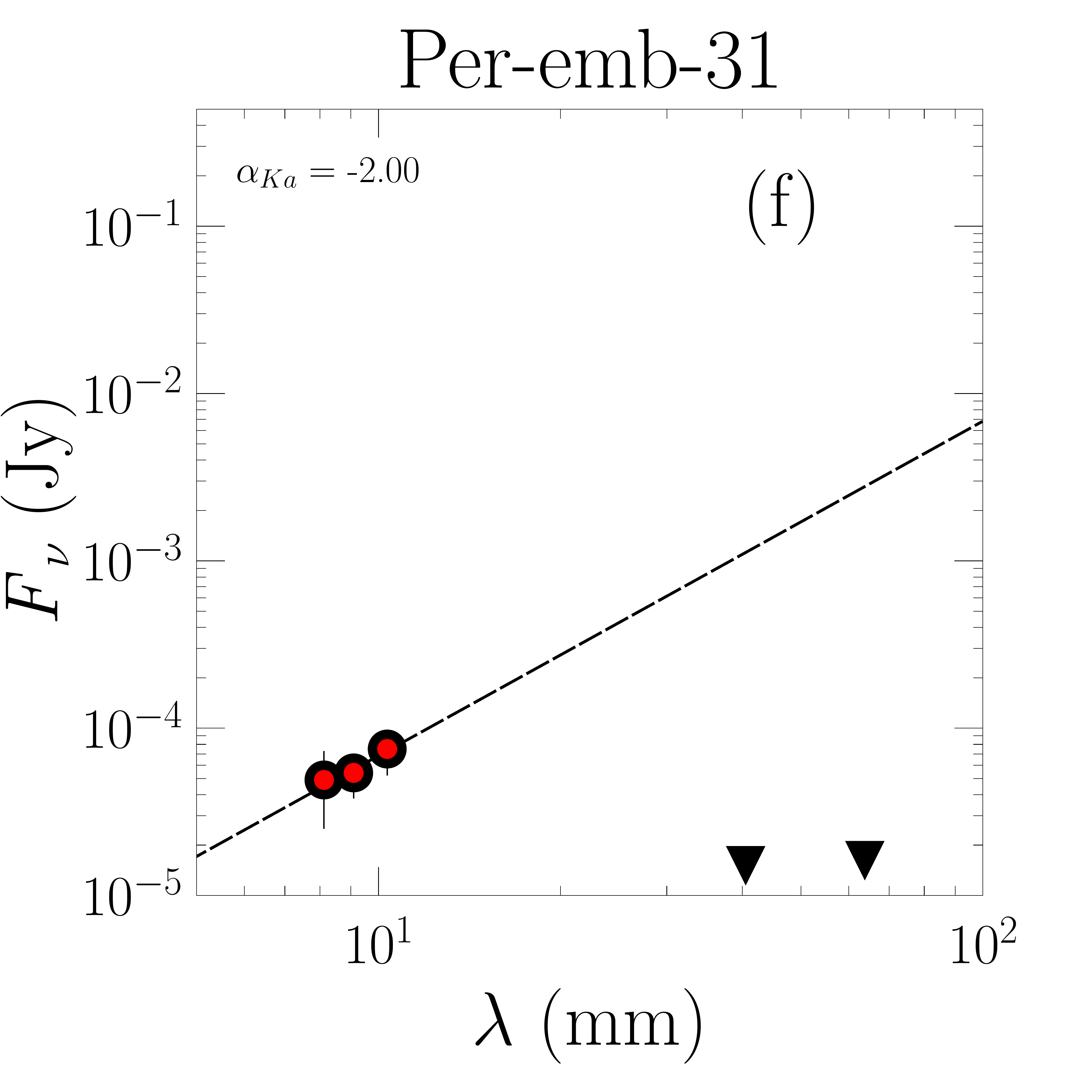}
  \includegraphics[width=0.30\linewidth]{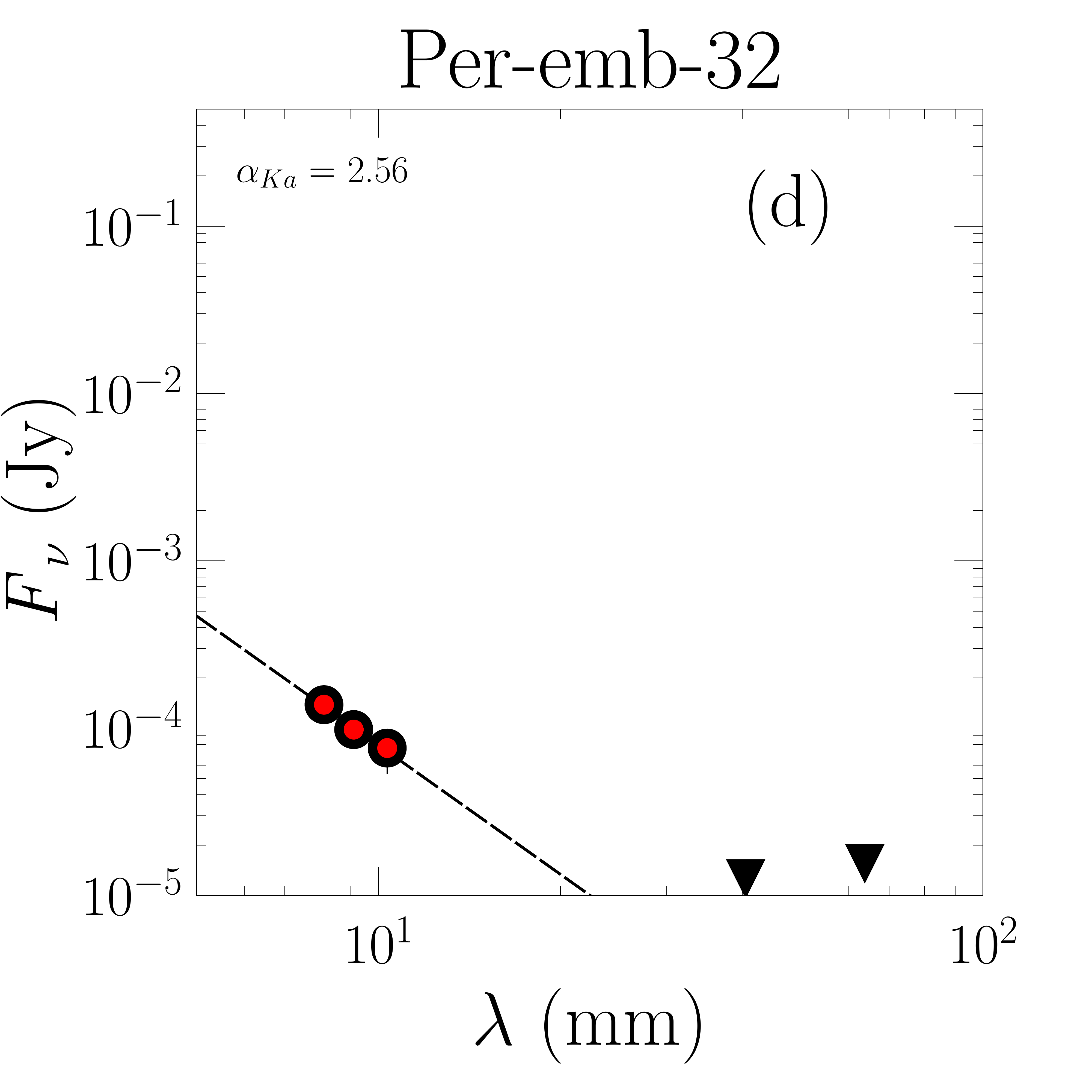}
  \includegraphics[width=0.30\linewidth]{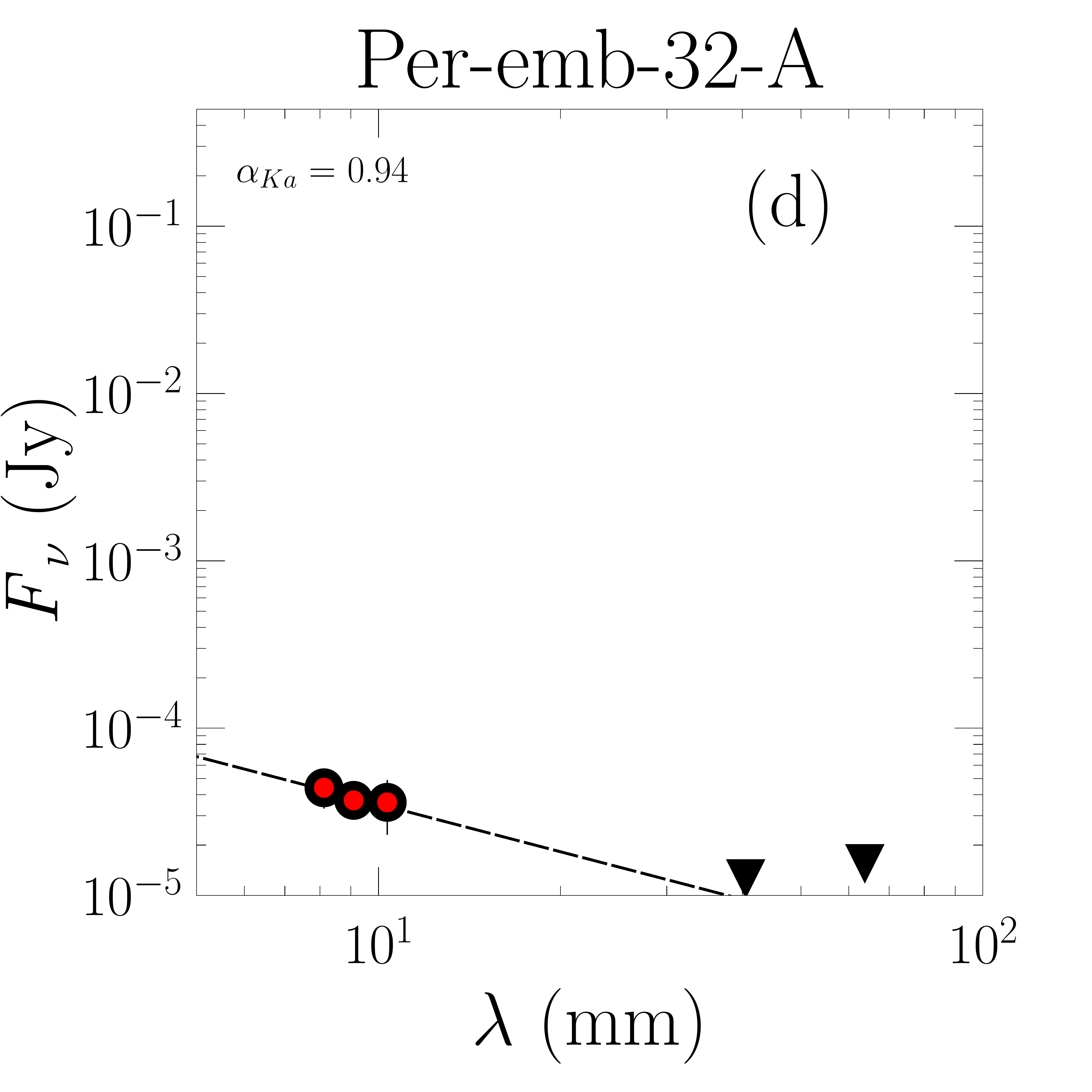}
  \includegraphics[width=0.30\linewidth]{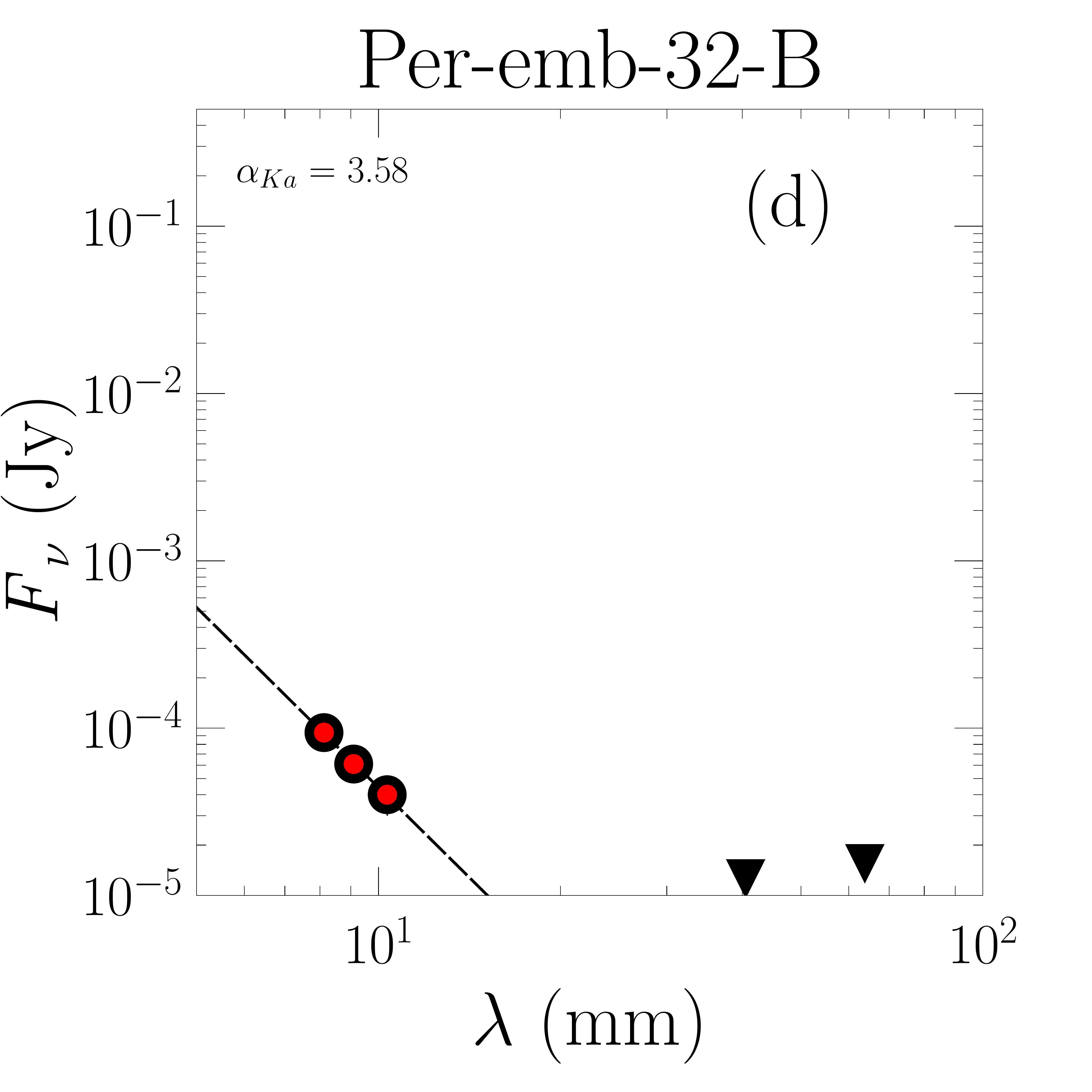}

\end{figure}
\begin{figure}[H]
\centering
  \includegraphics[width=0.30\linewidth]{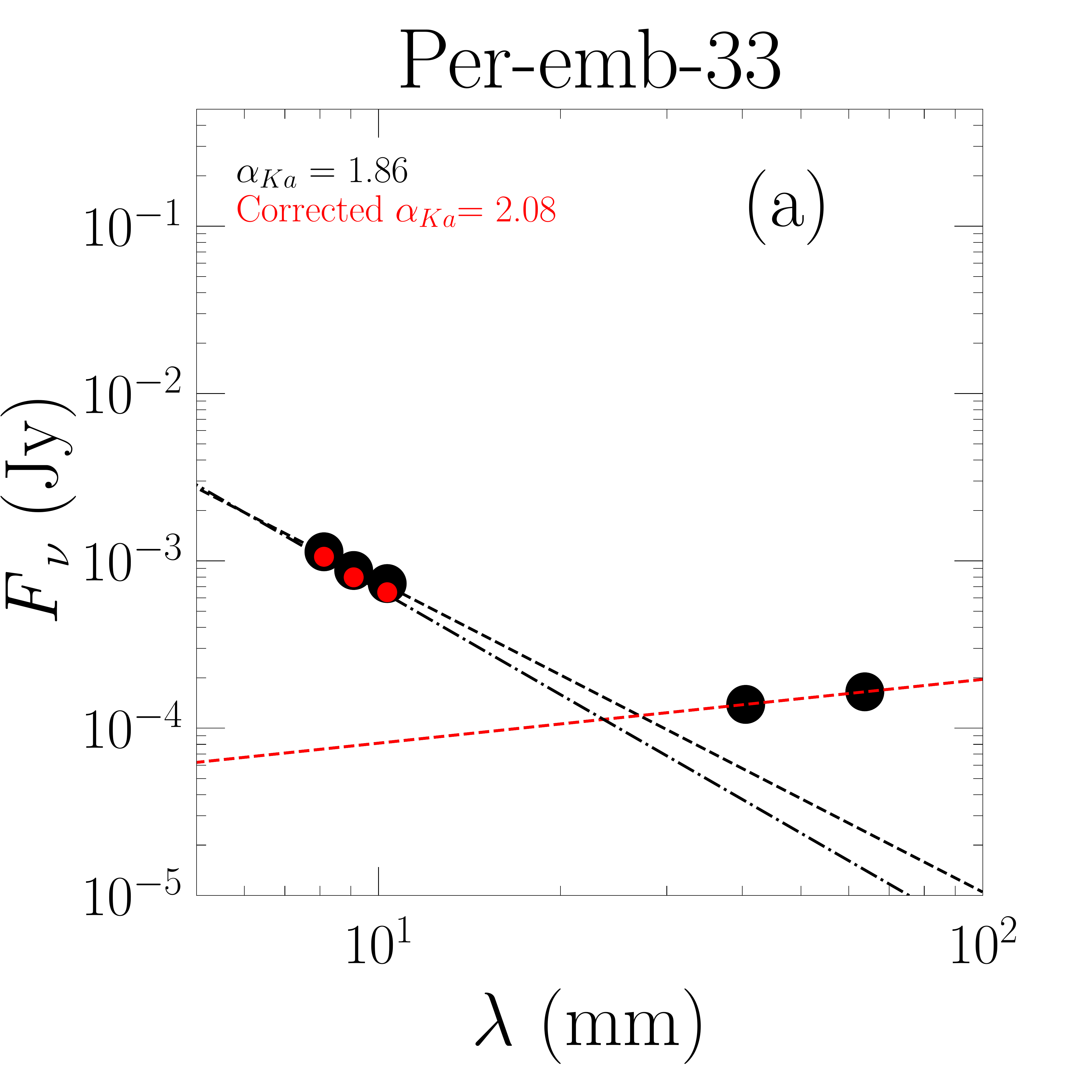}
  \includegraphics[width=0.30\linewidth]{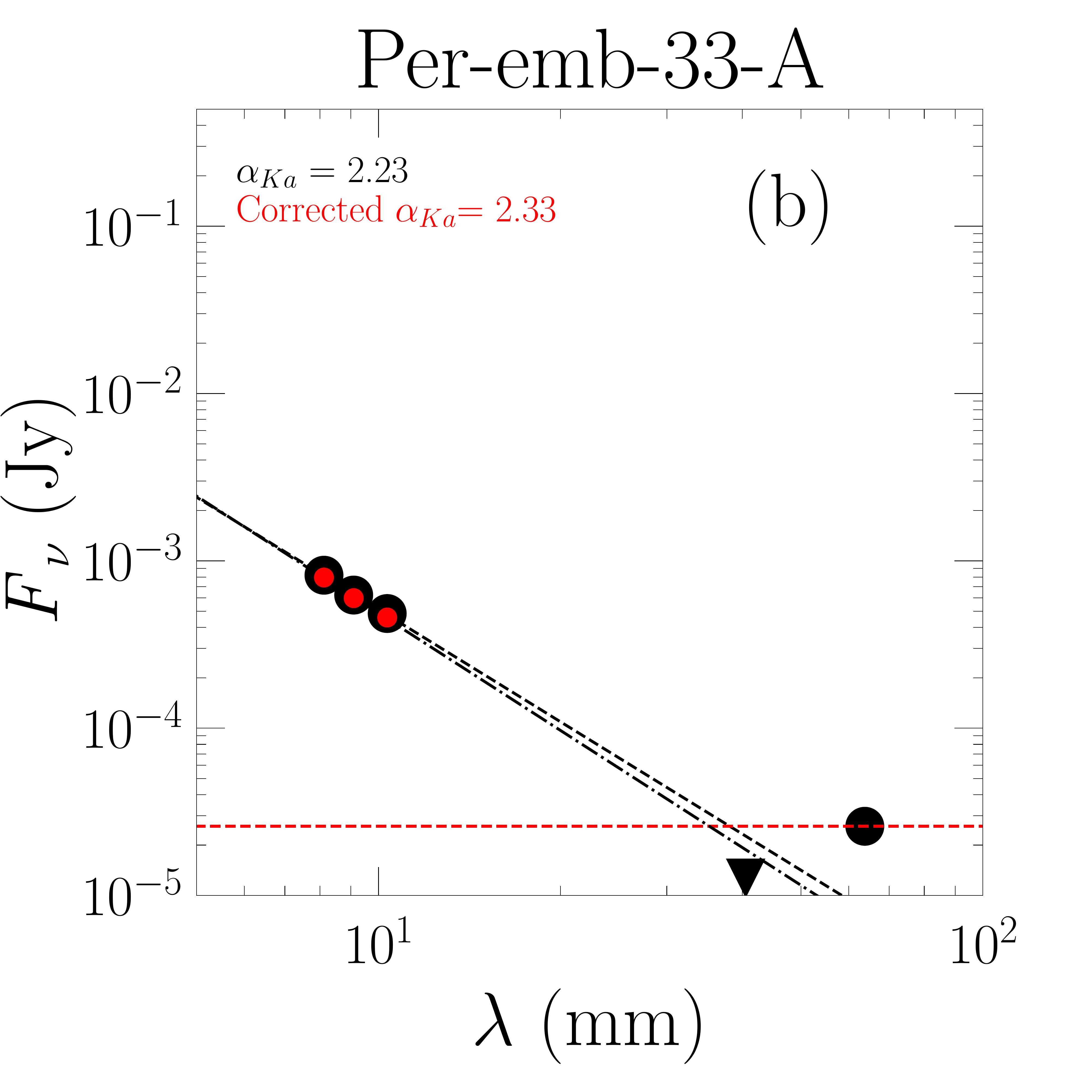}
  \includegraphics[width=0.30\linewidth]{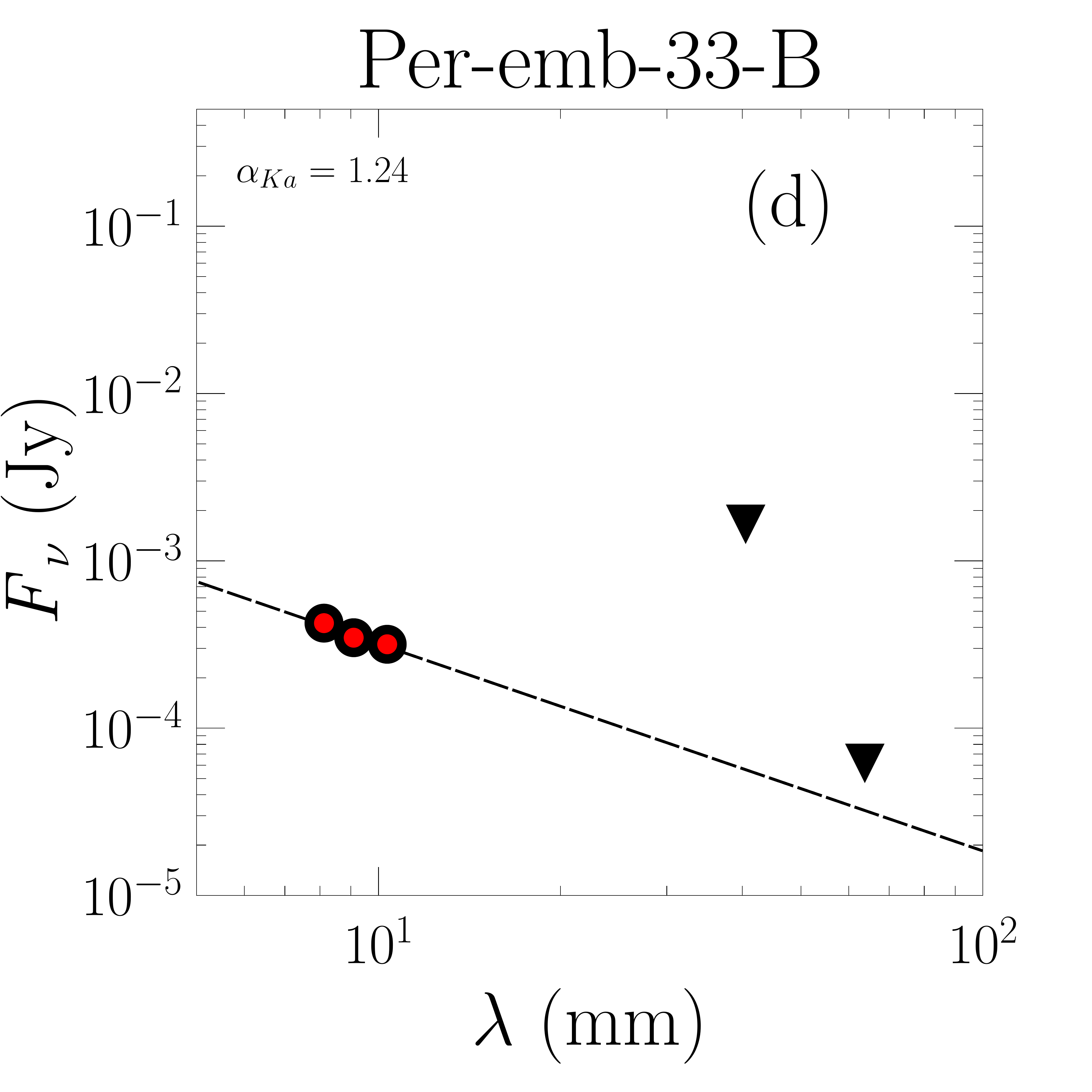}
  \includegraphics[width=0.30\linewidth]{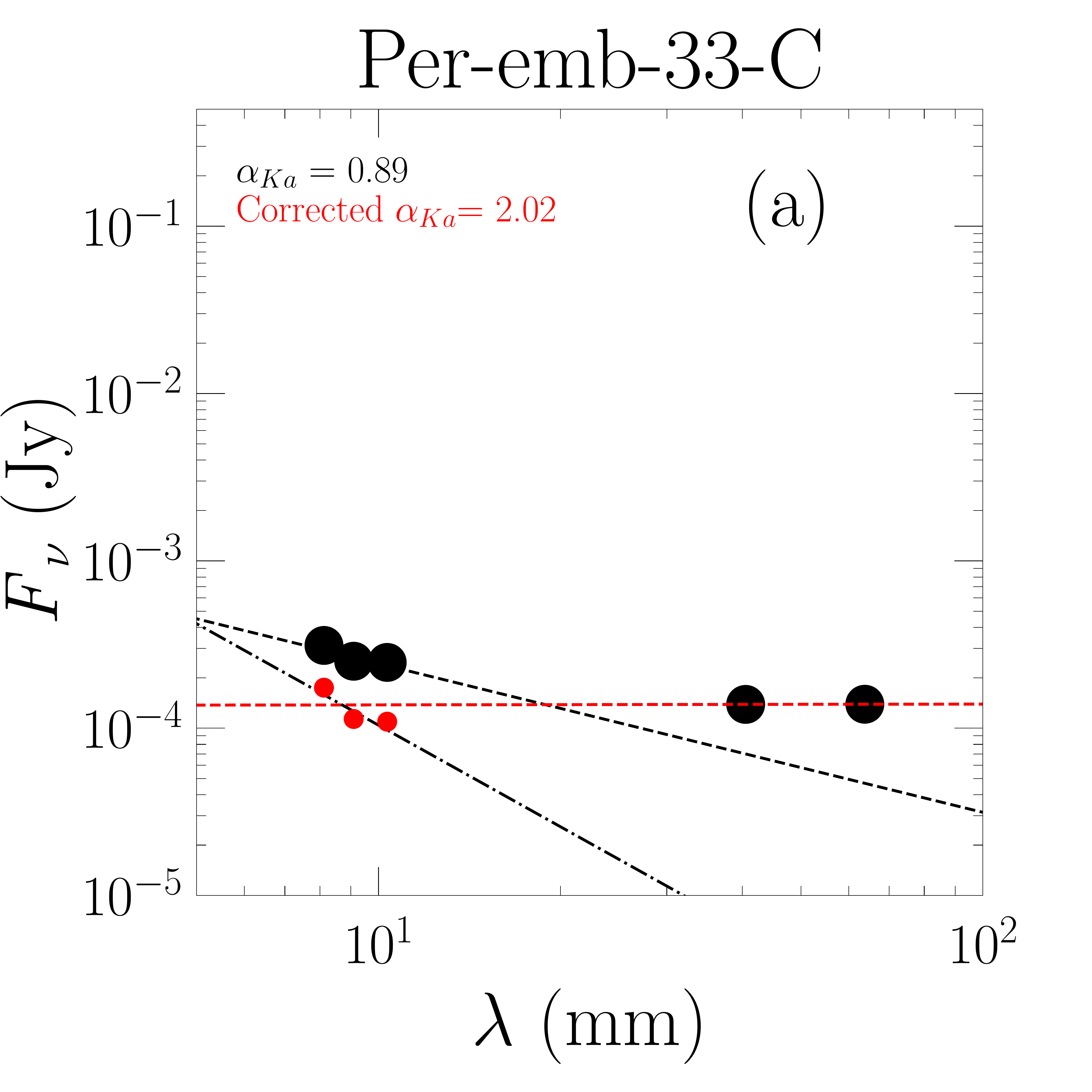}
  \includegraphics[width=0.30\linewidth]{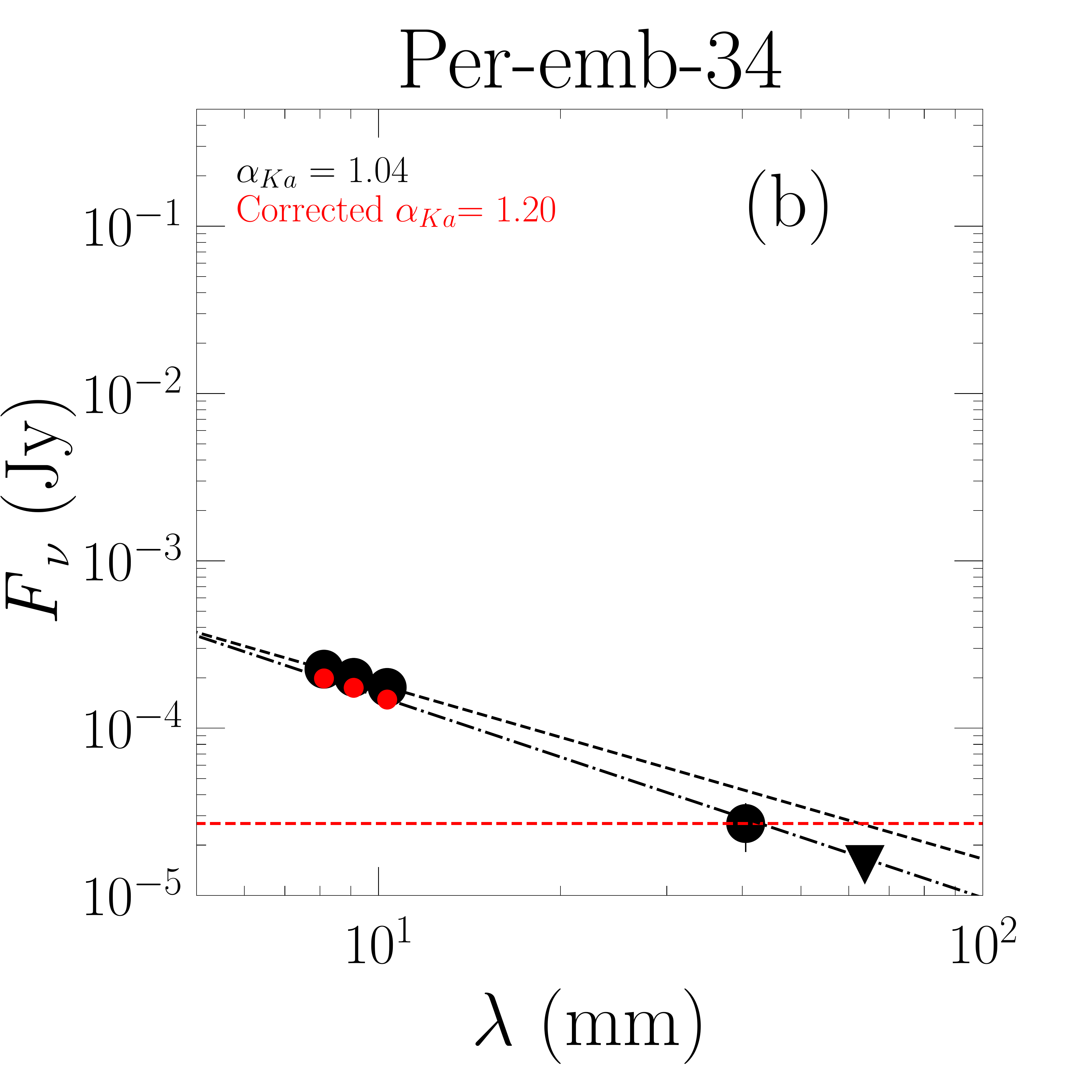}
  \includegraphics[width=0.30\linewidth]{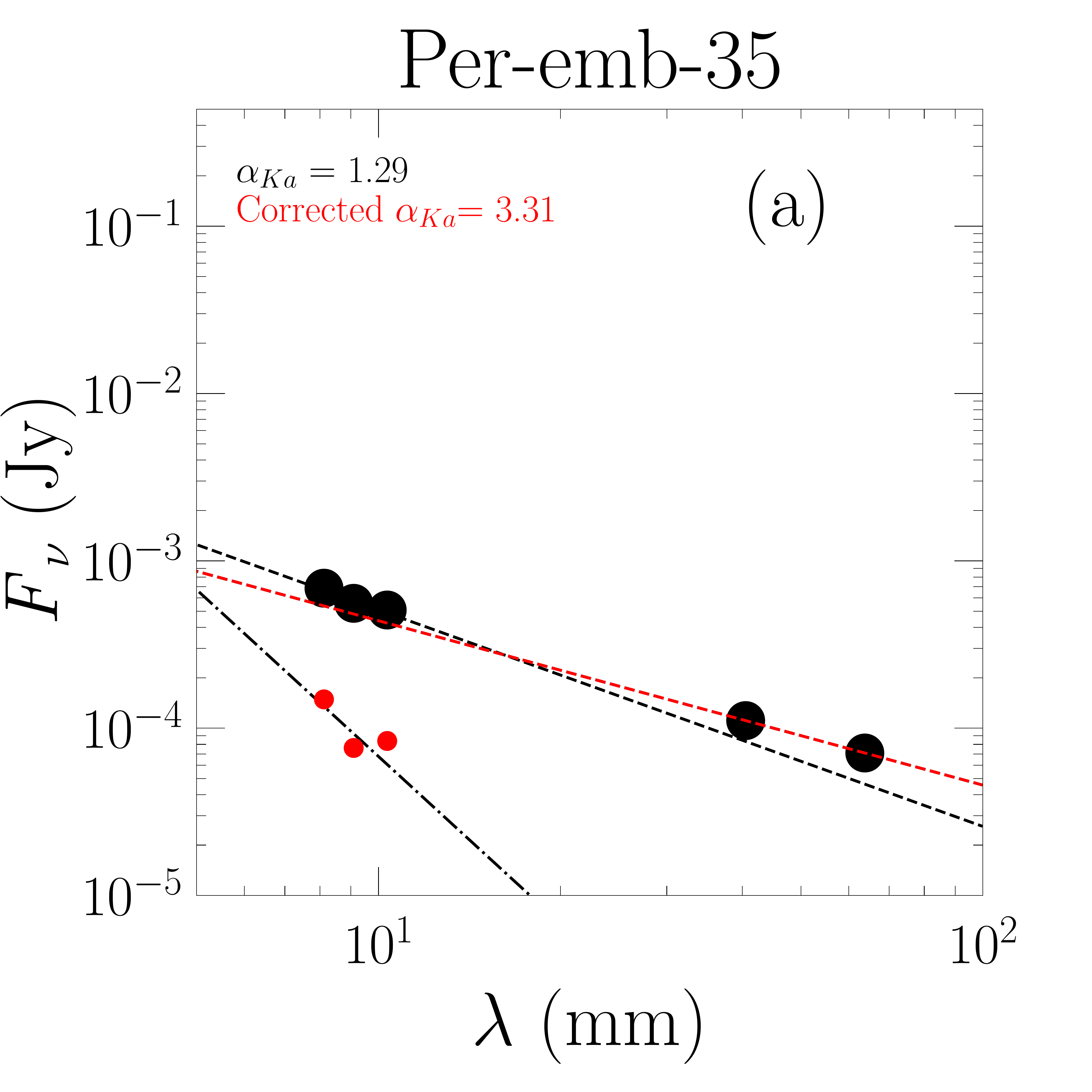}
  \includegraphics[width=0.30\linewidth]{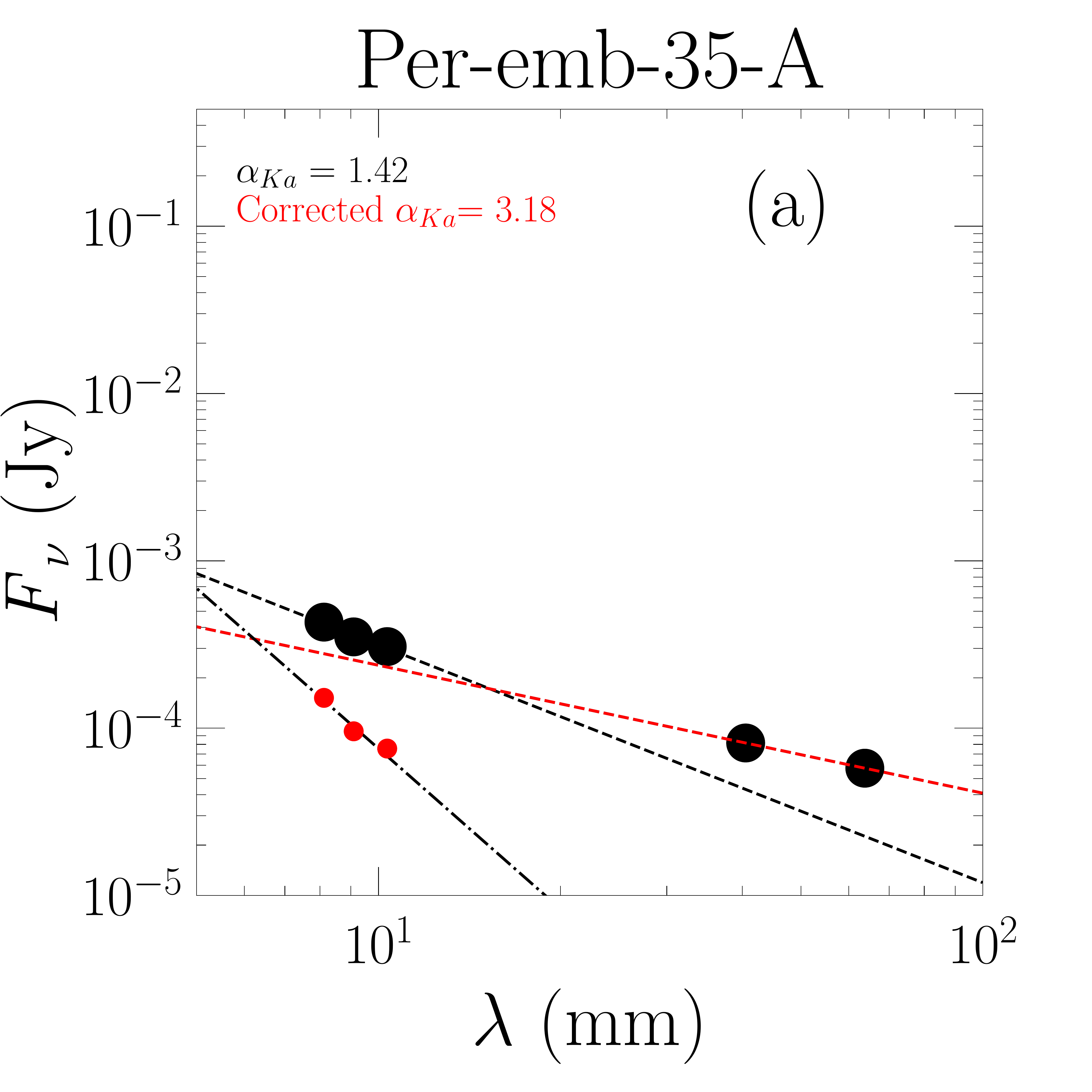}
  \includegraphics[width=0.30\linewidth]{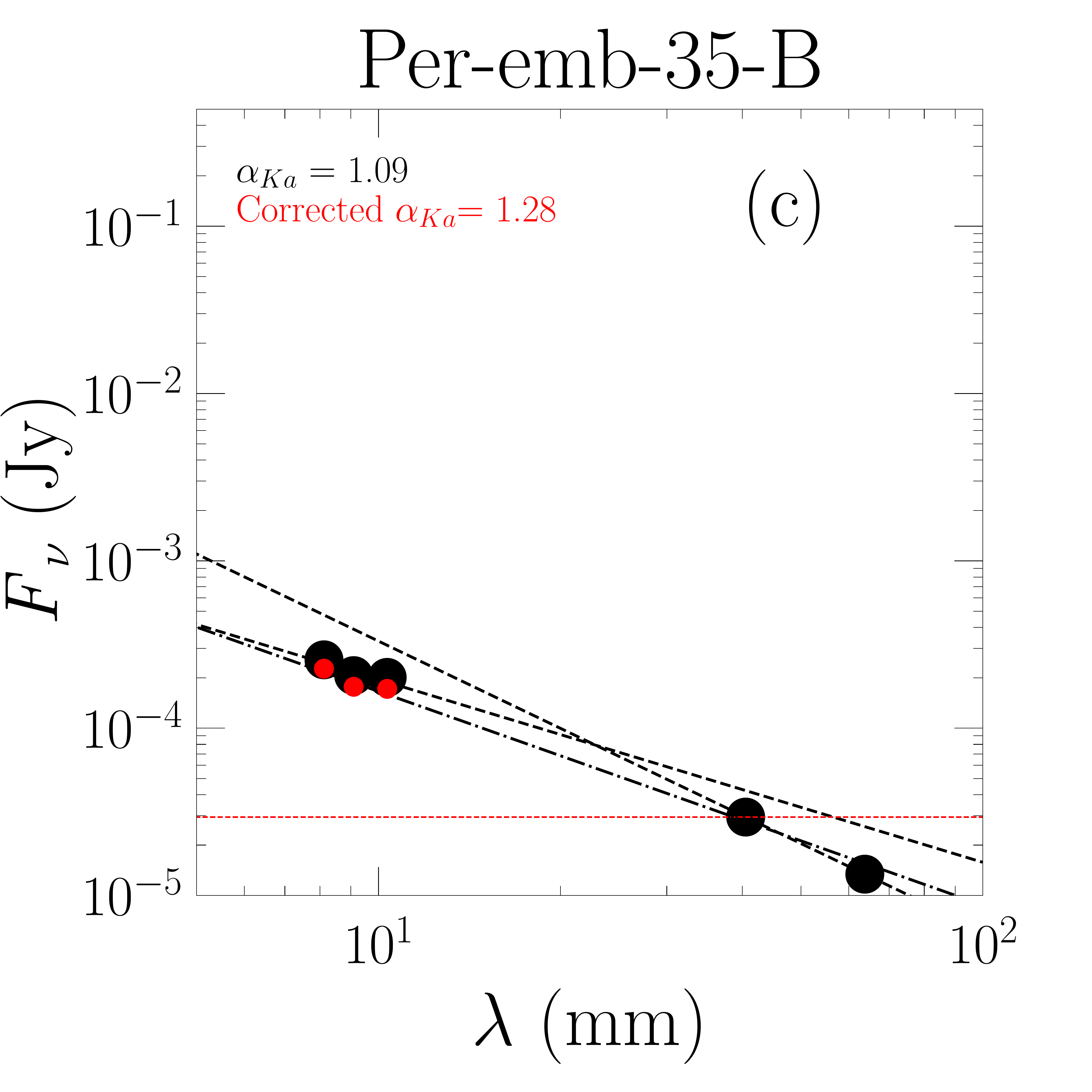}
  \includegraphics[width=0.30\linewidth]{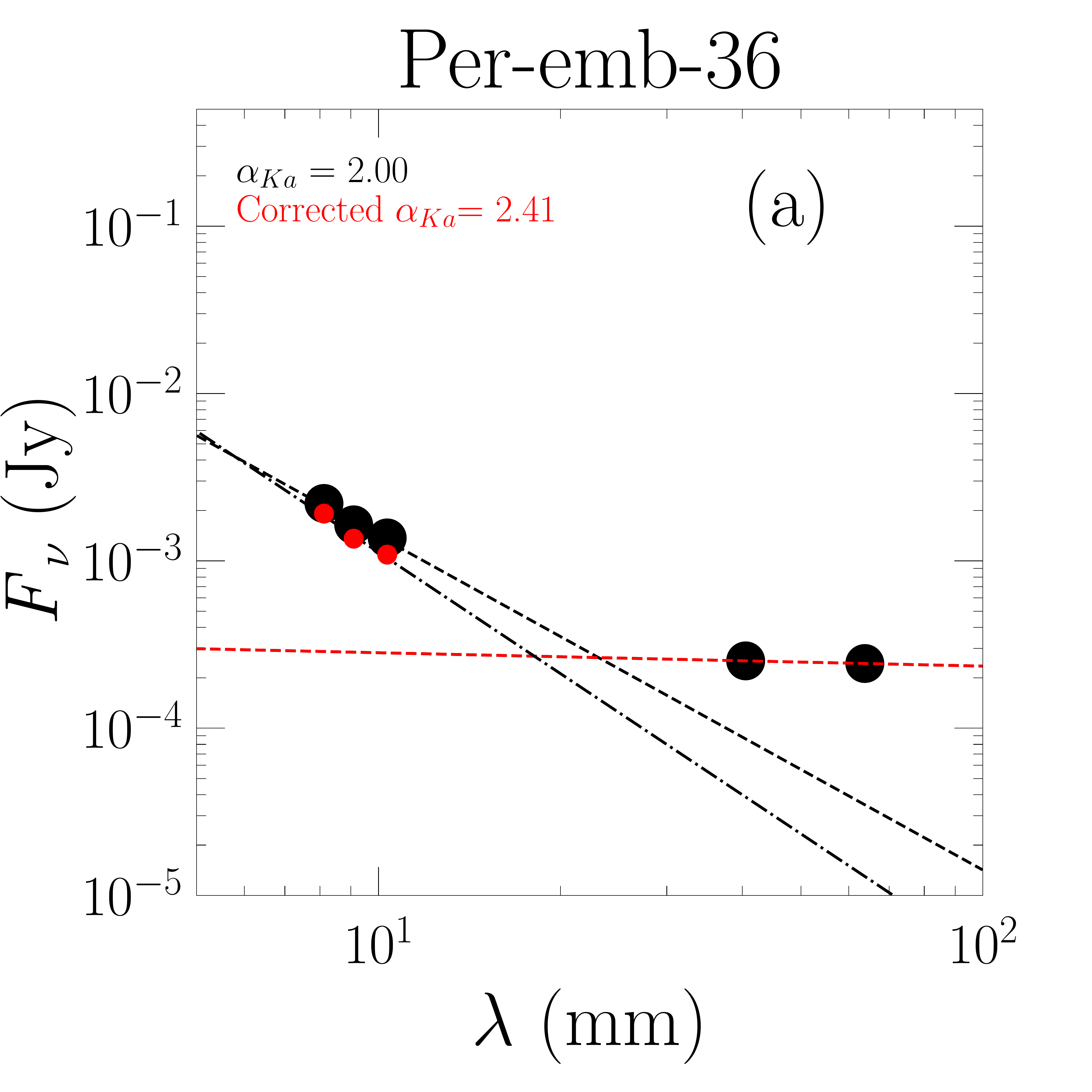}
  \includegraphics[width=0.30\linewidth]{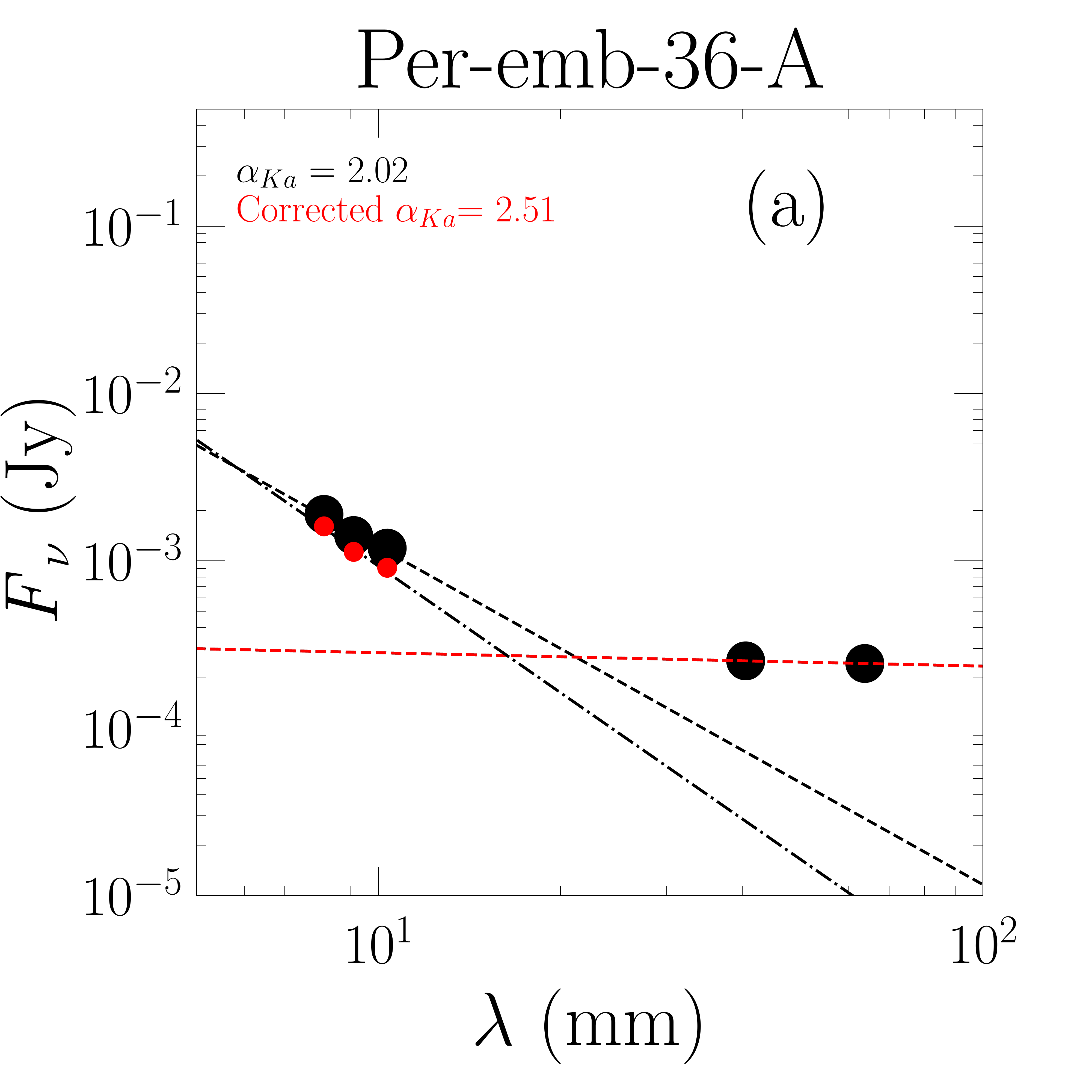}
  \includegraphics[width=0.30\linewidth]{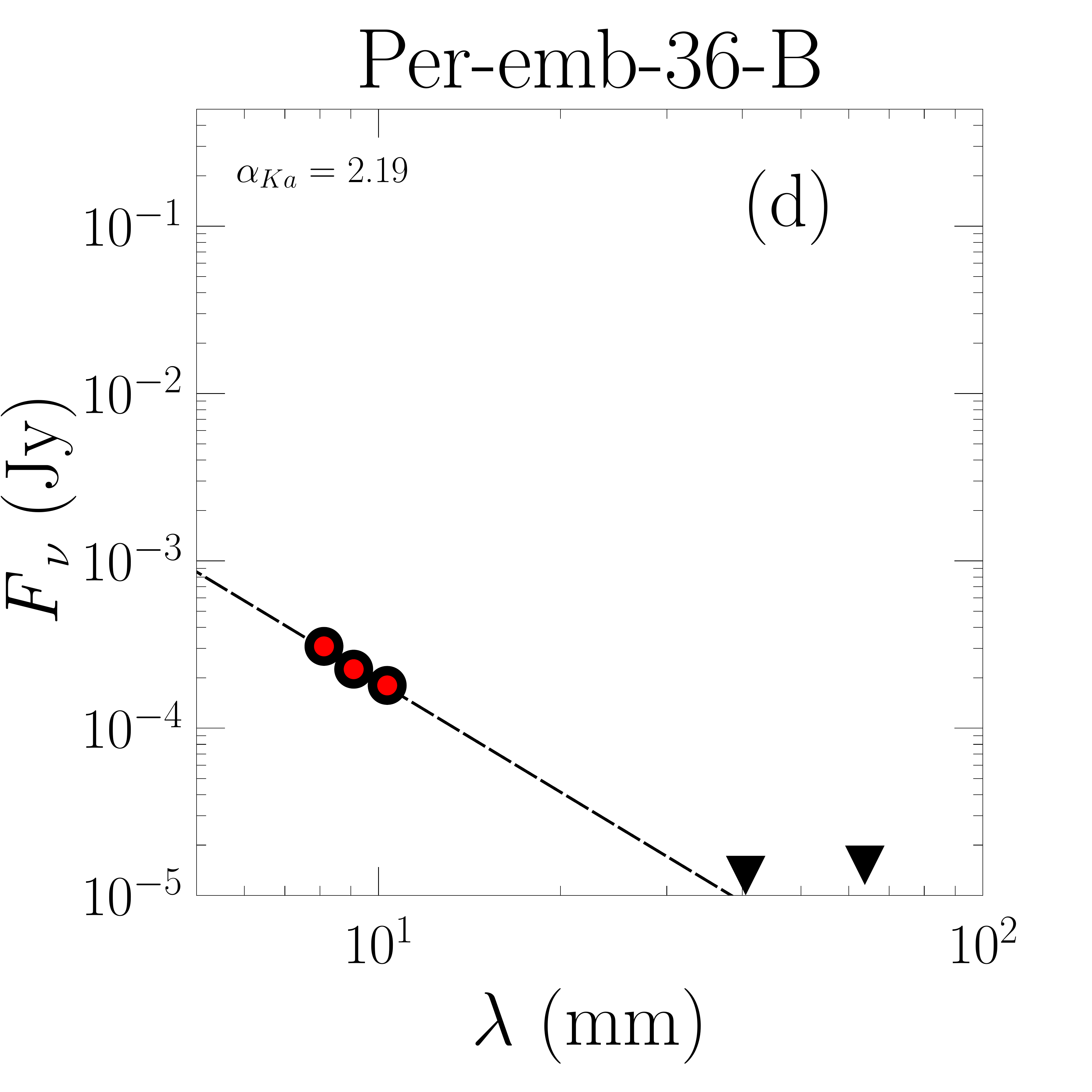}
  \includegraphics[width=0.30\linewidth]{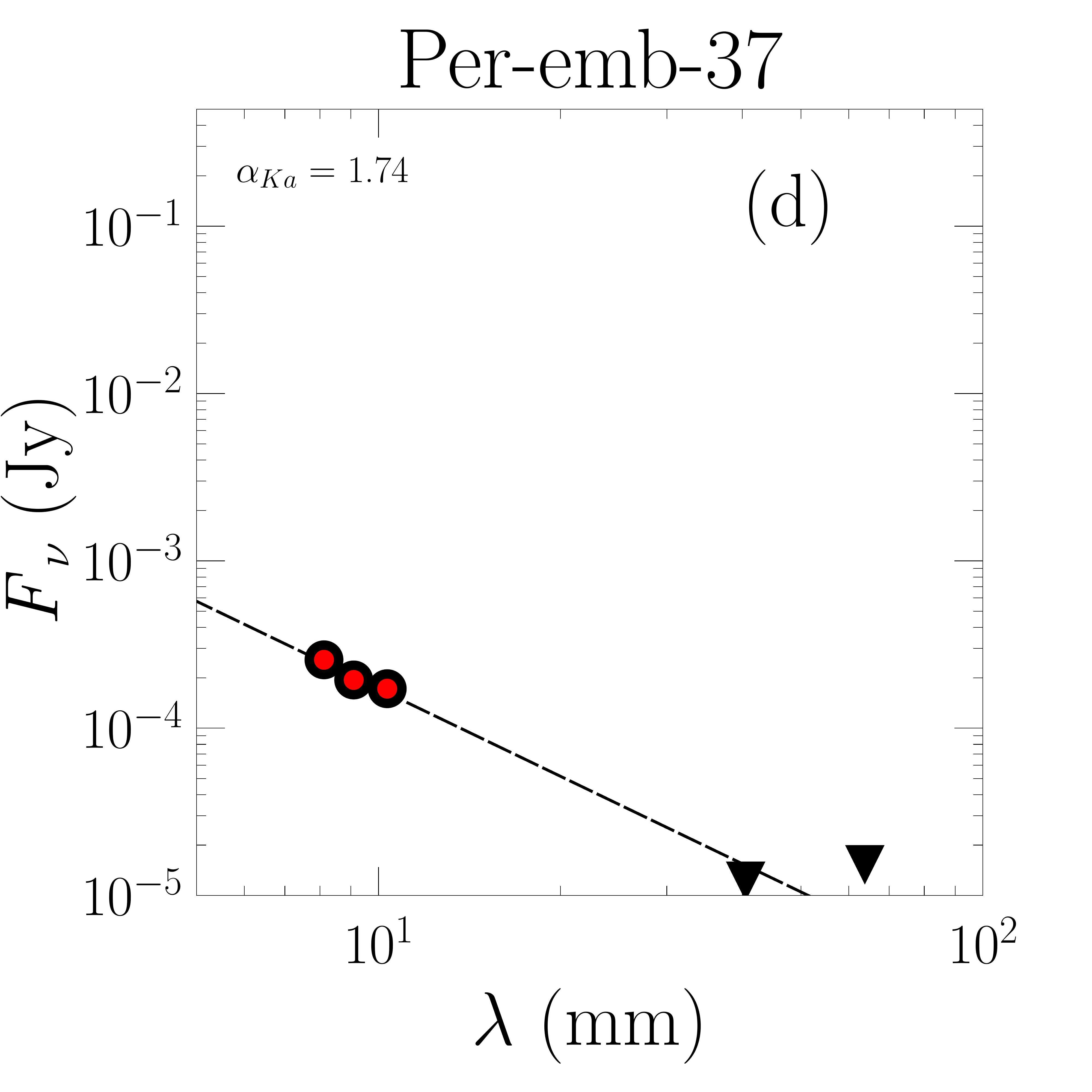}

\end{figure}
\begin{figure}[H]
\centering
  \includegraphics[width=0.30\linewidth]{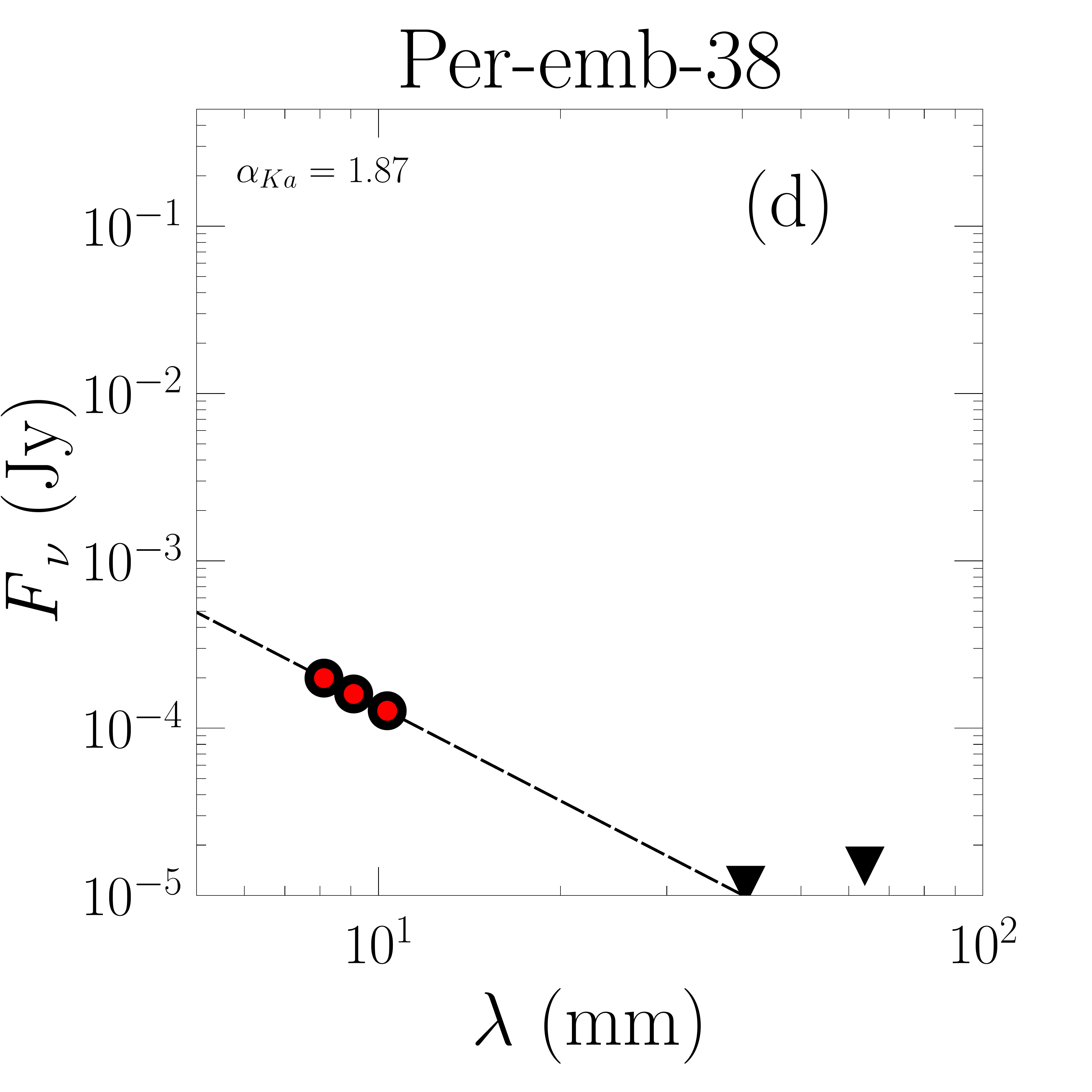}
  \includegraphics[width=0.30\linewidth]{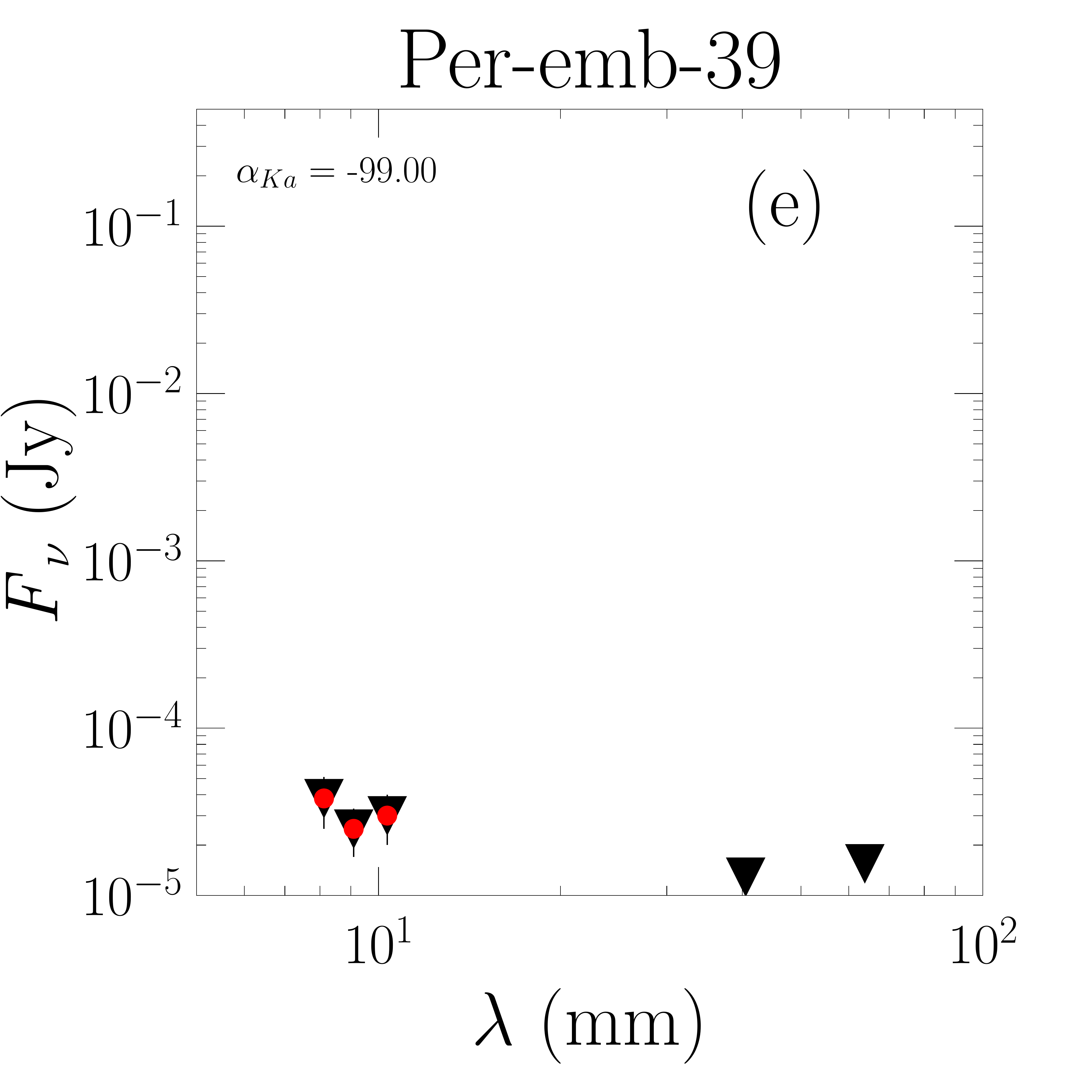}
  \includegraphics[width=0.30\linewidth]{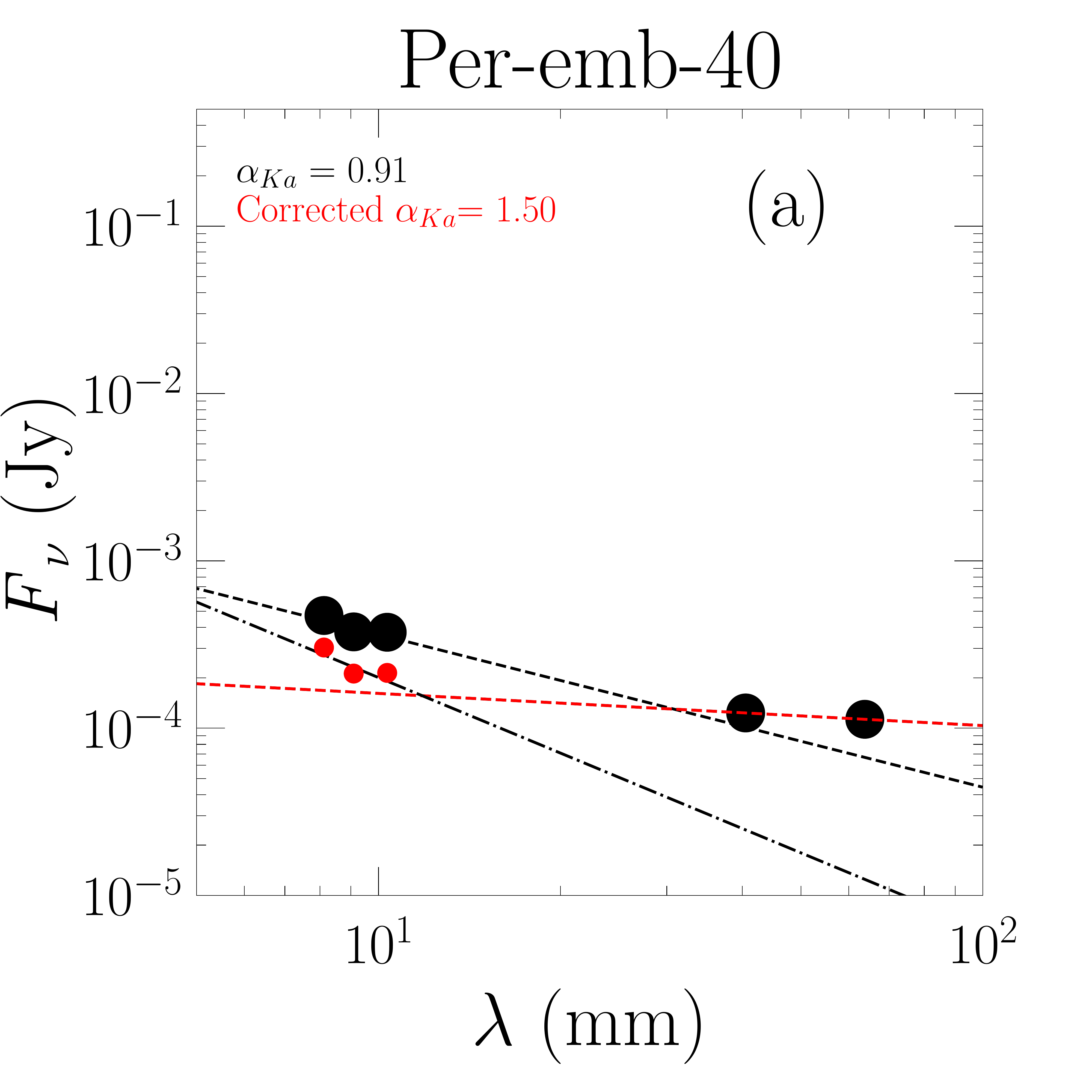}
  \includegraphics[width=0.30\linewidth]{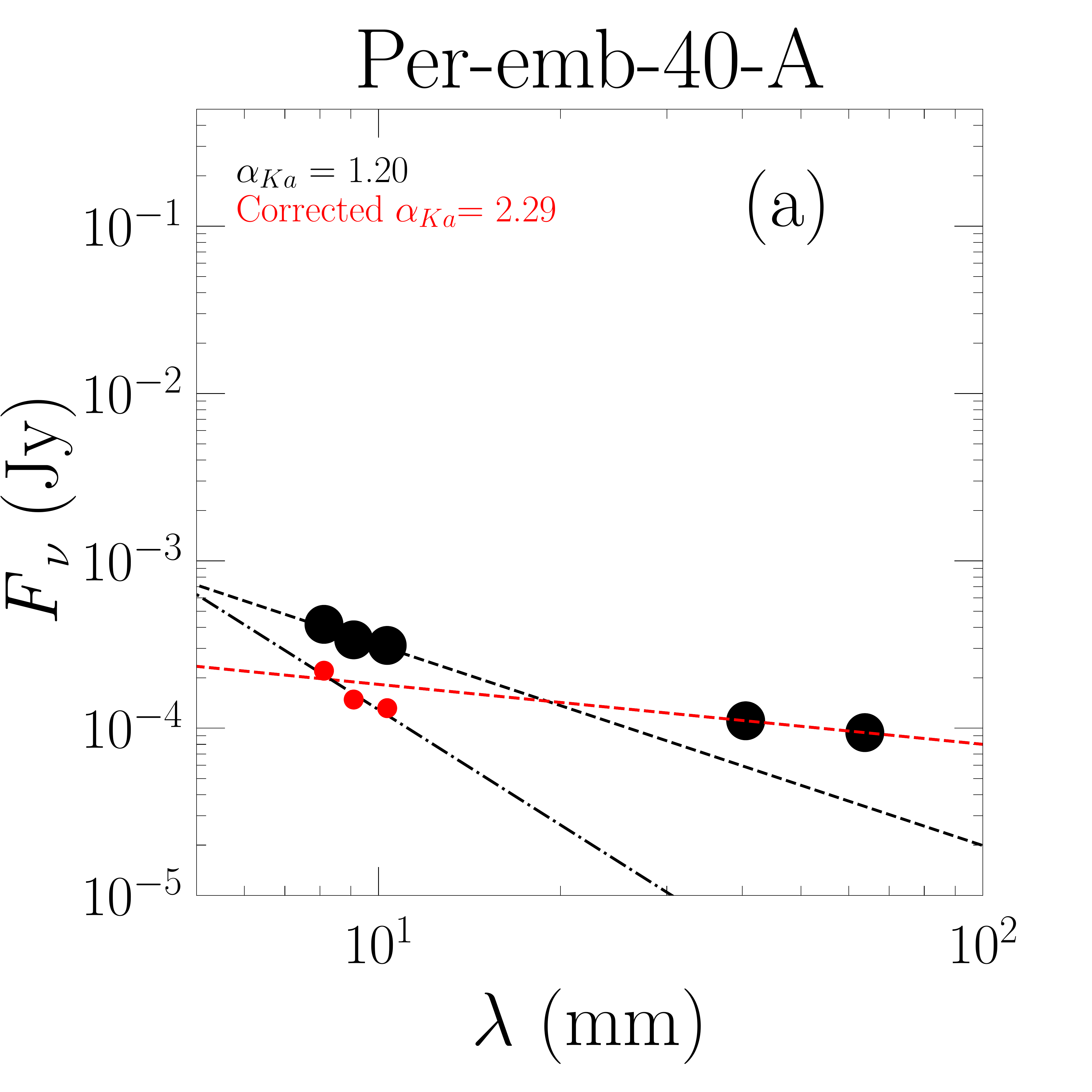}
  \includegraphics[width=0.30\linewidth]{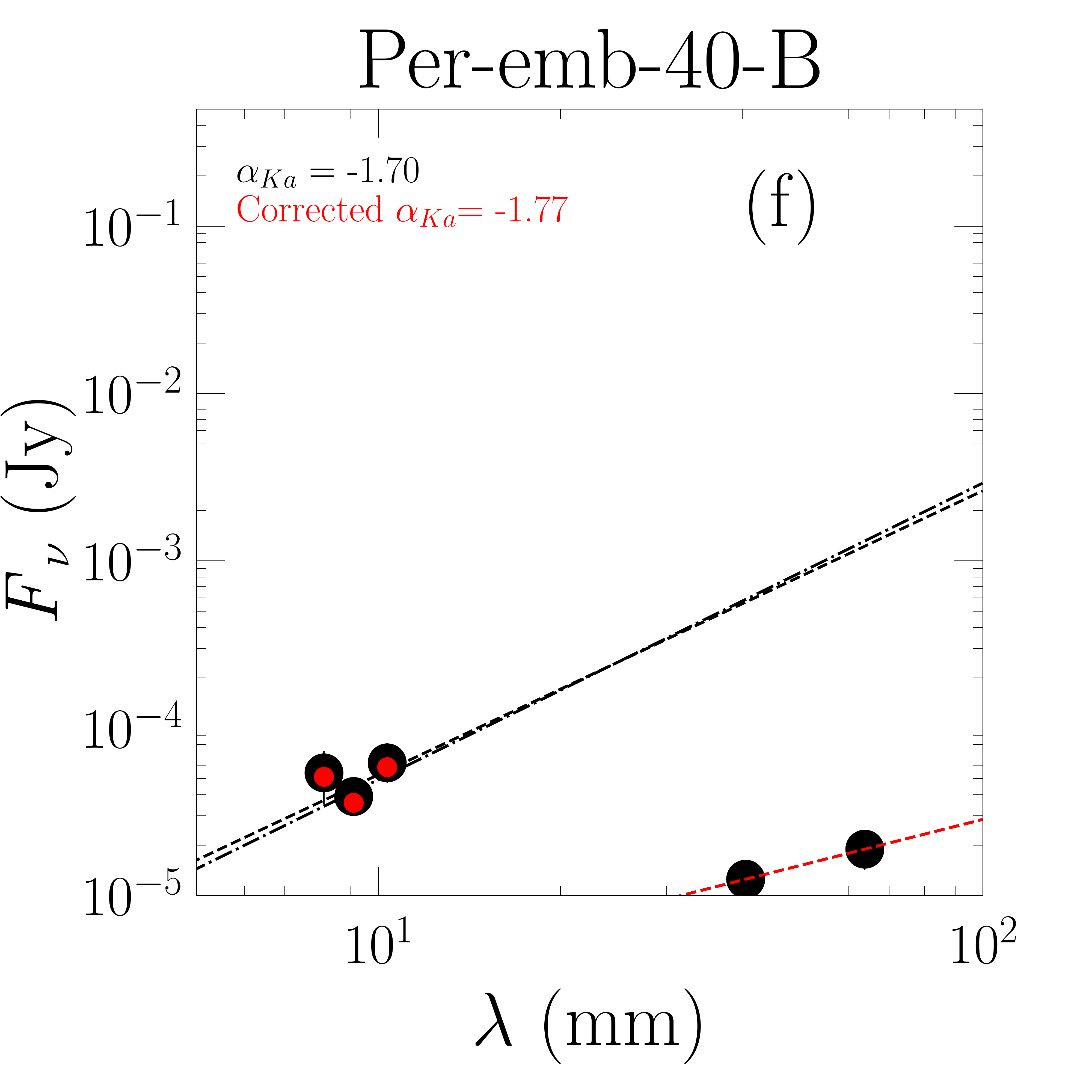}
  \includegraphics[width=0.30\linewidth]{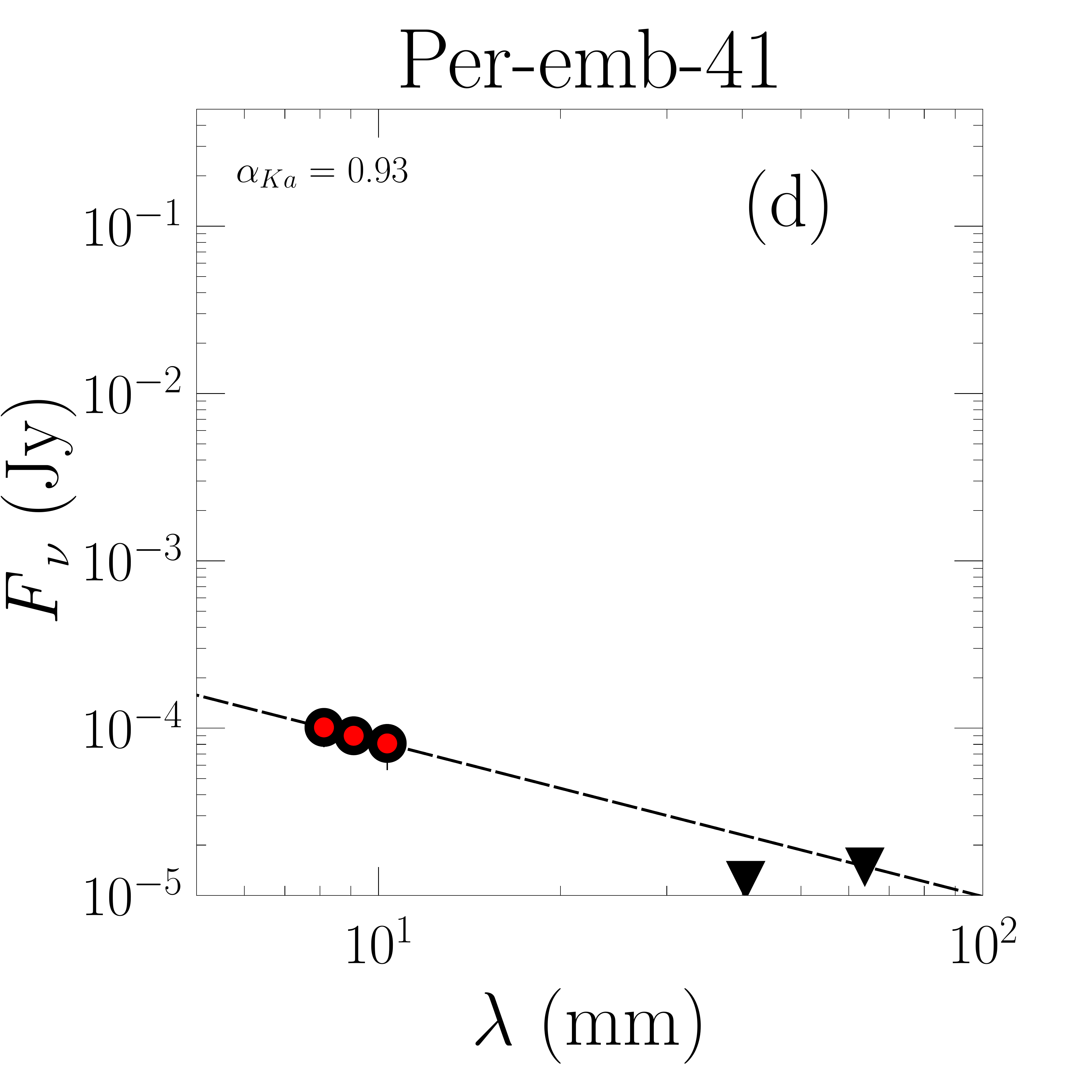}
  \includegraphics[width=0.30\linewidth]{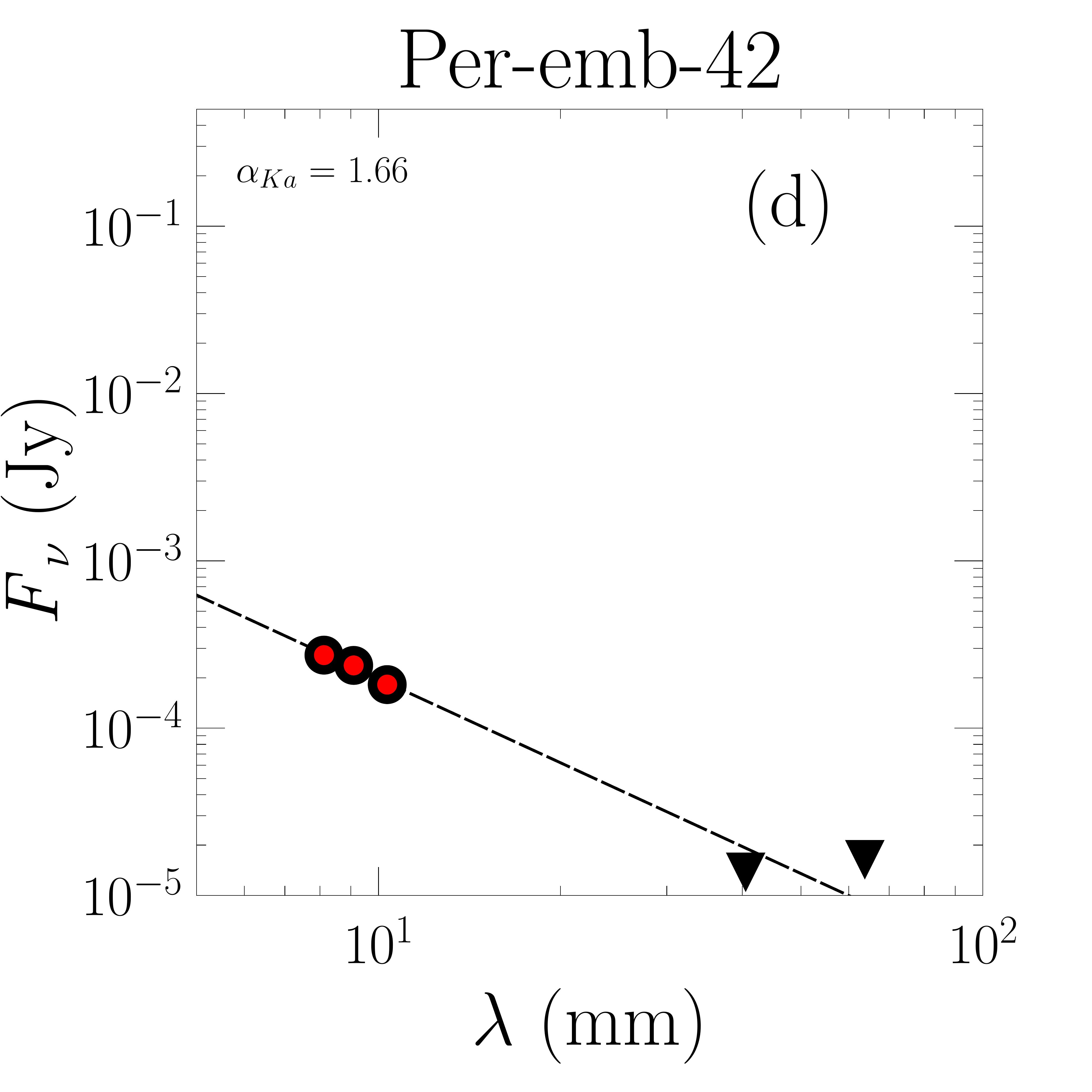}
  \includegraphics[width=0.30\linewidth]{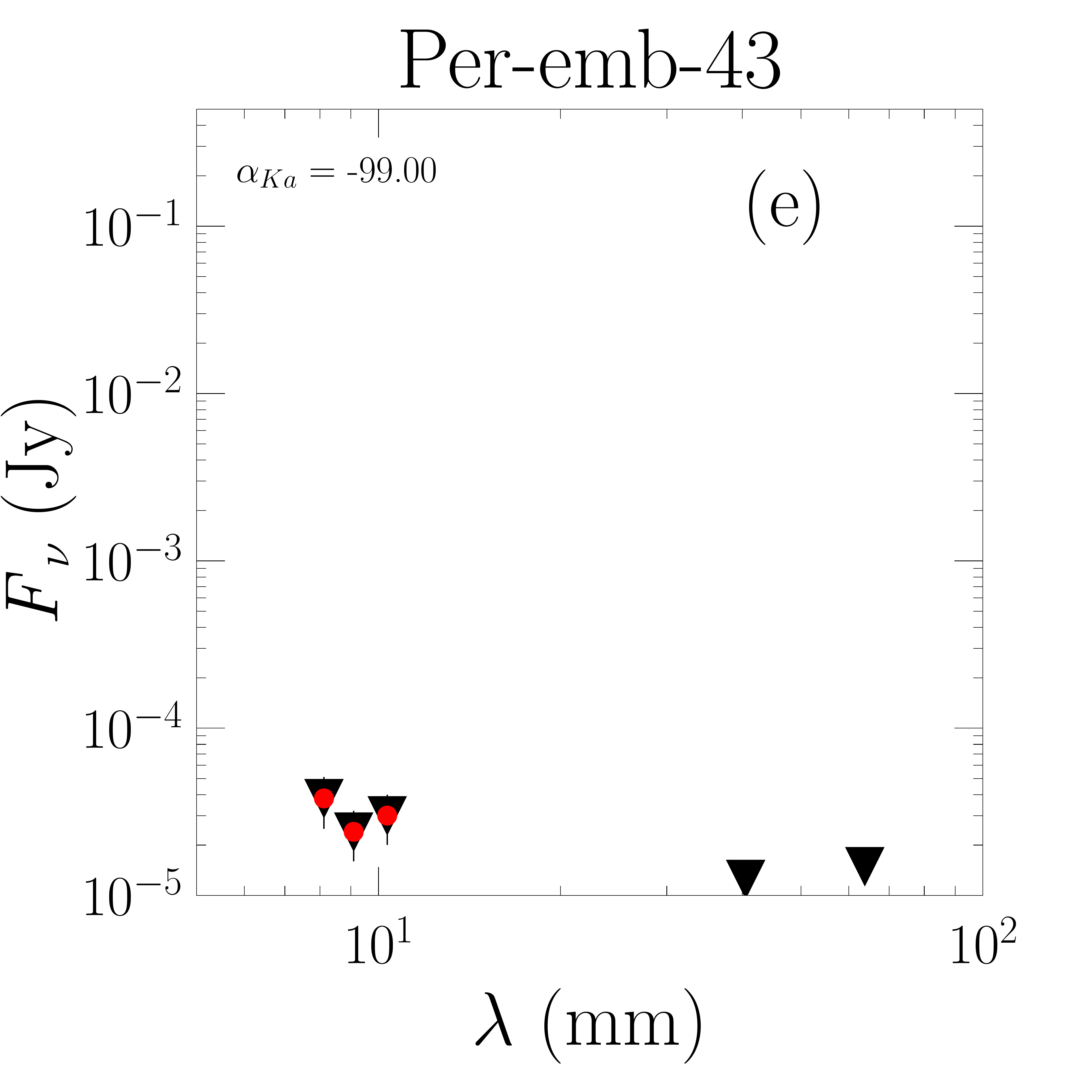}
  \includegraphics[width=0.30\linewidth]{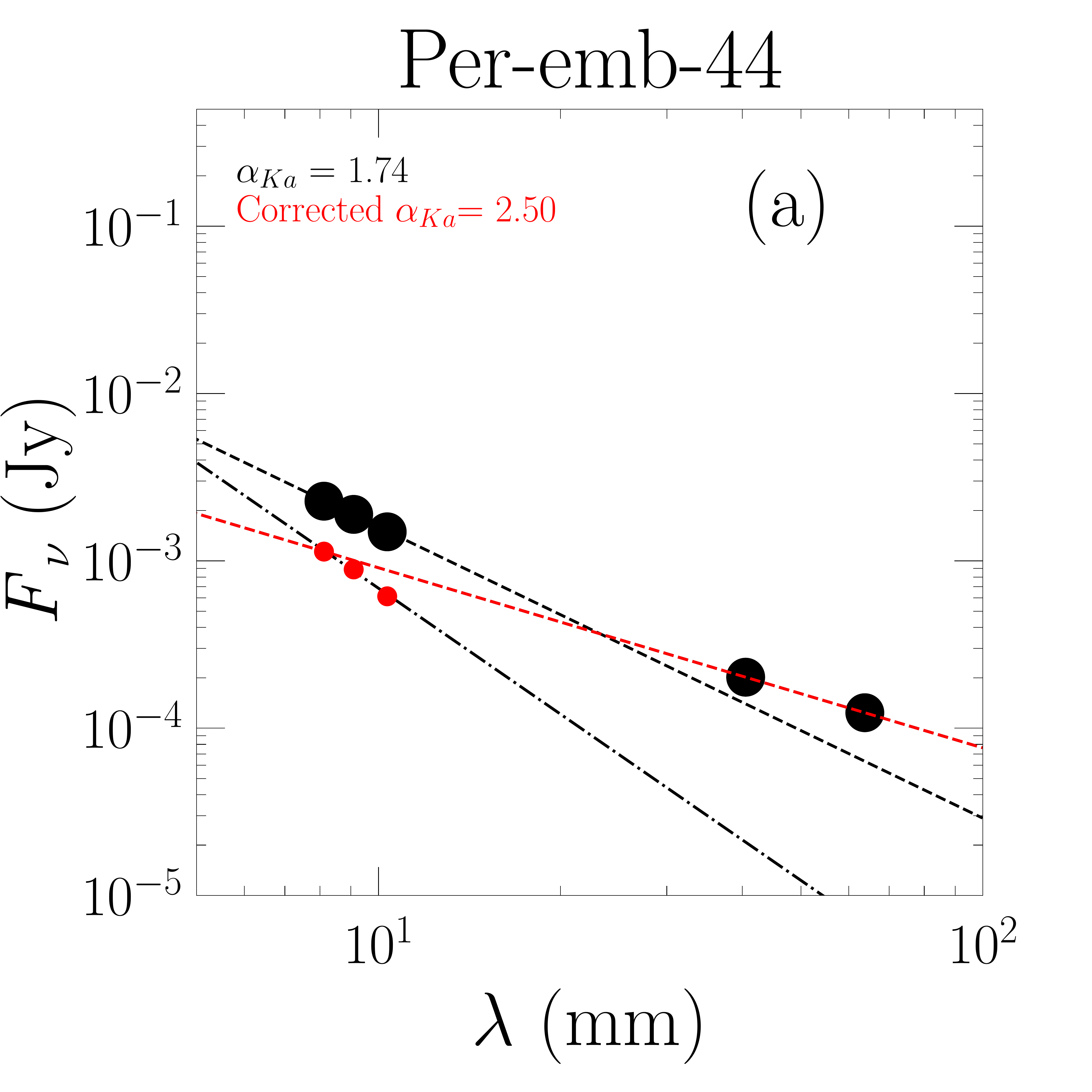}
  \includegraphics[width=0.30\linewidth]{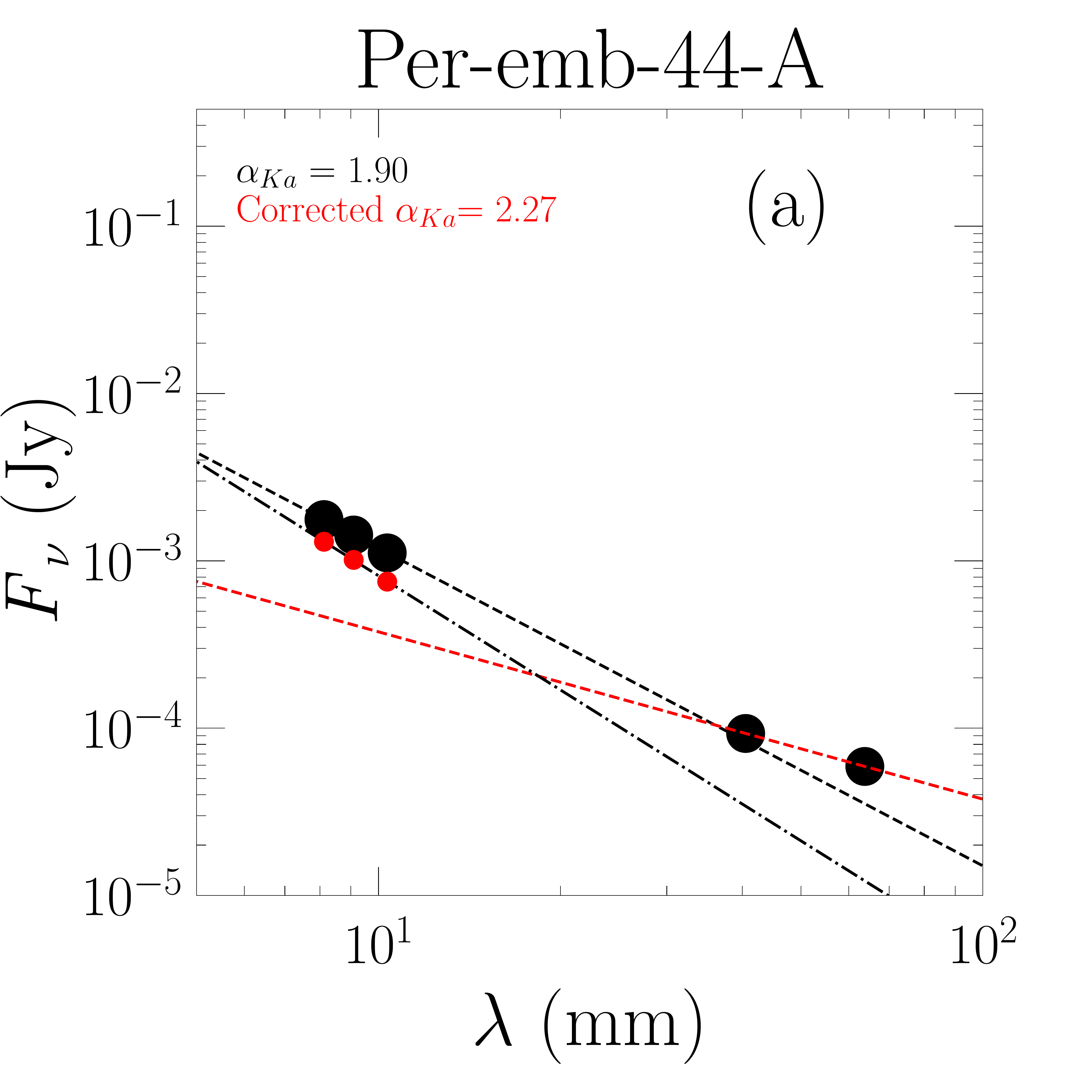}
  \includegraphics[width=0.30\linewidth]{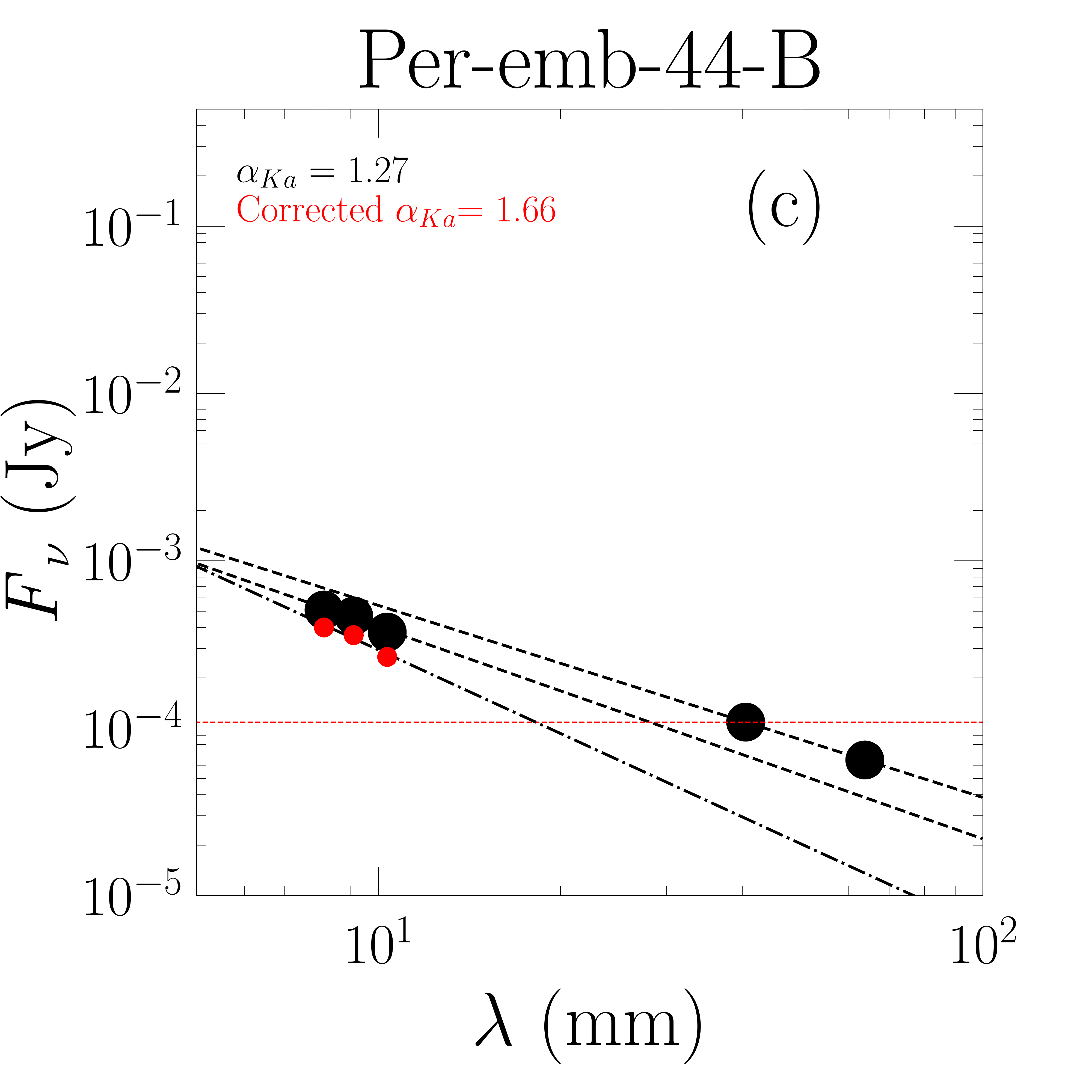}
  \includegraphics[width=0.30\linewidth]{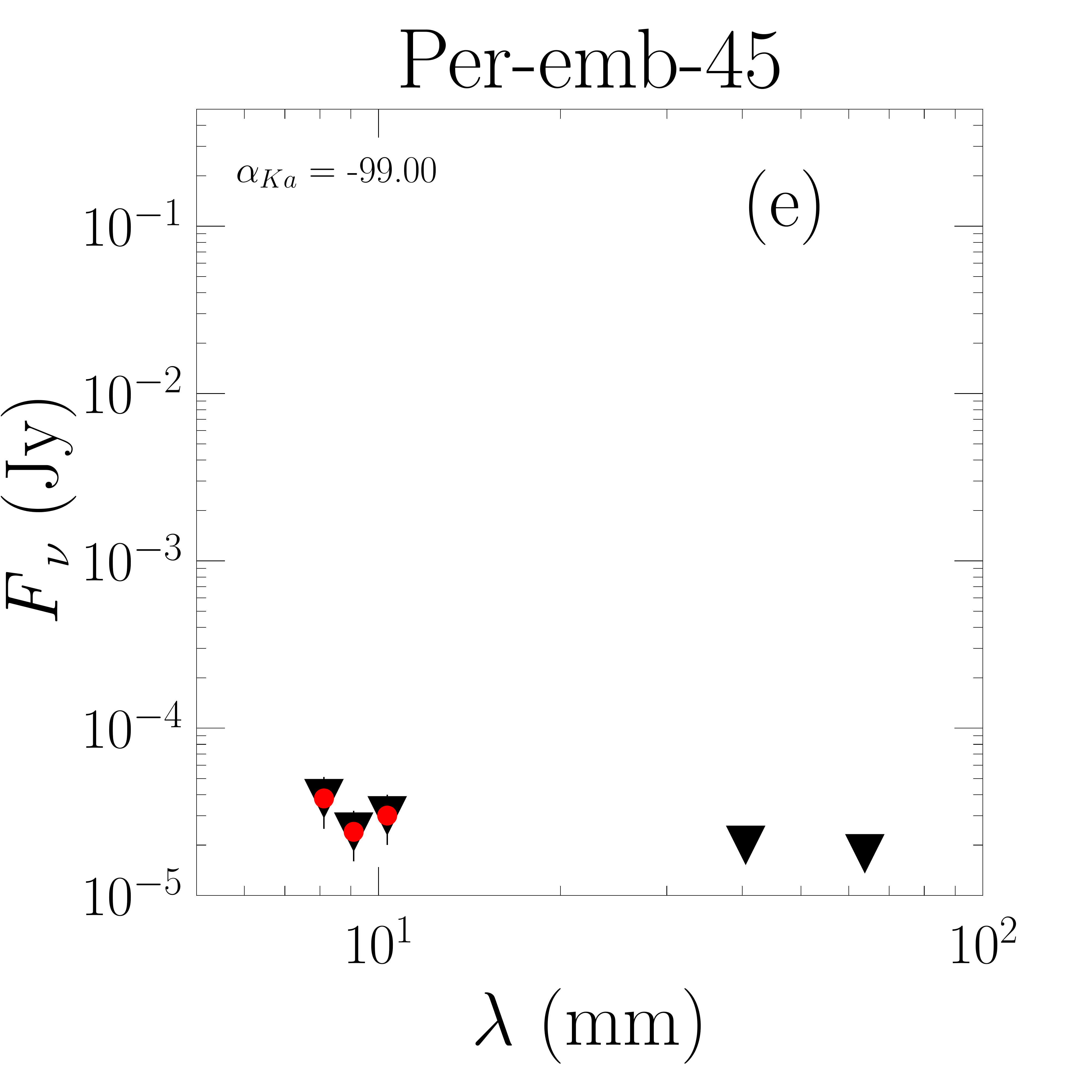}

\end{figure}
\begin{figure}[H]
\centering
  \includegraphics[width=0.30\linewidth]{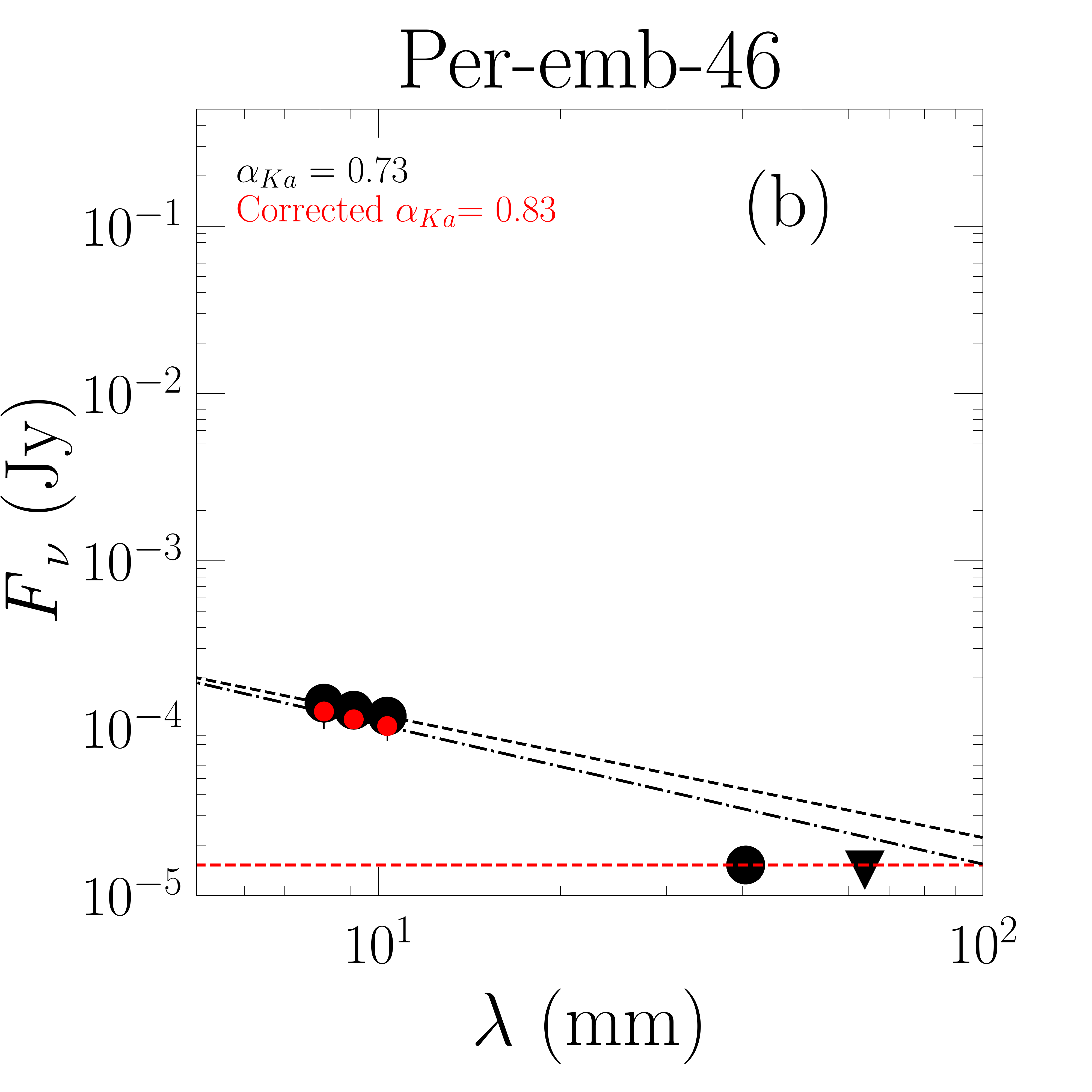}
  \includegraphics[width=0.30\linewidth]{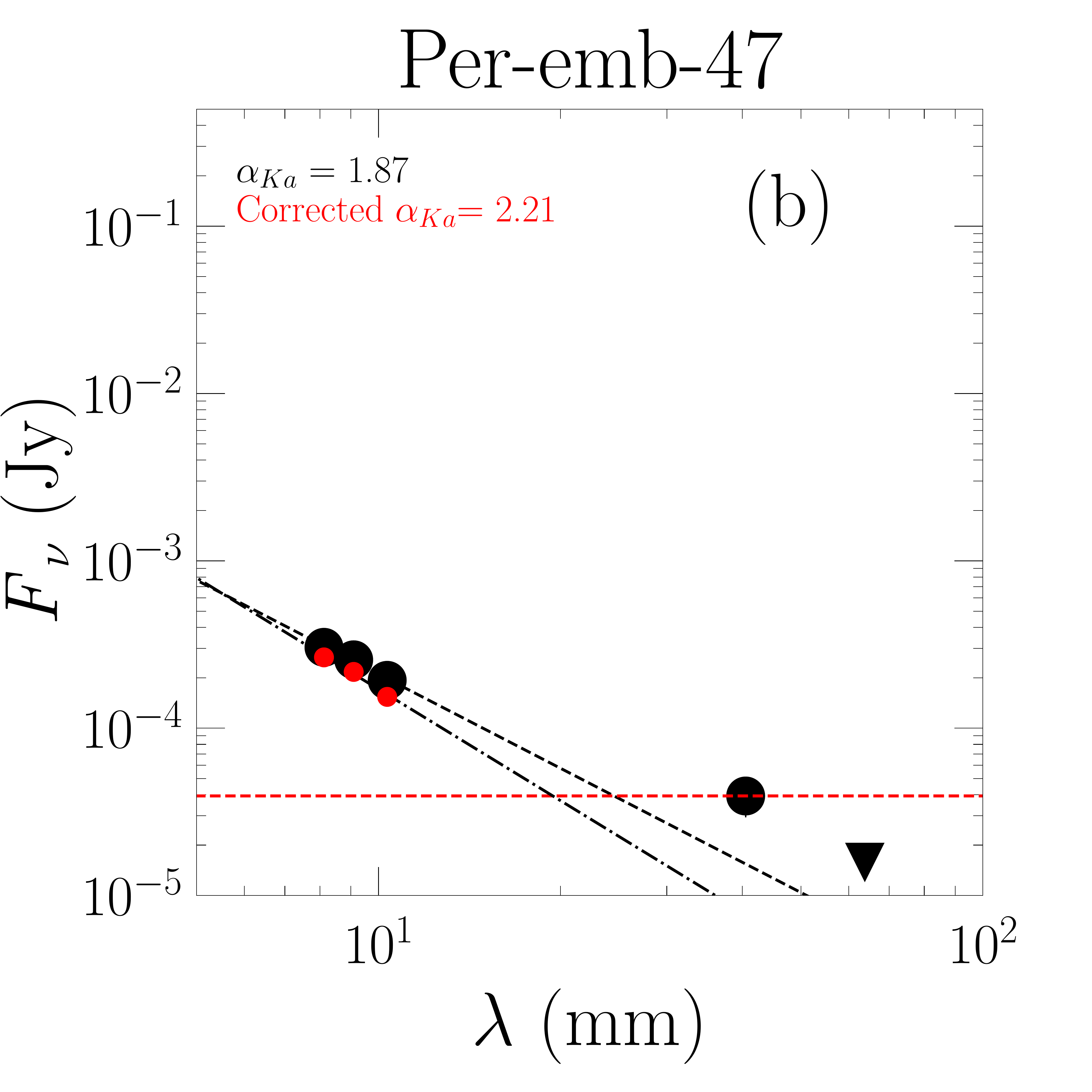}
  \includegraphics[width=0.30\linewidth]{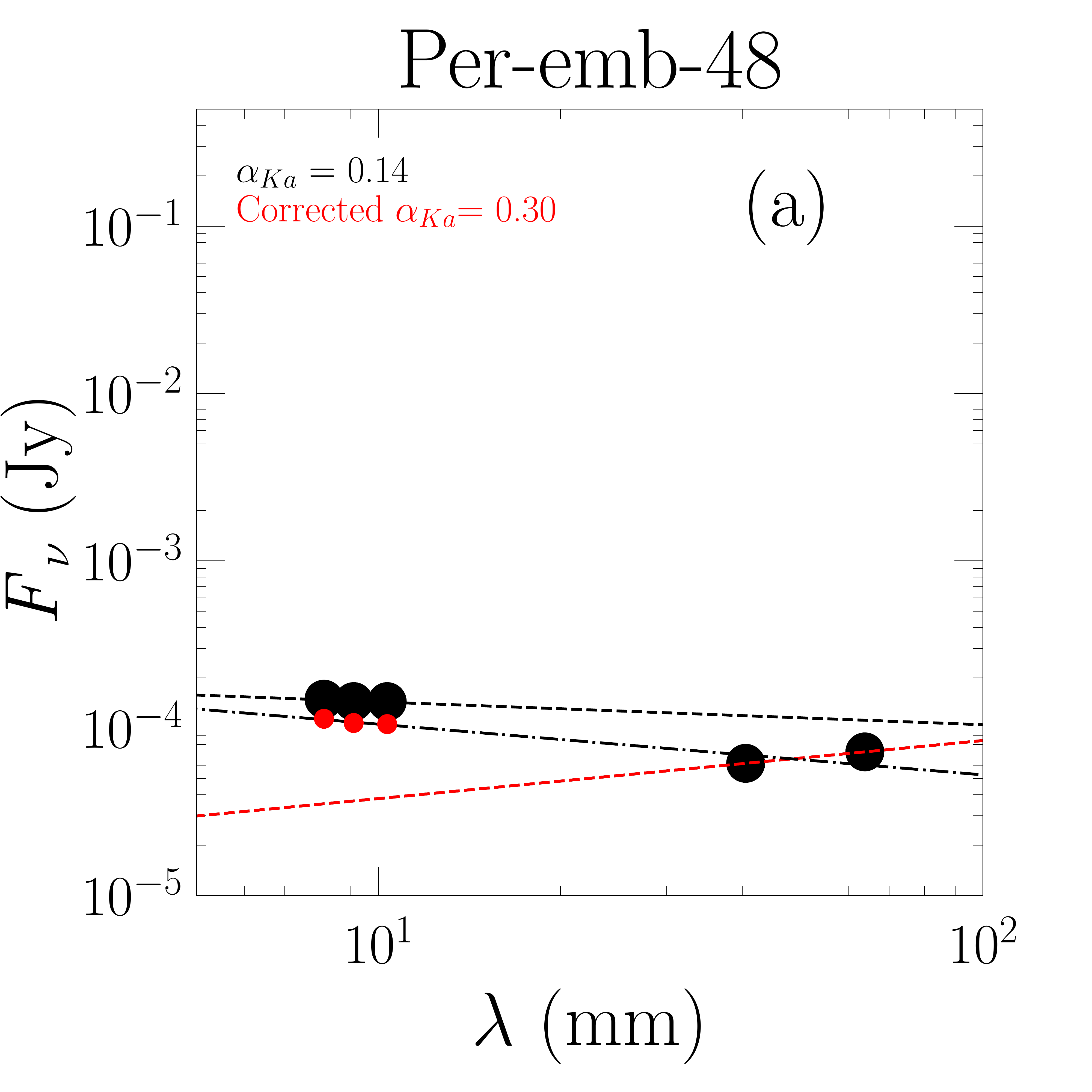}
  \includegraphics[width=0.30\linewidth]{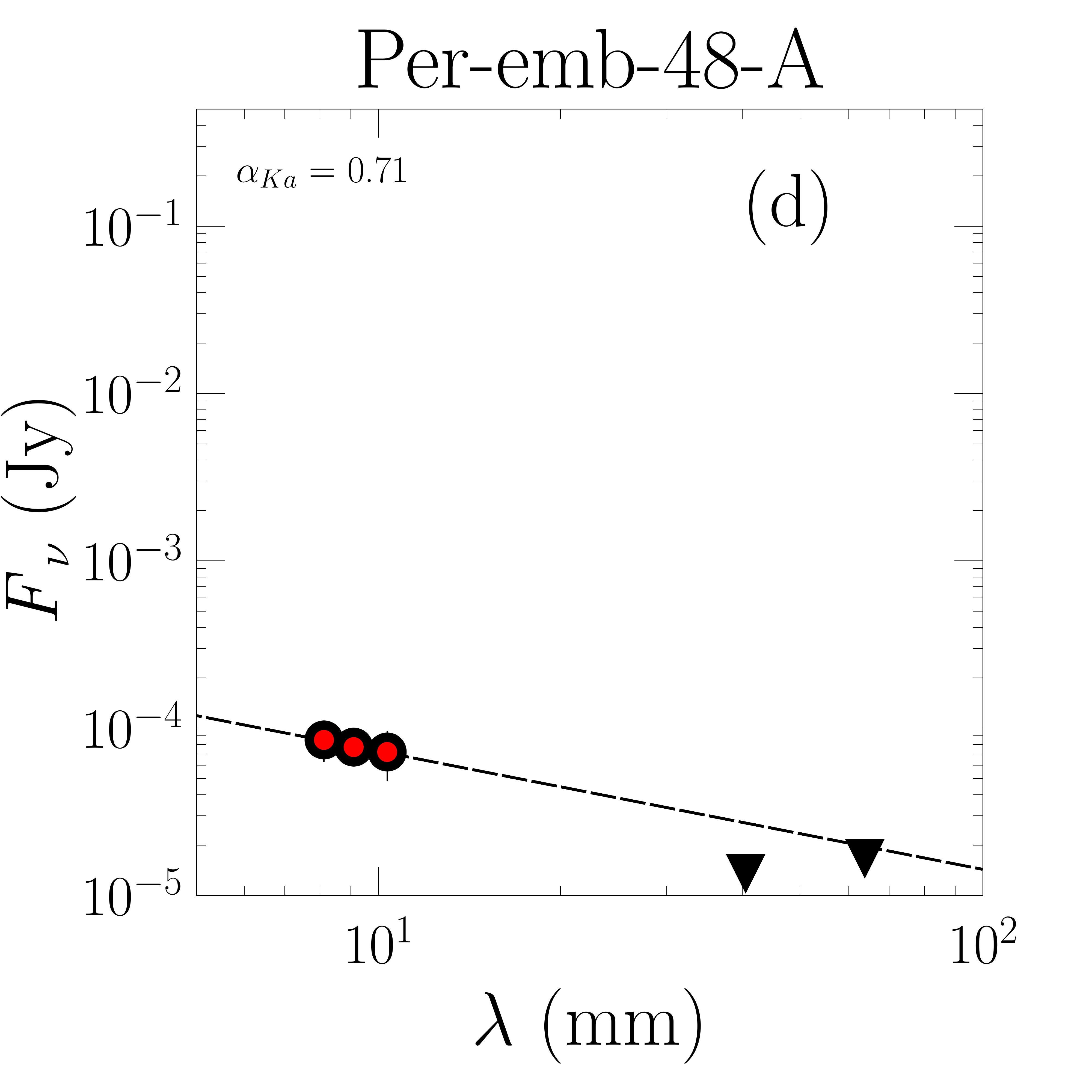}
  \includegraphics[width=0.30\linewidth]{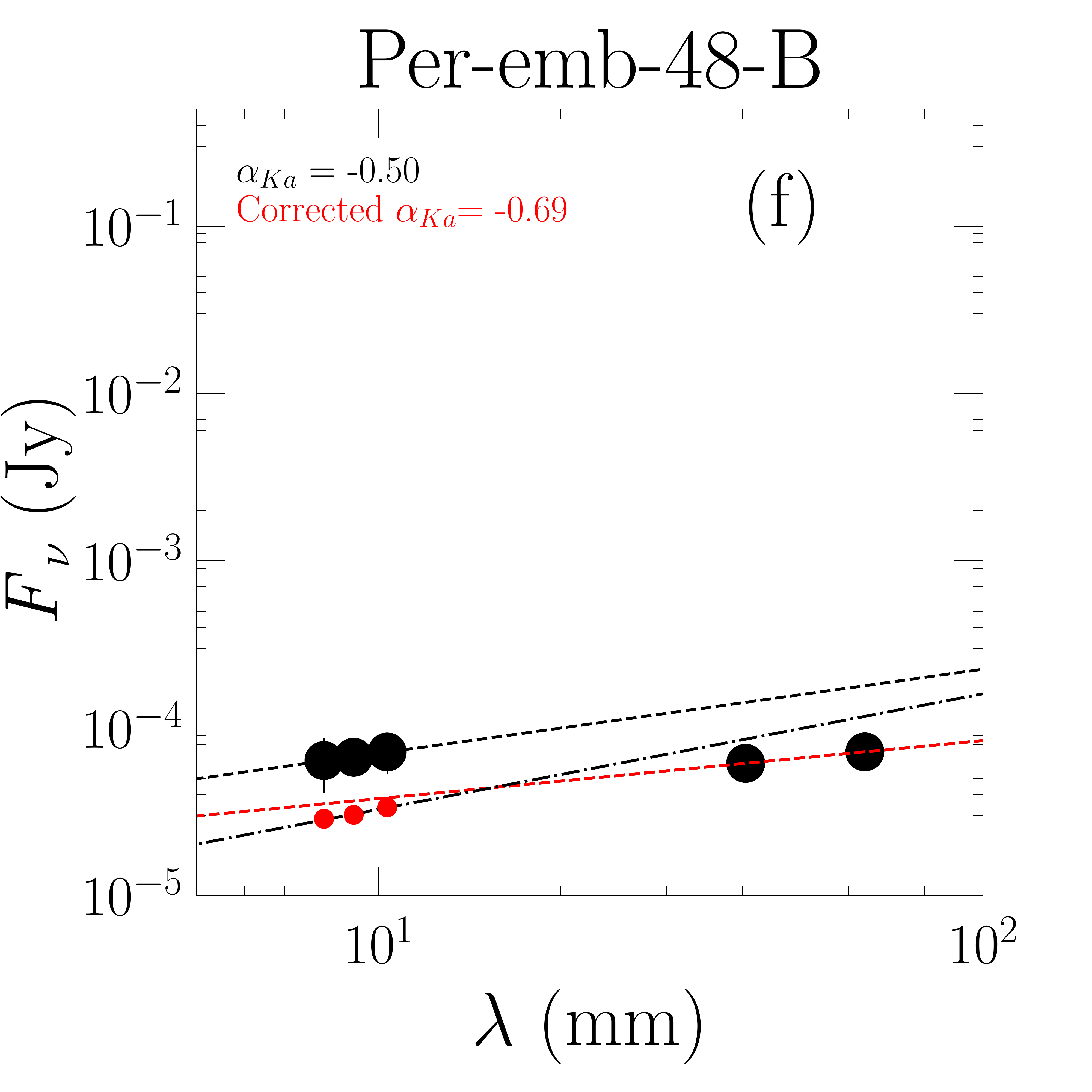}
  \includegraphics[width=0.30\linewidth]{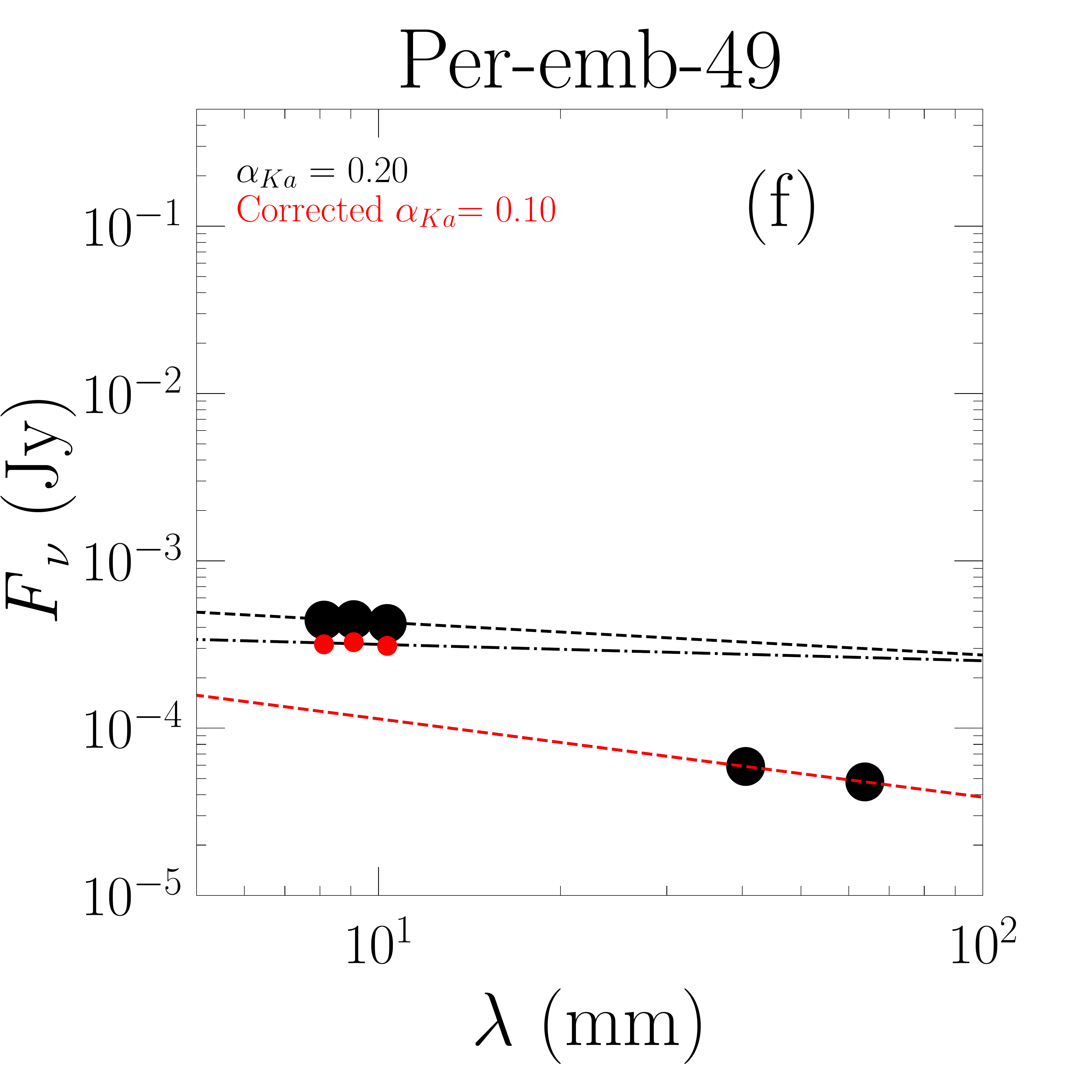}
  \includegraphics[width=0.30\linewidth]{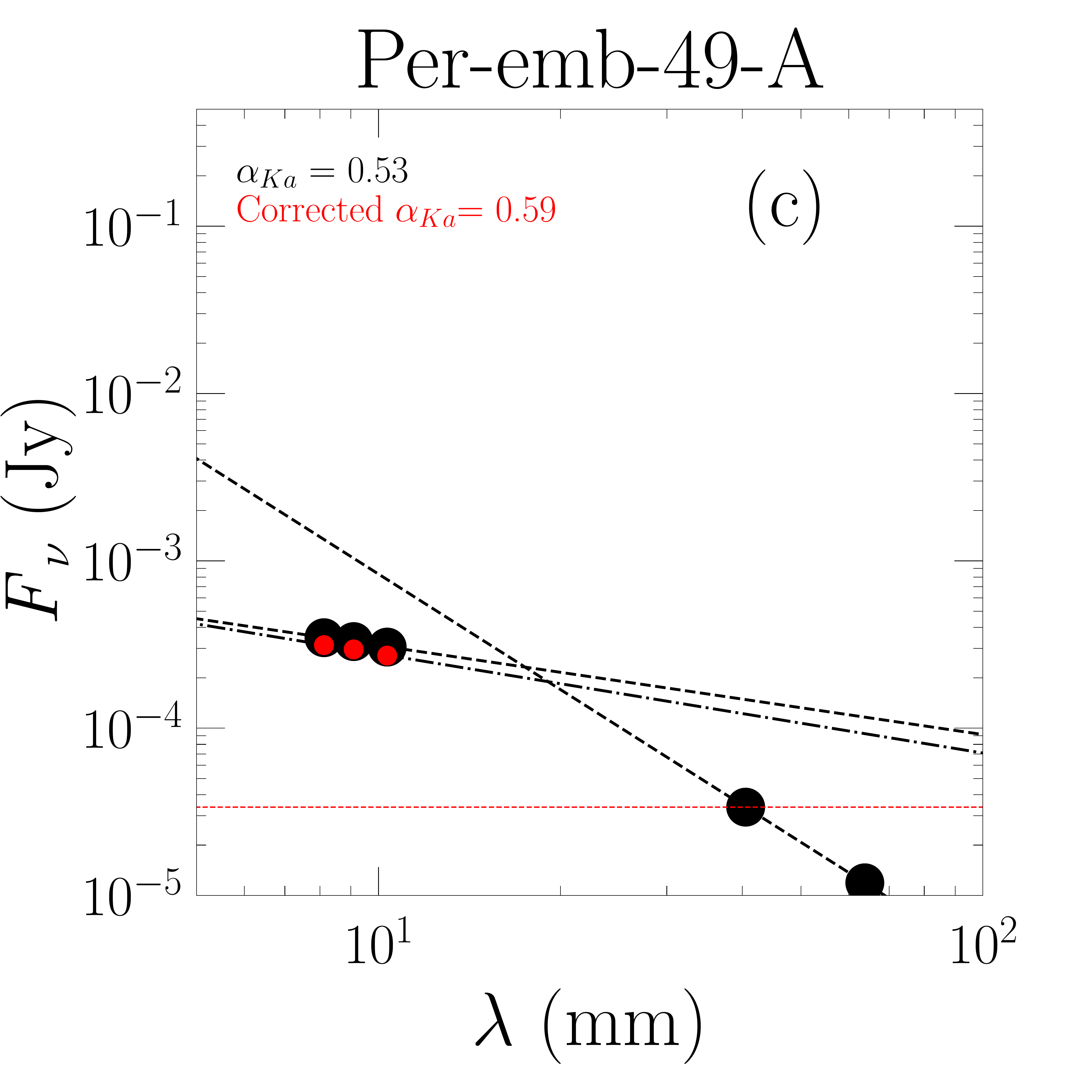}
  \includegraphics[width=0.30\linewidth]{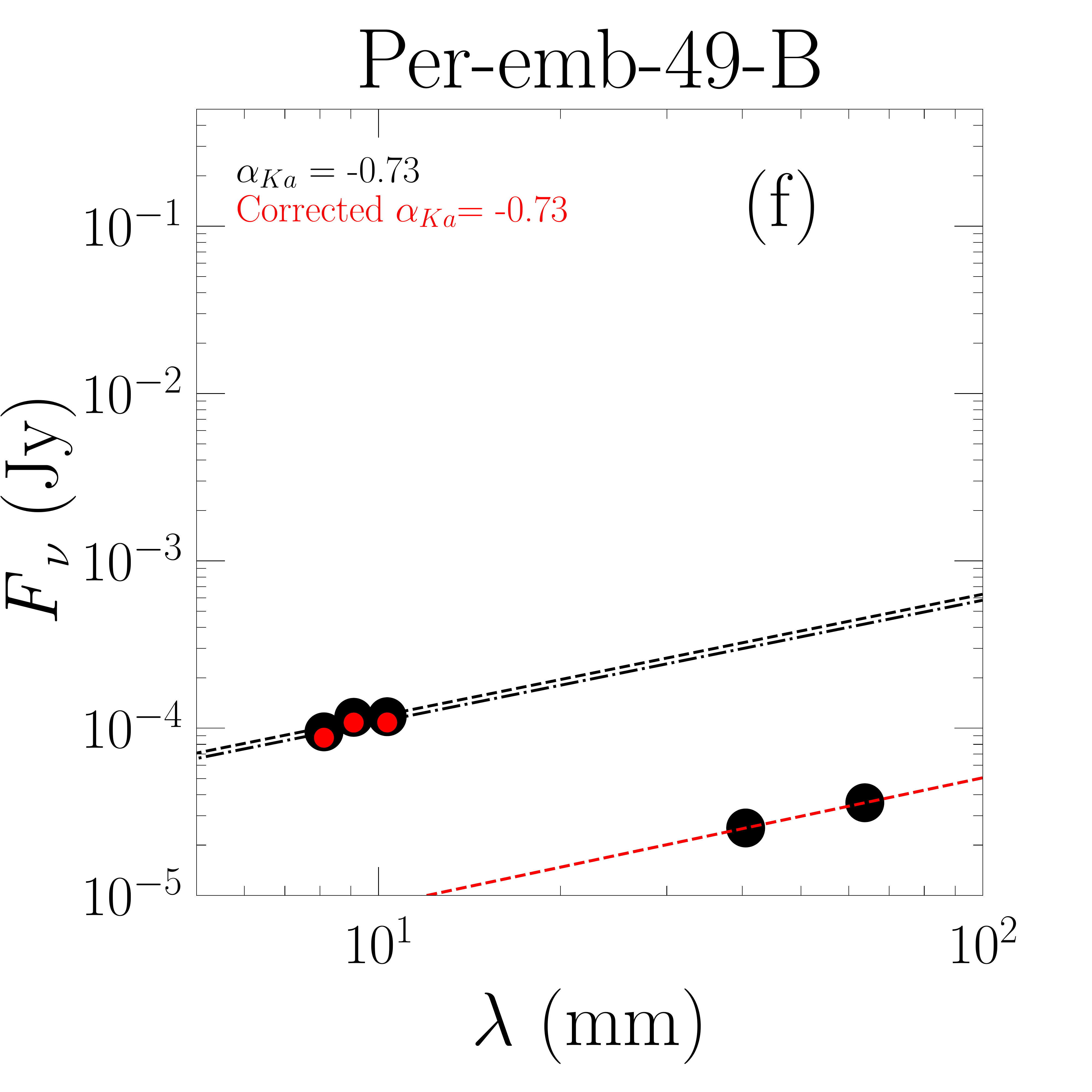}
  \includegraphics[width=0.30\linewidth]{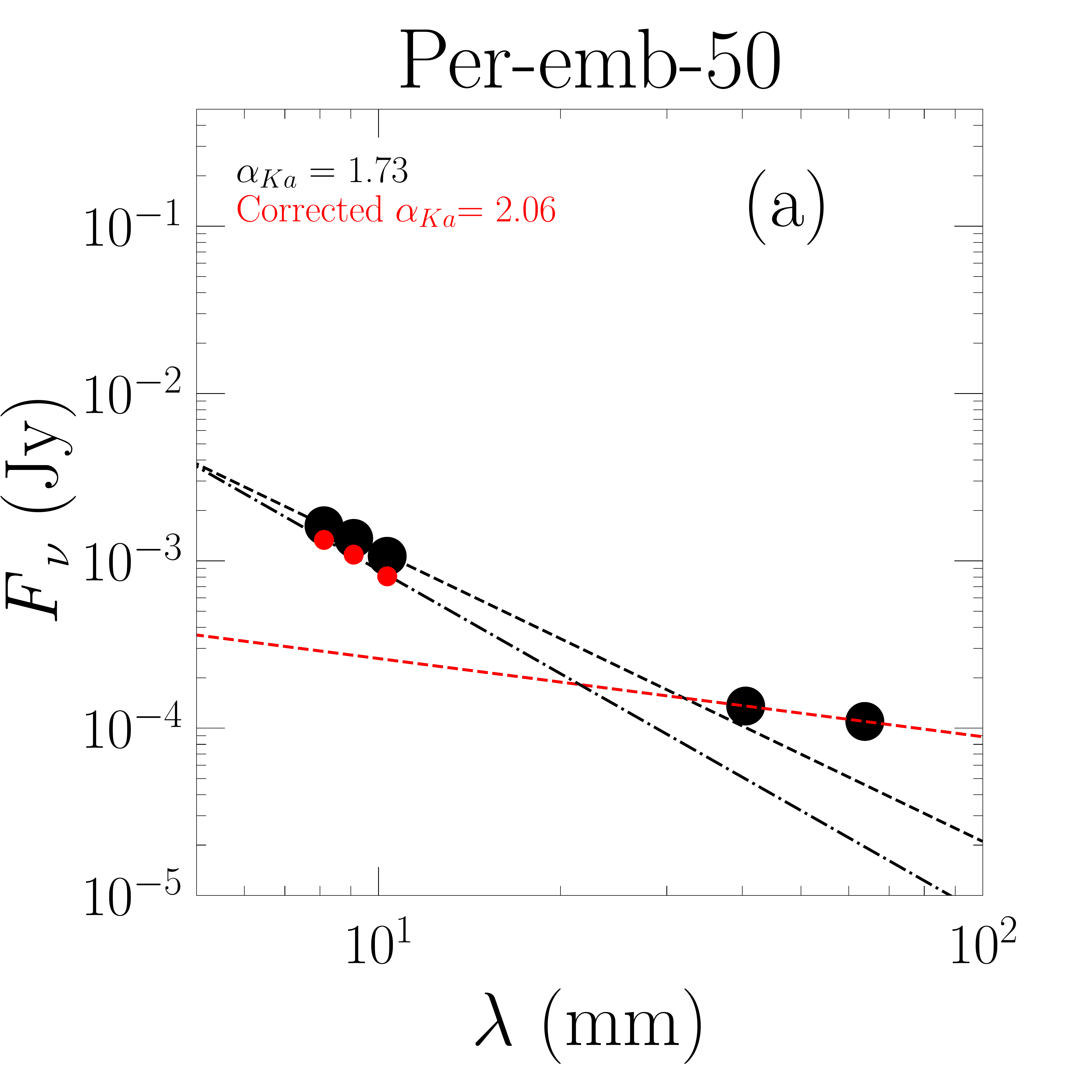}
  \includegraphics[width=0.30\linewidth]{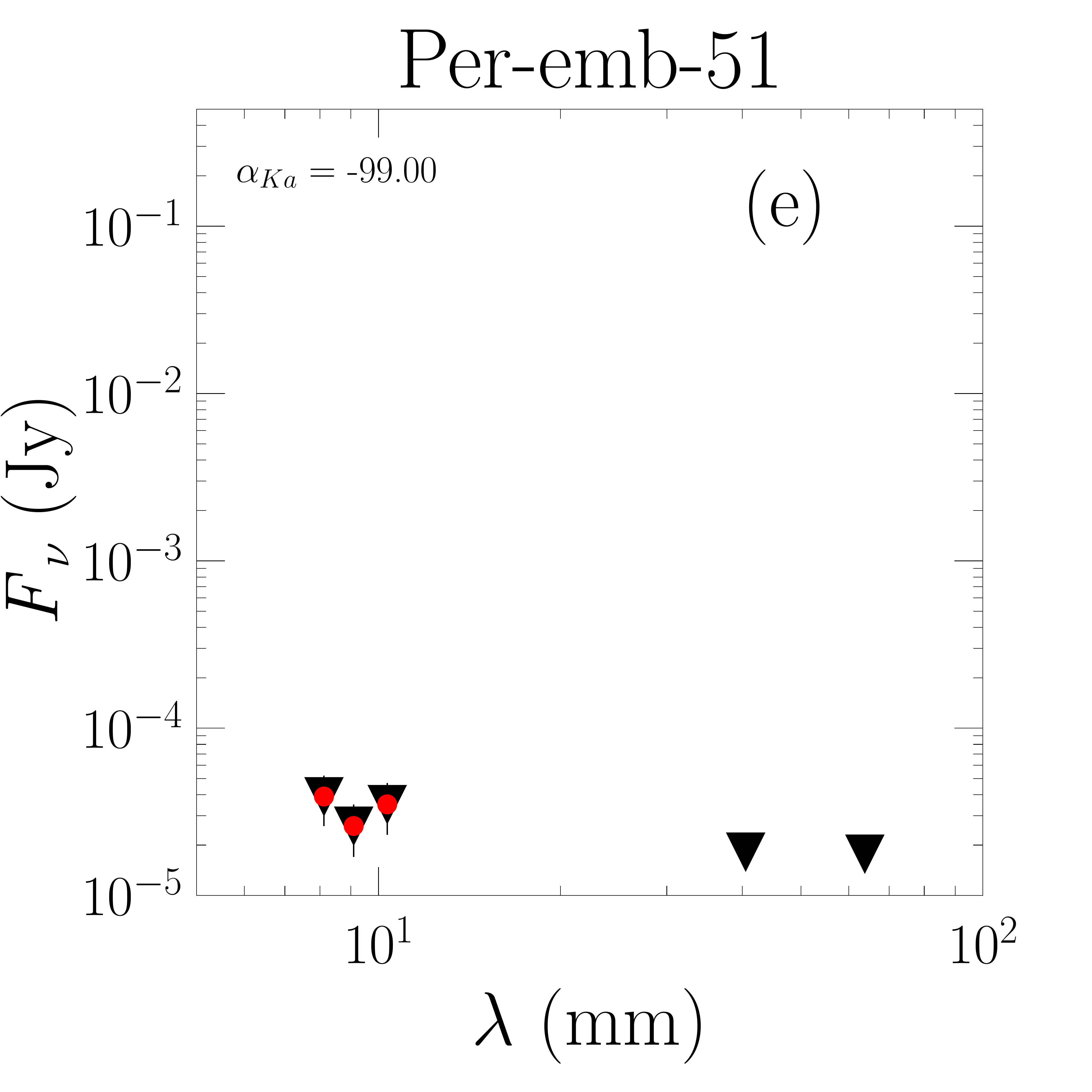}
  \includegraphics[width=0.30\linewidth]{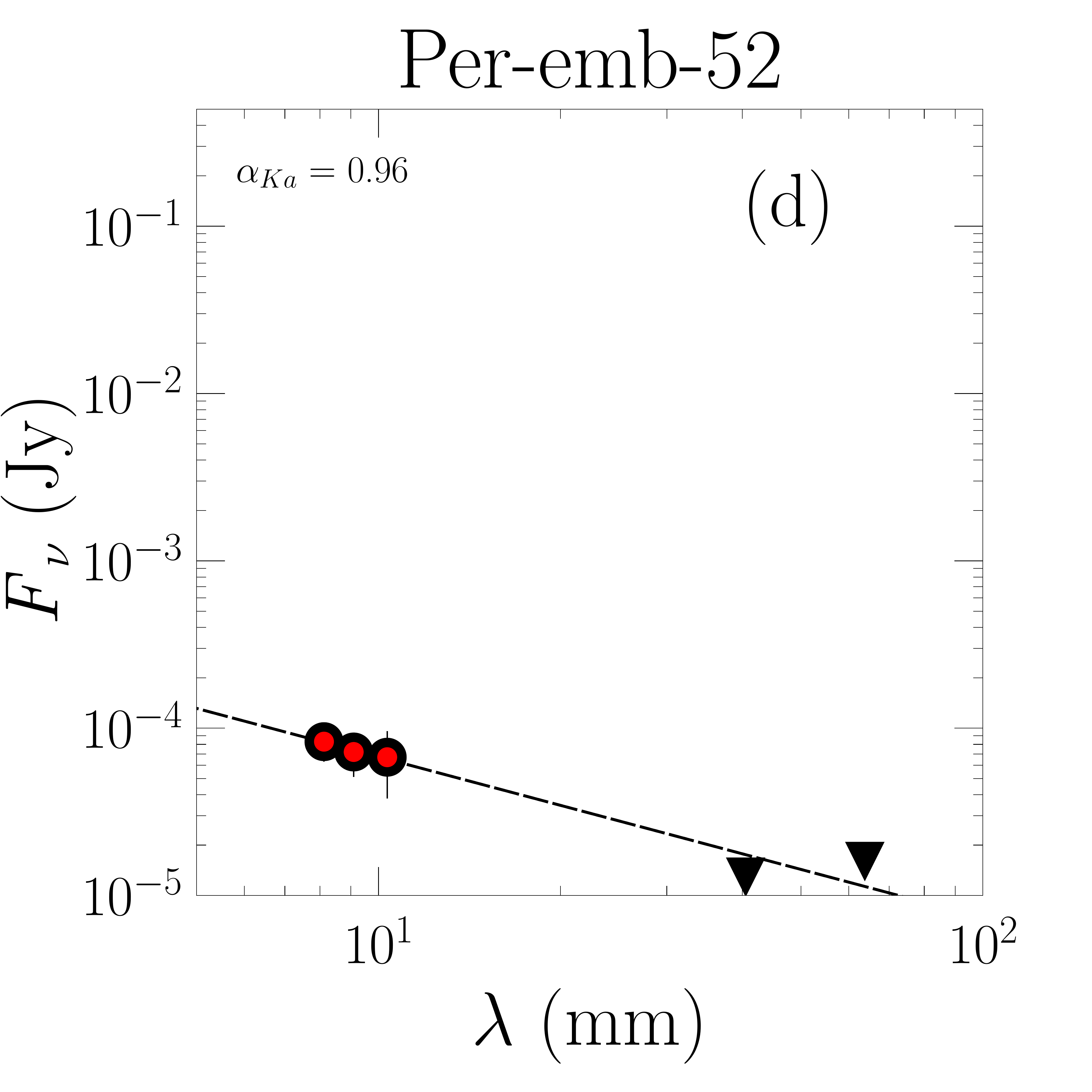}
  \includegraphics[width=0.30\linewidth]{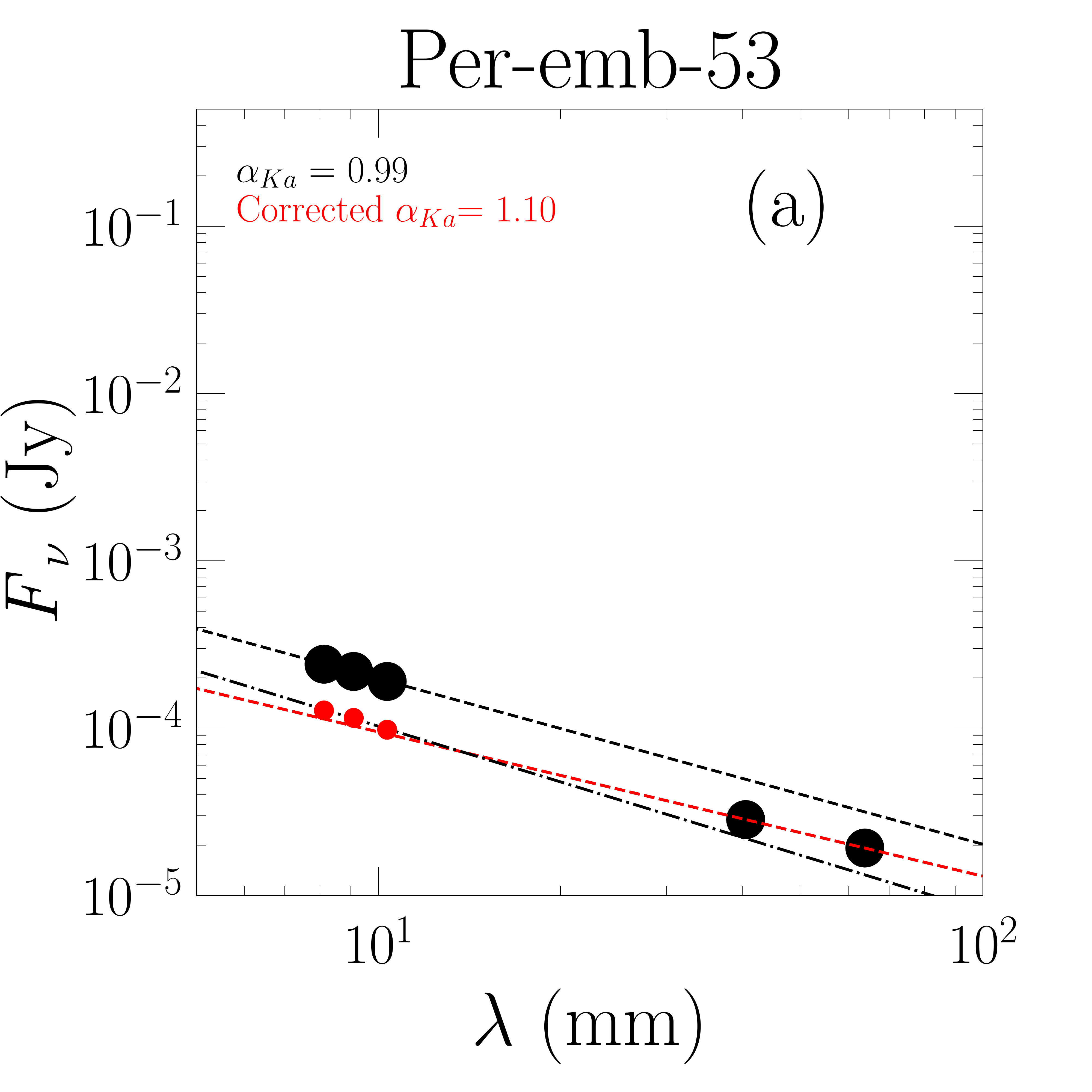}

\end{figure}
\begin{figure}[H]
\centering
  \includegraphics[width=0.30\linewidth]{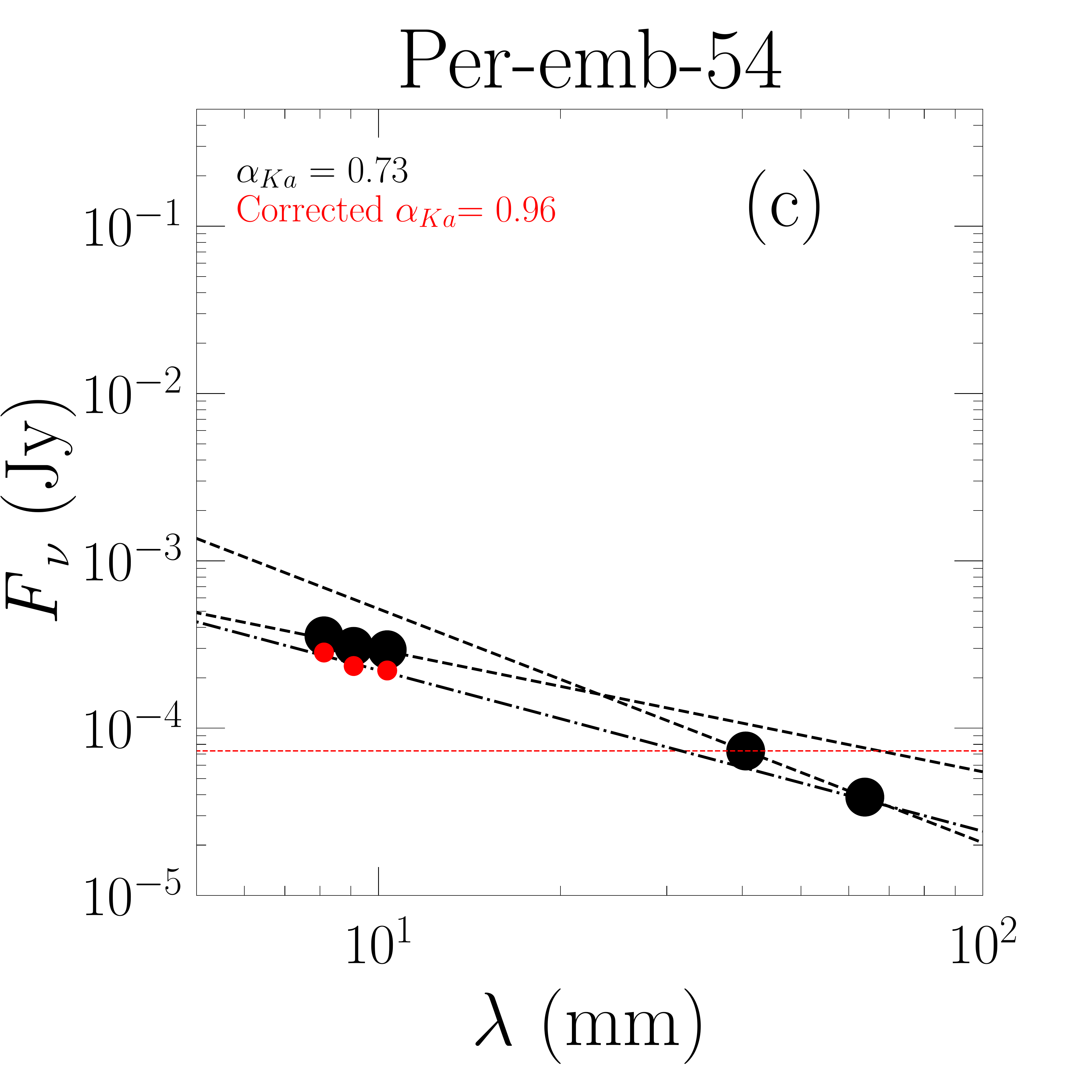}
  \includegraphics[width=0.30\linewidth]{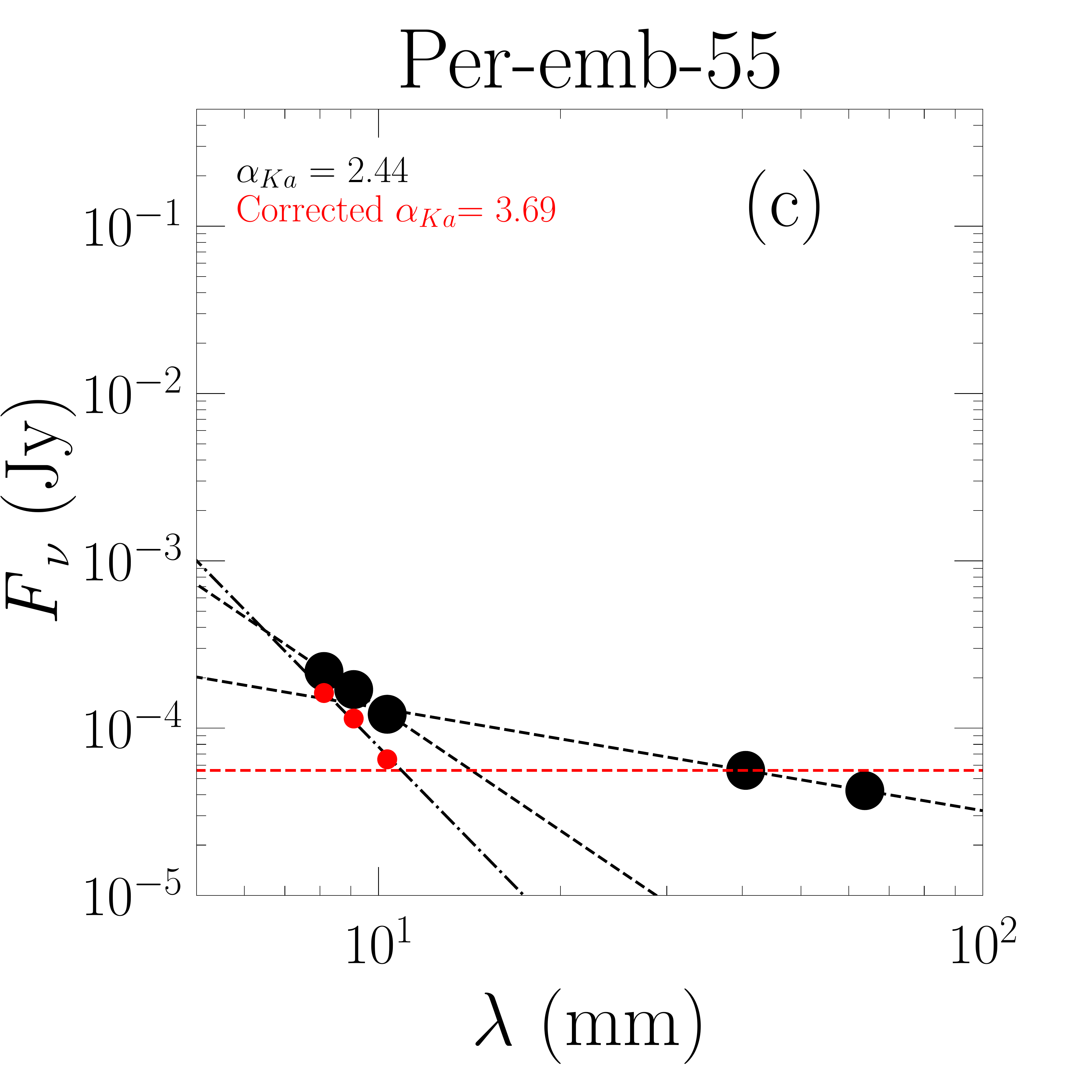}
  \includegraphics[width=0.30\linewidth]{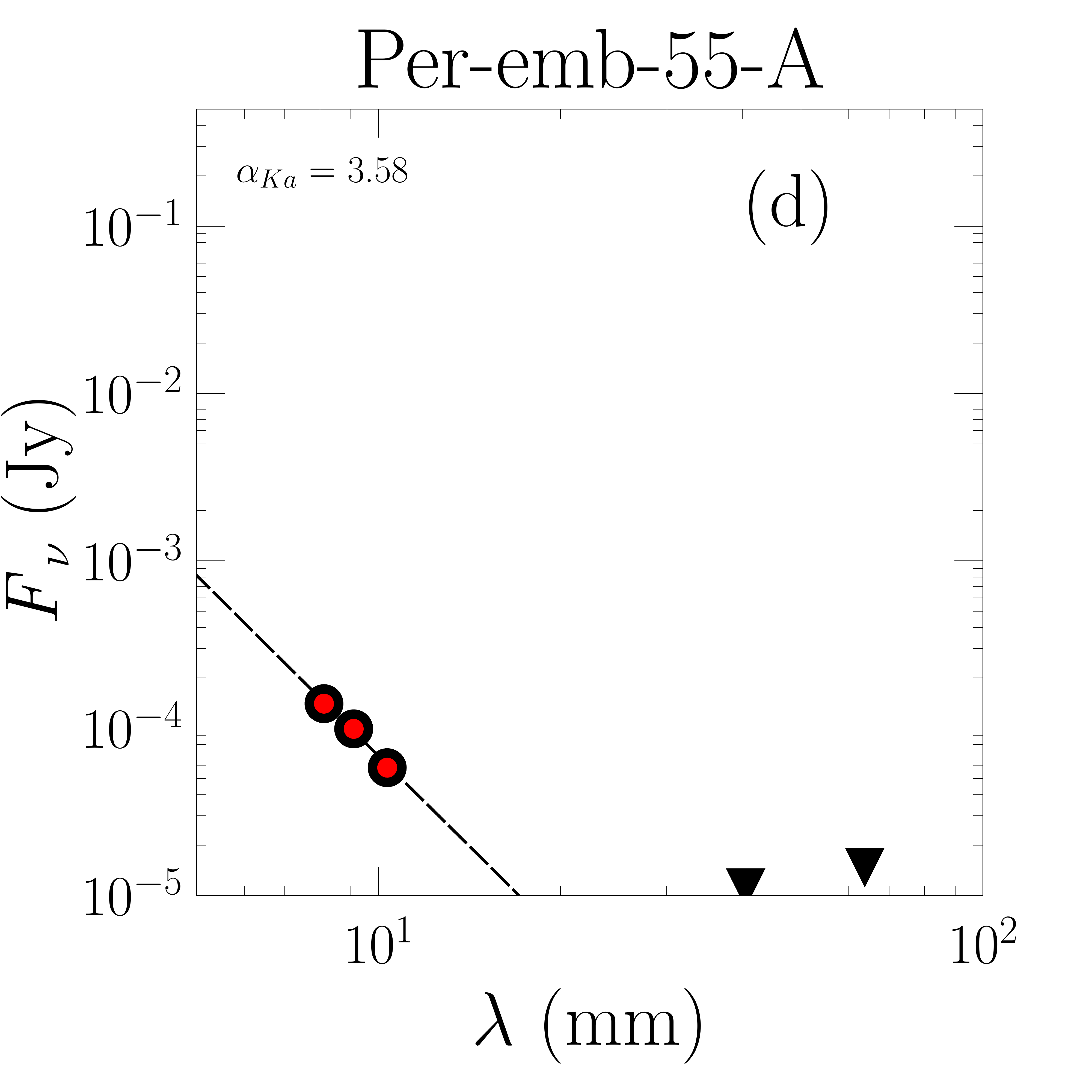}
  \includegraphics[width=0.30\linewidth]{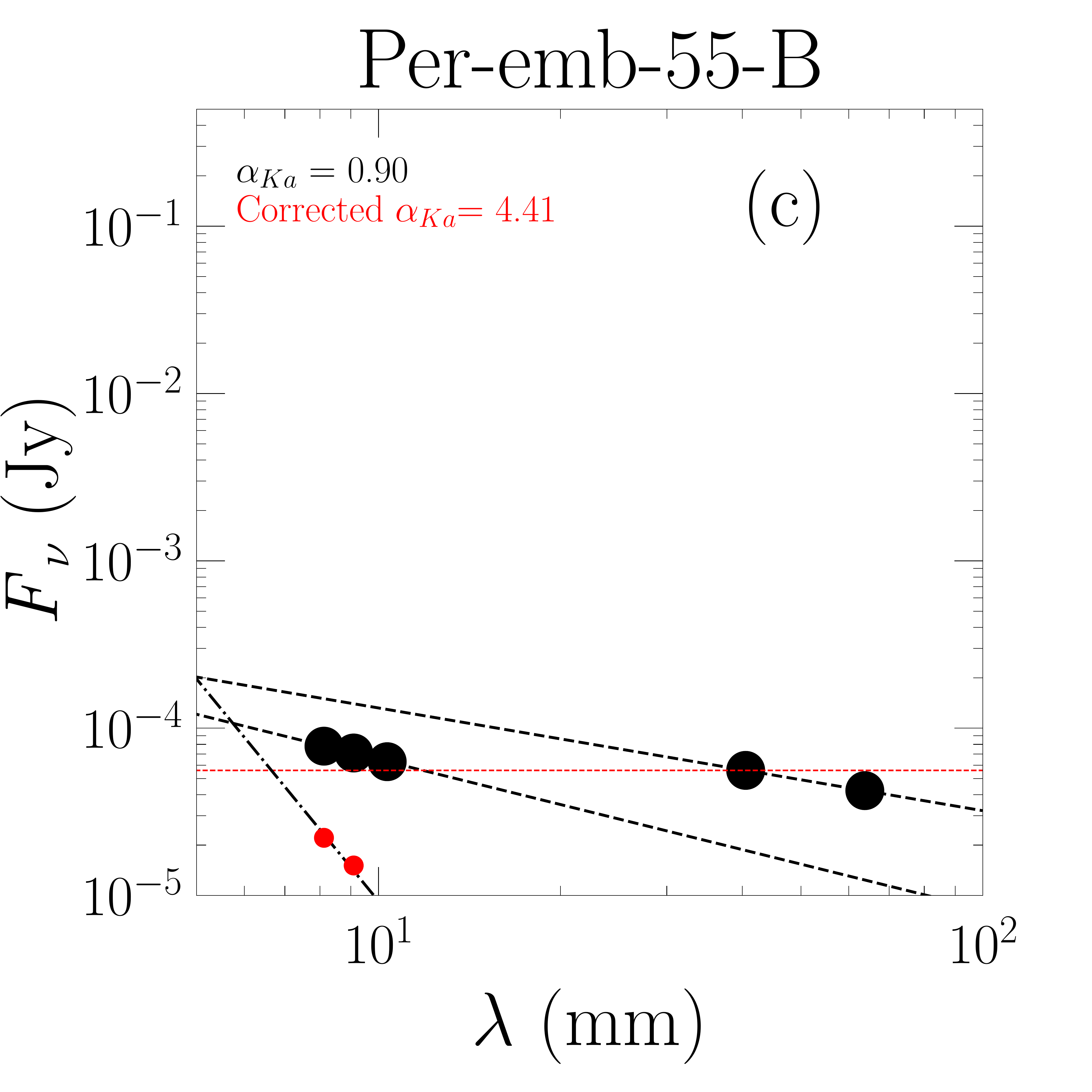}
  \includegraphics[width=0.30\linewidth]{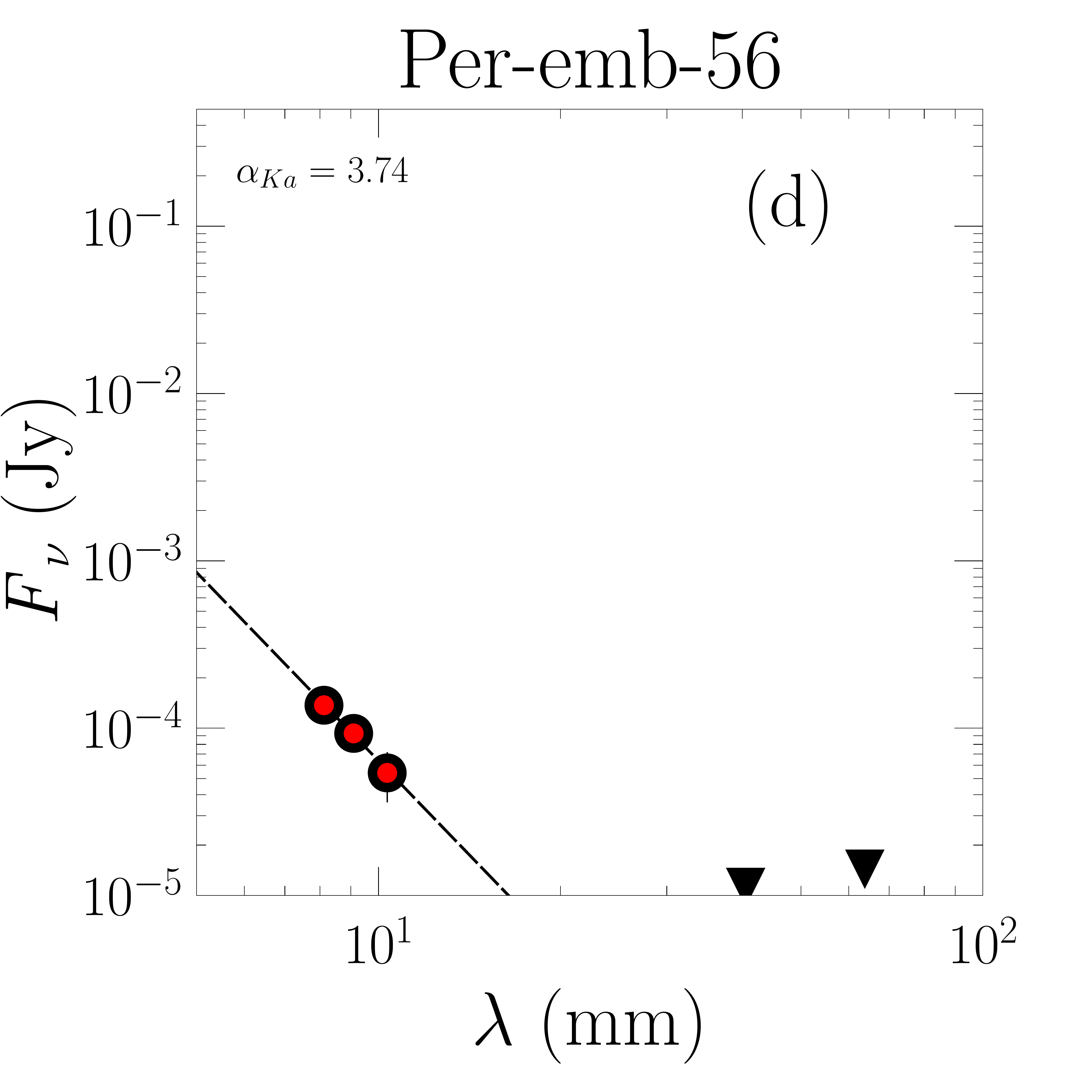}
  \includegraphics[width=0.30\linewidth]{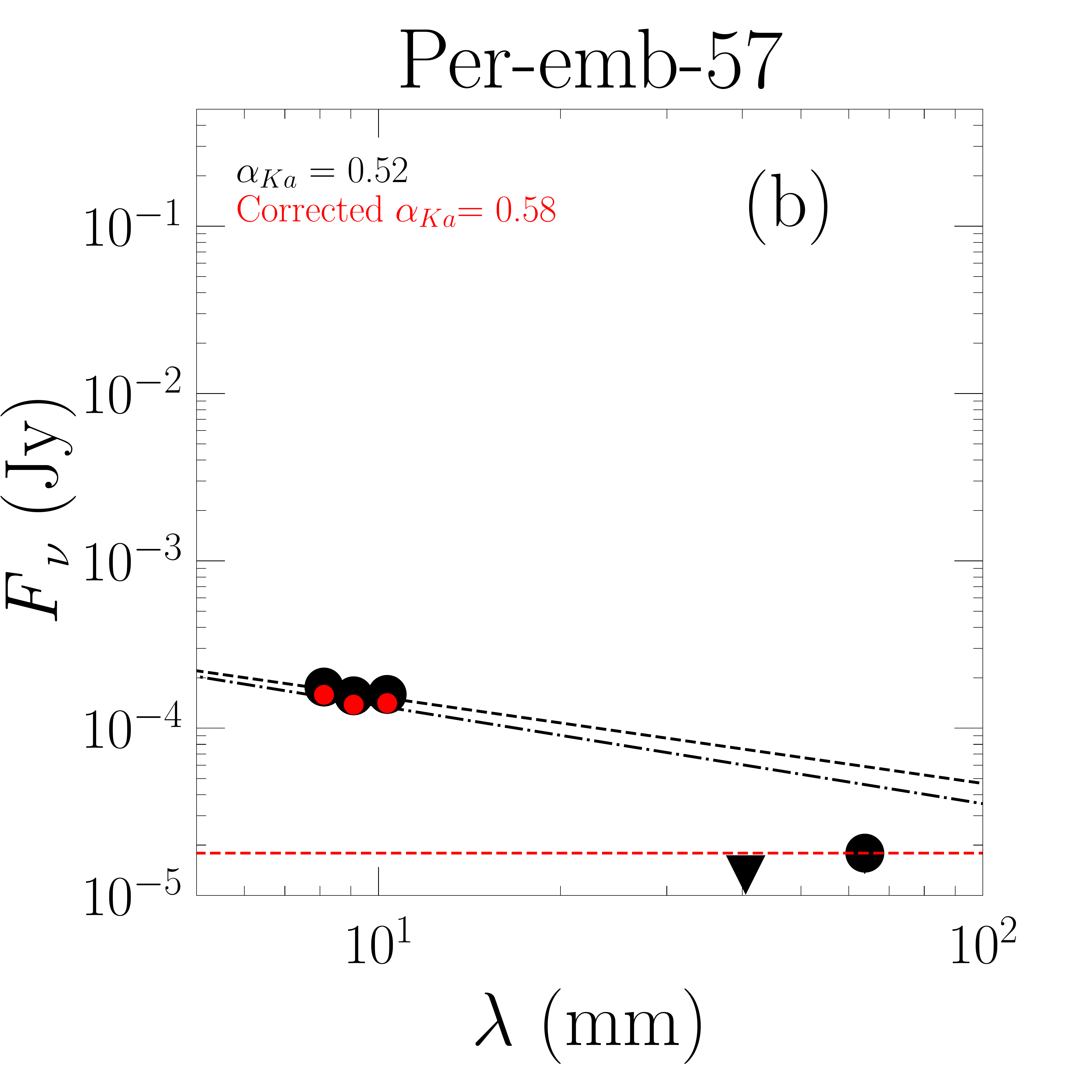}
  \includegraphics[width=0.30\linewidth]{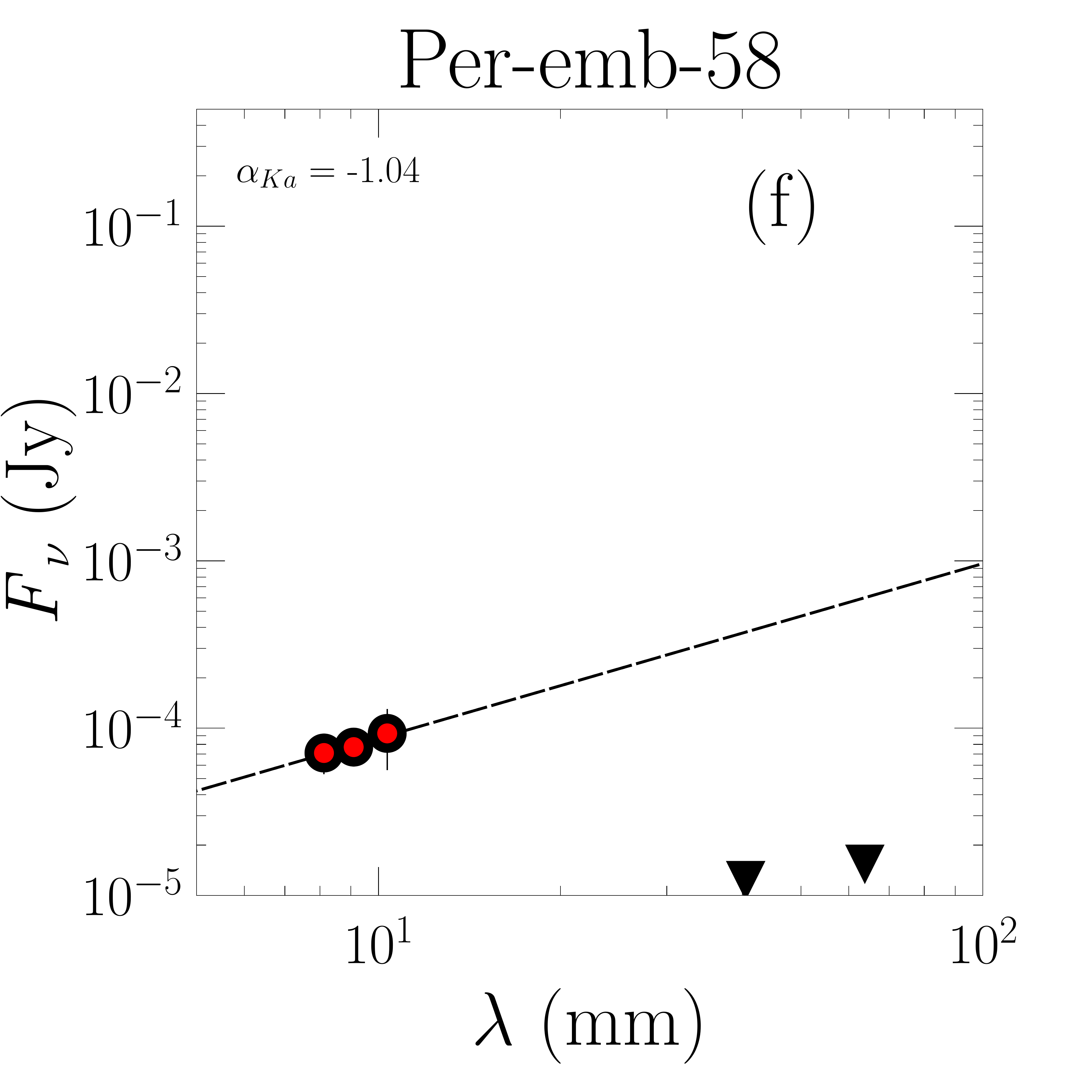}
  \includegraphics[width=0.30\linewidth]{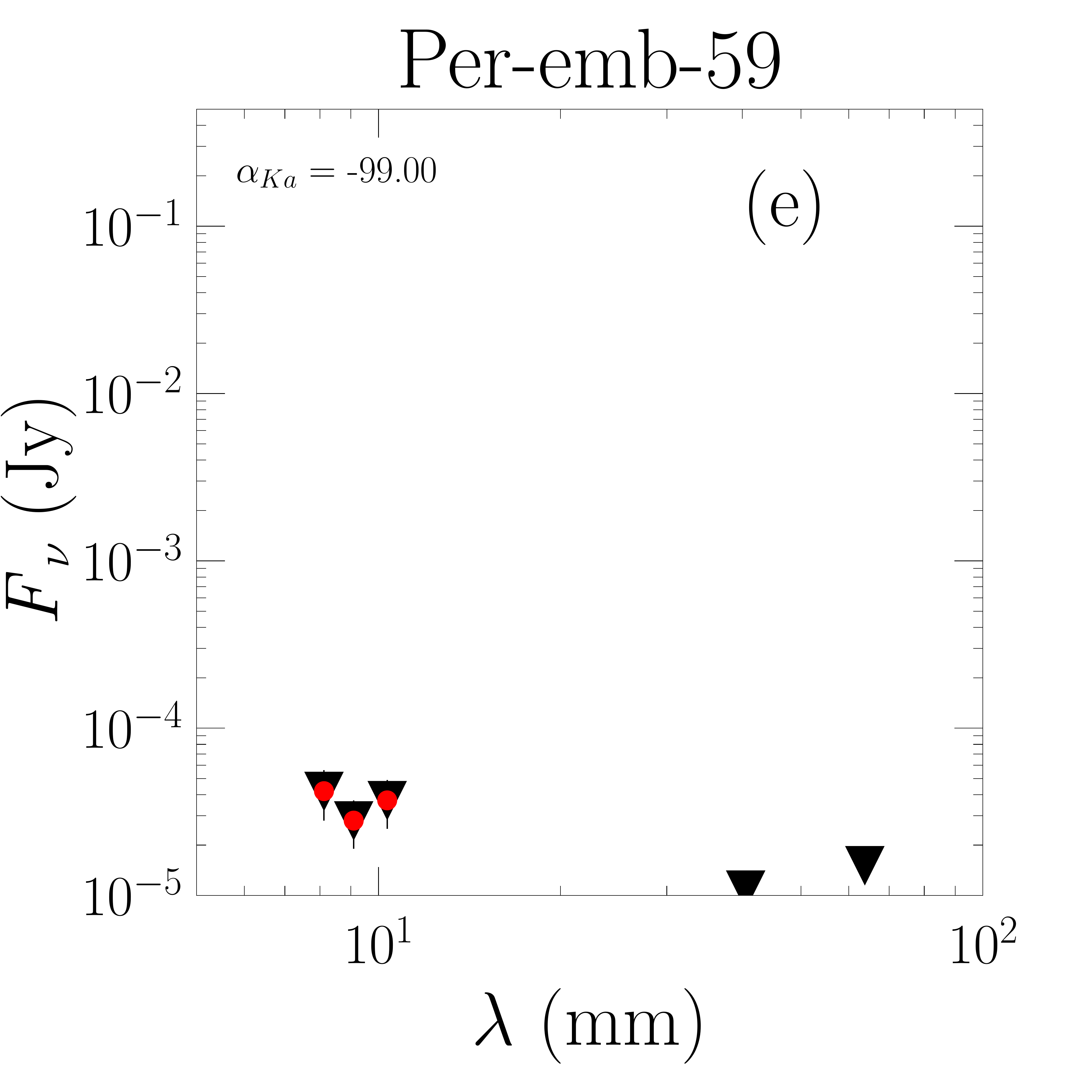}
  \includegraphics[width=0.30\linewidth]{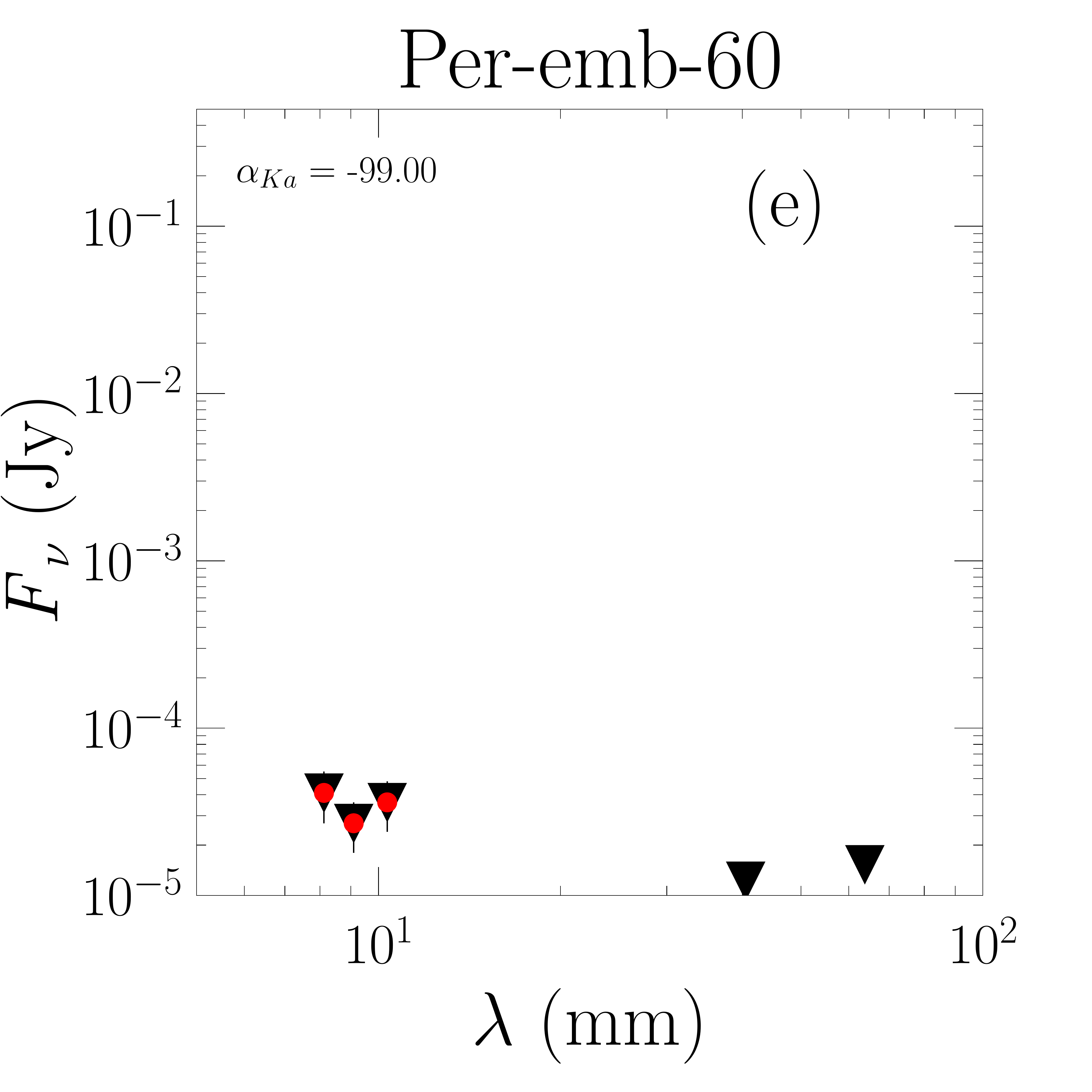}
  \includegraphics[width=0.30\linewidth]{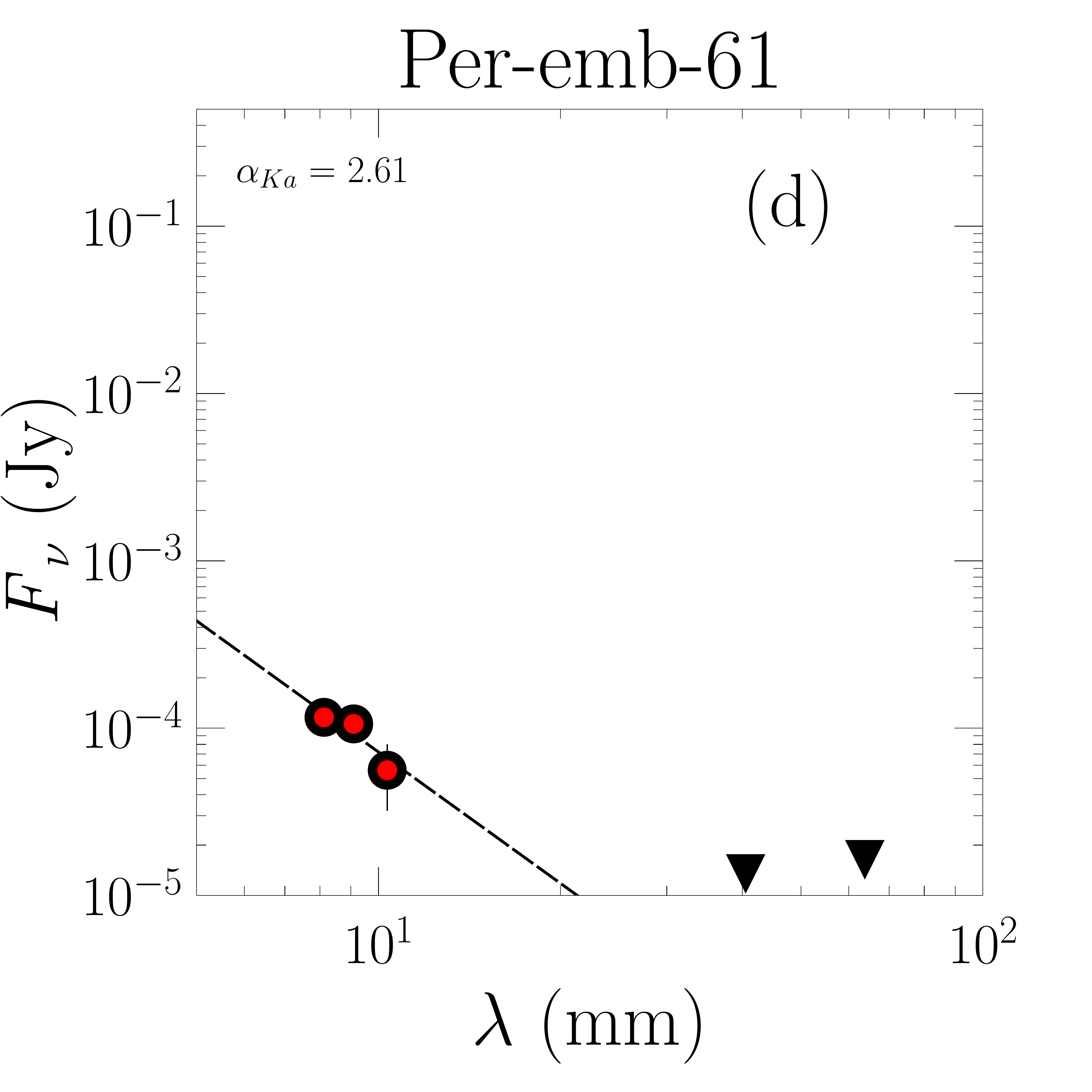}
  \includegraphics[width=0.30\linewidth]{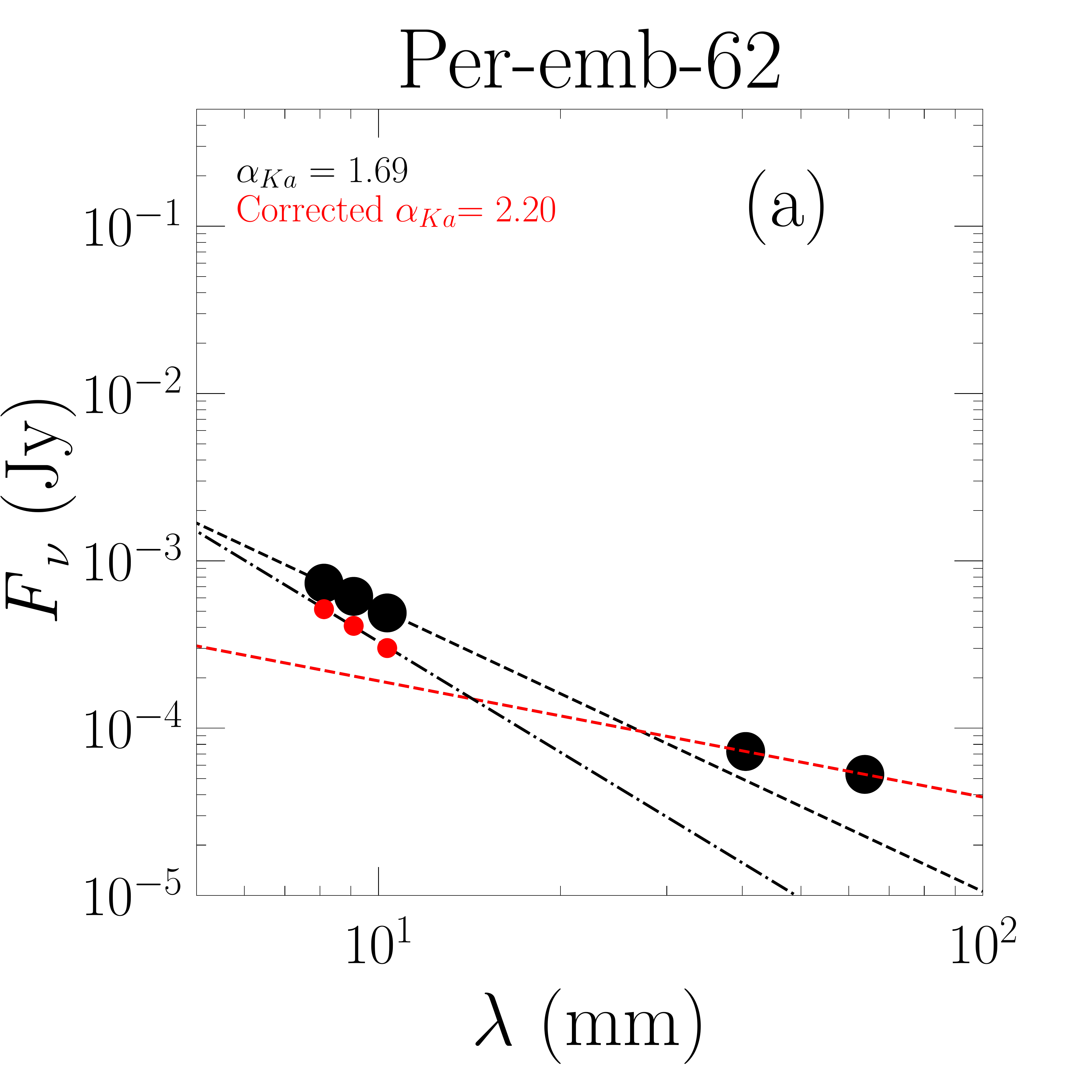}
  \includegraphics[width=0.30\linewidth]{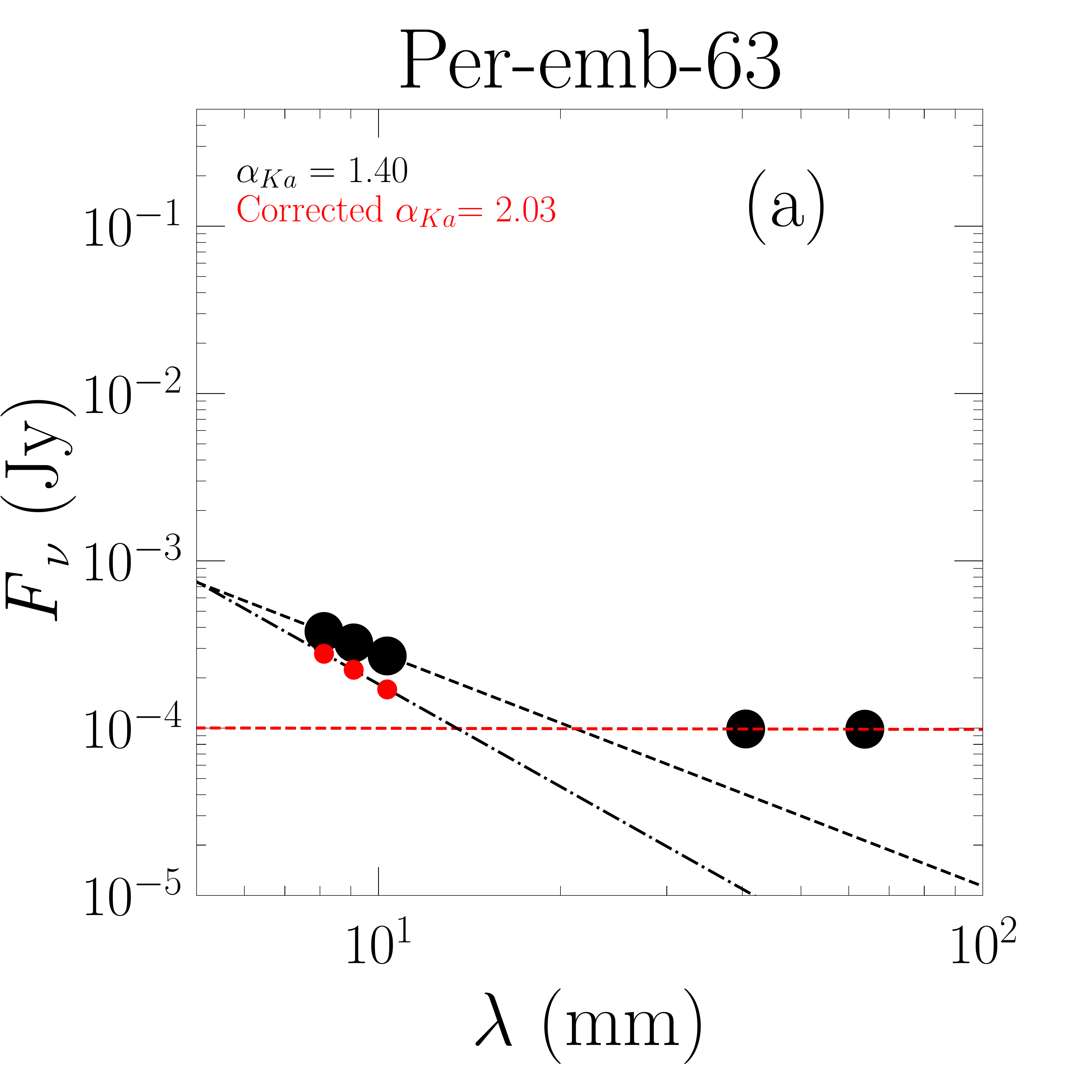}

\end{figure}

\begin{figure}[H]
\centering
  \includegraphics[width=0.30\linewidth]{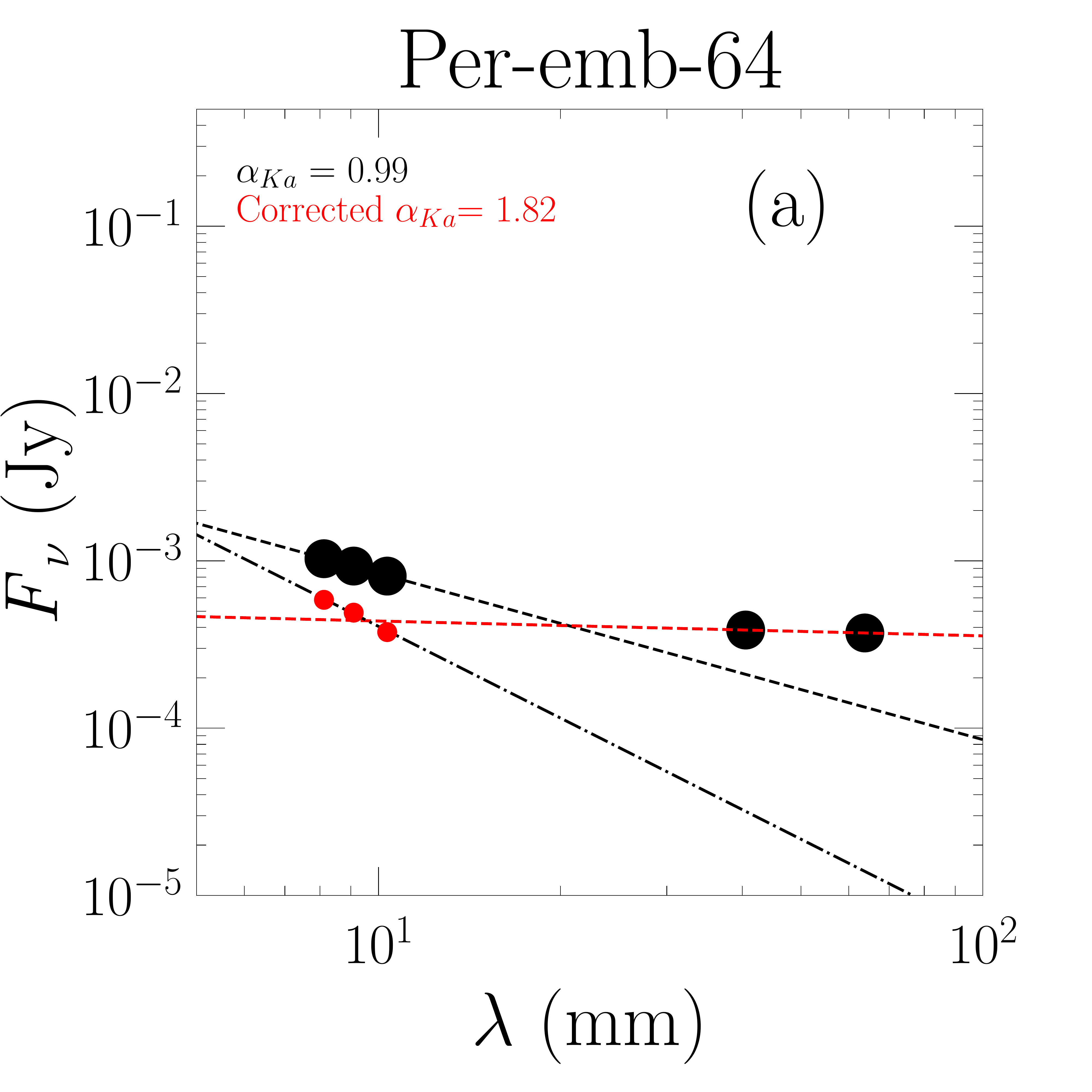}
  \includegraphics[width=0.30\linewidth]{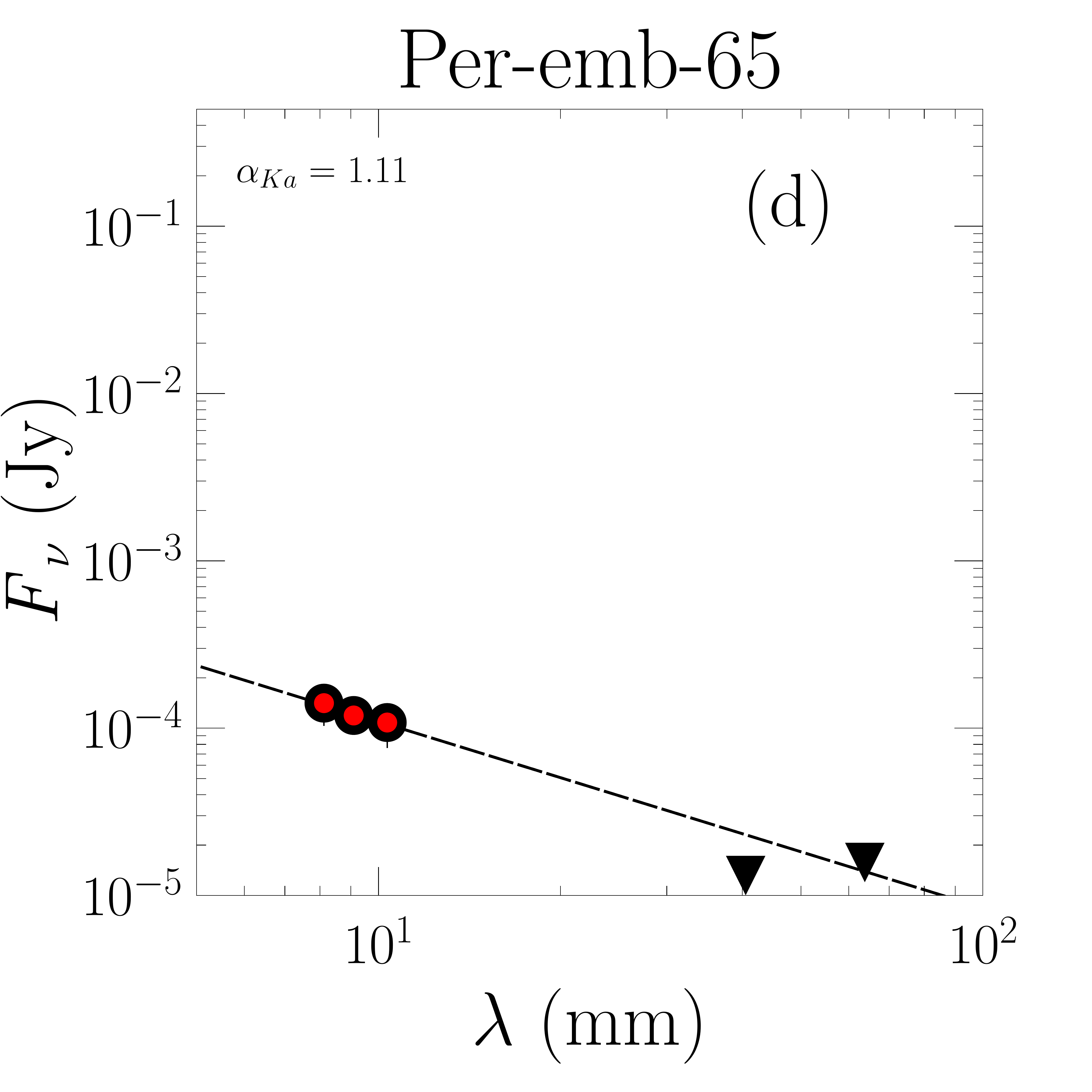}
  \includegraphics[width=0.30\linewidth]{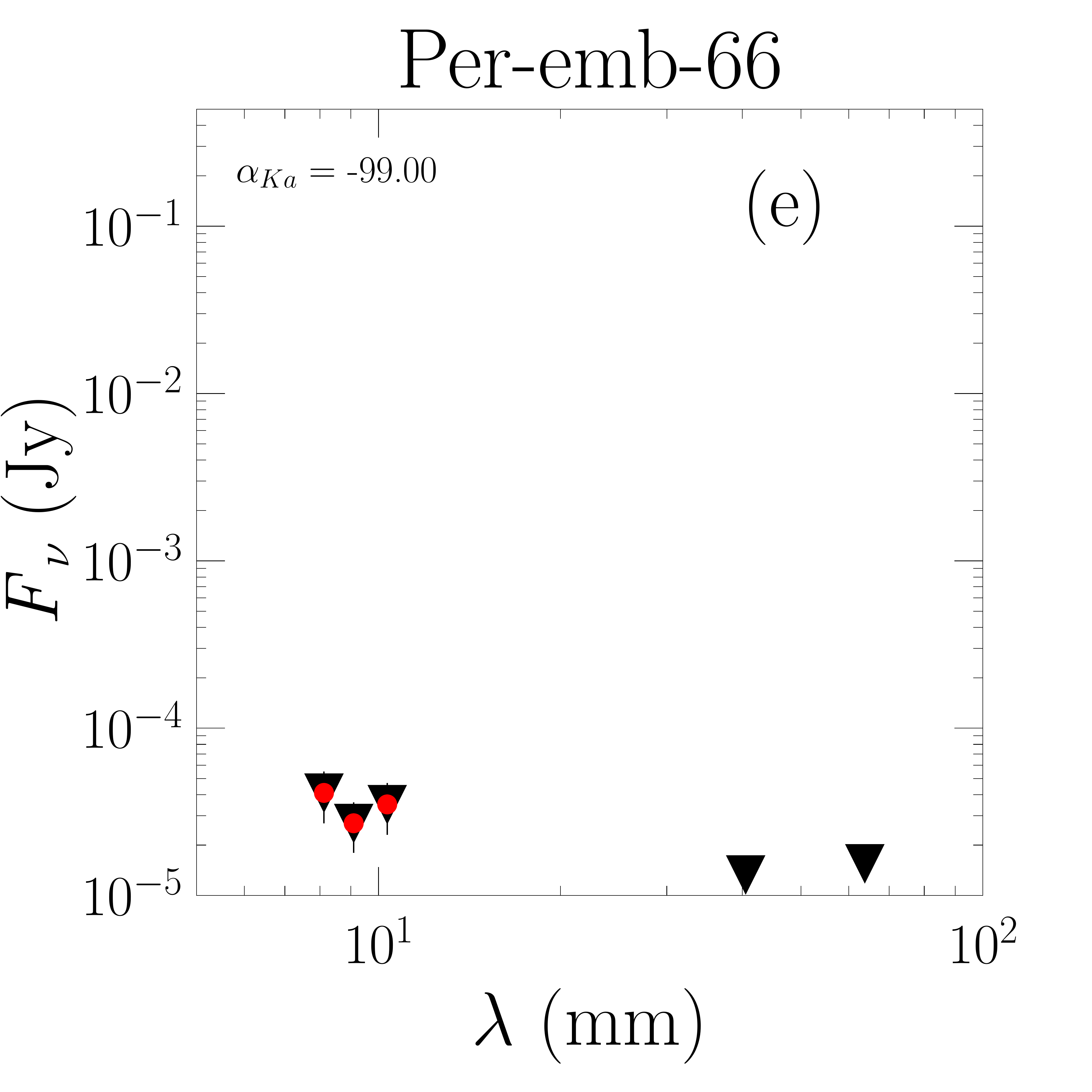}
  \includegraphics[width=0.30\linewidth]{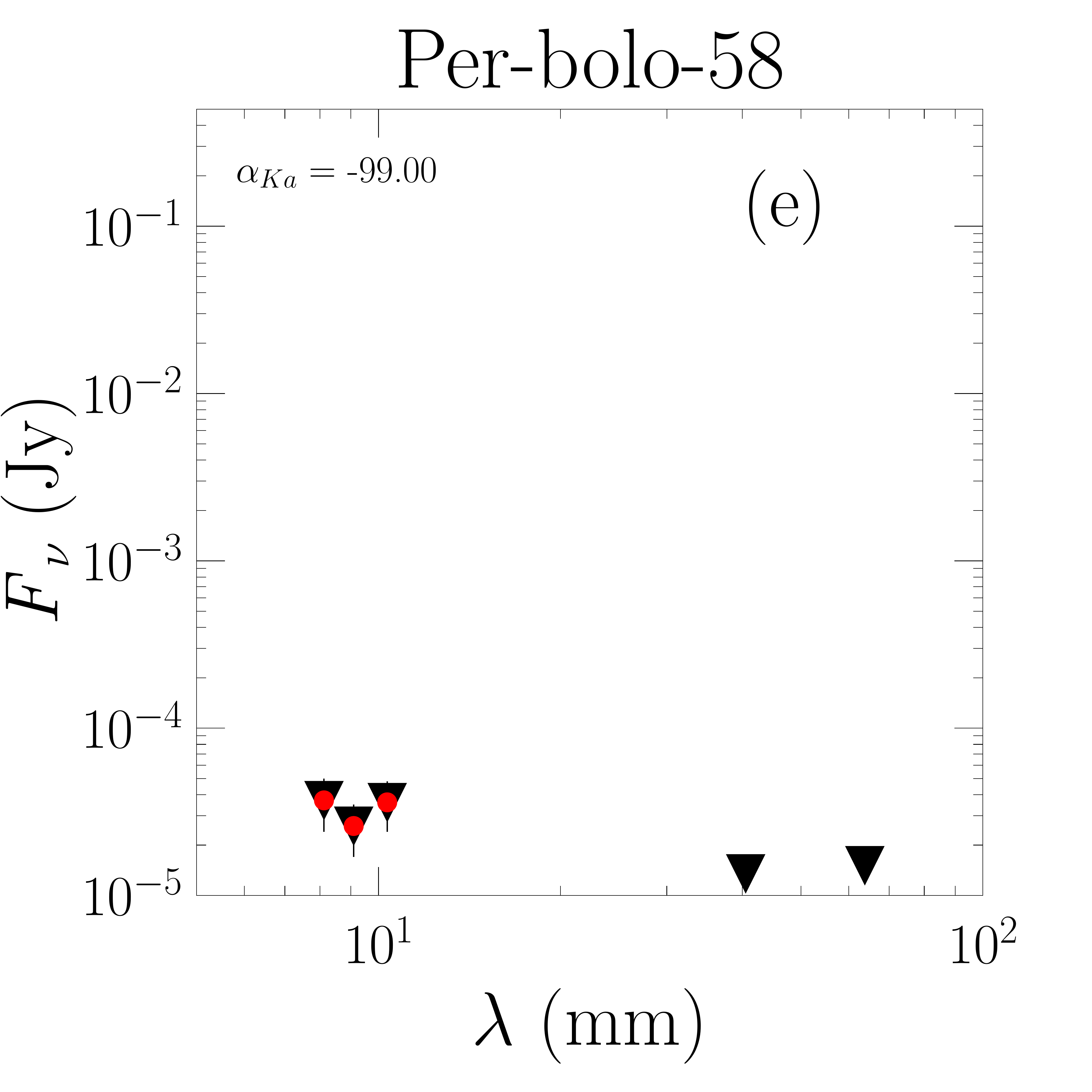}
  \includegraphics[width=0.30\linewidth]{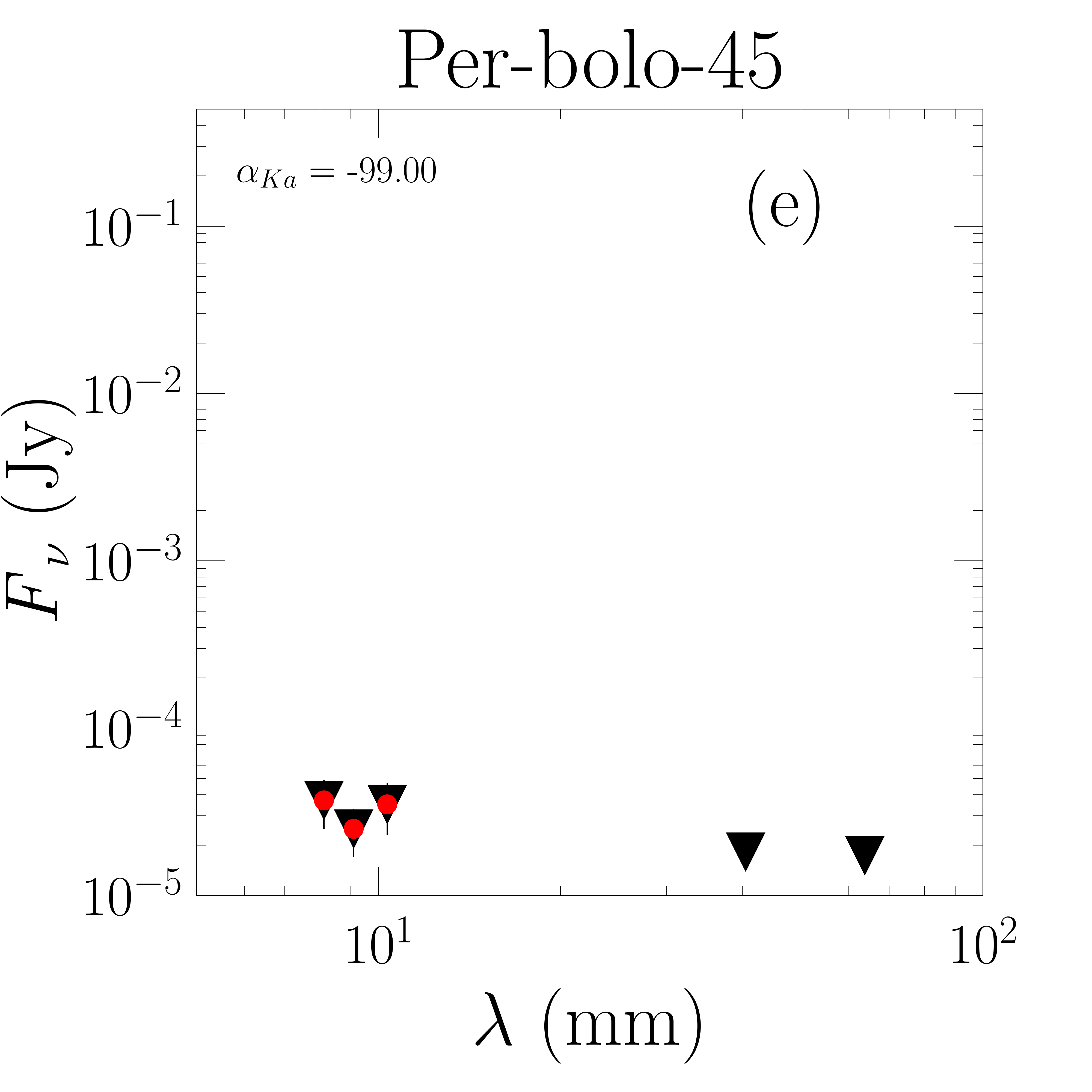}
  \includegraphics[width=0.30\linewidth]{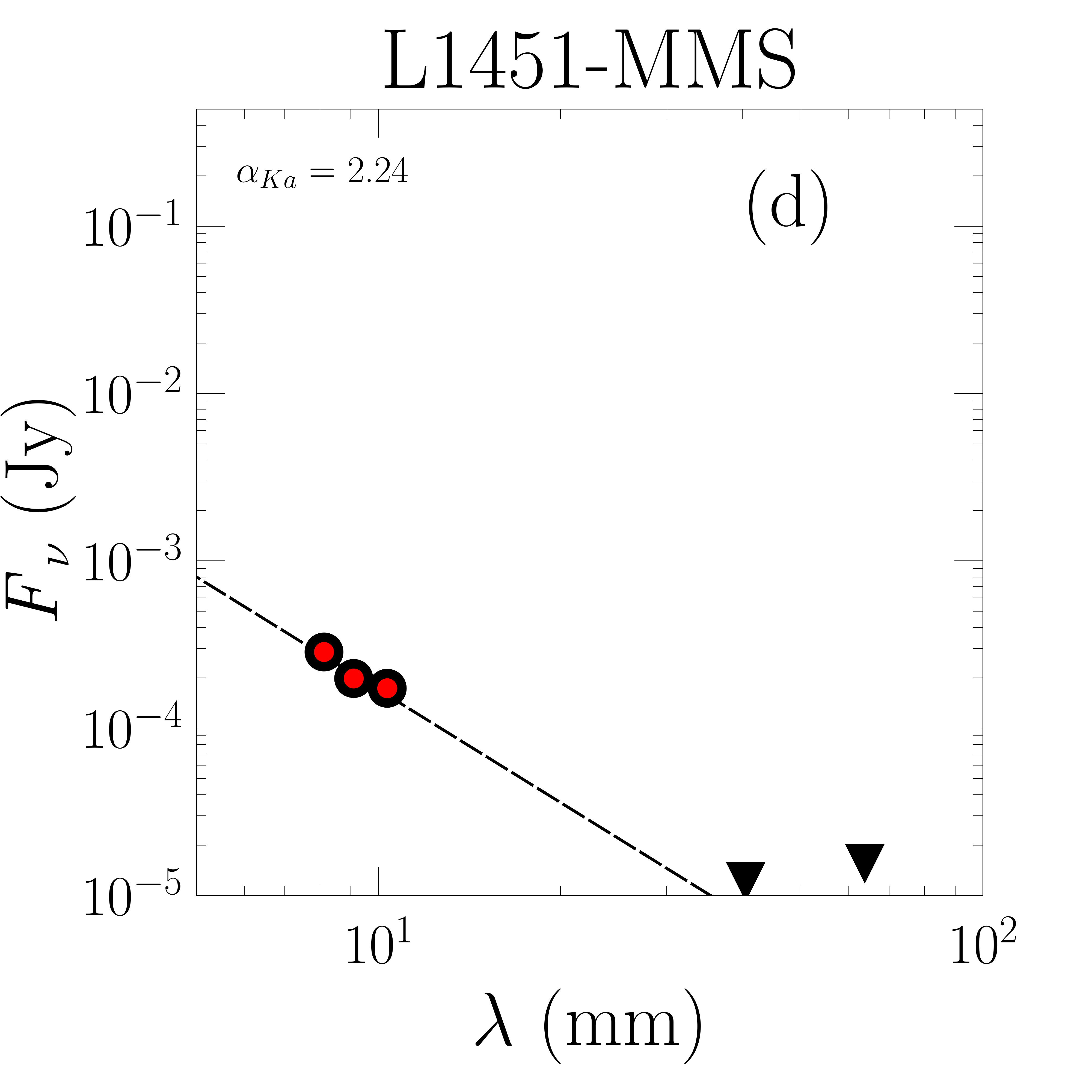}
  \includegraphics[width=0.30\linewidth]{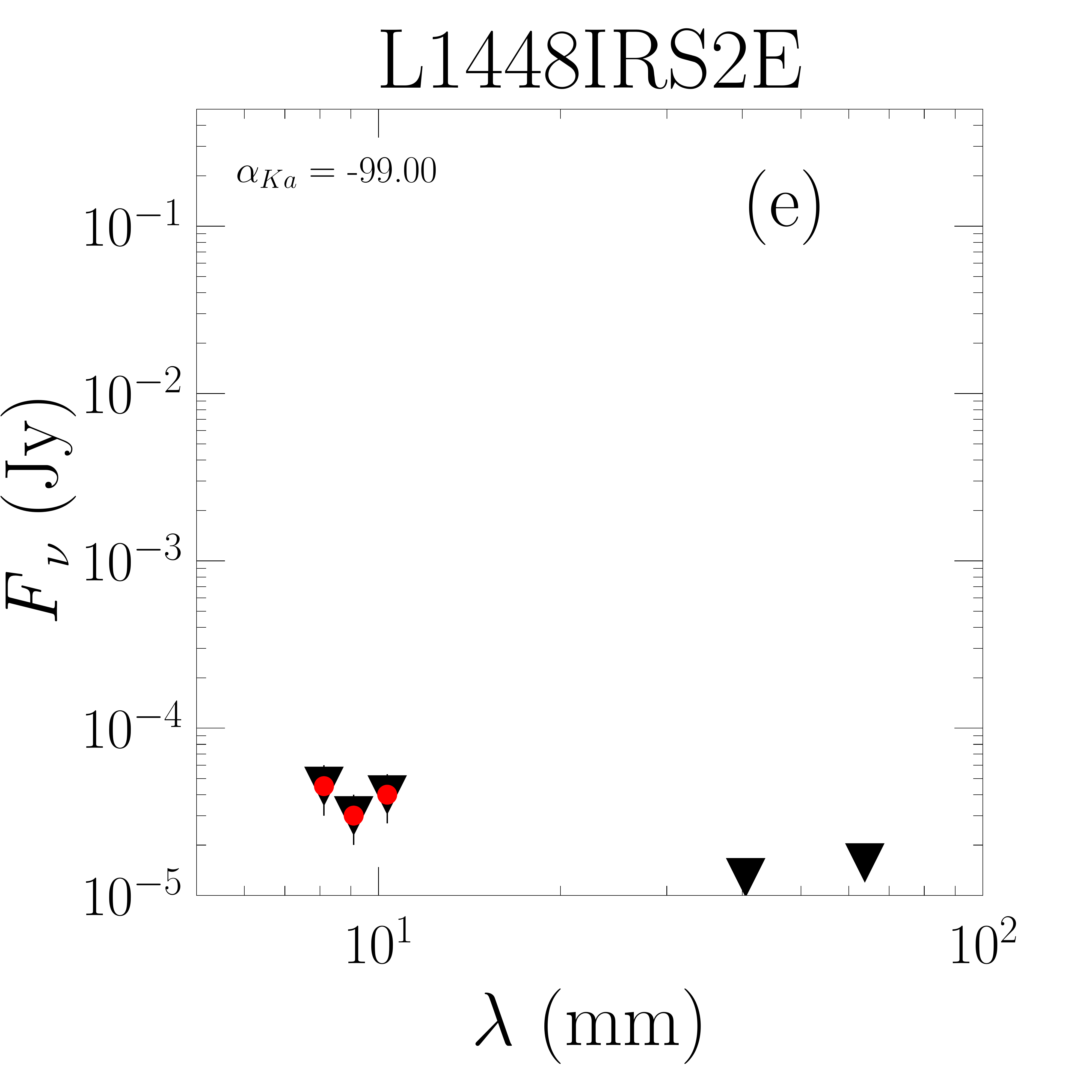}
  \includegraphics[width=0.30\linewidth]{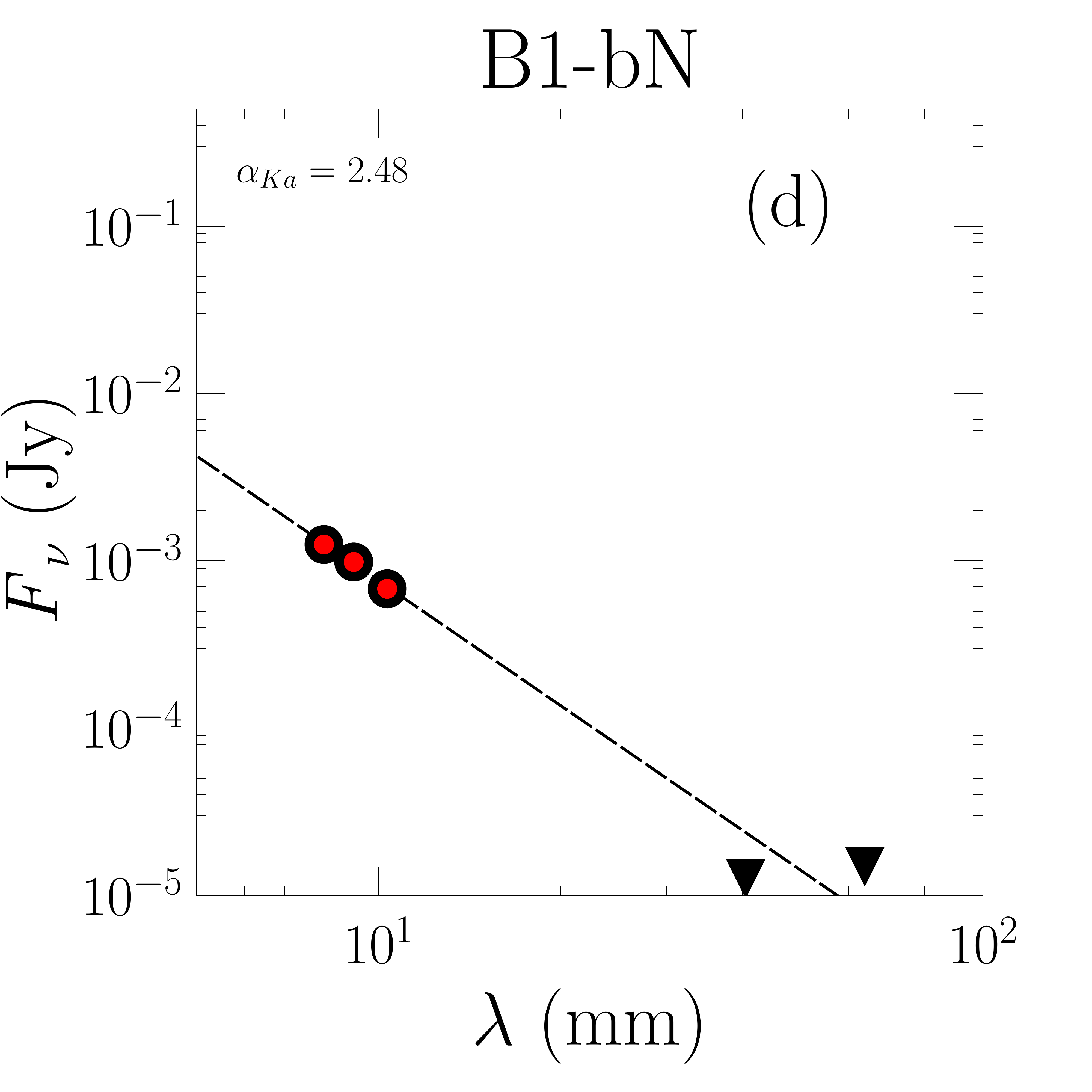}
  \includegraphics[width=0.30\linewidth]{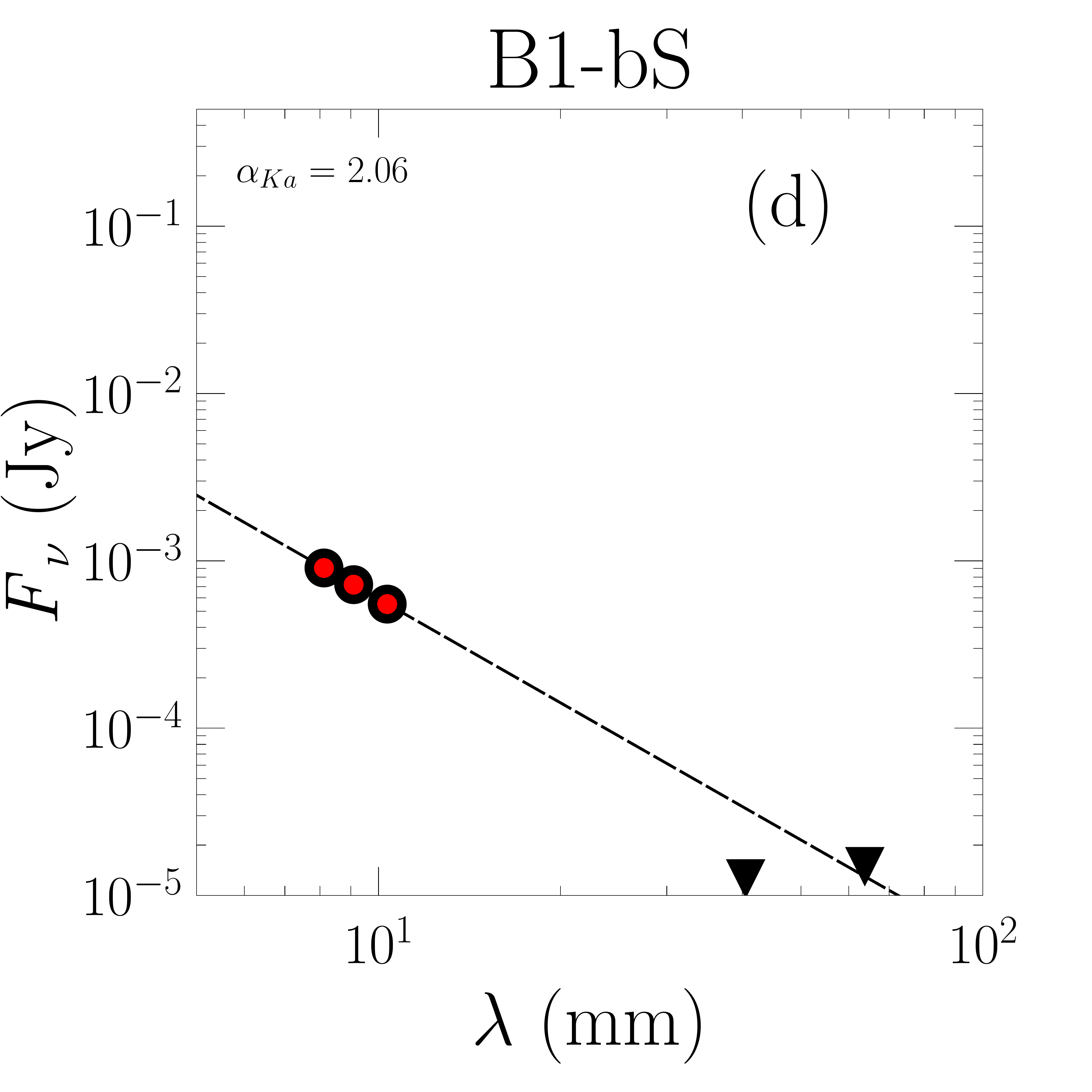}
  \includegraphics[width=0.30\linewidth]{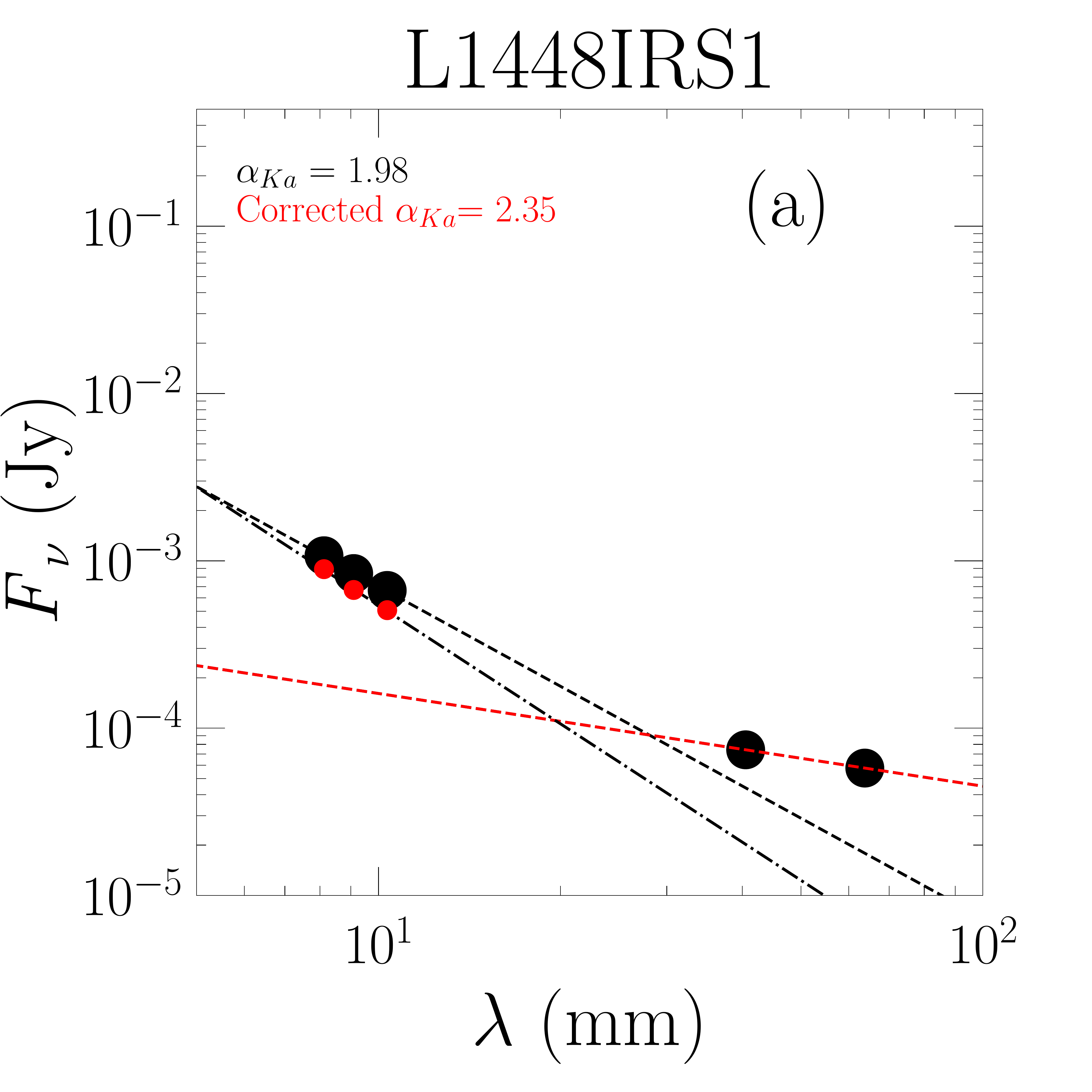}
  \includegraphics[width=0.30\linewidth]{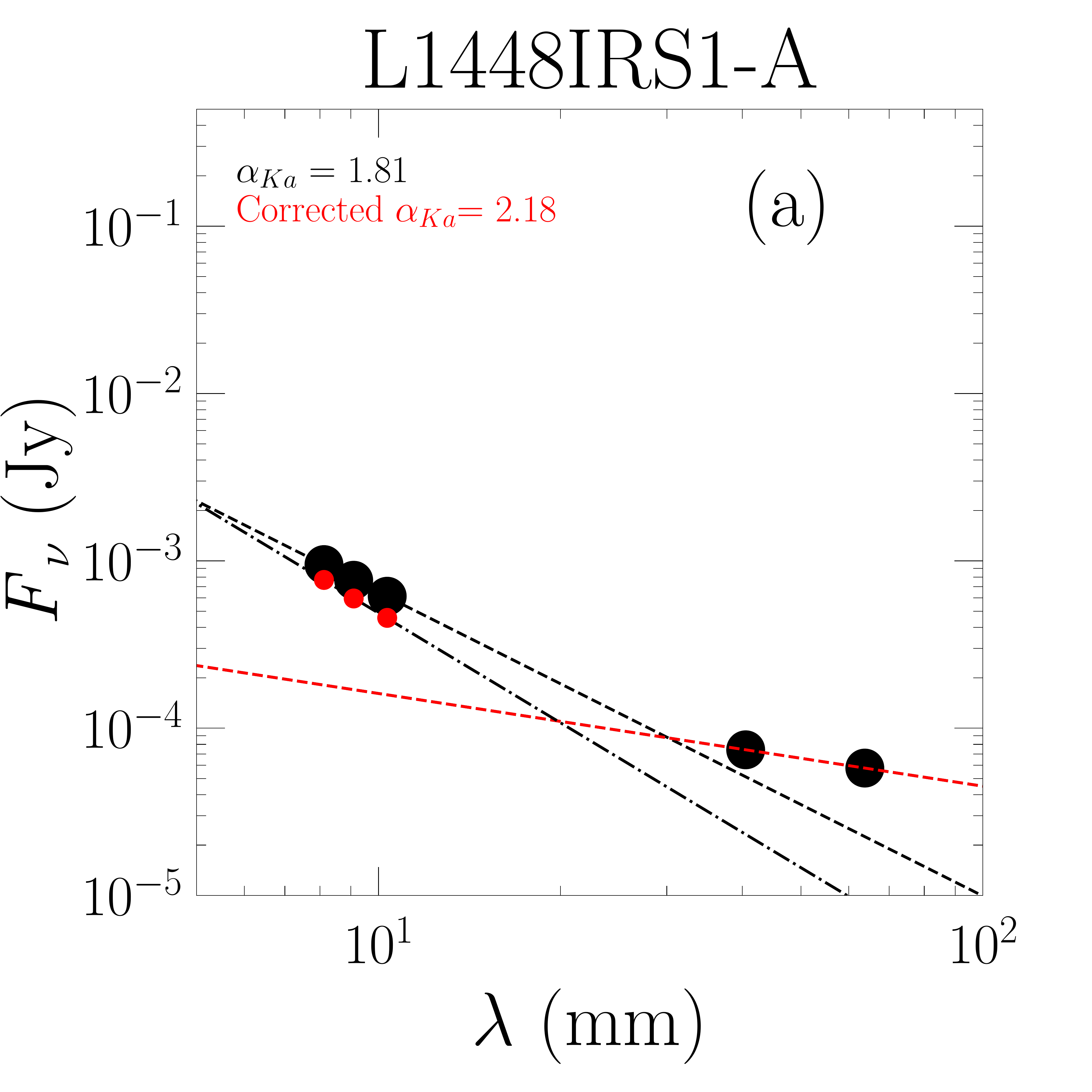}
  \includegraphics[width=0.30\linewidth]{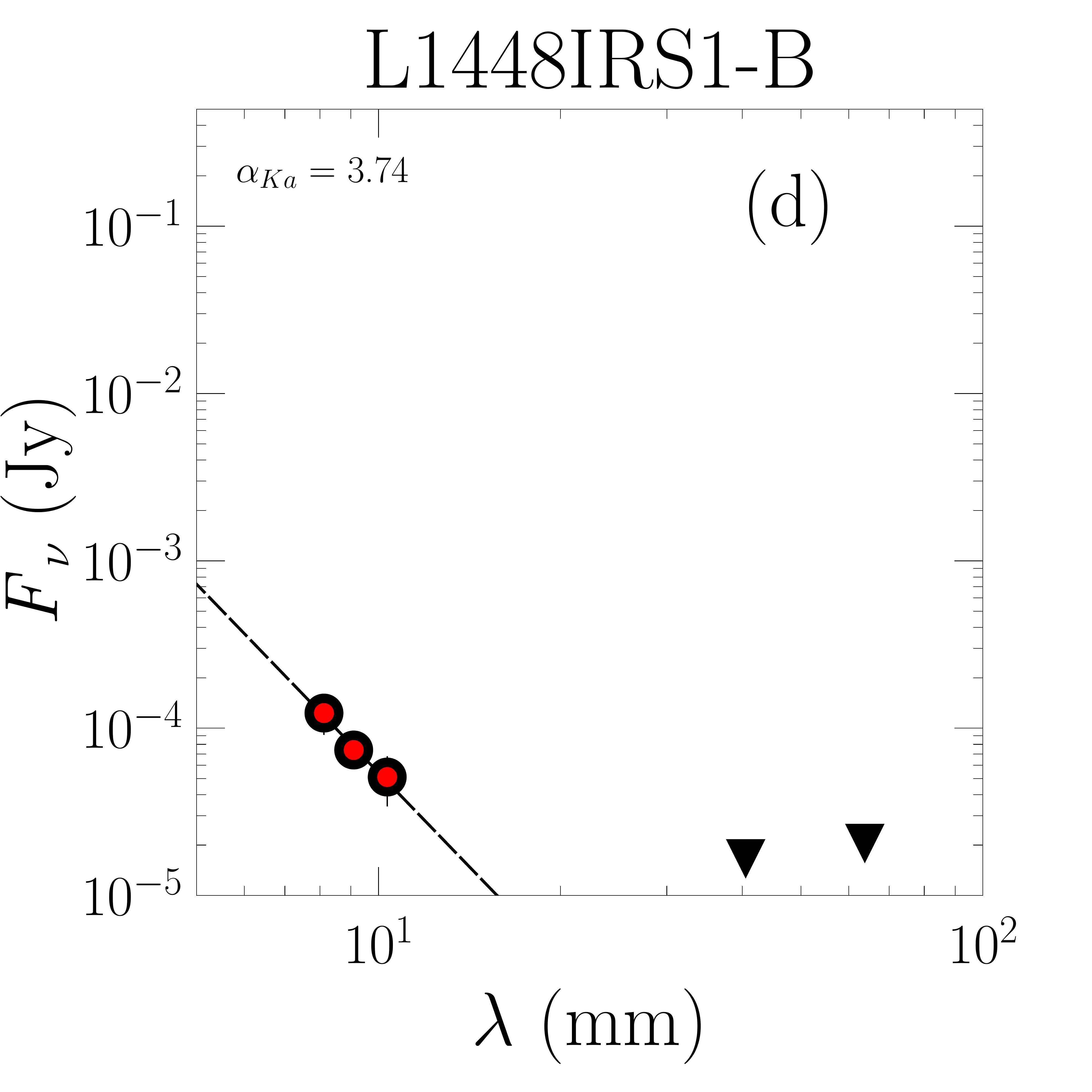}

\end{figure}

\begin{figure}[H]
\centering
  \includegraphics[width=0.30\linewidth]{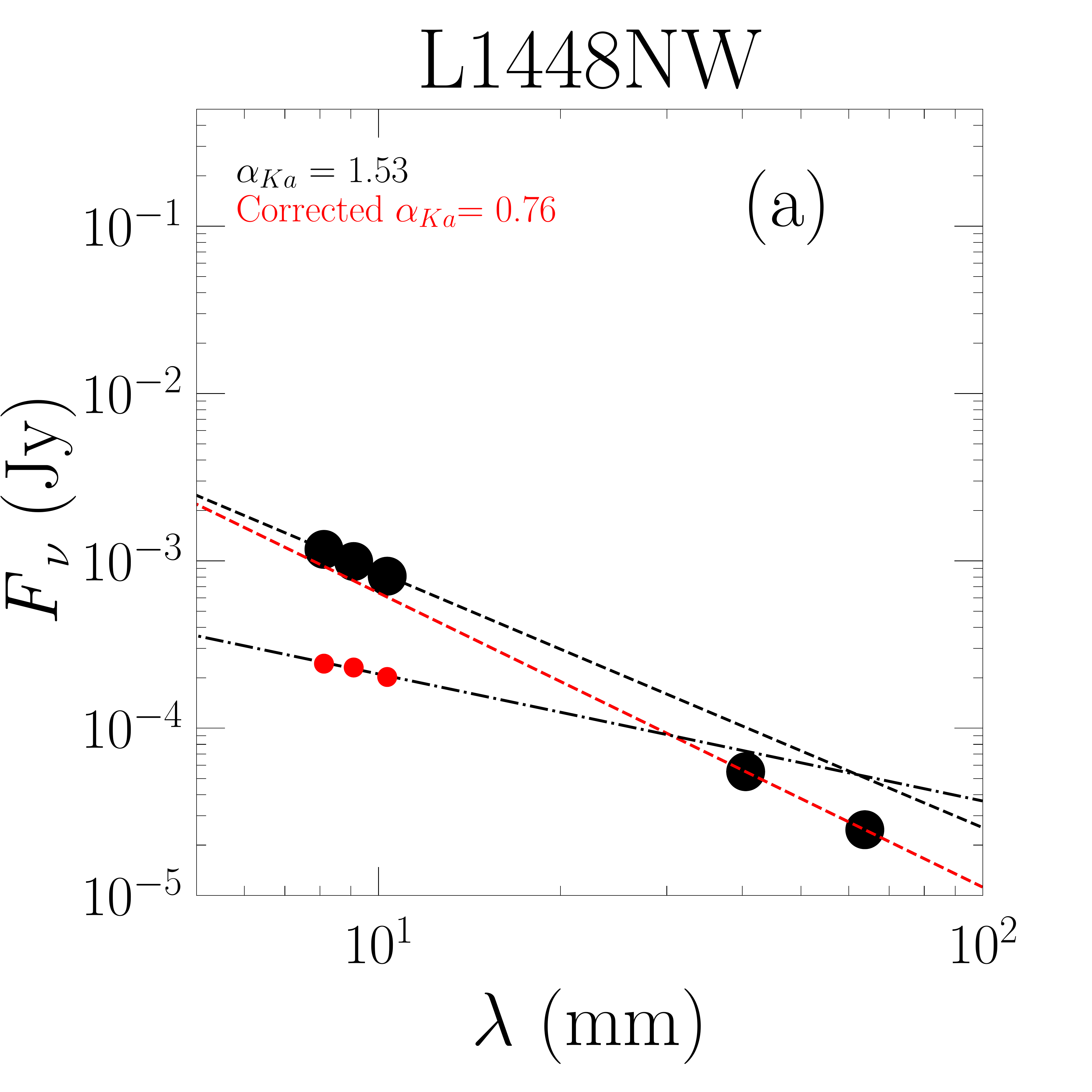}
  \includegraphics[width=0.30\linewidth]{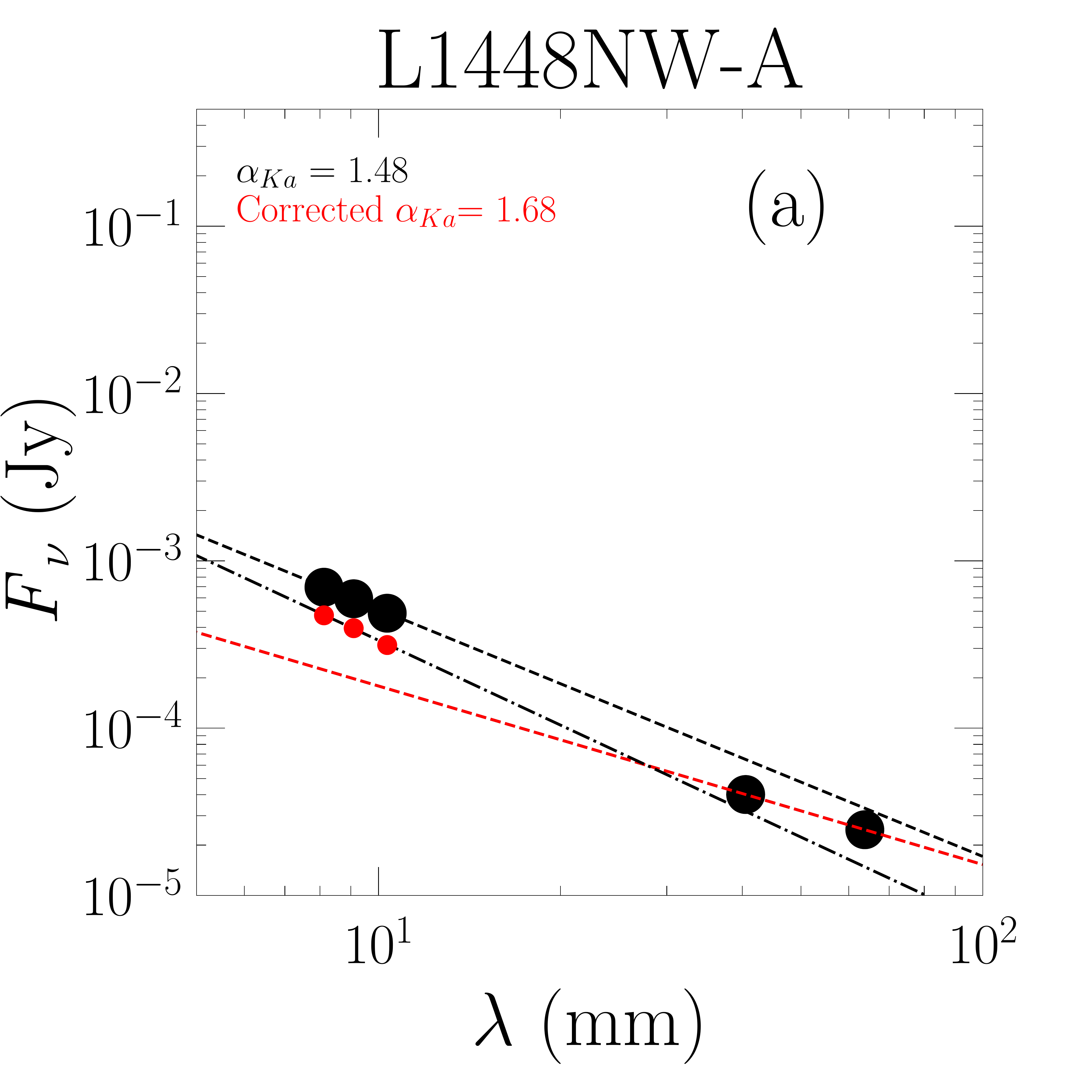}
  \includegraphics[width=0.30\linewidth]{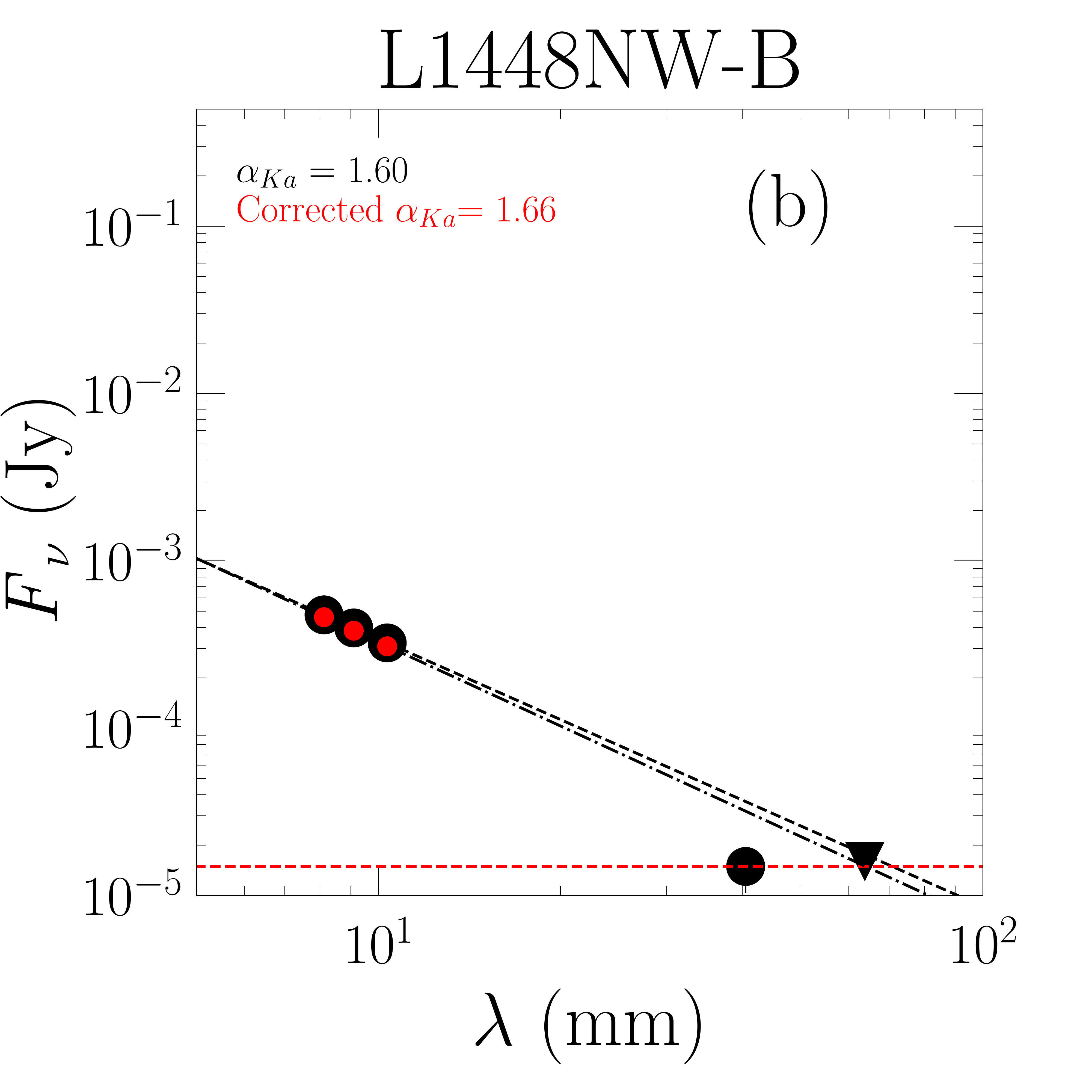}
  \includegraphics[width=0.30\linewidth]{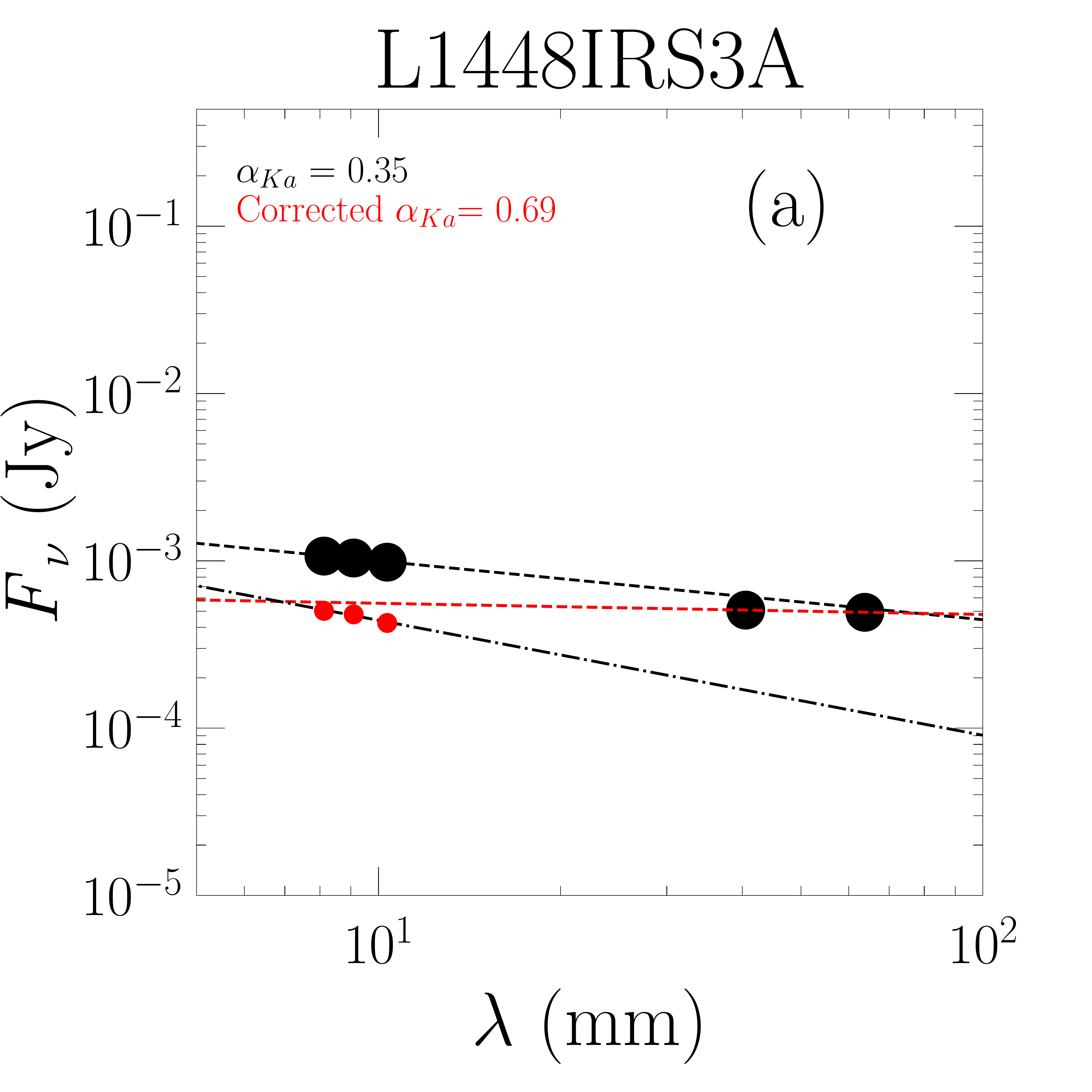}
  \includegraphics[width=0.30\linewidth]{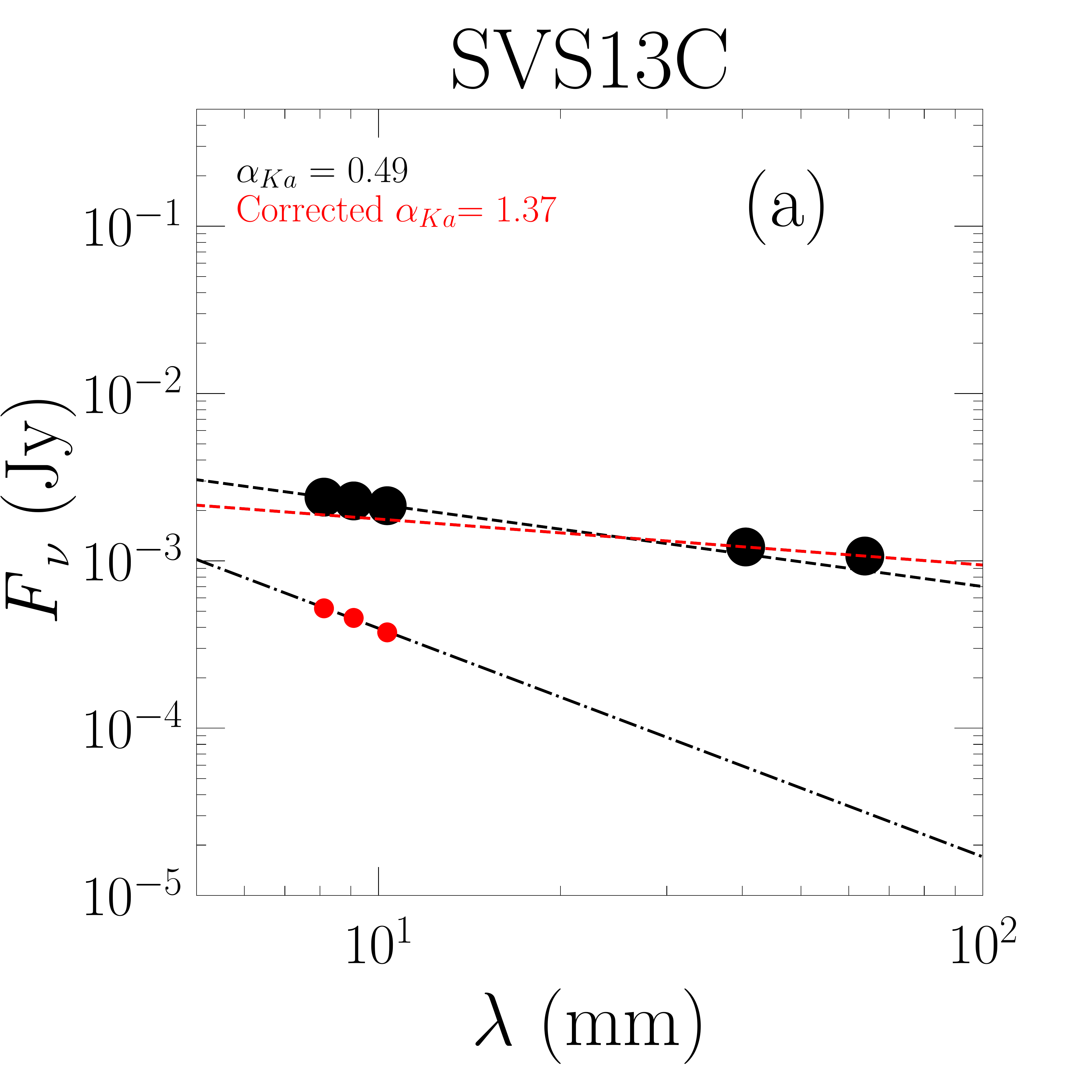}
  \includegraphics[width=0.30\linewidth]{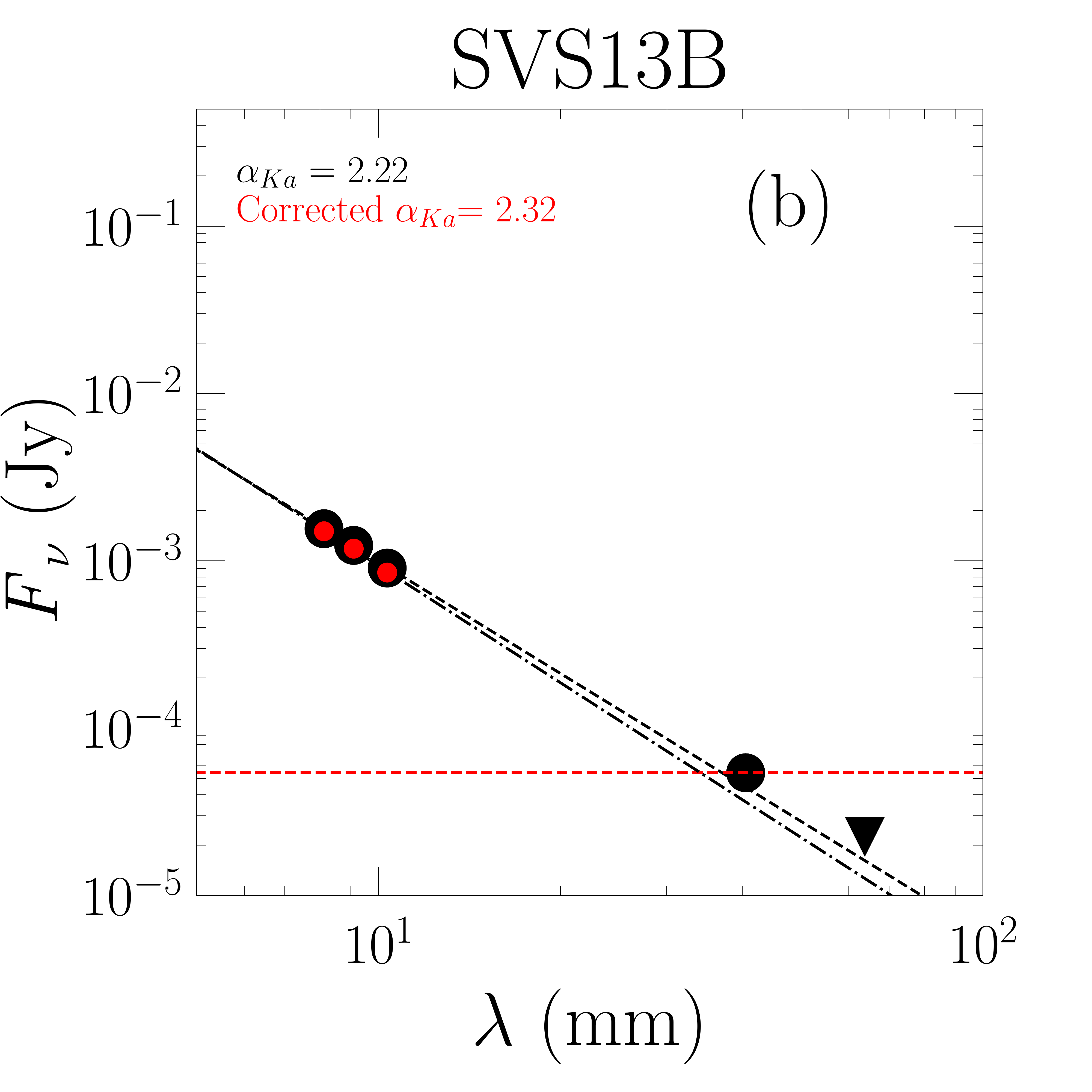}
  \includegraphics[width=0.30\linewidth]{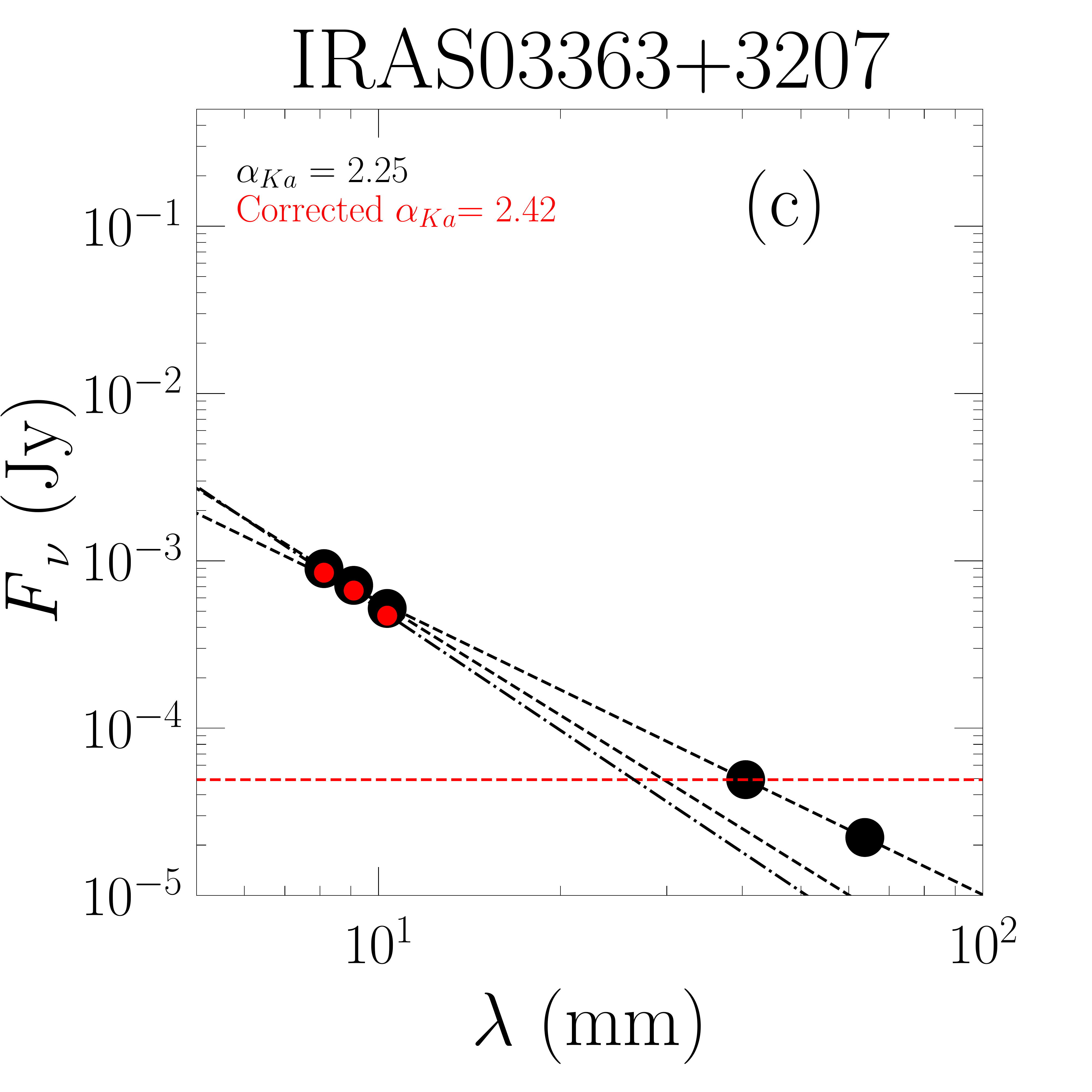}
  \includegraphics[width=0.30\linewidth]{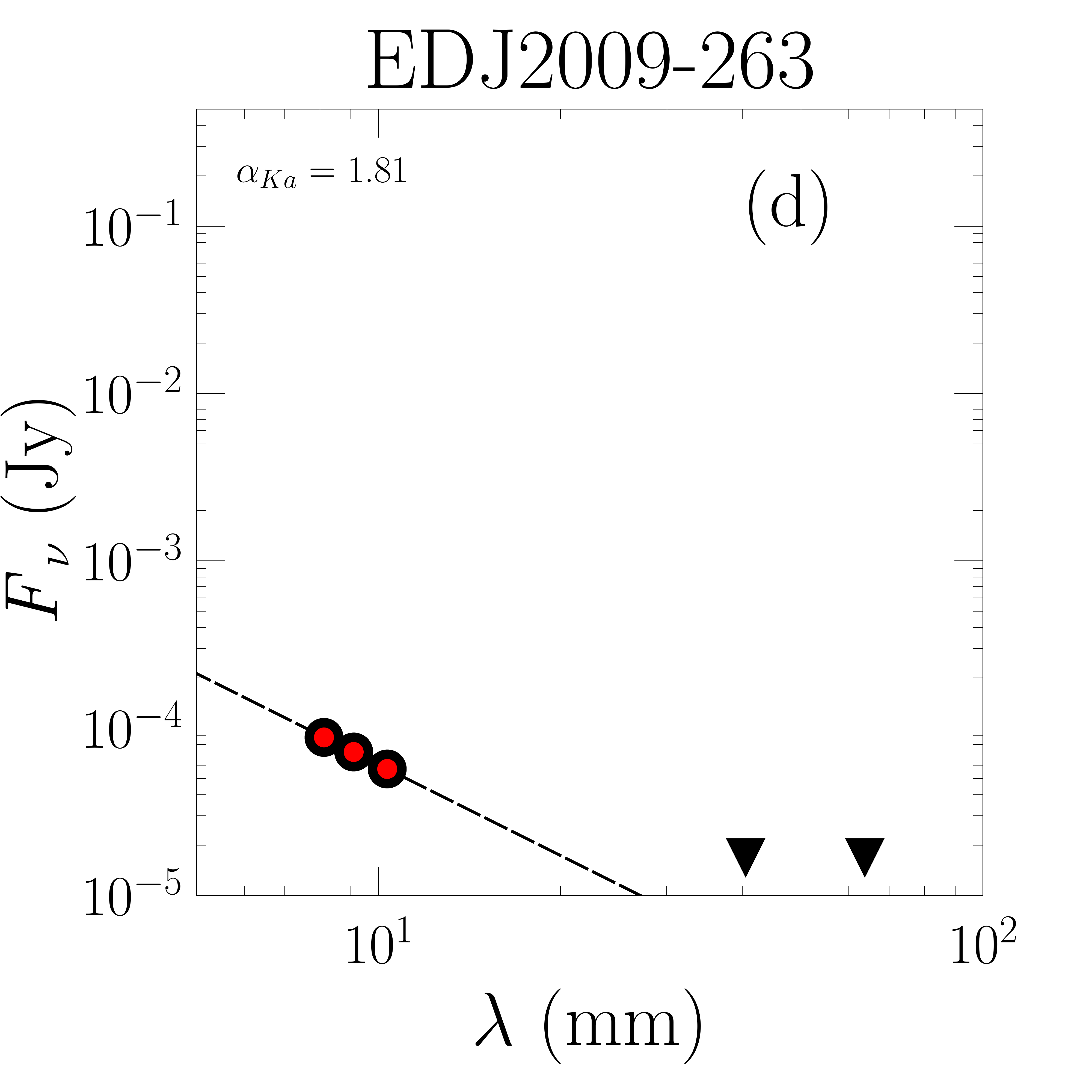}
  \includegraphics[width=0.30\linewidth]{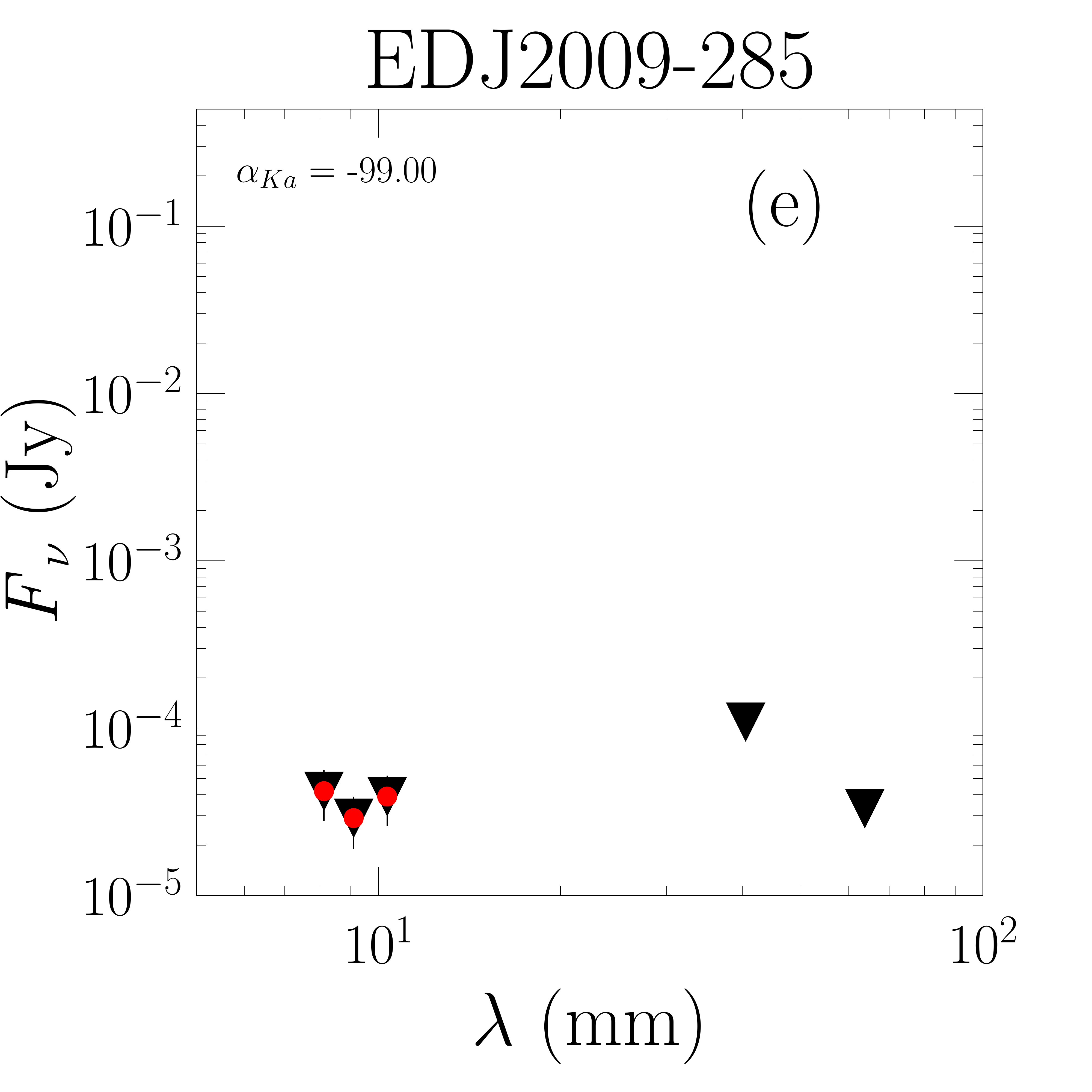}
  \includegraphics[width=0.30\linewidth]{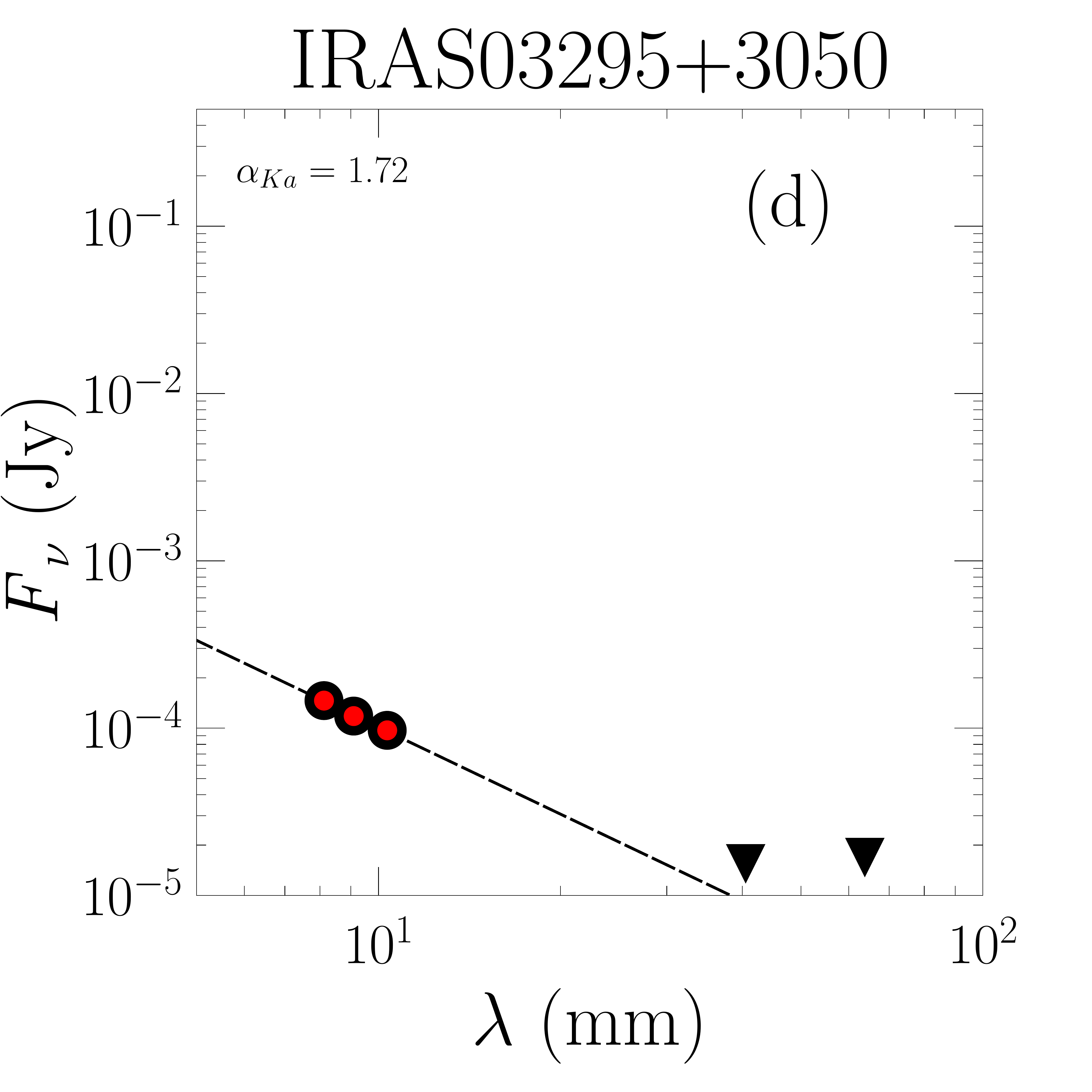}
  \includegraphics[width=0.30\linewidth]{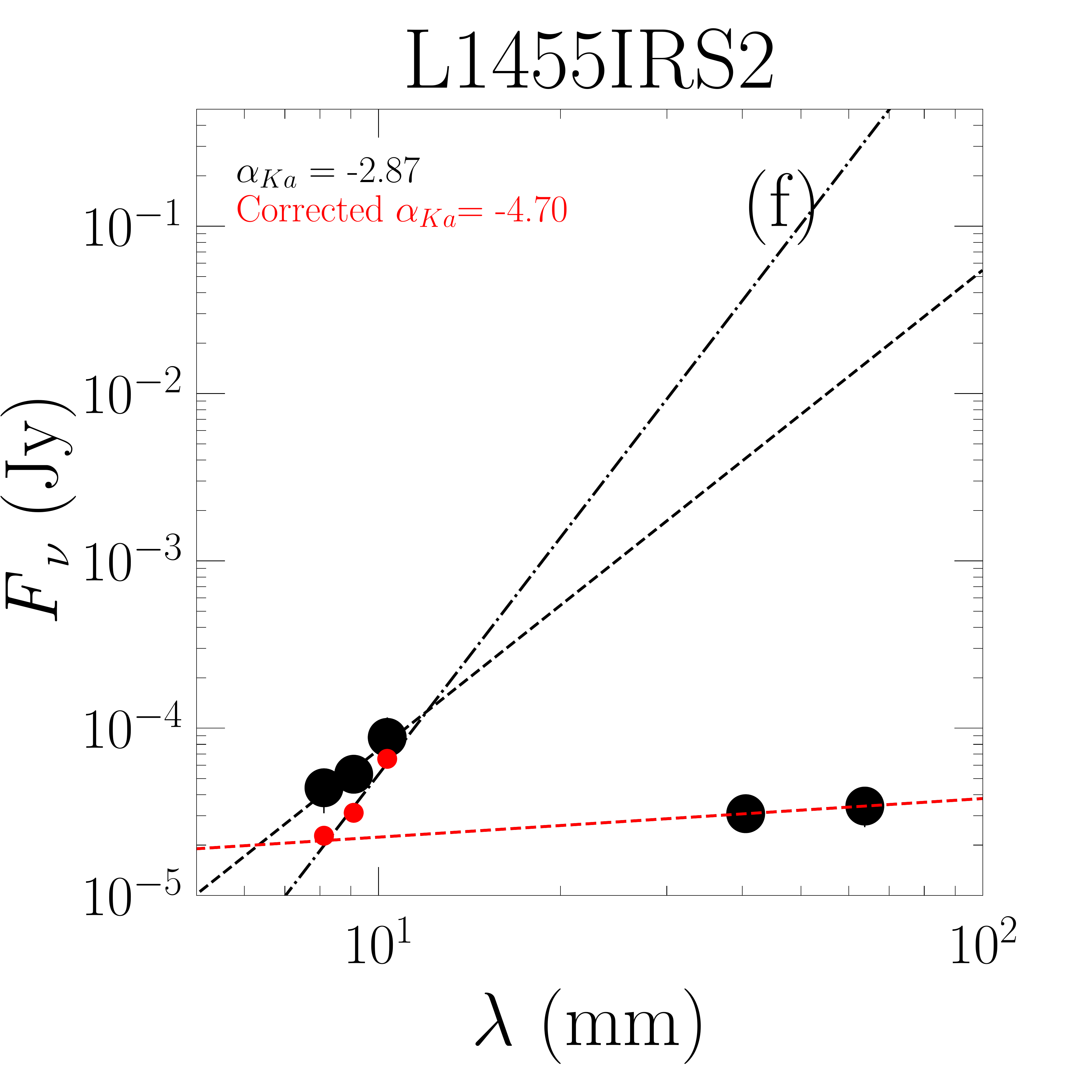}
  \includegraphics[width=0.30\linewidth]{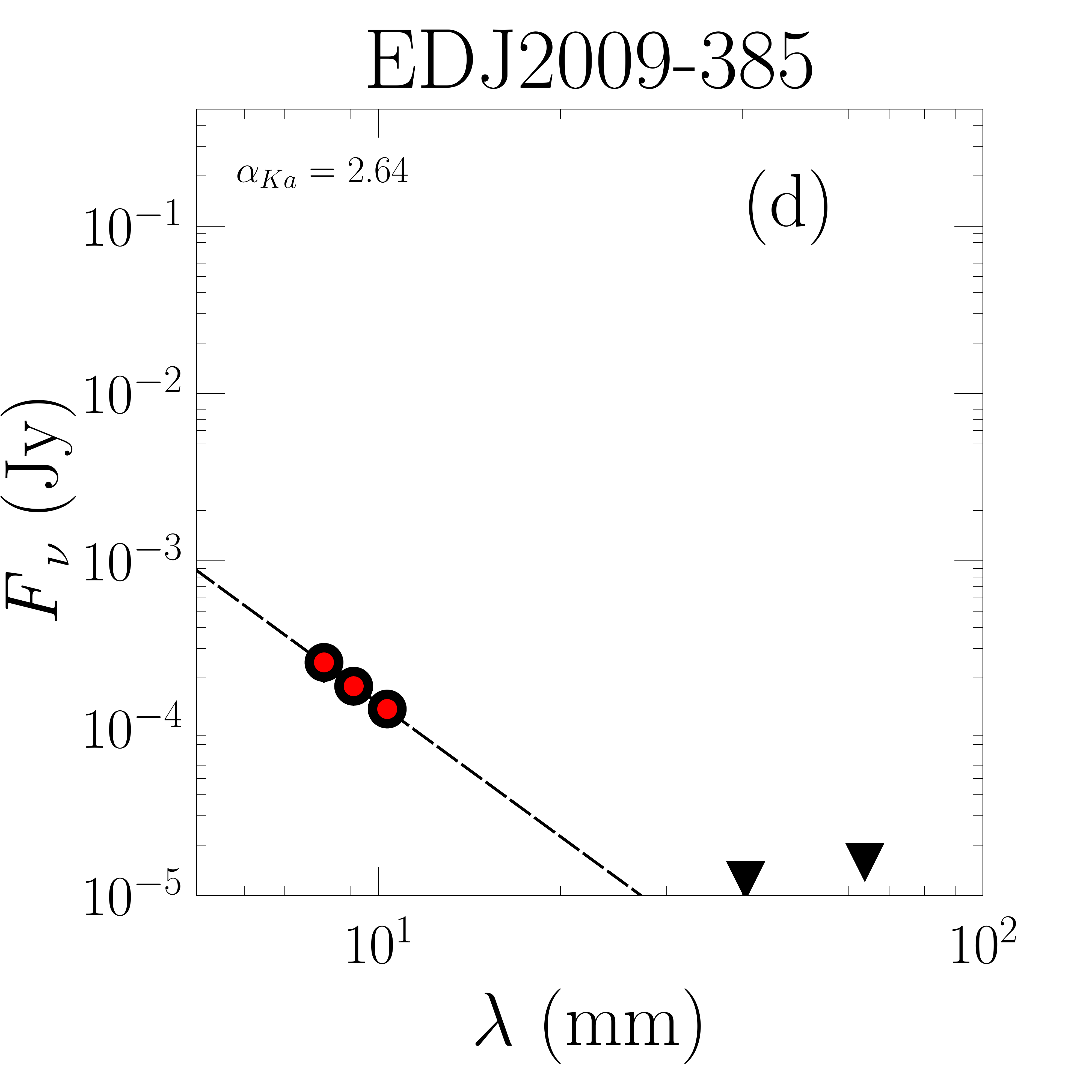}

\end{figure}

\begin{figure}
\centering
  \includegraphics[width=0.30\linewidth]{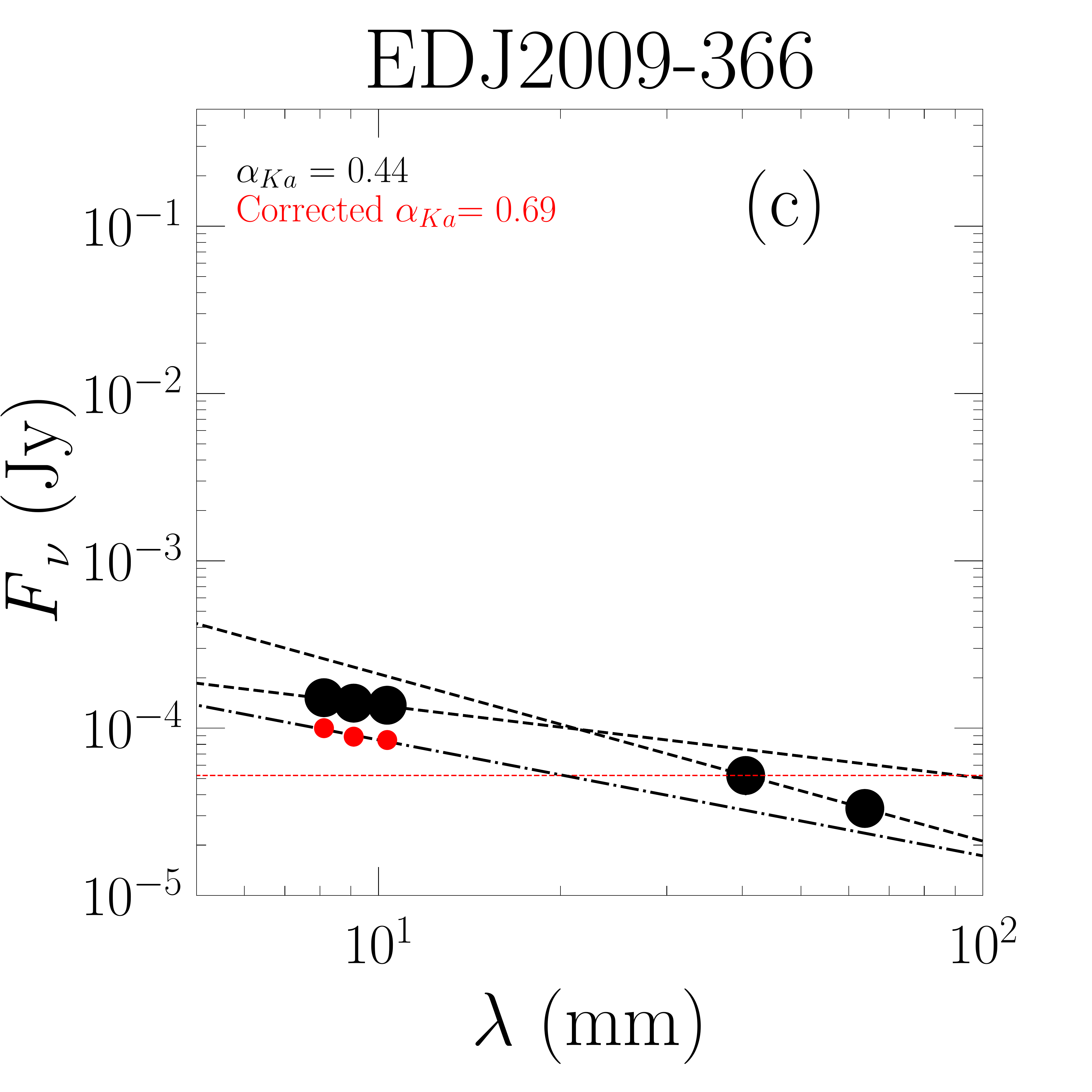}
  \includegraphics[width=0.30\linewidth]{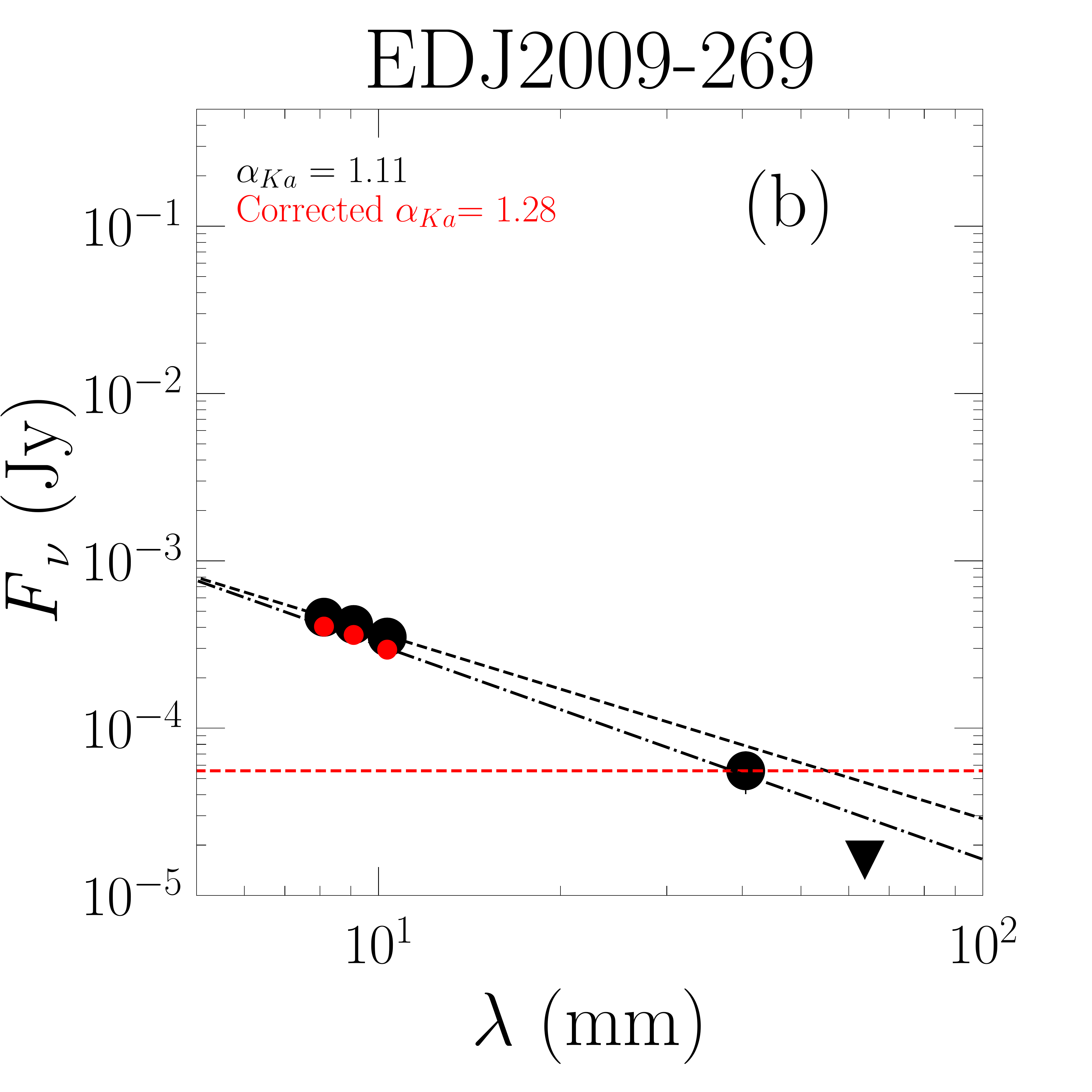}
  \includegraphics[width=0.30\linewidth]{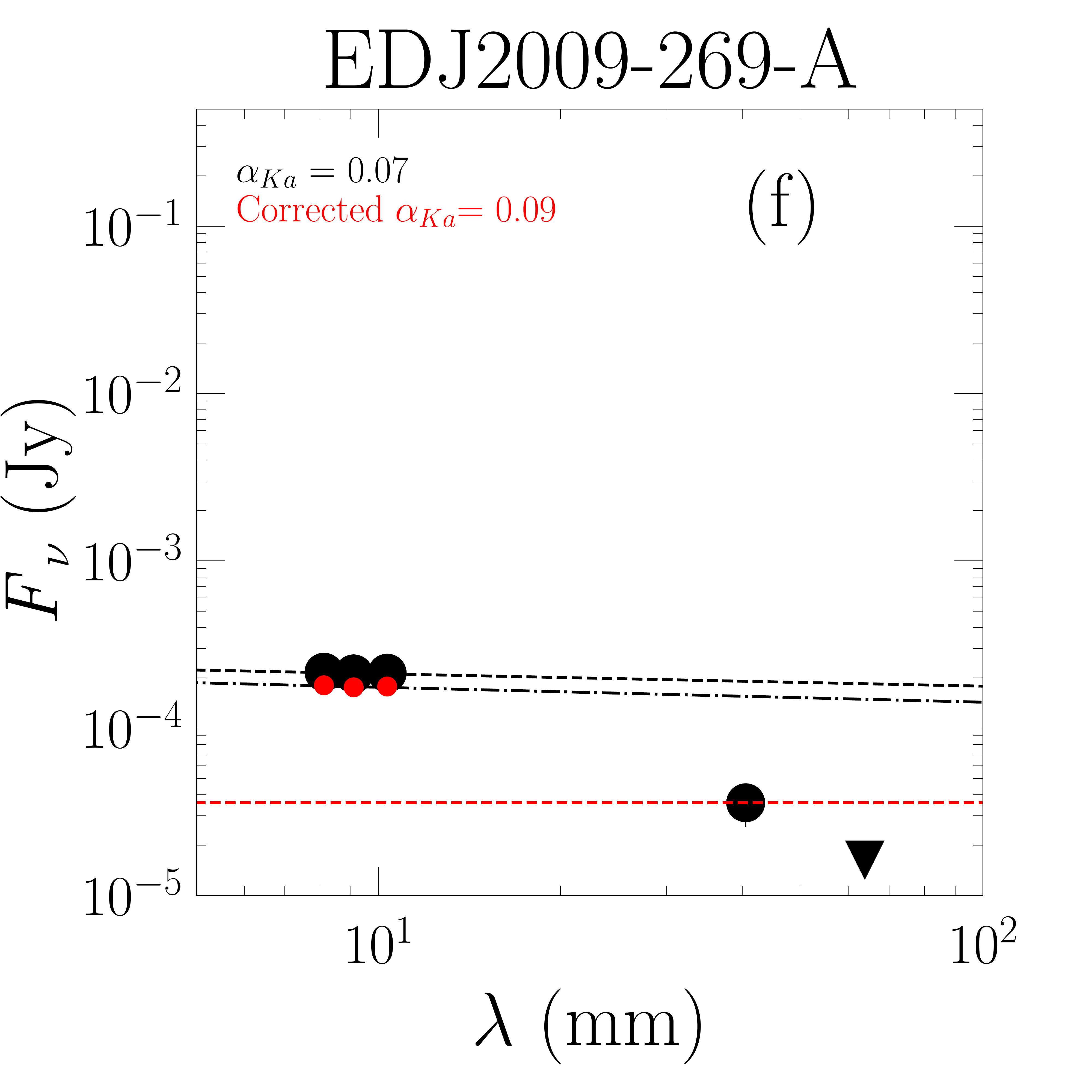}
  \includegraphics[width=0.30\linewidth]{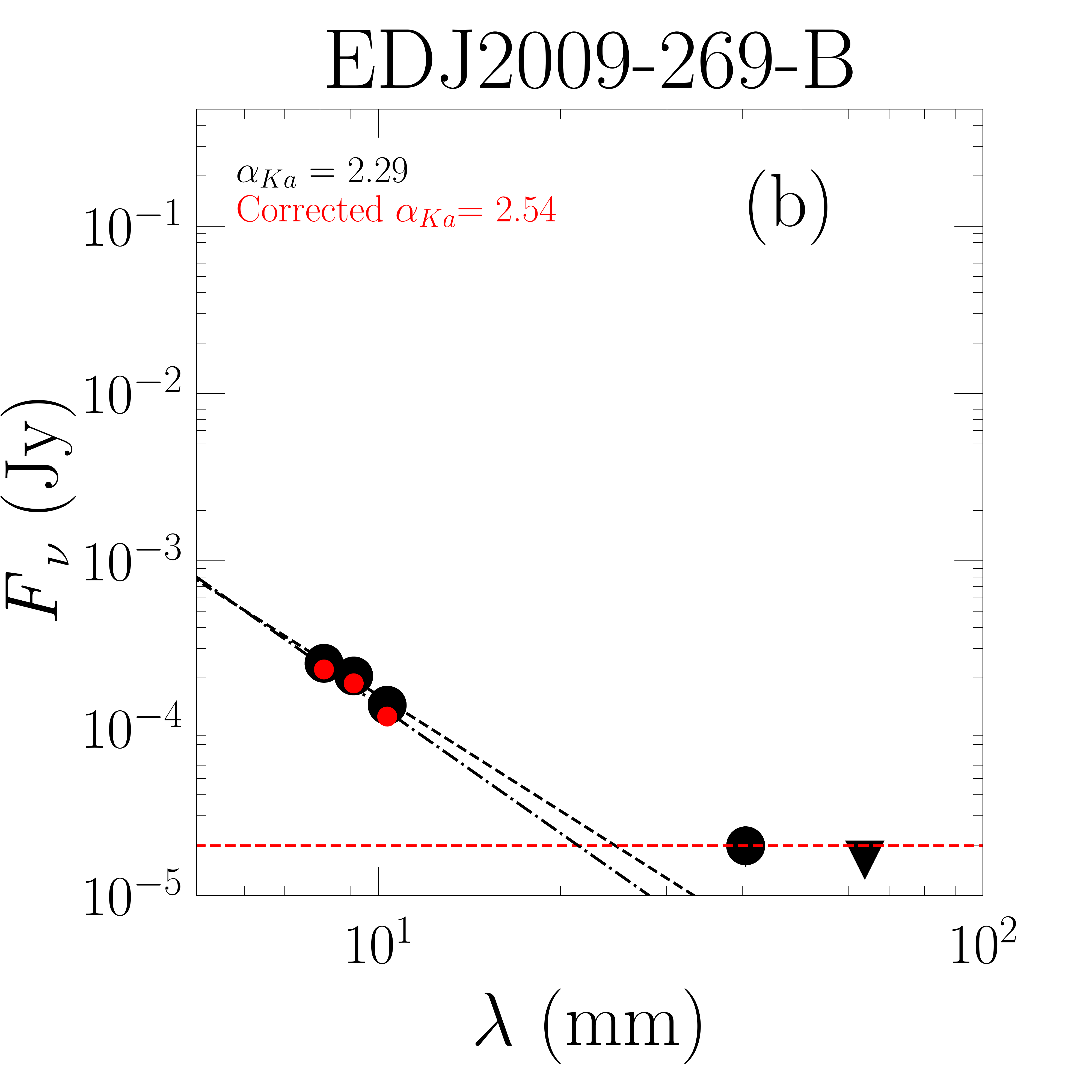}
  \includegraphics[width=0.30\linewidth]{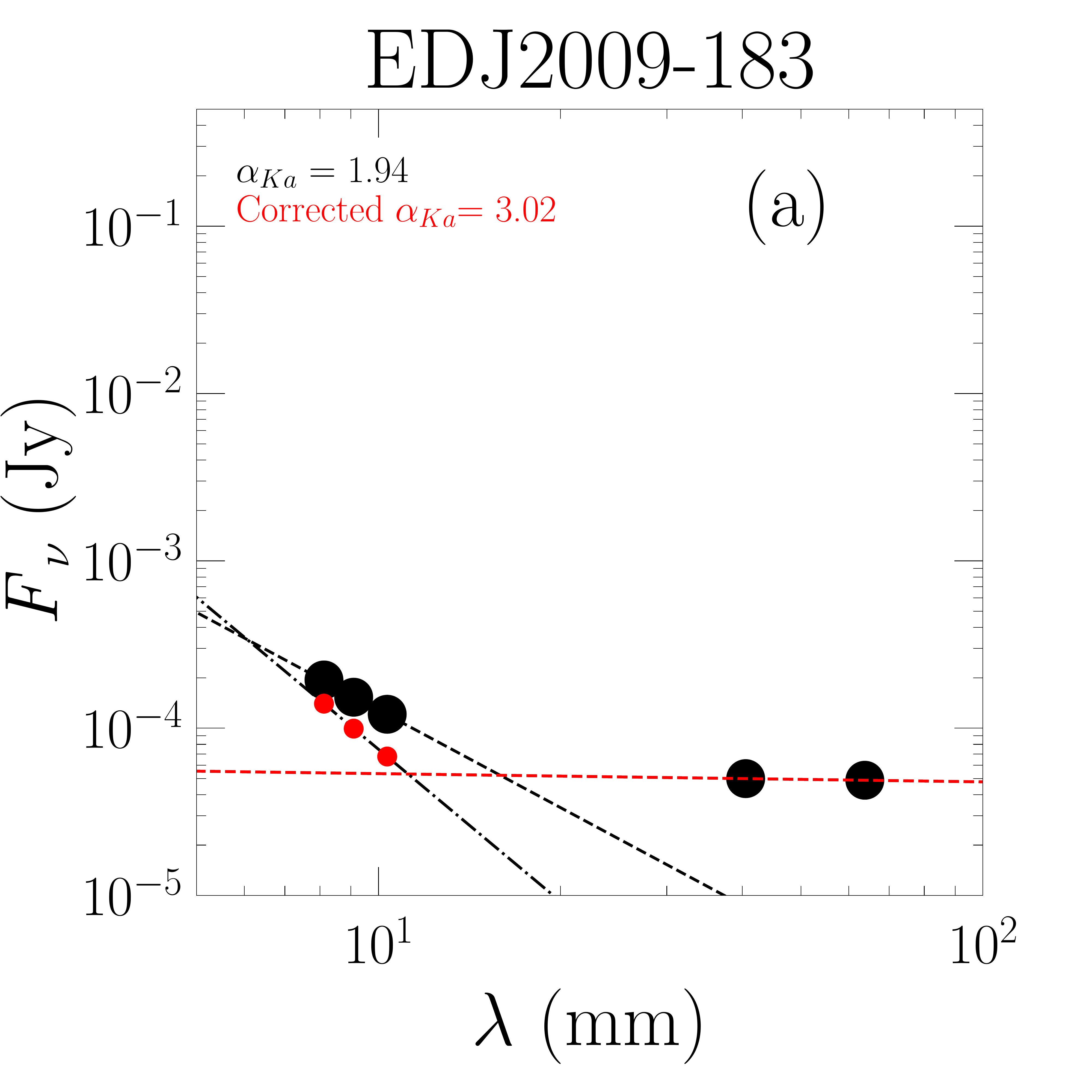}
  \includegraphics[width=0.30\linewidth]{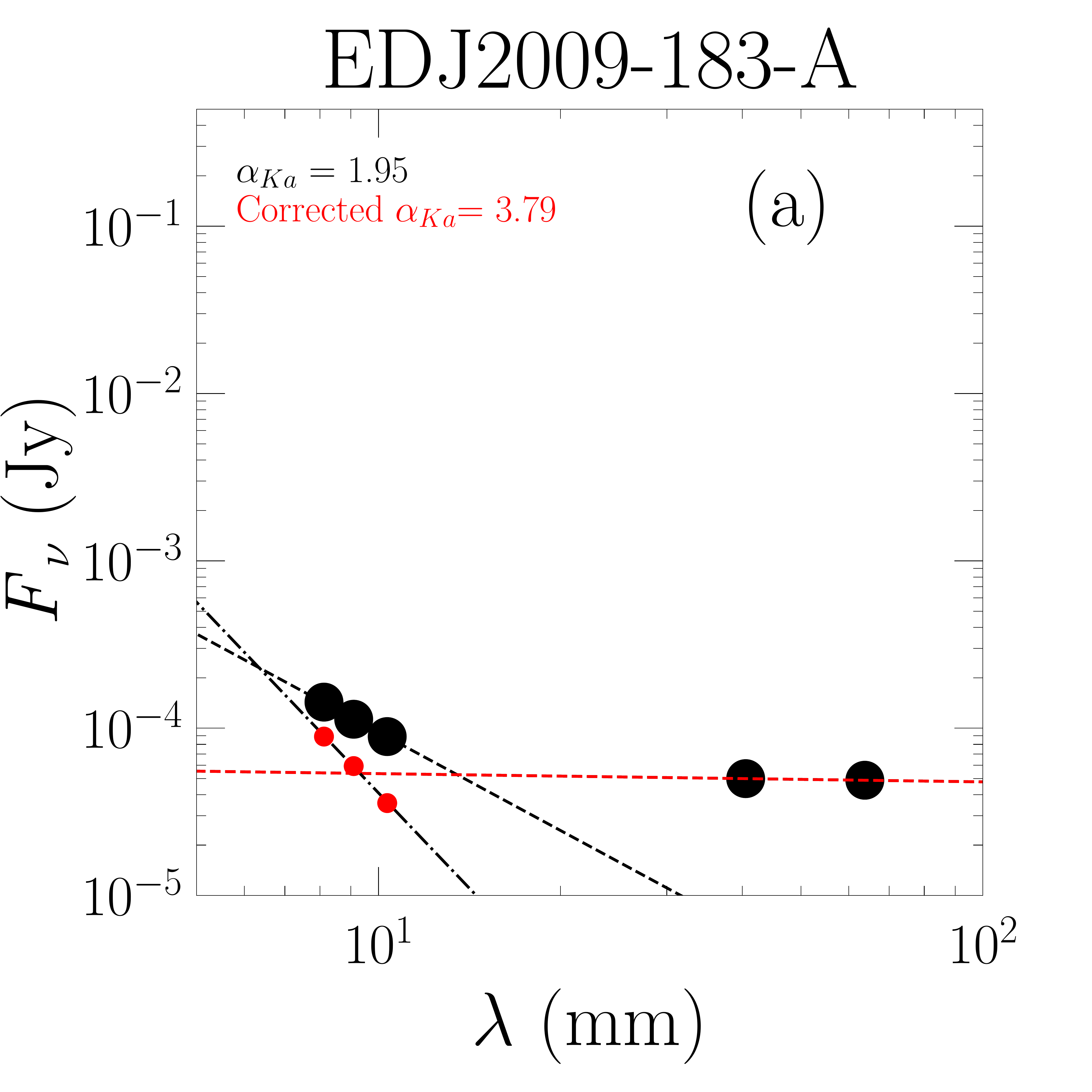}
  \includegraphics[width=0.30\linewidth]{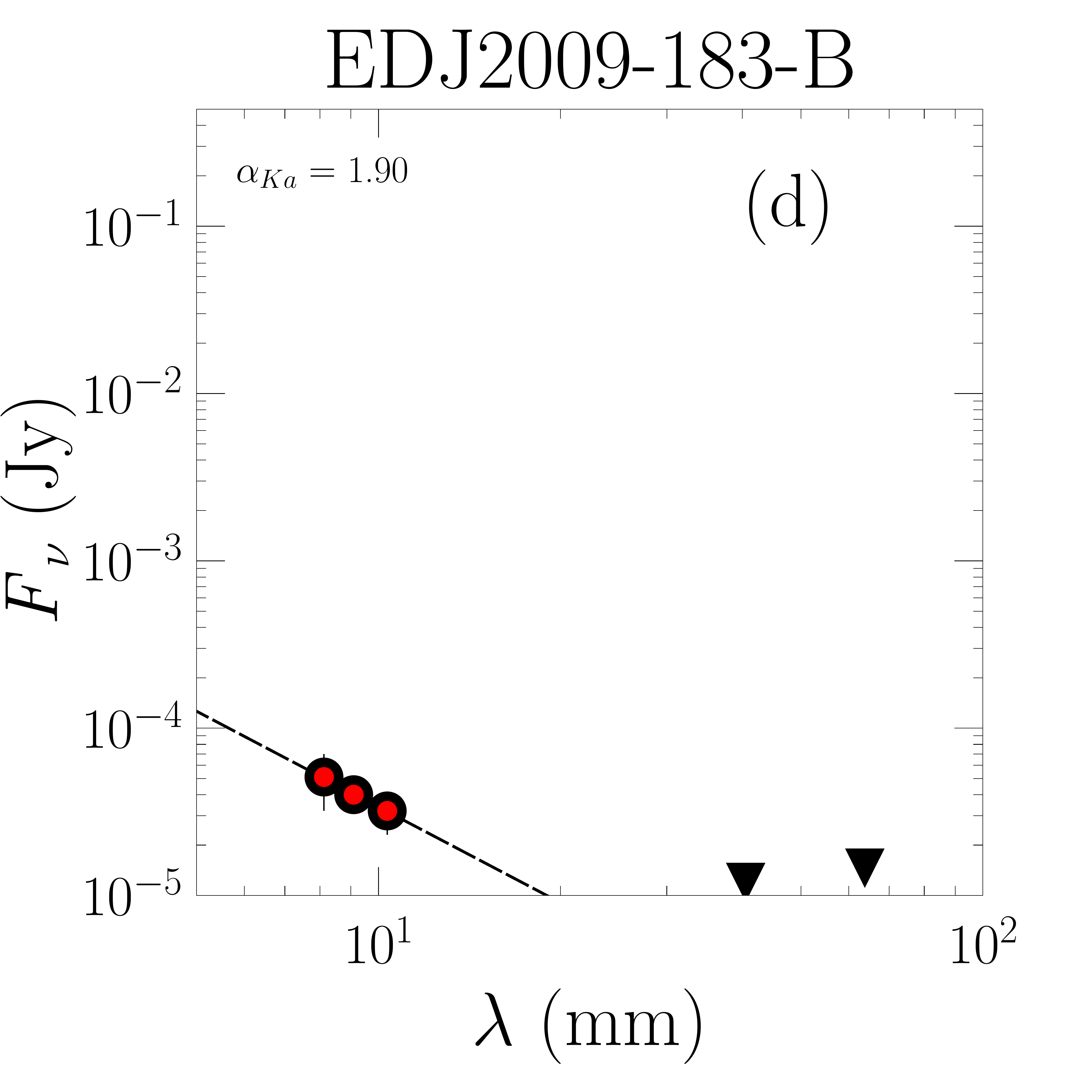}
  \includegraphics[width=0.30\linewidth]{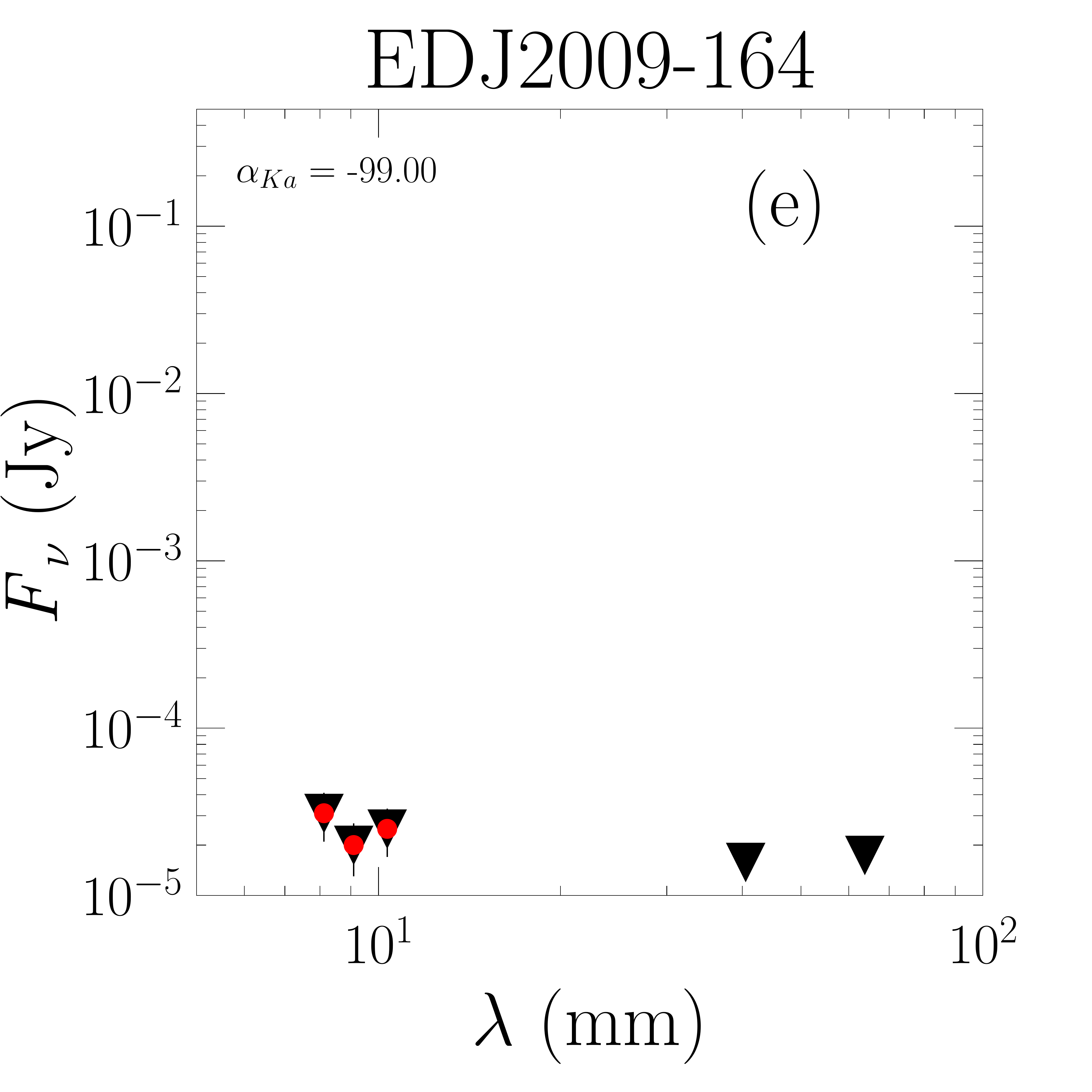}
  \includegraphics[width=0.30\linewidth]{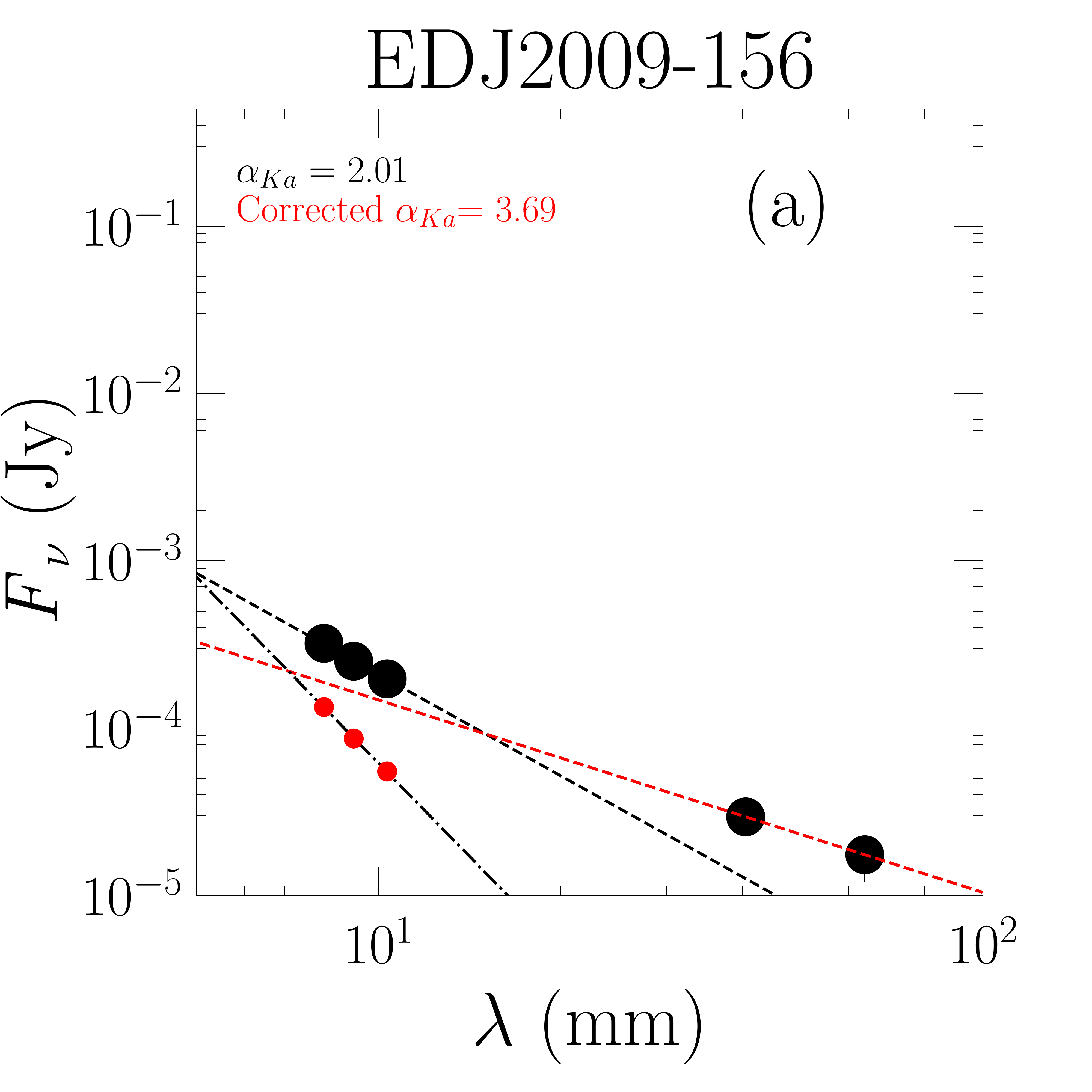}
  \includegraphics[width=0.30\linewidth]{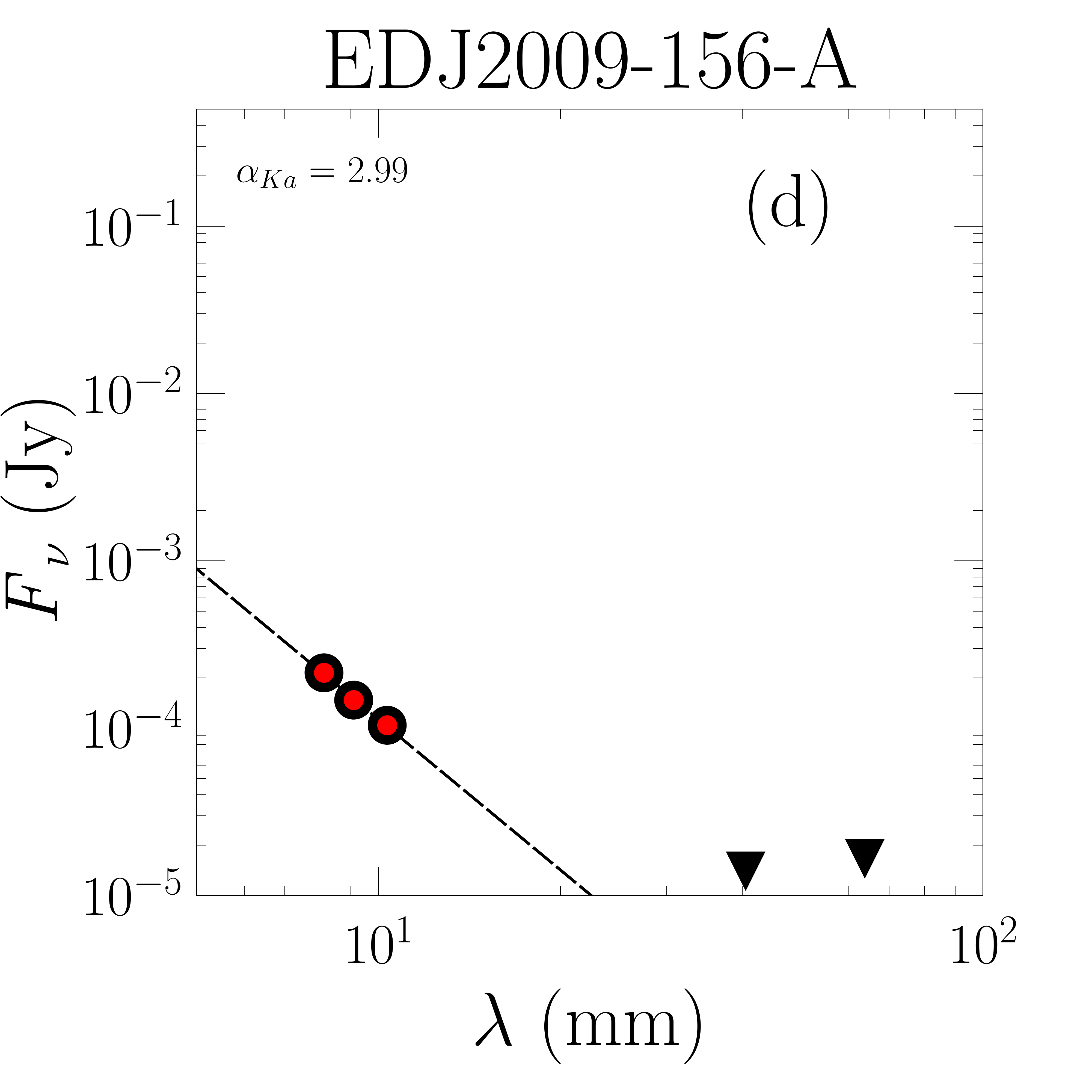}
  \includegraphics[width=0.30\linewidth]{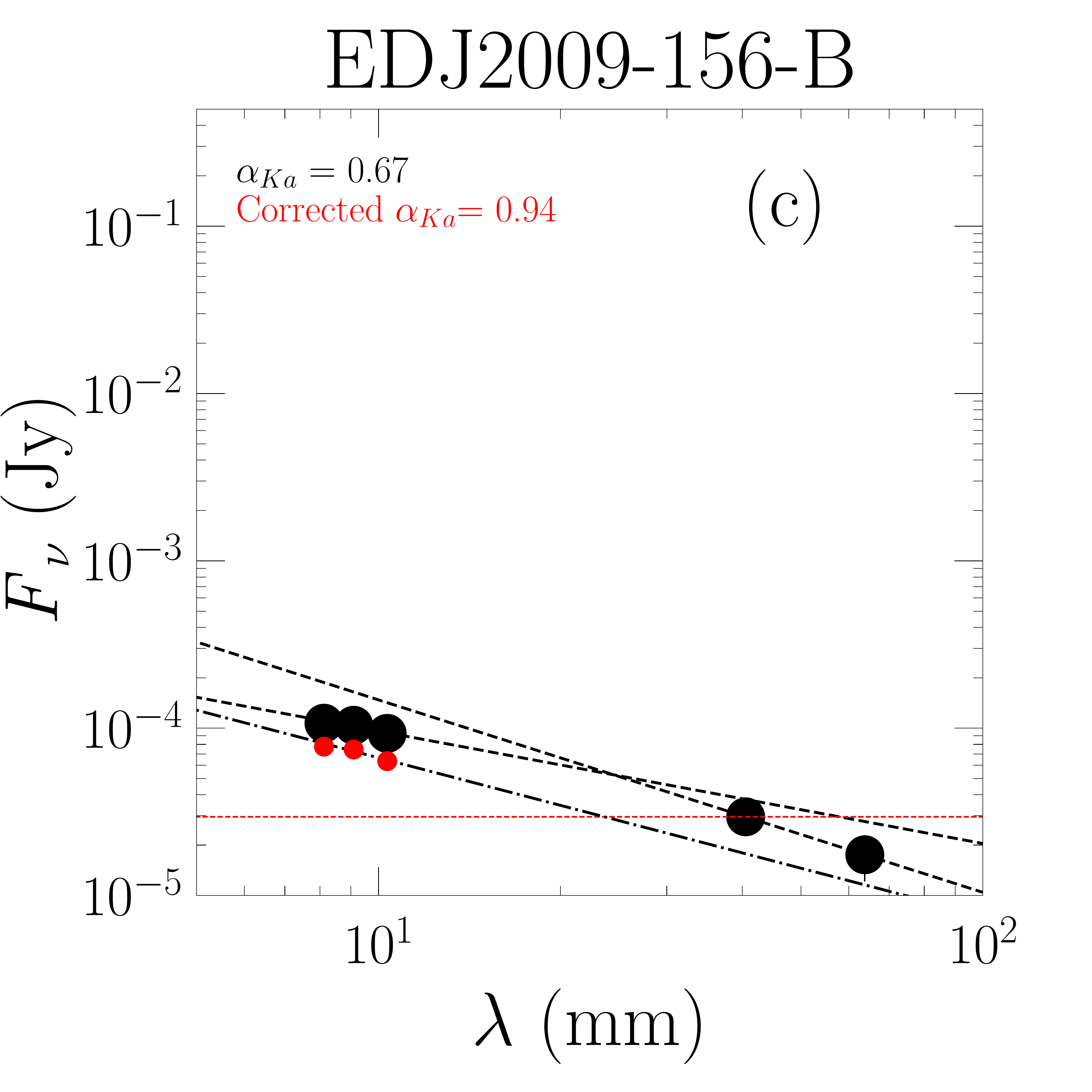}
  \includegraphics[width=0.30\linewidth]{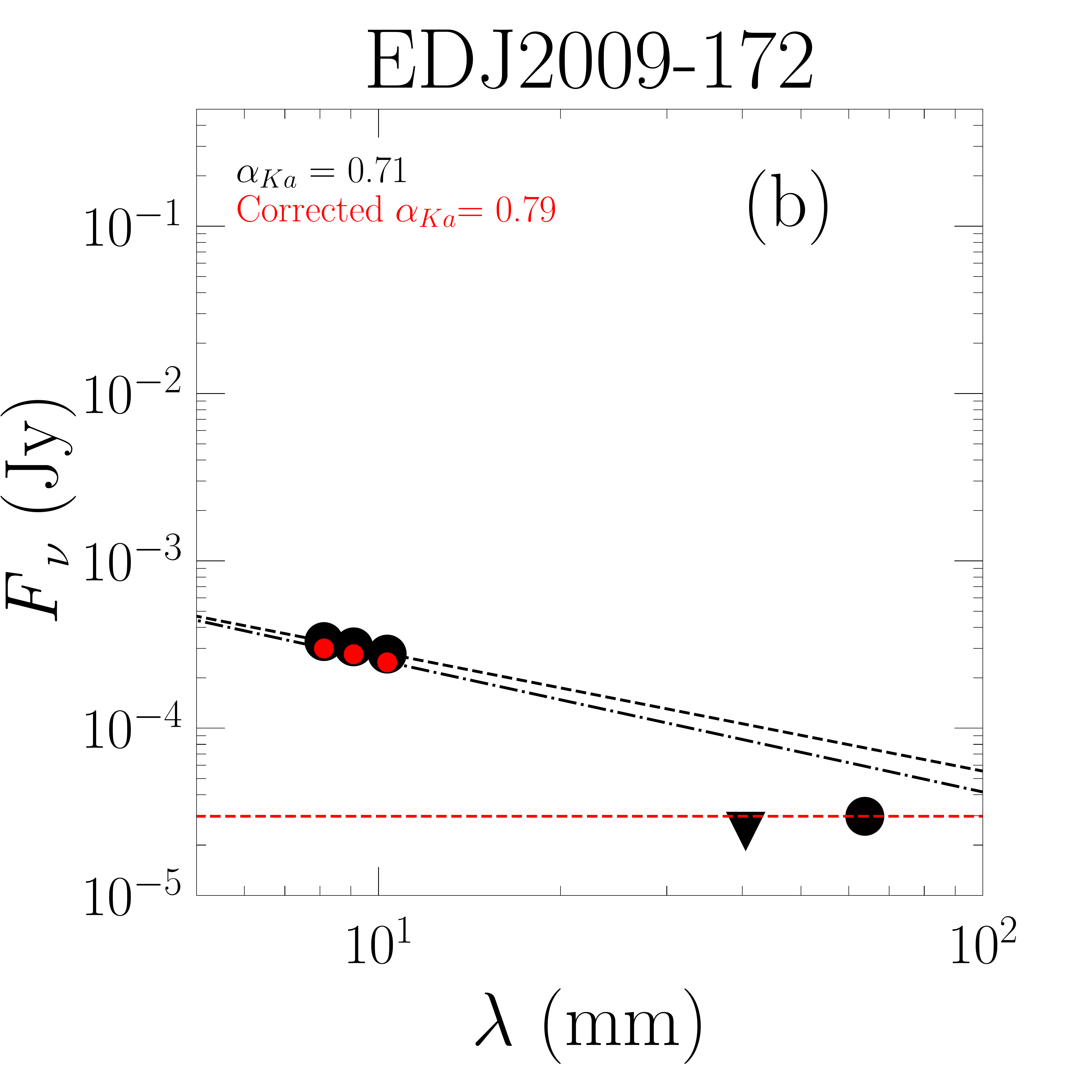}

\end{figure}

\begin{figure}[H]
\centering

  \includegraphics[width=0.30\linewidth]{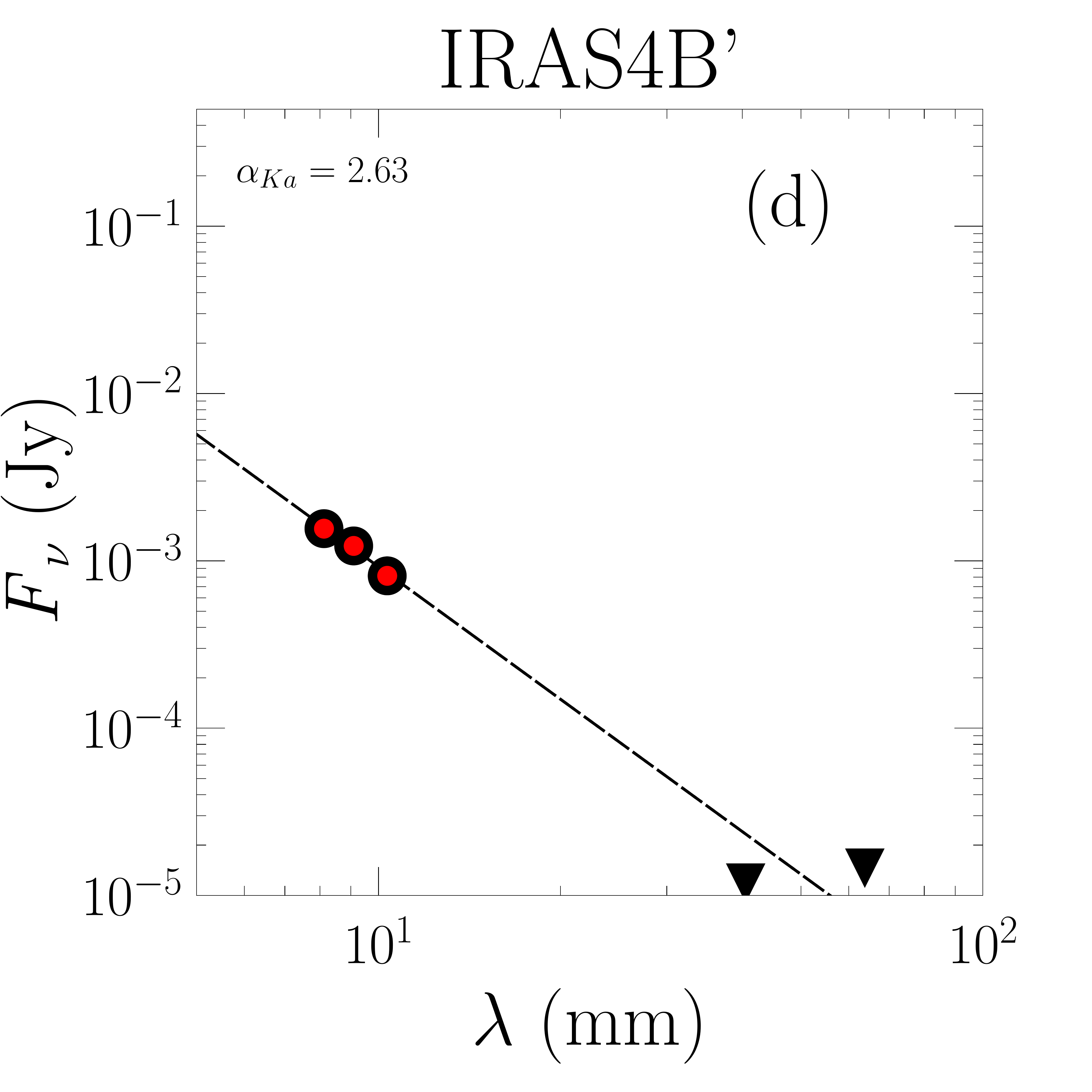}
  \includegraphics[width=0.30\linewidth]{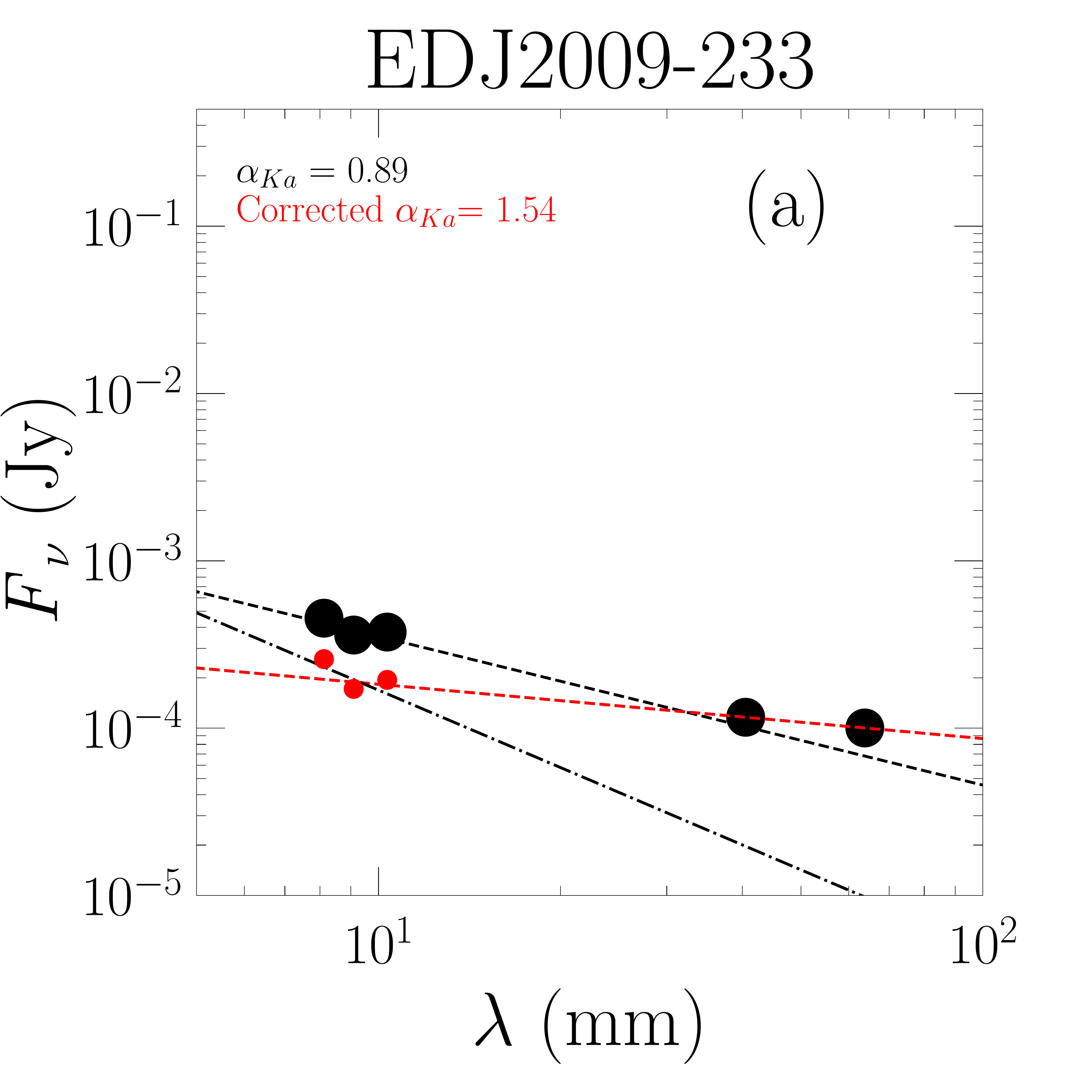}
  \includegraphics[width=0.30\linewidth]{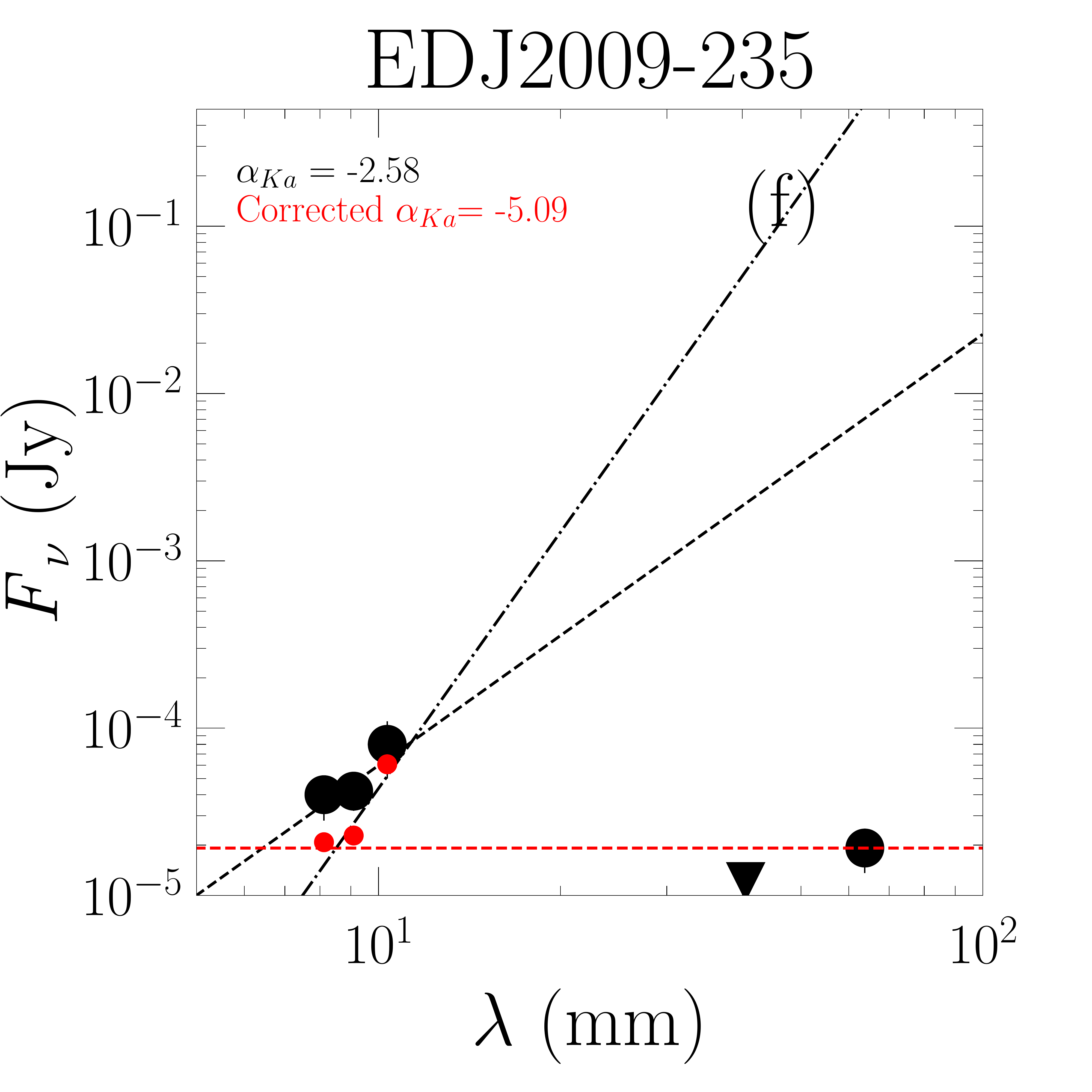}
  \includegraphics[width=0.30\linewidth]{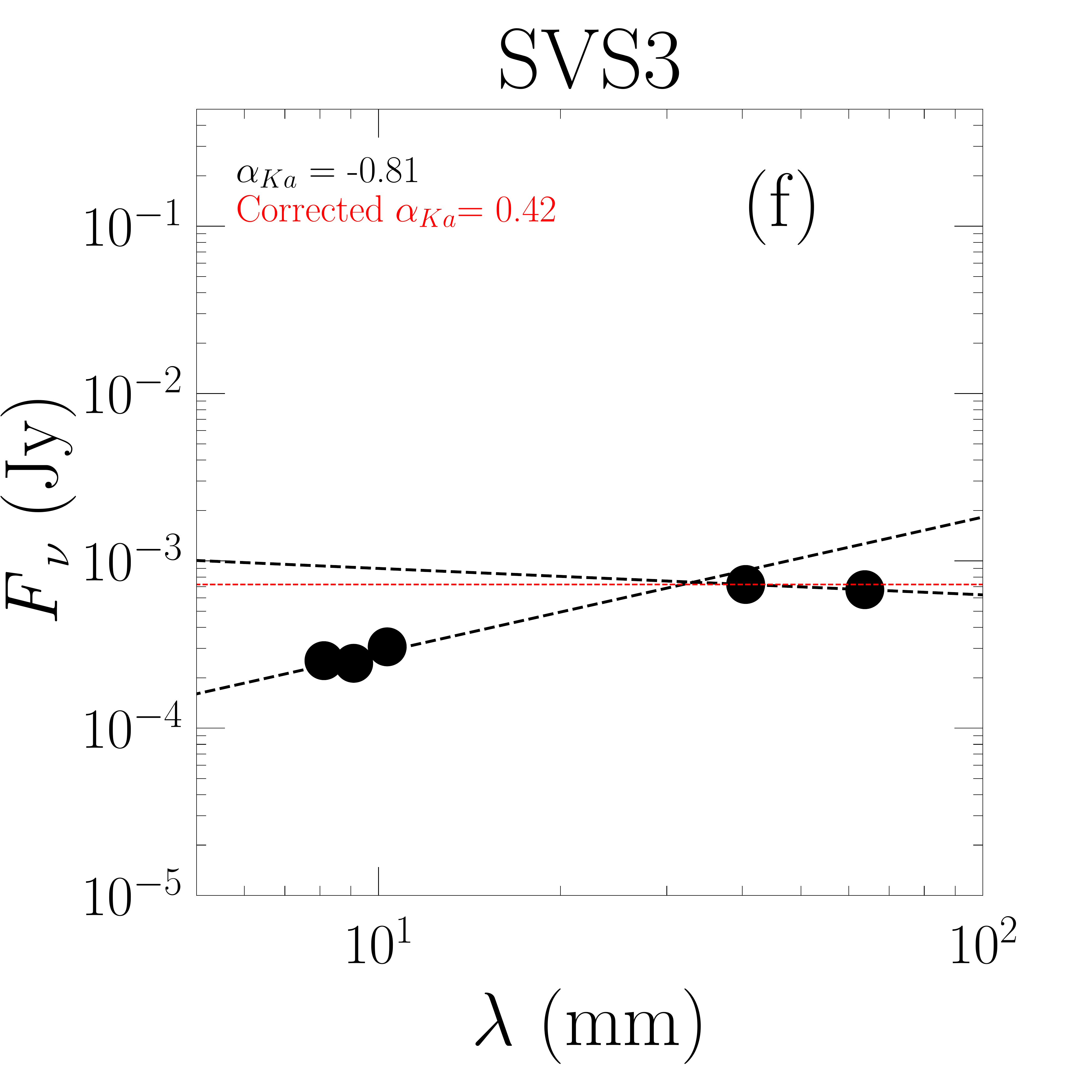}
  \includegraphics[width=0.30\linewidth]{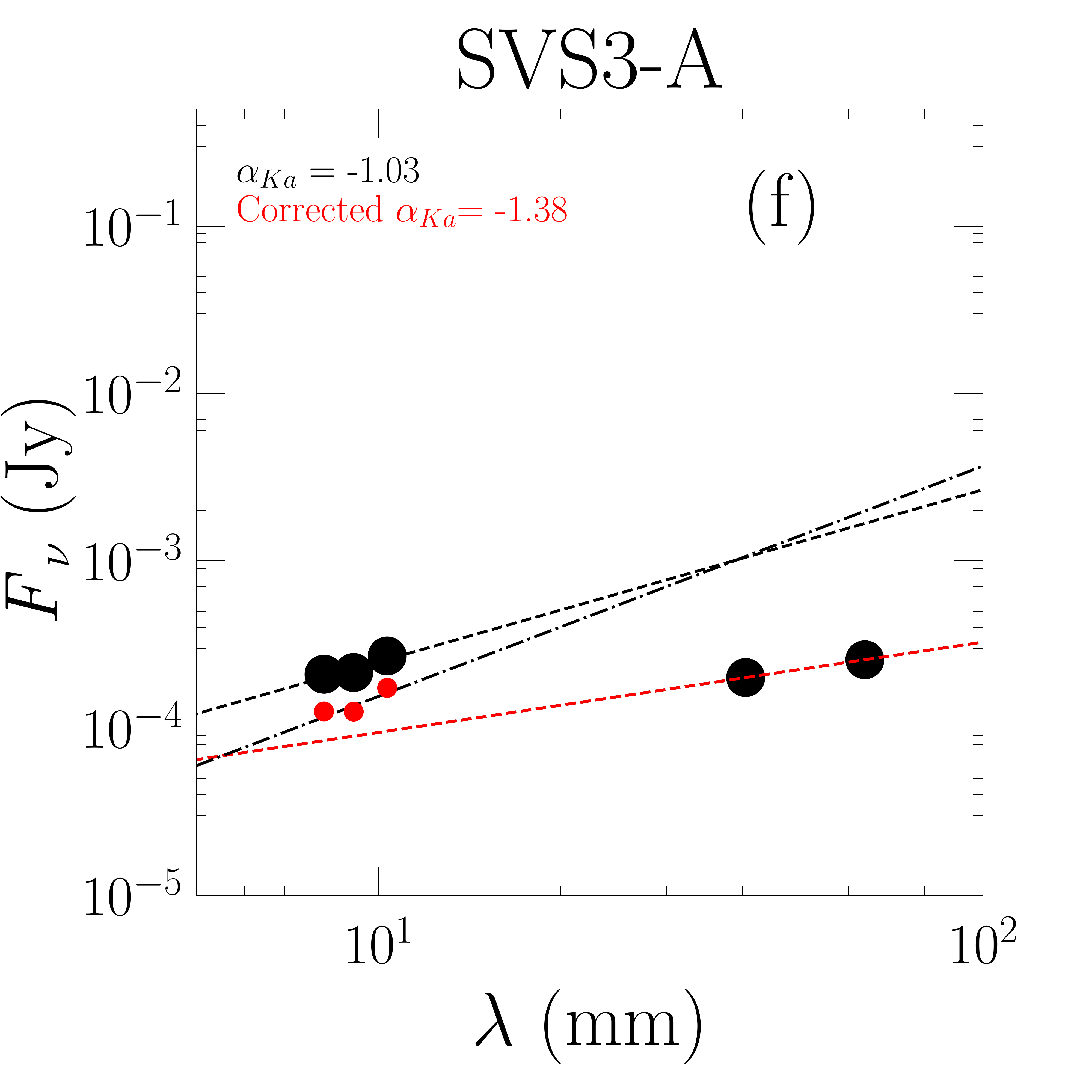}
  \includegraphics[width=0.30\linewidth]{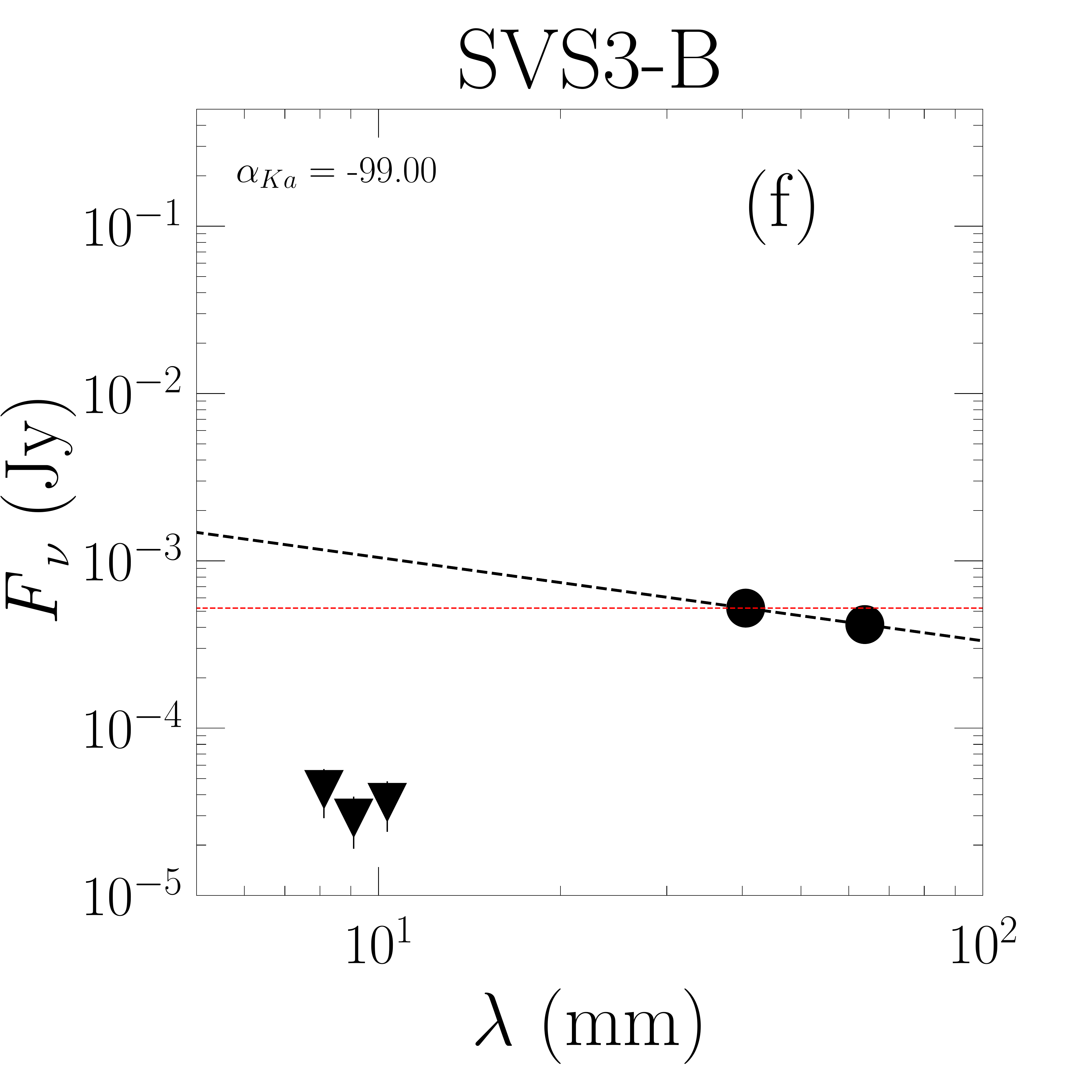}
  \includegraphics[width=0.30\linewidth]{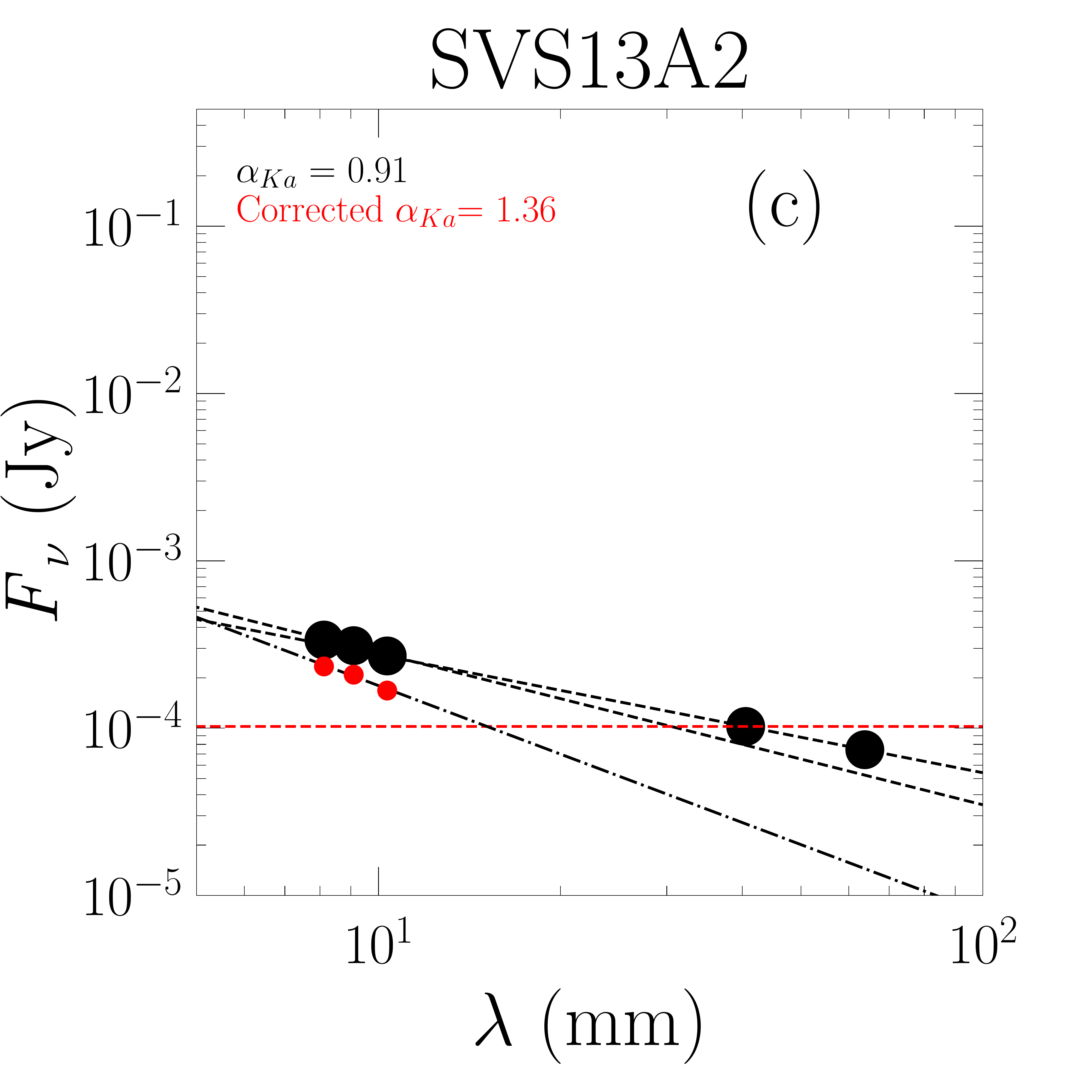}
  \includegraphics[width=0.30\linewidth]{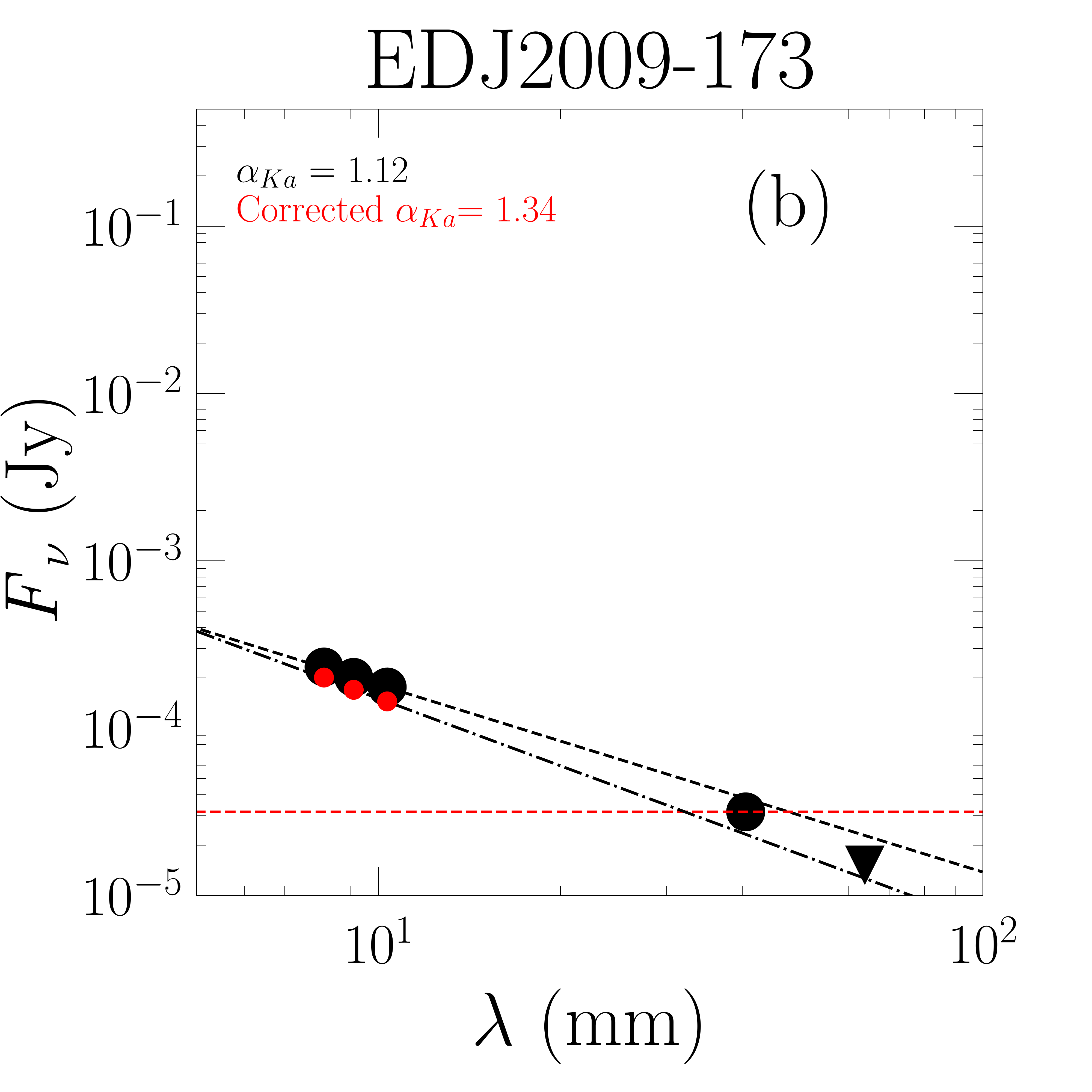}

\label{fig:appendixx}
\end{figure}

\clearpage

\section{Protostars of the VANDAM survey in Ka-band and C-band}
\begin{figure}[H]
 \centering
  \includegraphics[width=0.24\linewidth]{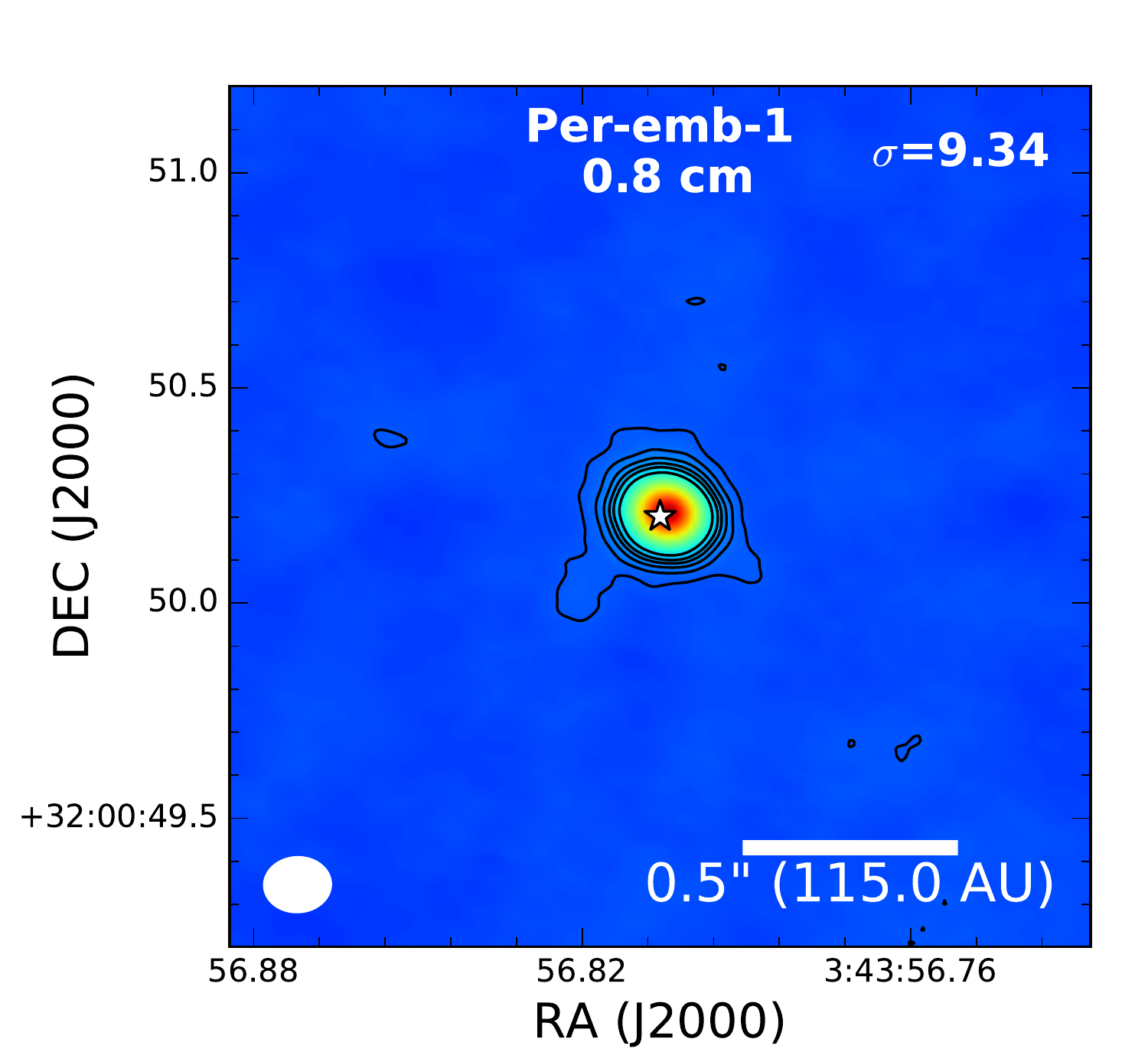}
  \includegraphics[width=0.24\linewidth]{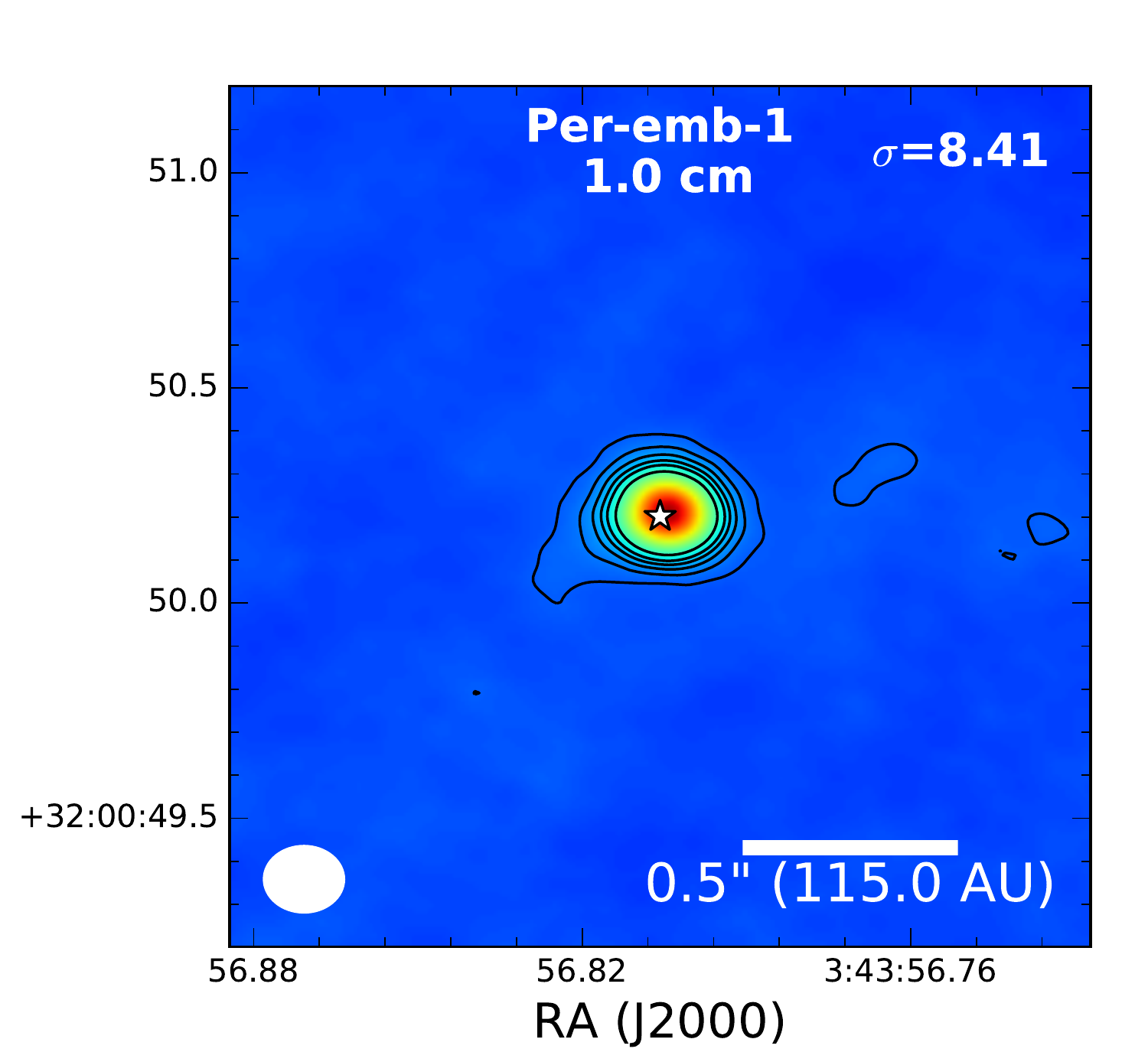}
  \includegraphics[width=0.24\linewidth]{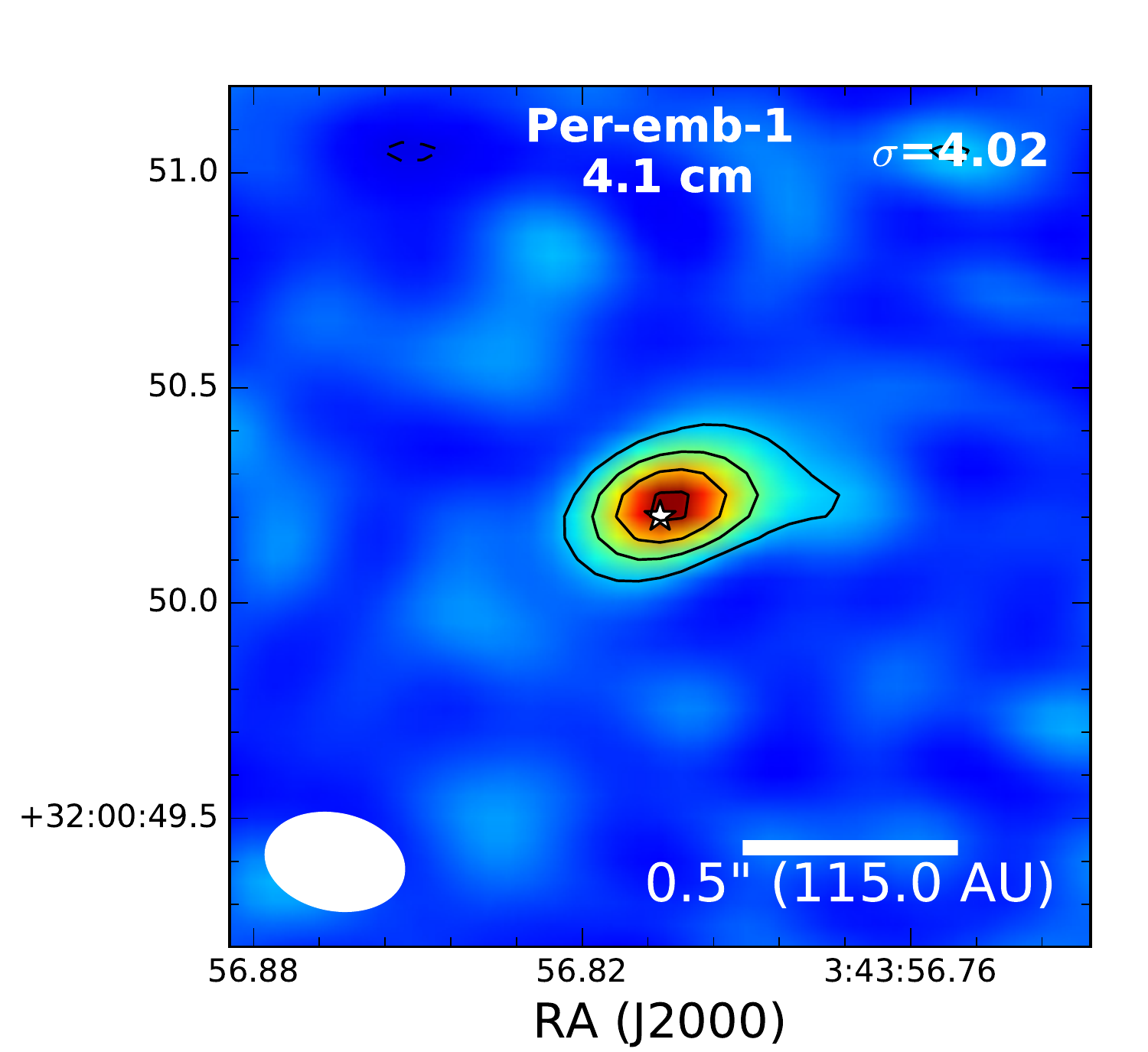}
  \includegraphics[width=0.24\linewidth]{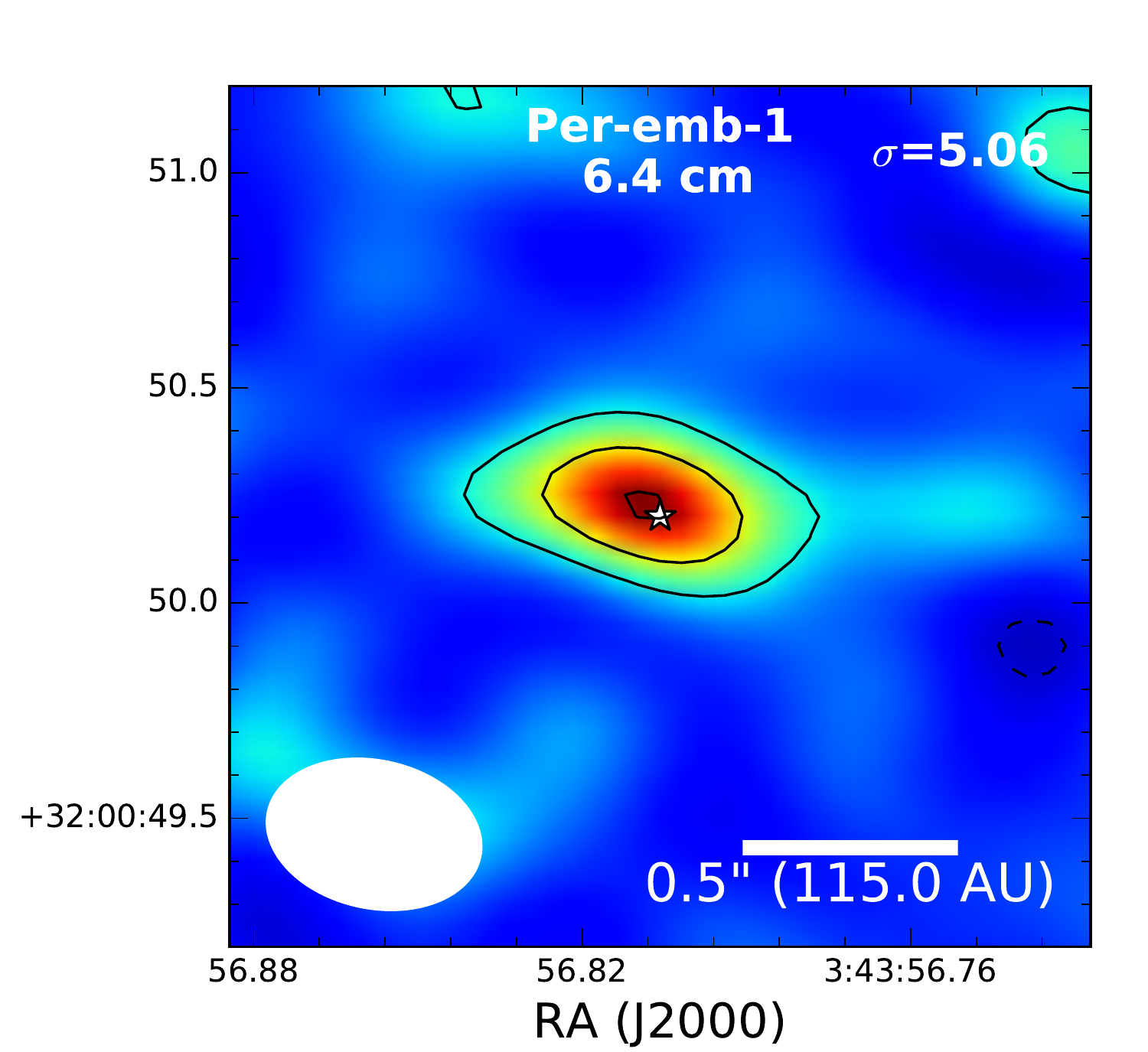}

  \includegraphics[width=0.24\linewidth]{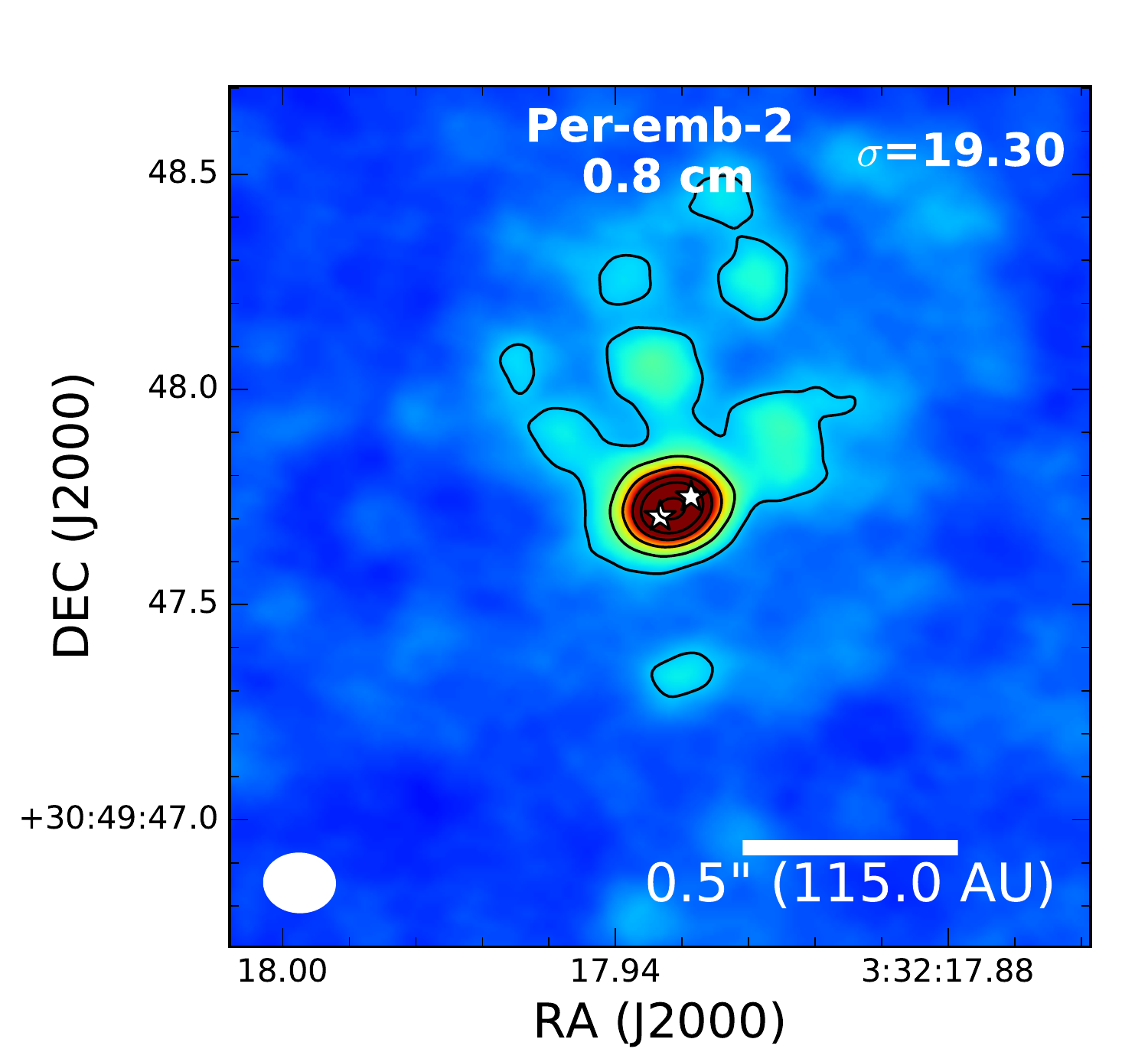}
  \includegraphics[width=0.24\linewidth]{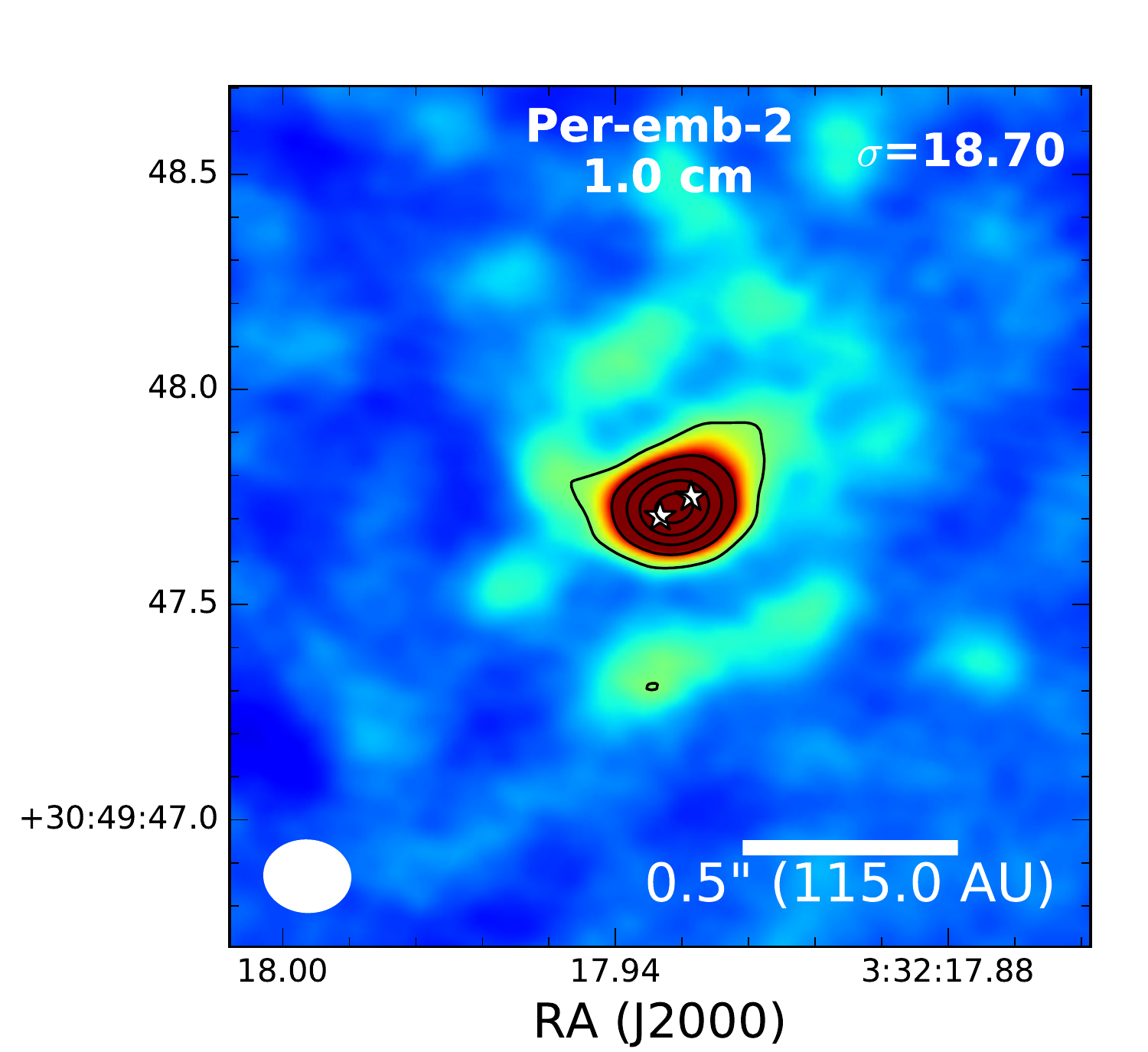}
  \includegraphics[width=0.24\linewidth]{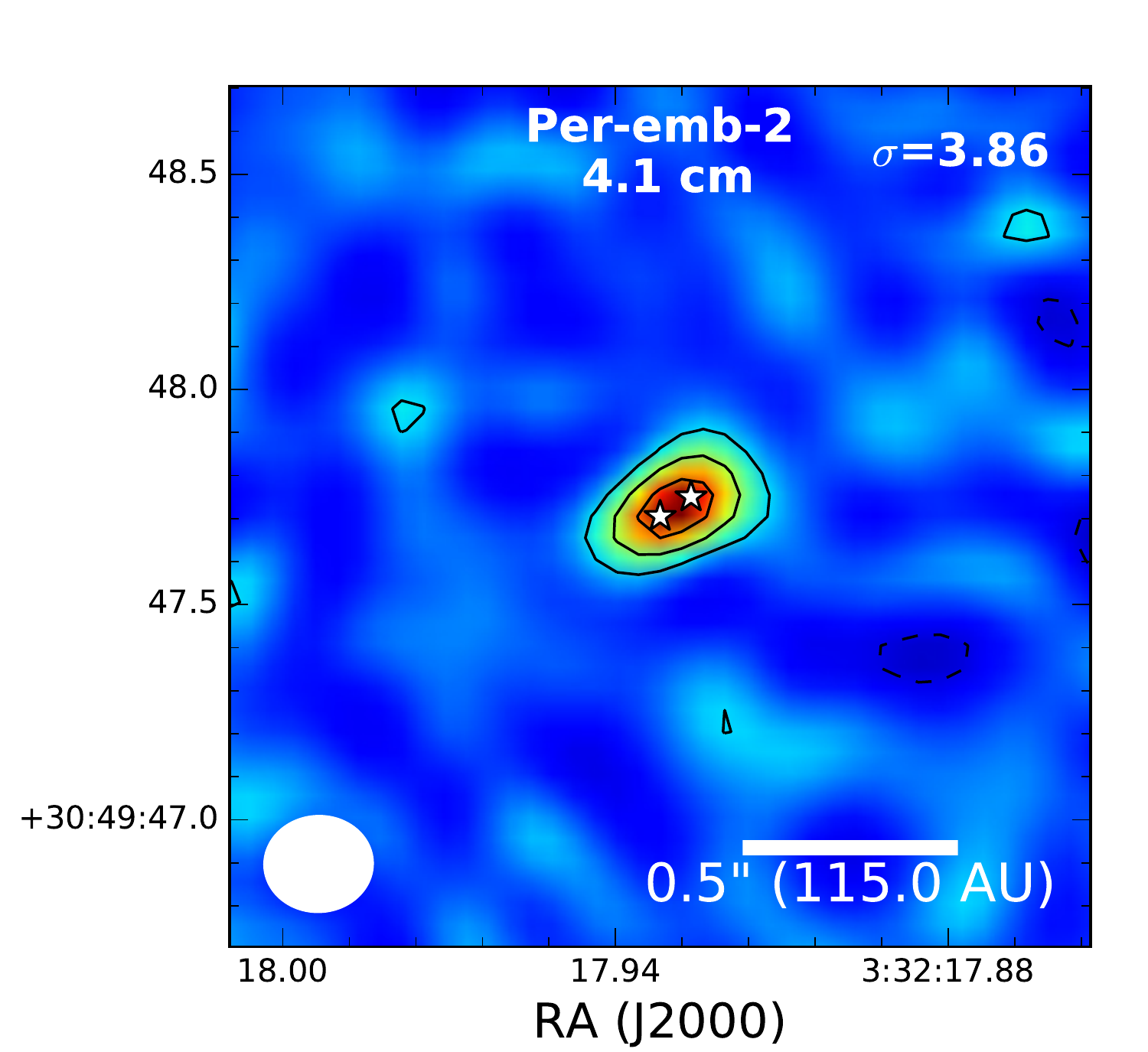}
  \includegraphics[width=0.24\linewidth]{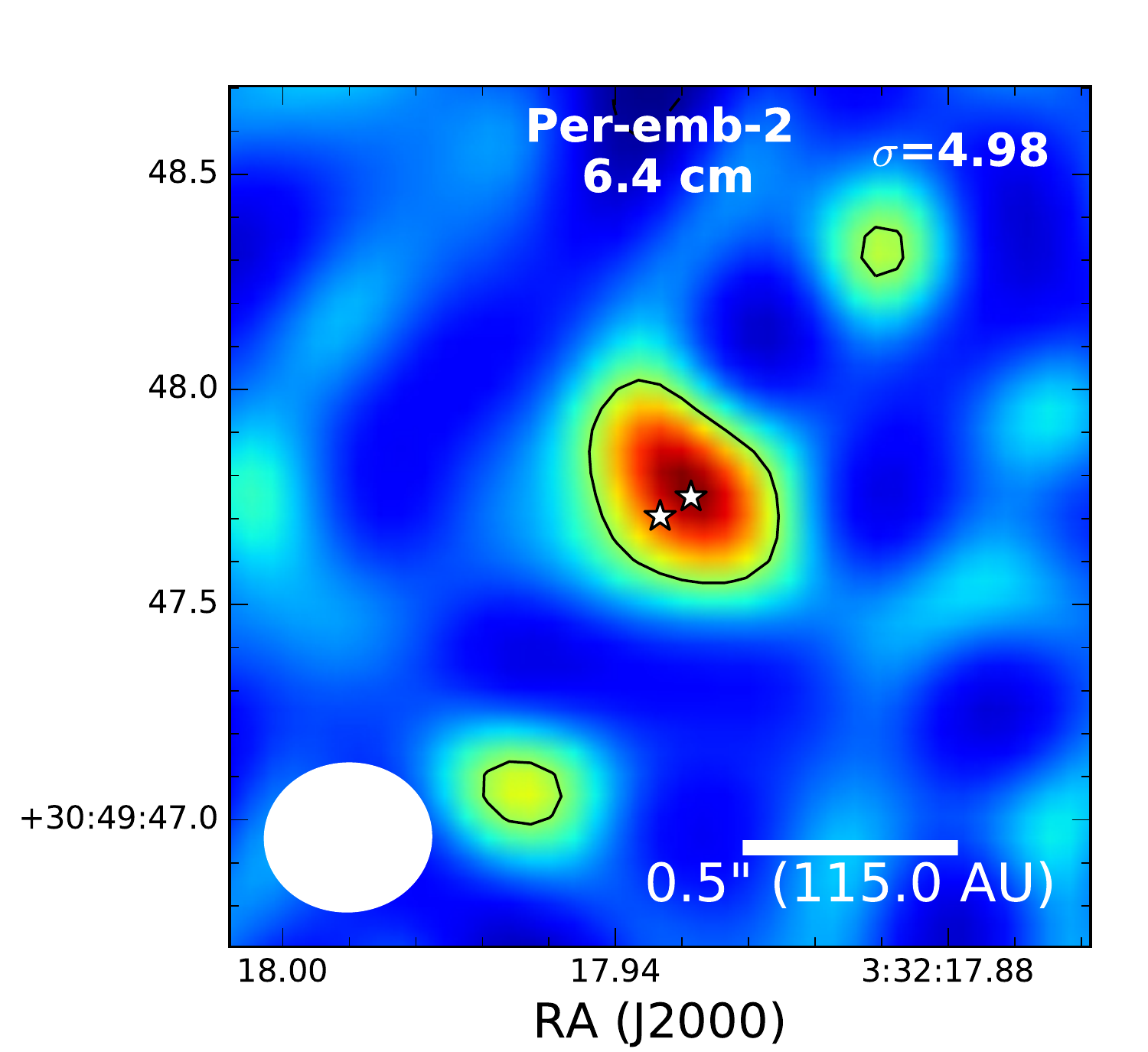}

  \includegraphics[width=0.24\linewidth]{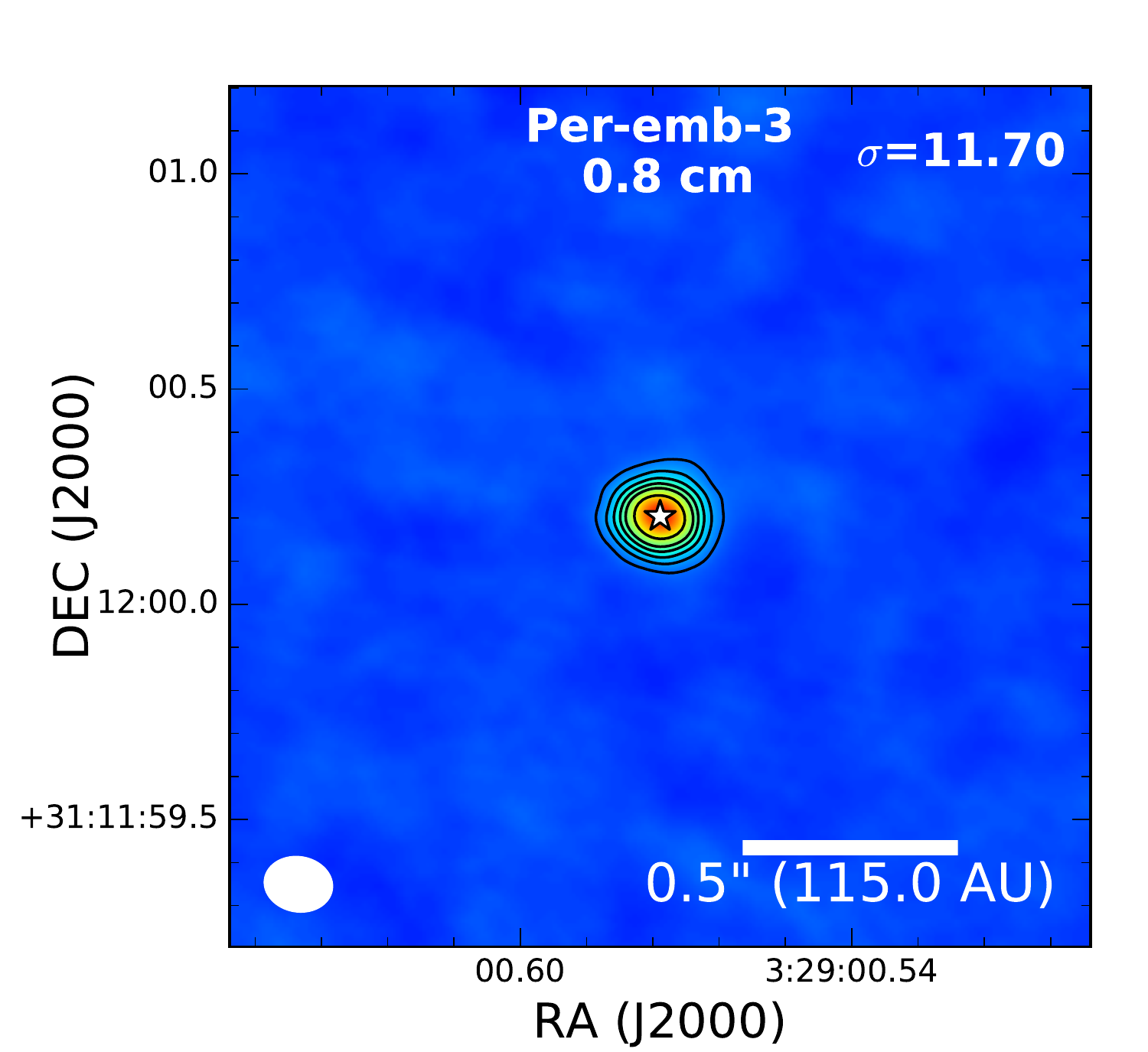}
  \includegraphics[width=0.24\linewidth]{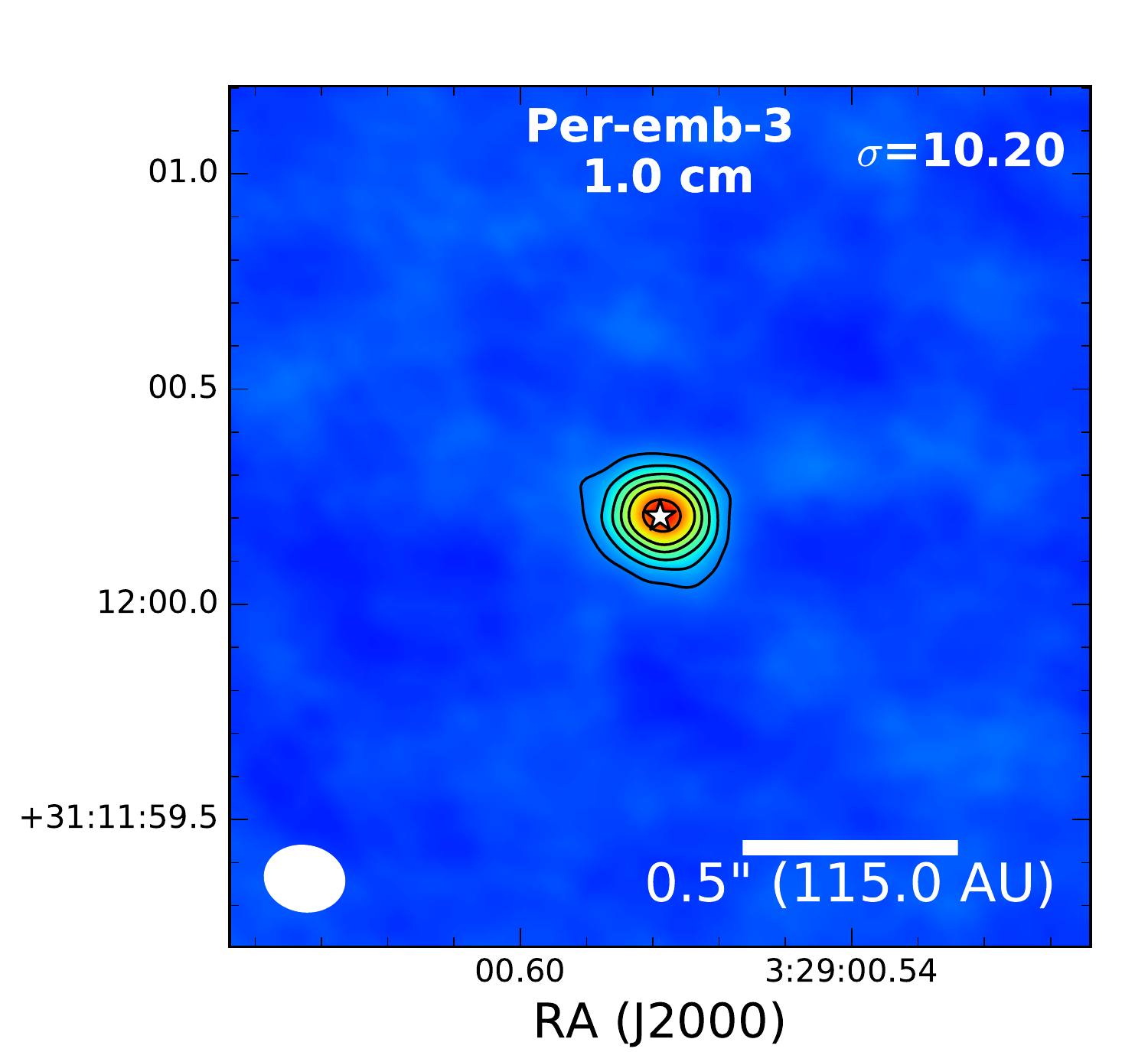}
  \includegraphics[width=0.24\linewidth]{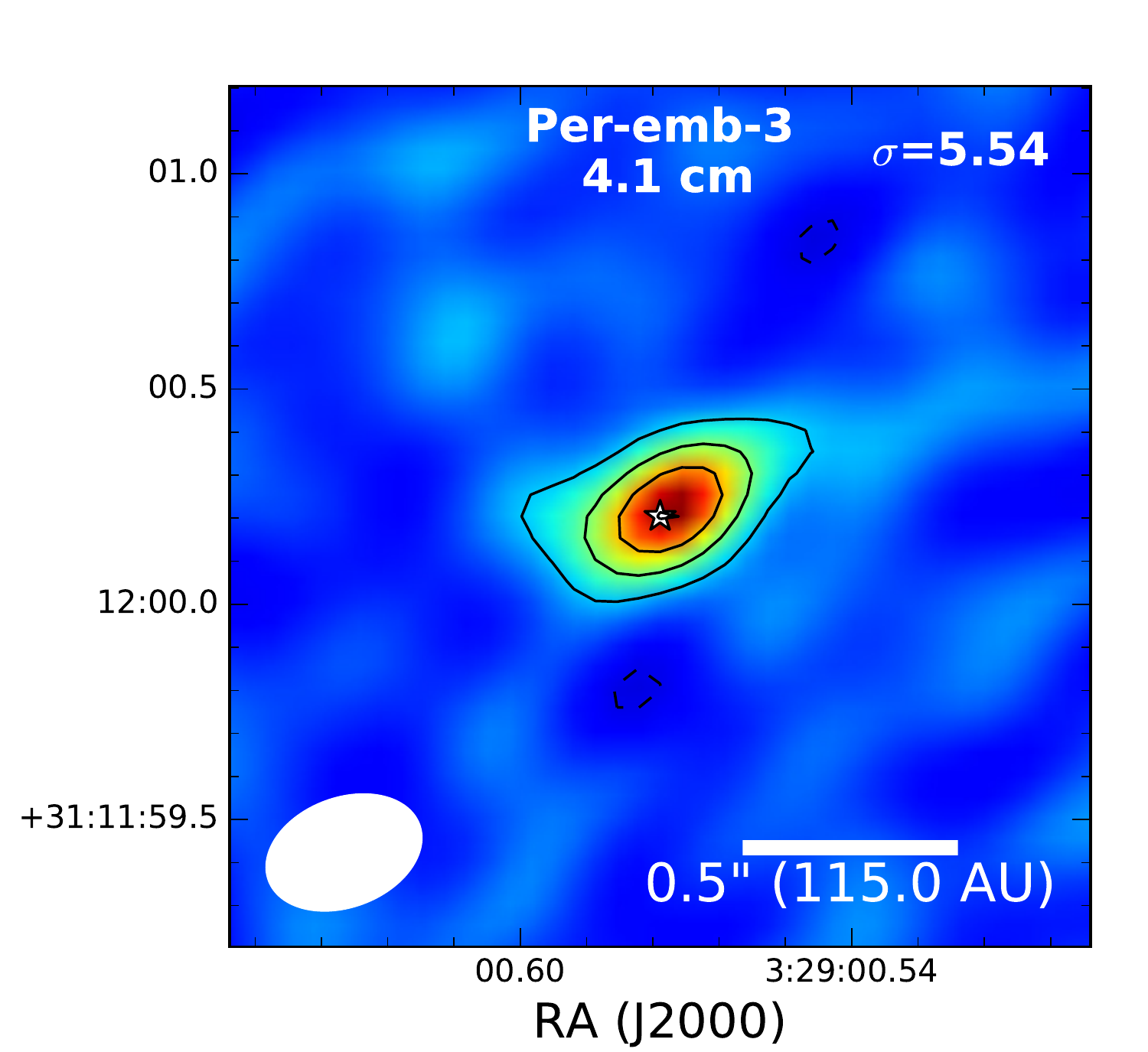}
  \includegraphics[width=0.24\linewidth]{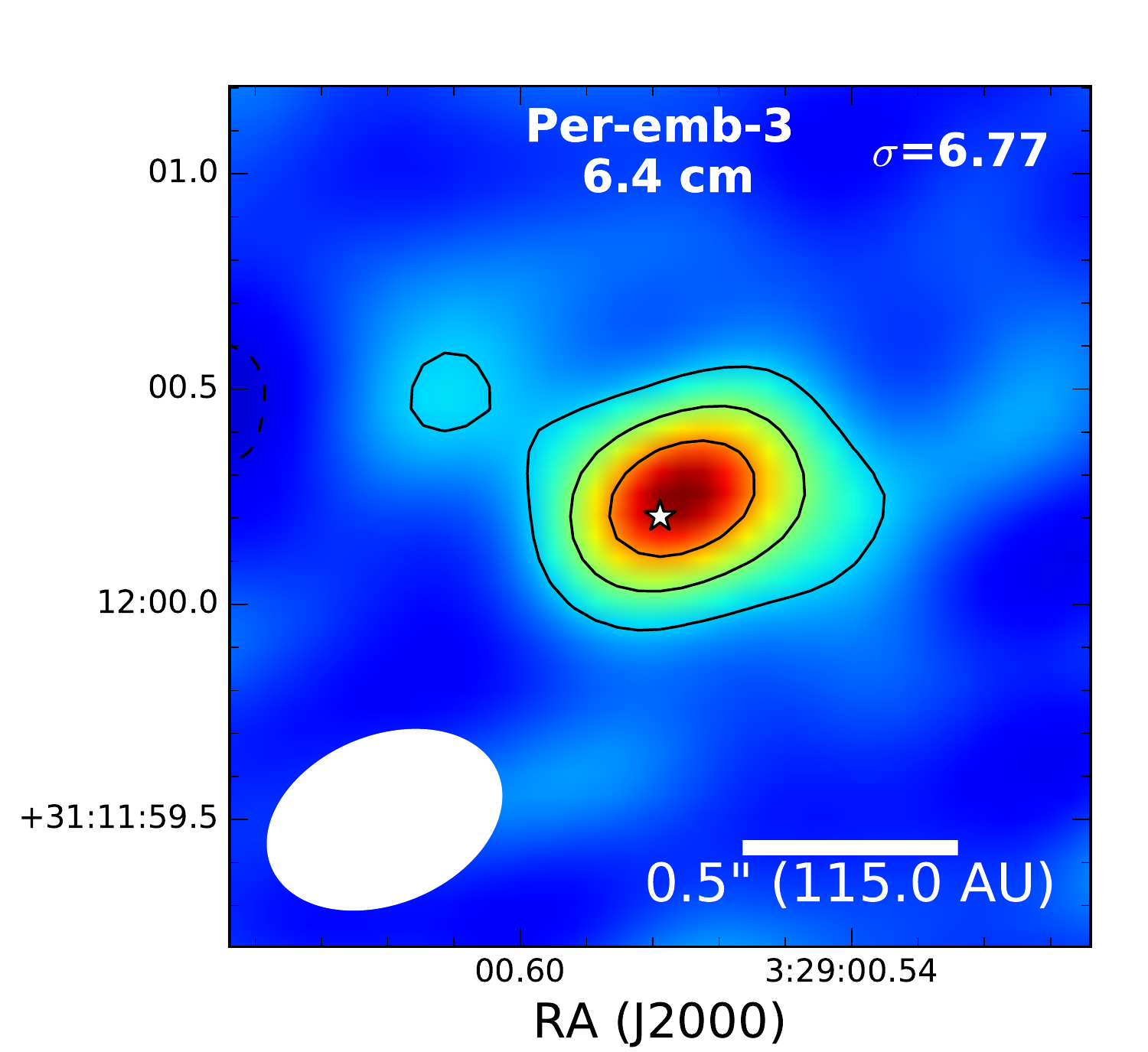}

  \includegraphics[width=0.24\linewidth]{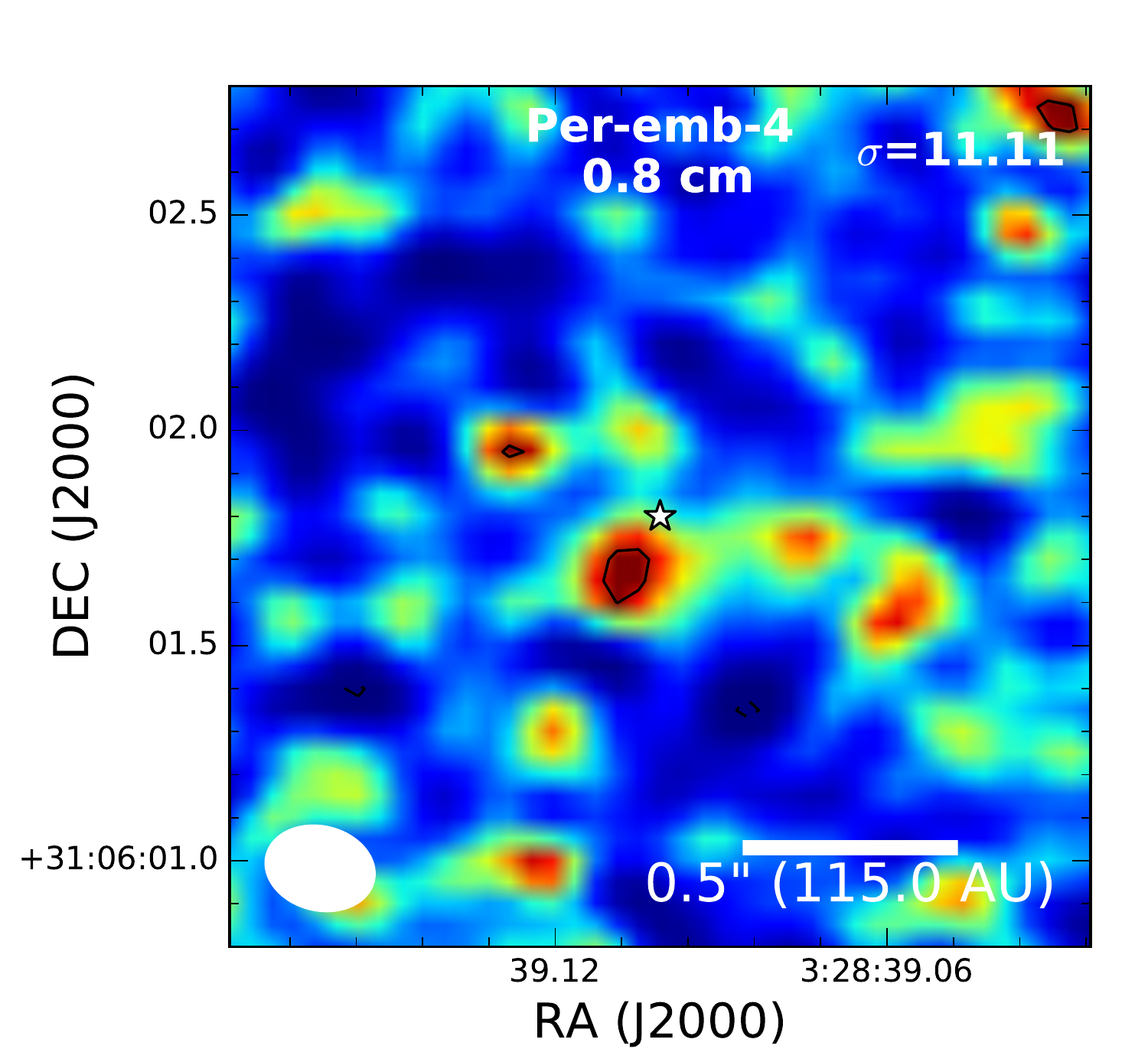}
  \includegraphics[width=0.24\linewidth]{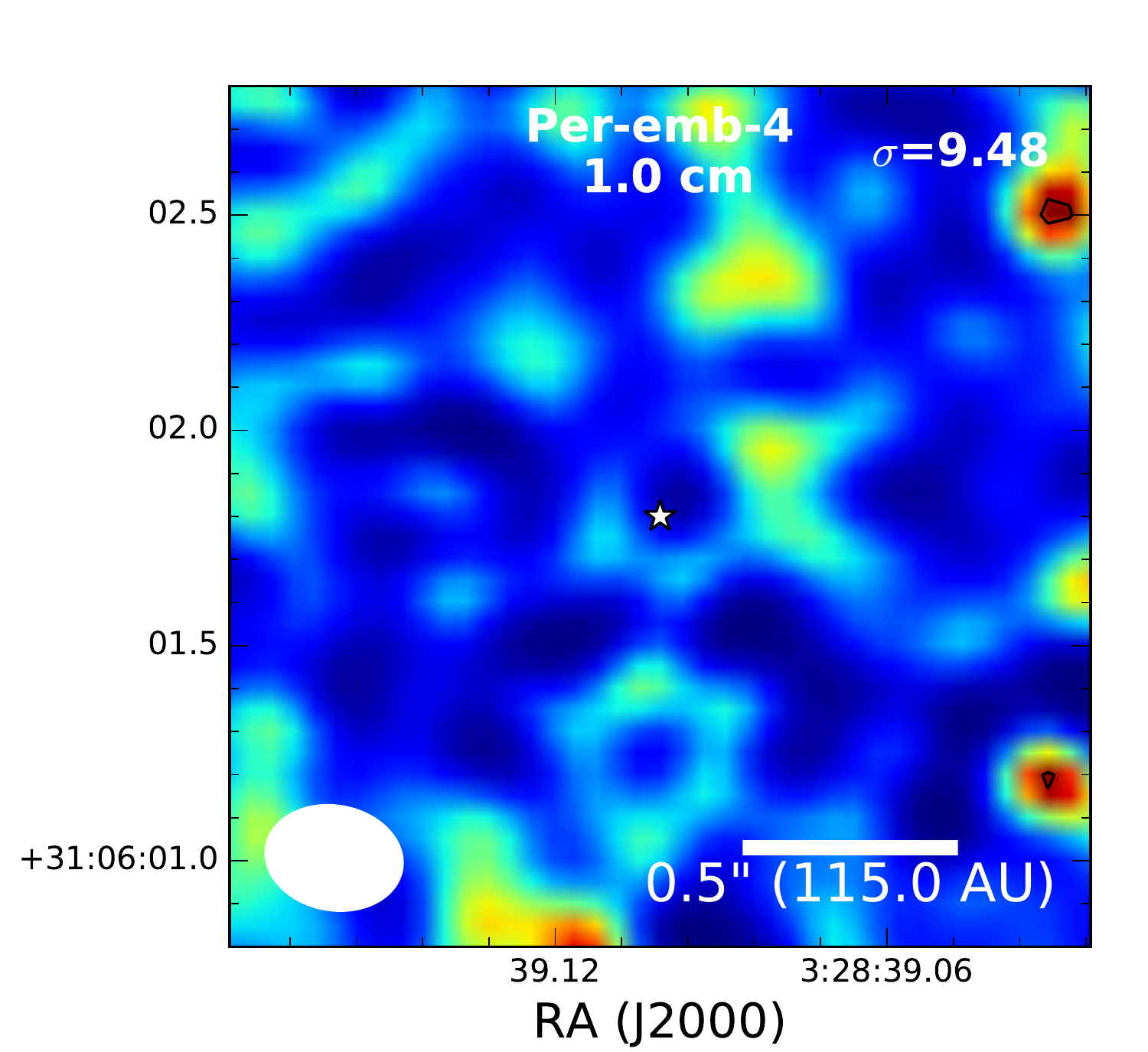}
  \includegraphics[width=0.24\linewidth]{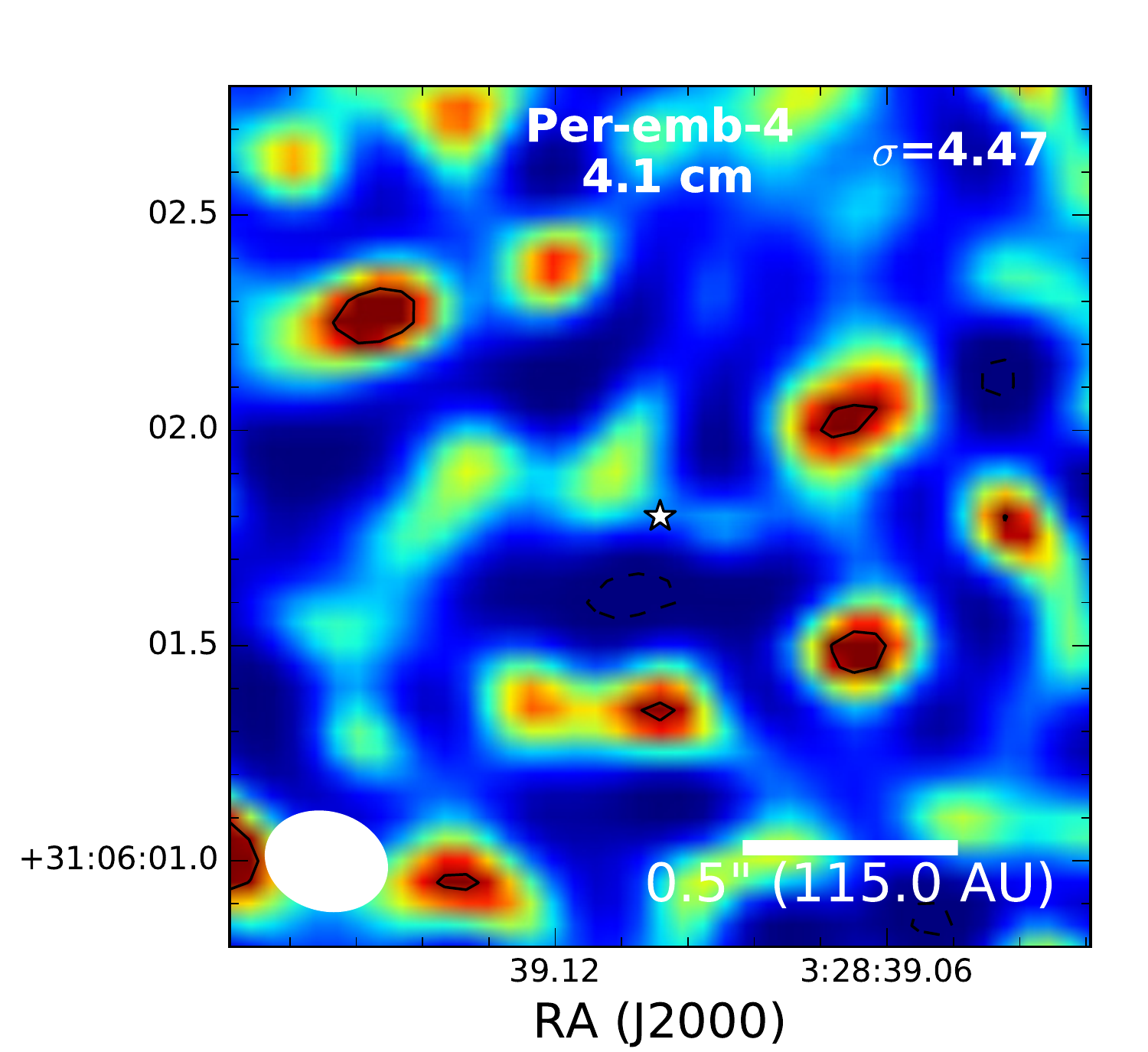}
  \includegraphics[width=0.24\linewidth]{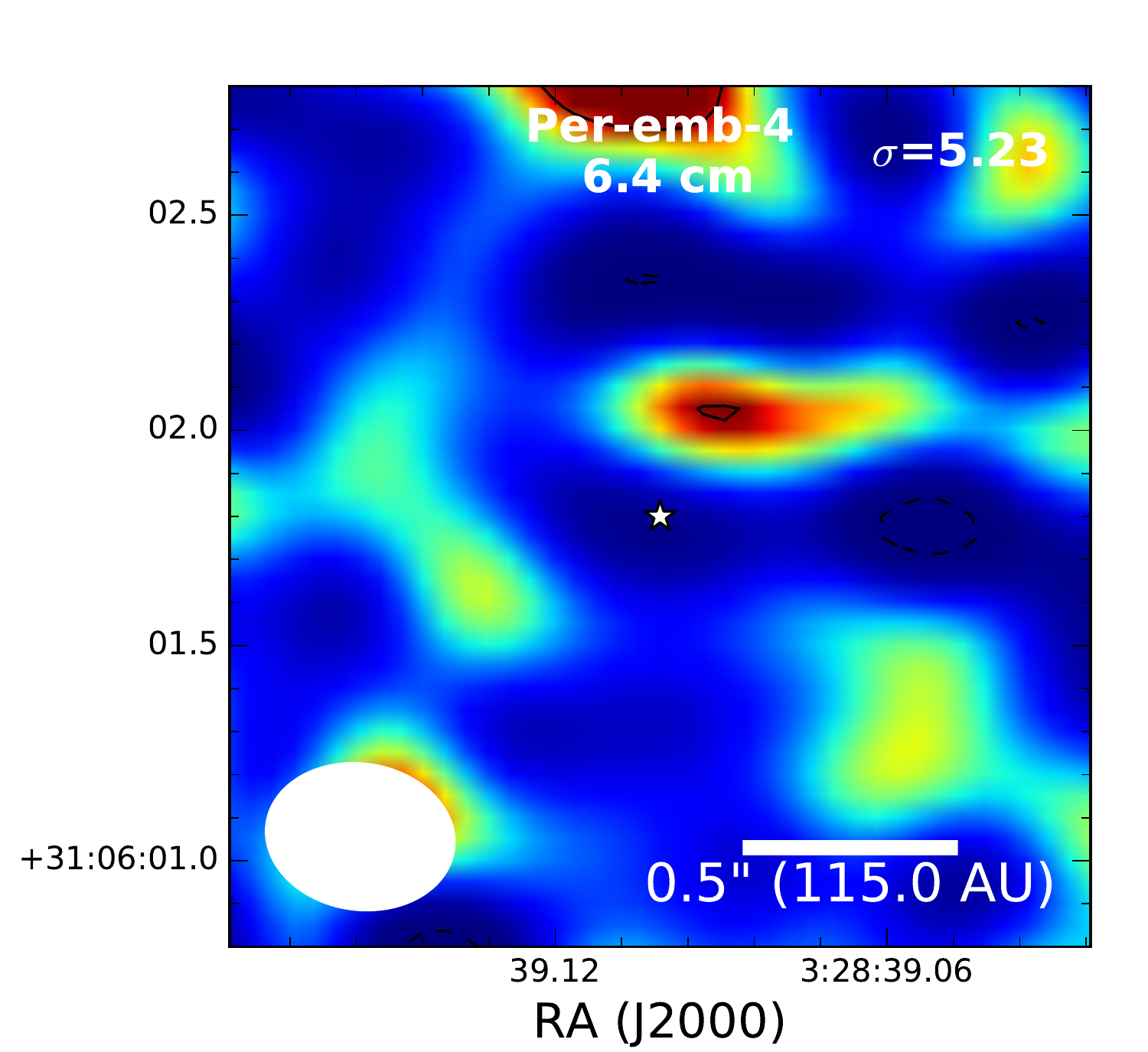}

\caption{Images of all protostars targeted by both Ka-band (0.8 \& 1.0 cm) and C-band (4.1 \& 6.4 cm) observations with increasing wavelength from left to right.
 The contours are [-3, 3, 6, 9, 12, 15, 20] $\times\  \sigma$ where $\sigma$ in mJy for each map is provided in top-right corner.}
\label{fig:all1}
\end{figure}

Here we present the images of all protostars targeted in VANDAM survey. Figure \ref{fig:all1} shows four images of each protostar obtained in Ka-band (0.8 \& 1.0 cm) and in C-band (4.1 \& 6.4 cm).

\begin{figure}[H]
 \centering
 \includegraphics[width=0.24\linewidth]{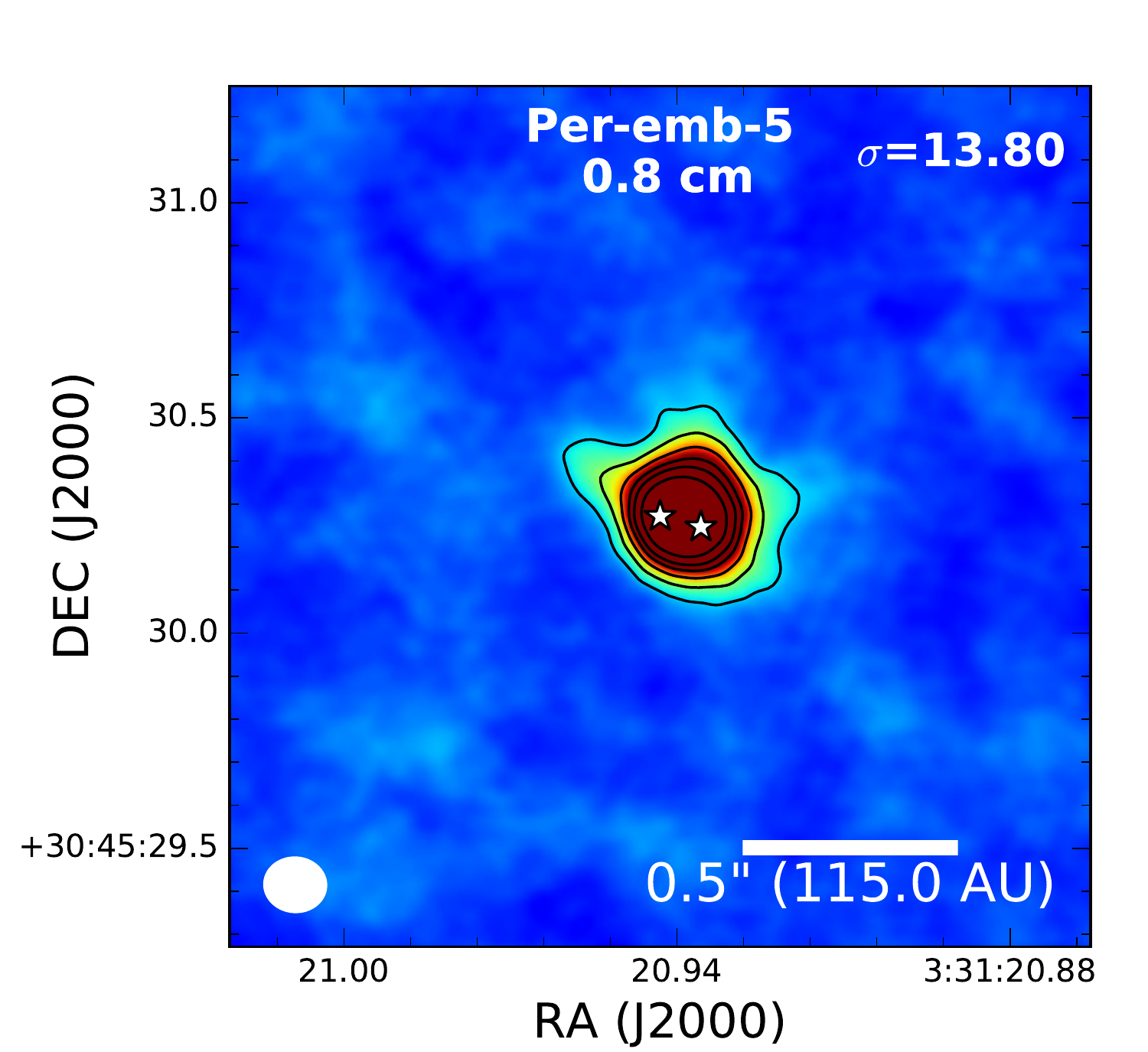}
  \includegraphics[width=0.24\linewidth]{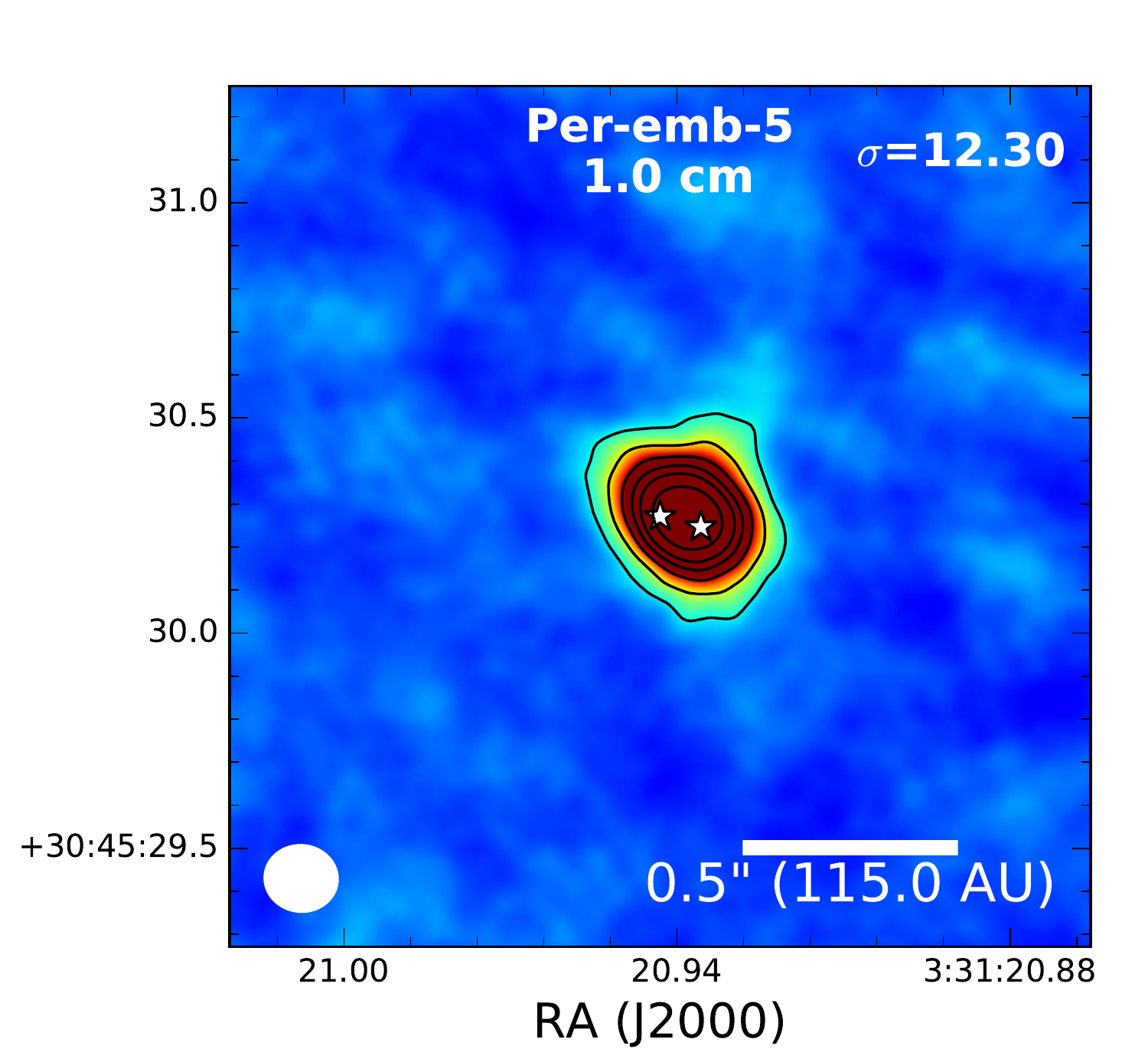}
  \includegraphics[width=0.24\linewidth]{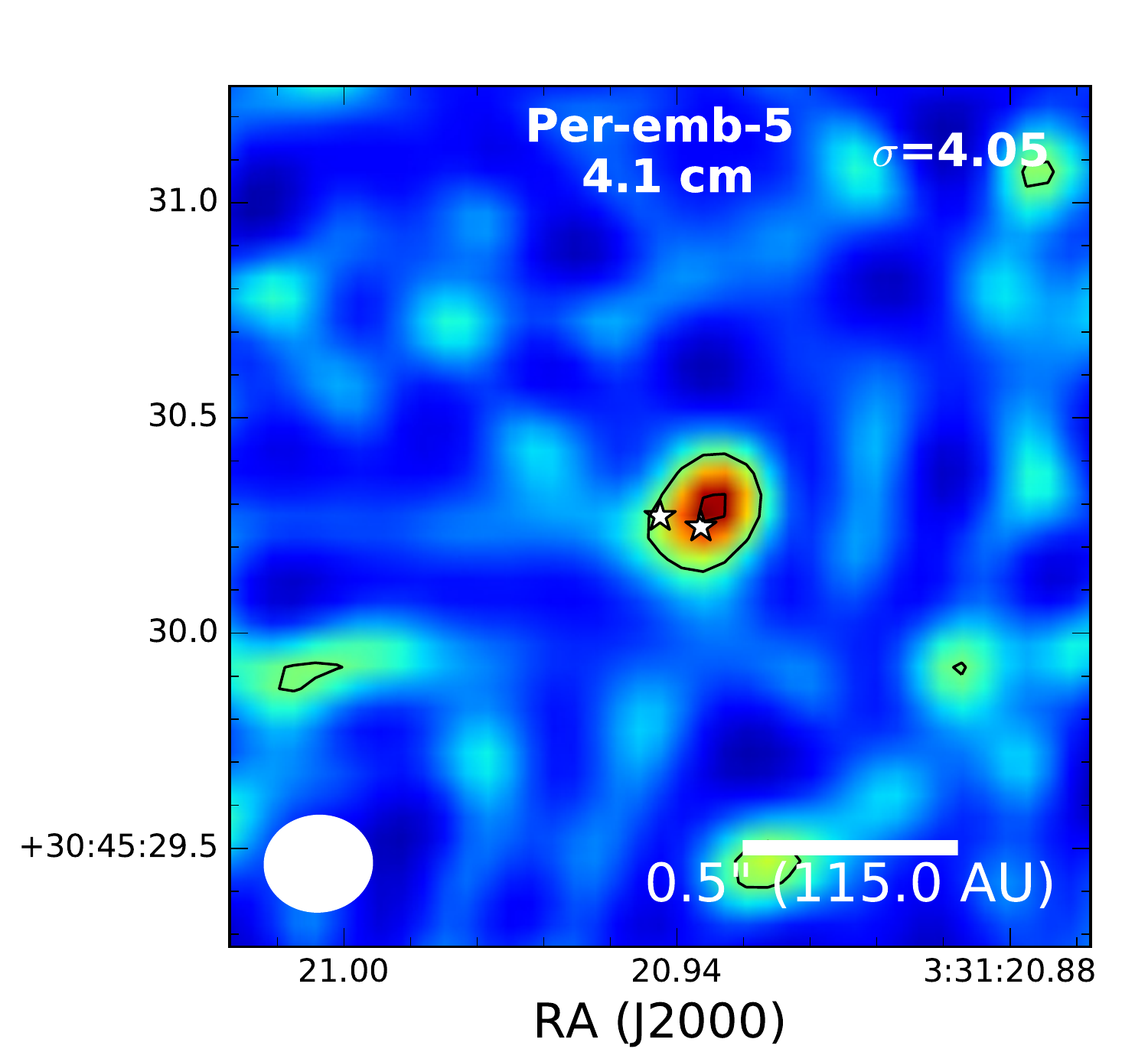}
  \includegraphics[width=0.24\linewidth]{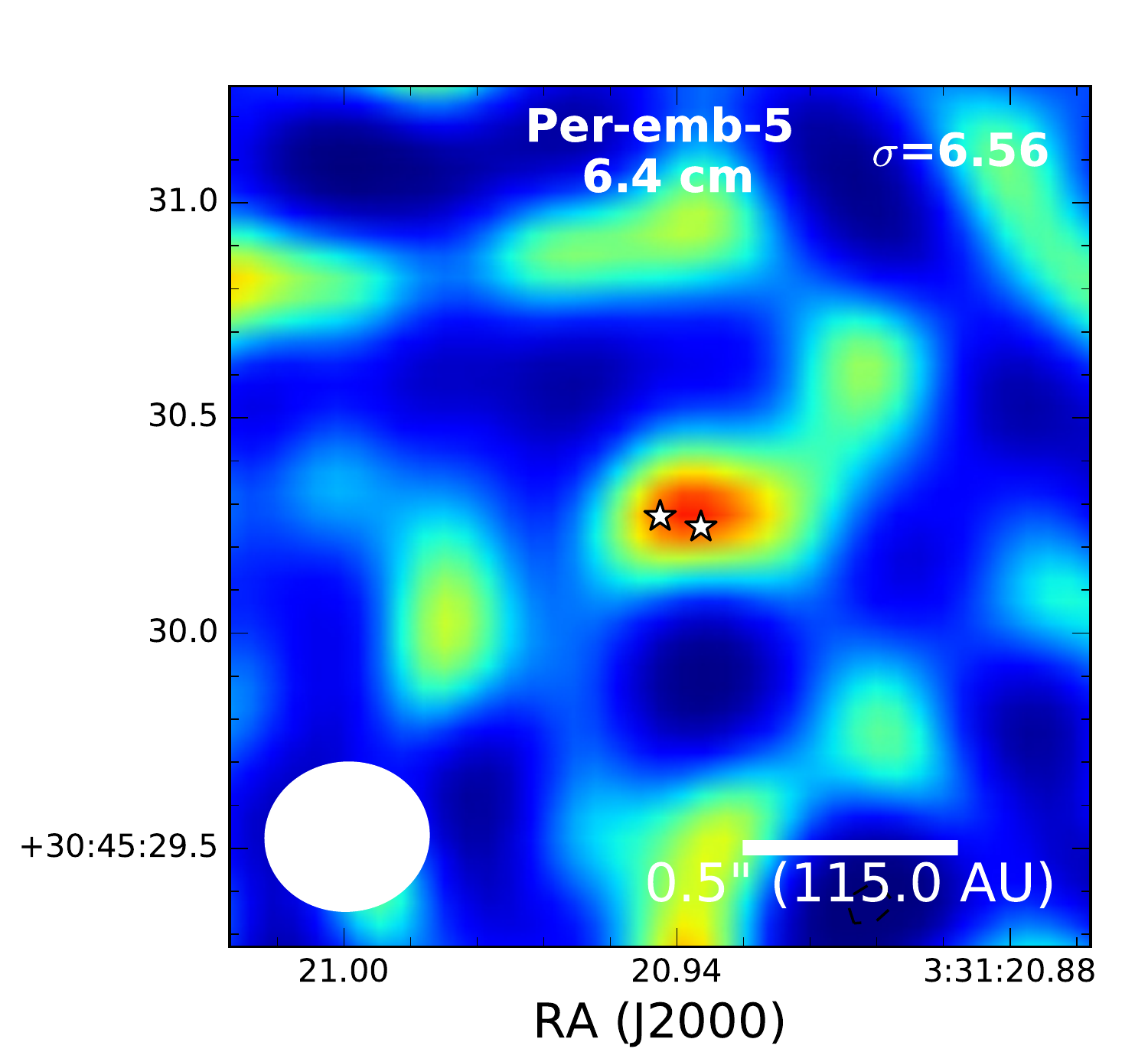}

  \includegraphics[width=0.24\linewidth]{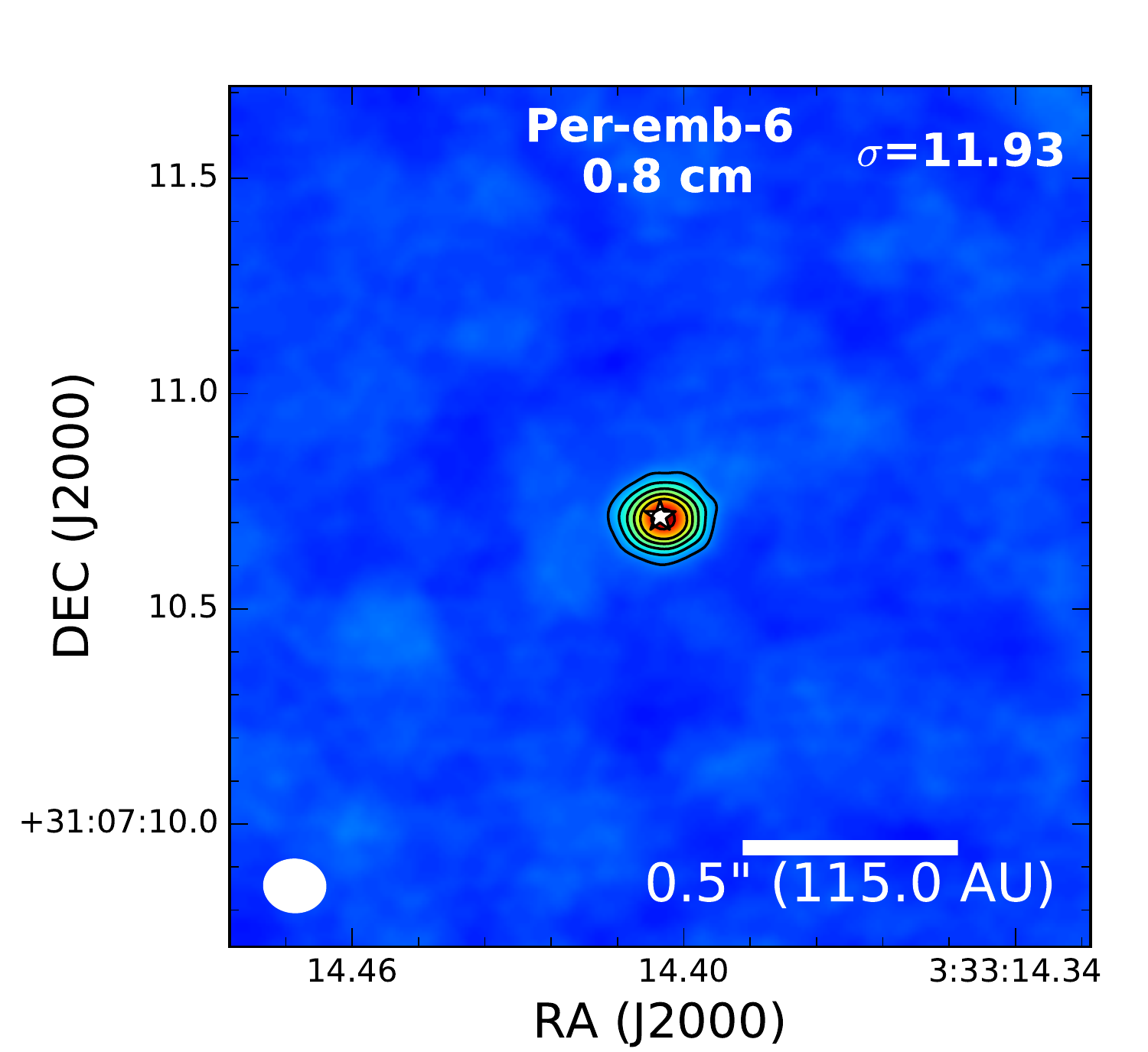}
  \includegraphics[width=0.24\linewidth]{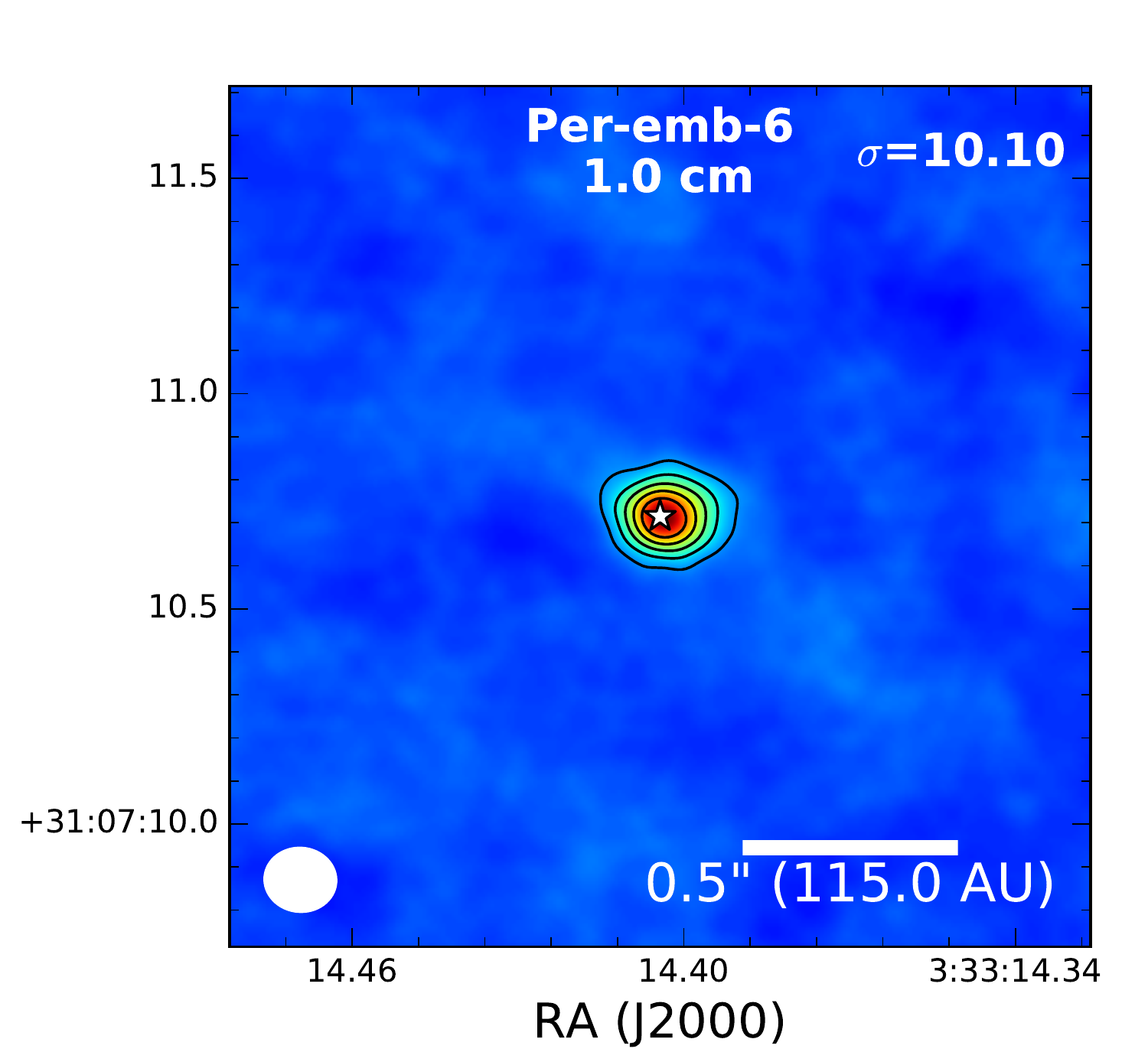}
  \includegraphics[width=0.24\linewidth]{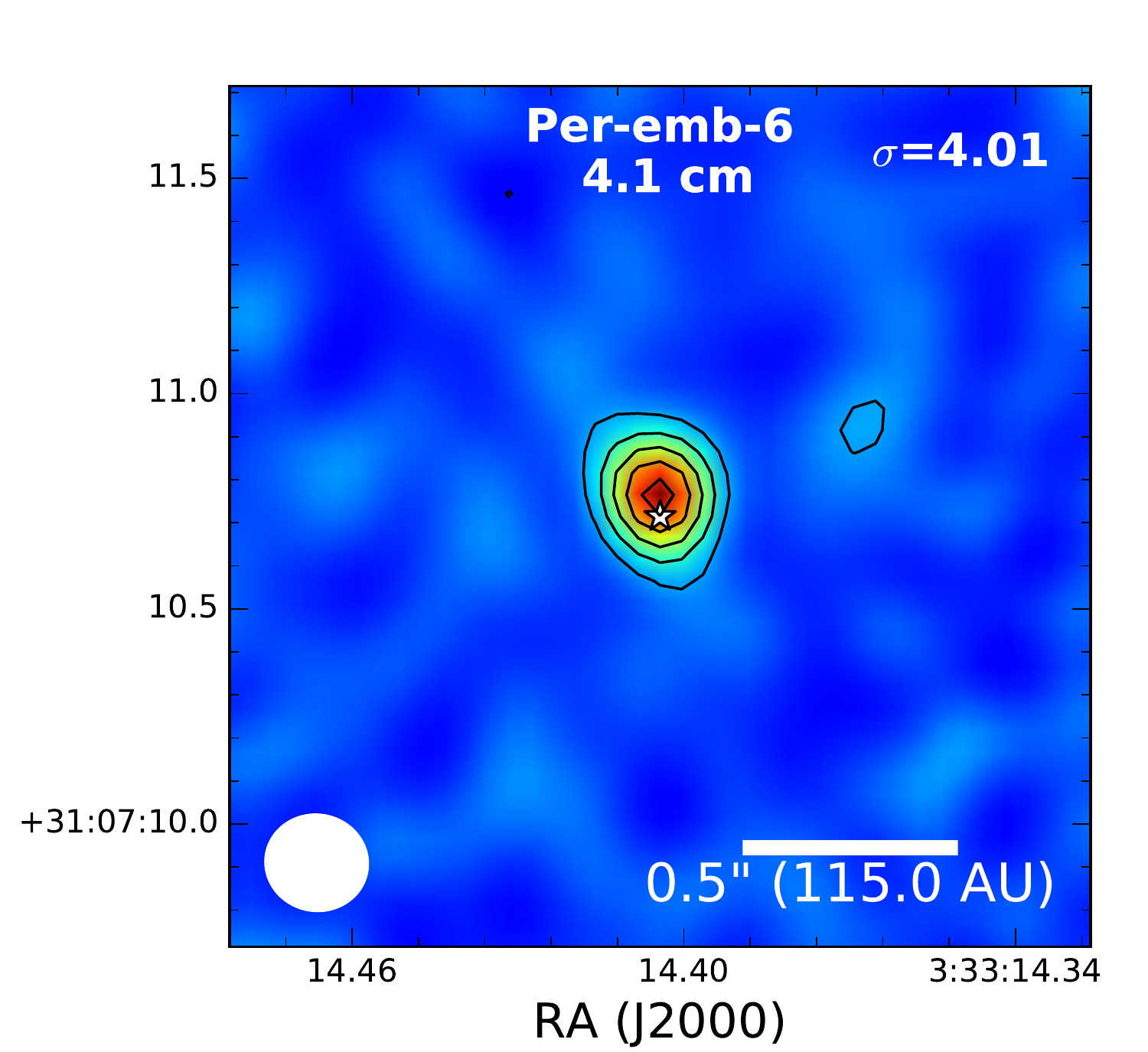}
  \includegraphics[width=0.24\linewidth]{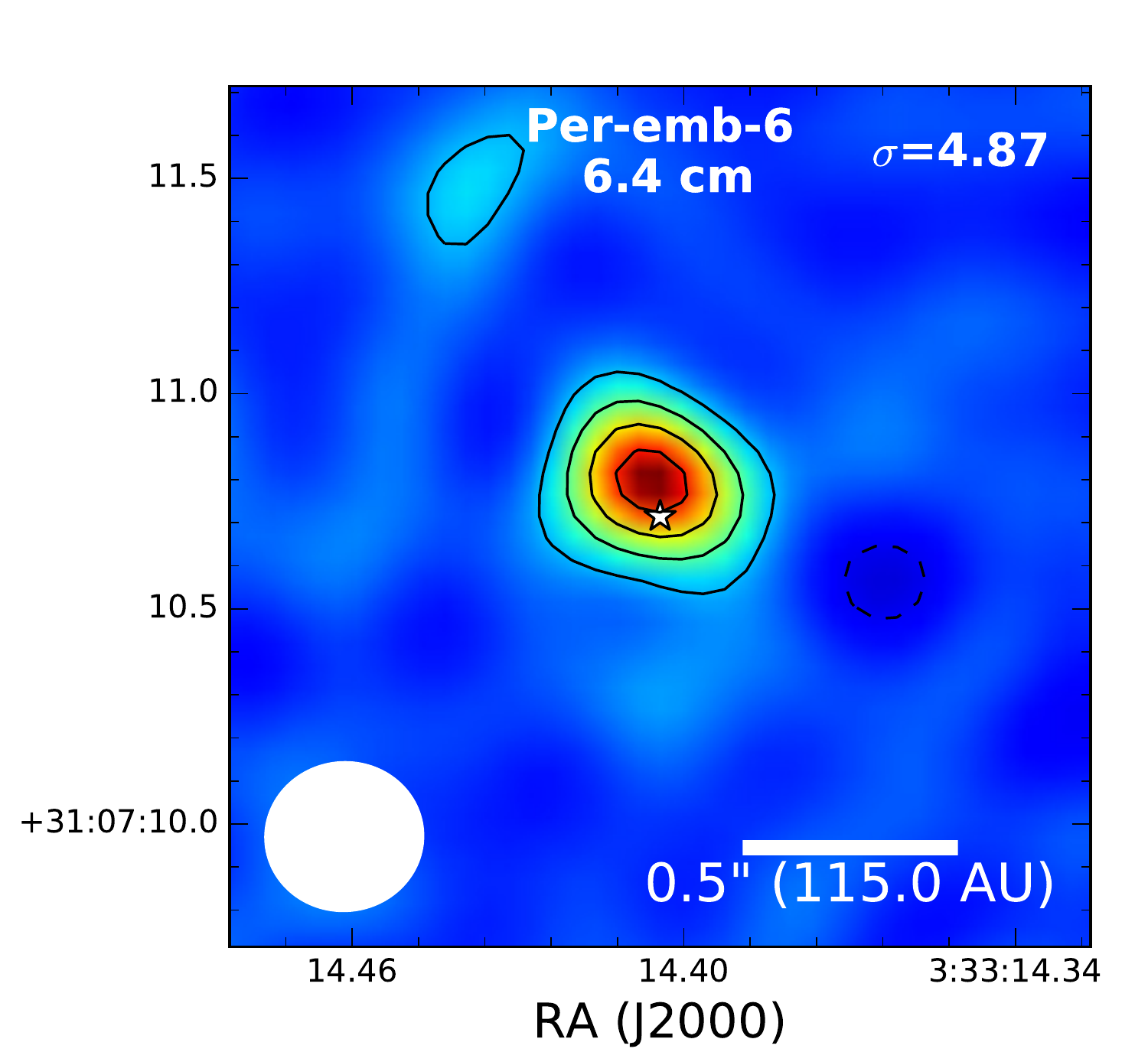}

  \includegraphics[width=0.24\linewidth]{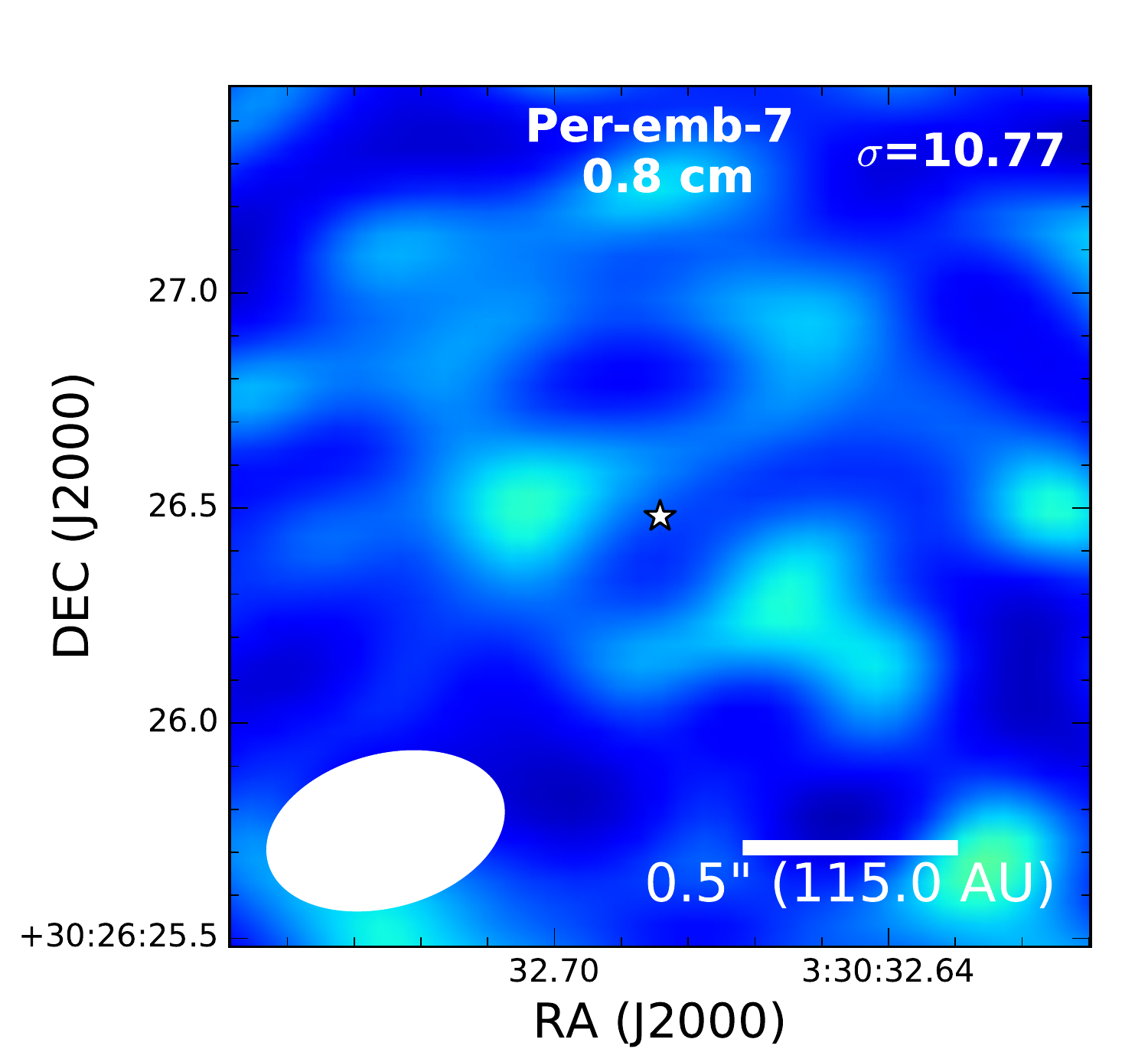}
  \includegraphics[width=0.24\linewidth]{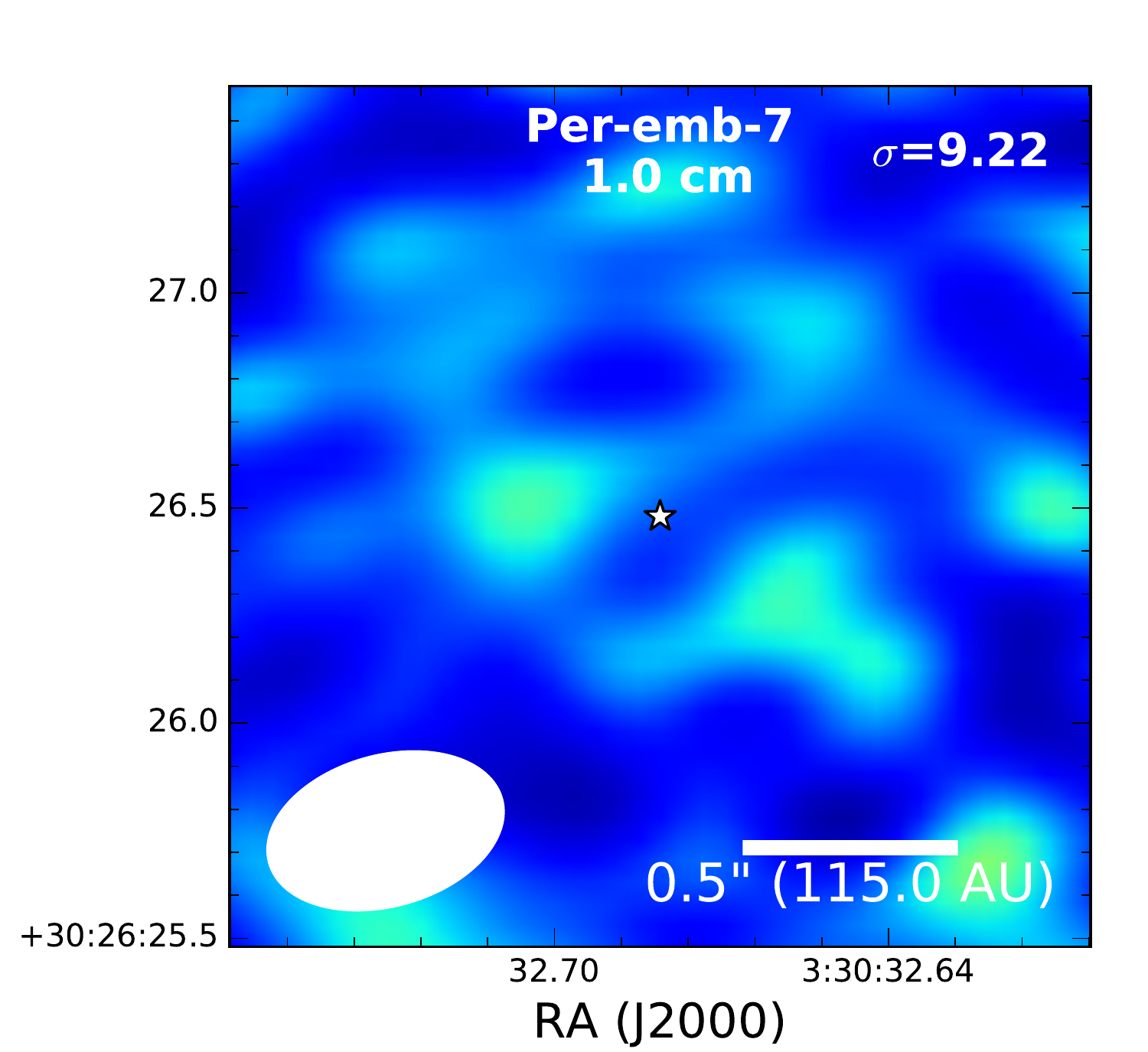}
  \includegraphics[width=0.24\linewidth]{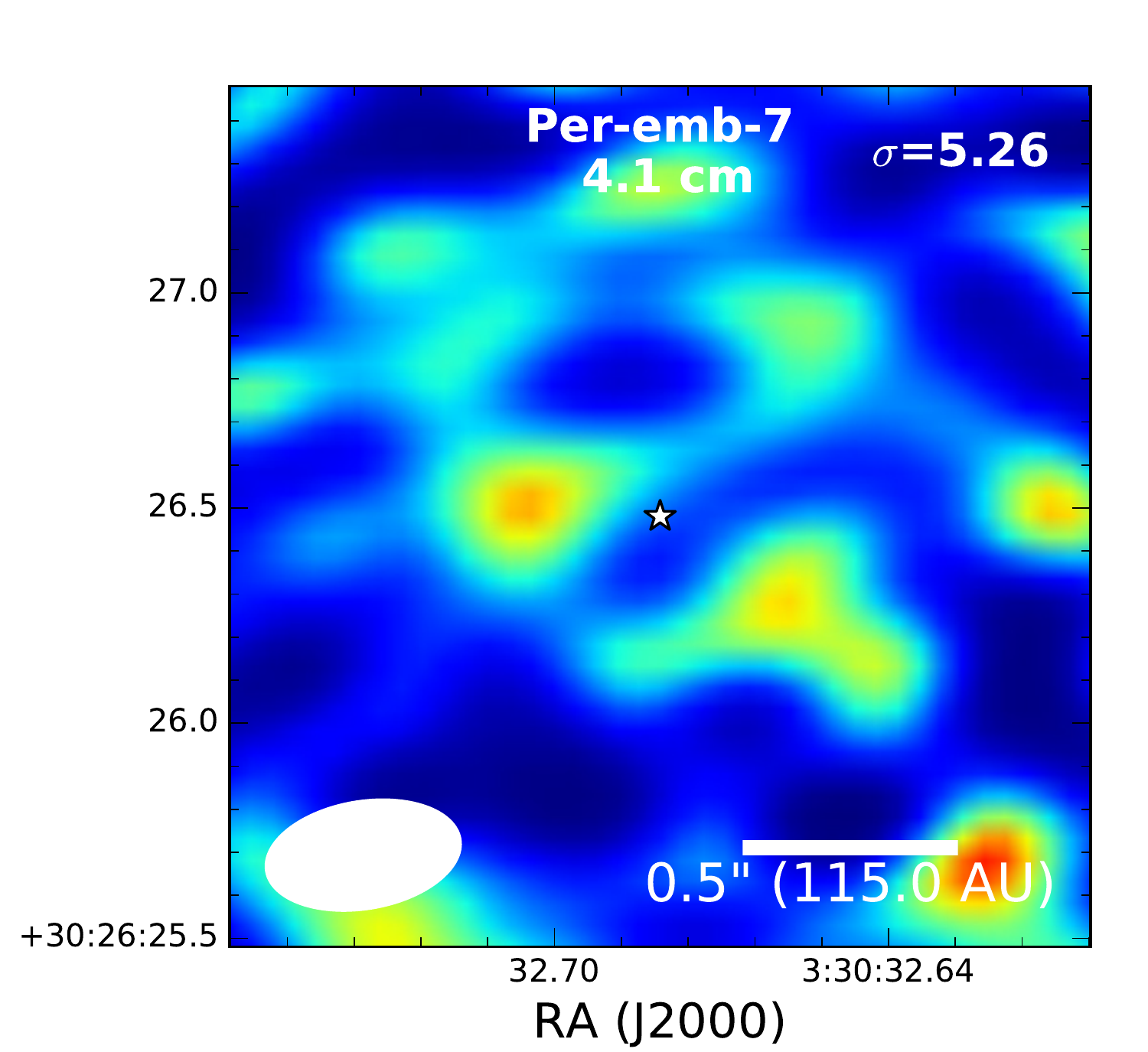}
  \includegraphics[width=0.24\linewidth]{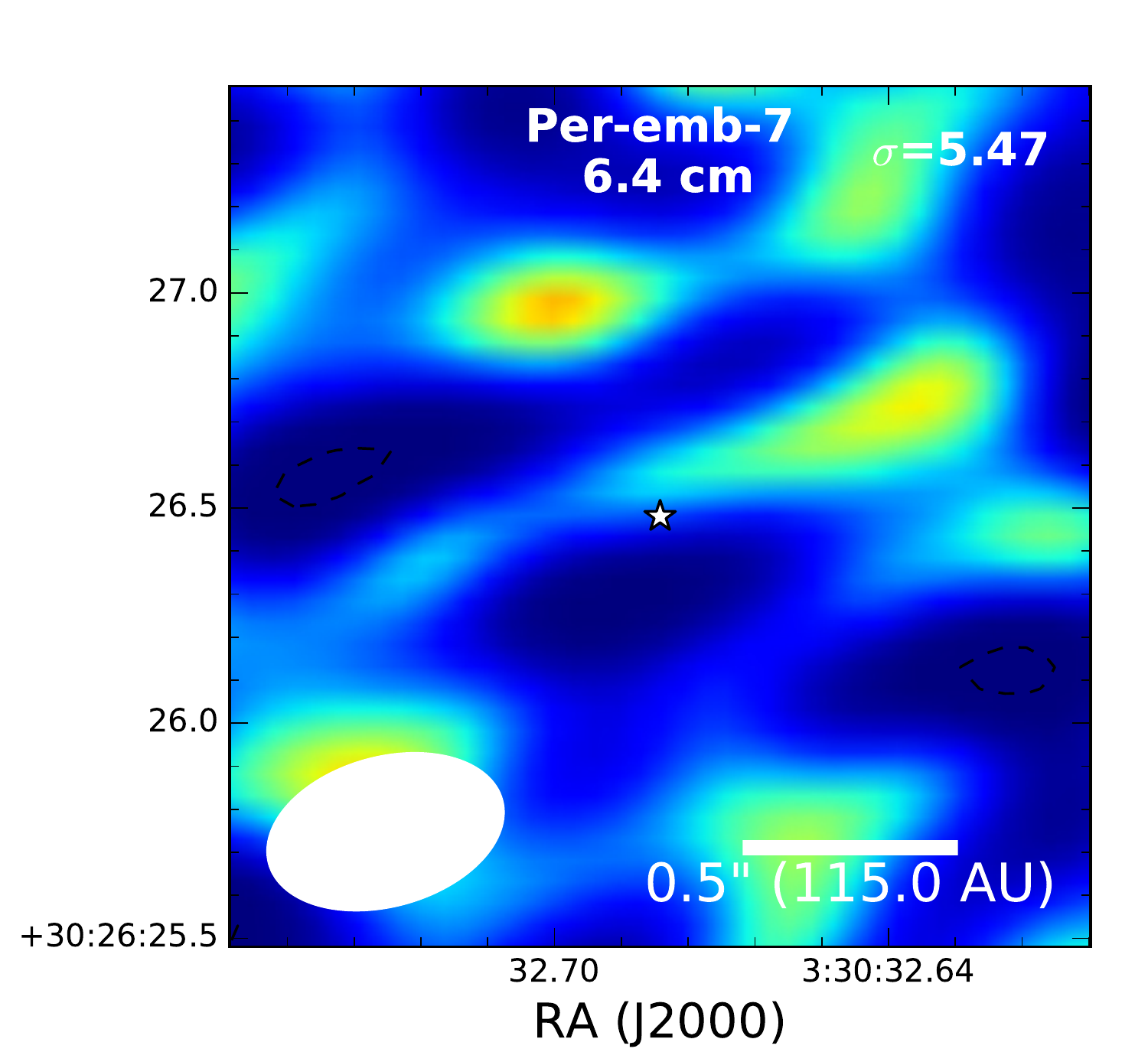}

  \includegraphics[width=0.24\linewidth]{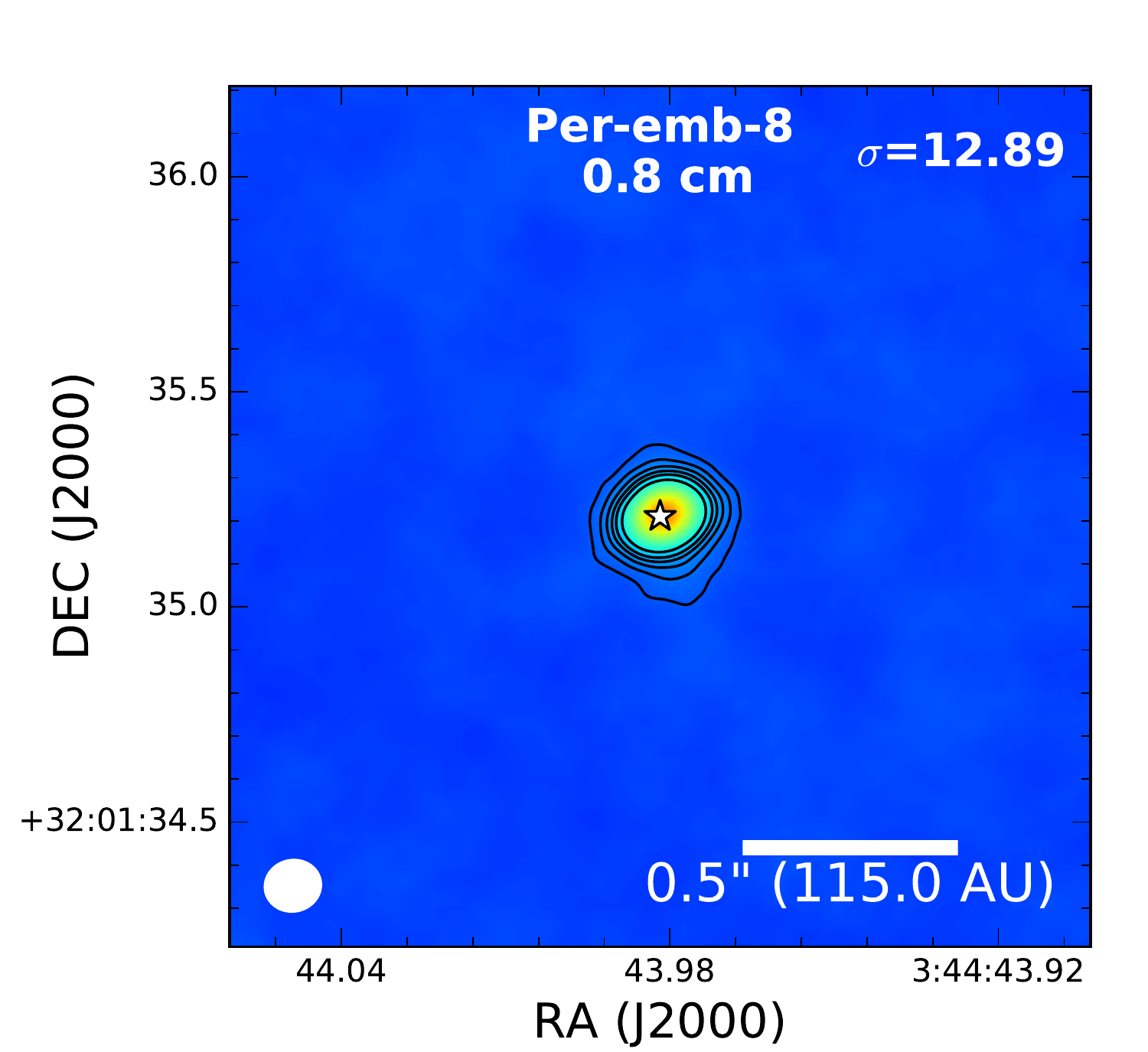}
  \includegraphics[width=0.24\linewidth]{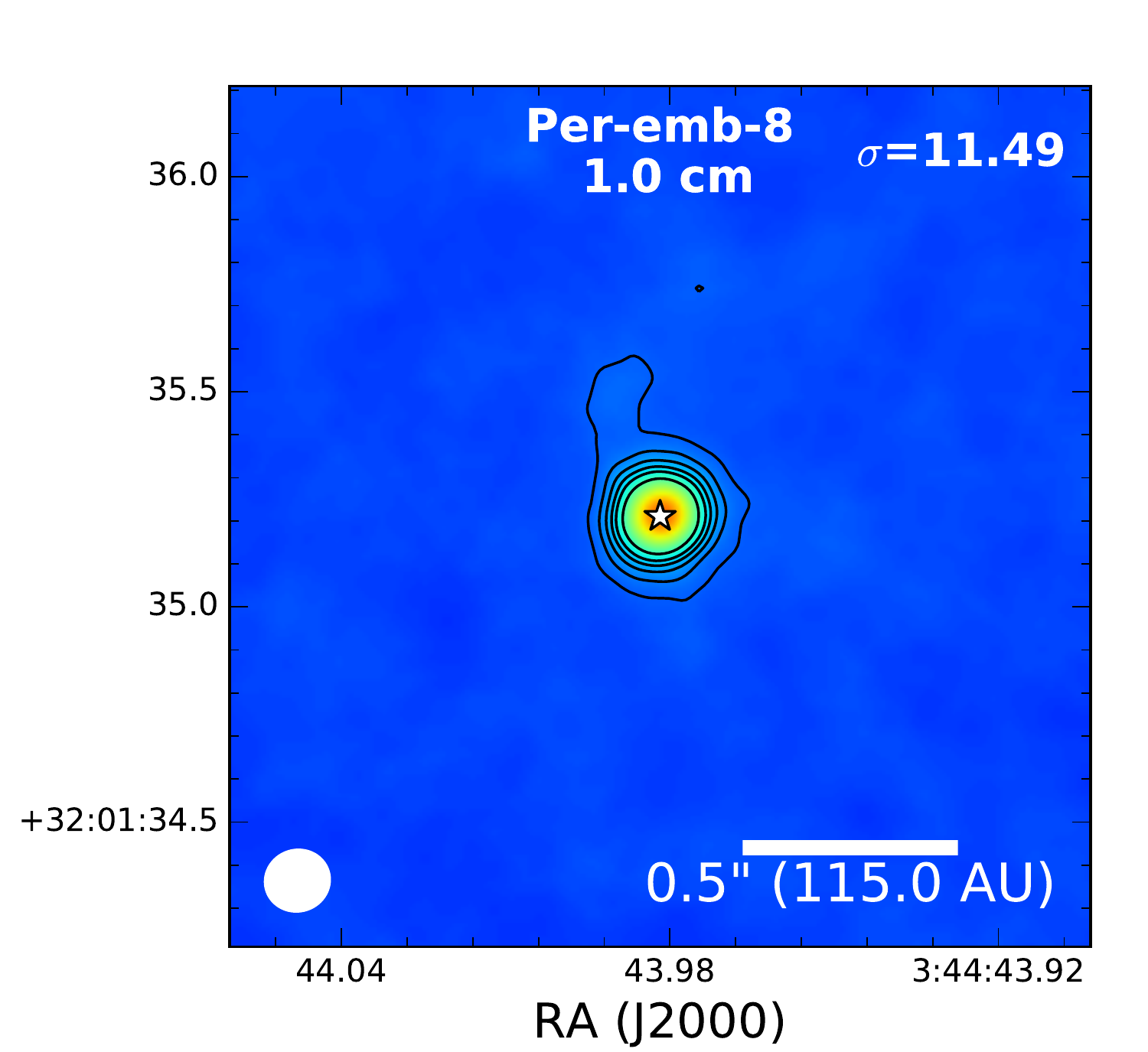}
  \includegraphics[width=0.24\linewidth]{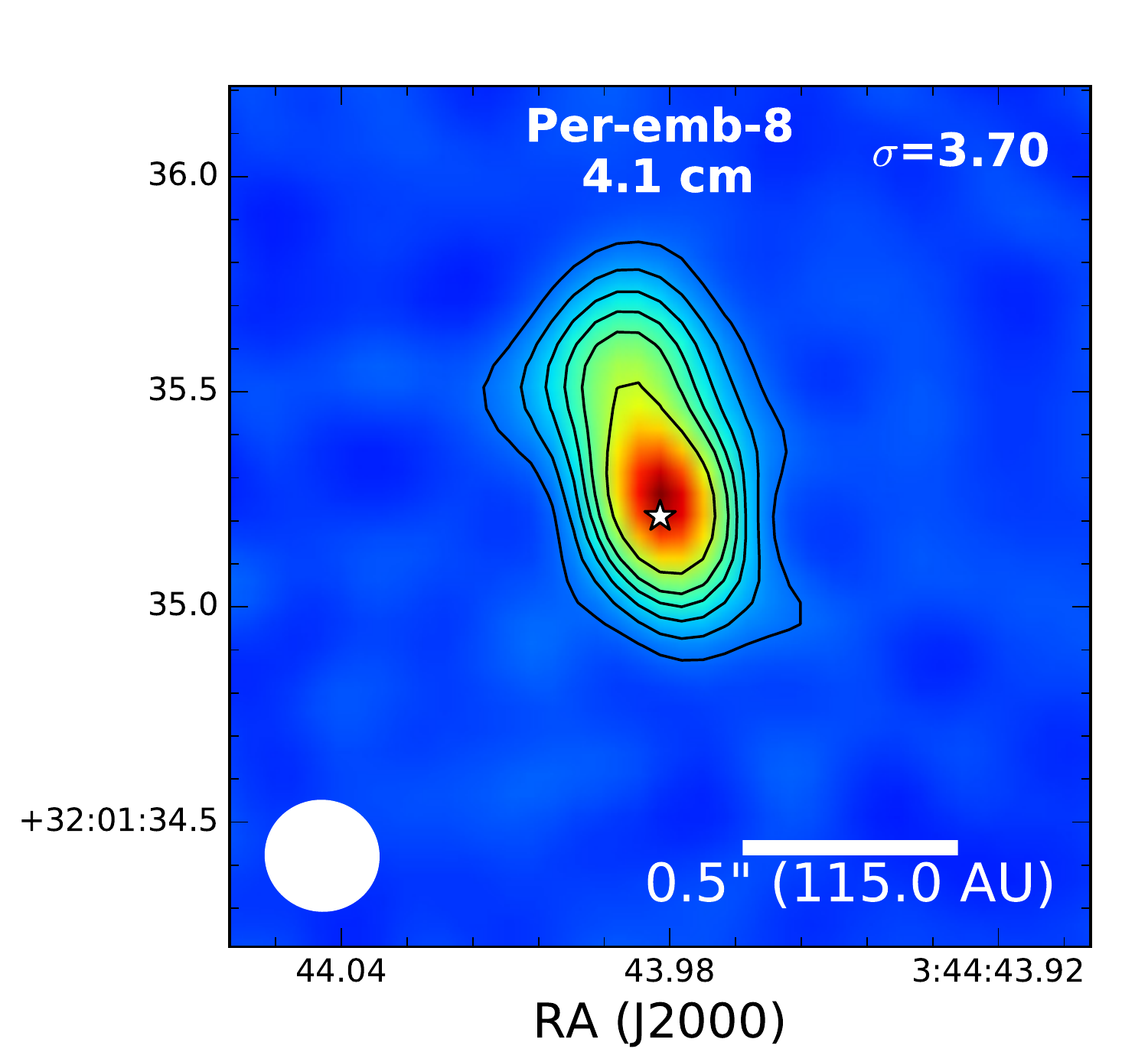}
  \includegraphics[width=0.24\linewidth]{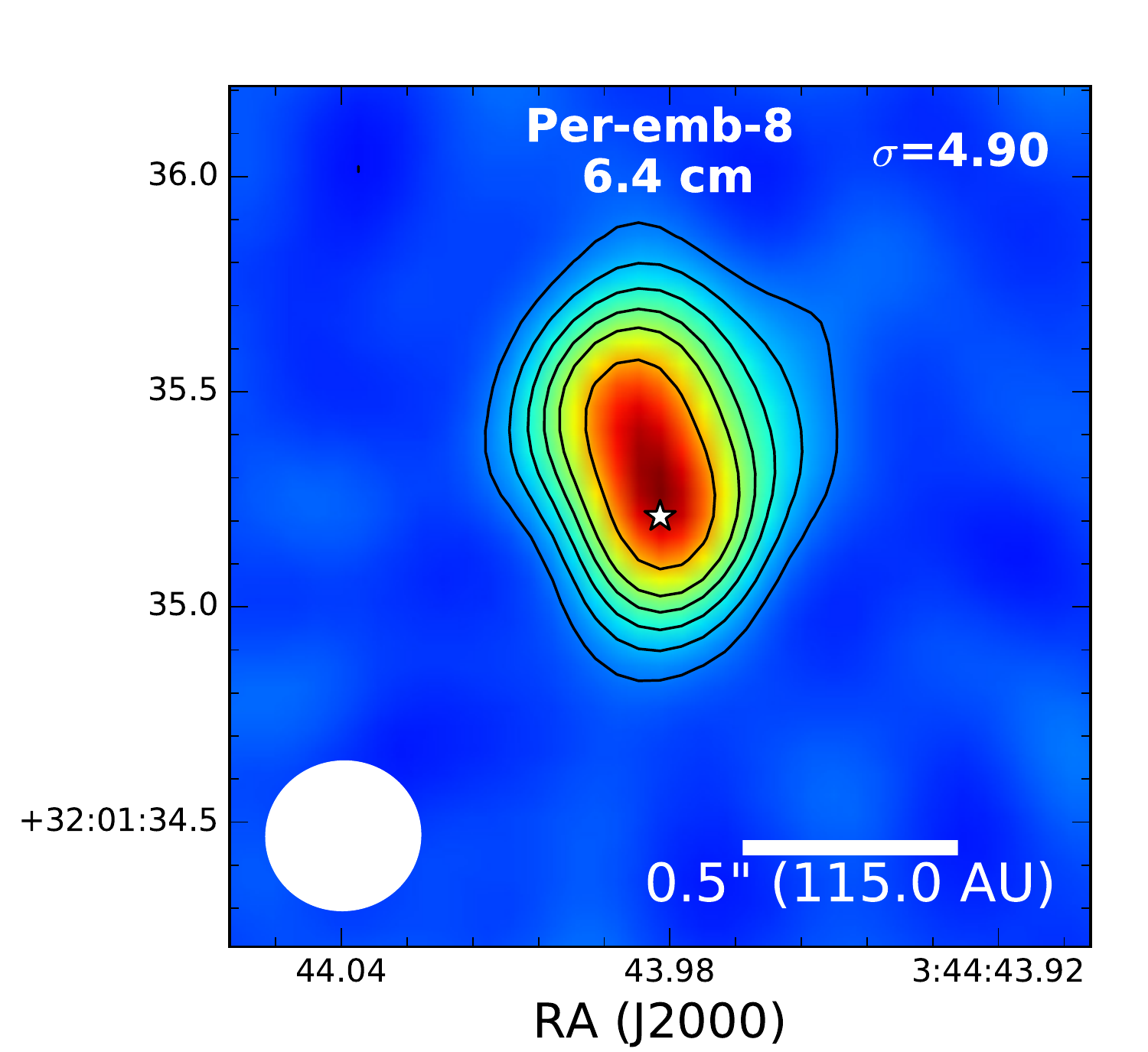}

  \includegraphics[width=0.24\linewidth]{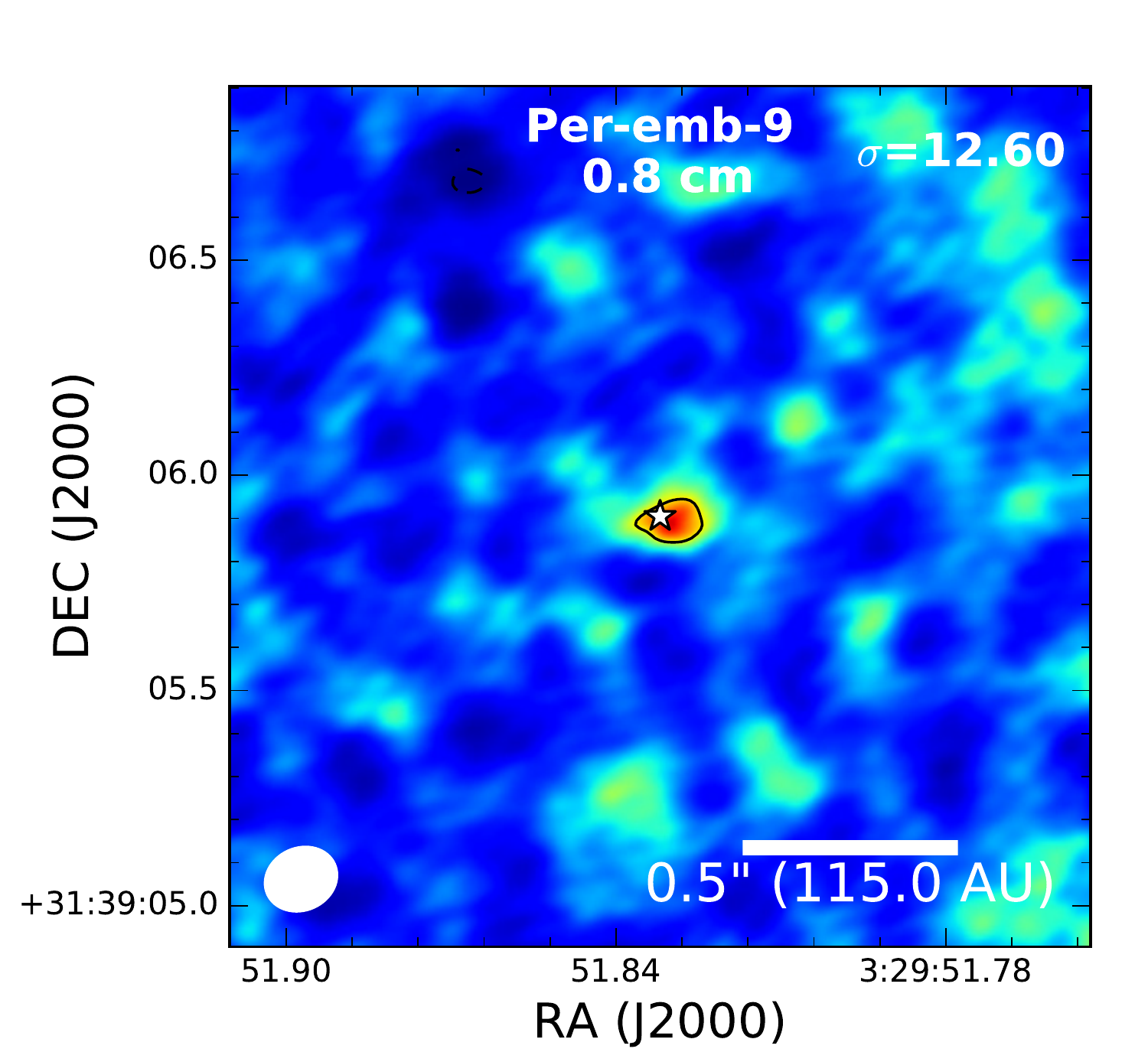}
  \includegraphics[width=0.24\linewidth]{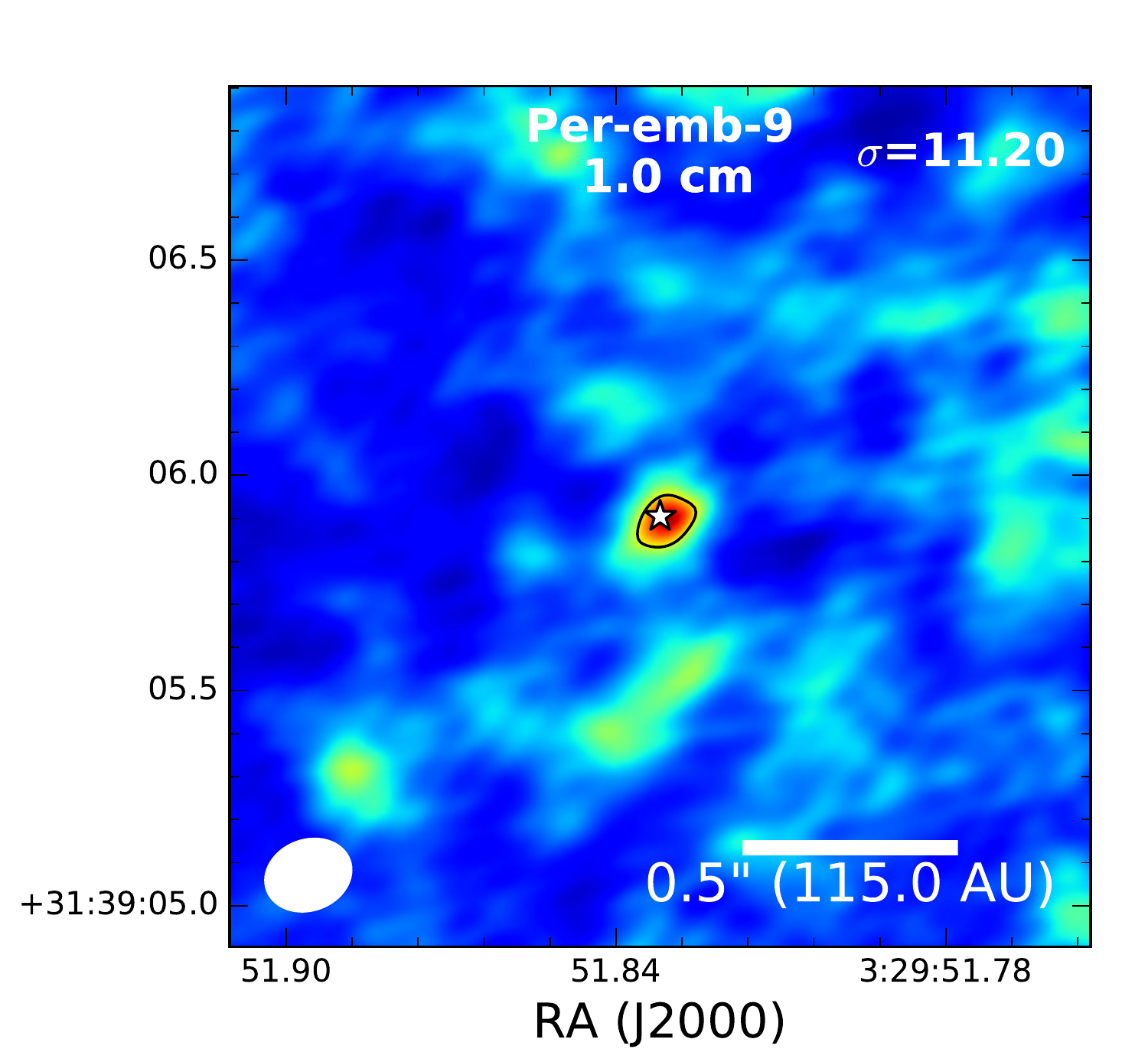}
  \includegraphics[width=0.24\linewidth]{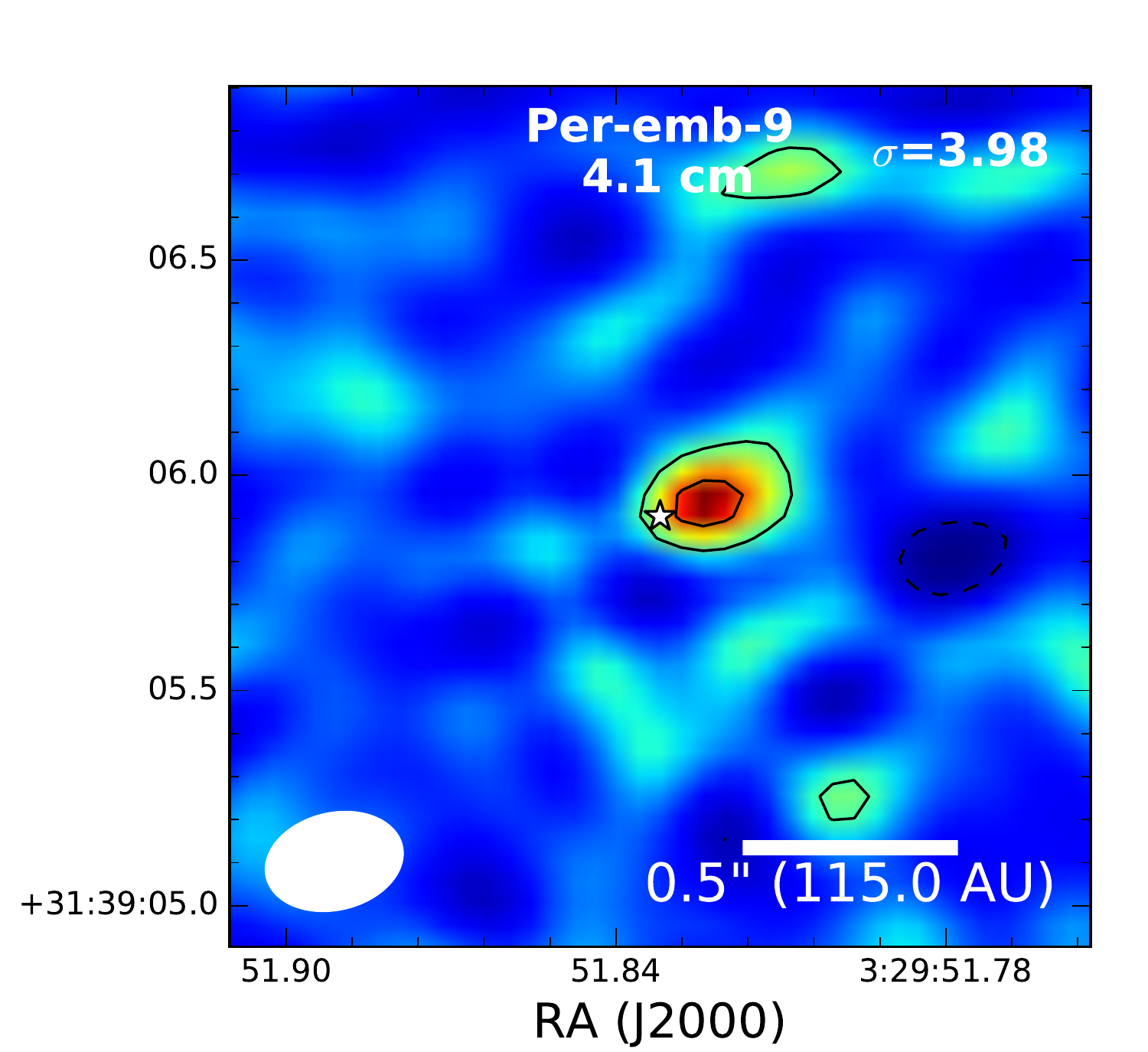}
  \includegraphics[width=0.24\linewidth]{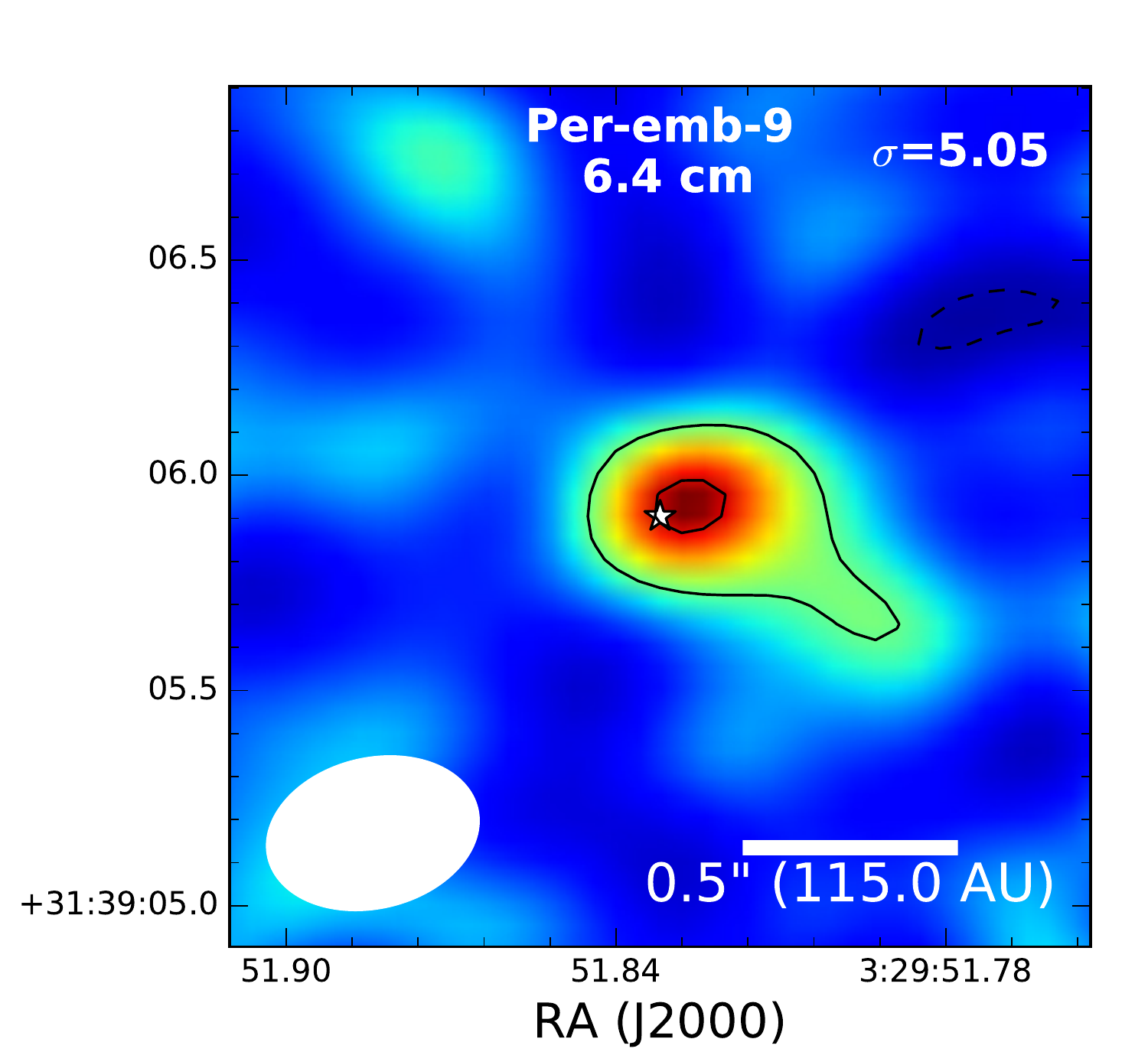}

  \includegraphics[width=0.24\linewidth]{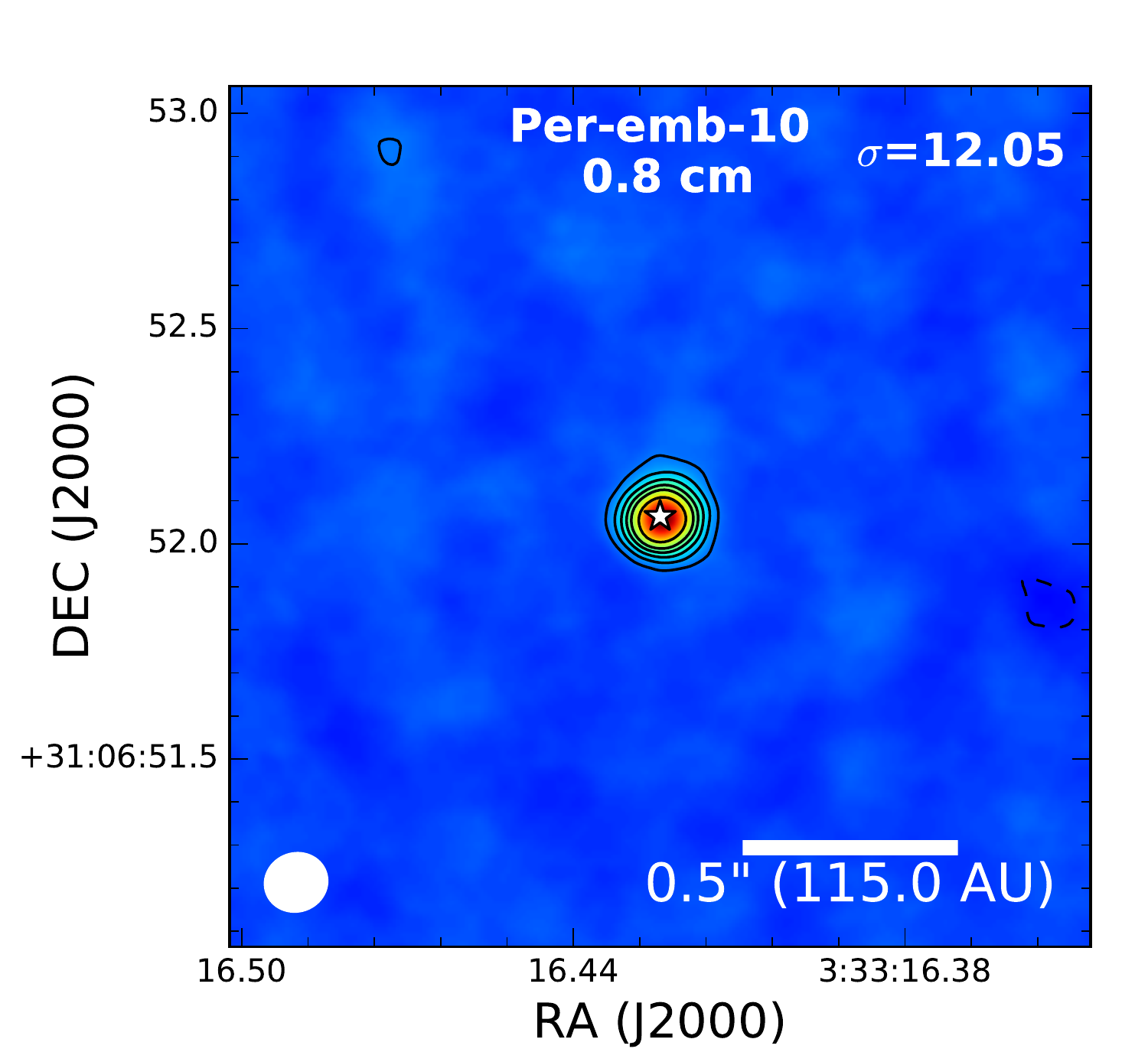}
  \includegraphics[width=0.24\linewidth]{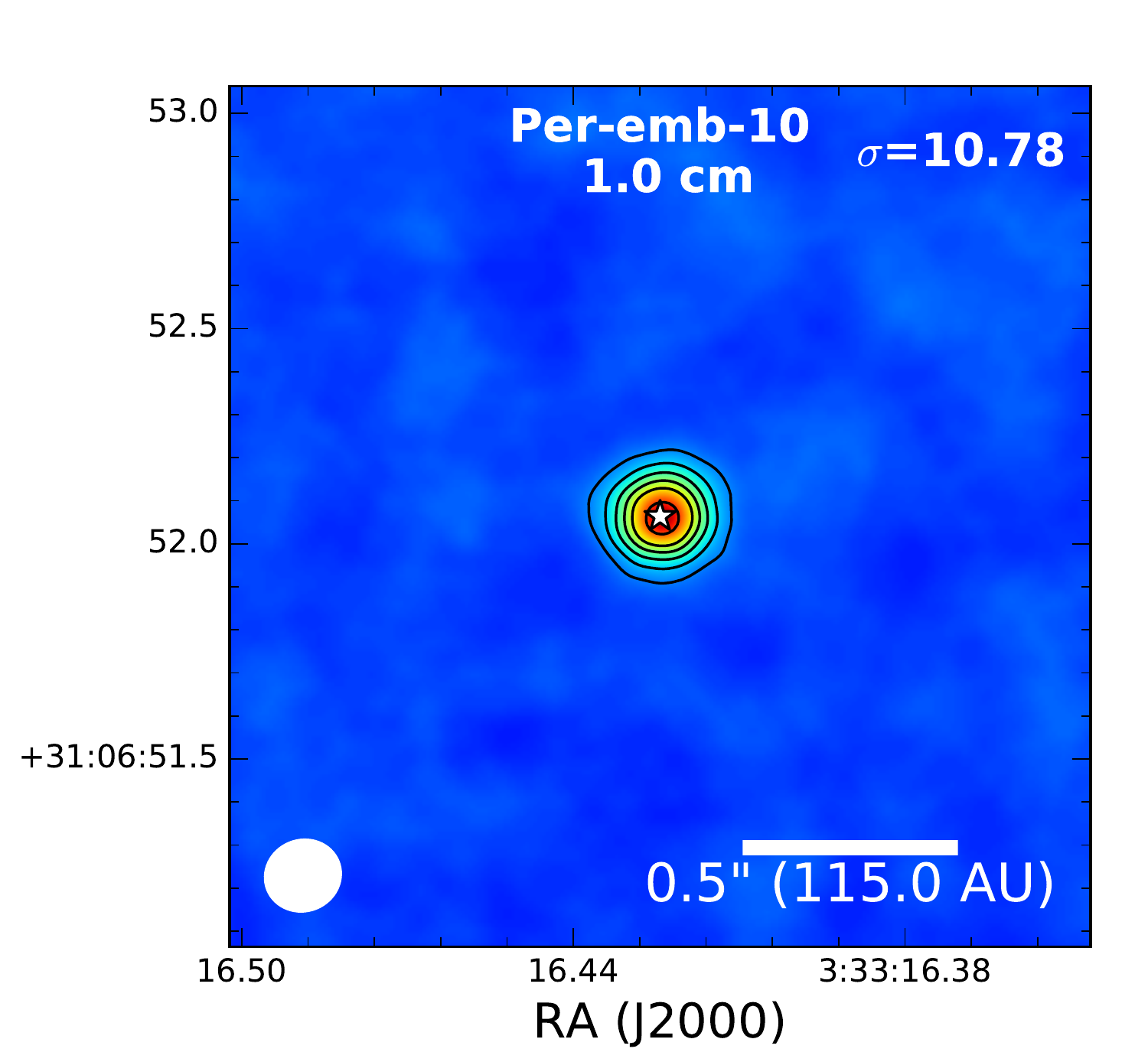}
  \includegraphics[width=0.24\linewidth]{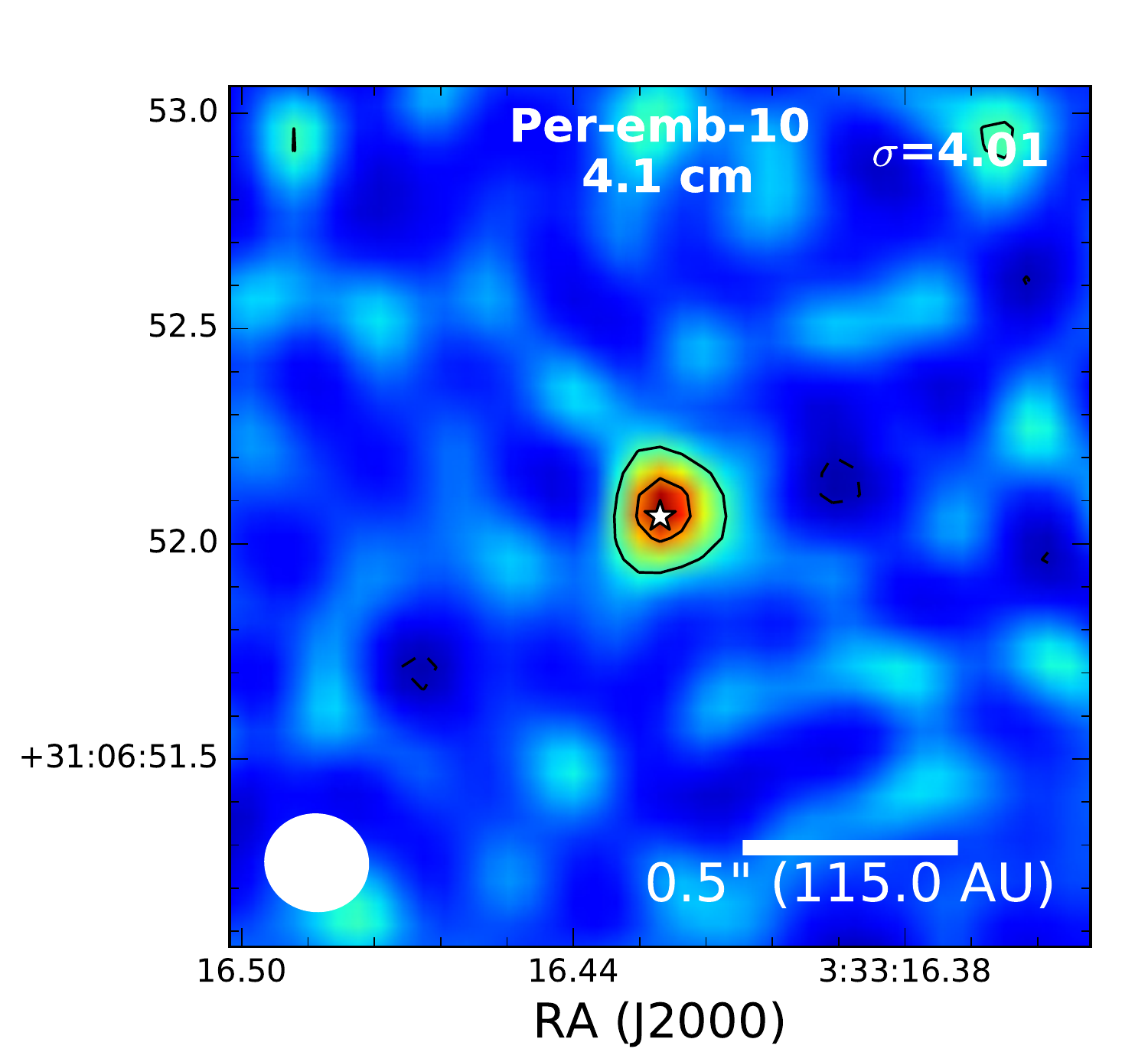}
  \includegraphics[width=0.24\linewidth]{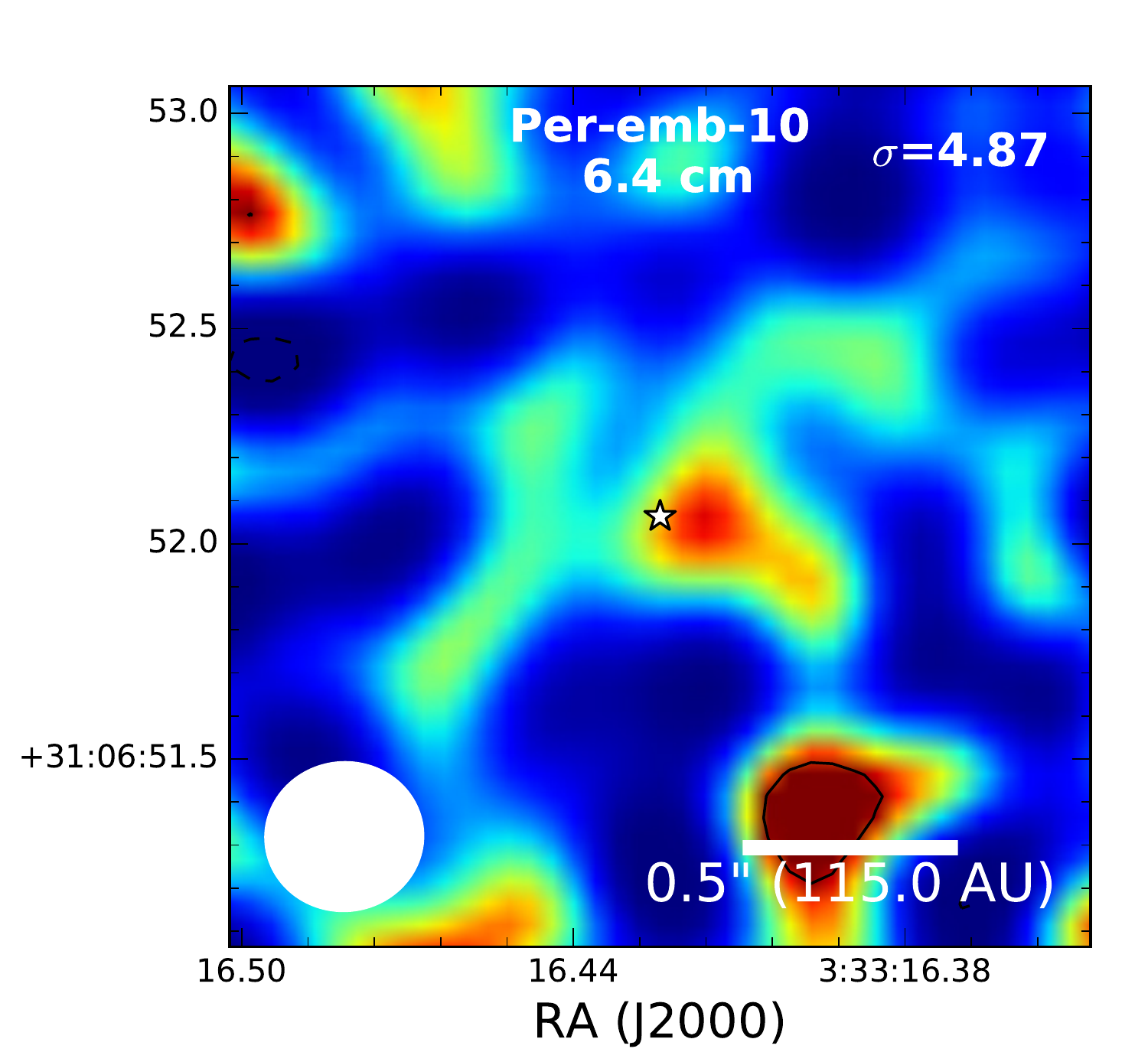}

\end{figure}

\begin{figure}[H]

  \includegraphics[width=0.24\linewidth]{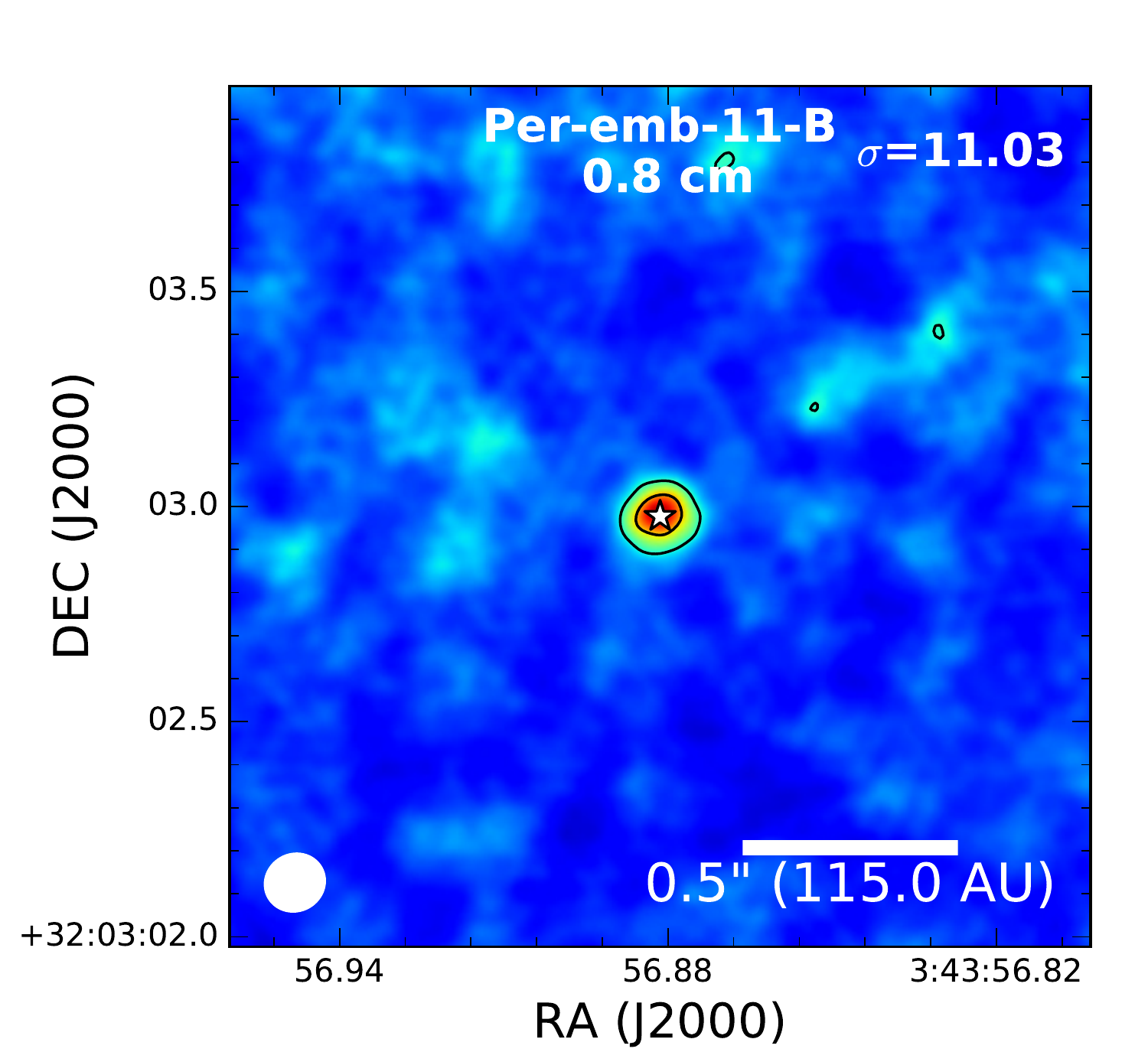}
  \includegraphics[width=0.24\linewidth]{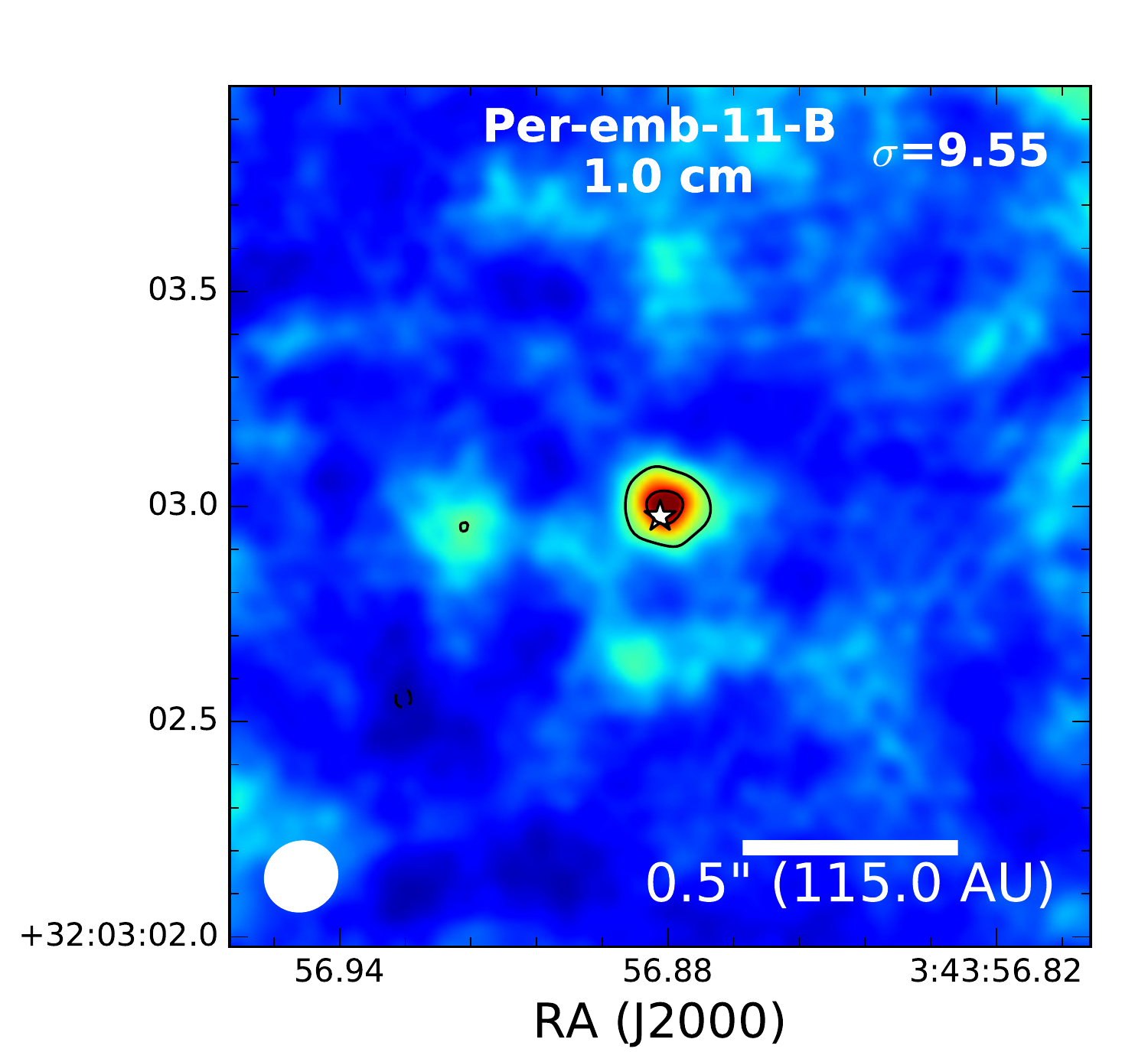}
  \includegraphics[width=0.24\linewidth]{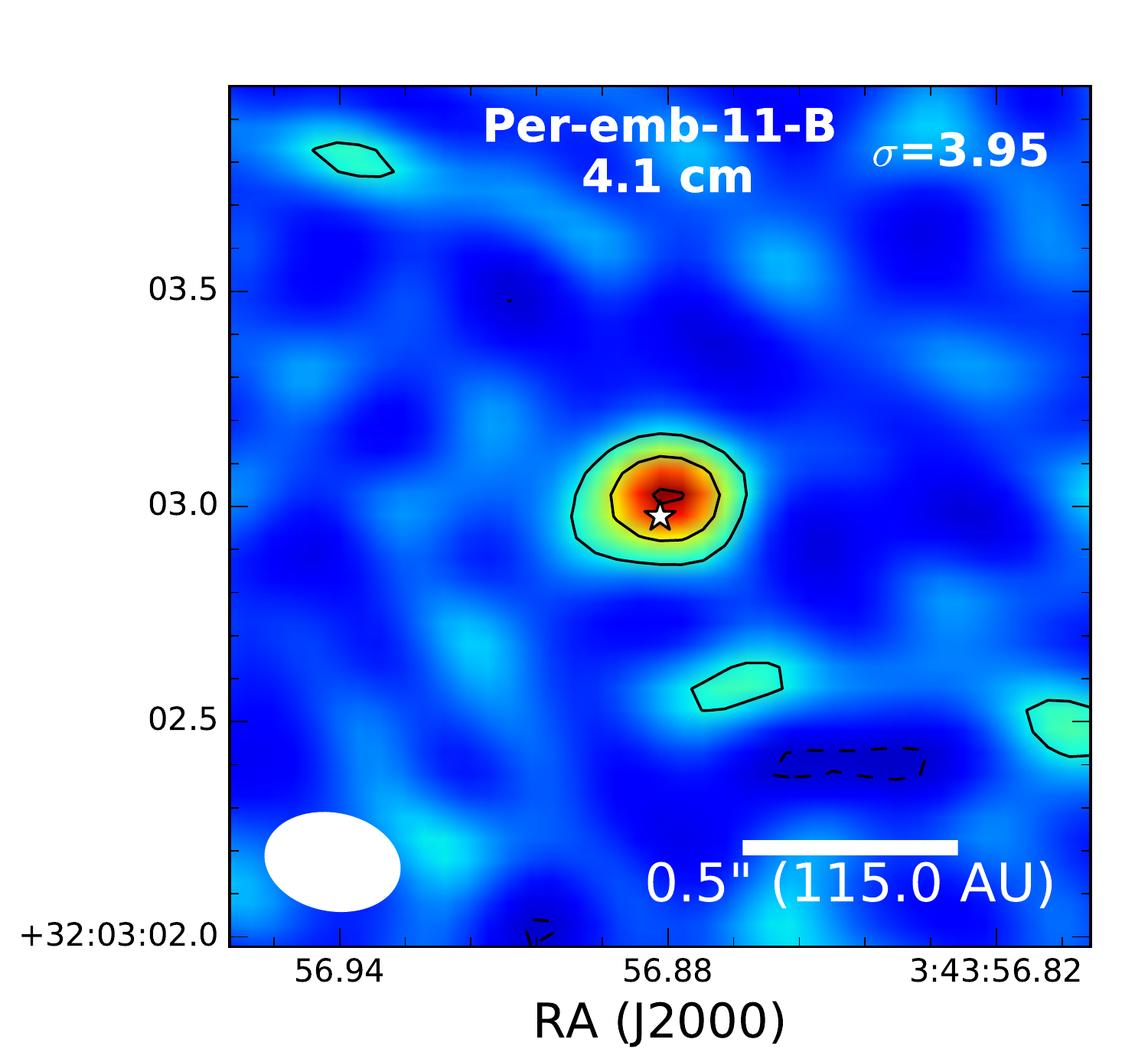}
  \includegraphics[width=0.24\linewidth]{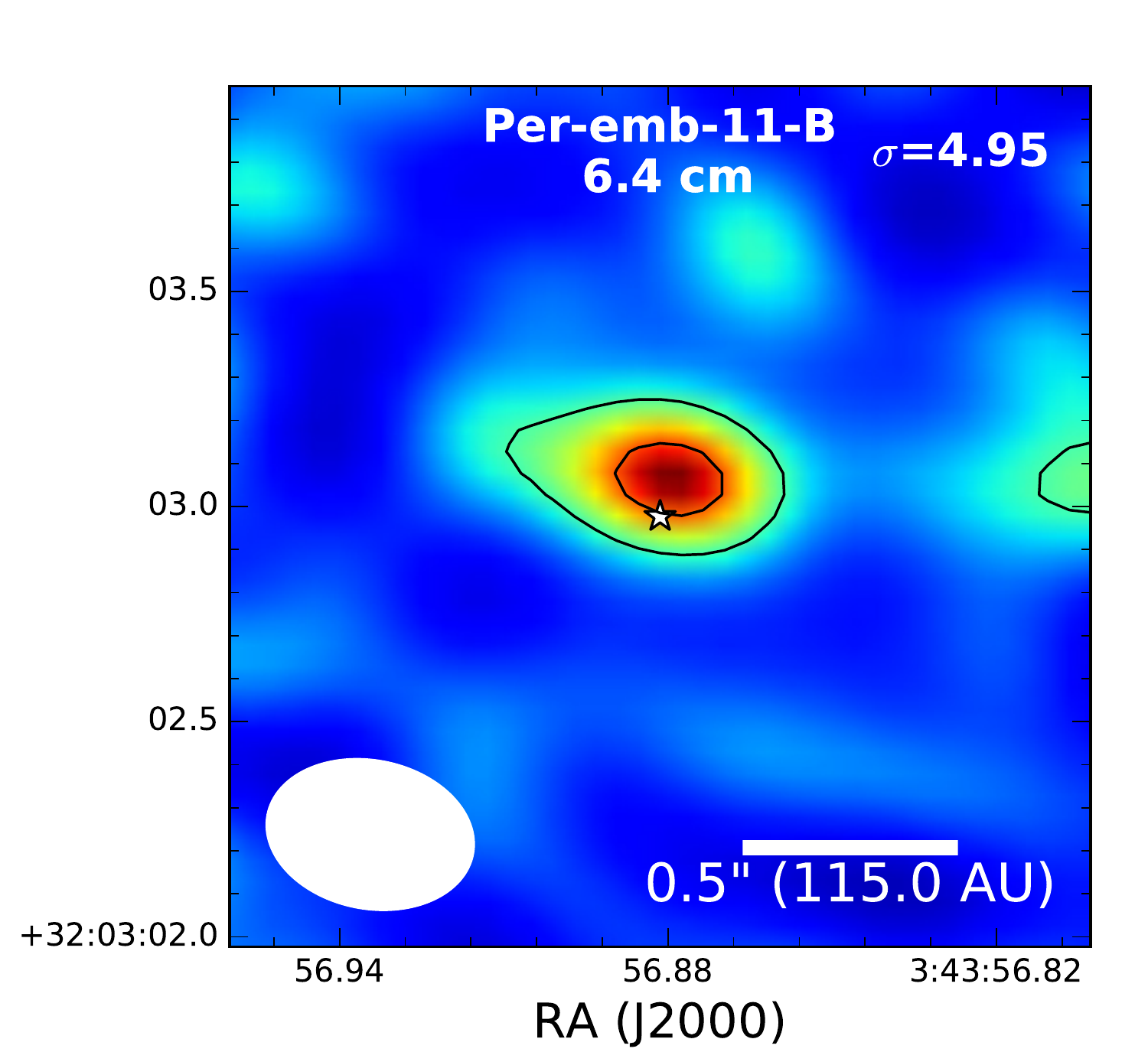}

  \includegraphics[width=0.24\linewidth]{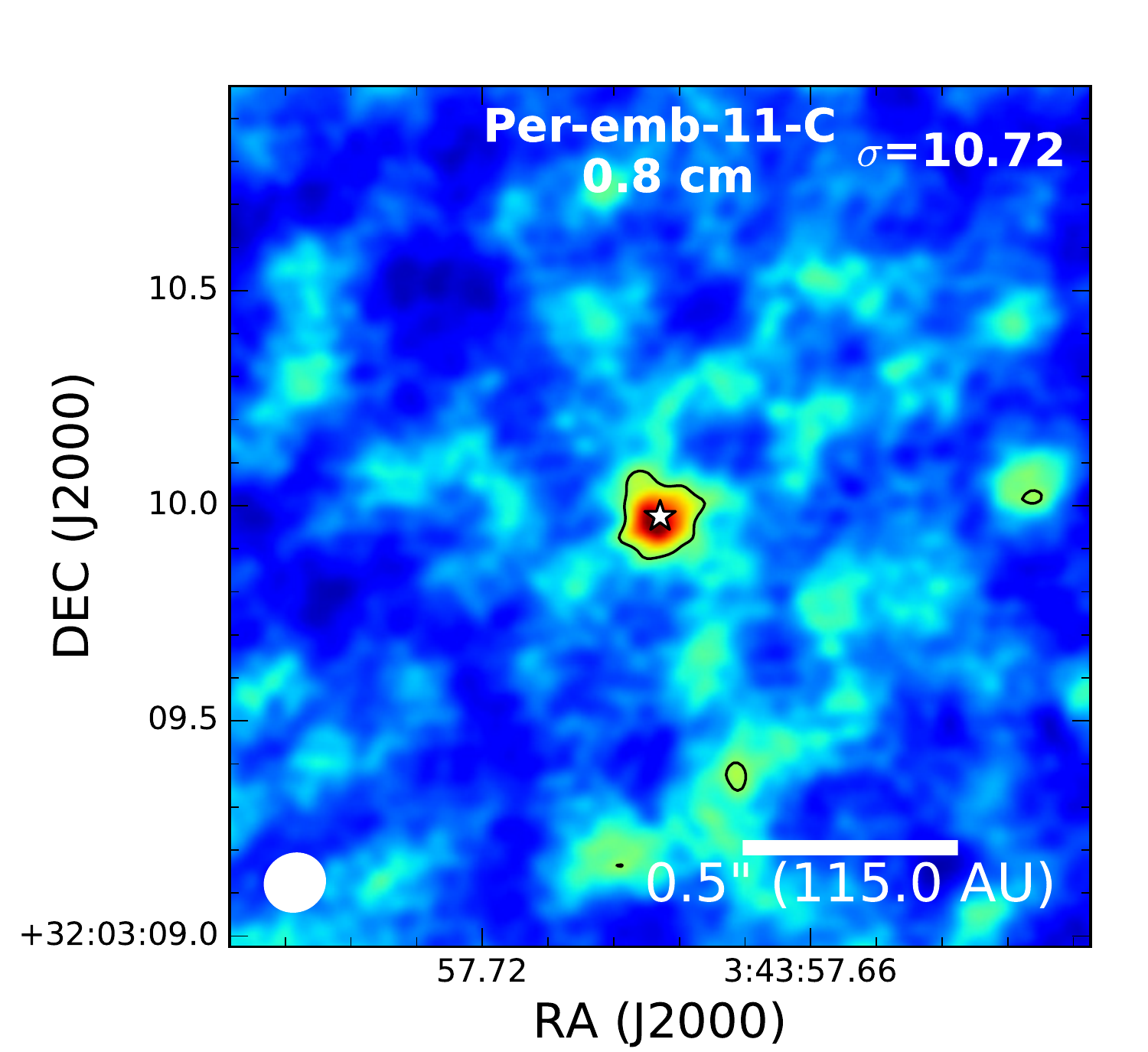}
  \includegraphics[width=0.24\linewidth]{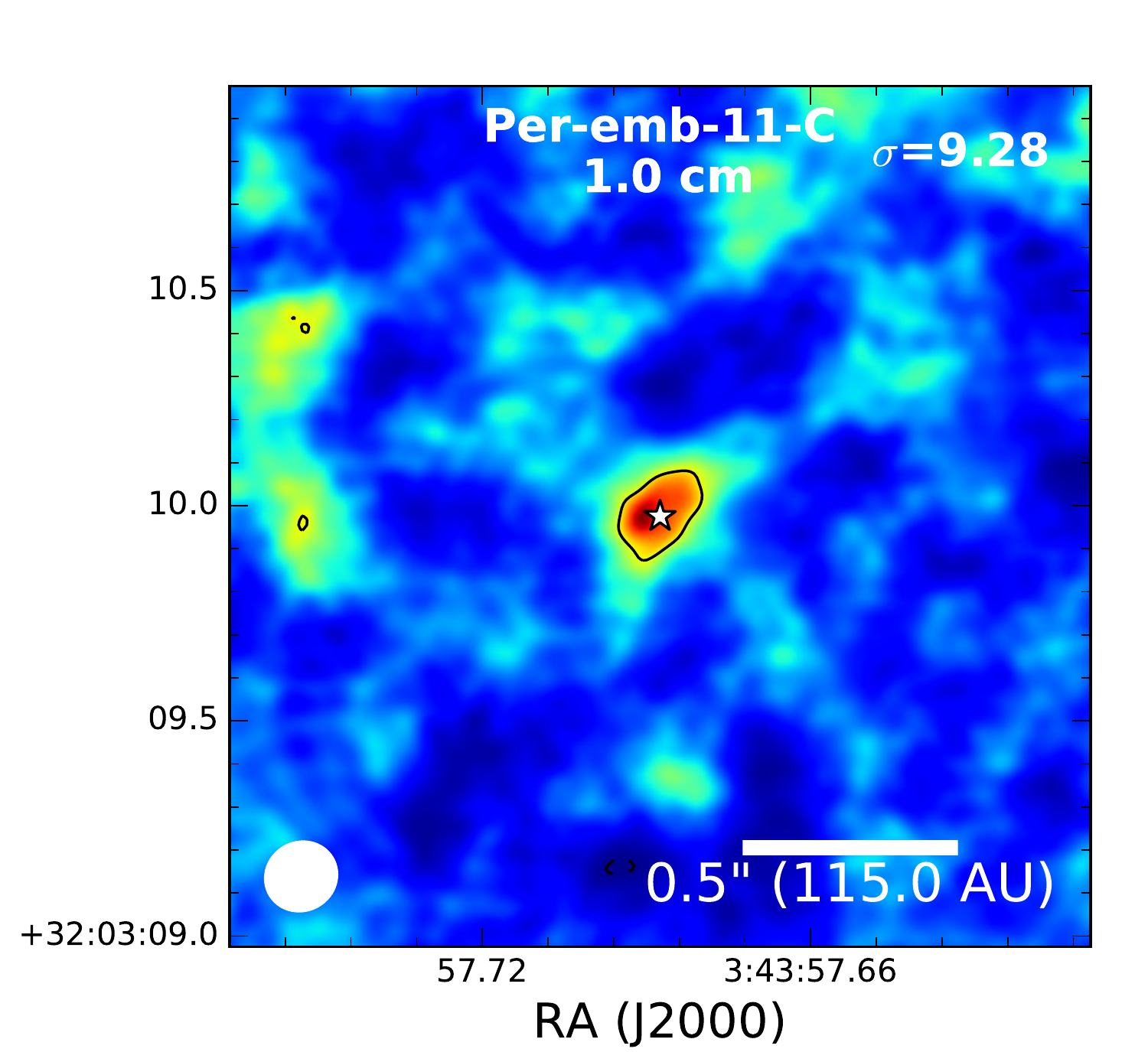}
  \includegraphics[width=0.24\linewidth]{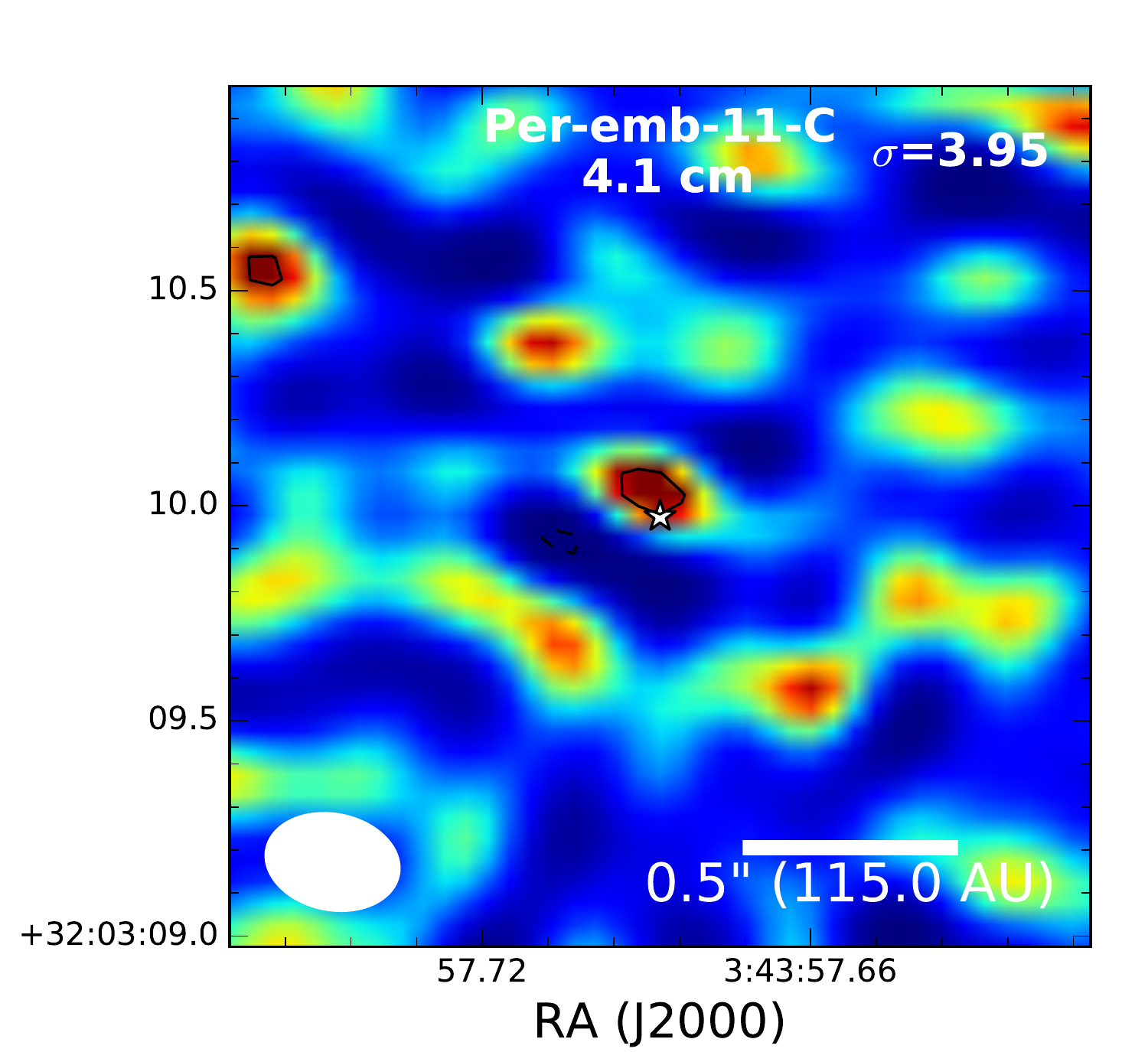}
  \includegraphics[width=0.24\linewidth]{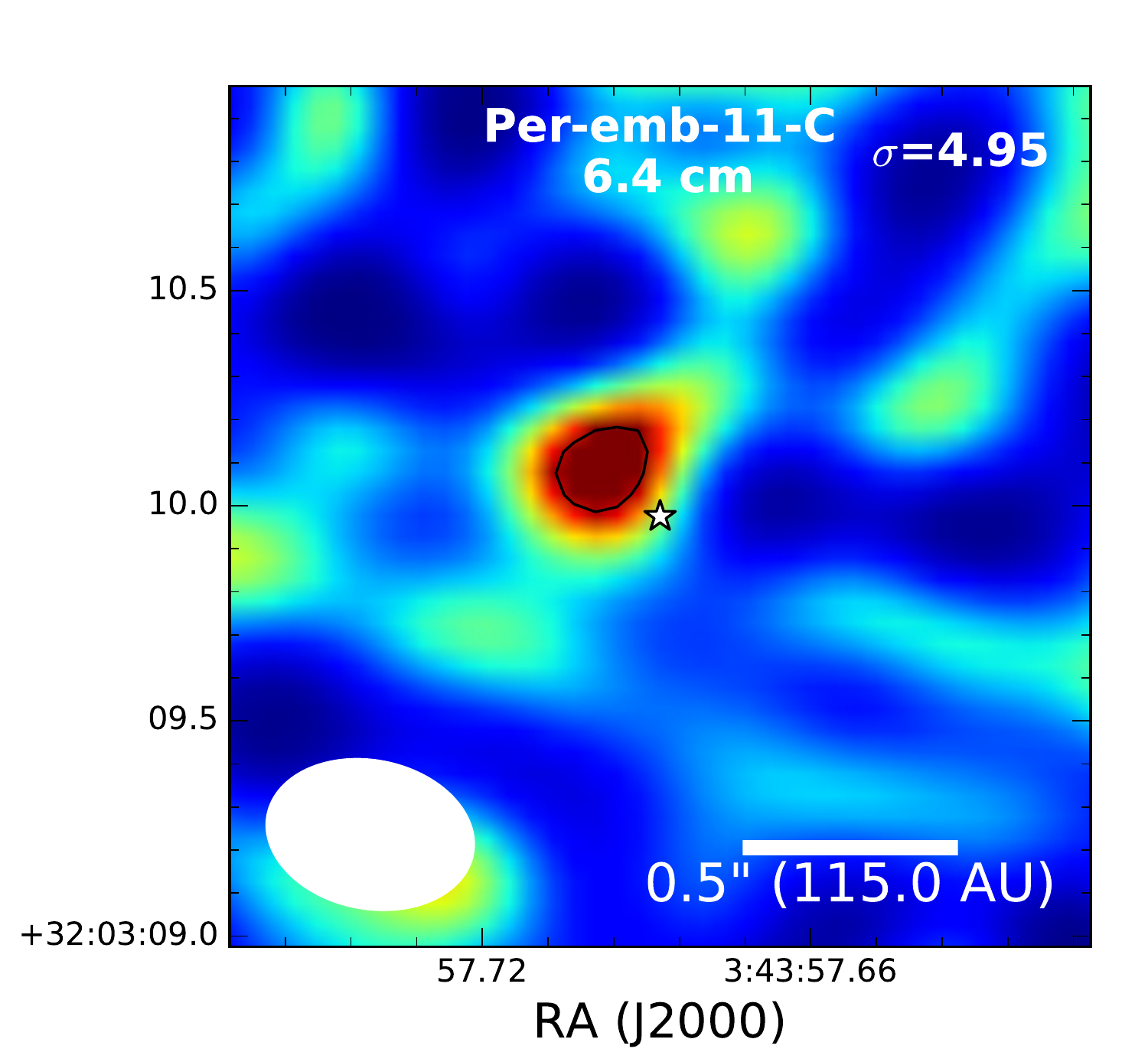}

  \includegraphics[width=0.24\linewidth]{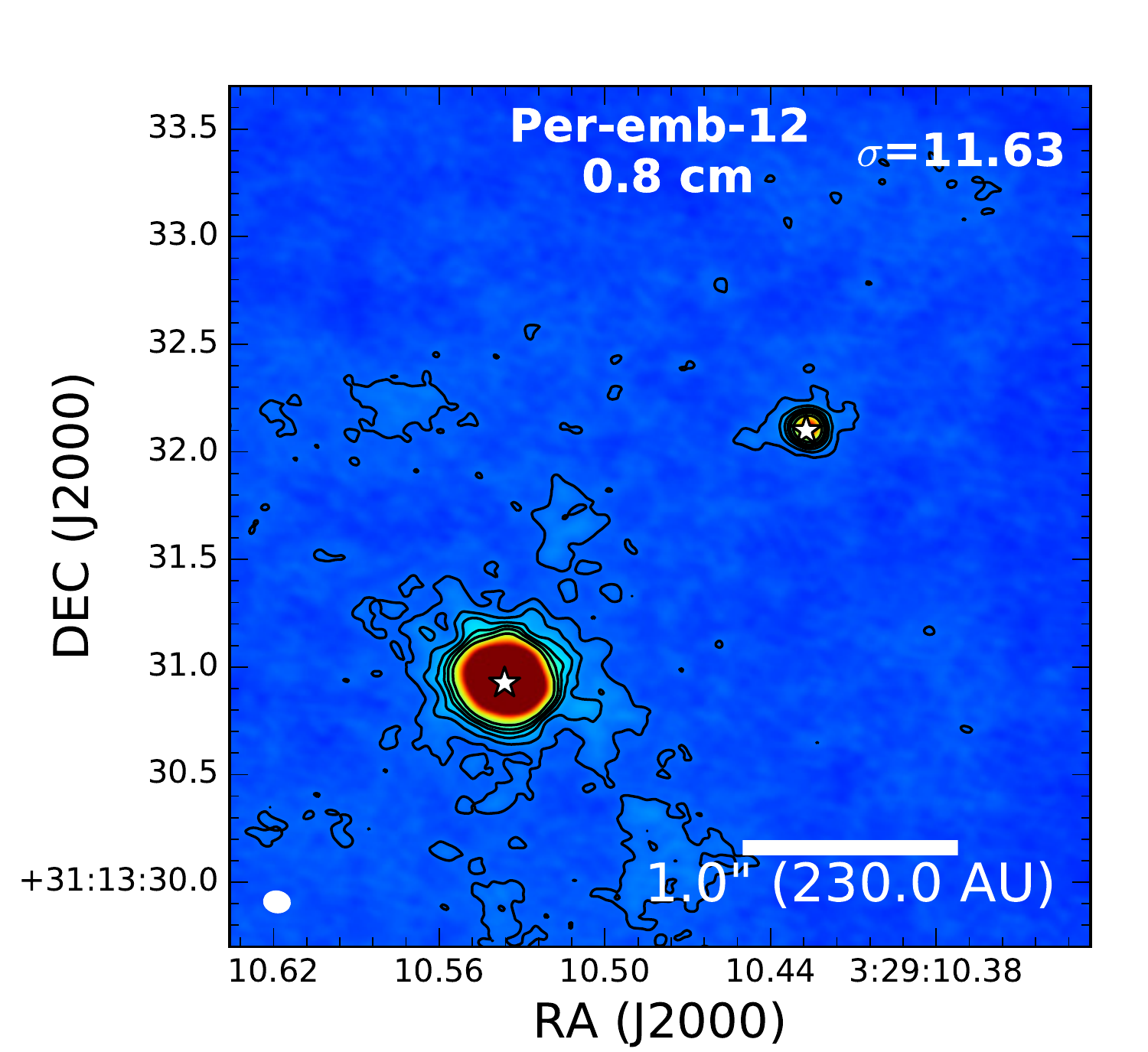}
  \includegraphics[width=0.24\linewidth]{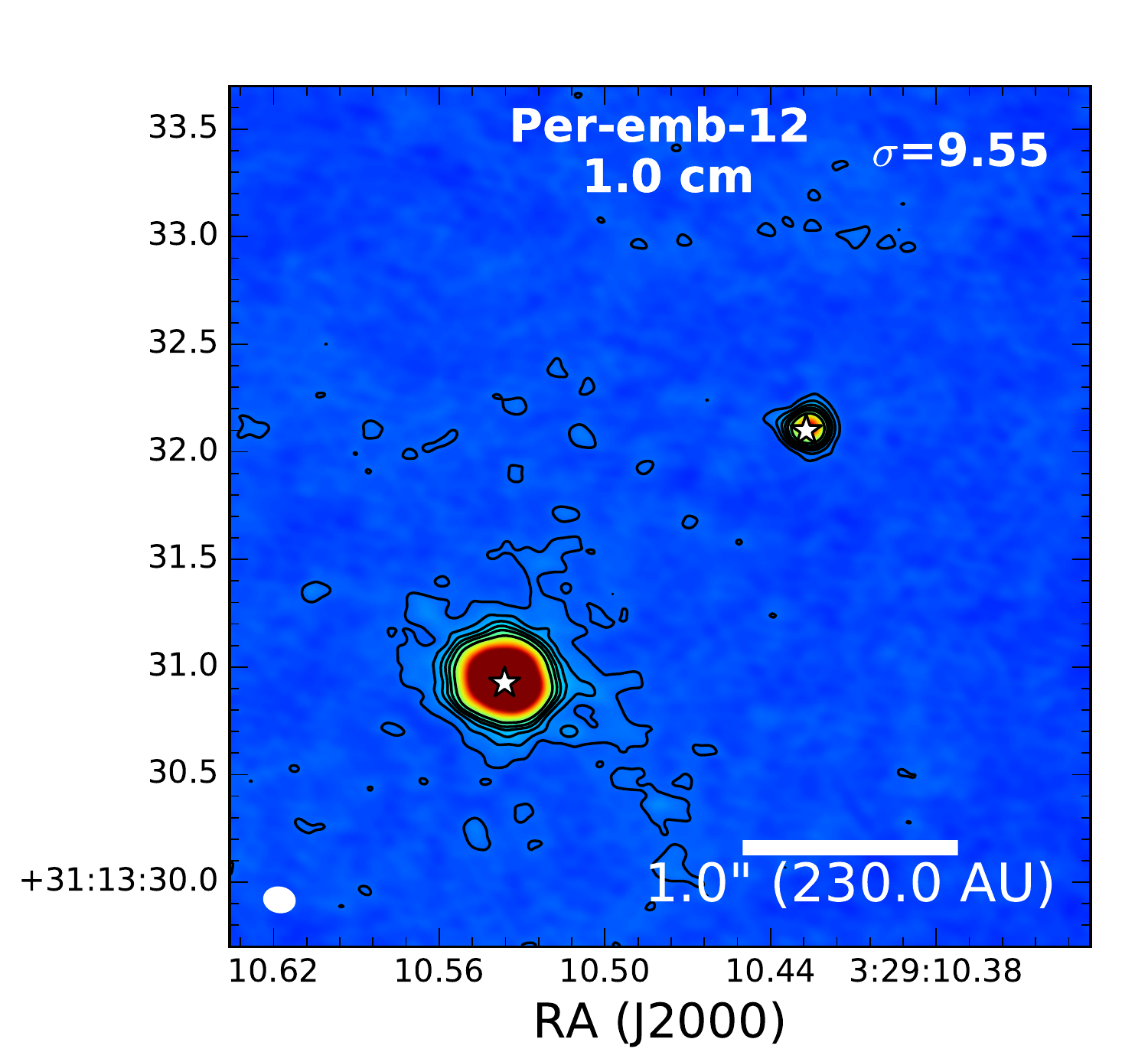}
  \includegraphics[width=0.24\linewidth]{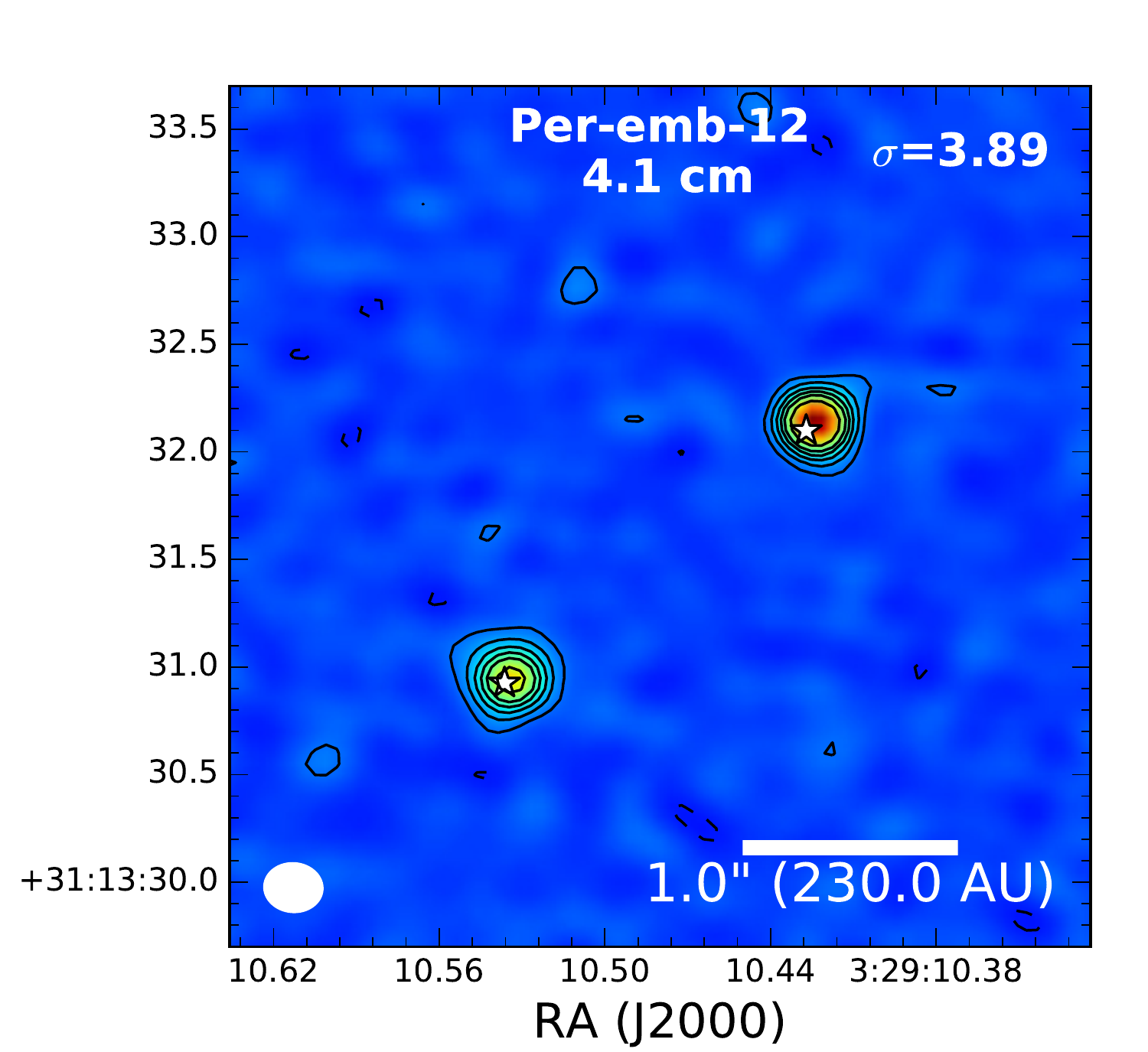}
  \includegraphics[width=0.24\linewidth]{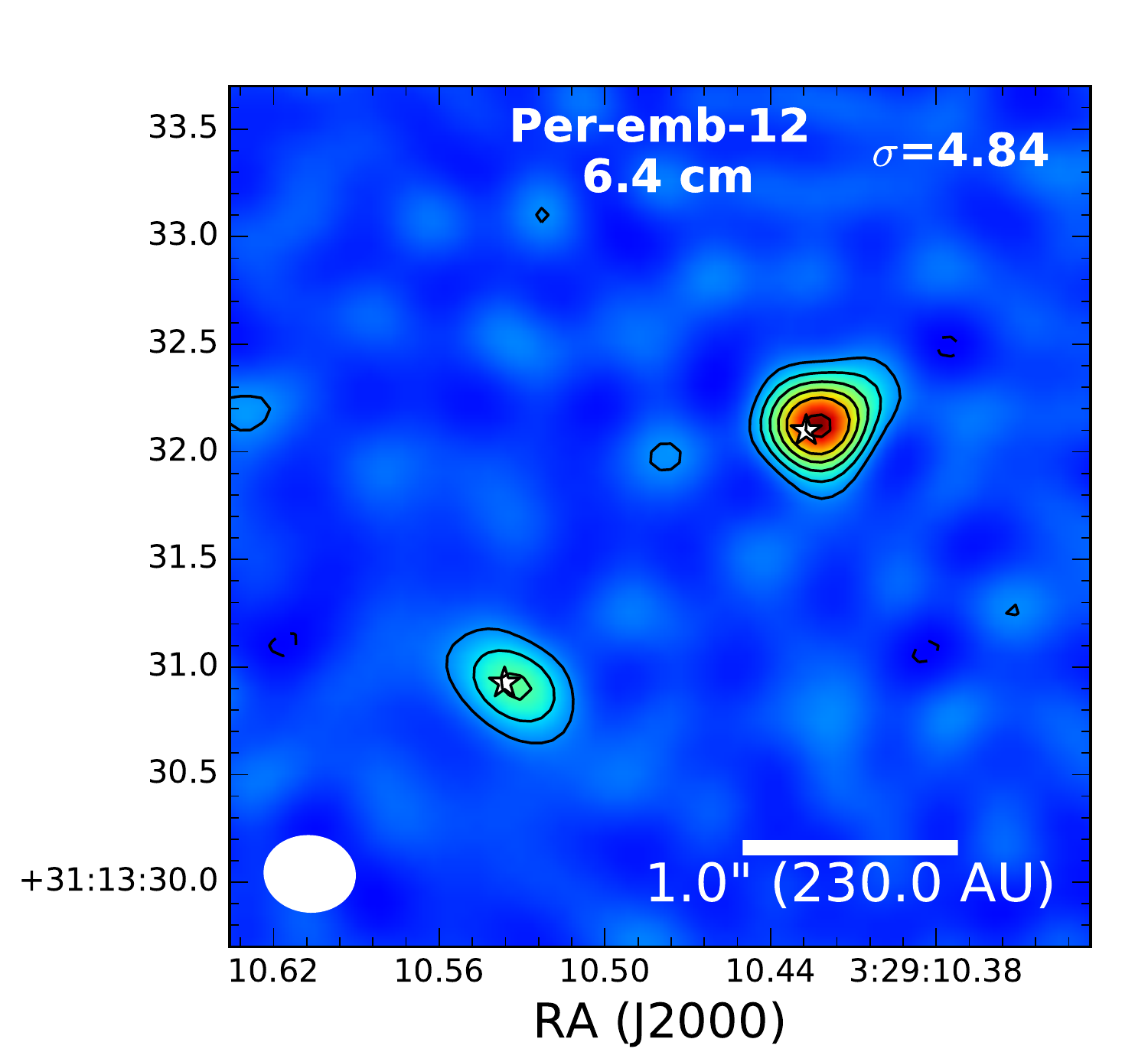}

  \includegraphics[width=0.24\linewidth]{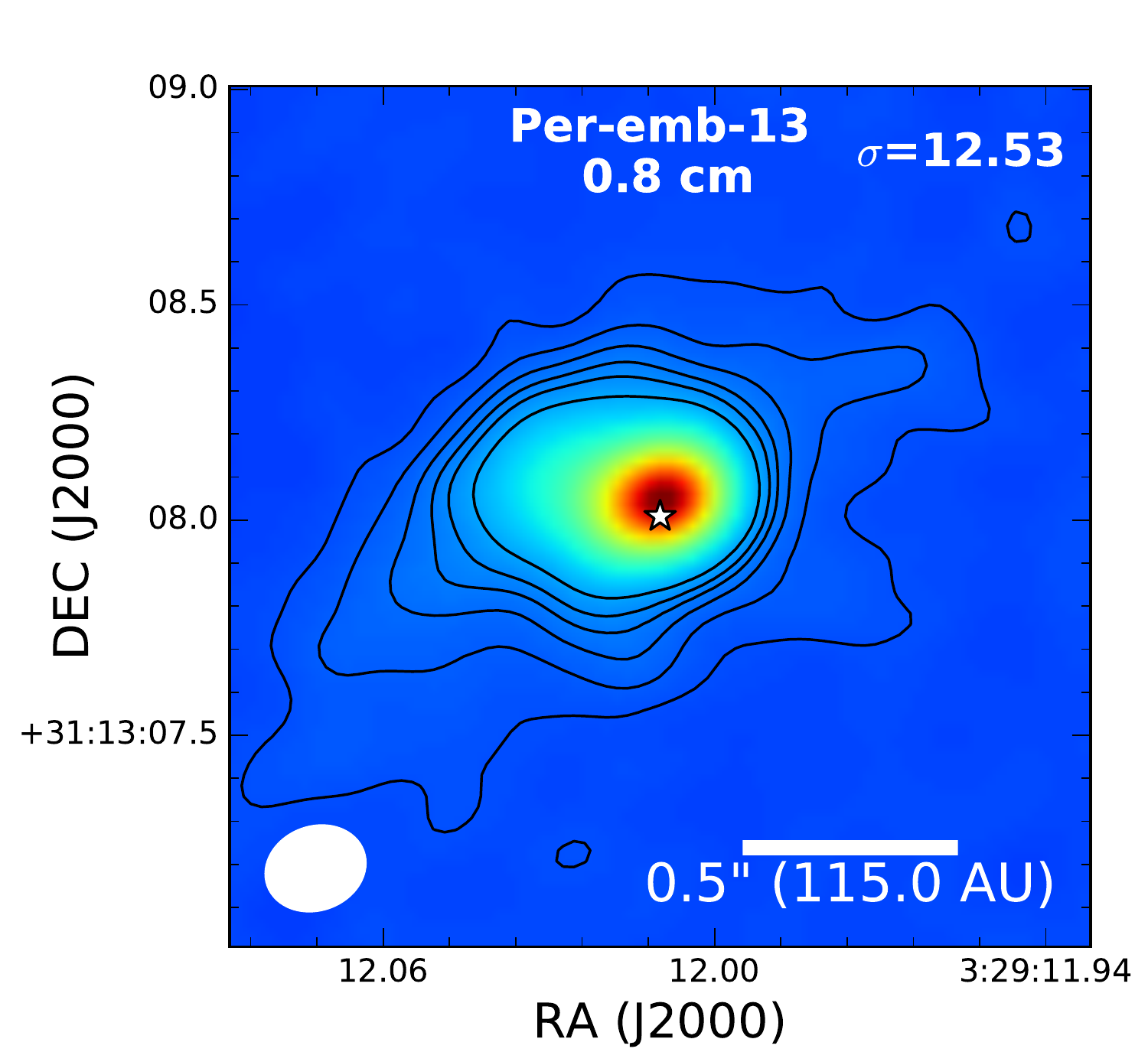}
  \includegraphics[width=0.24\linewidth]{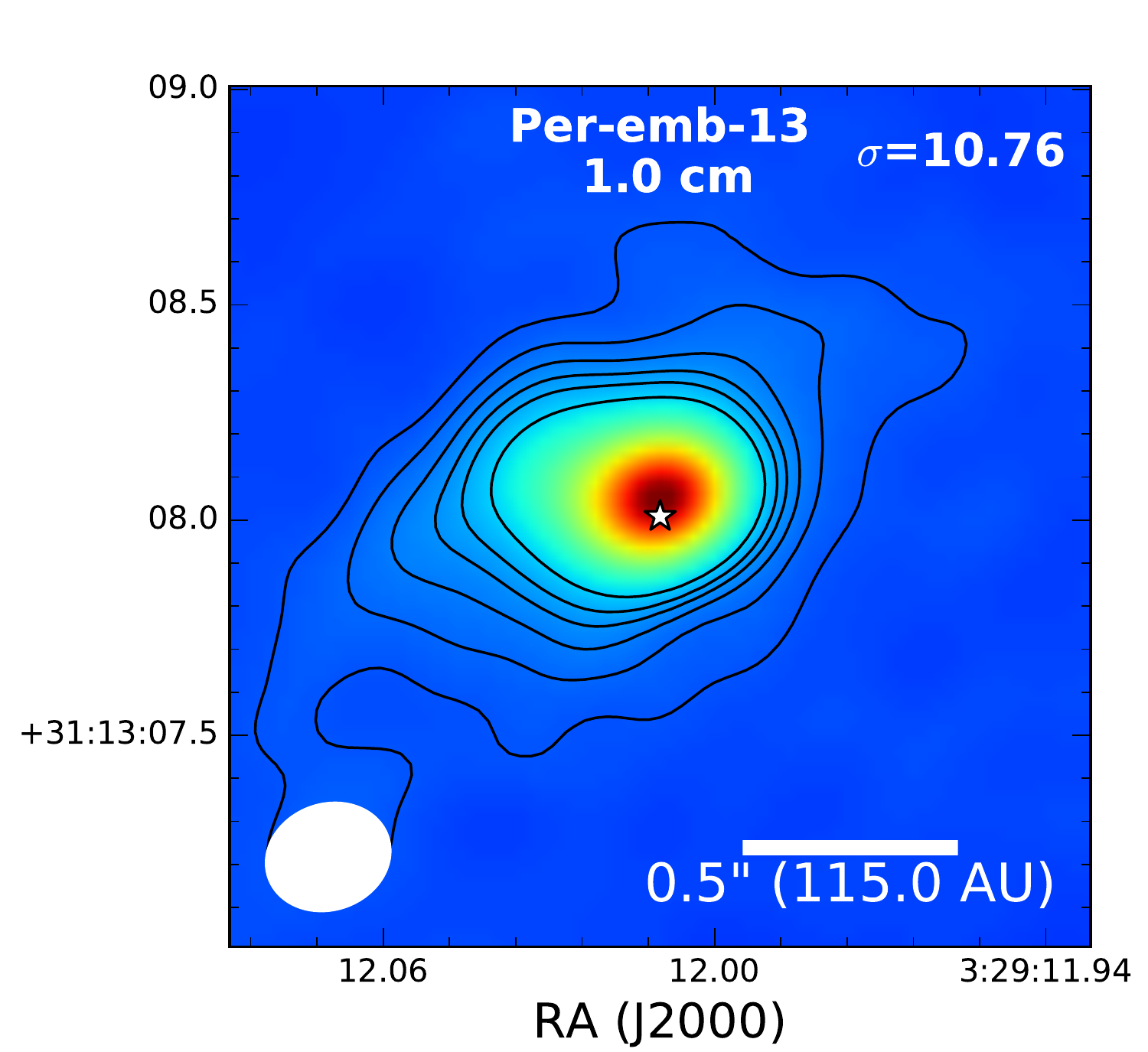}
  \includegraphics[width=0.24\linewidth]{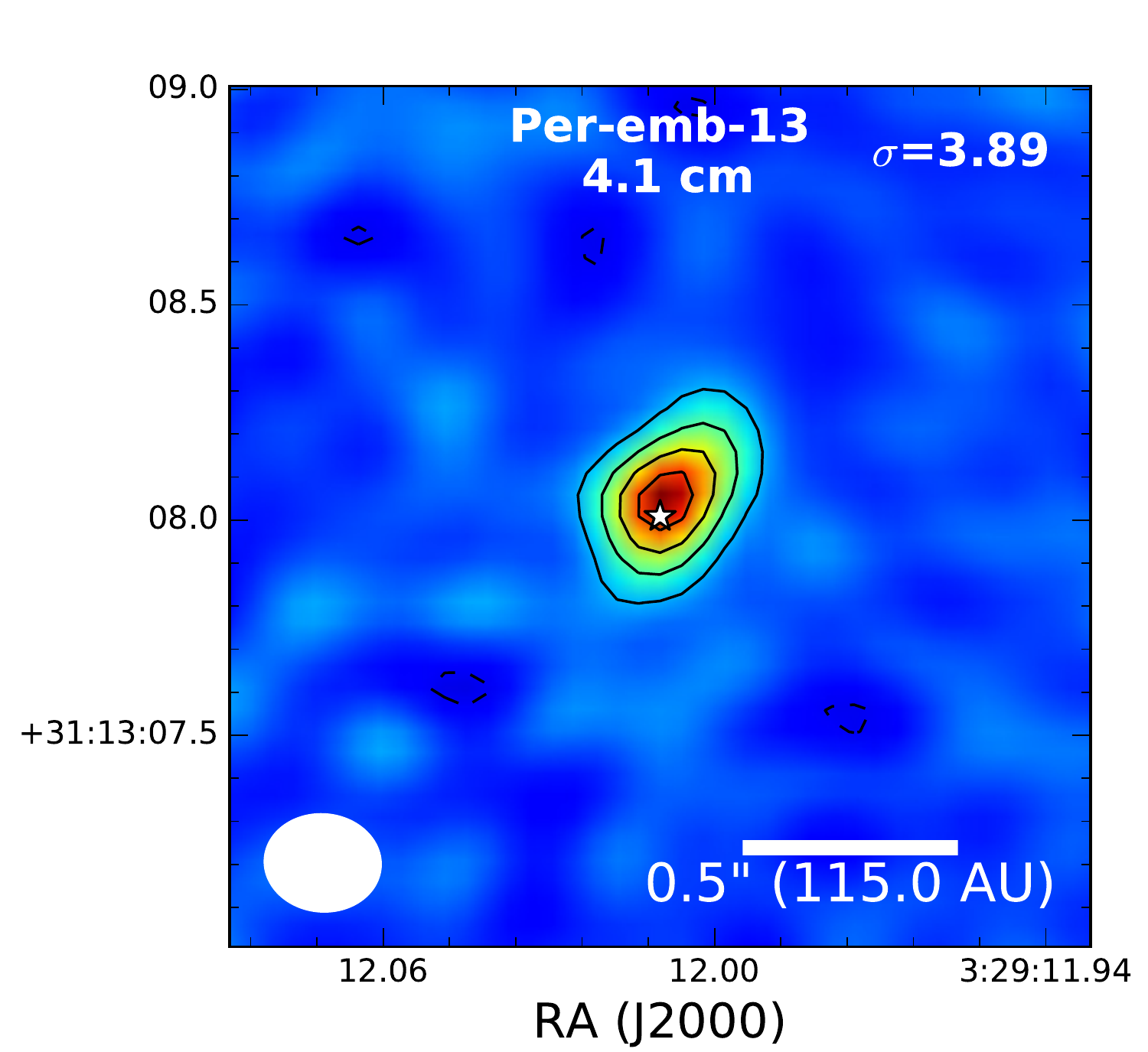}
  \includegraphics[width=0.24\linewidth]{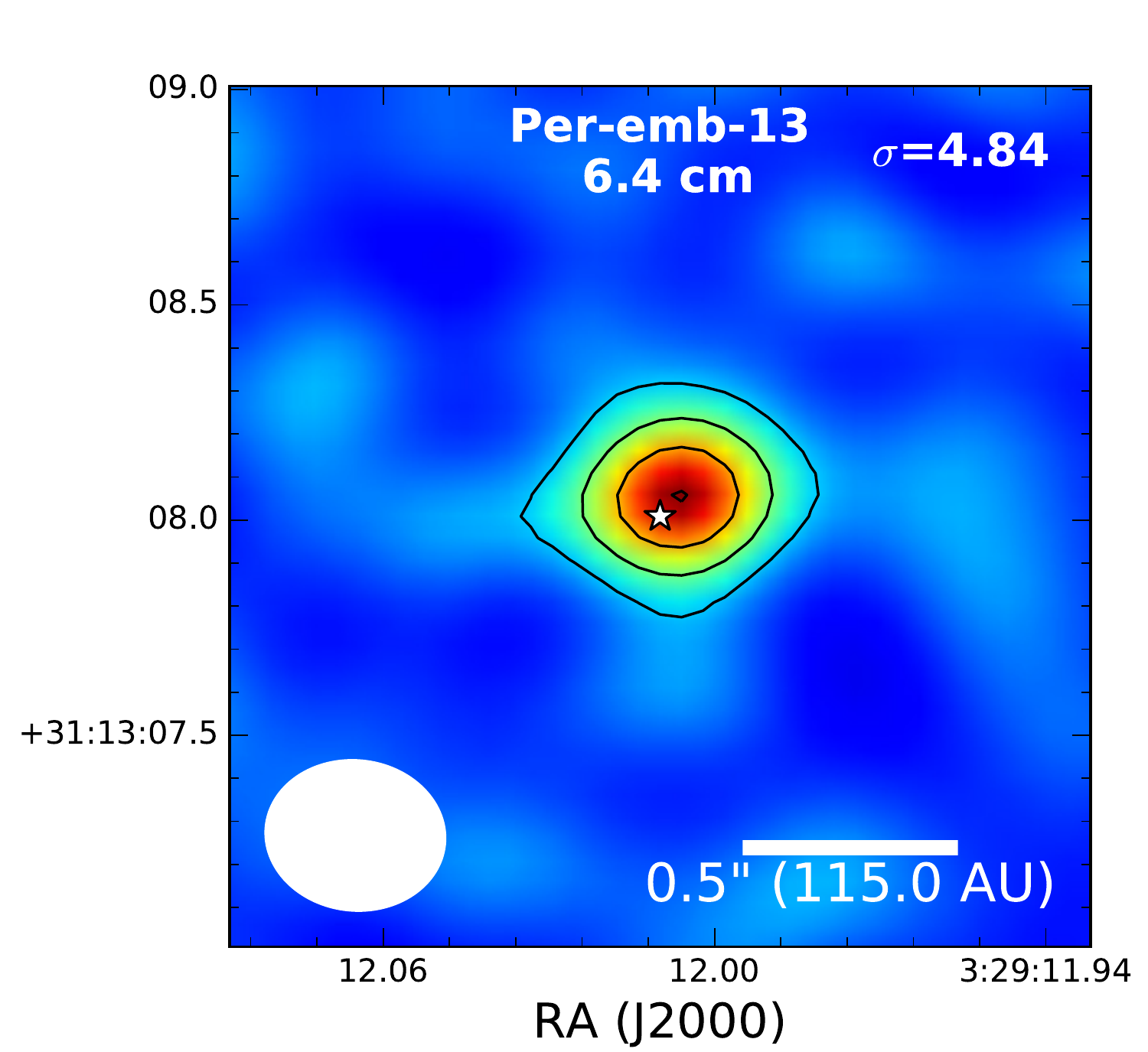}

  \includegraphics[width=0.24\linewidth]{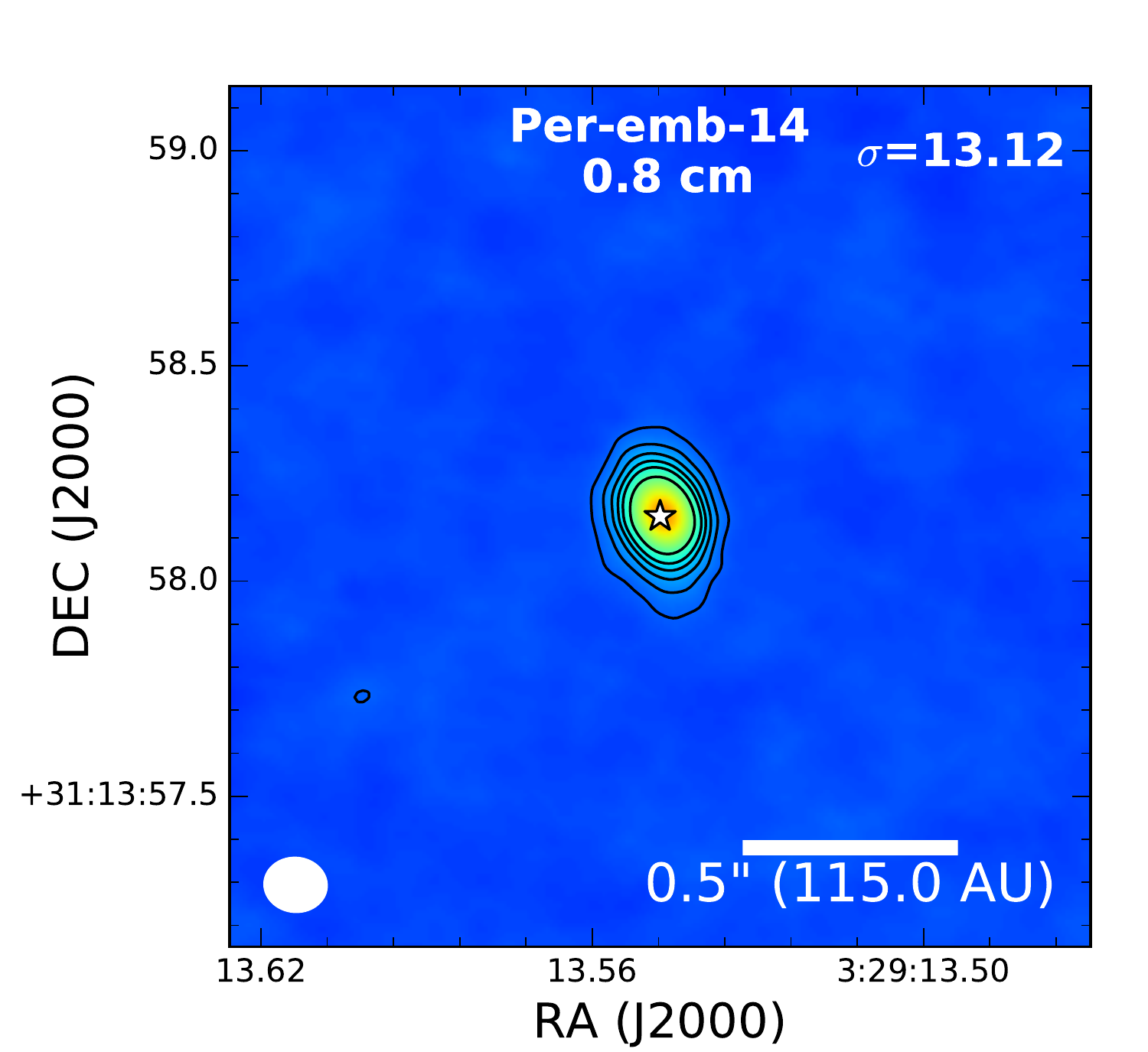}
  \includegraphics[width=0.24\linewidth]{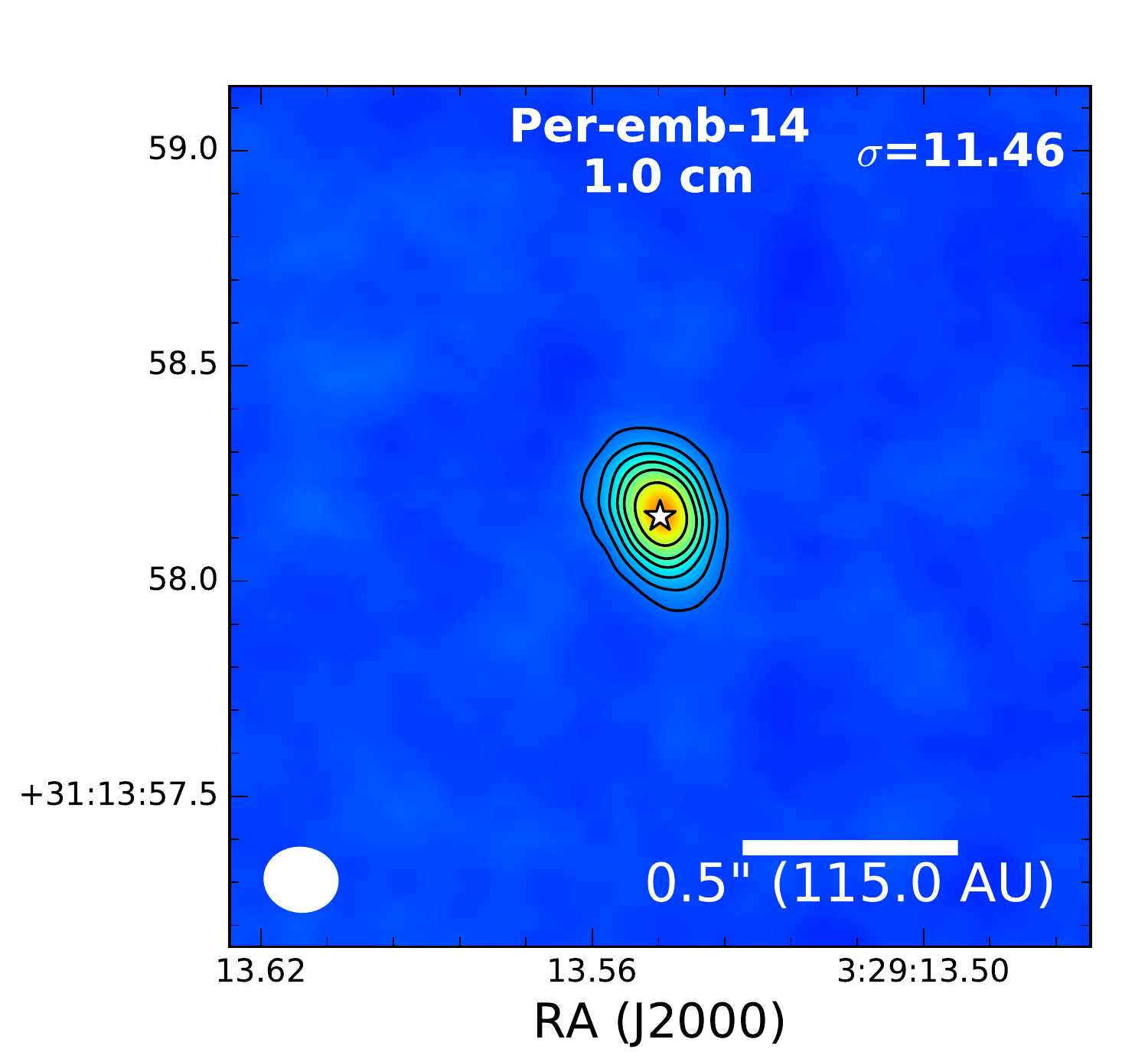}
  \includegraphics[width=0.24\linewidth]{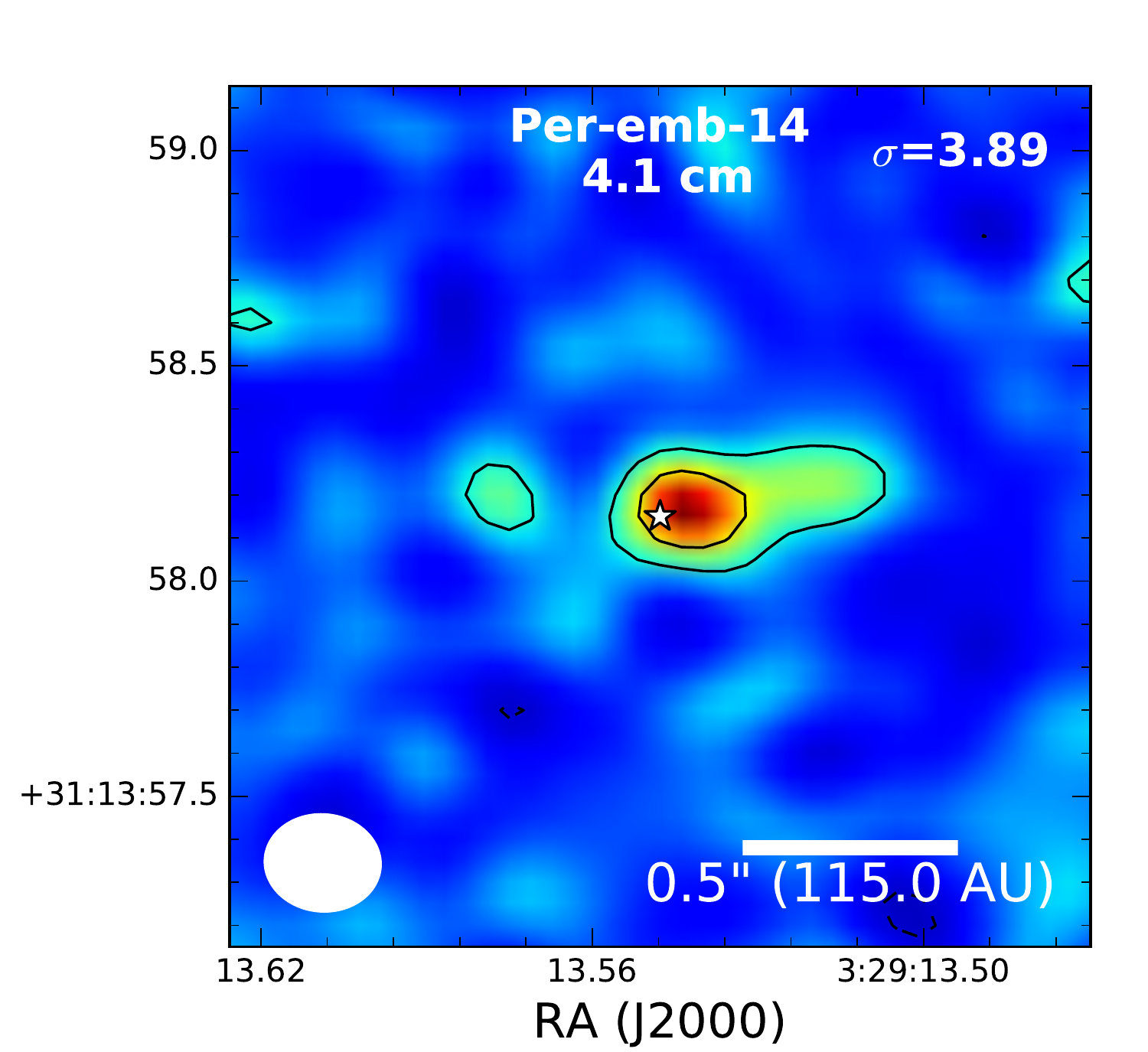}
  \includegraphics[width=0.24\linewidth]{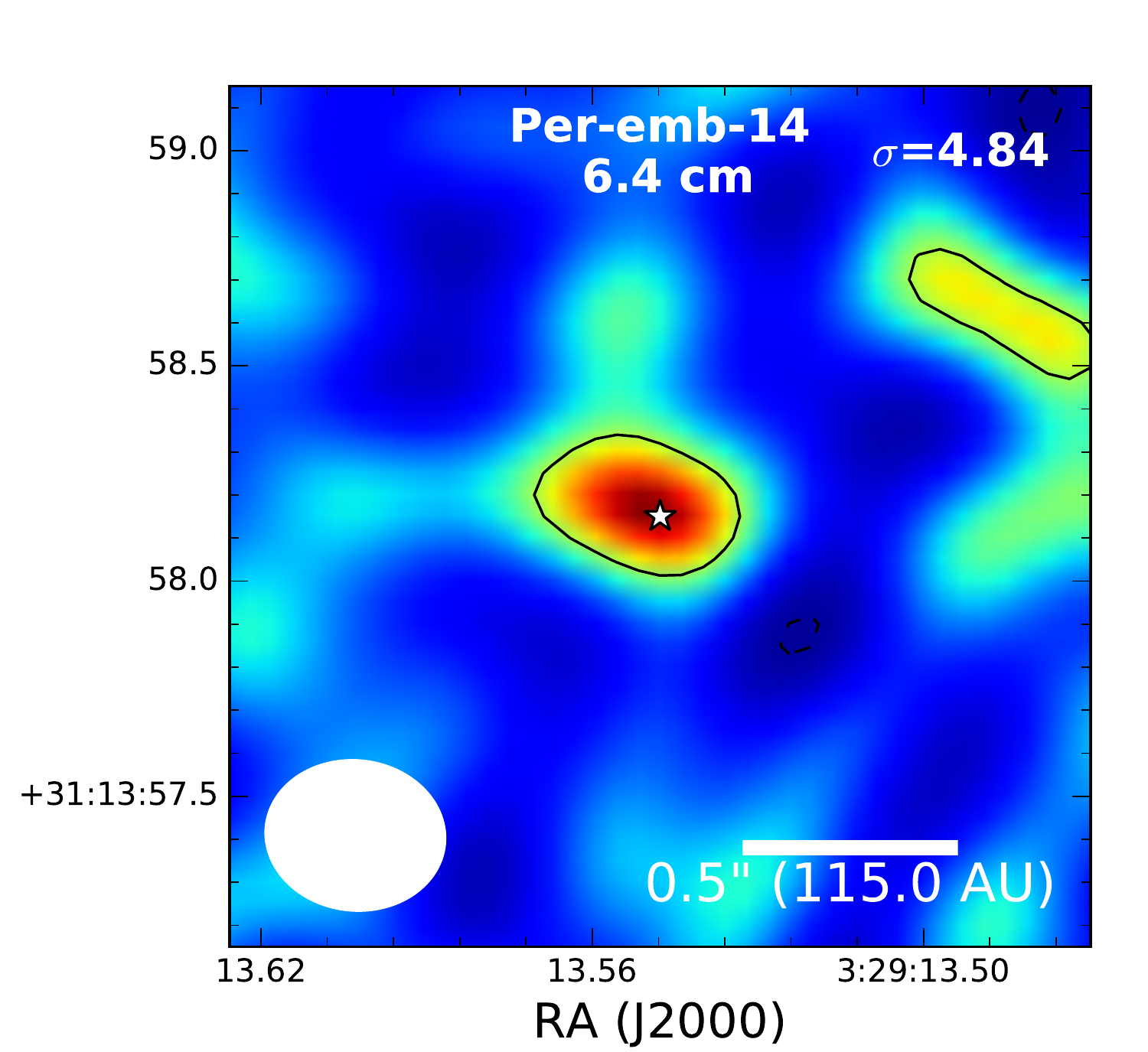}

  \includegraphics[width=0.24\linewidth]{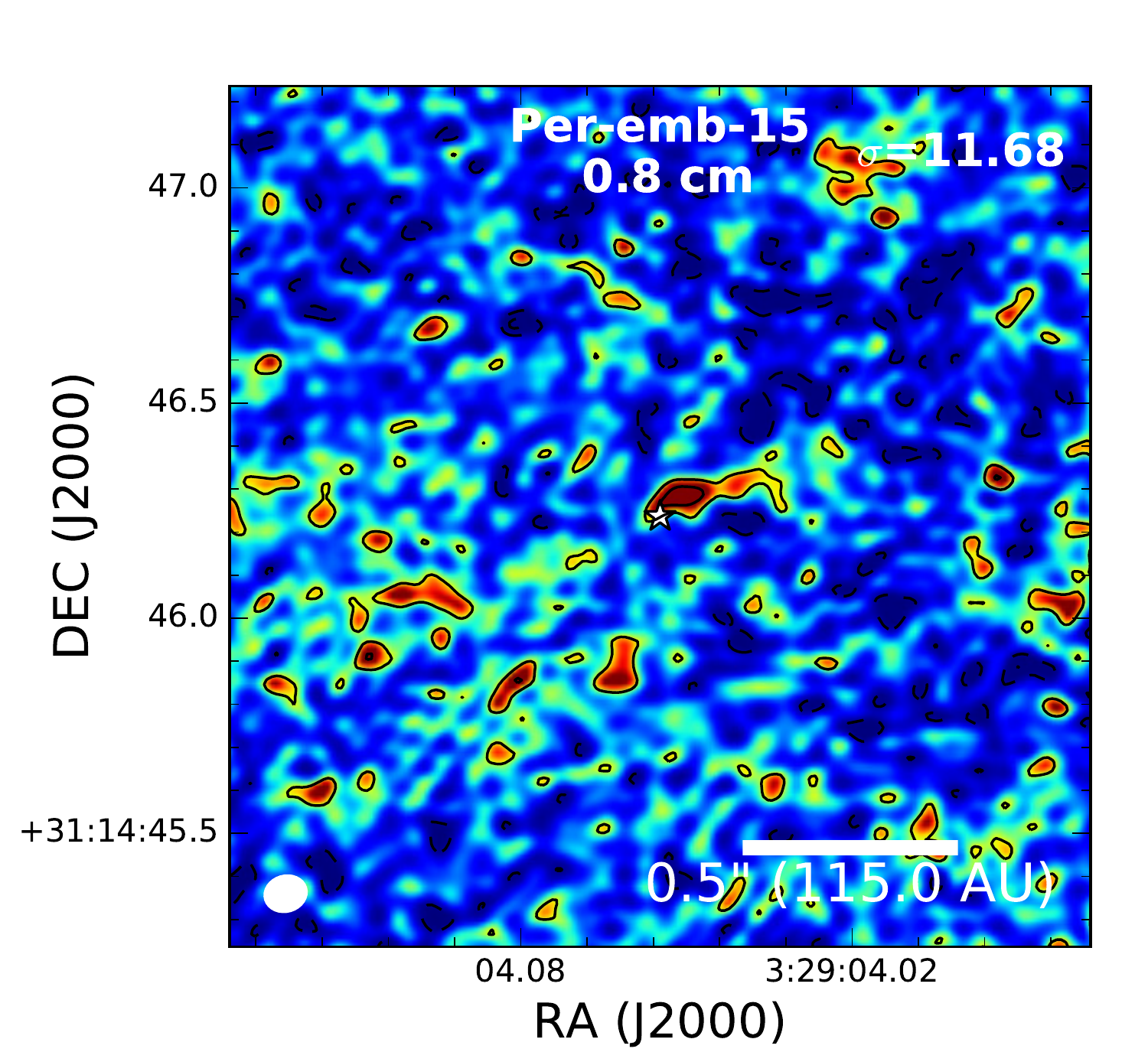}
  \includegraphics[width=0.24\linewidth]{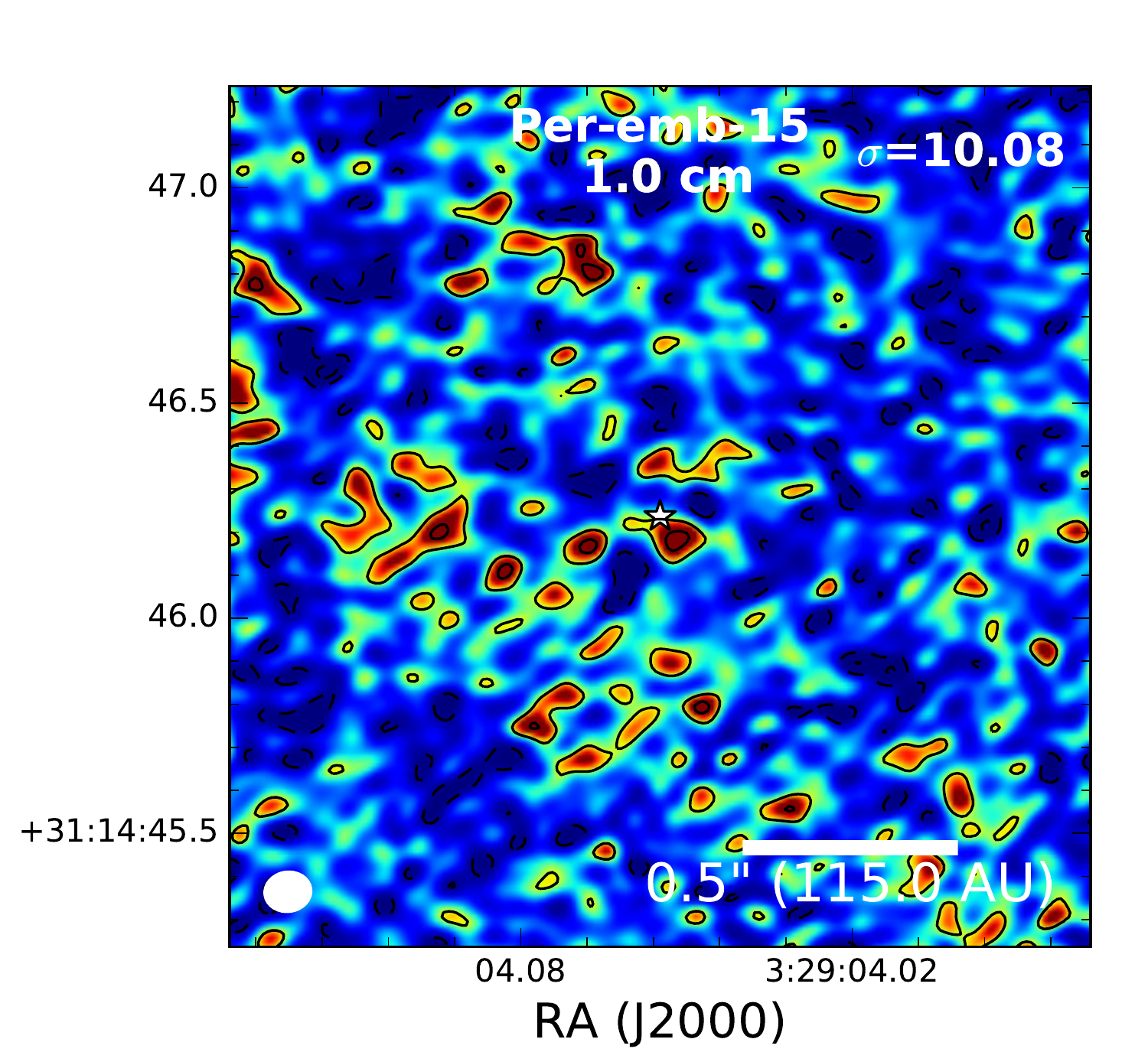}
  \includegraphics[width=0.24\linewidth]{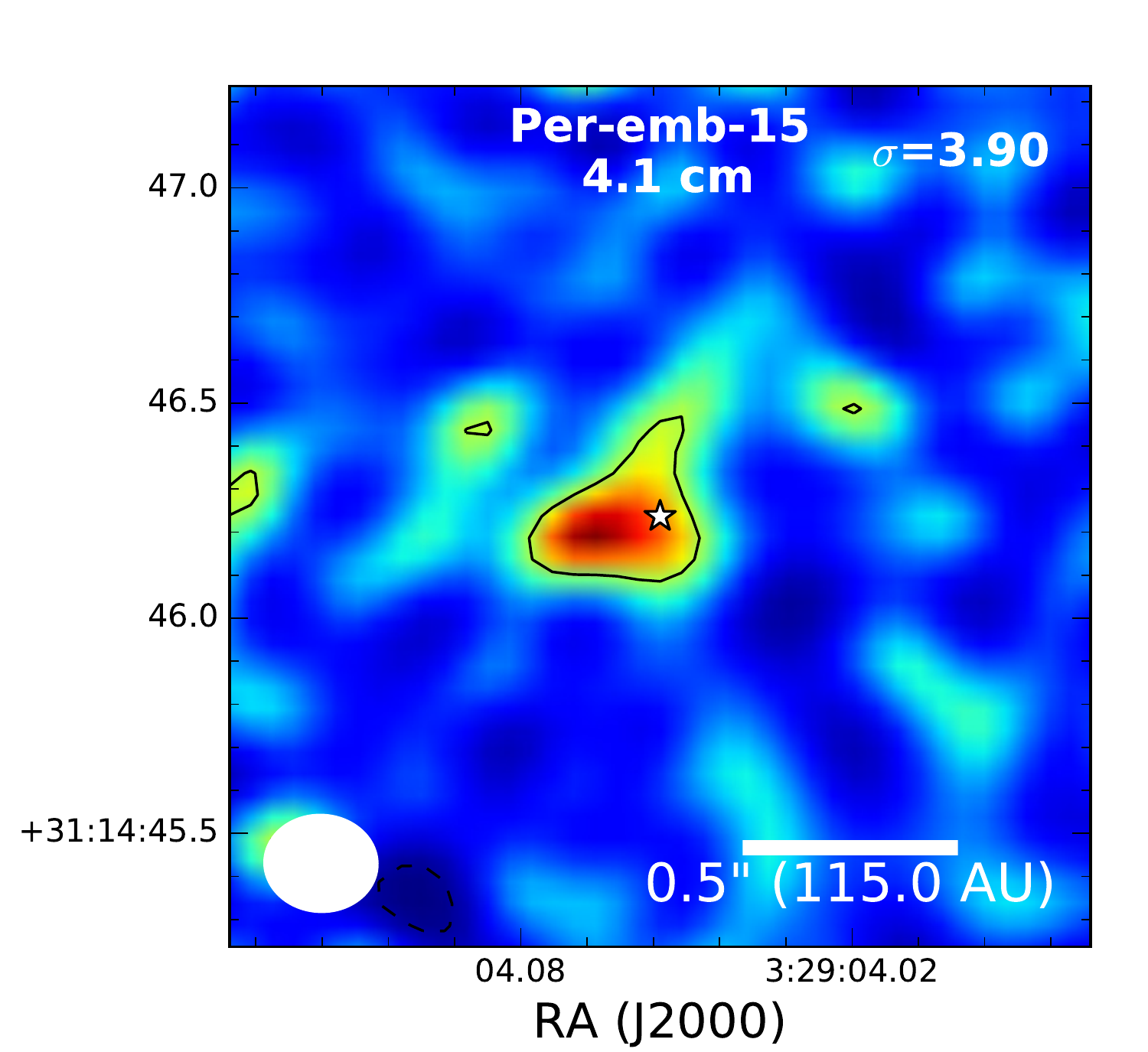}
  \includegraphics[width=0.24\linewidth]{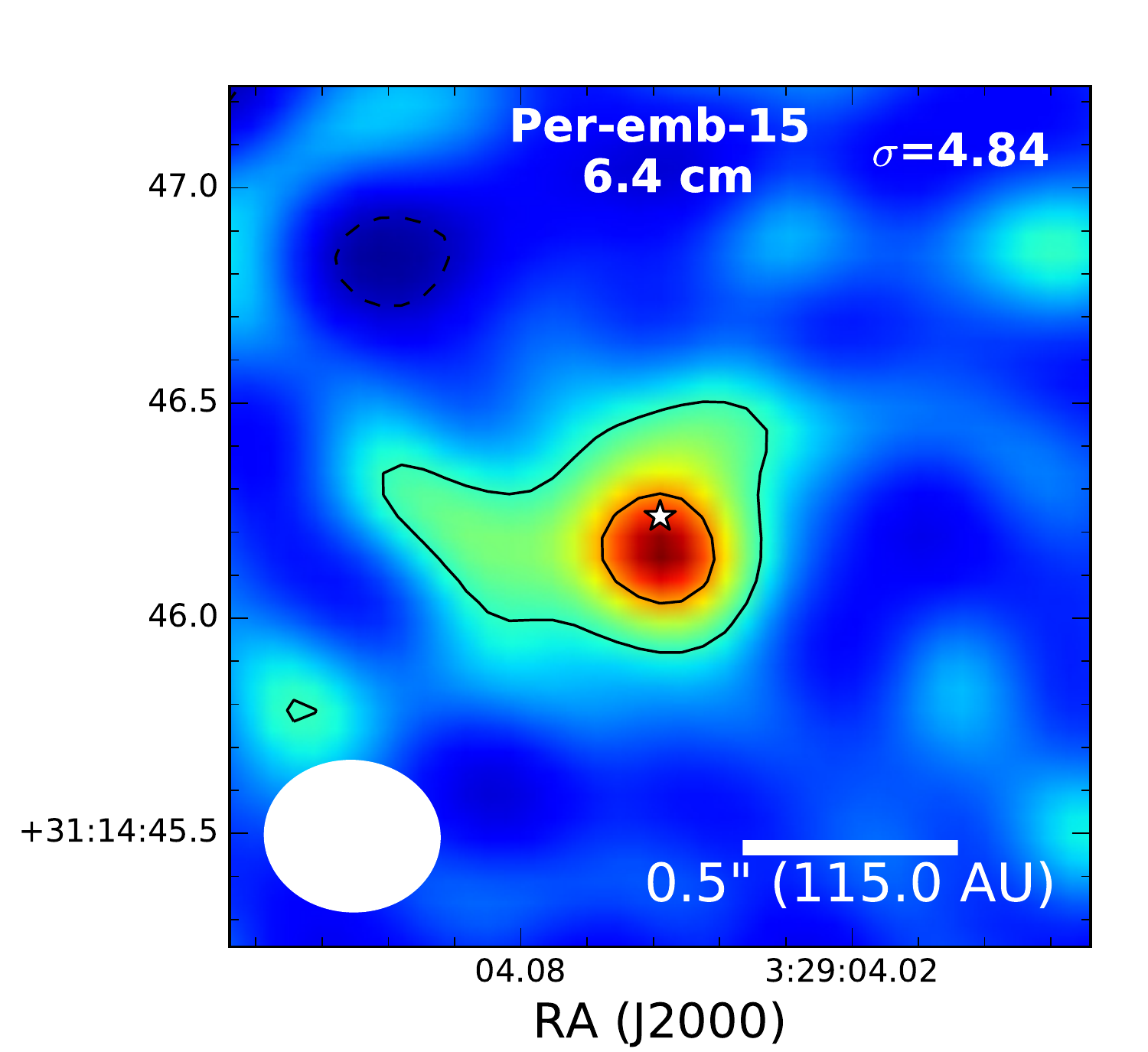}

\end{figure}

\begin{figure}

  \includegraphics[width=0.24\linewidth]{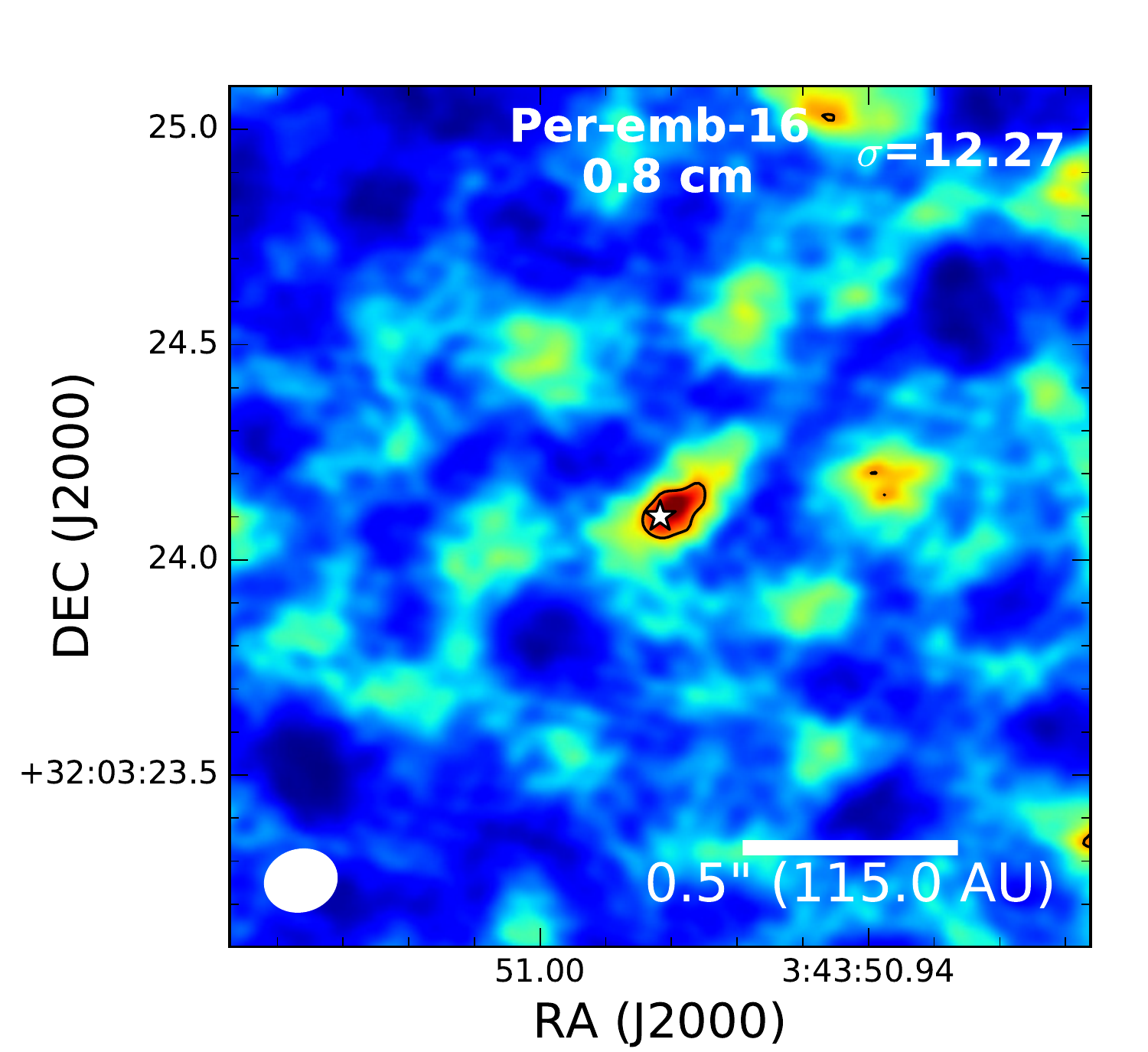}
  \includegraphics[width=0.24\linewidth]{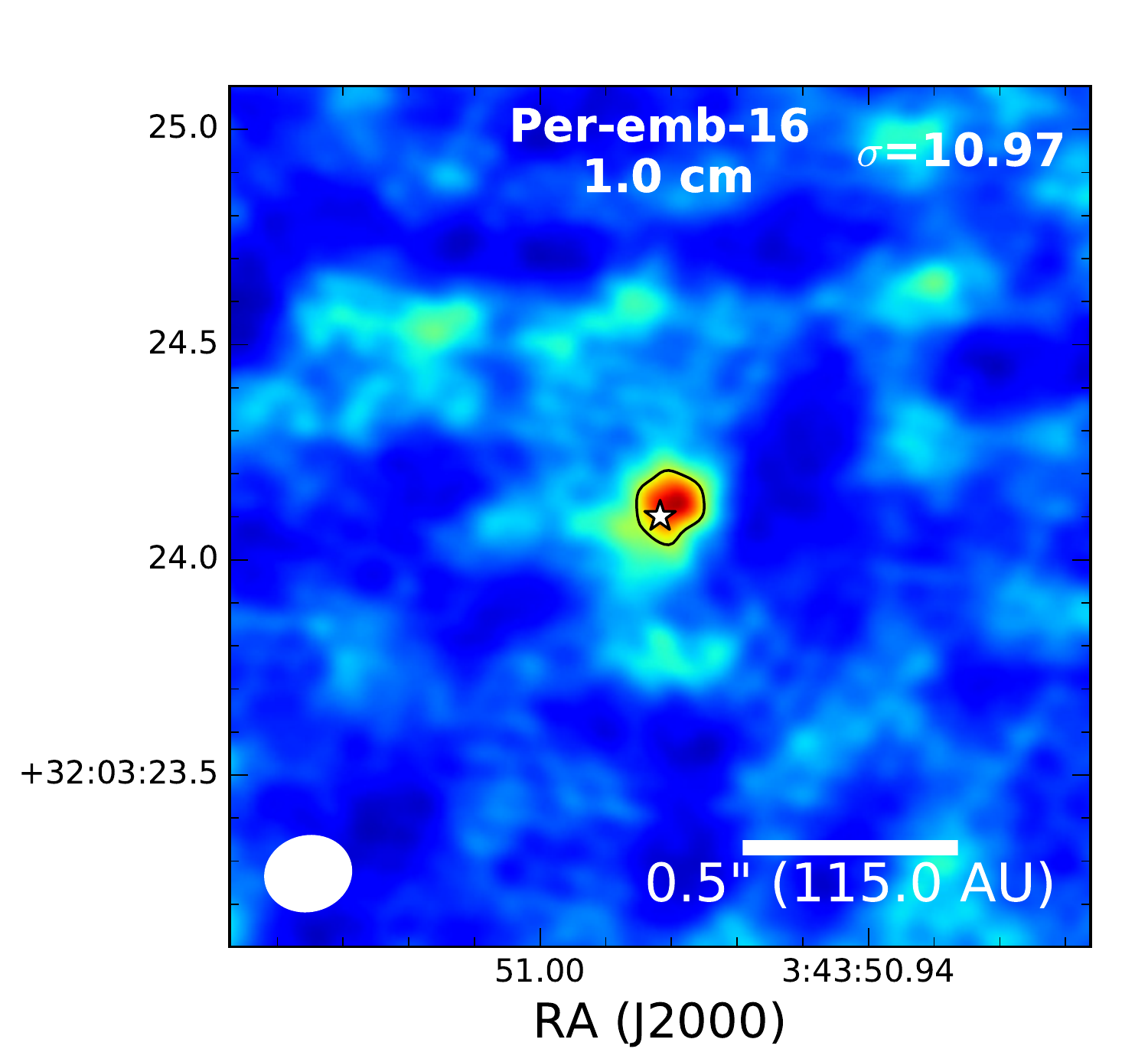}
  \includegraphics[width=0.24\linewidth]{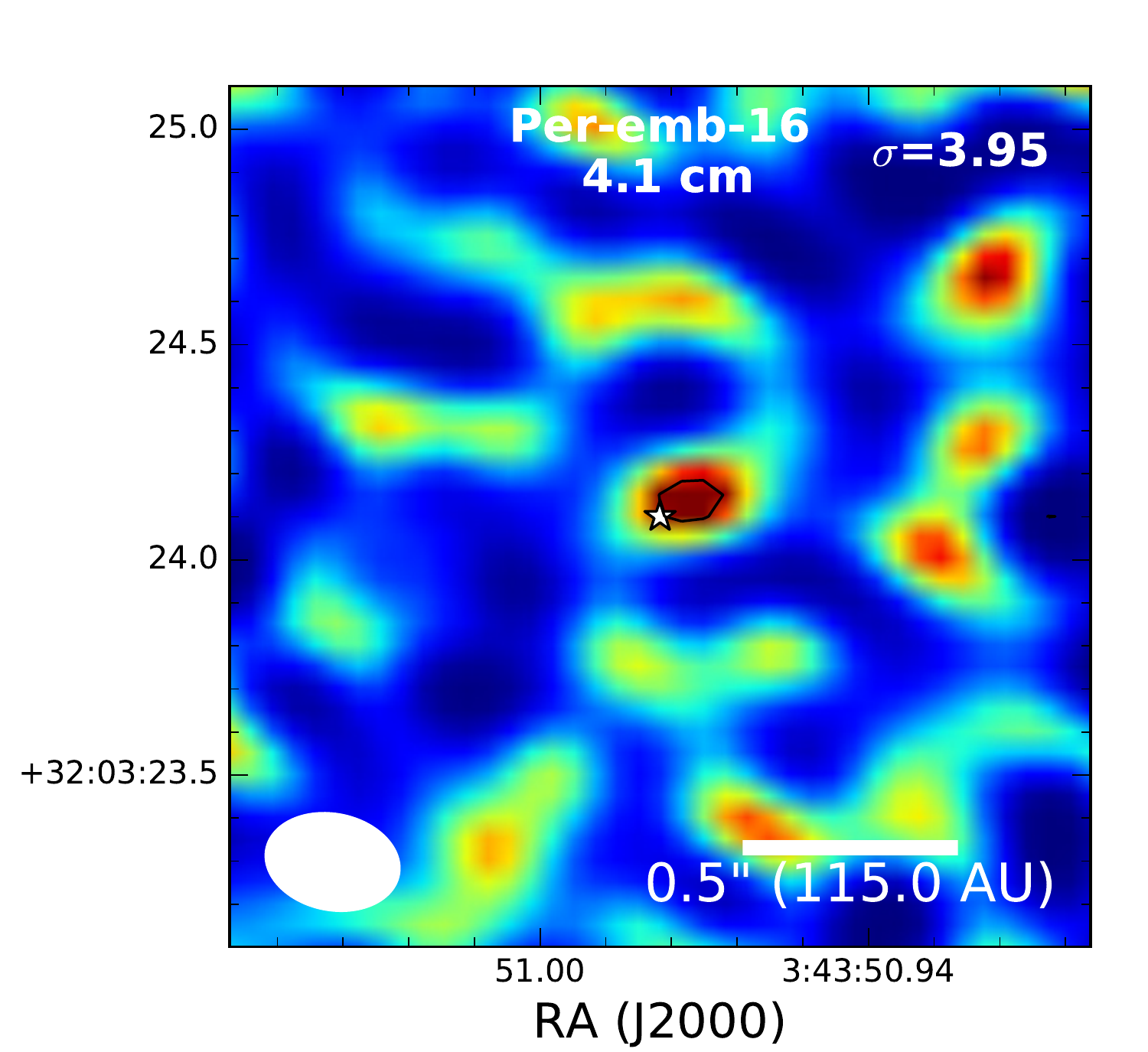}
  \includegraphics[width=0.24\linewidth]{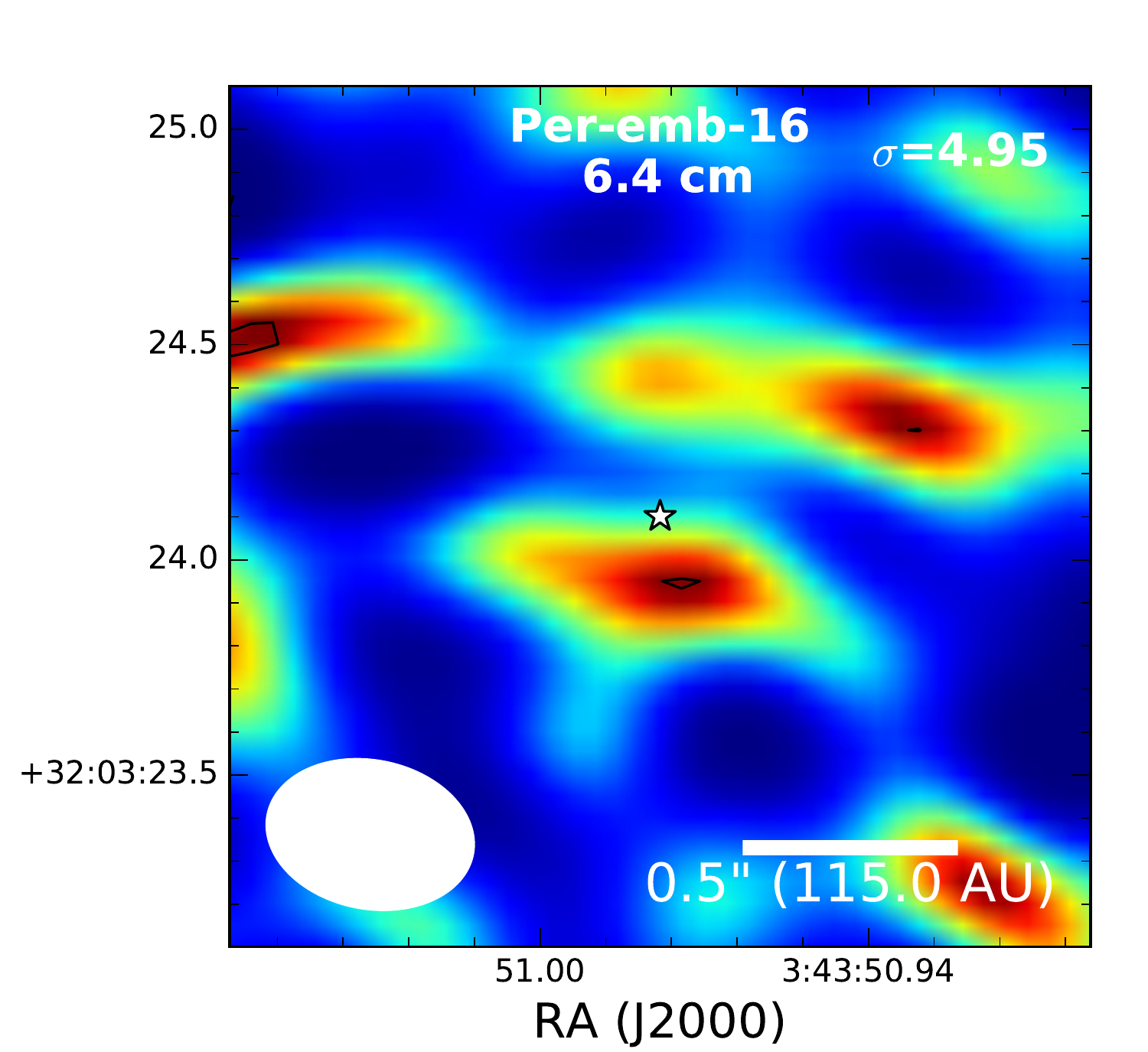}

  \includegraphics[width=0.24\linewidth]{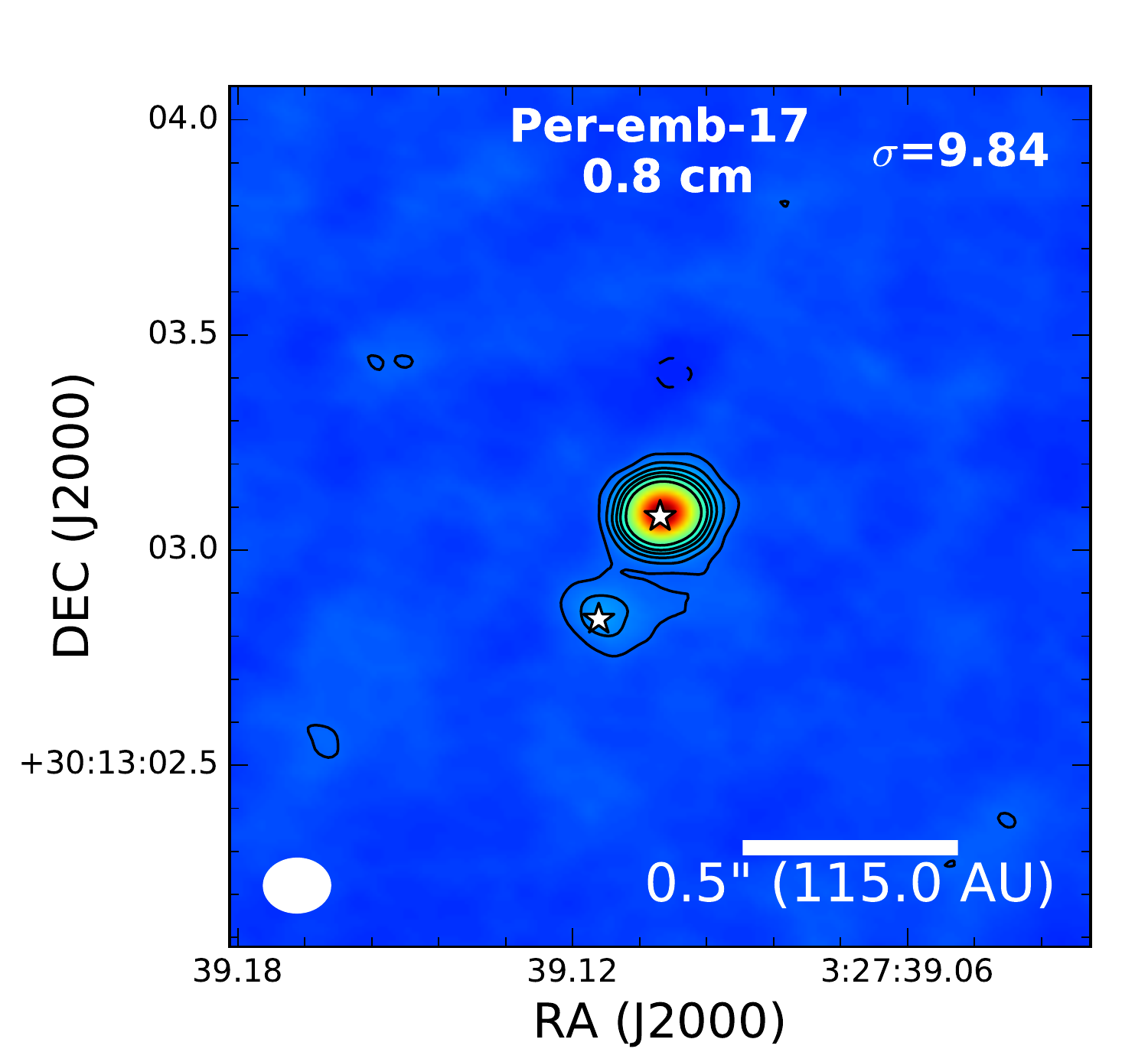}
  \includegraphics[width=0.24\linewidth]{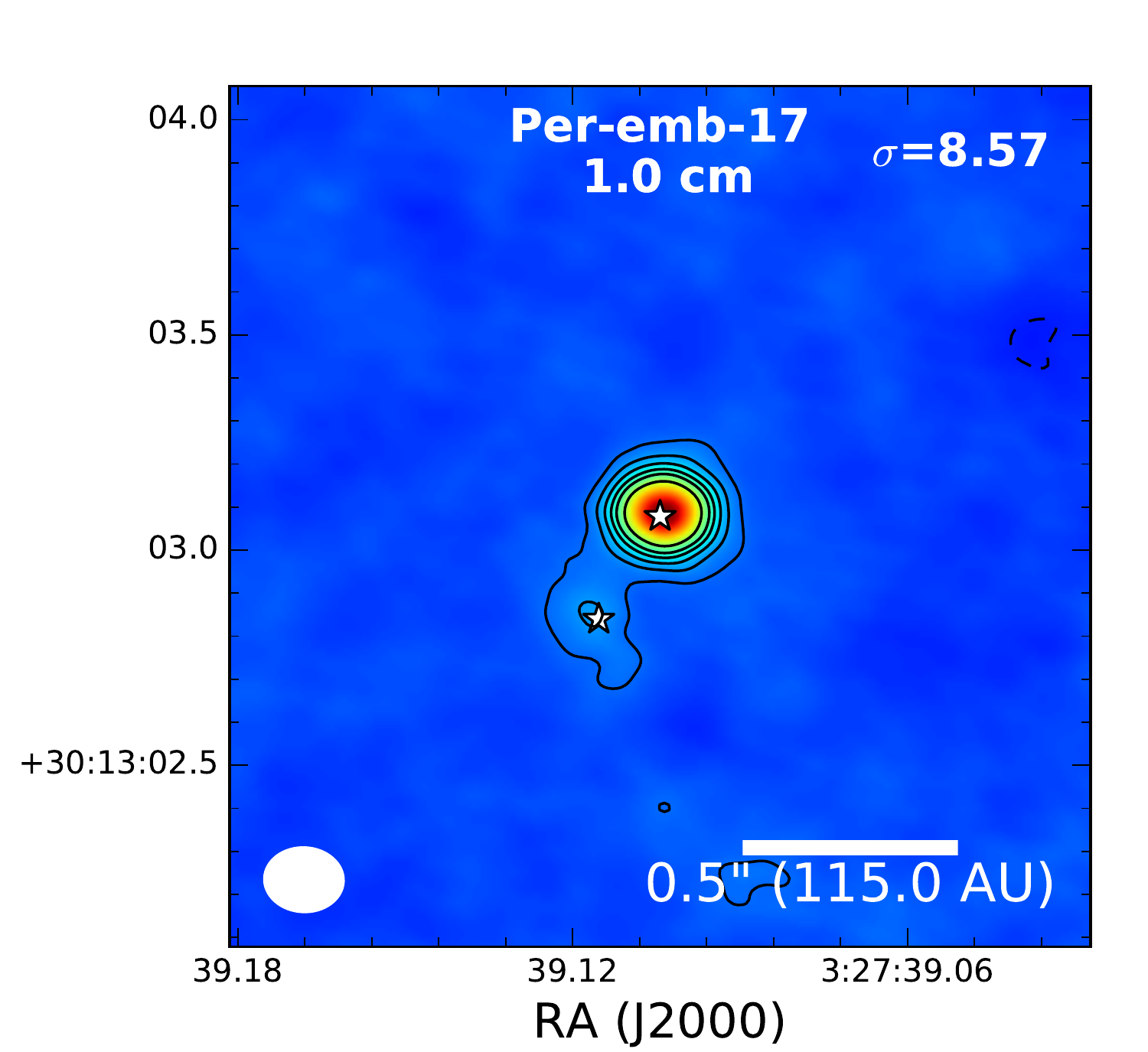}
  \includegraphics[width=0.24\linewidth]{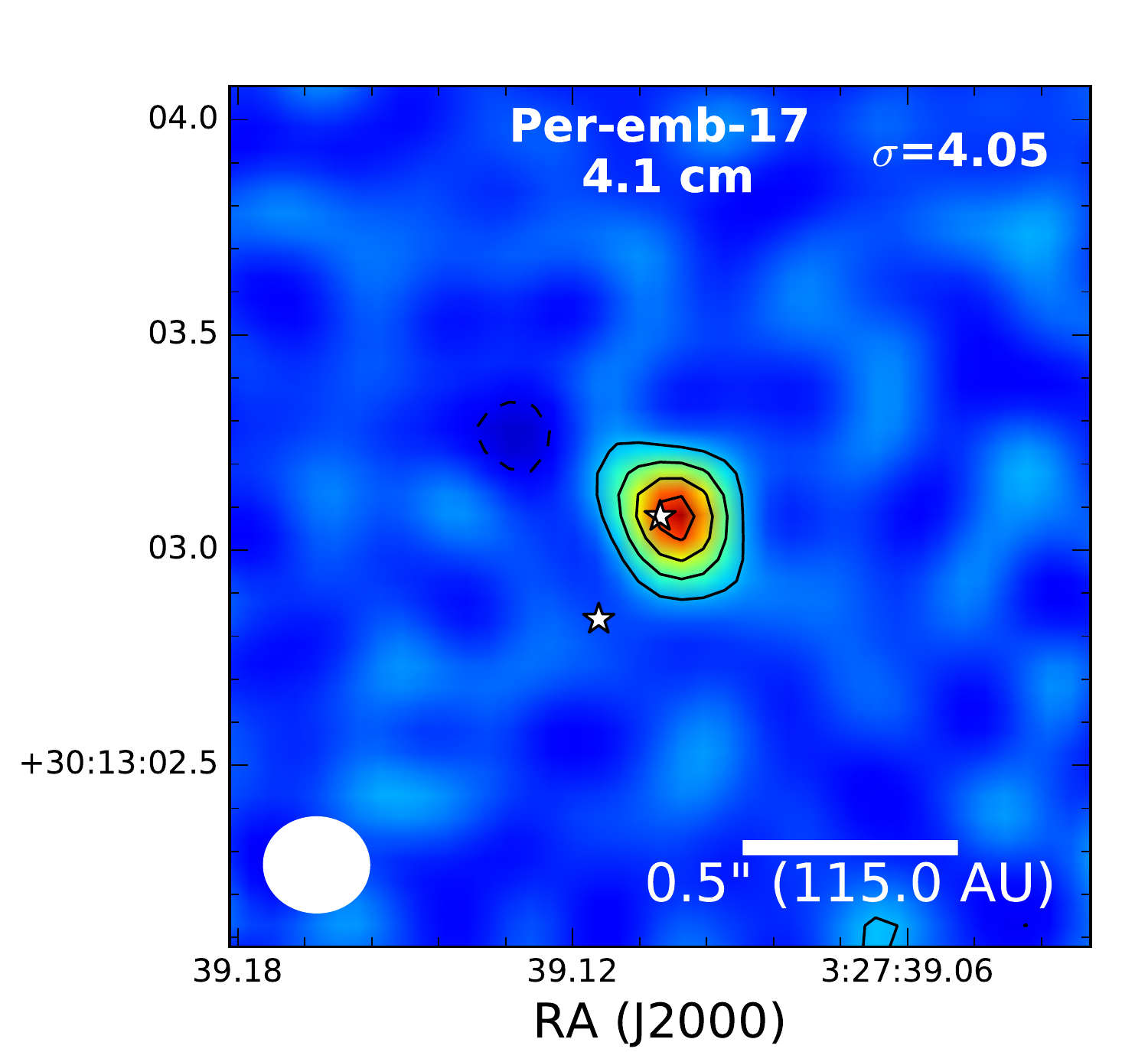}
  \includegraphics[width=0.24\linewidth]{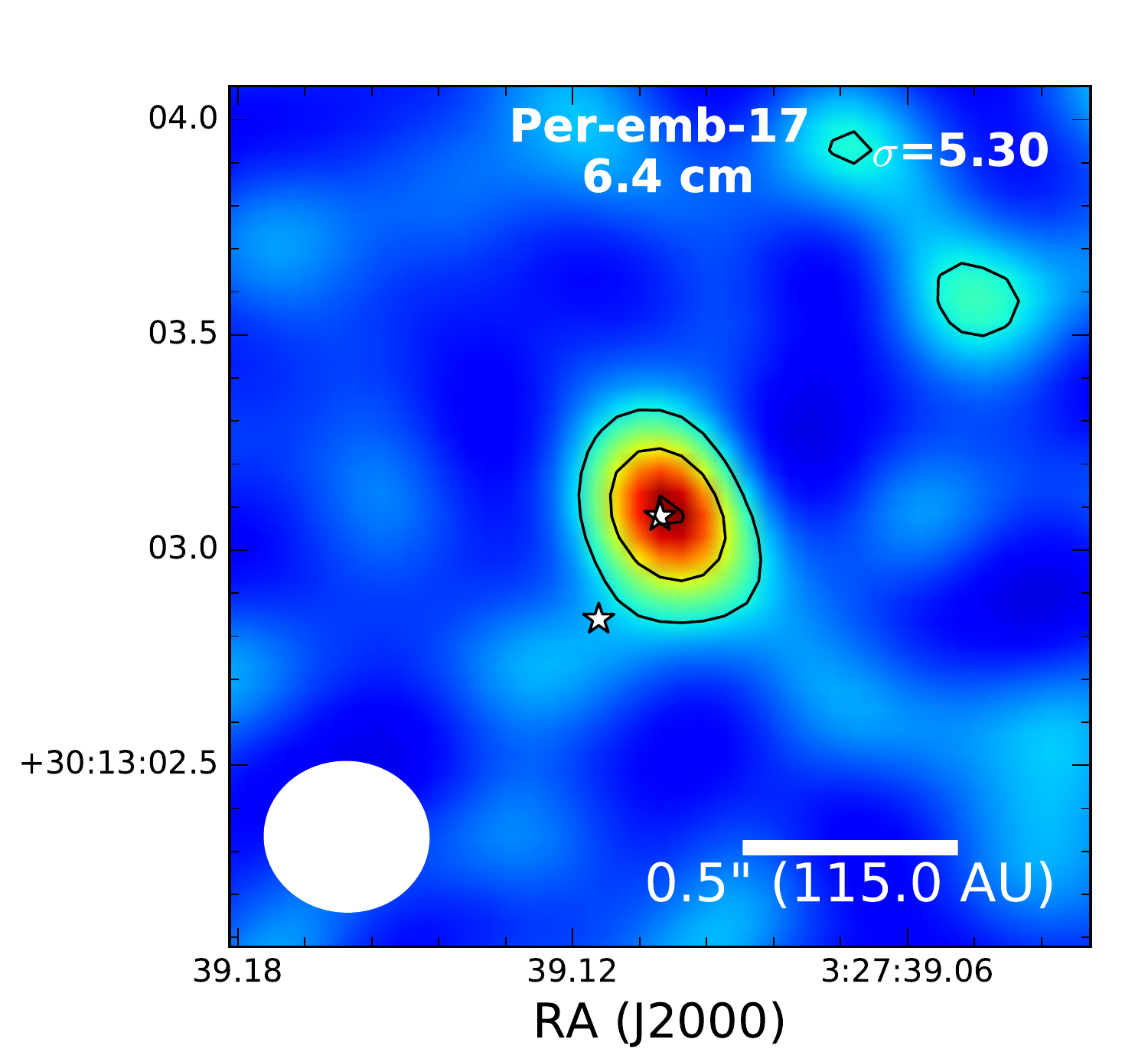}

  \includegraphics[width=0.24\linewidth]{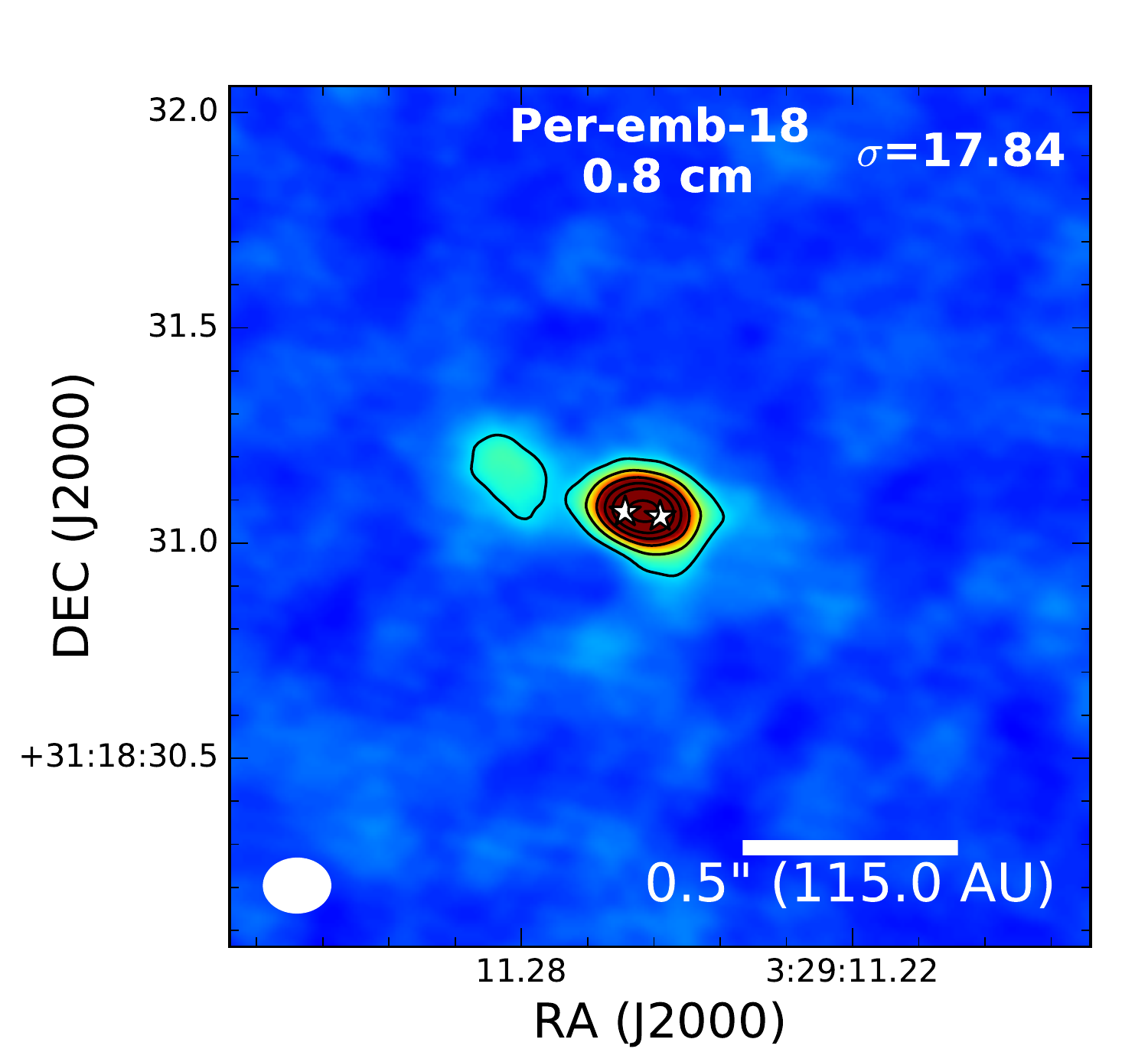}
  \includegraphics[width=0.24\linewidth]{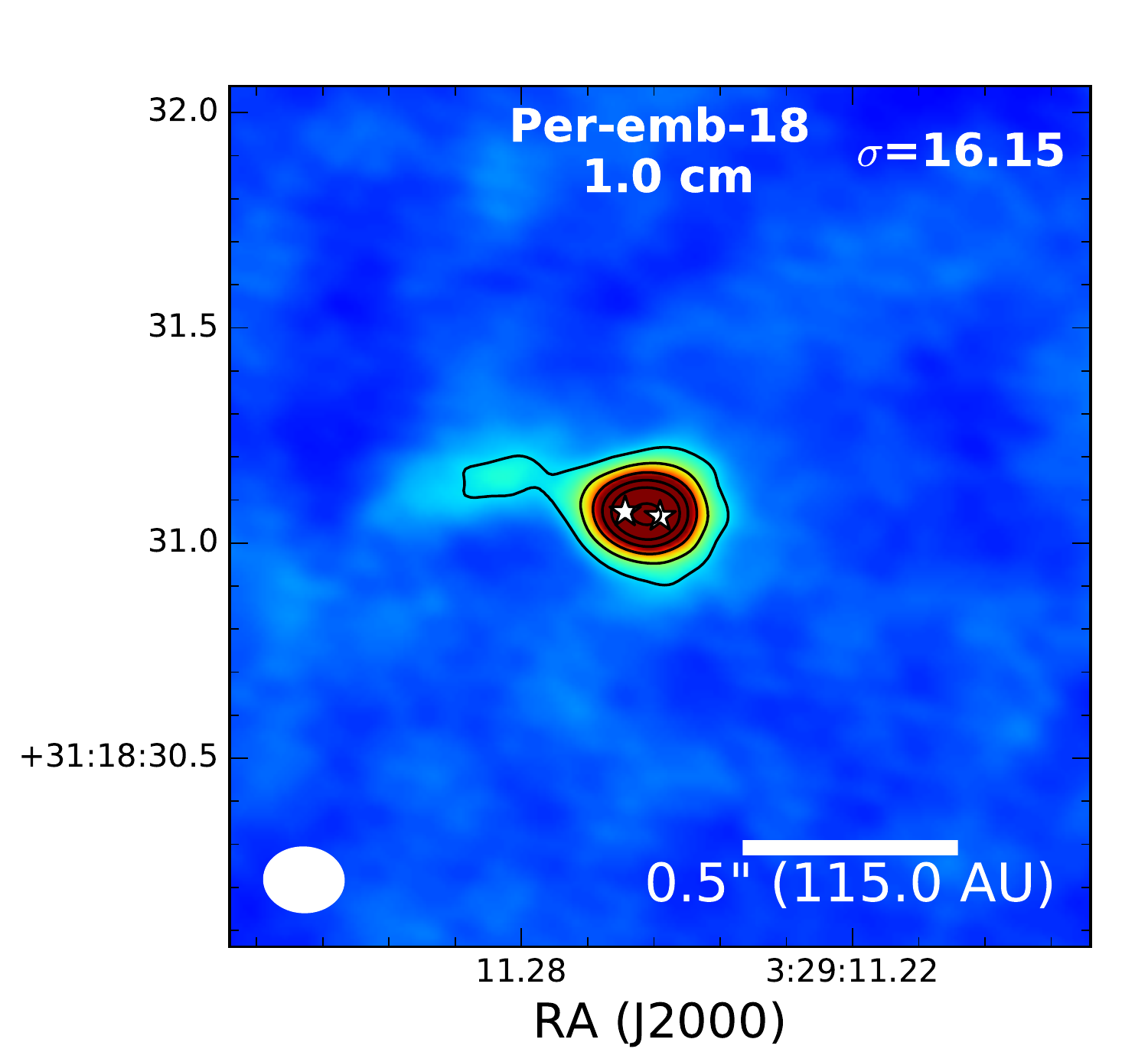}
  \includegraphics[width=0.24\linewidth]{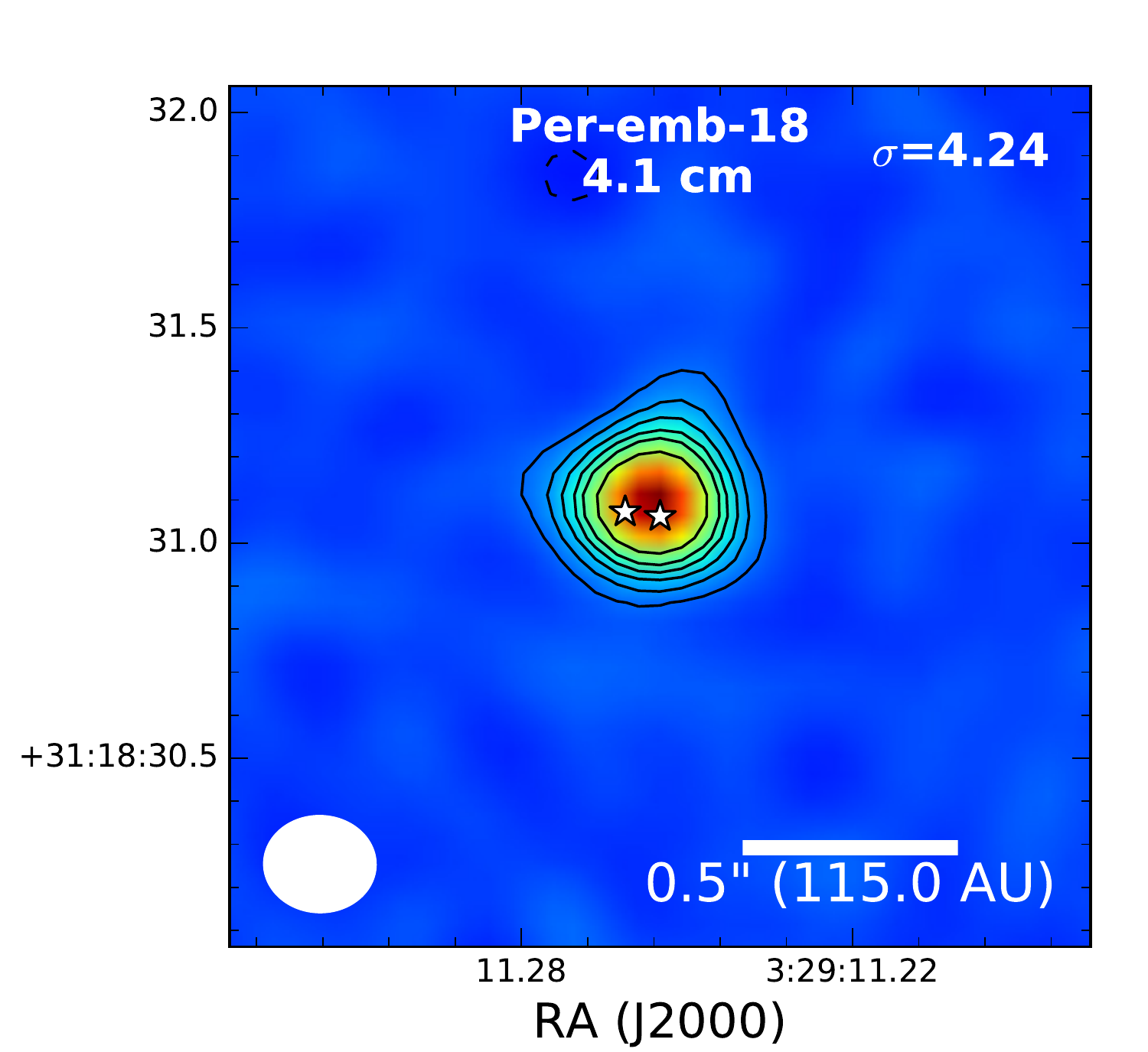}
  \includegraphics[width=0.24\linewidth]{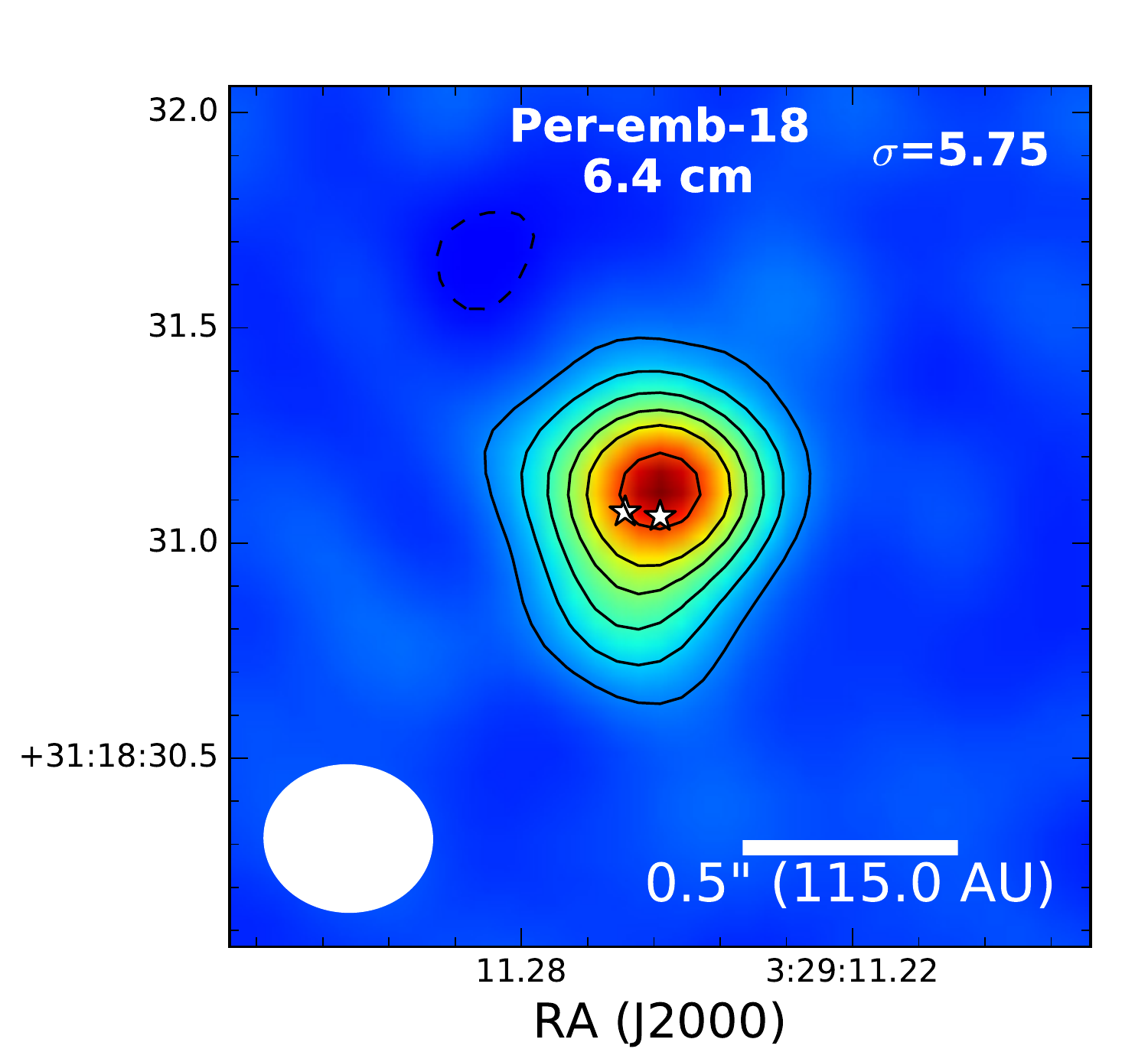}

  \includegraphics[width=0.24\linewidth]{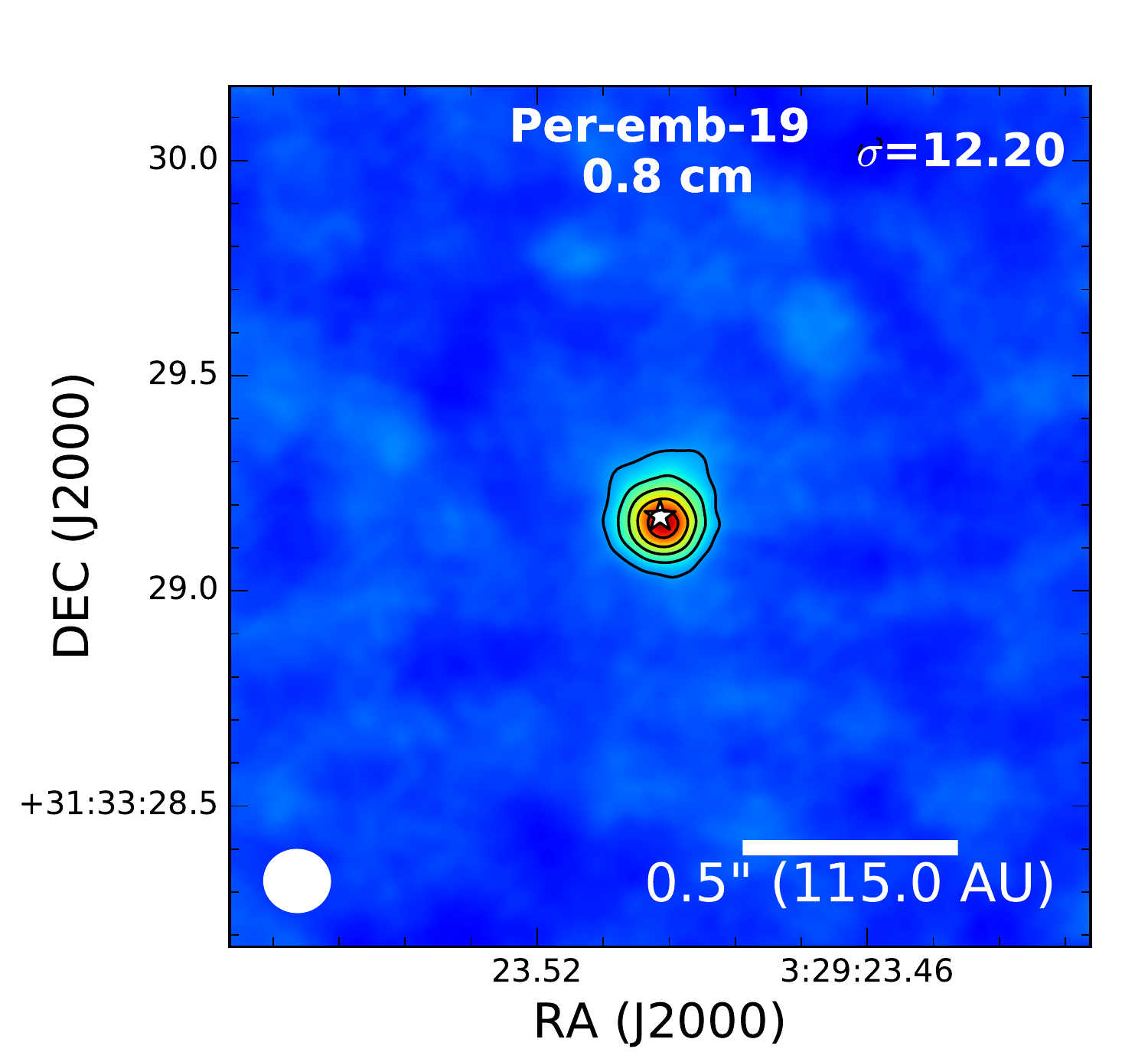}
  \includegraphics[width=0.24\linewidth]{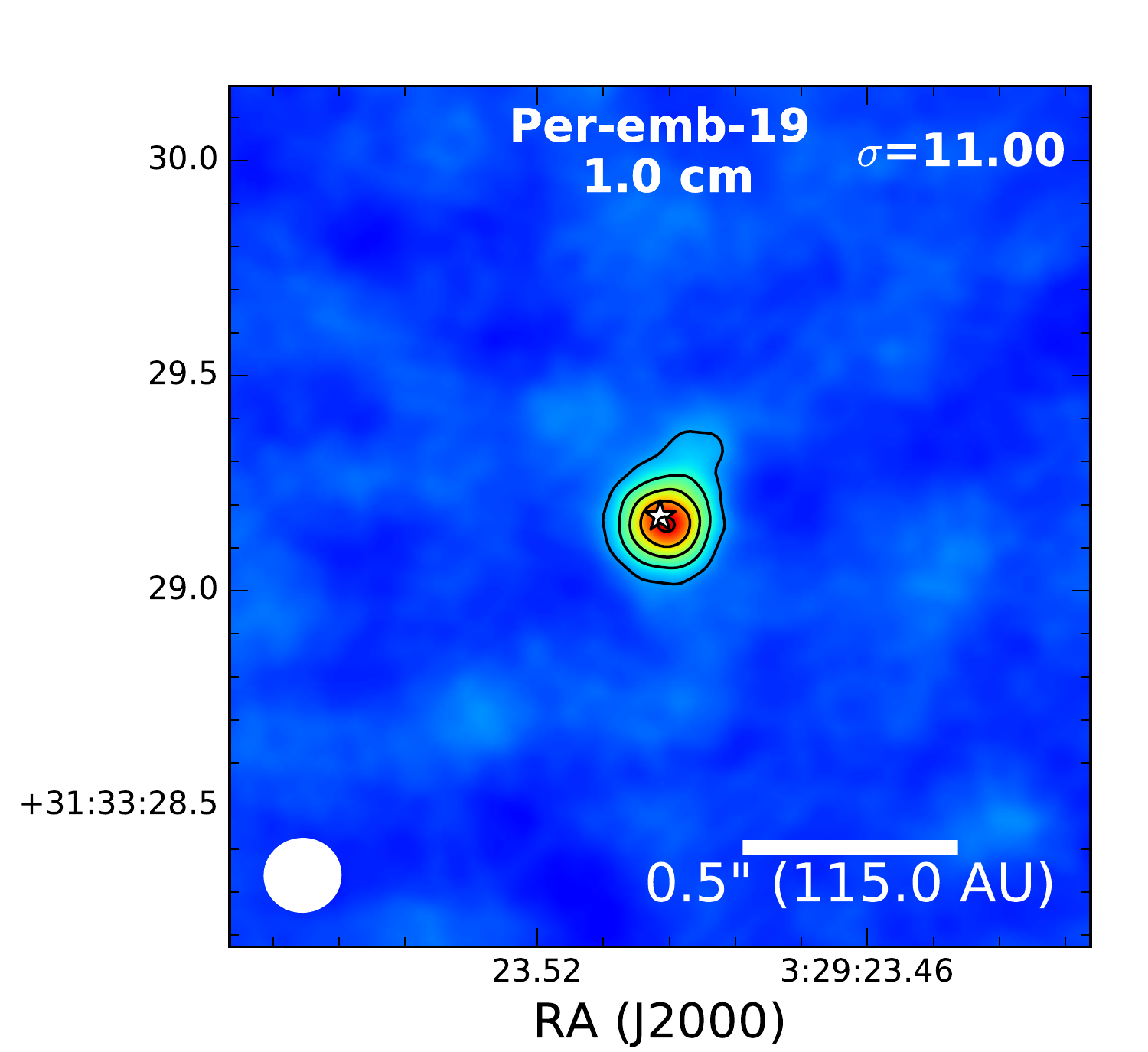}
  \includegraphics[width=0.24\linewidth]{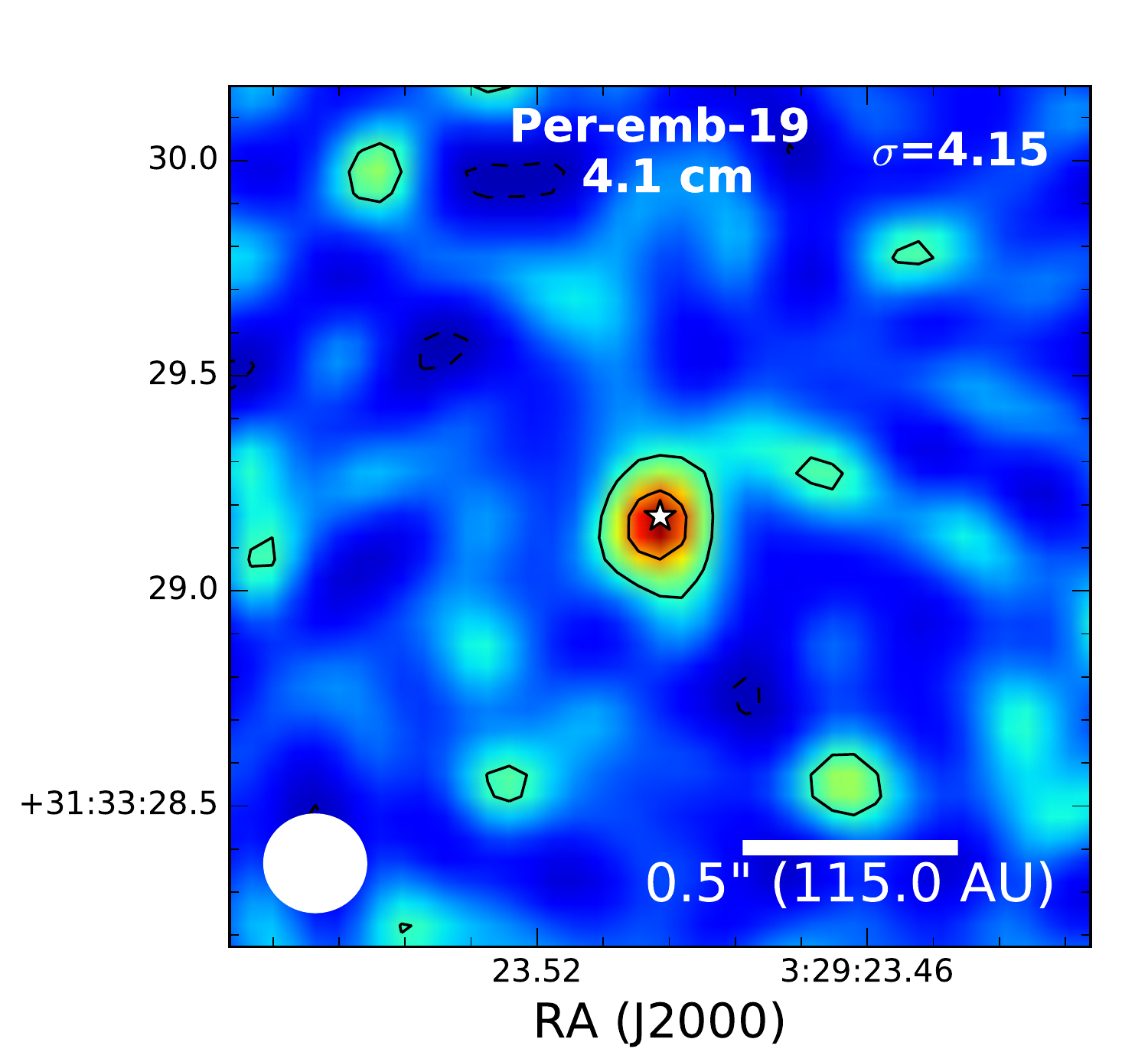}
  \includegraphics[width=0.24\linewidth]{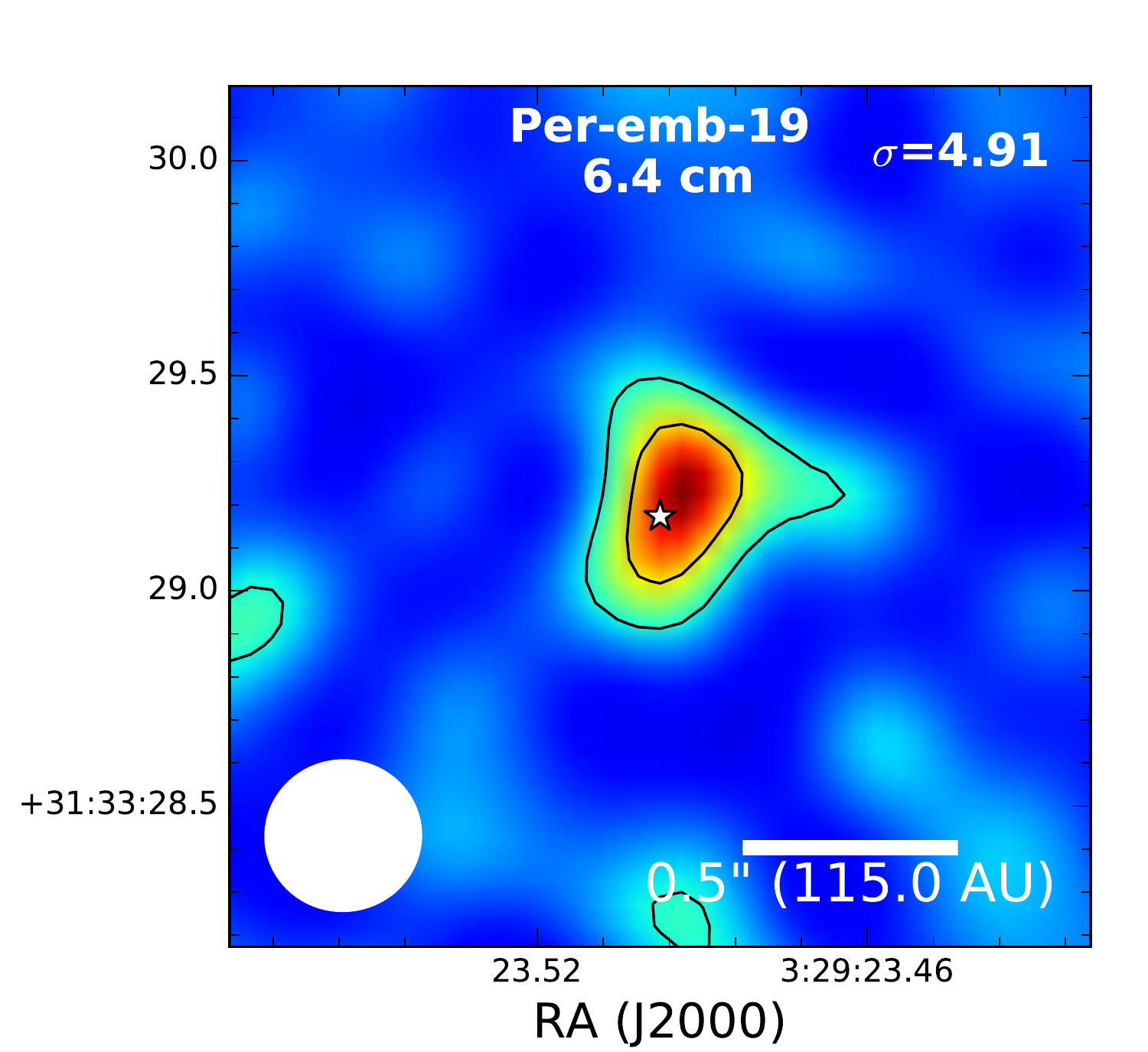}

  \includegraphics[width=0.24\linewidth]{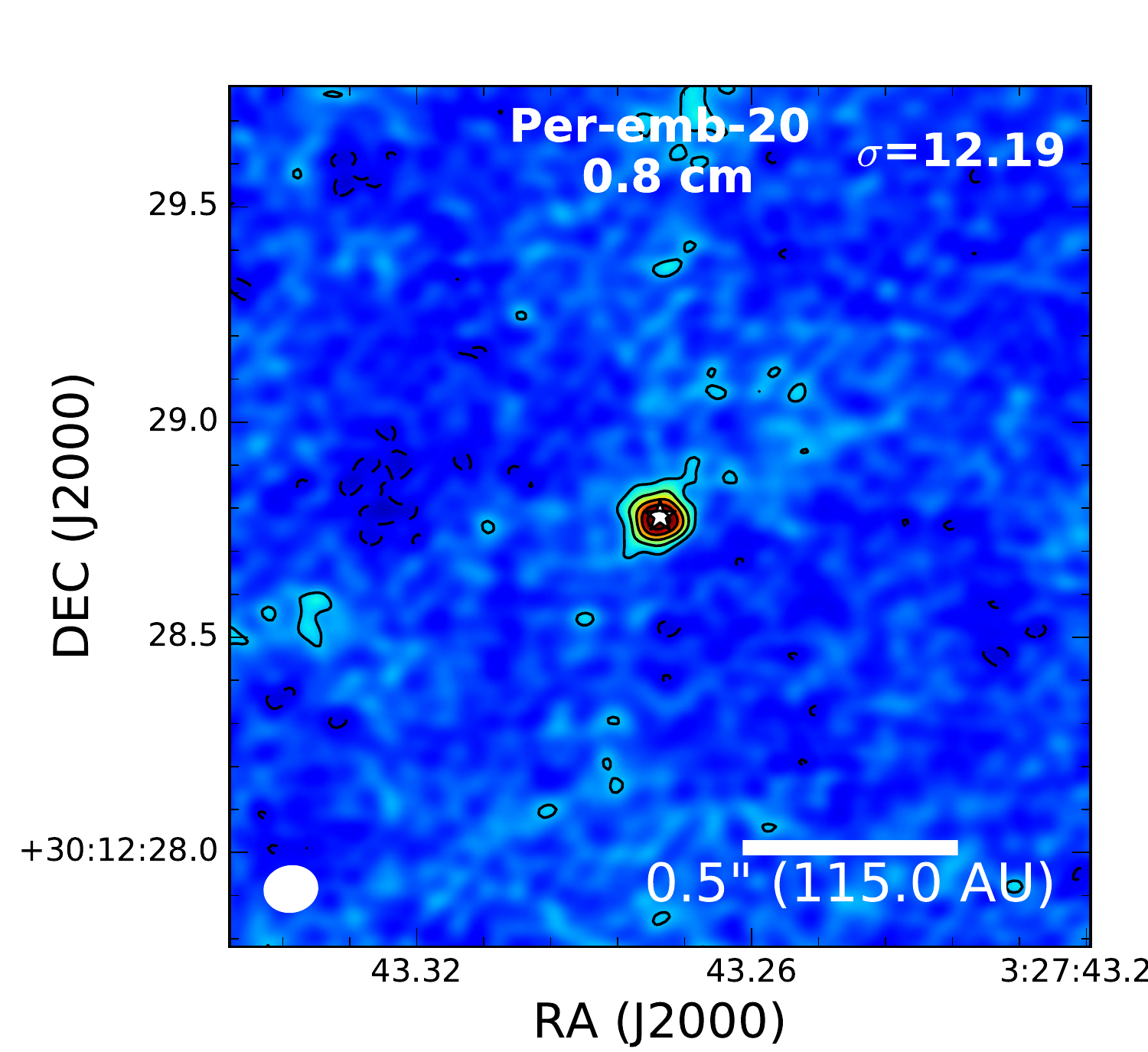}
  \includegraphics[width=0.24\linewidth]{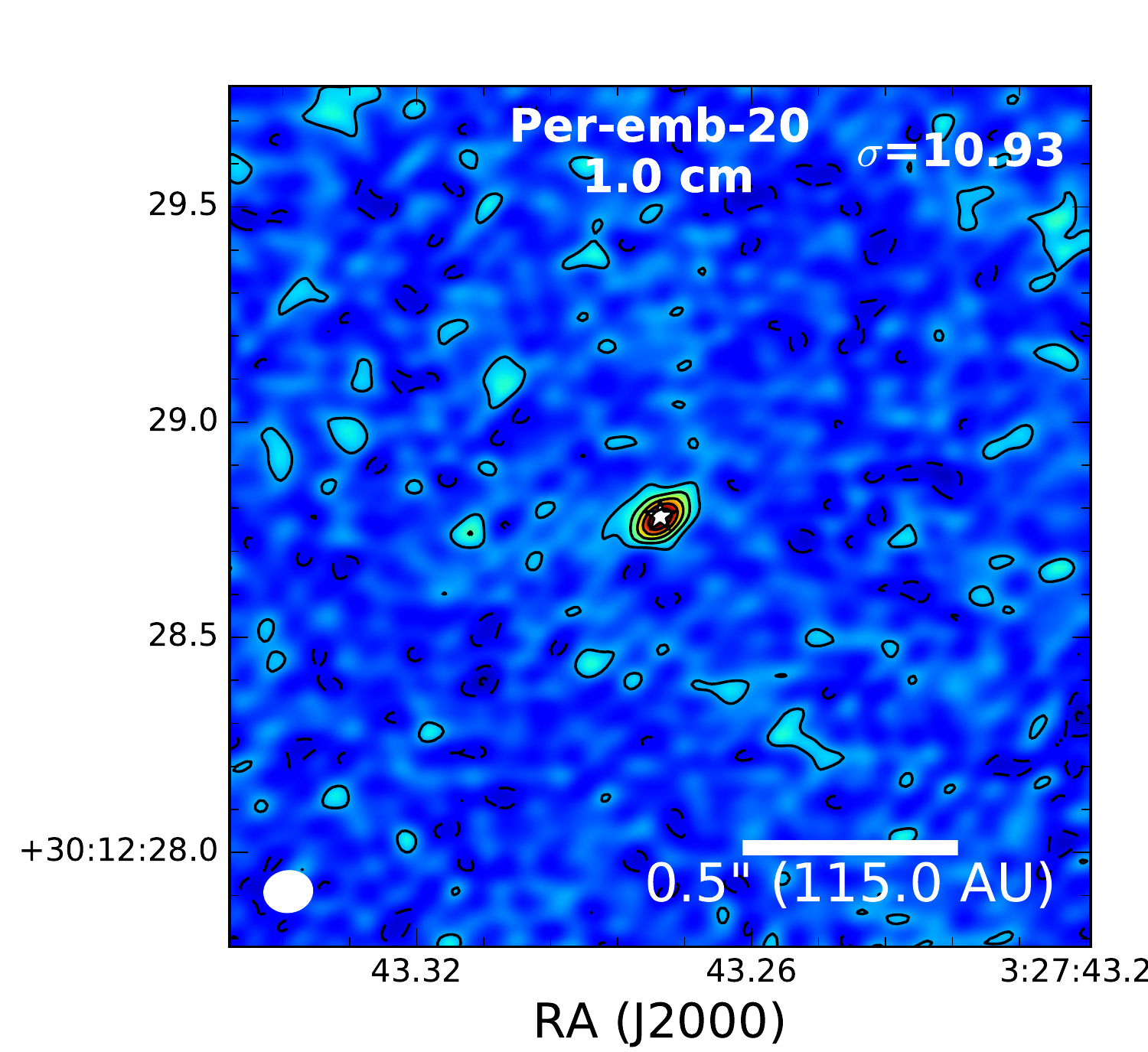}
  \includegraphics[width=0.24\linewidth]{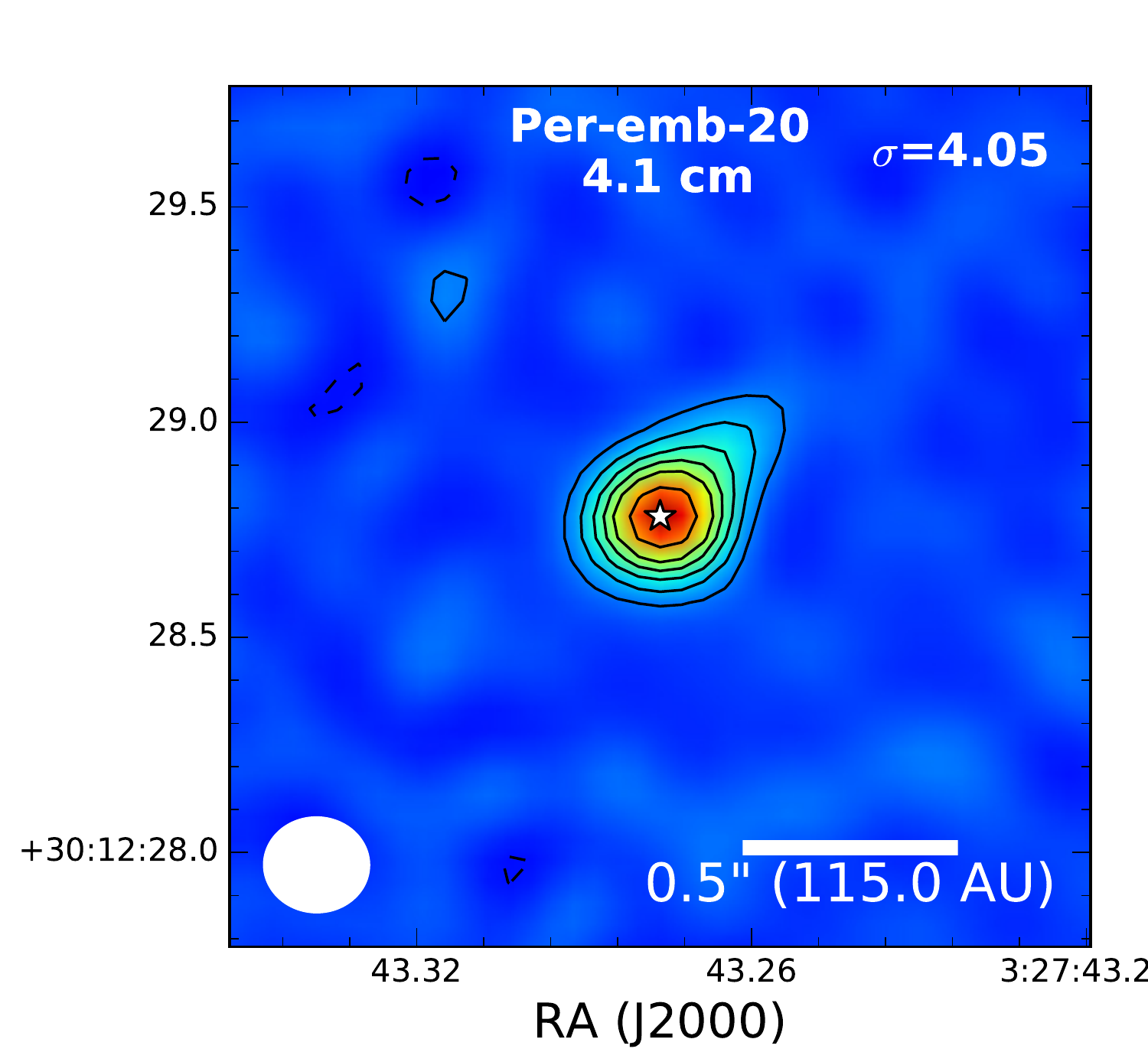}
  \includegraphics[width=0.24\linewidth]{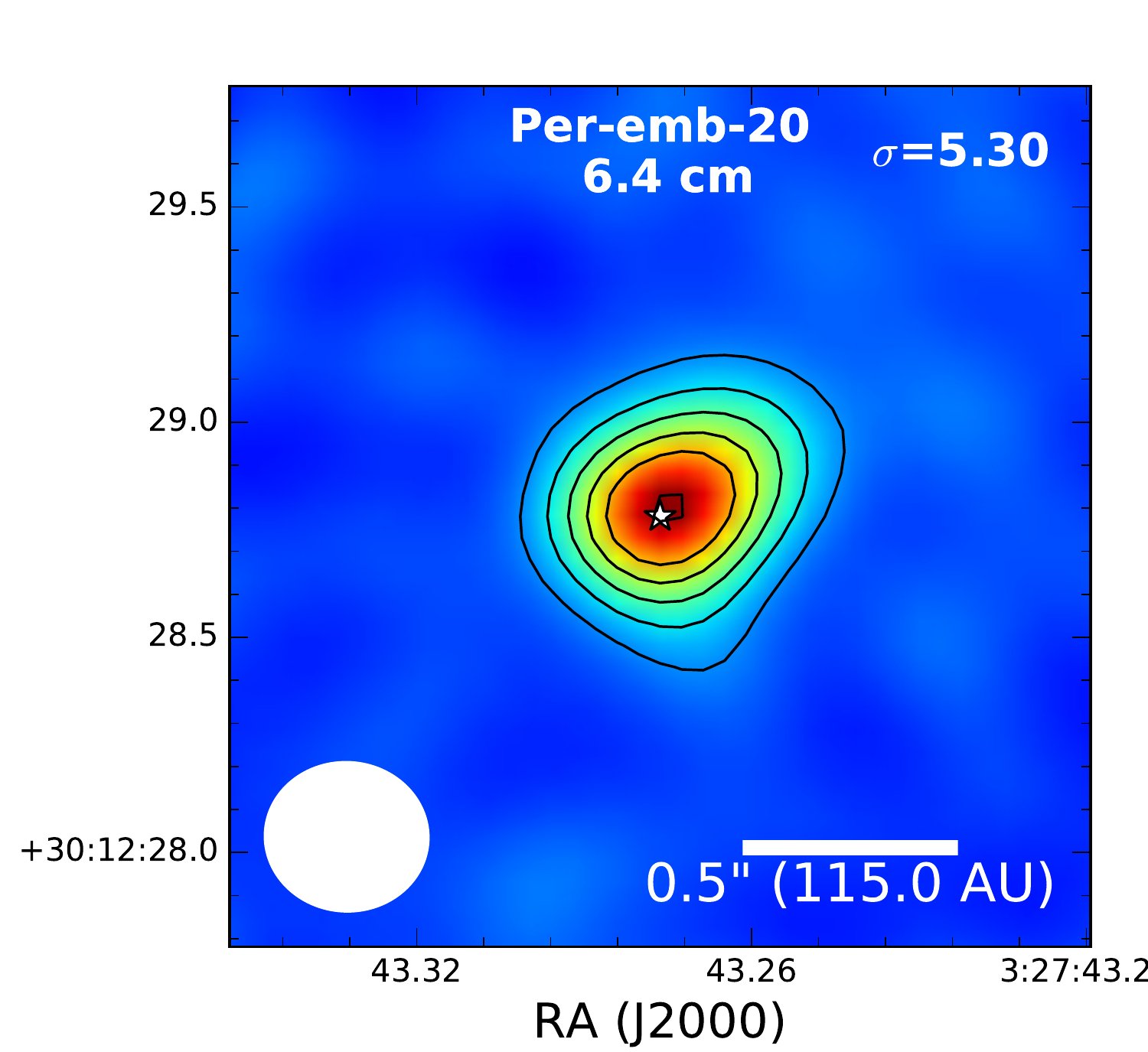}

  \includegraphics[width=0.24\linewidth]{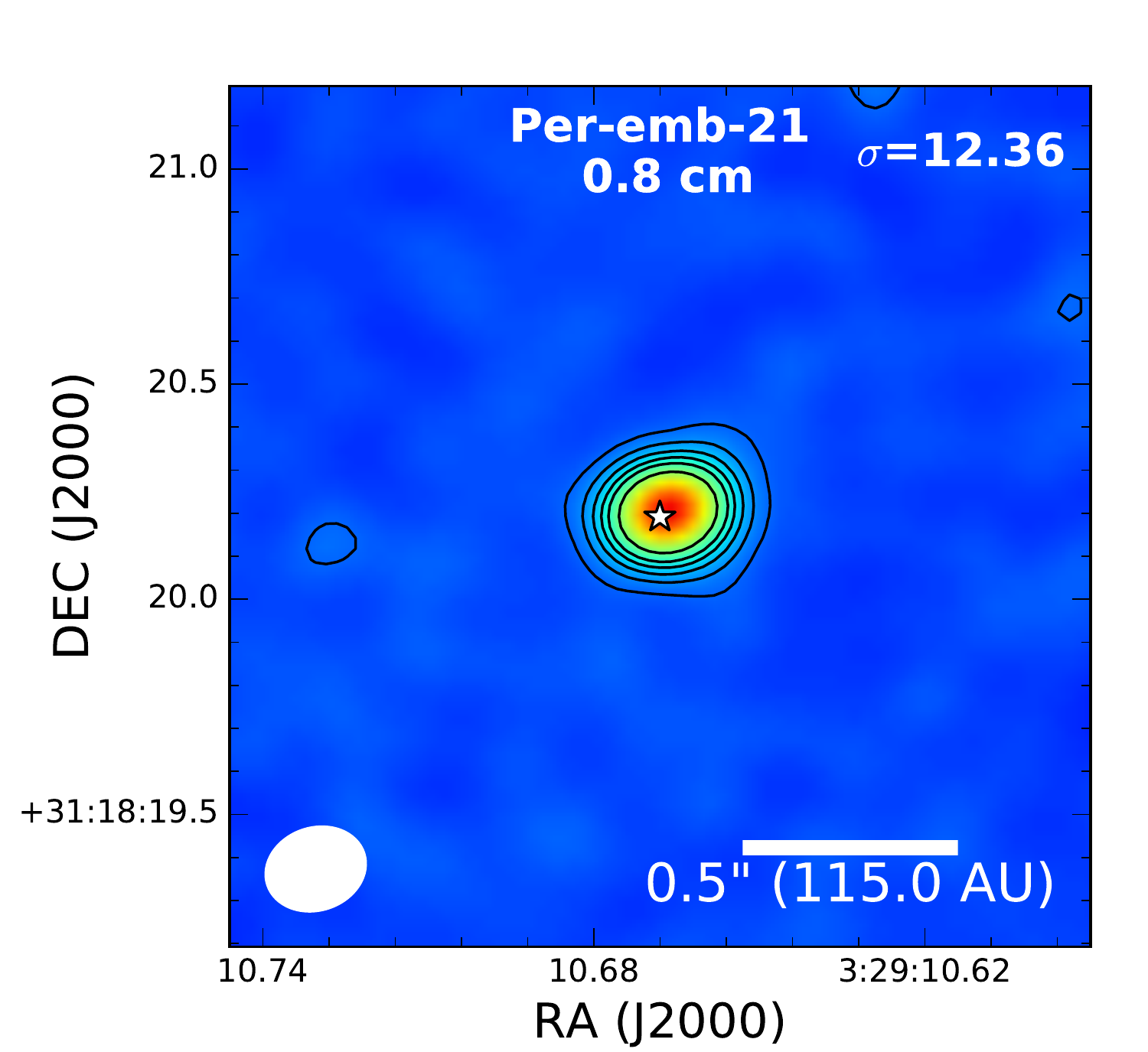}
  \includegraphics[width=0.24\linewidth]{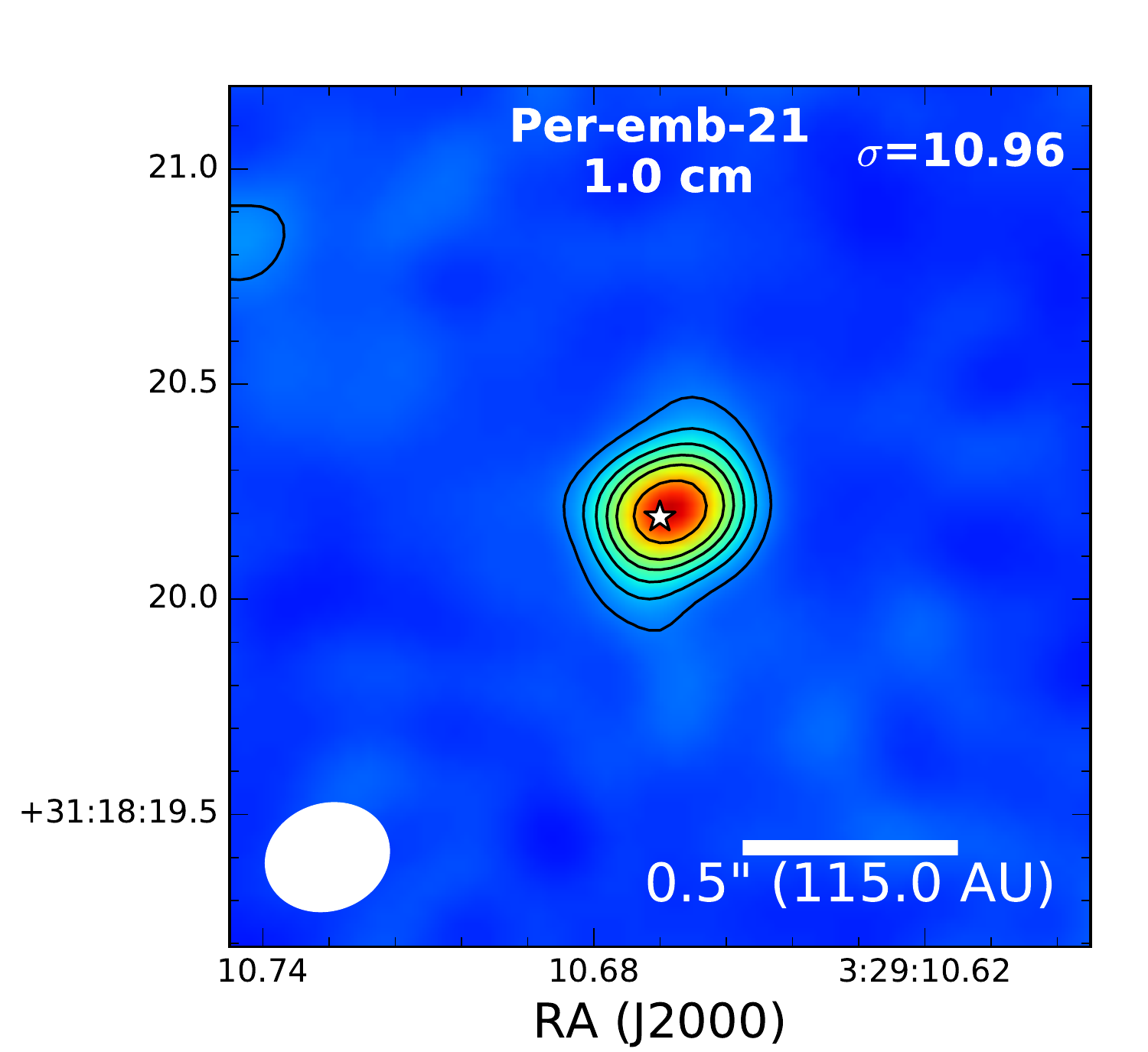}
  \includegraphics[width=0.24\linewidth]{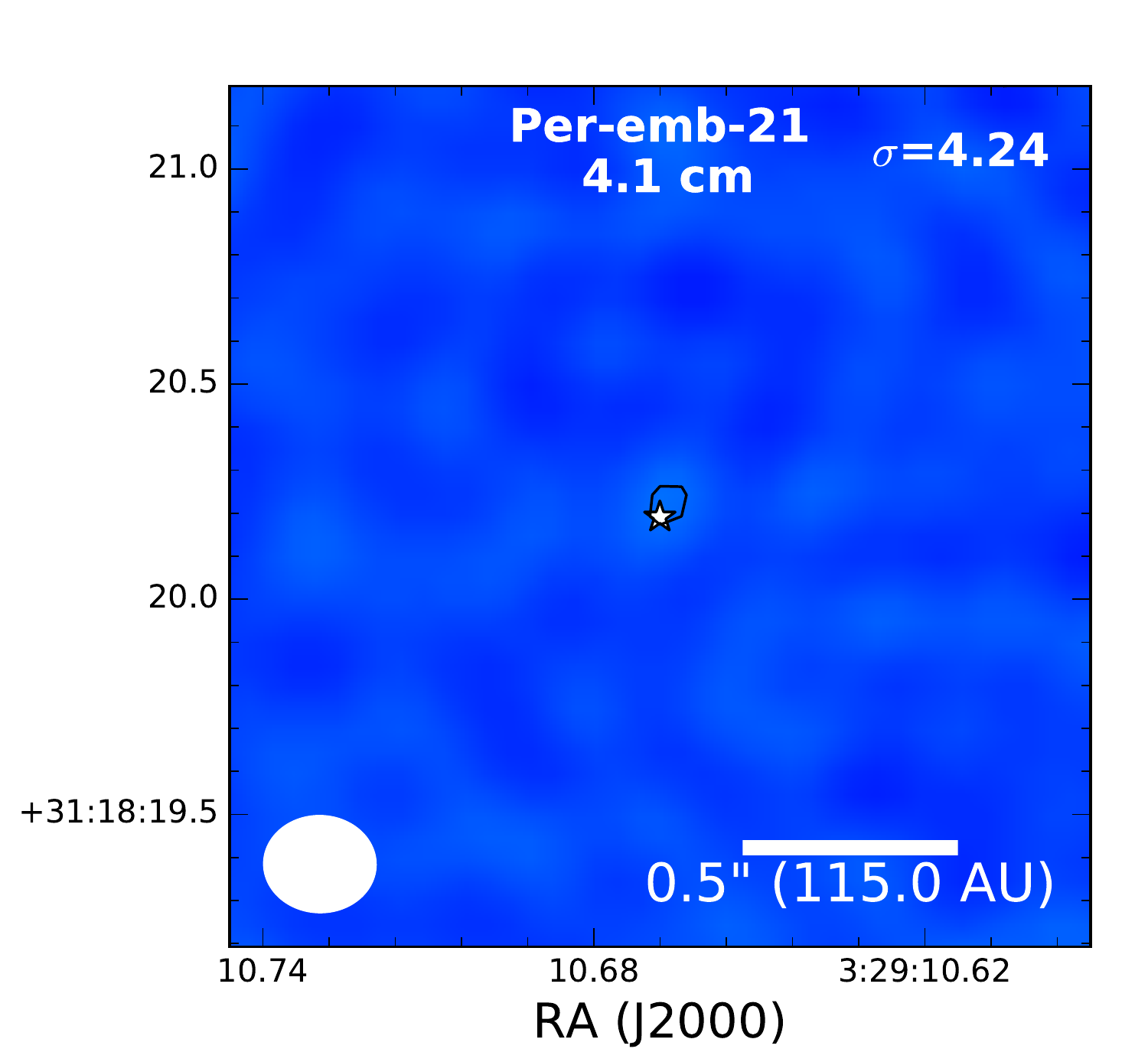}
  \includegraphics[width=0.24\linewidth]{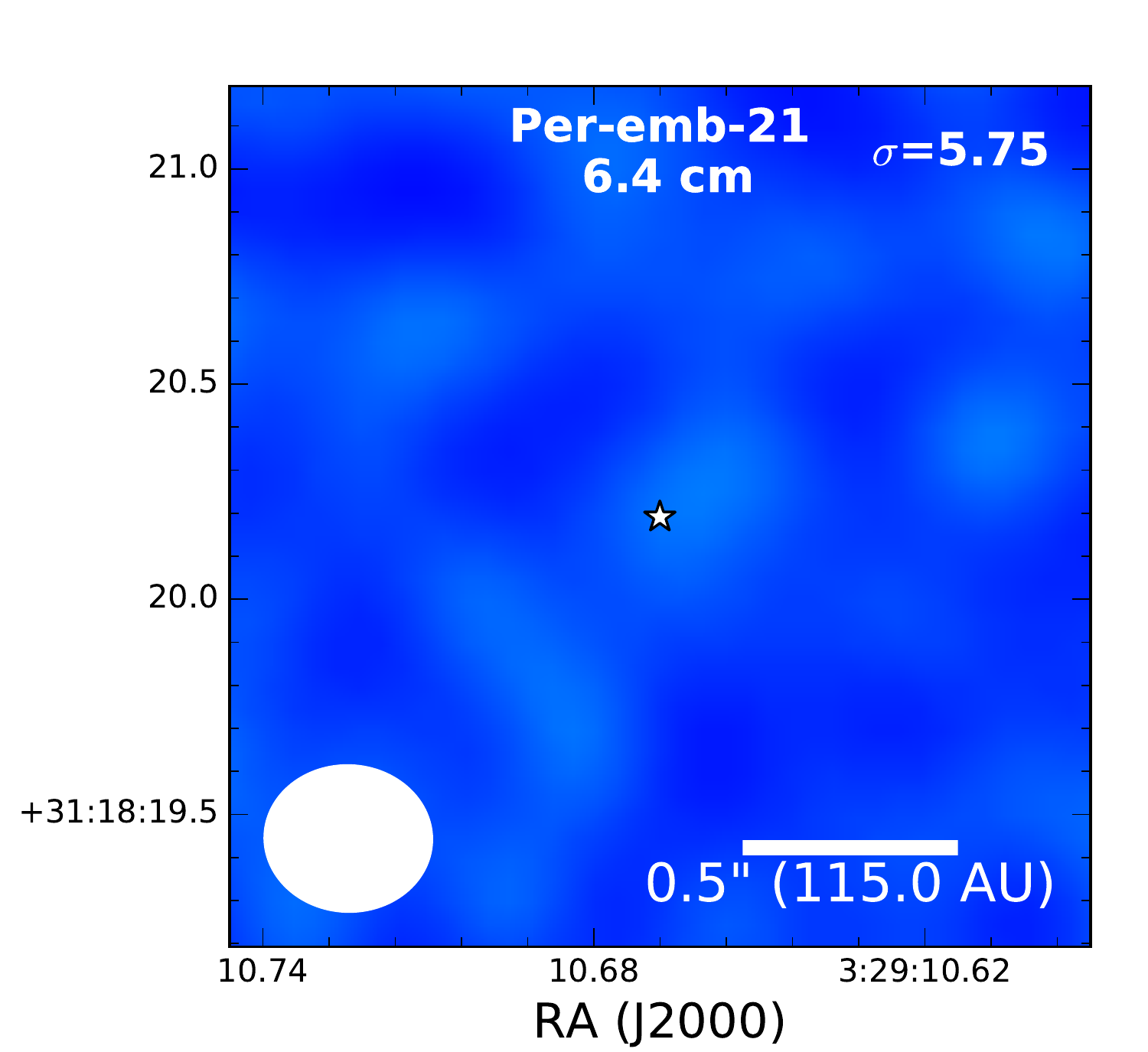}

\end{figure}

\begin{figure}

  \includegraphics[width=0.24\linewidth]{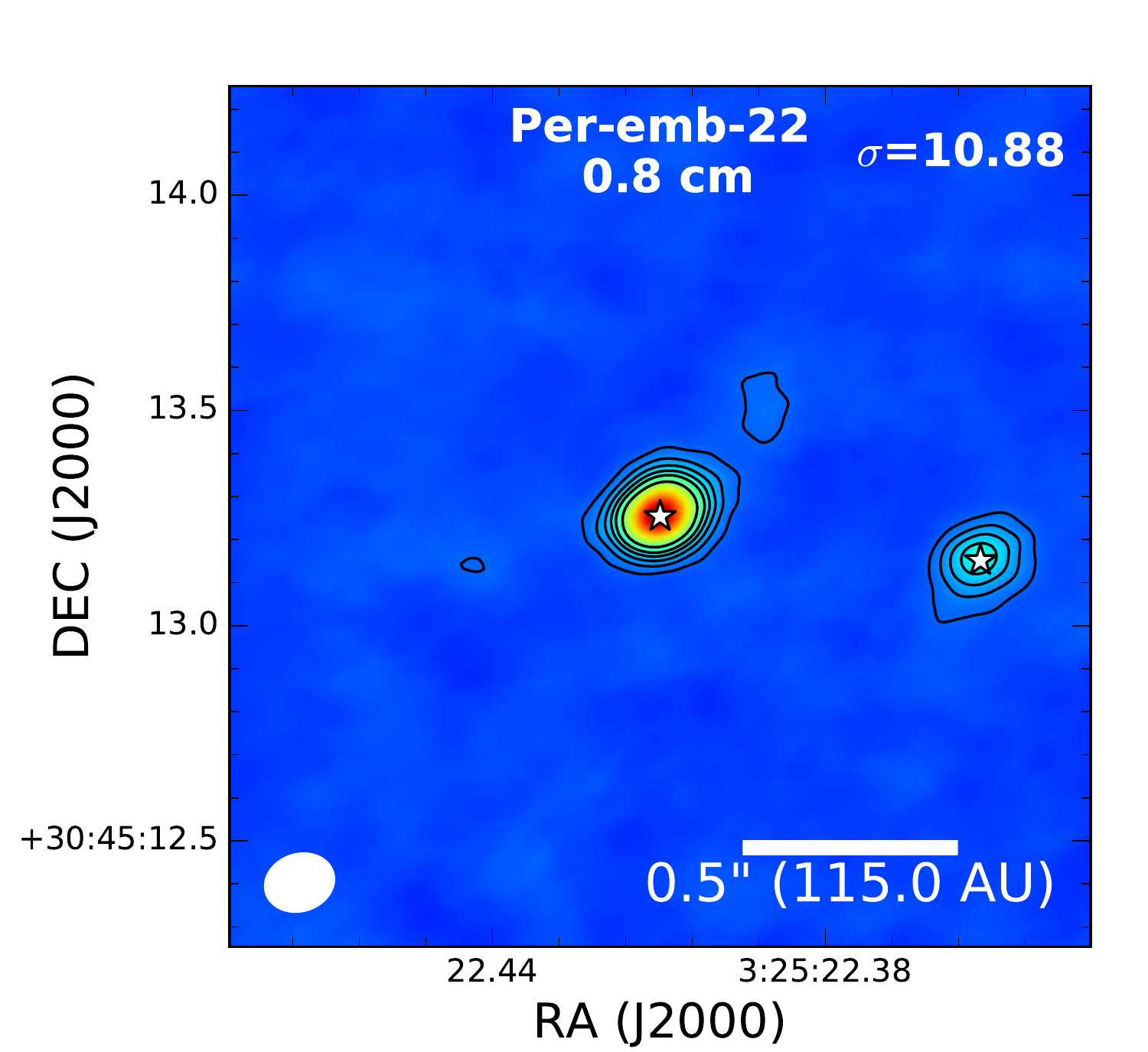}
  \includegraphics[width=0.24\linewidth]{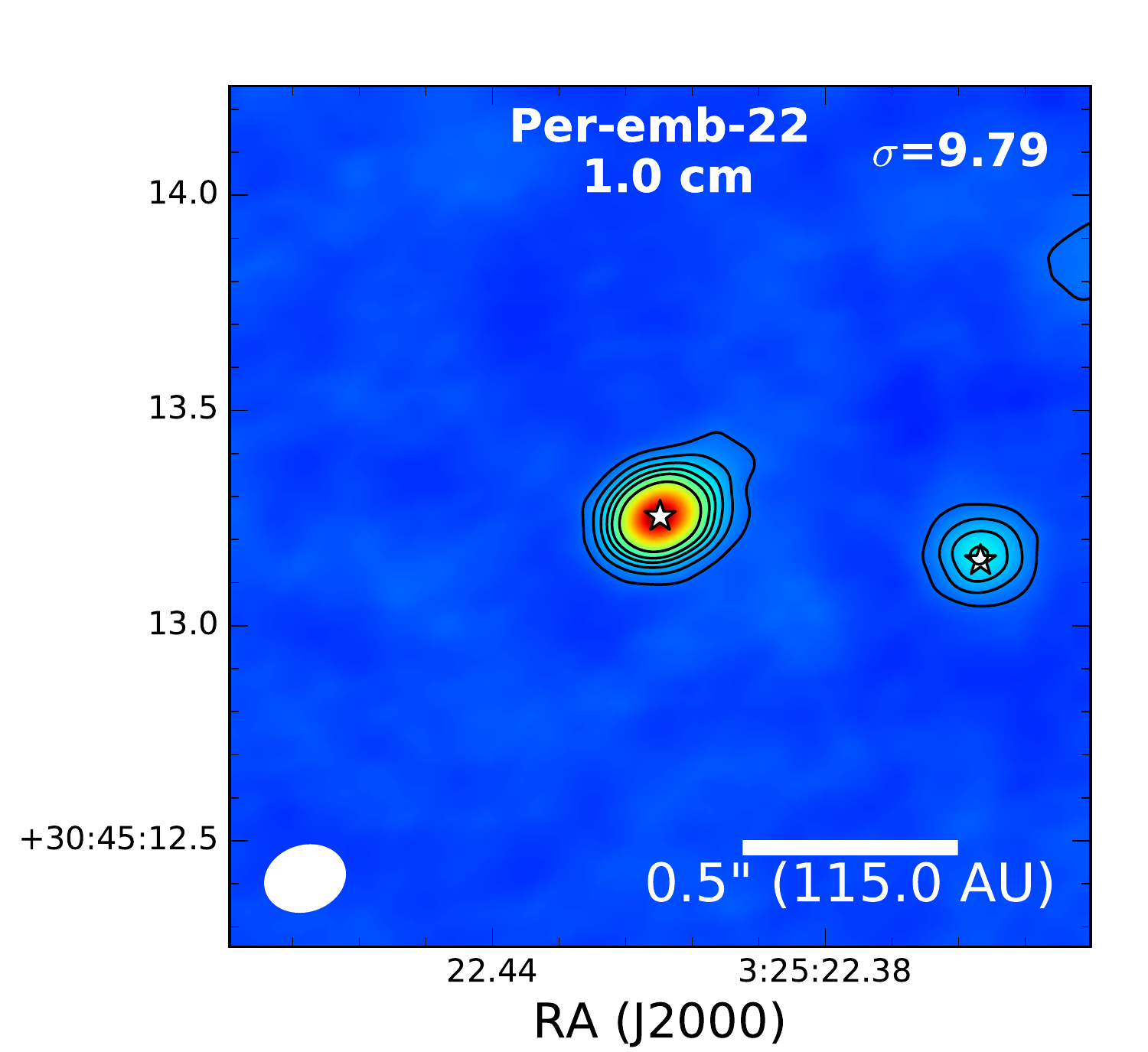}
  \includegraphics[width=0.24\linewidth]{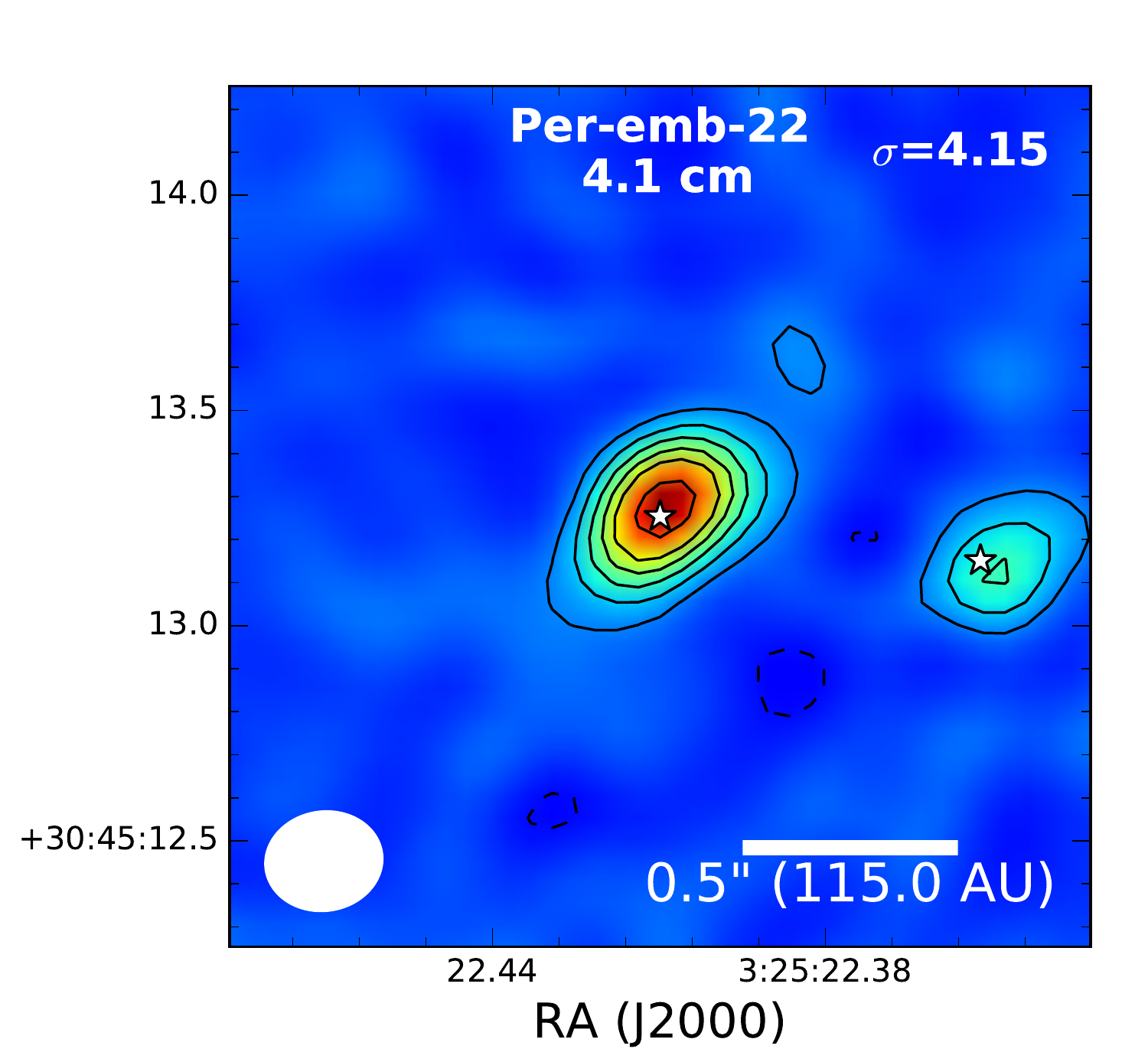}
  \includegraphics[width=0.24\linewidth]{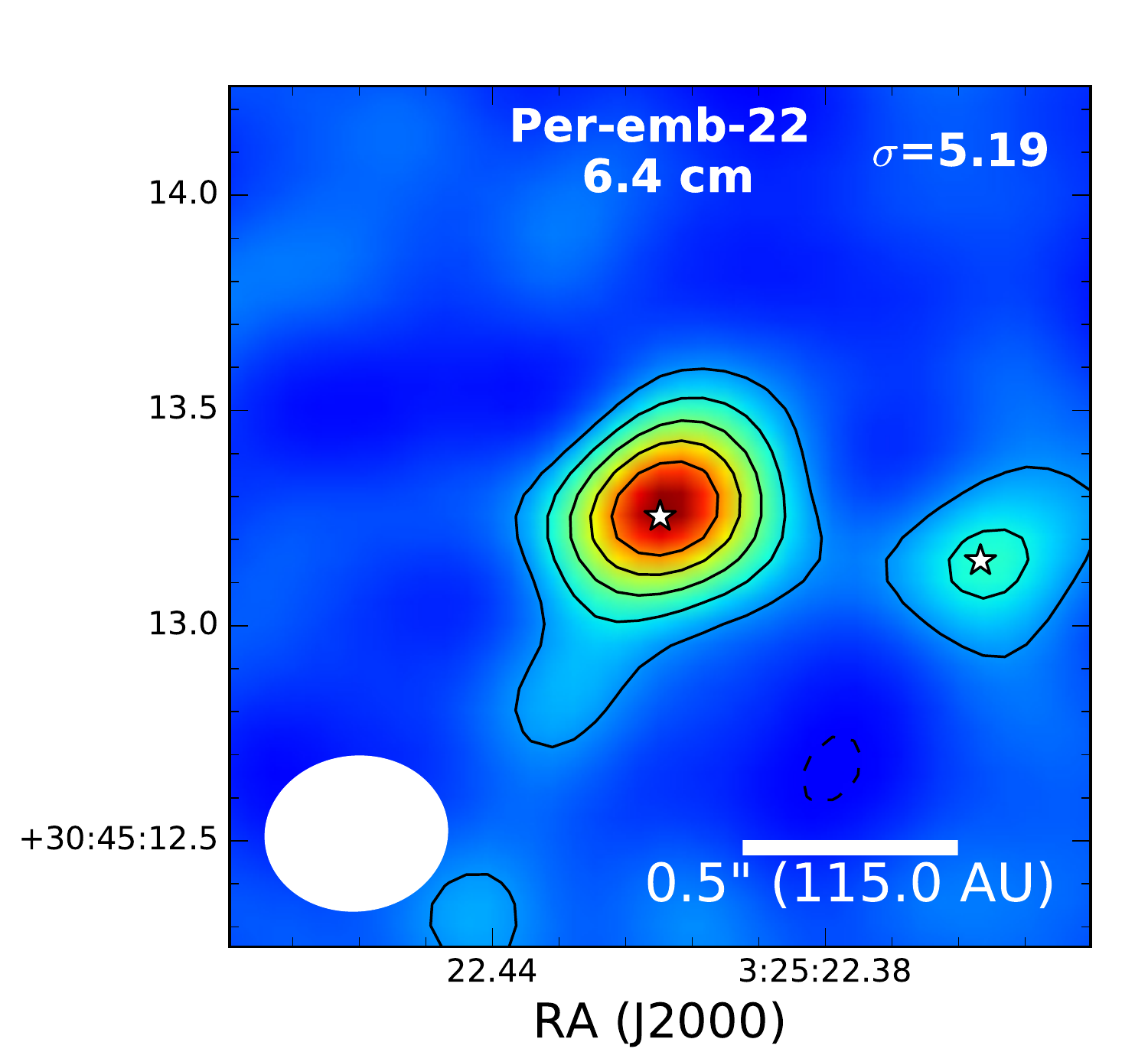}

  \includegraphics[width=0.24\linewidth]{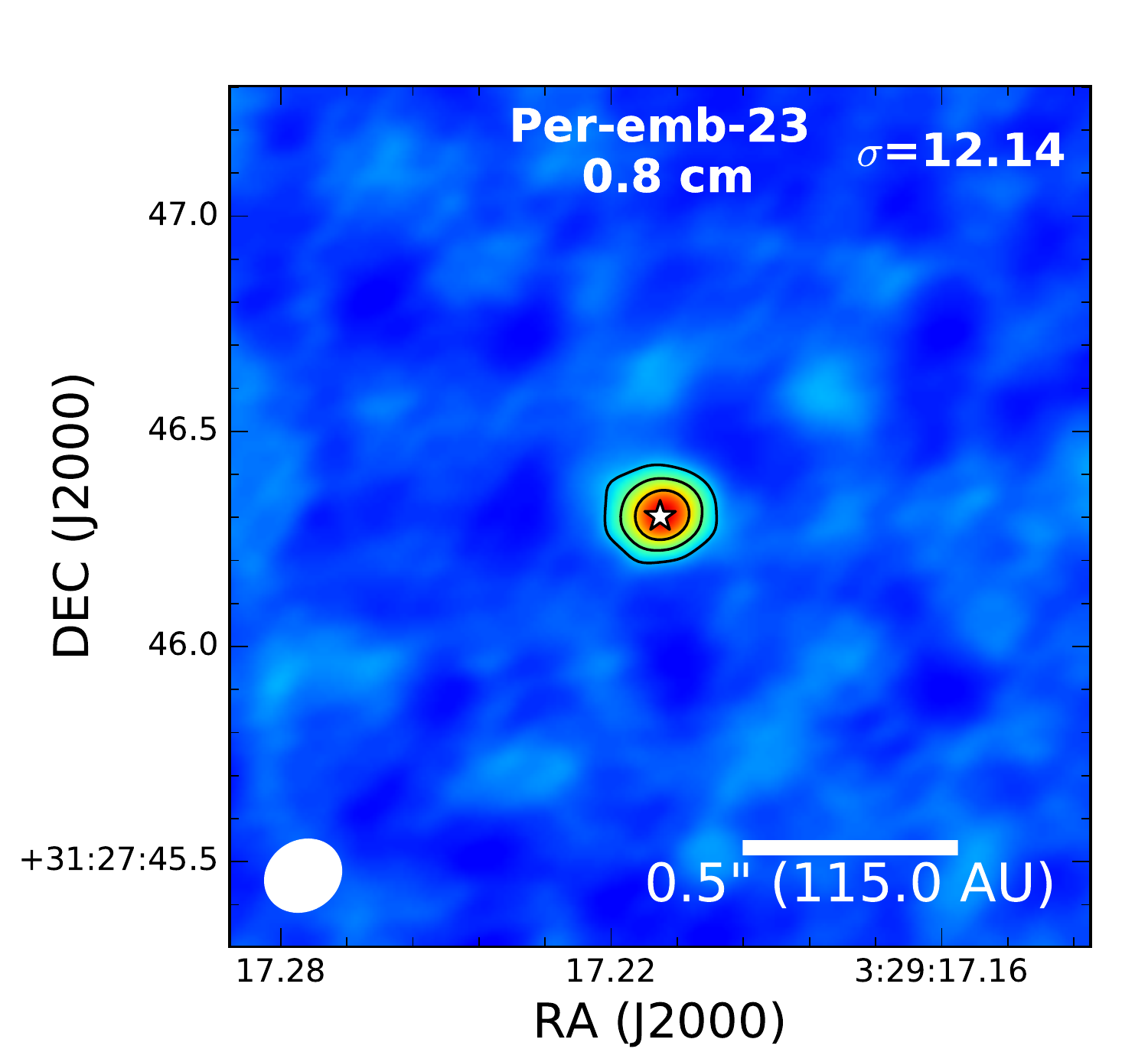}
  \includegraphics[width=0.24\linewidth]{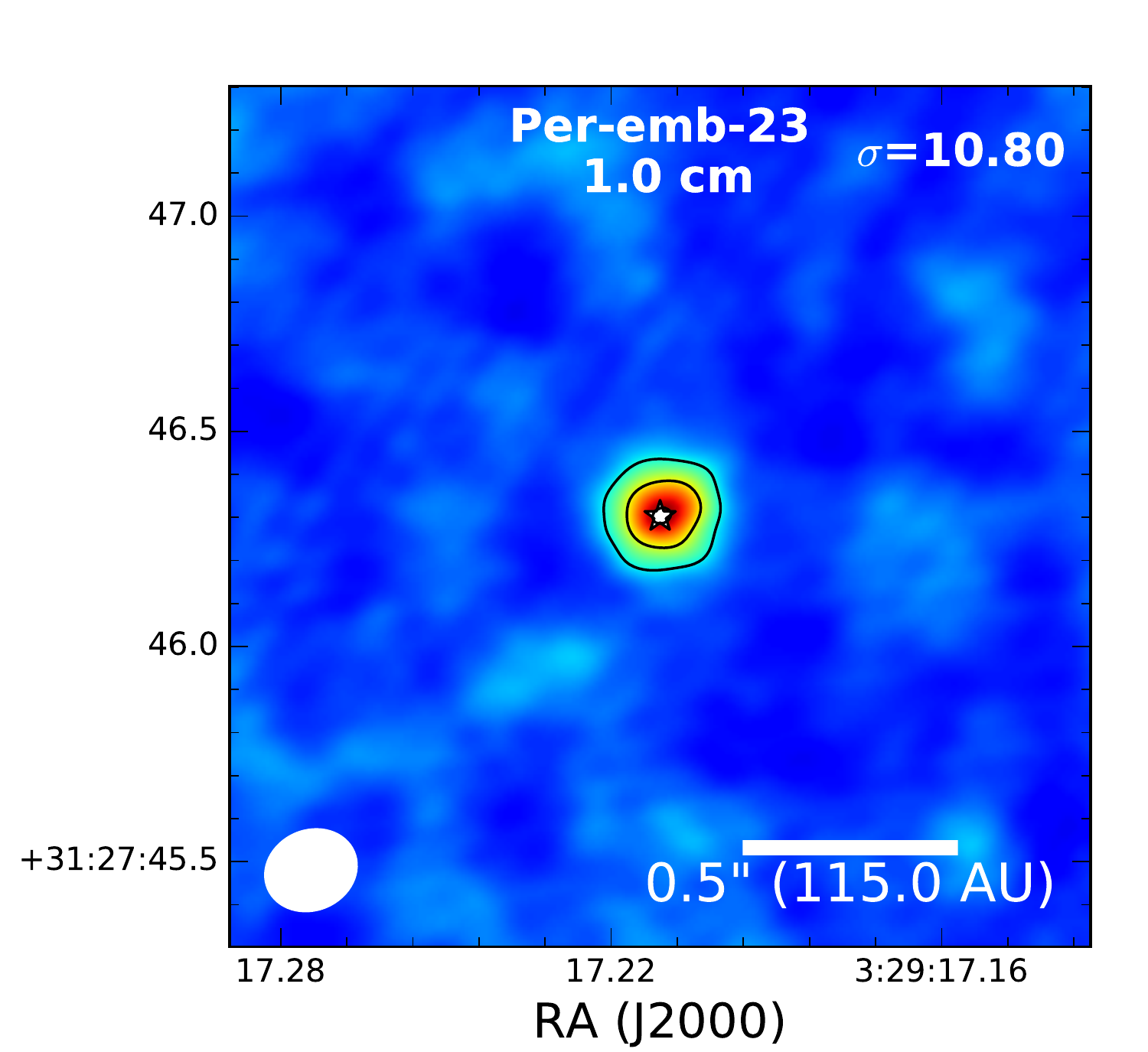}
  \includegraphics[width=0.24\linewidth]{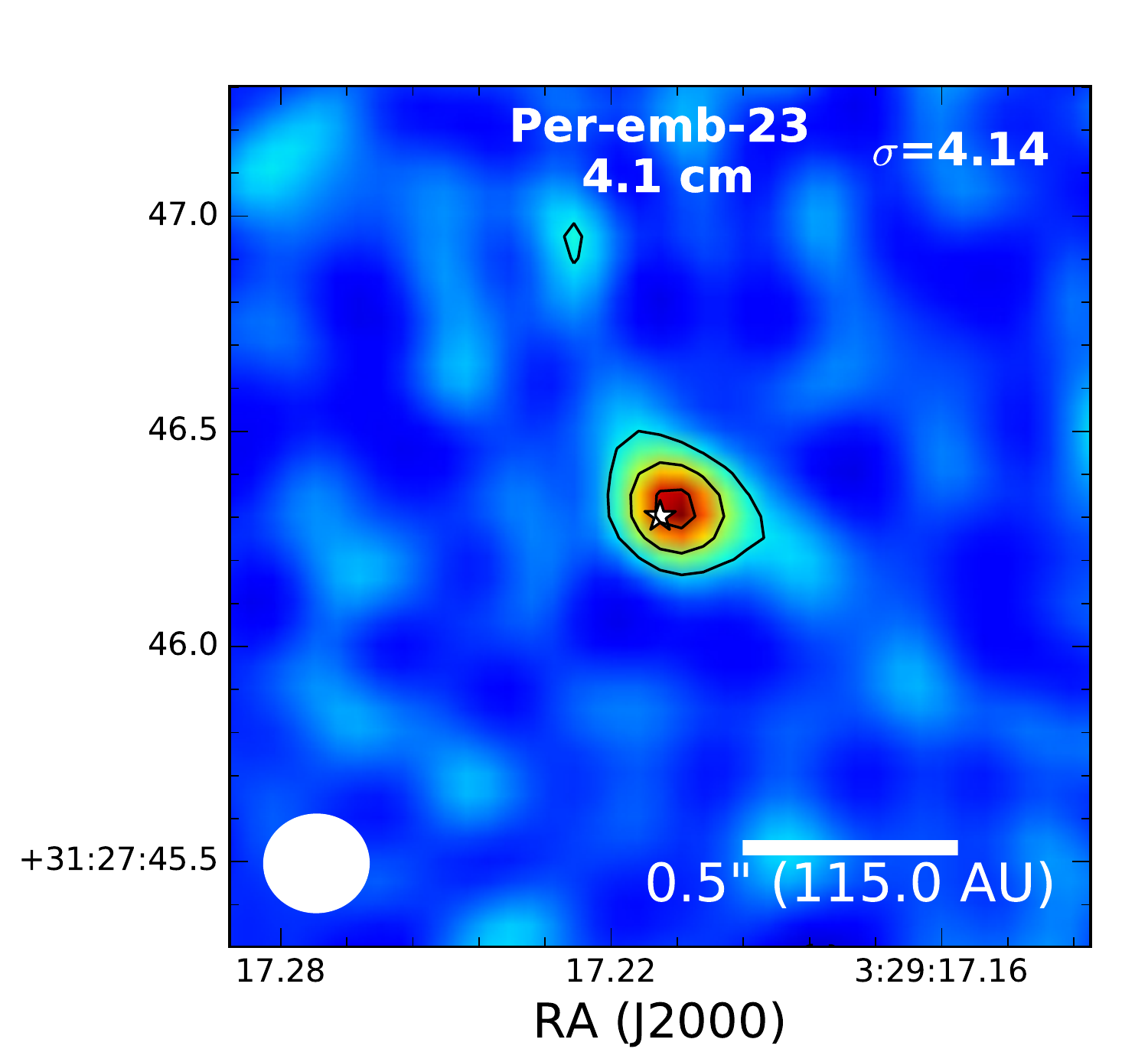}
  \includegraphics[width=0.24\linewidth]{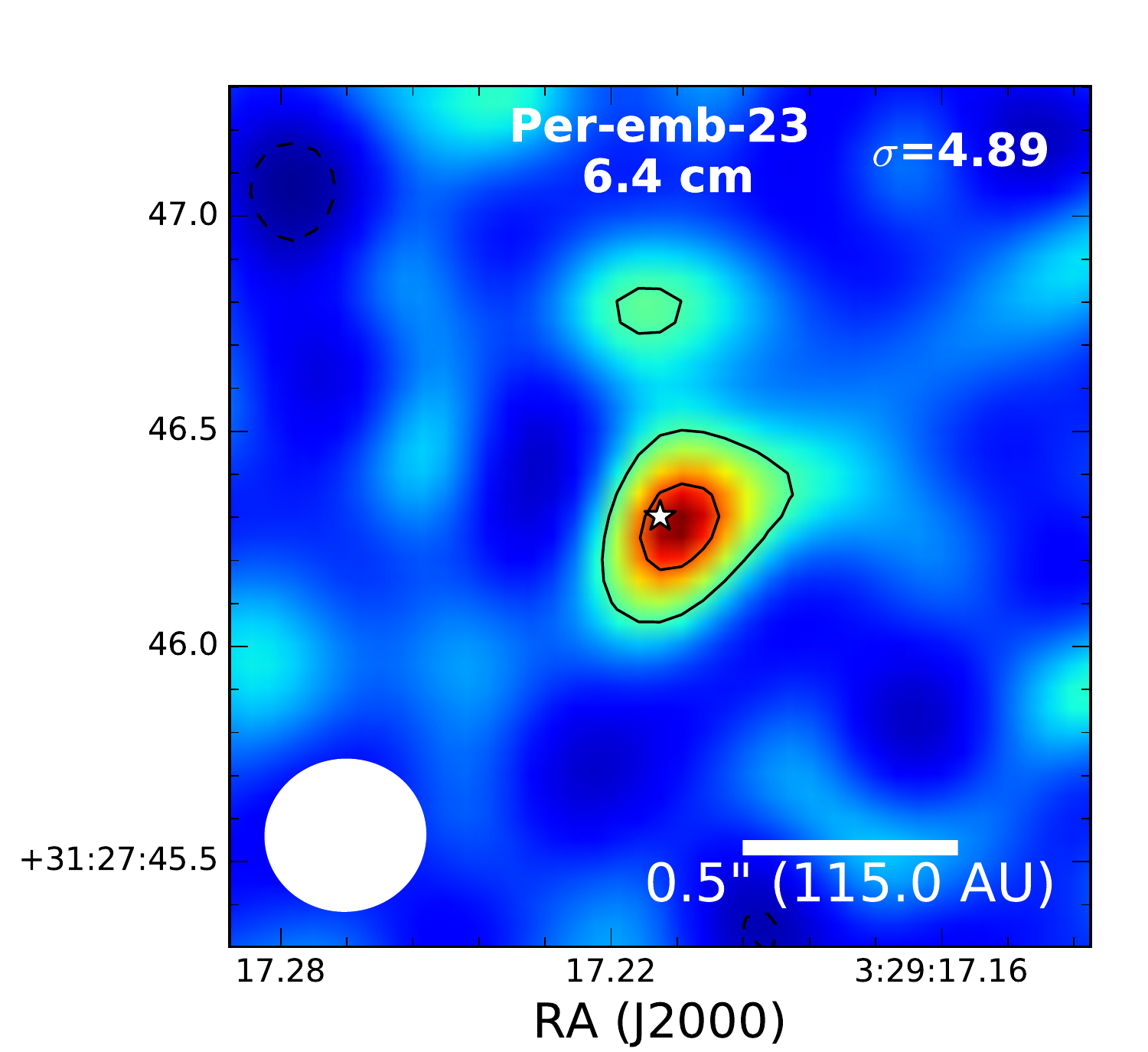}

  \includegraphics[width=0.24\linewidth]{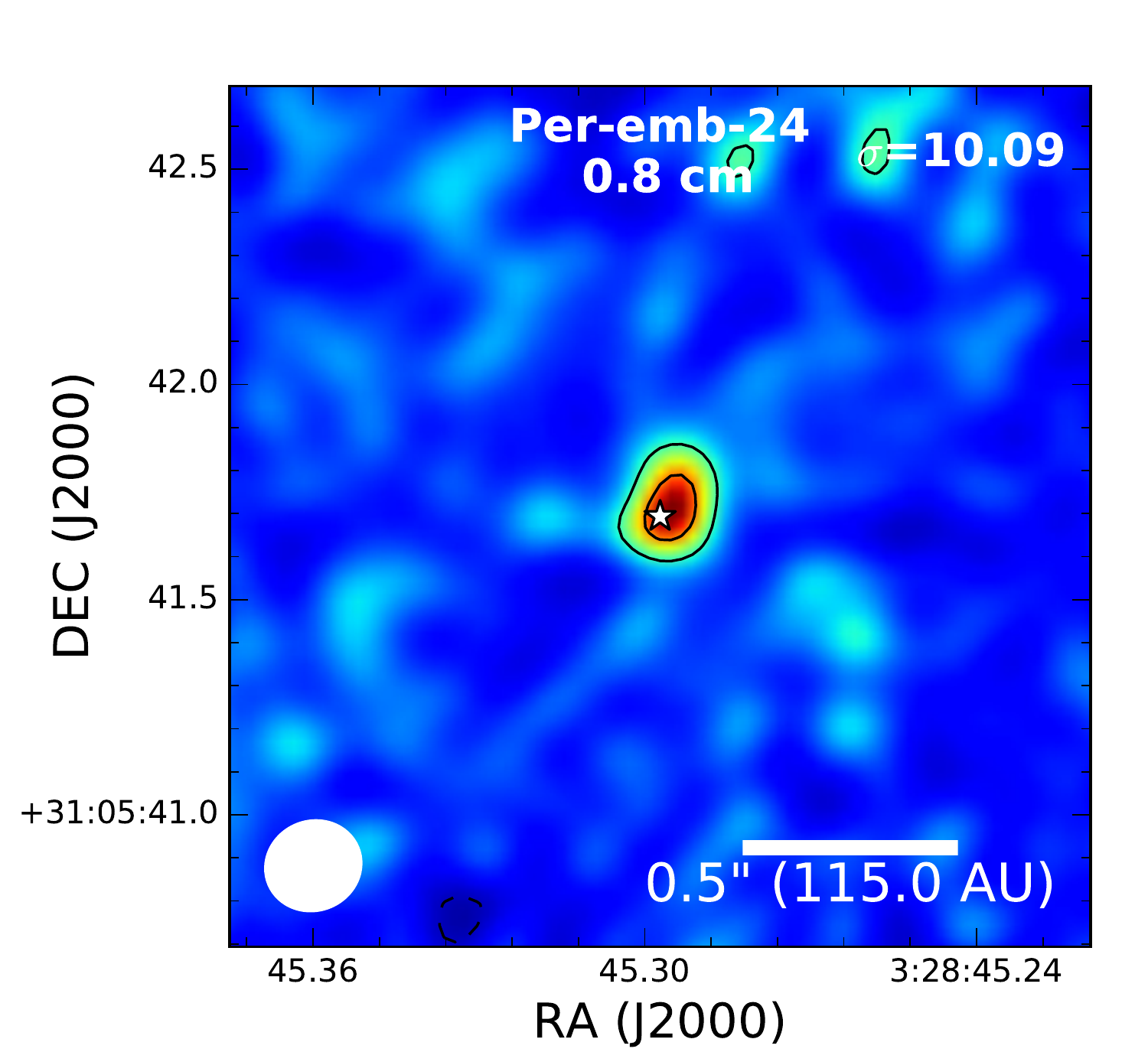}
  \includegraphics[width=0.24\linewidth]{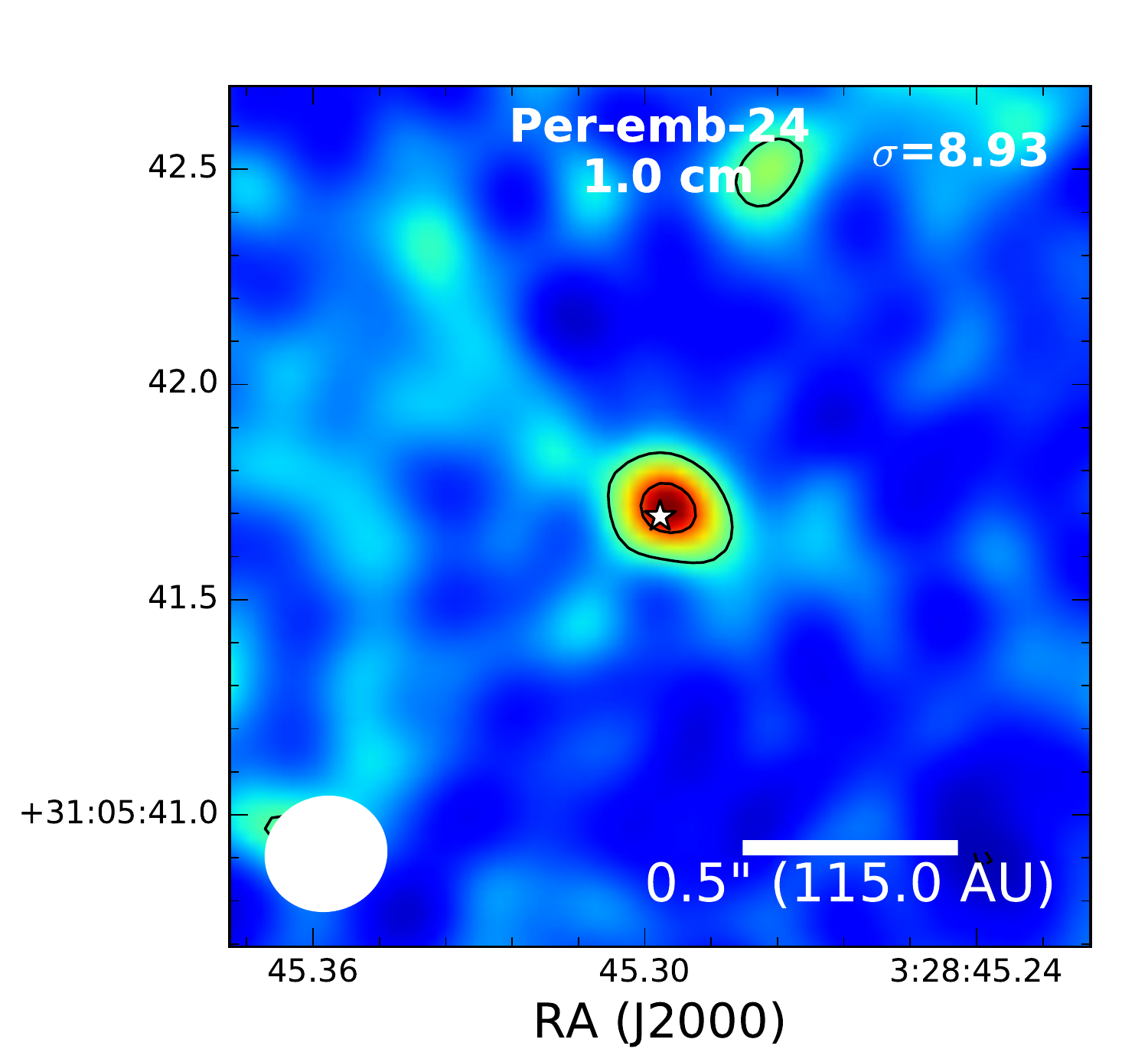}
  \includegraphics[width=0.24\linewidth]{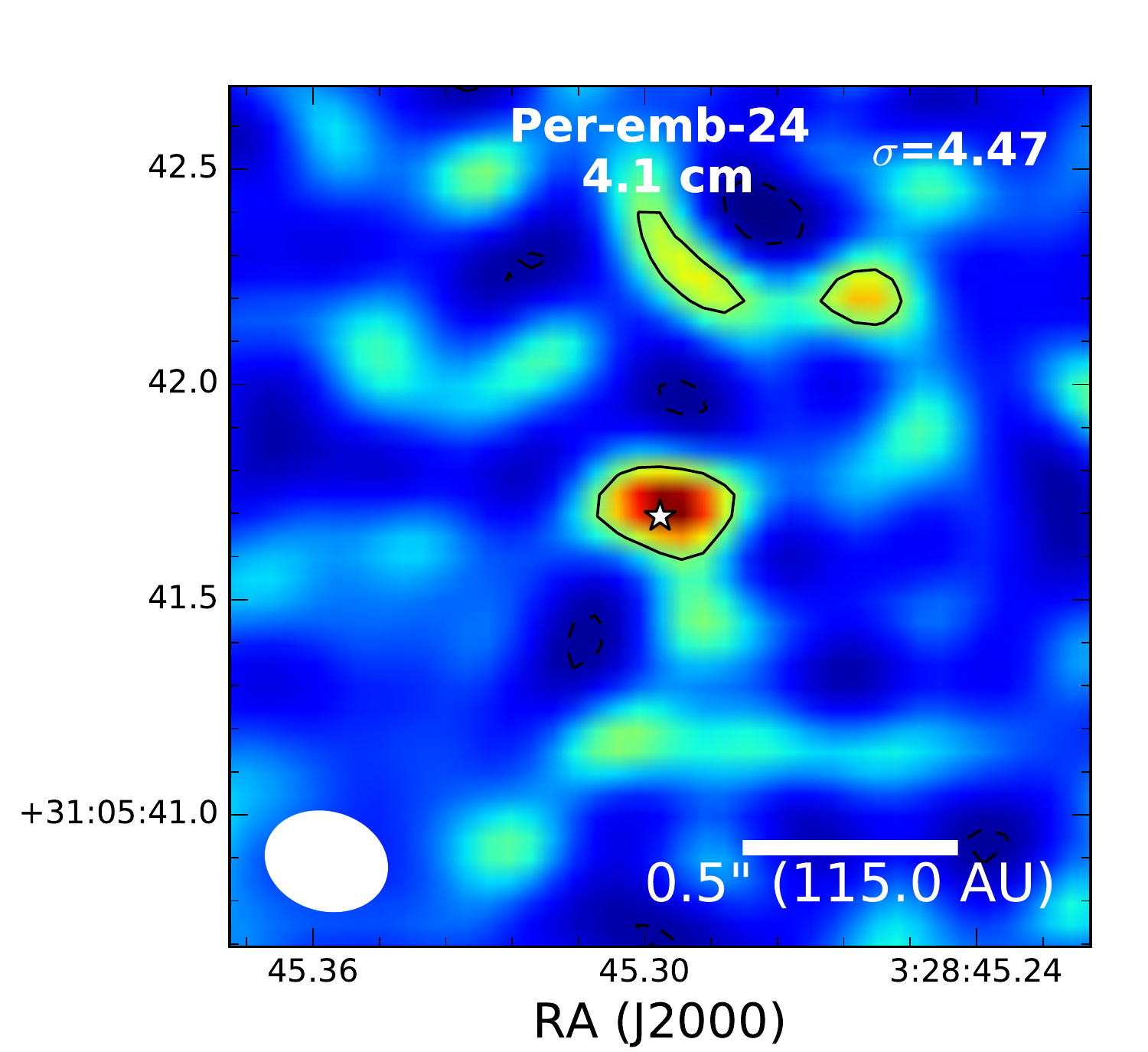}
  \includegraphics[width=0.24\linewidth]{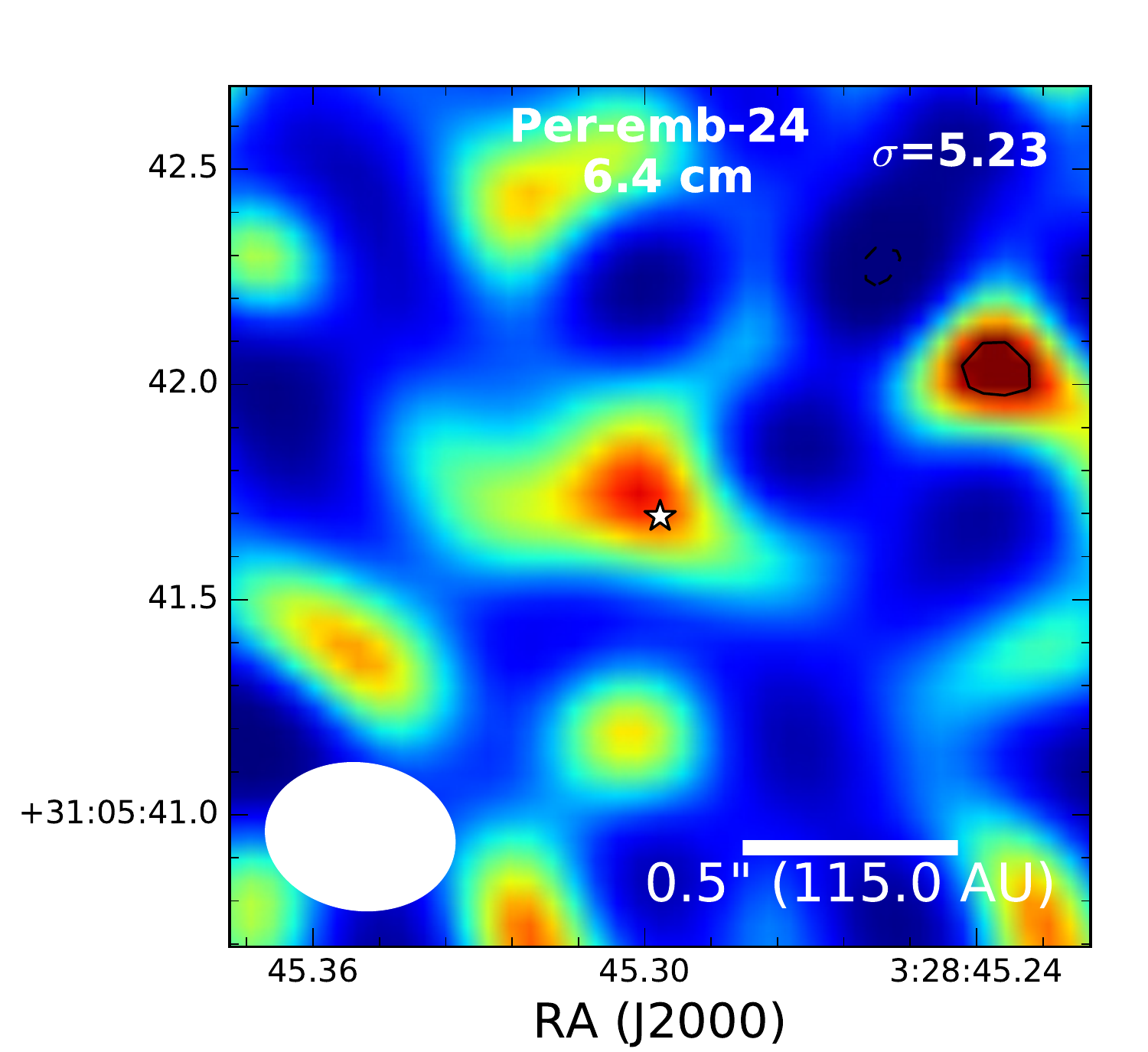}

  \includegraphics[width=0.24\linewidth]{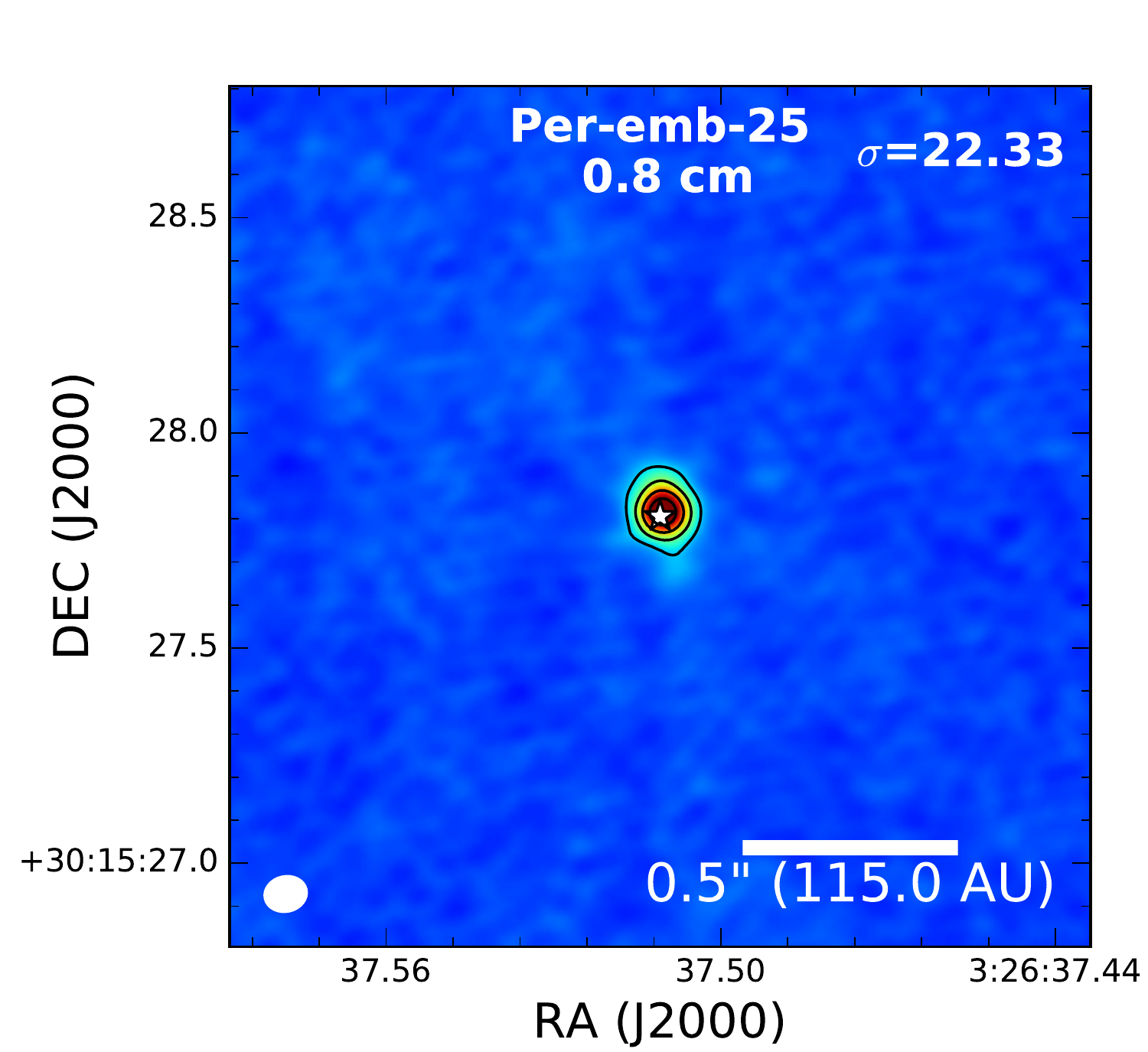}
  \includegraphics[width=0.24\linewidth]{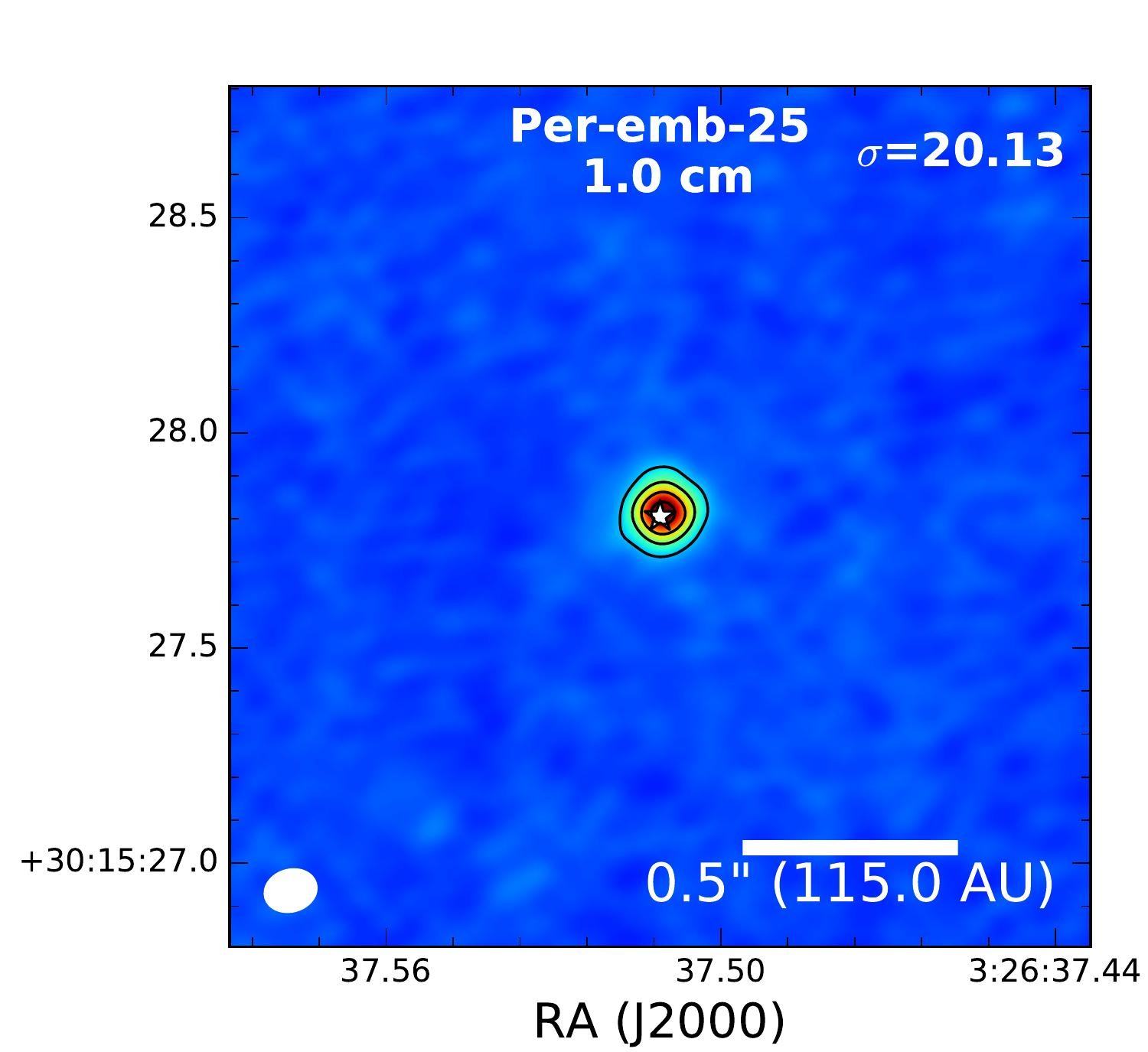}
  \includegraphics[width=0.24\linewidth]{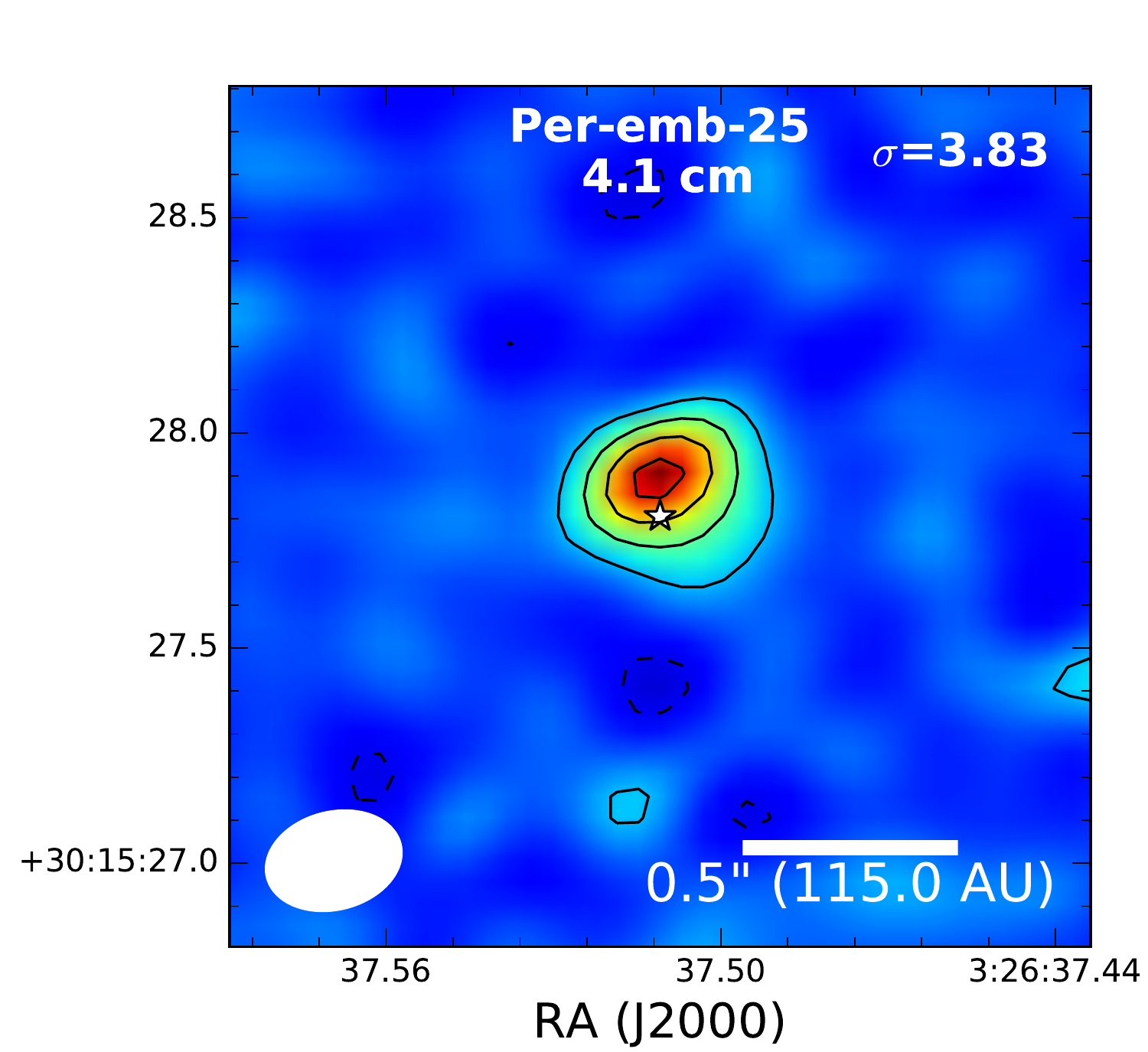}
  \includegraphics[width=0.24\linewidth]{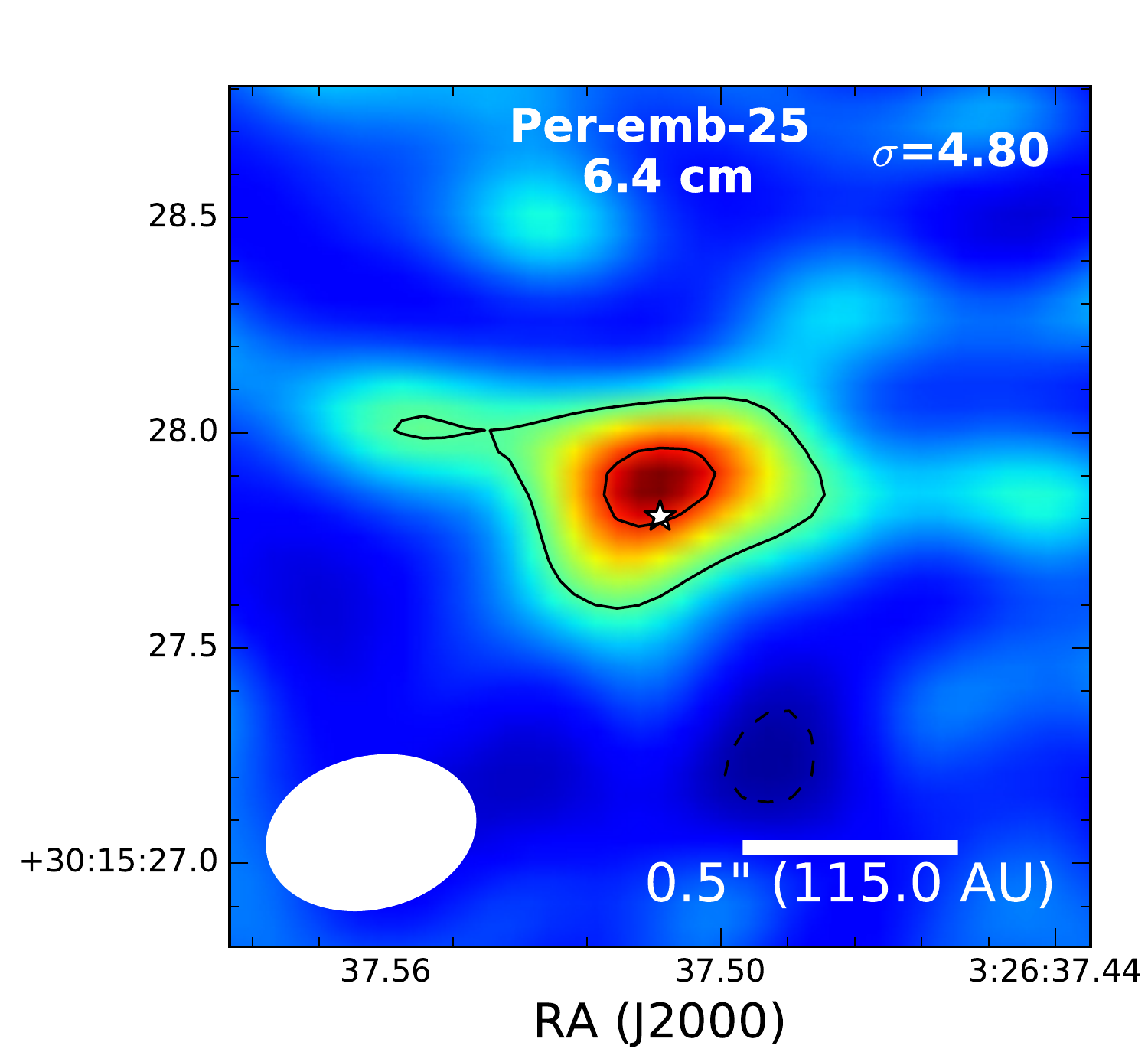}

  \includegraphics[width=0.24\linewidth]{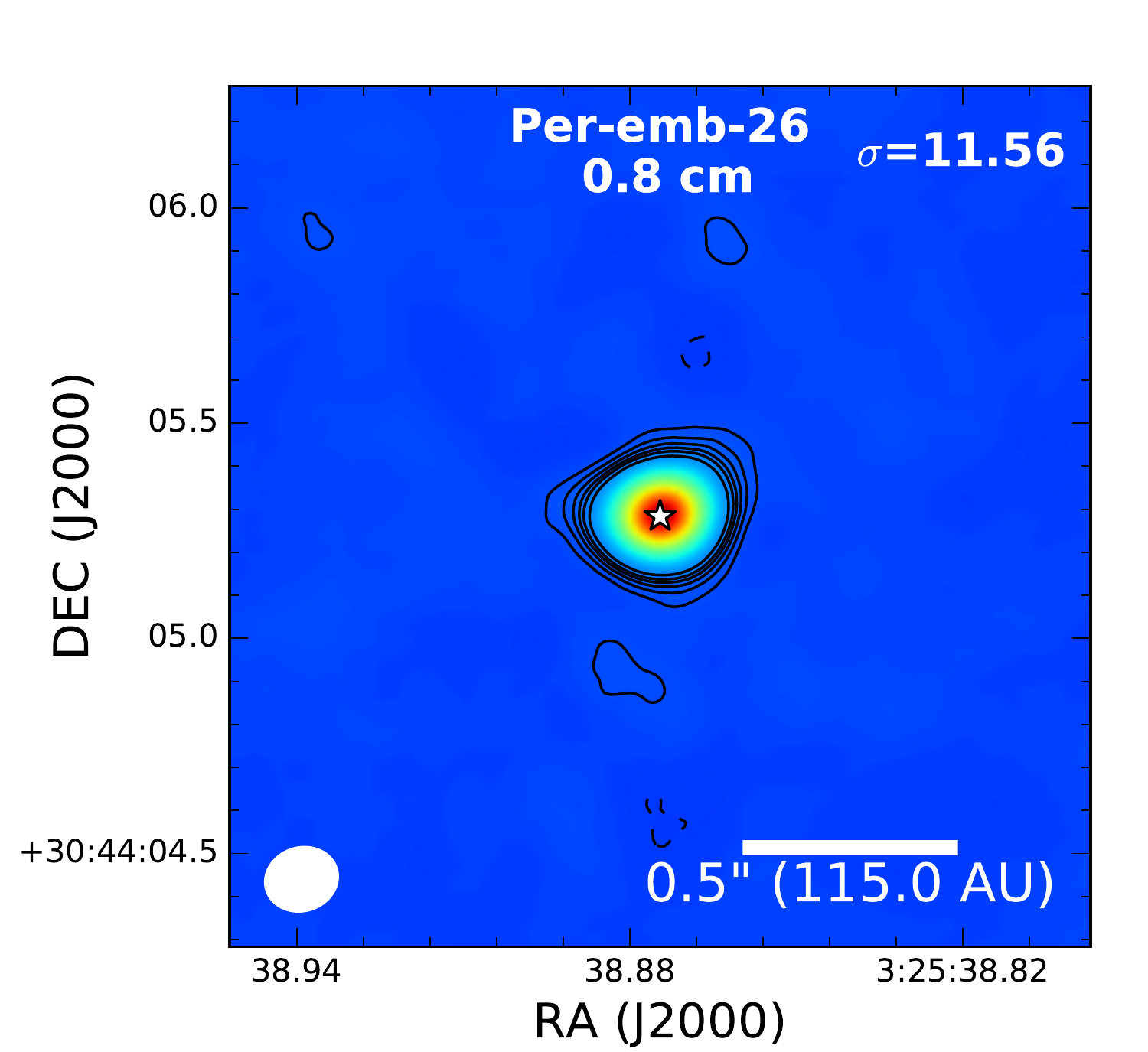}
  \includegraphics[width=0.24\linewidth]{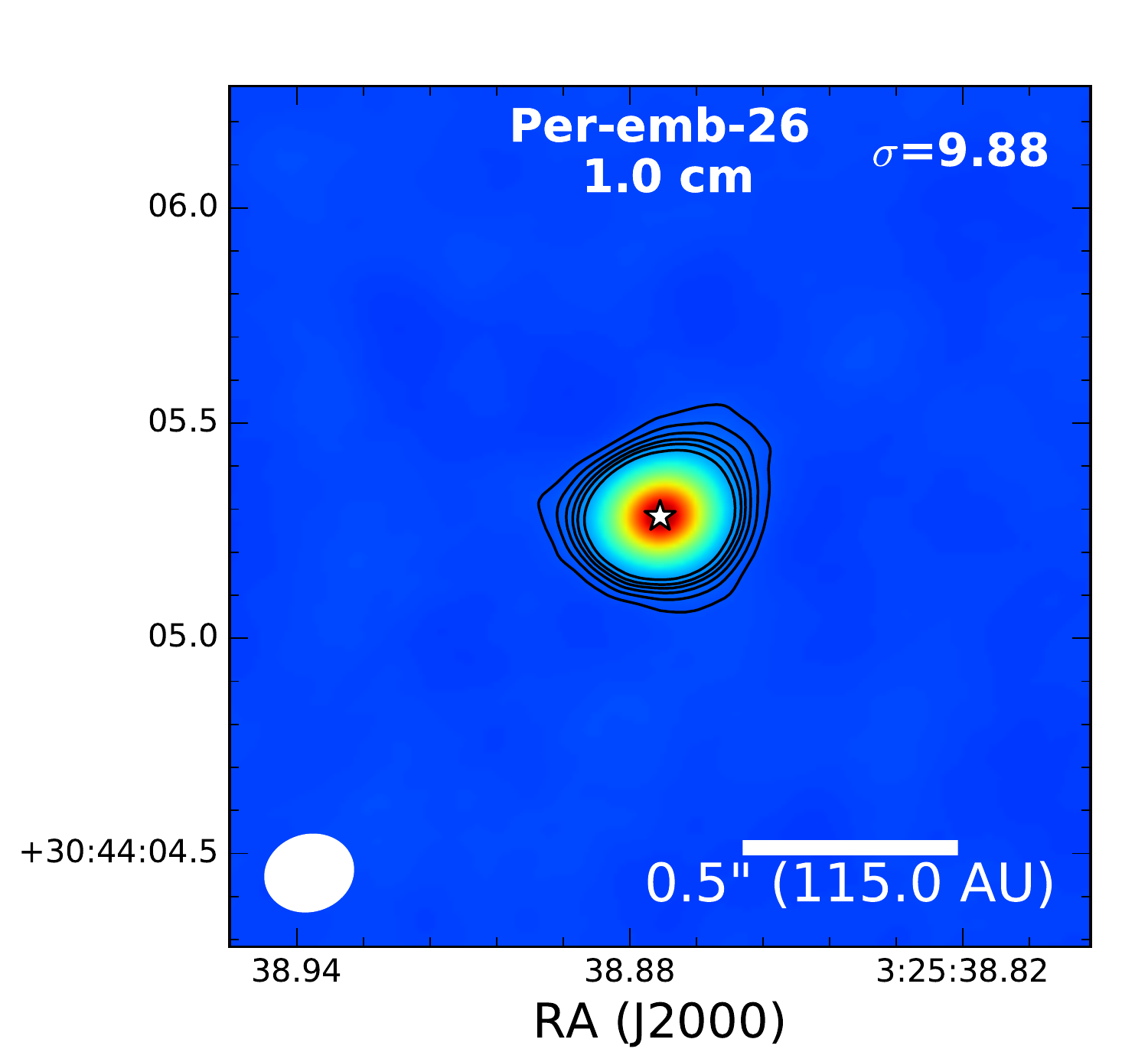}
  \includegraphics[width=0.24\linewidth]{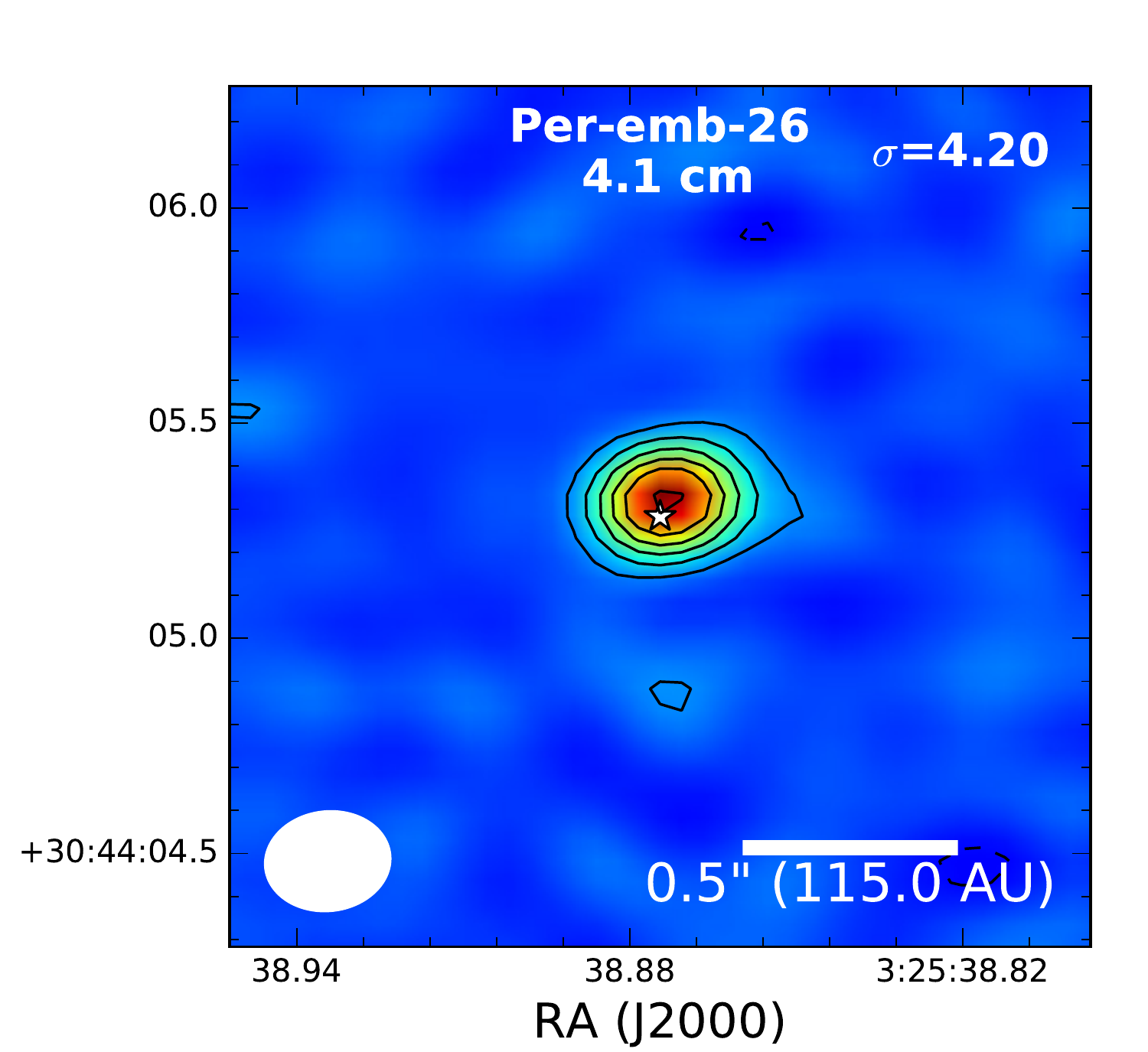}
  \includegraphics[width=0.24\linewidth]{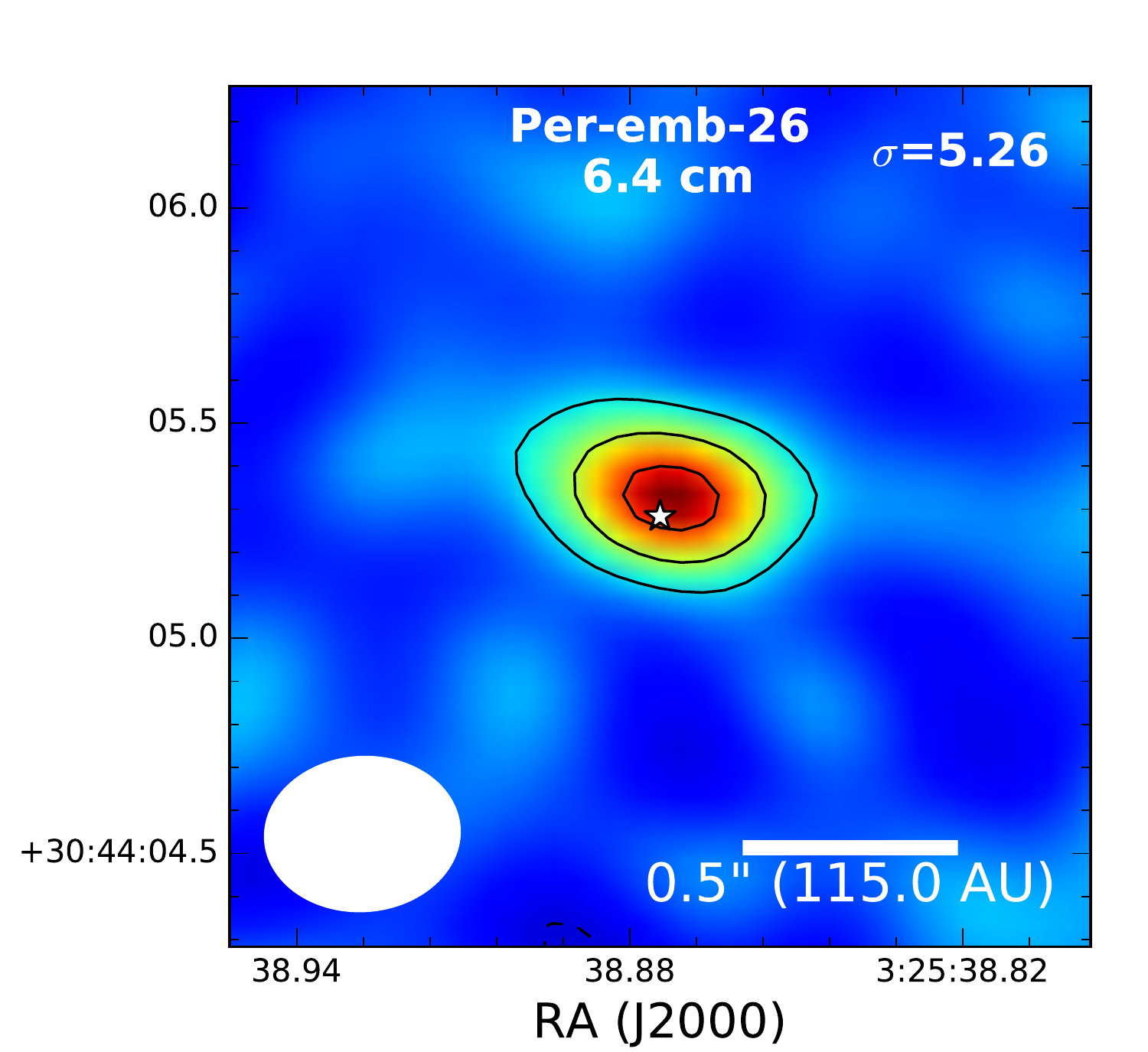}

  \includegraphics[width=0.24\linewidth]{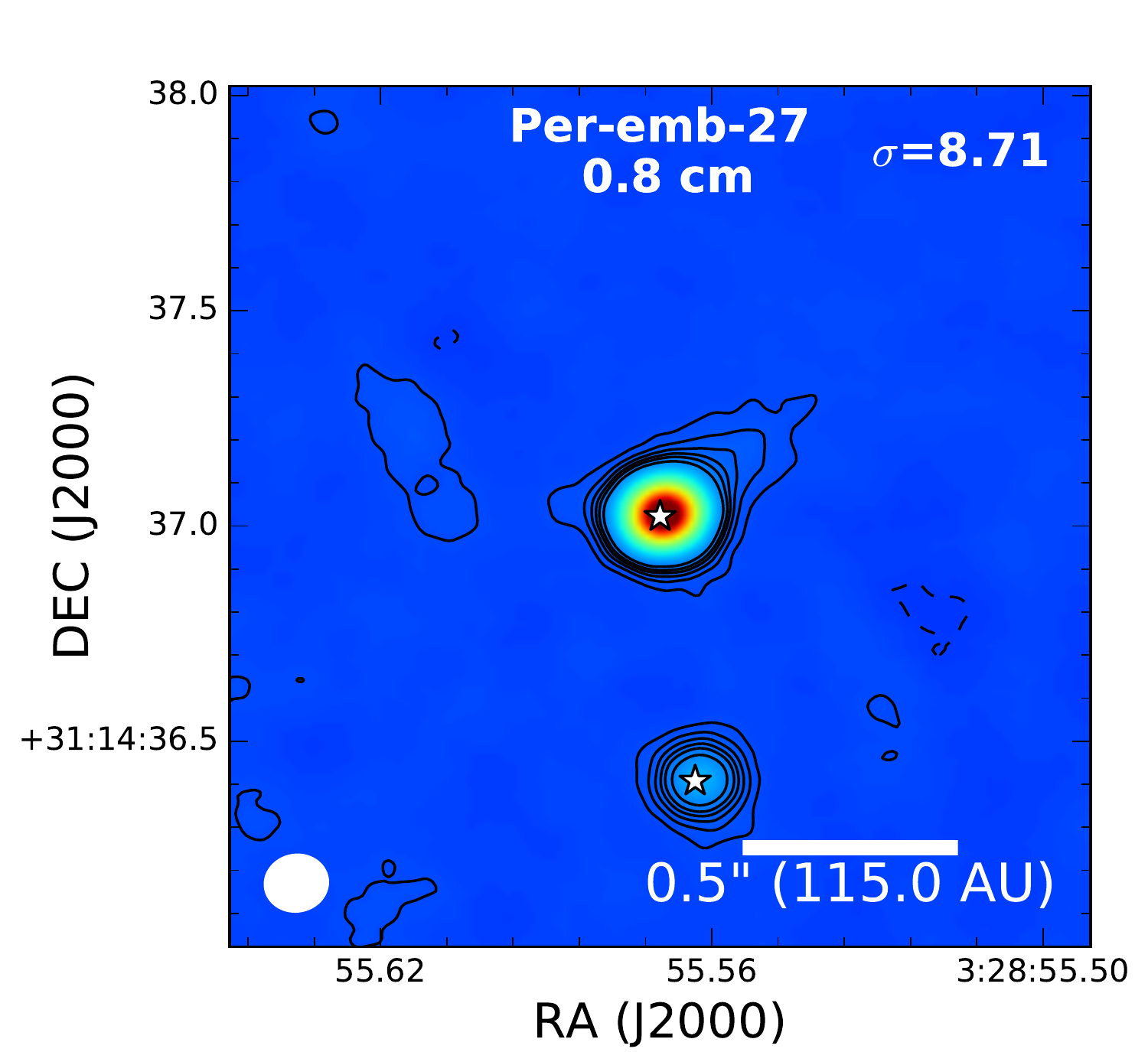}
  \includegraphics[width=0.24\linewidth]{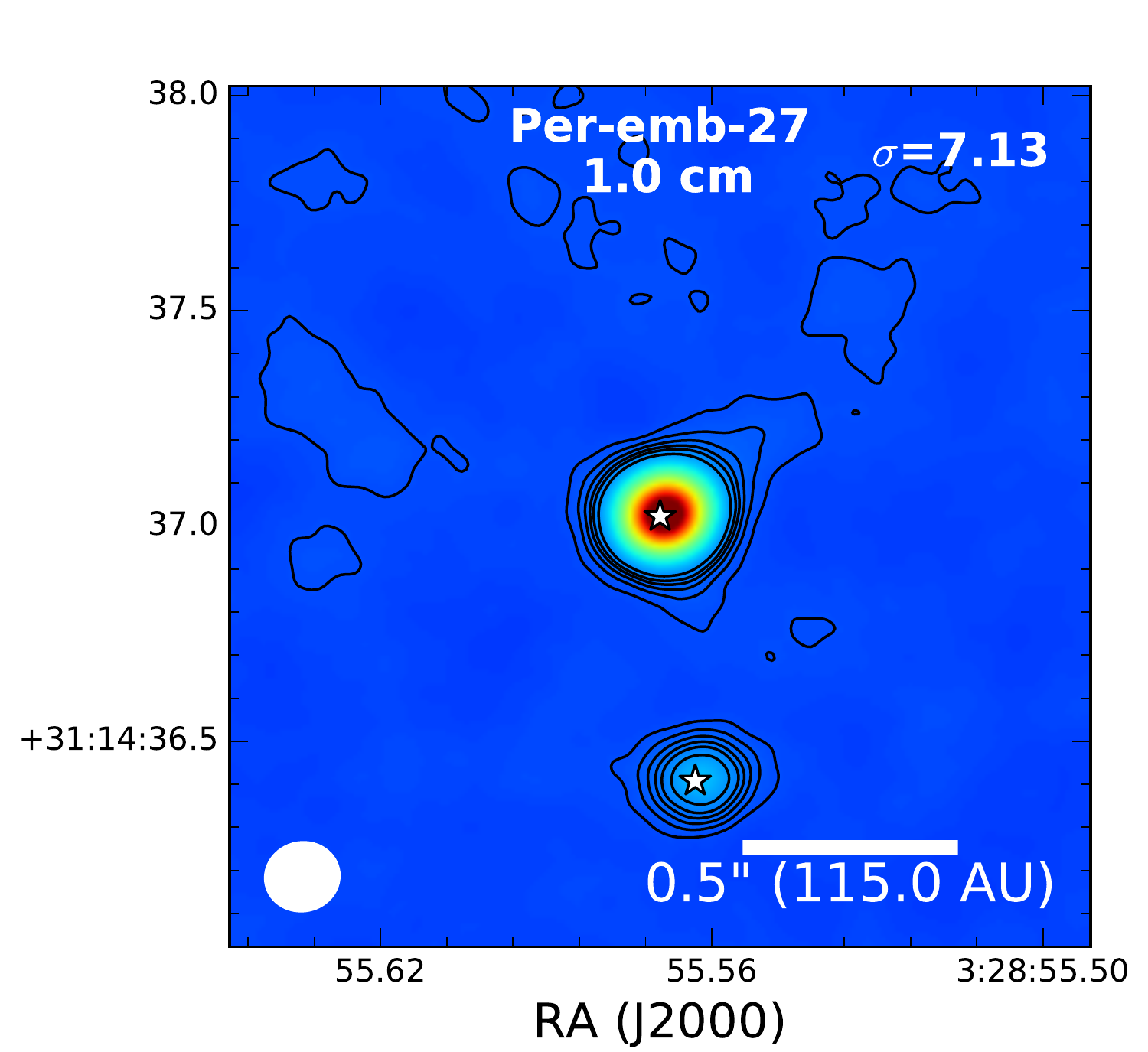}
  \includegraphics[width=0.24\linewidth]{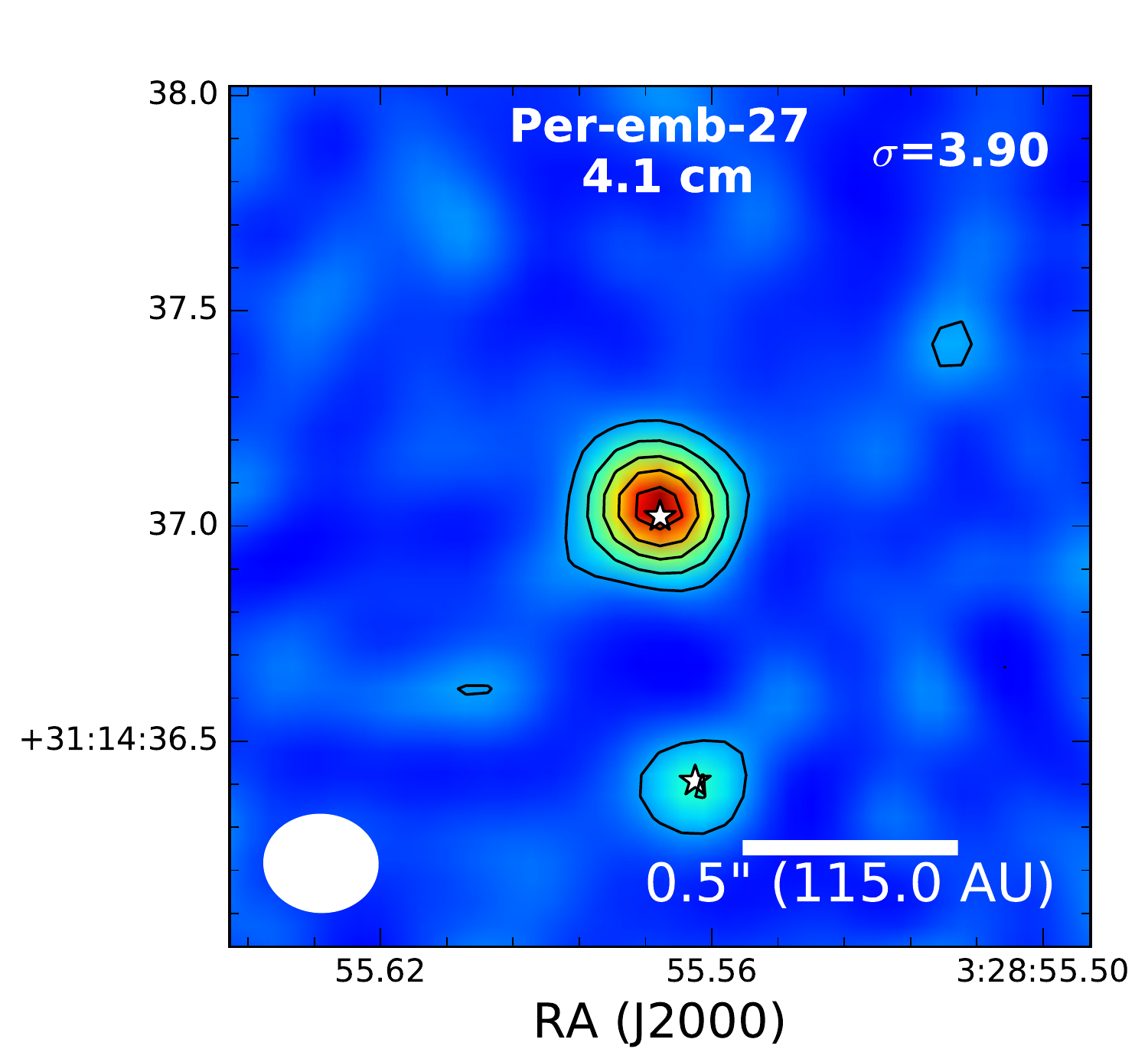}
  \includegraphics[width=0.24\linewidth]{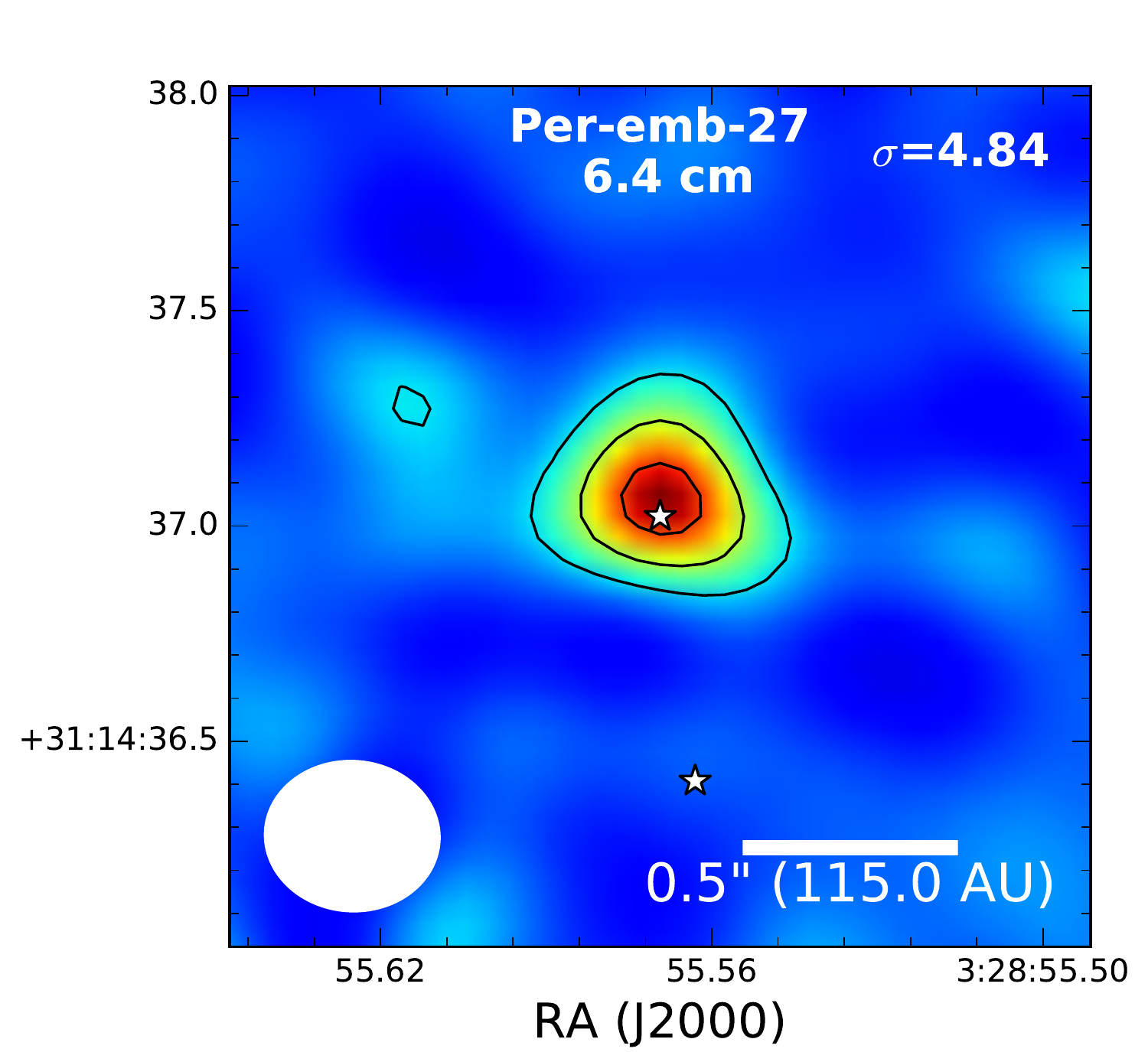}

\end{figure}

\begin{figure}

  \includegraphics[width=0.24\linewidth]{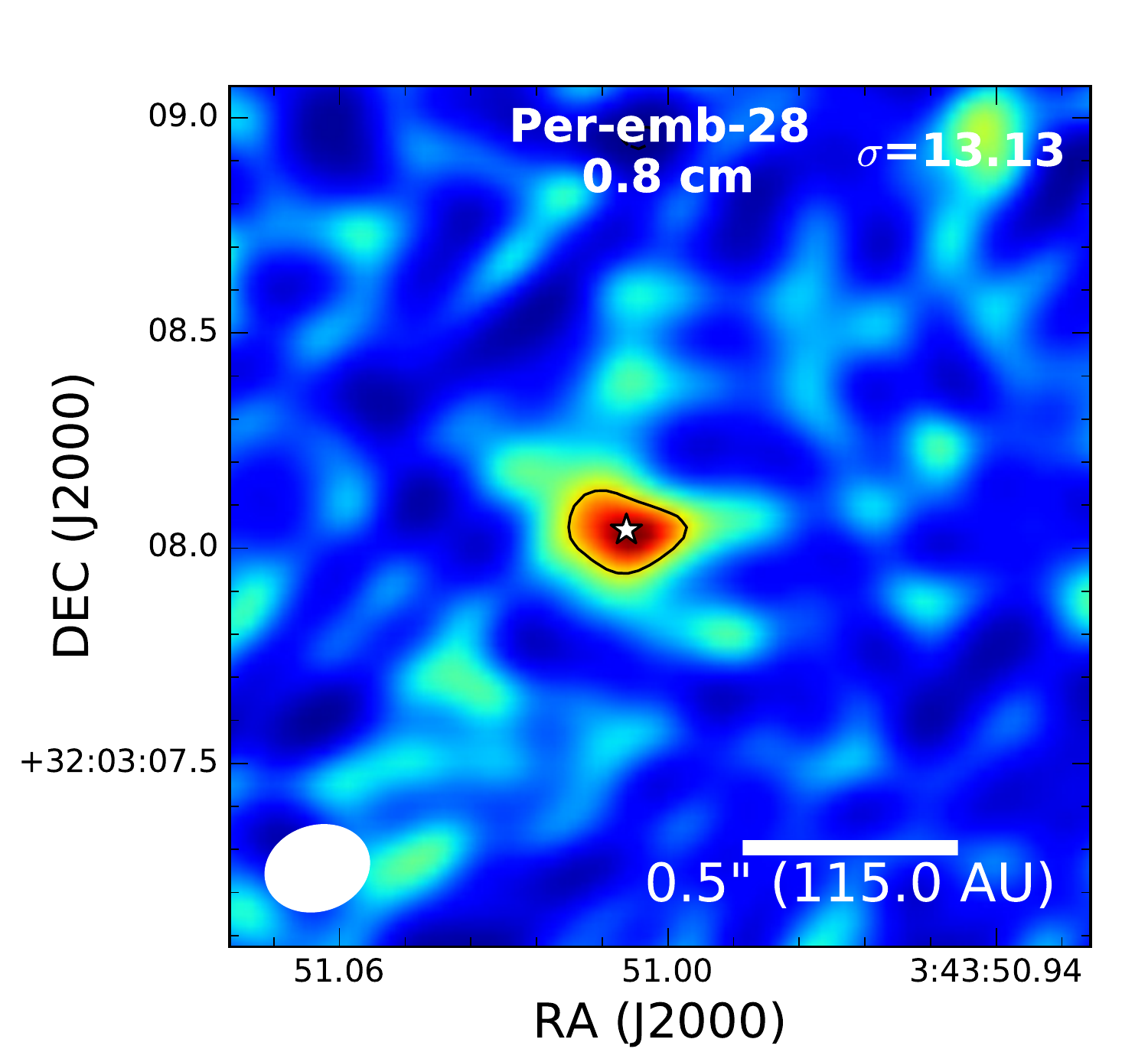}
  \includegraphics[width=0.24\linewidth]{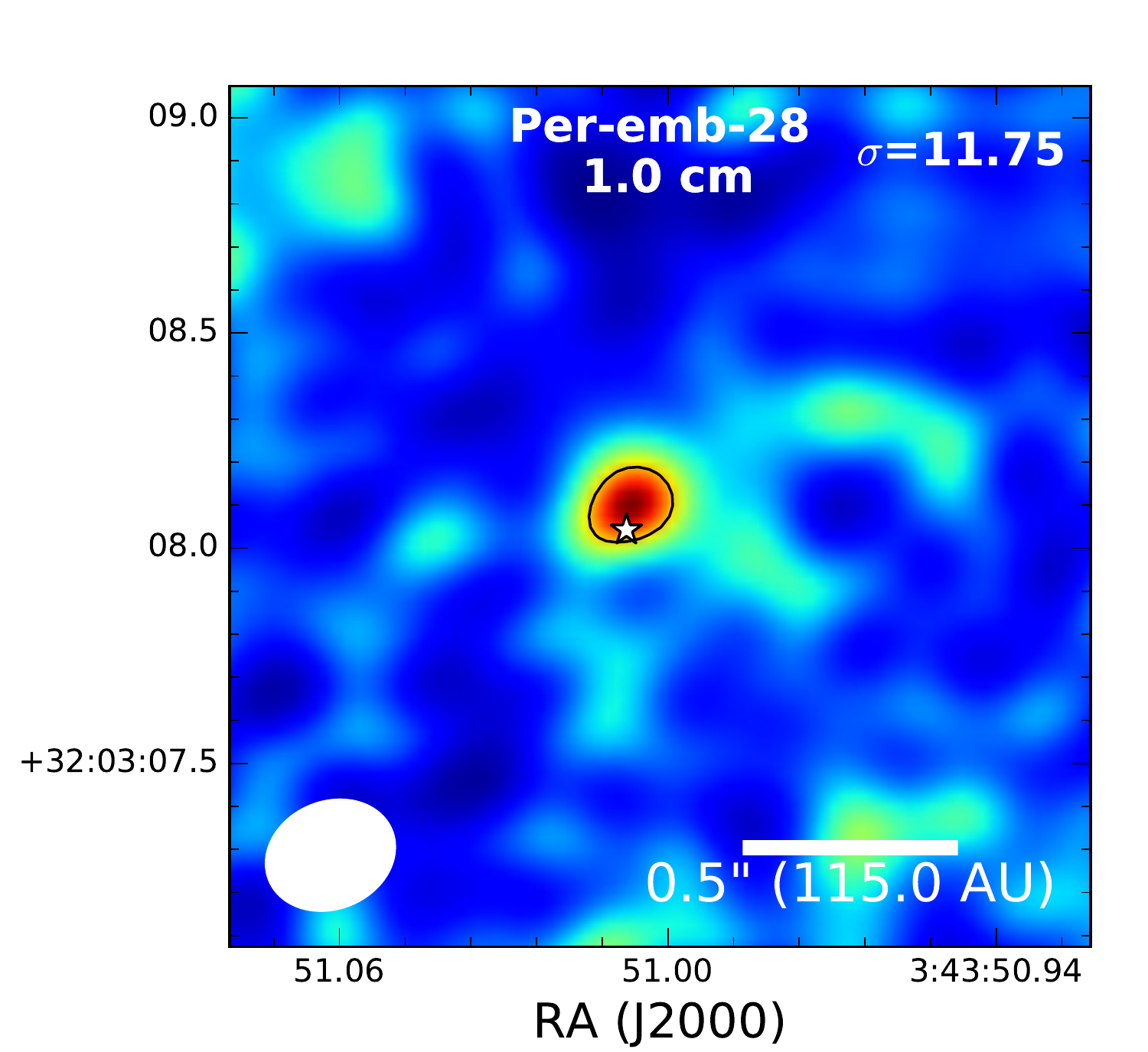}
  \includegraphics[width=0.24\linewidth]{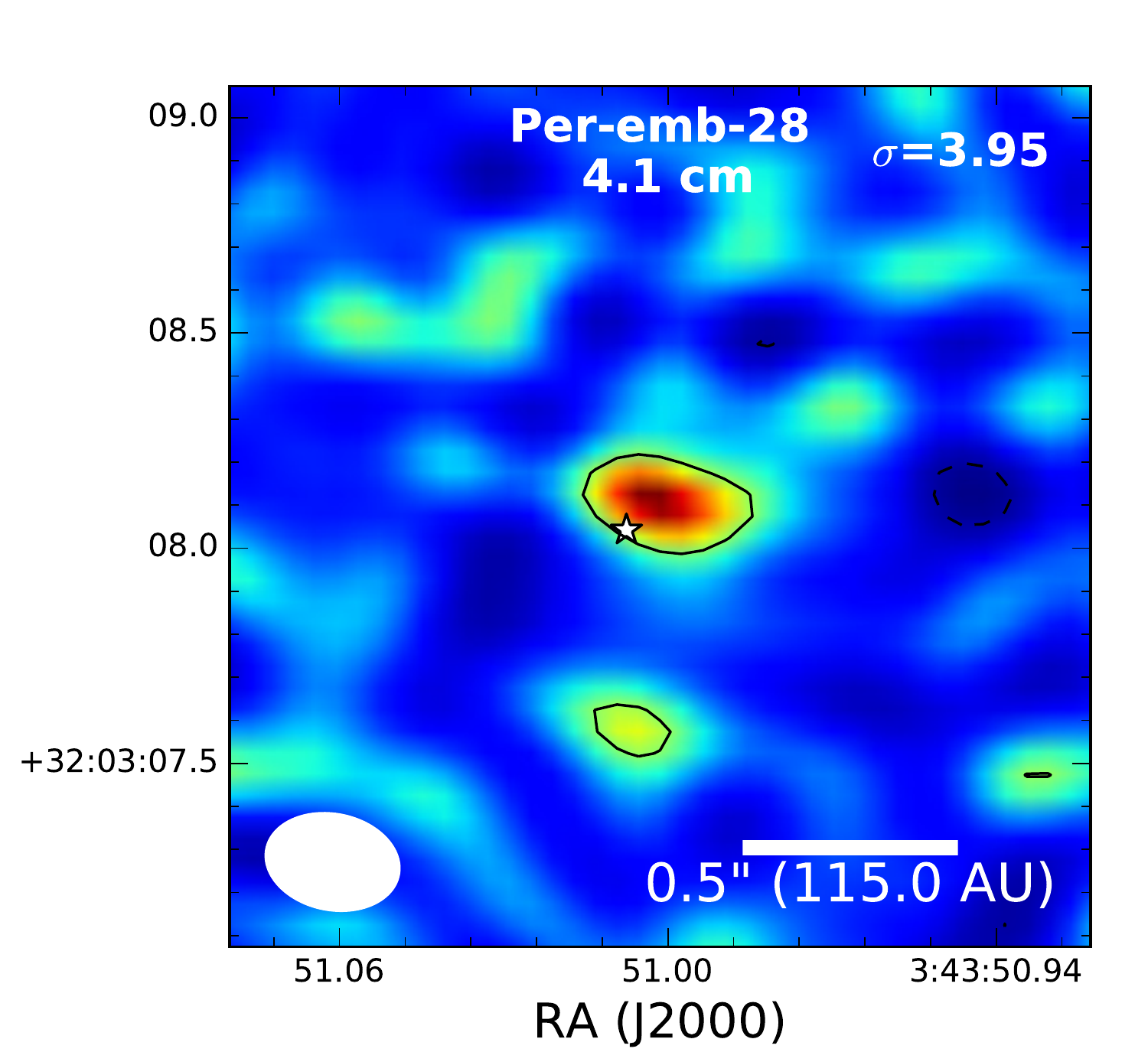}
  \includegraphics[width=0.24\linewidth]{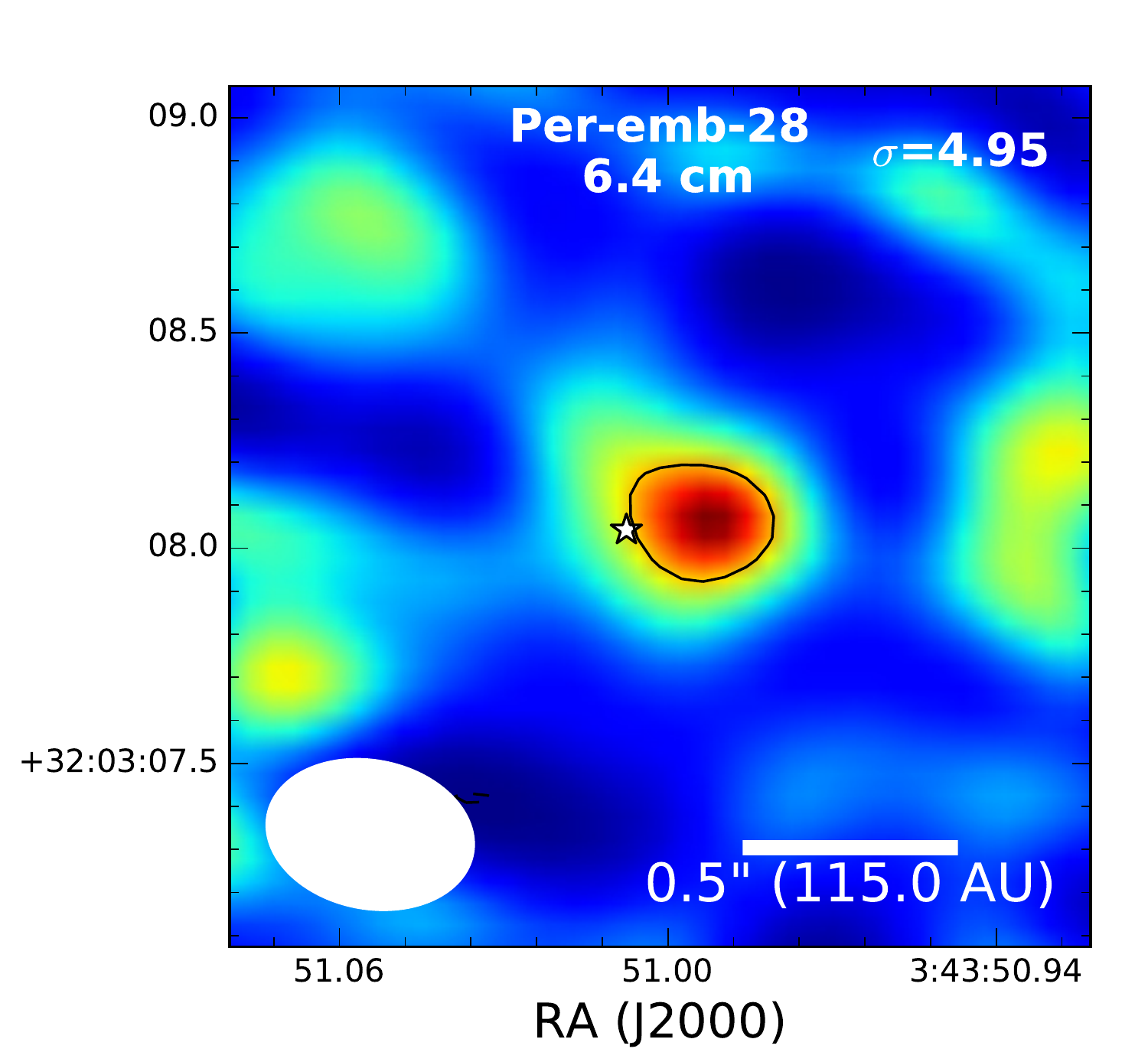}

  \includegraphics[width=0.24\linewidth]{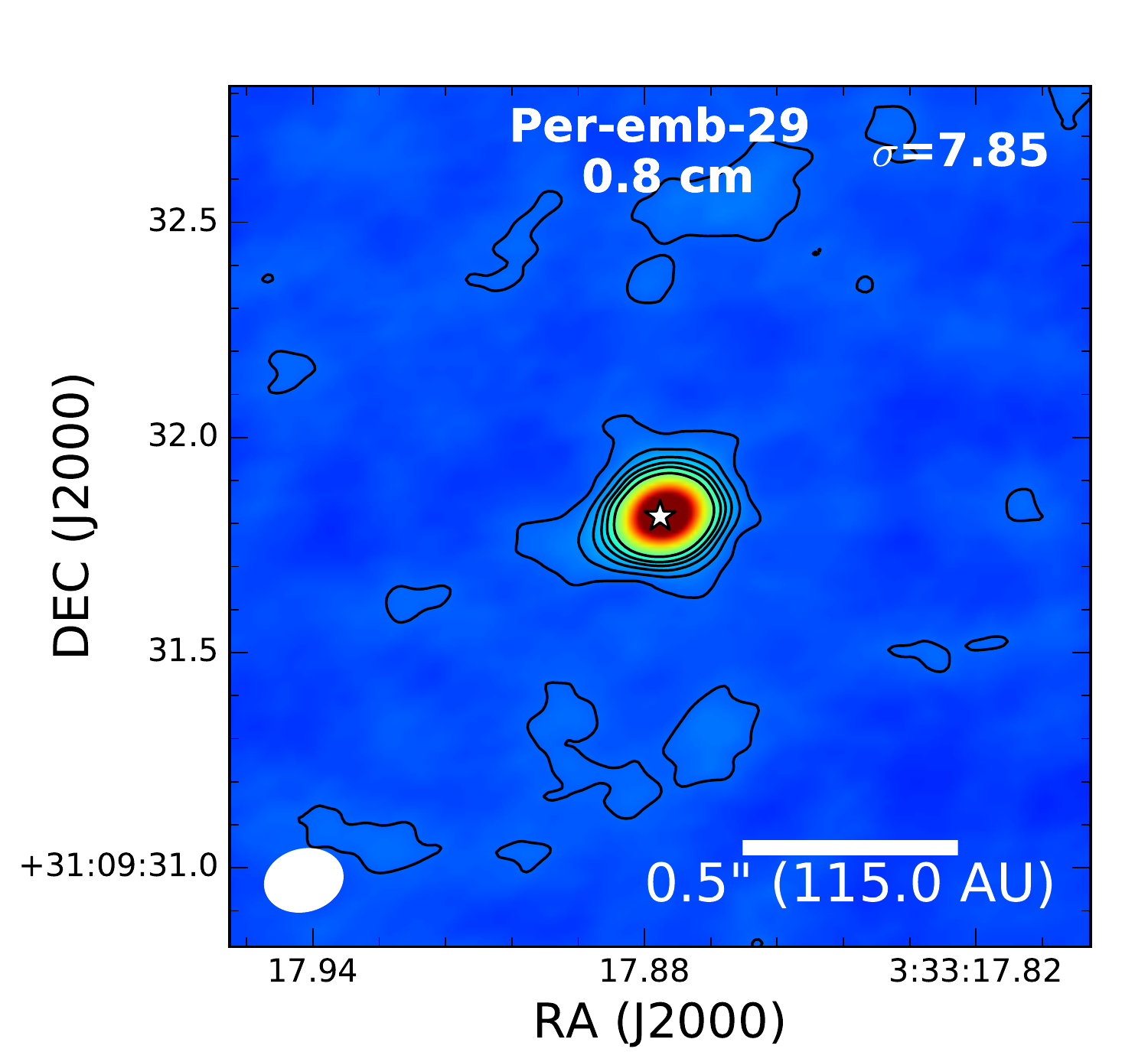}
  \includegraphics[width=0.24\linewidth]{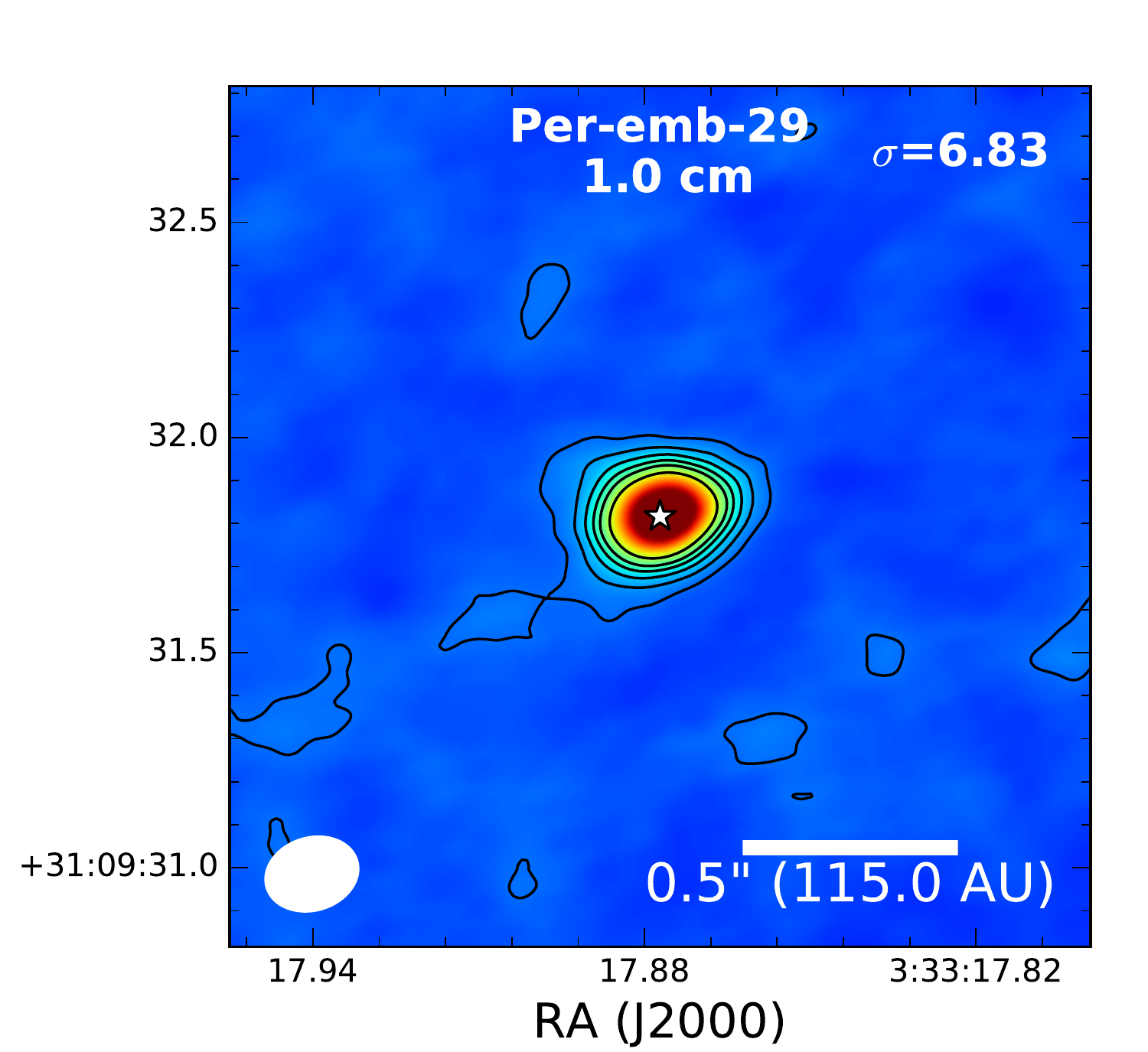}
  \includegraphics[width=0.24\linewidth]{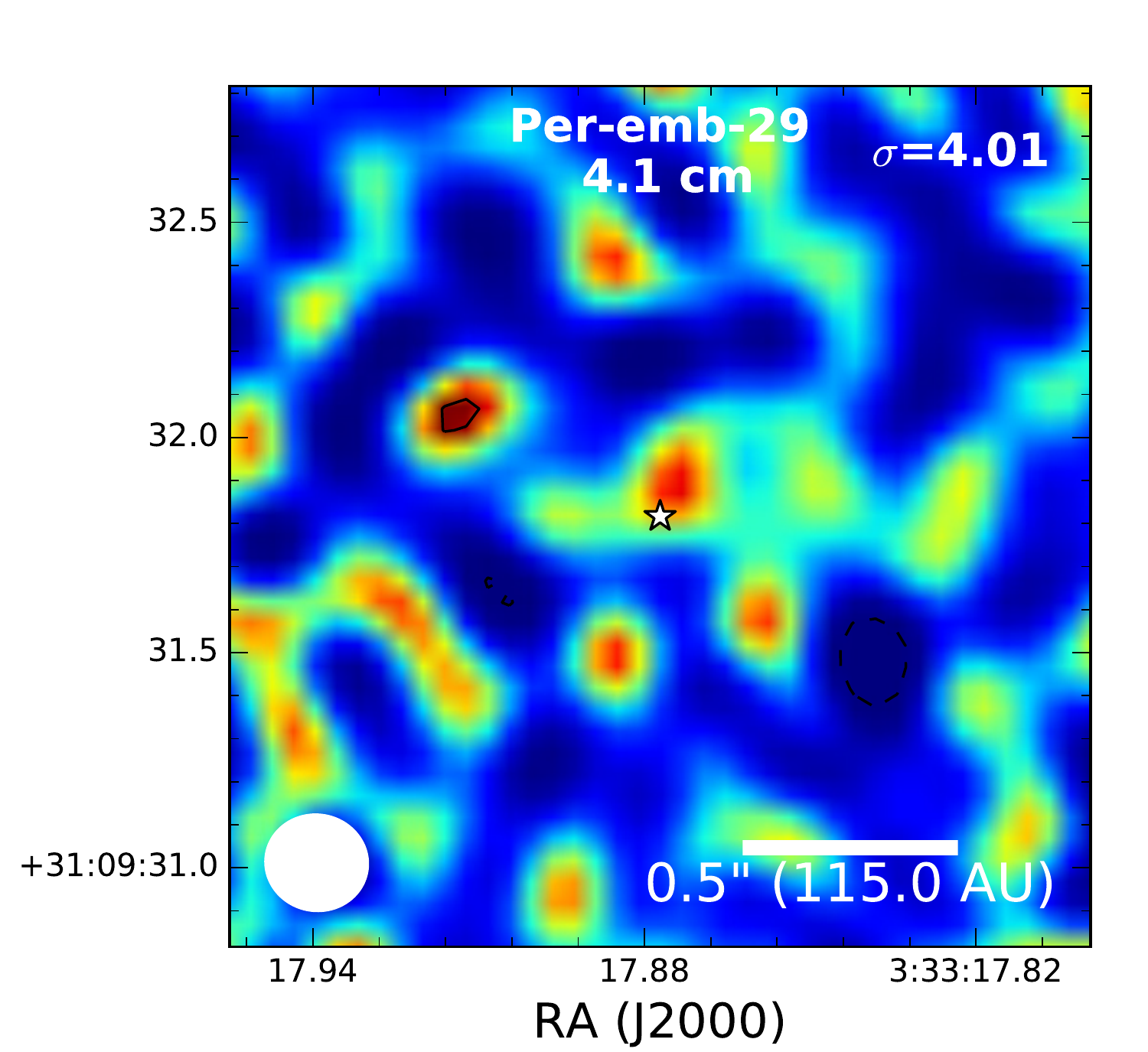}
  \includegraphics[width=0.24\linewidth]{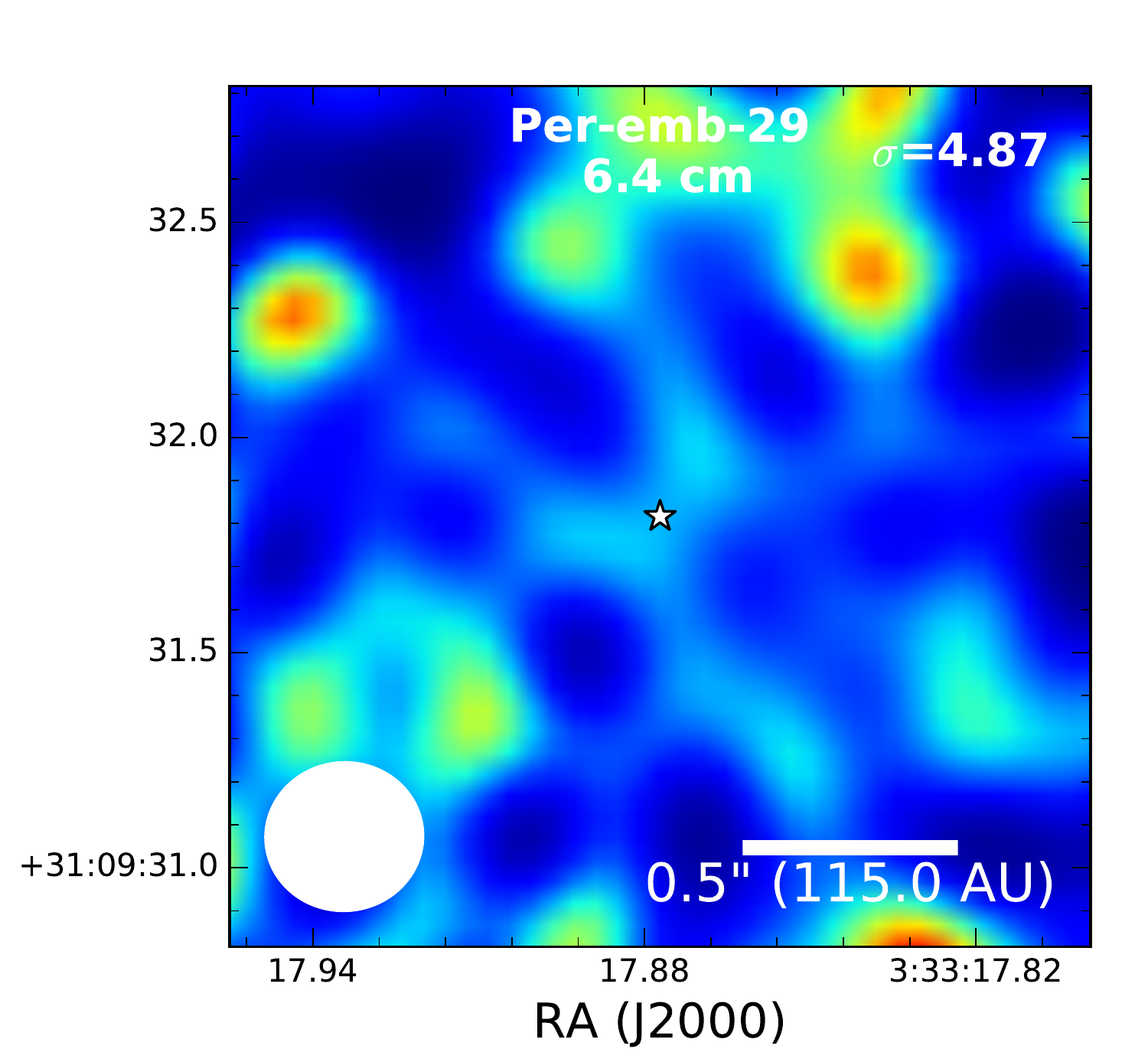}

  \includegraphics[width=0.24\linewidth]{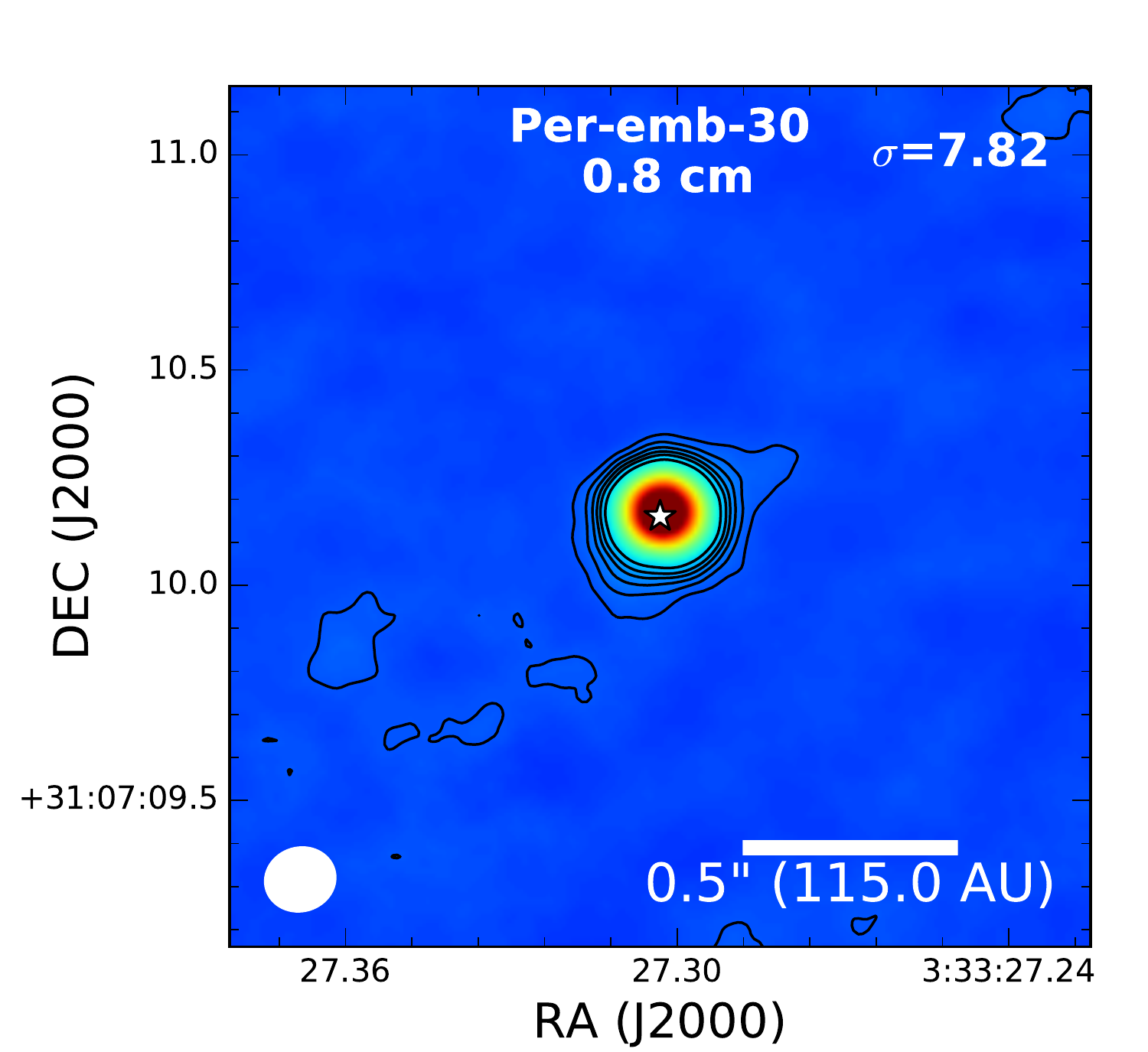}
  \includegraphics[width=0.24\linewidth]{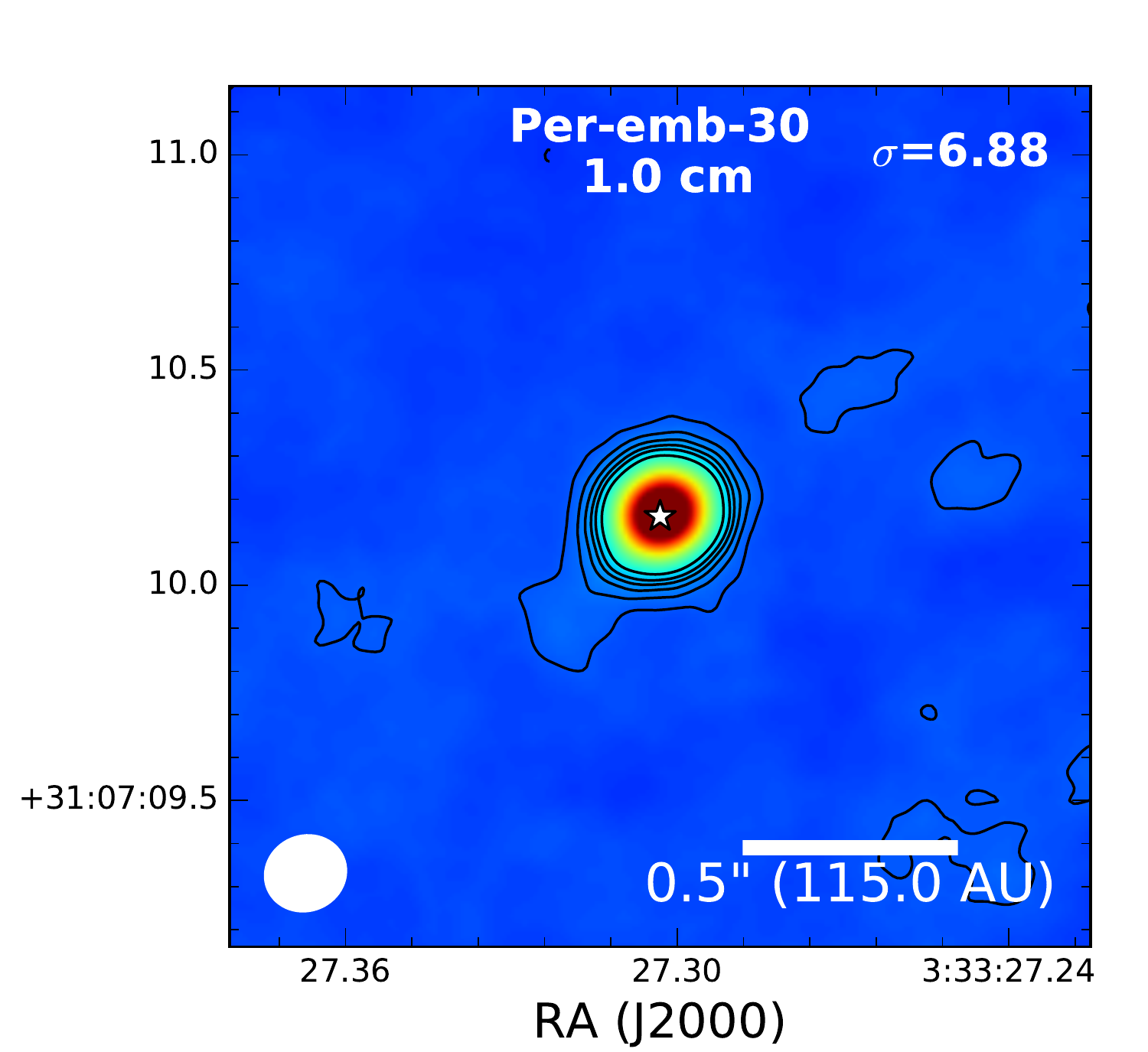}
  \includegraphics[width=0.24\linewidth]{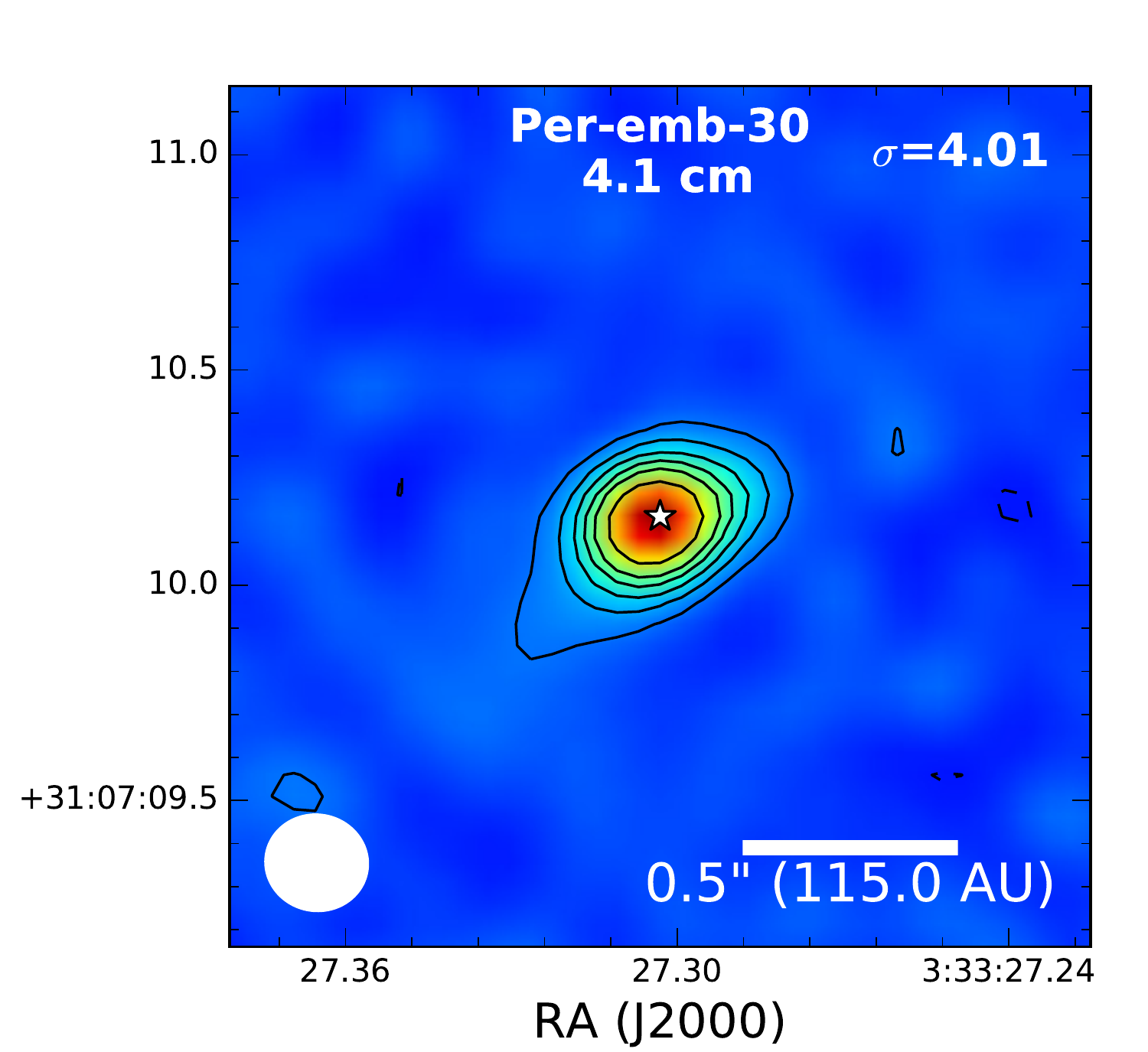}
  \includegraphics[width=0.24\linewidth]{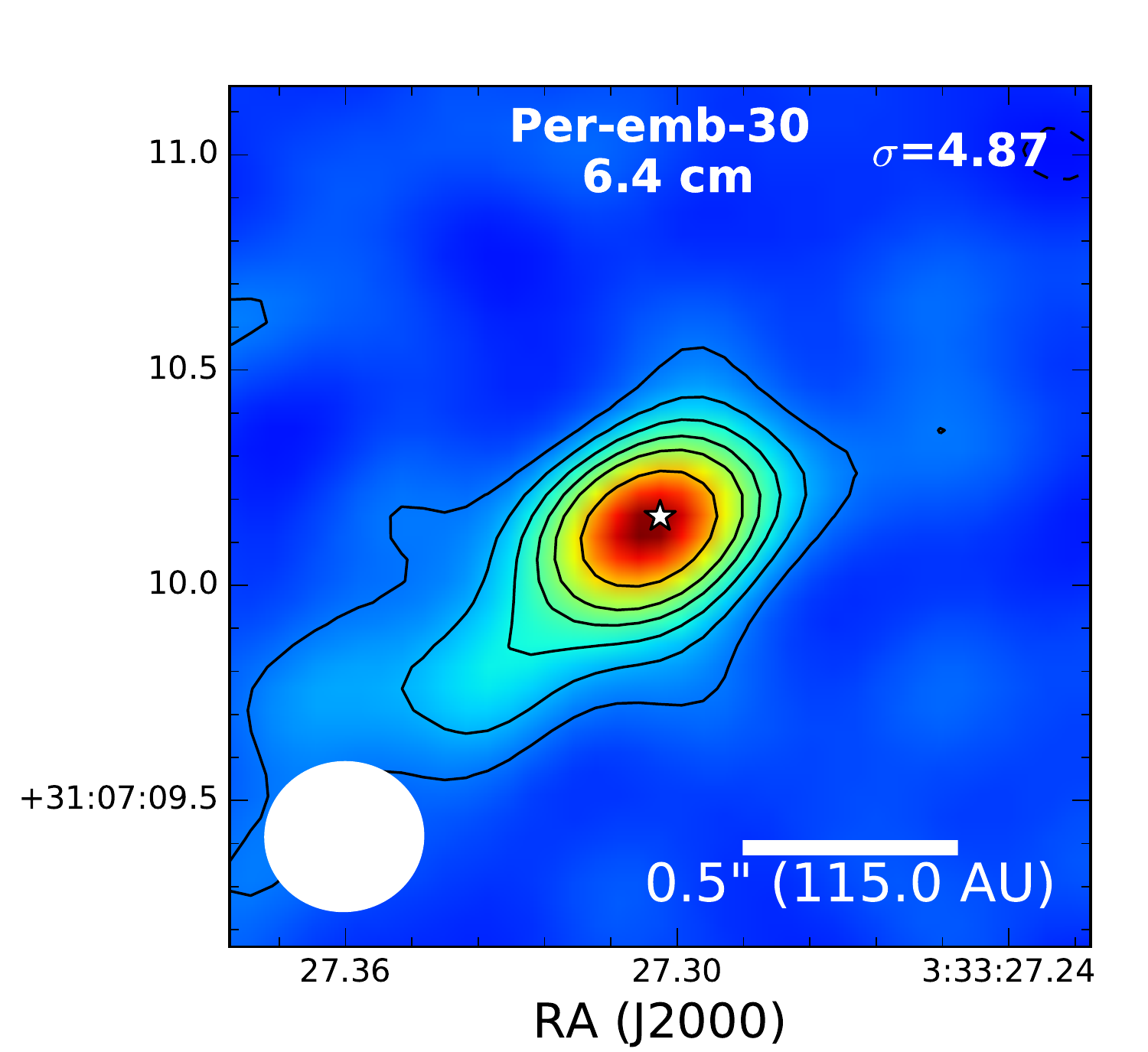}

  \includegraphics[width=0.24\linewidth]{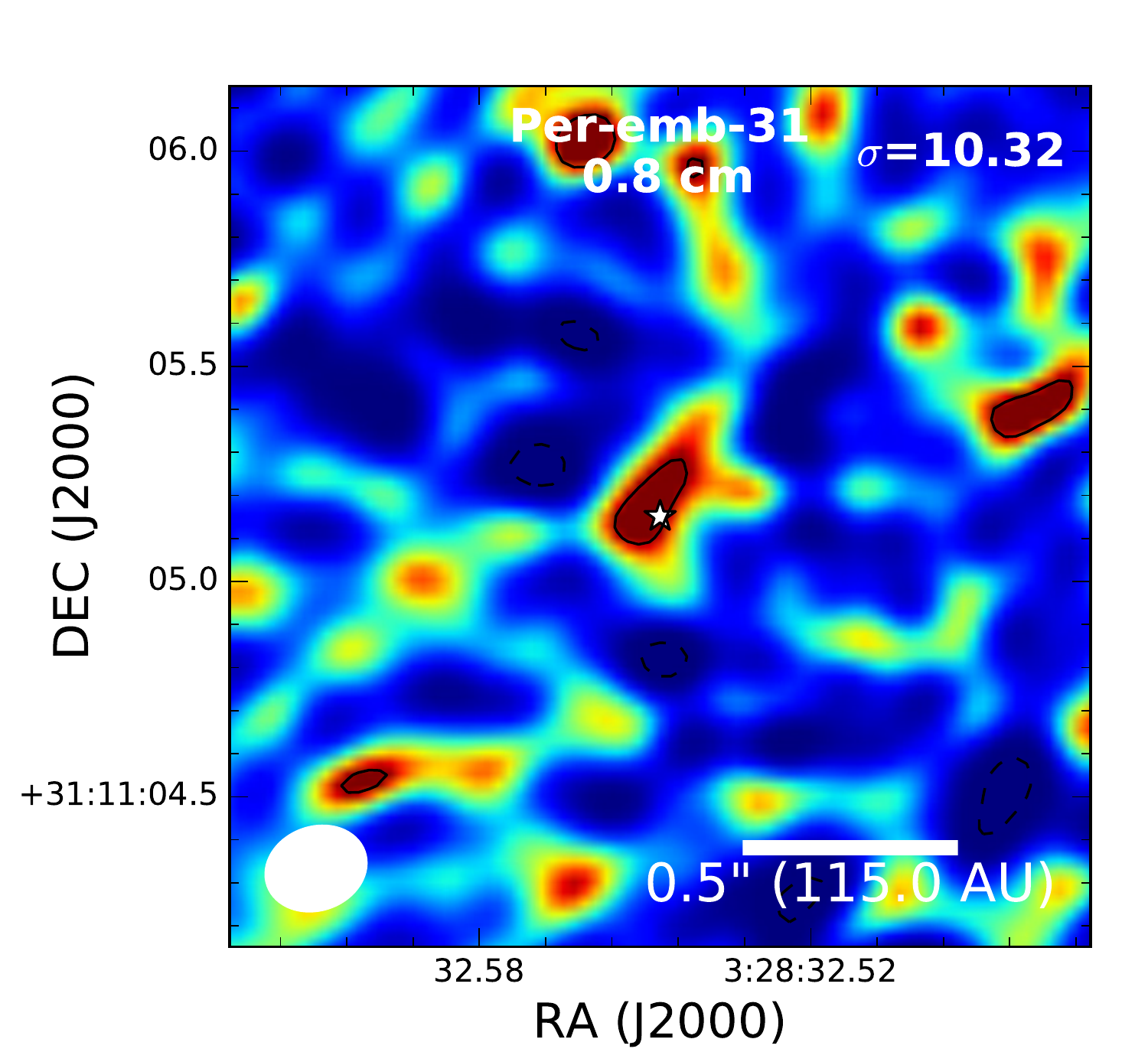}
  \includegraphics[width=0.24\linewidth]{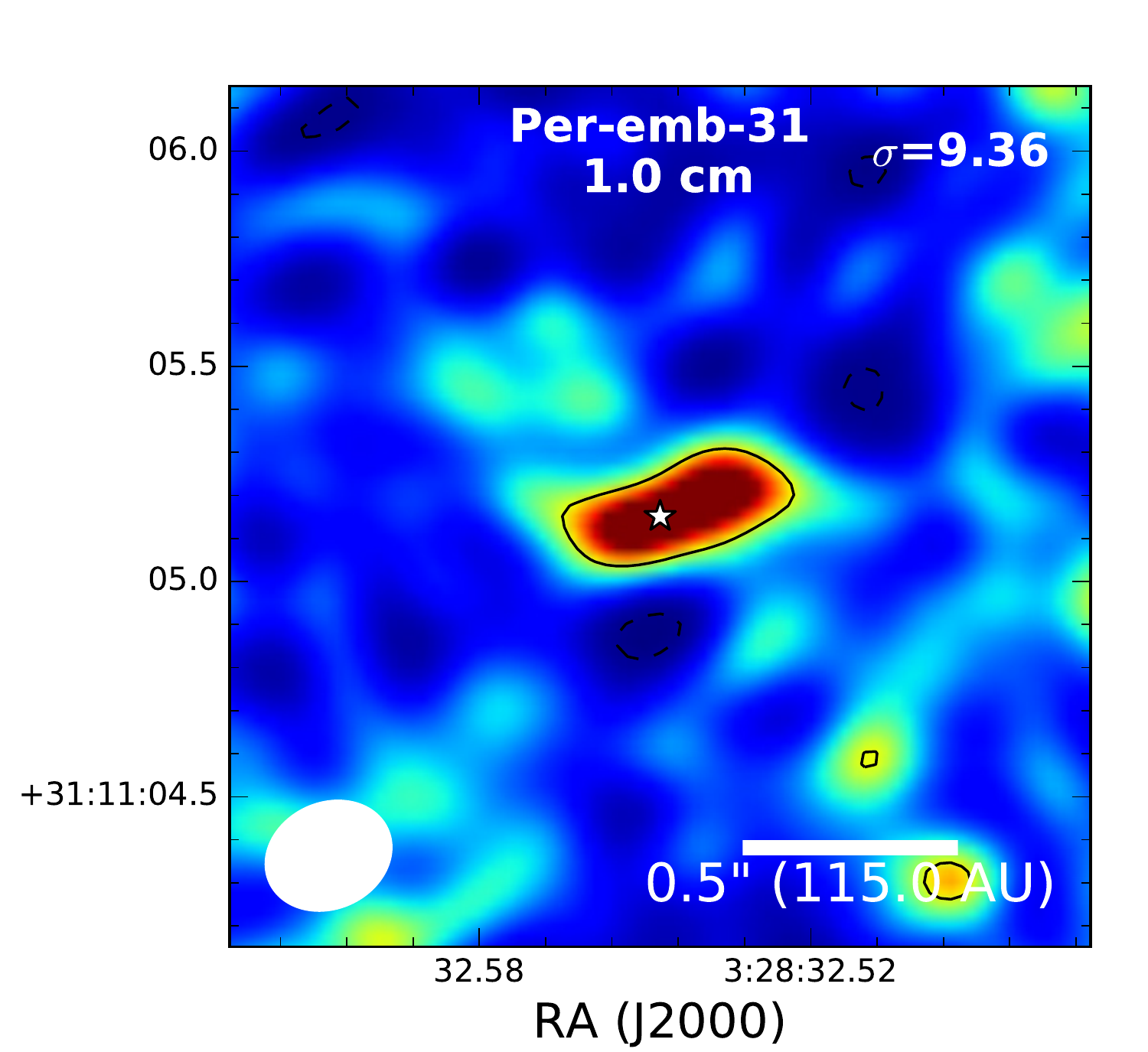}
  \includegraphics[width=0.24\linewidth]{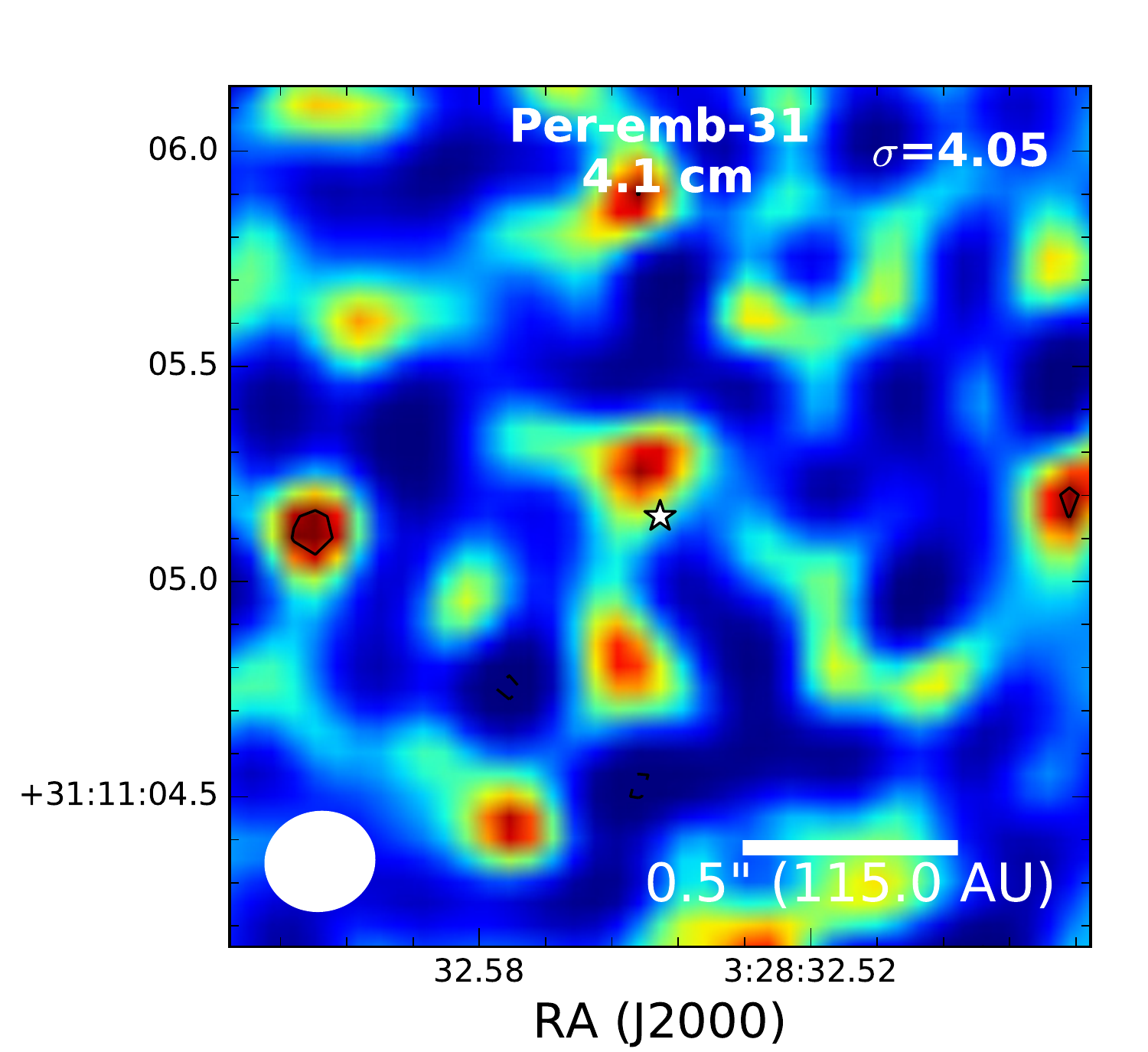}
  \includegraphics[width=0.24\linewidth]{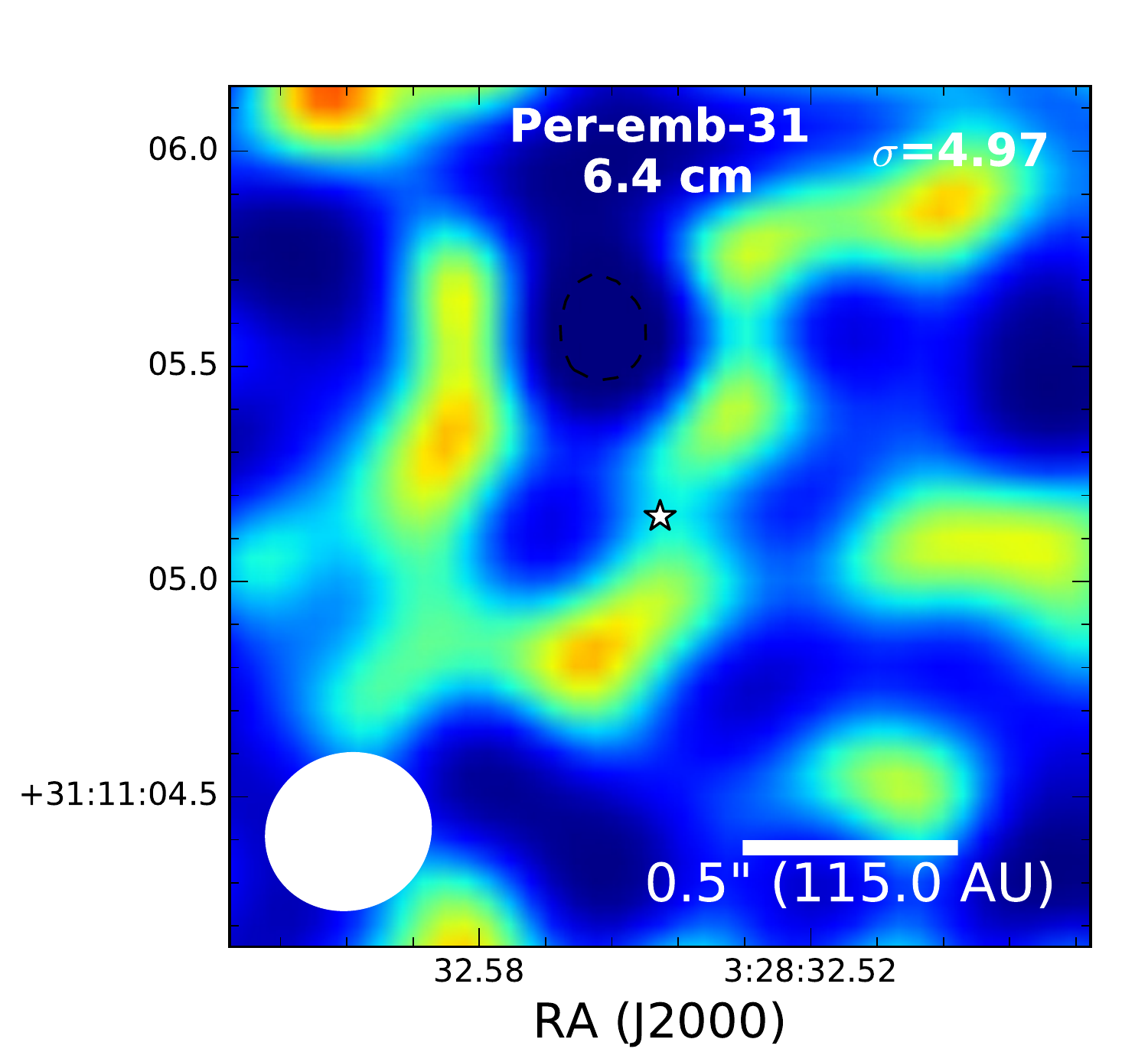}

  \includegraphics[width=0.24\linewidth]{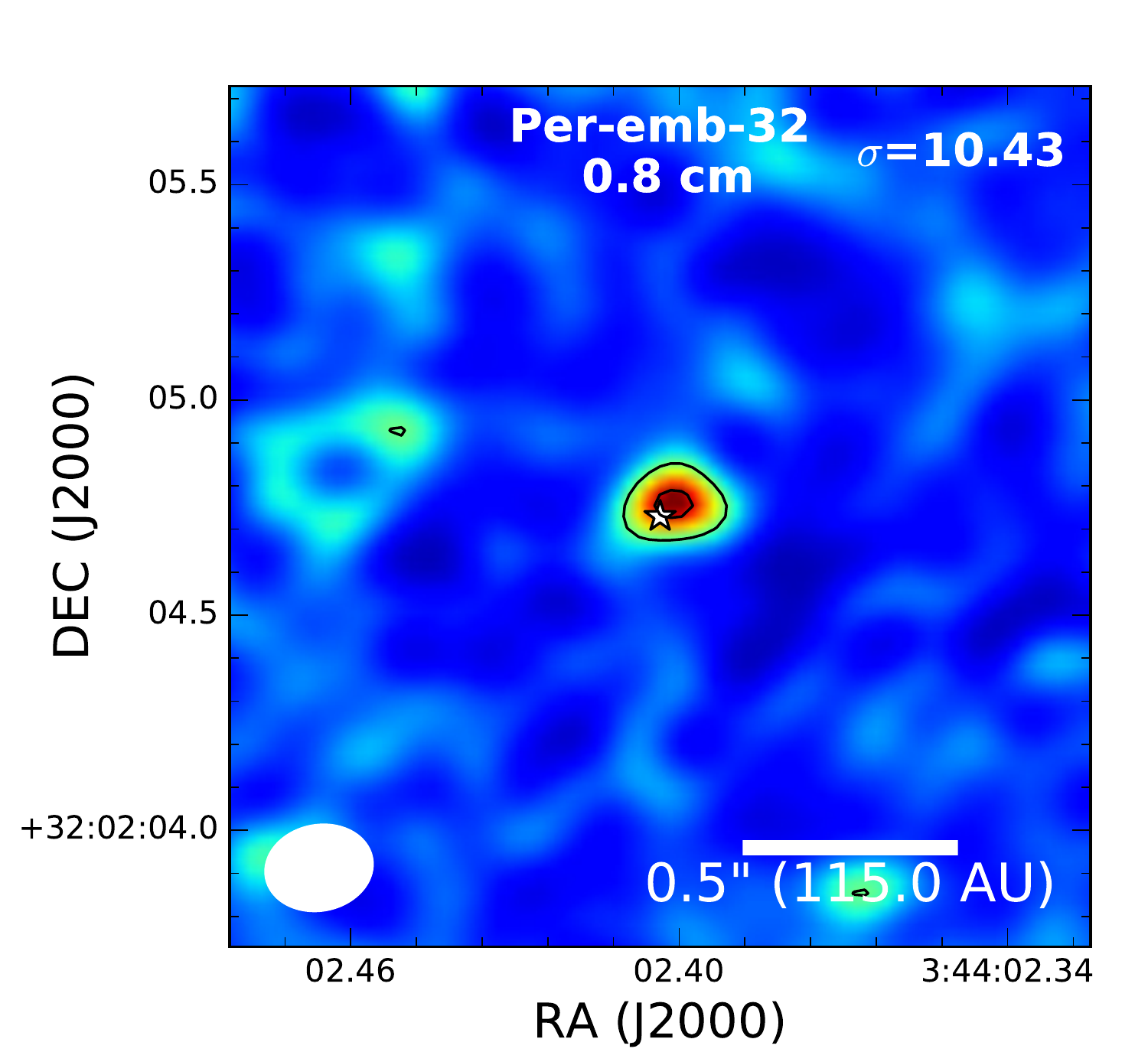}
  \includegraphics[width=0.24\linewidth]{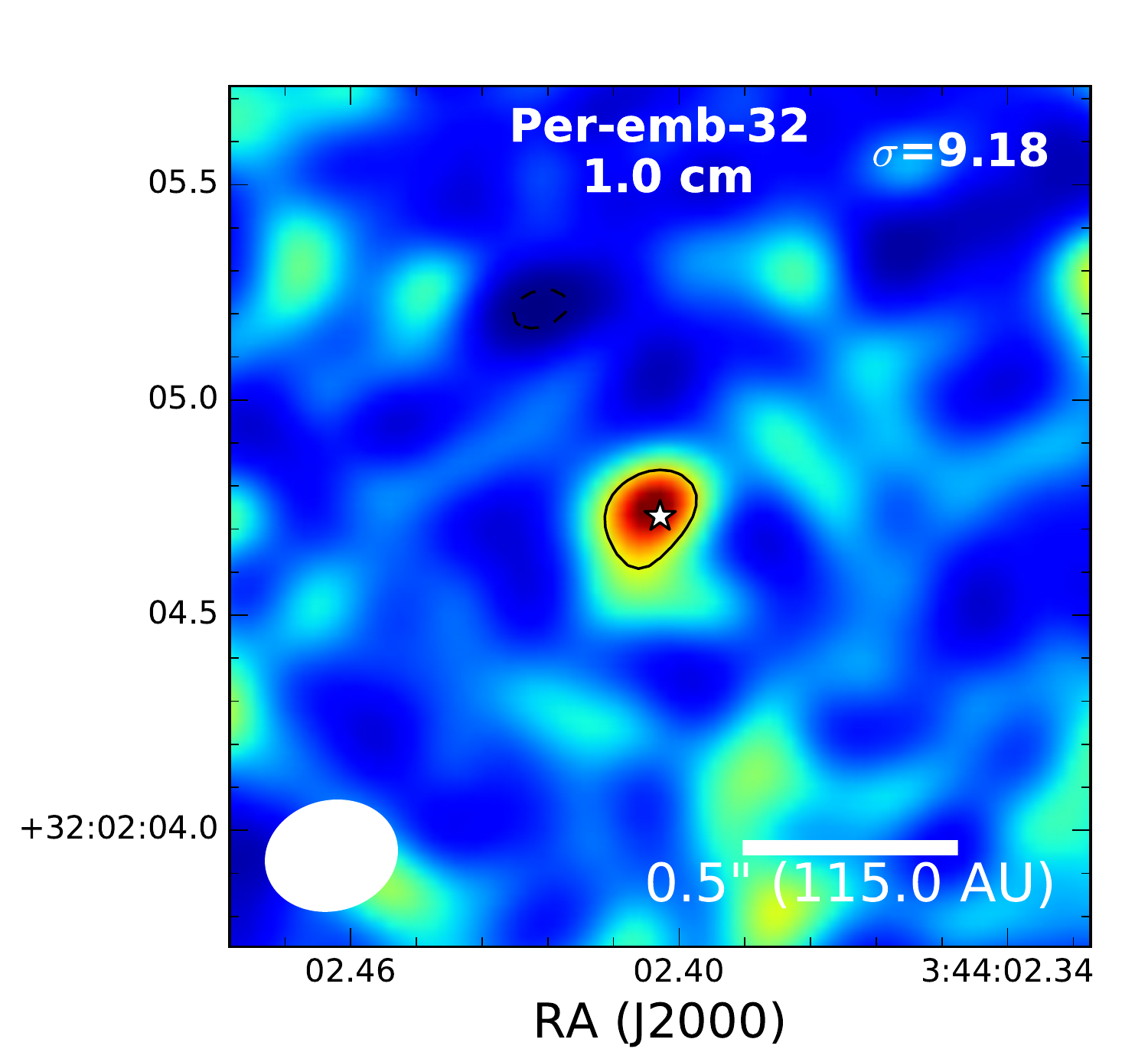}
  \includegraphics[width=0.24\linewidth]{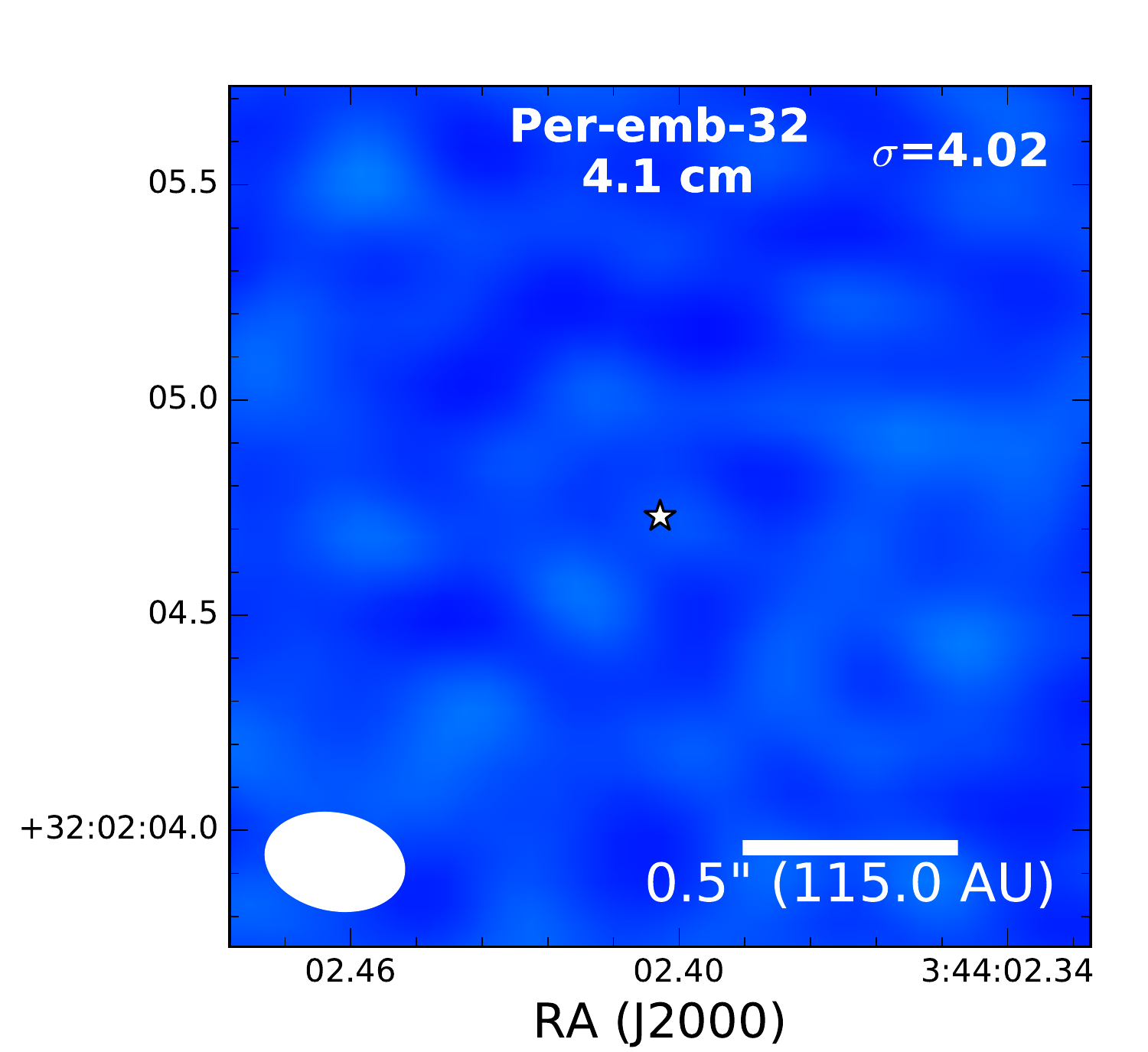}
  \includegraphics[width=0.24\linewidth]{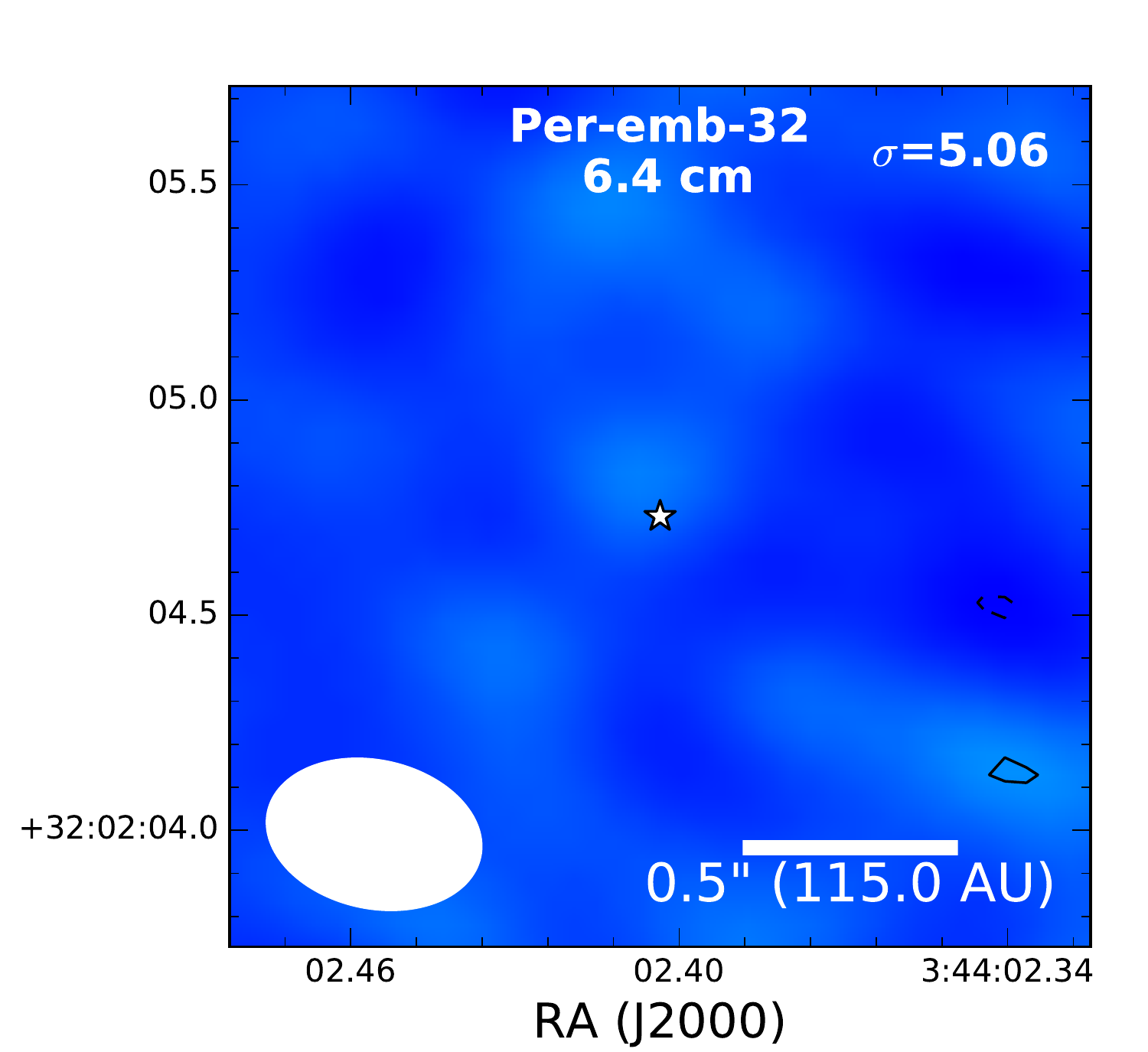}

  \includegraphics[width=0.24\linewidth]{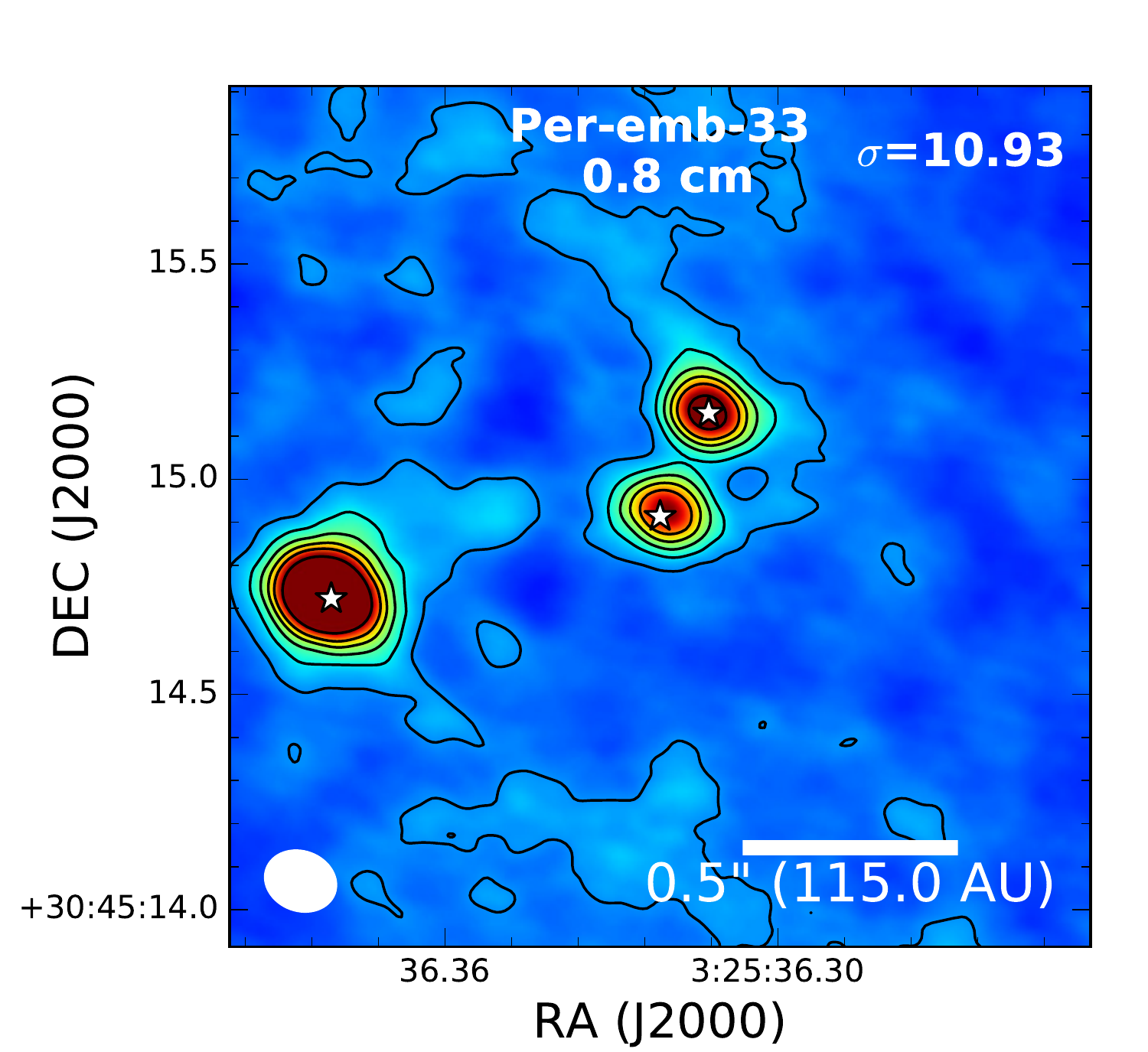}
  \includegraphics[width=0.24\linewidth]{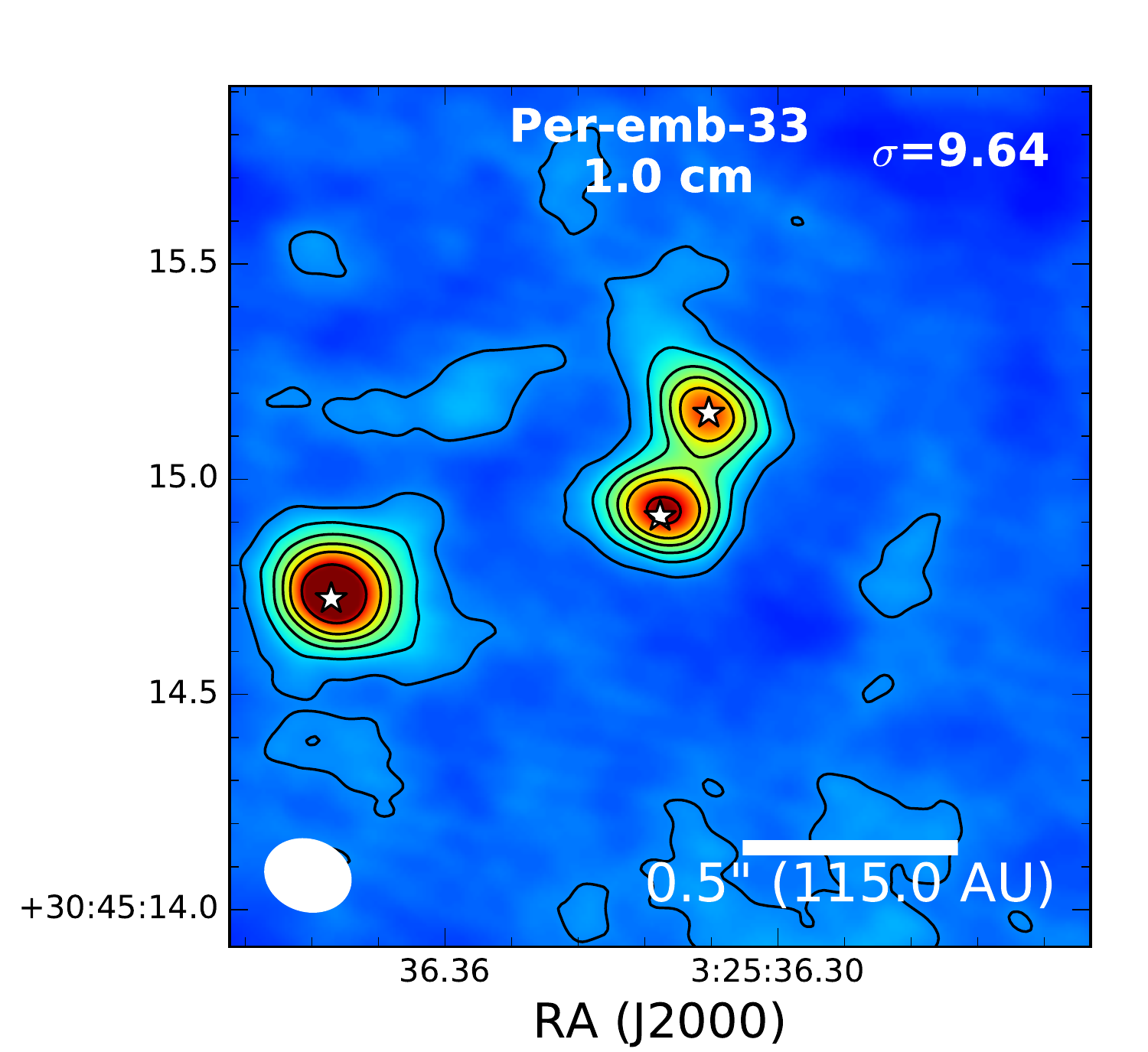}
  \includegraphics[width=0.24\linewidth]{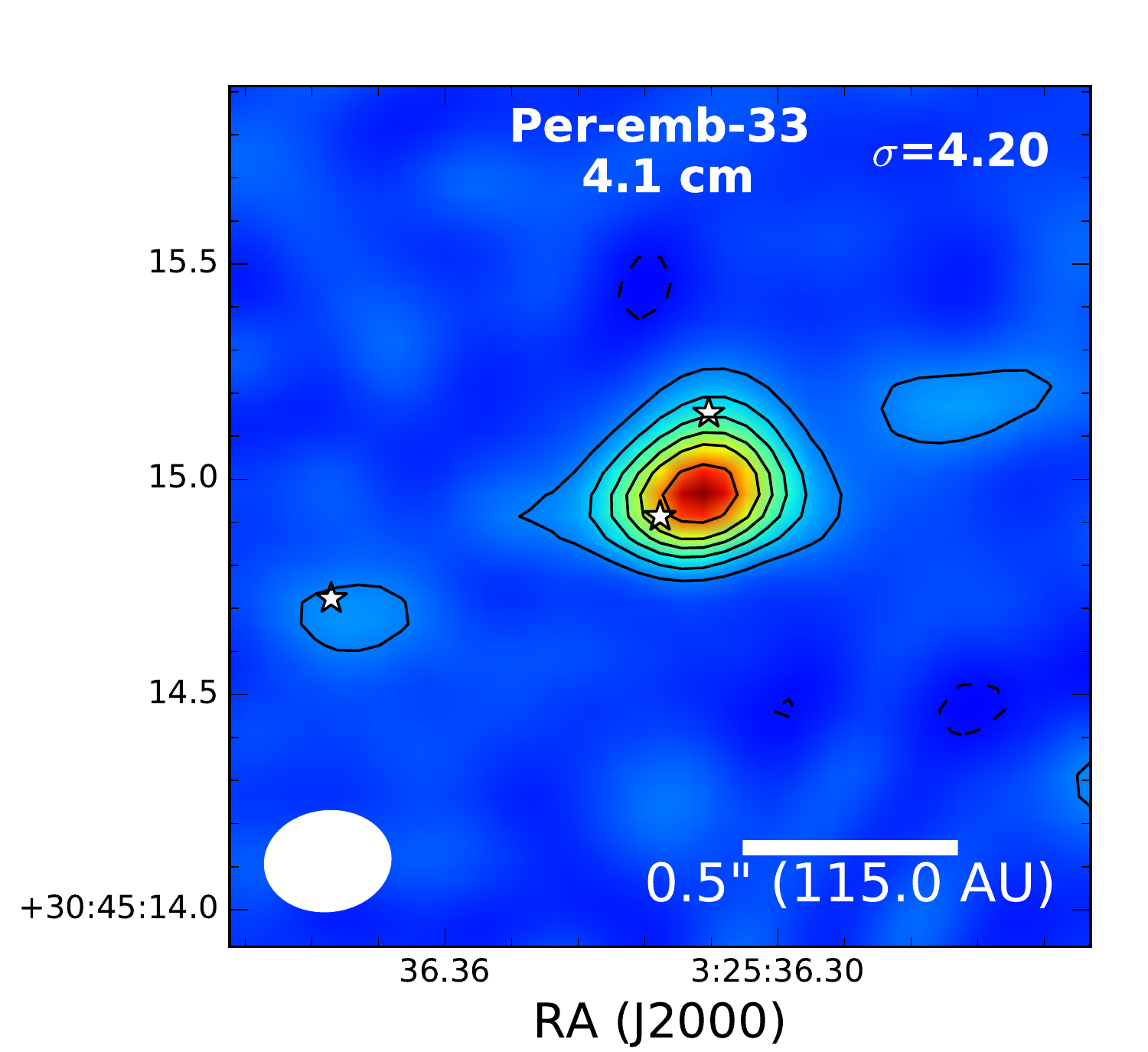}
  \includegraphics[width=0.24\linewidth]{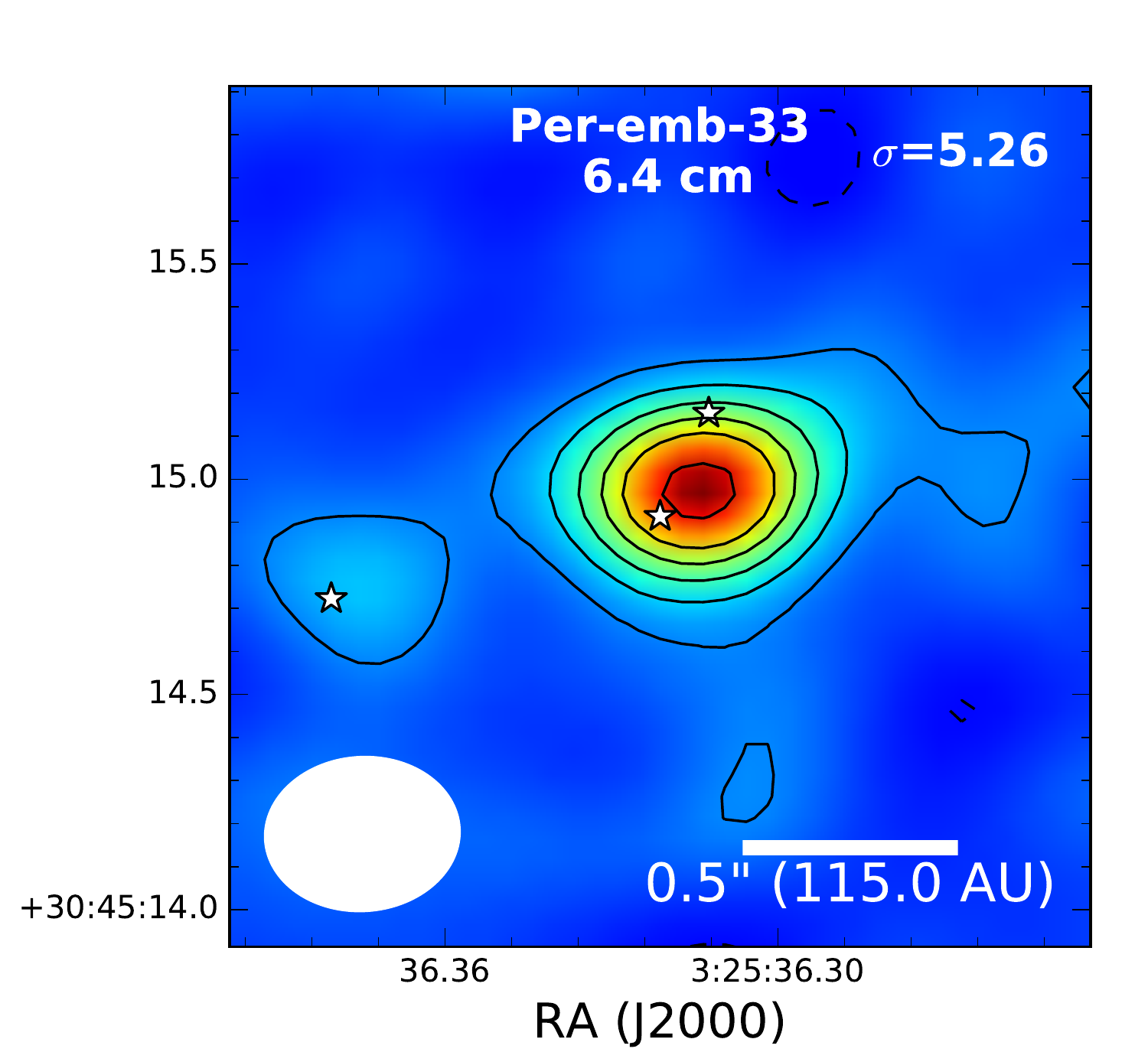}

\end{figure}

\begin{figure}

  \includegraphics[width=0.24\linewidth]{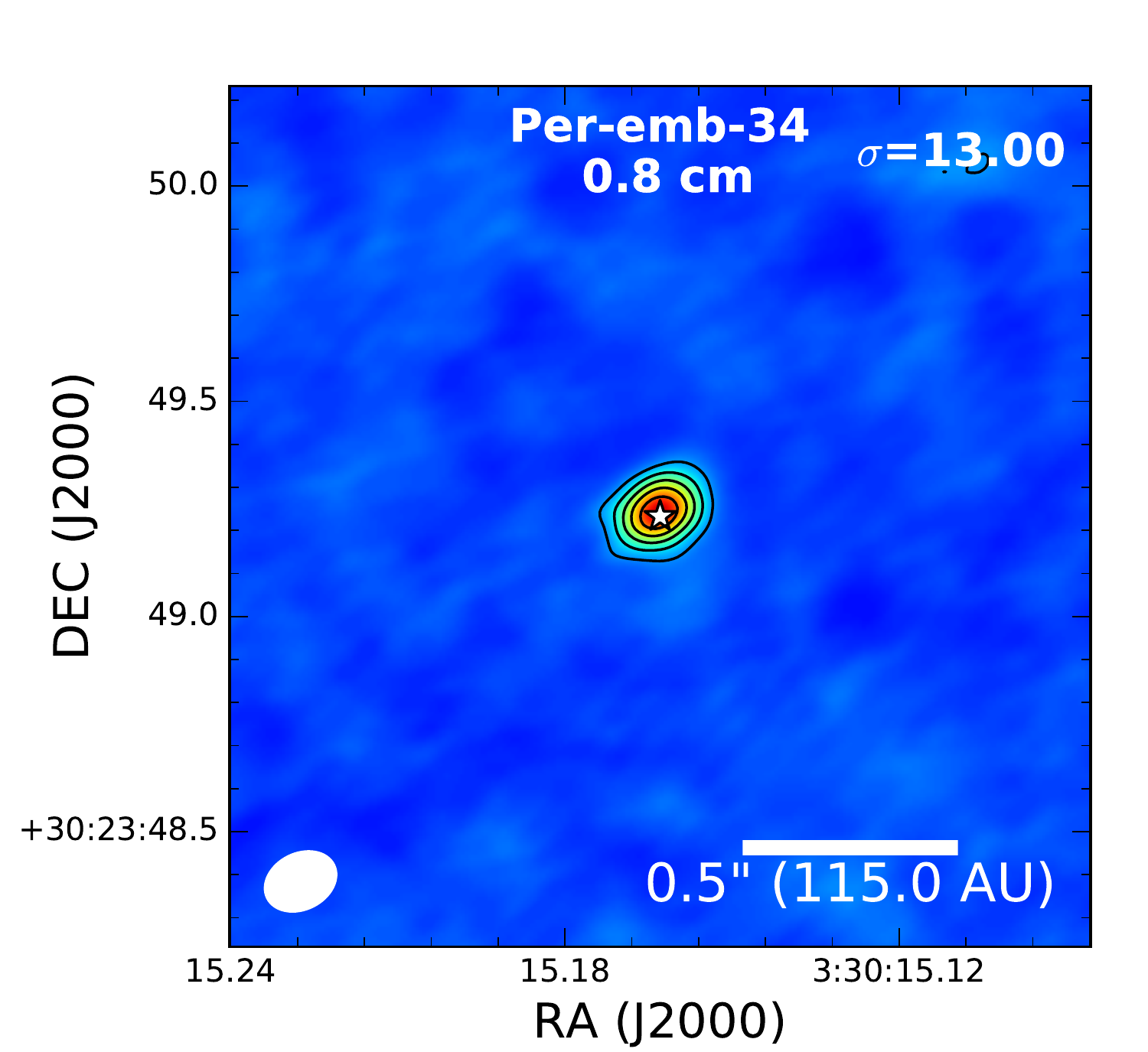}
  \includegraphics[width=0.24\linewidth]{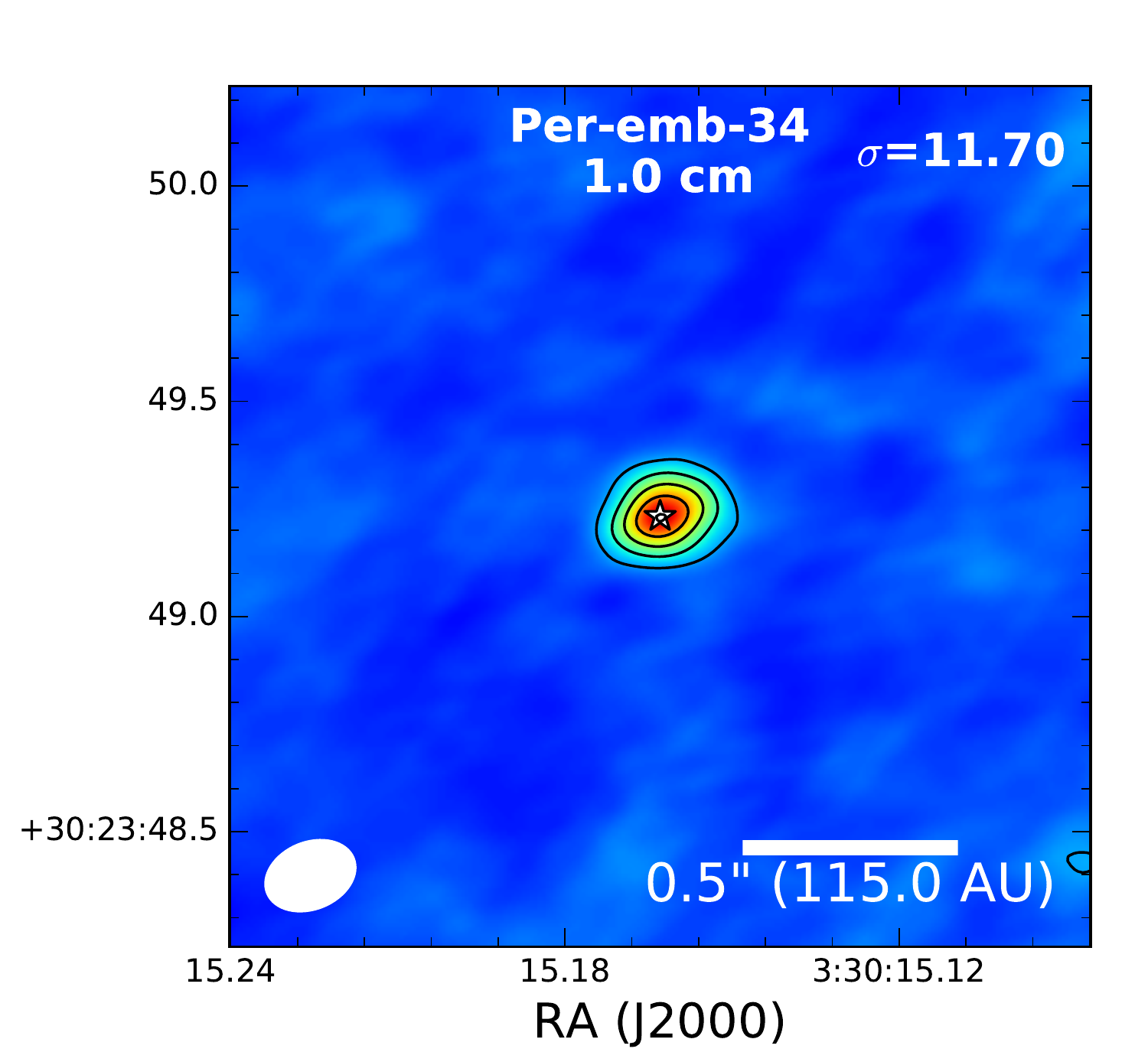}
  \includegraphics[width=0.24\linewidth]{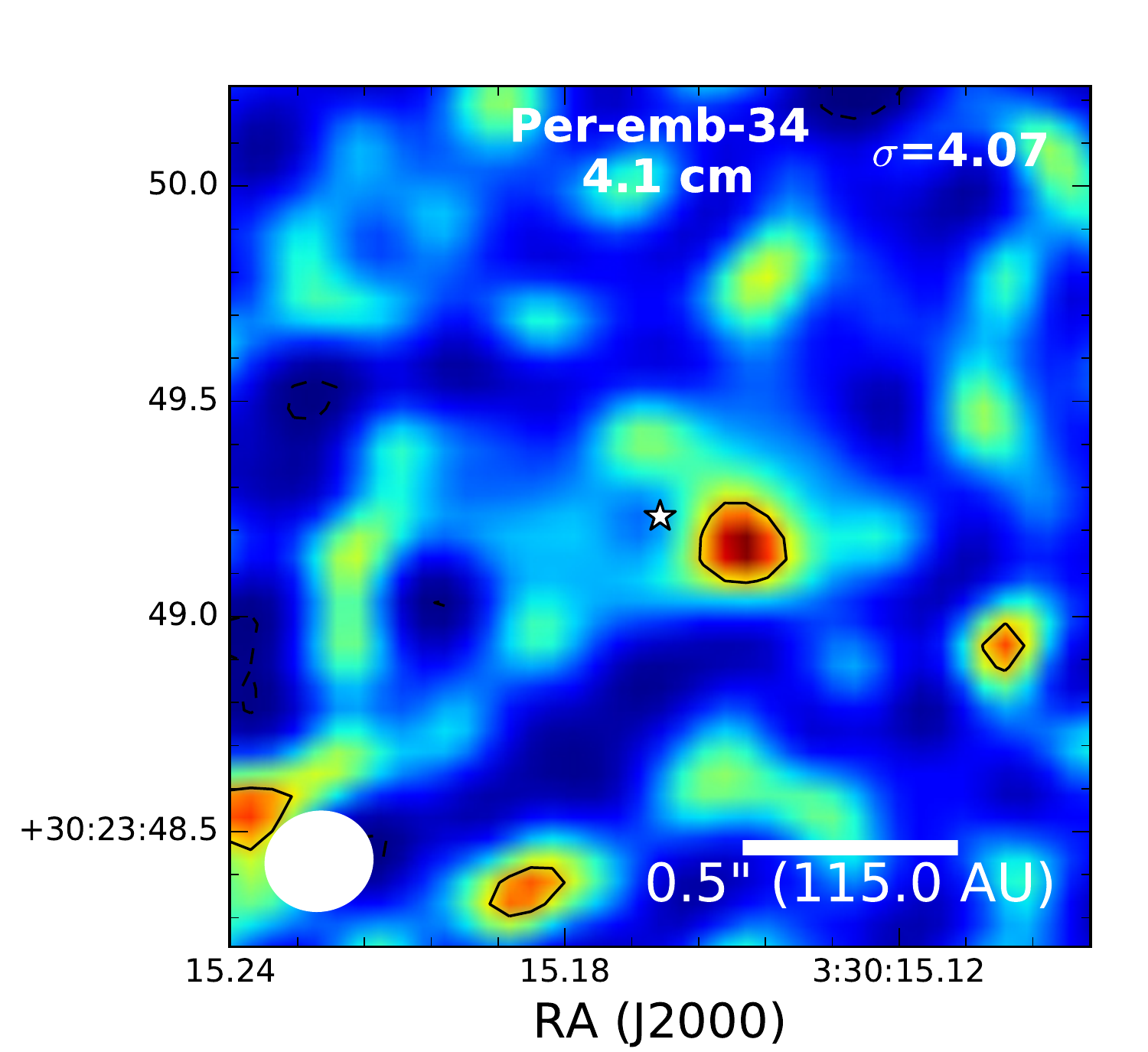}
  \includegraphics[width=0.24\linewidth]{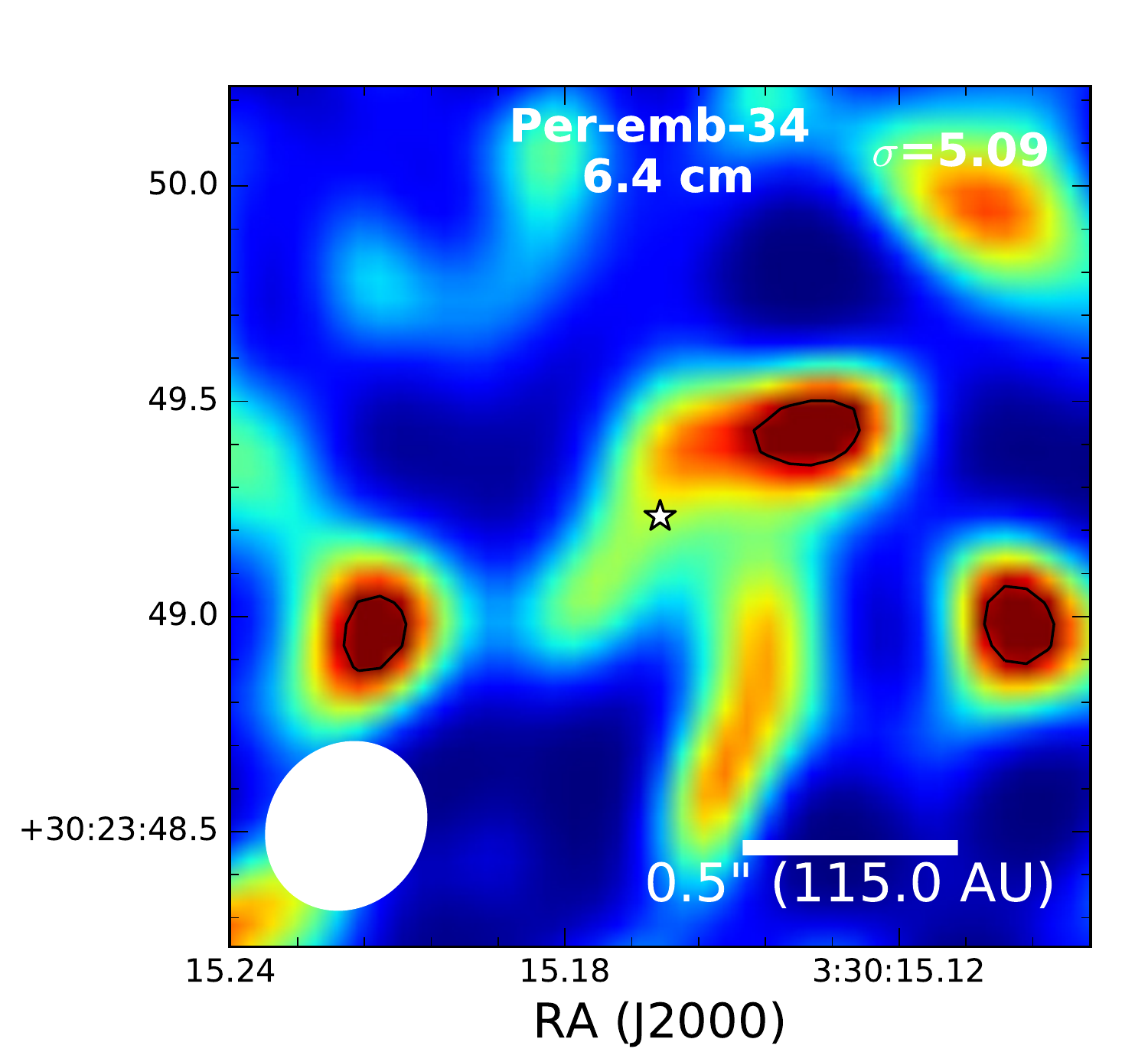}

  \includegraphics[width=0.24\linewidth]{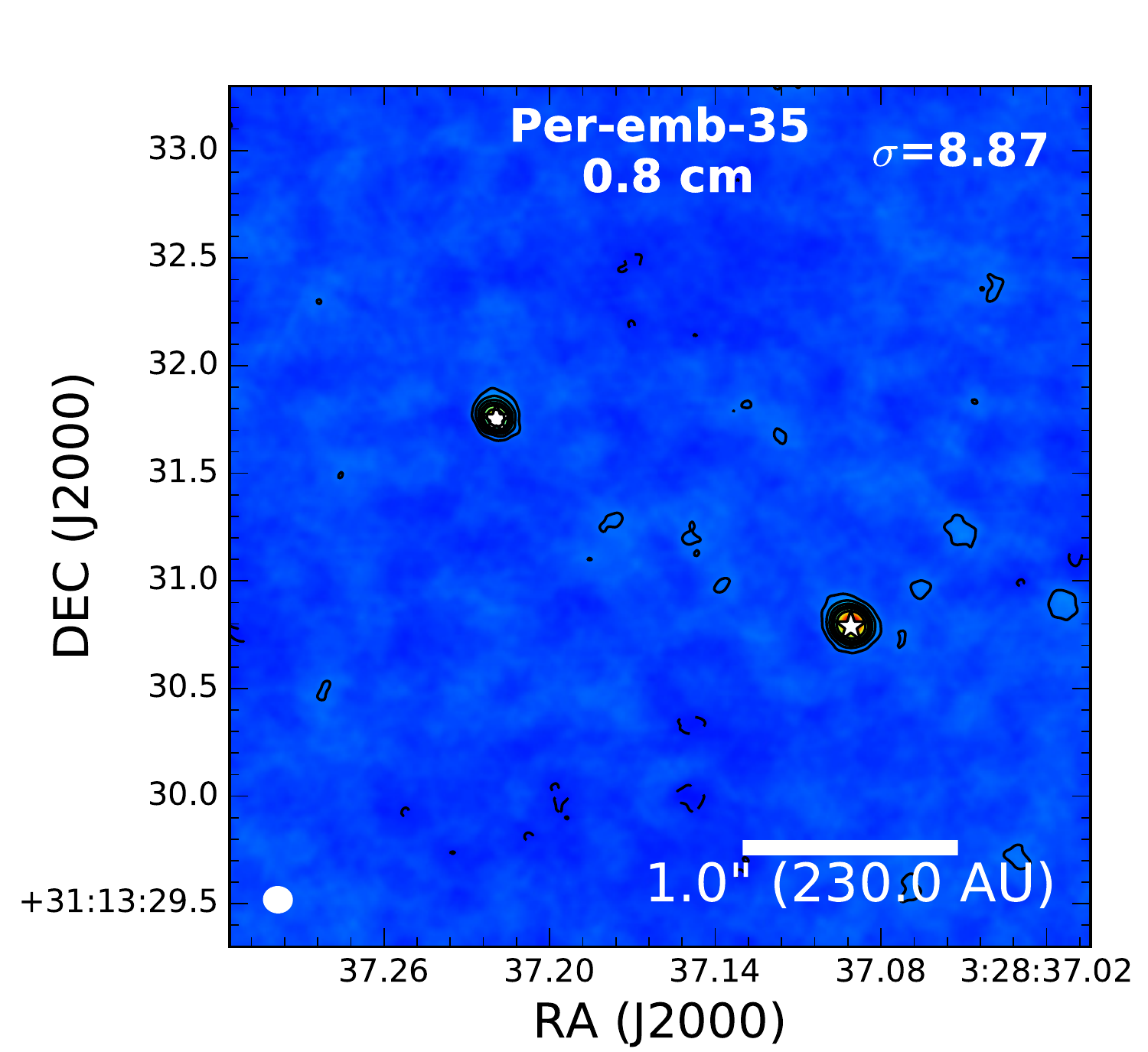}
  \includegraphics[width=0.24\linewidth]{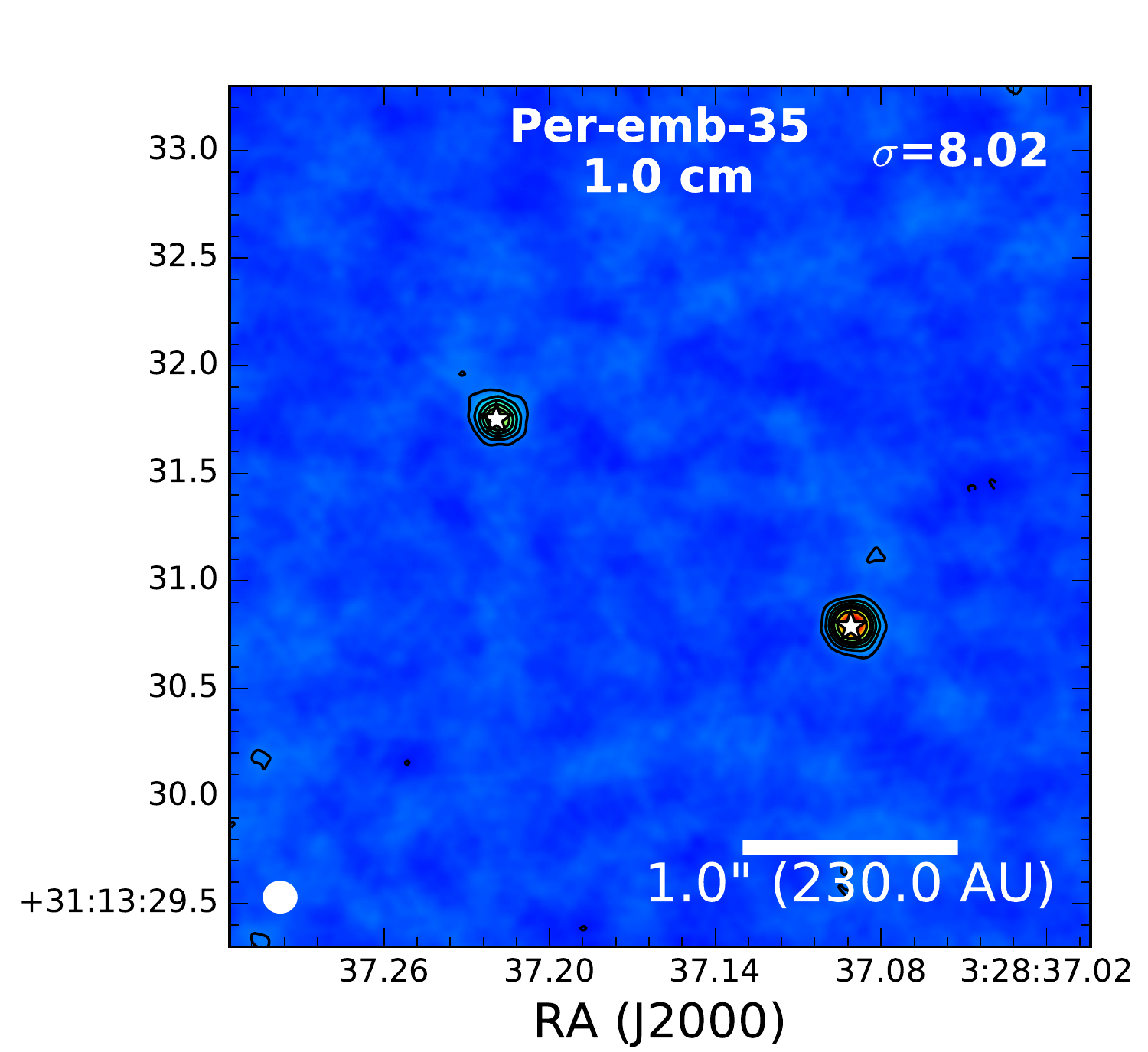}
  \includegraphics[width=0.24\linewidth]{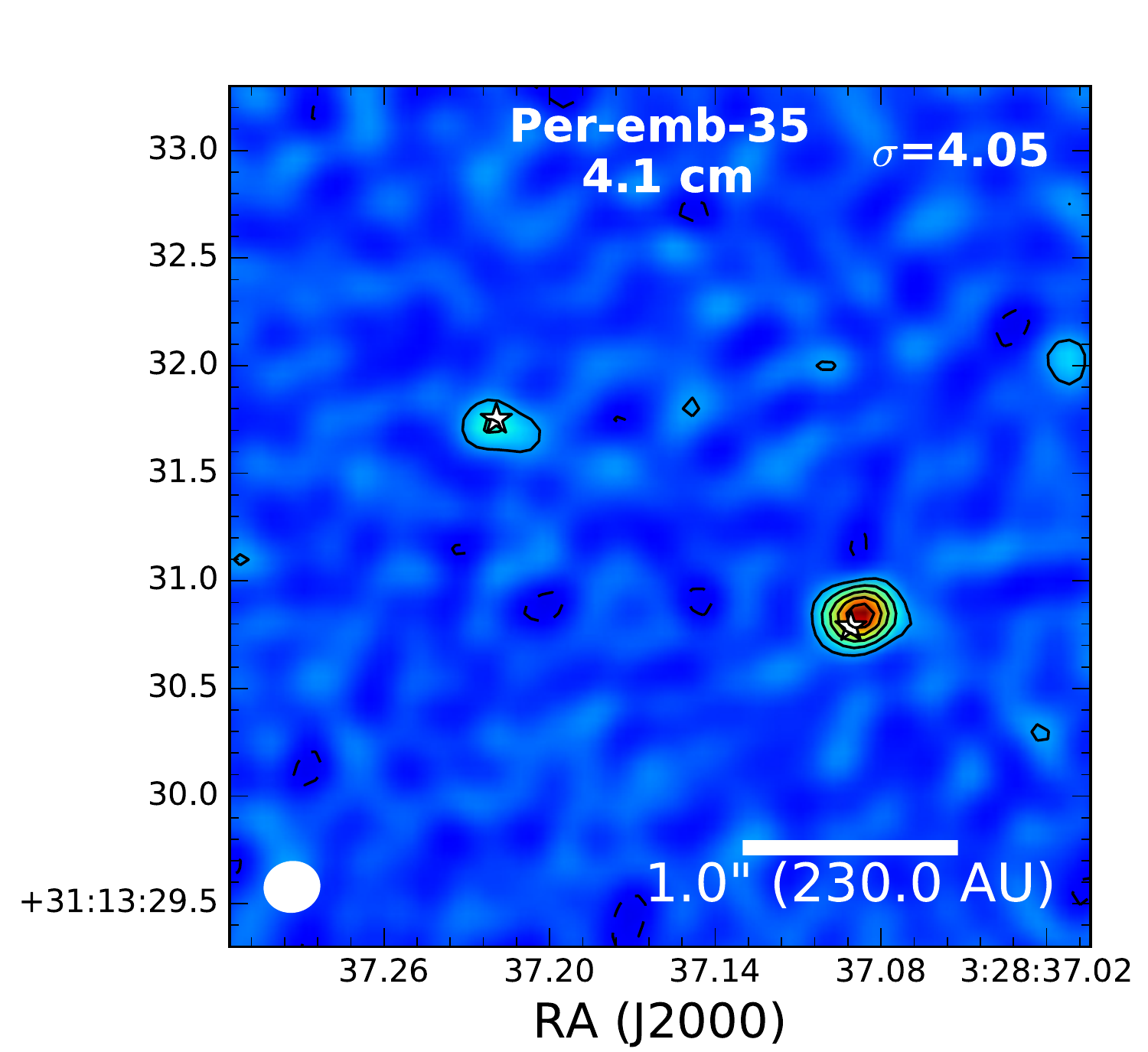}
  \includegraphics[width=0.24\linewidth]{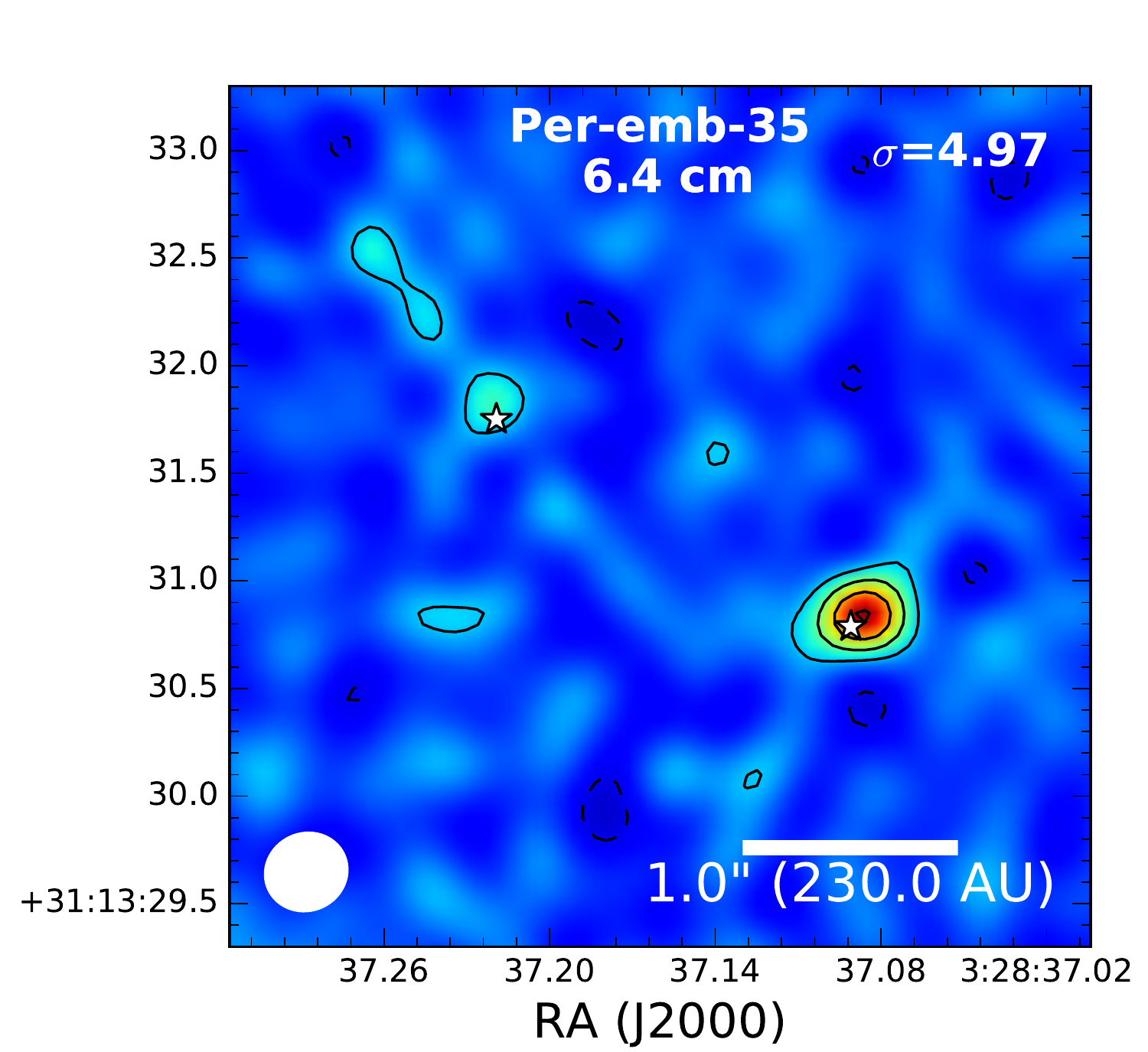}

  \includegraphics[width=0.24\linewidth]{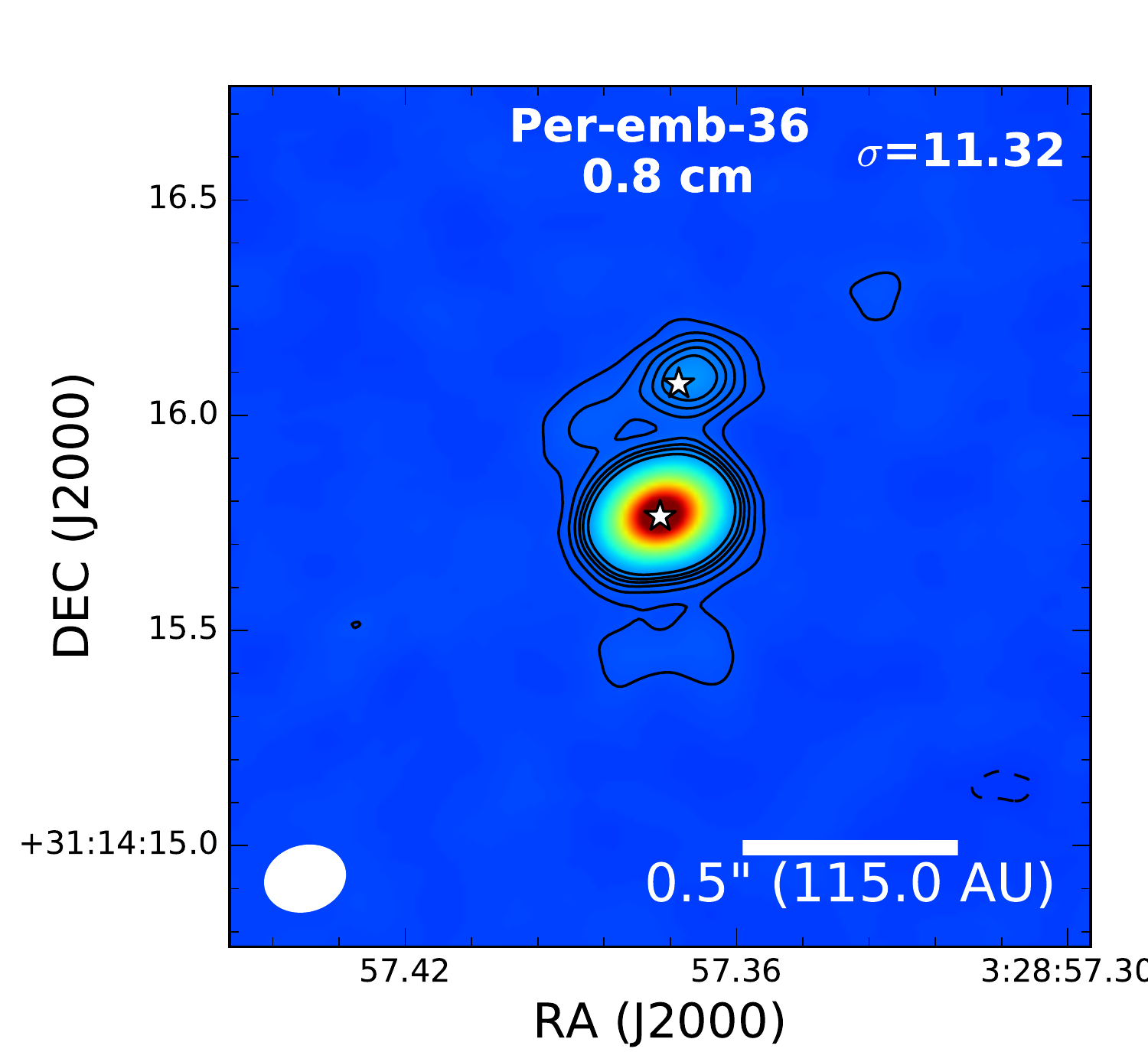}
  \includegraphics[width=0.24\linewidth]{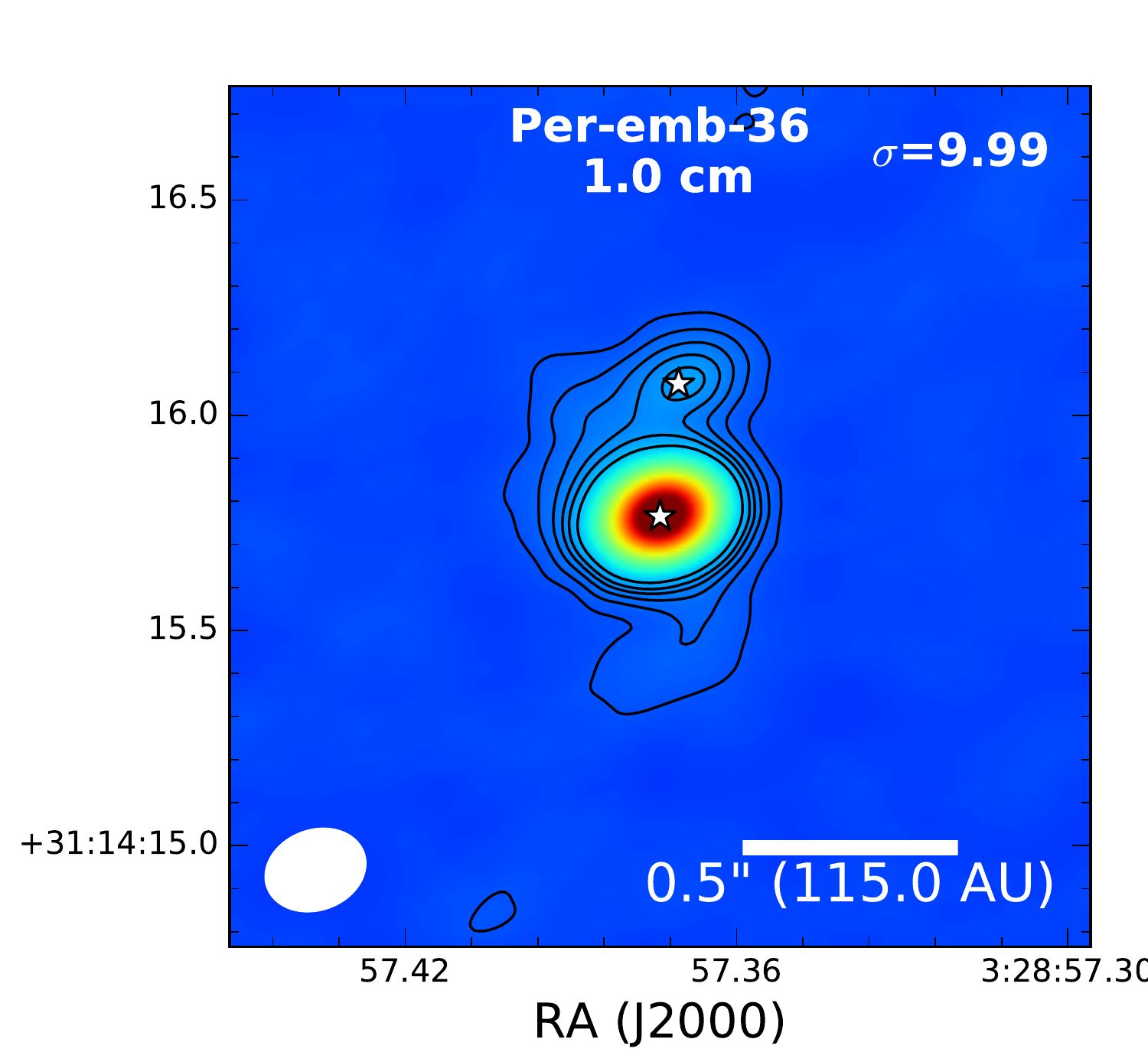}
  \includegraphics[width=0.24\linewidth]{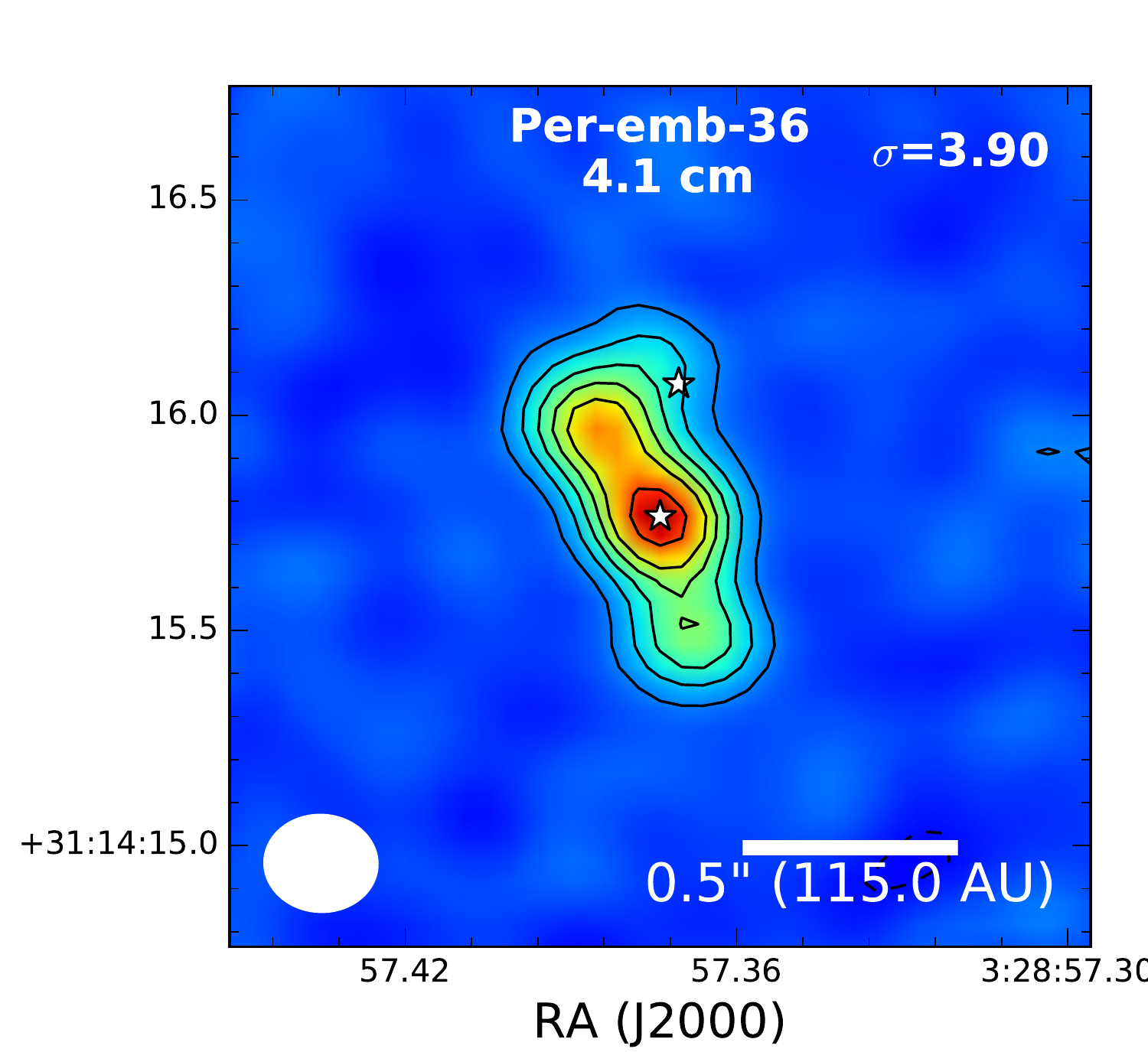}
  \includegraphics[width=0.24\linewidth]{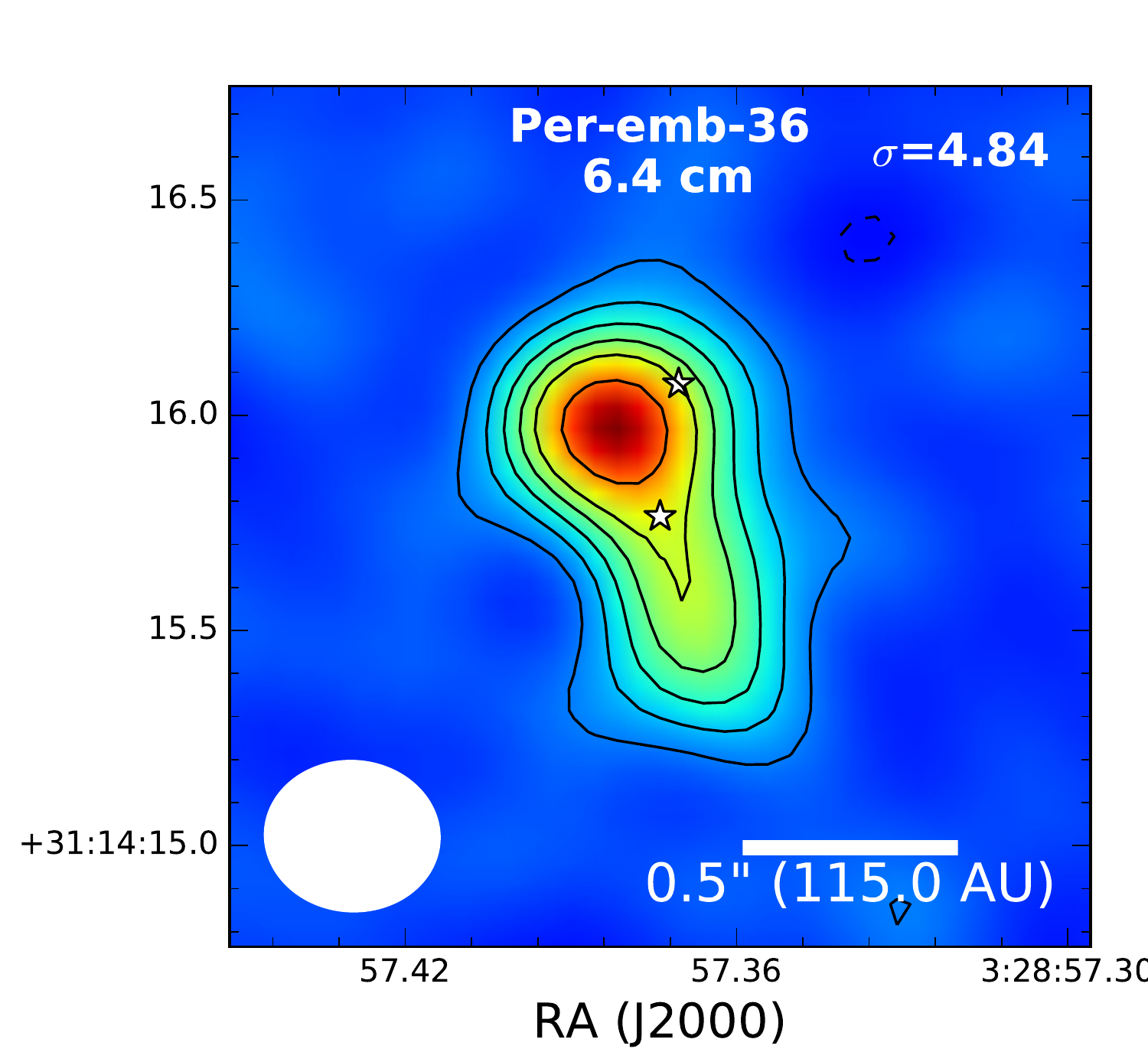}

  \includegraphics[width=0.24\linewidth]{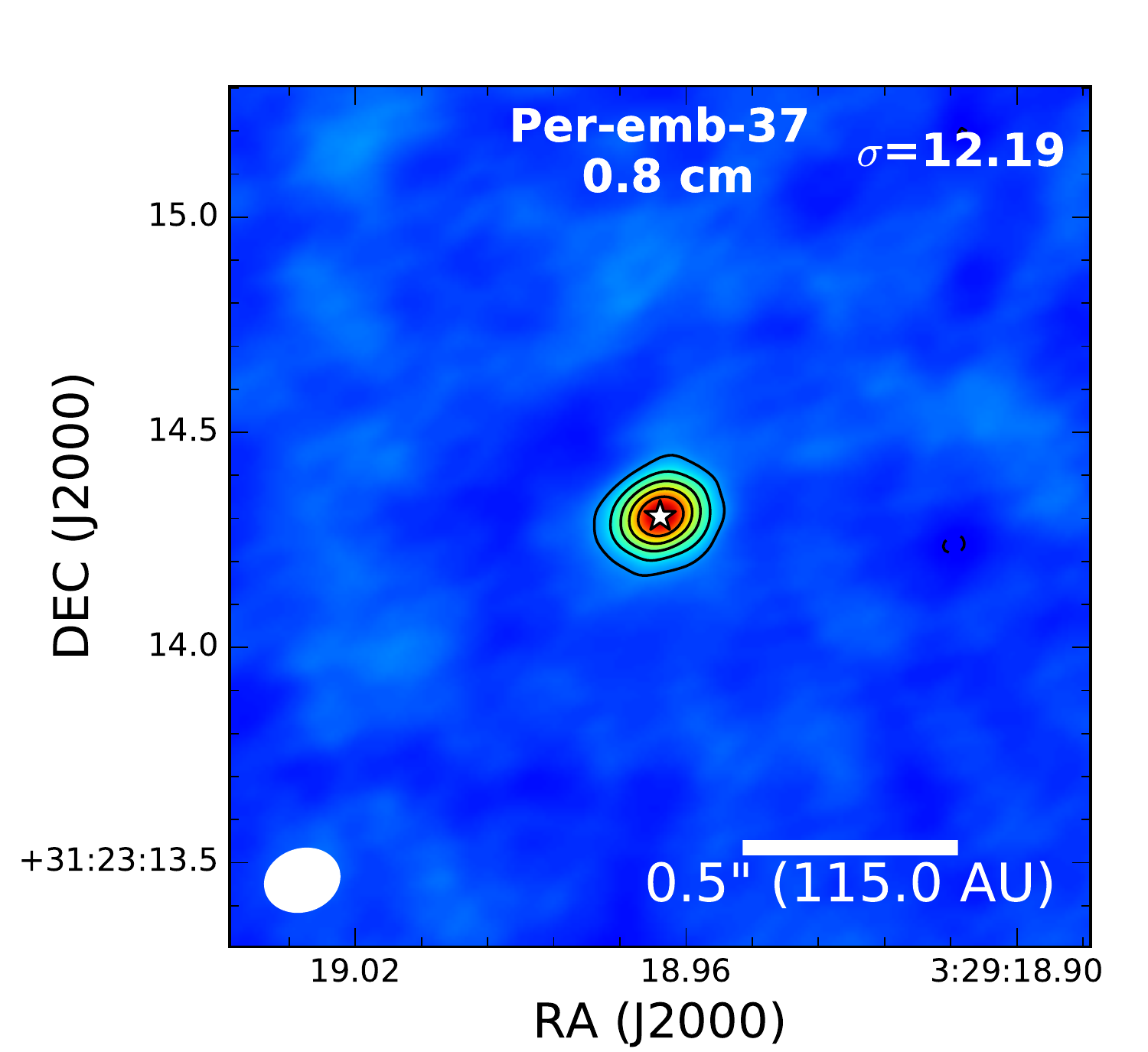}
  \includegraphics[width=0.24\linewidth]{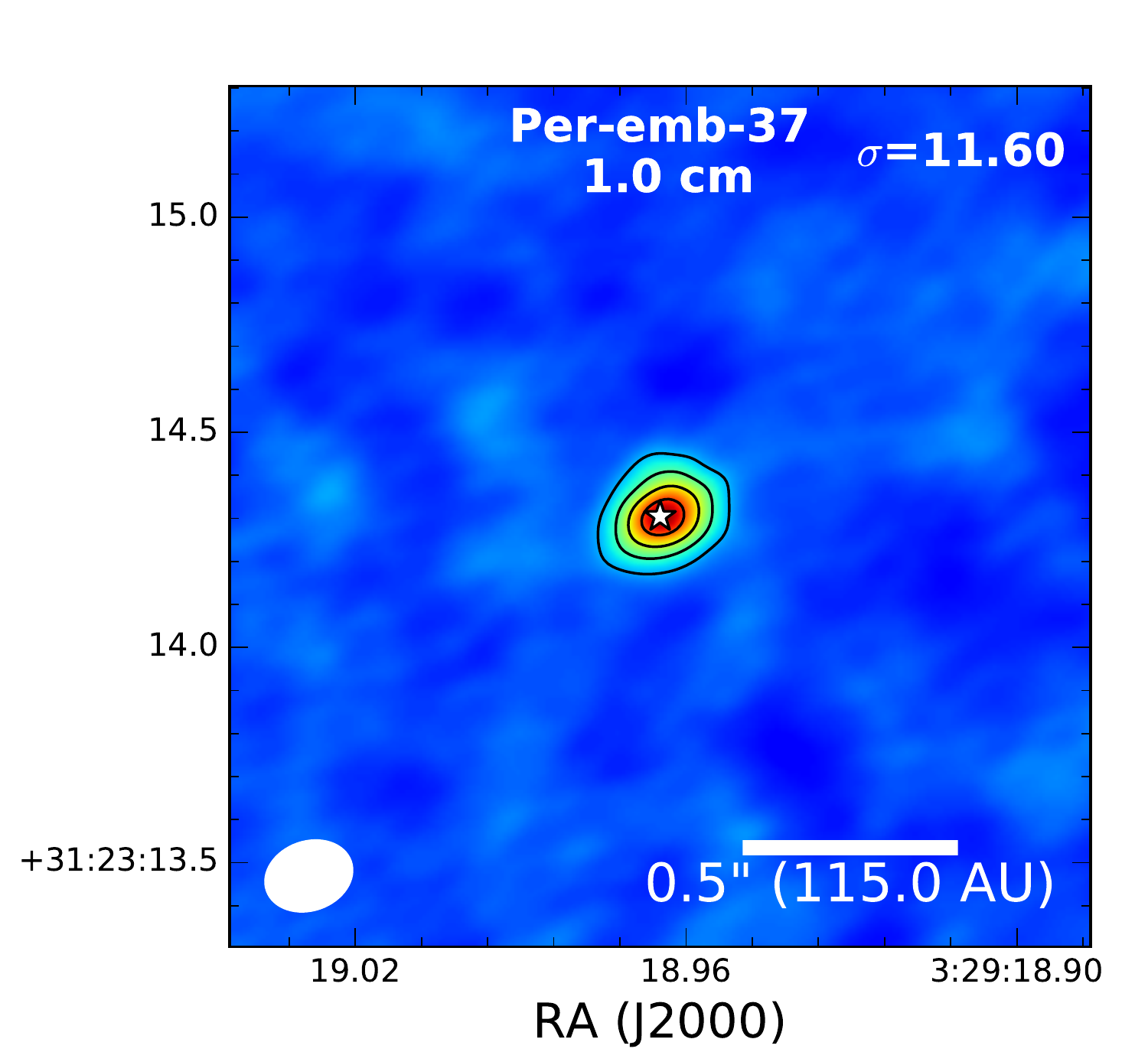}
  \includegraphics[width=0.24\linewidth]{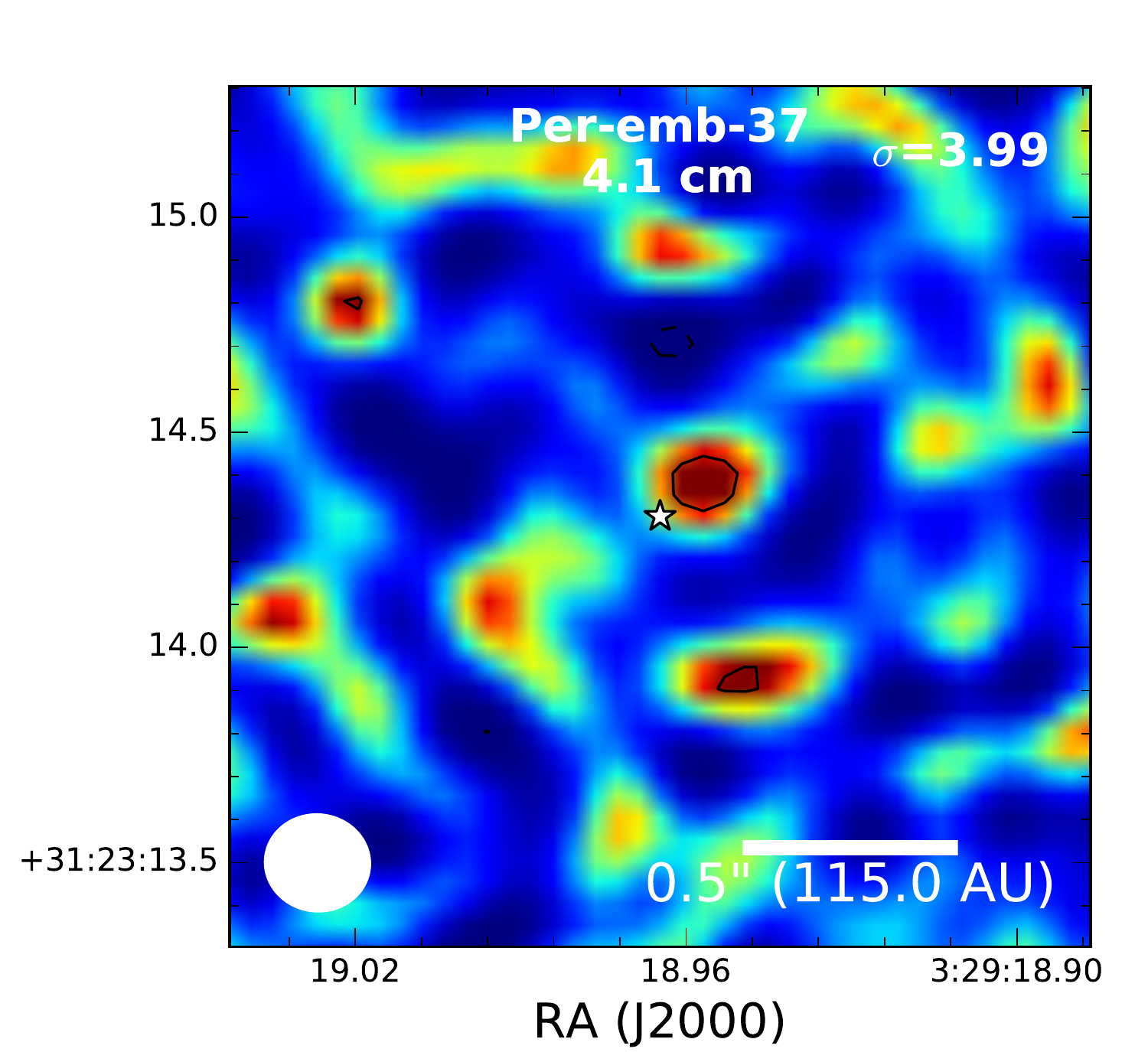}
  \includegraphics[width=0.24\linewidth]{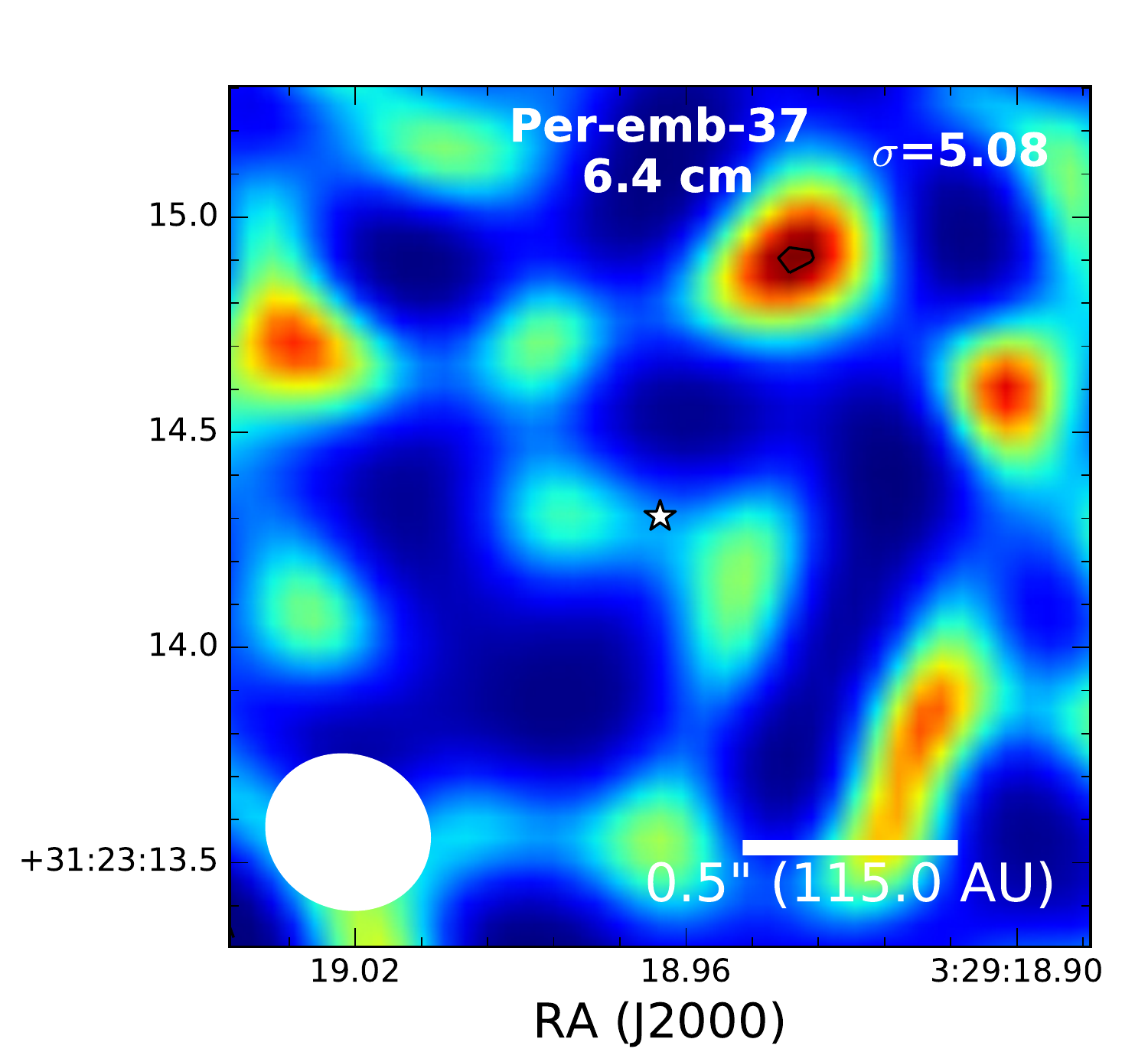}

  \includegraphics[width=0.24\linewidth]{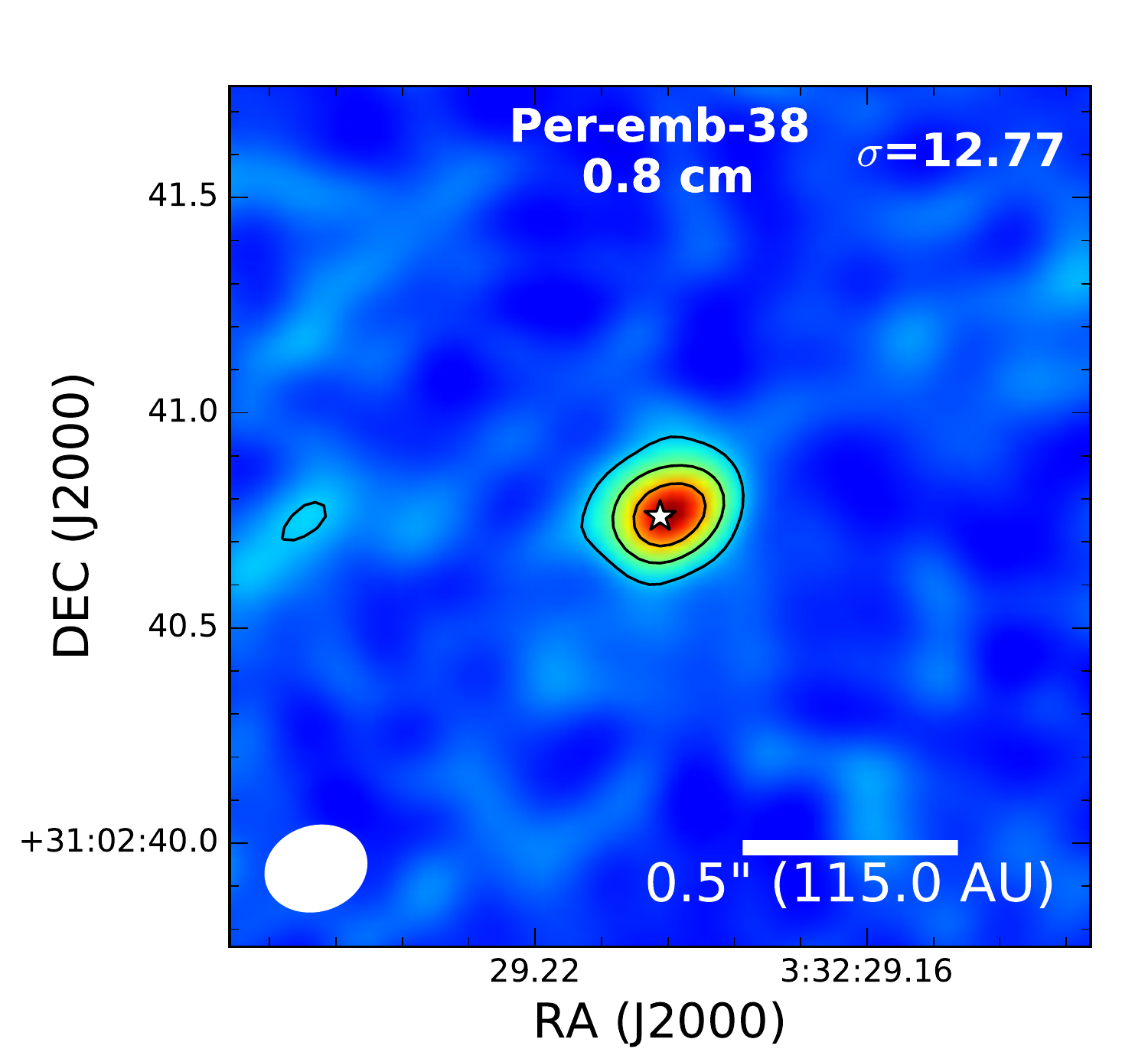}
  \includegraphics[width=0.24\linewidth]{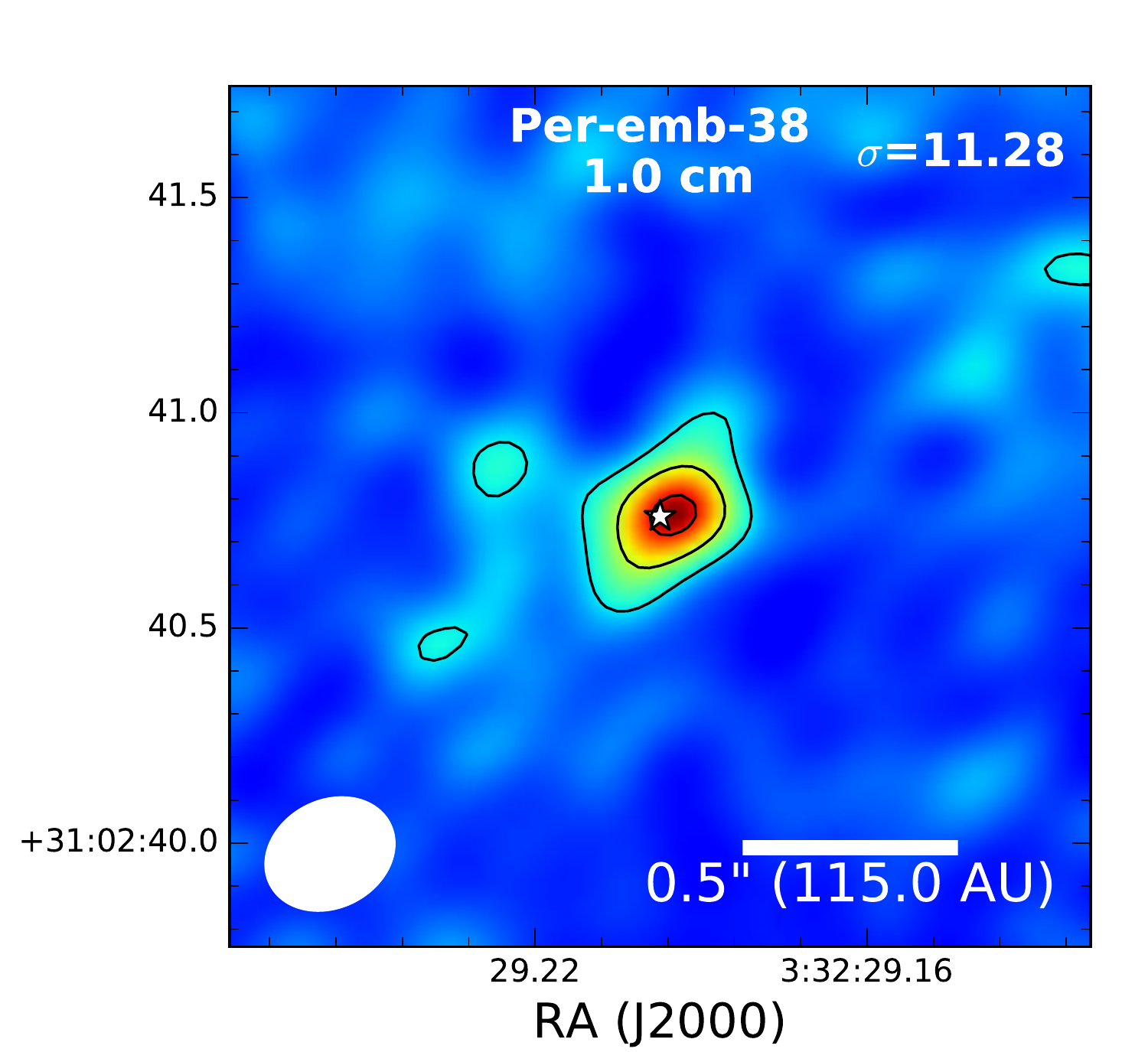}
  \includegraphics[width=0.24\linewidth]{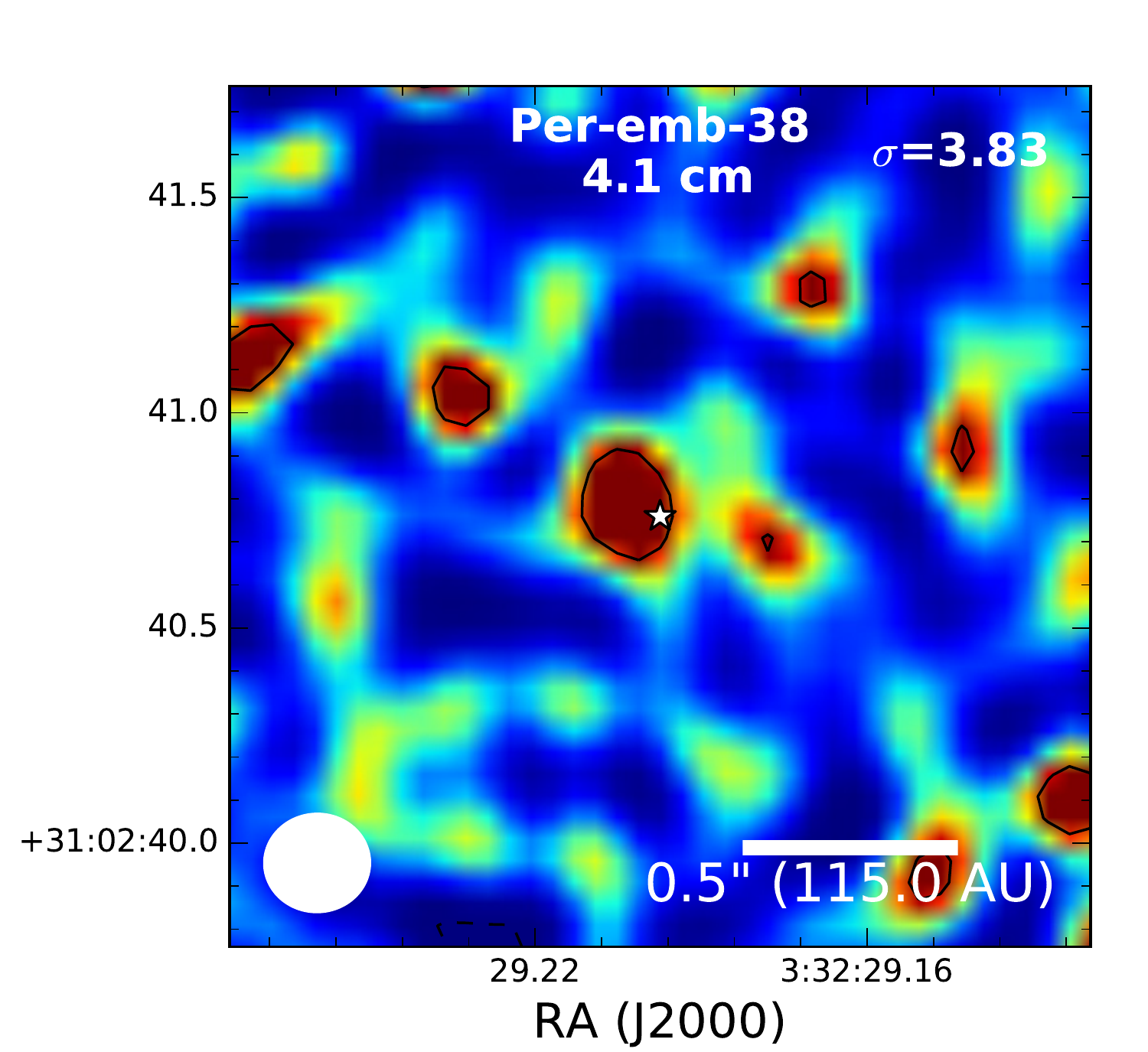}
  \includegraphics[width=0.24\linewidth]{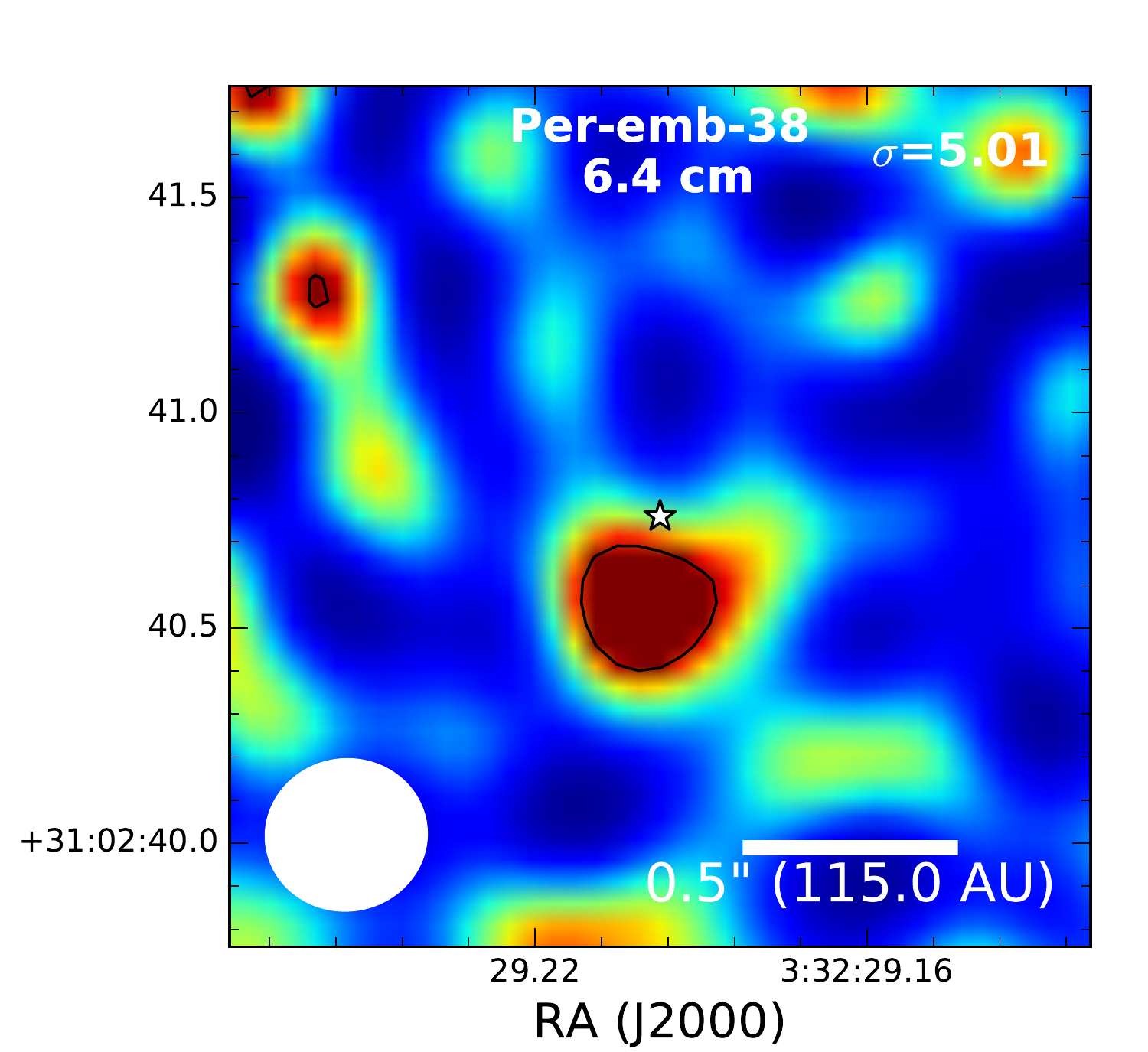}

  \includegraphics[width=0.24\linewidth]{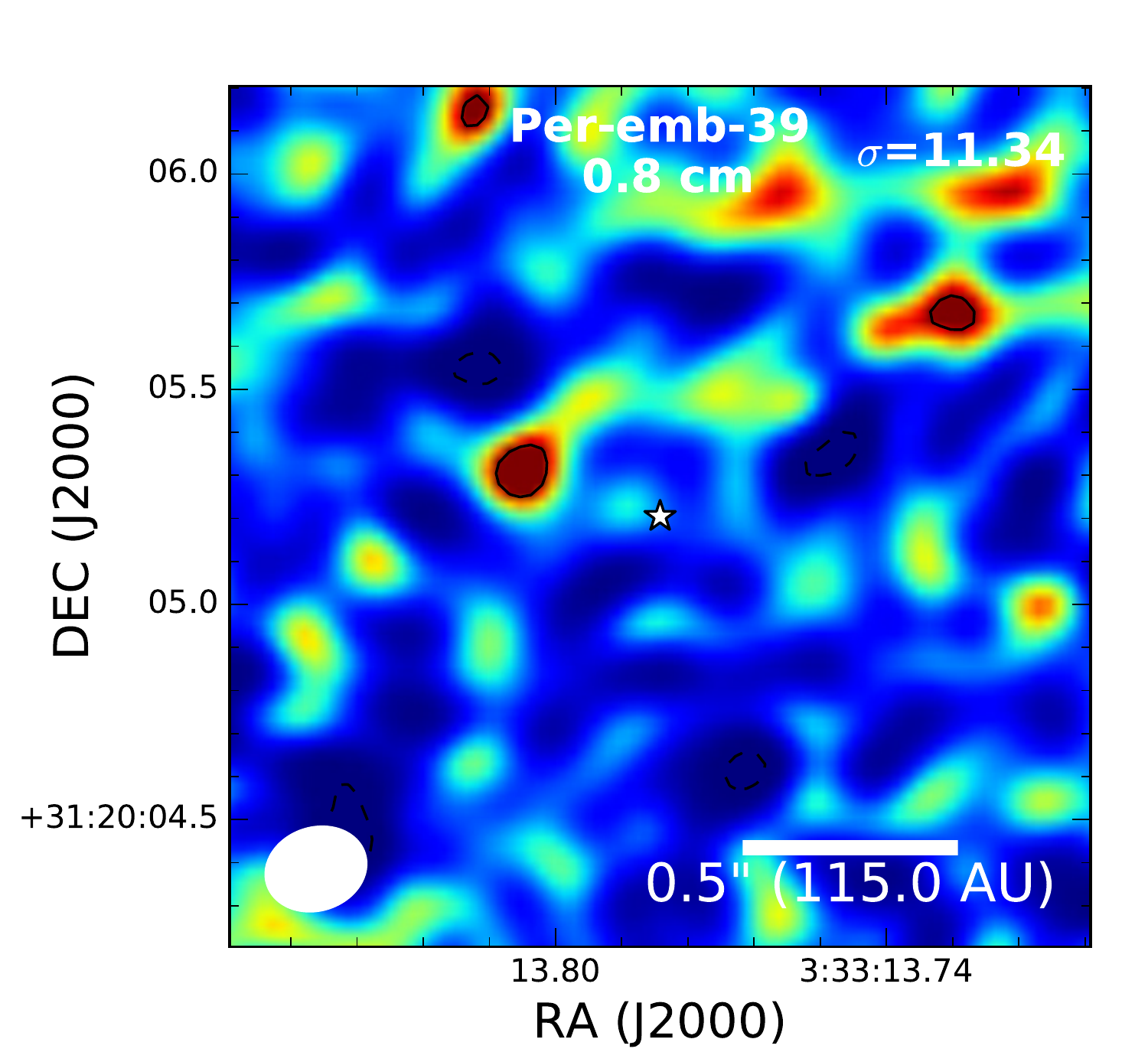}
  \includegraphics[width=0.24\linewidth]{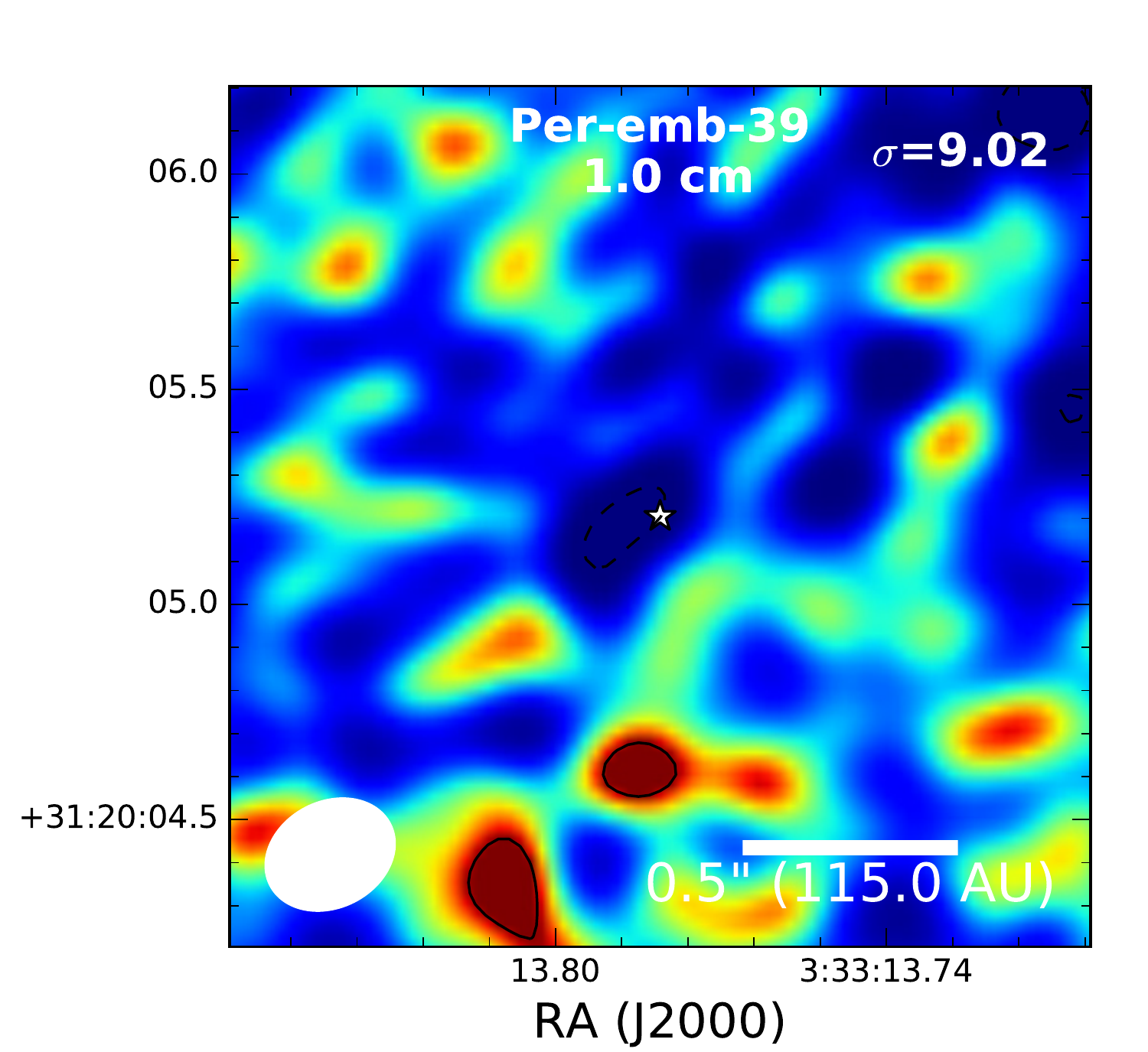}
  \includegraphics[width=0.24\linewidth]{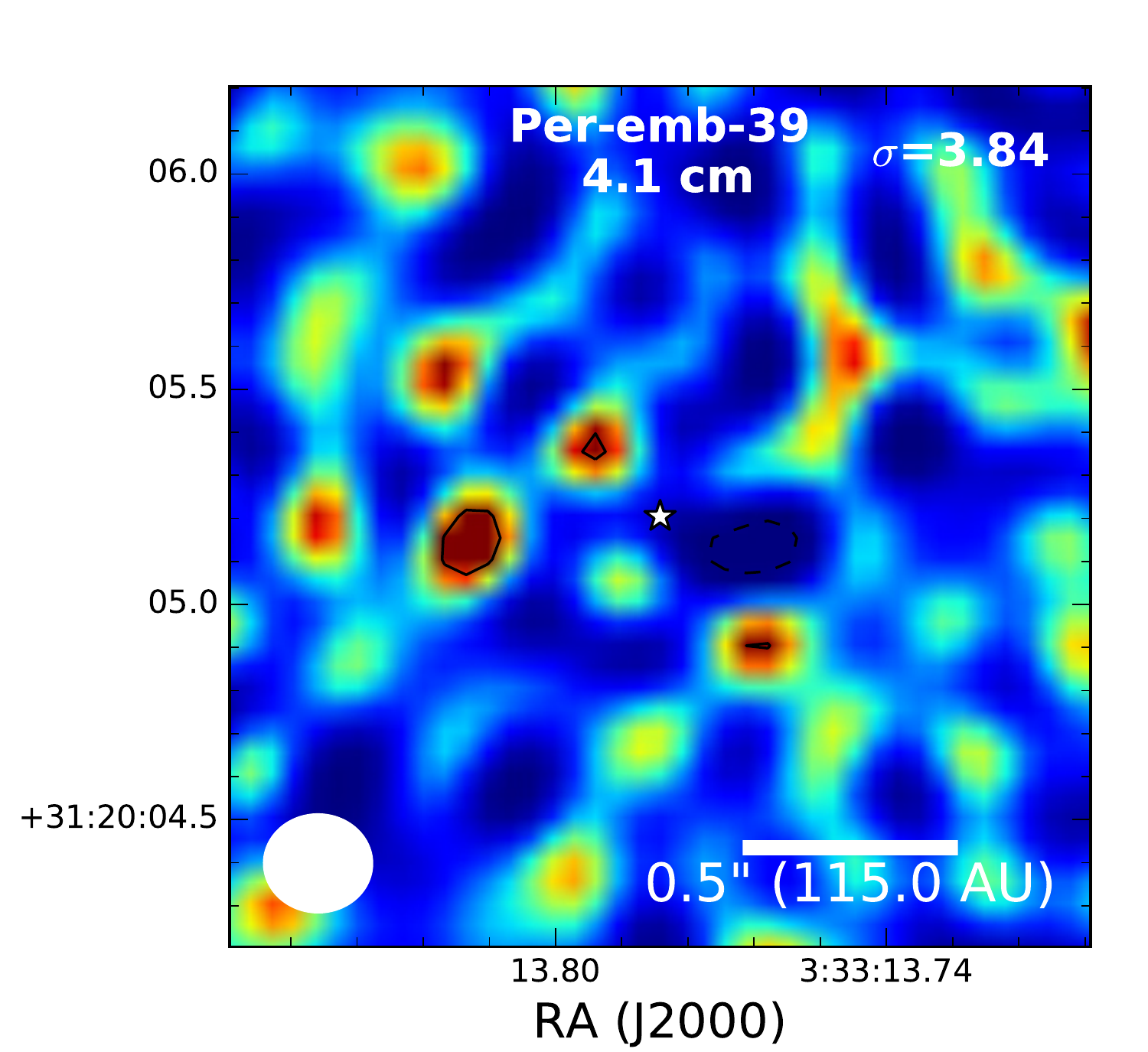}
  \includegraphics[width=0.24\linewidth]{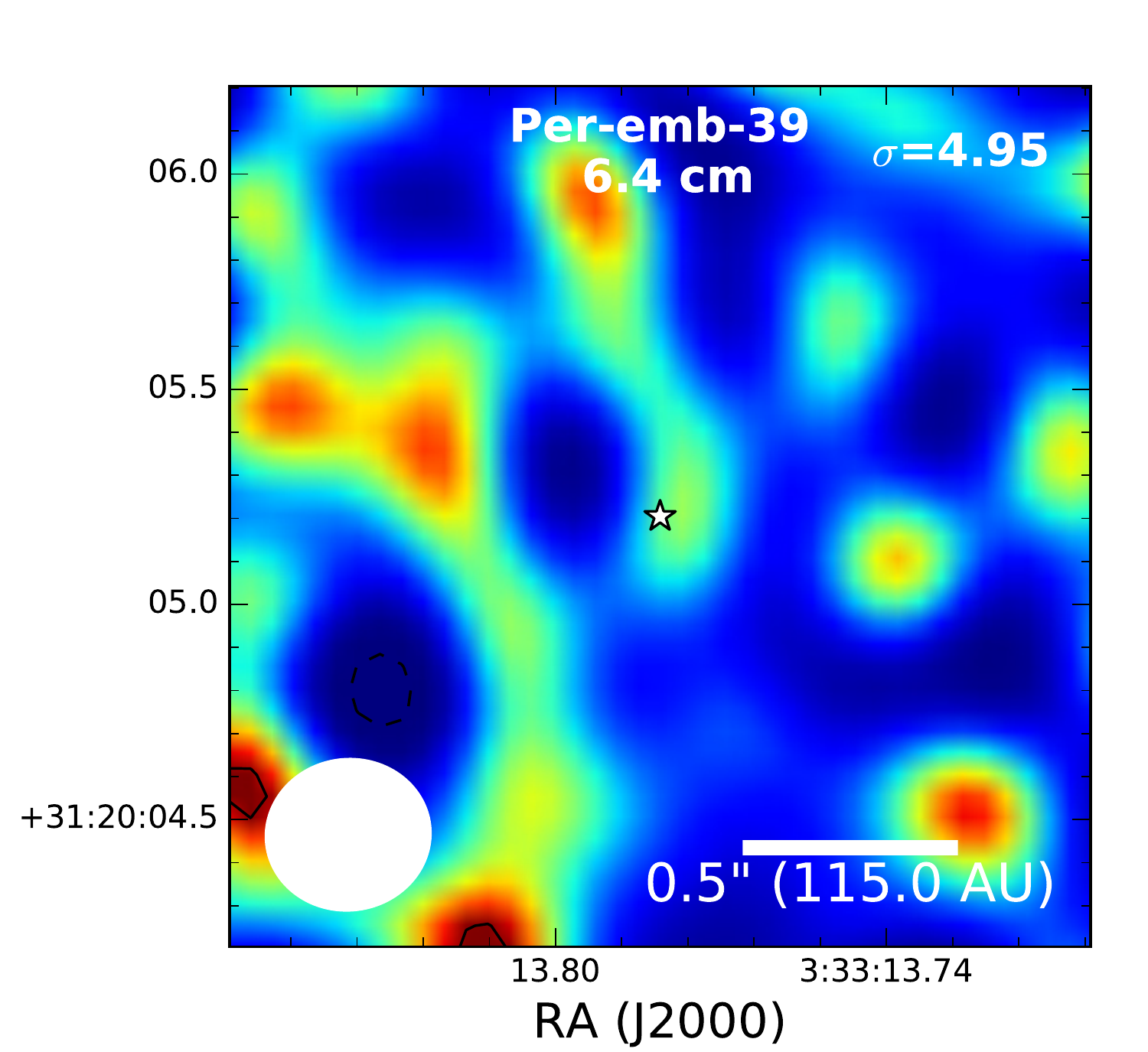}

\end{figure}

\begin{figure}

  \includegraphics[width=0.24\linewidth]{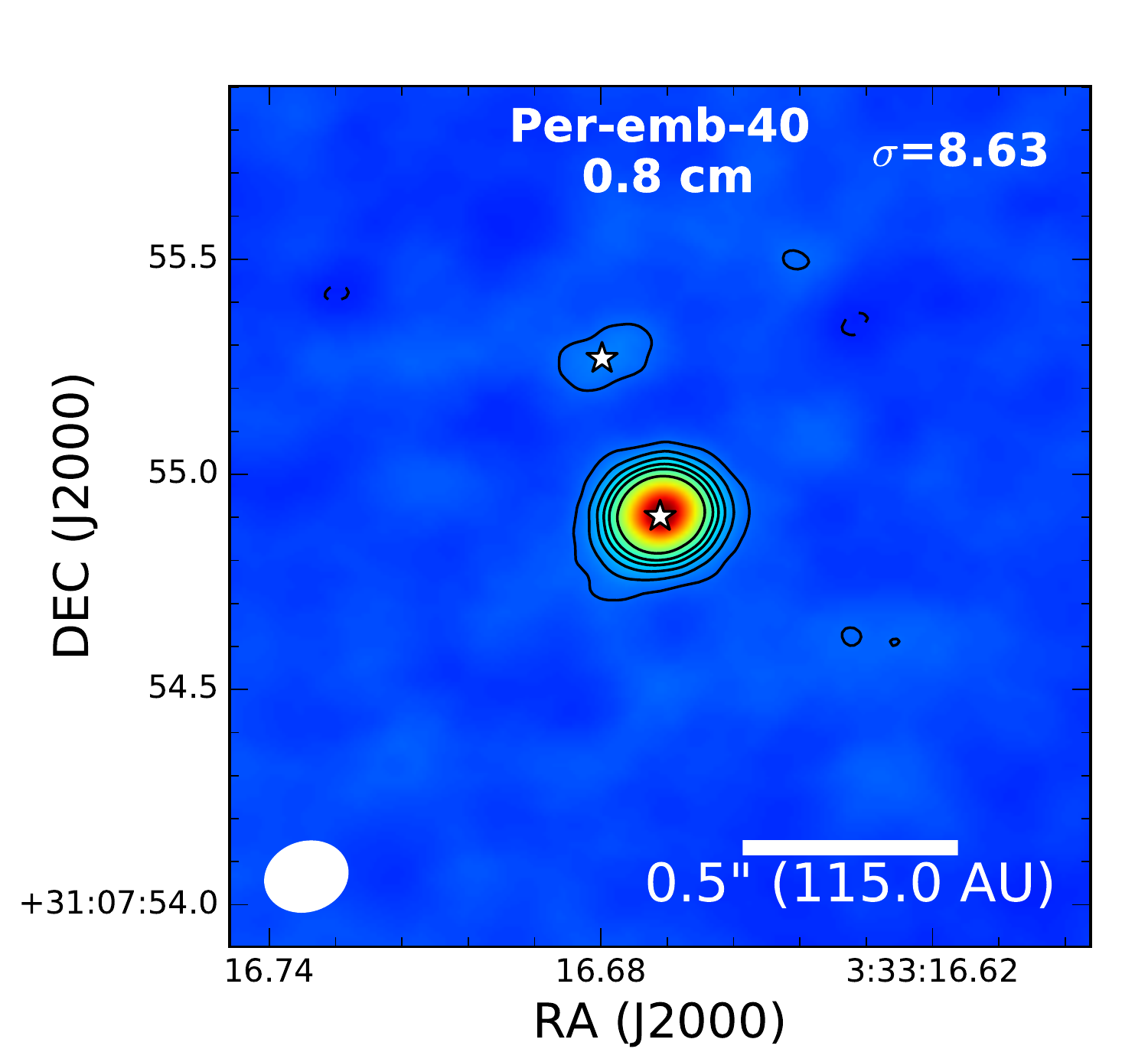}
  \includegraphics[width=0.24\linewidth]{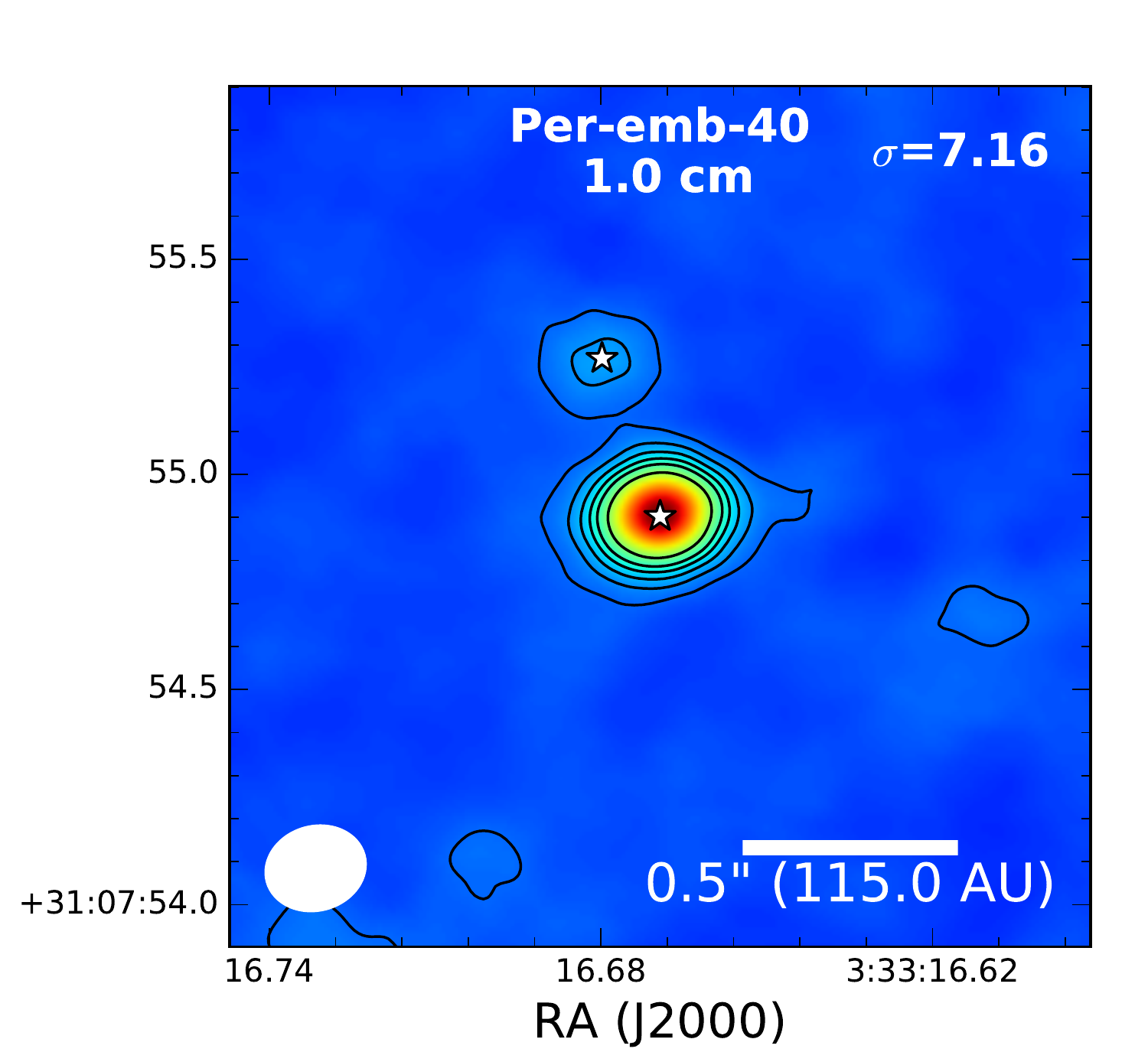}
  \includegraphics[width=0.24\linewidth]{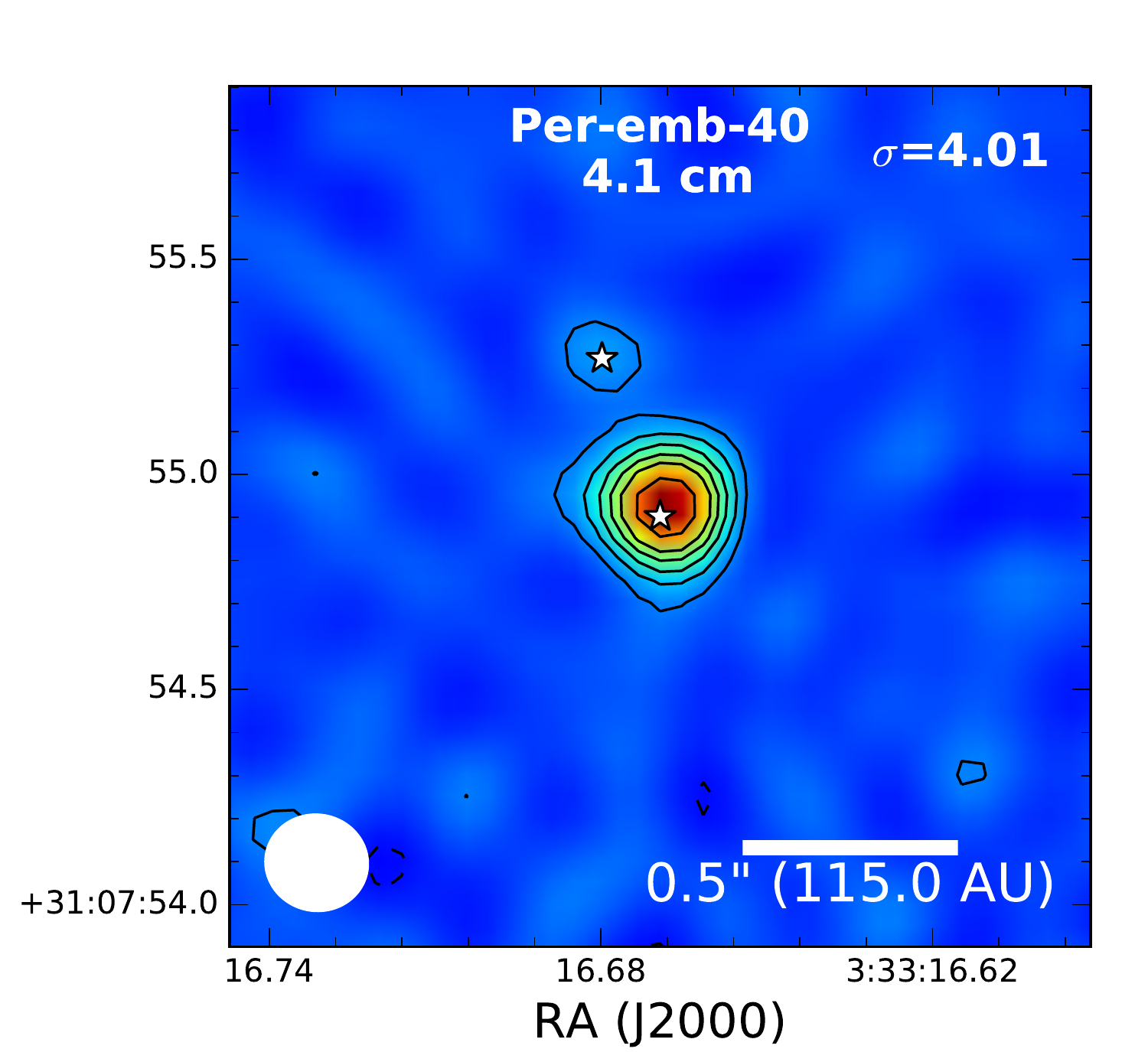}
  \includegraphics[width=0.24\linewidth]{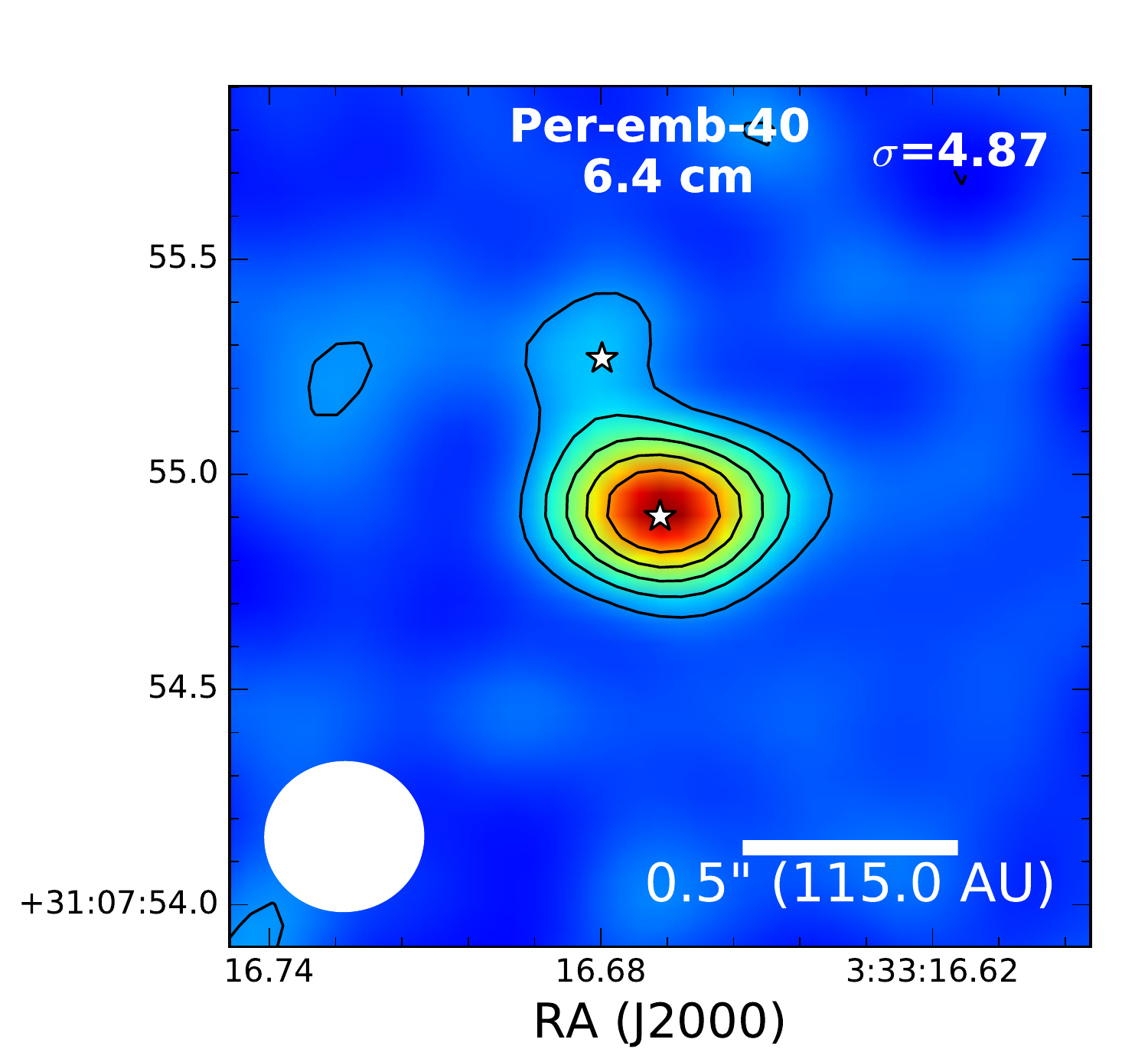}

  \includegraphics[width=0.24\linewidth]{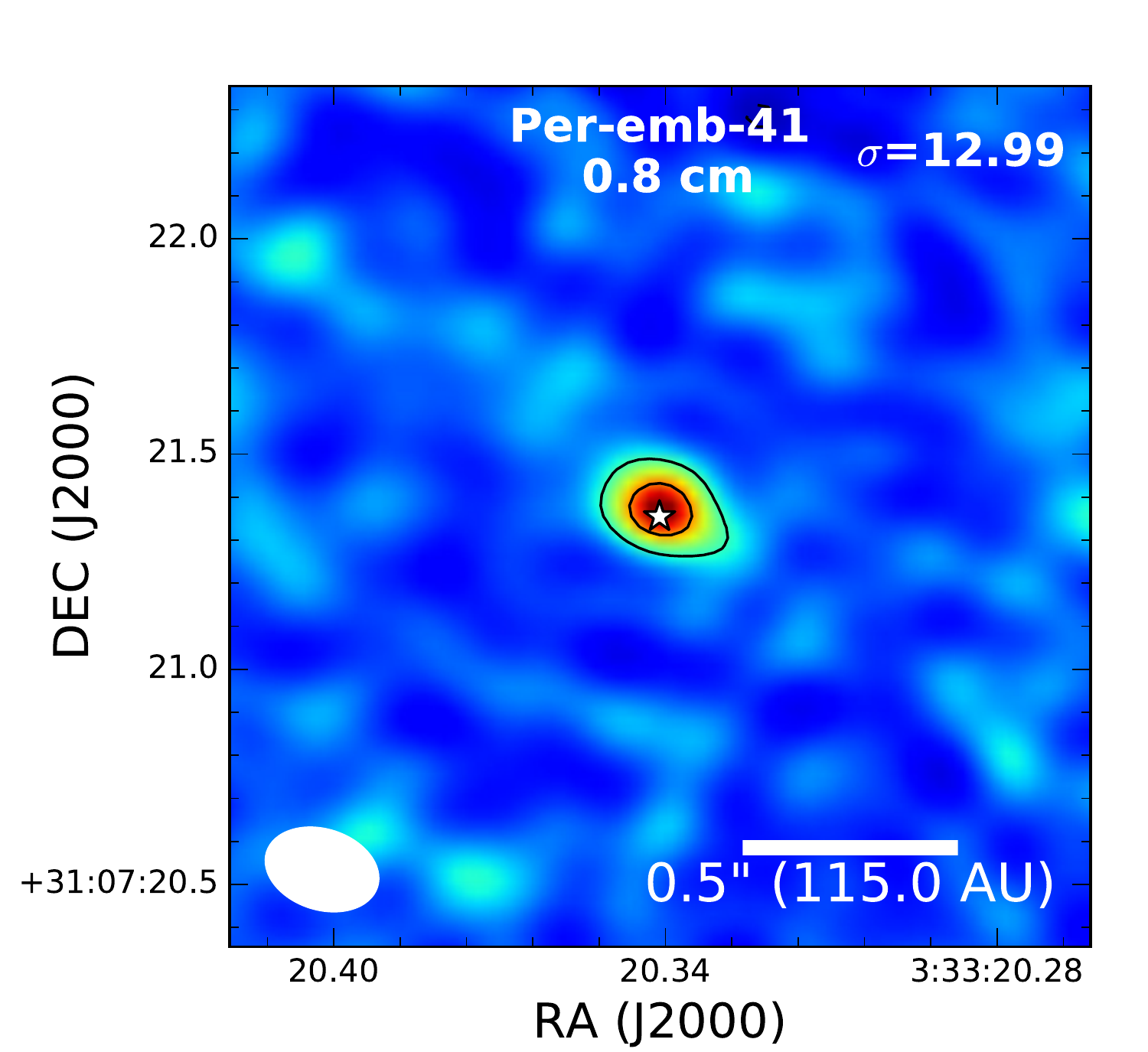}
  \includegraphics[width=0.24\linewidth]{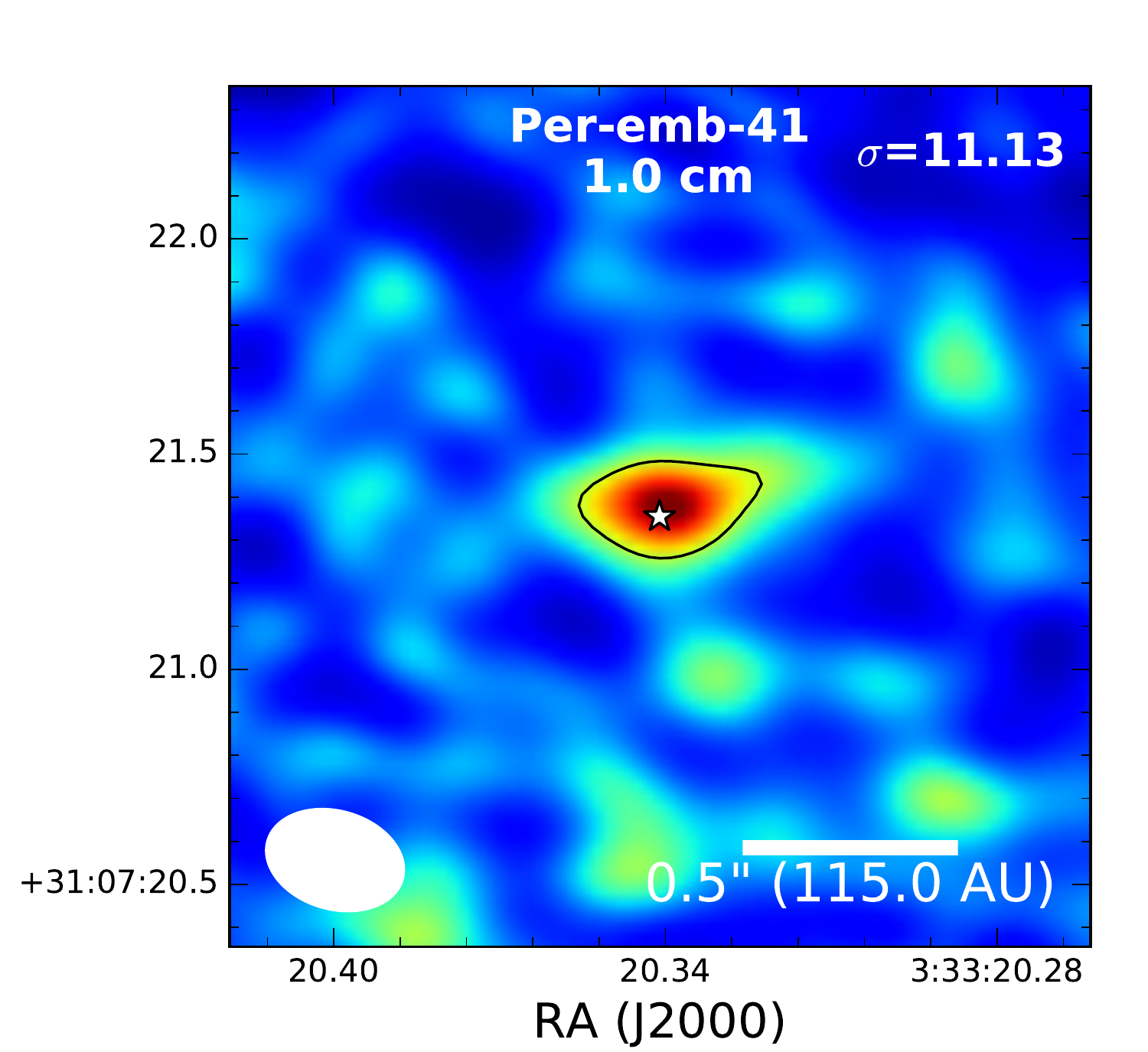}
  \includegraphics[width=0.24\linewidth]{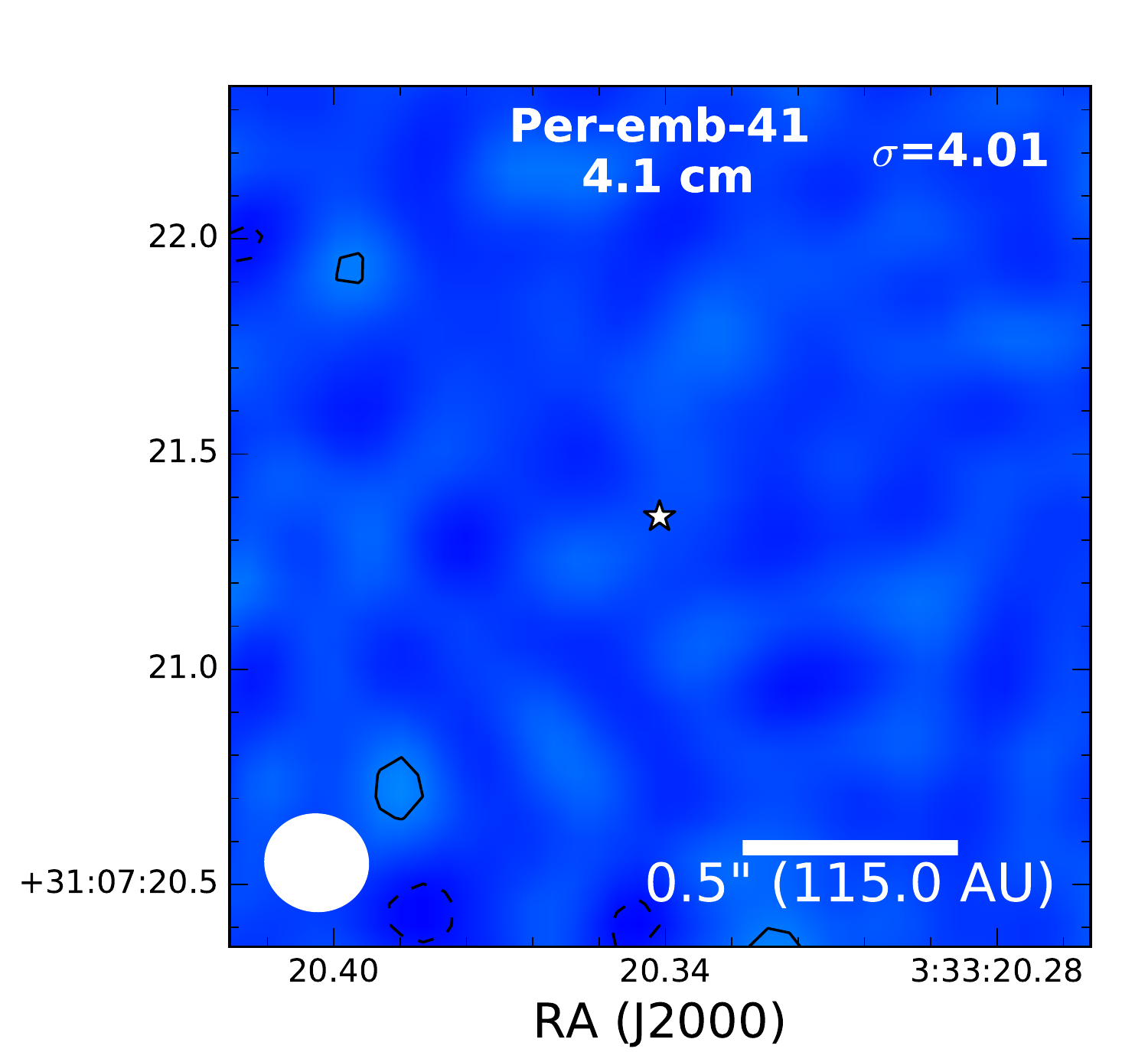}
  \includegraphics[width=0.24\linewidth]{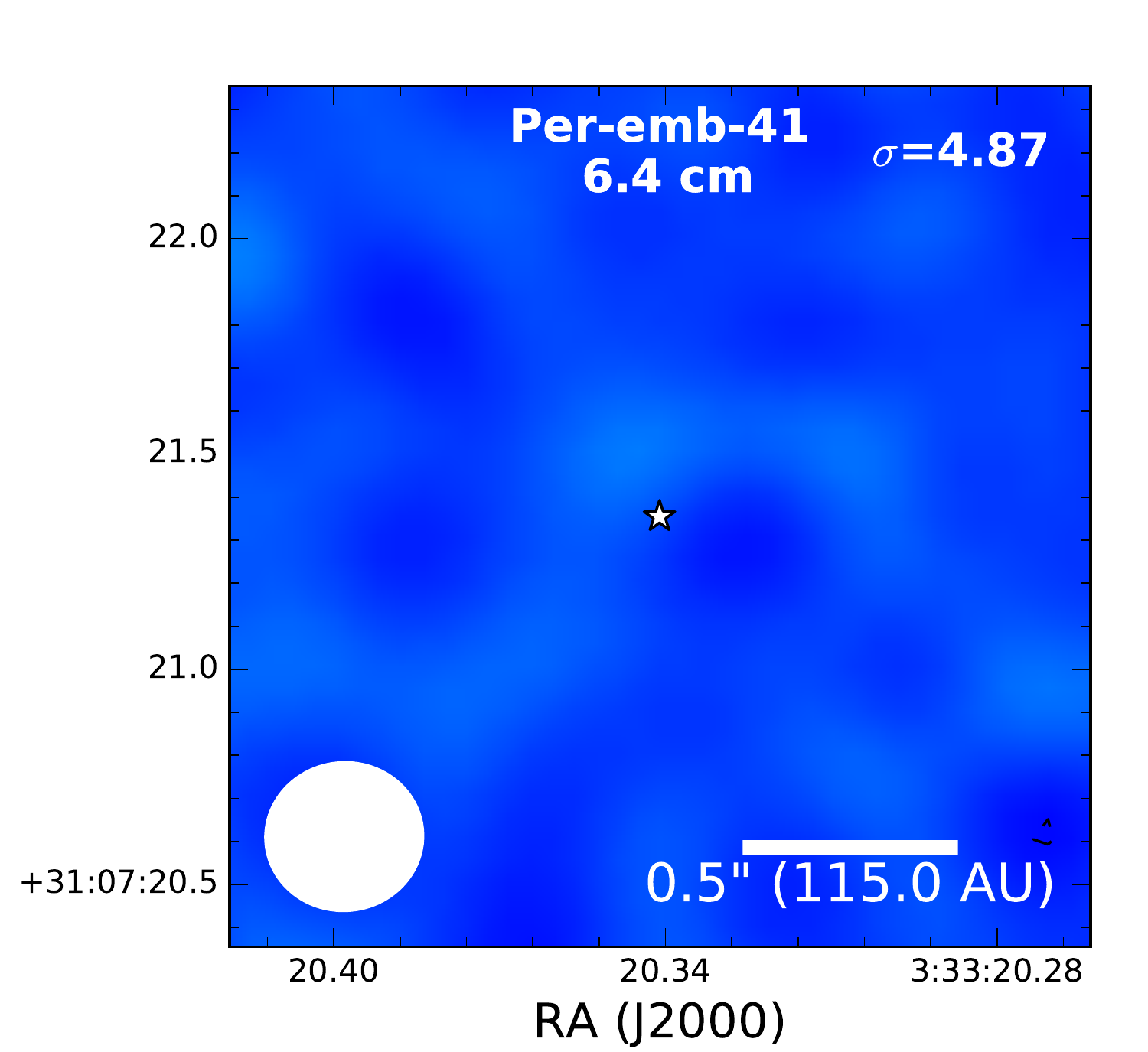}

  \includegraphics[width=0.24\linewidth]{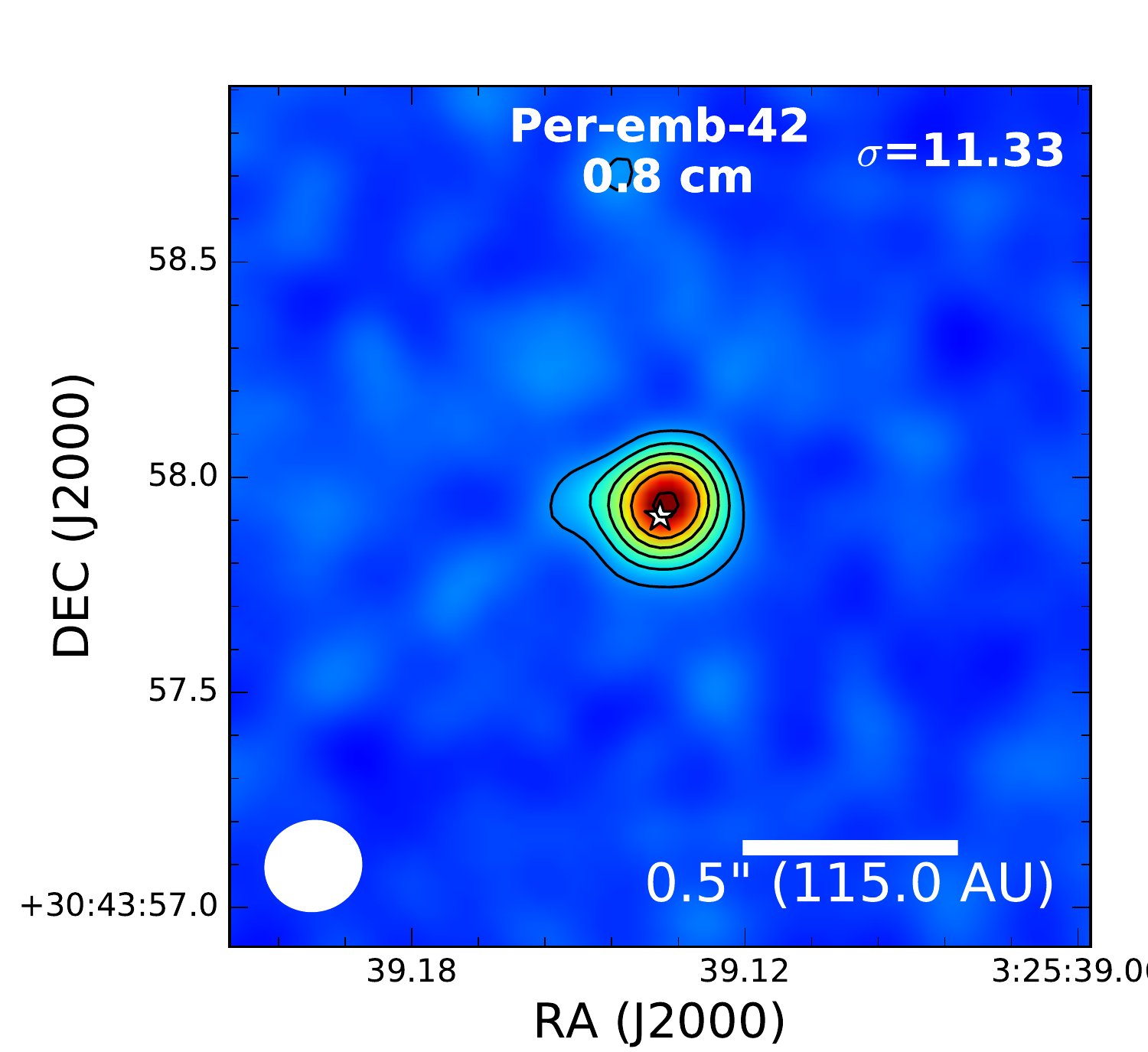}
  \includegraphics[width=0.24\linewidth]{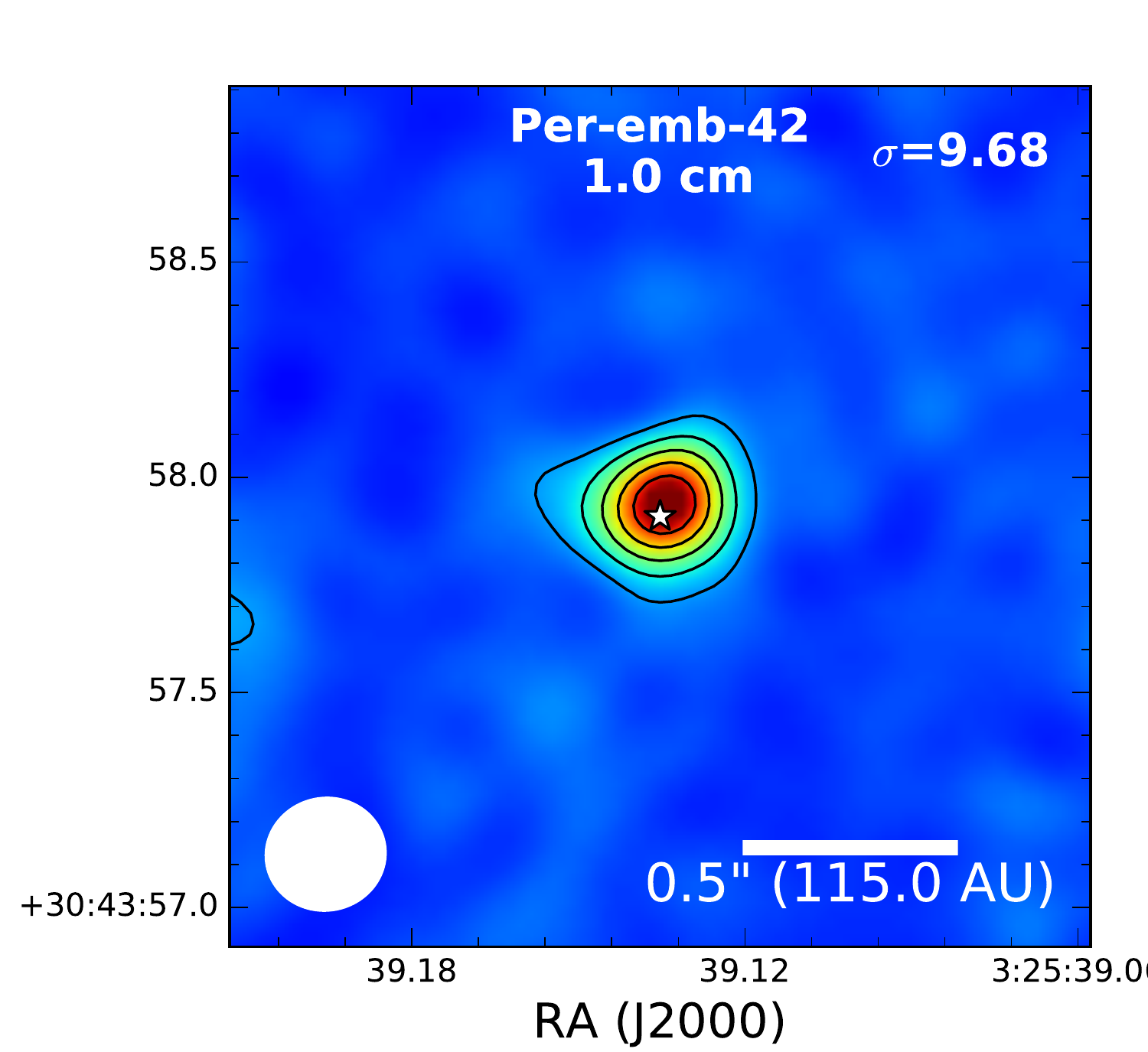}
  \includegraphics[width=0.24\linewidth]{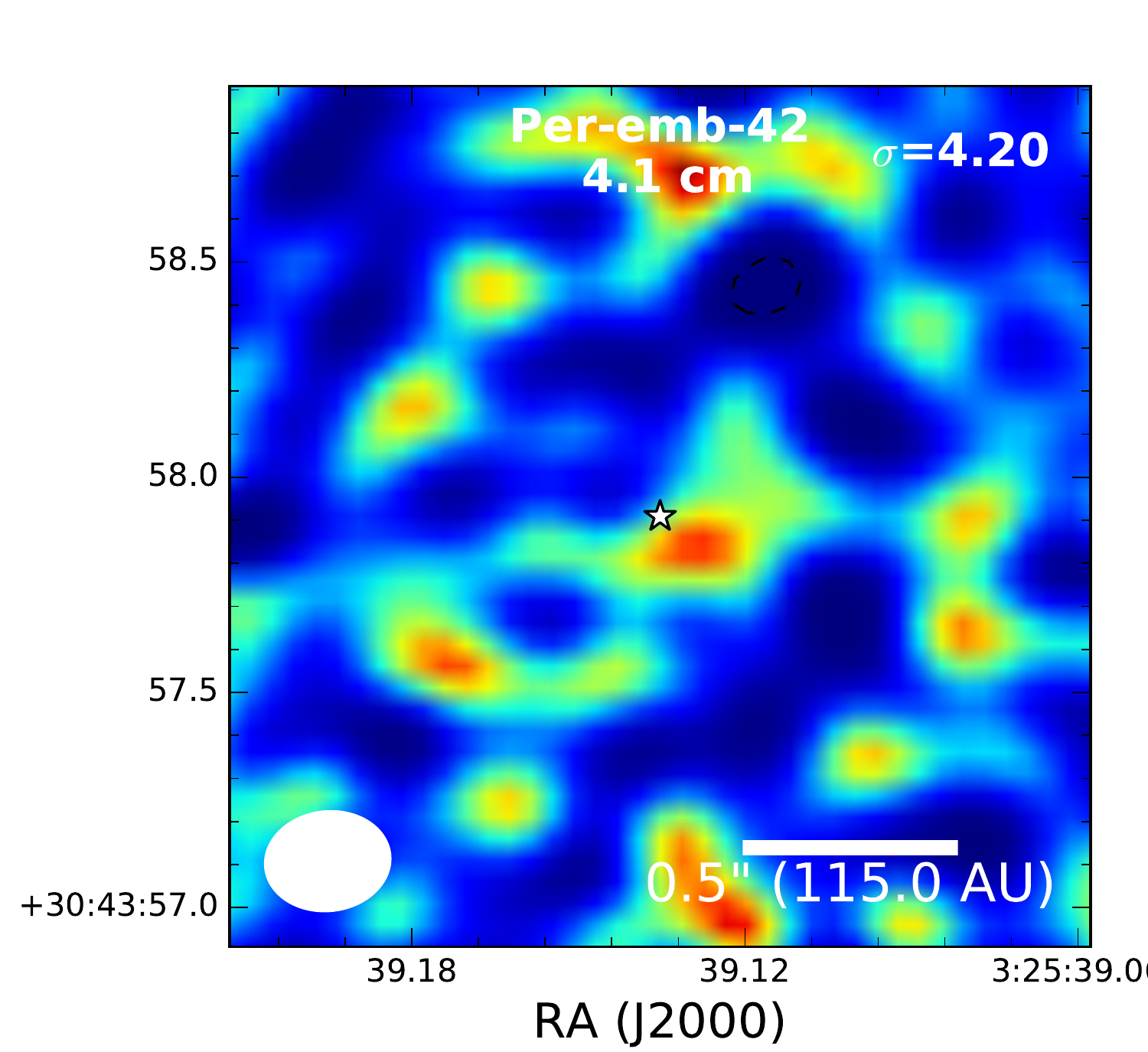}
  \includegraphics[width=0.24\linewidth]{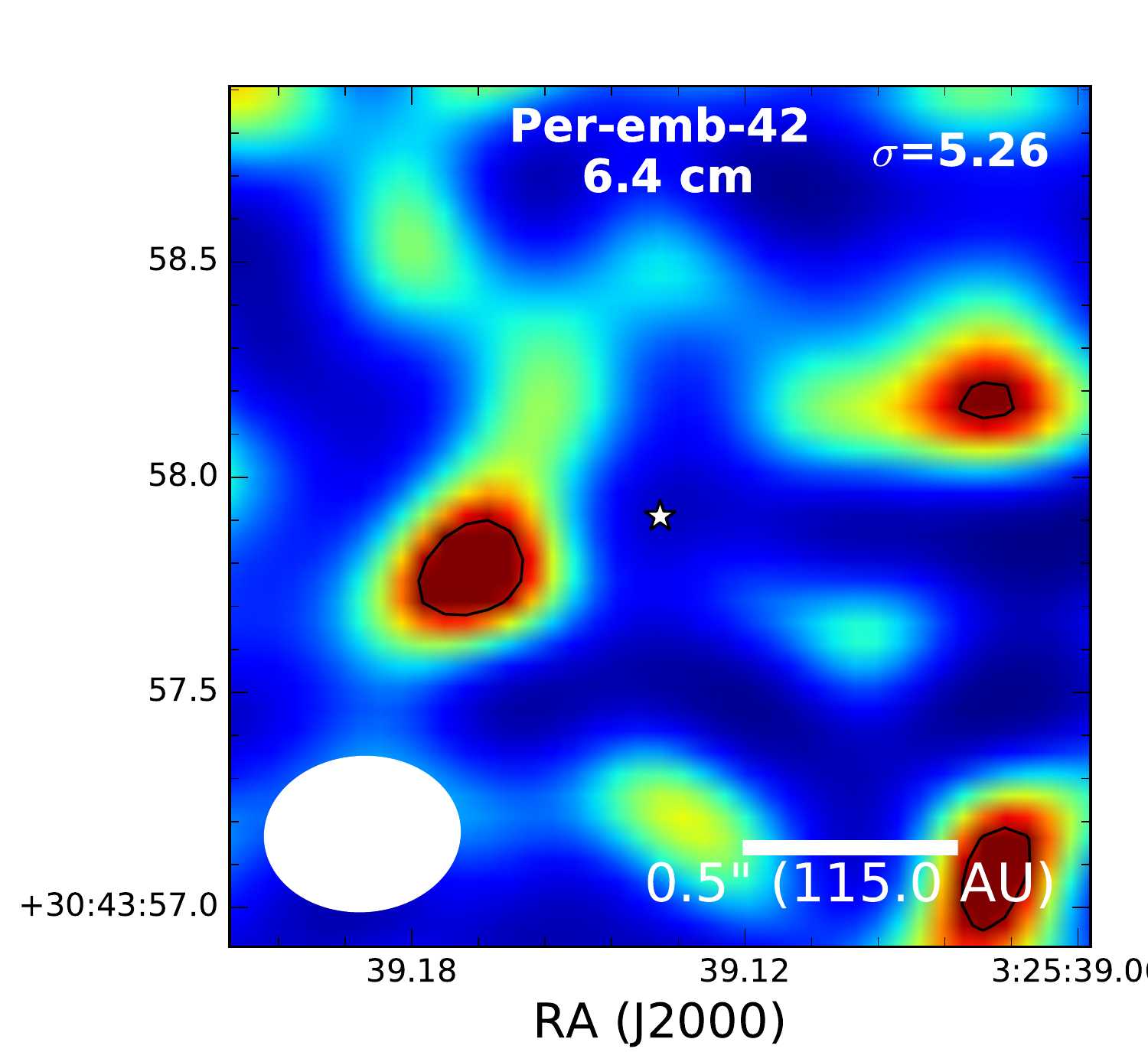}

  \includegraphics[width=0.24\linewidth]{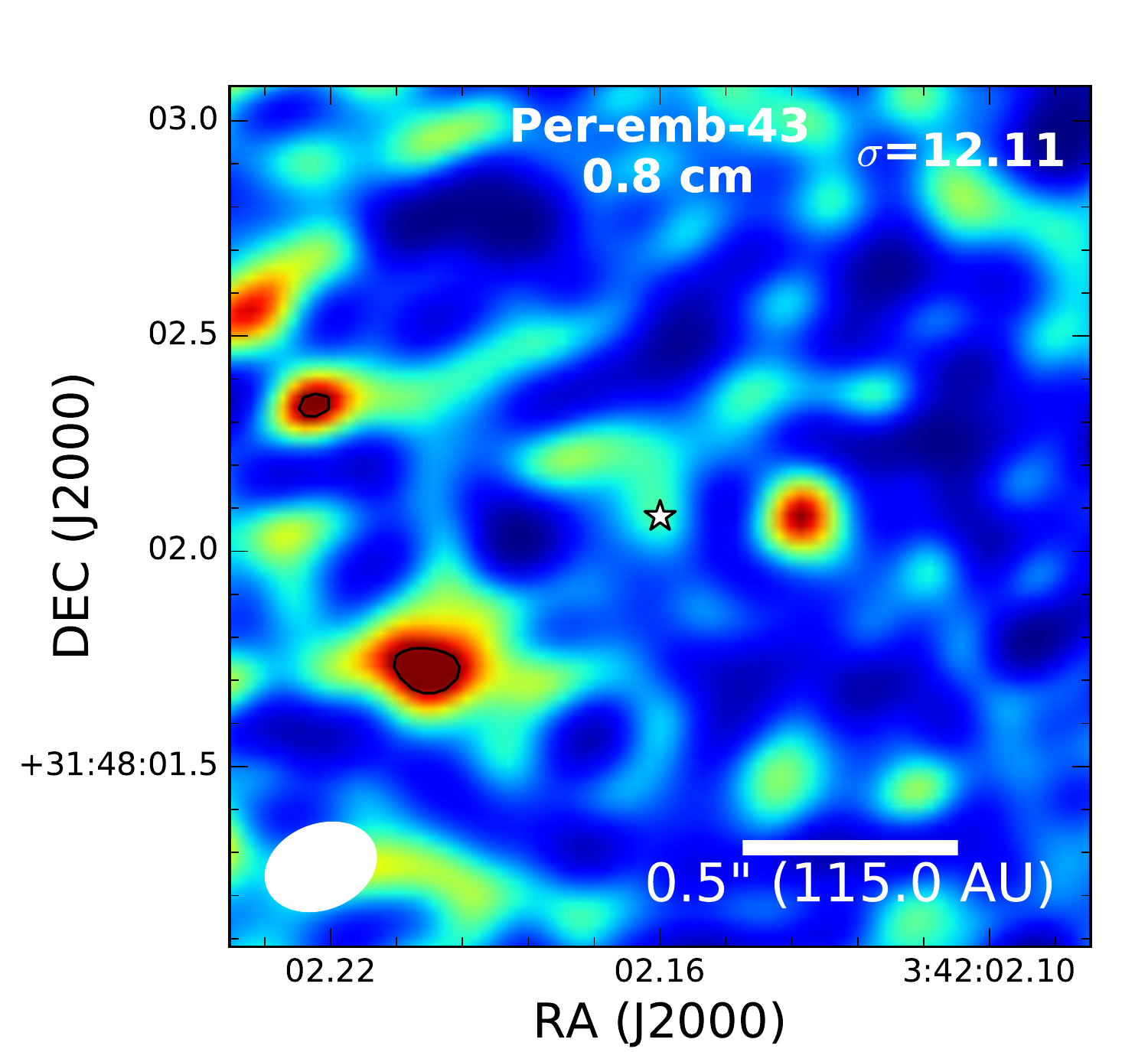}
  \includegraphics[width=0.24\linewidth]{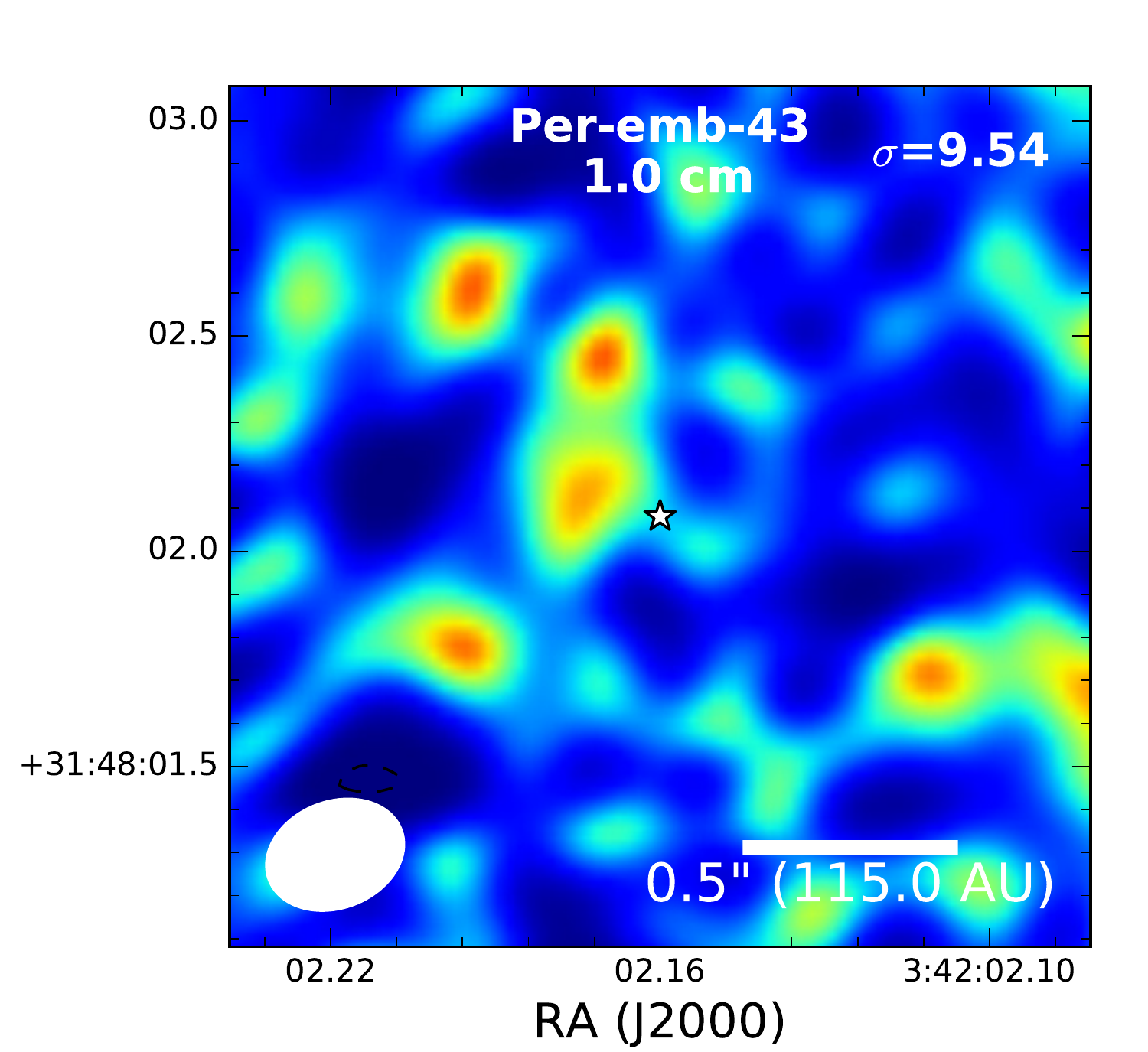}
  \includegraphics[width=0.24\linewidth]{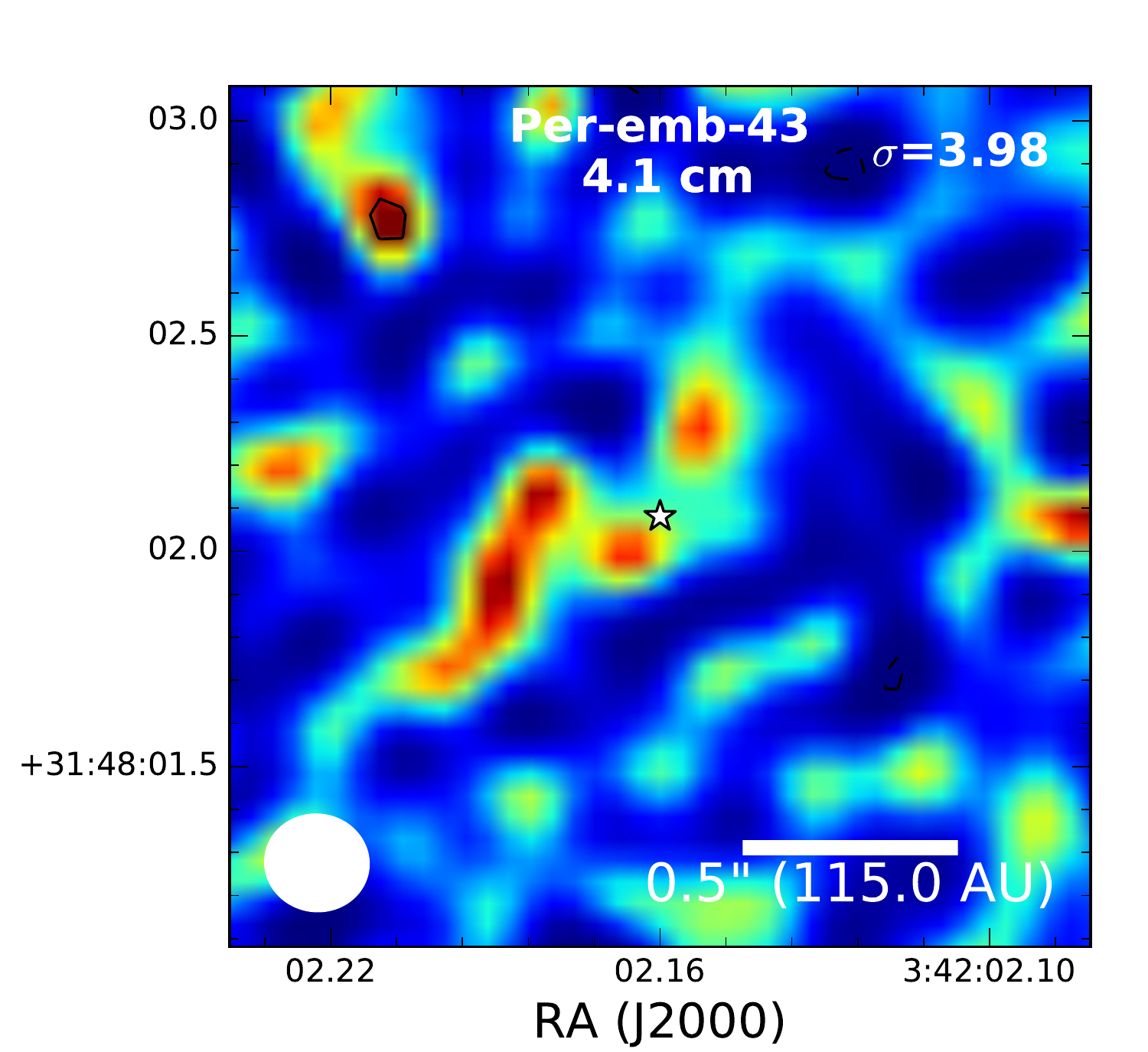}
  \includegraphics[width=0.24\linewidth]{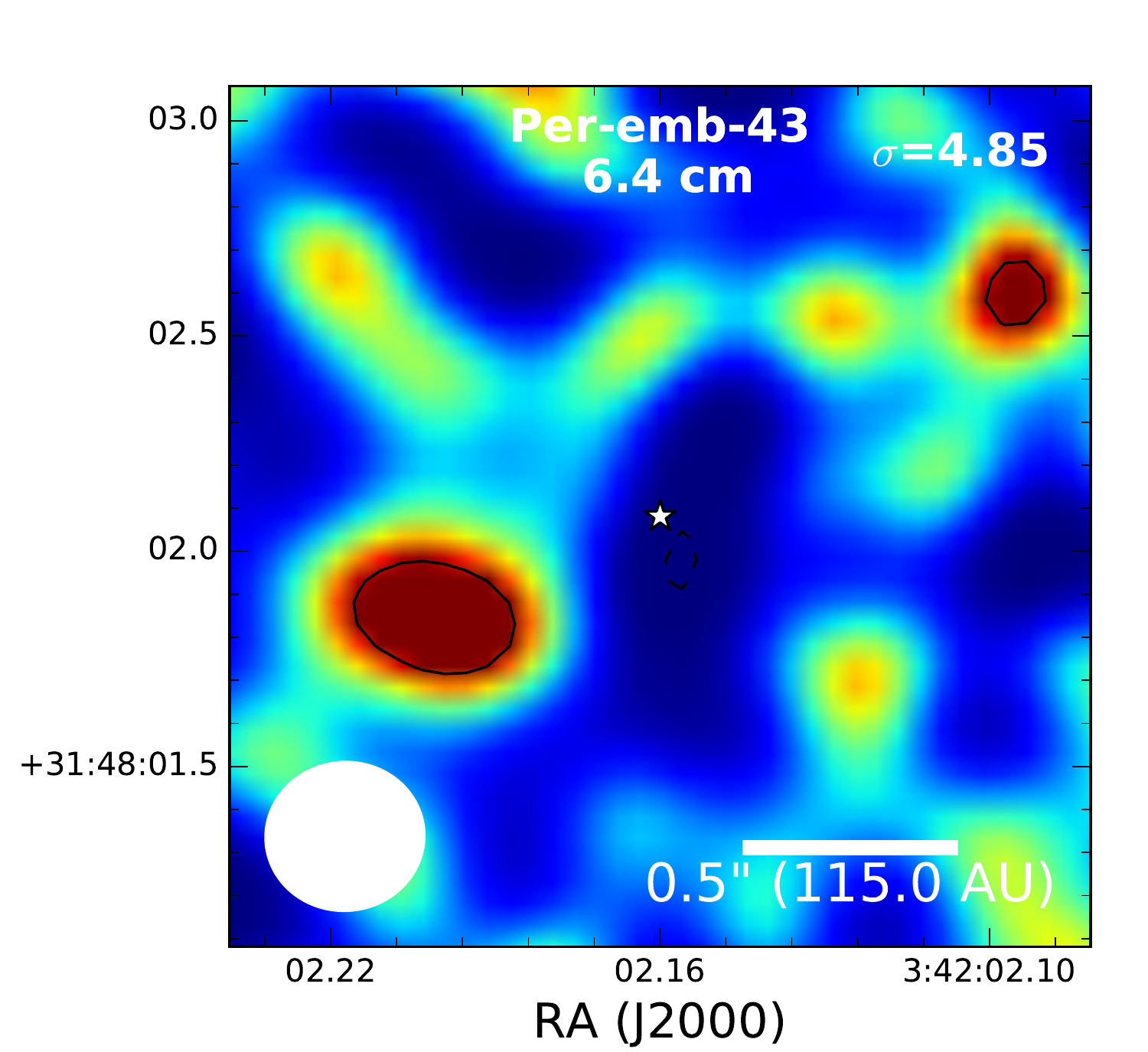}

  \includegraphics[width=0.24\linewidth]{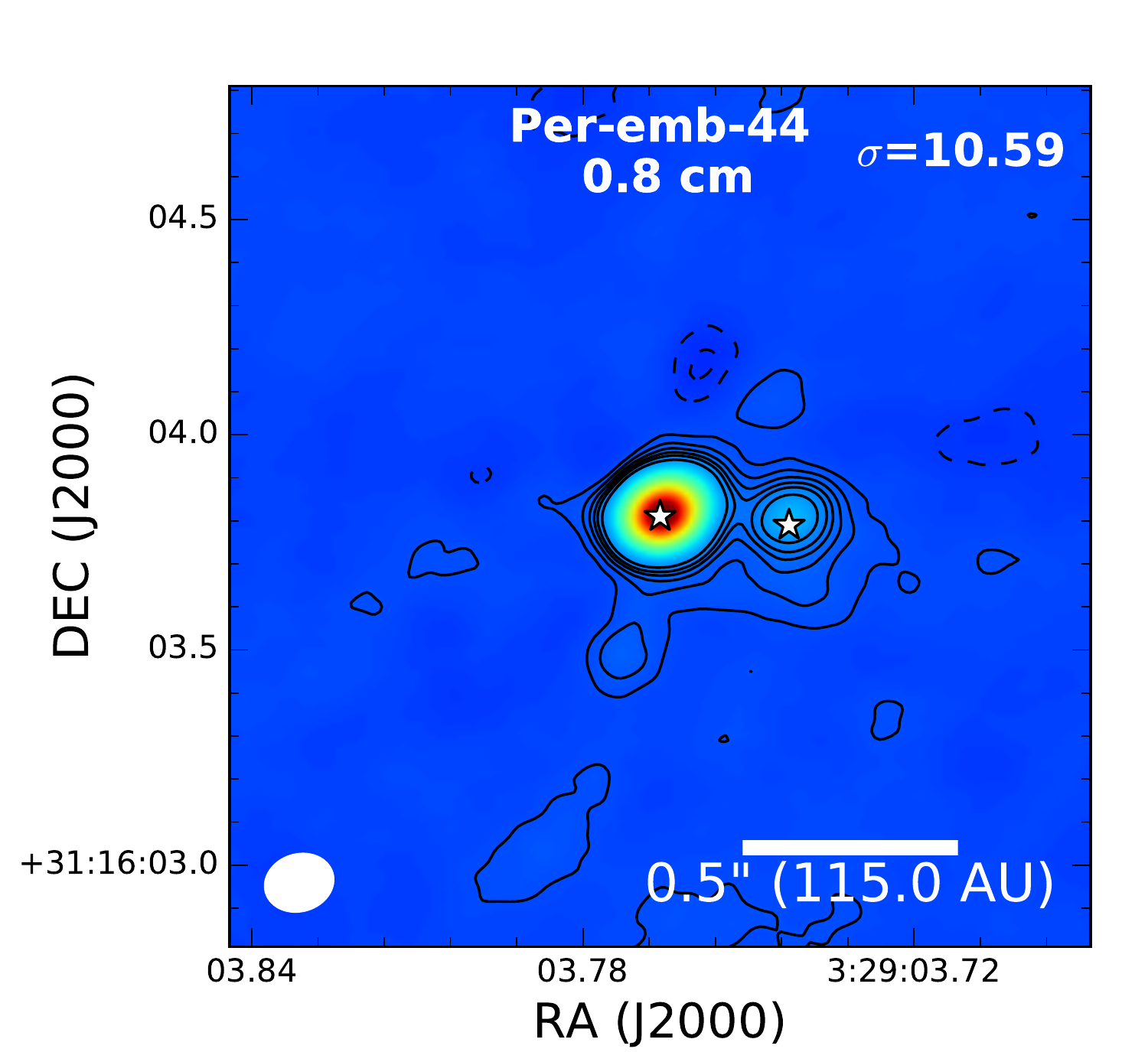}
  \includegraphics[width=0.24\linewidth]{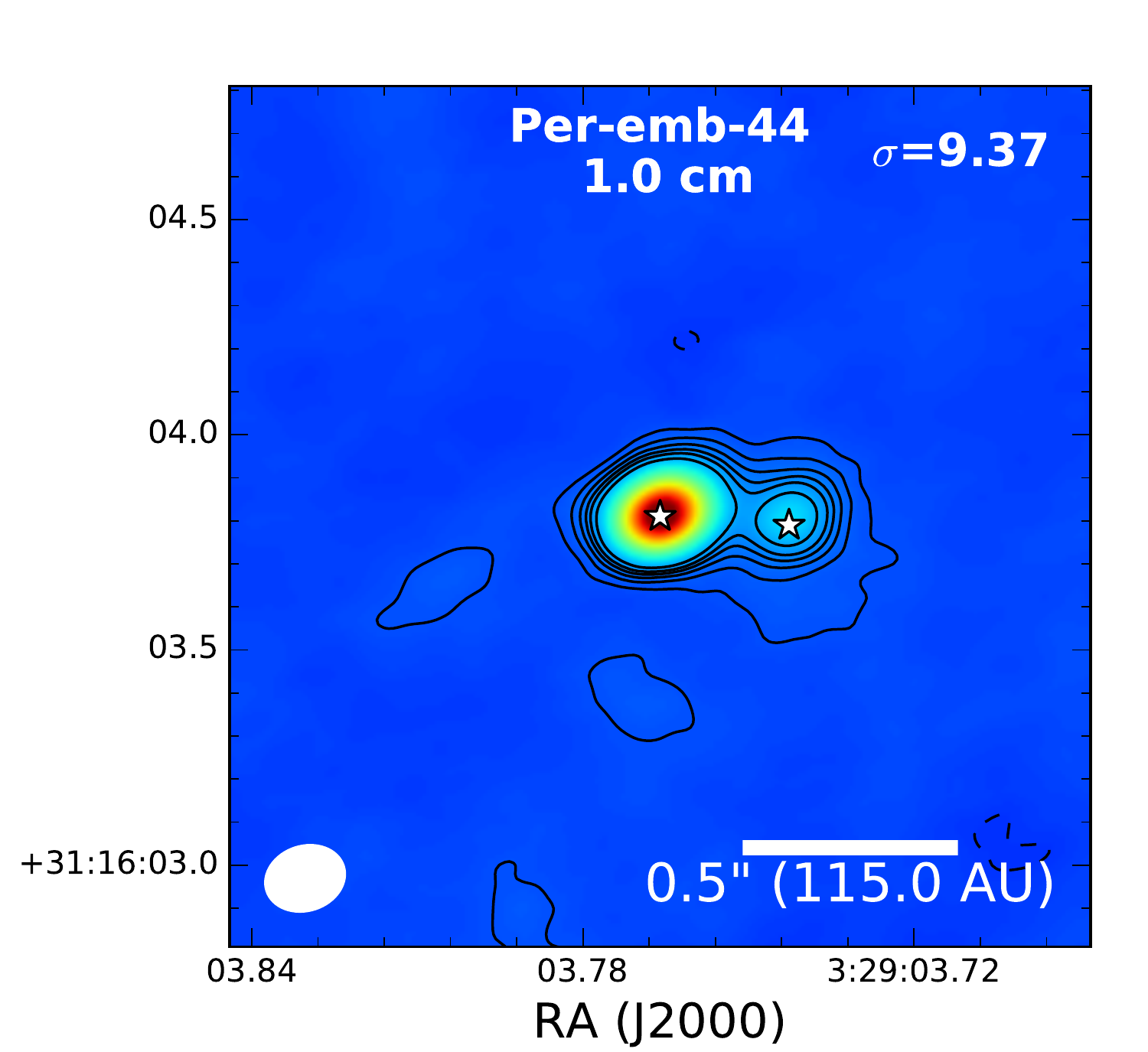}
  \includegraphics[width=0.24\linewidth]{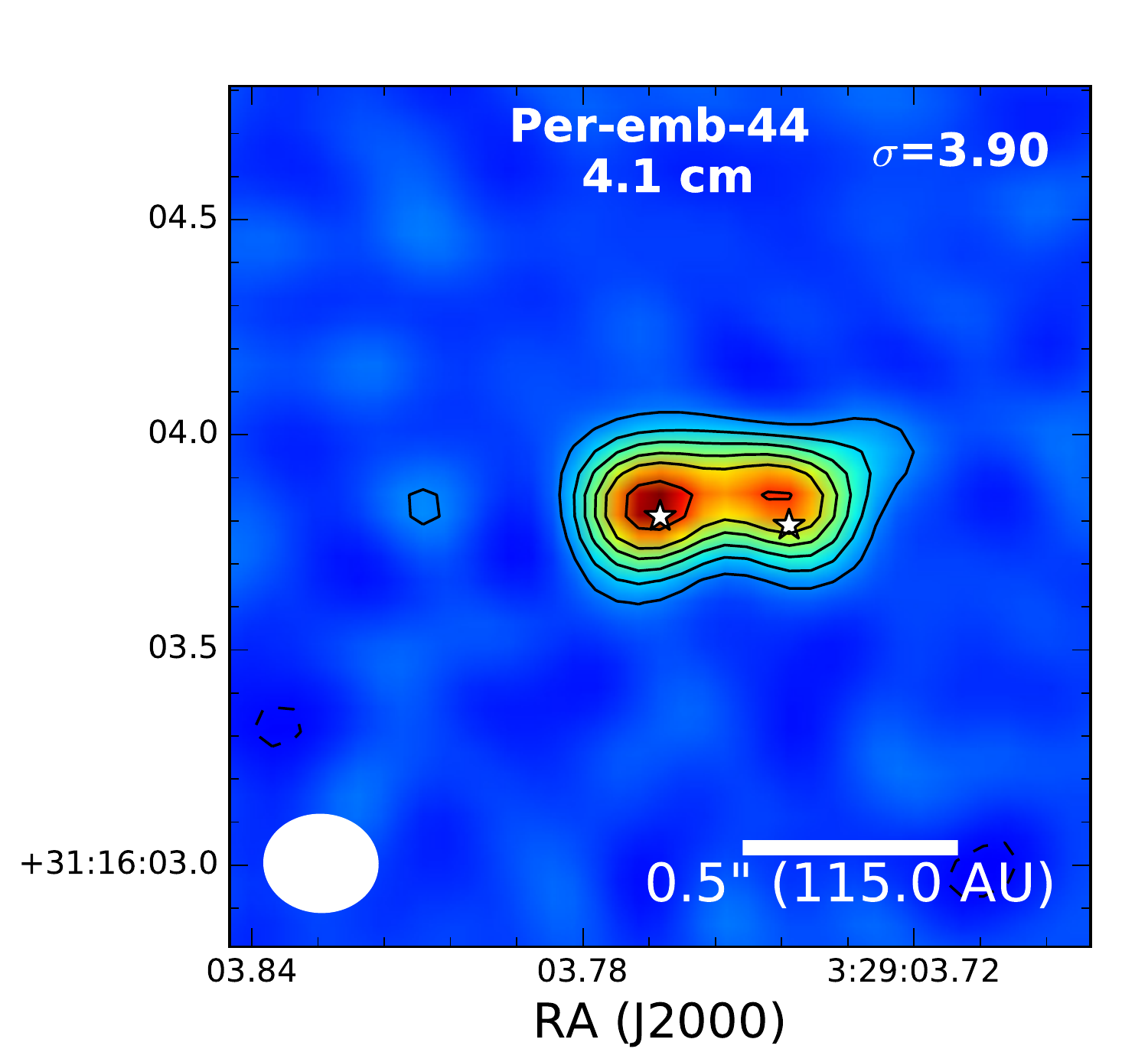}
  \includegraphics[width=0.24\linewidth]{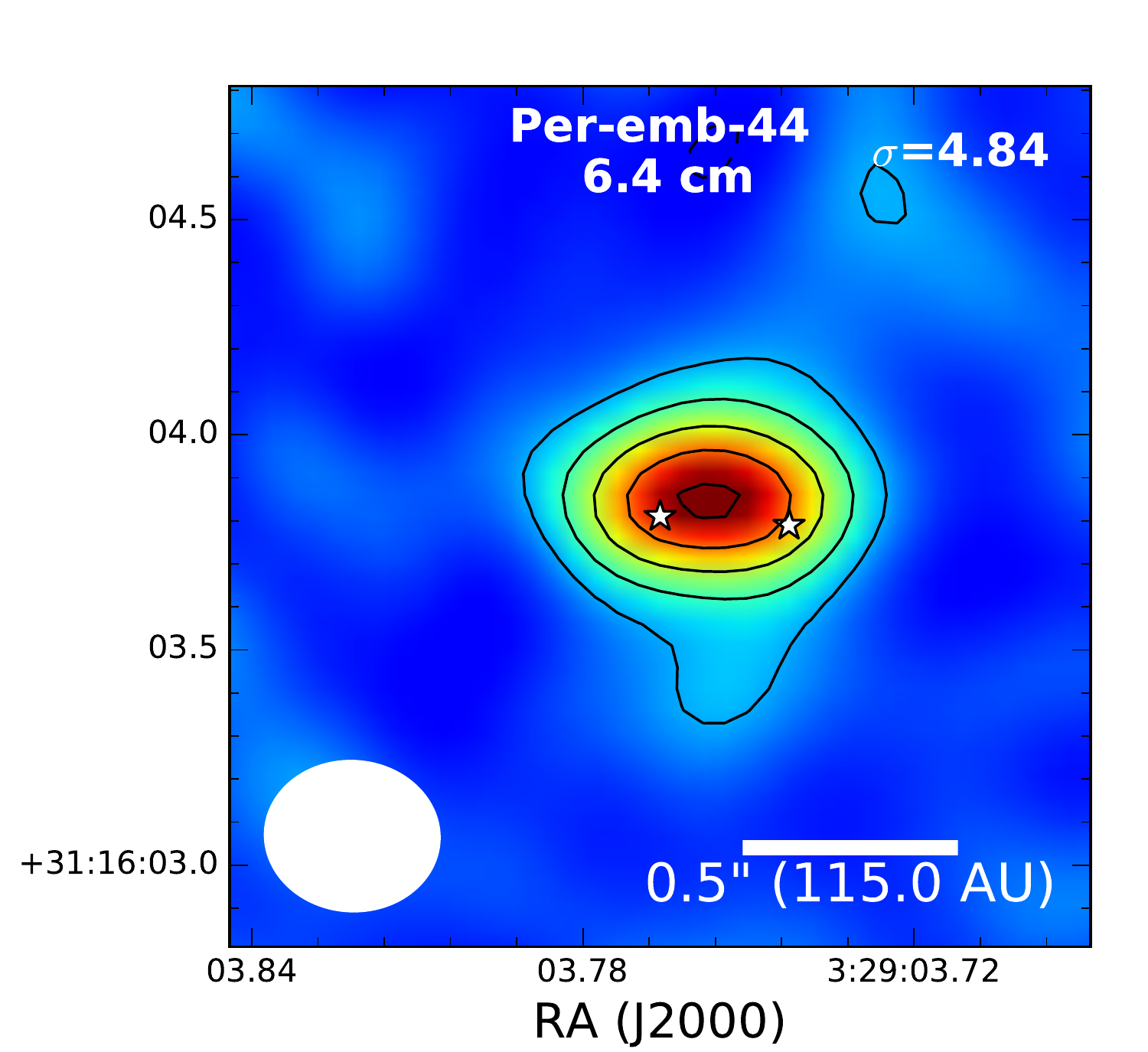}

  \includegraphics[width=0.24\linewidth]{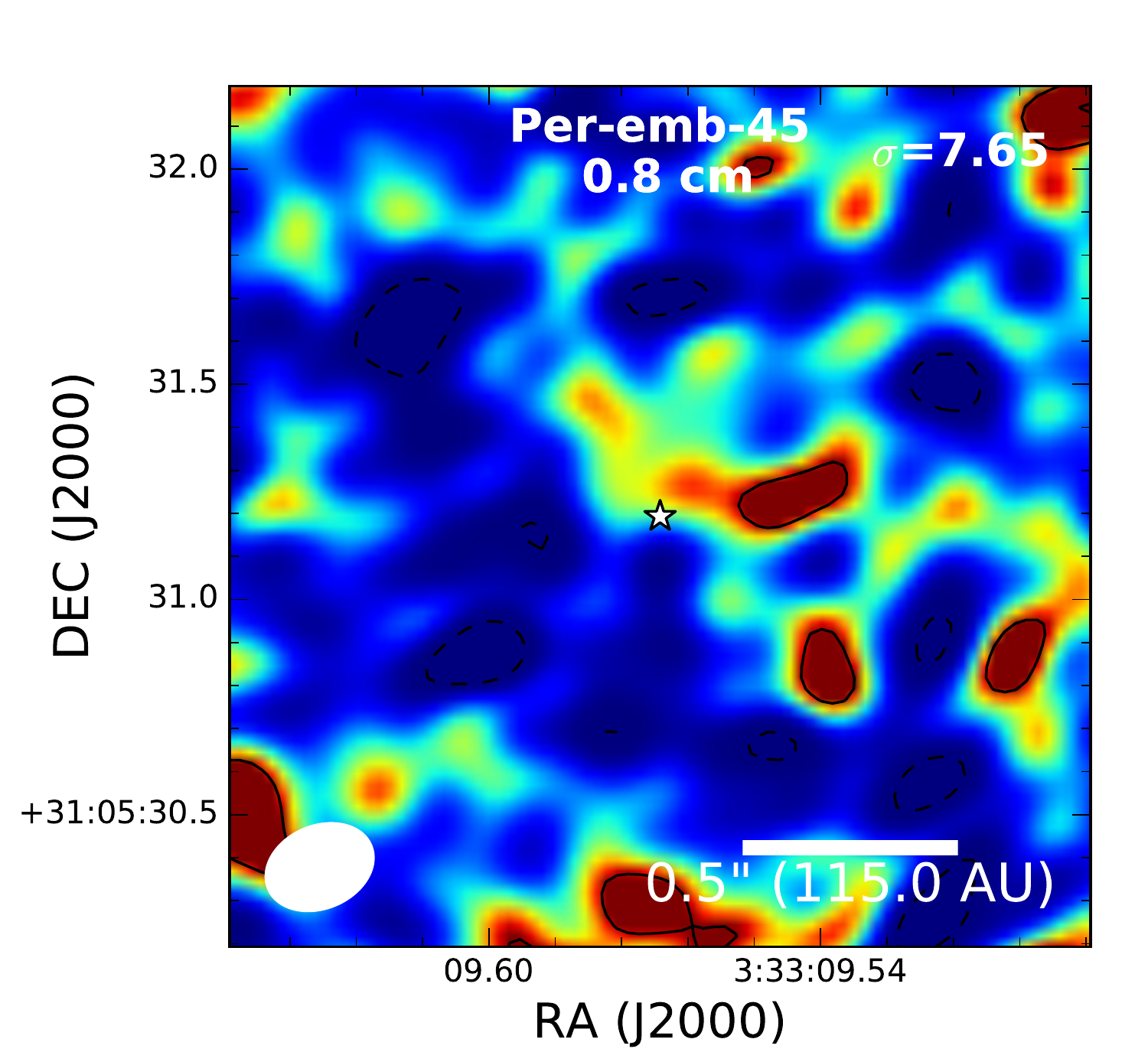}
  \includegraphics[width=0.24\linewidth]{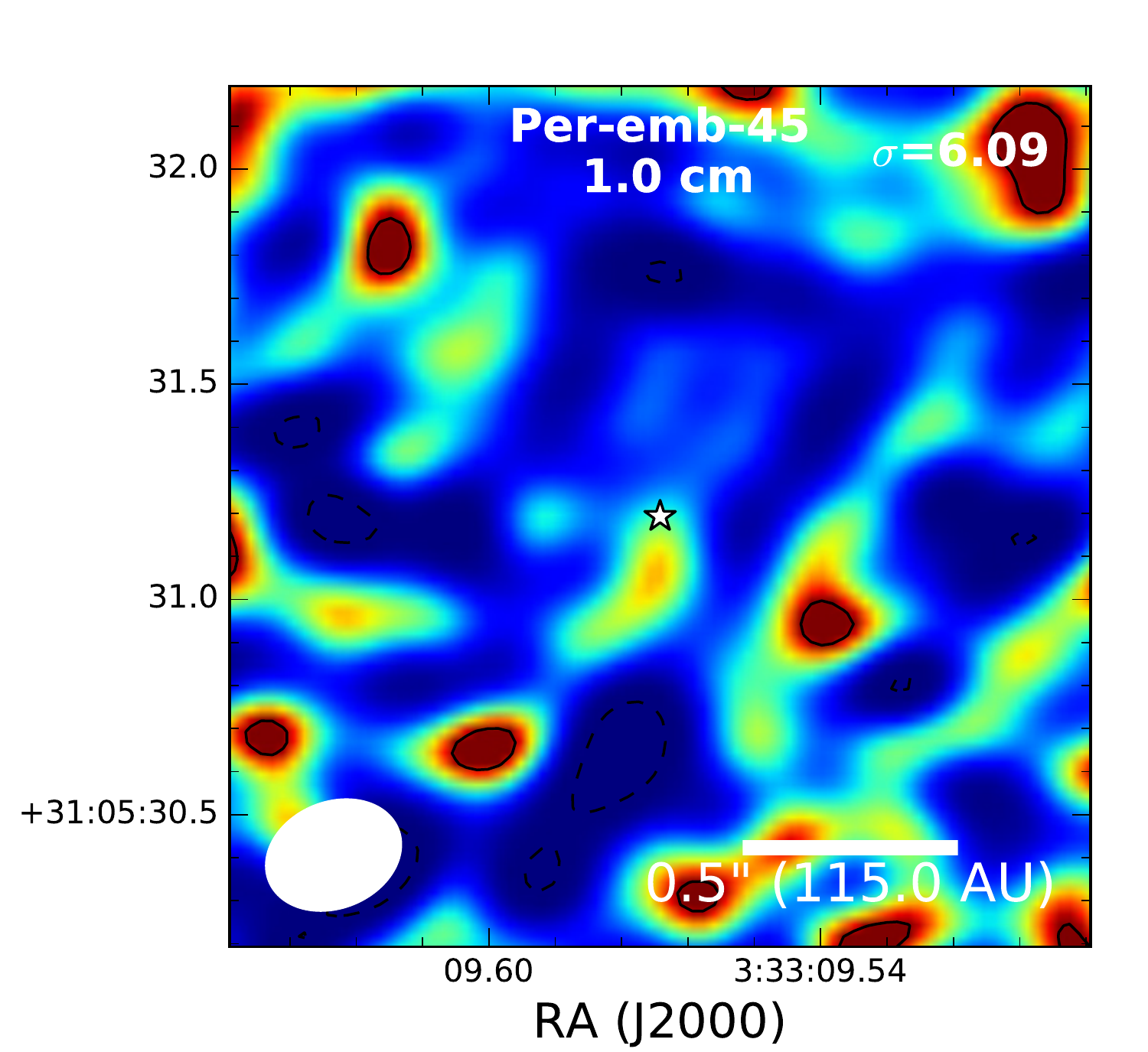}
  \includegraphics[width=0.24\linewidth]{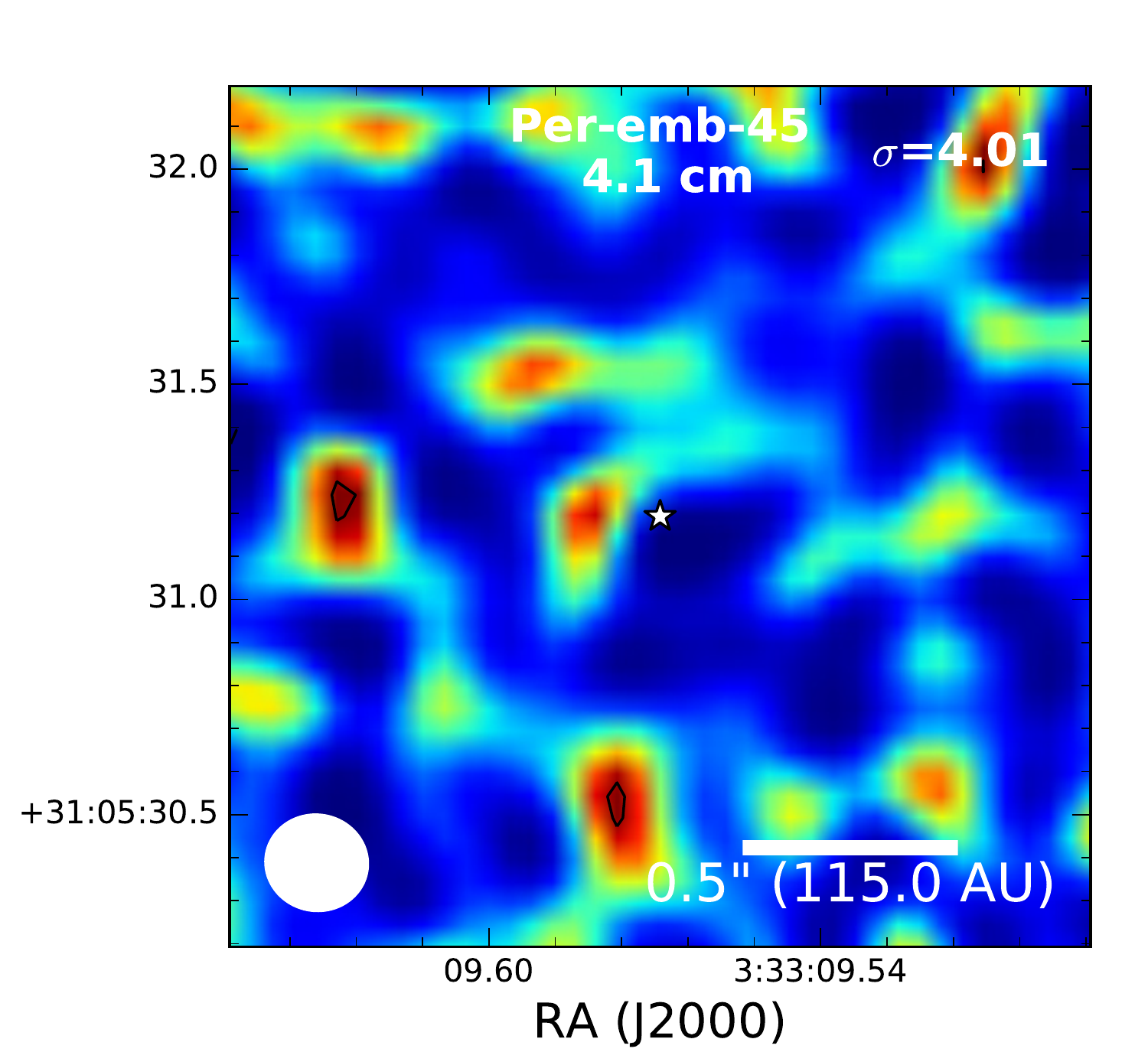}
  \includegraphics[width=0.24\linewidth]{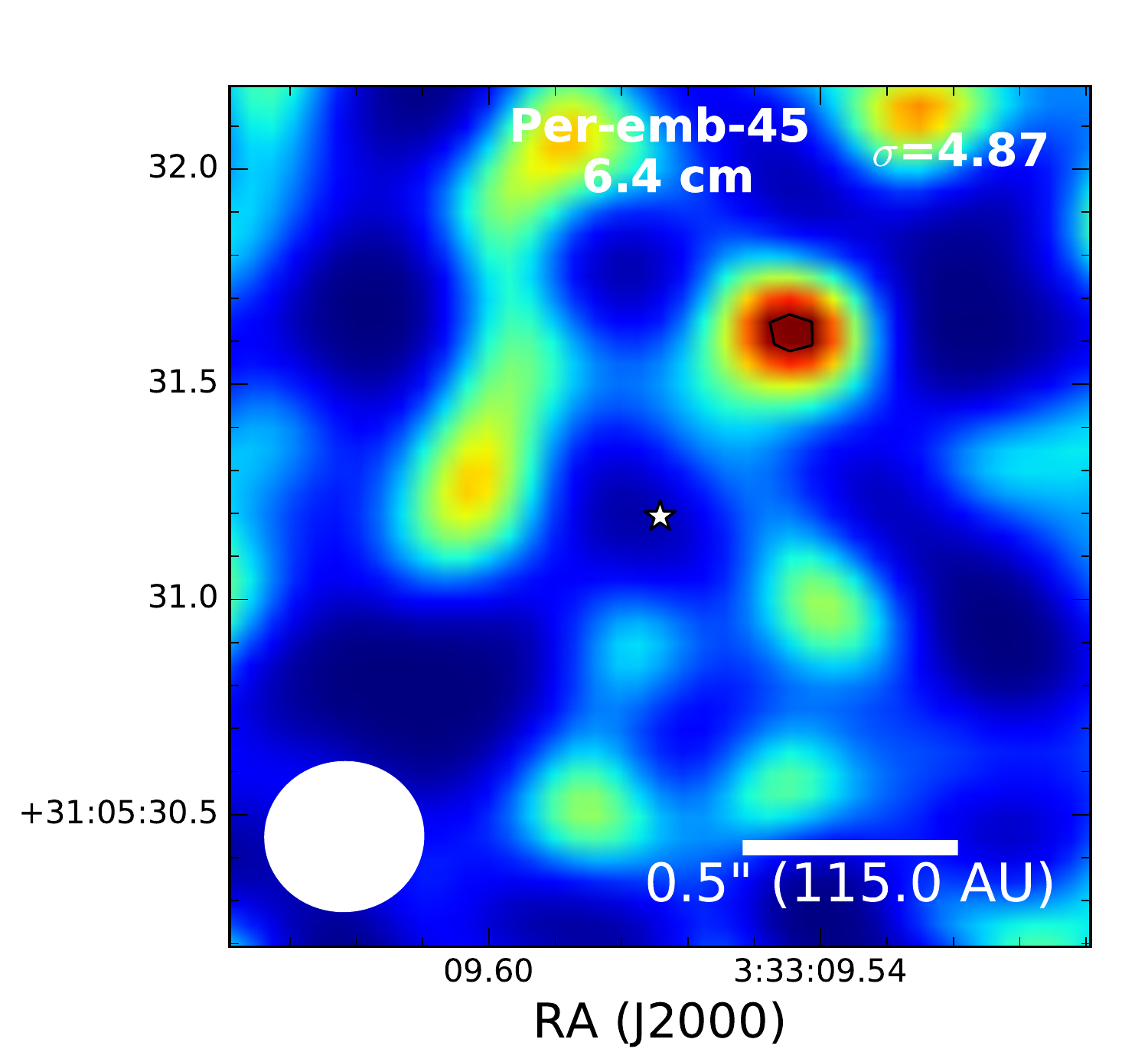}

\end{figure}

\begin{figure}

  \includegraphics[width=0.24\linewidth]{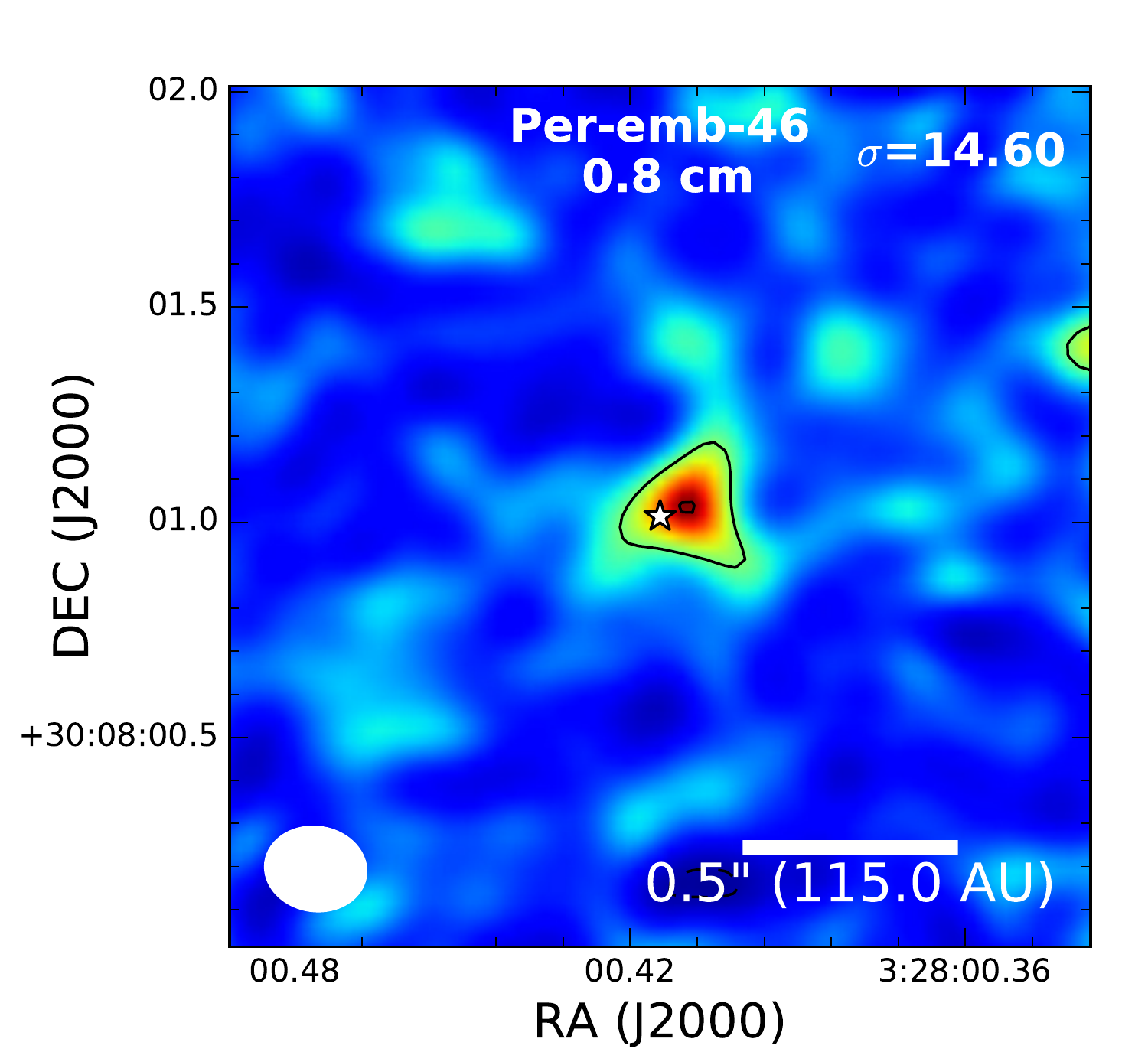}
  \includegraphics[width=0.24\linewidth]{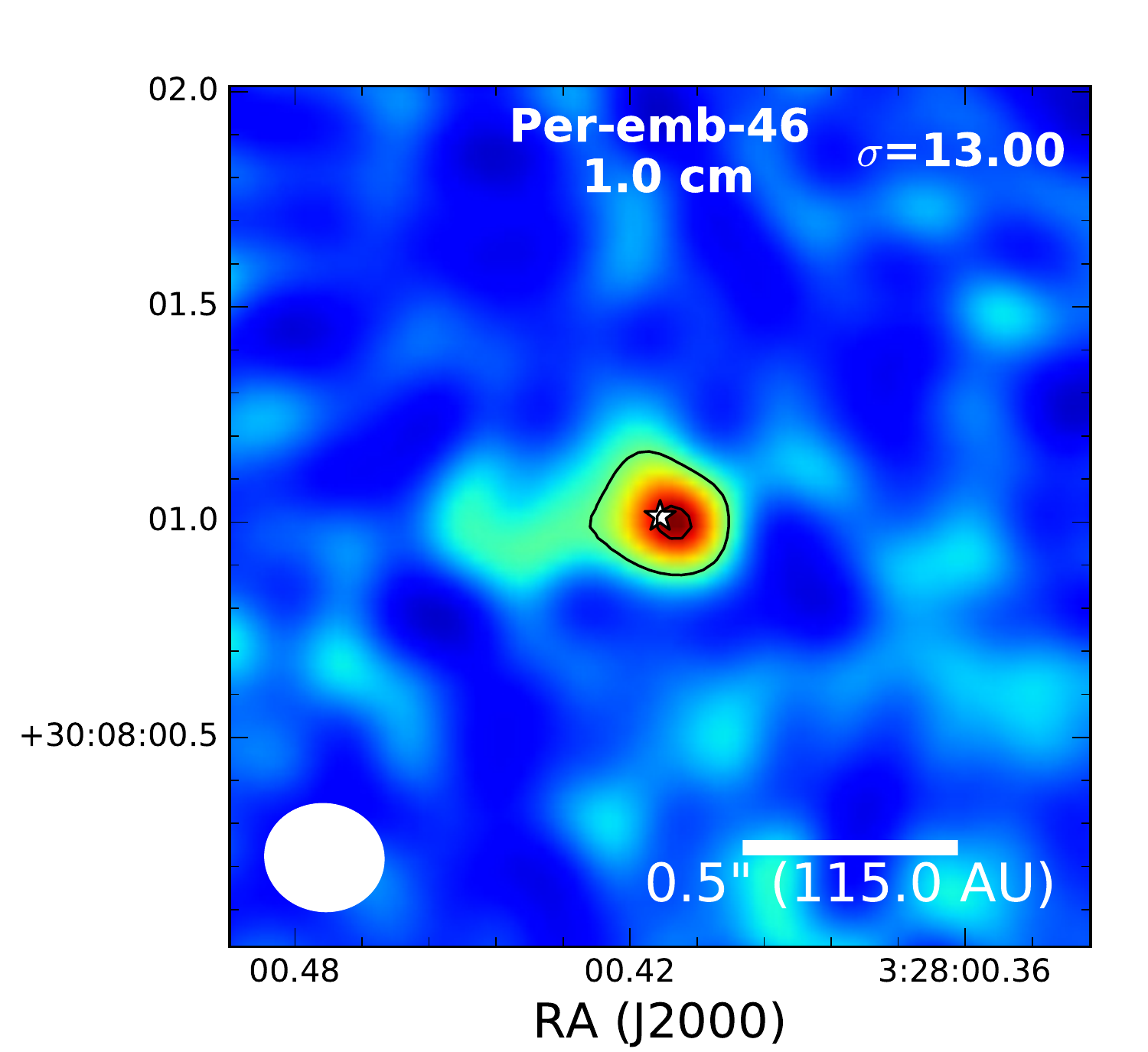}
  \includegraphics[width=0.24\linewidth]{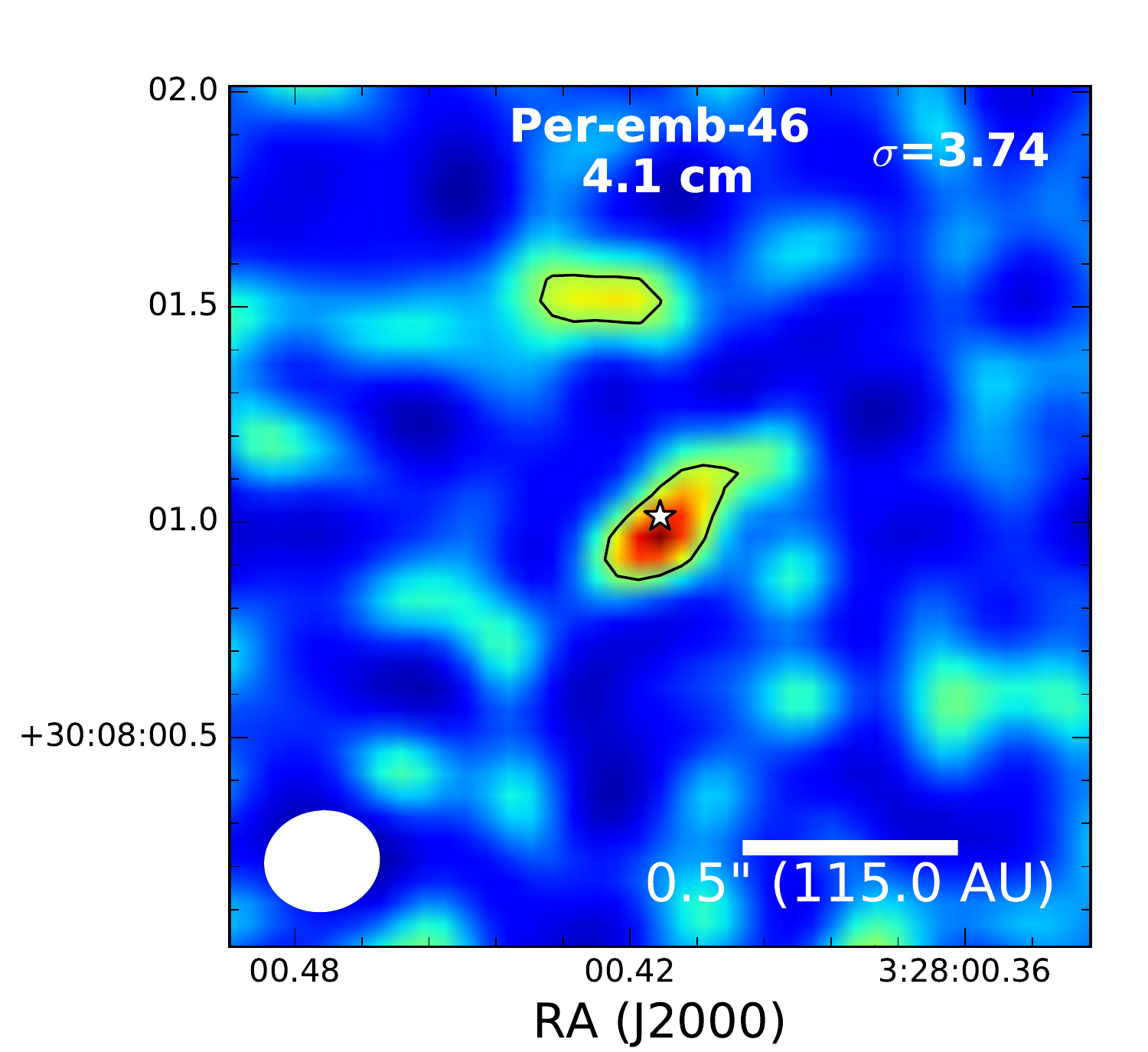}
  \includegraphics[width=0.24\linewidth]{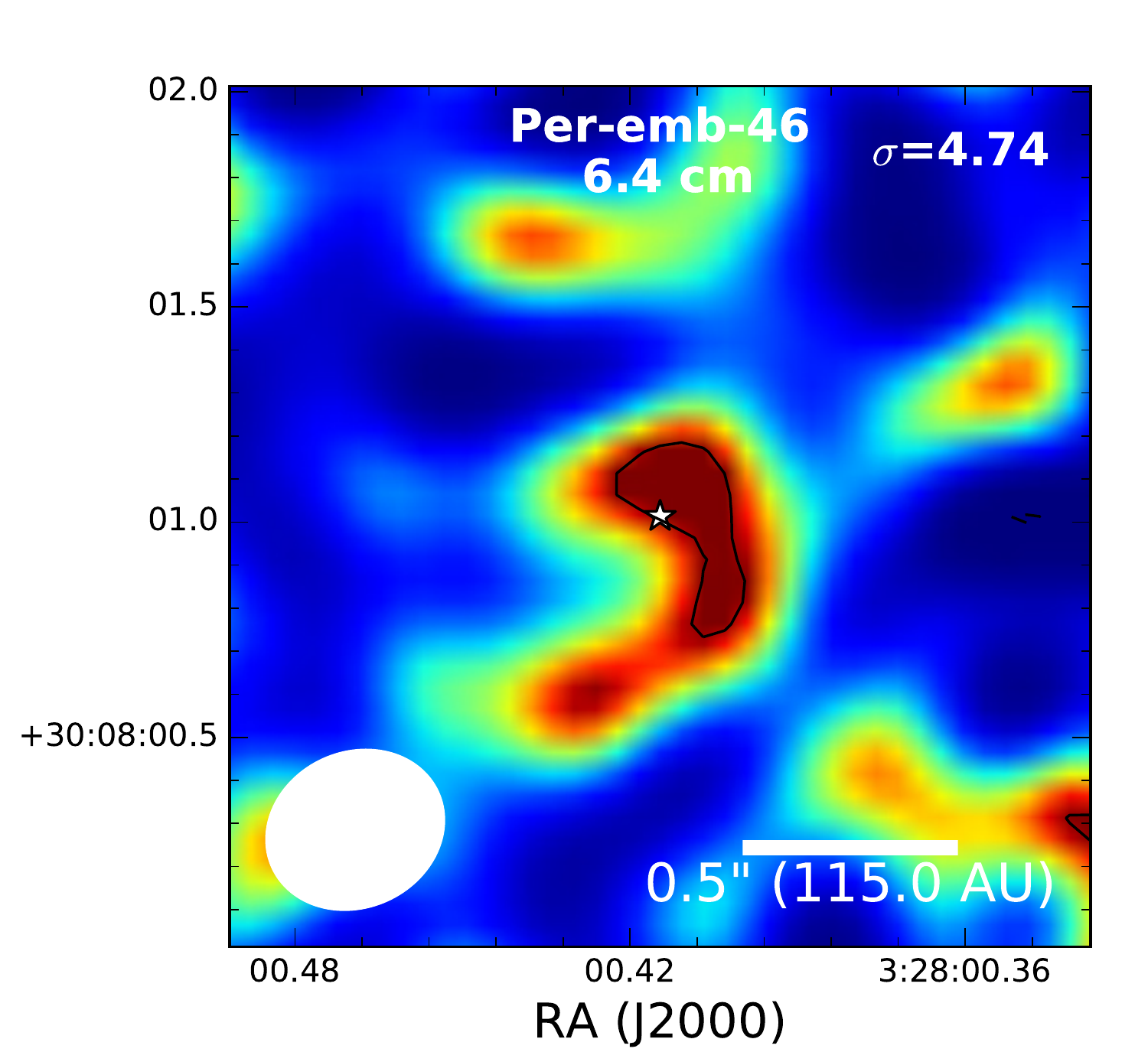}

  \includegraphics[width=0.24\linewidth]{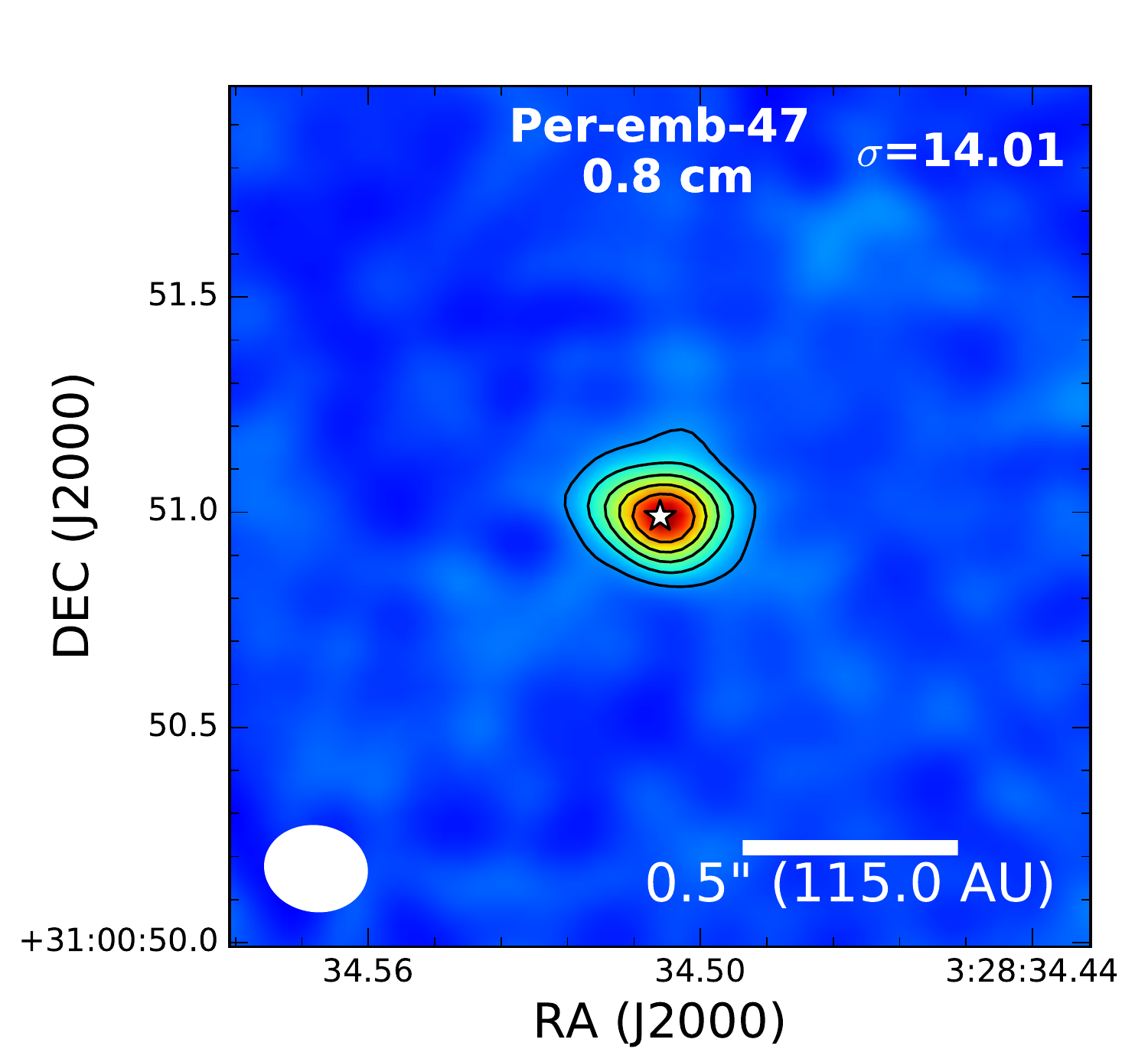}
  \includegraphics[width=0.24\linewidth]{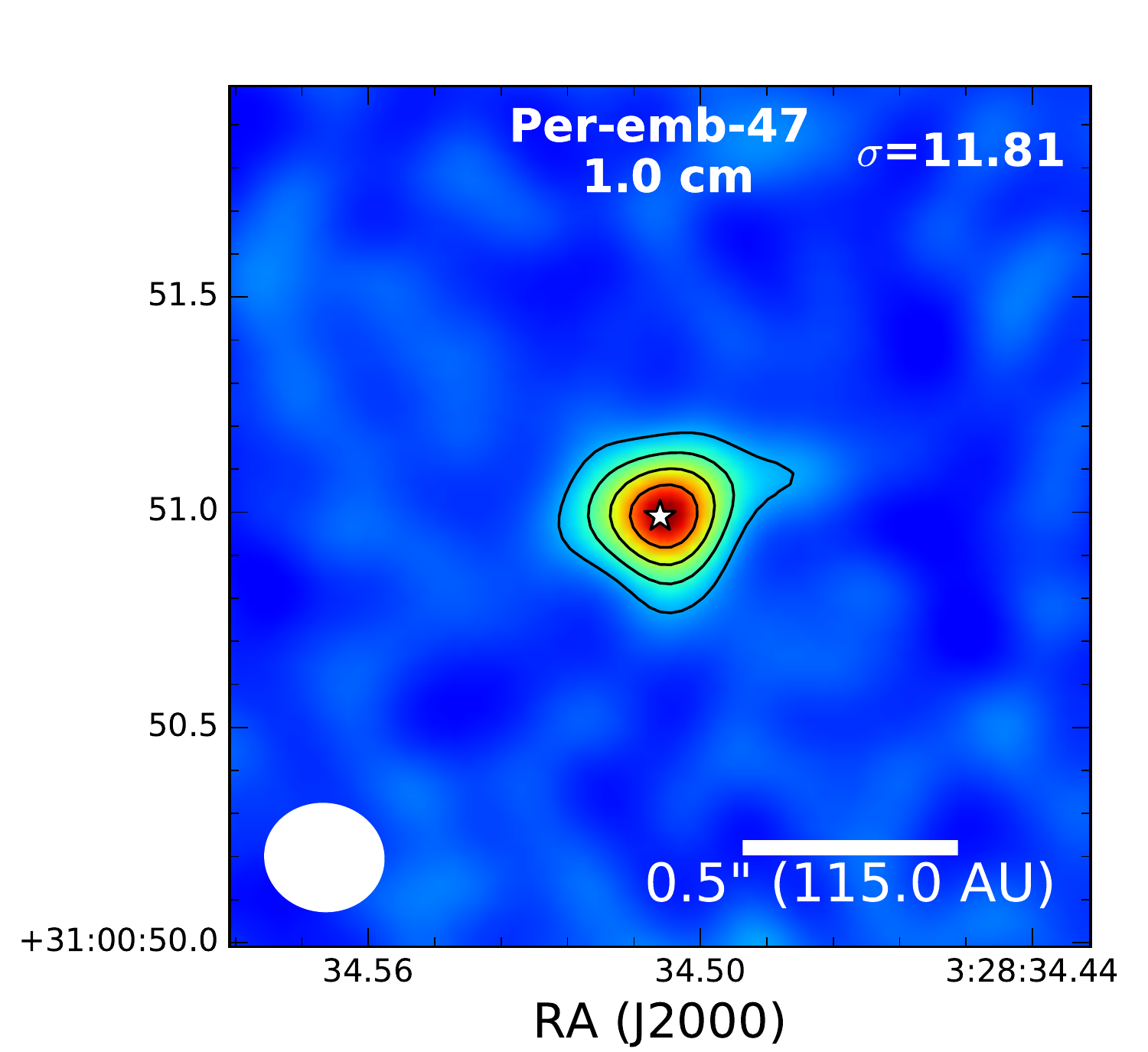}
  \includegraphics[width=0.24\linewidth]{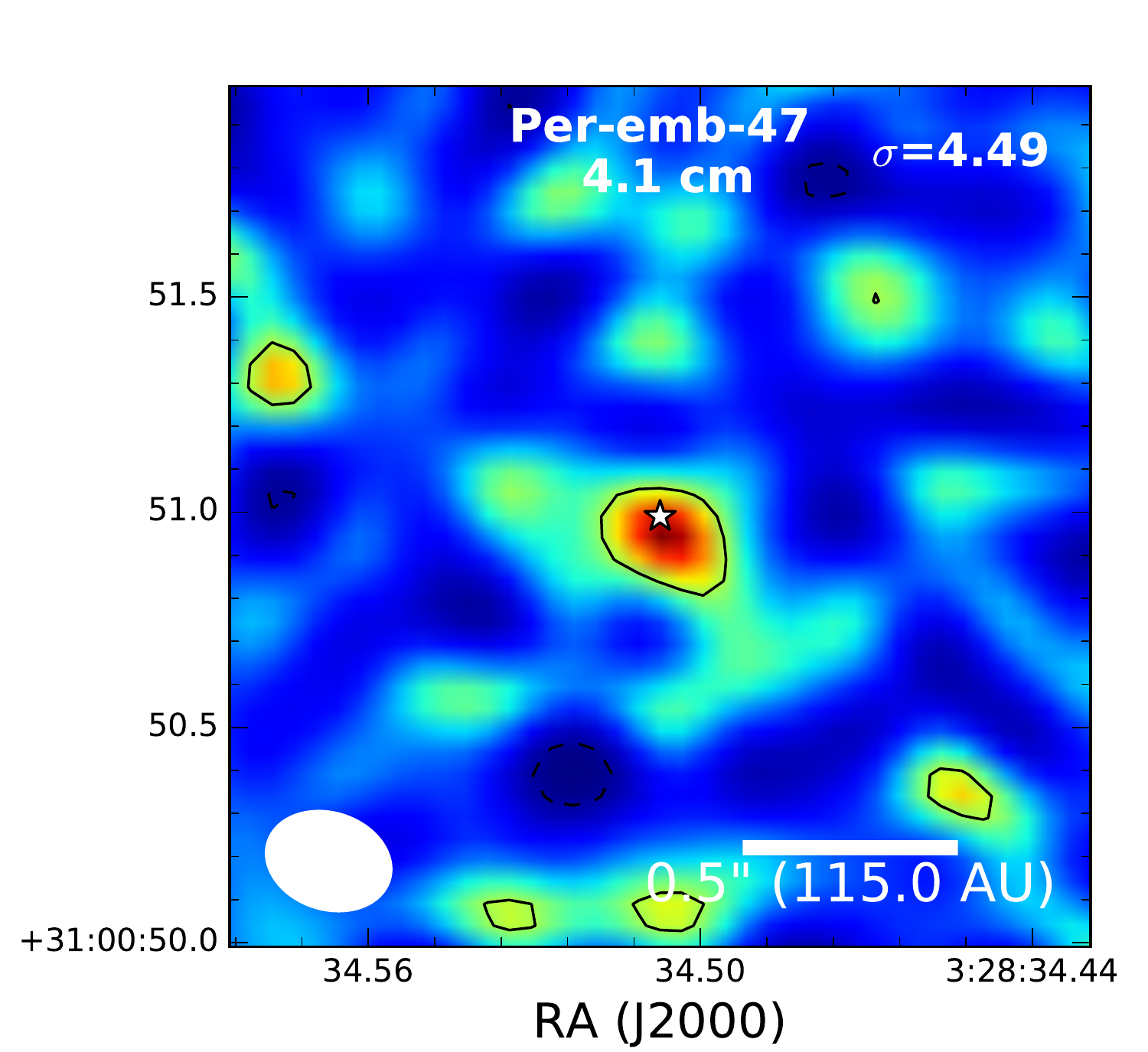}
  \includegraphics[width=0.24\linewidth]{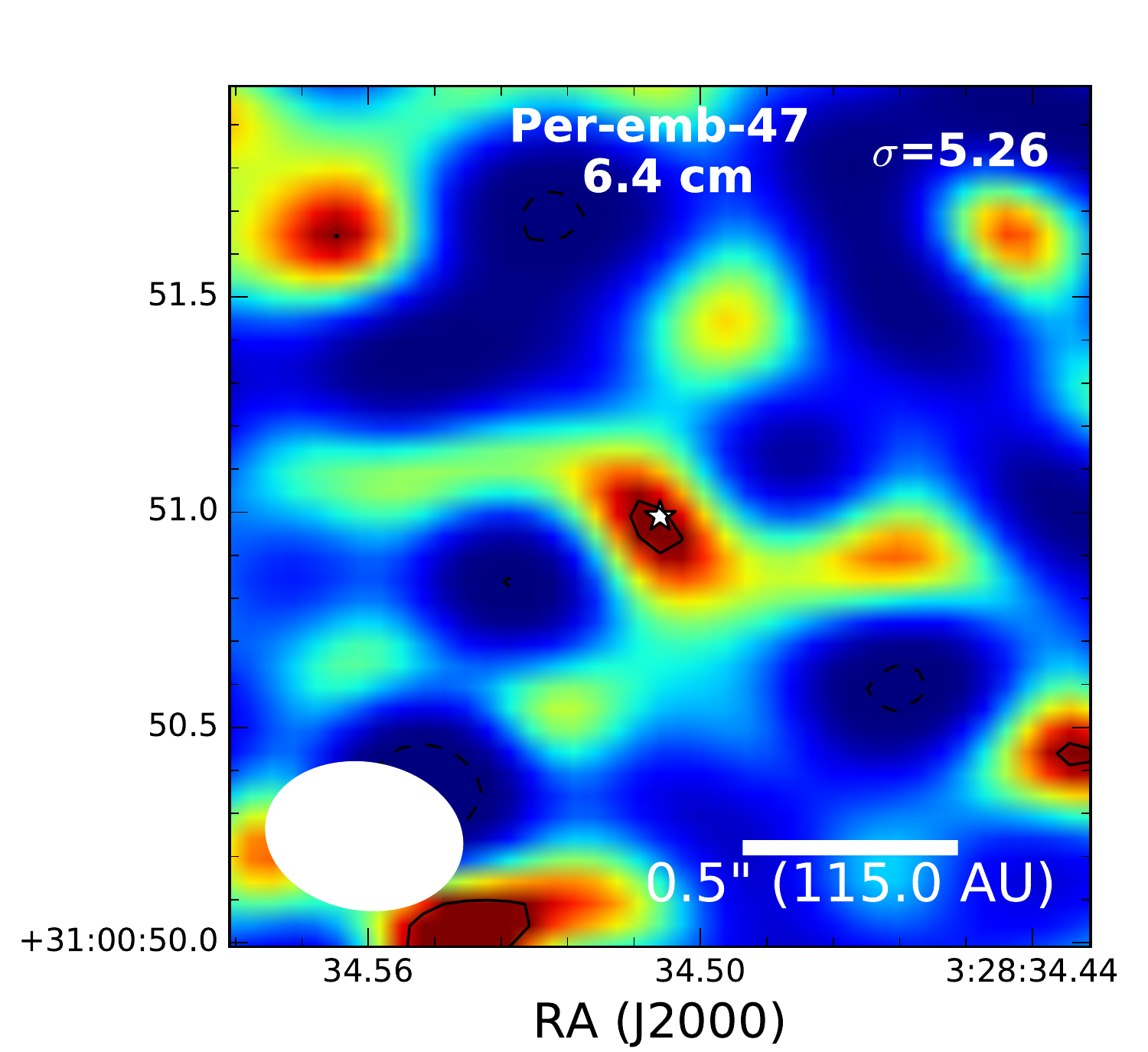}

  \includegraphics[width=0.24\linewidth]{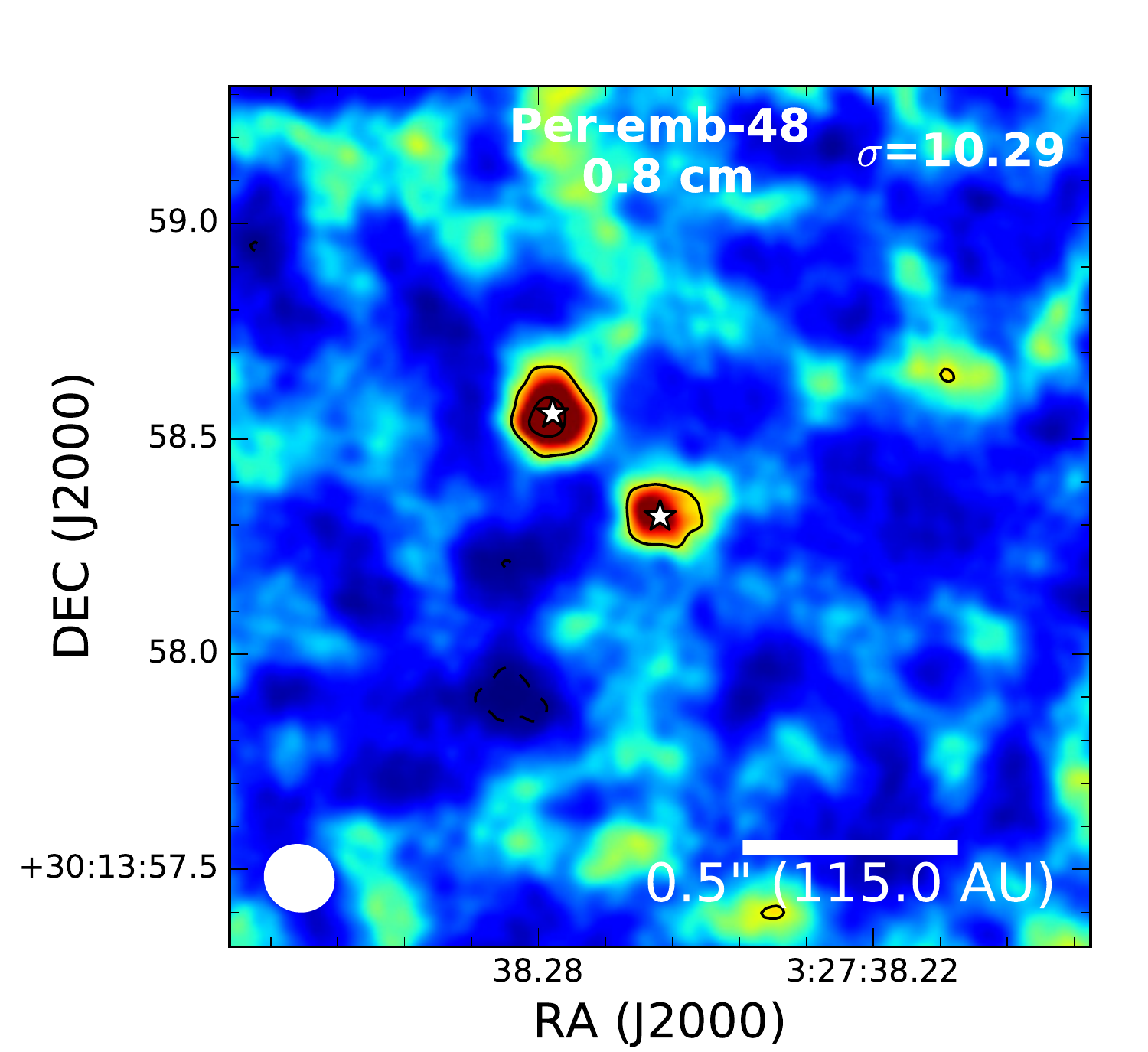}
  \includegraphics[width=0.24\linewidth]{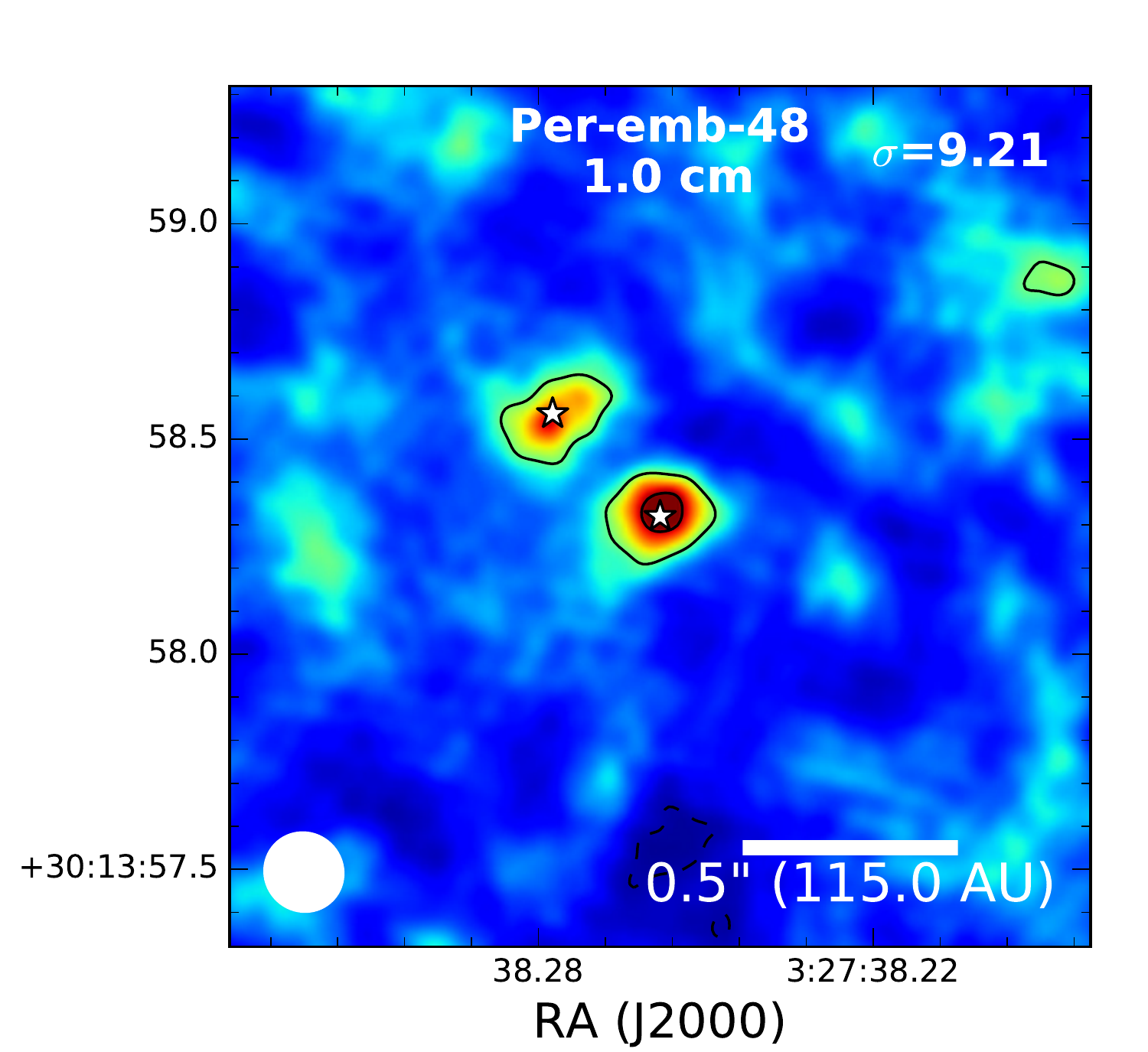}
  \includegraphics[width=0.24\linewidth]{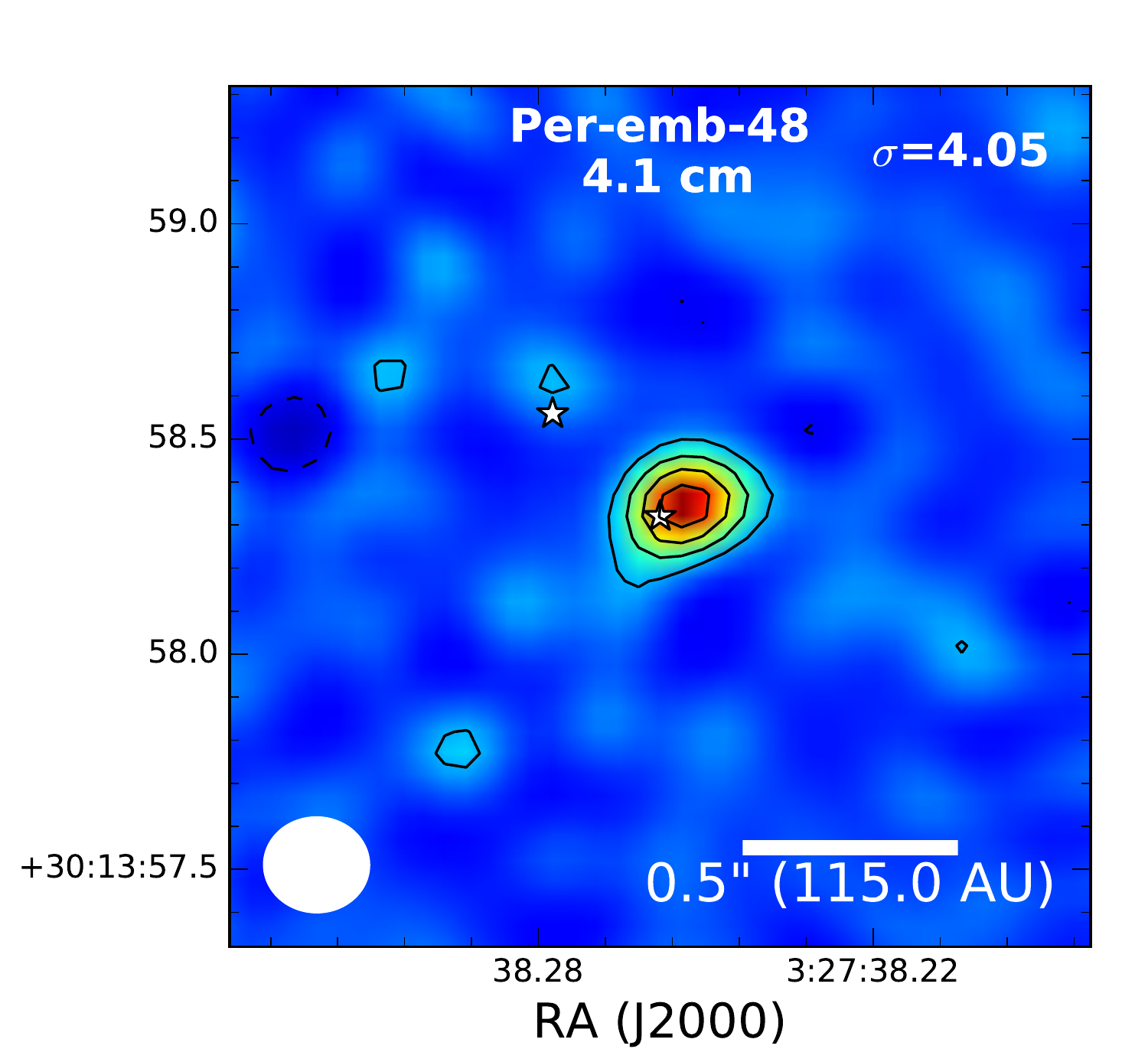}
  \includegraphics[width=0.24\linewidth]{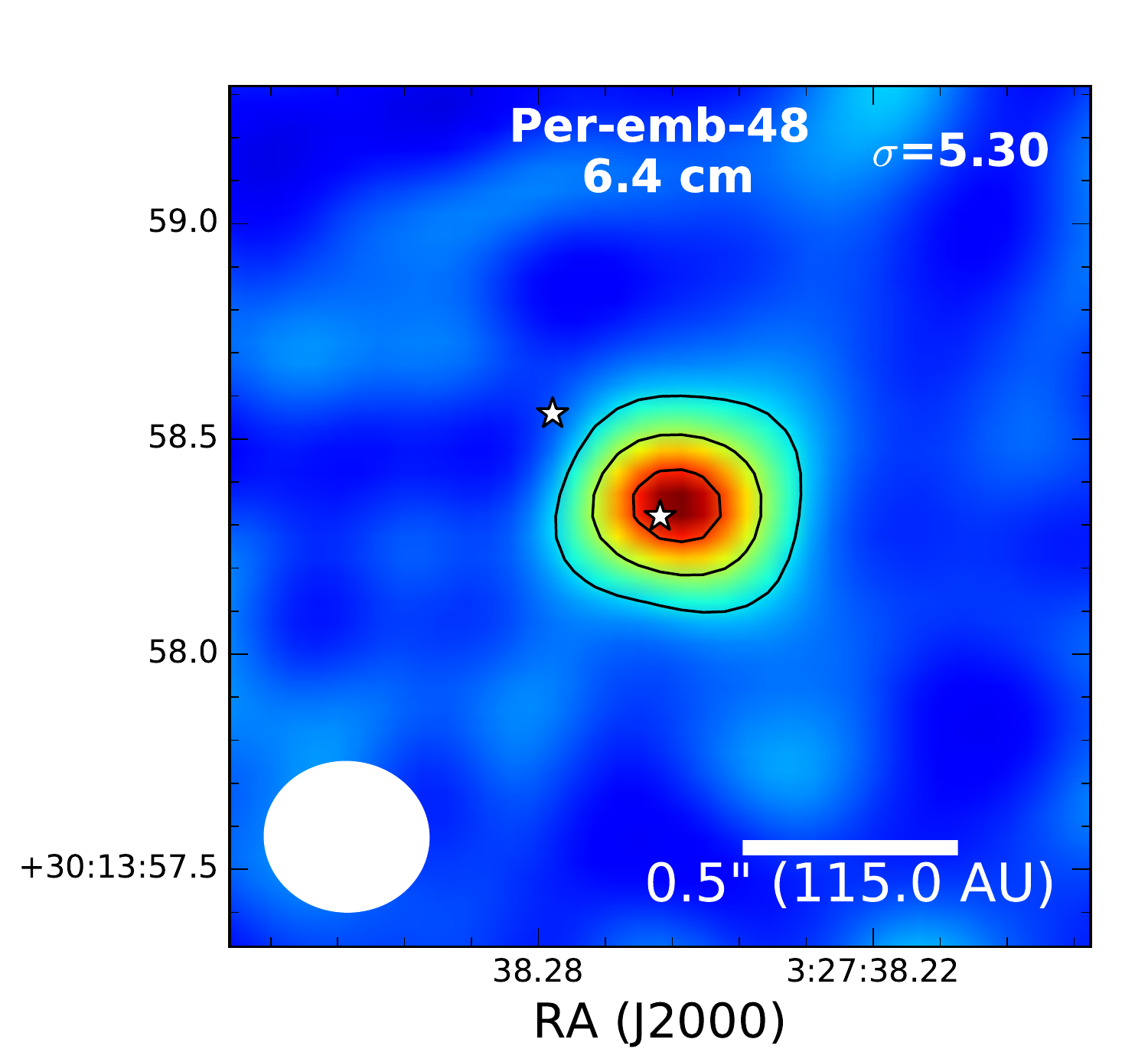}

  \includegraphics[width=0.24\linewidth]{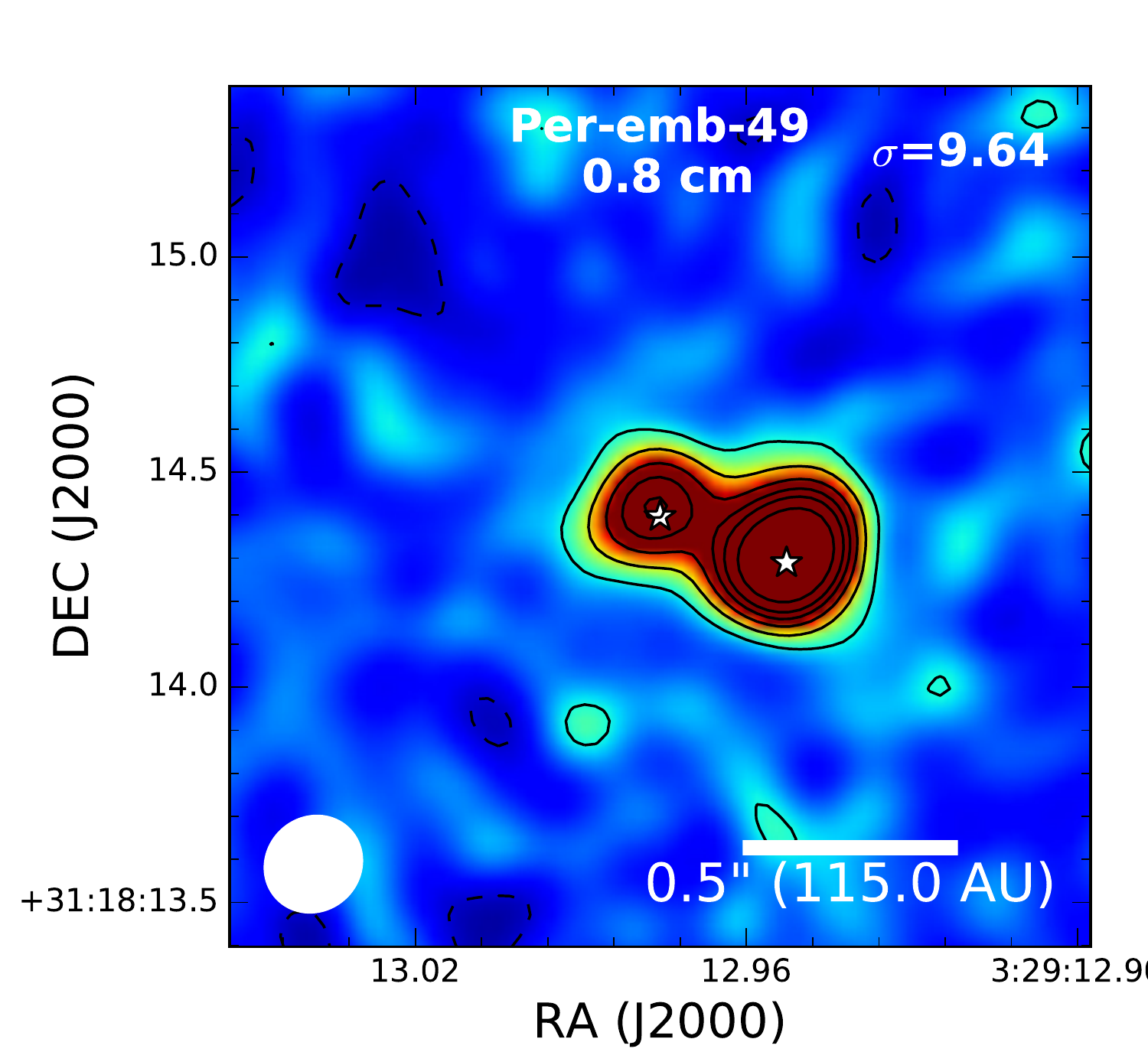}
  \includegraphics[width=0.24\linewidth]{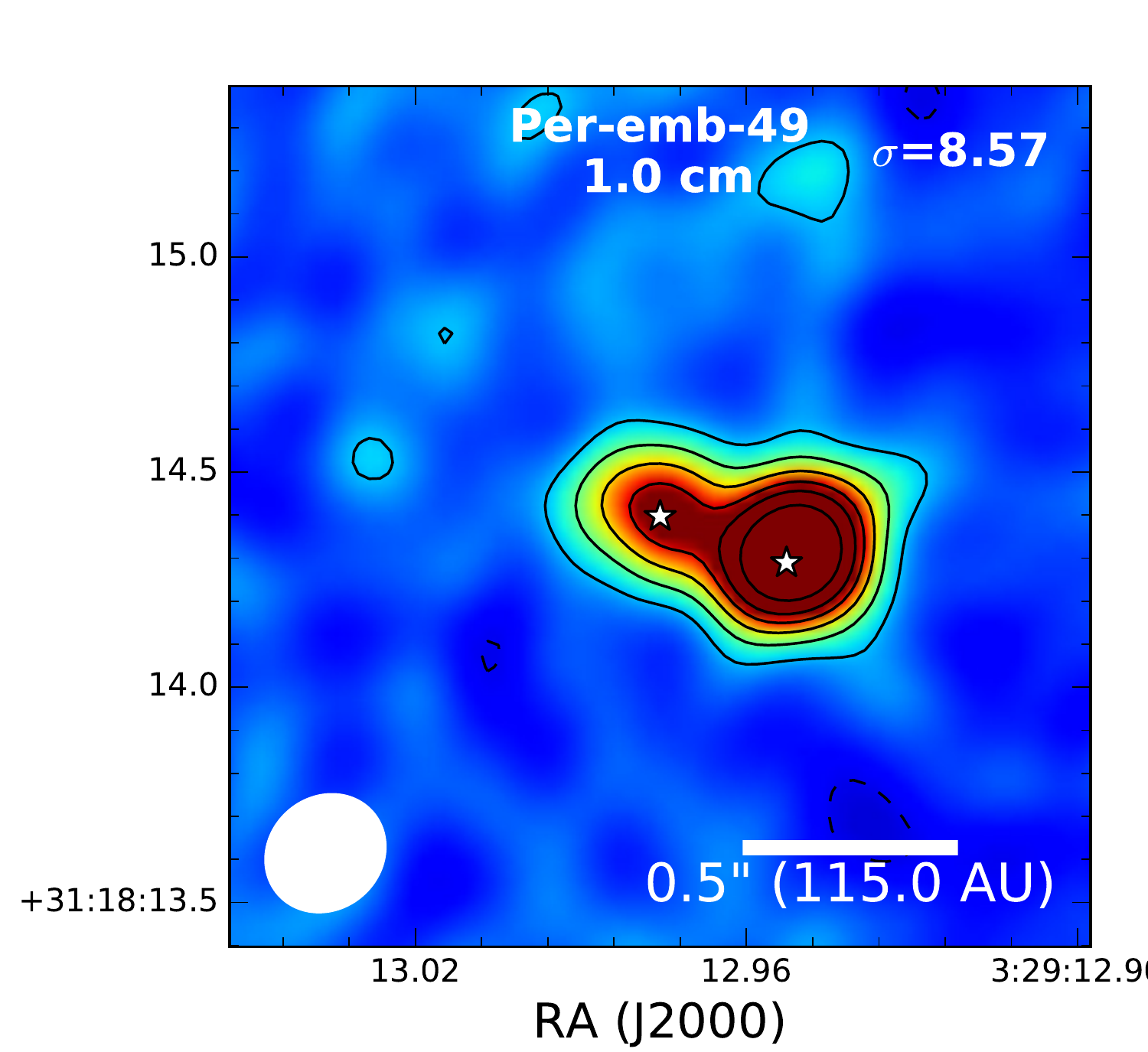}
  \includegraphics[width=0.24\linewidth]{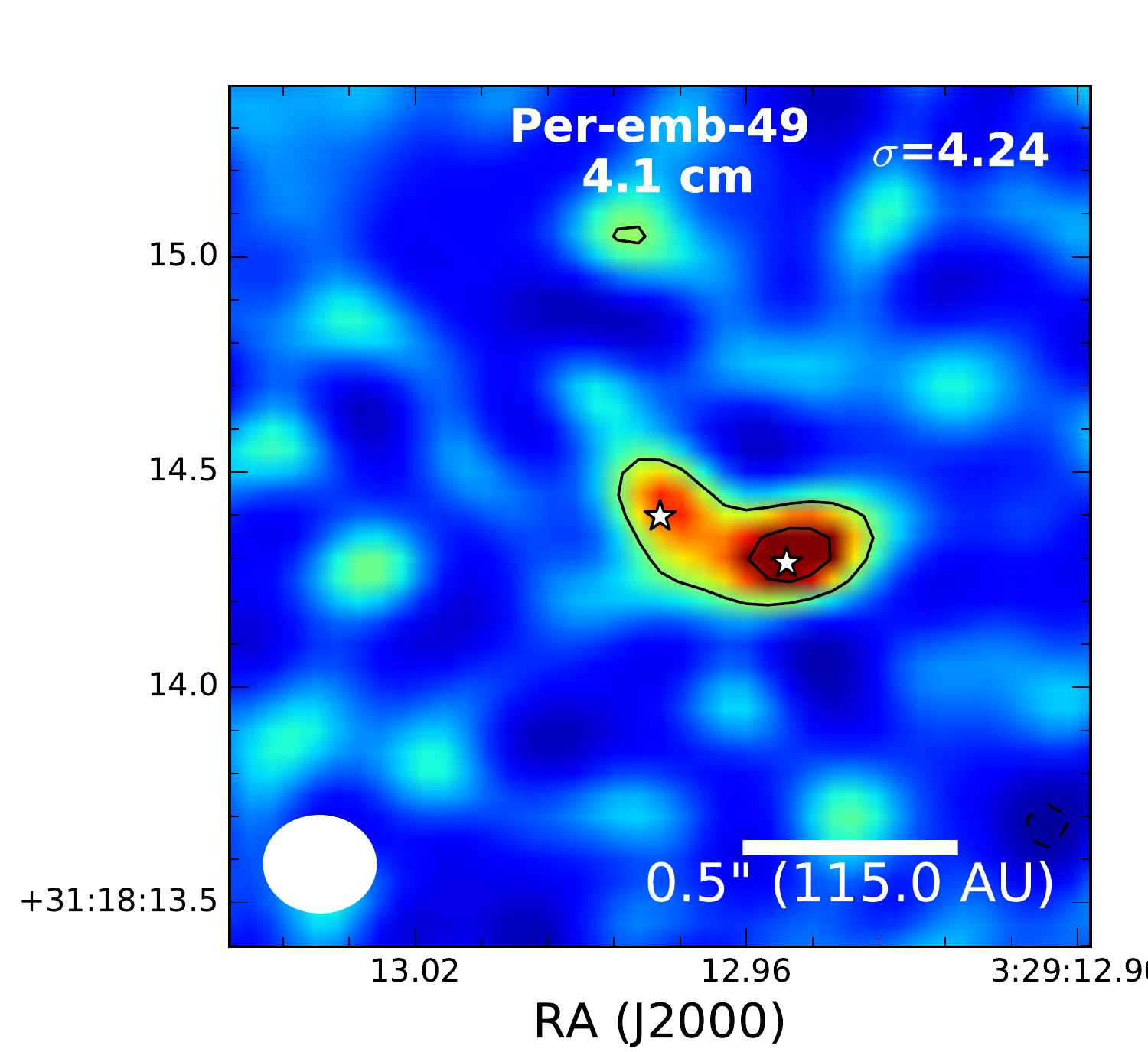}
  \includegraphics[width=0.24\linewidth]{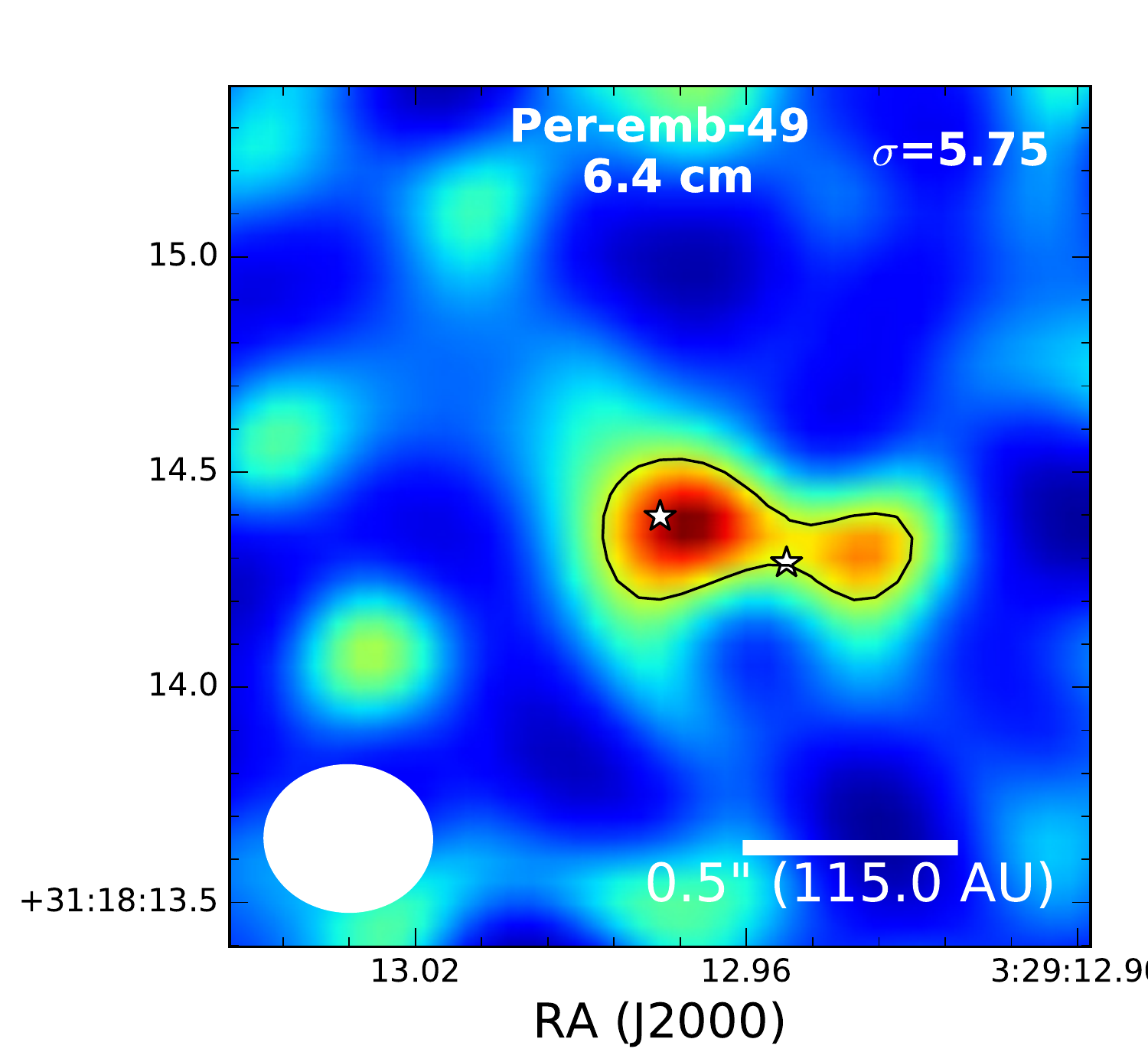}

  \includegraphics[width=0.24\linewidth]{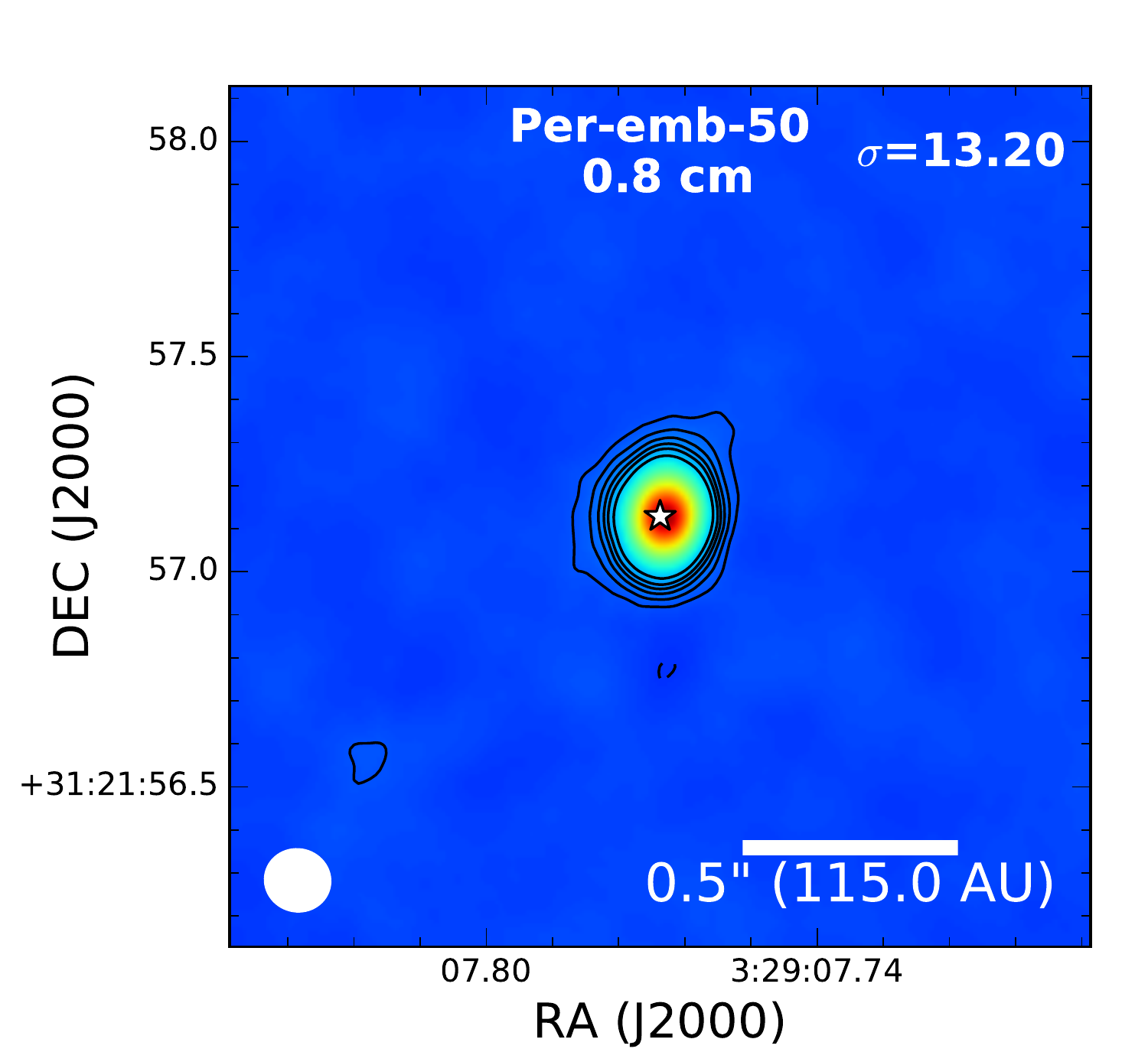}
  \includegraphics[width=0.24\linewidth]{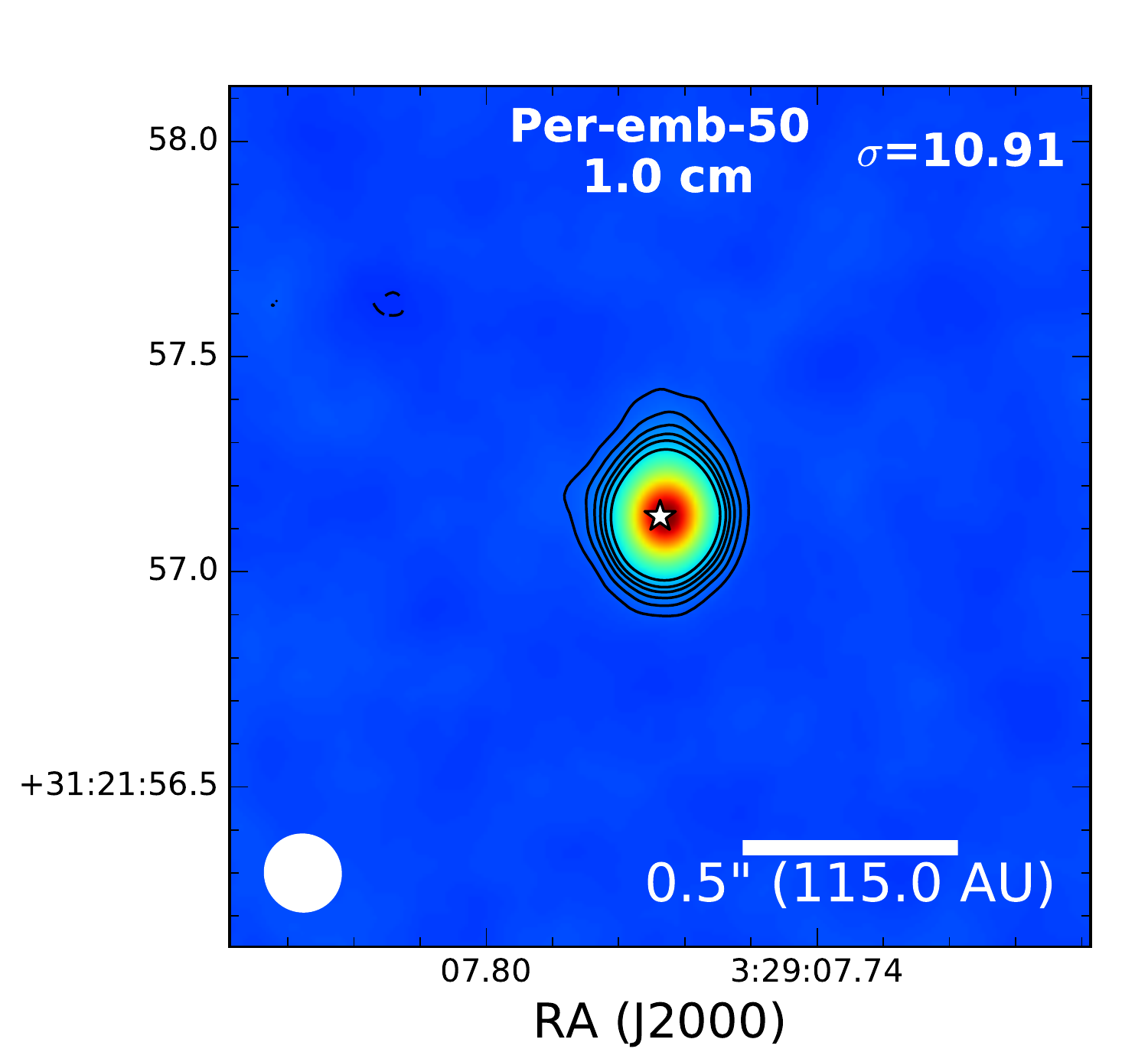}
  \includegraphics[width=0.24\linewidth]{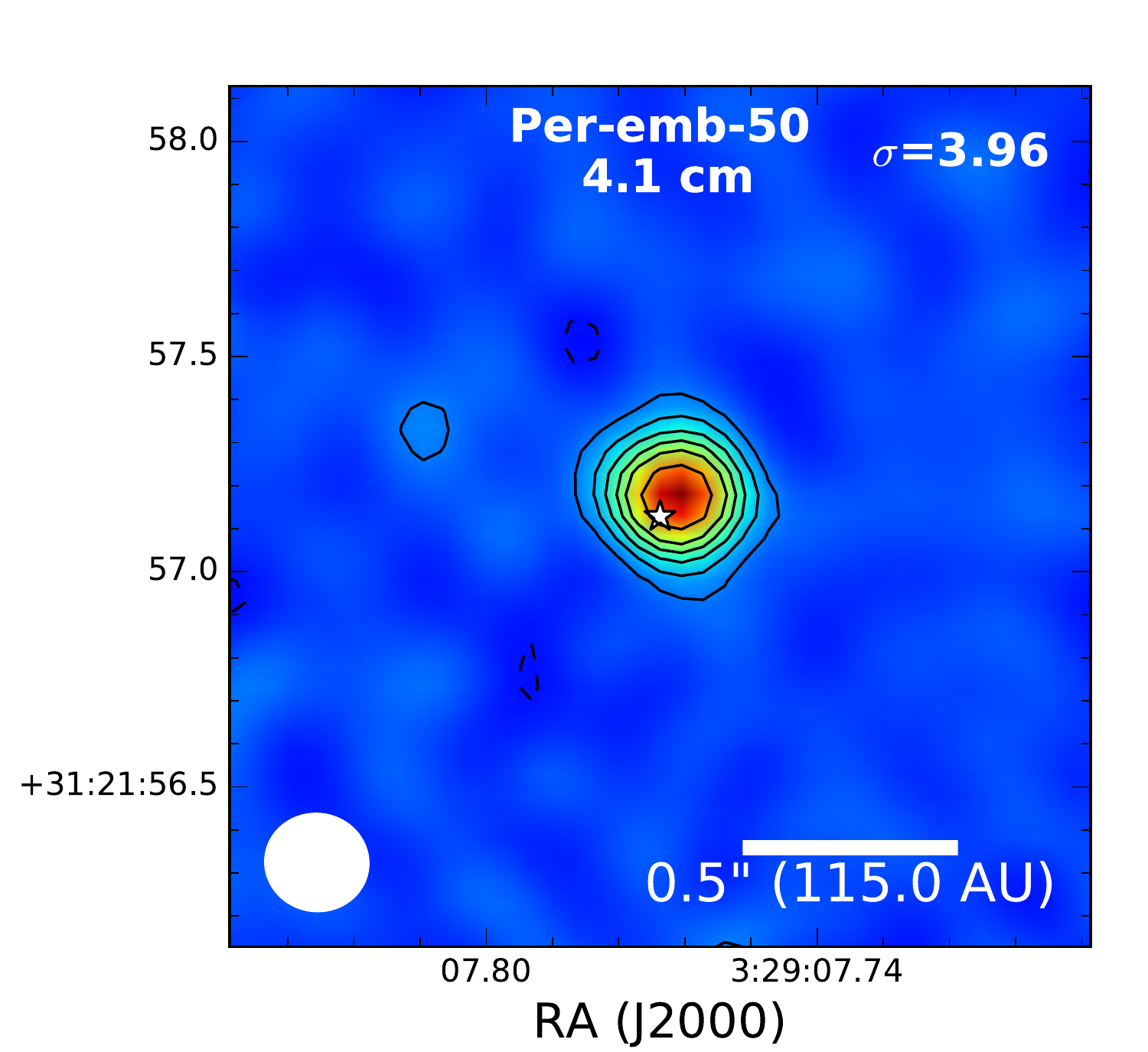}
  \includegraphics[width=0.24\linewidth]{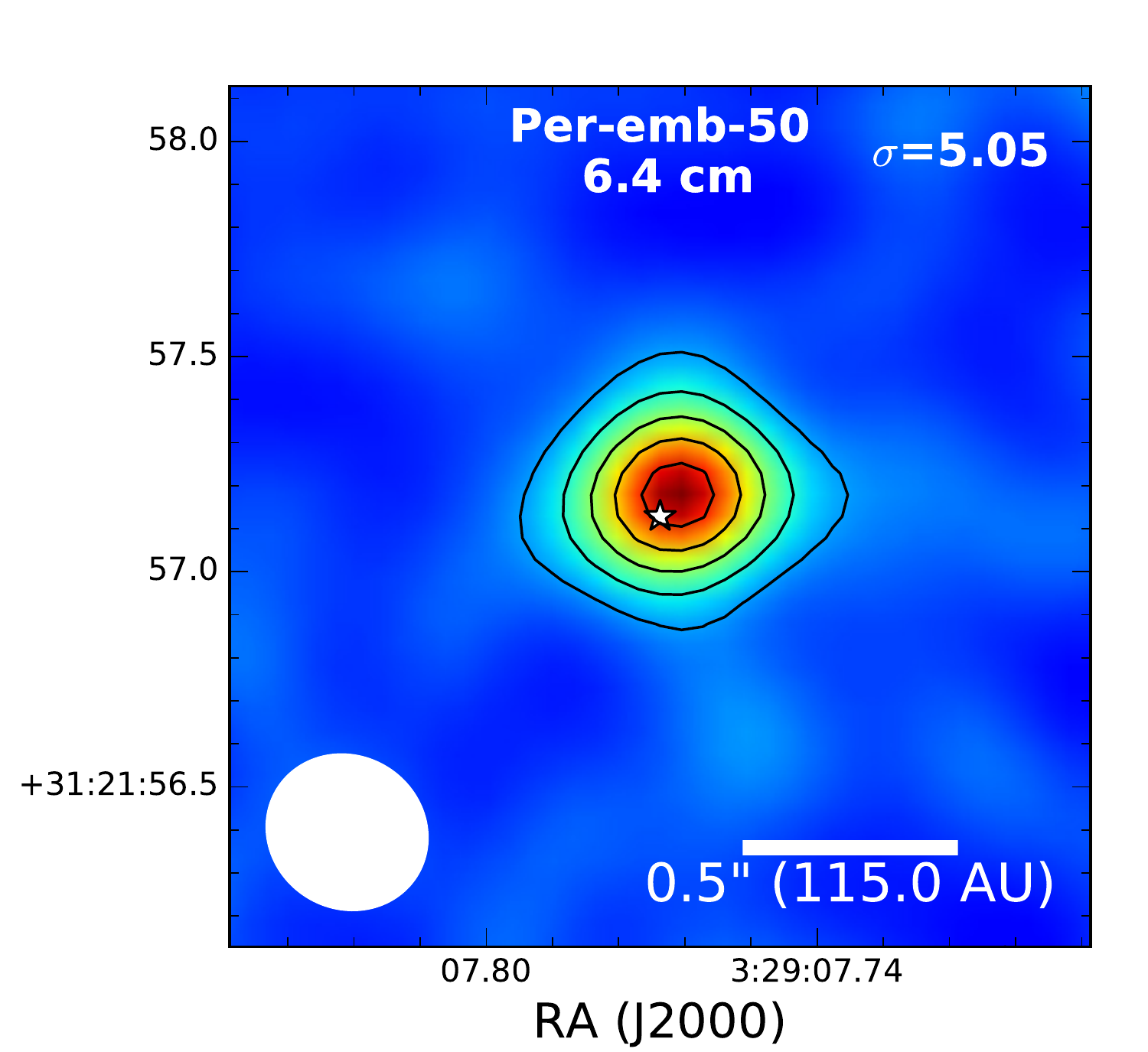}

  \includegraphics[width=0.24\linewidth]{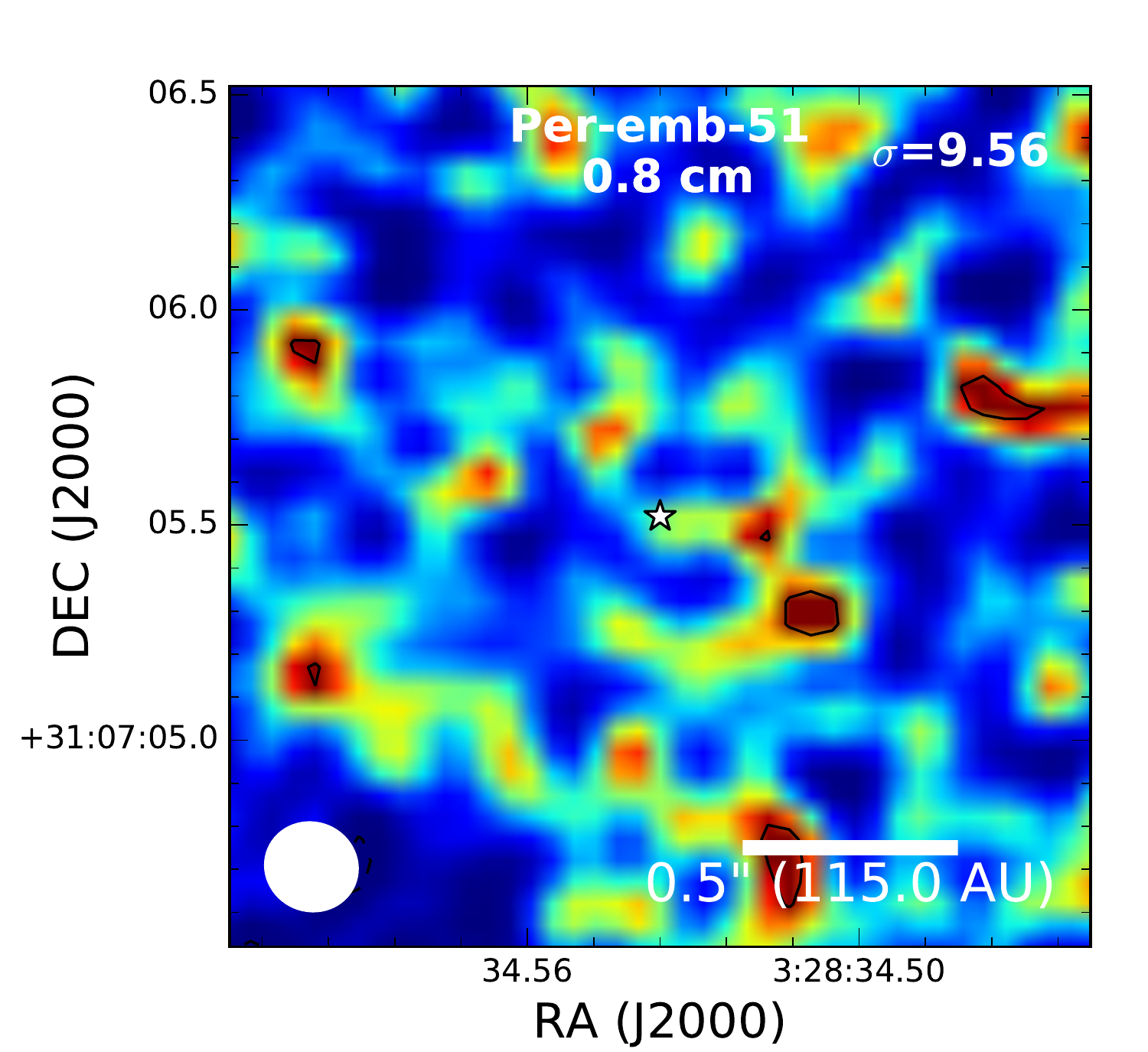}
  \includegraphics[width=0.24\linewidth]{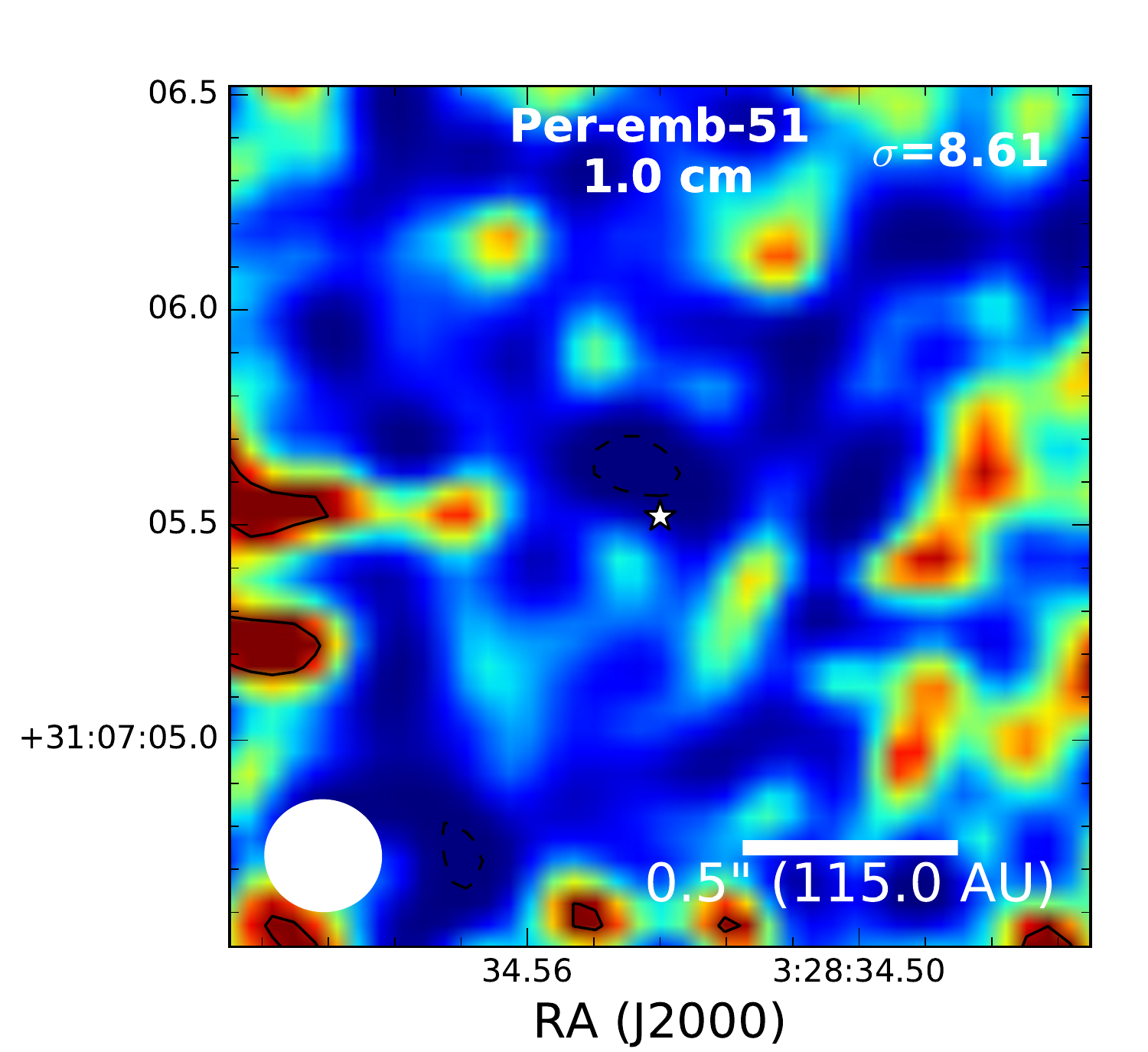}
  \includegraphics[width=0.24\linewidth]{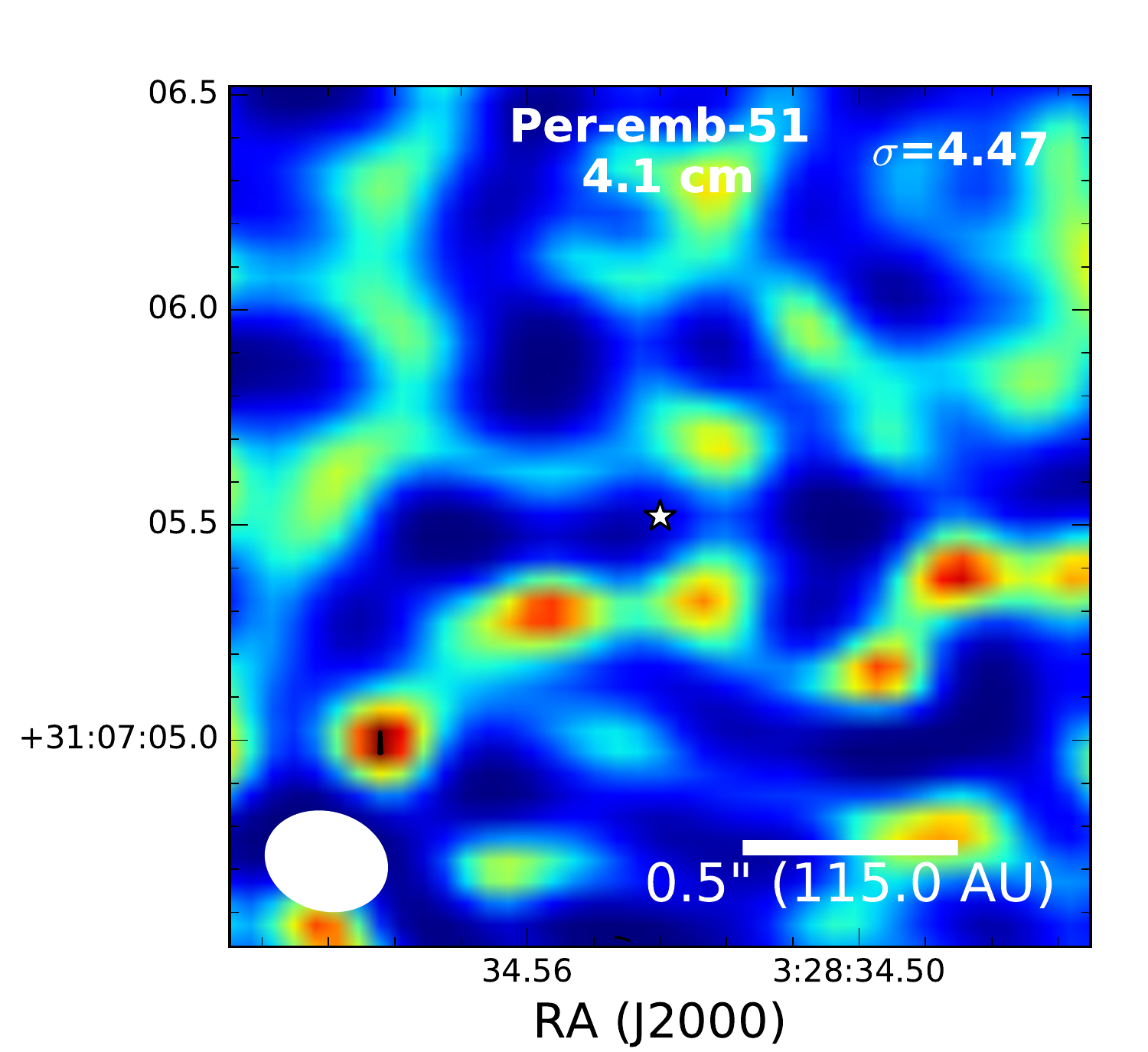}
  \includegraphics[width=0.24\linewidth]{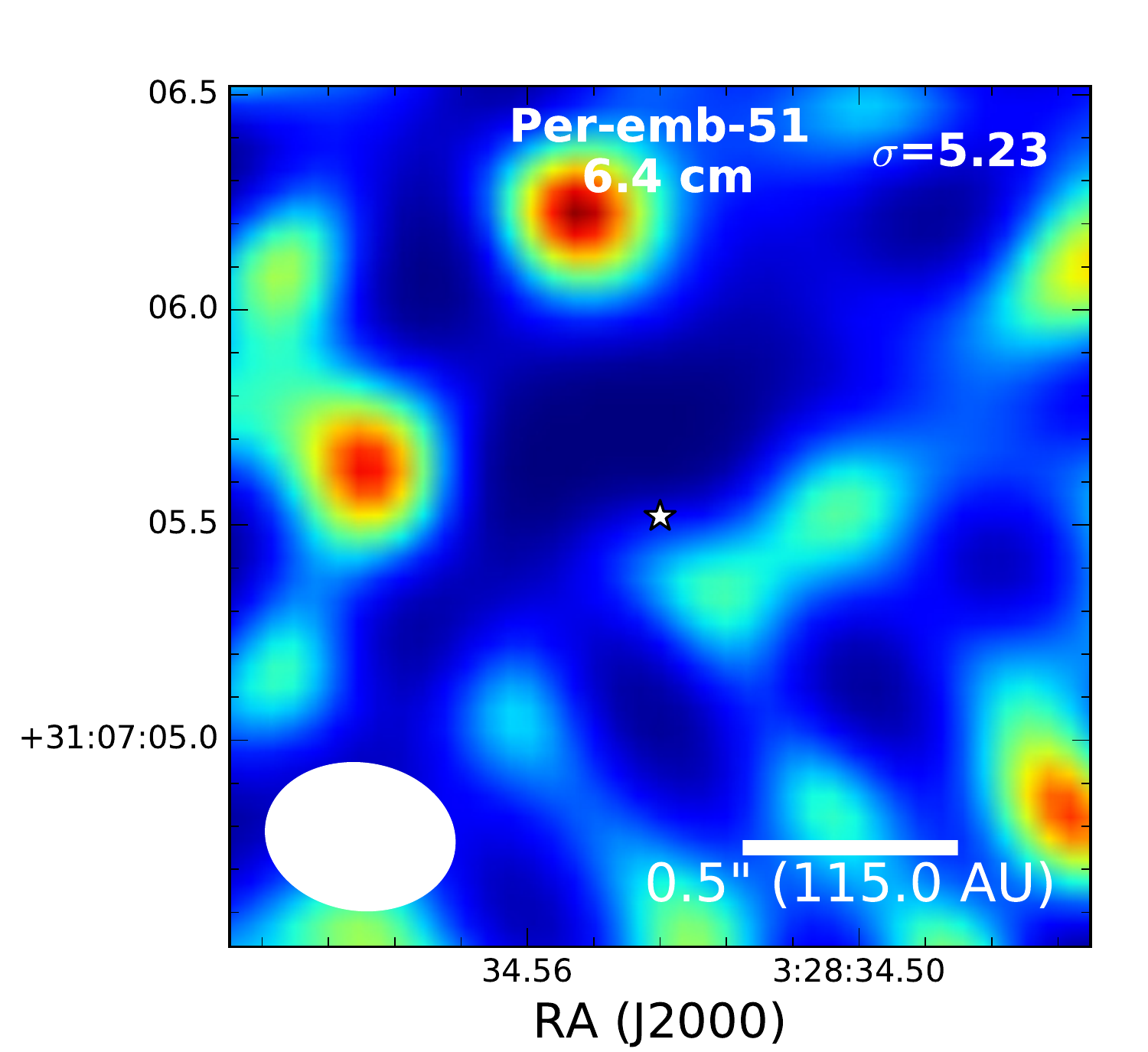}

\end{figure}

\begin{figure}

  \includegraphics[width=0.24\linewidth]{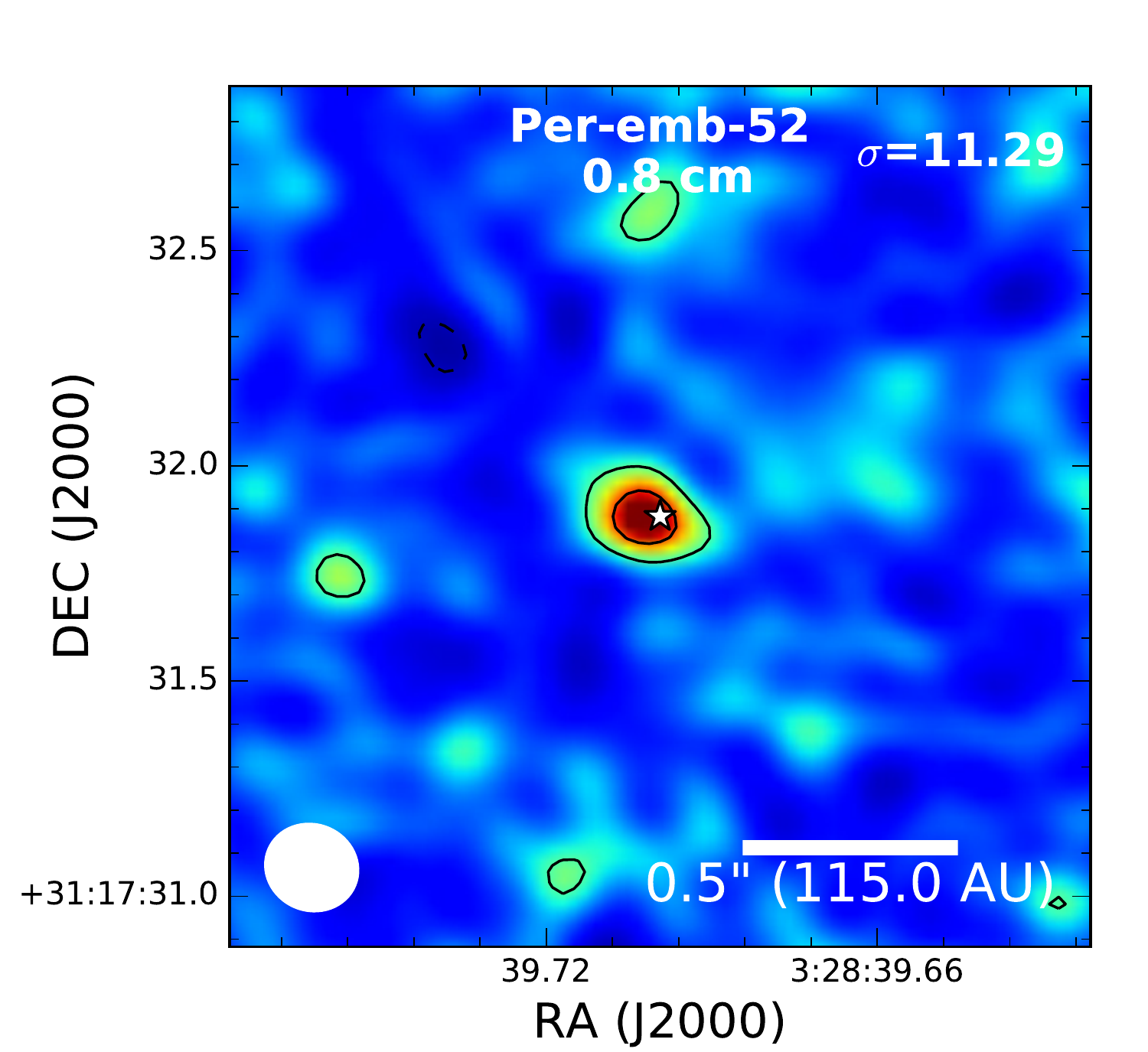}
  \includegraphics[width=0.24\linewidth]{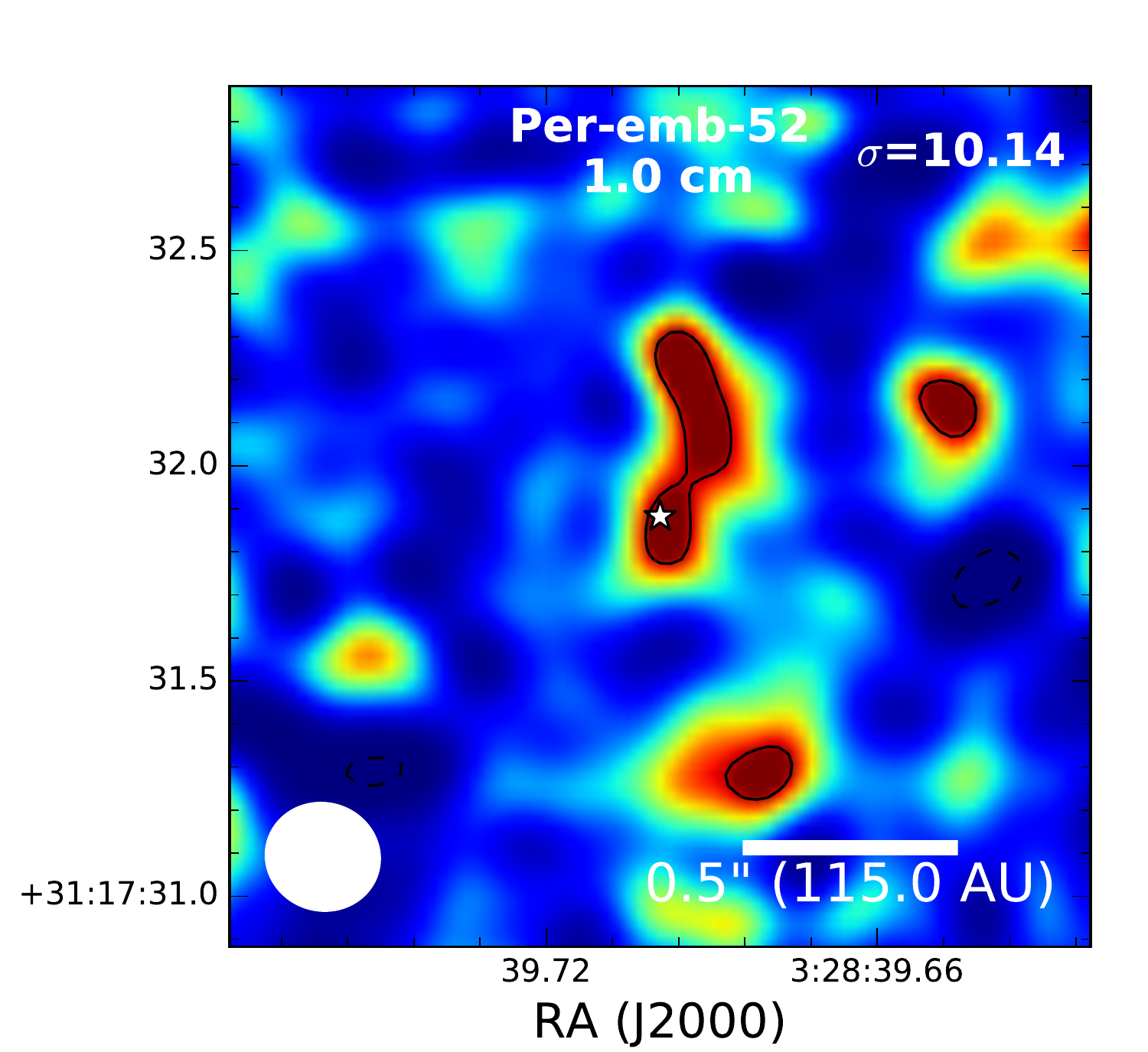}
  \includegraphics[width=0.24\linewidth]{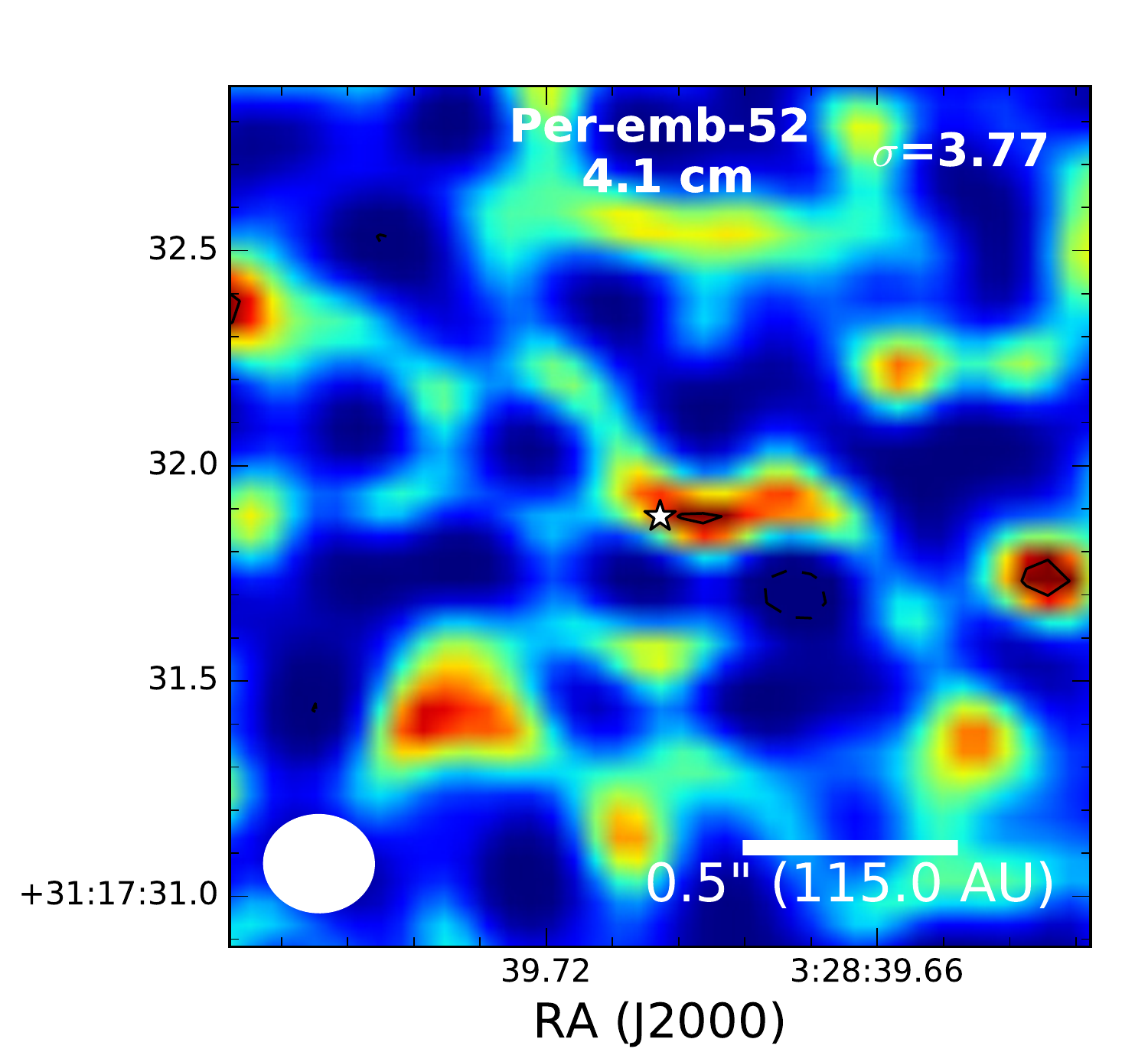}
  \includegraphics[width=0.24\linewidth]{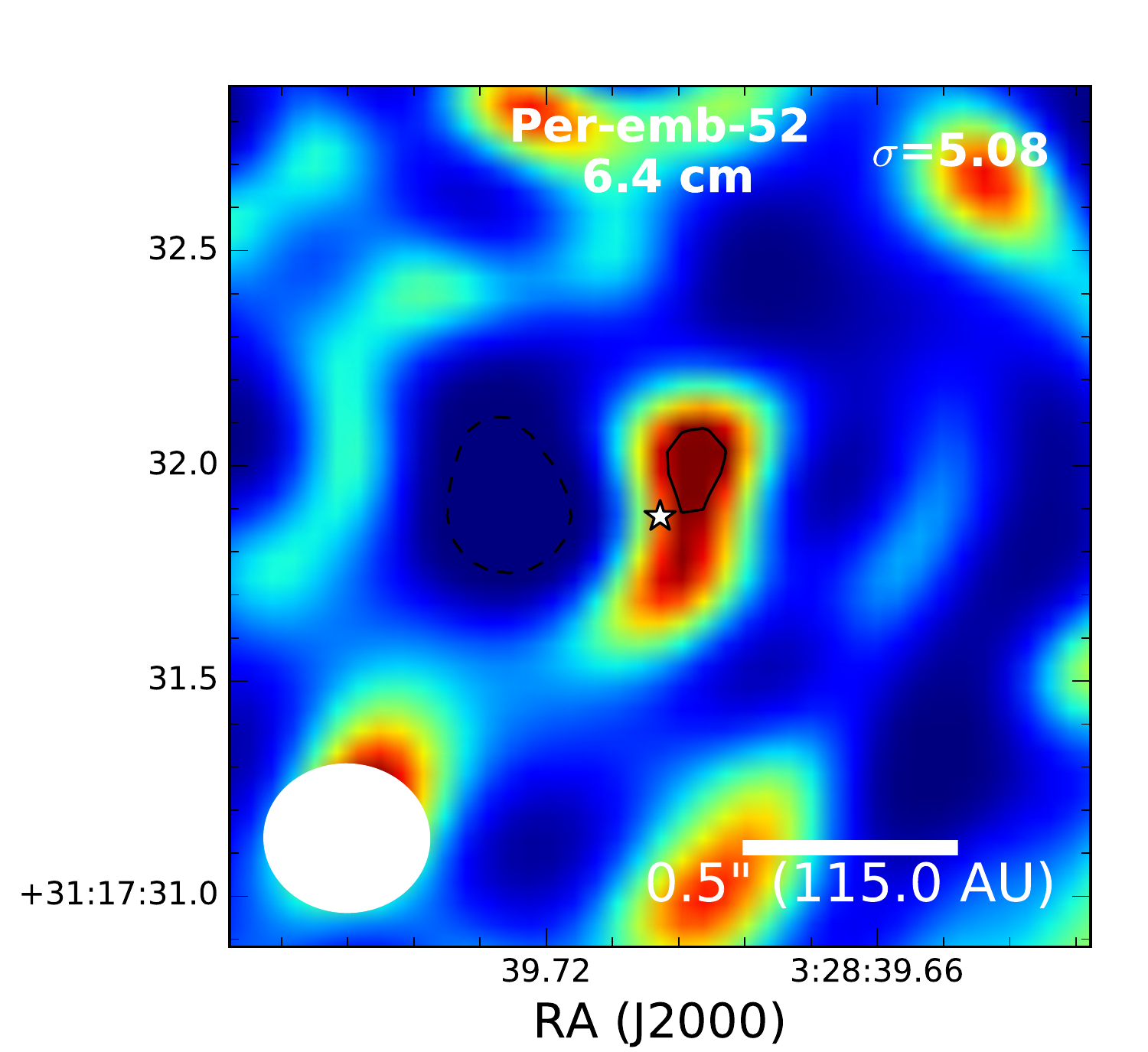}

  \includegraphics[width=0.24\linewidth]{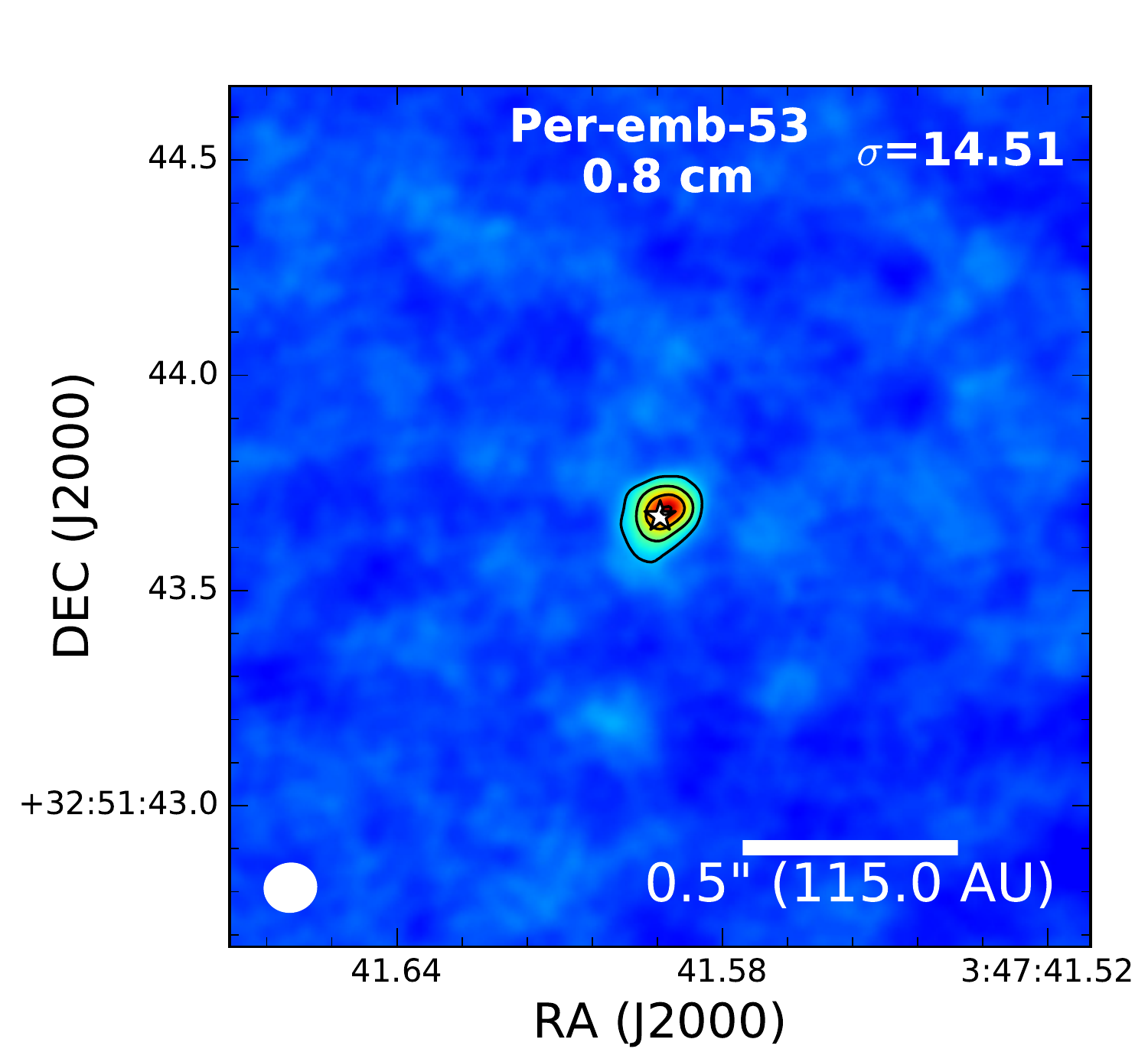}
  \includegraphics[width=0.24\linewidth]{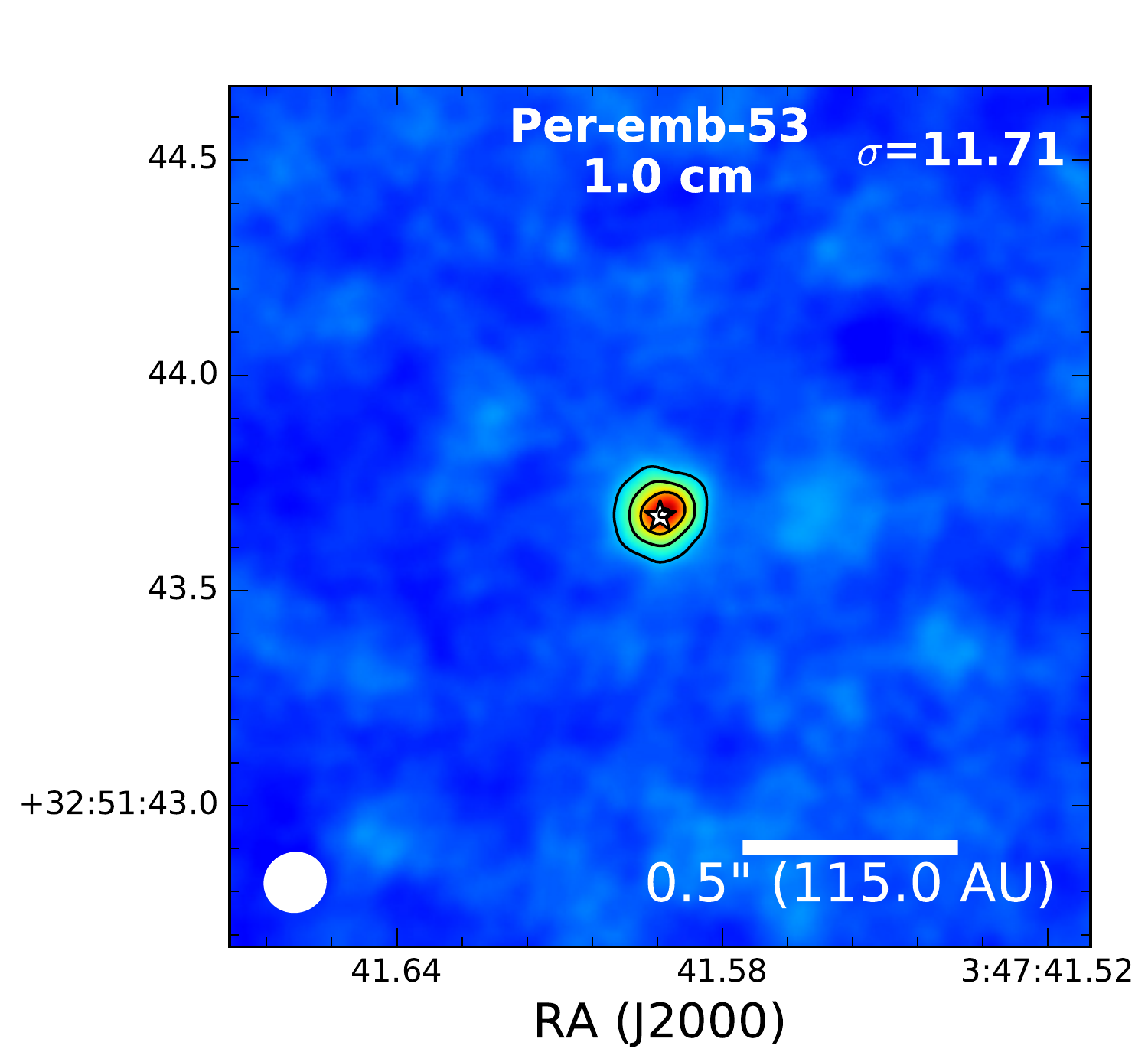}
  \includegraphics[width=0.24\linewidth]{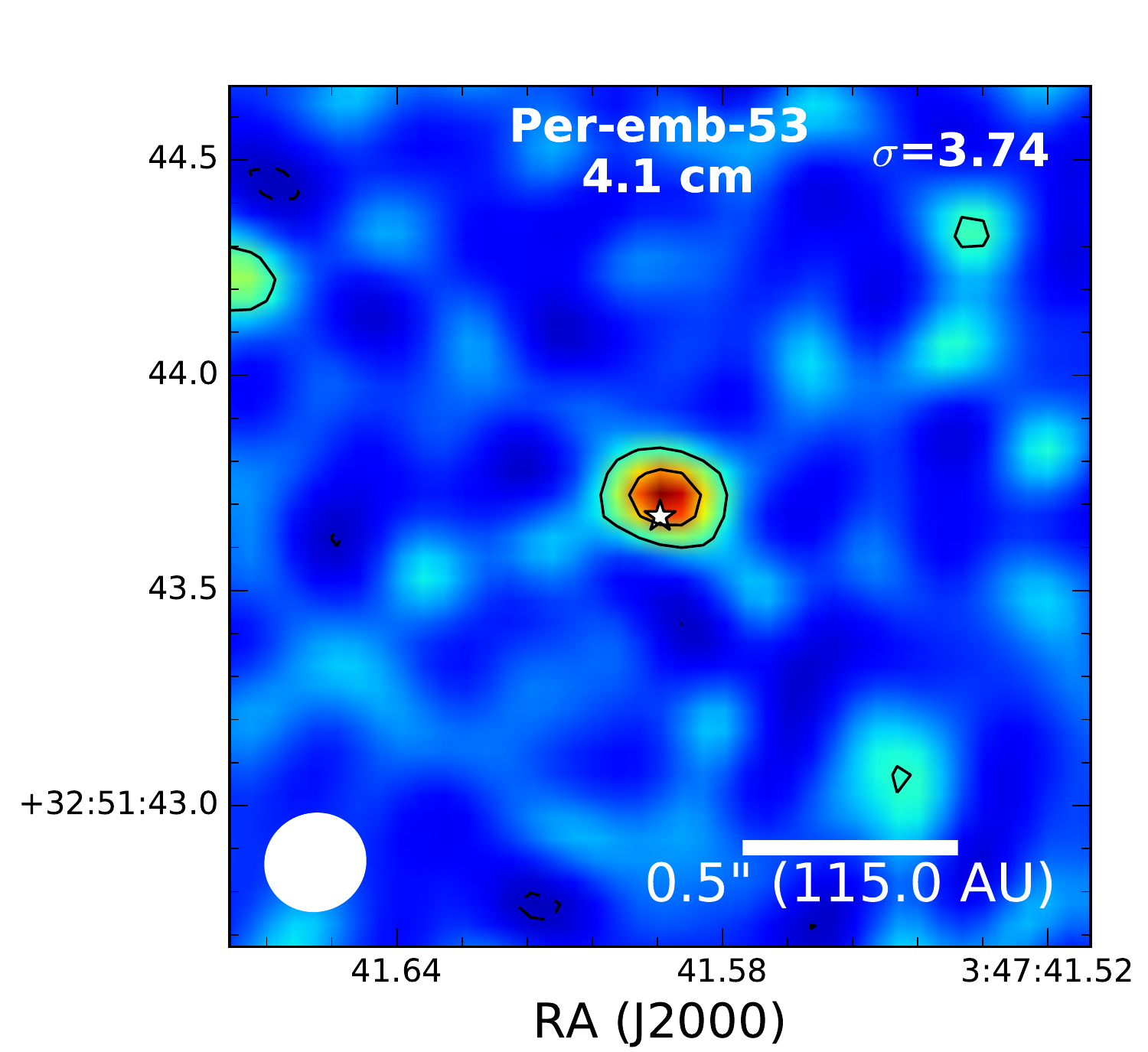}
  \includegraphics[width=0.24\linewidth]{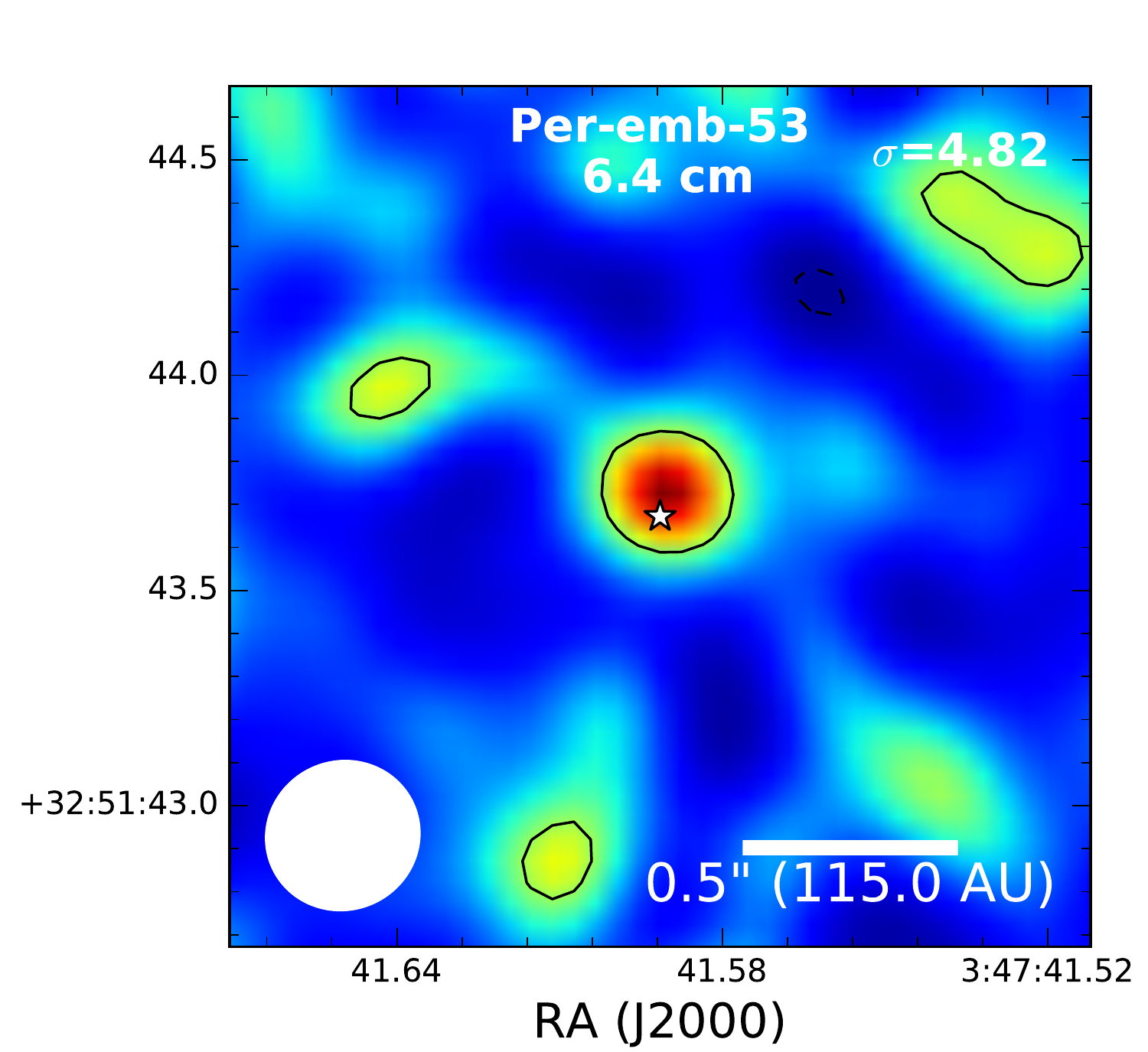}

  \includegraphics[width=0.24\linewidth]{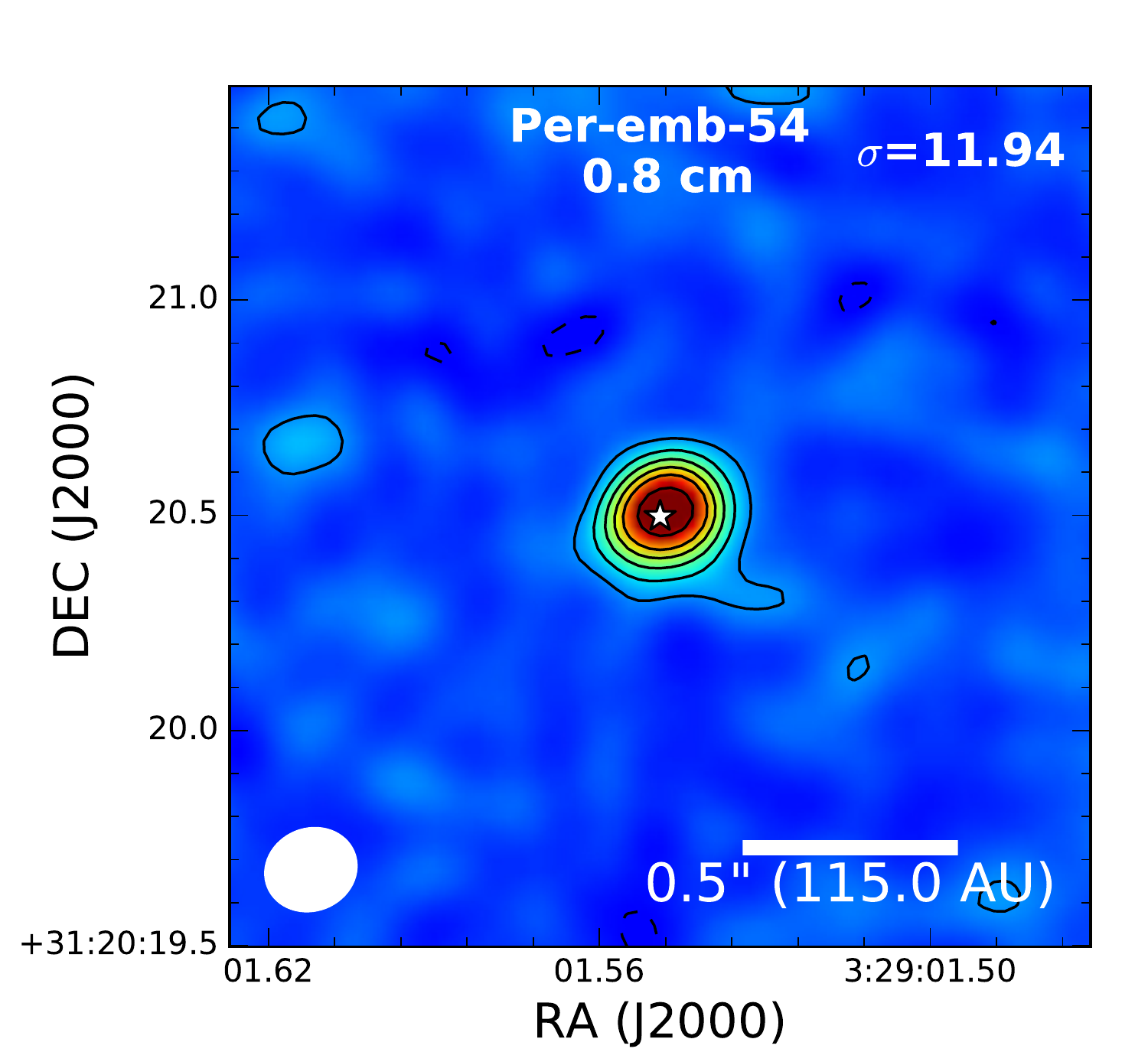}
  \includegraphics[width=0.24\linewidth]{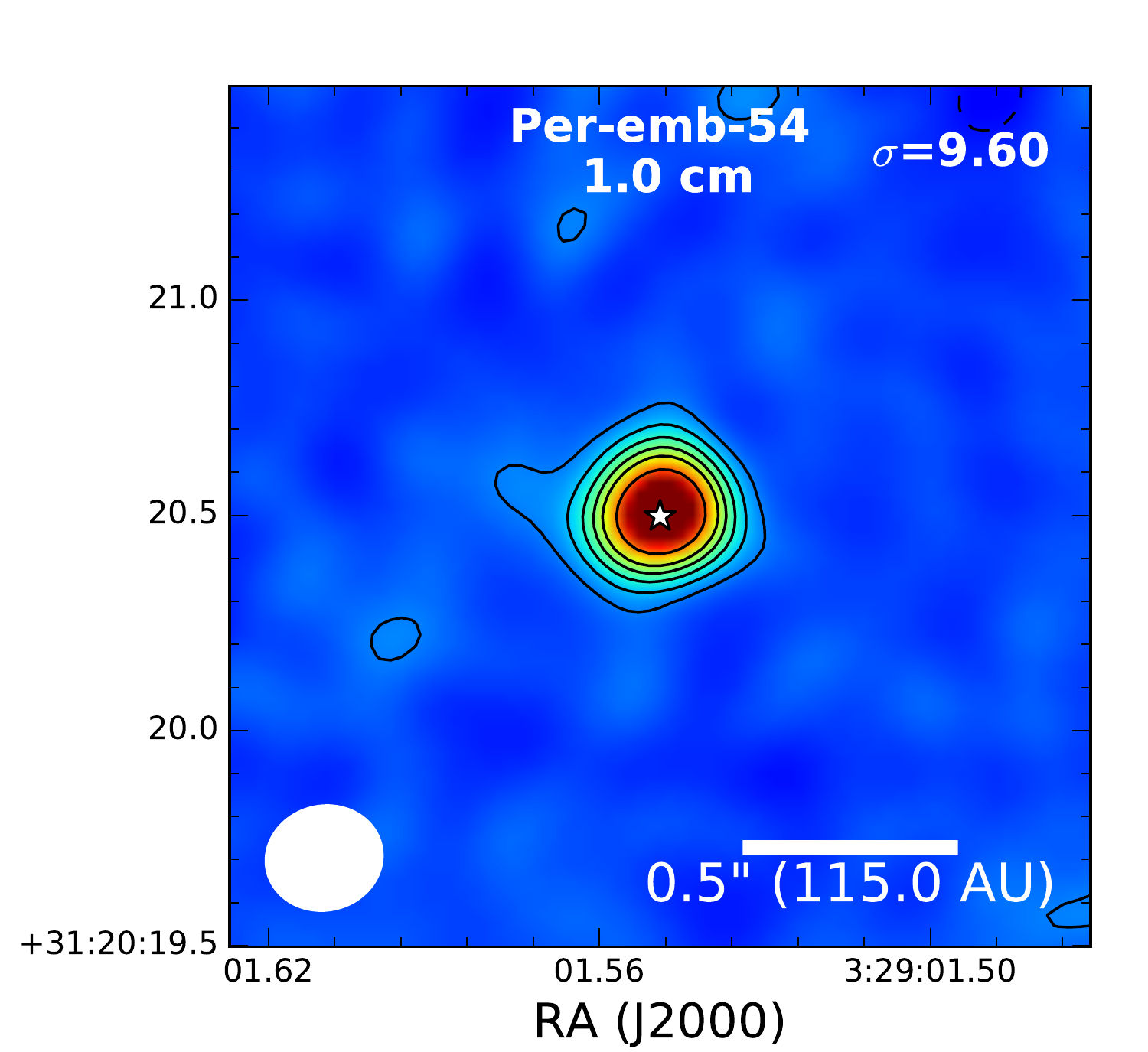}
  \includegraphics[width=0.24\linewidth]{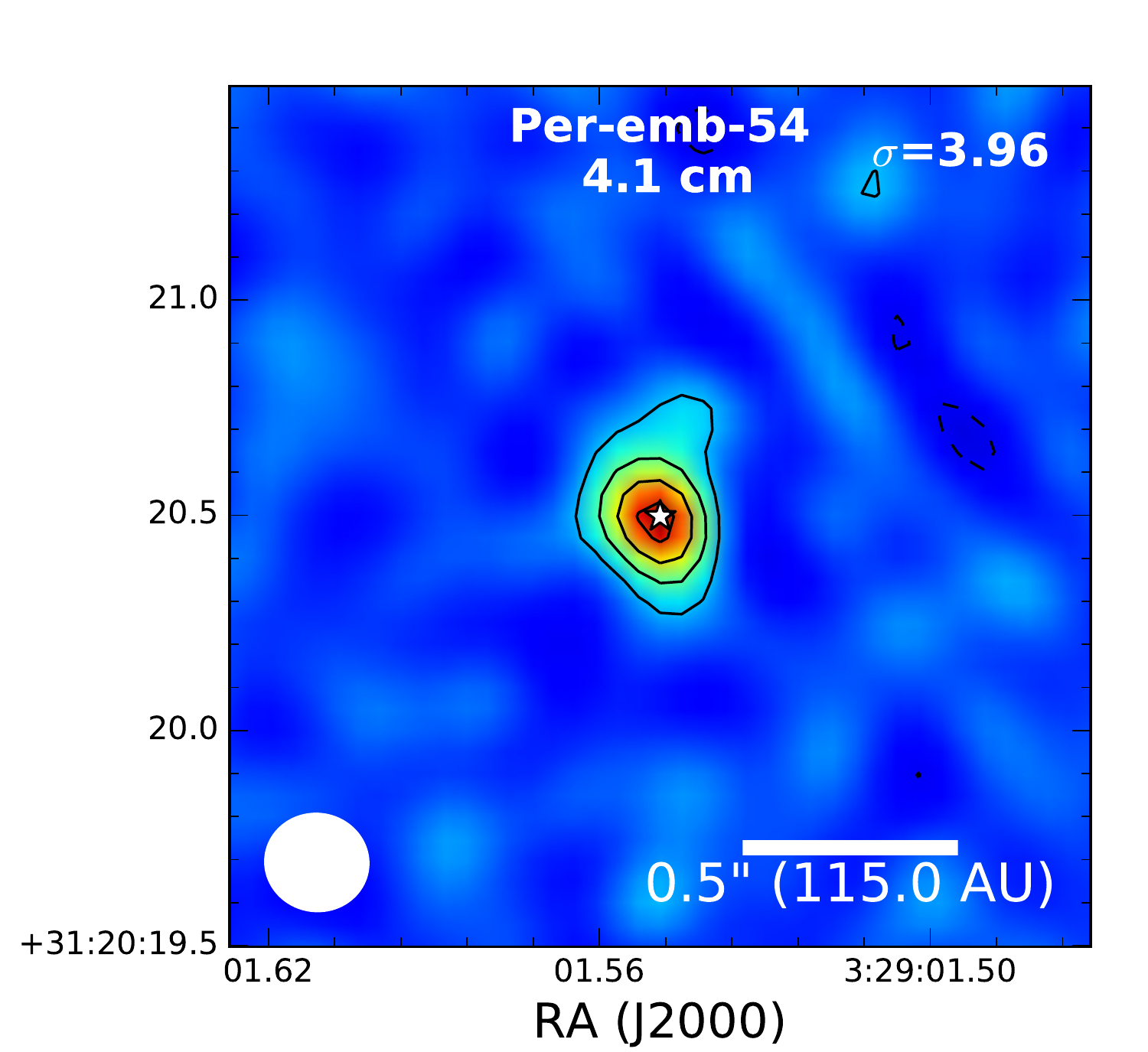}
  \includegraphics[width=0.24\linewidth]{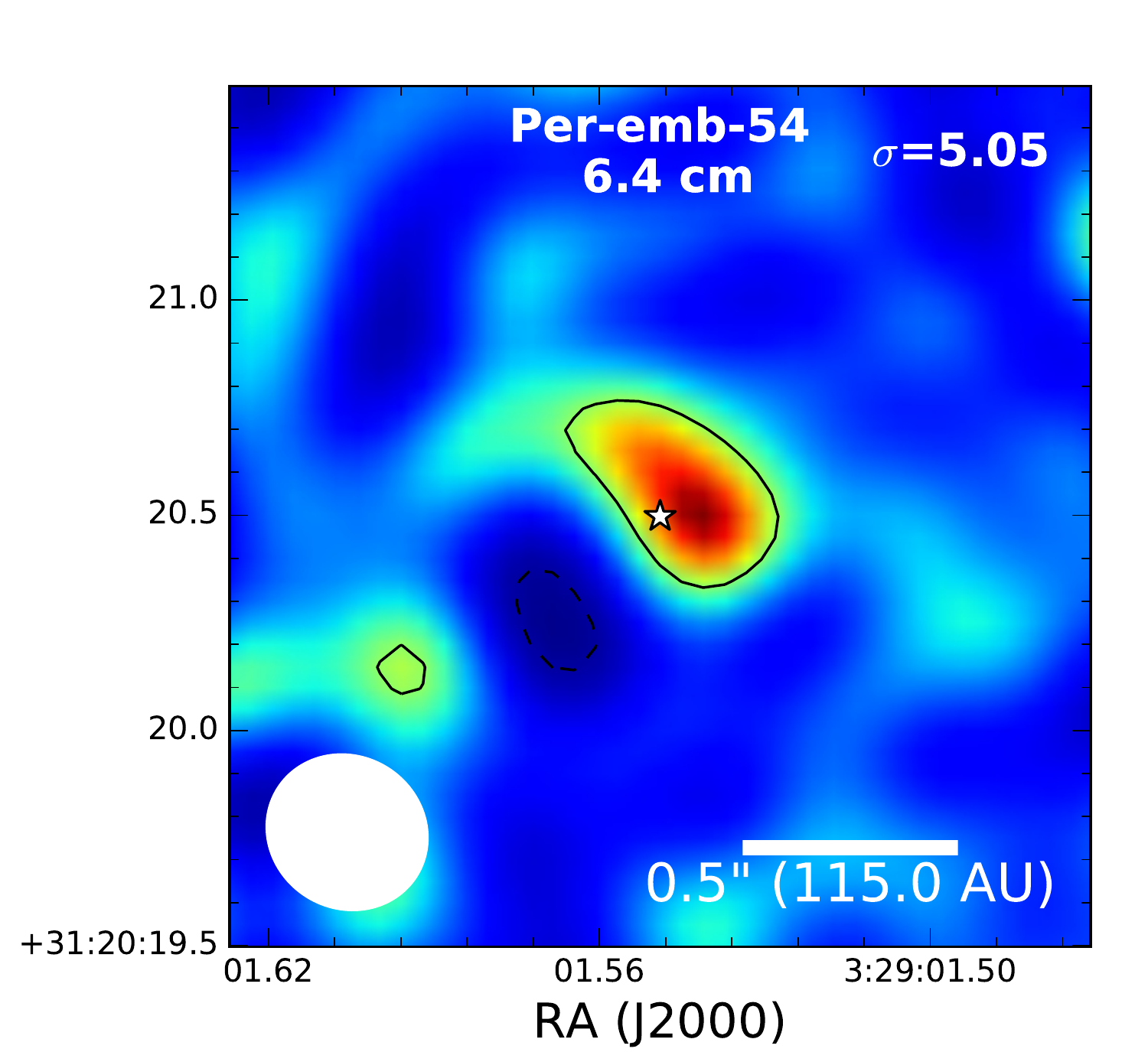}

  \includegraphics[width=0.24\linewidth]{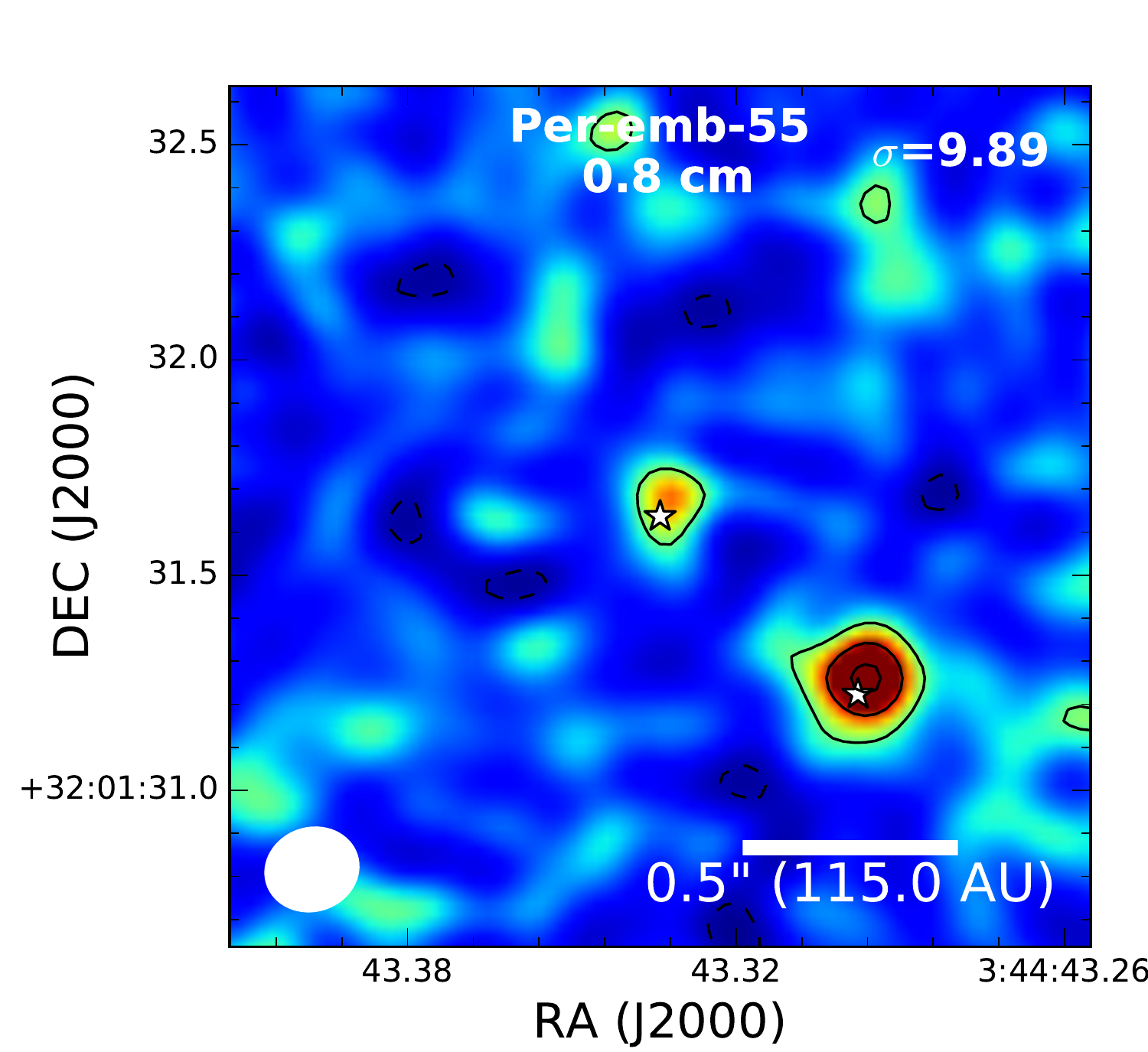}
  \includegraphics[width=0.24\linewidth]{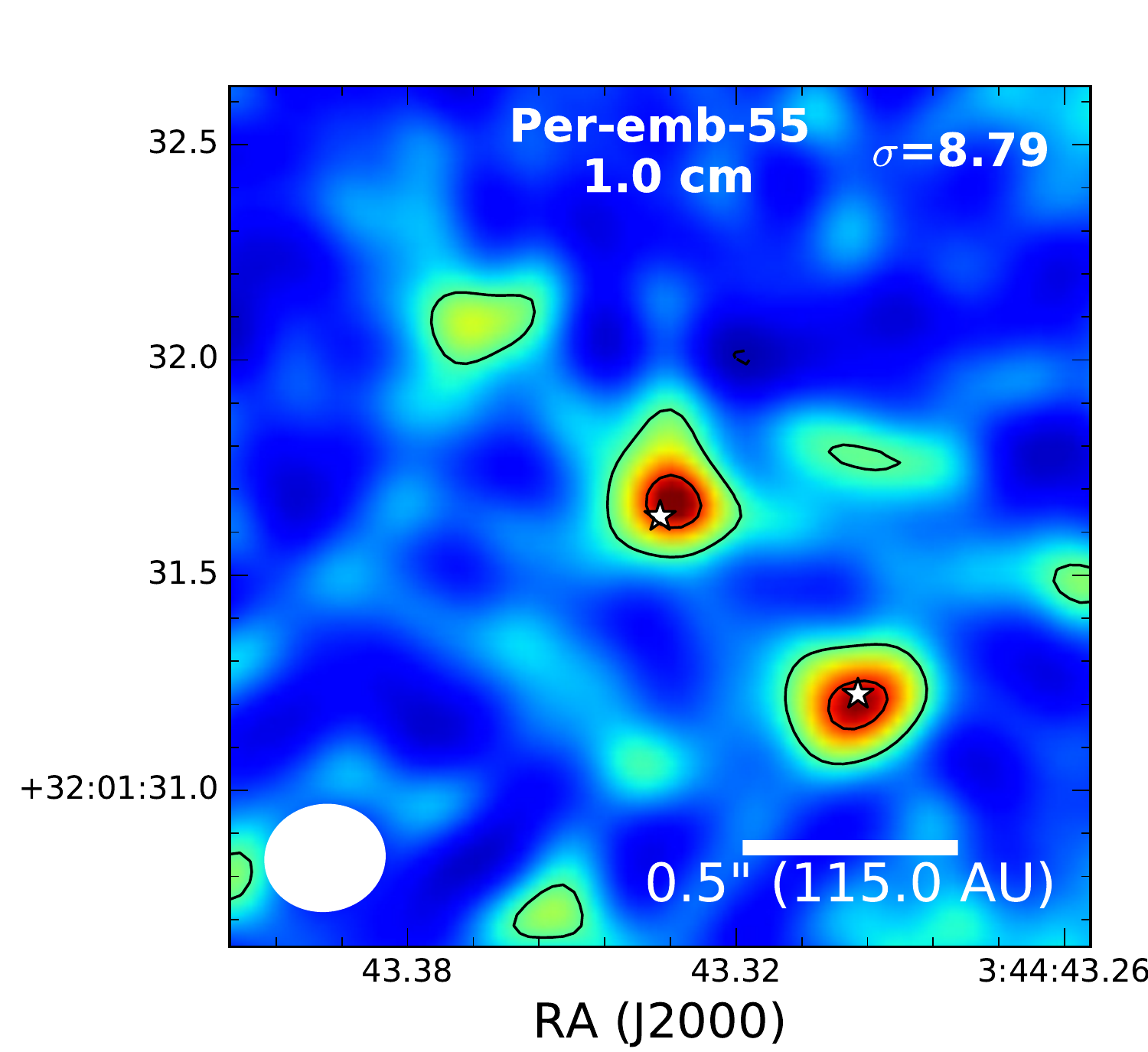}
  \includegraphics[width=0.24\linewidth]{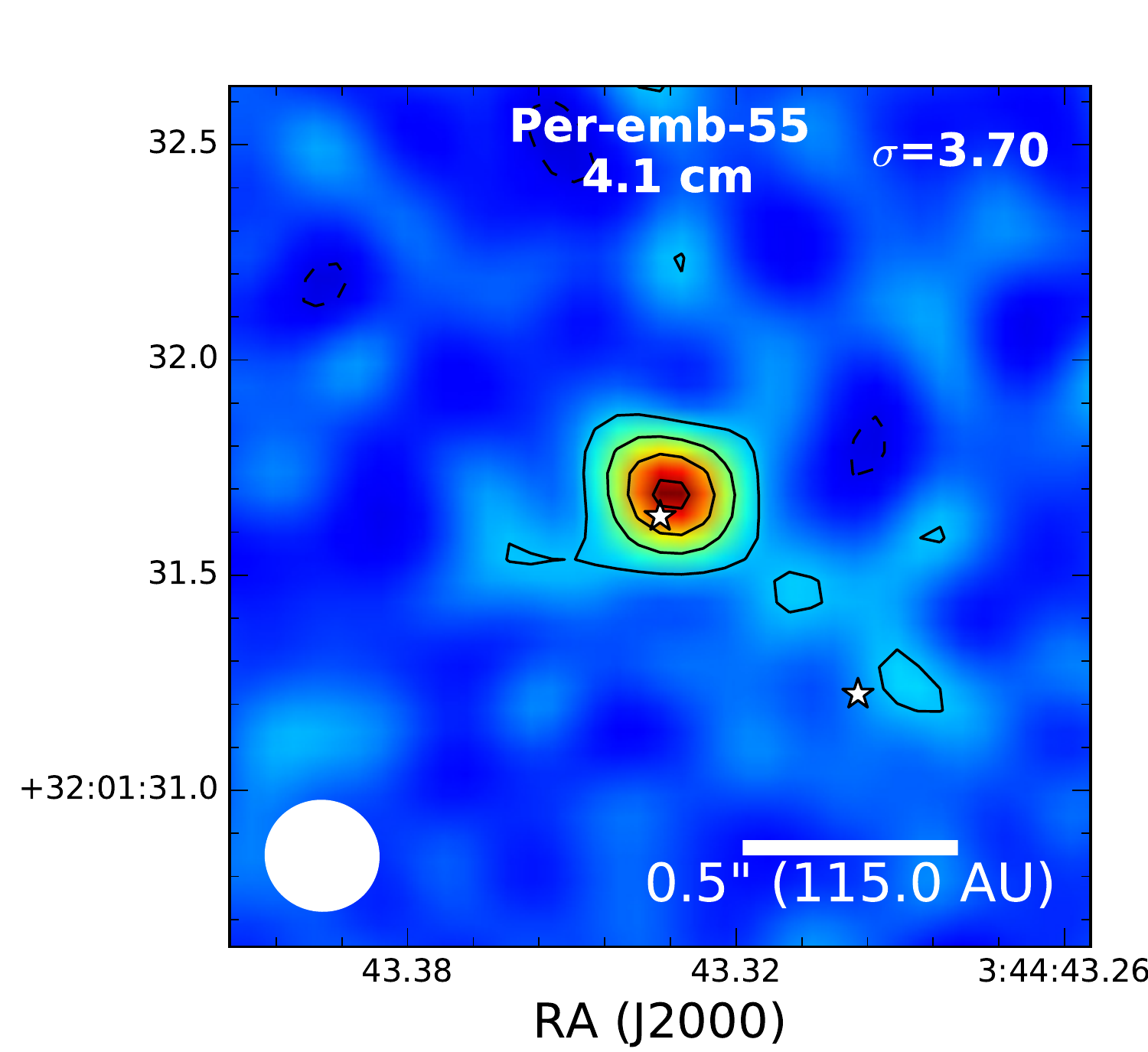}
  \includegraphics[width=0.24\linewidth]{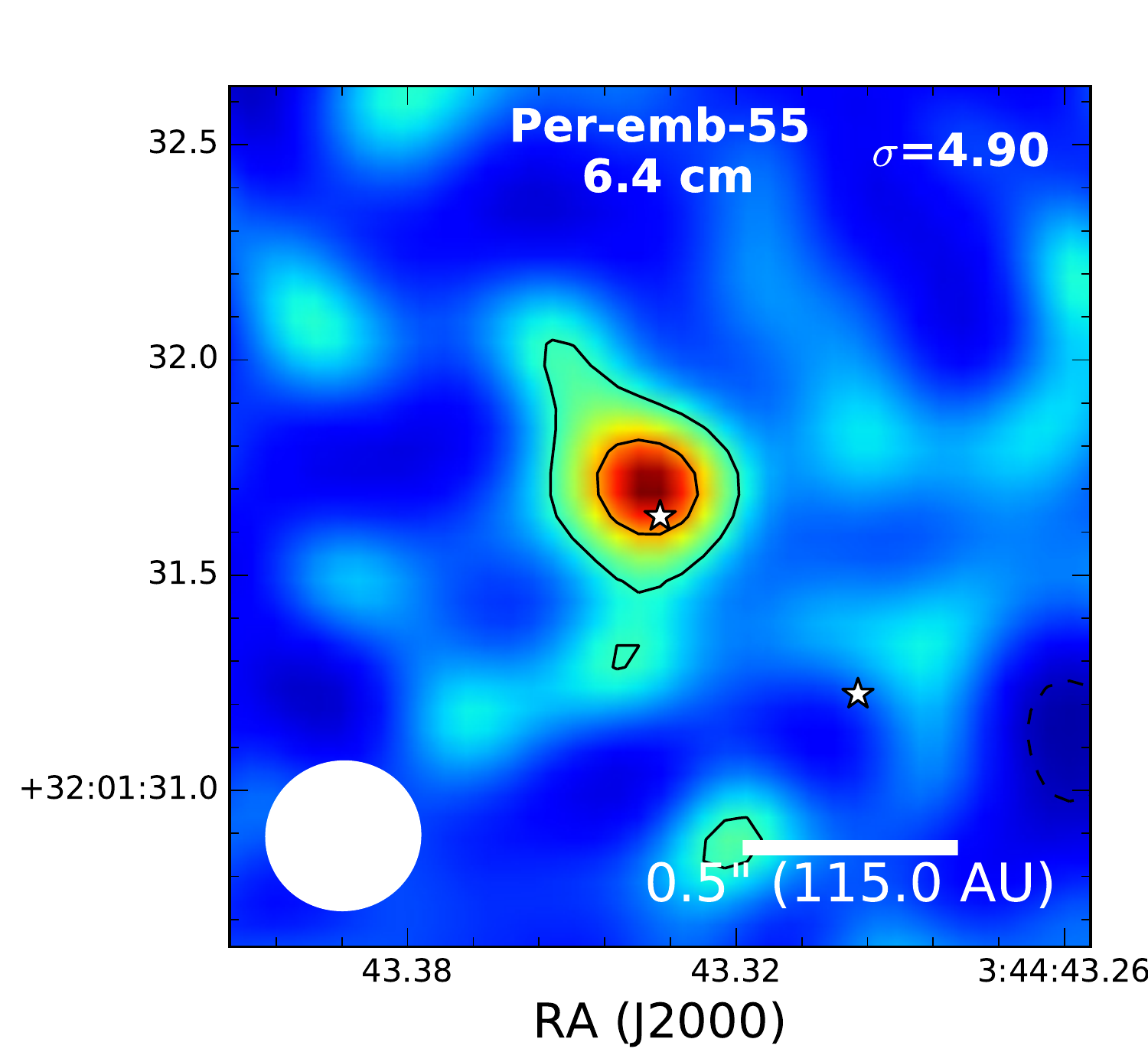}

  \includegraphics[width=0.24\linewidth]{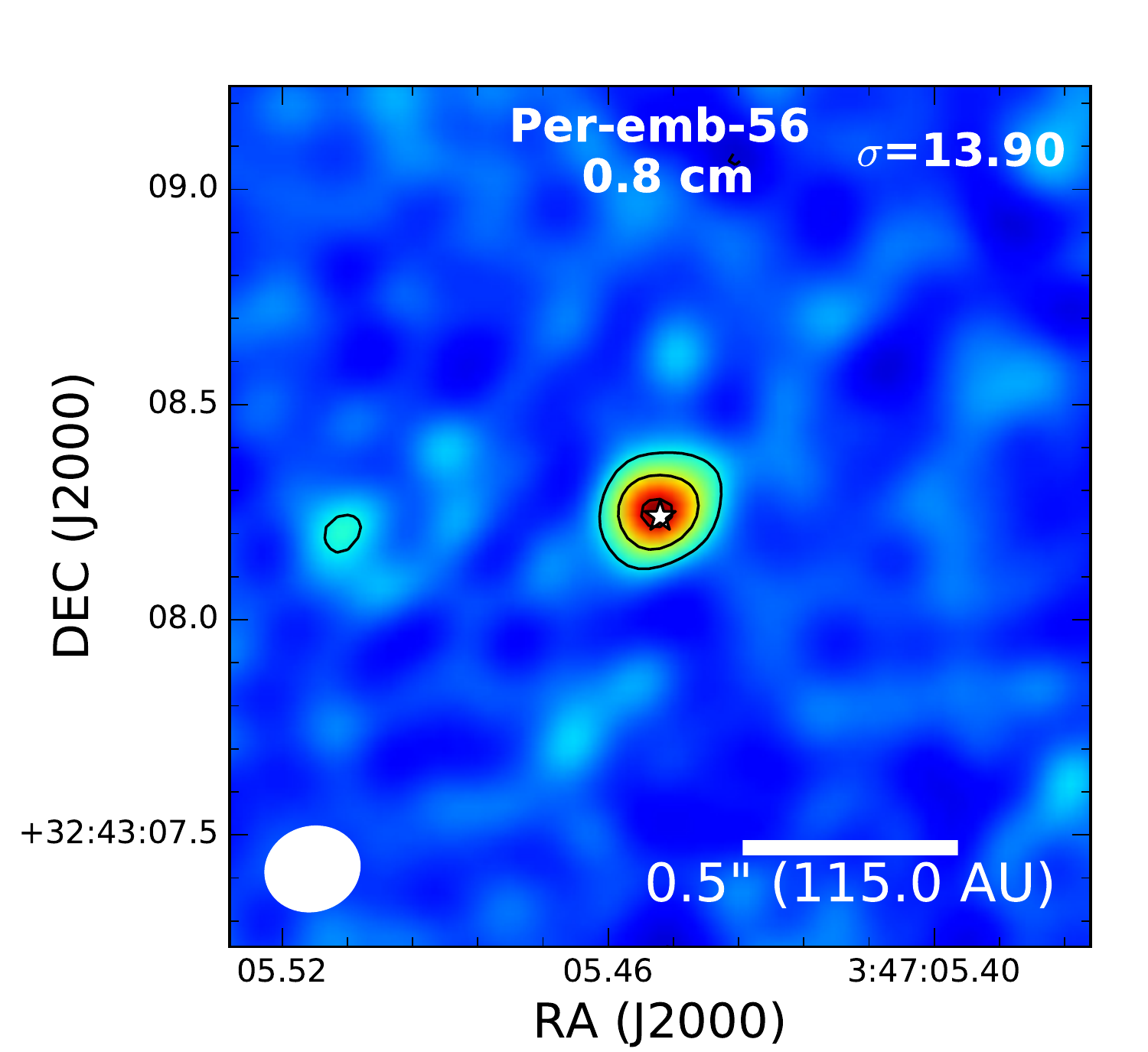}
  \includegraphics[width=0.24\linewidth]{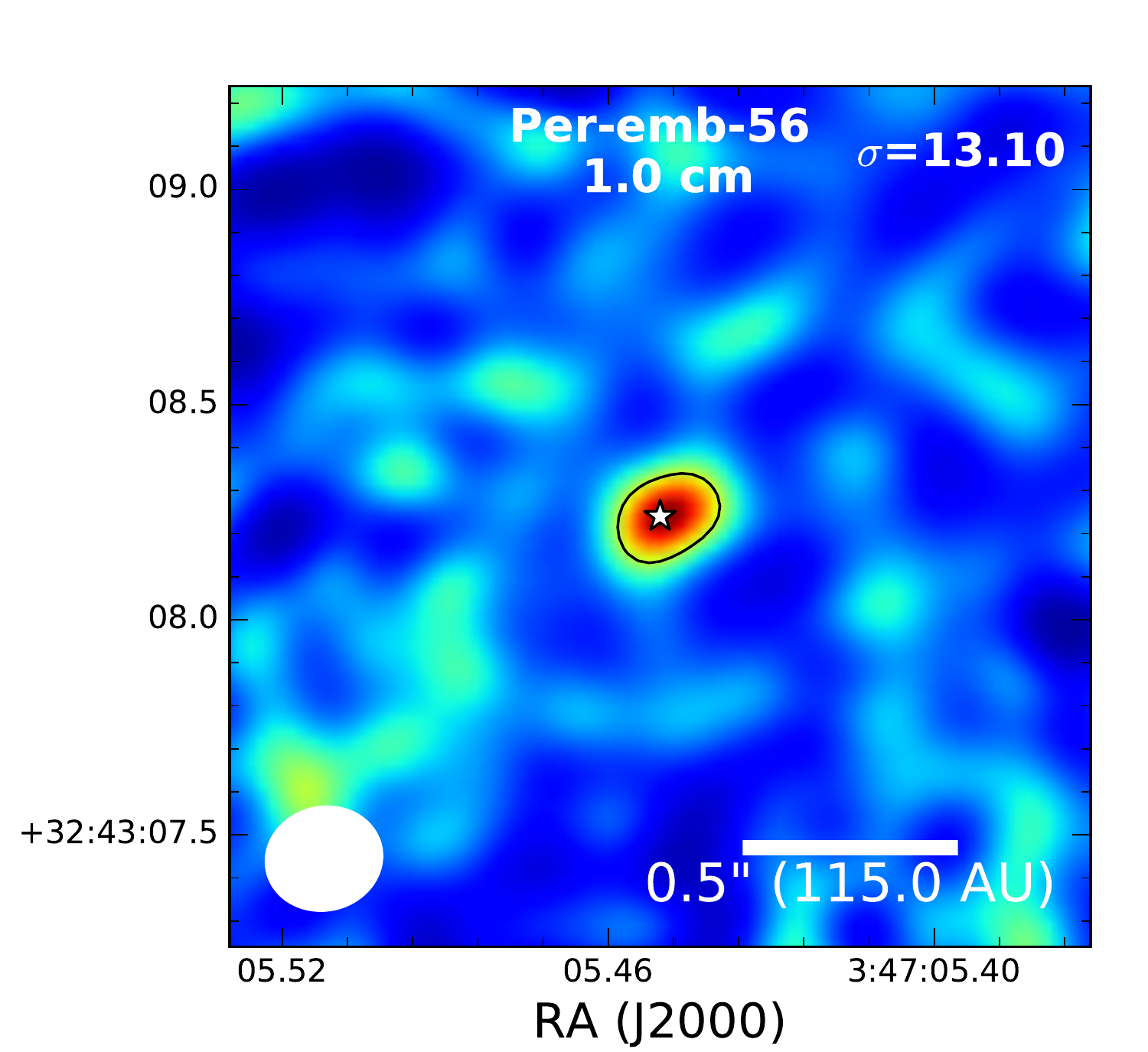}
  \includegraphics[width=0.24\linewidth]{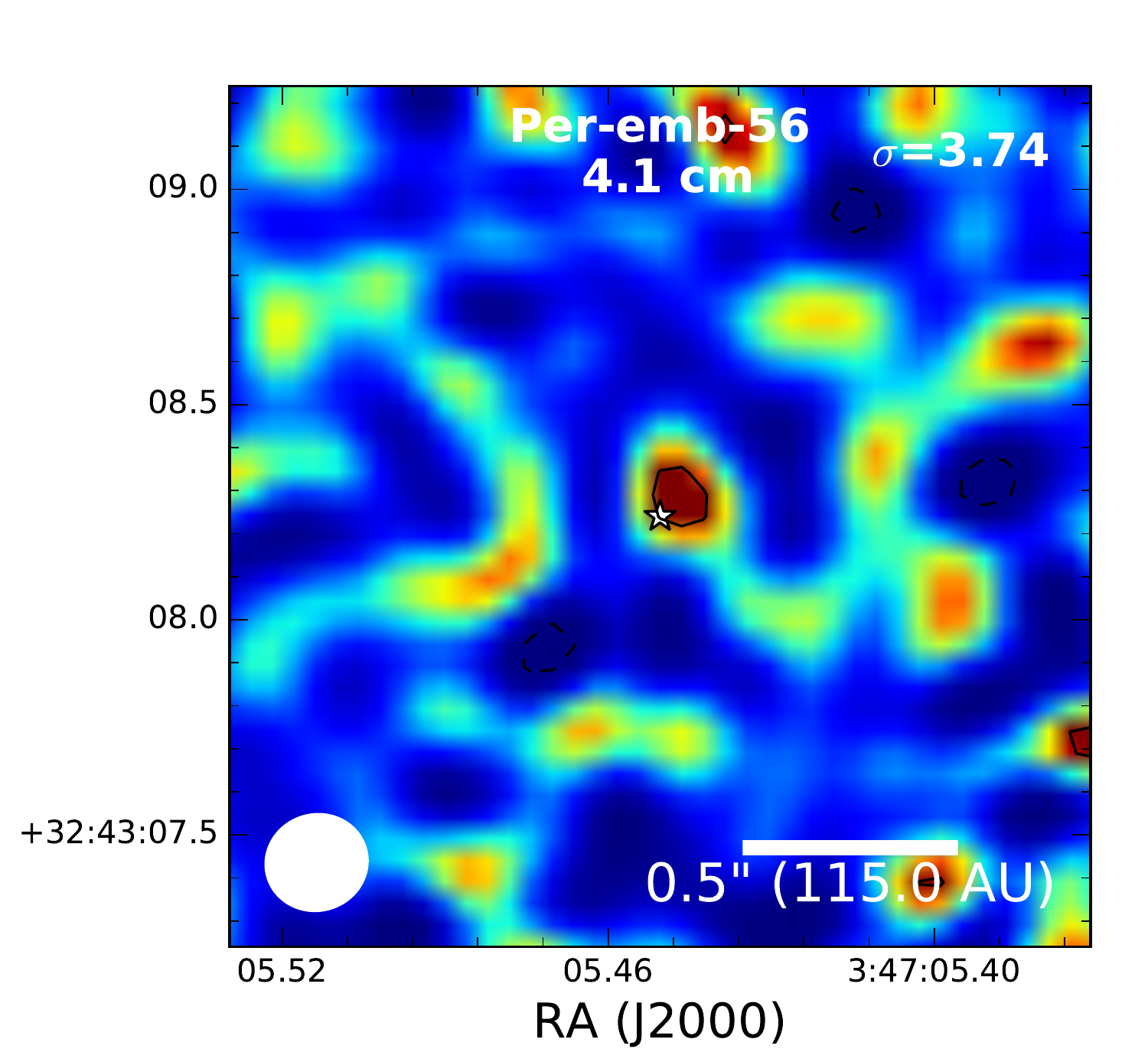}
  \includegraphics[width=0.24\linewidth]{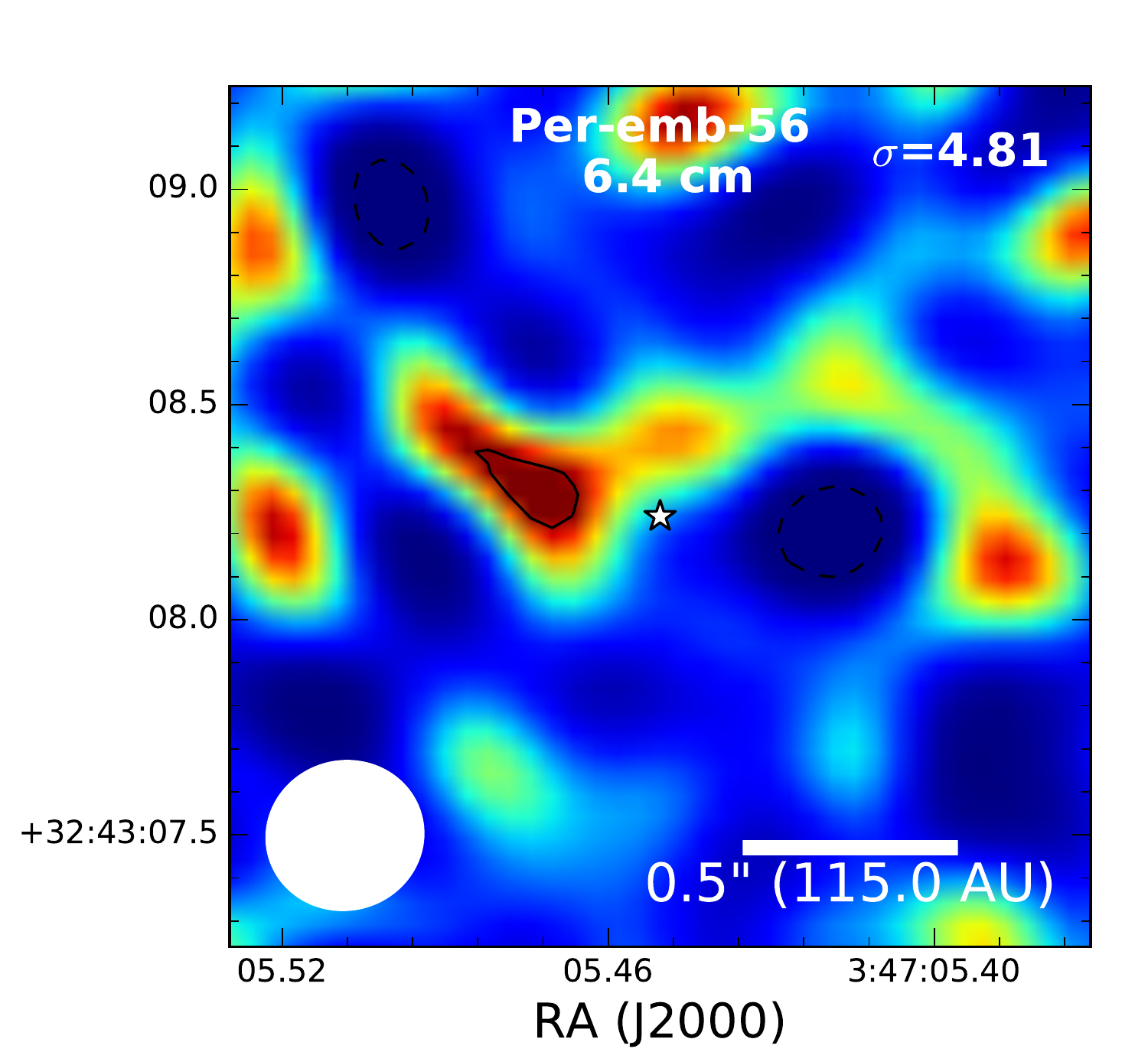}

  \includegraphics[width=0.24\linewidth]{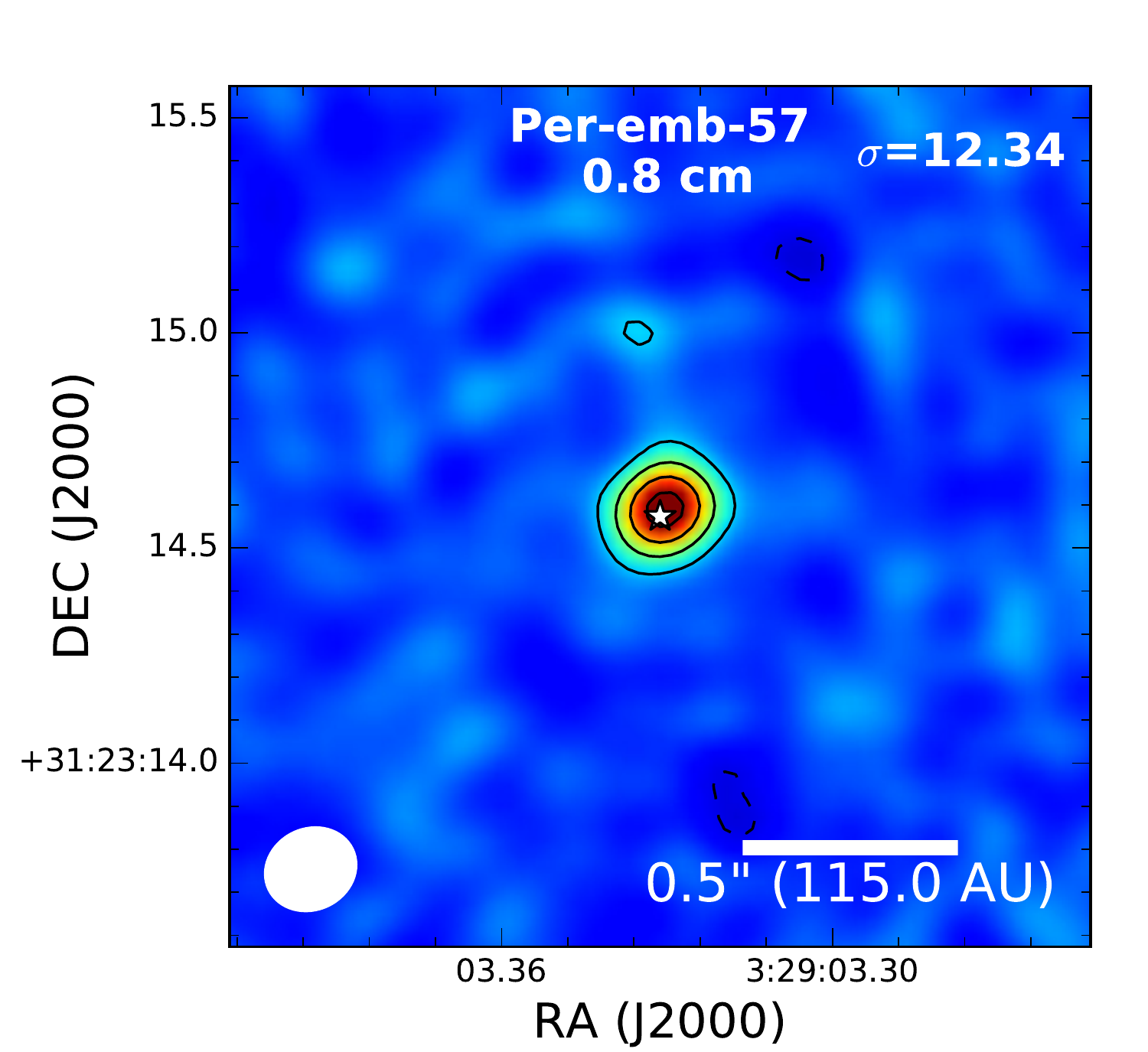}
  \includegraphics[width=0.24\linewidth]{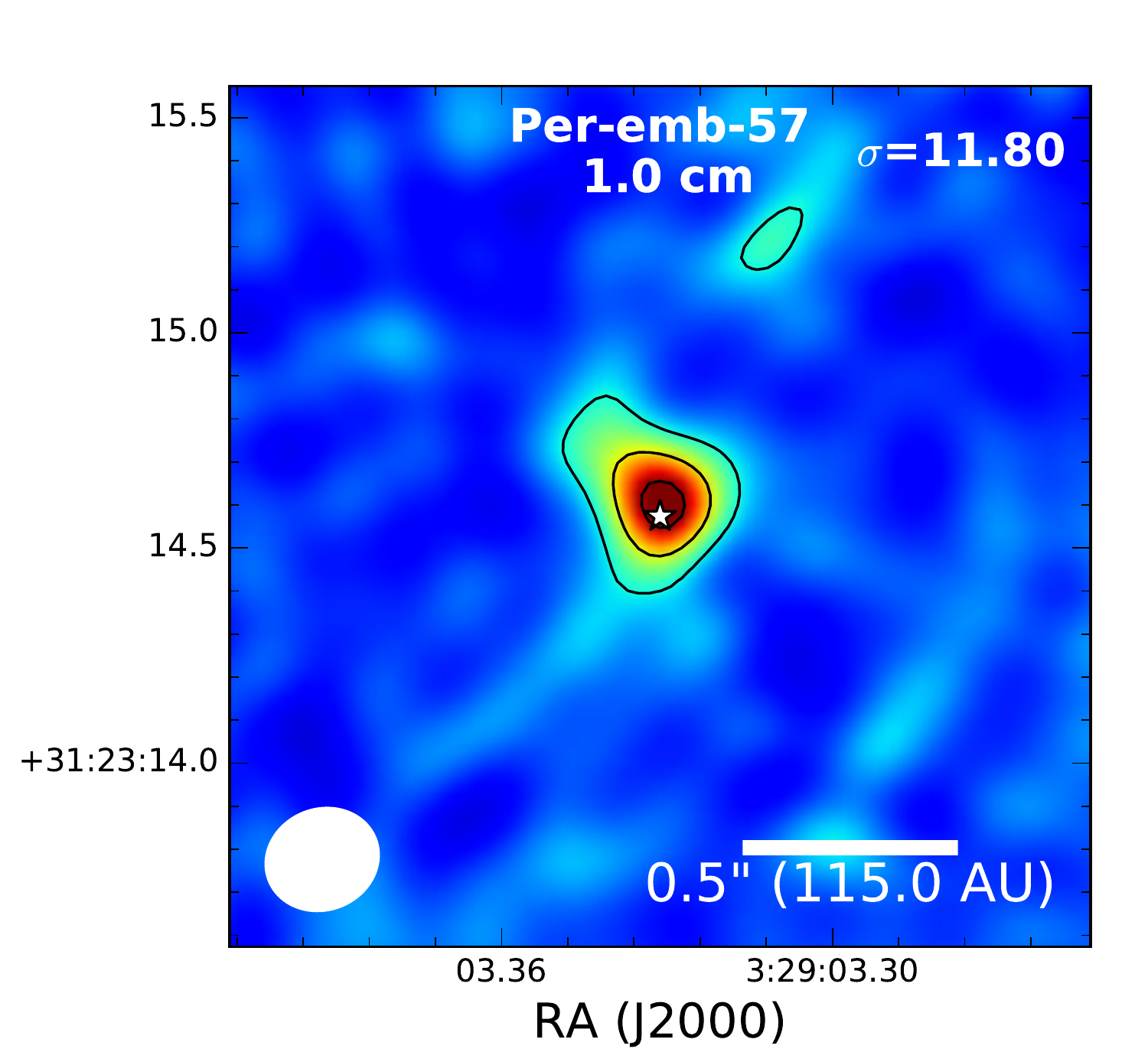}
  \includegraphics[width=0.24\linewidth]{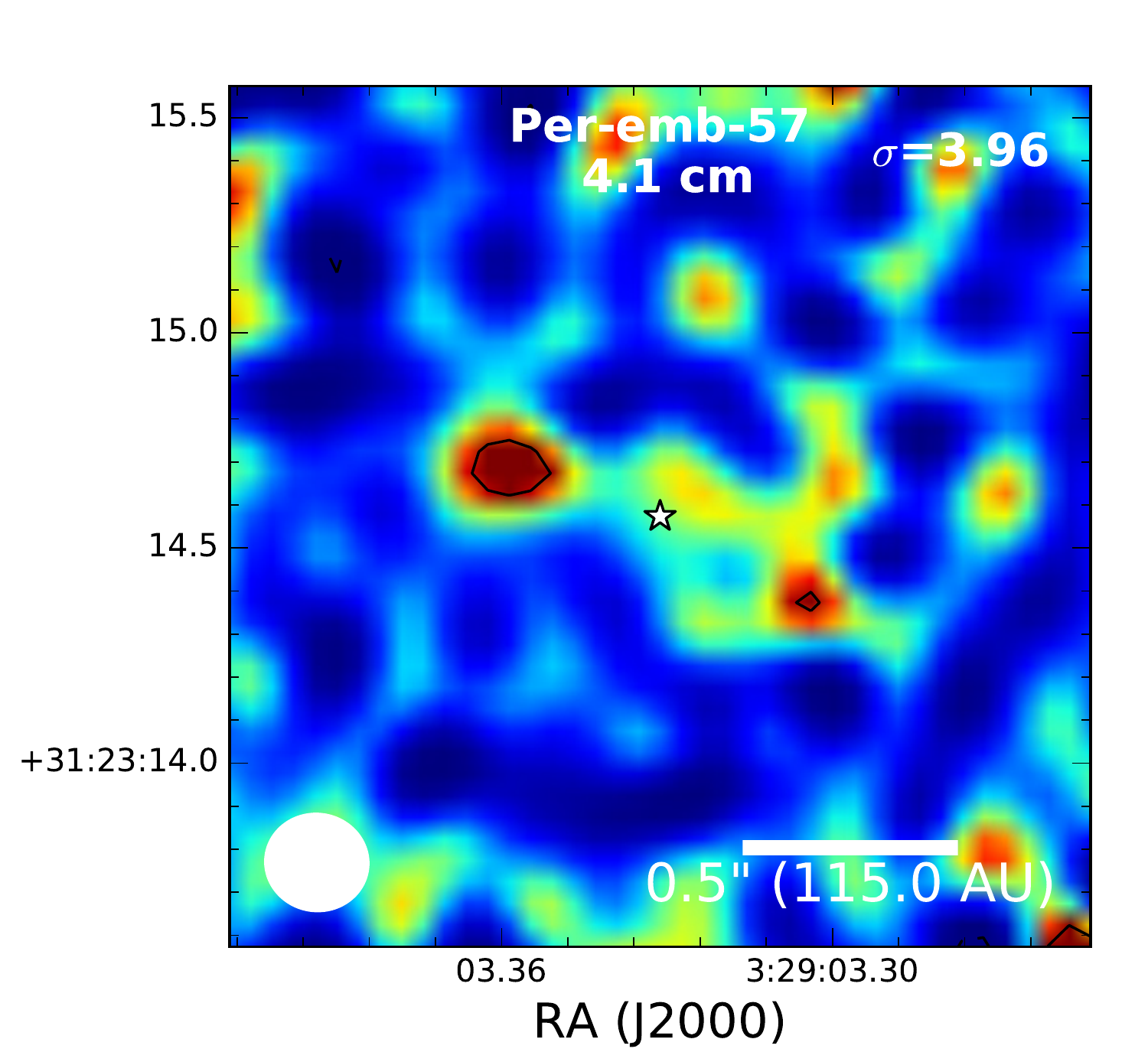}
  \includegraphics[width=0.24\linewidth]{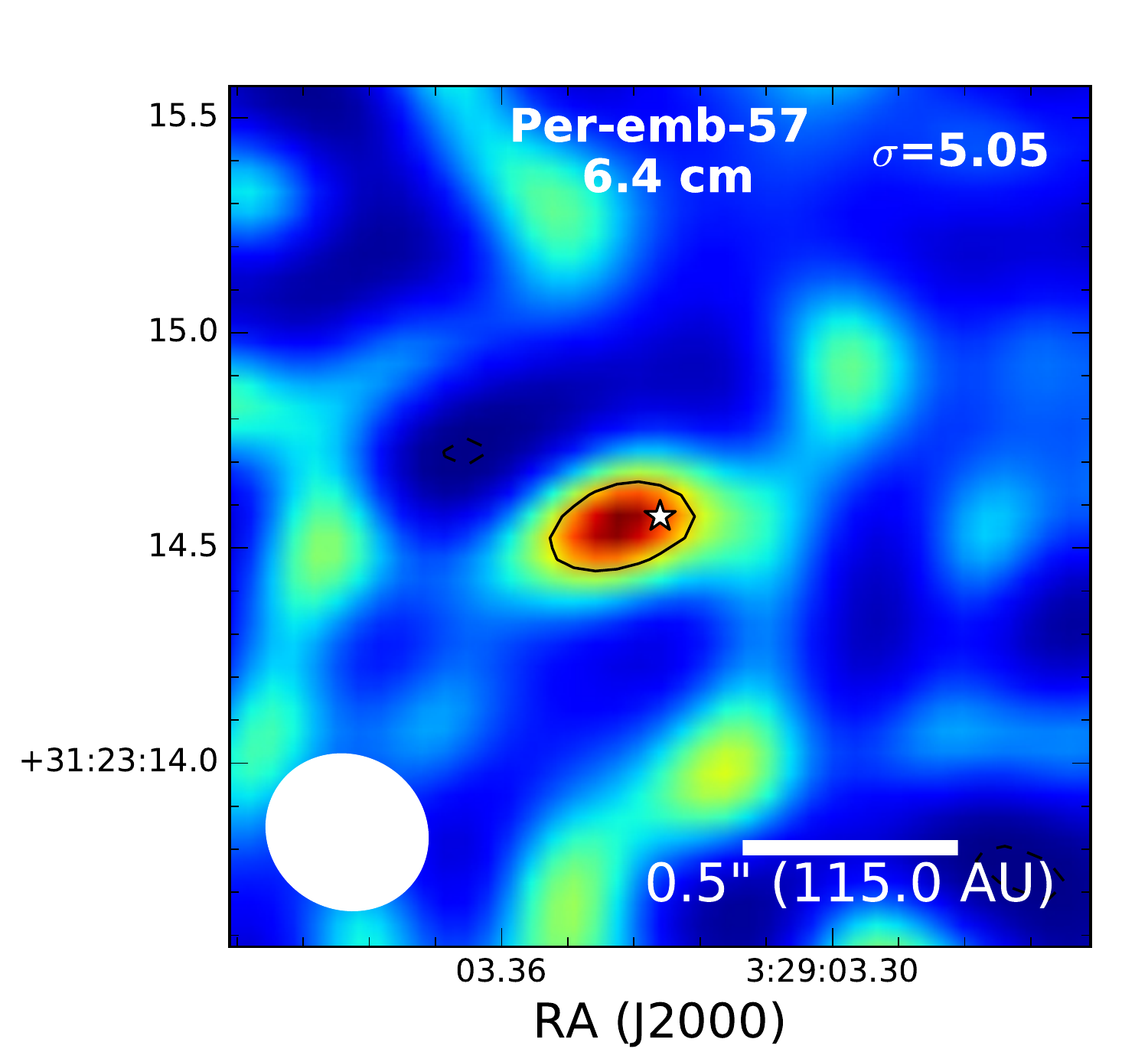}

\end{figure}

\begin{figure}

  \includegraphics[width=0.24\linewidth]{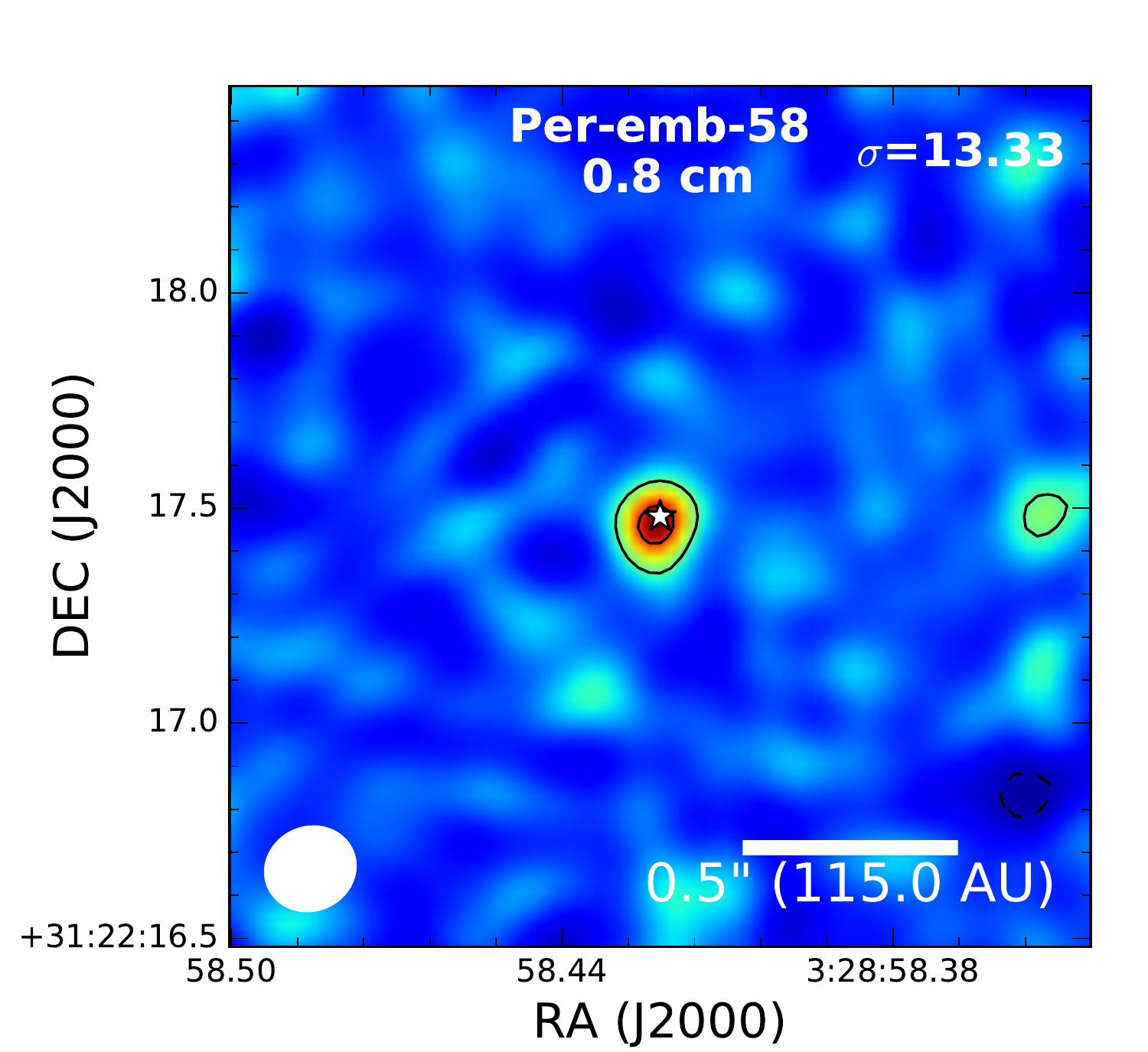}
  \includegraphics[width=0.24\linewidth]{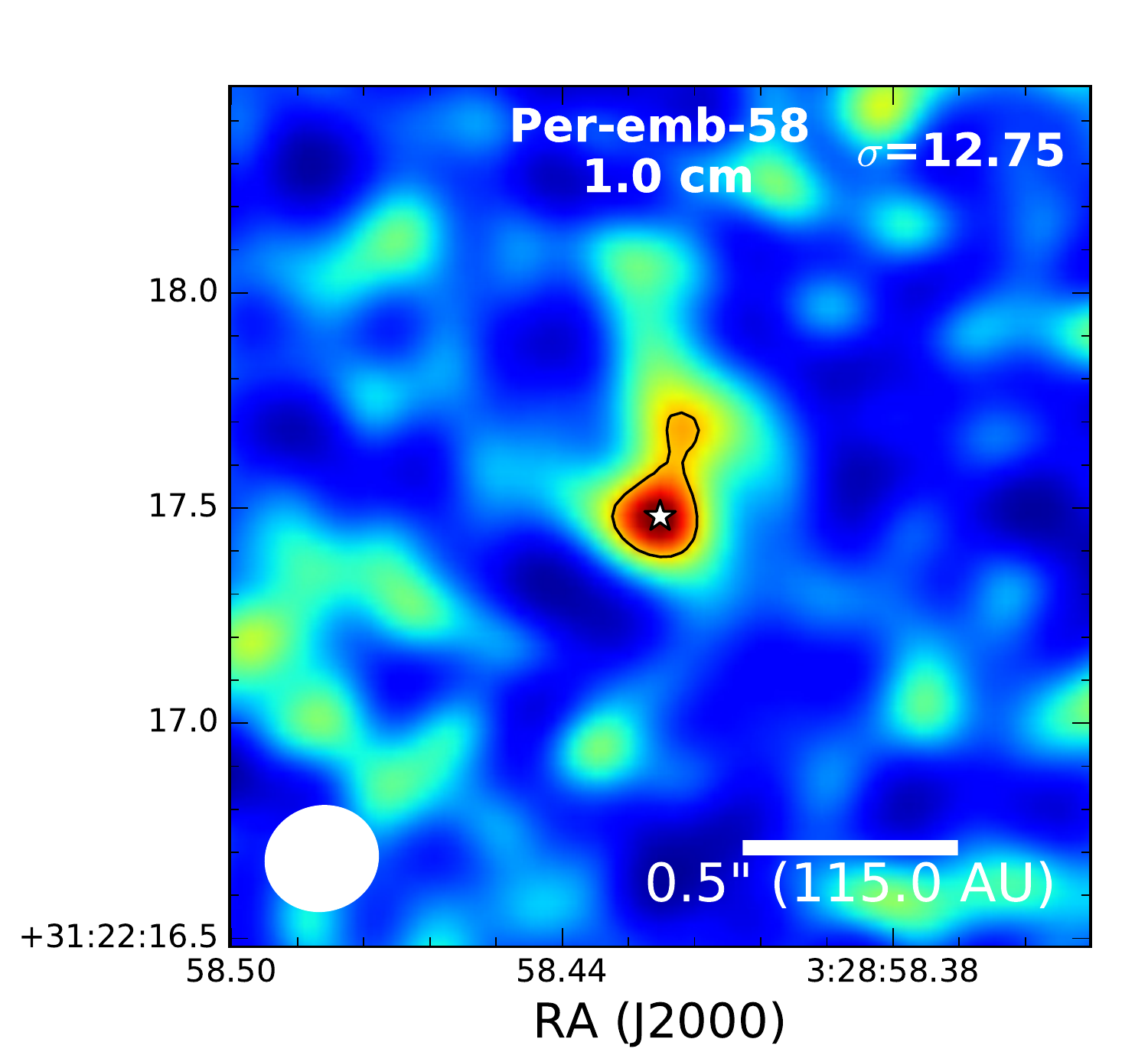}
  \includegraphics[width=0.24\linewidth]{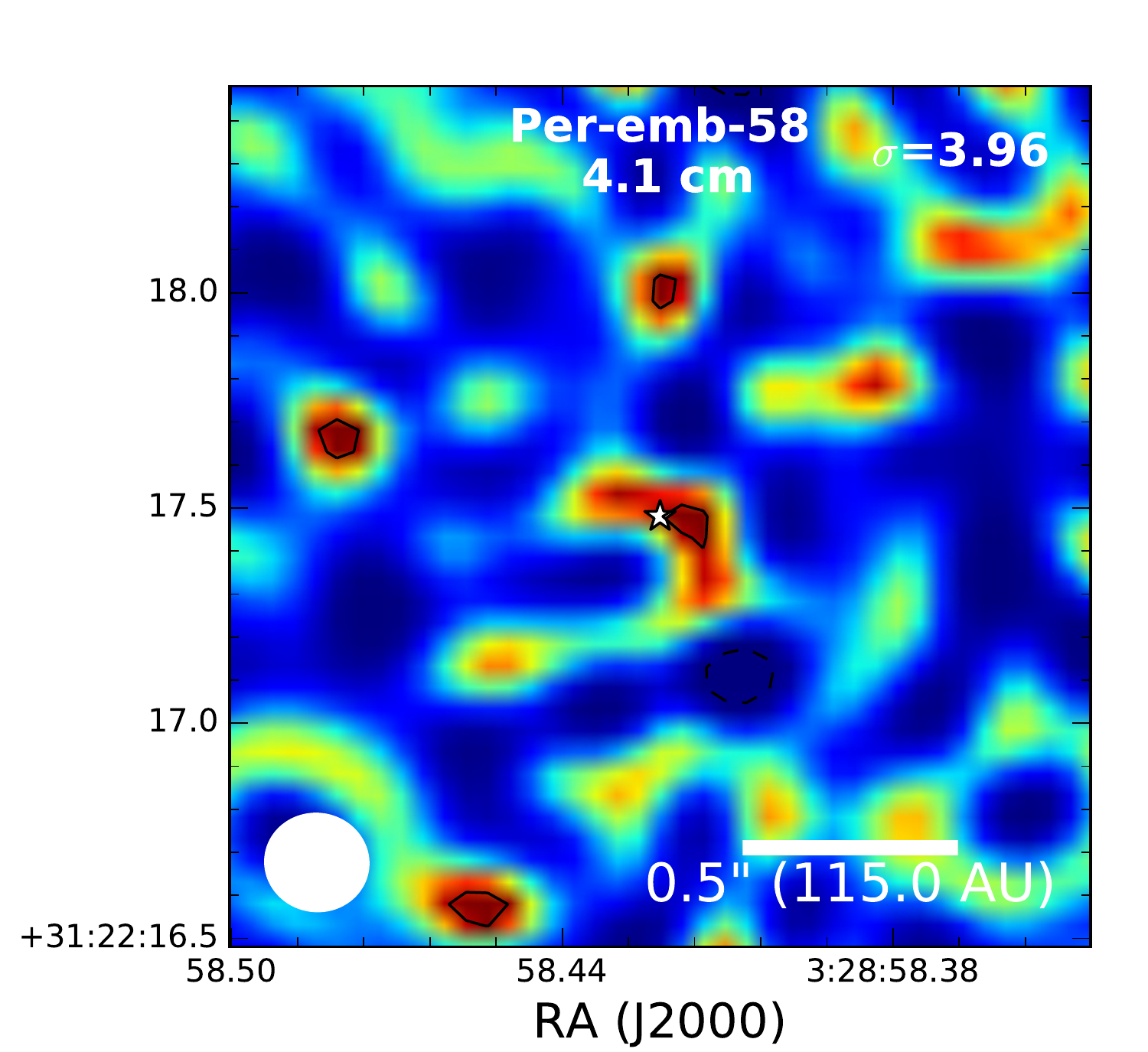}
  \includegraphics[width=0.24\linewidth]{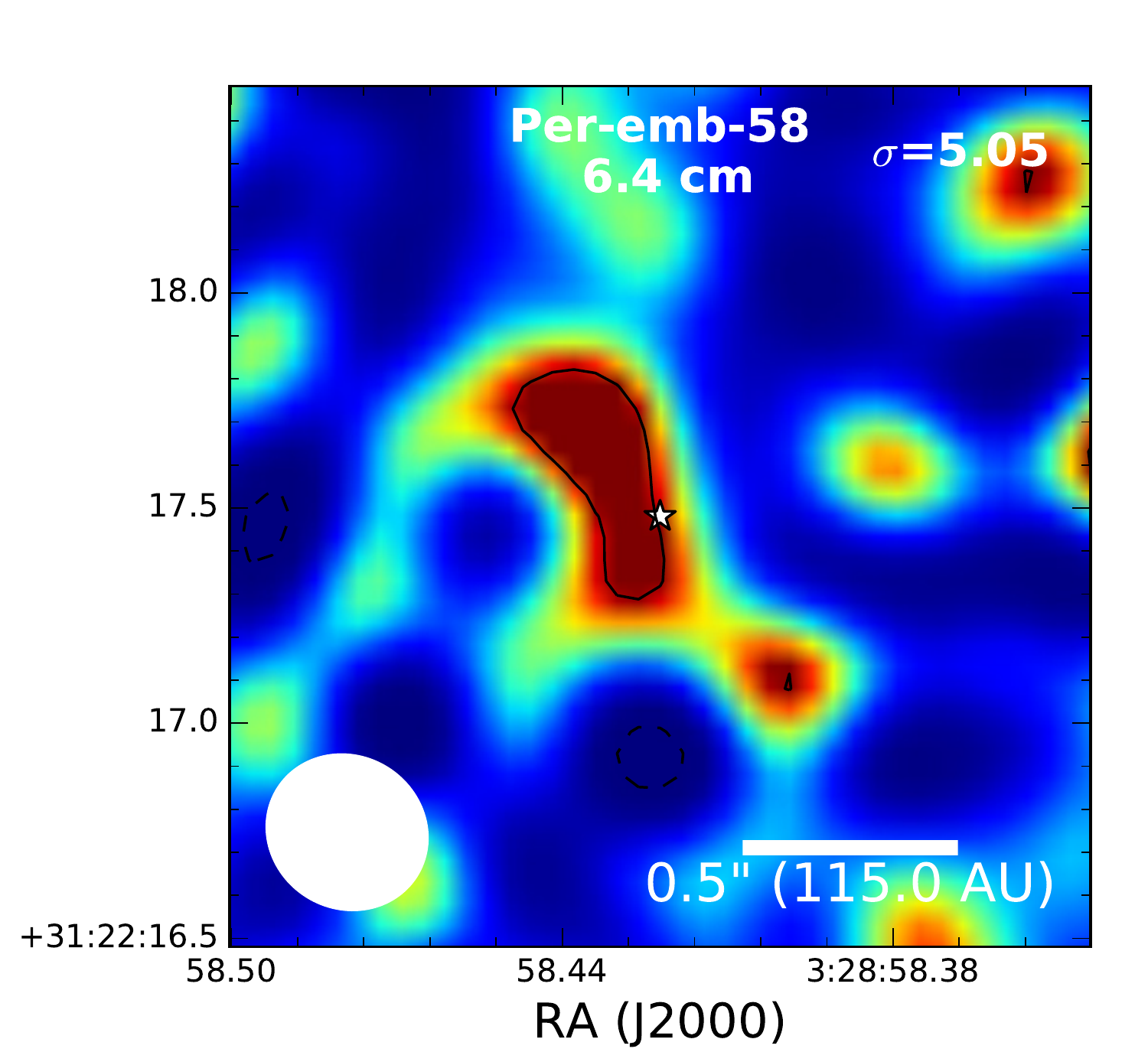}

  \includegraphics[width=0.24\linewidth]{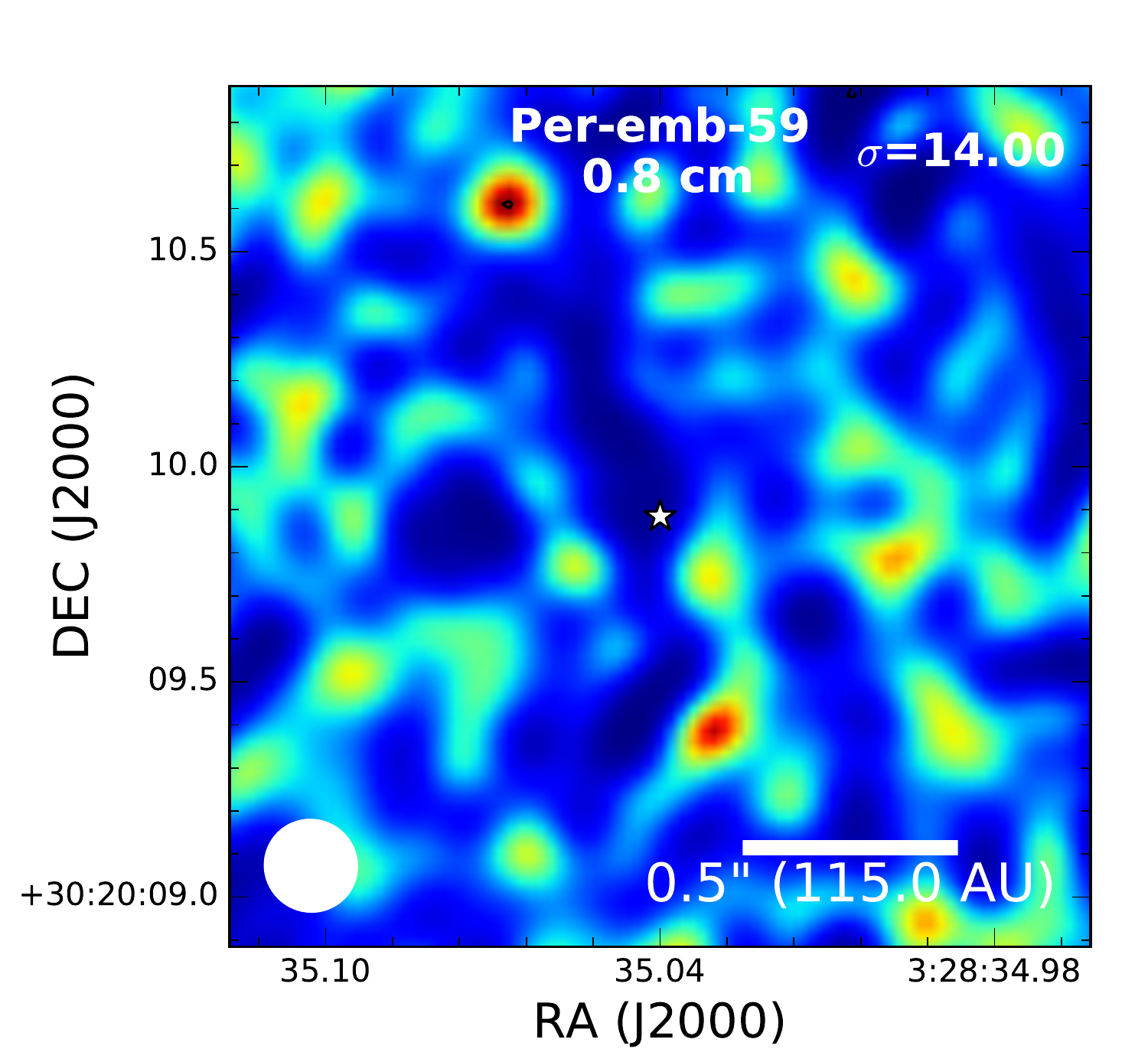}
  \includegraphics[width=0.24\linewidth]{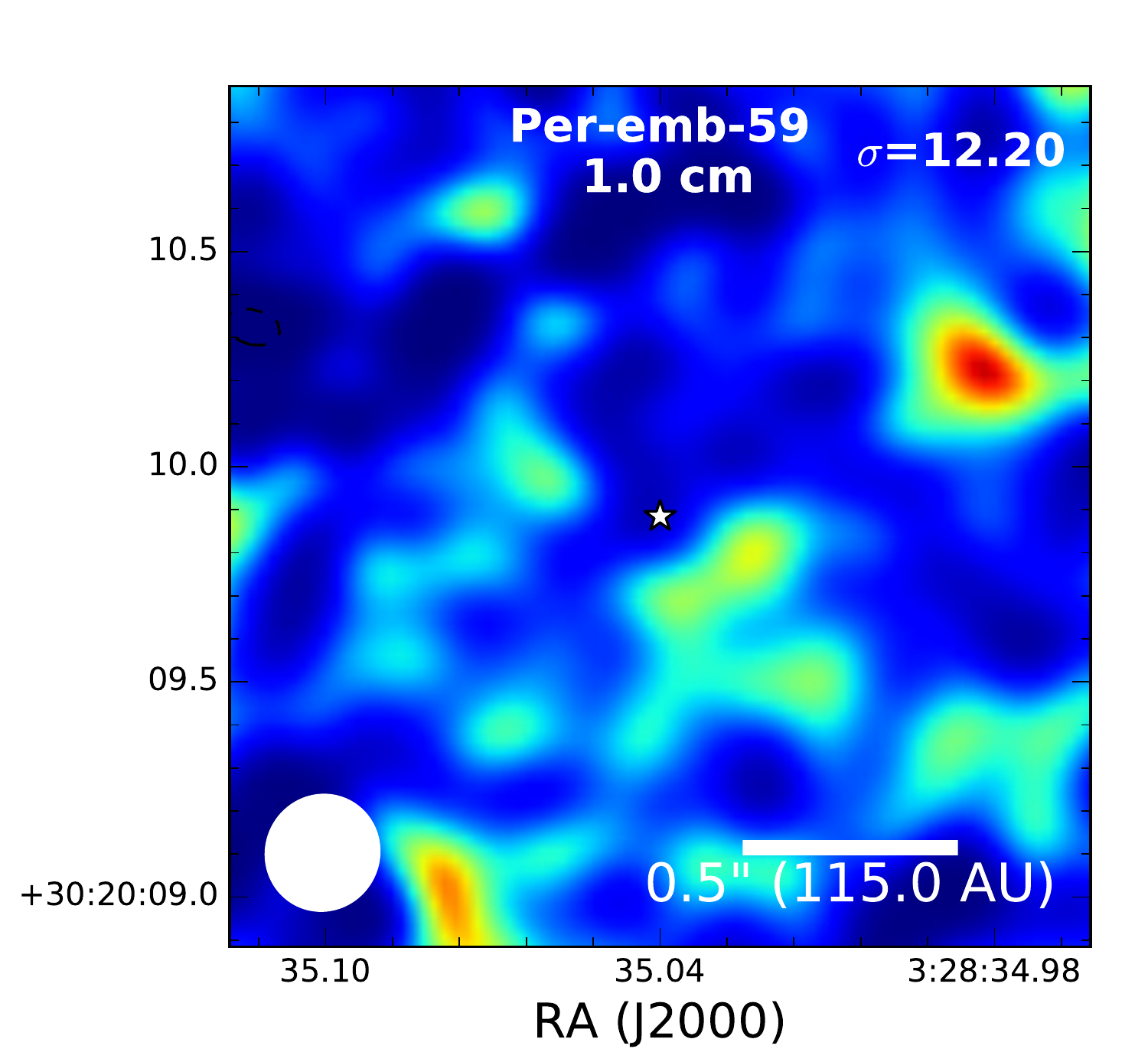}
  \includegraphics[width=0.24\linewidth]{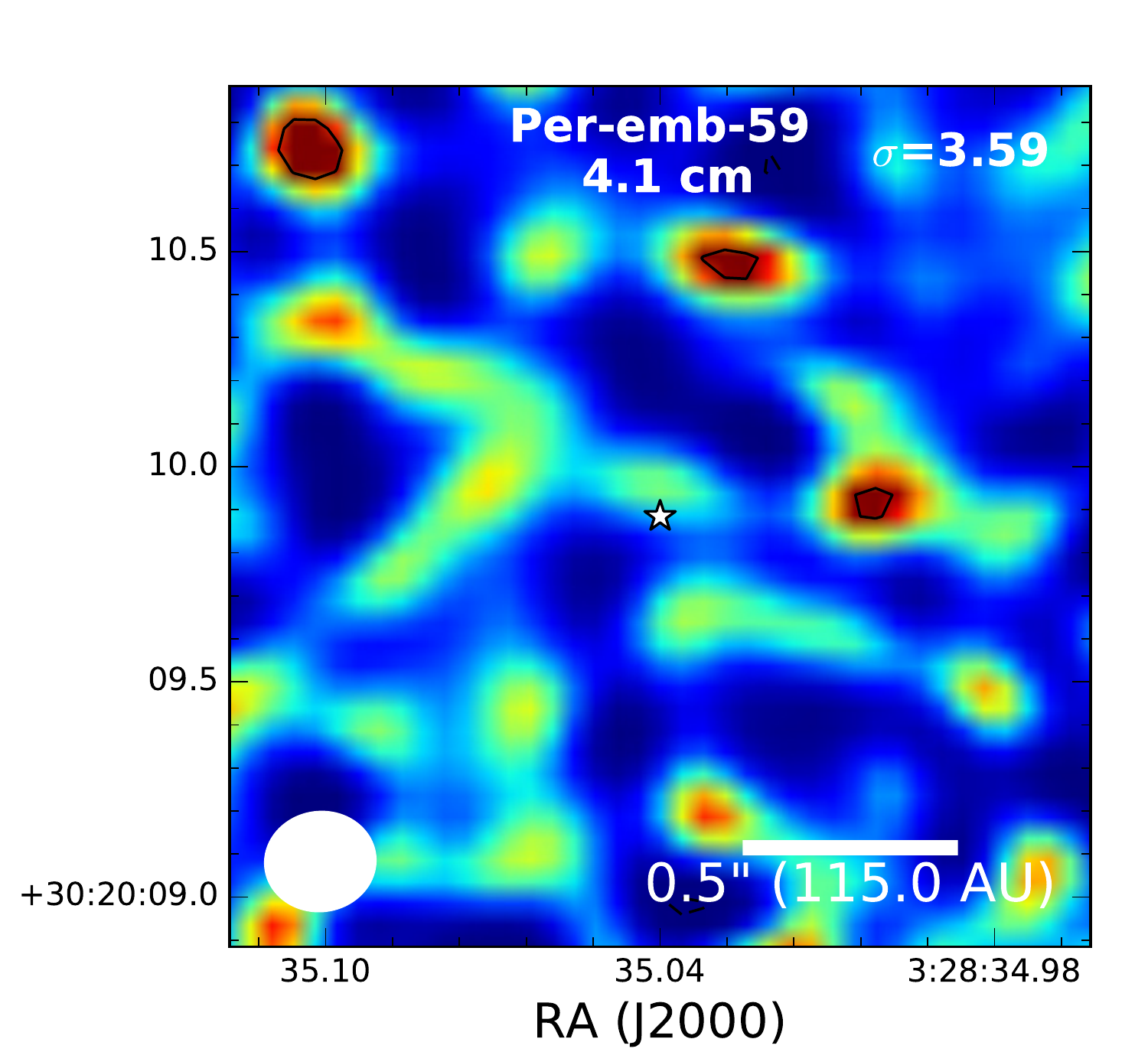}
  \includegraphics[width=0.24\linewidth]{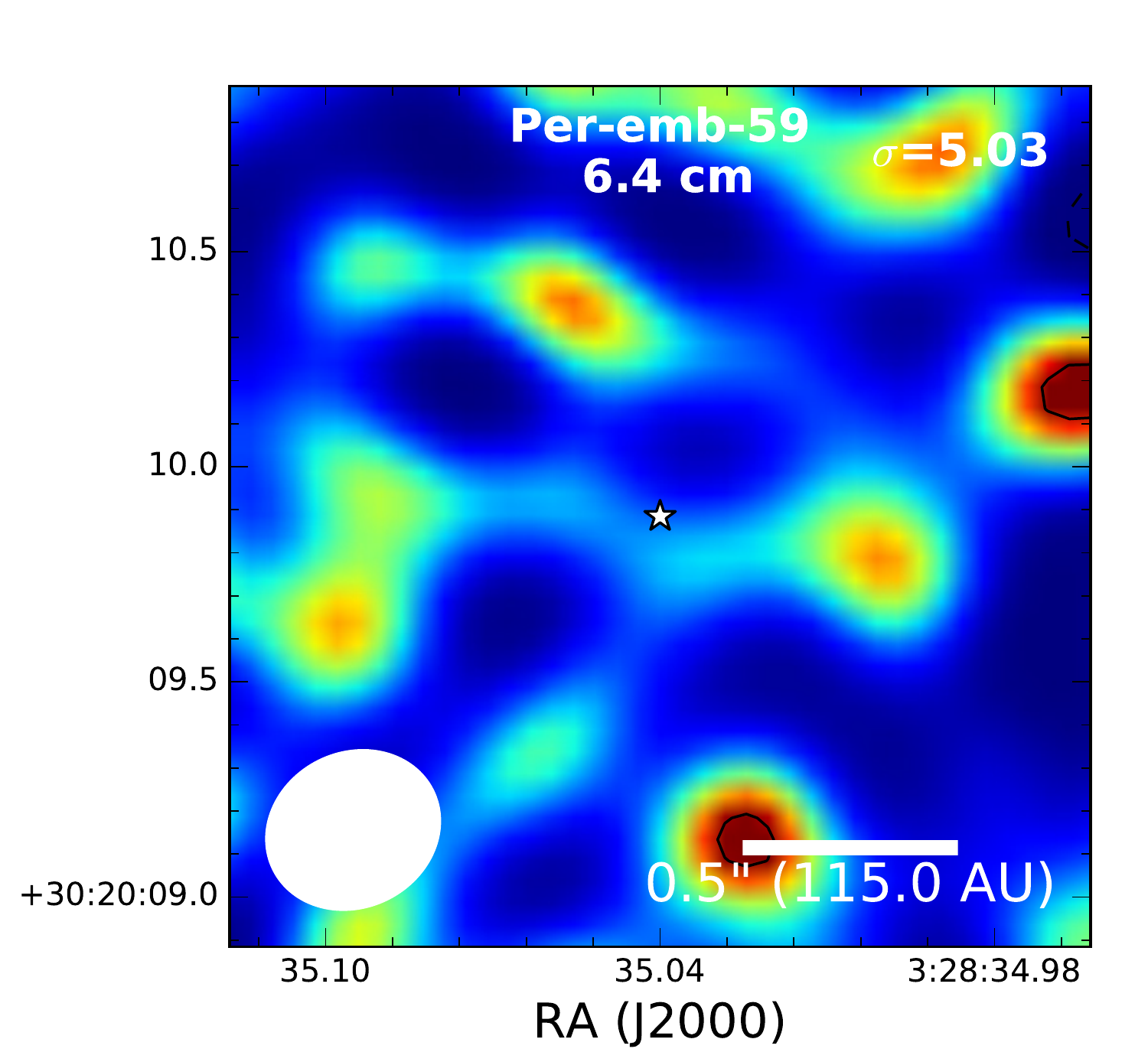}

  \includegraphics[width=0.24\linewidth]{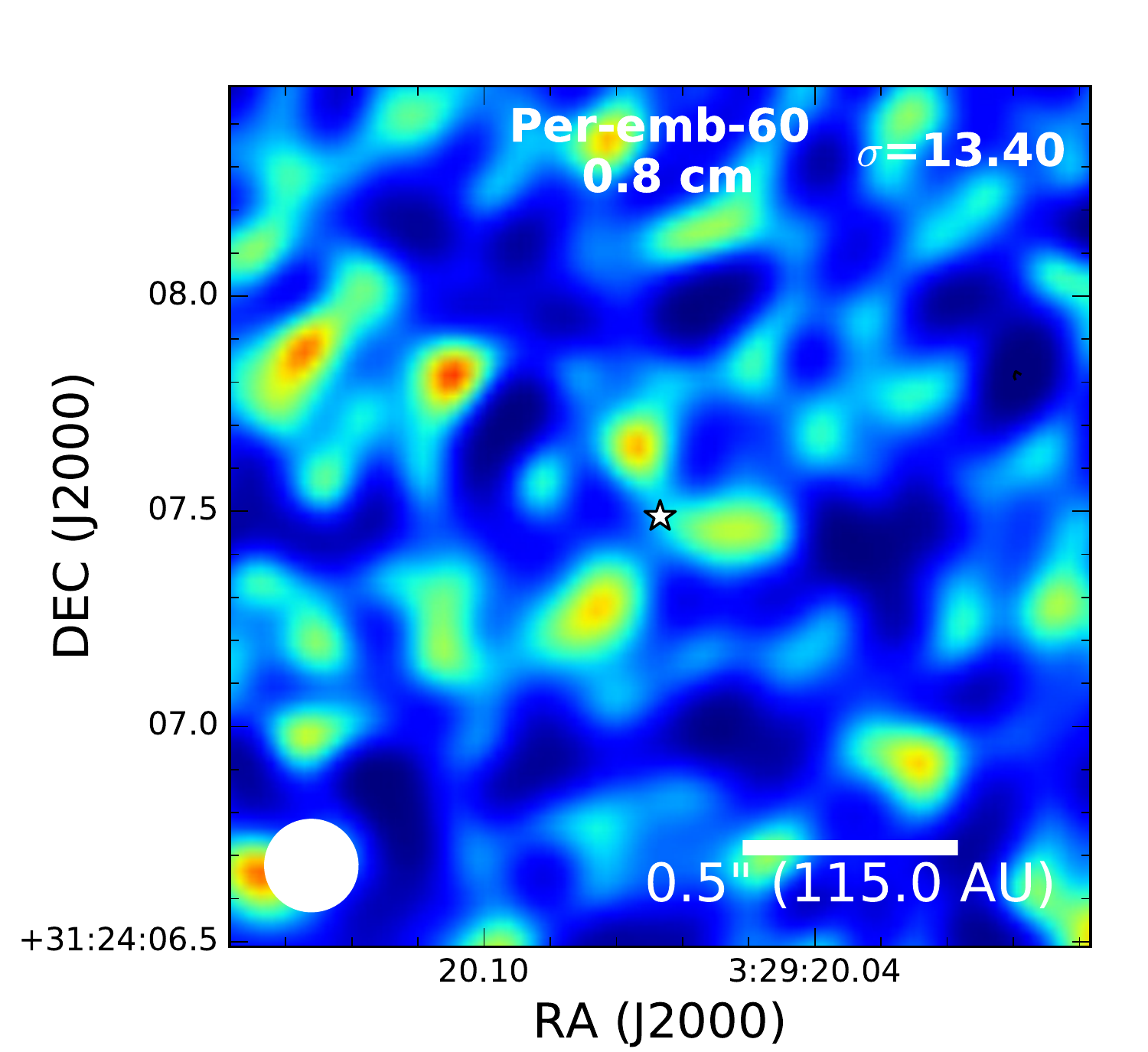}
  \includegraphics[width=0.24\linewidth]{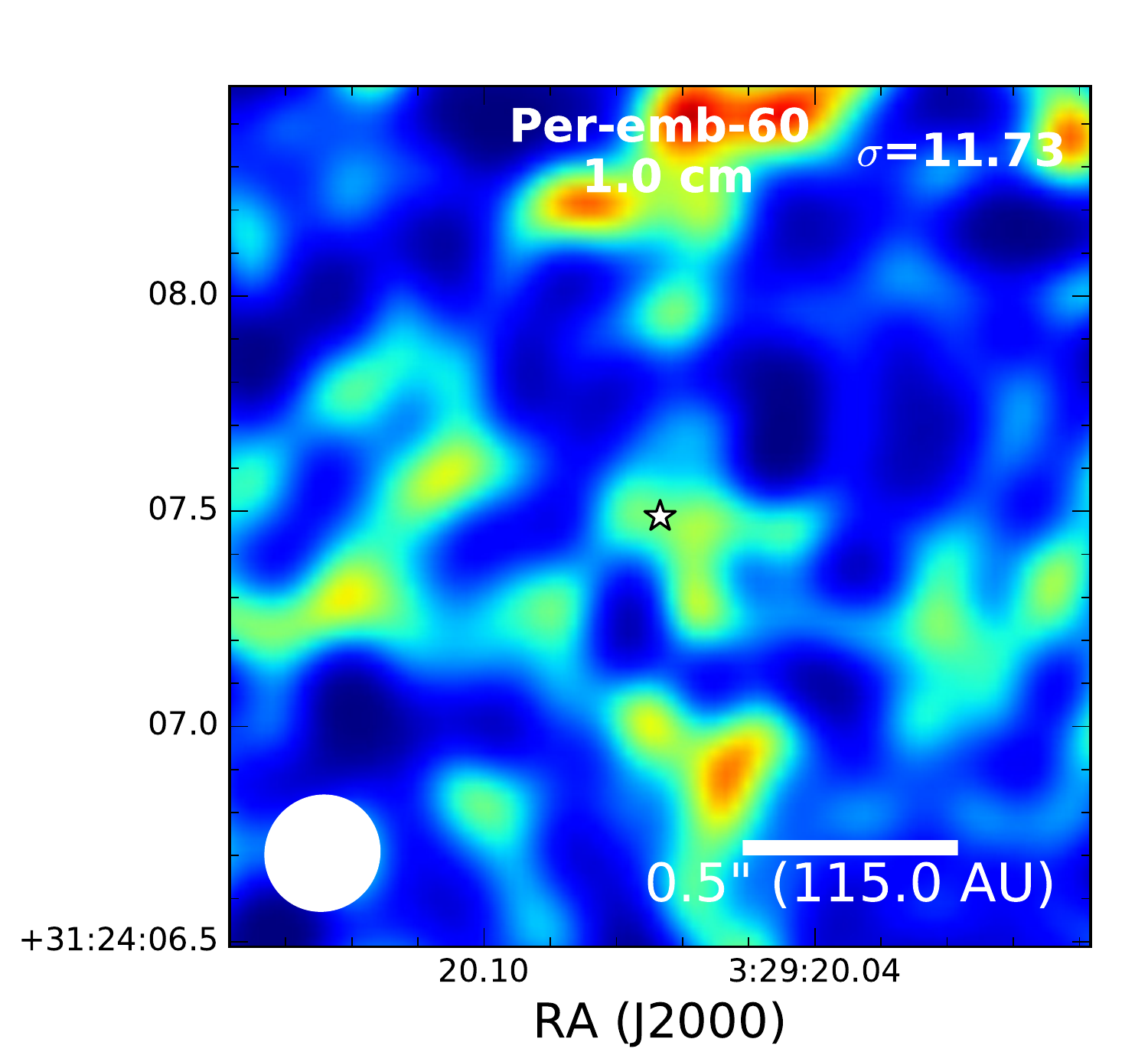}
  \includegraphics[width=0.24\linewidth]{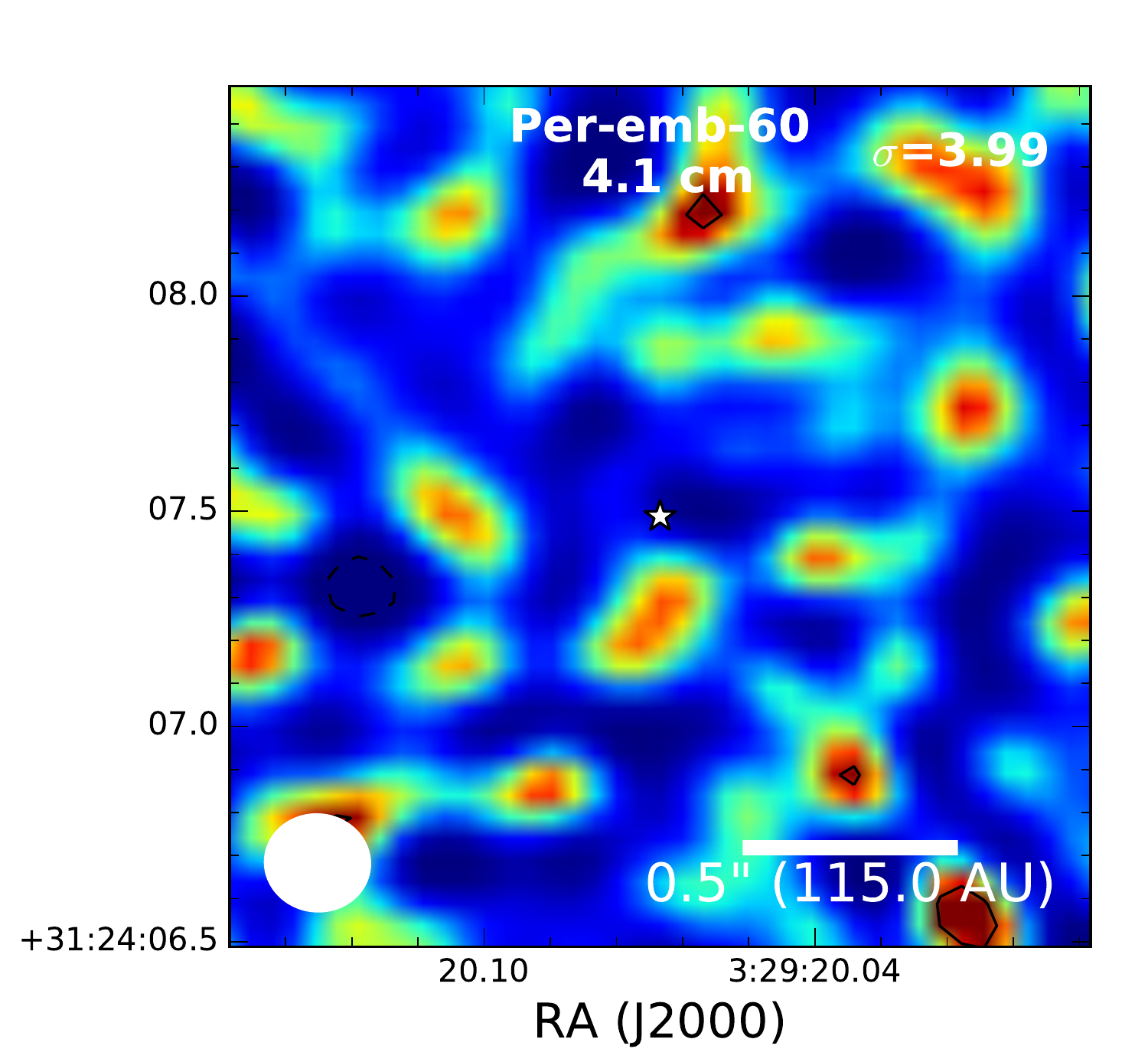}
  \includegraphics[width=0.24\linewidth]{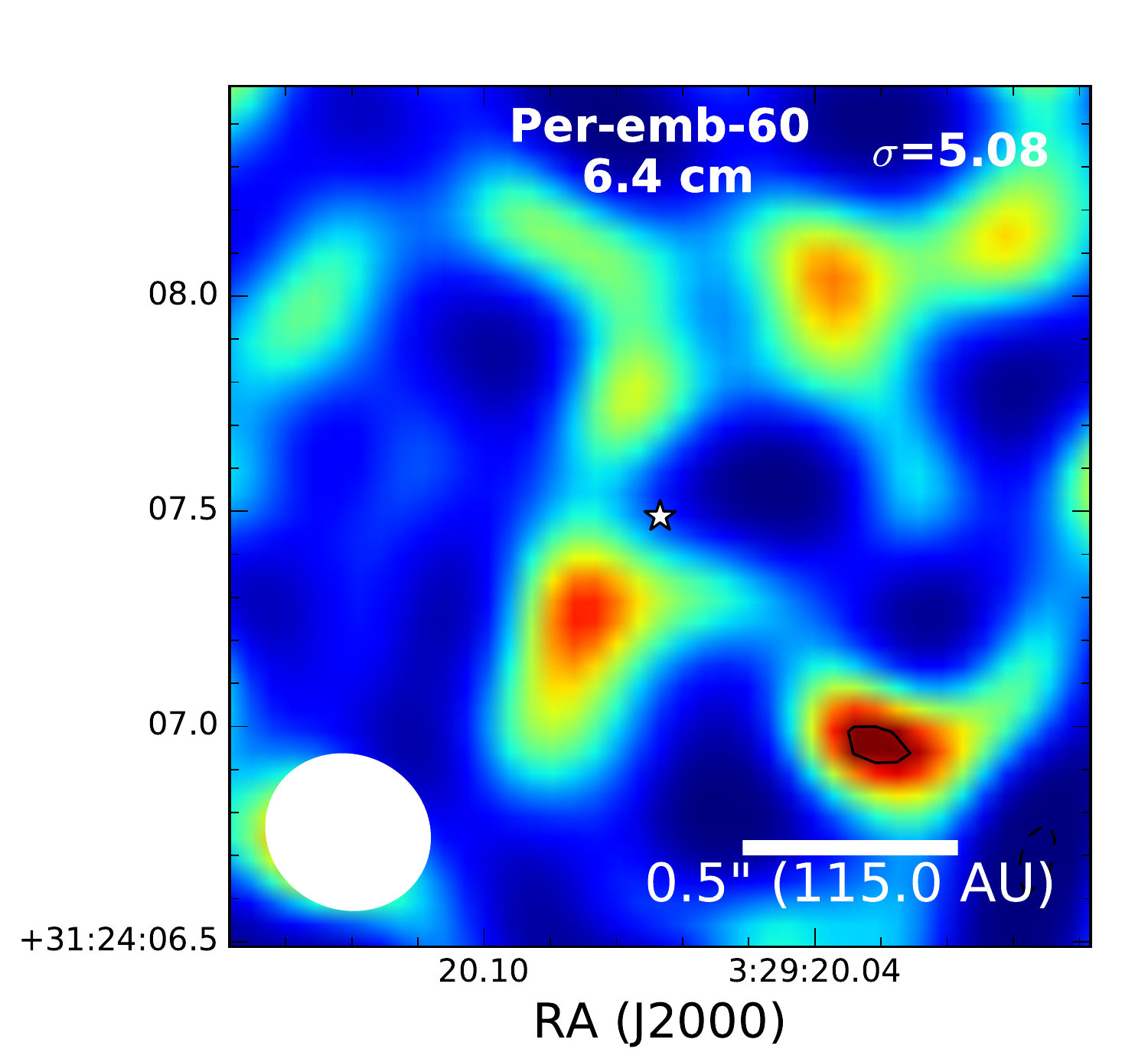}

  \includegraphics[width=0.24\linewidth]{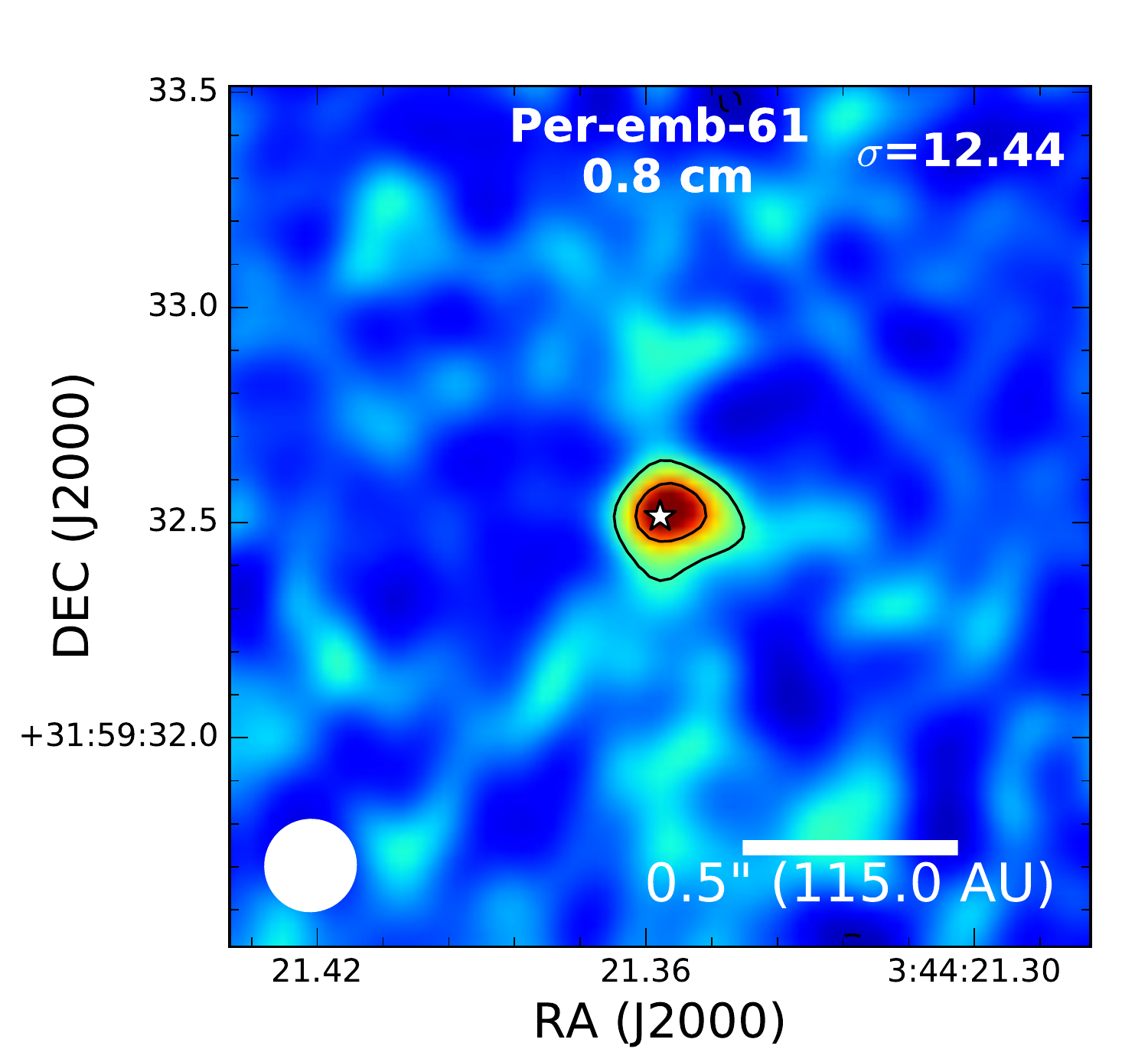}
  \includegraphics[width=0.24\linewidth]{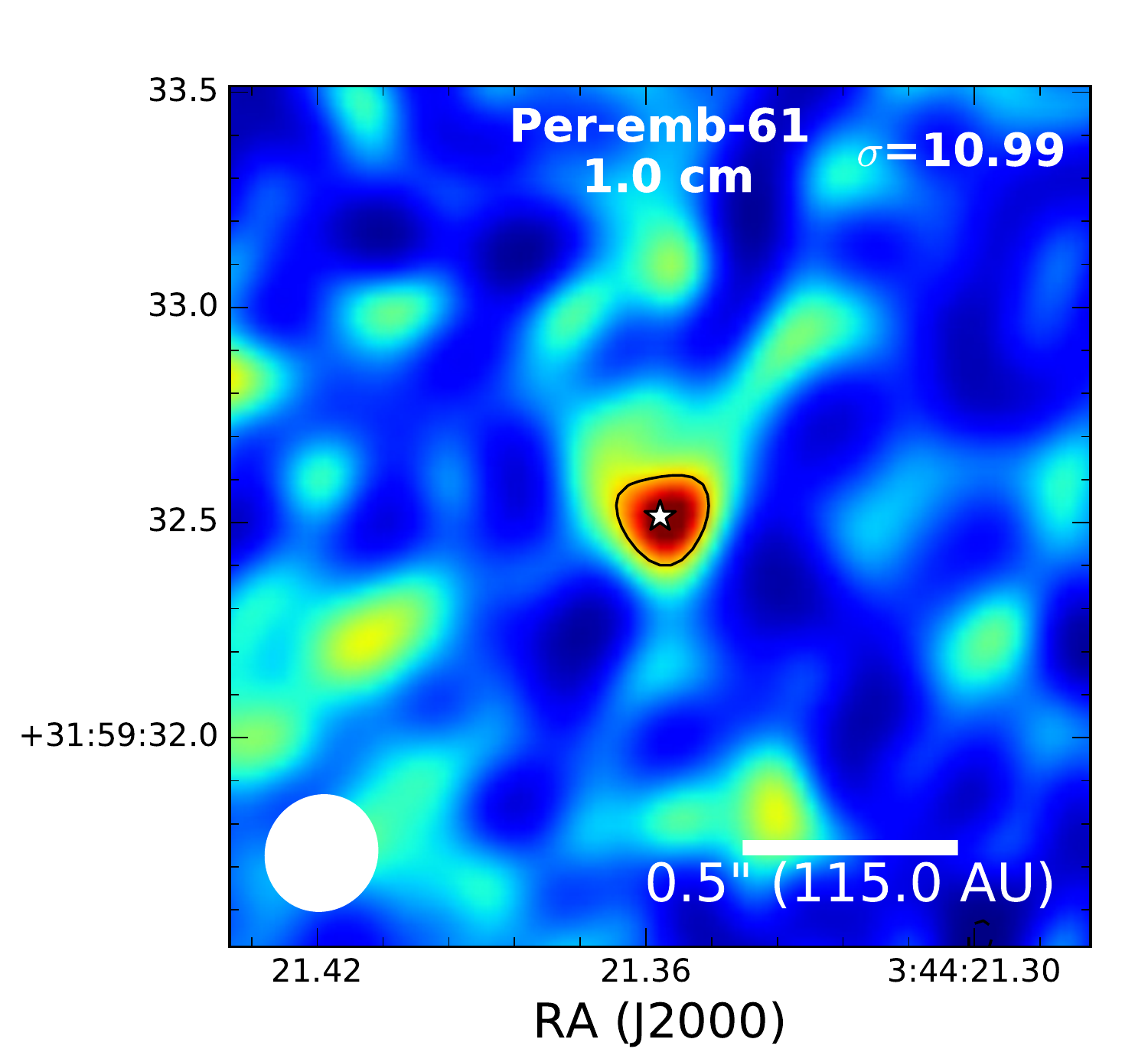}
  \includegraphics[width=0.24\linewidth]{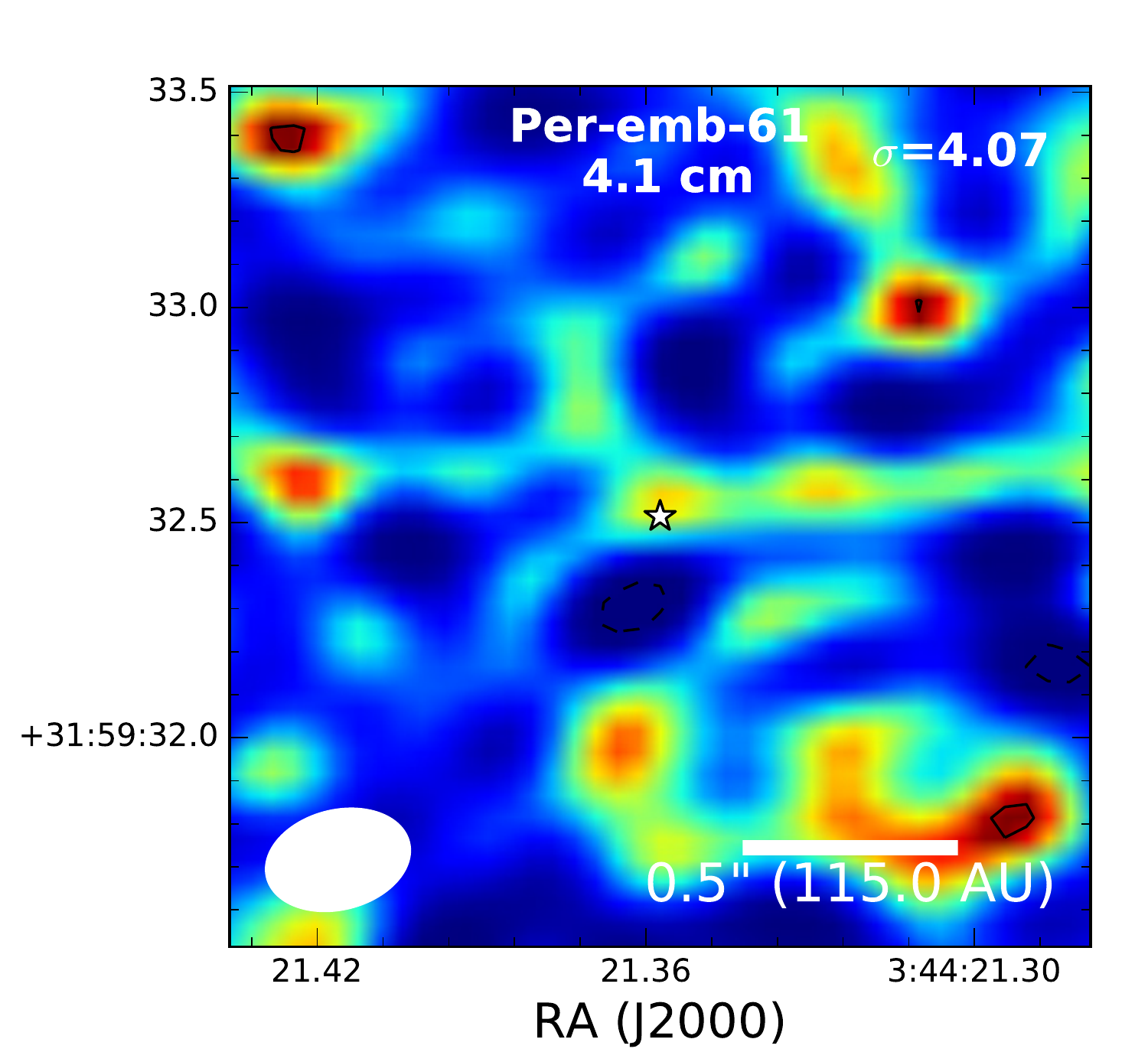}
  \includegraphics[width=0.24\linewidth]{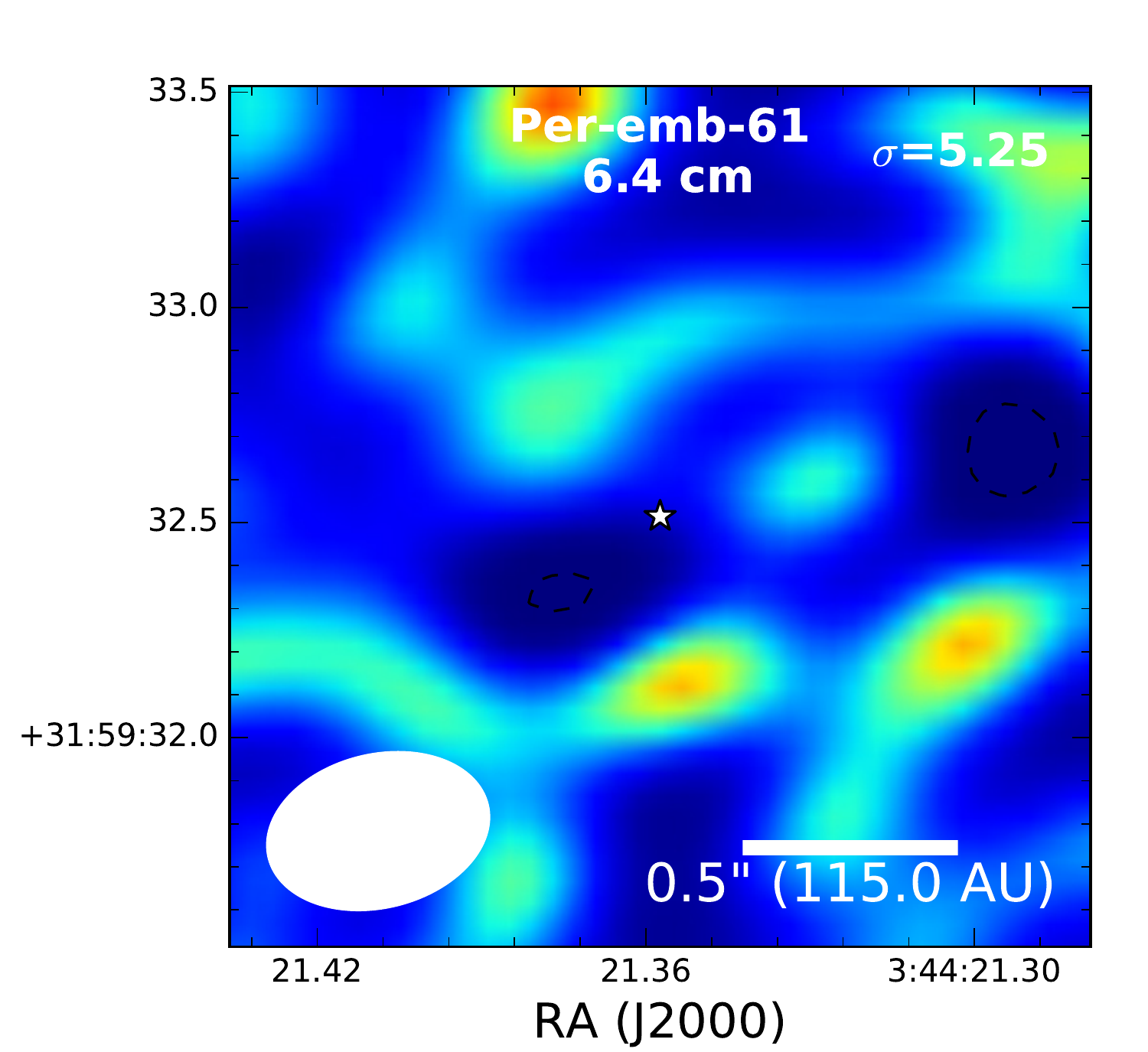}

  \includegraphics[width=0.24\linewidth]{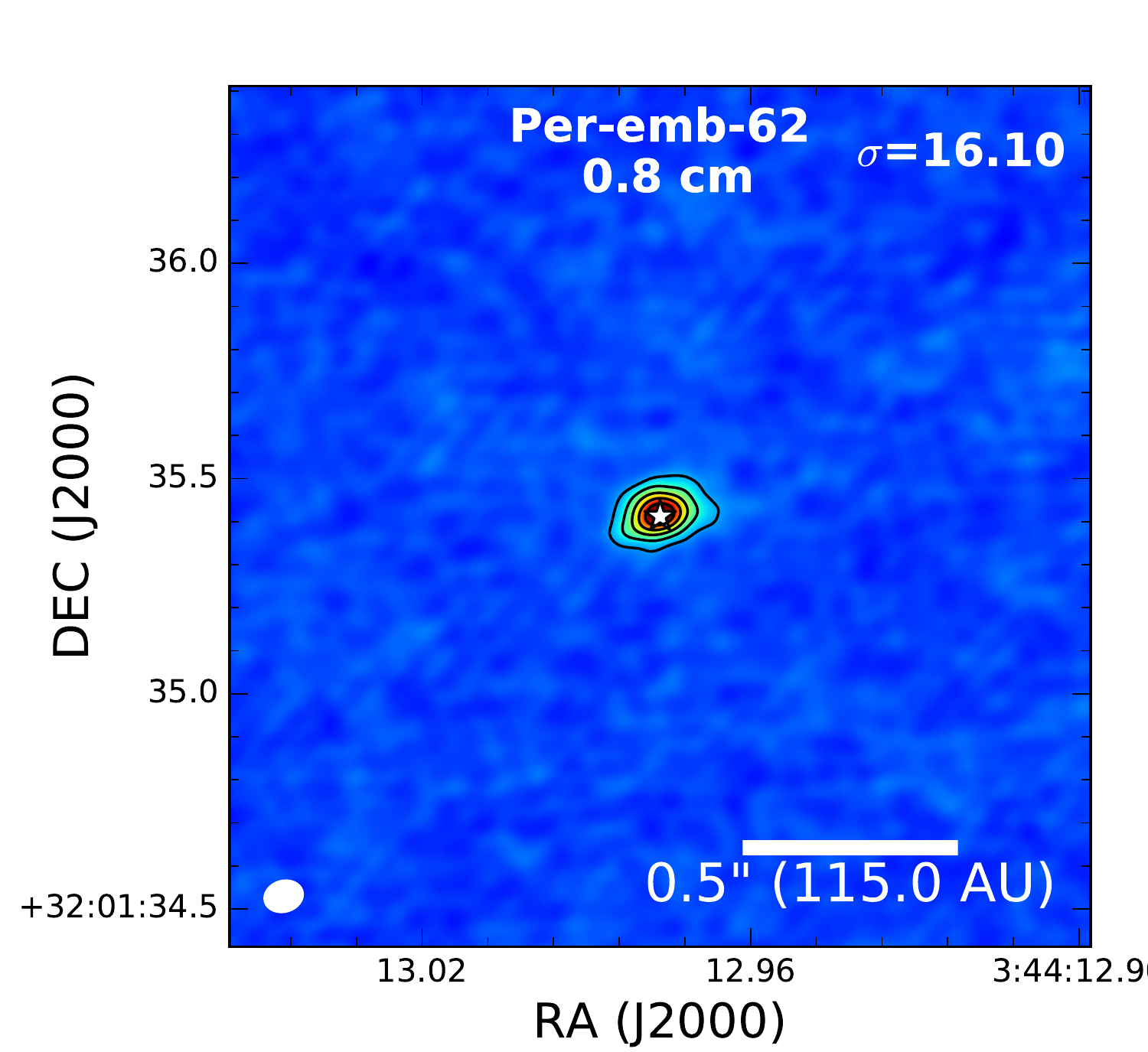}
  \includegraphics[width=0.24\linewidth]{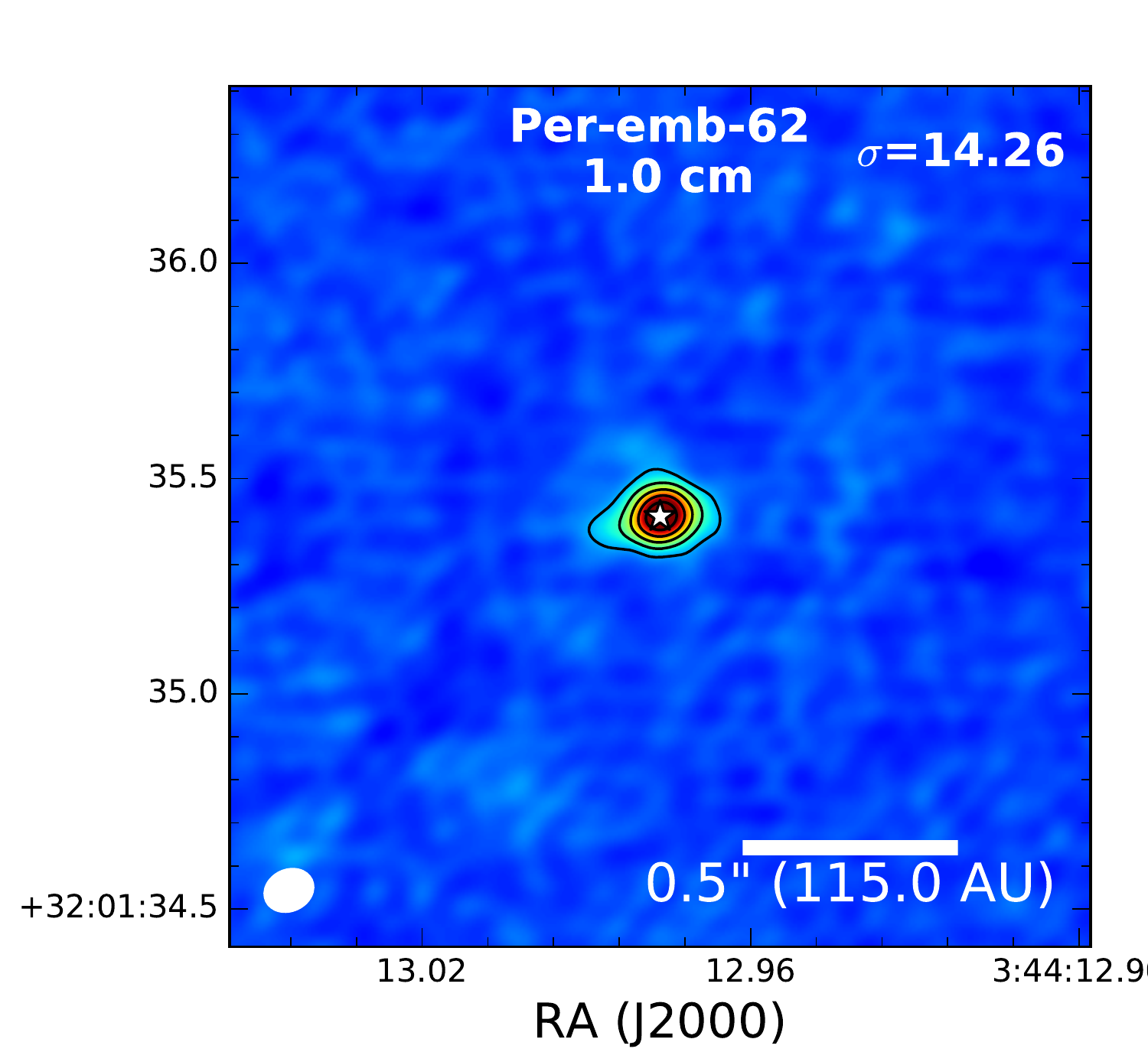}
  \includegraphics[width=0.24\linewidth]{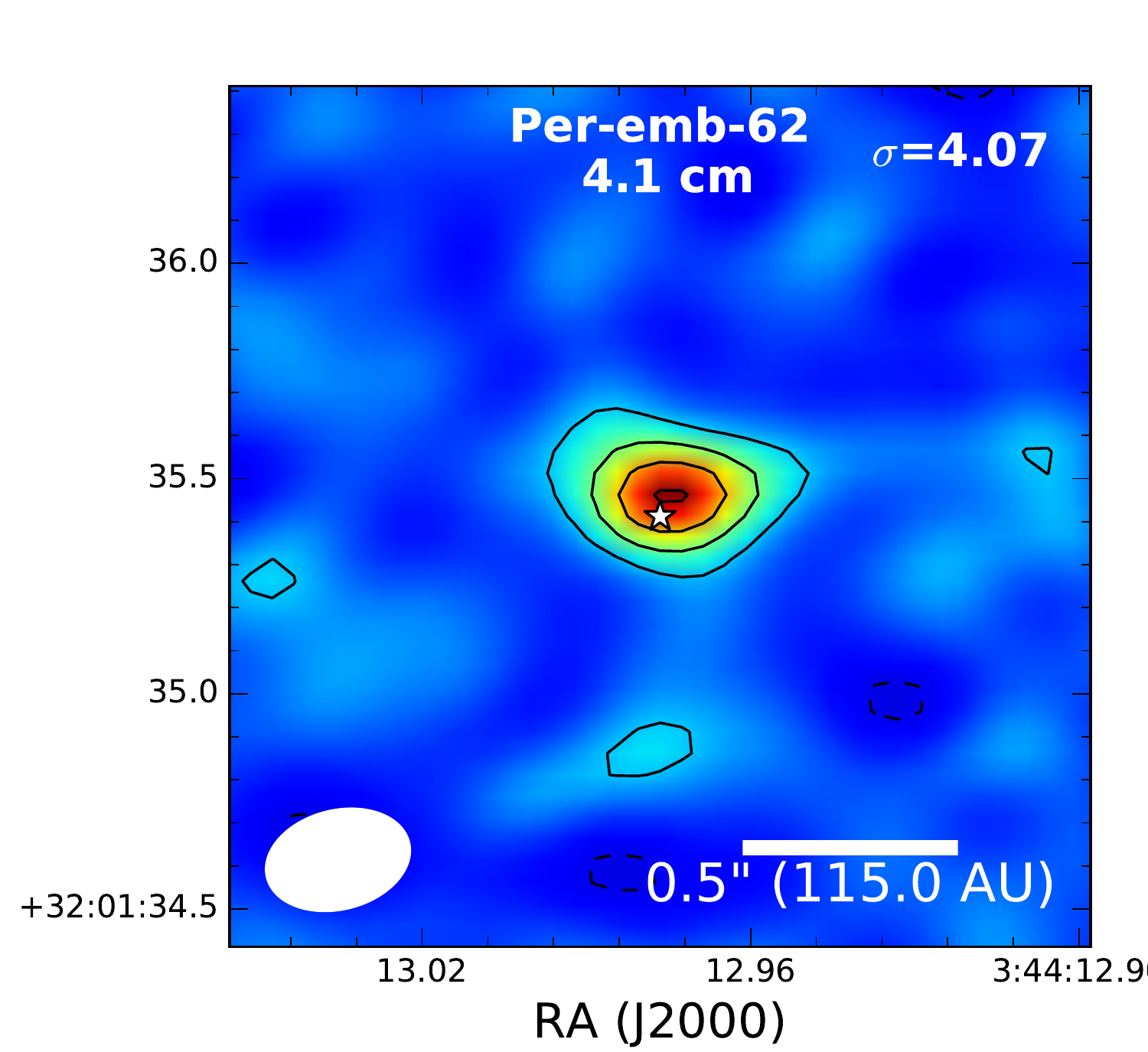}
  \includegraphics[width=0.24\linewidth]{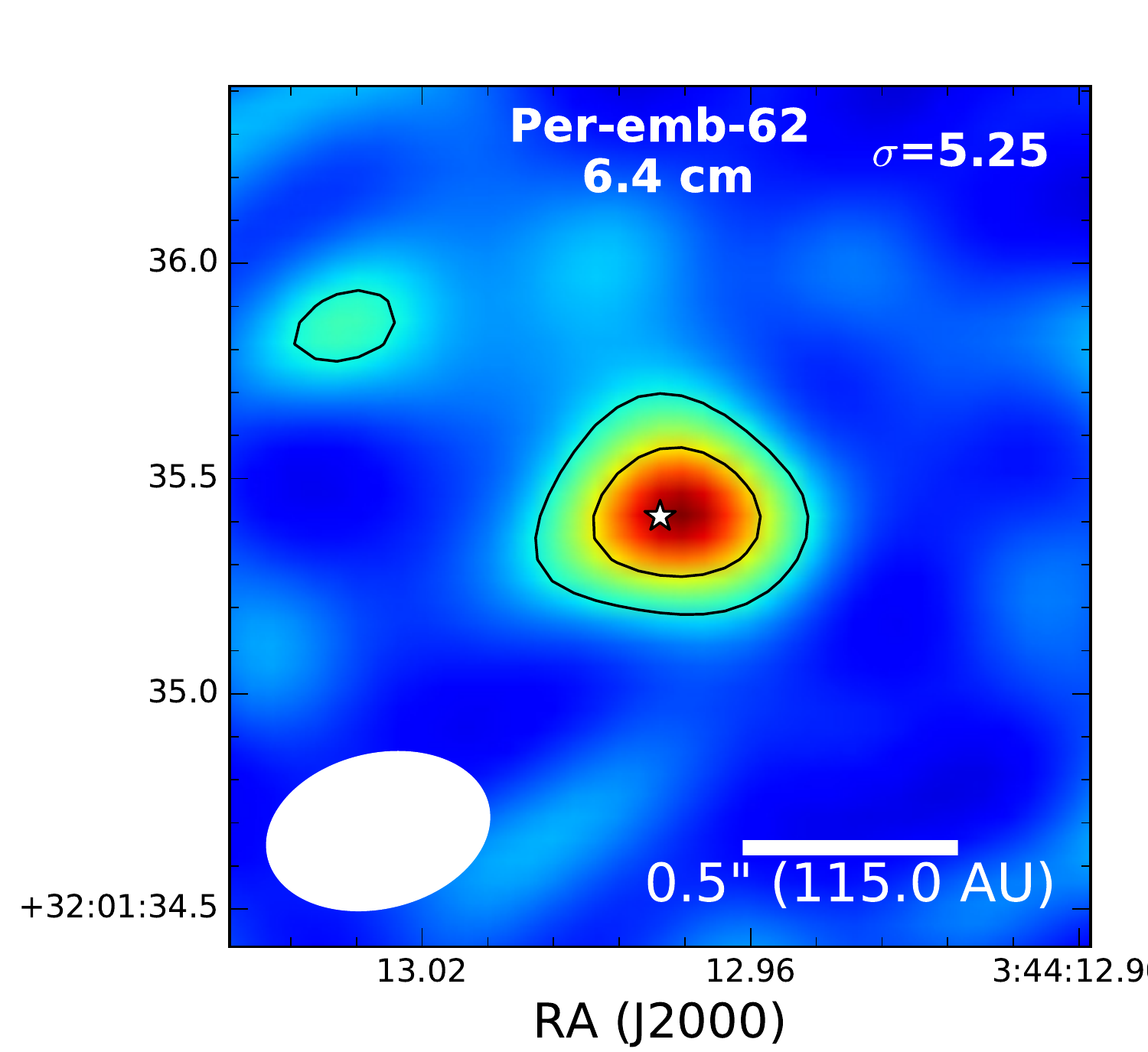}

  \includegraphics[width=0.24\linewidth]{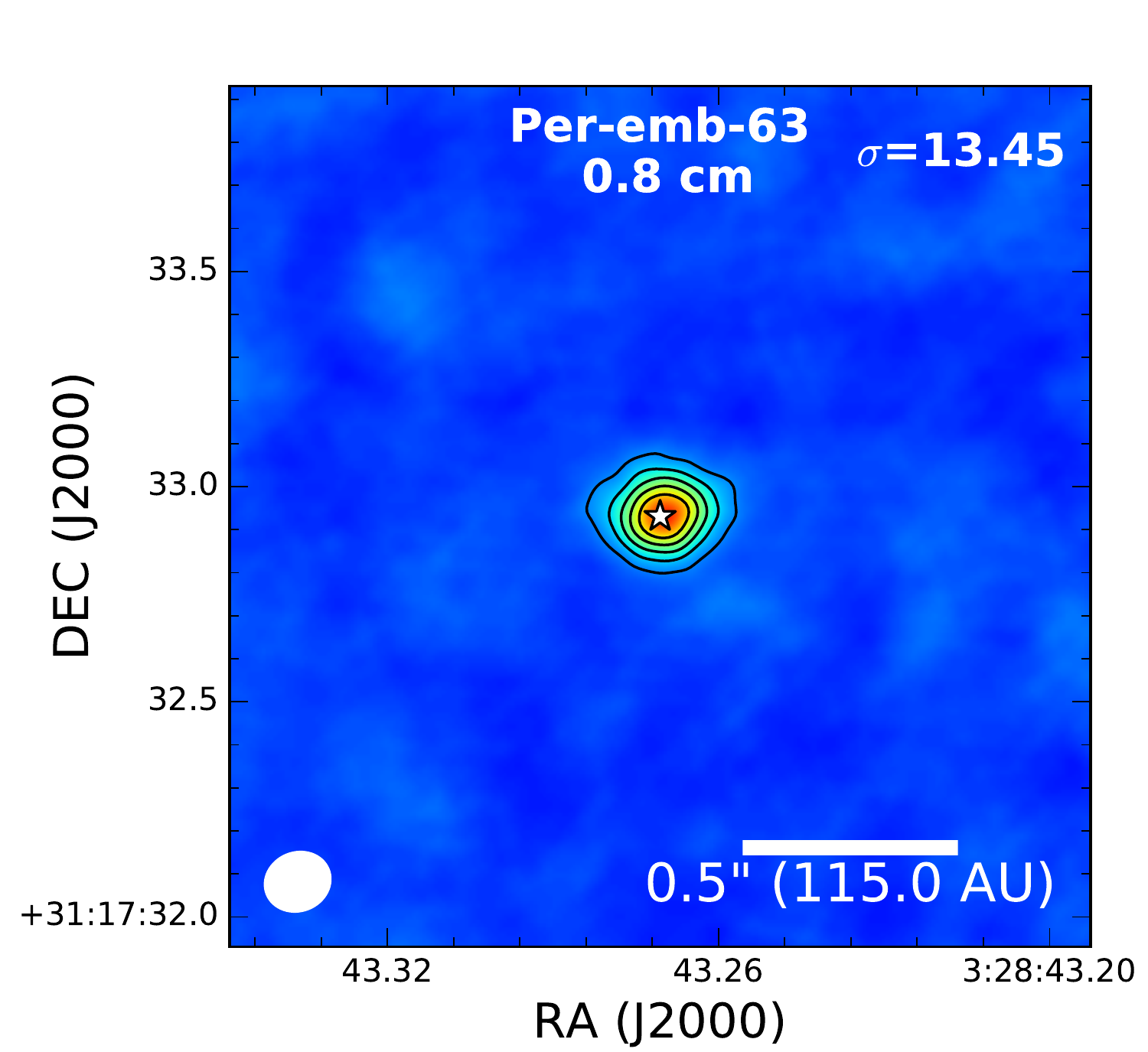}
  \includegraphics[width=0.24\linewidth]{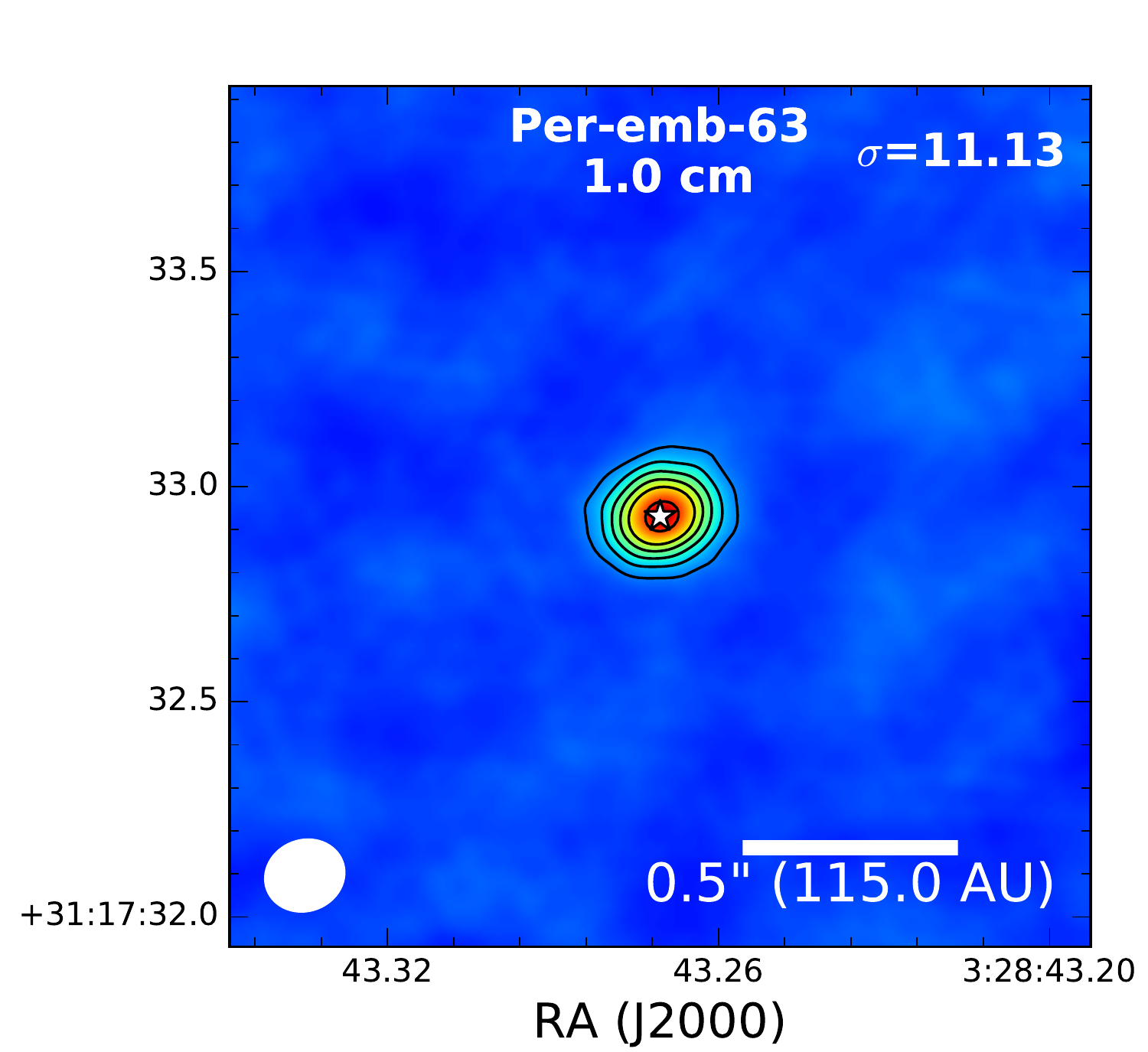}
  \includegraphics[width=0.24\linewidth]{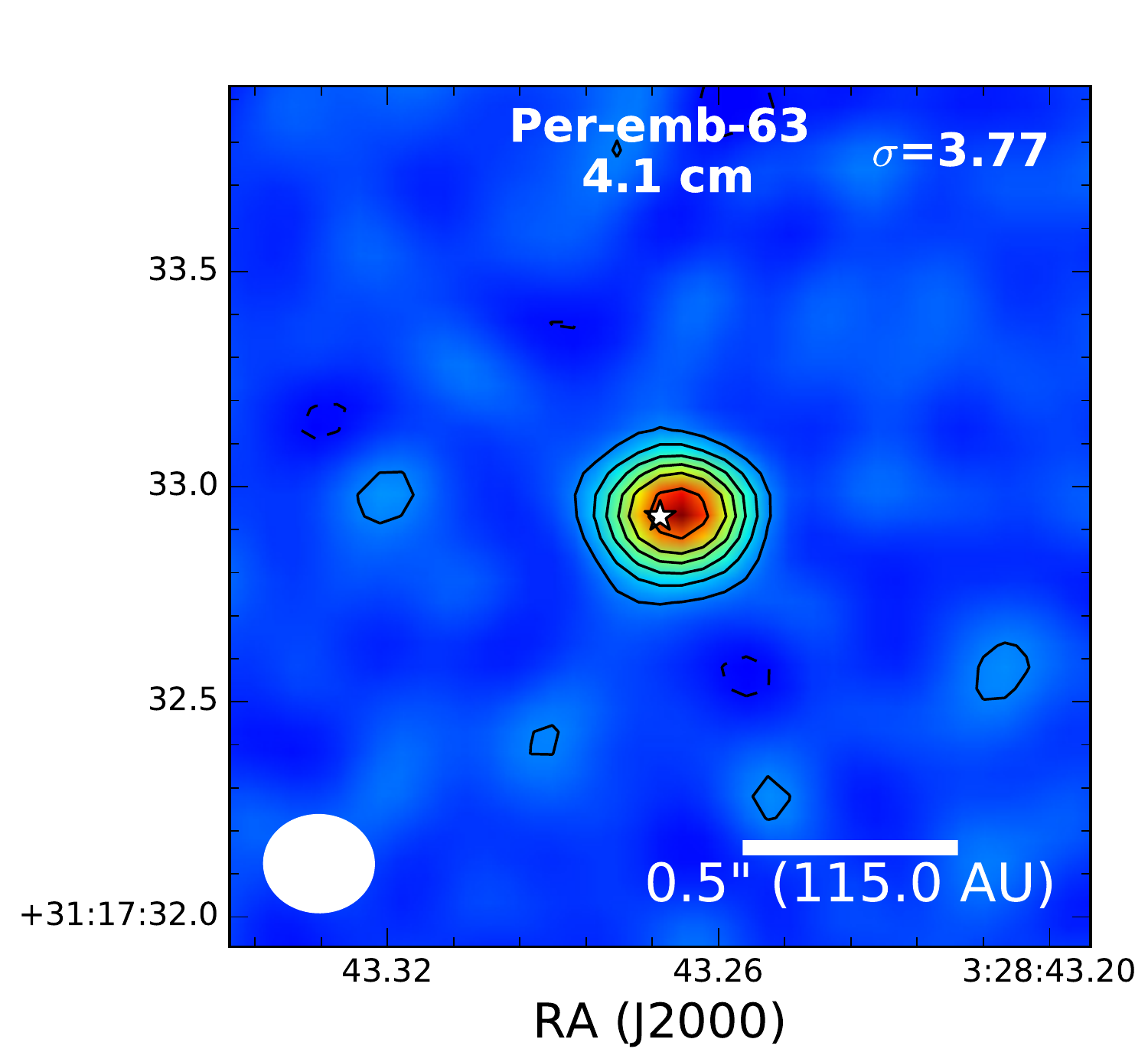}
  \includegraphics[width=0.24\linewidth]{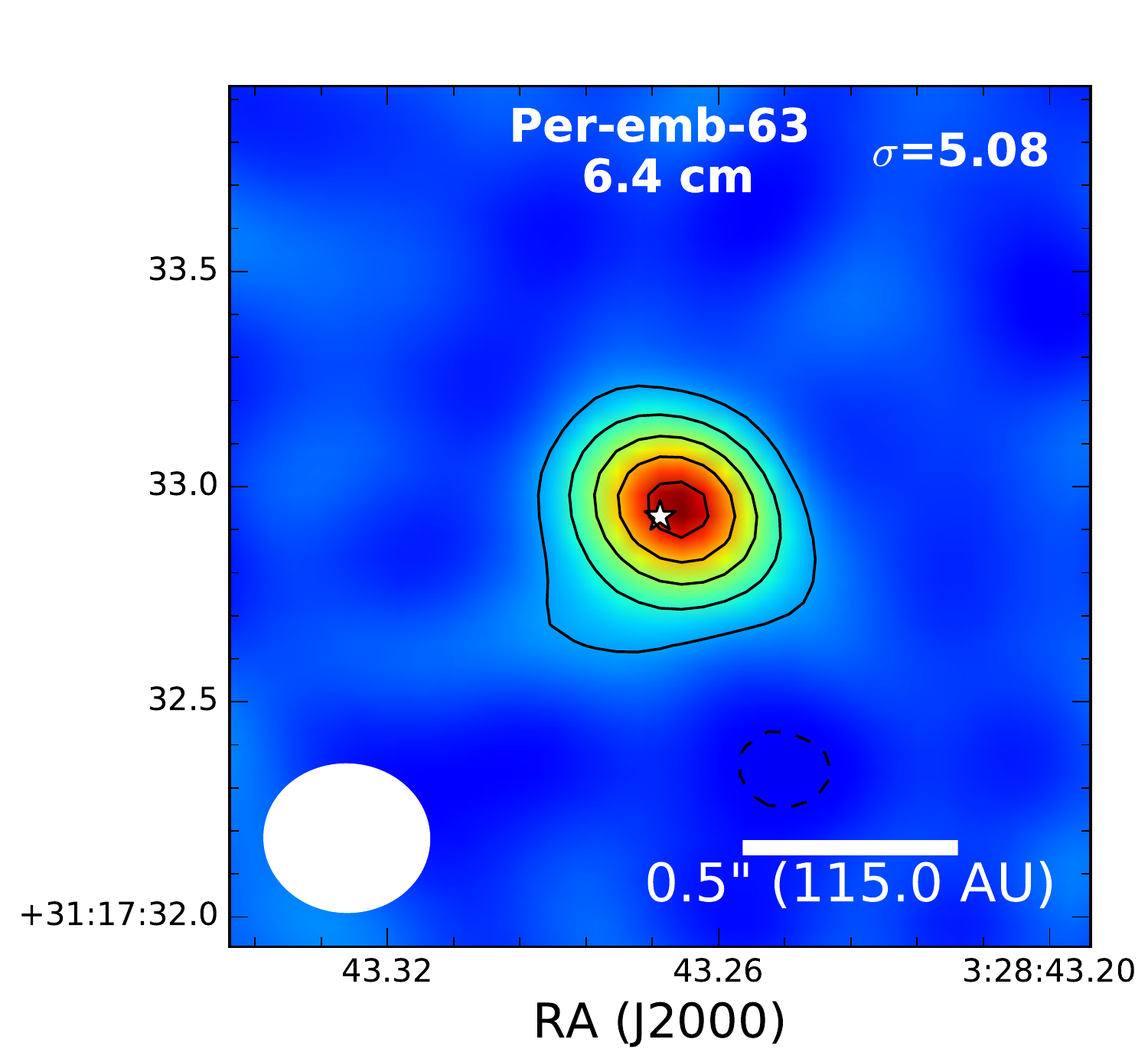}

\end{figure}
\begin{figure}

  \includegraphics[width=0.24\linewidth]{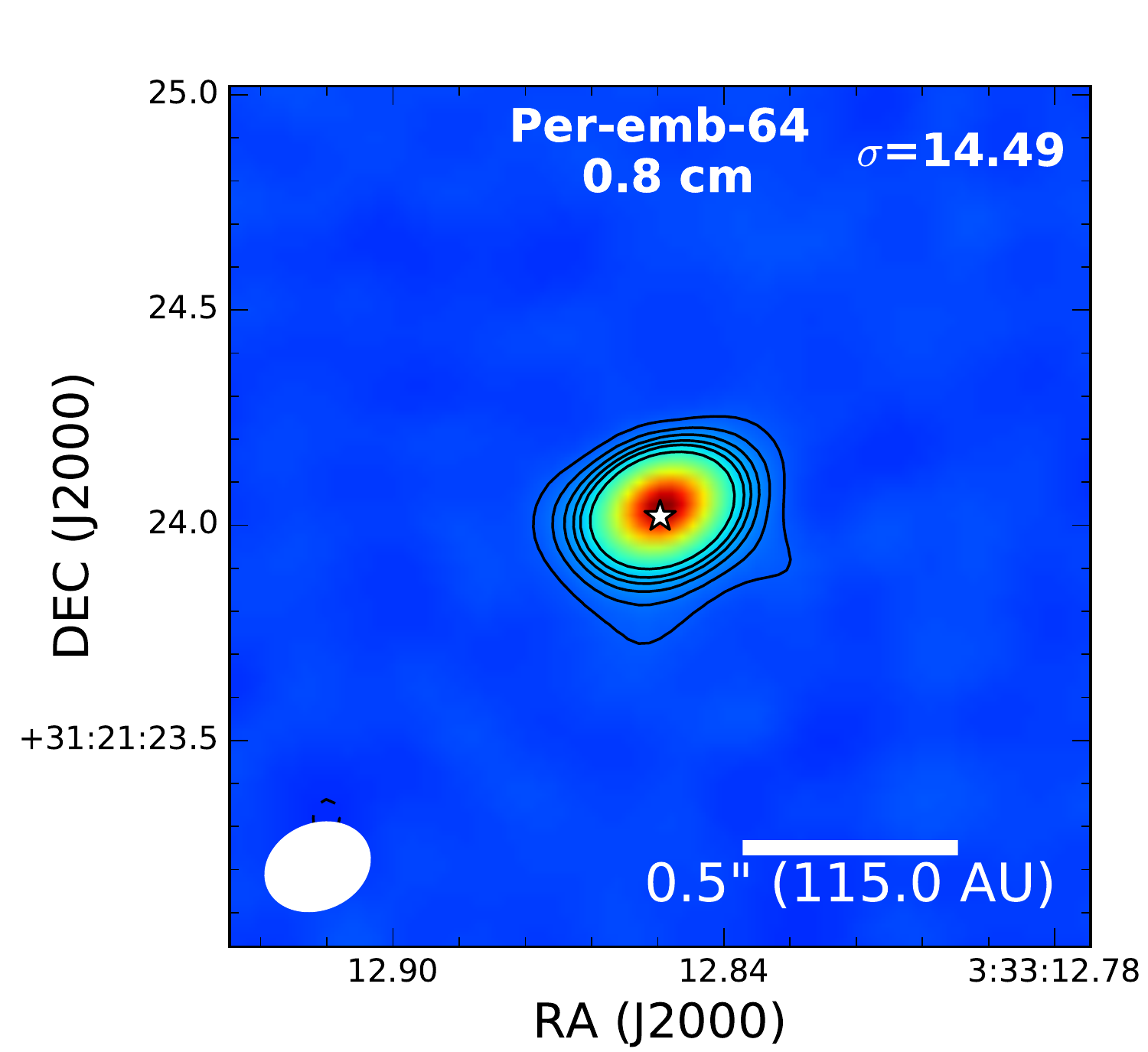}
  \includegraphics[width=0.24\linewidth]{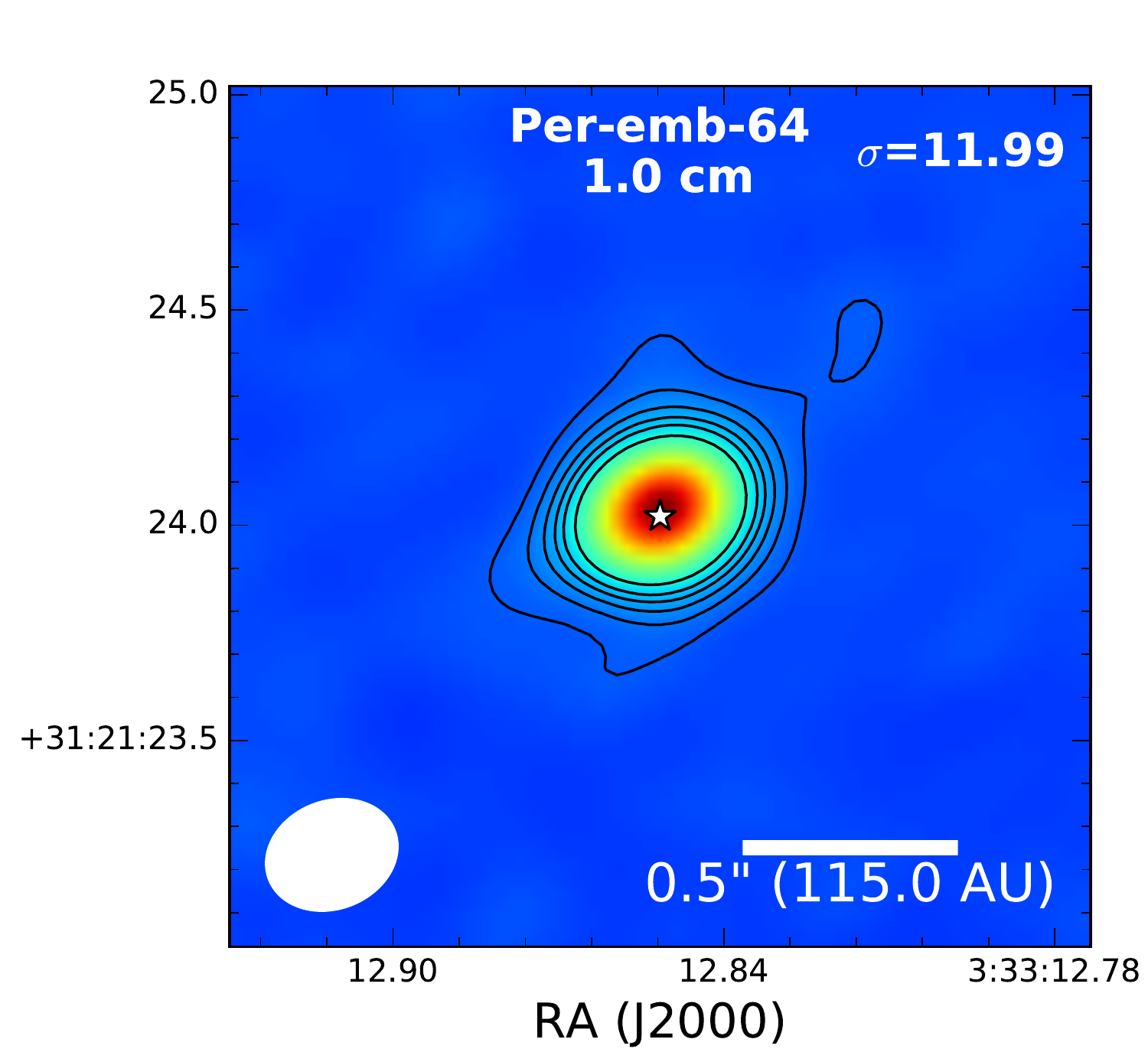}
  \includegraphics[width=0.24\linewidth]{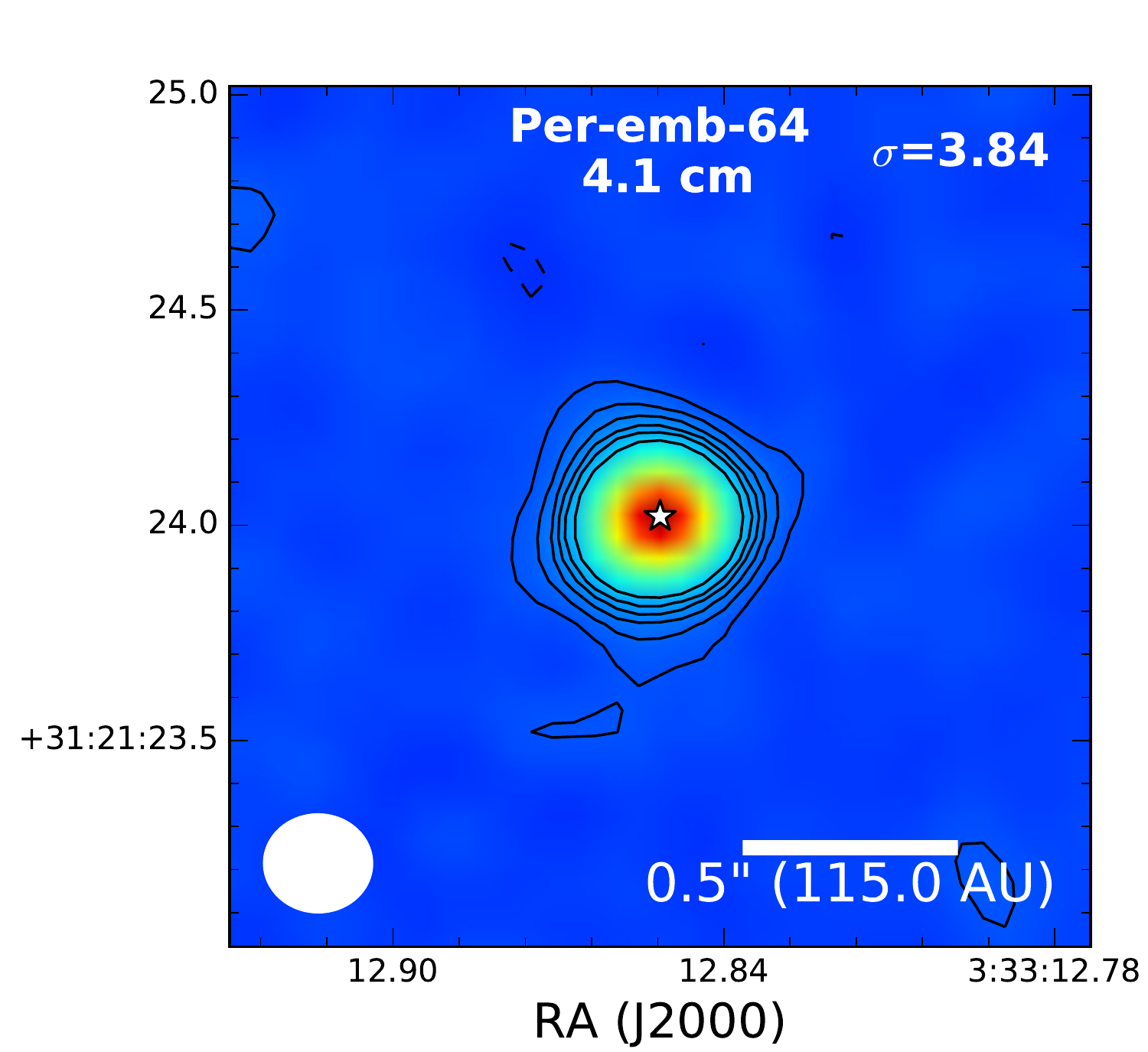}
  \includegraphics[width=0.24\linewidth]{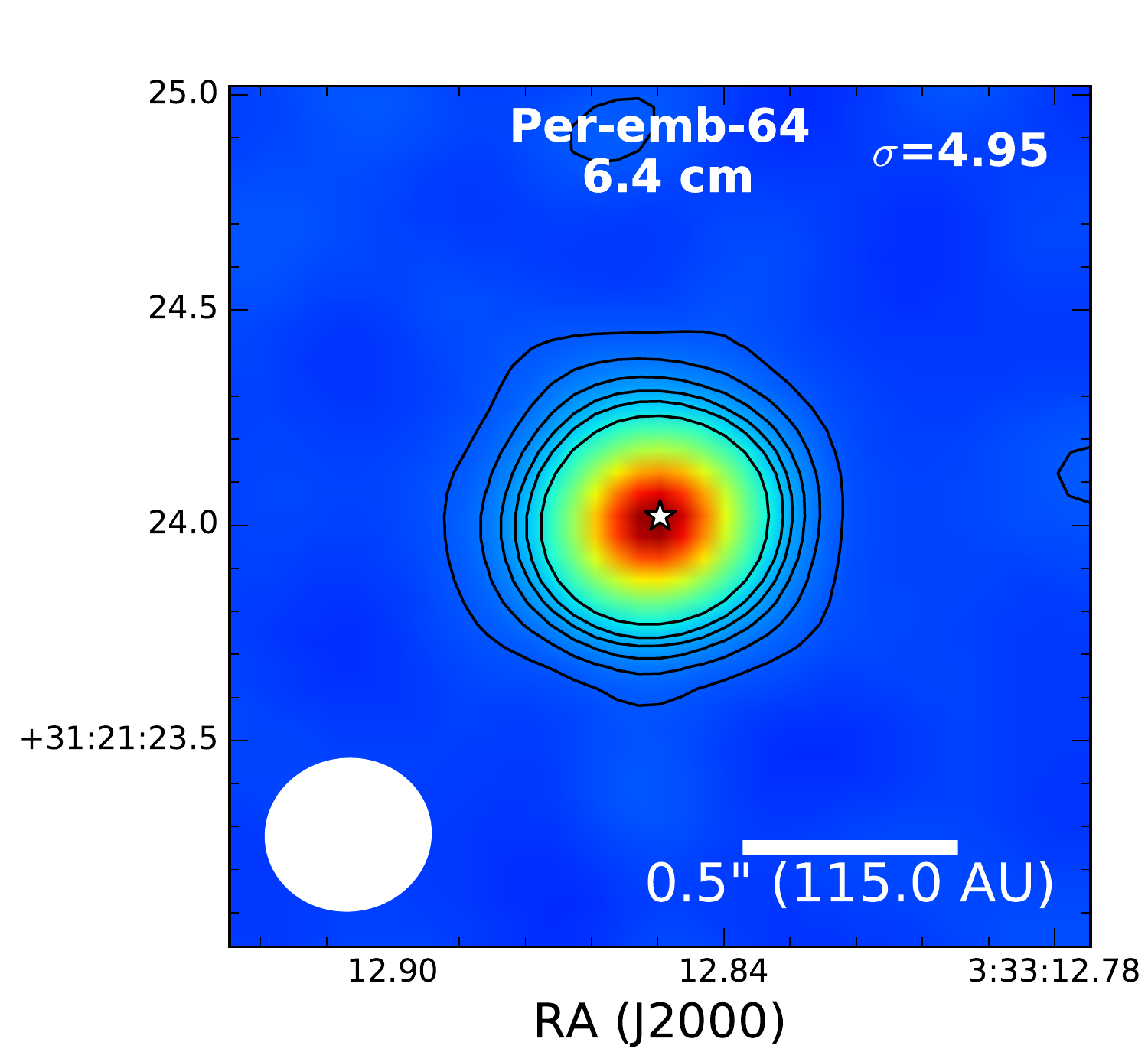}

  \includegraphics[width=0.24\linewidth]{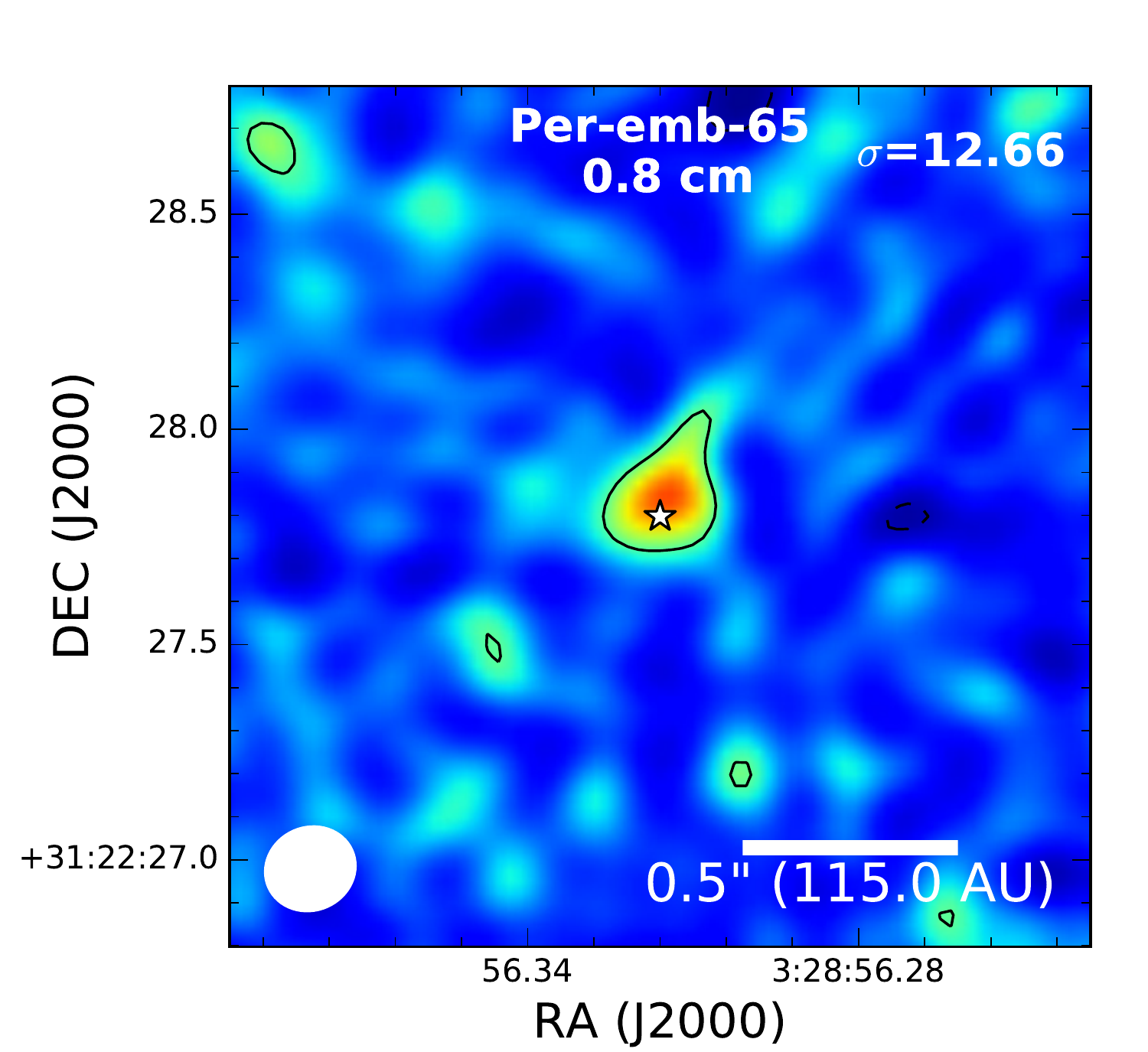}
  \includegraphics[width=0.24\linewidth]{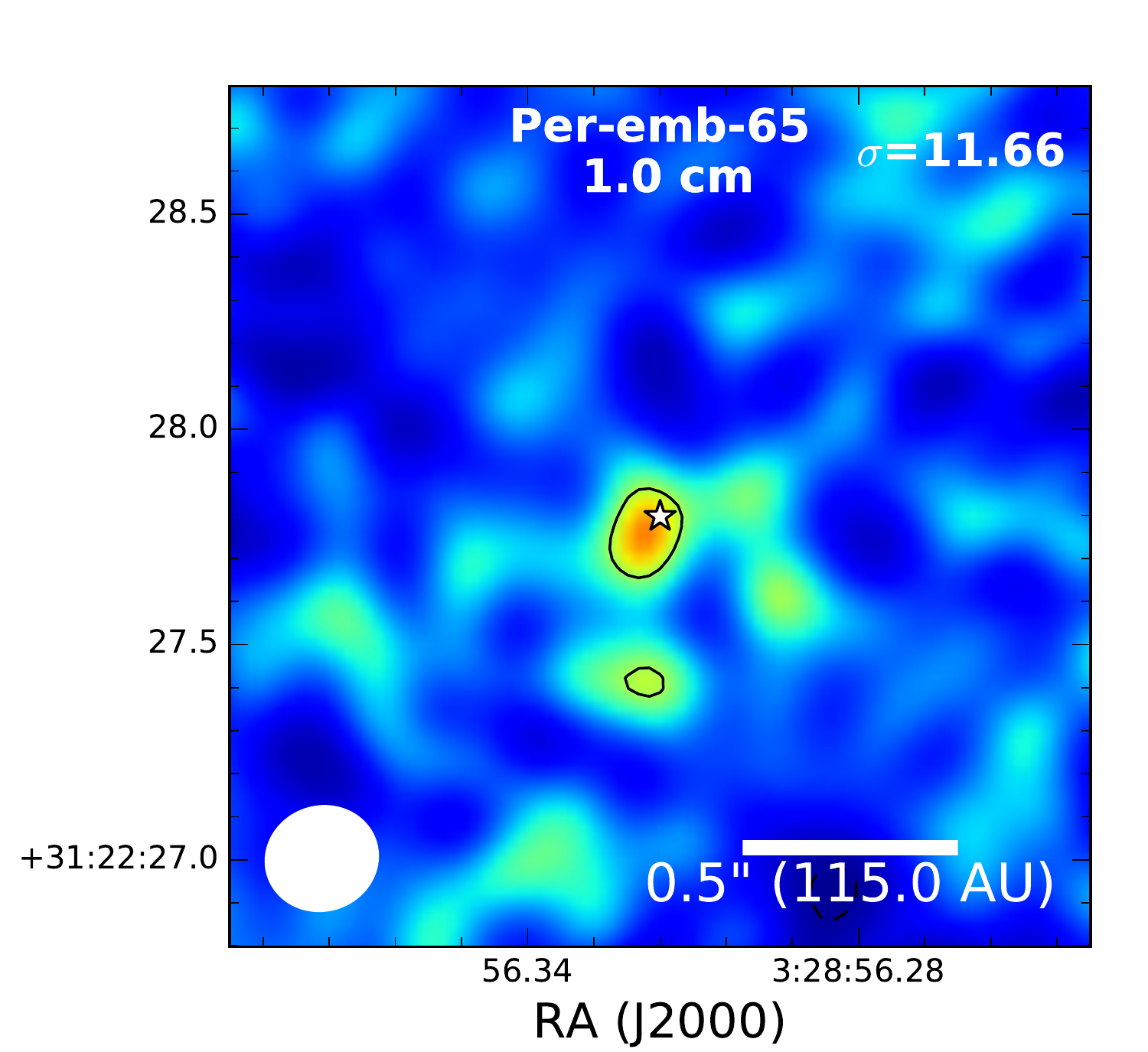}
  \includegraphics[width=0.24\linewidth]{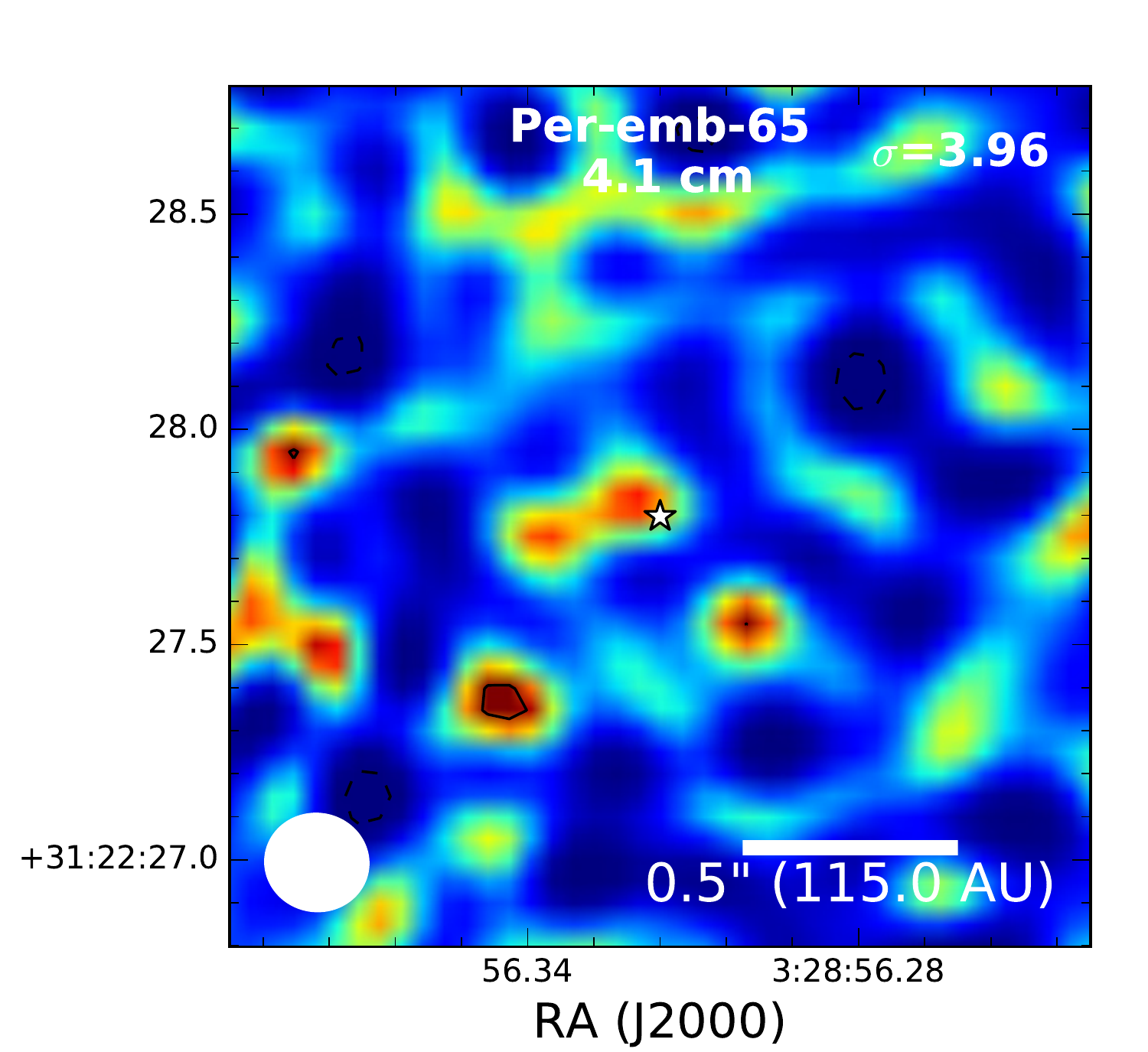}
  \includegraphics[width=0.24\linewidth]{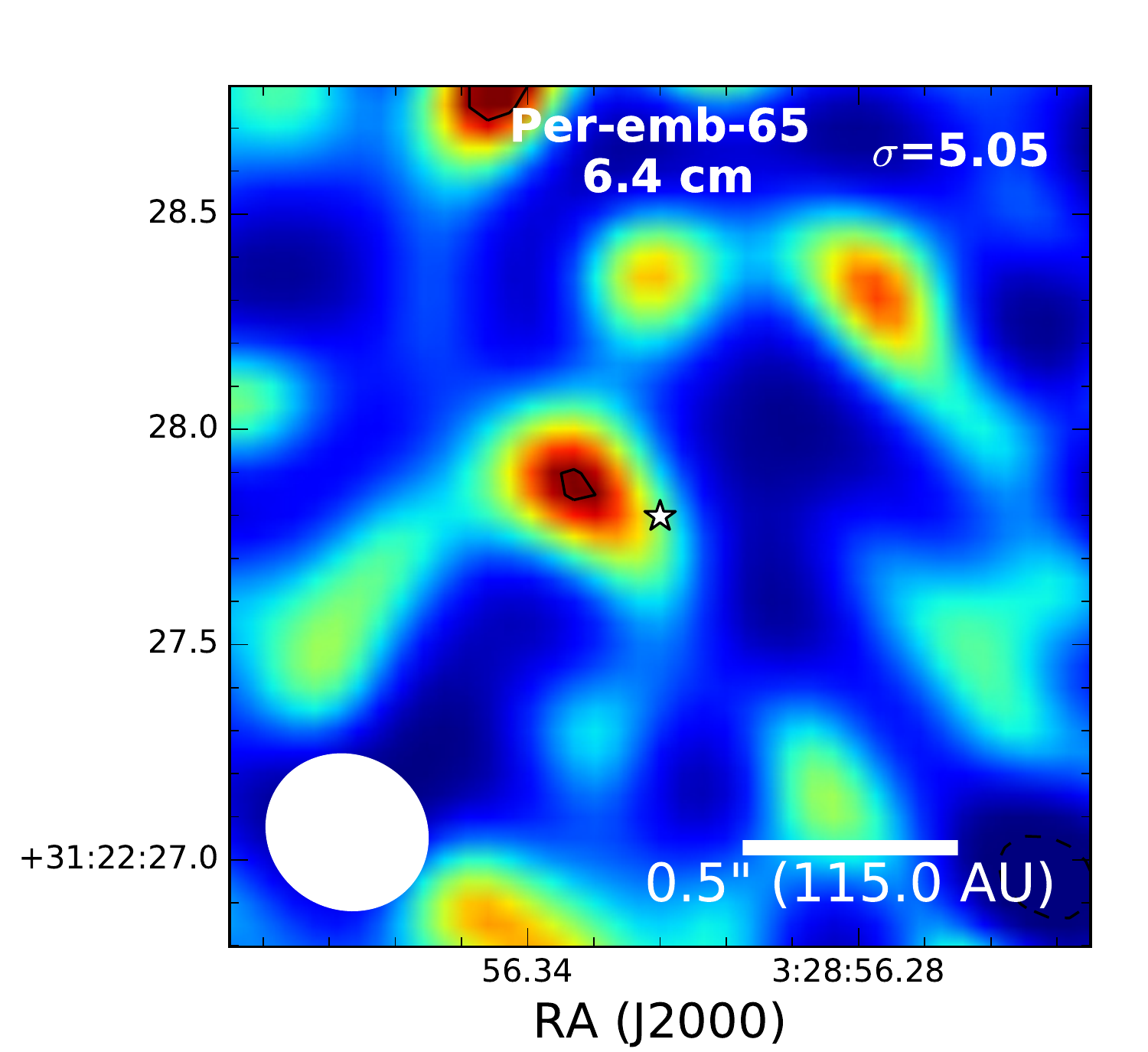}

  \includegraphics[width=0.24\linewidth]{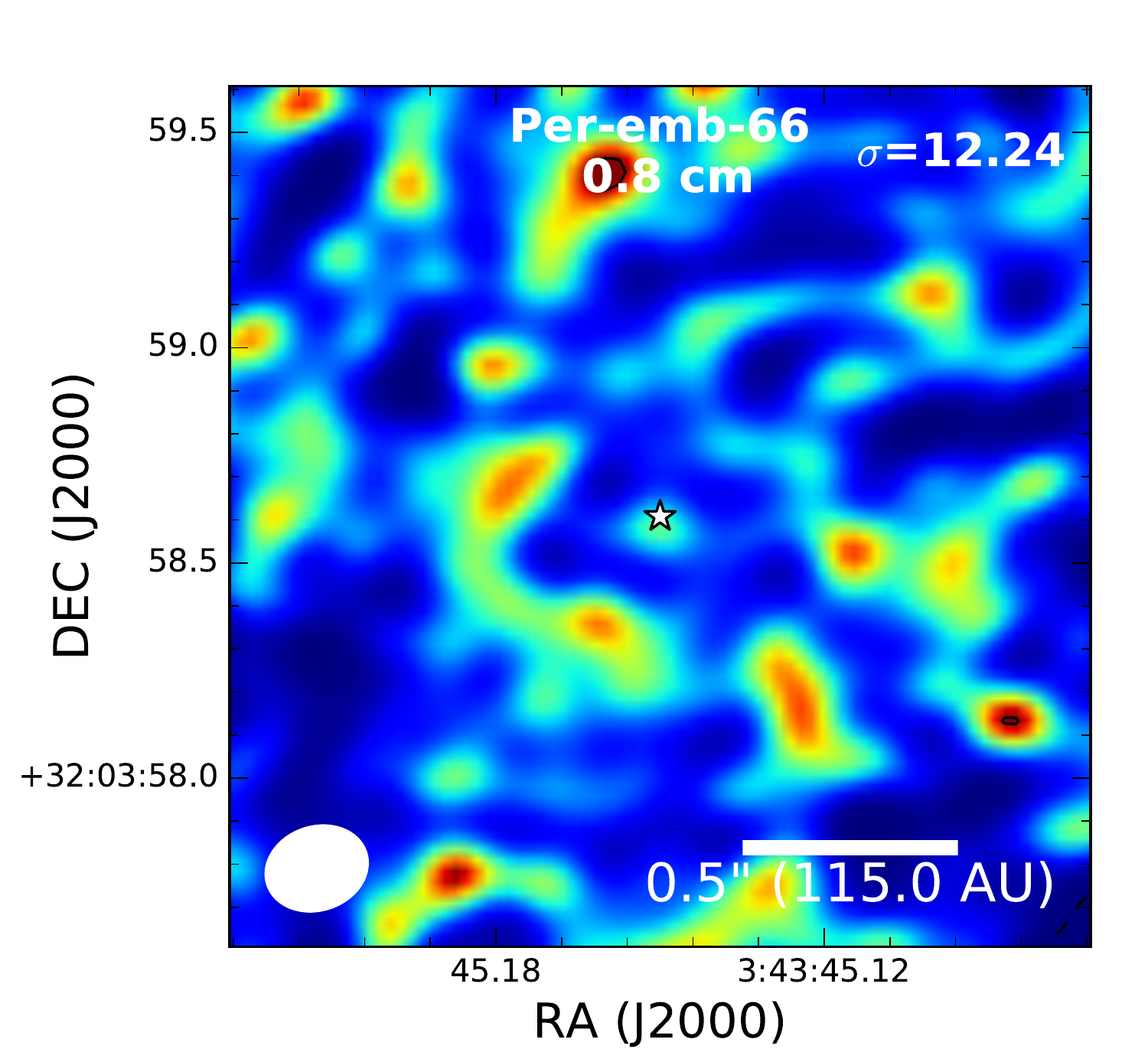}
  \includegraphics[width=0.24\linewidth]{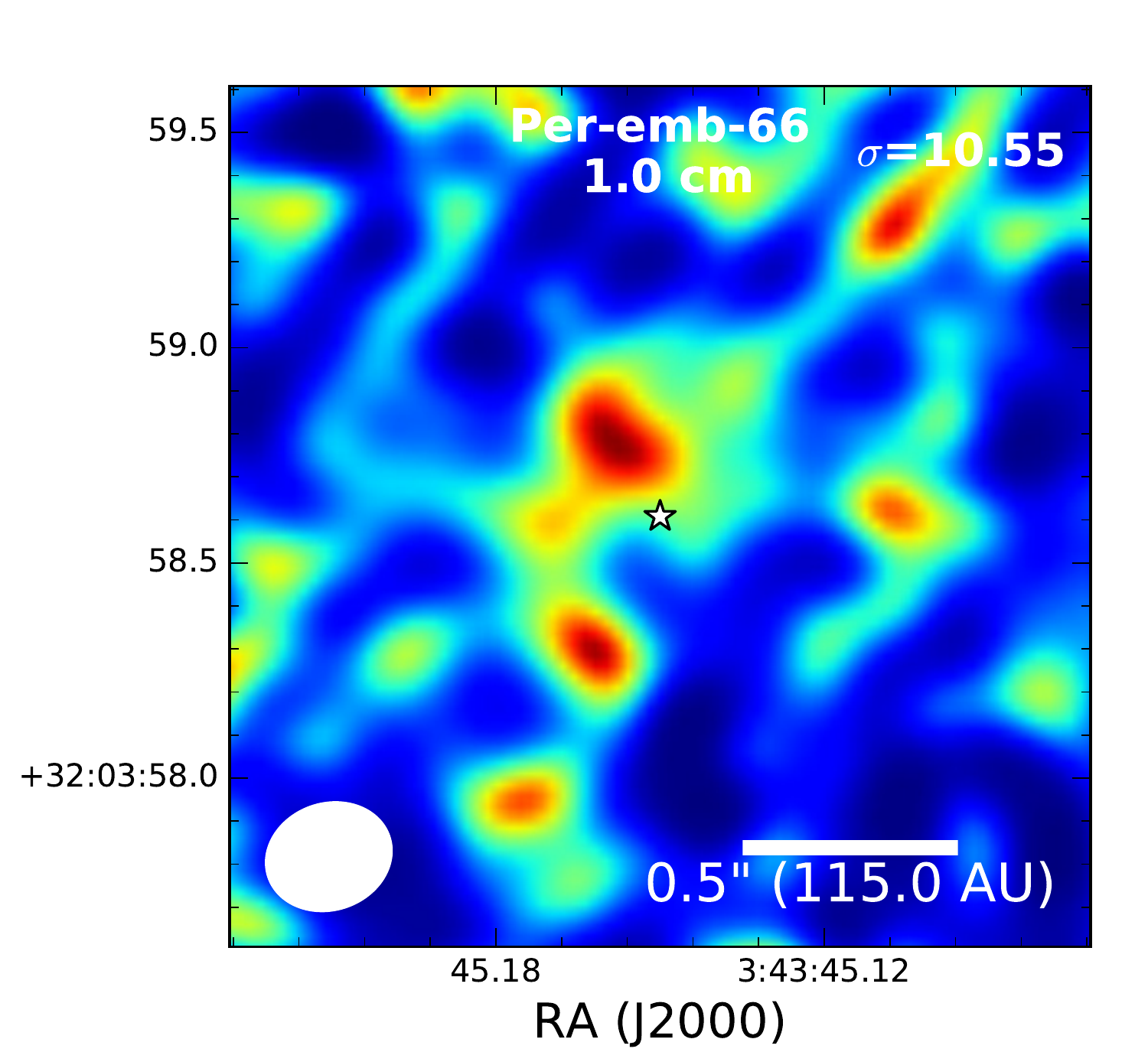}
  \includegraphics[width=0.24\linewidth]{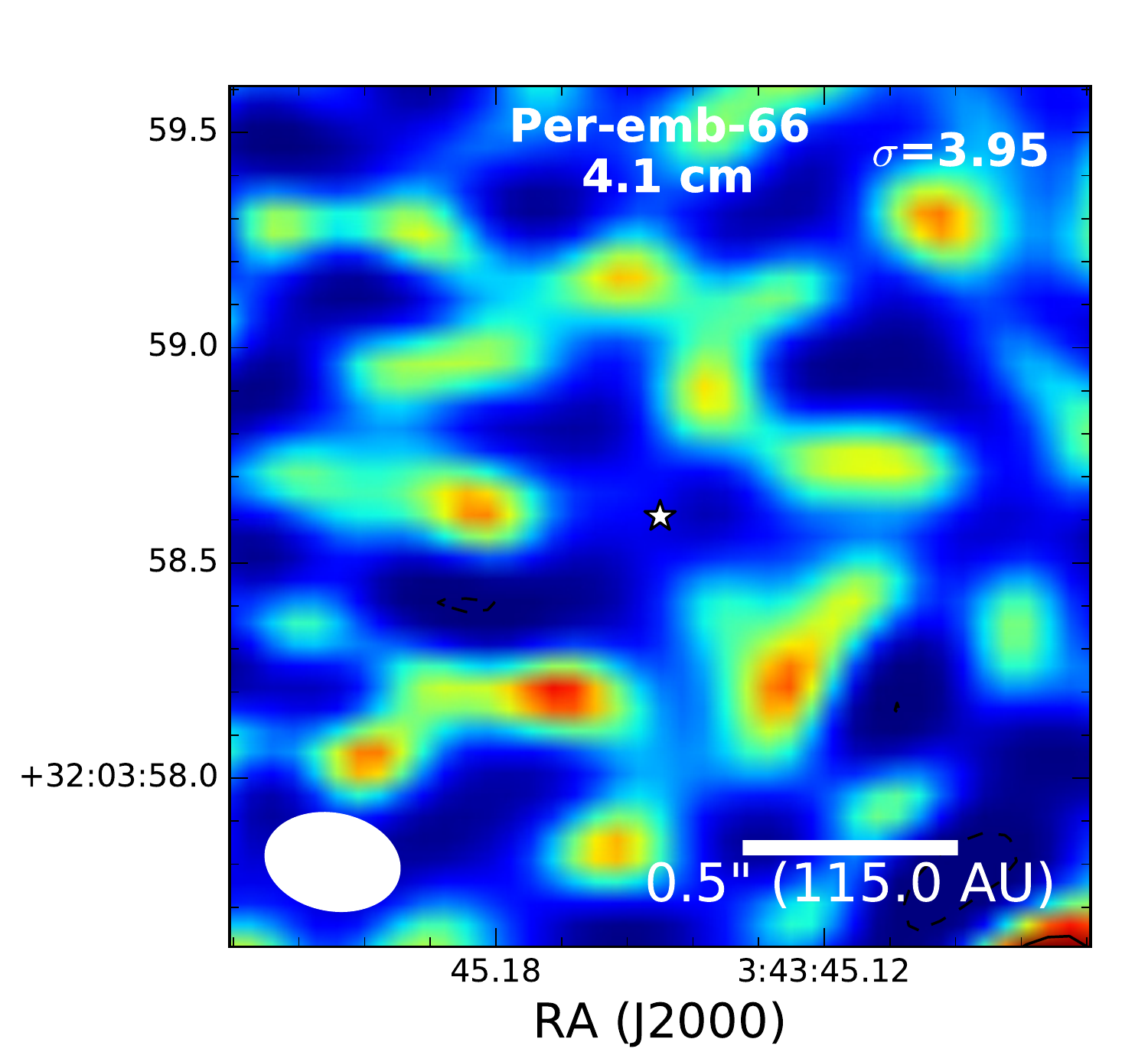}
  \includegraphics[width=0.24\linewidth]{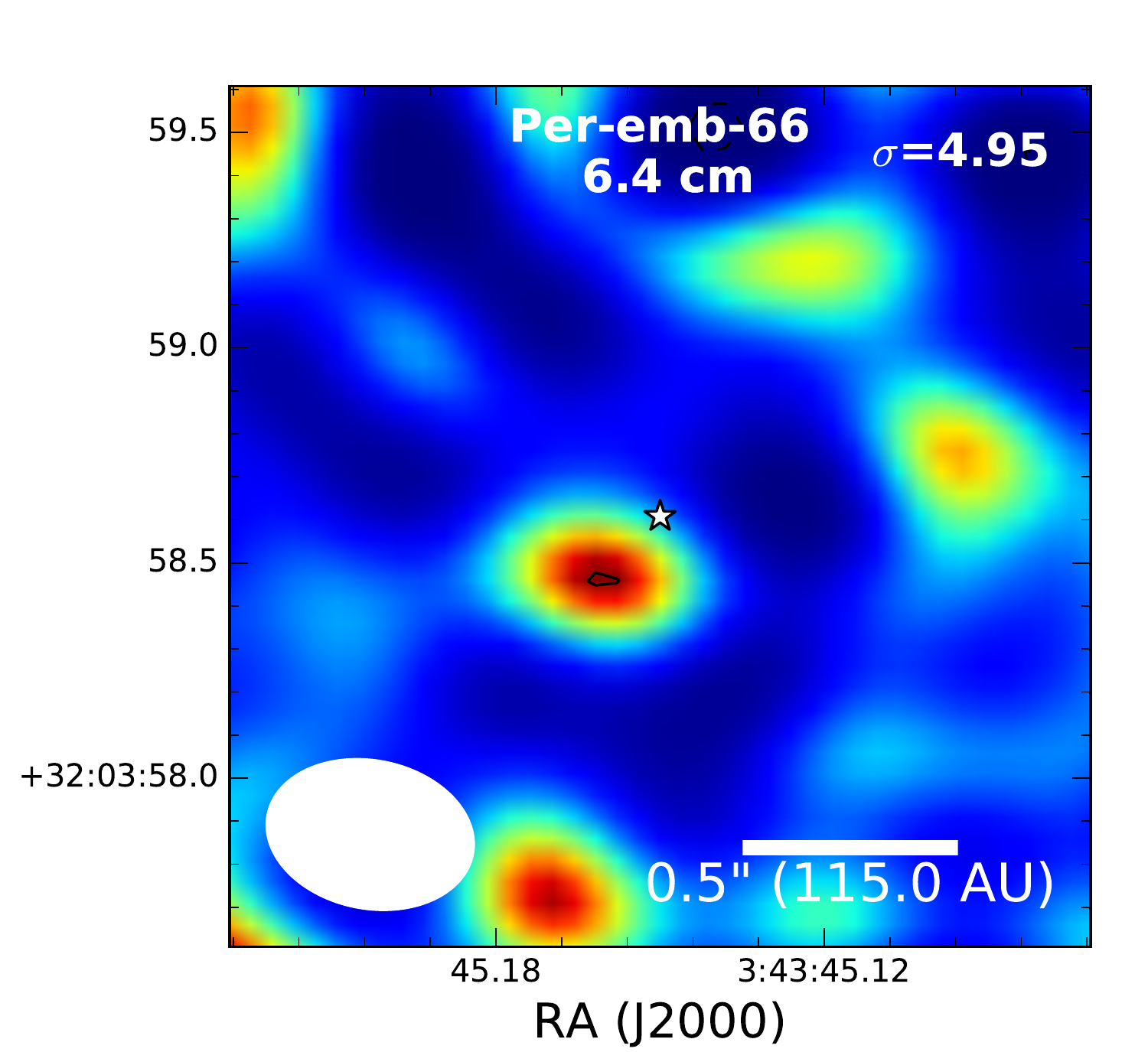}

  \includegraphics[width=0.24\linewidth]{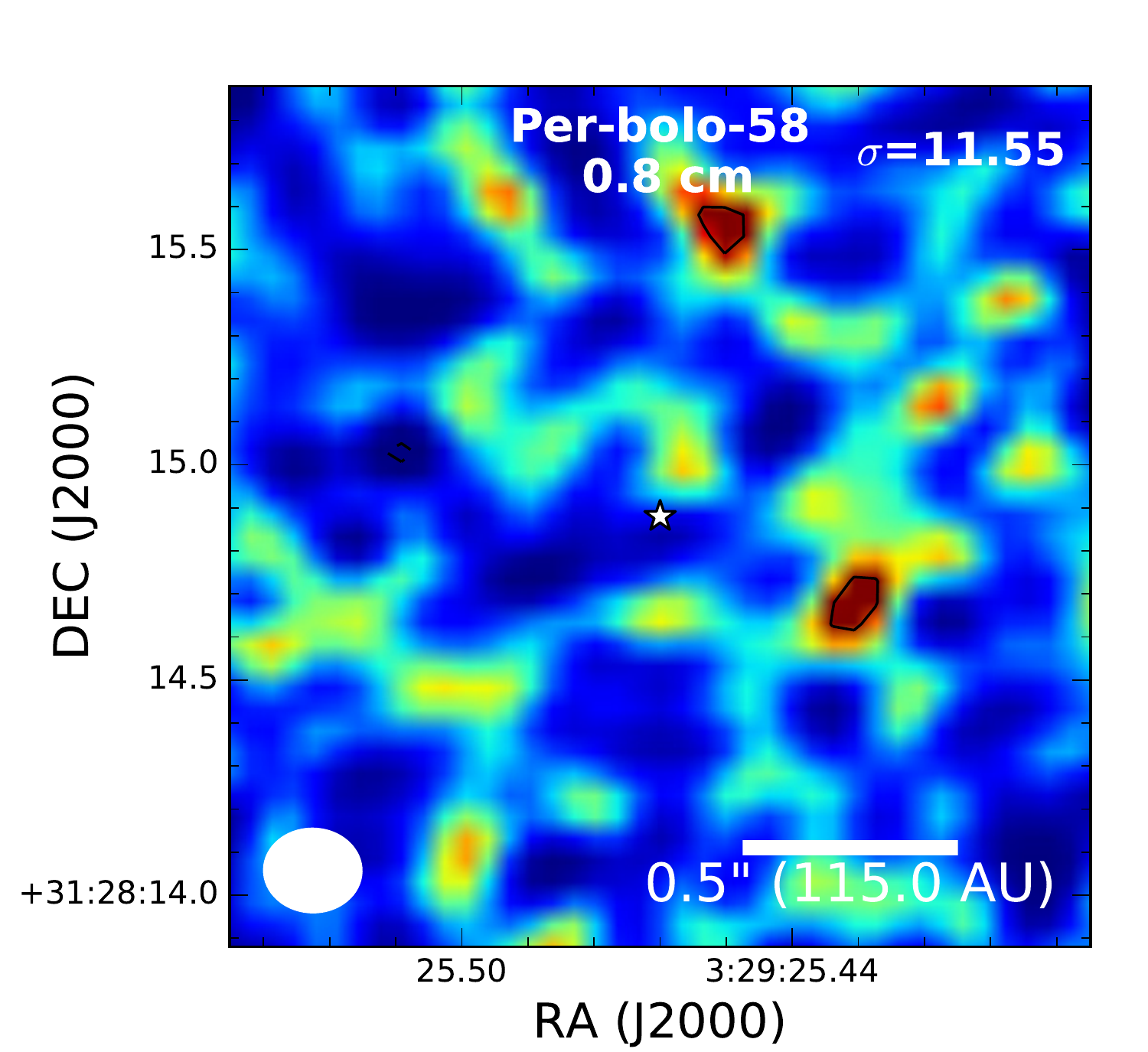}
  \includegraphics[width=0.24\linewidth]{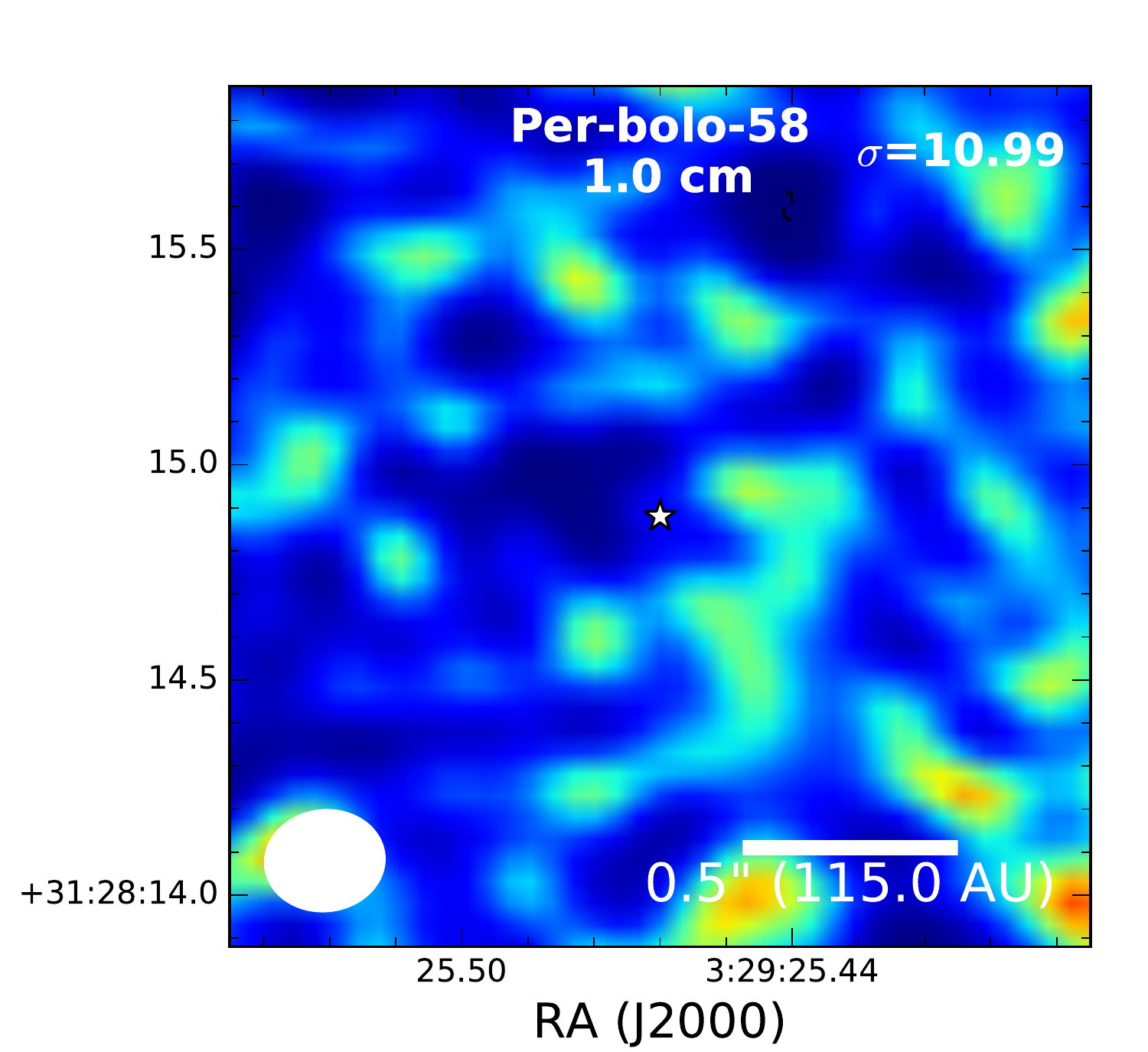}
  \includegraphics[width=0.24\linewidth]{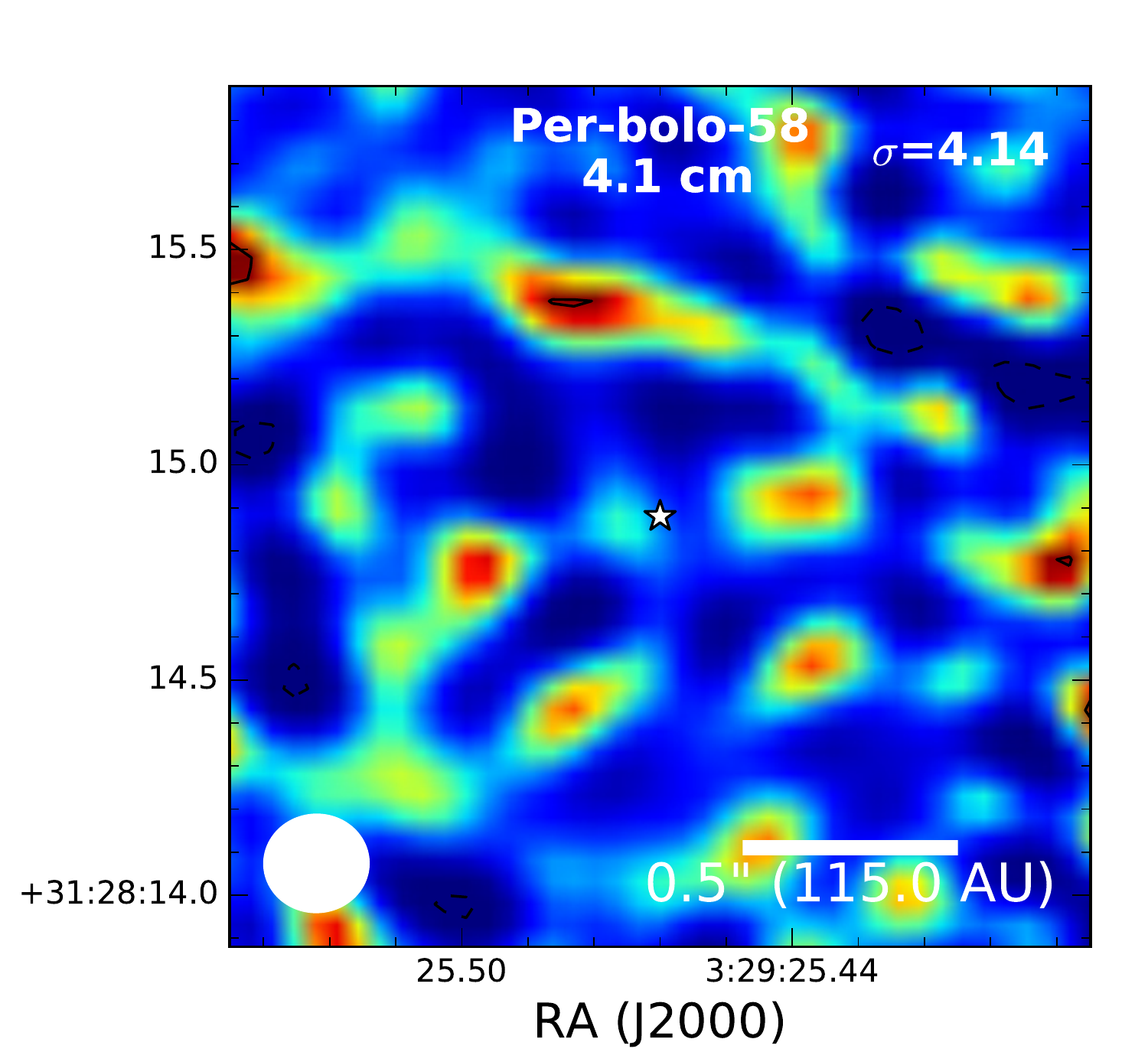}
  \includegraphics[width=0.24\linewidth]{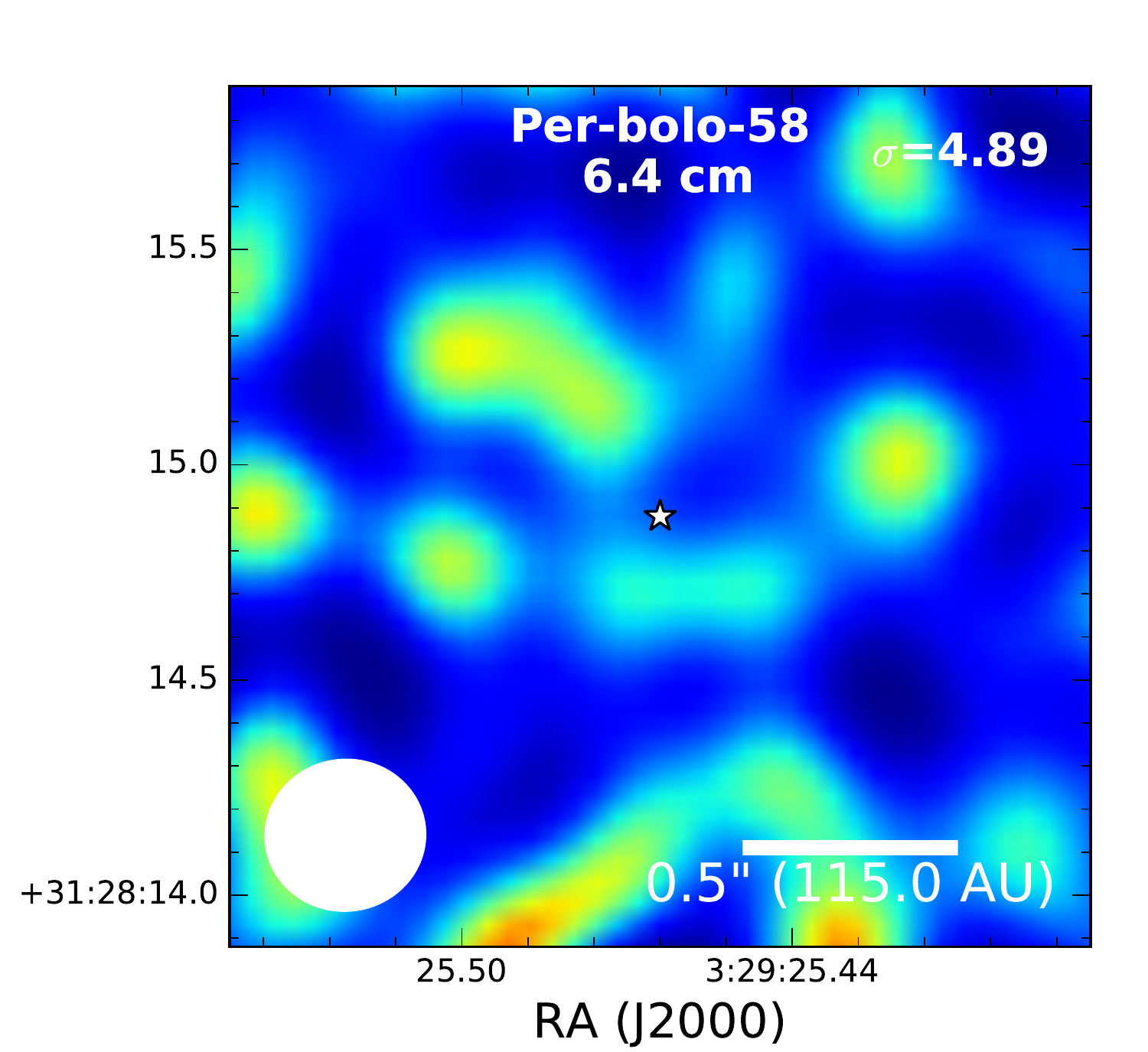}

  \includegraphics[width=0.24\linewidth]{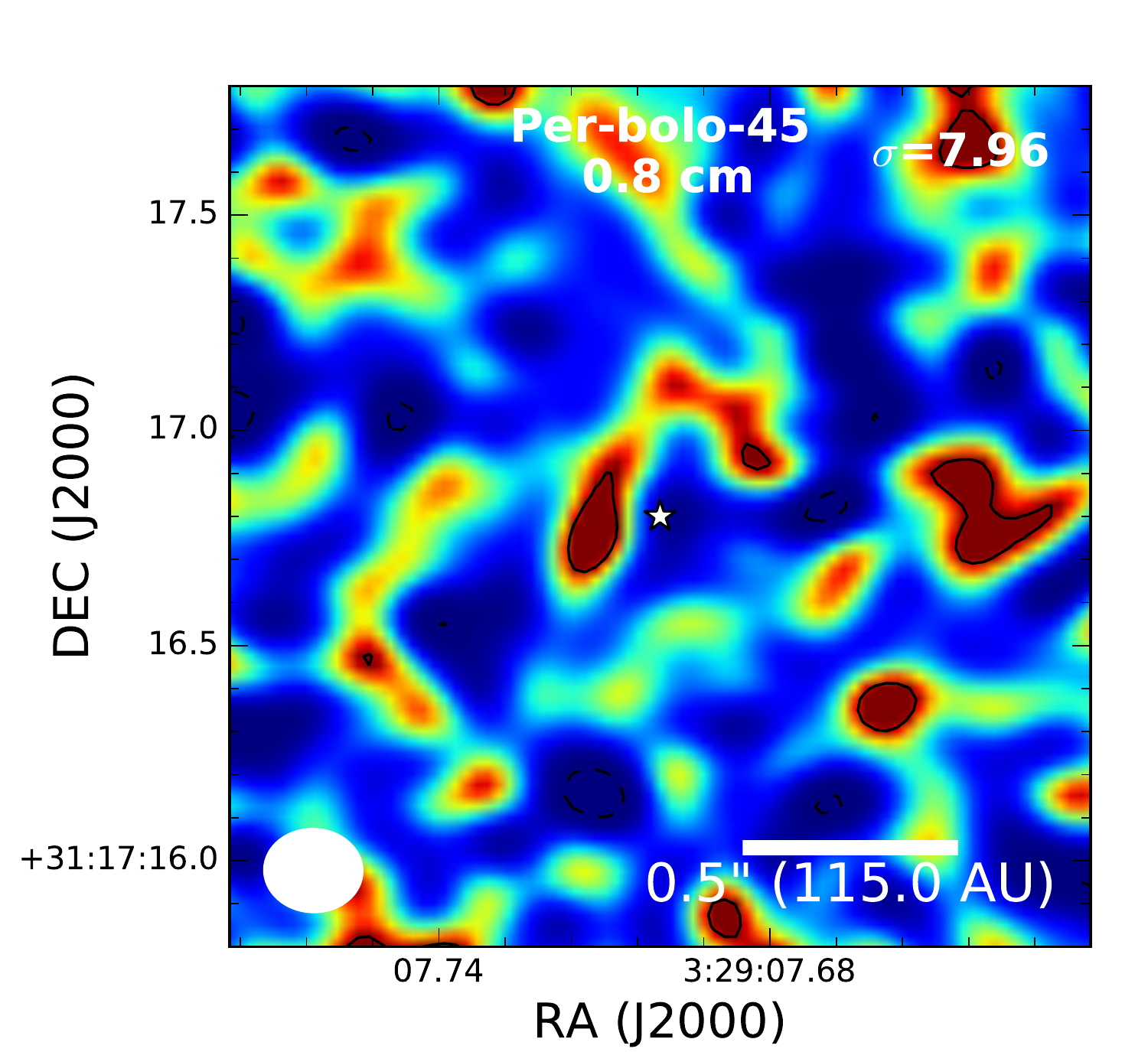}
  \includegraphics[width=0.24\linewidth]{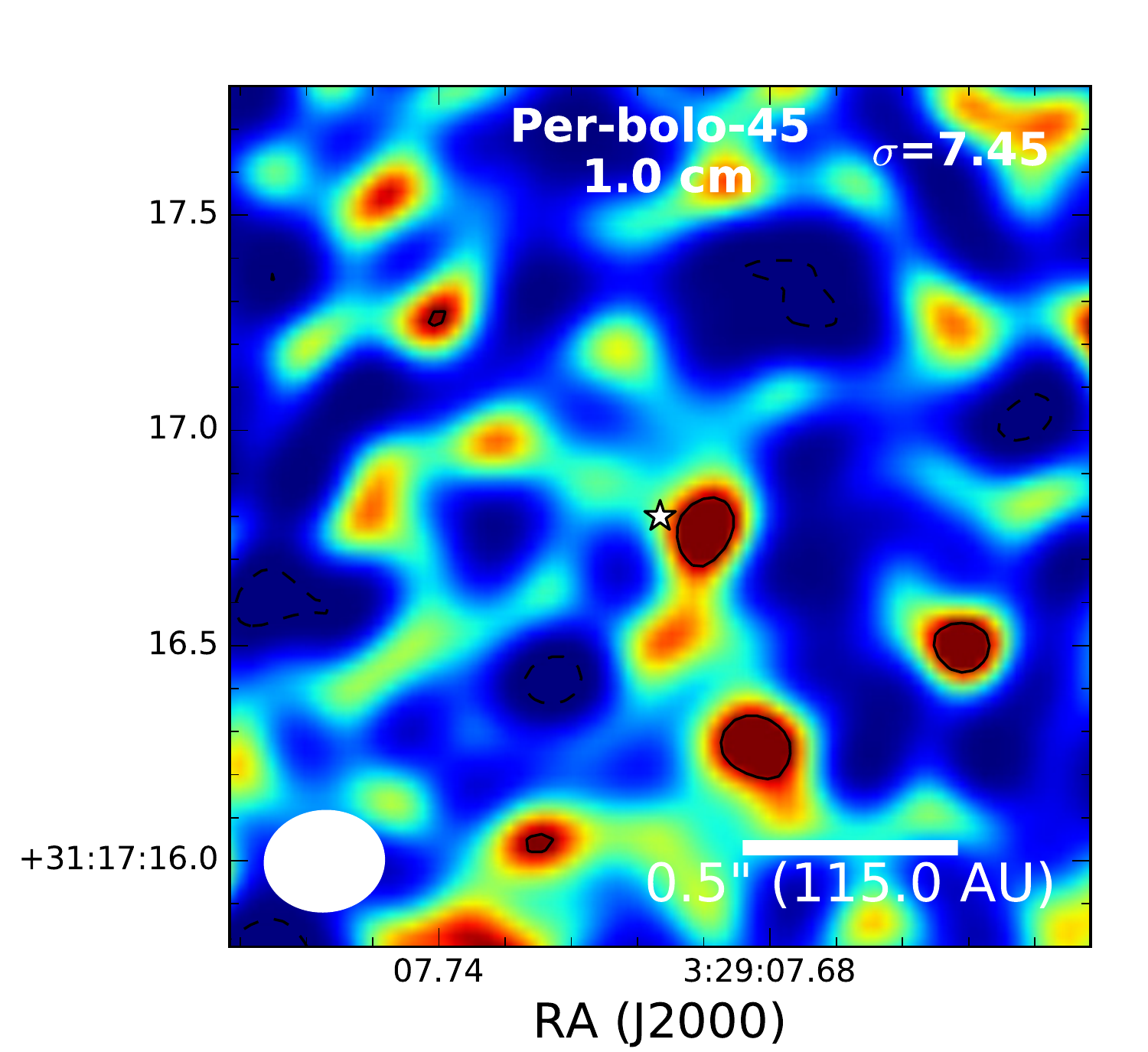}
  \includegraphics[width=0.24\linewidth]{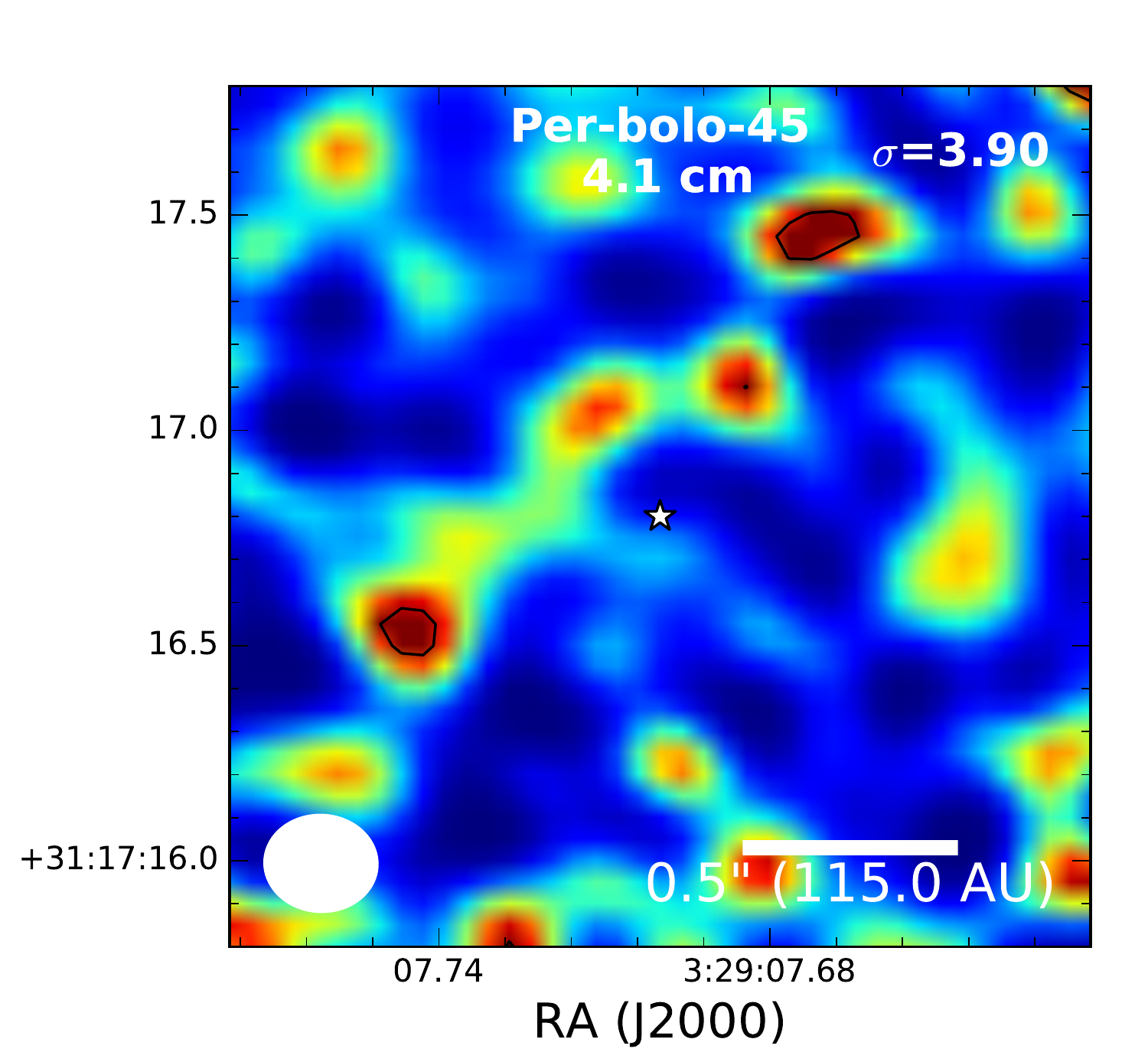}
  \includegraphics[width=0.24\linewidth]{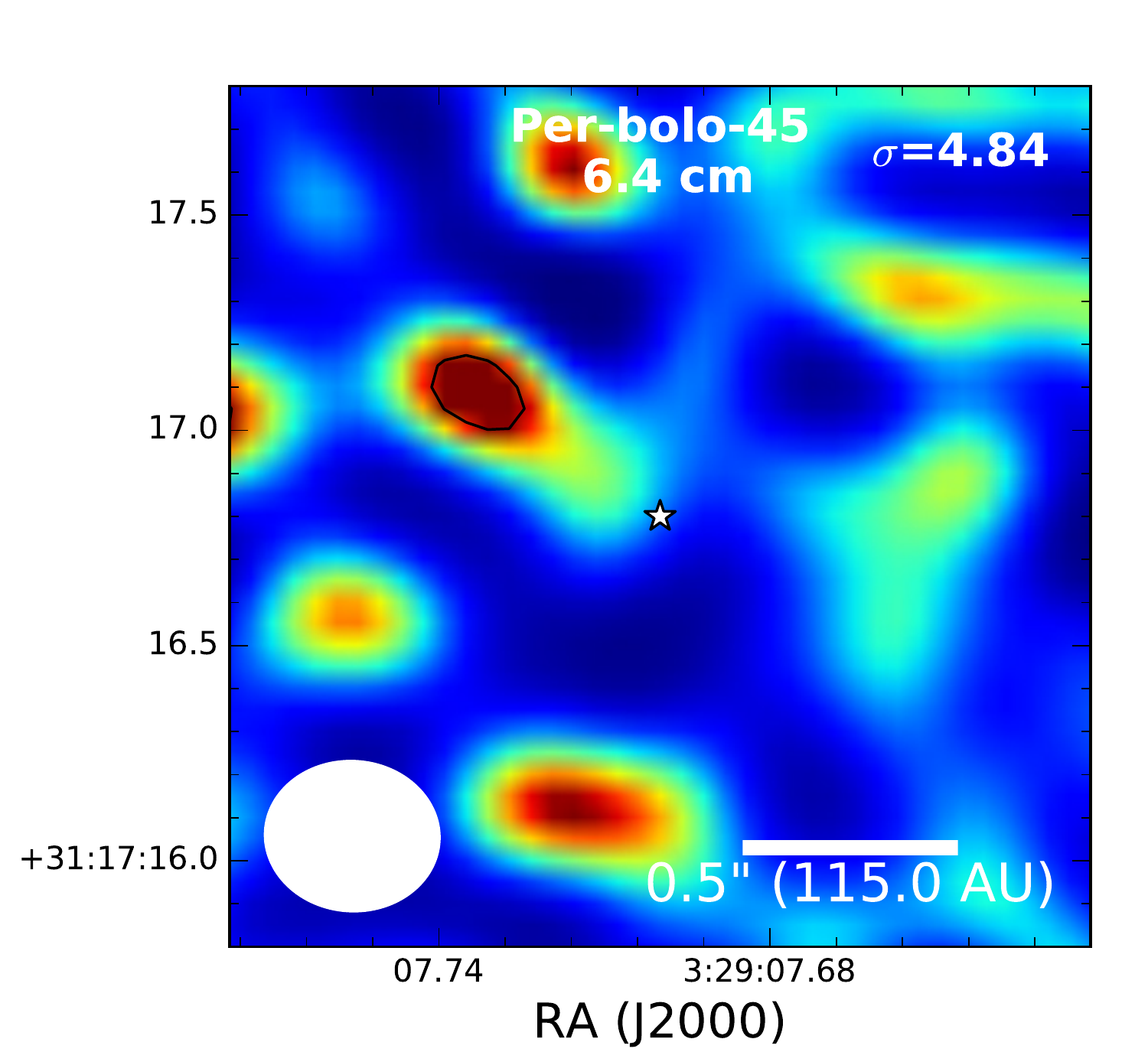}

  \includegraphics[width=0.24\linewidth]{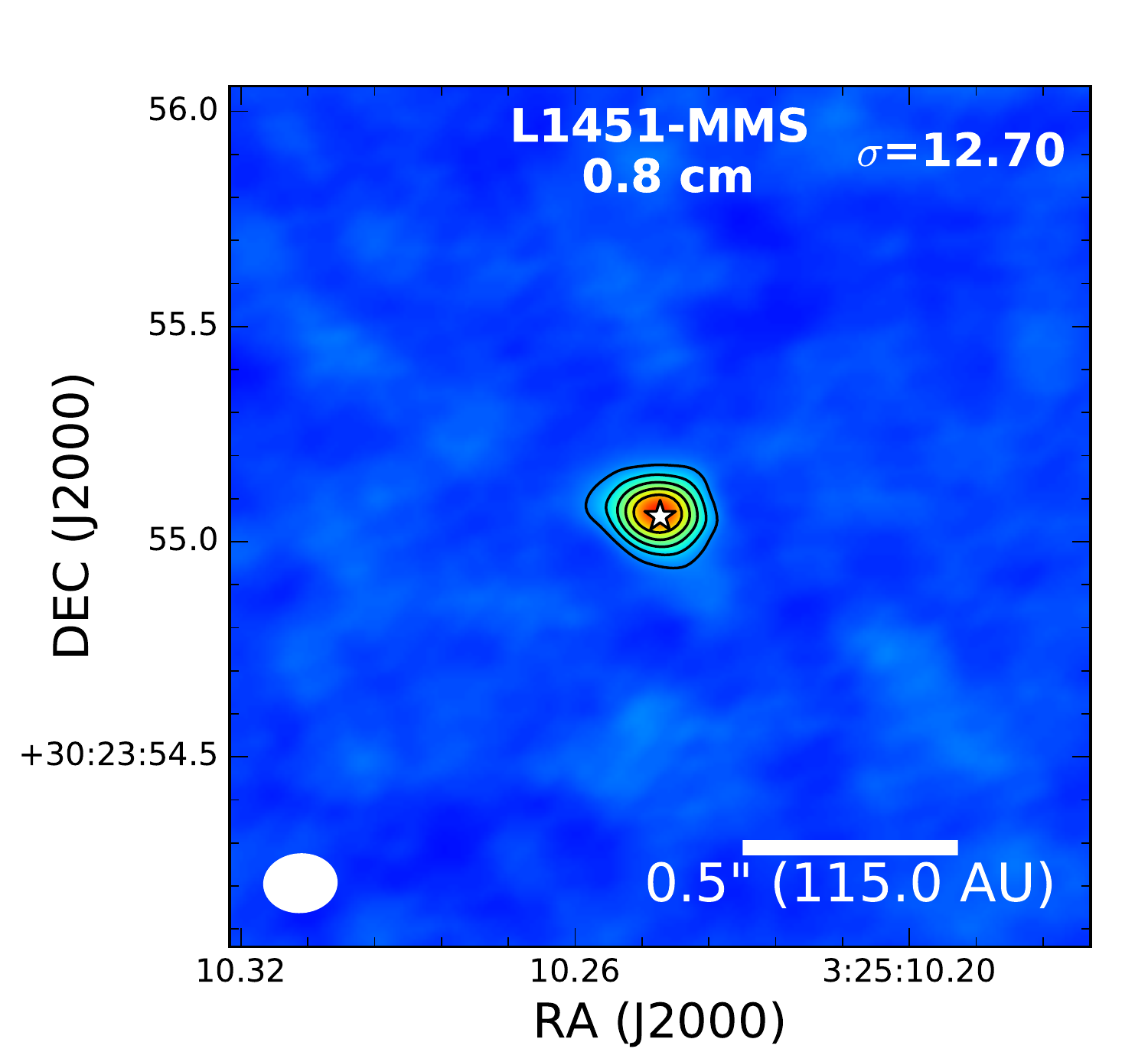}
  \includegraphics[width=0.24\linewidth]{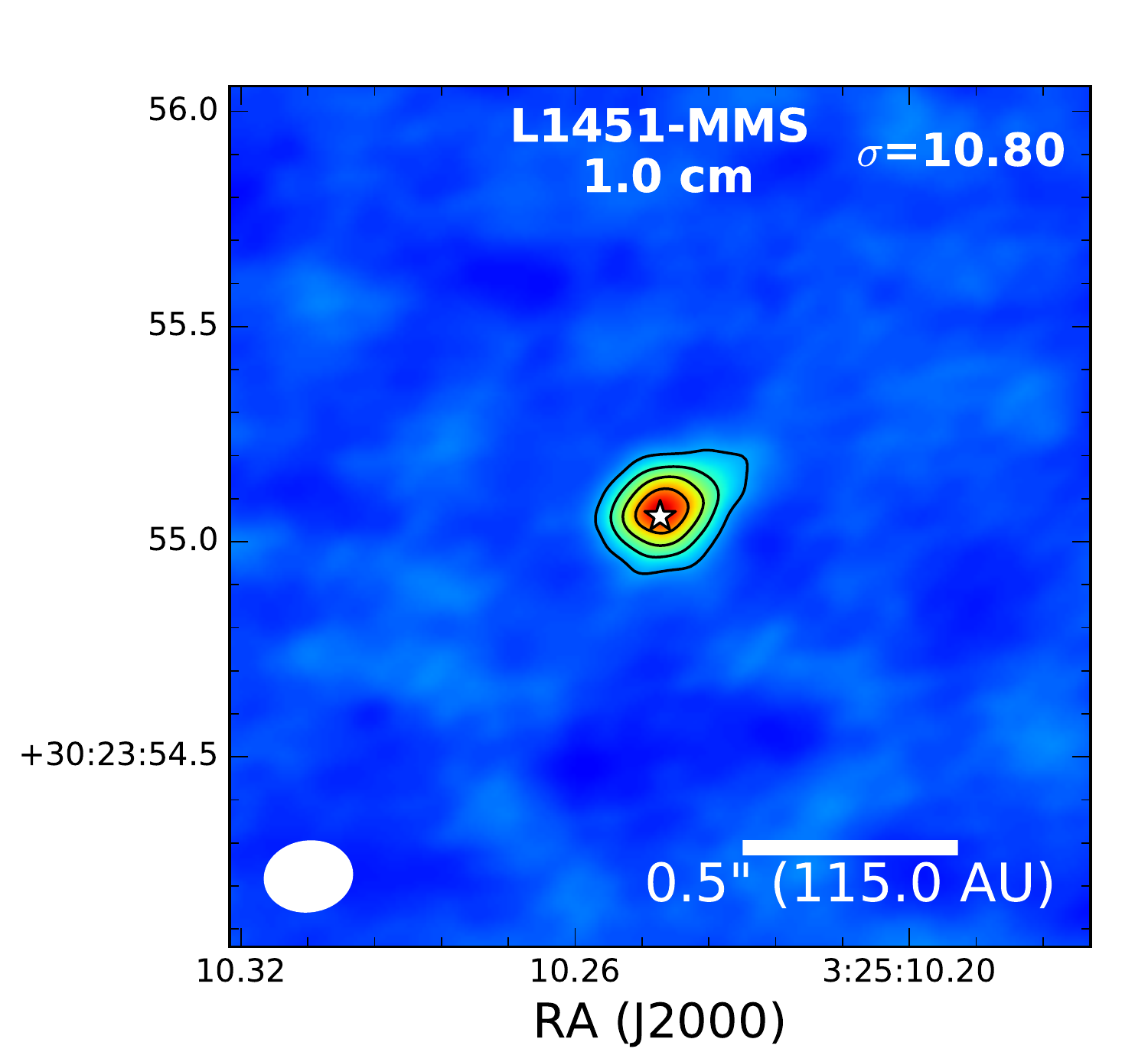}
  \includegraphics[width=0.24\linewidth]{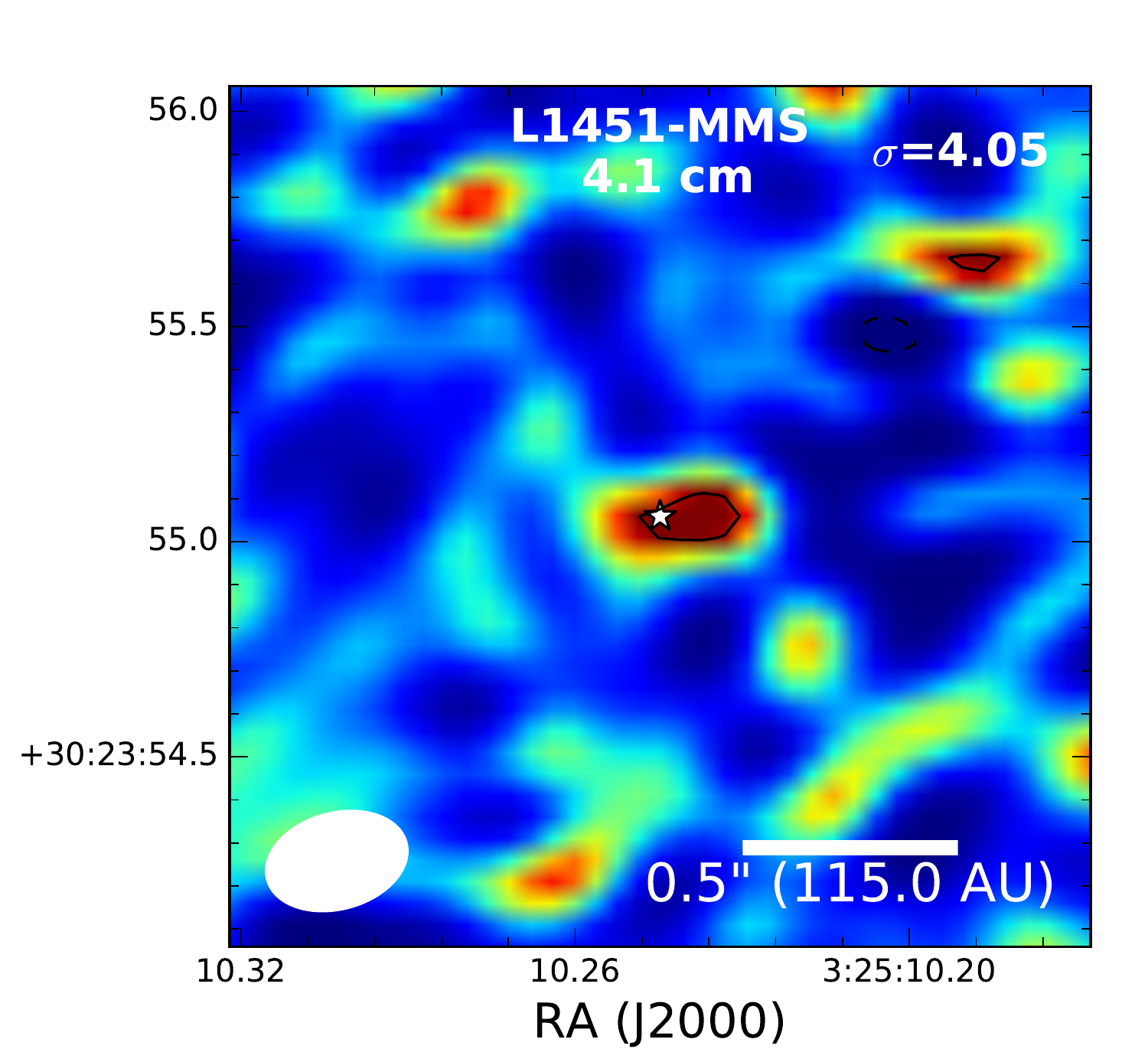}
  \includegraphics[width=0.24\linewidth]{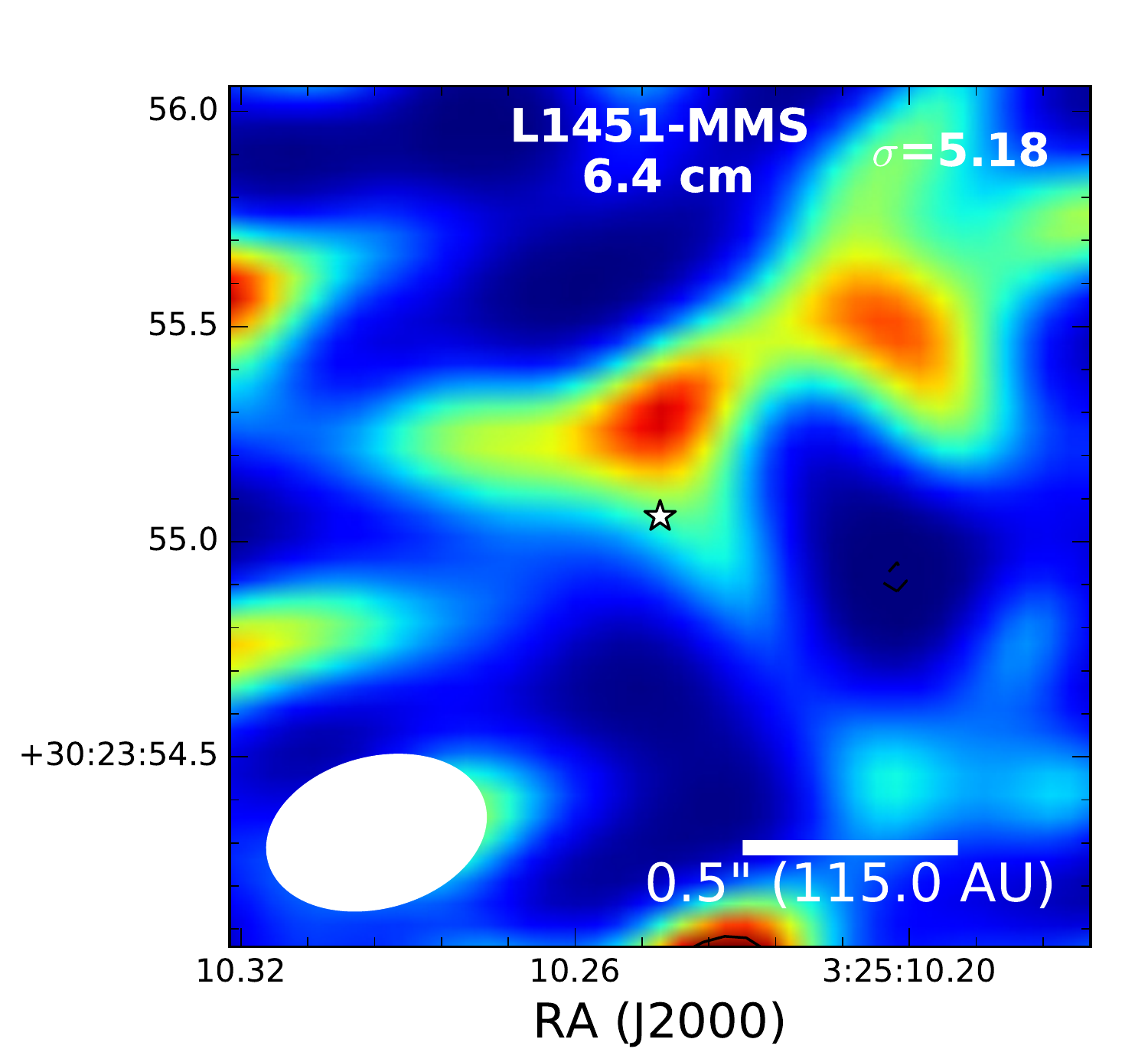}

\end{figure}
\begin{figure}

  \includegraphics[width=0.24\linewidth]{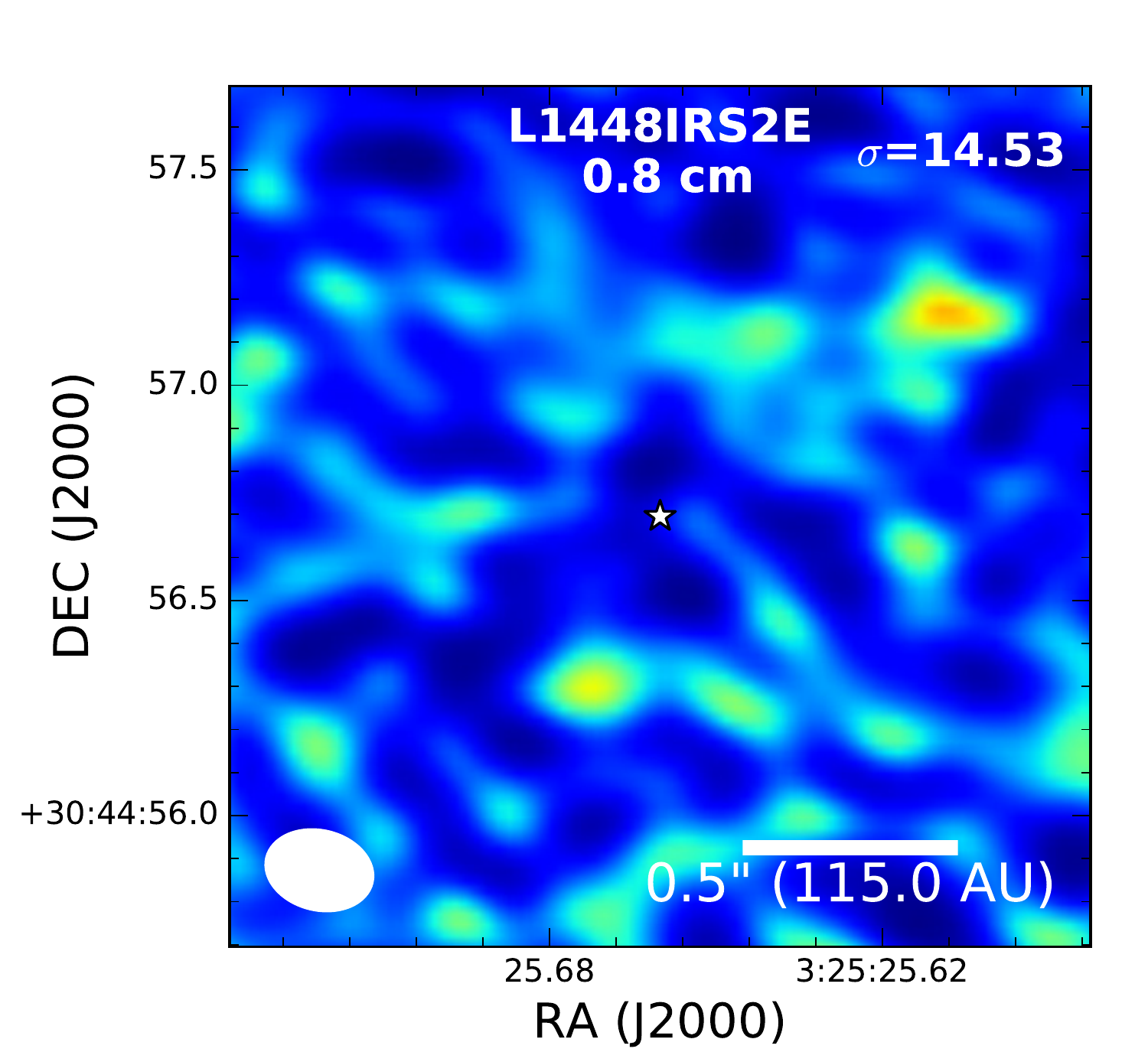}
  \includegraphics[width=0.24\linewidth]{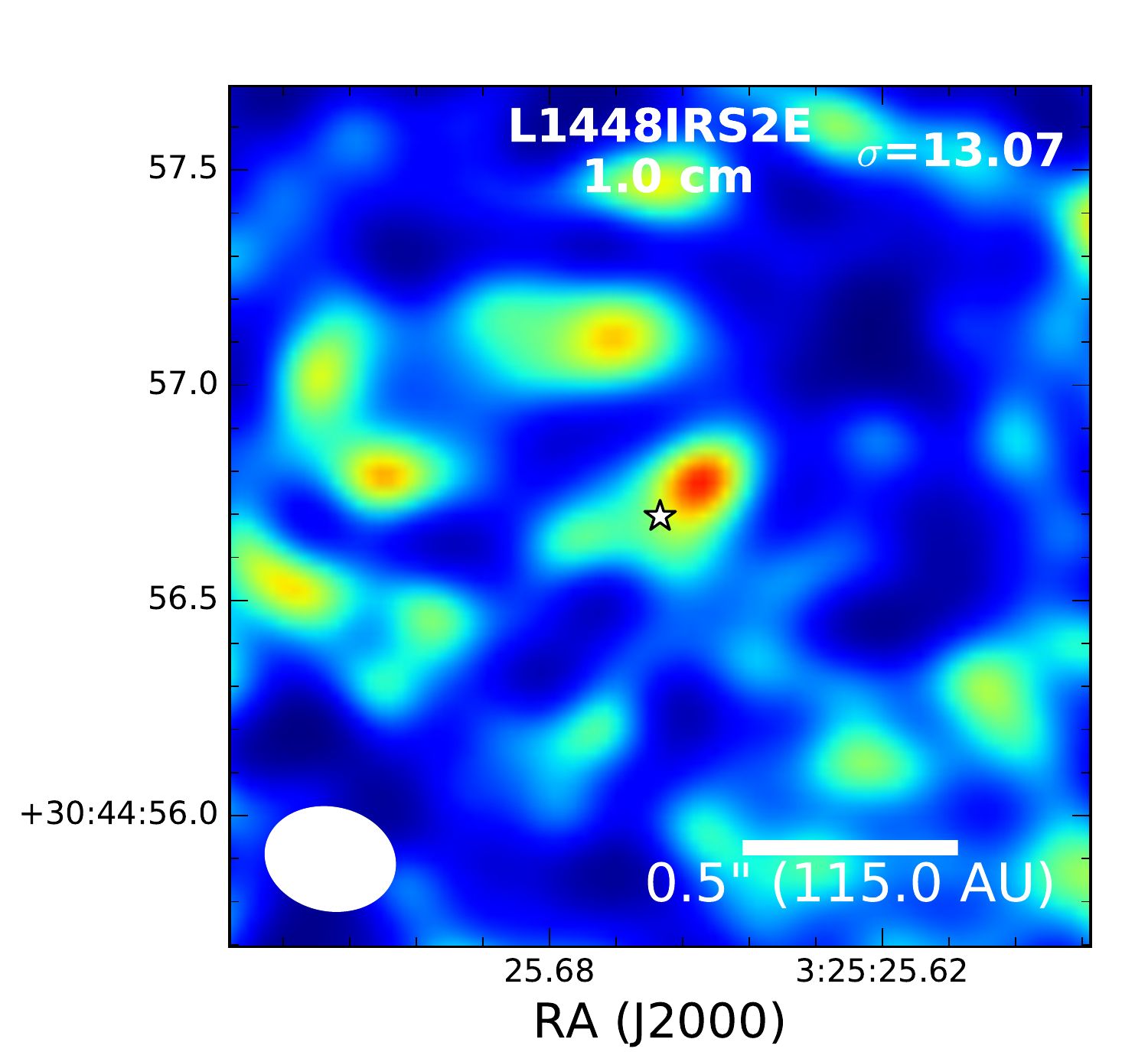}
  \includegraphics[width=0.24\linewidth]{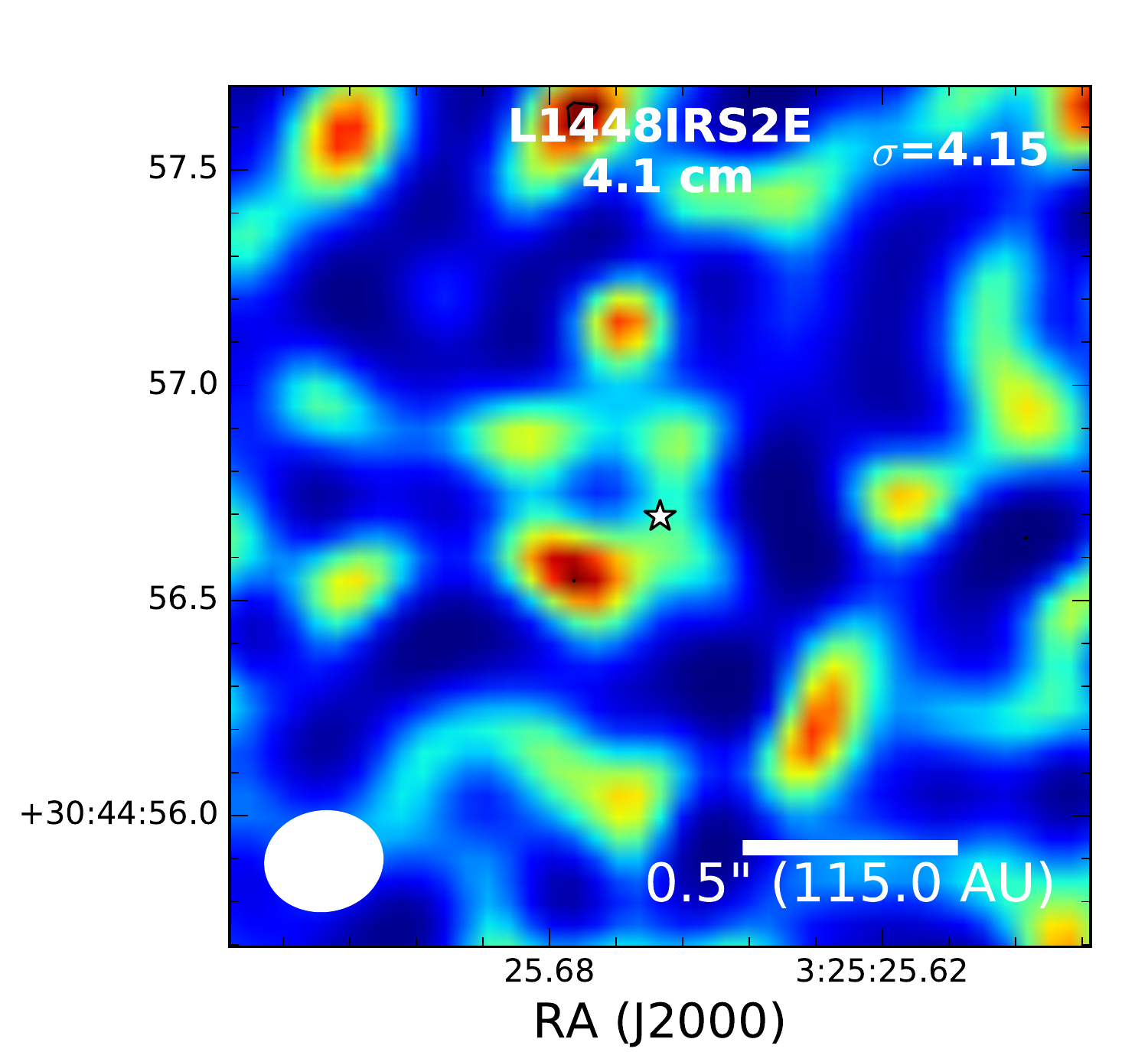}
  \includegraphics[width=0.24\linewidth]{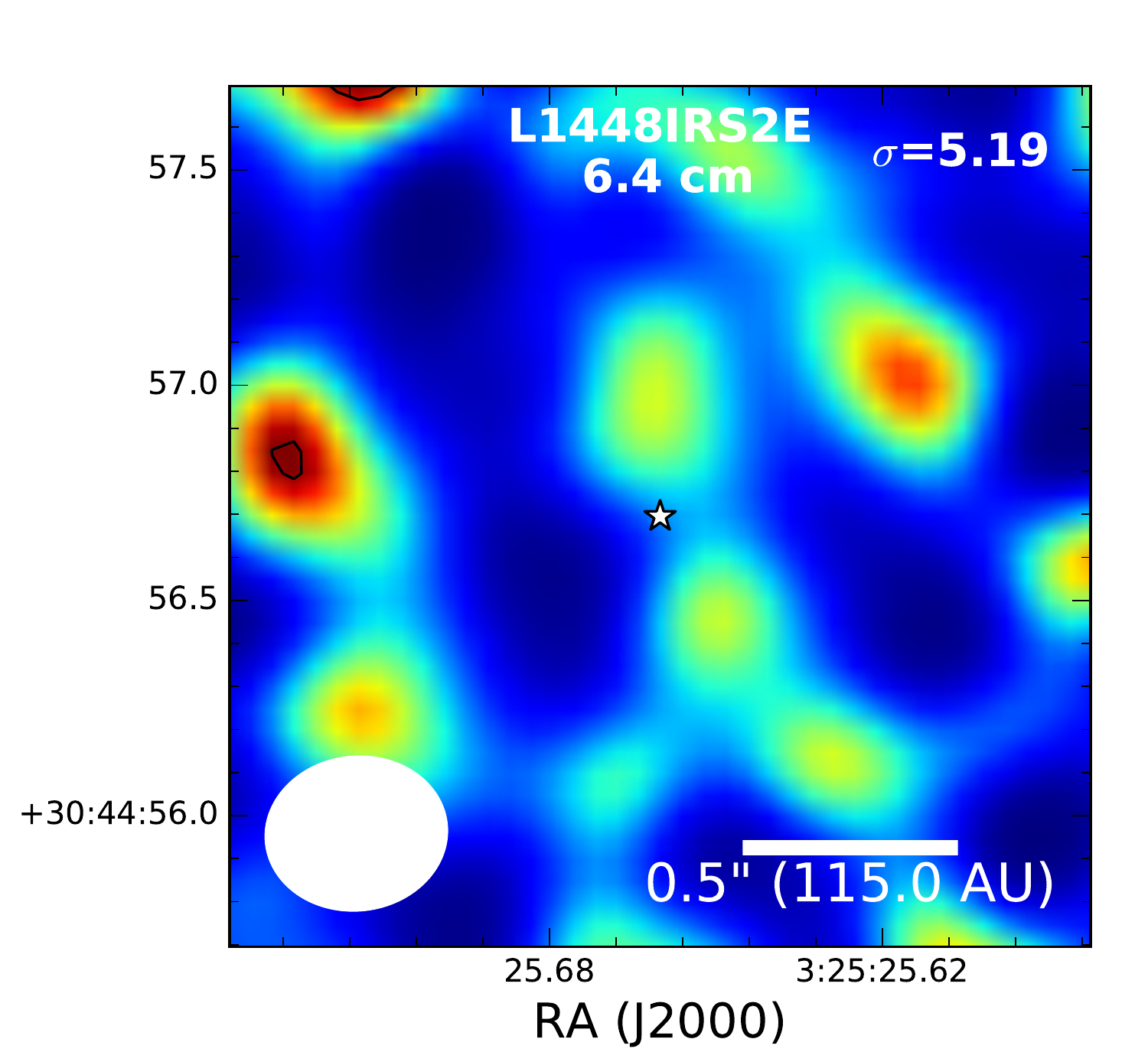}

  \includegraphics[width=0.24\linewidth]{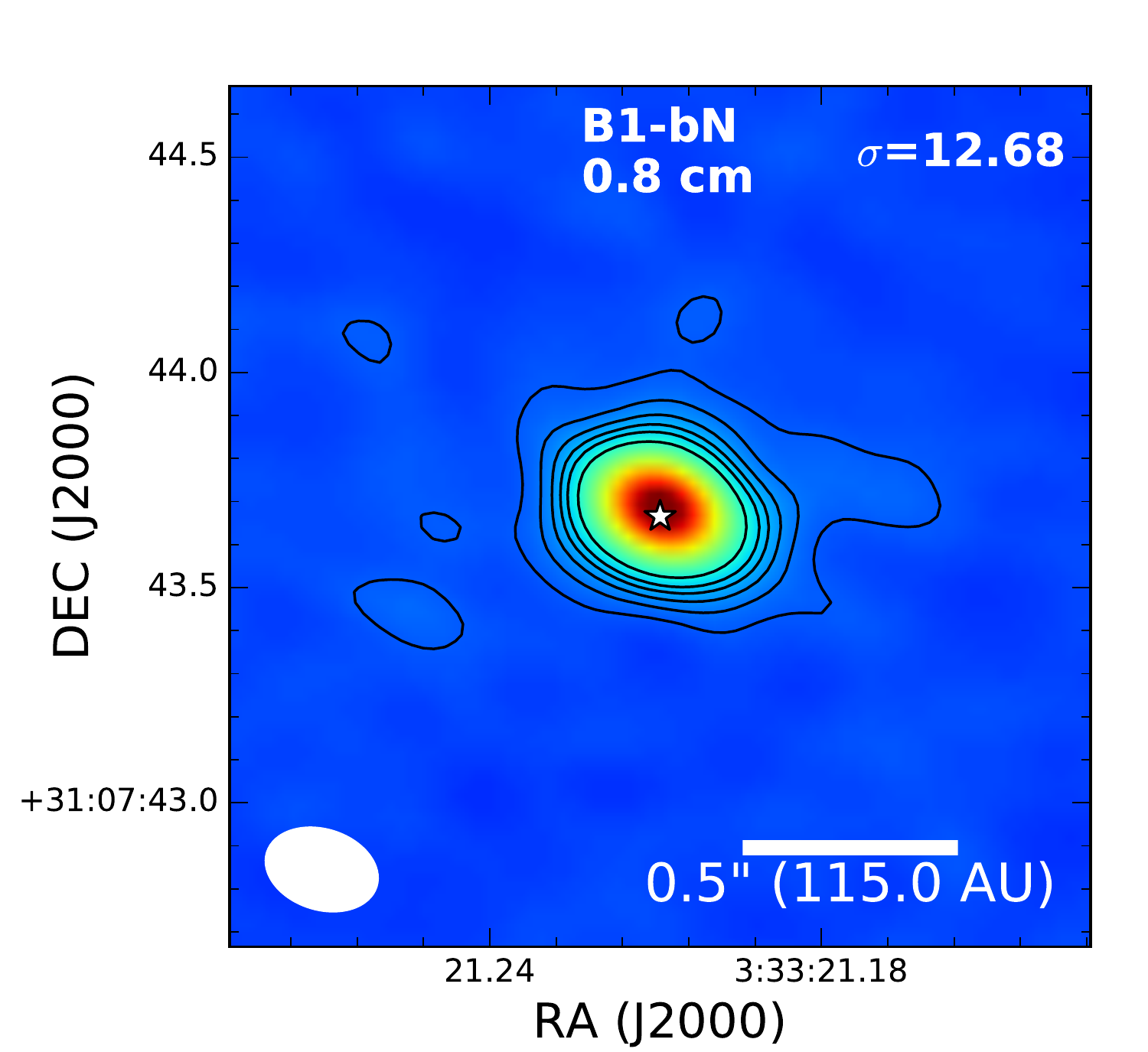}
  \includegraphics[width=0.24\linewidth]{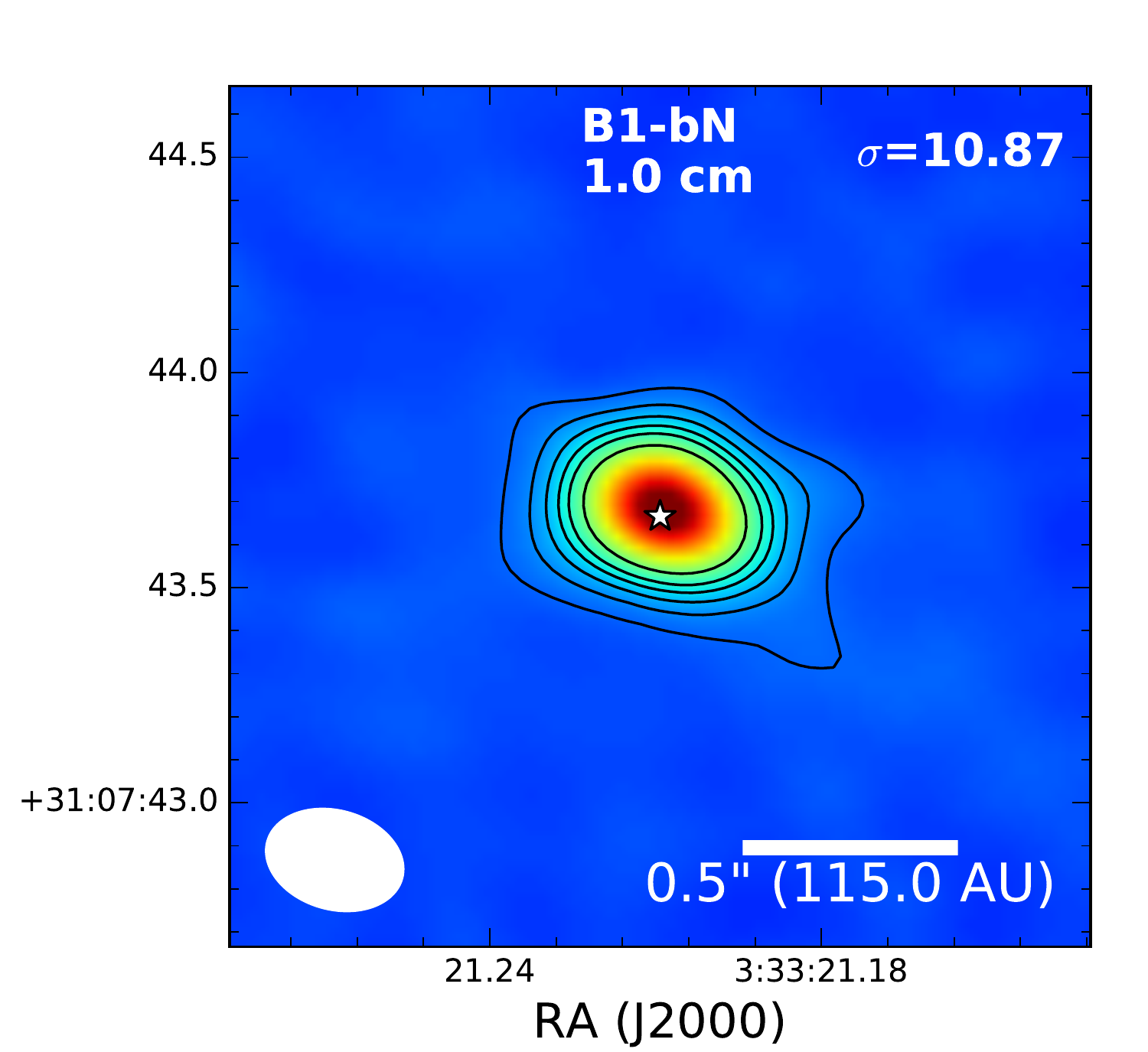}
  \includegraphics[width=0.24\linewidth]{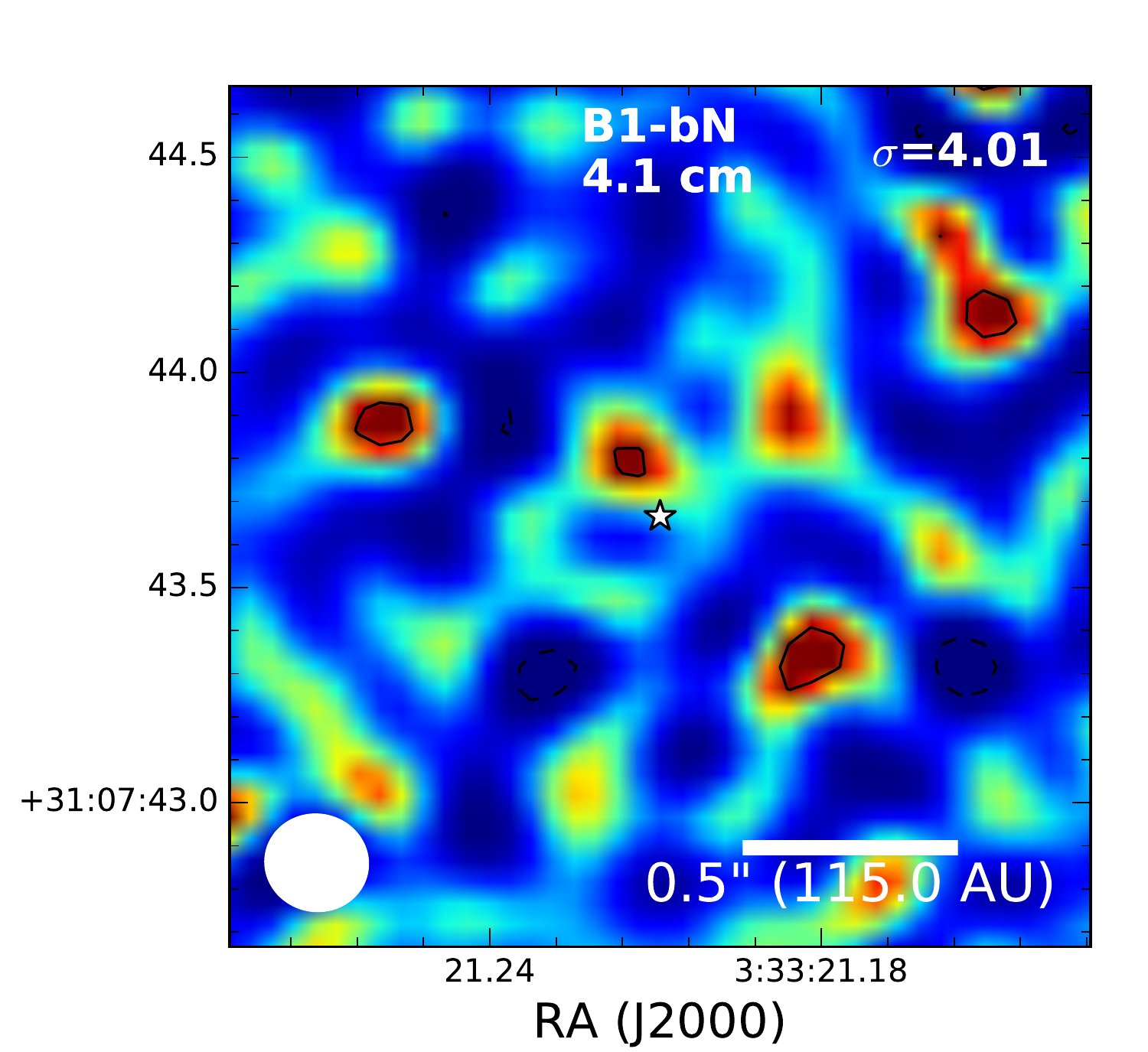}
  \includegraphics[width=0.24\linewidth]{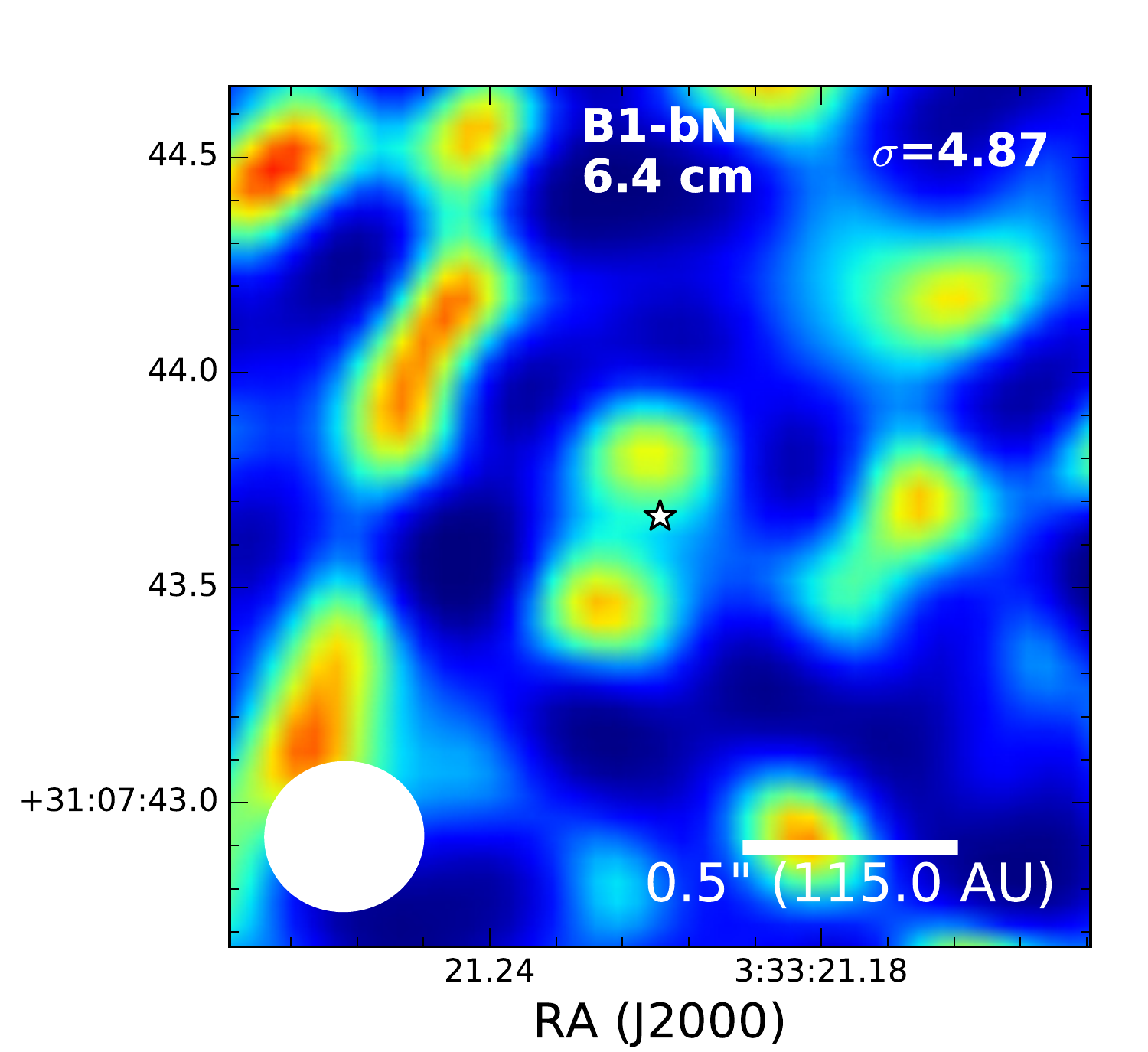}

  \includegraphics[width=0.24\linewidth]{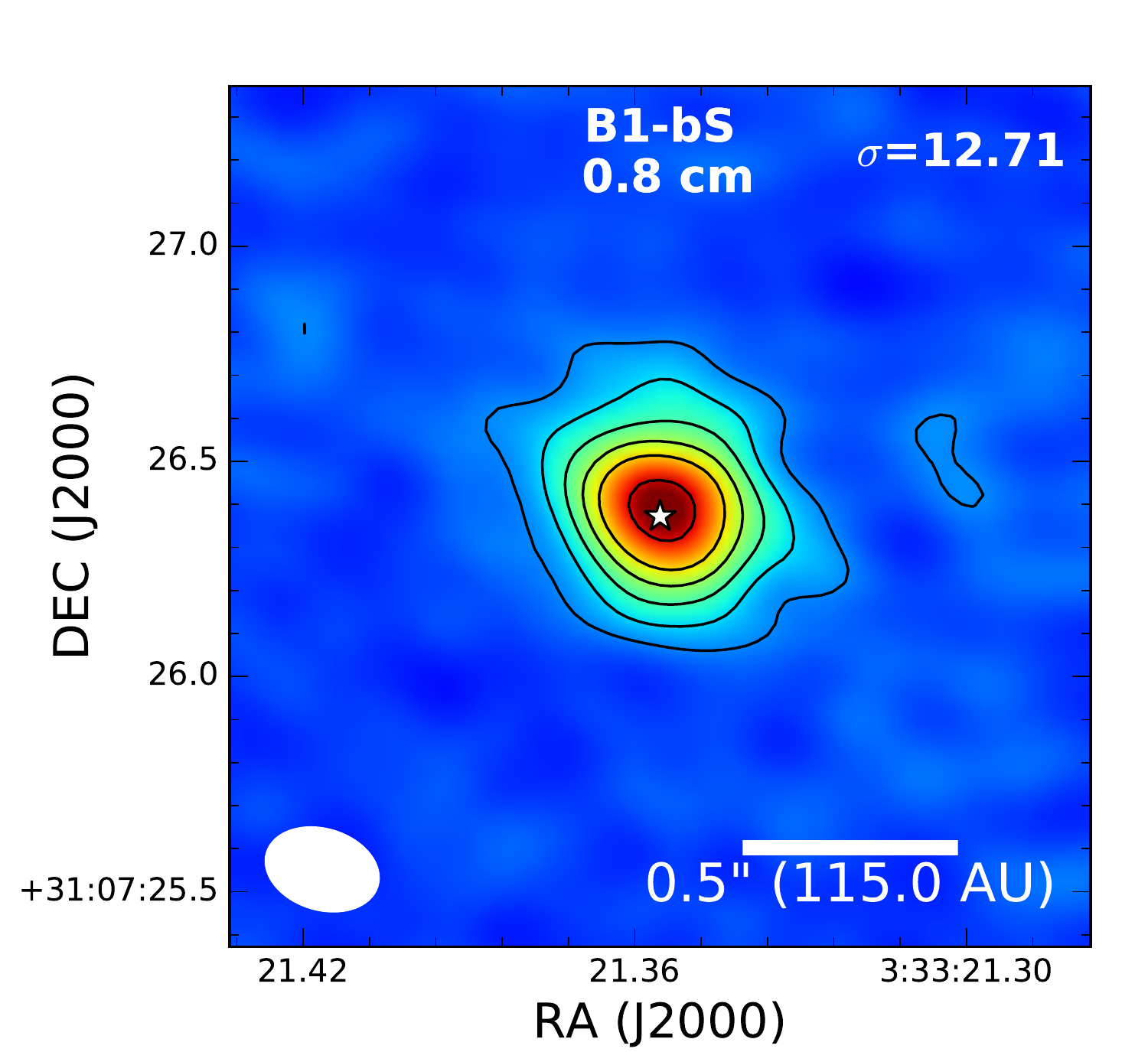}
  \includegraphics[width=0.24\linewidth]{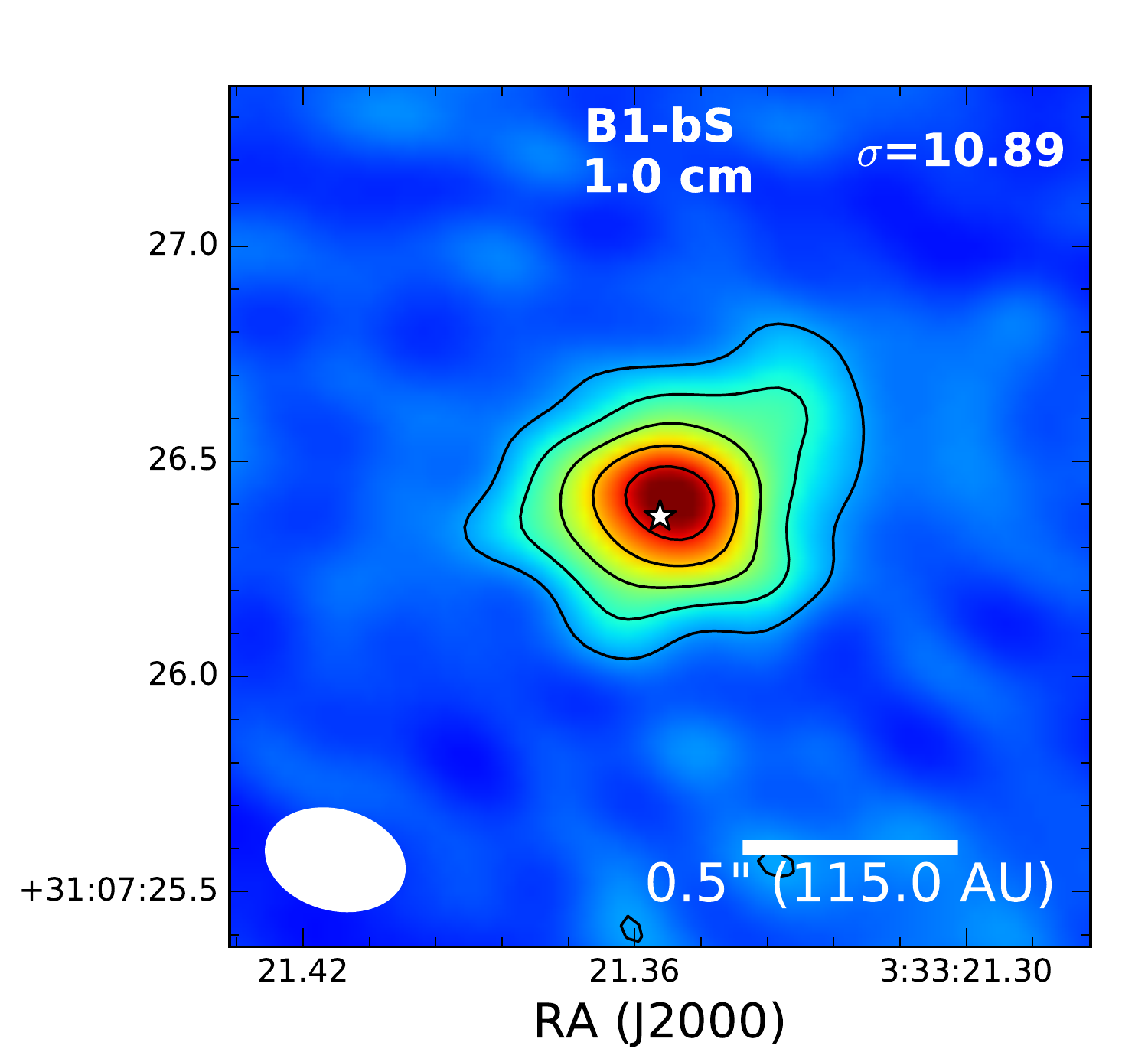}
  \includegraphics[width=0.24\linewidth]{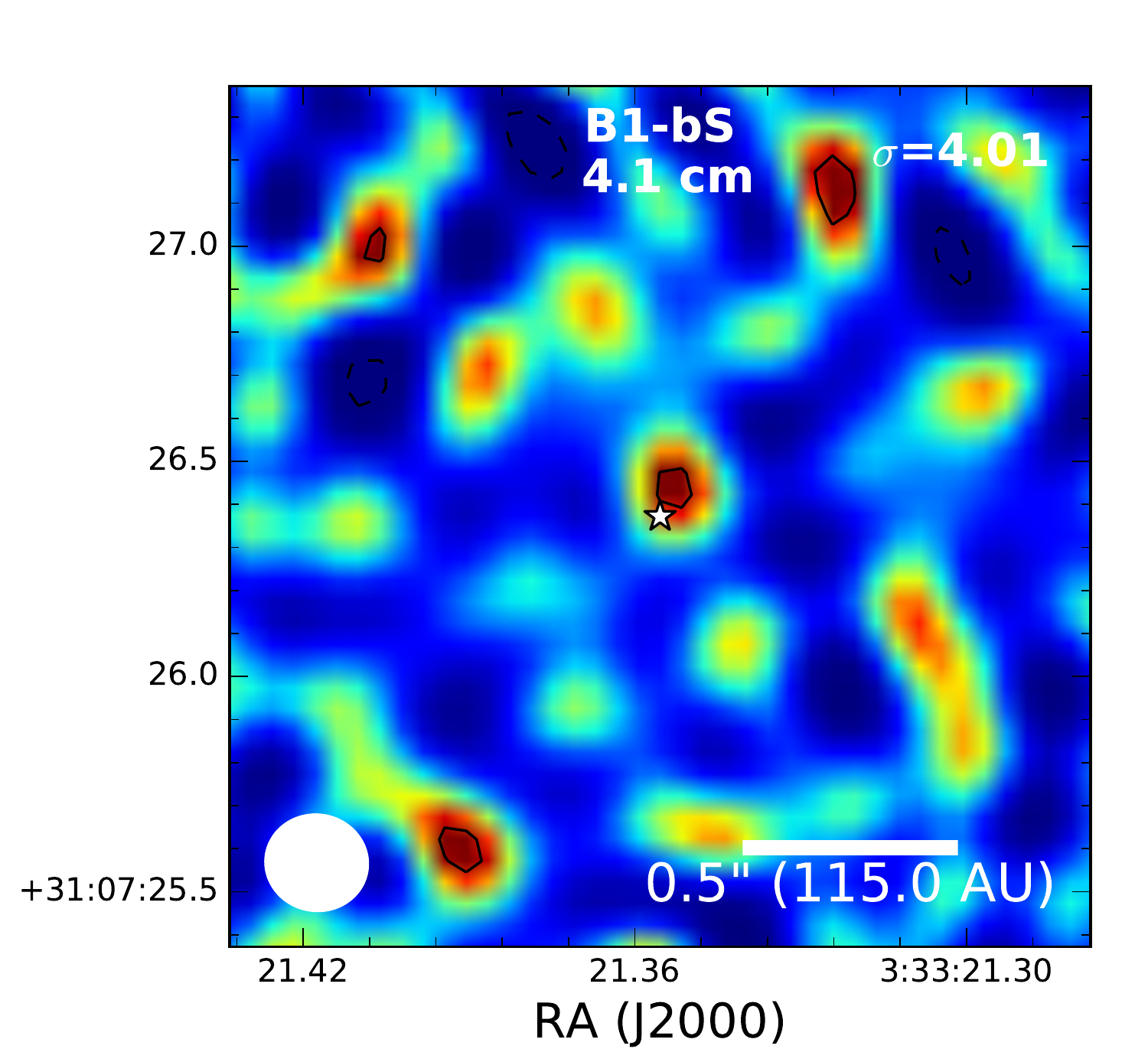}
  \includegraphics[width=0.24\linewidth]{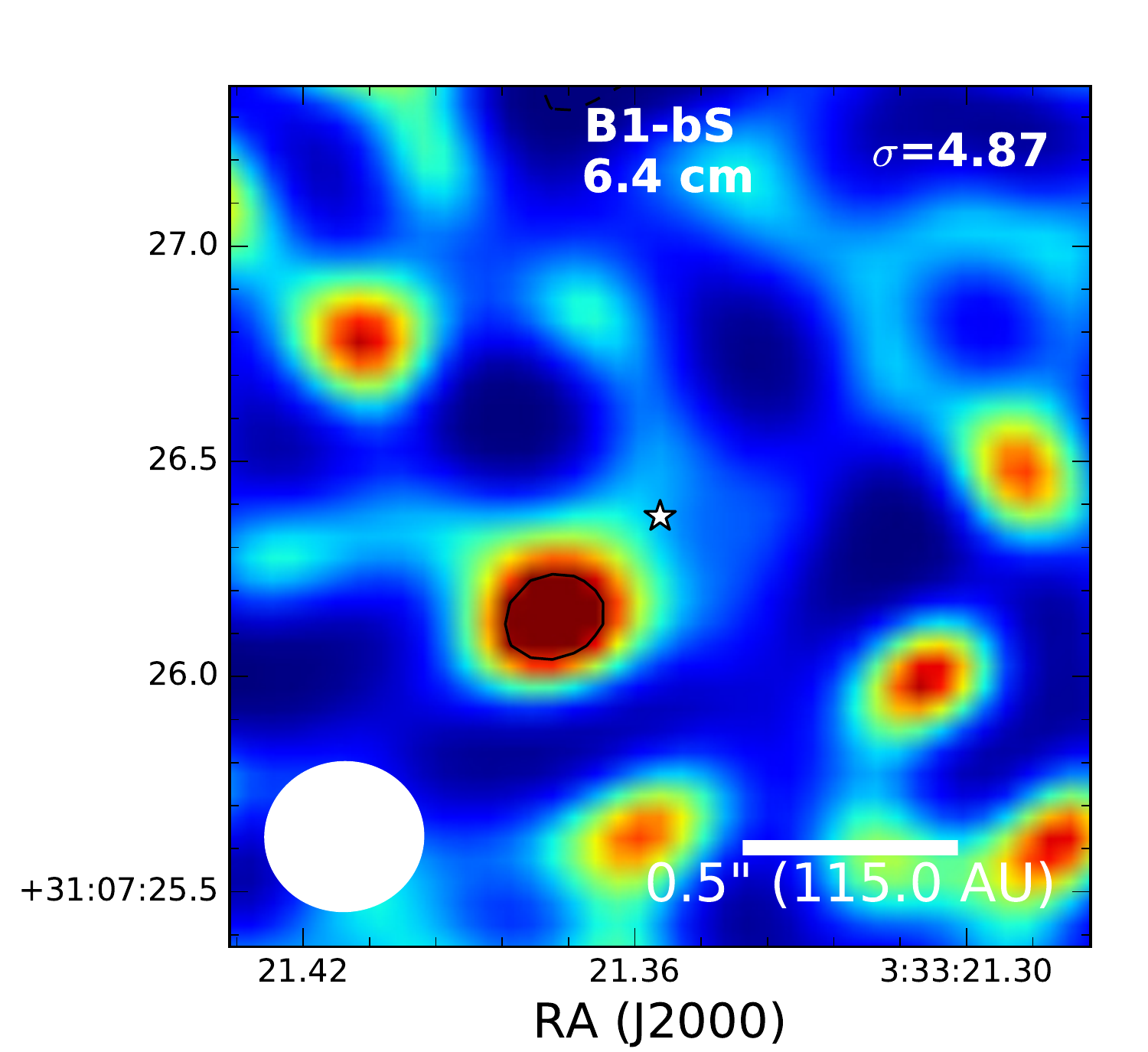}

  \includegraphics[width=0.24\linewidth]{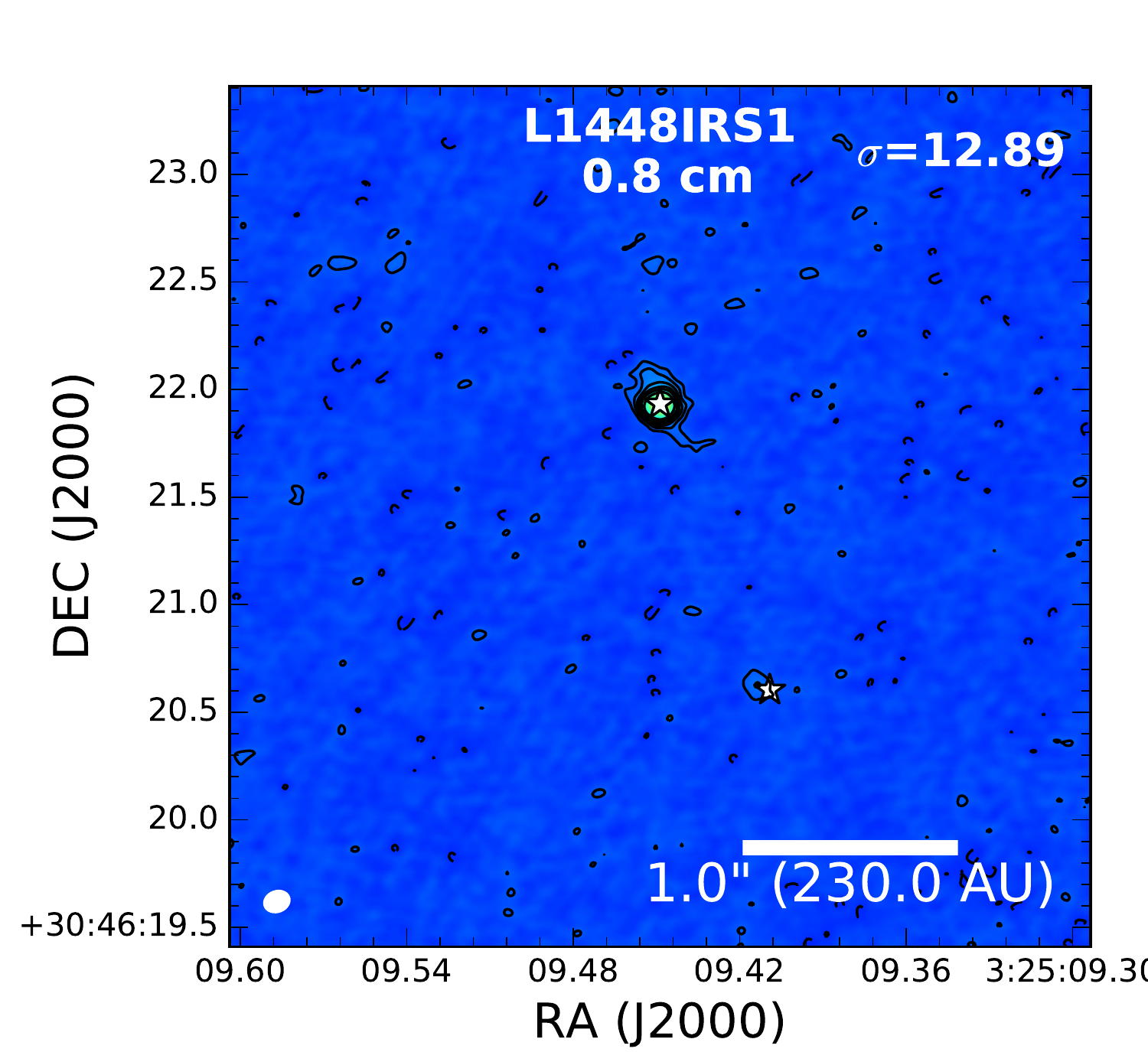}
  \includegraphics[width=0.24\linewidth]{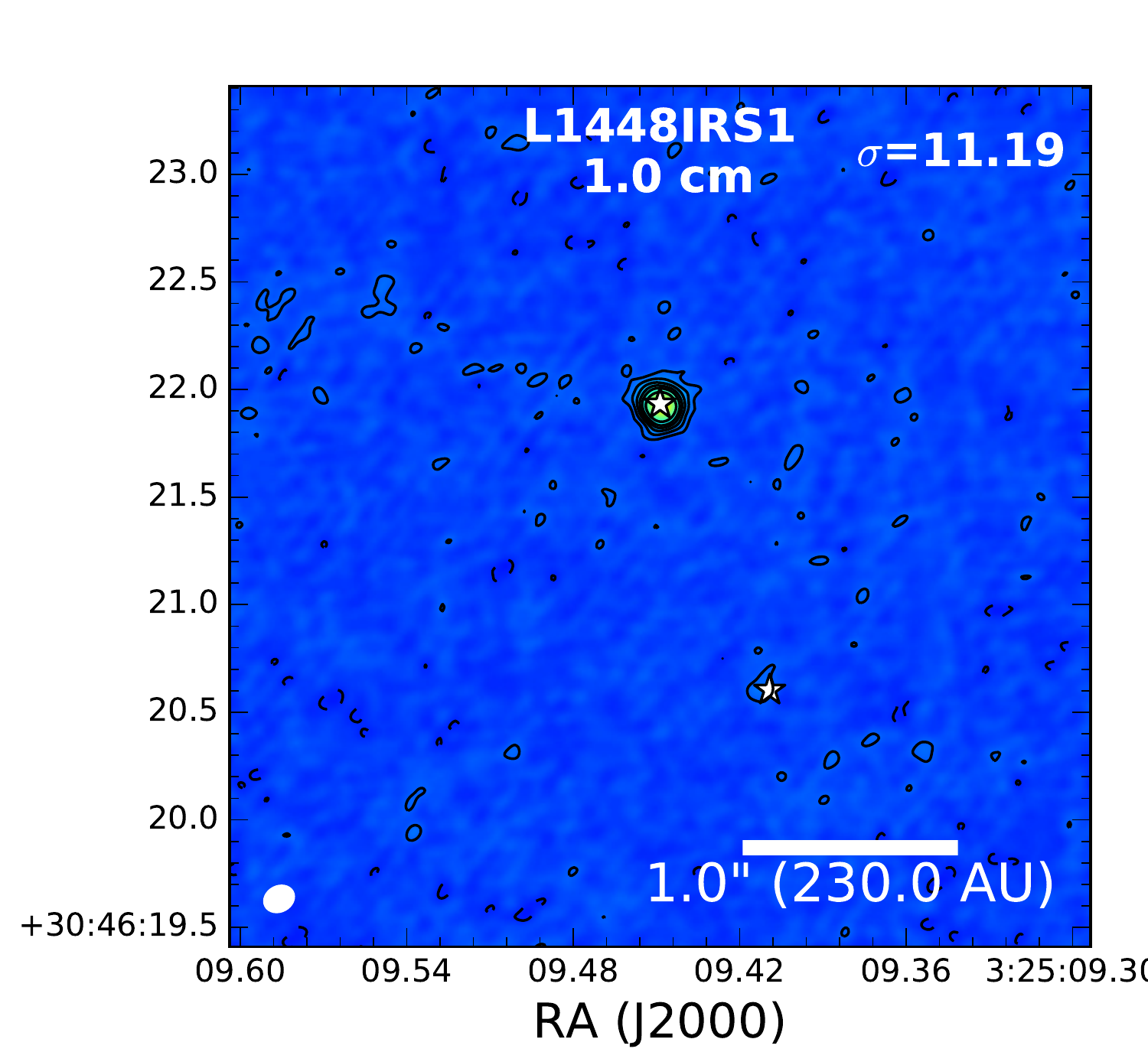}
  \includegraphics[width=0.24\linewidth]{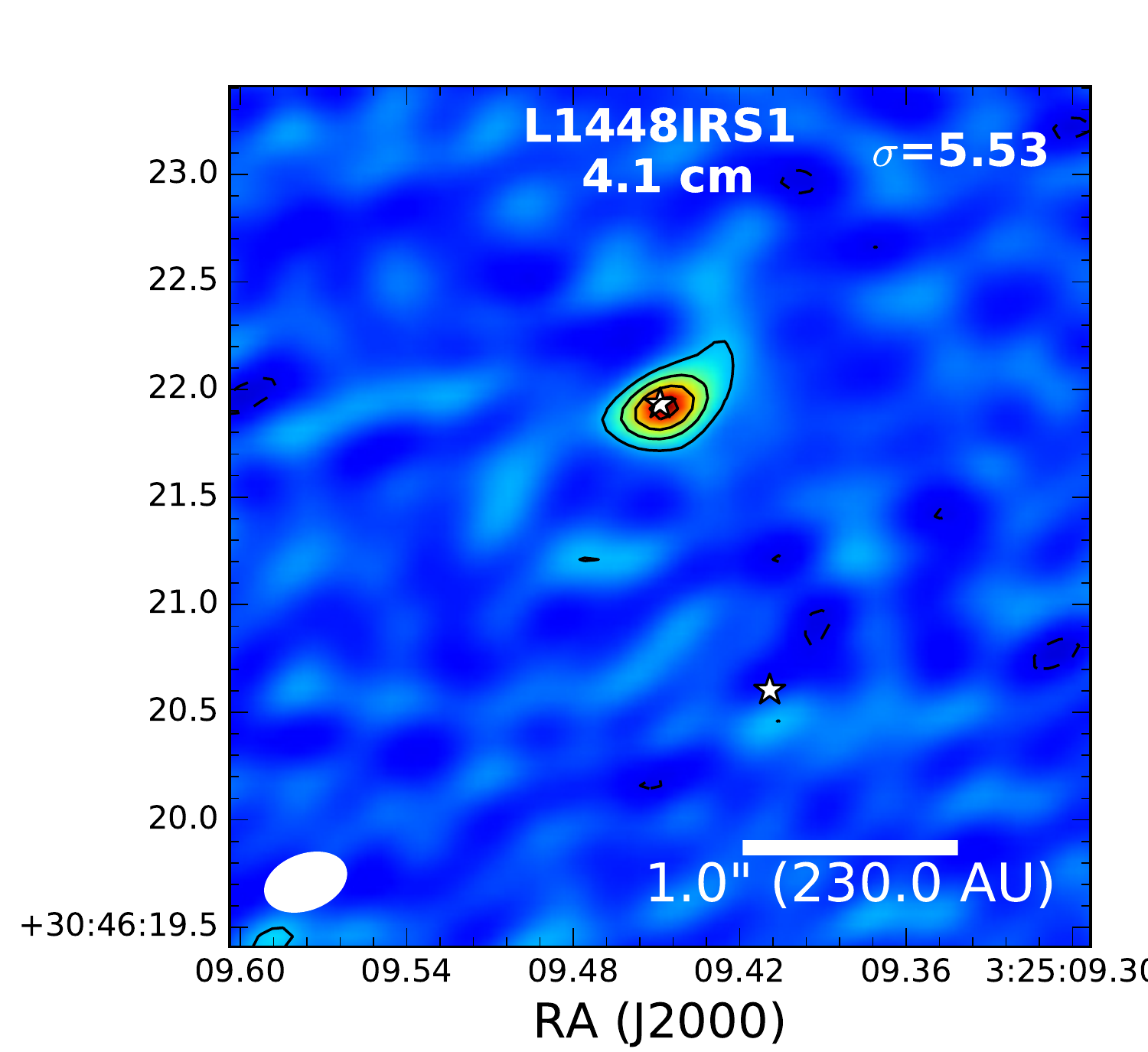}
  \includegraphics[width=0.24\linewidth]{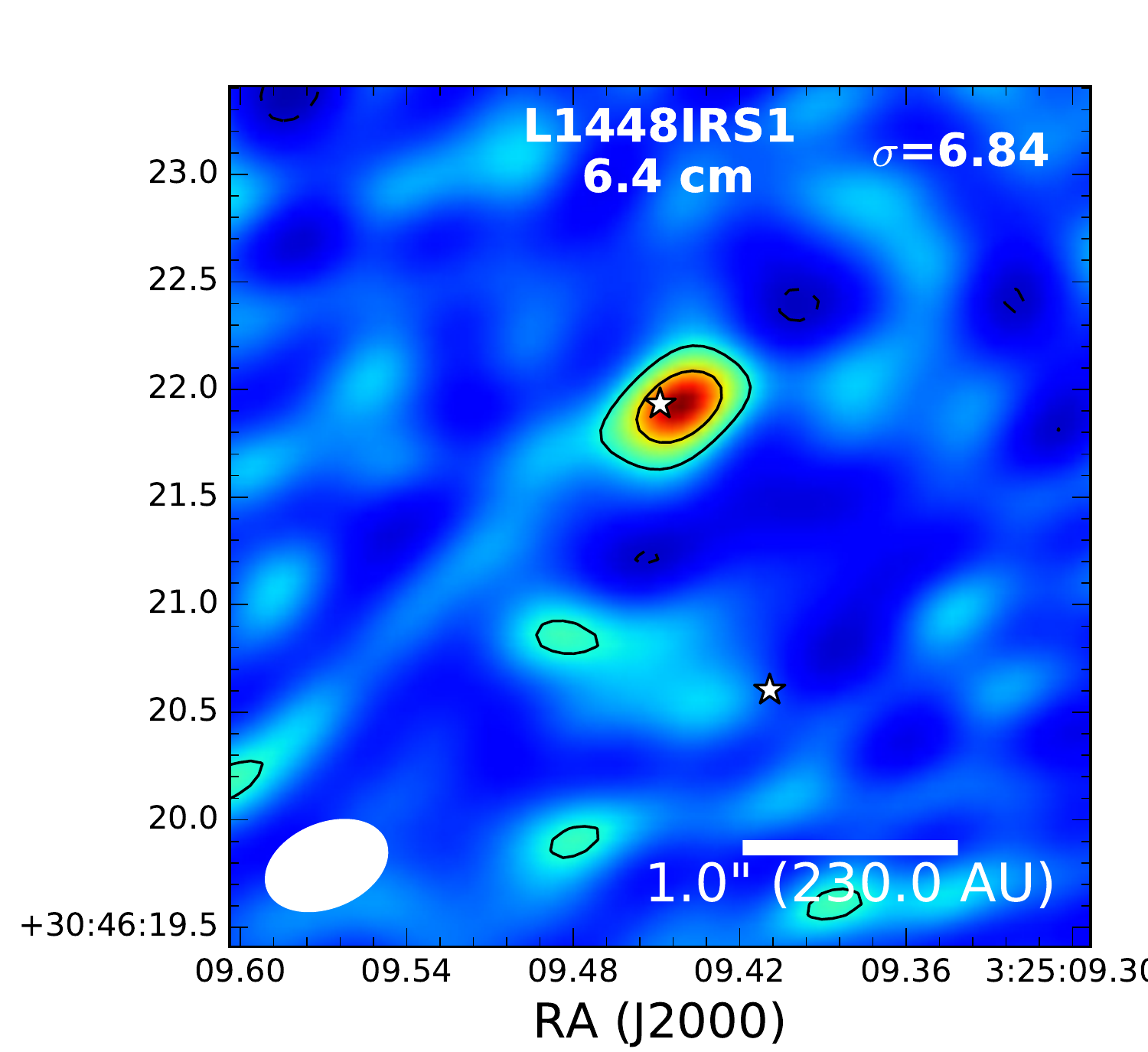}

  \includegraphics[width=0.24\linewidth]{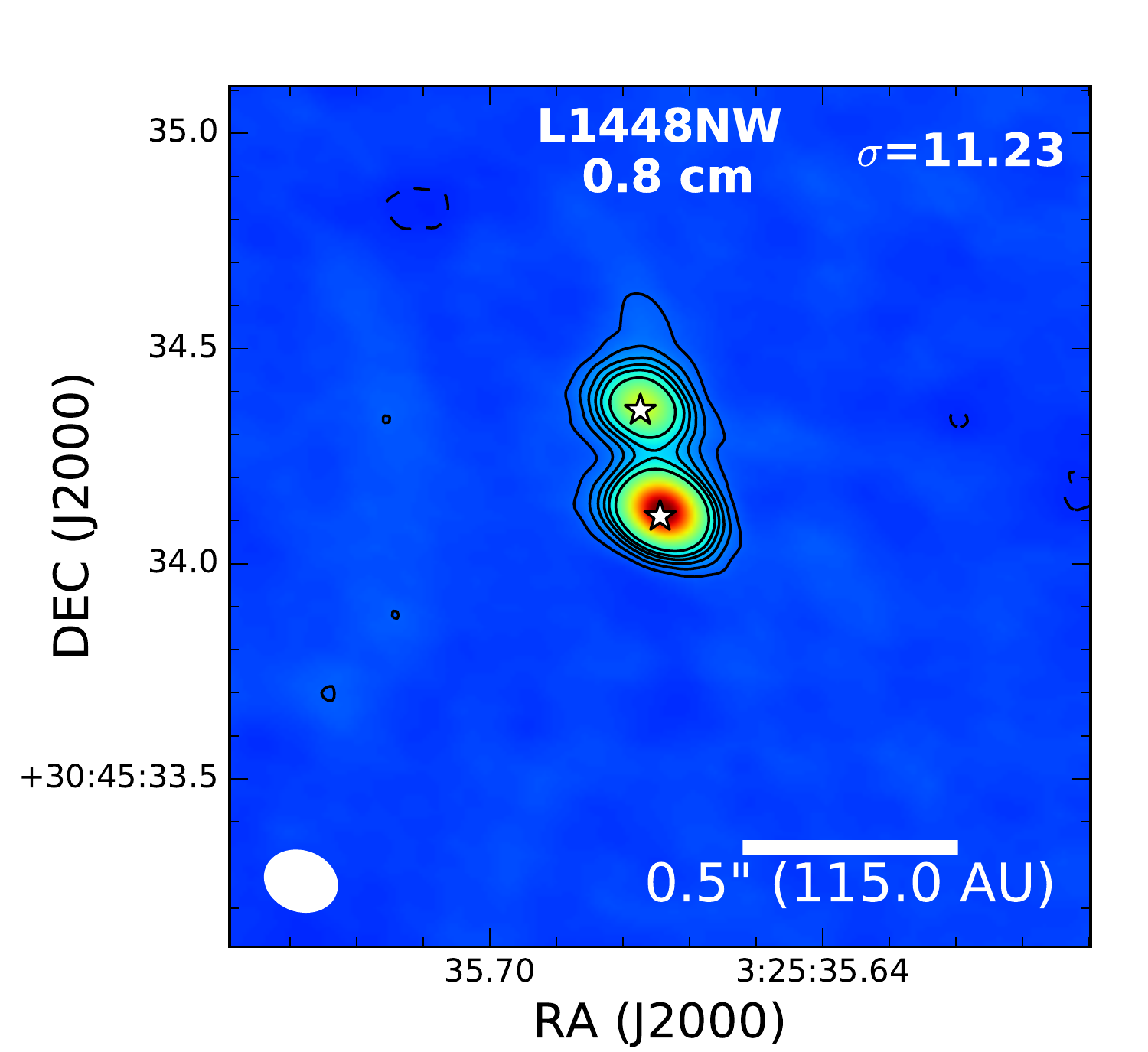}
  \includegraphics[width=0.24\linewidth]{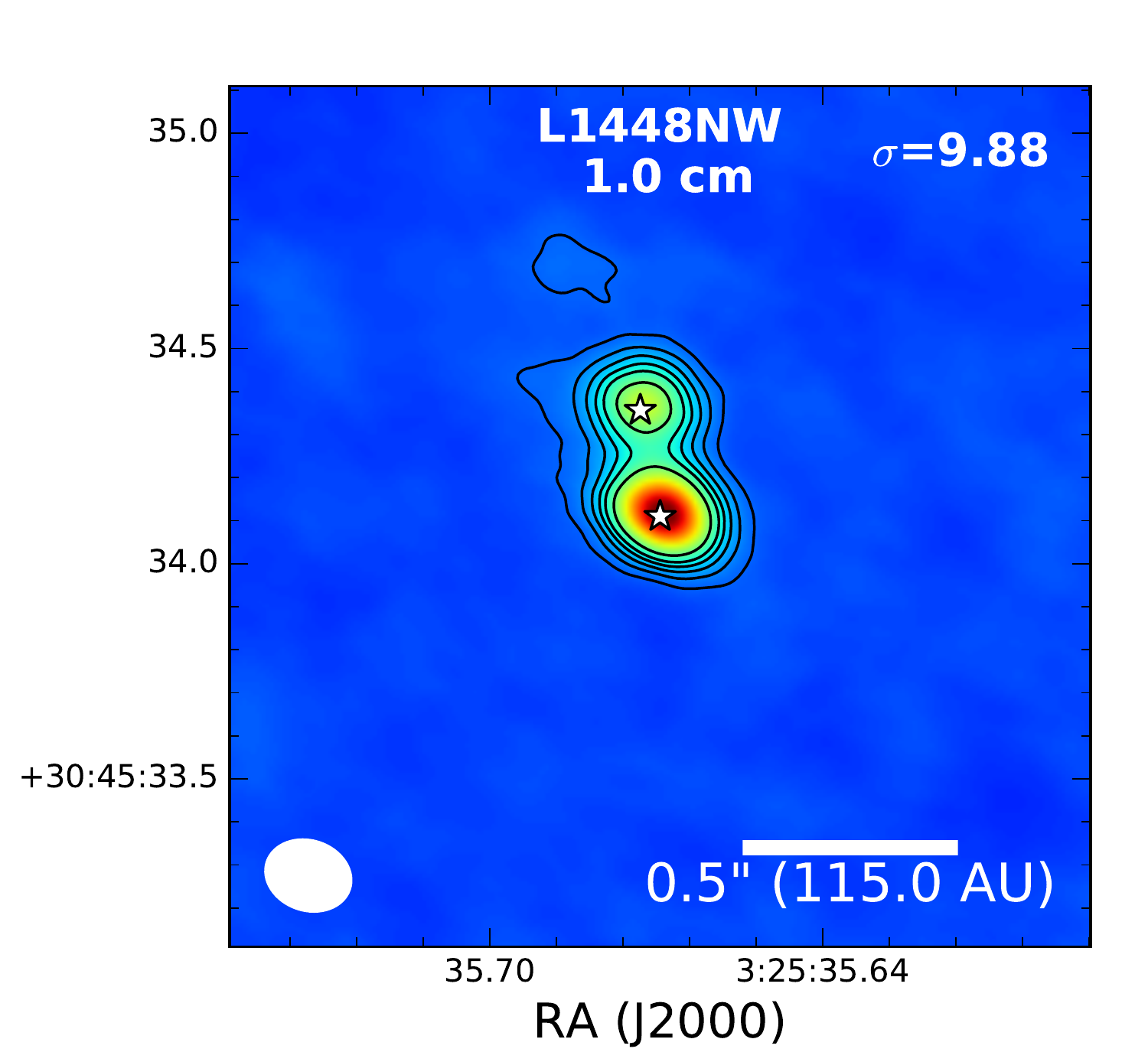}
  \includegraphics[width=0.24\linewidth]{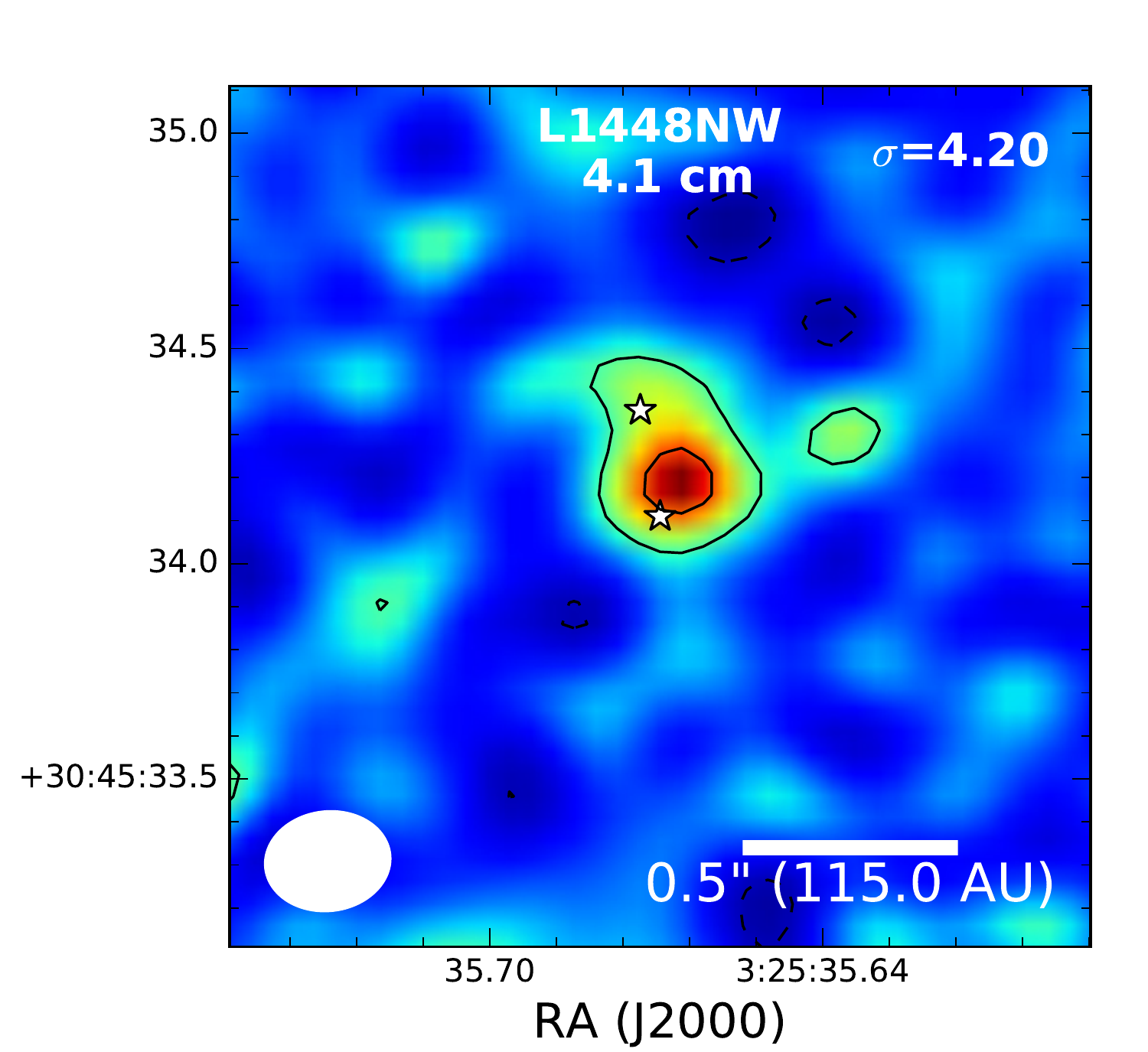}
  \includegraphics[width=0.24\linewidth]{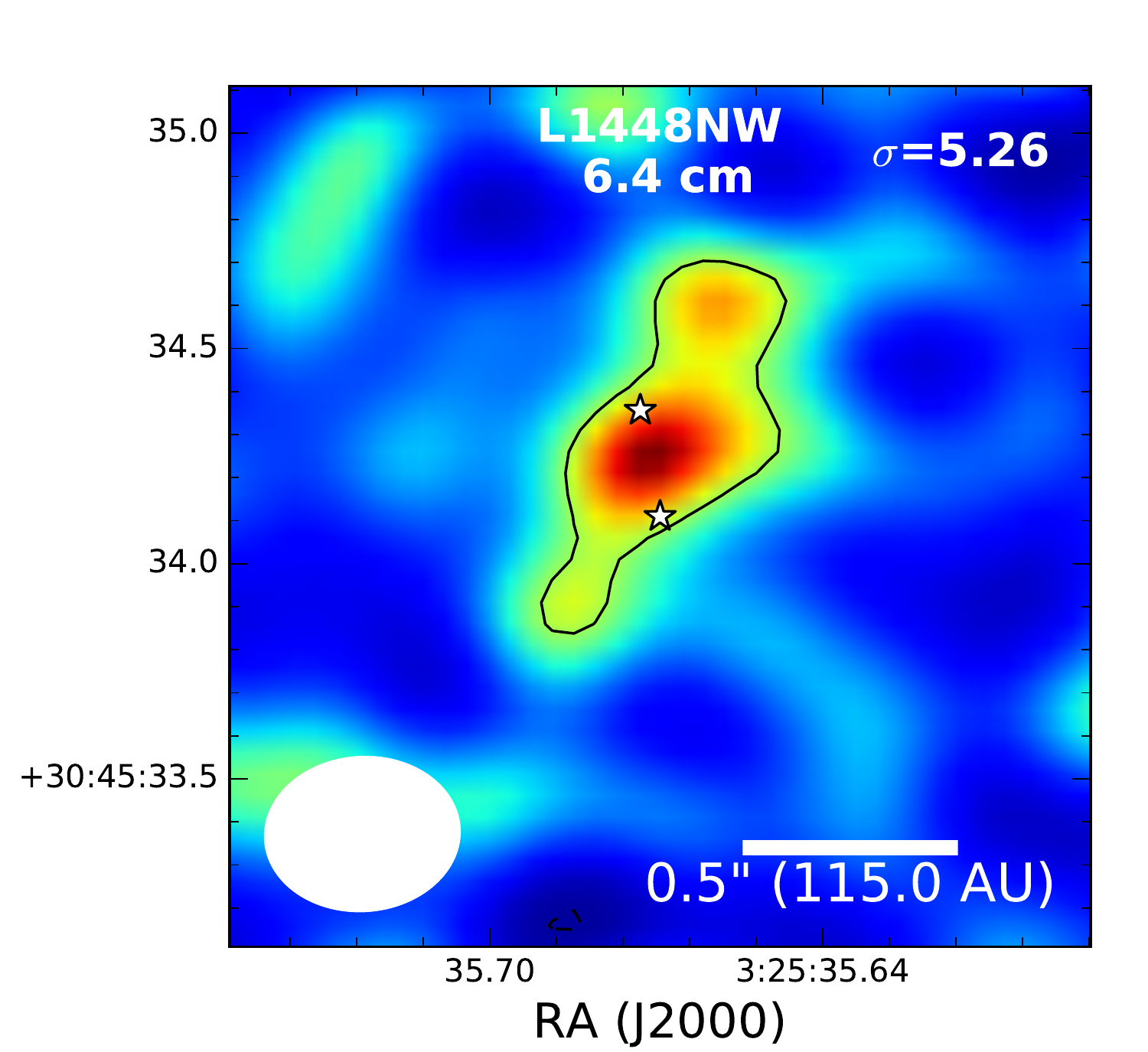}

  \includegraphics[width=0.24\linewidth]{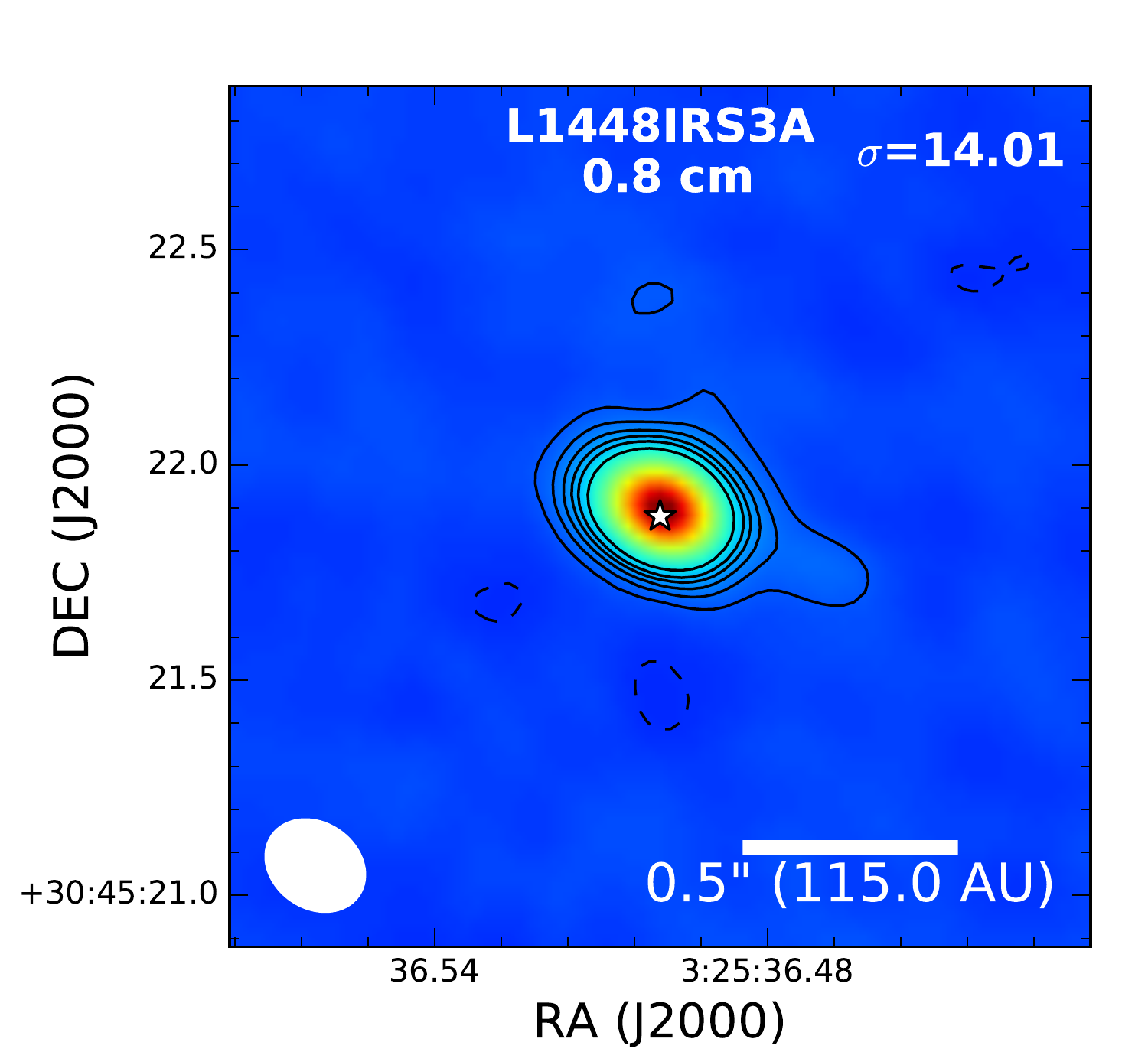}
  \includegraphics[width=0.24\linewidth]{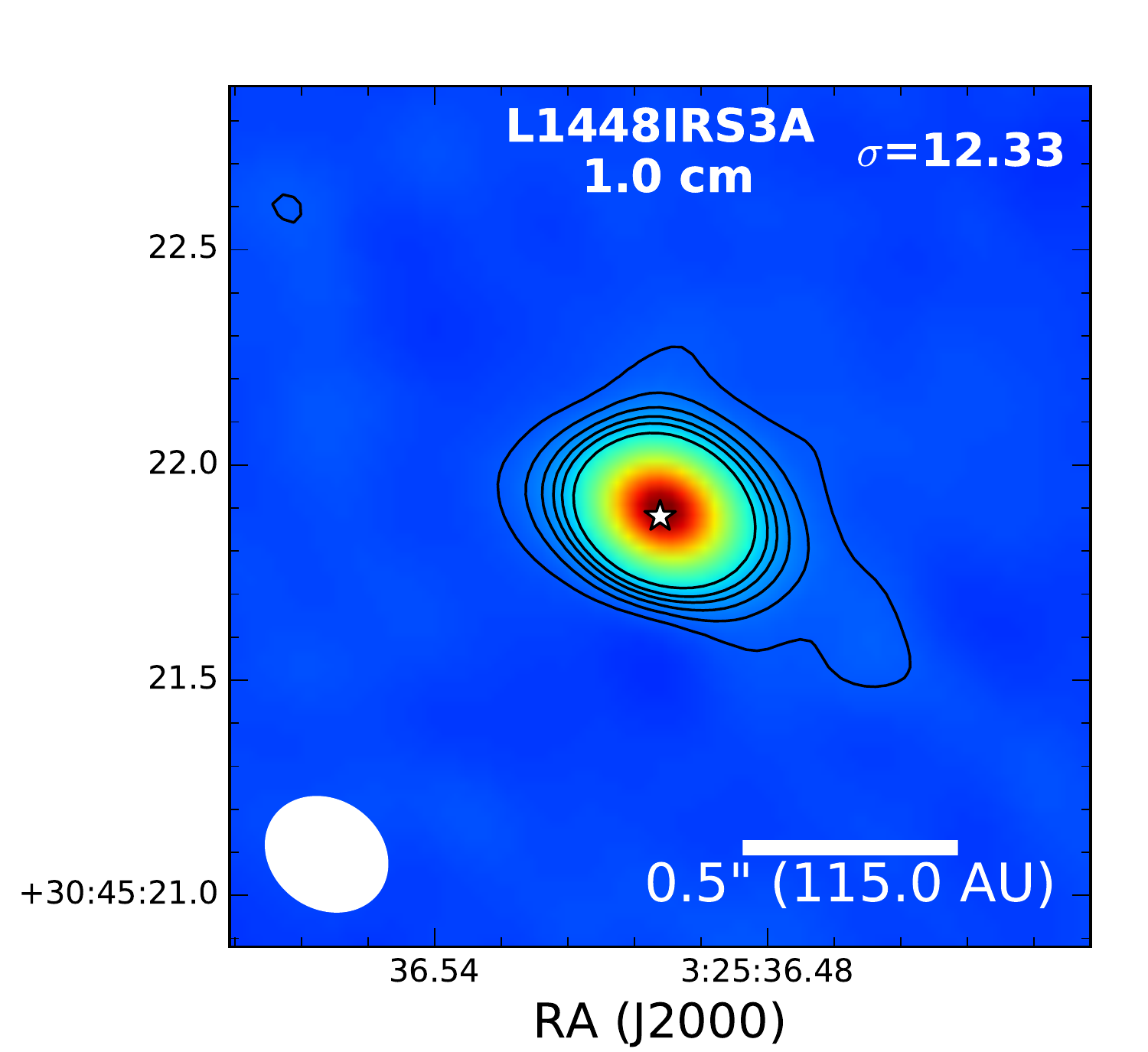}
  \includegraphics[width=0.24\linewidth]{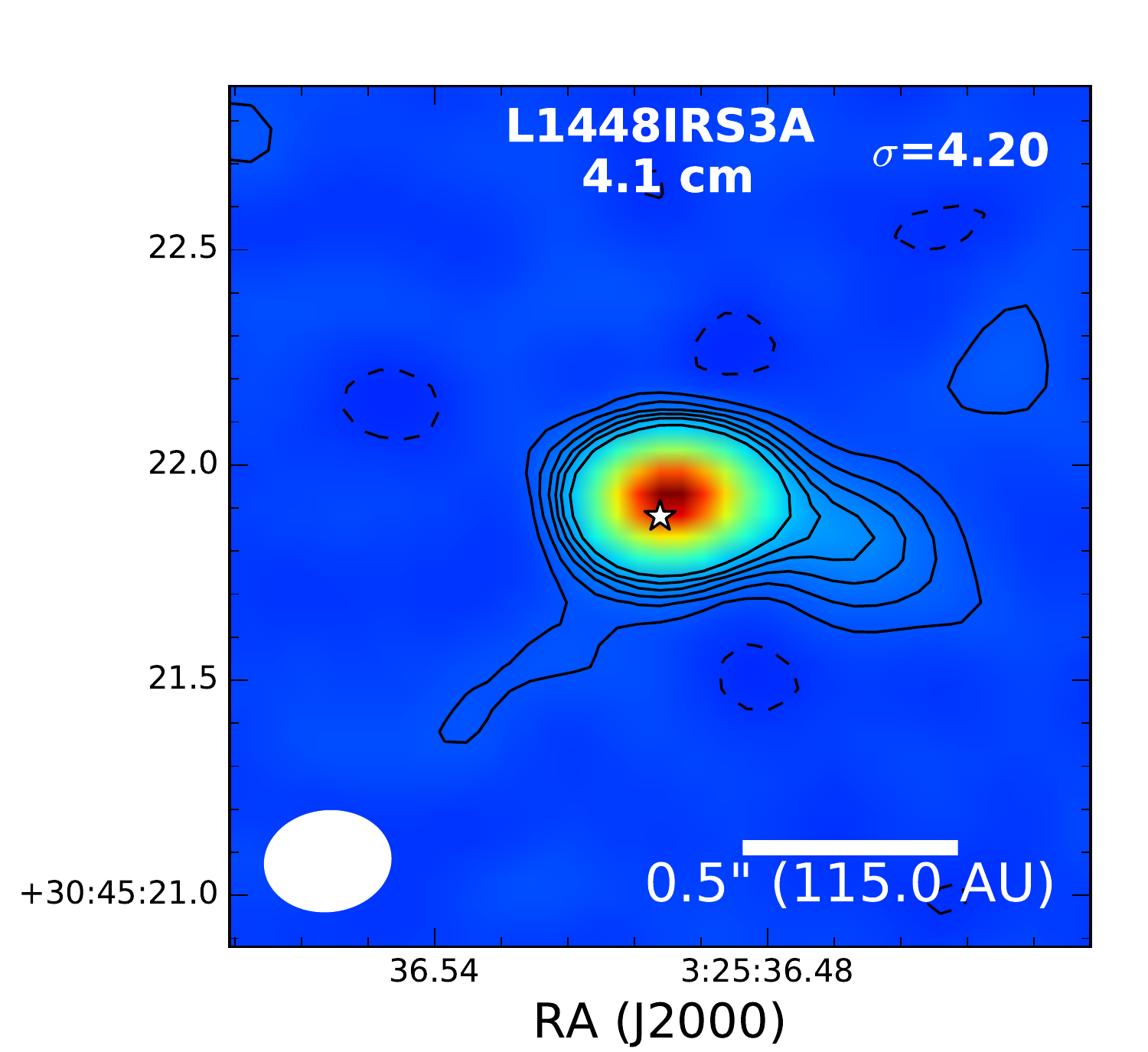}
  \includegraphics[width=0.24\linewidth]{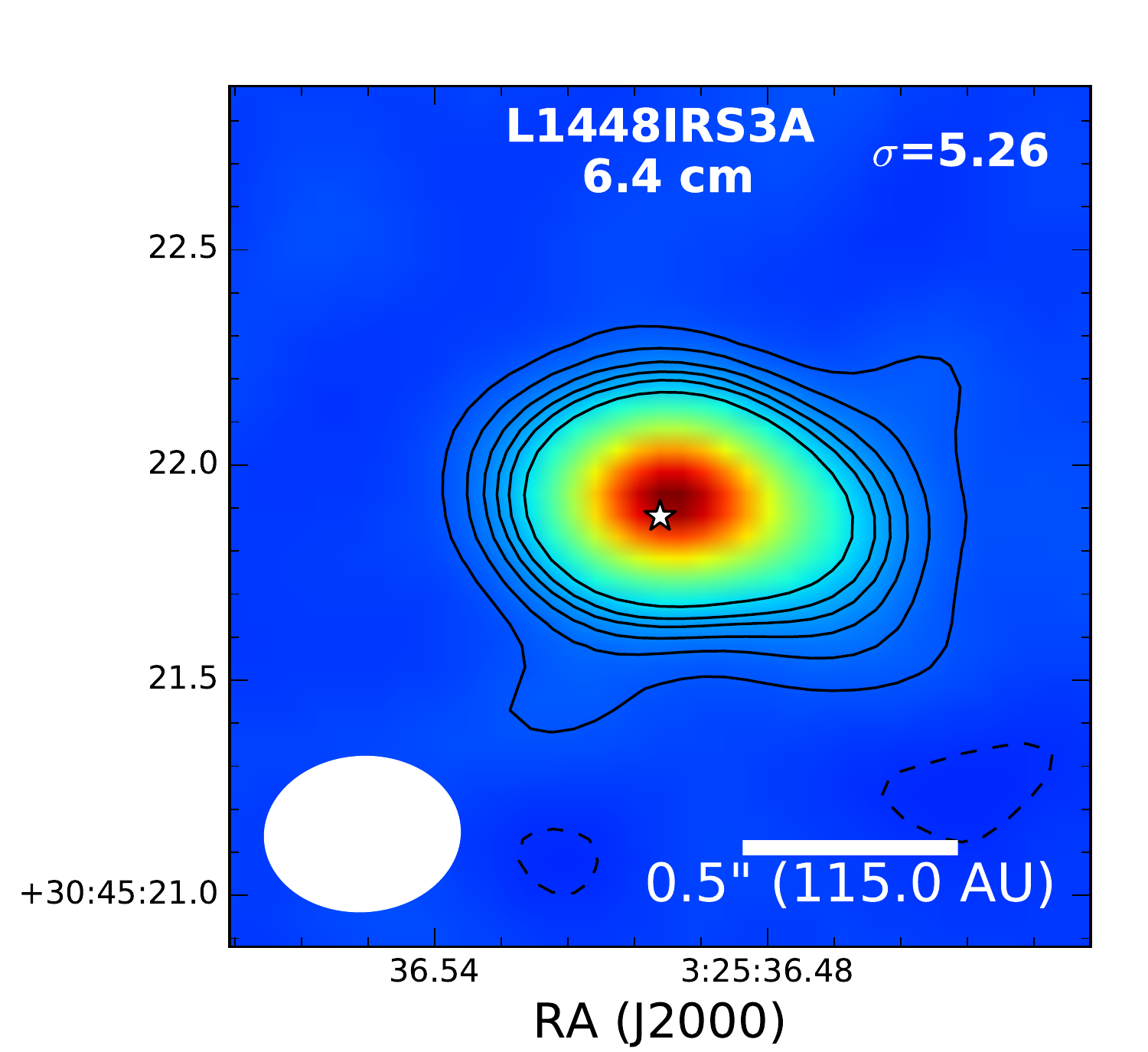}

\end{figure}

\begin{figure}

  \includegraphics[width=0.24\linewidth]{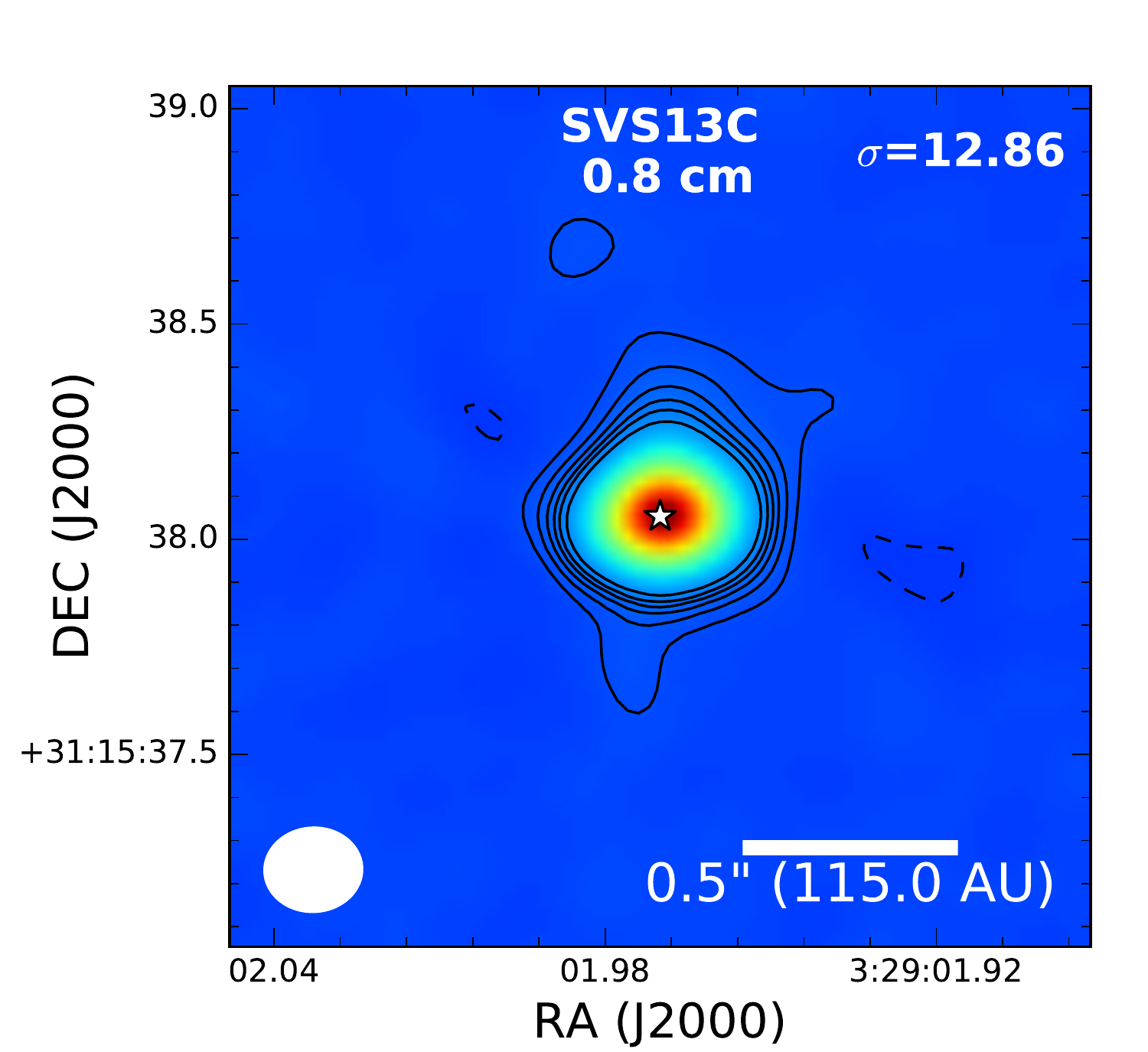}
  \includegraphics[width=0.24\linewidth]{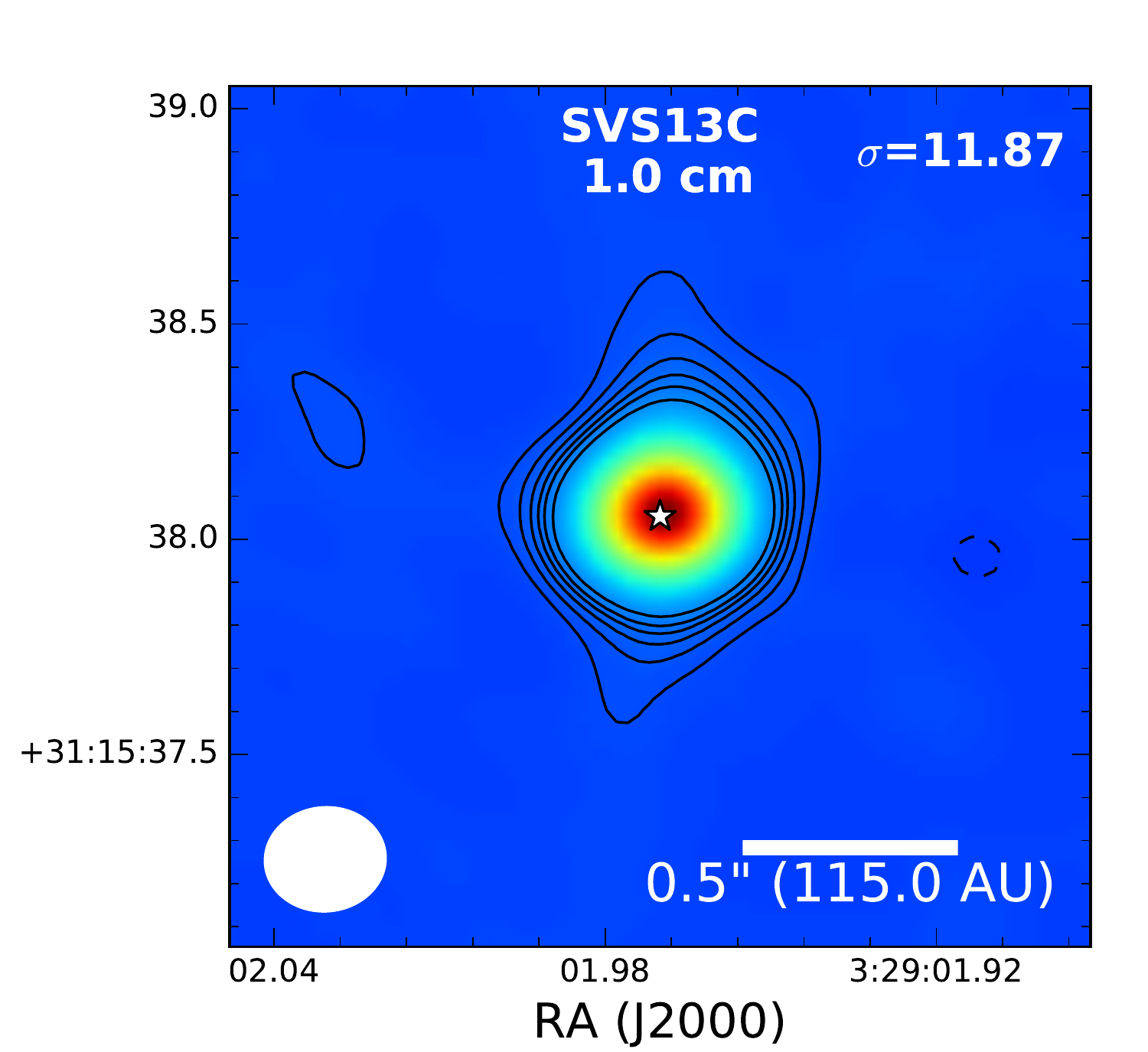}
  \includegraphics[width=0.24\linewidth]{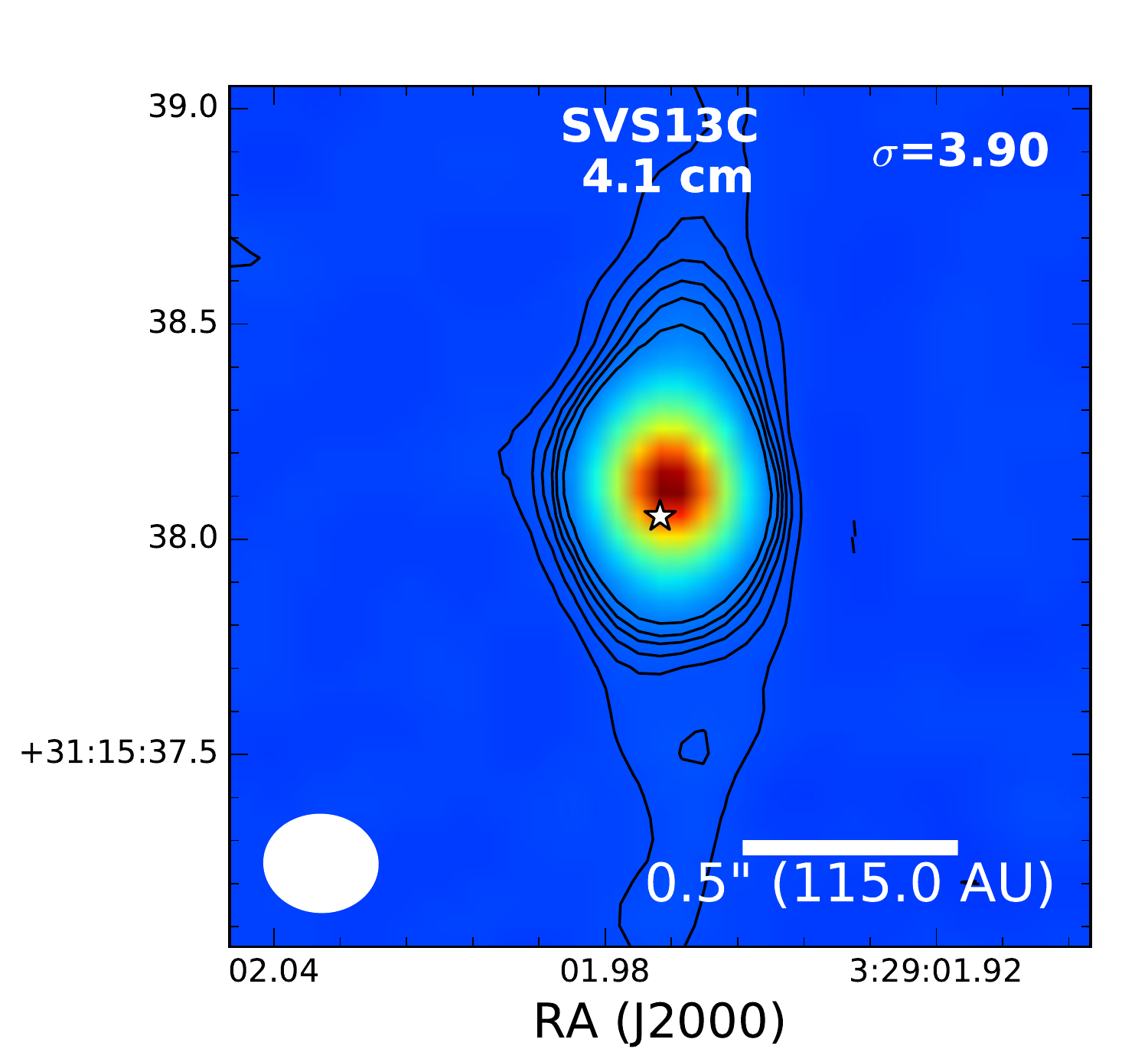}
  \includegraphics[width=0.24\linewidth]{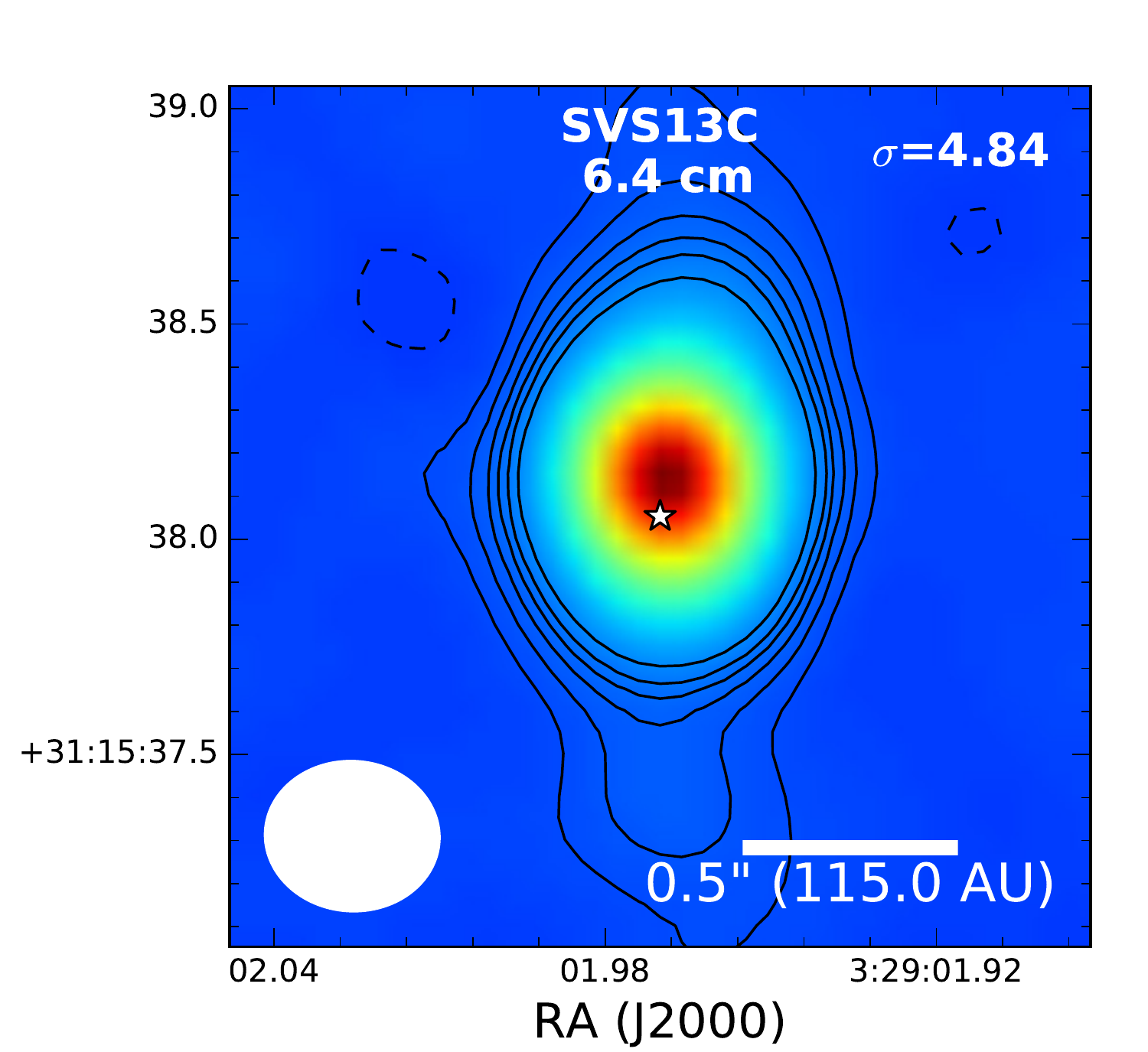}

  \includegraphics[width=0.24\linewidth]{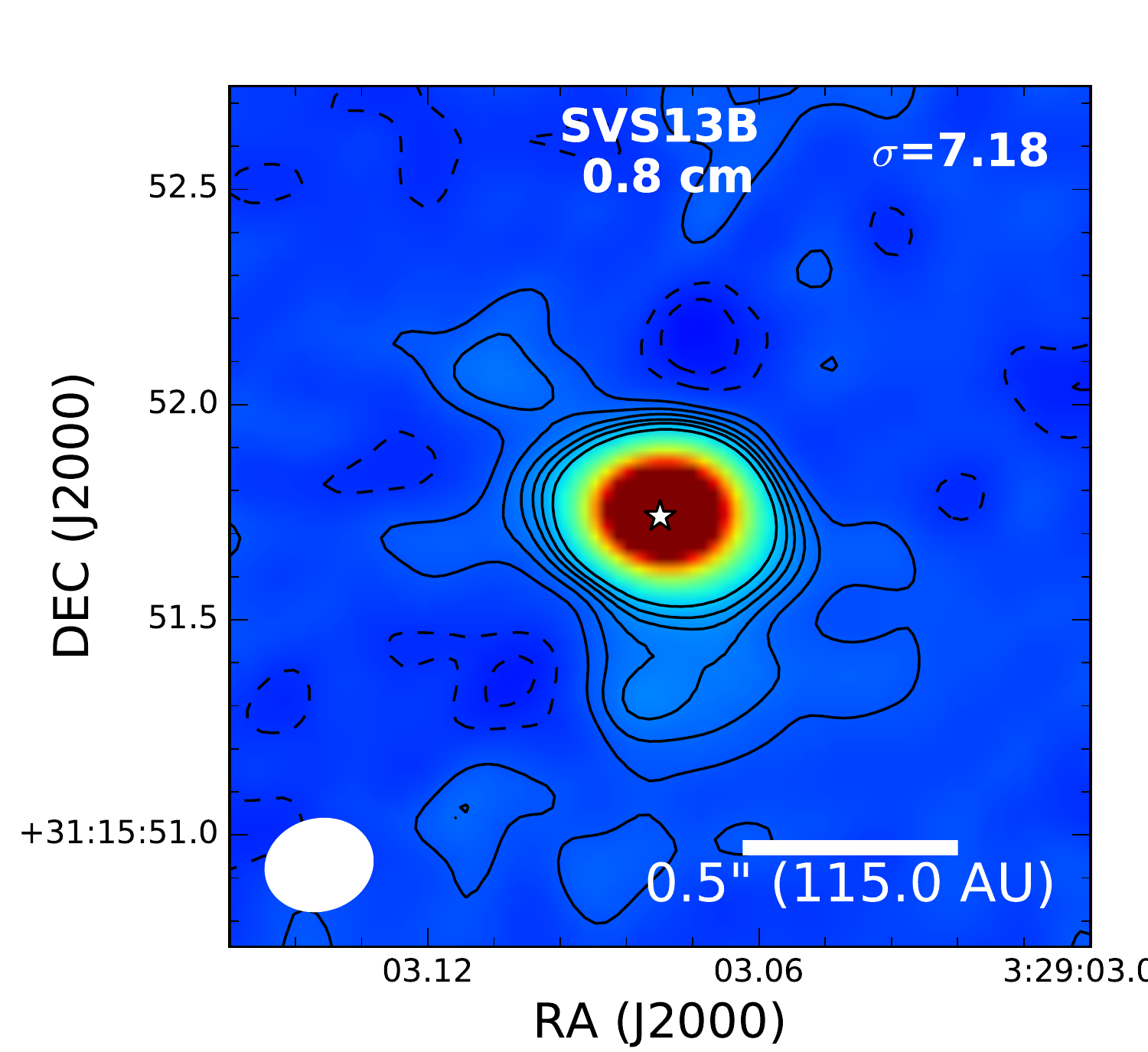}
  \includegraphics[width=0.24\linewidth]{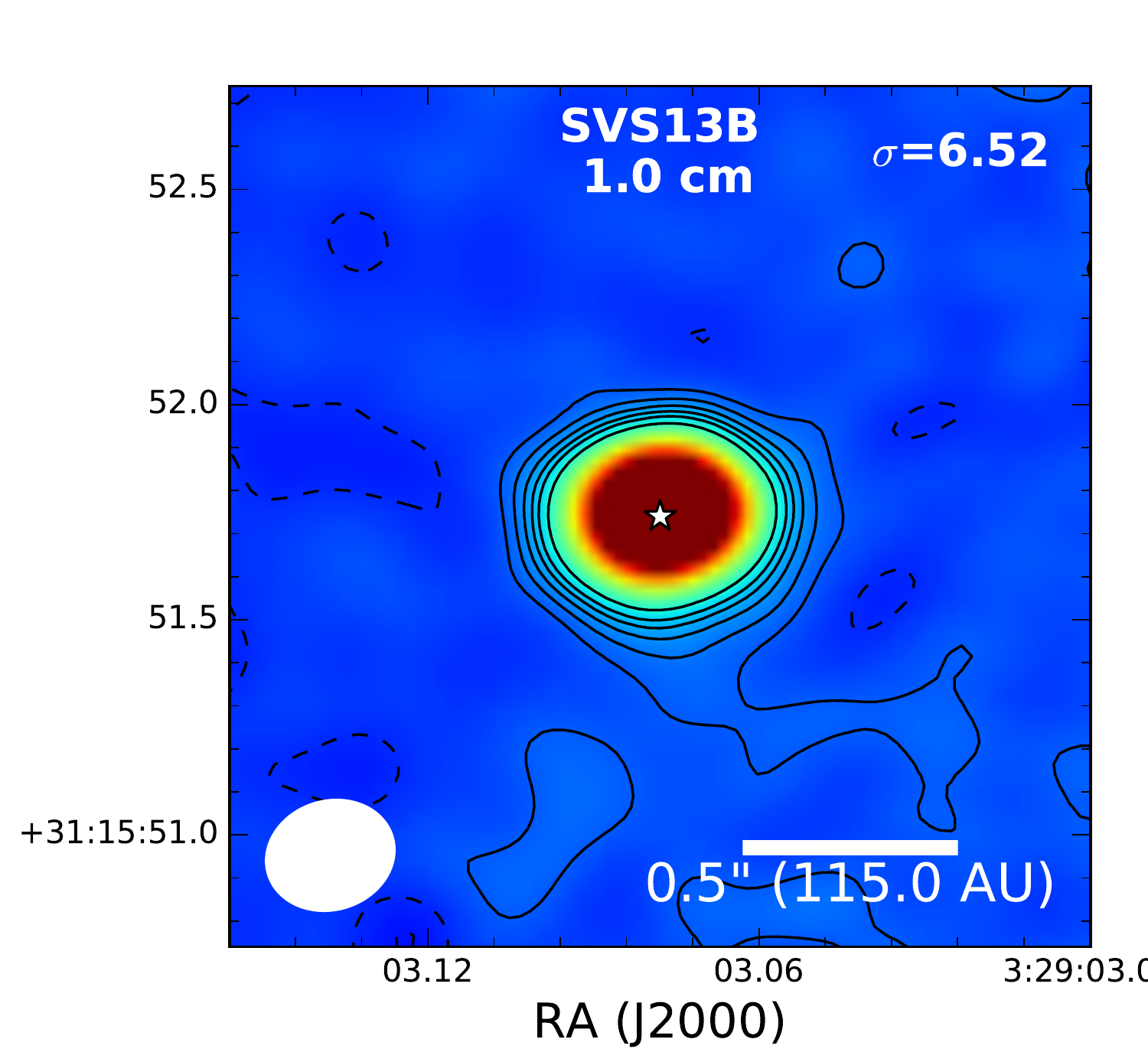}
  \includegraphics[width=0.24\linewidth]{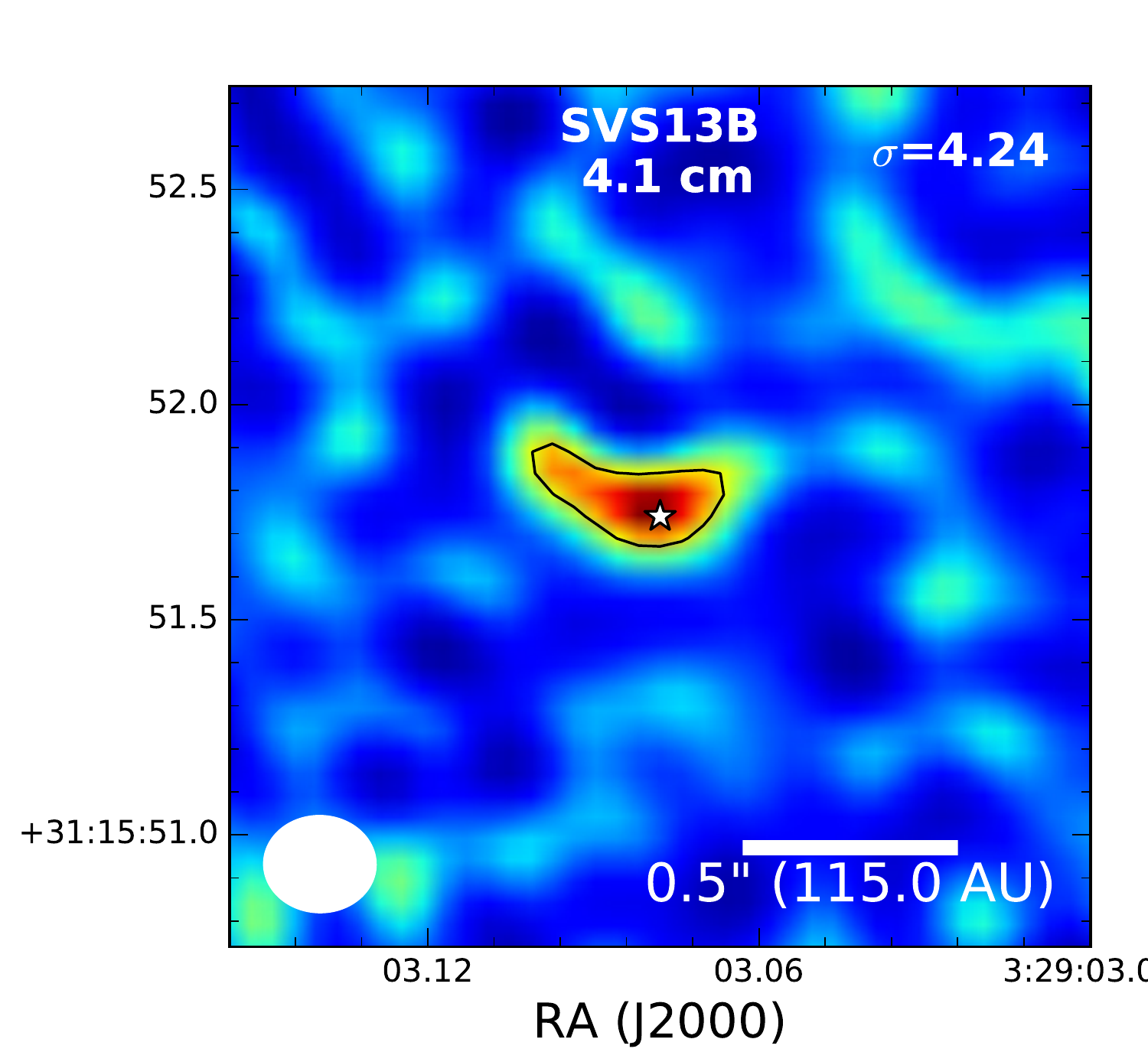}
  \includegraphics[width=0.24\linewidth]{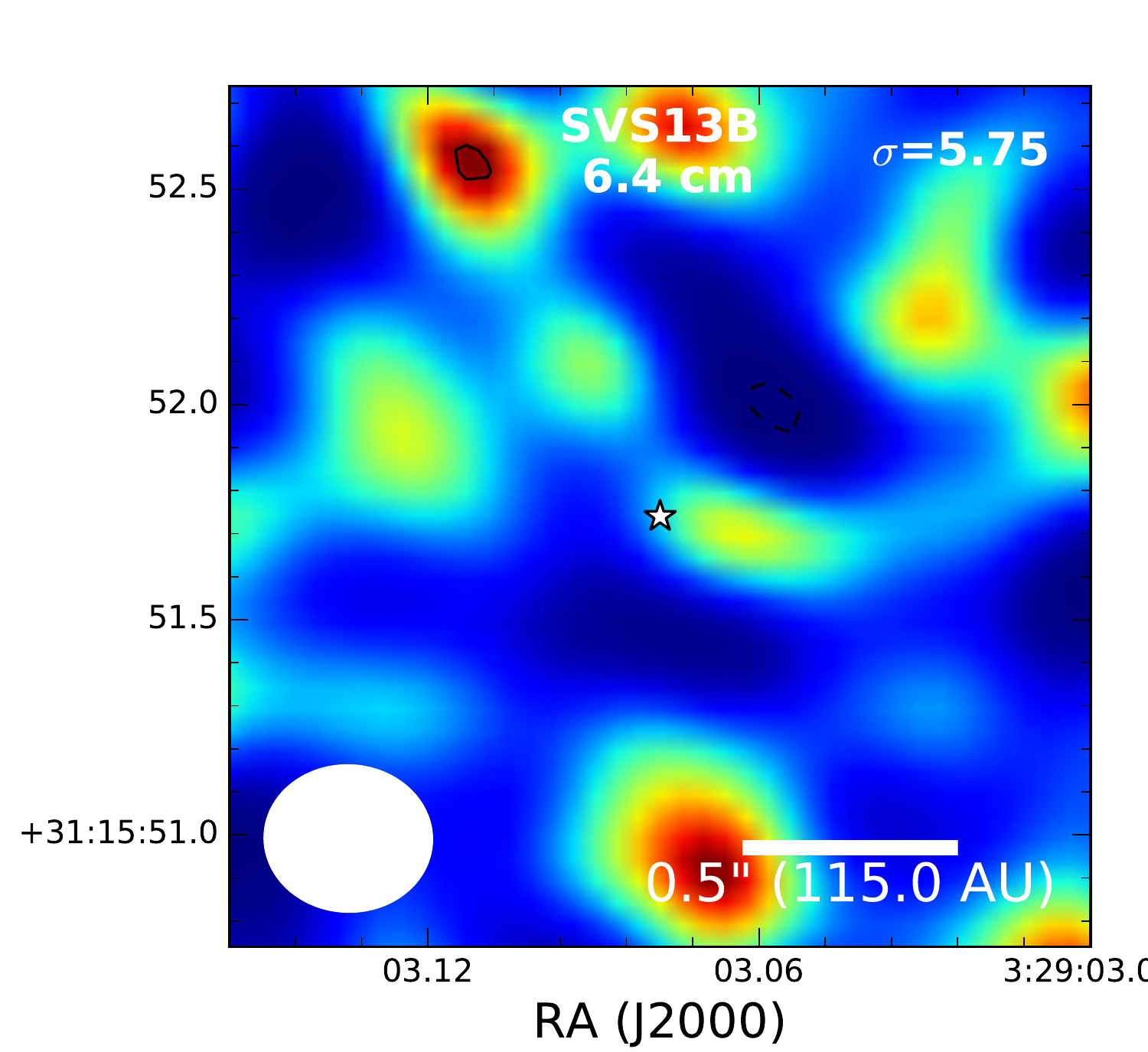}

  \includegraphics[width=0.24\linewidth]{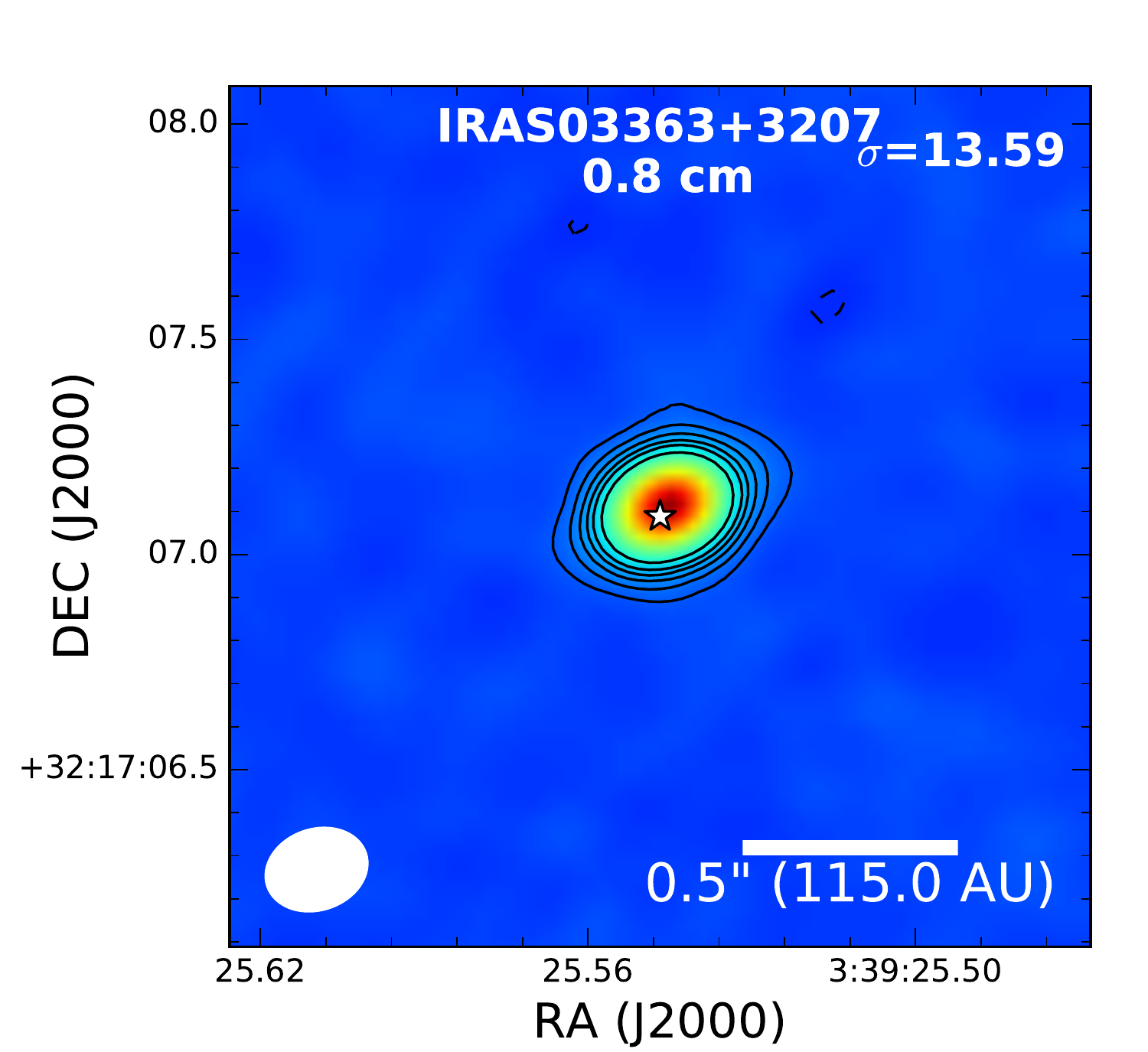}
  \includegraphics[width=0.24\linewidth]{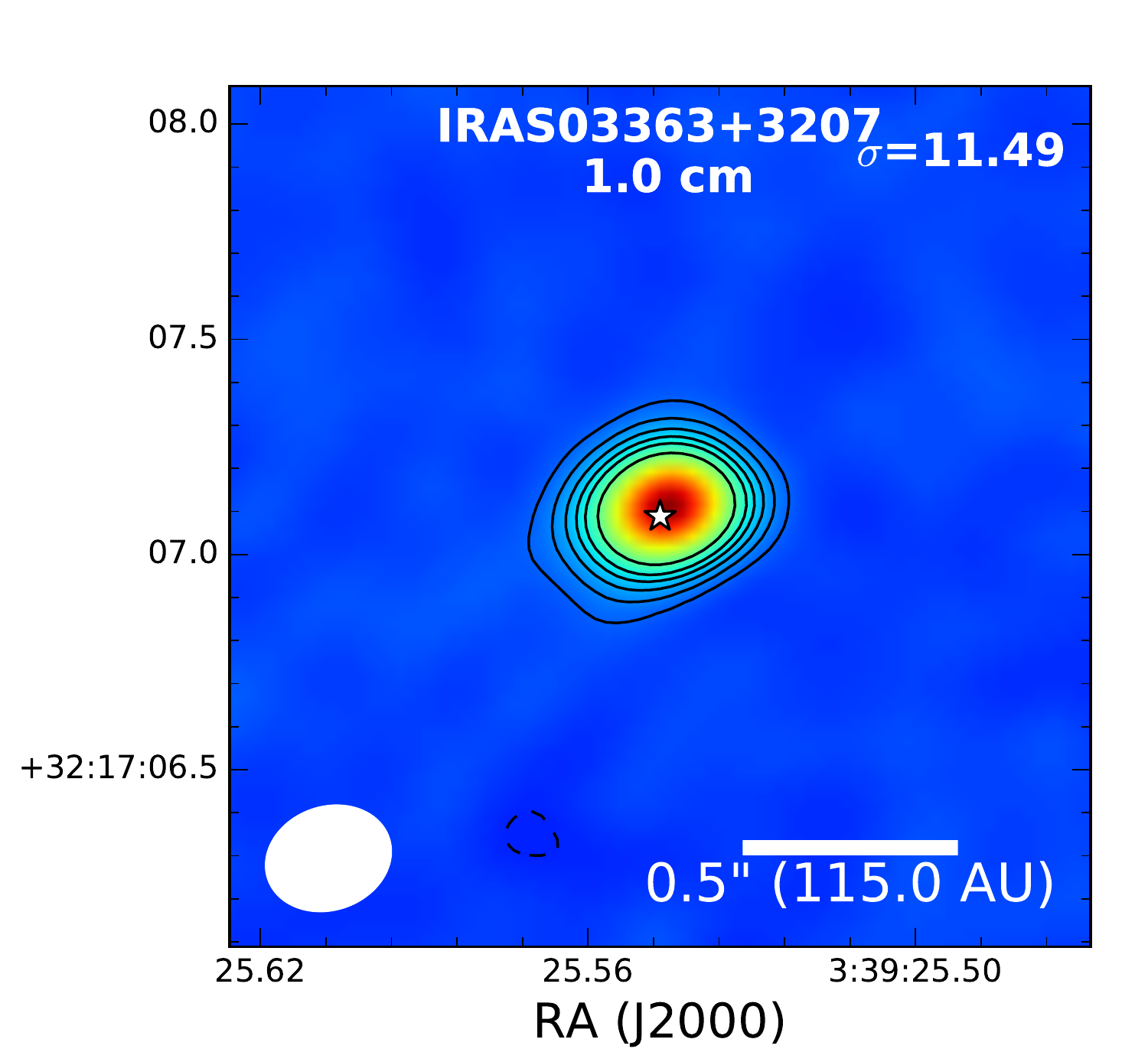}
  \includegraphics[width=0.24\linewidth]{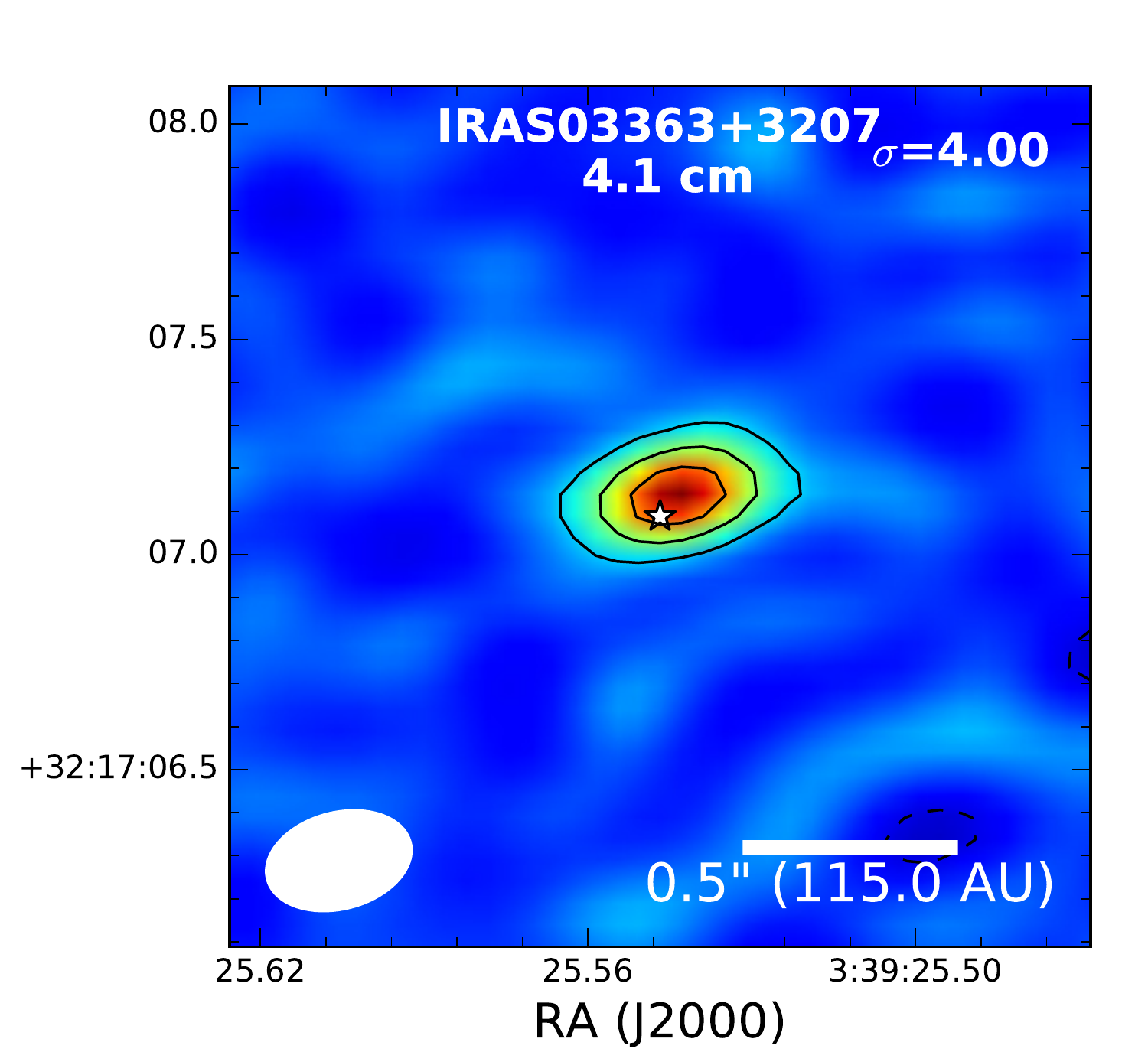}
  \includegraphics[width=0.24\linewidth]{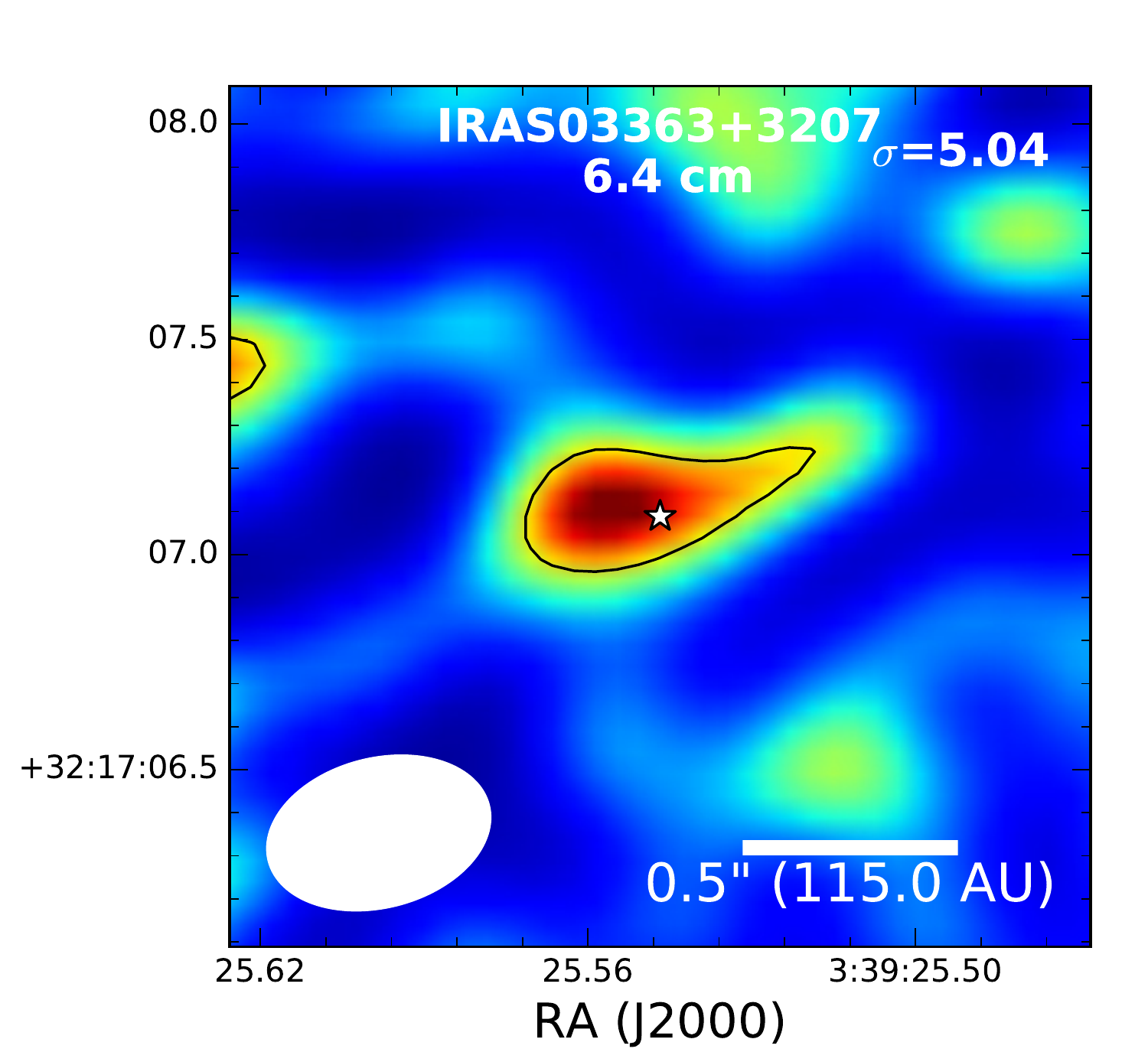}

  \includegraphics[width=0.24\linewidth]{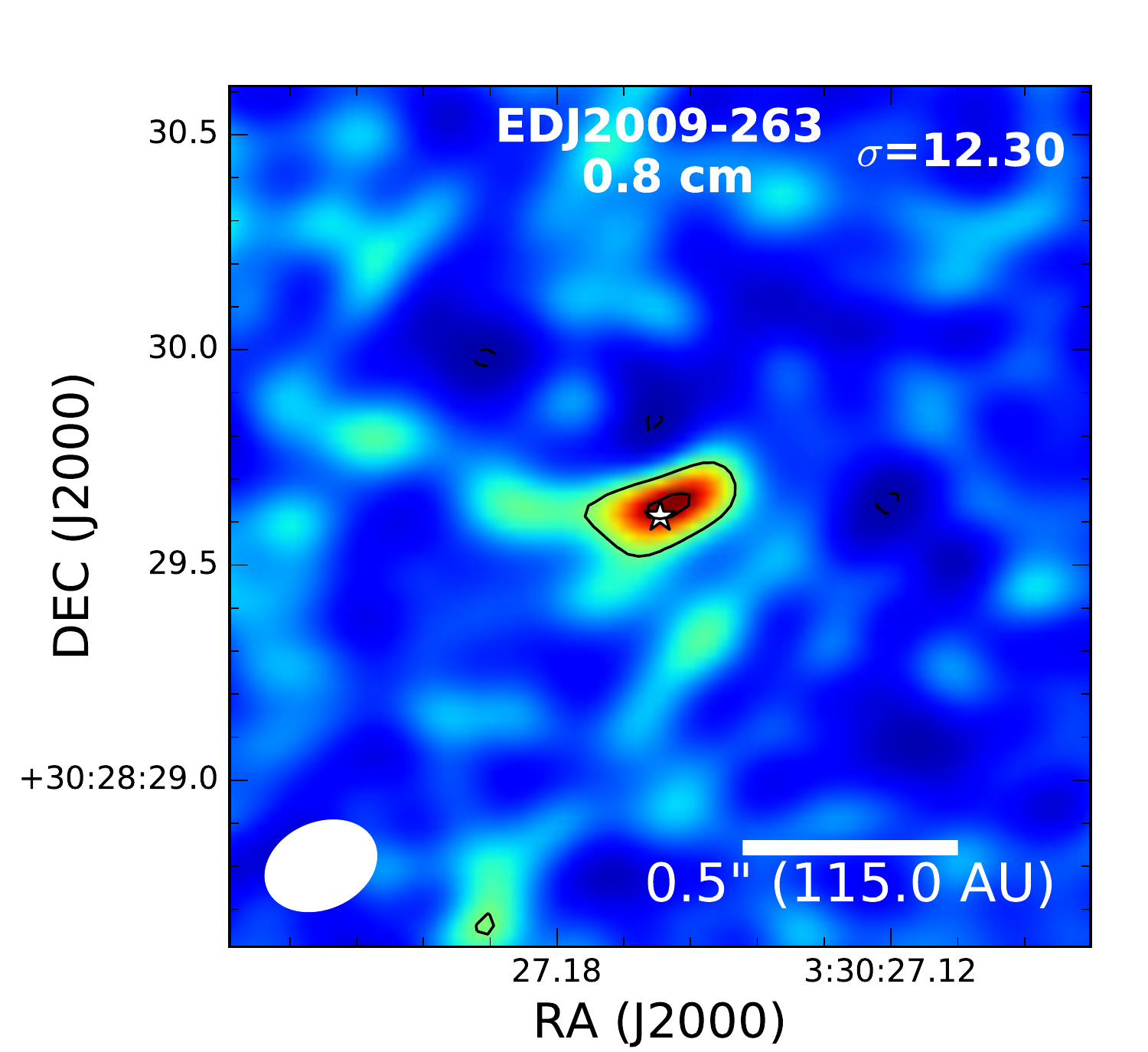}
  \includegraphics[width=0.24\linewidth]{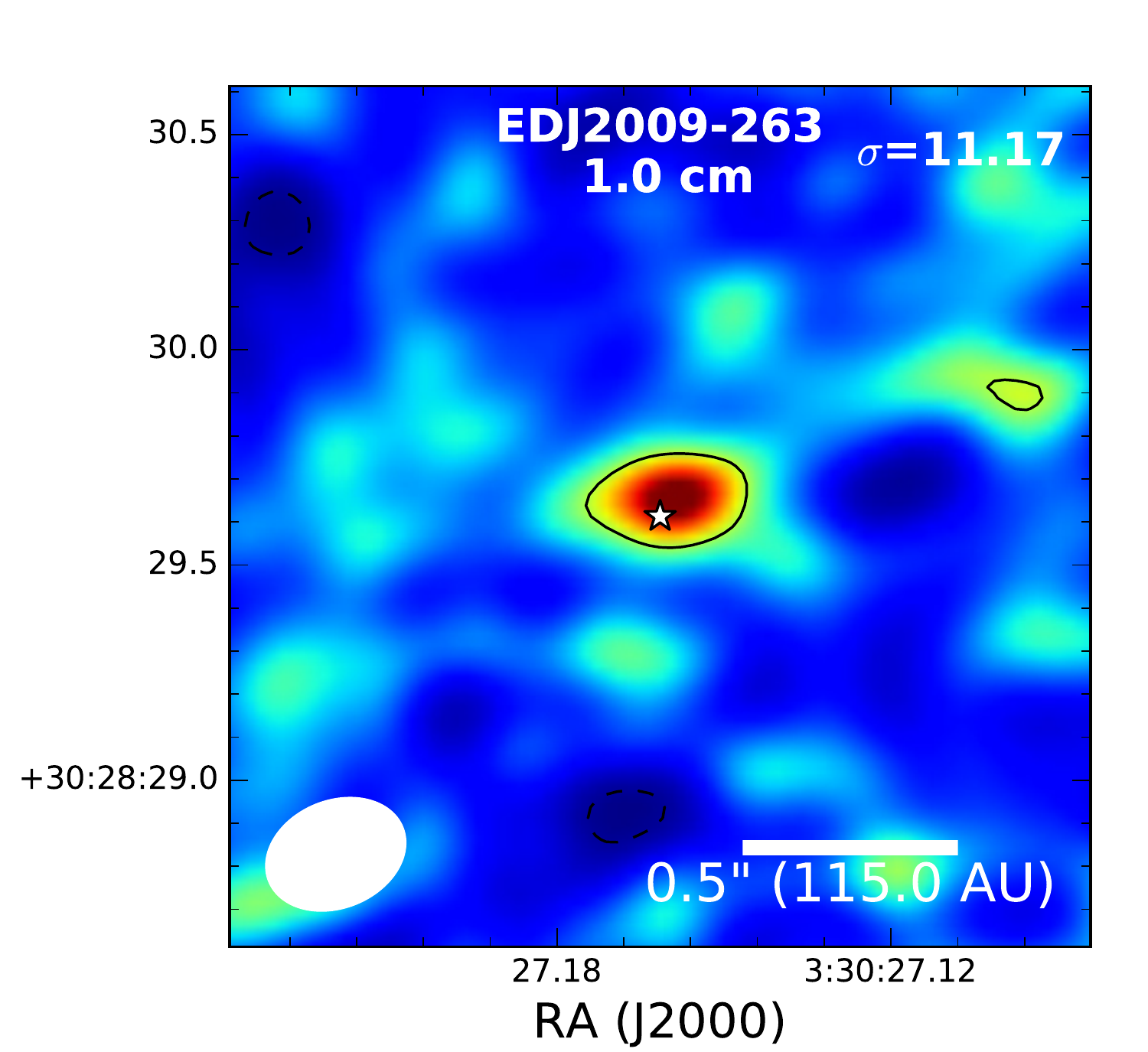}
  \includegraphics[width=0.24\linewidth]{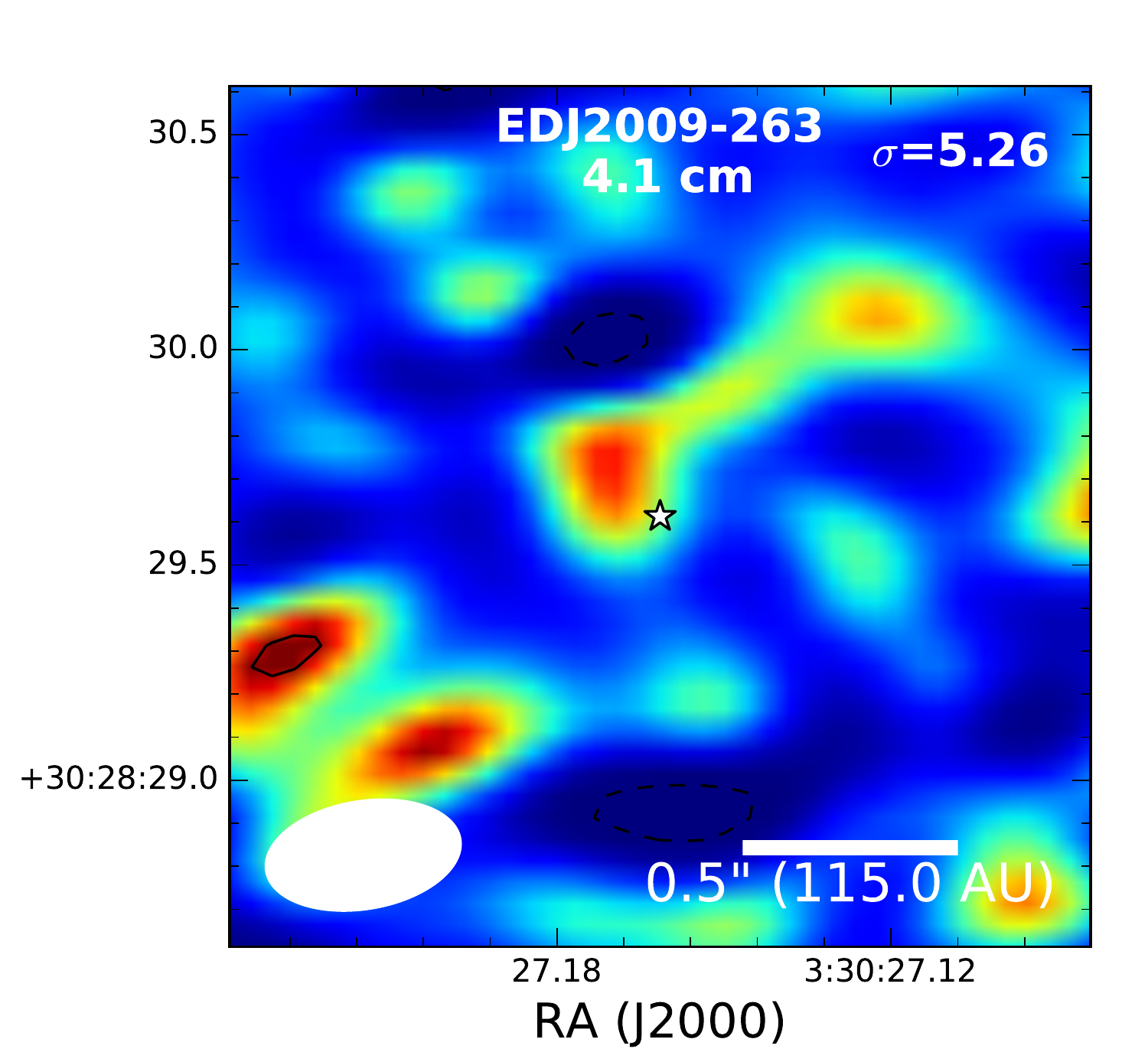}
  \includegraphics[width=0.24\linewidth]{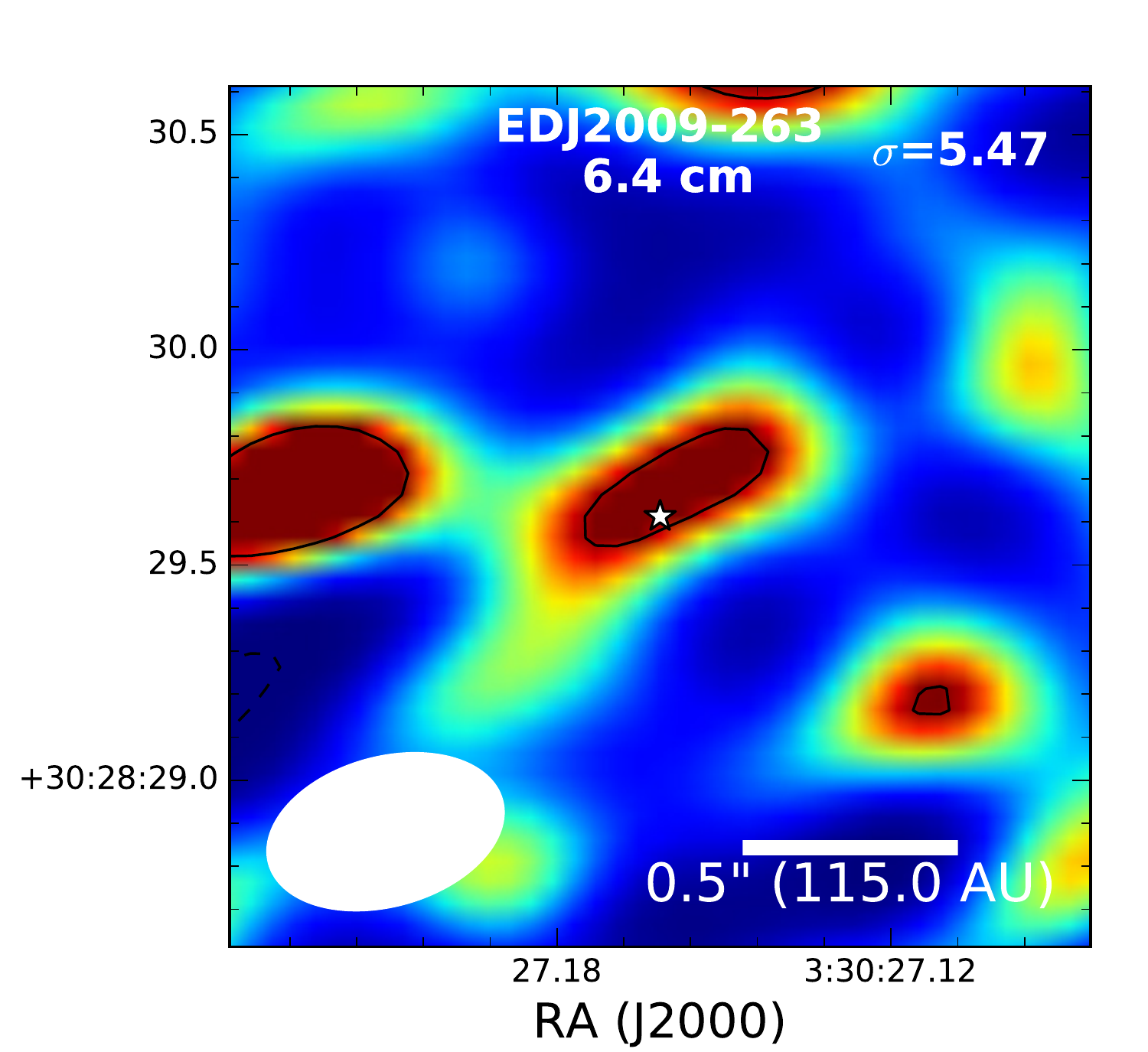}

  \includegraphics[width=0.24\linewidth]{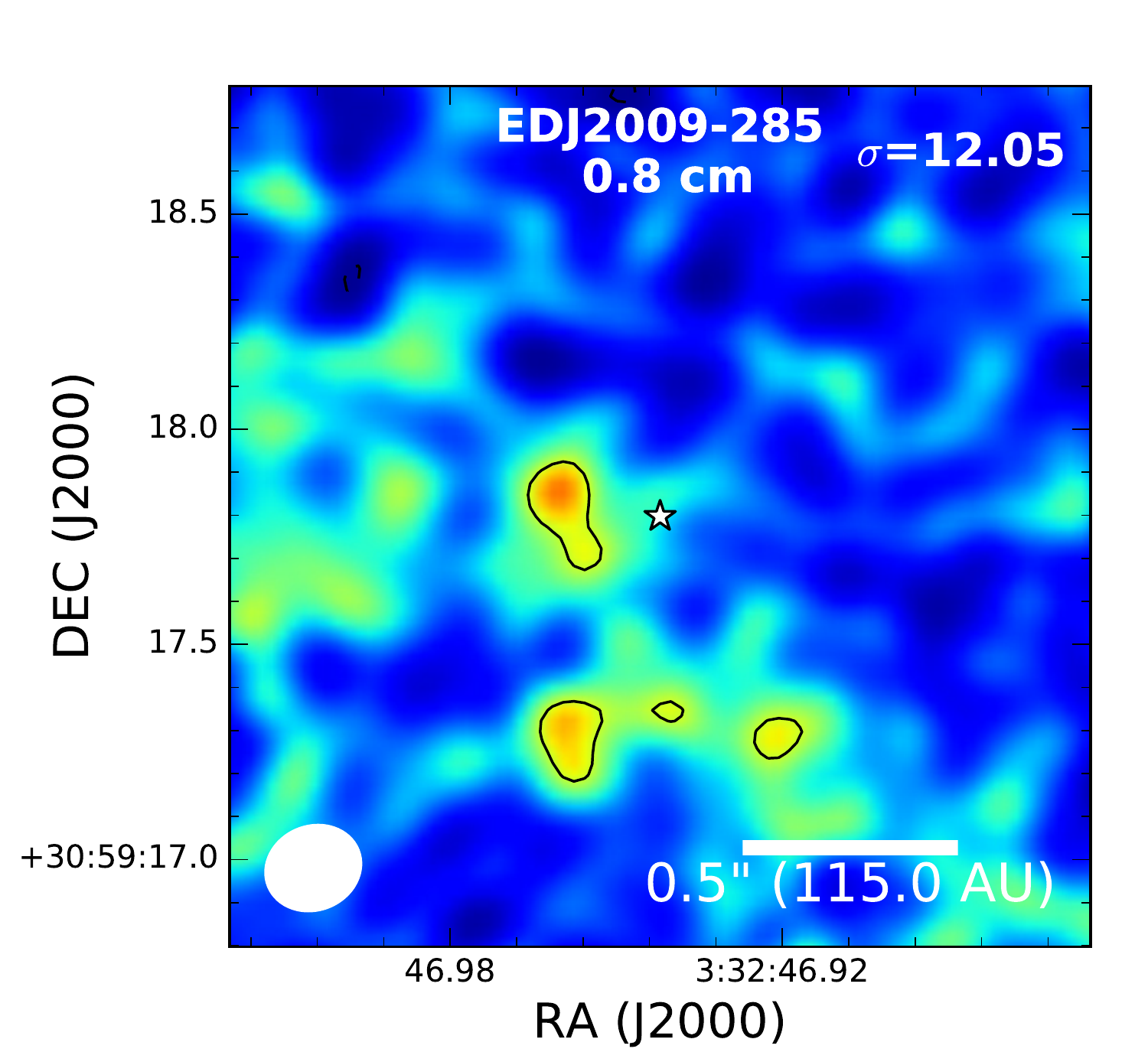}
  \includegraphics[width=0.24\linewidth]{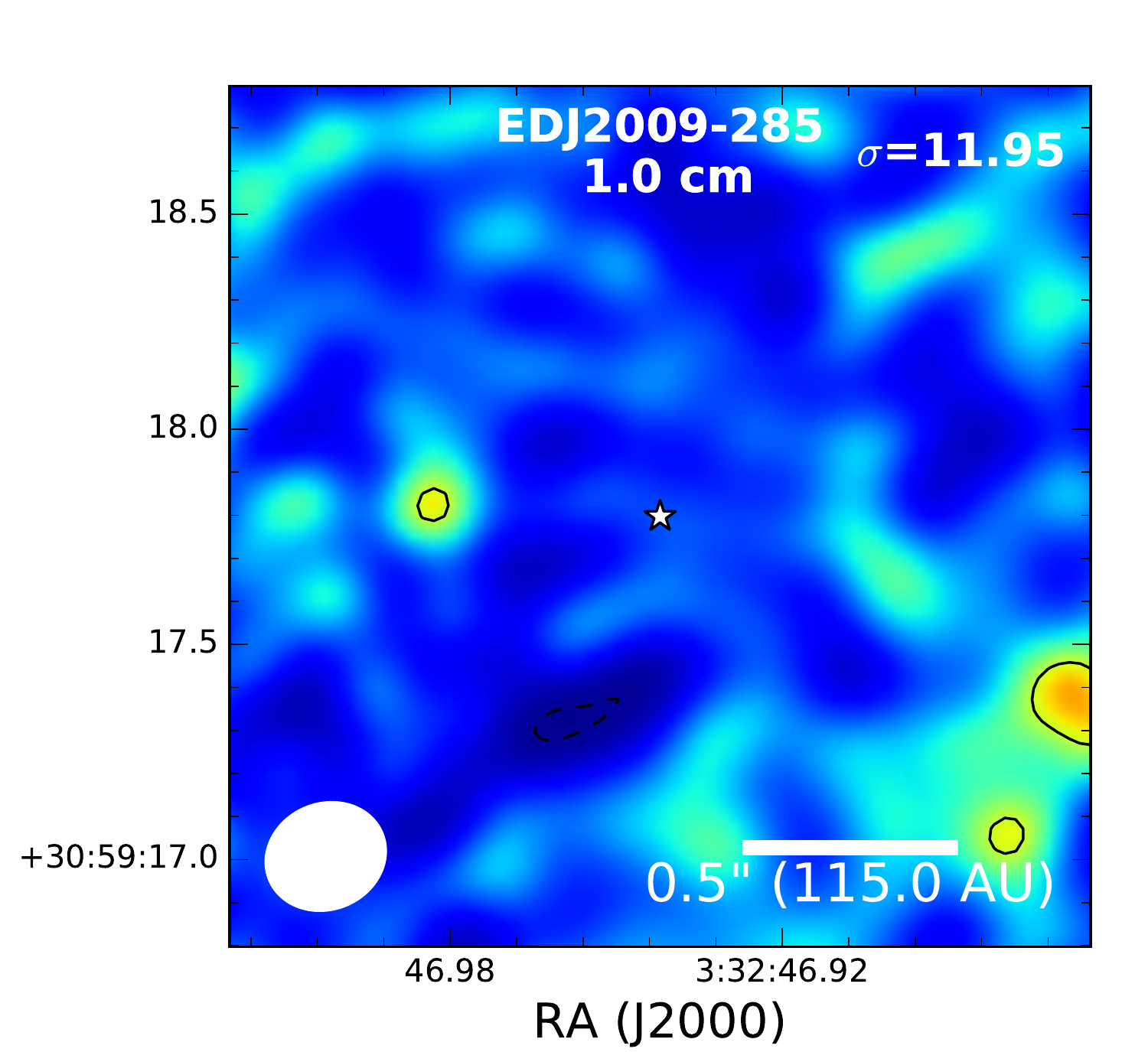}
  \includegraphics[width=0.24\linewidth]{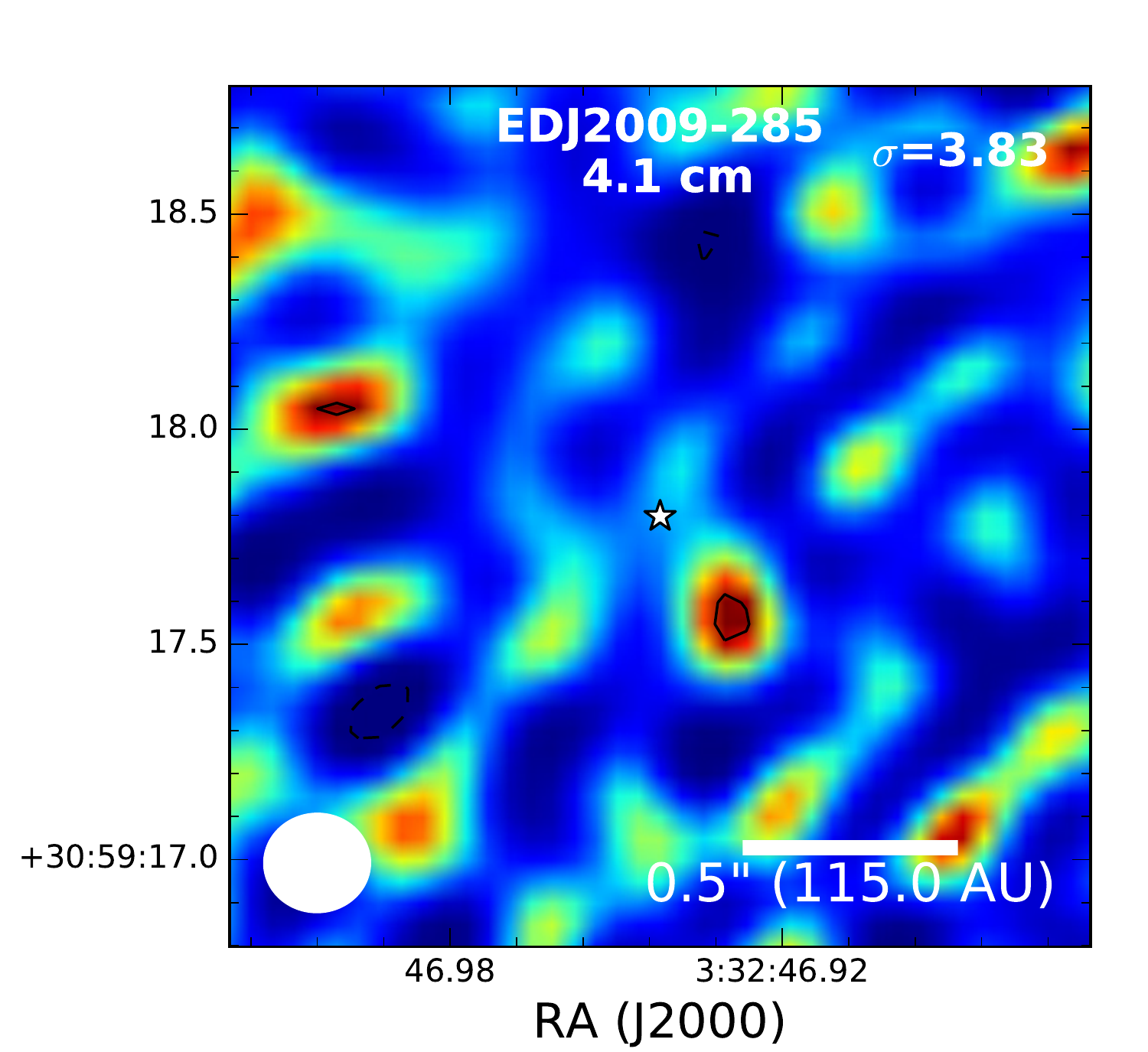}
  \includegraphics[width=0.24\linewidth]{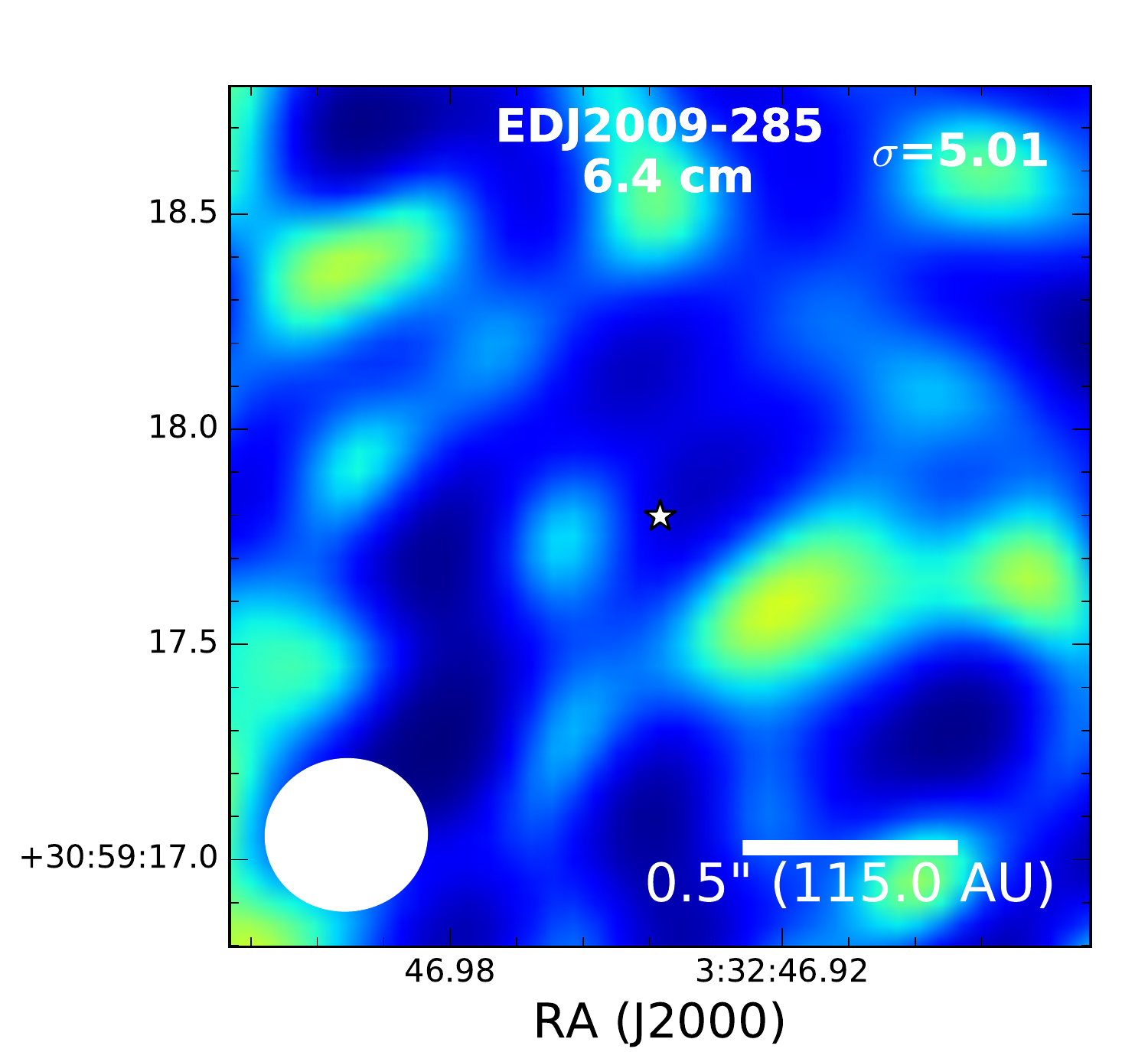}

  \includegraphics[width=0.24\linewidth]{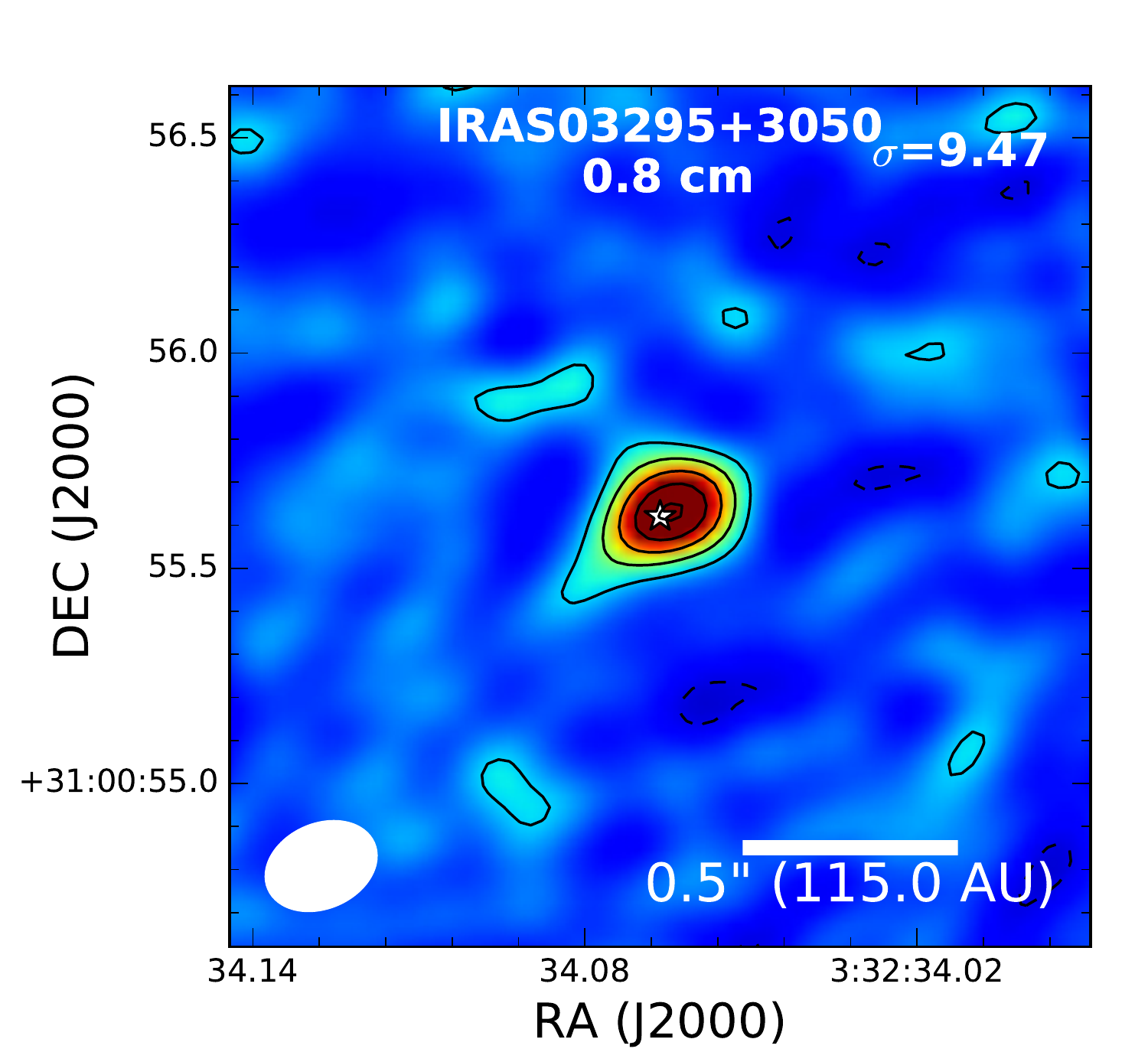}
  \includegraphics[width=0.24\linewidth]{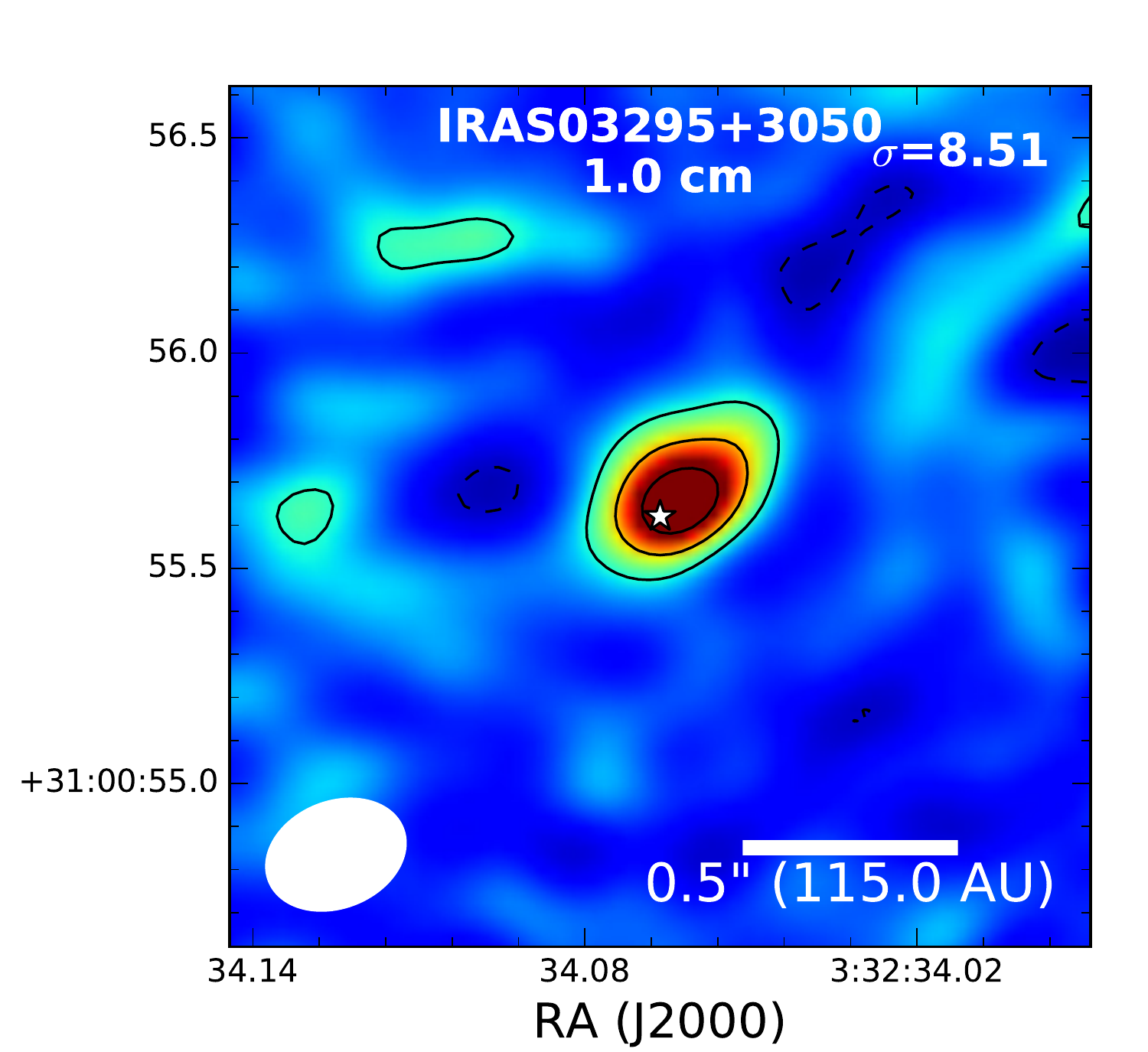}
  \includegraphics[width=0.24\linewidth]{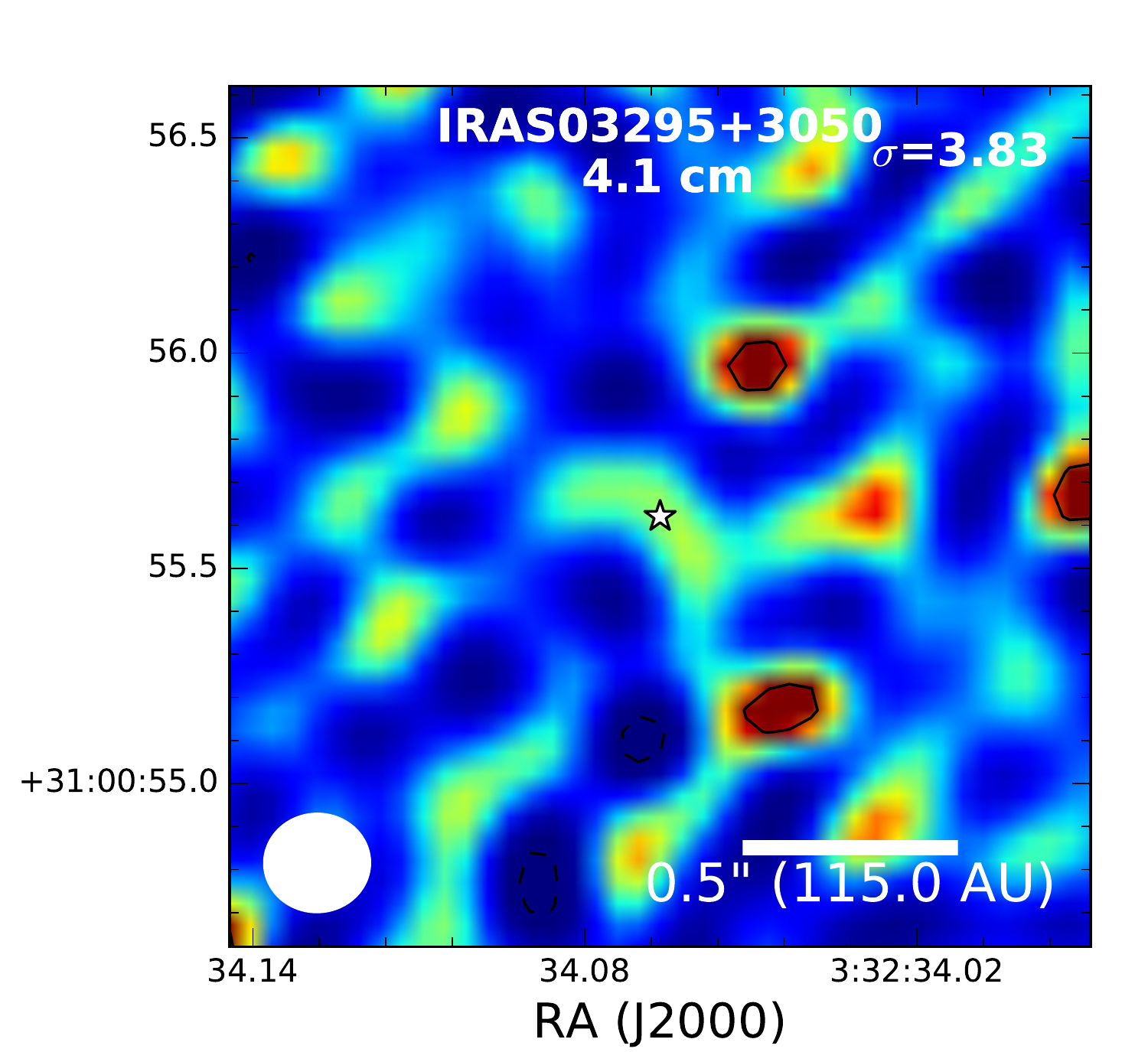}
  \includegraphics[width=0.24\linewidth]{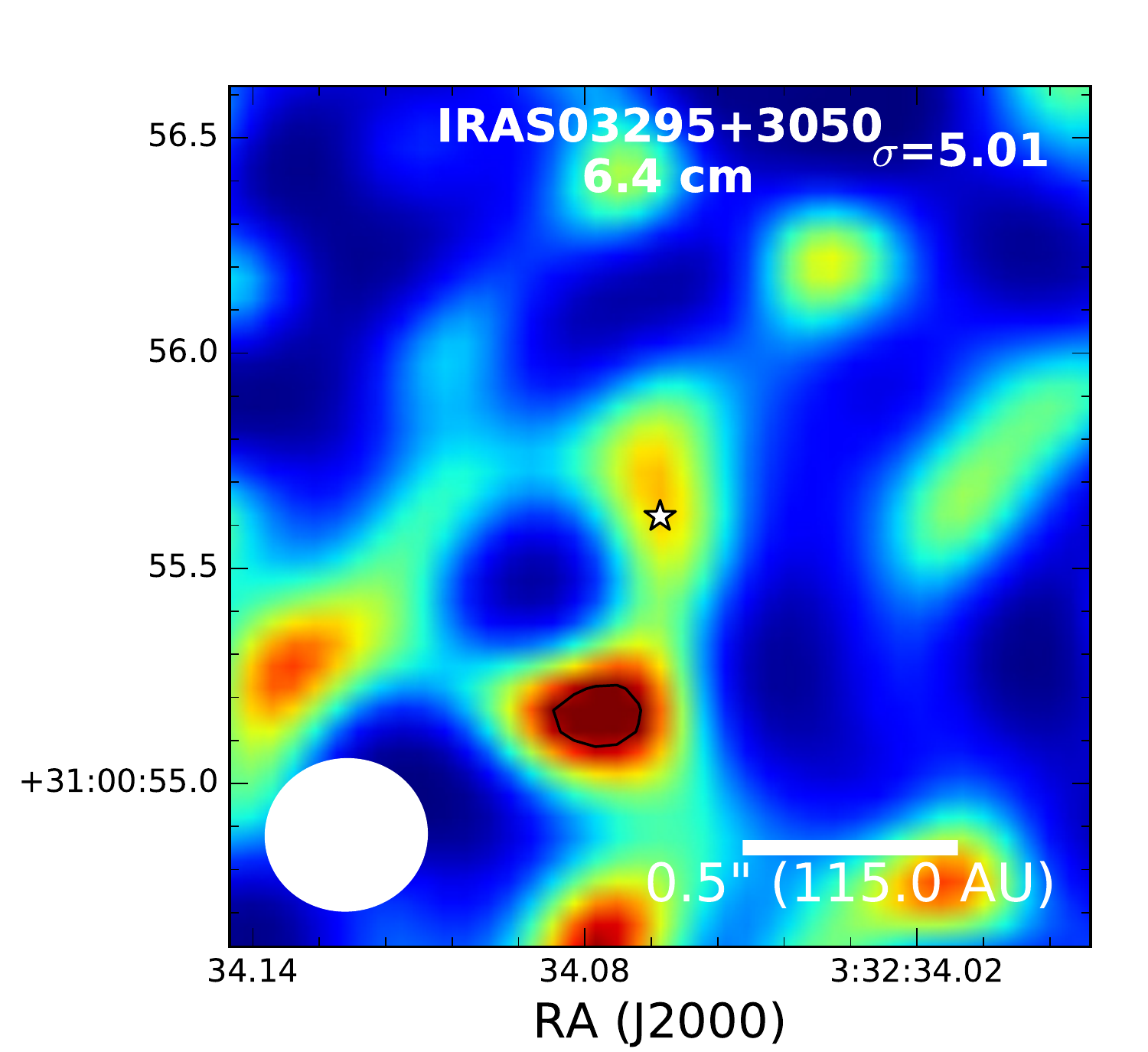}

\end{figure}
\begin{figure}

  \includegraphics[width=0.24\linewidth]{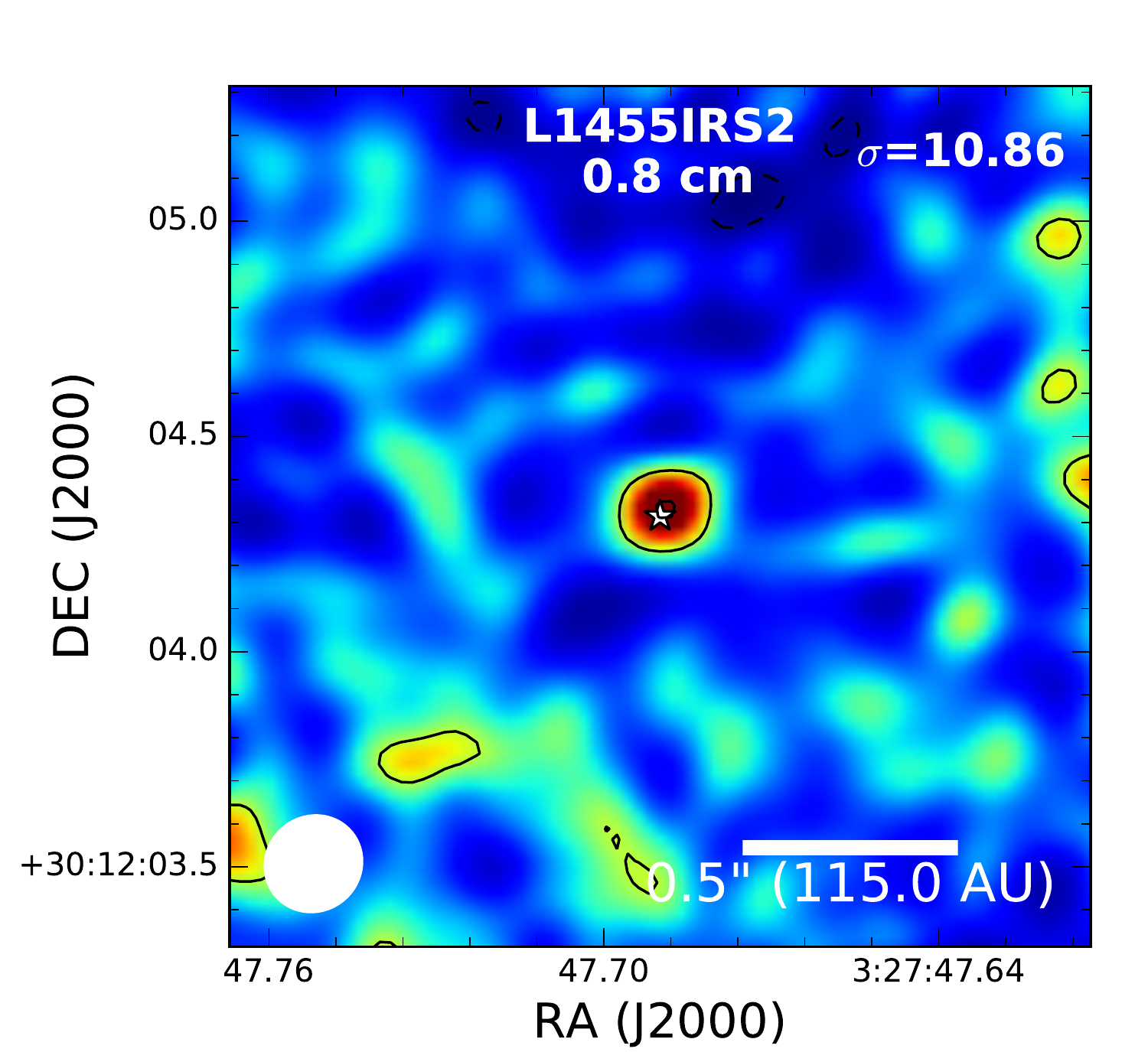}
  \includegraphics[width=0.24\linewidth]{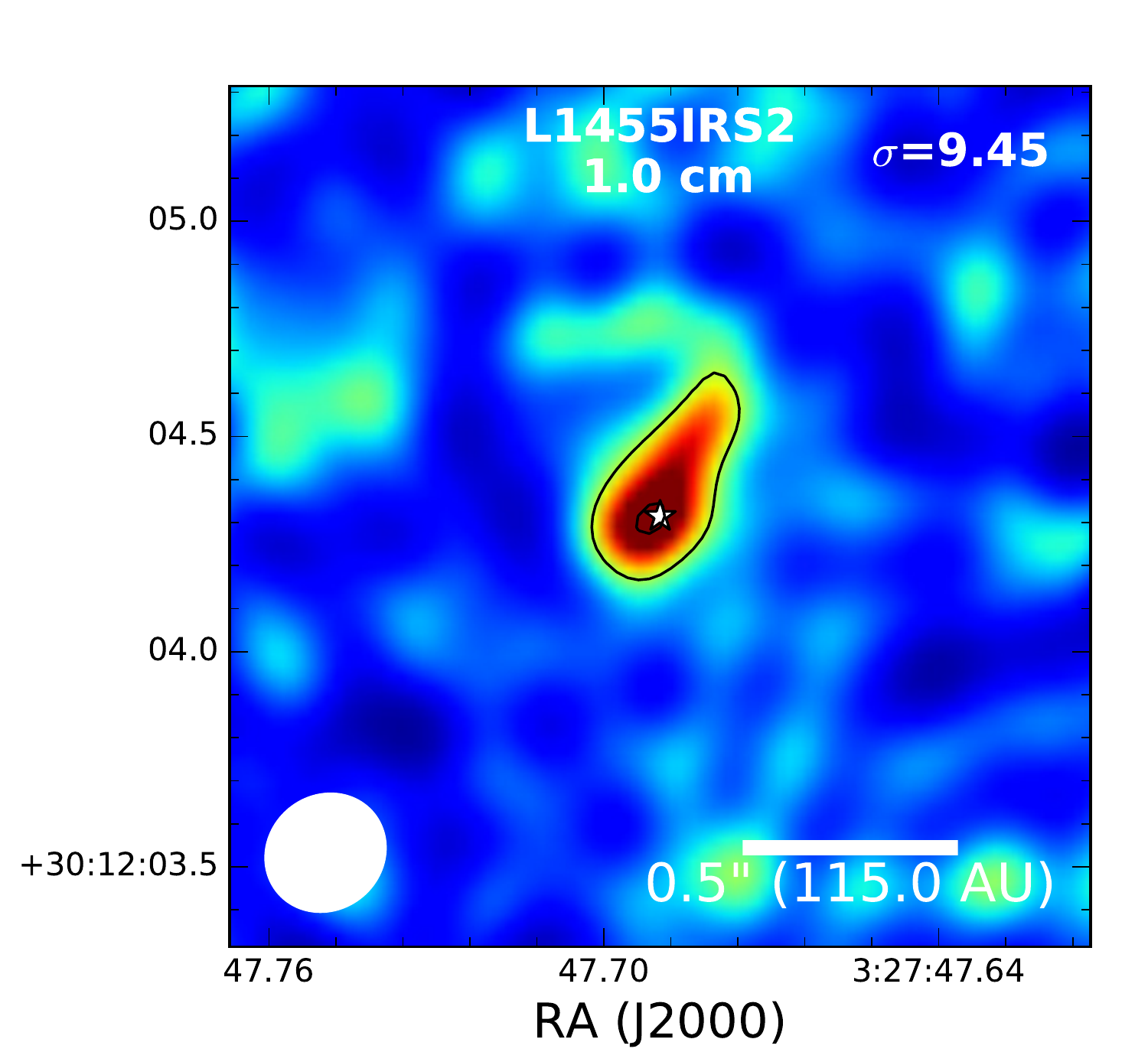}
  \includegraphics[width=0.24\linewidth]{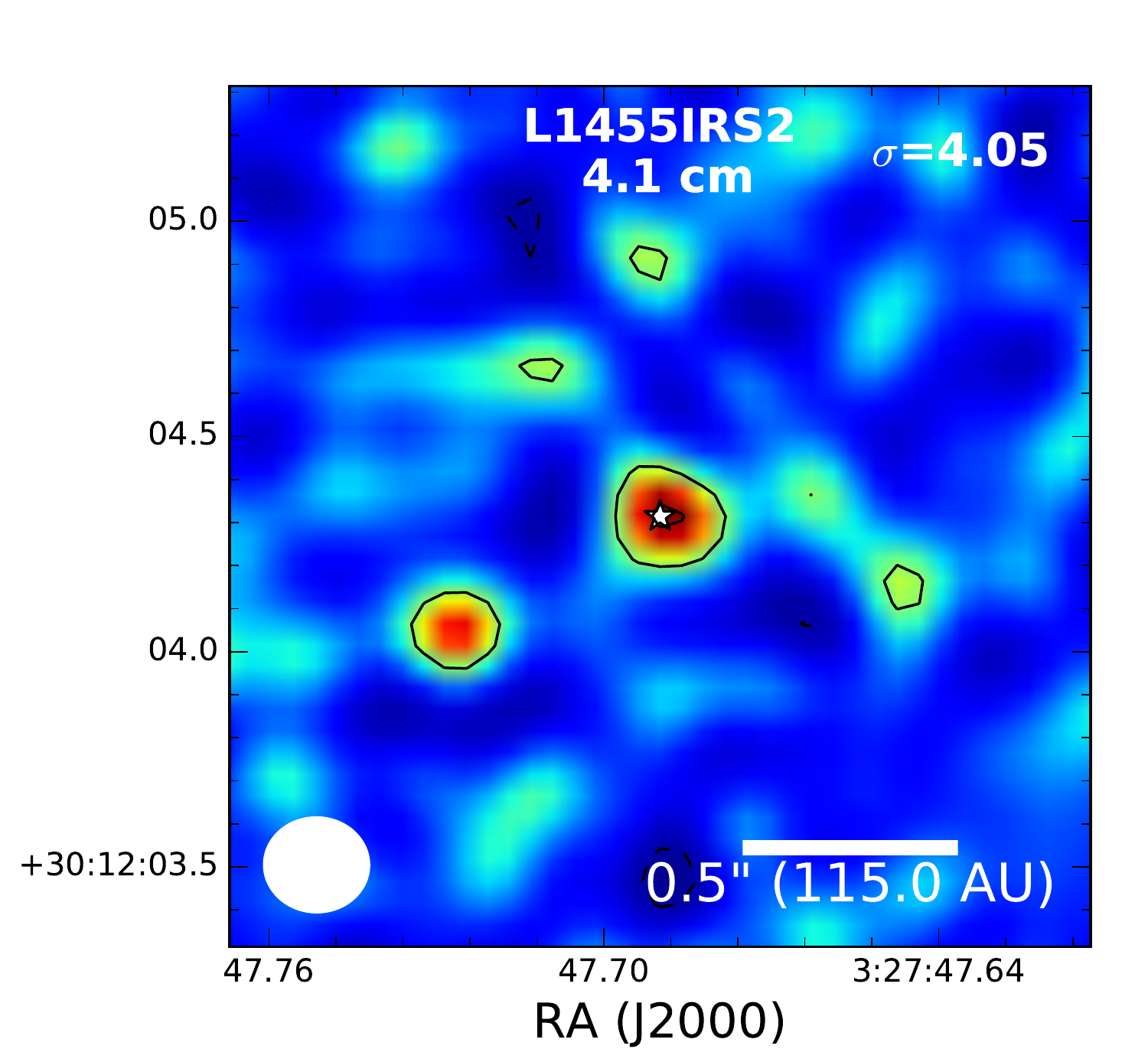}
  \includegraphics[width=0.24\linewidth]{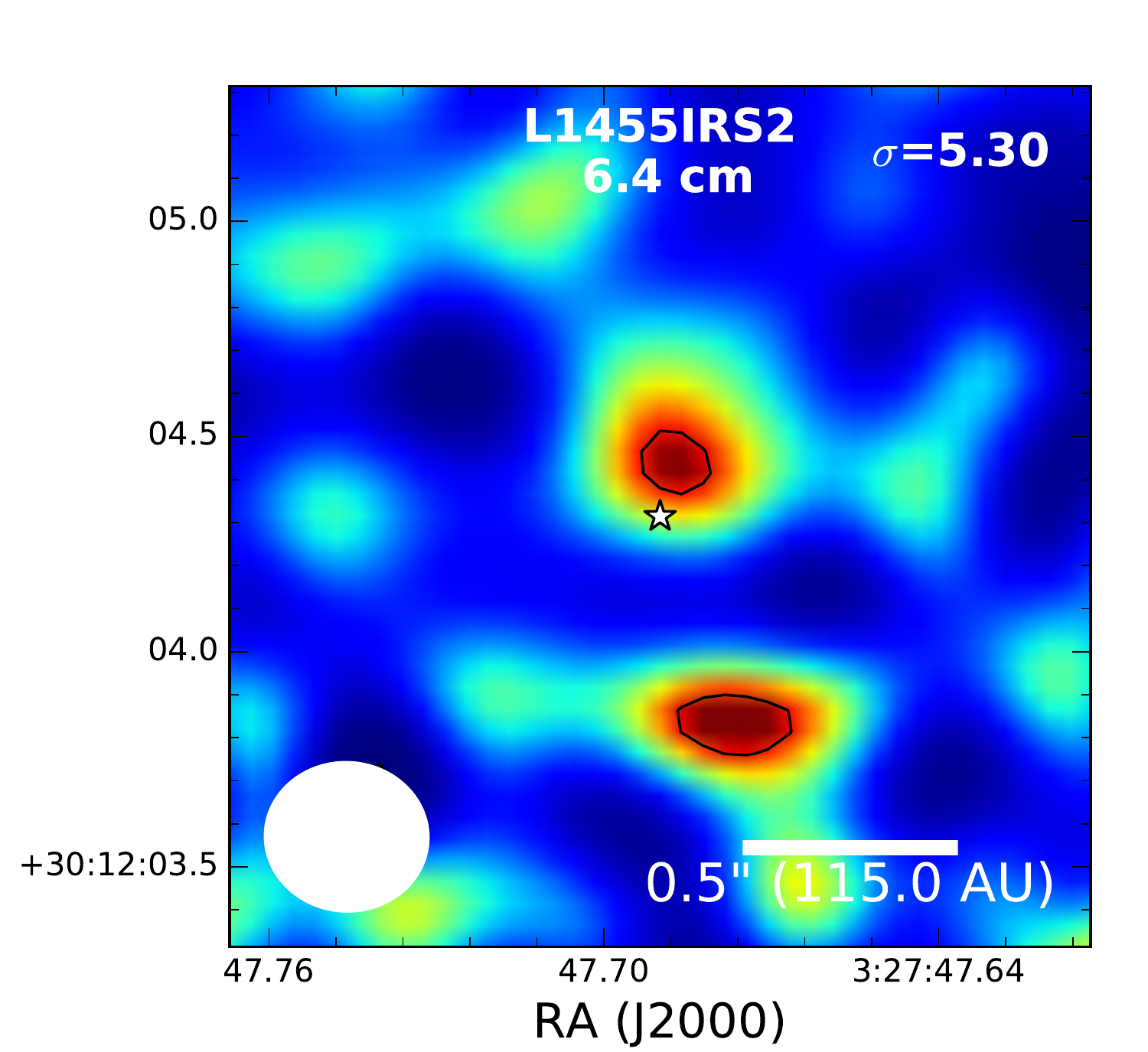}

  \includegraphics[width=0.24\linewidth]{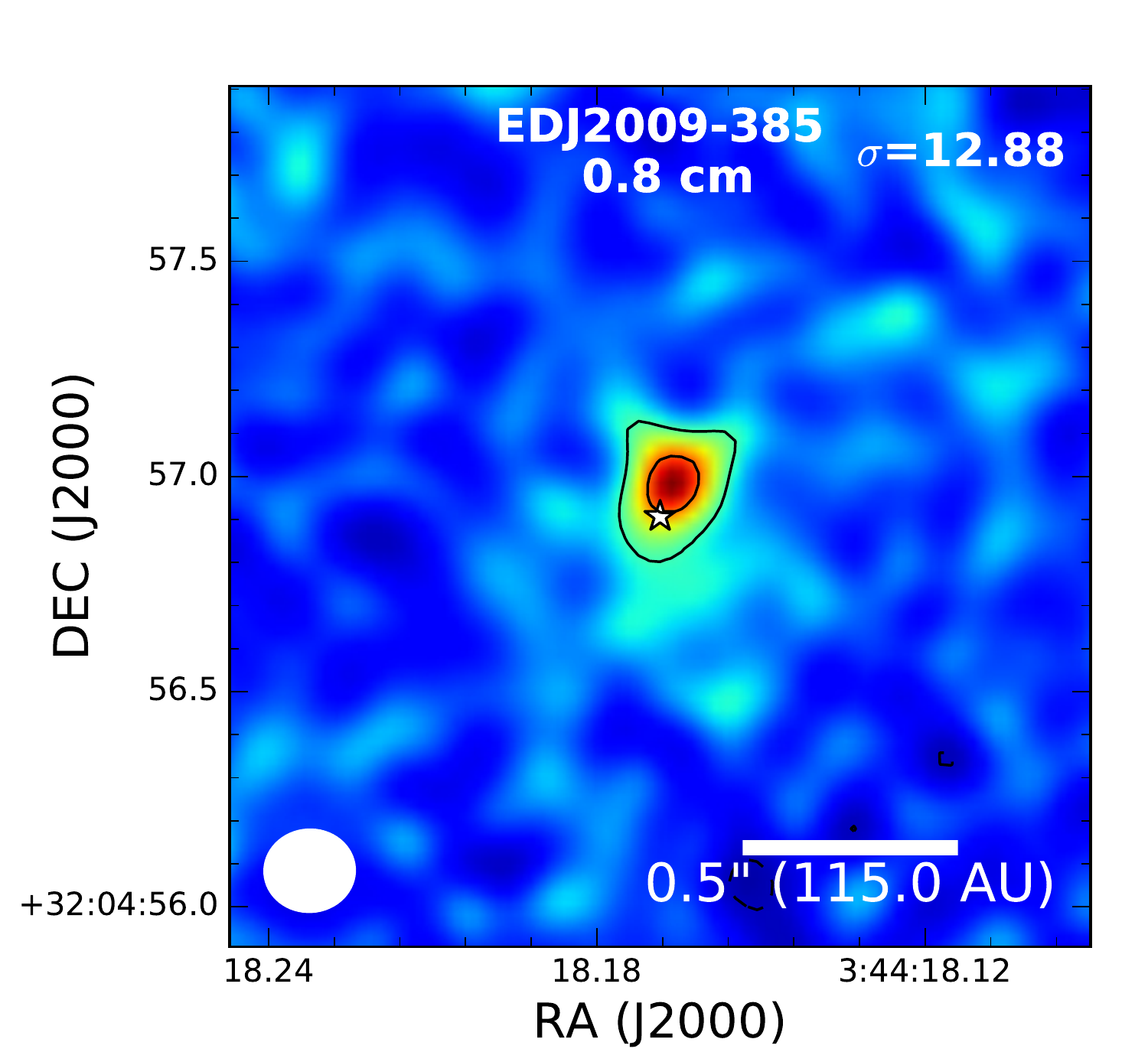}
  \includegraphics[width=0.24\linewidth]{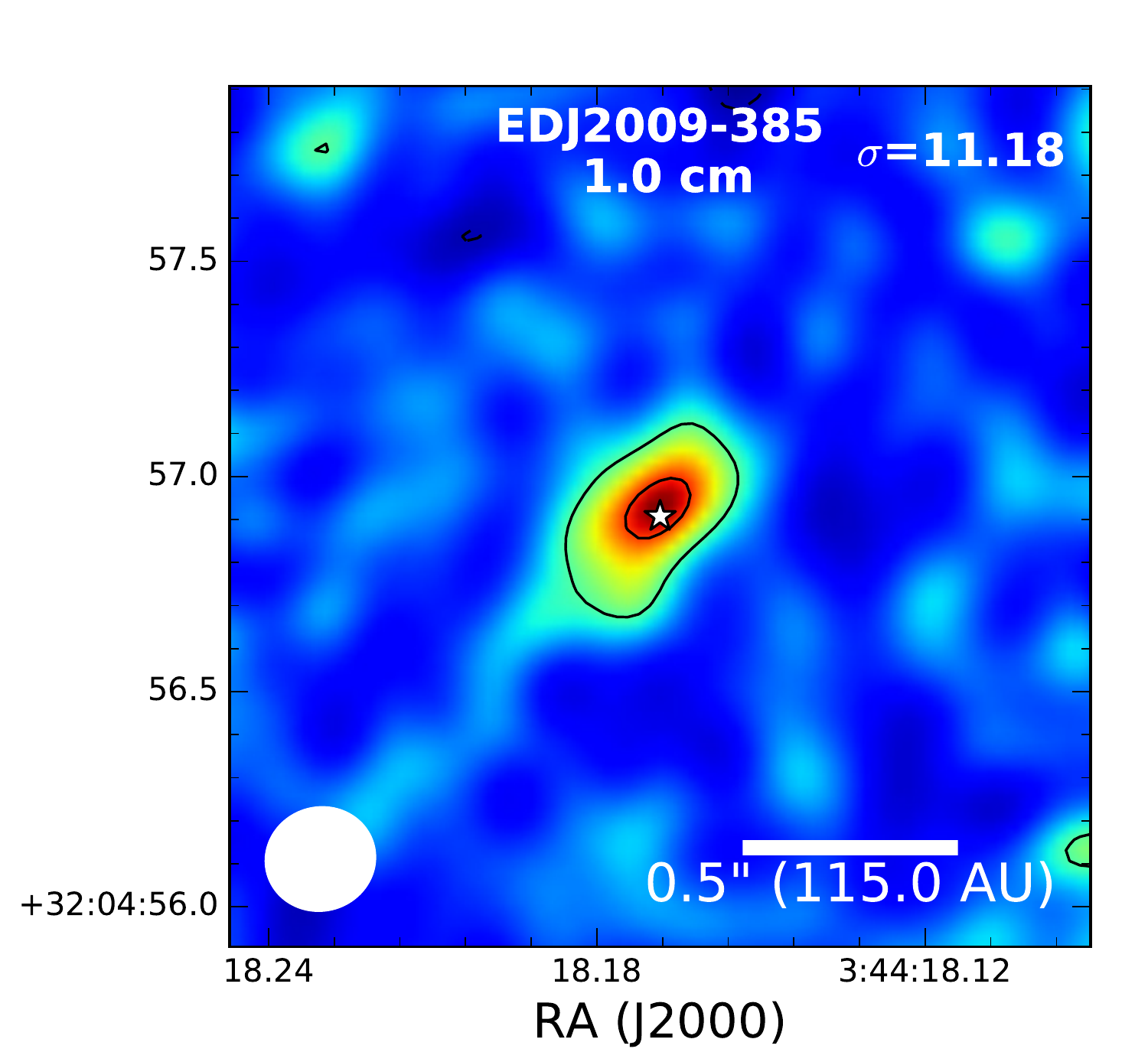}
  \includegraphics[width=0.24\linewidth]{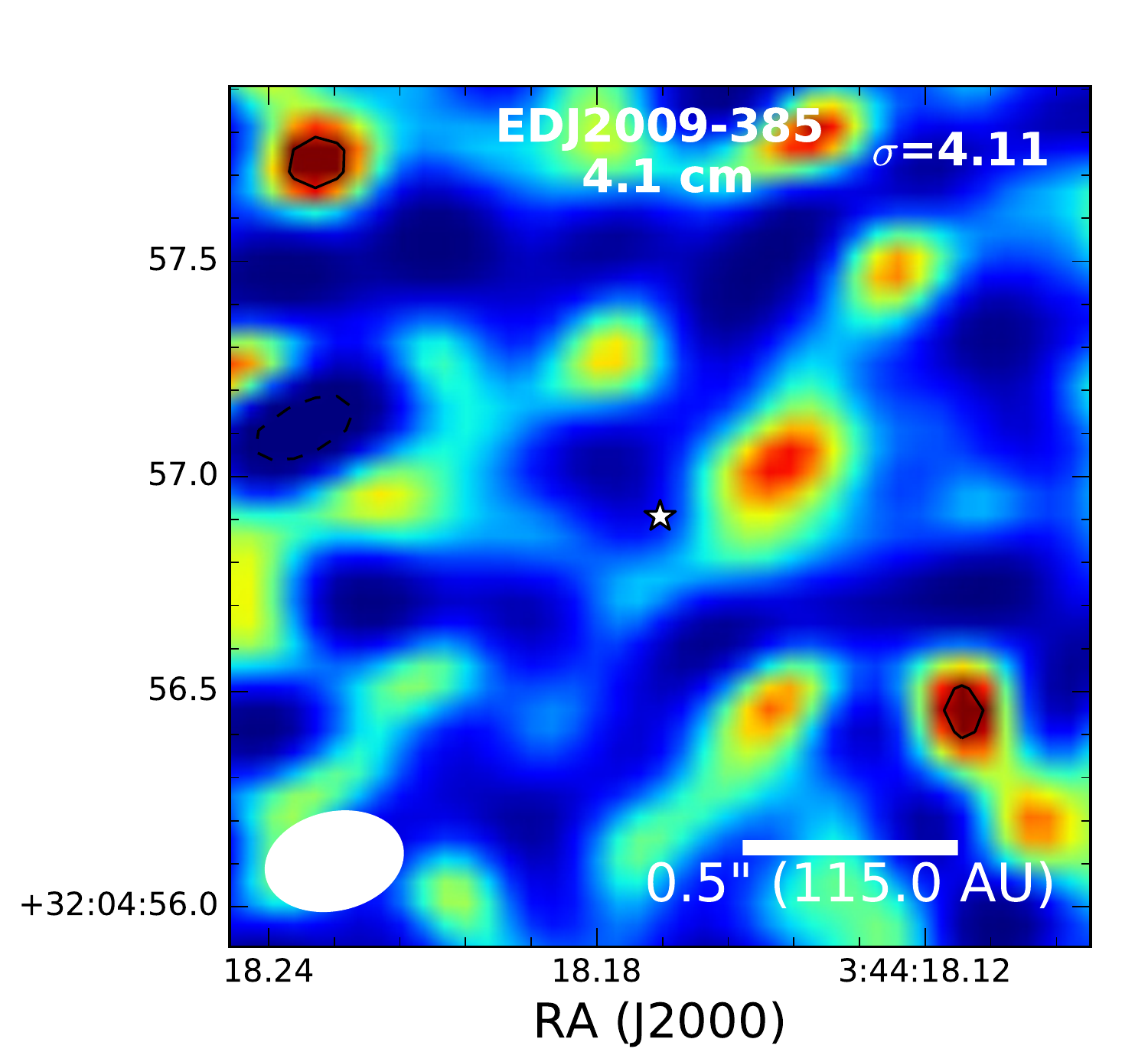}
  \includegraphics[width=0.24\linewidth]{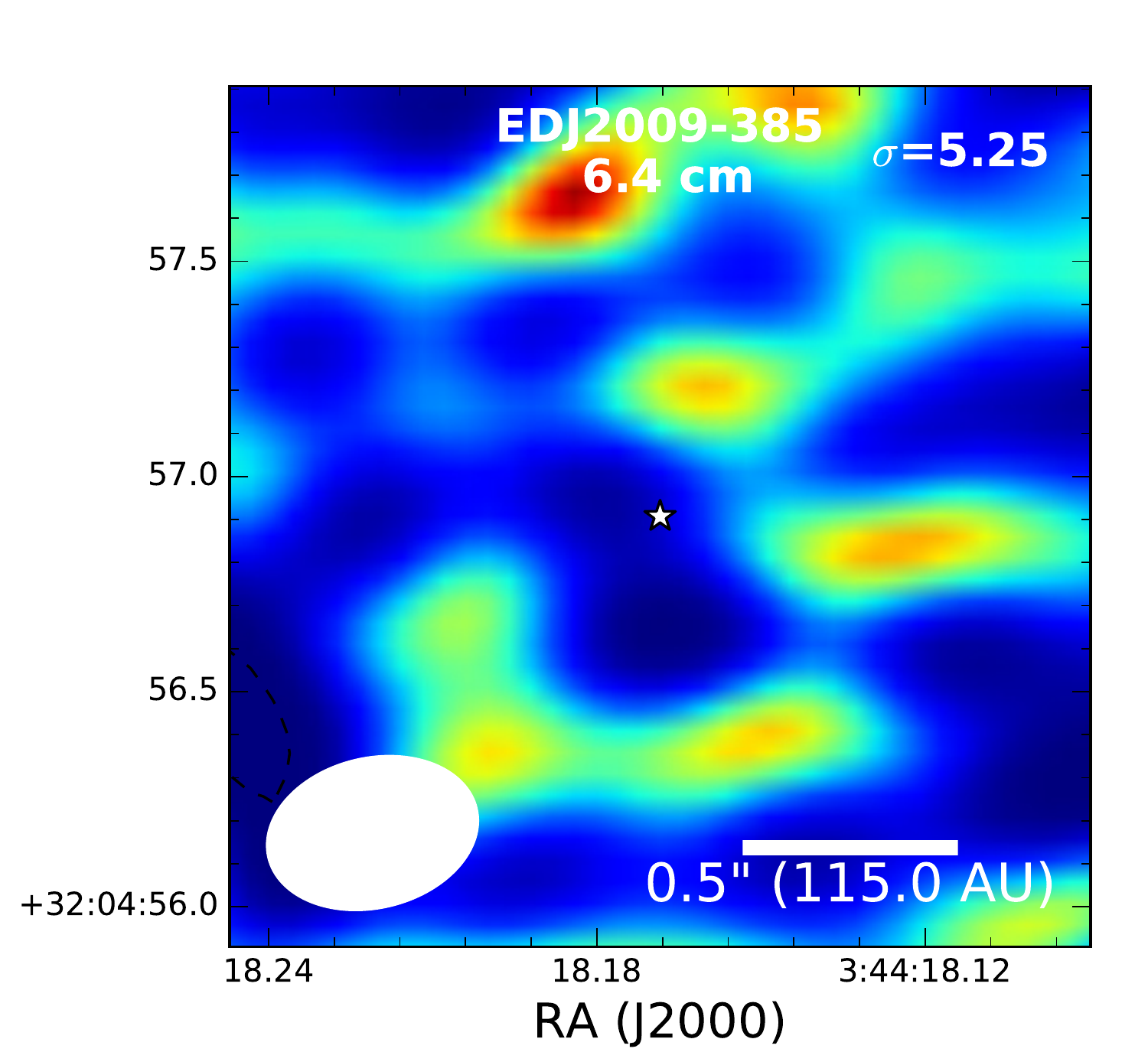}

  \includegraphics[width=0.24\linewidth]{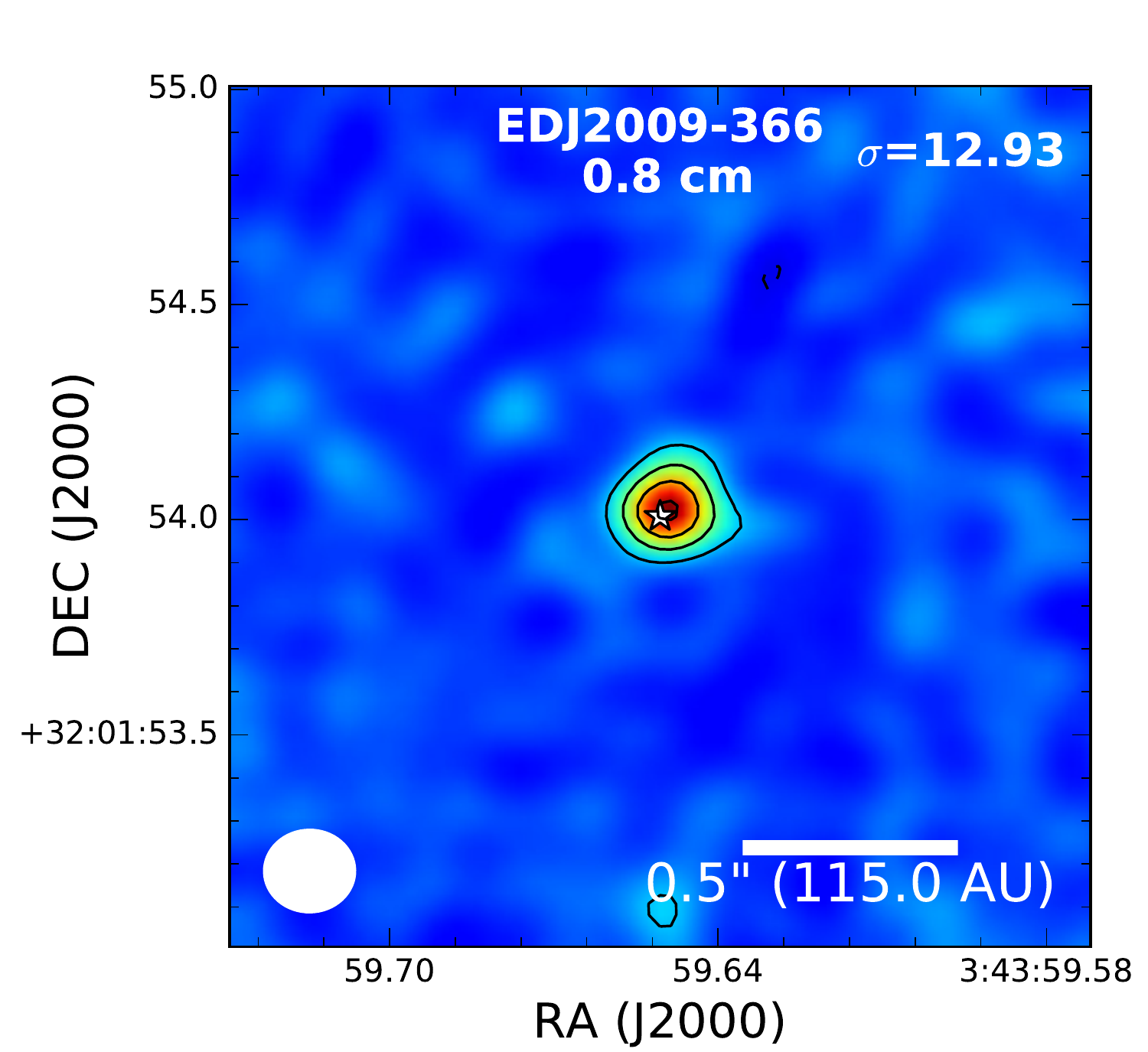}
  \includegraphics[width=0.24\linewidth]{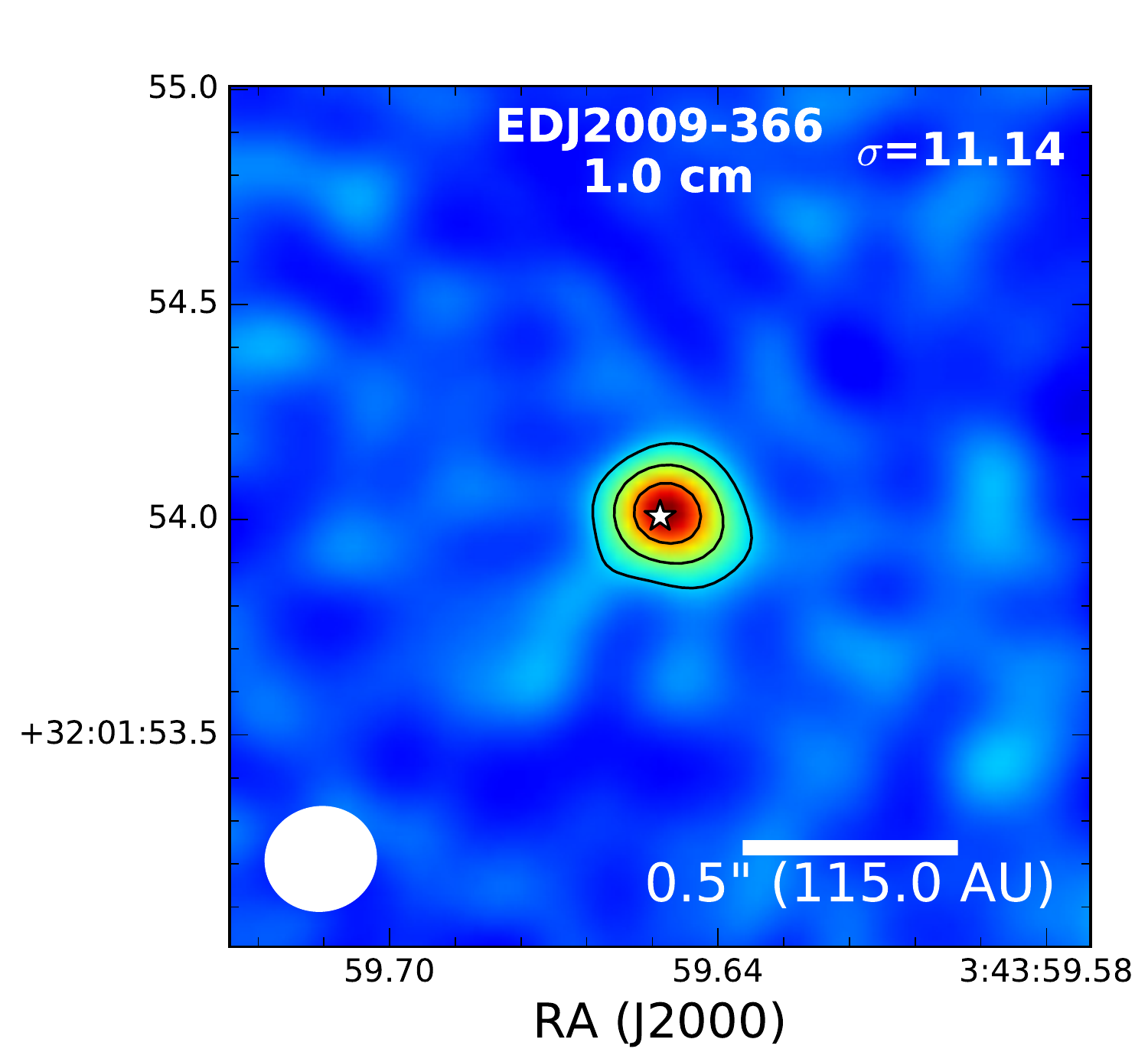}
  \includegraphics[width=0.24\linewidth]{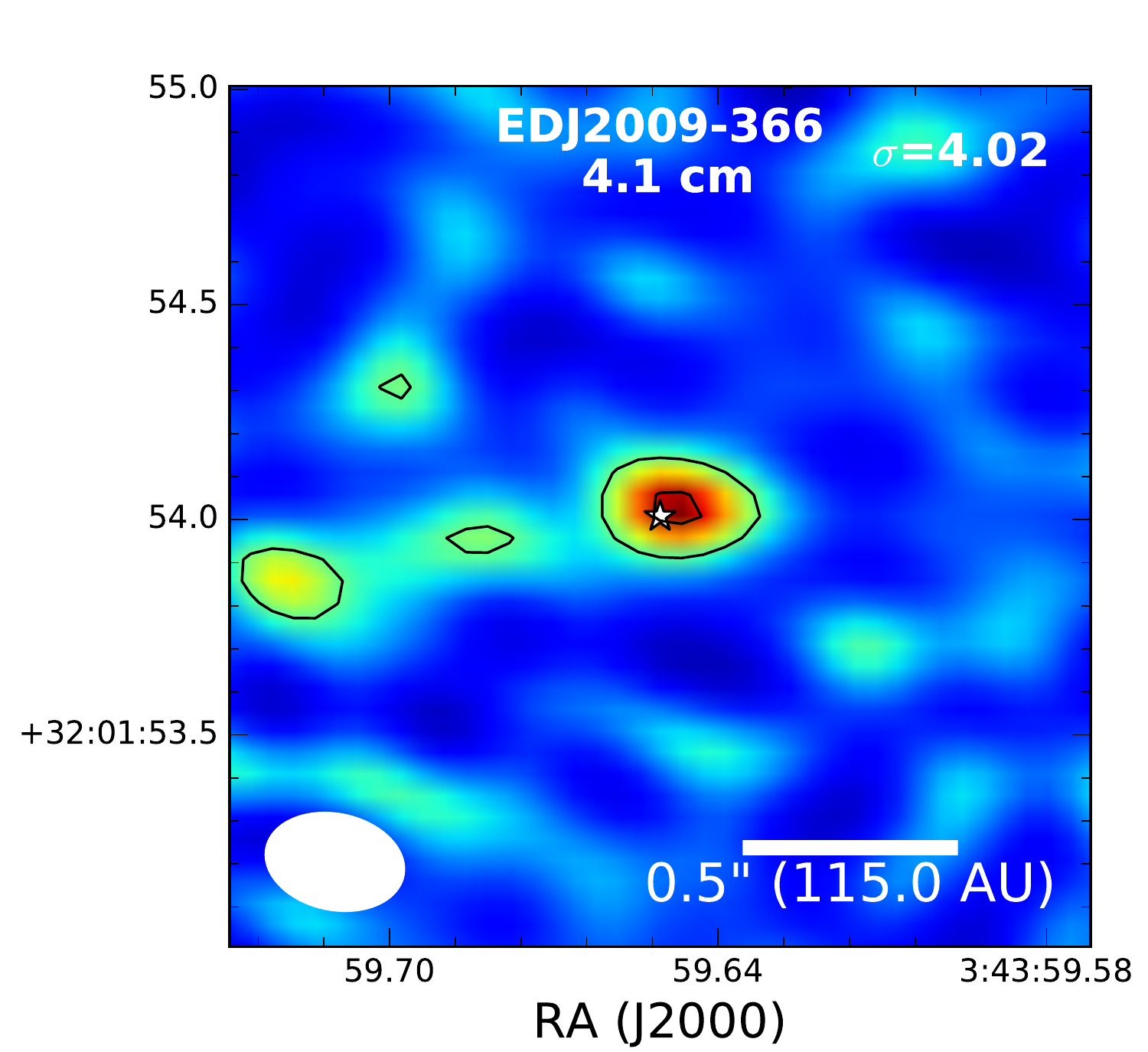}
  \includegraphics[width=0.24\linewidth]{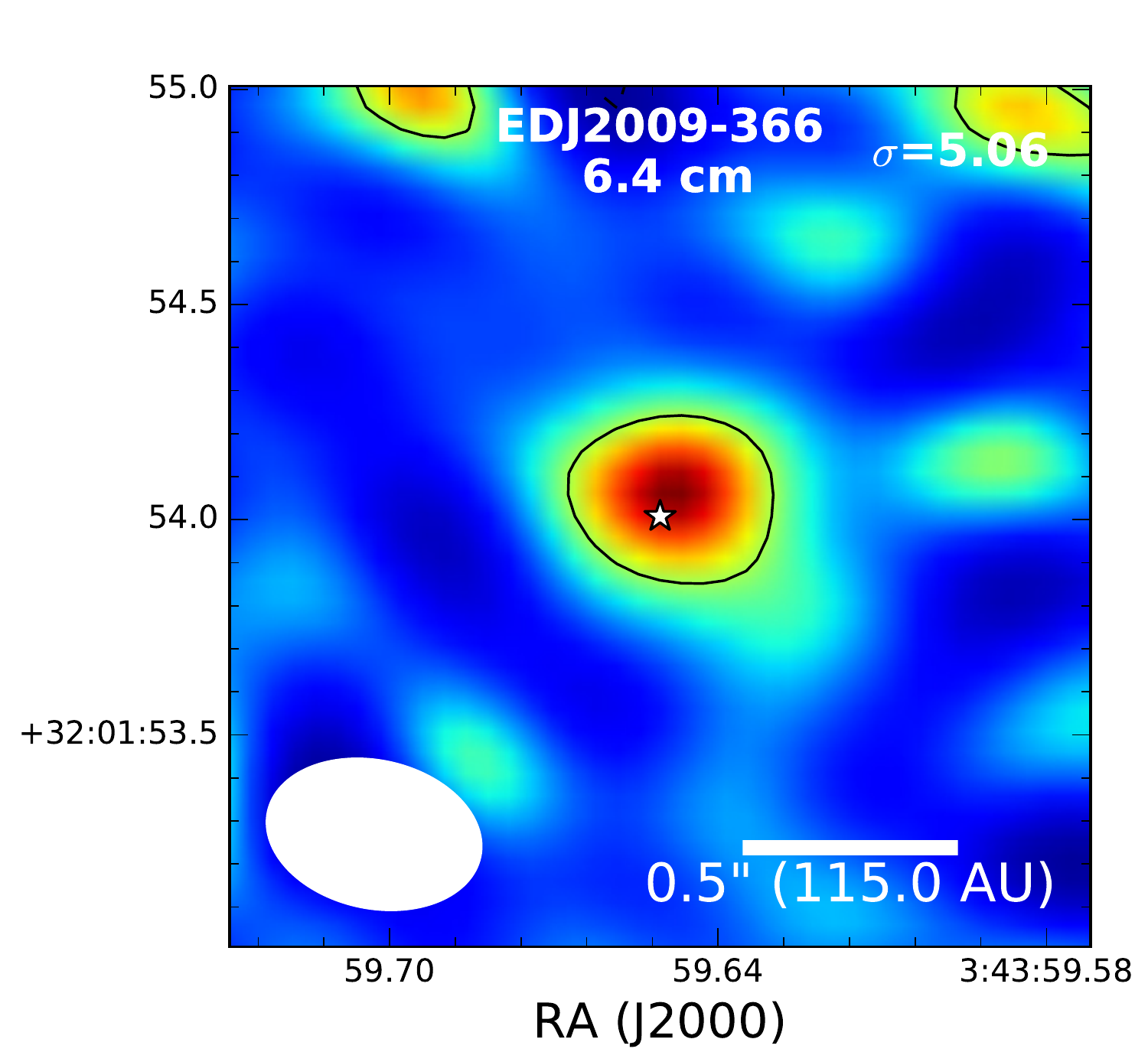}

  \includegraphics[width=0.24\linewidth]{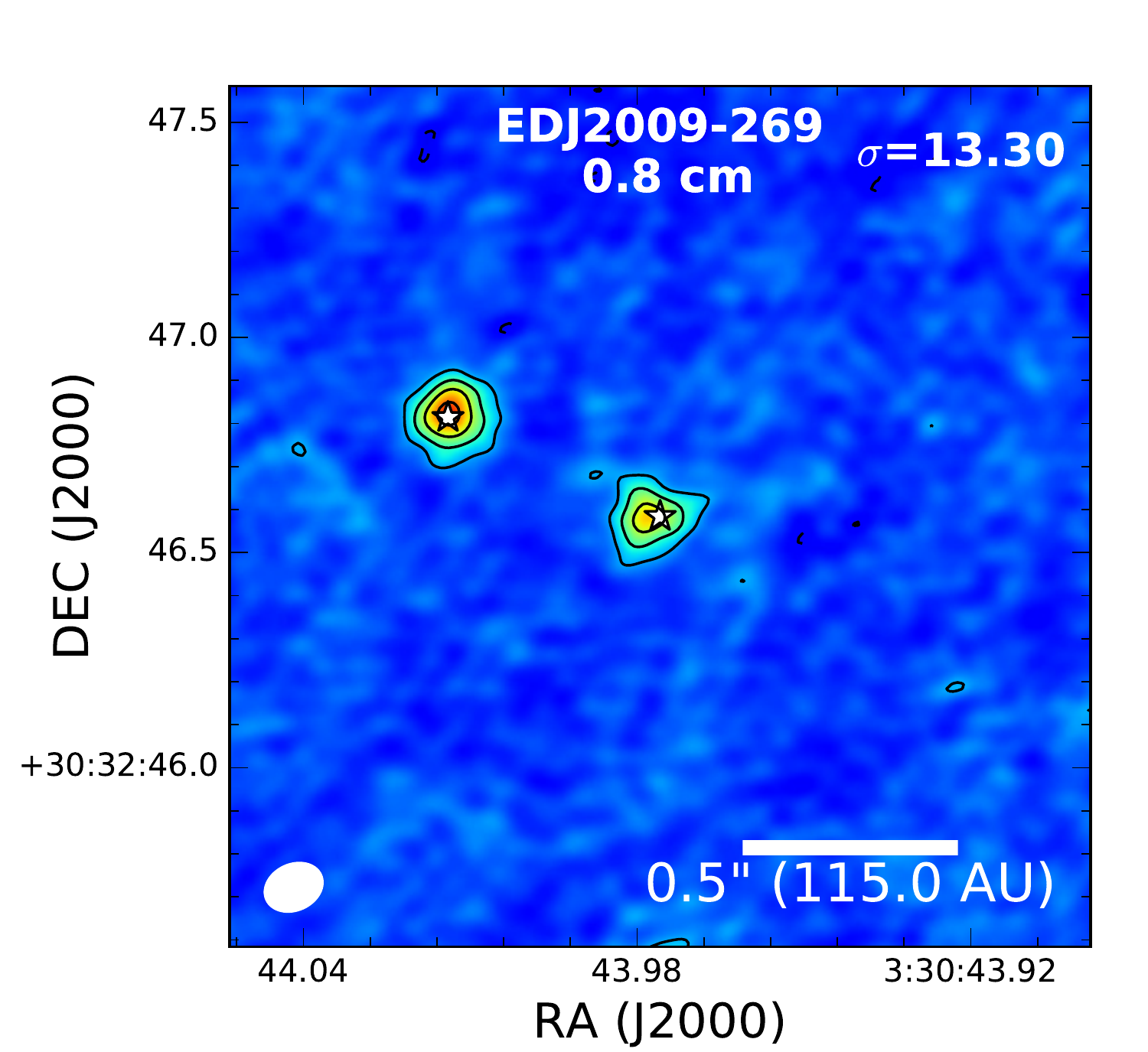}
  \includegraphics[width=0.24\linewidth]{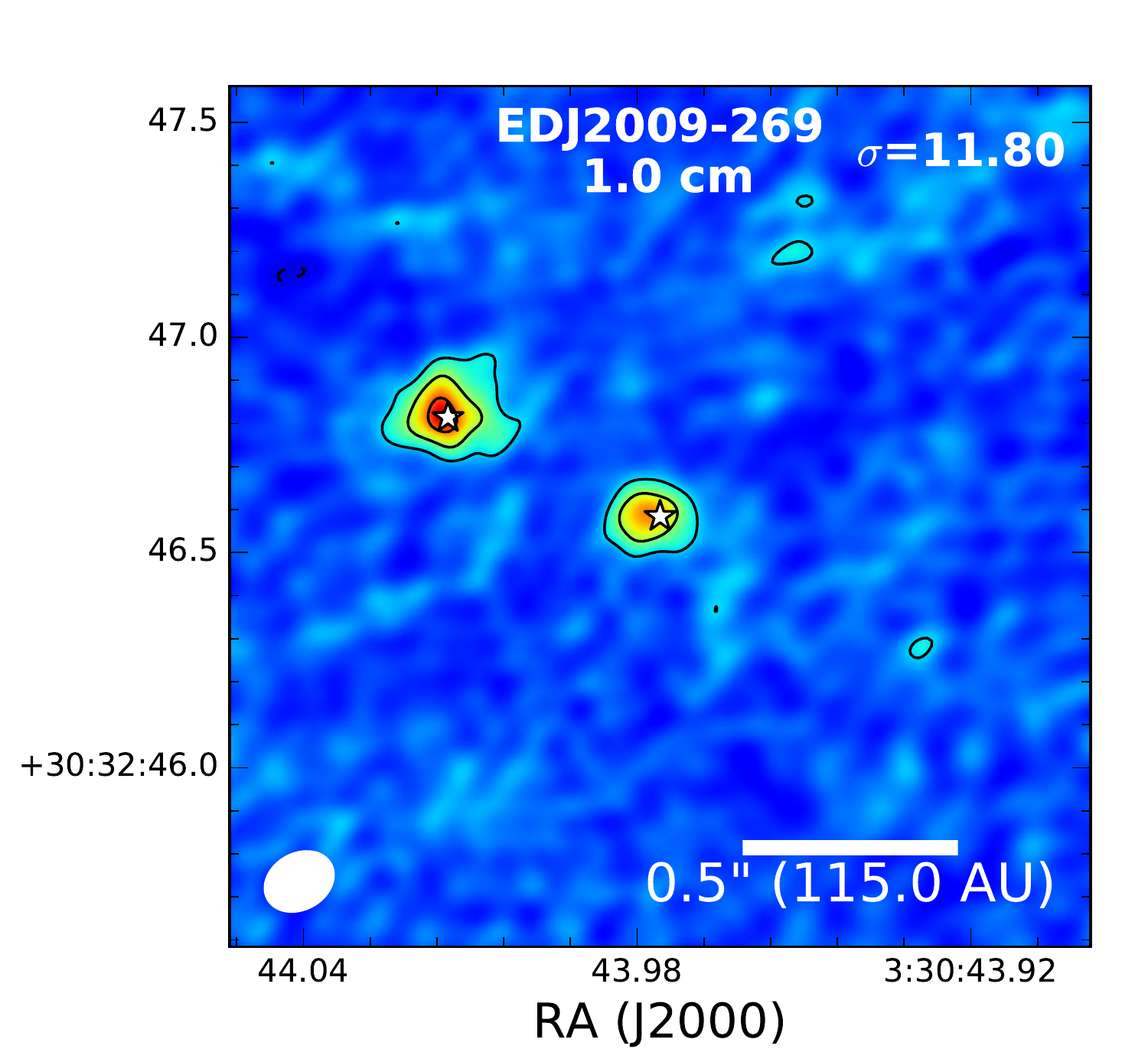}
  \includegraphics[width=0.24\linewidth]{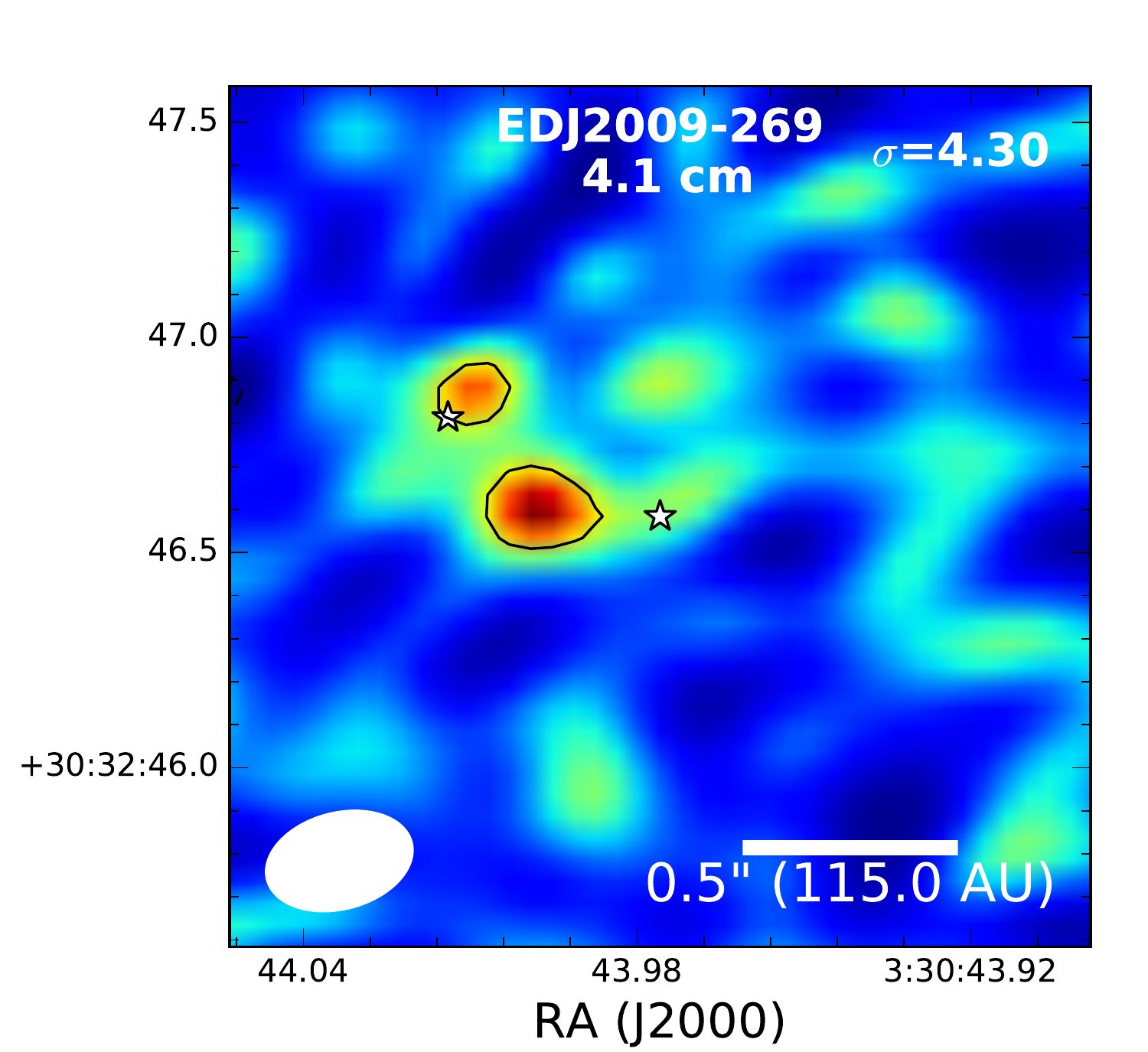}
  \includegraphics[width=0.24\linewidth]{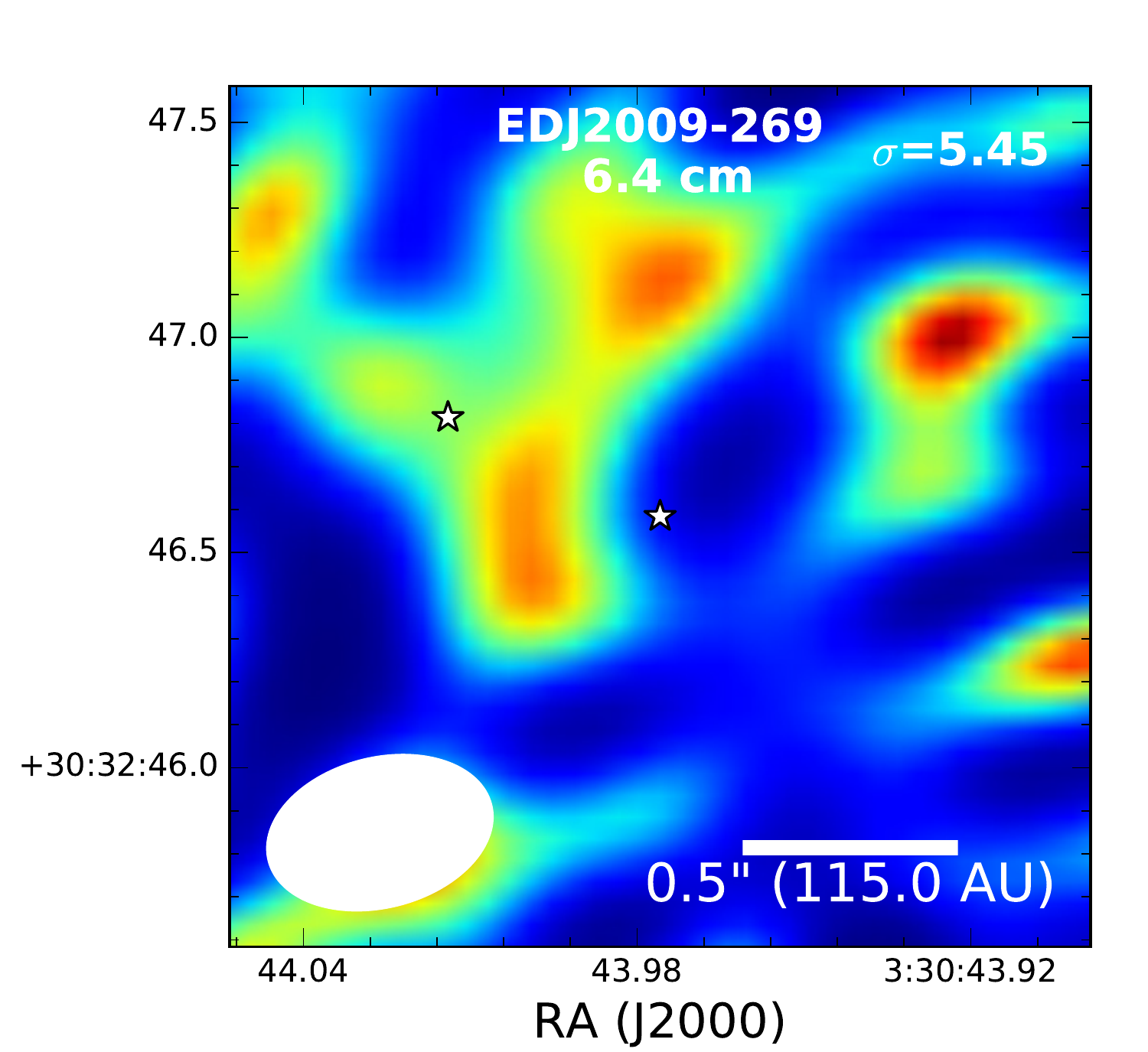}

  \includegraphics[width=0.24\linewidth]{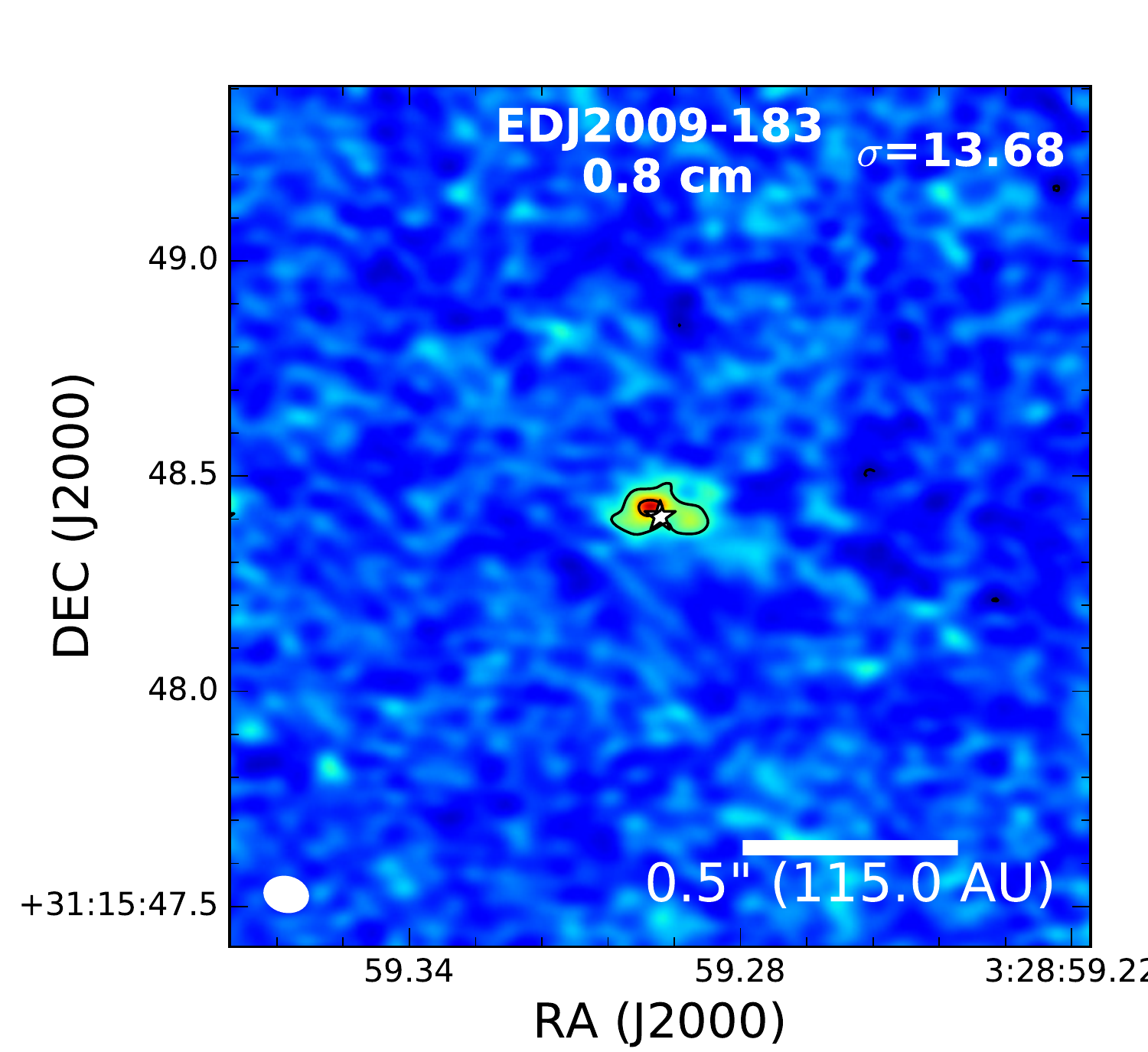}
  \includegraphics[width=0.24\linewidth]{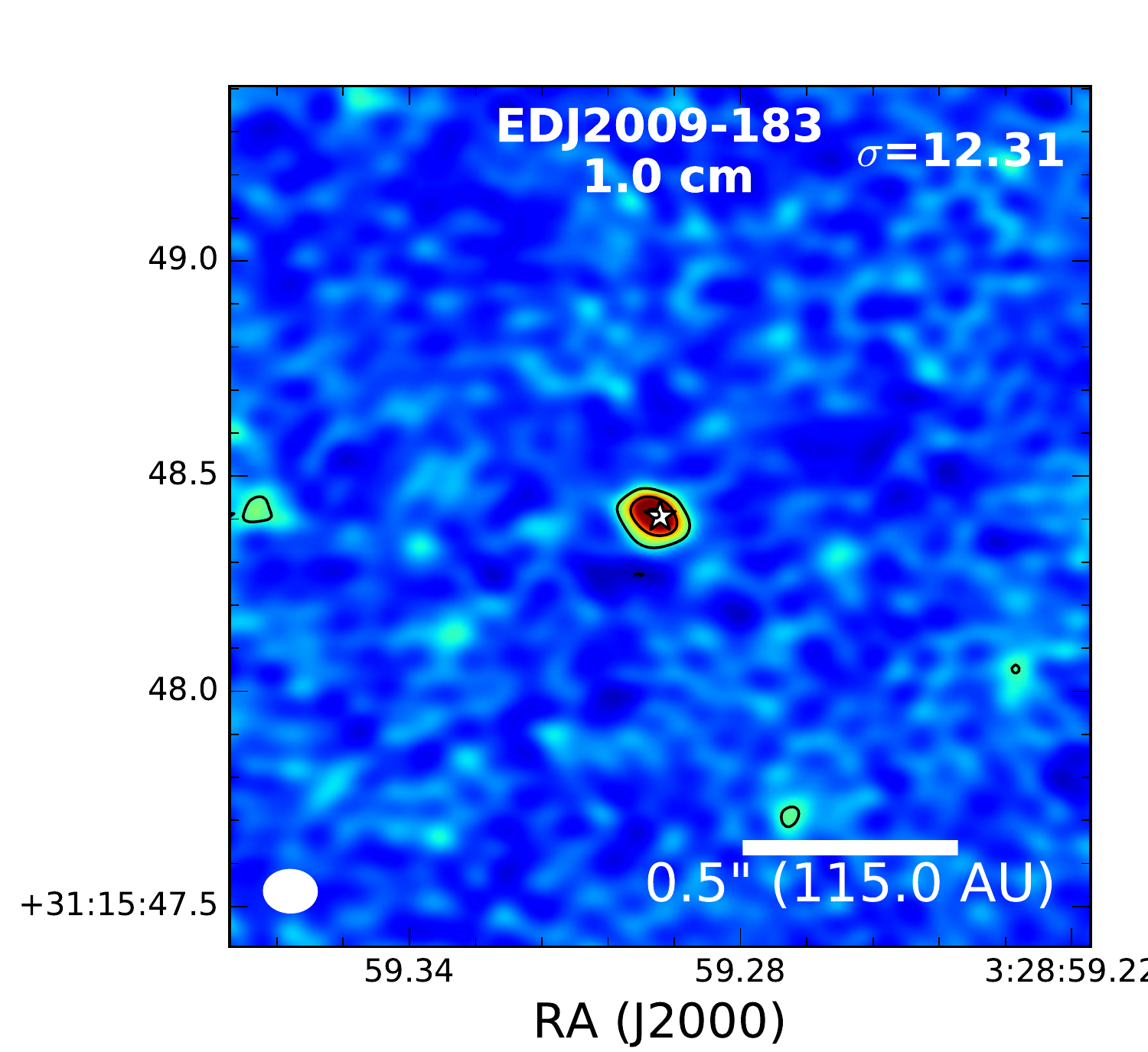}
  \includegraphics[width=0.24\linewidth]{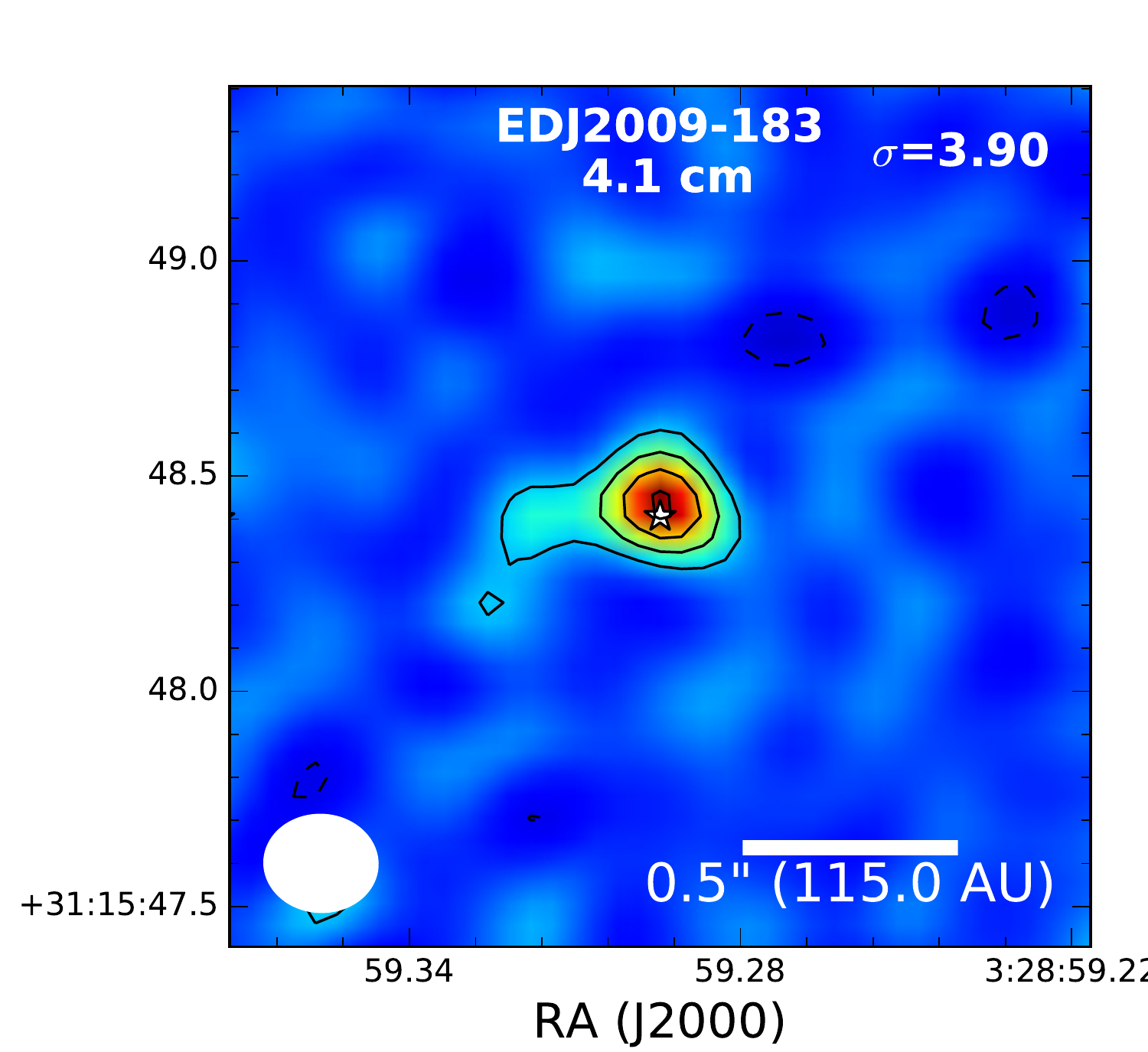}
  \includegraphics[width=0.24\linewidth]{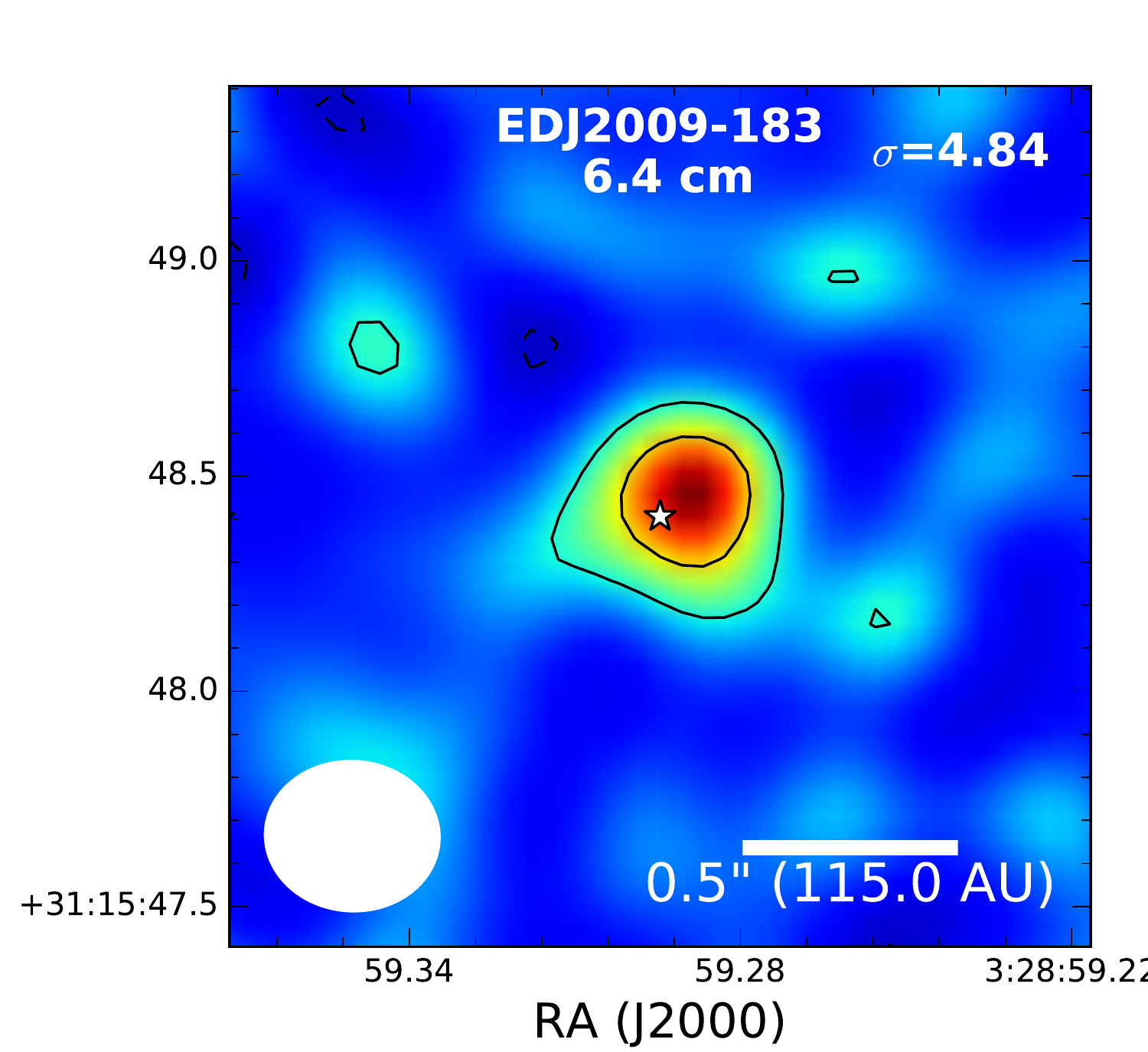}

  \includegraphics[width=0.24\linewidth]{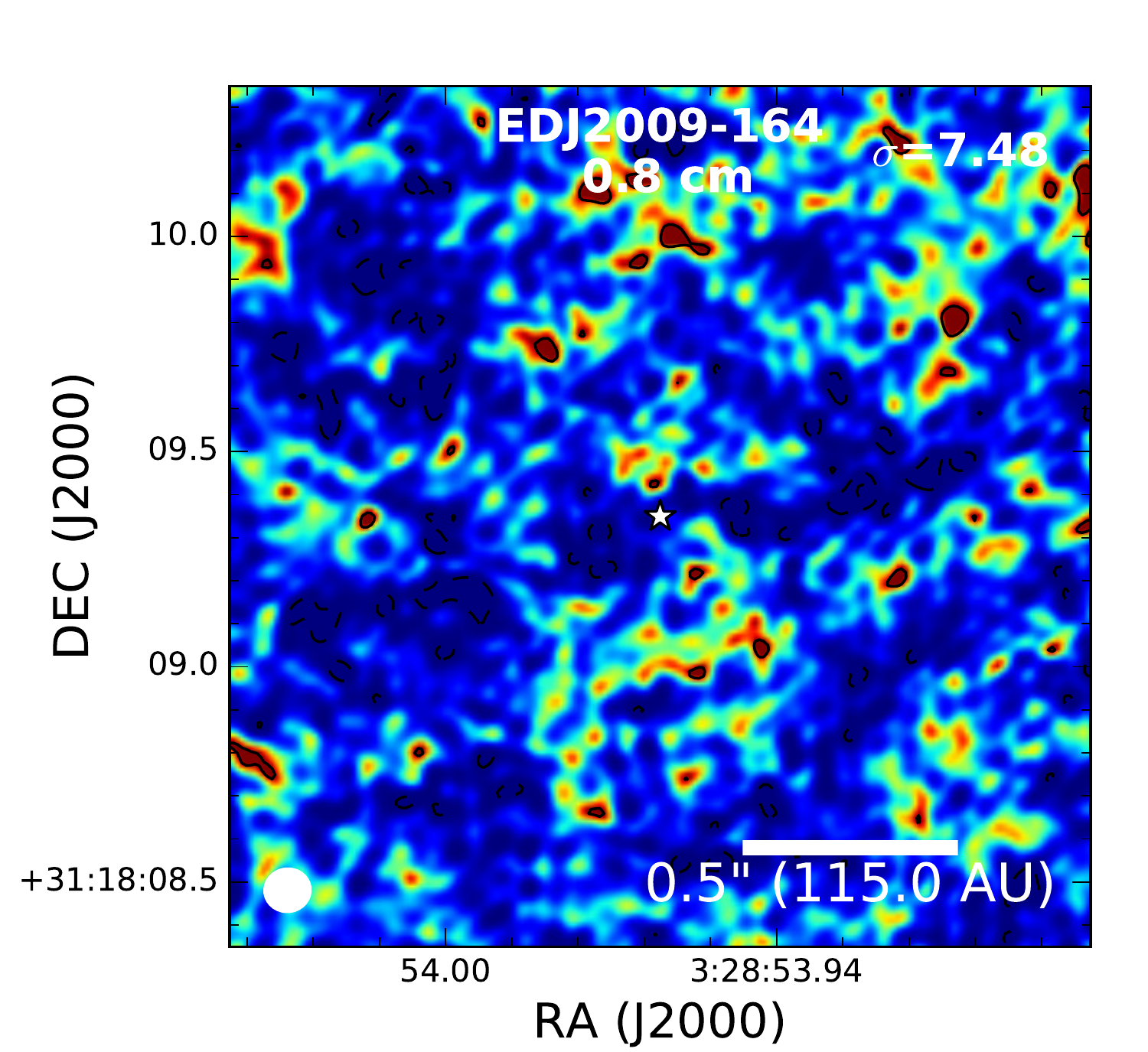}
  \includegraphics[width=0.24\linewidth]{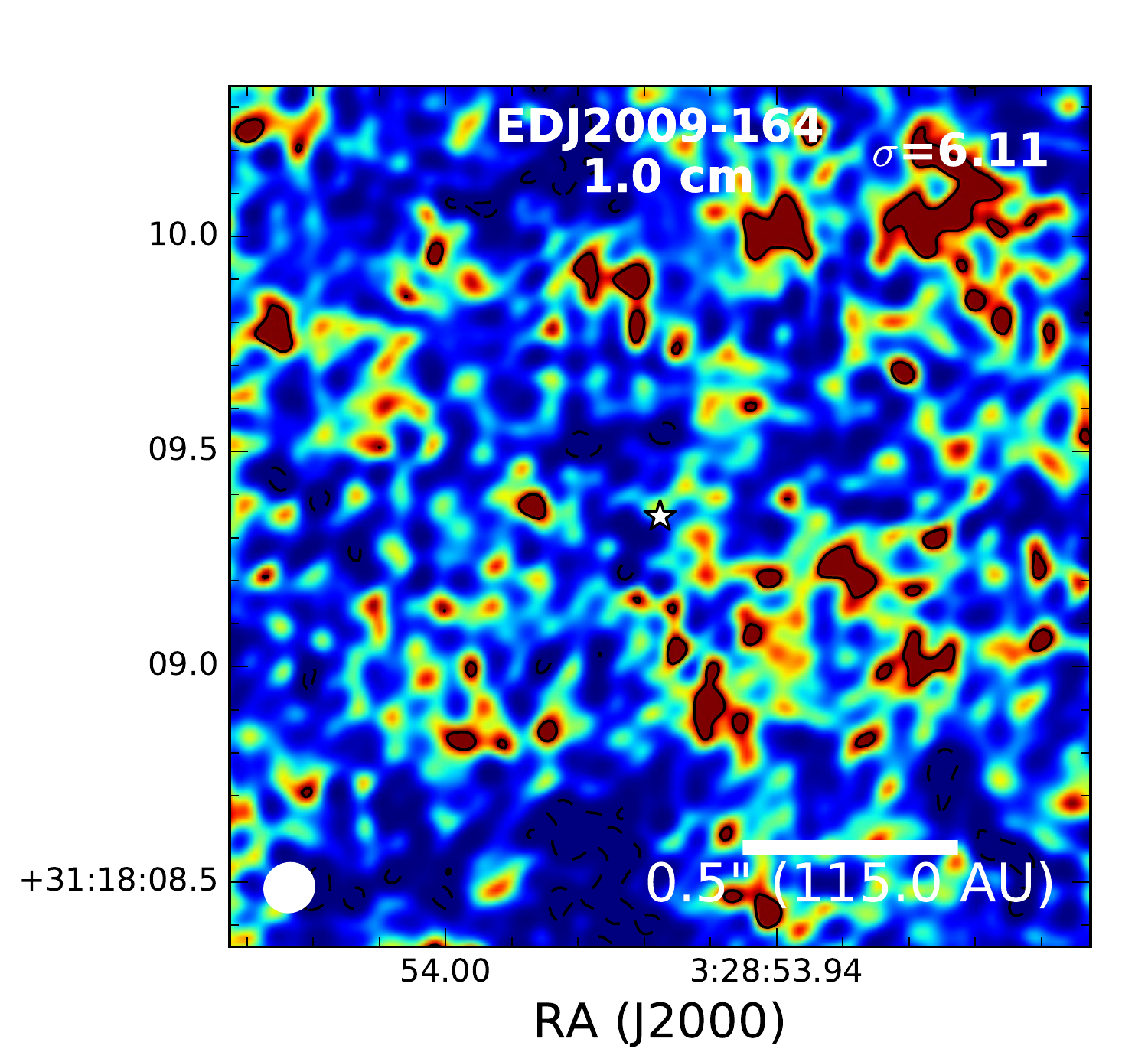}
  \includegraphics[width=0.24\linewidth]{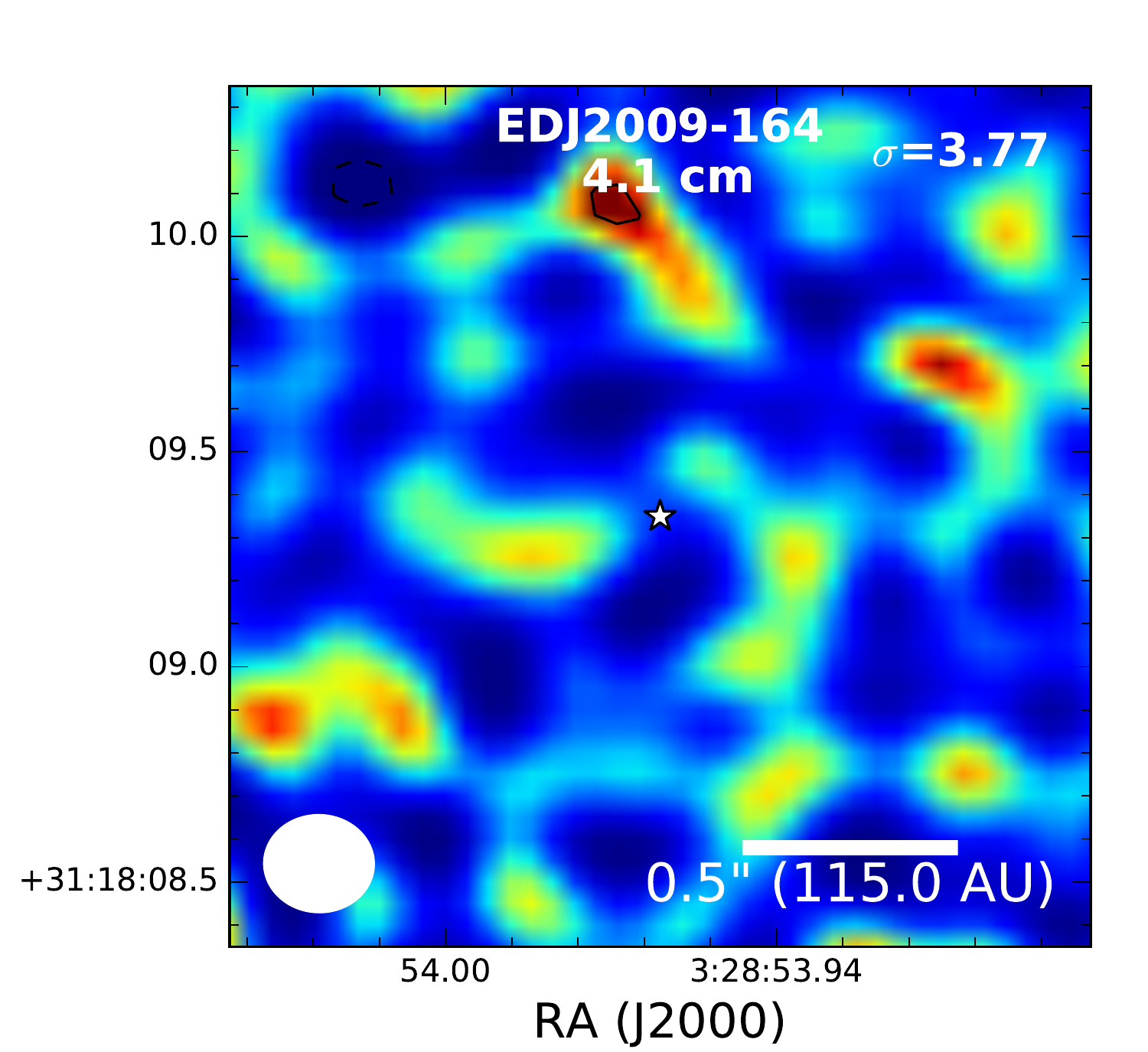}
  \includegraphics[width=0.24\linewidth]{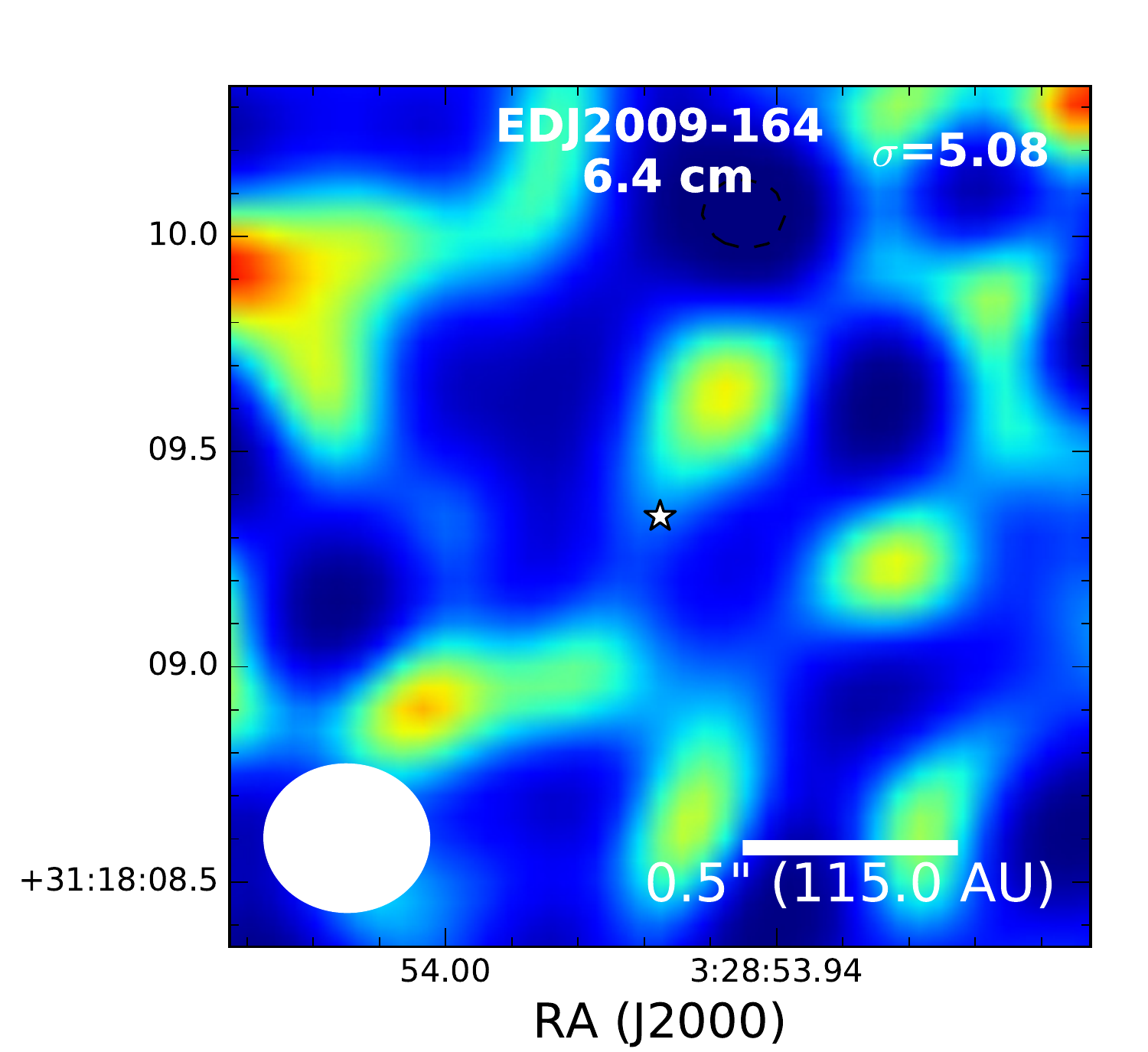}

\end{figure}
\begin{figure}

  \includegraphics[width=0.24\linewidth]{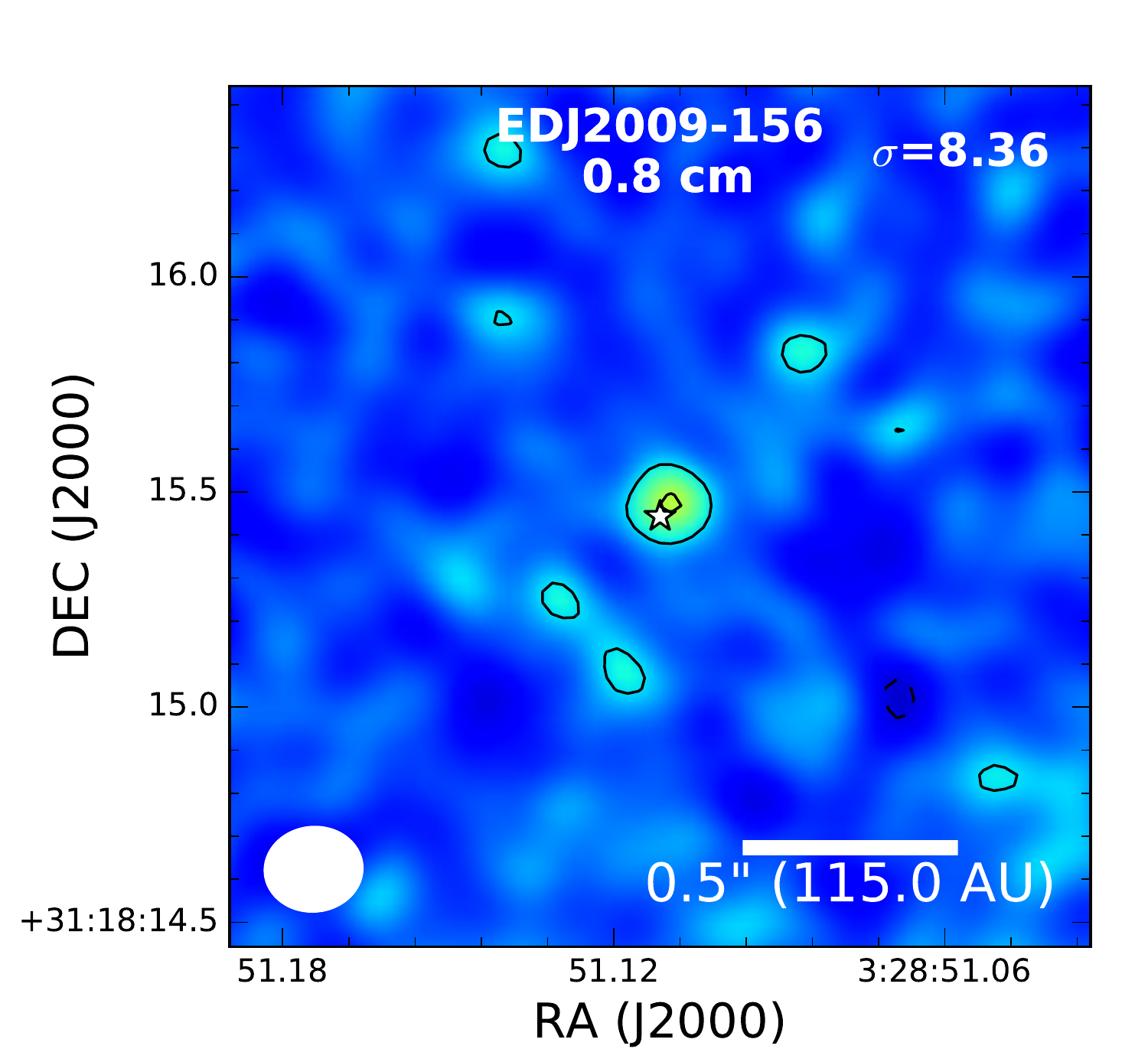}
  \includegraphics[width=0.24\linewidth]{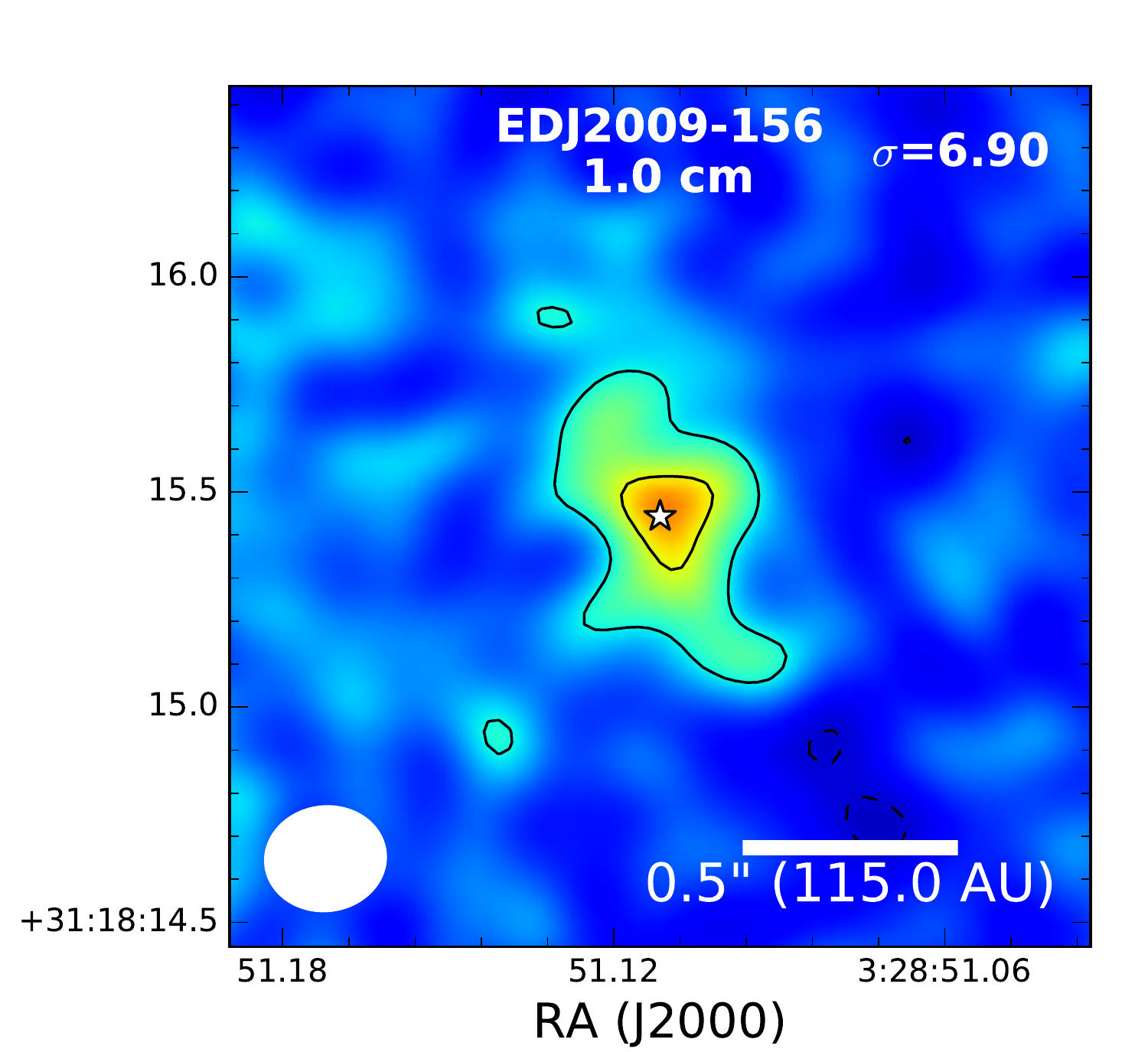}
  \includegraphics[width=0.24\linewidth]{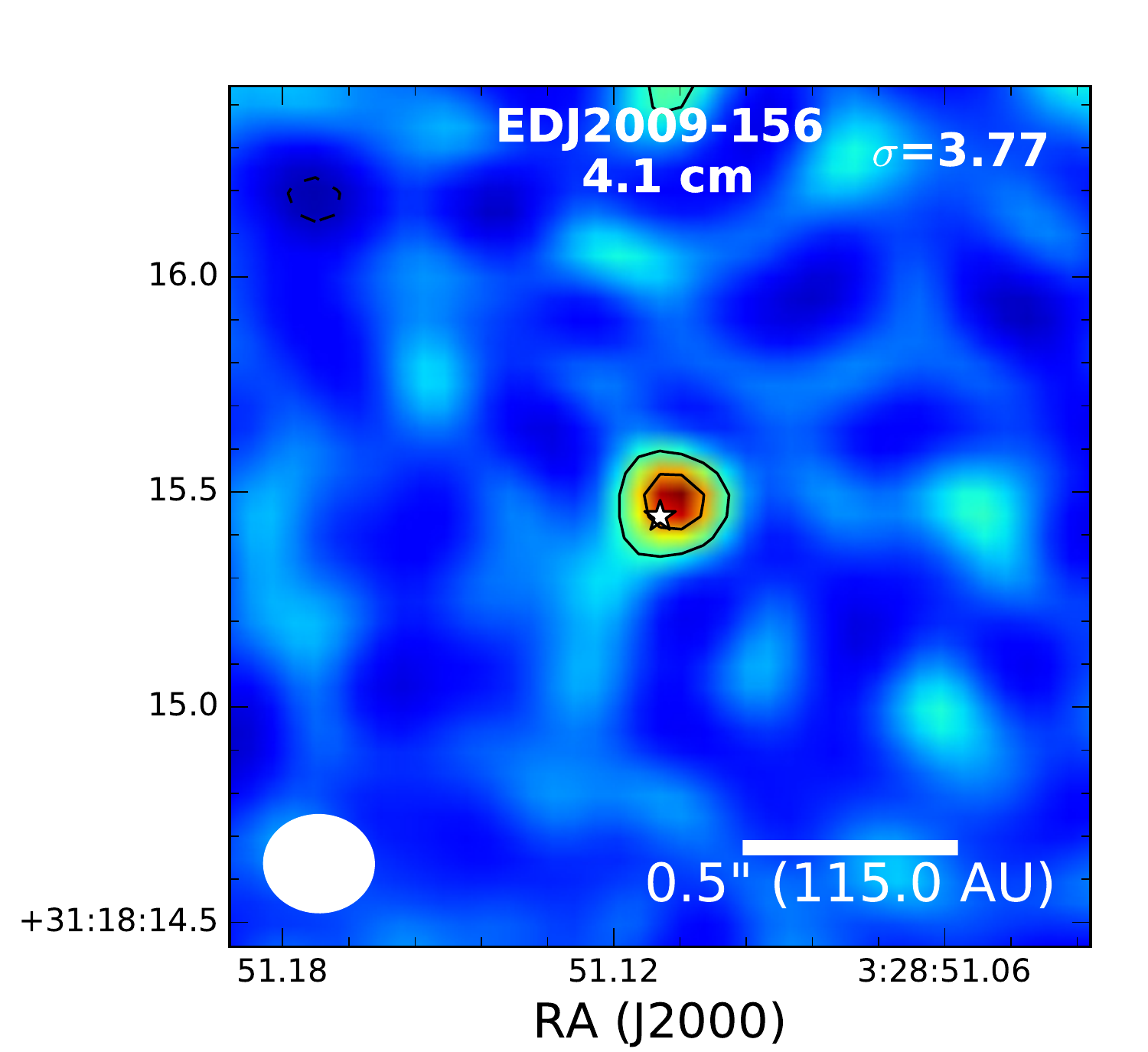}
  \includegraphics[width=0.24\linewidth]{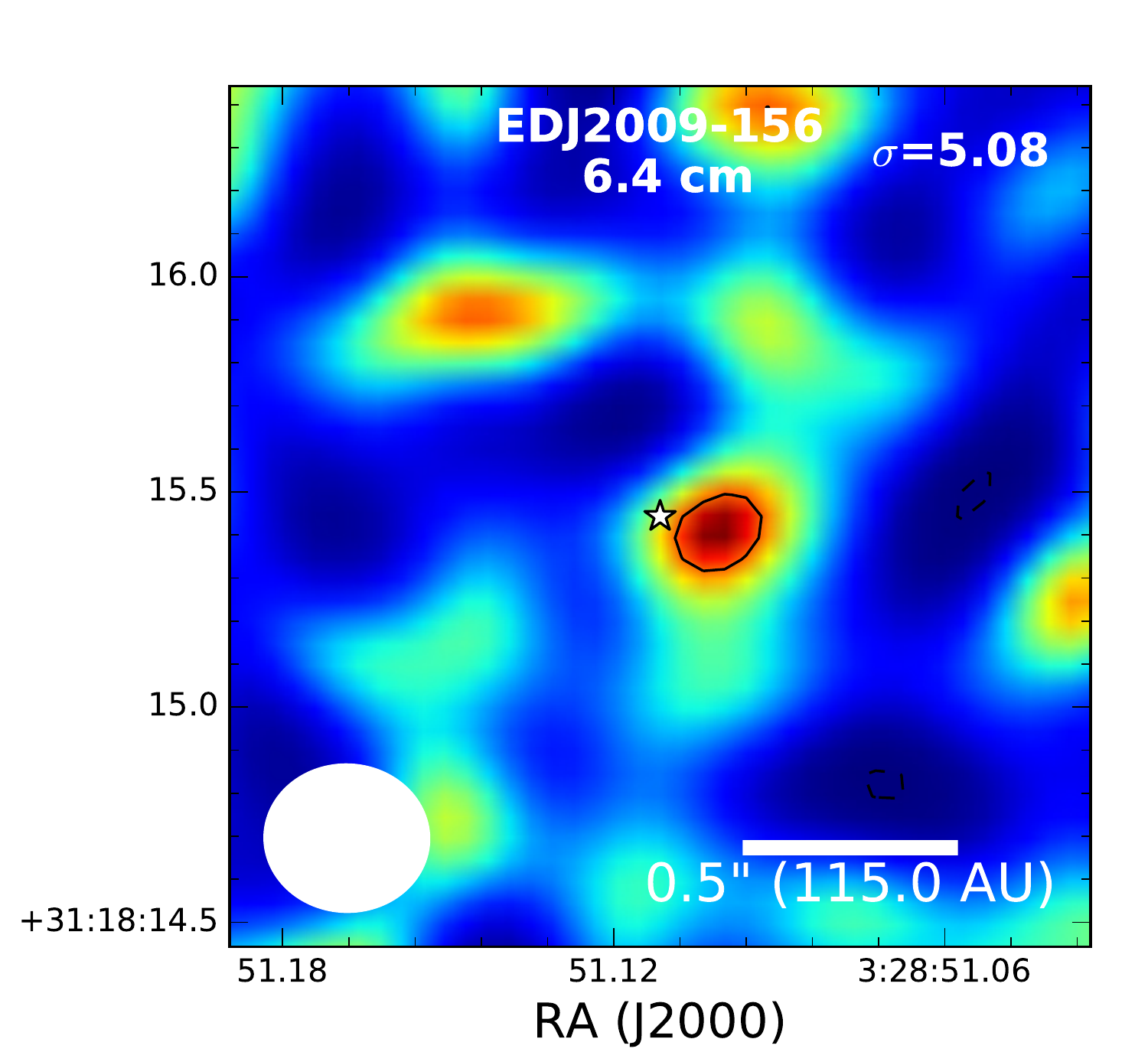}

  \includegraphics[width=0.24\linewidth]{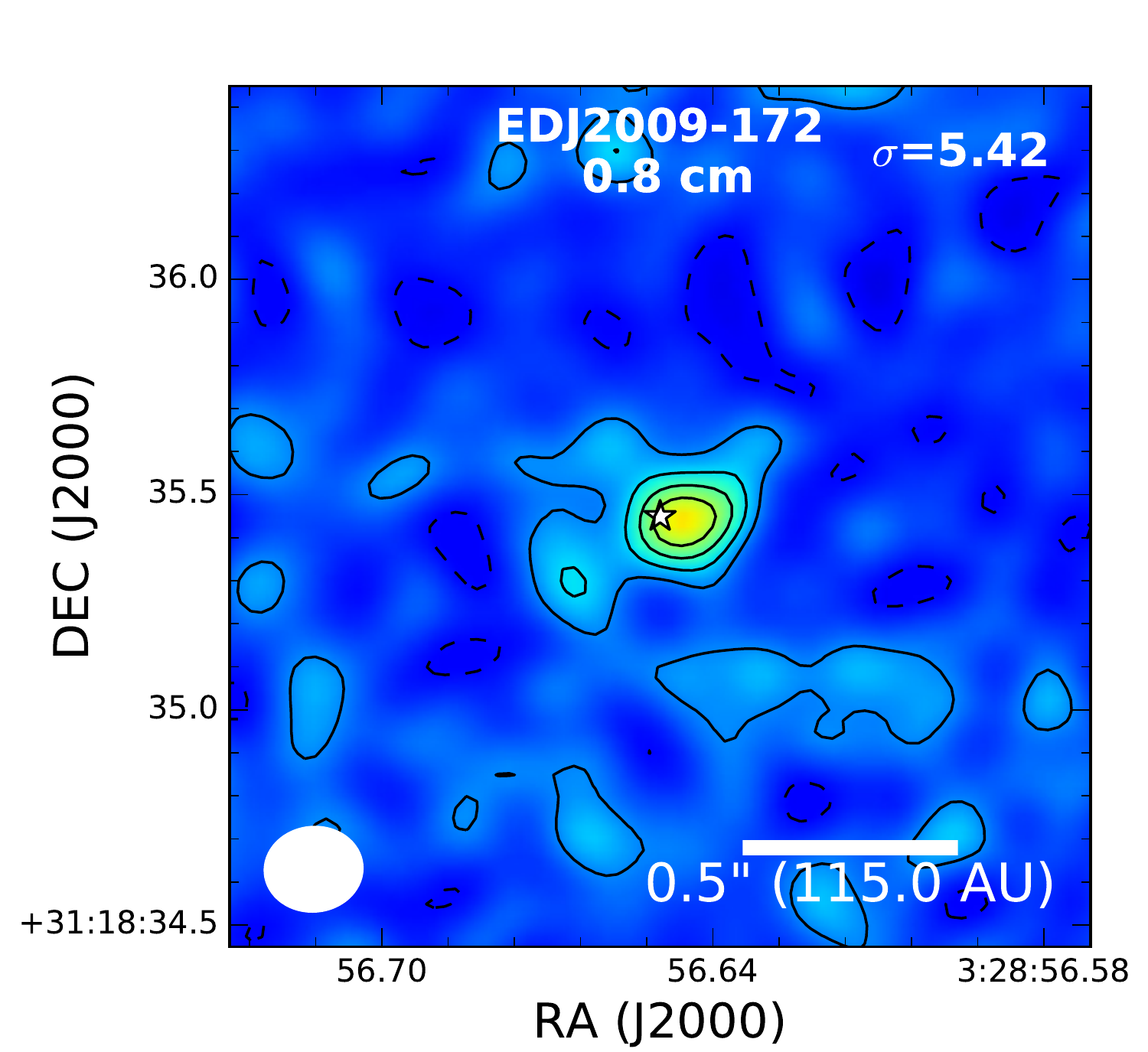}
  \includegraphics[width=0.24\linewidth]{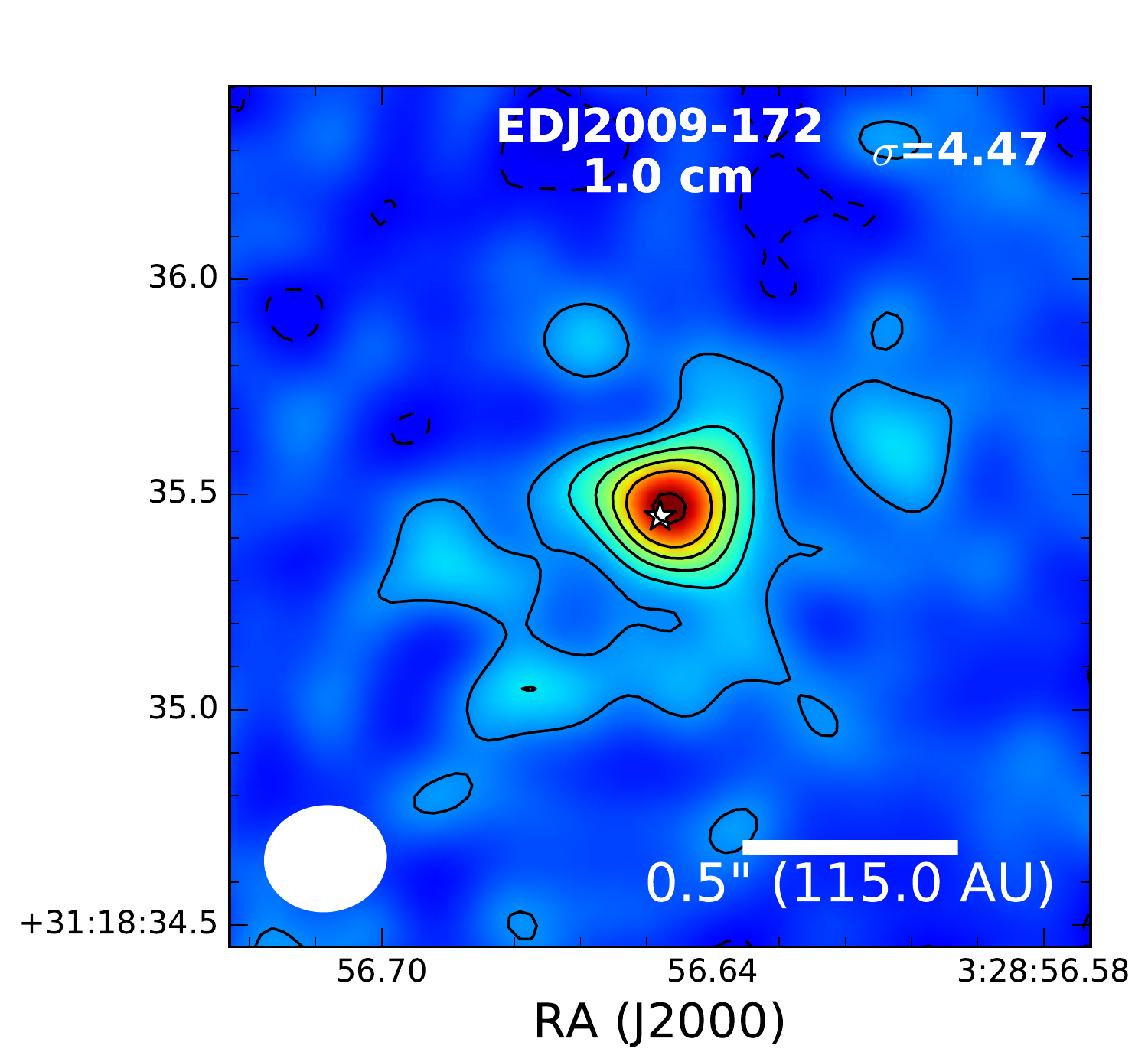}
  \includegraphics[width=0.24\linewidth]{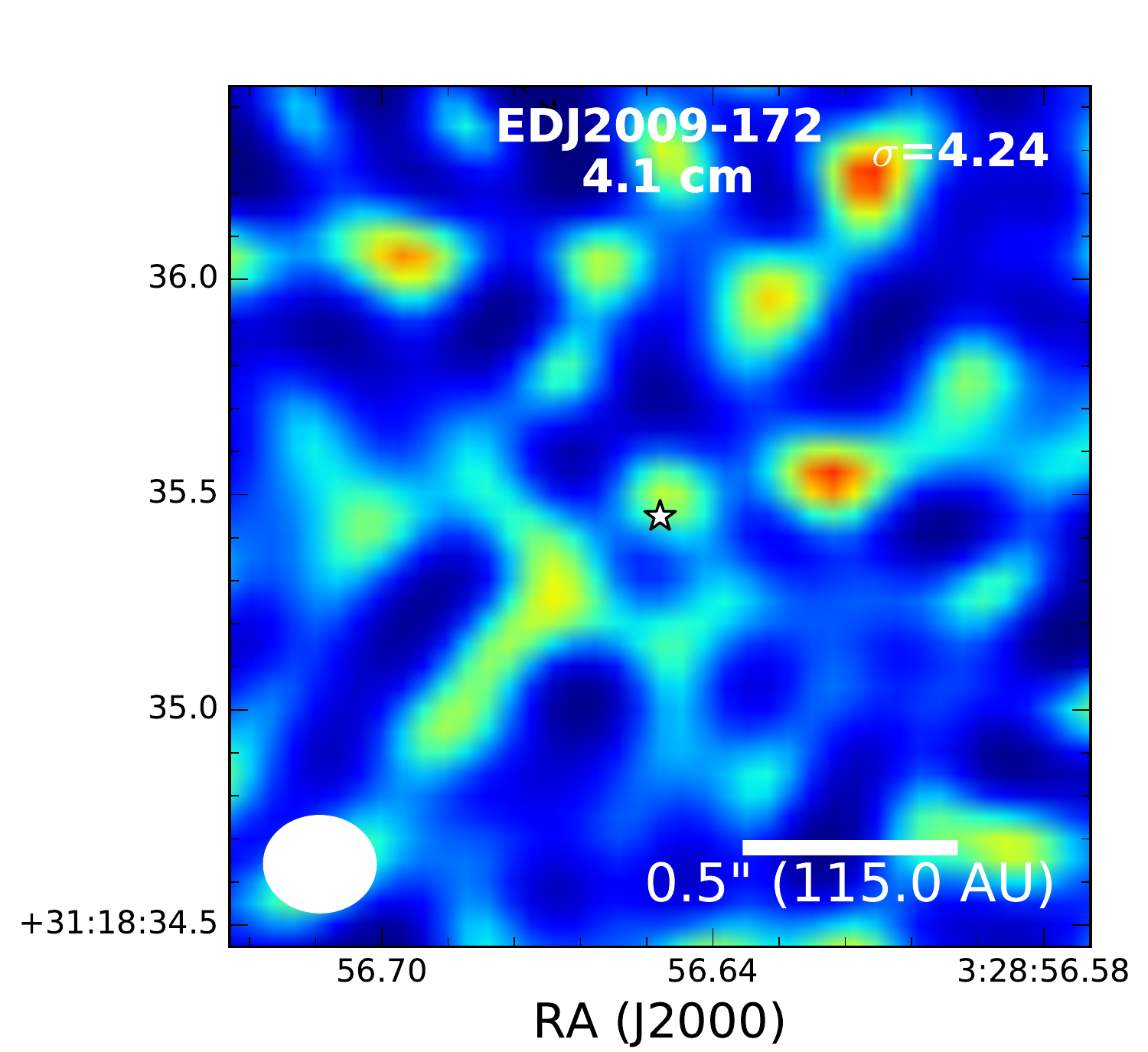}
  \includegraphics[width=0.24\linewidth]{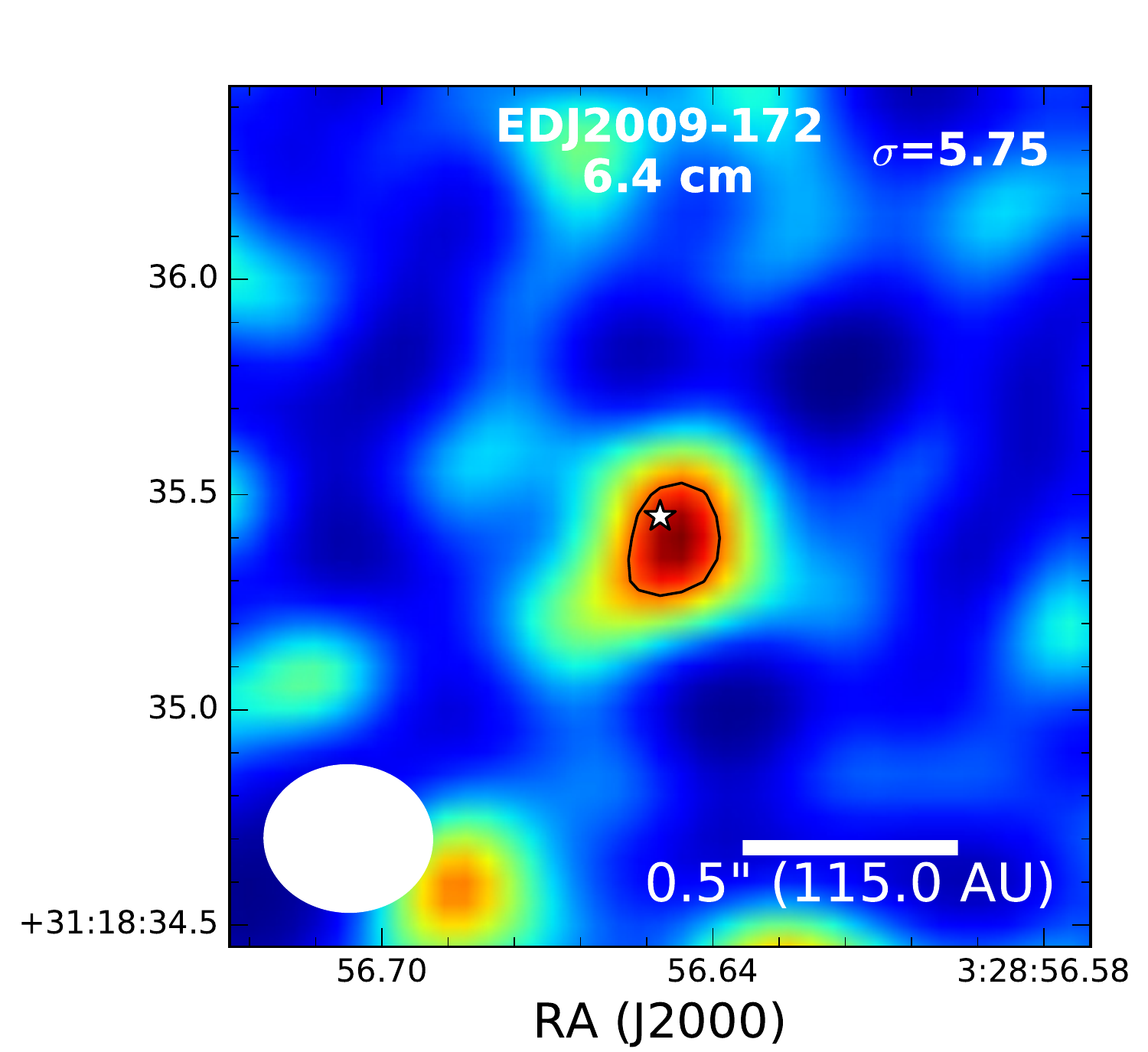}

  \includegraphics[width=0.24\linewidth]{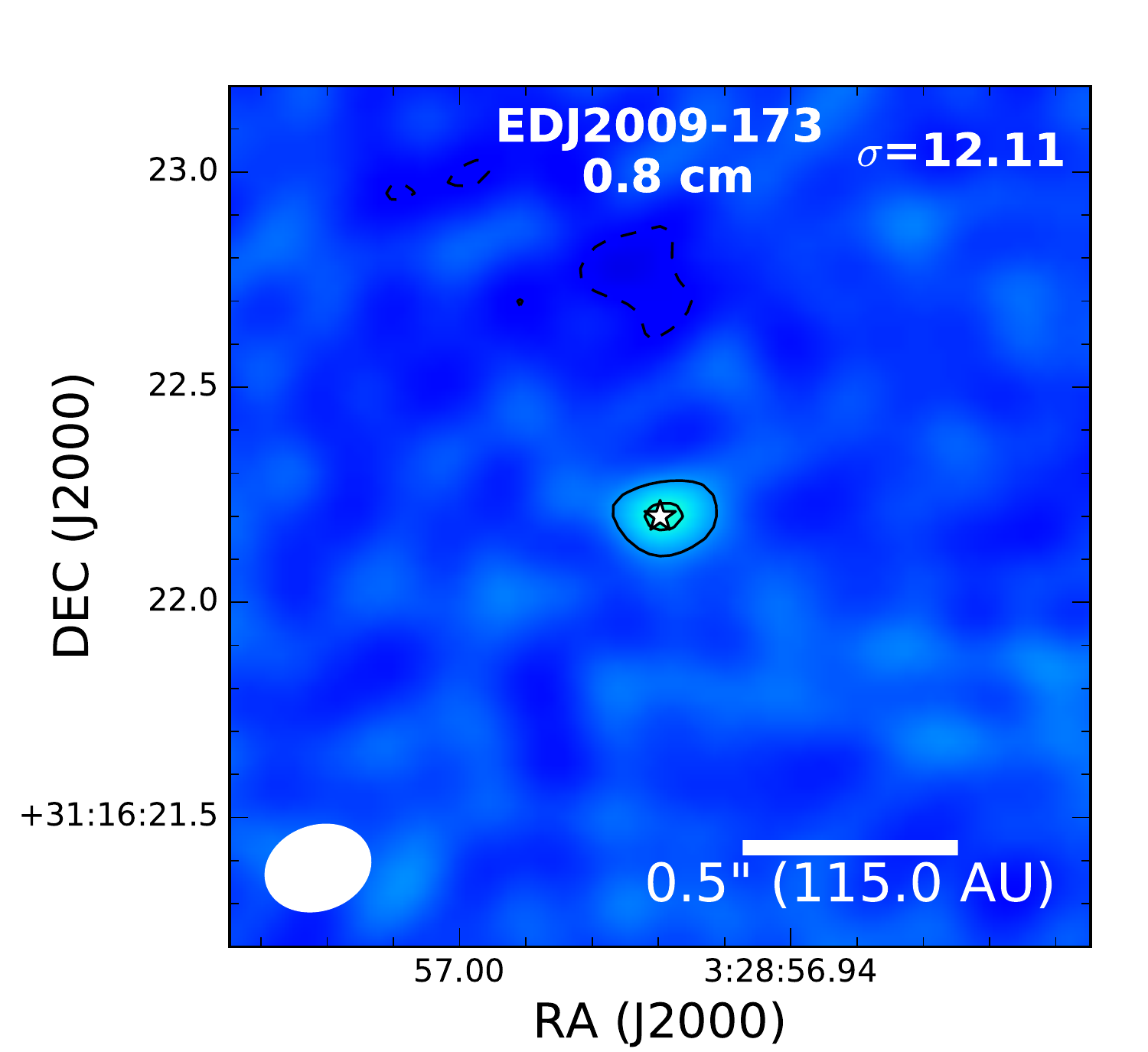}
  \includegraphics[width=0.24\linewidth]{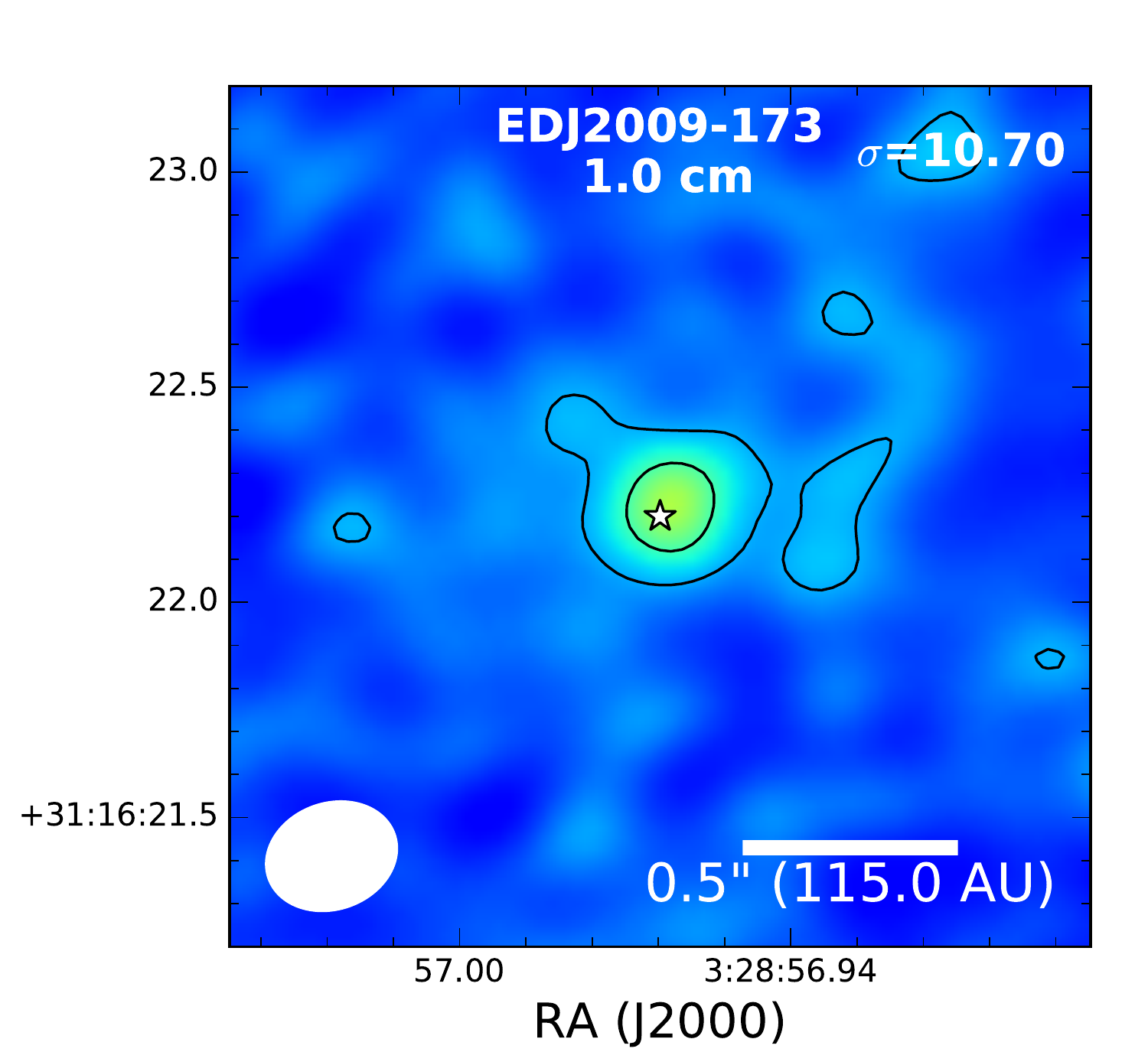}
  \includegraphics[width=0.24\linewidth]{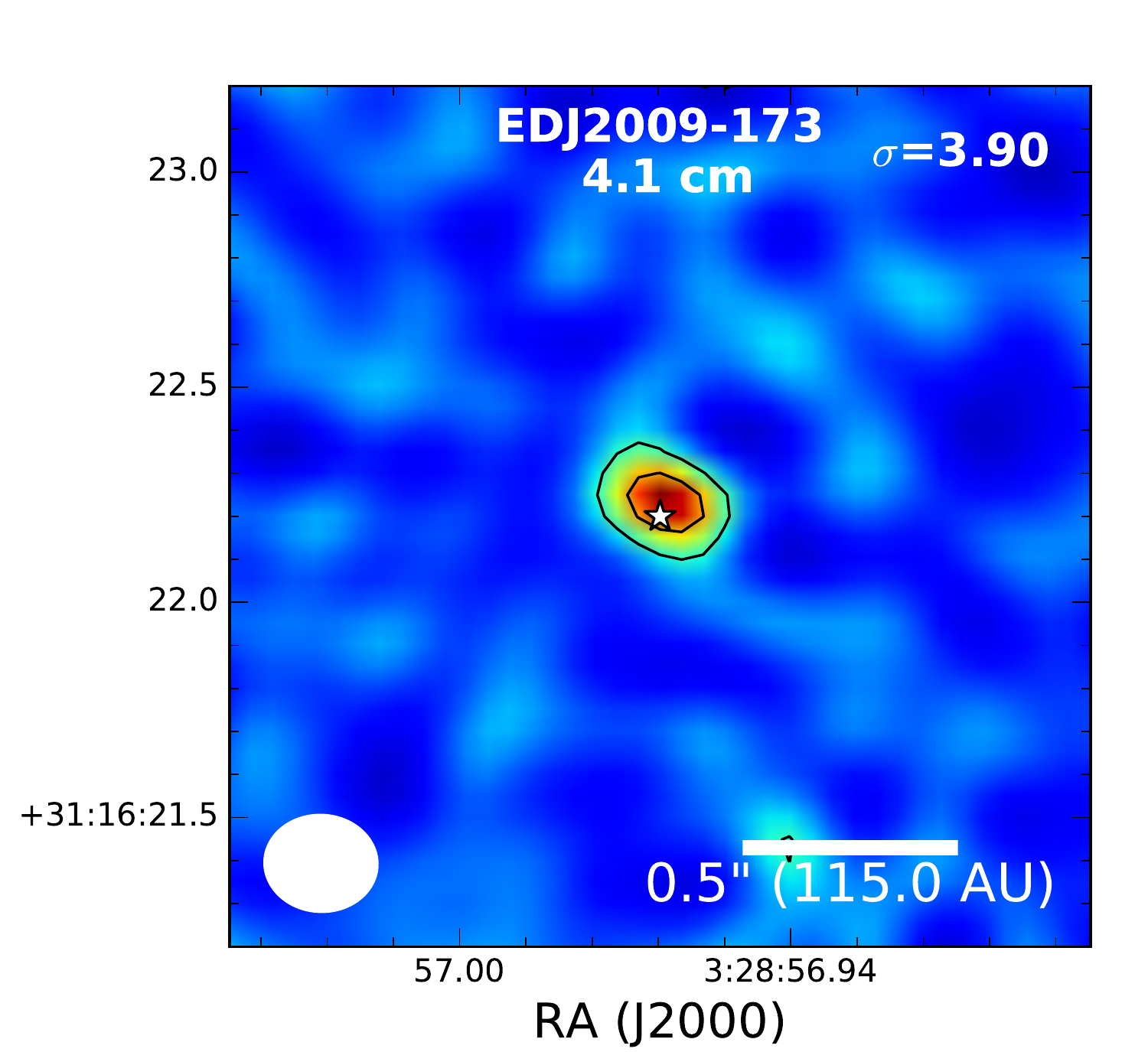}
  \includegraphics[width=0.24\linewidth]{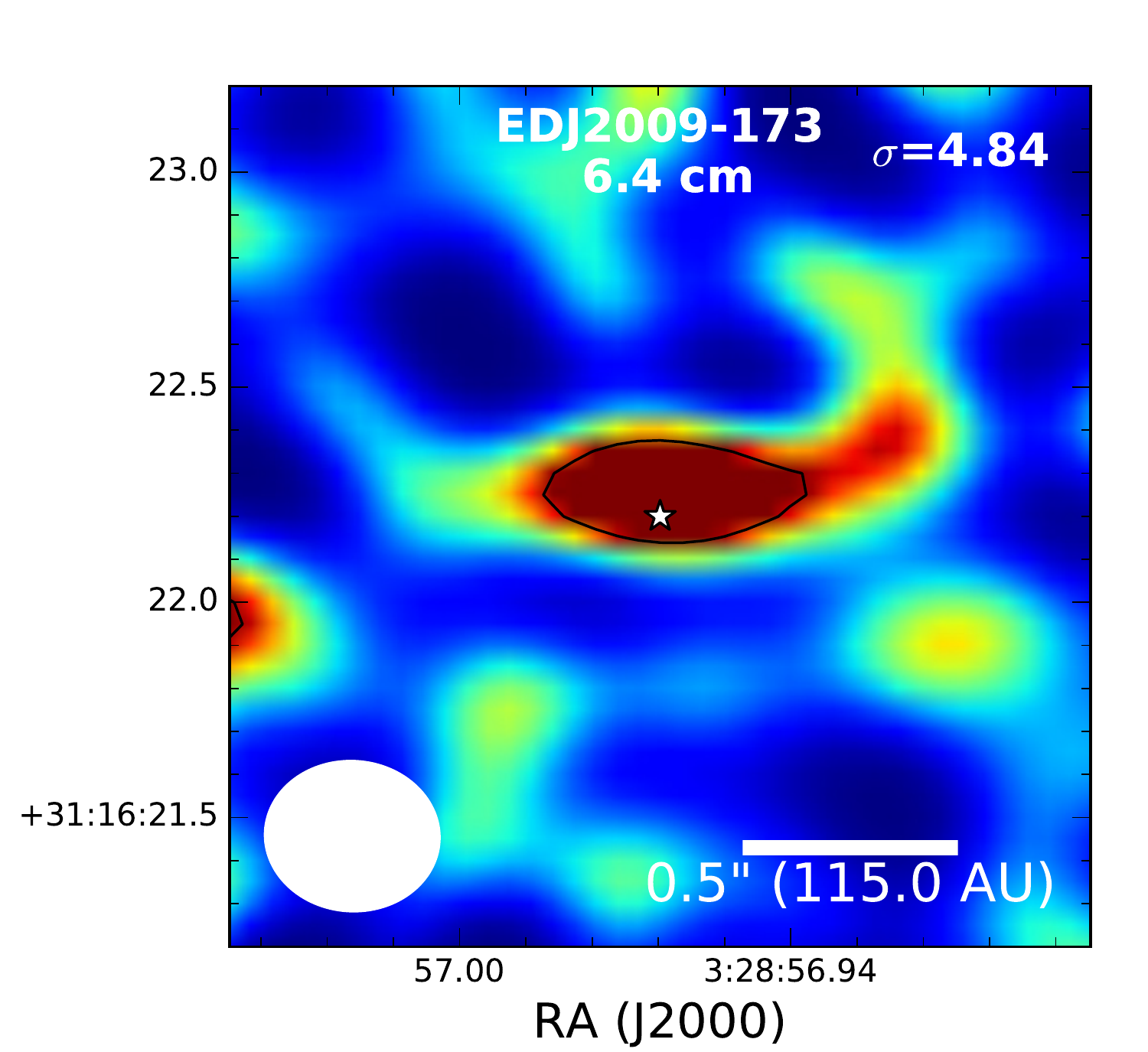}

  \includegraphics[width=0.24\linewidth]{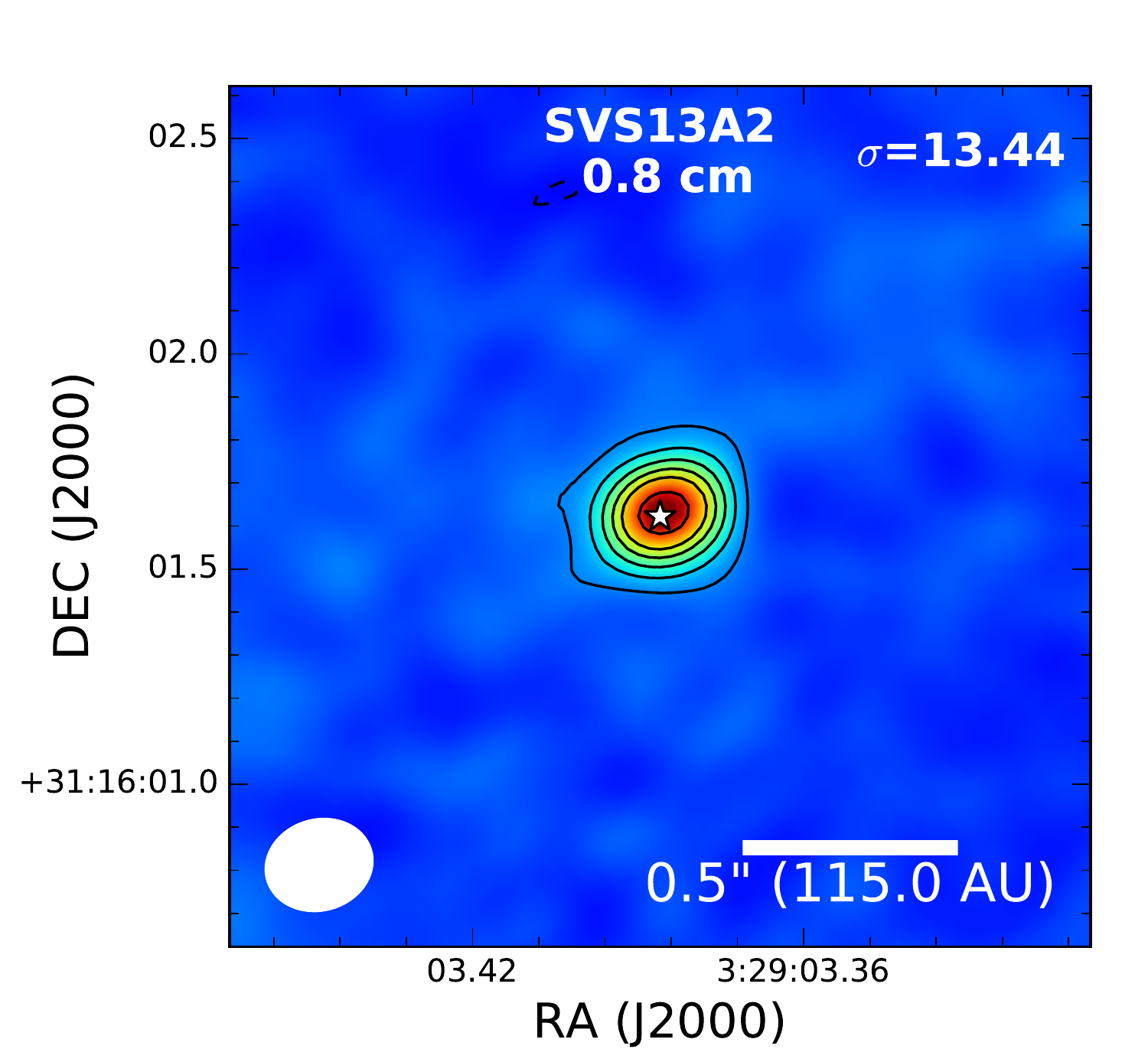}
  \includegraphics[width=0.24\linewidth]{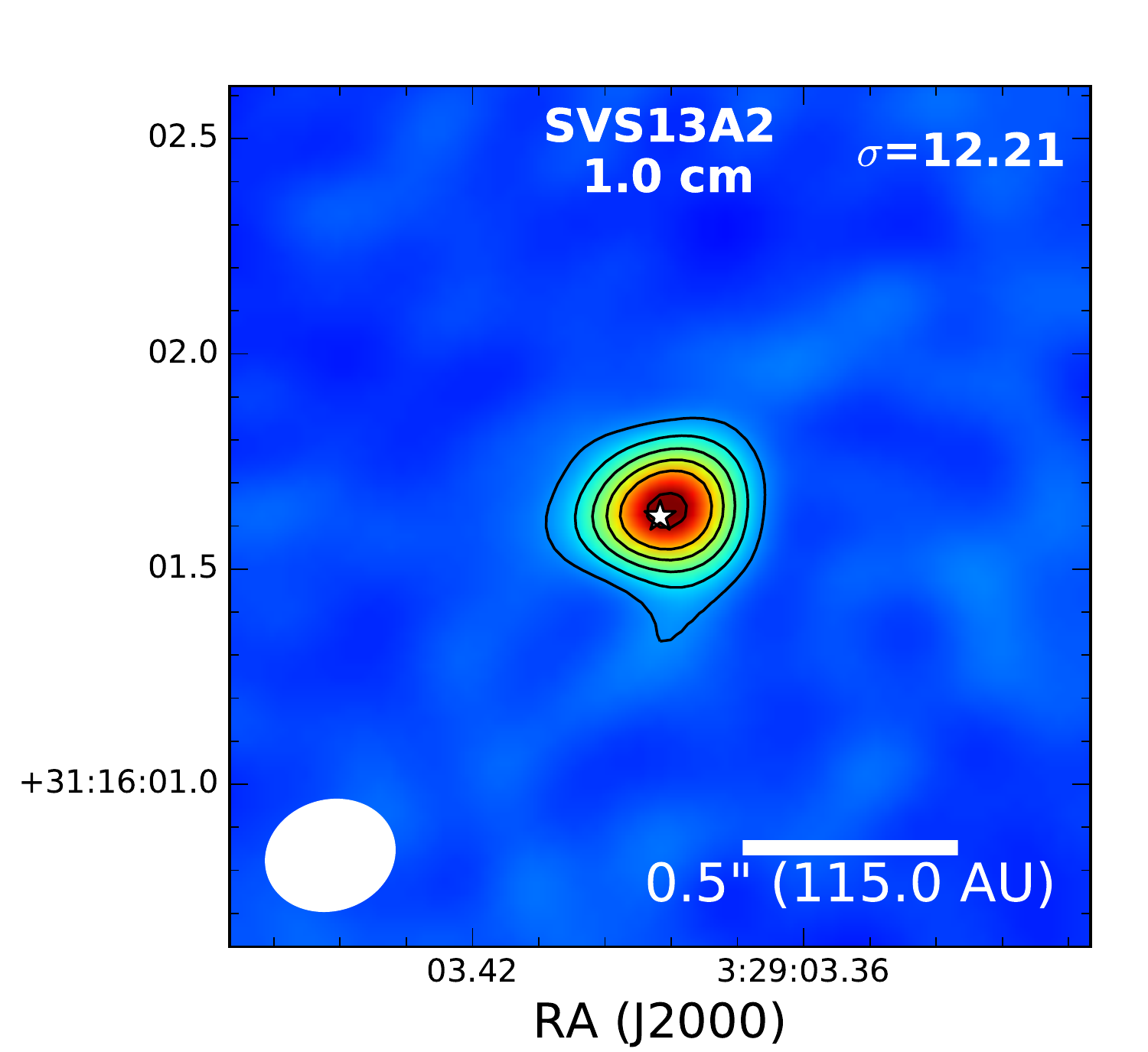}
  \includegraphics[width=0.24\linewidth]{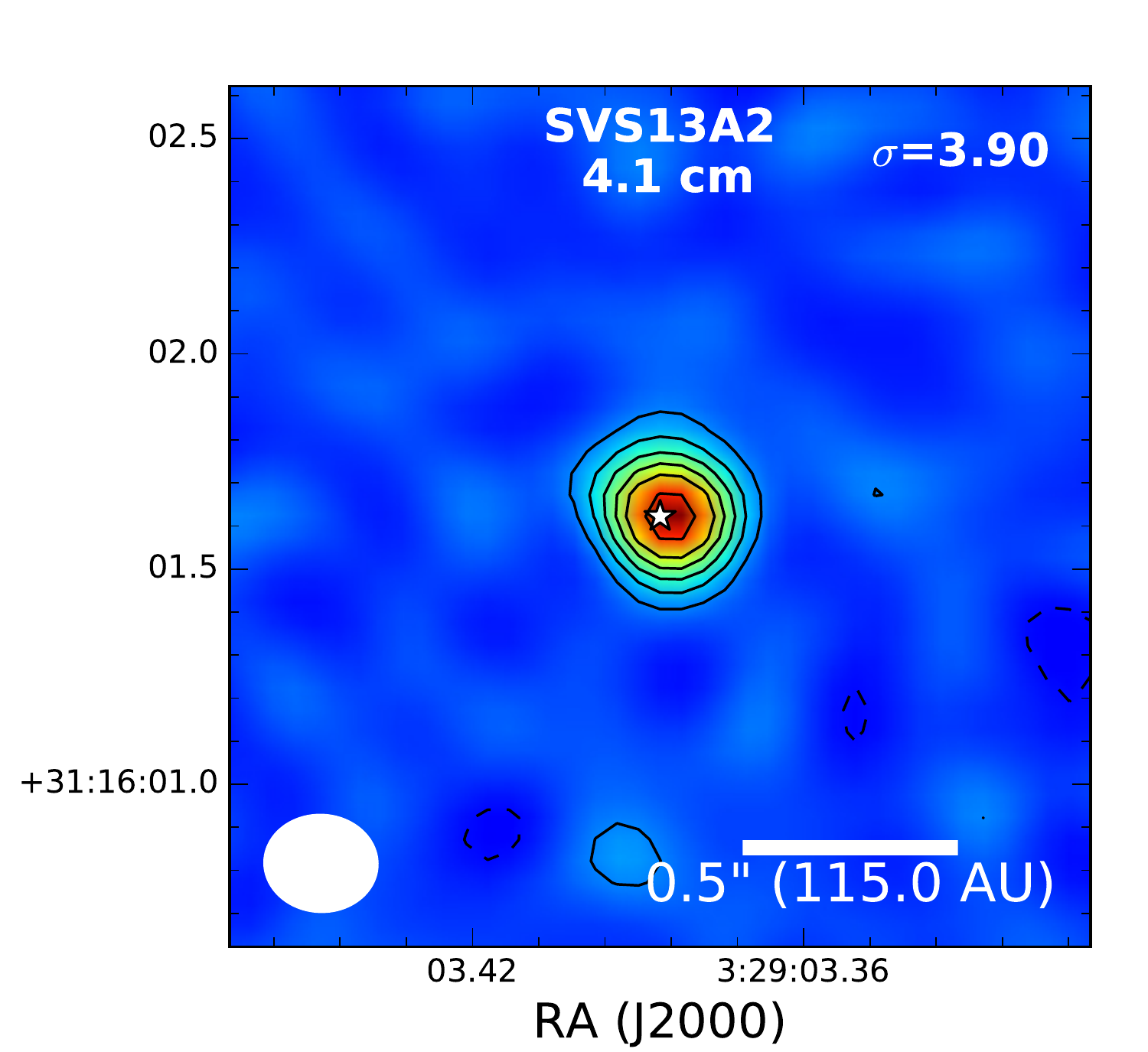}
  \includegraphics[width=0.24\linewidth]{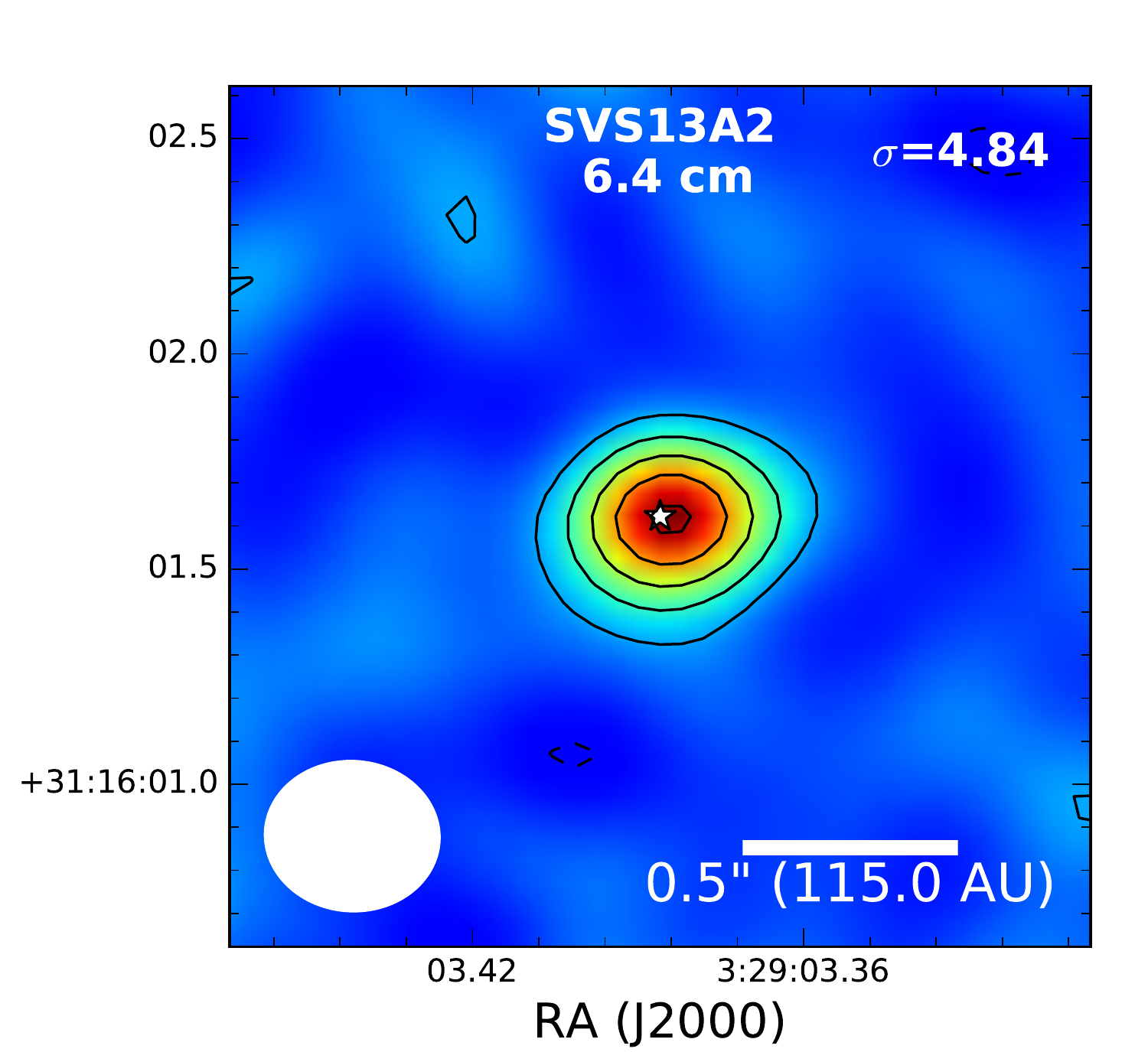}

  \includegraphics[width=0.24\linewidth]{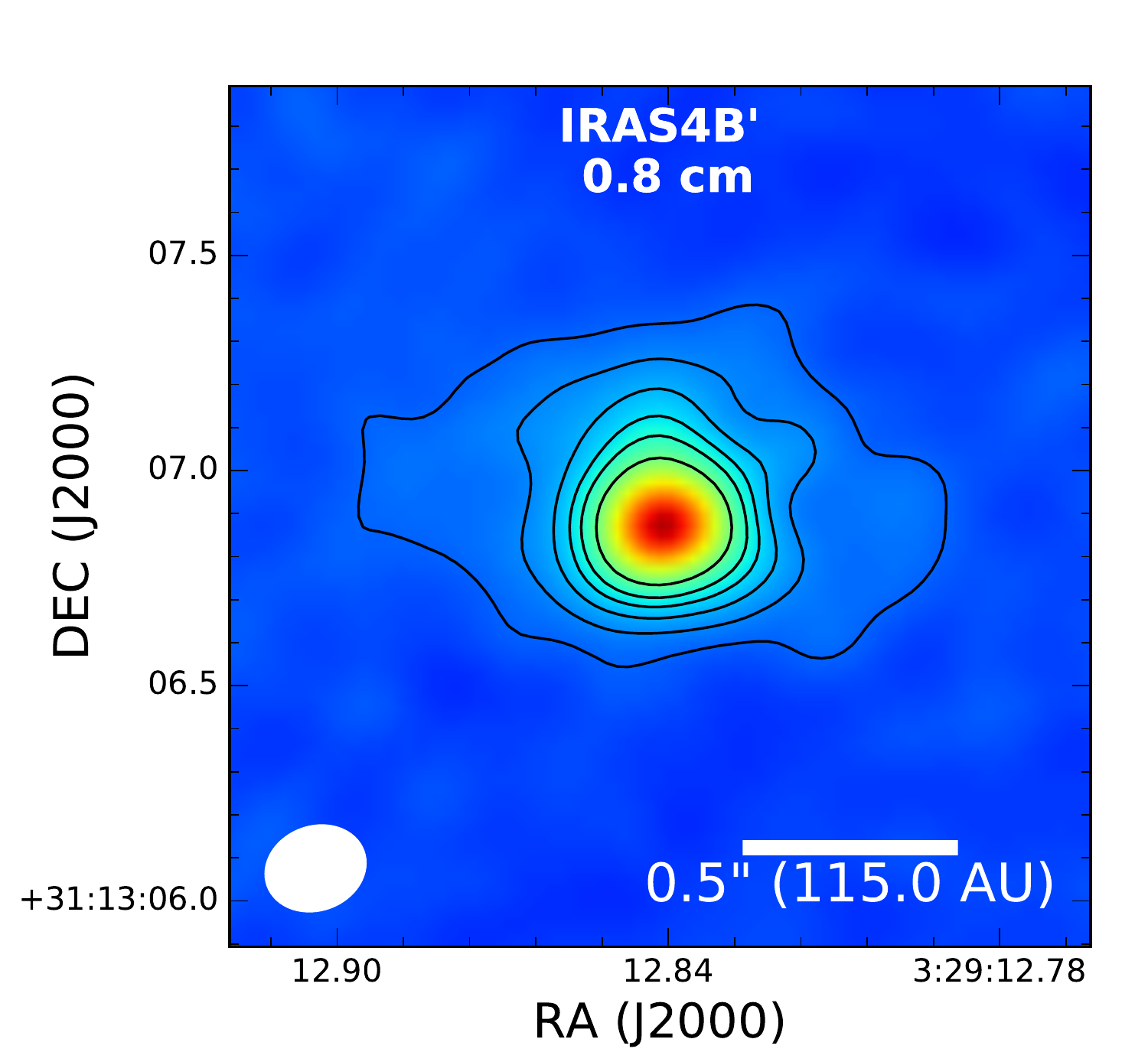}
  \includegraphics[width=0.24\linewidth]{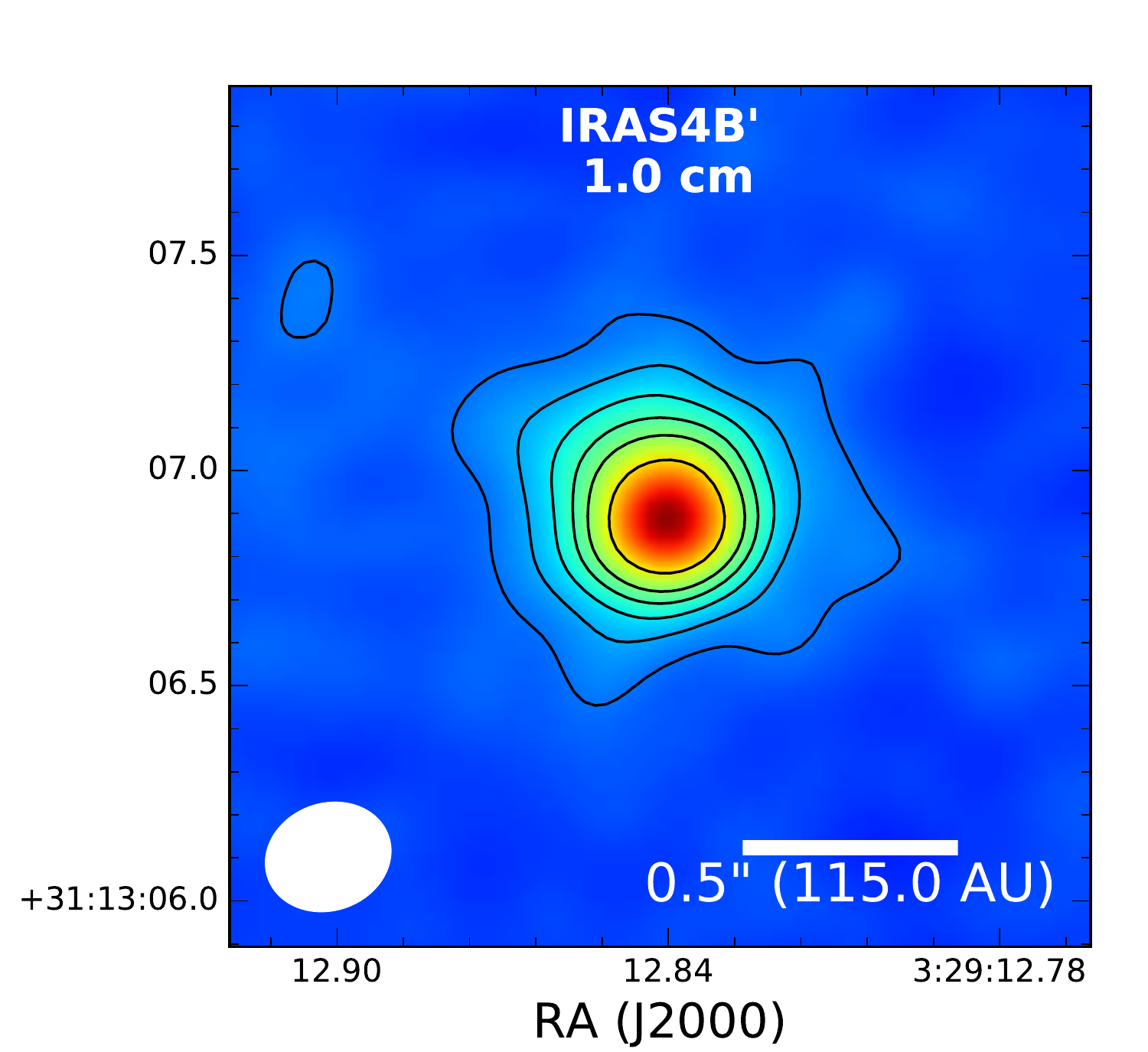}
  \includegraphics[width=0.24\linewidth]{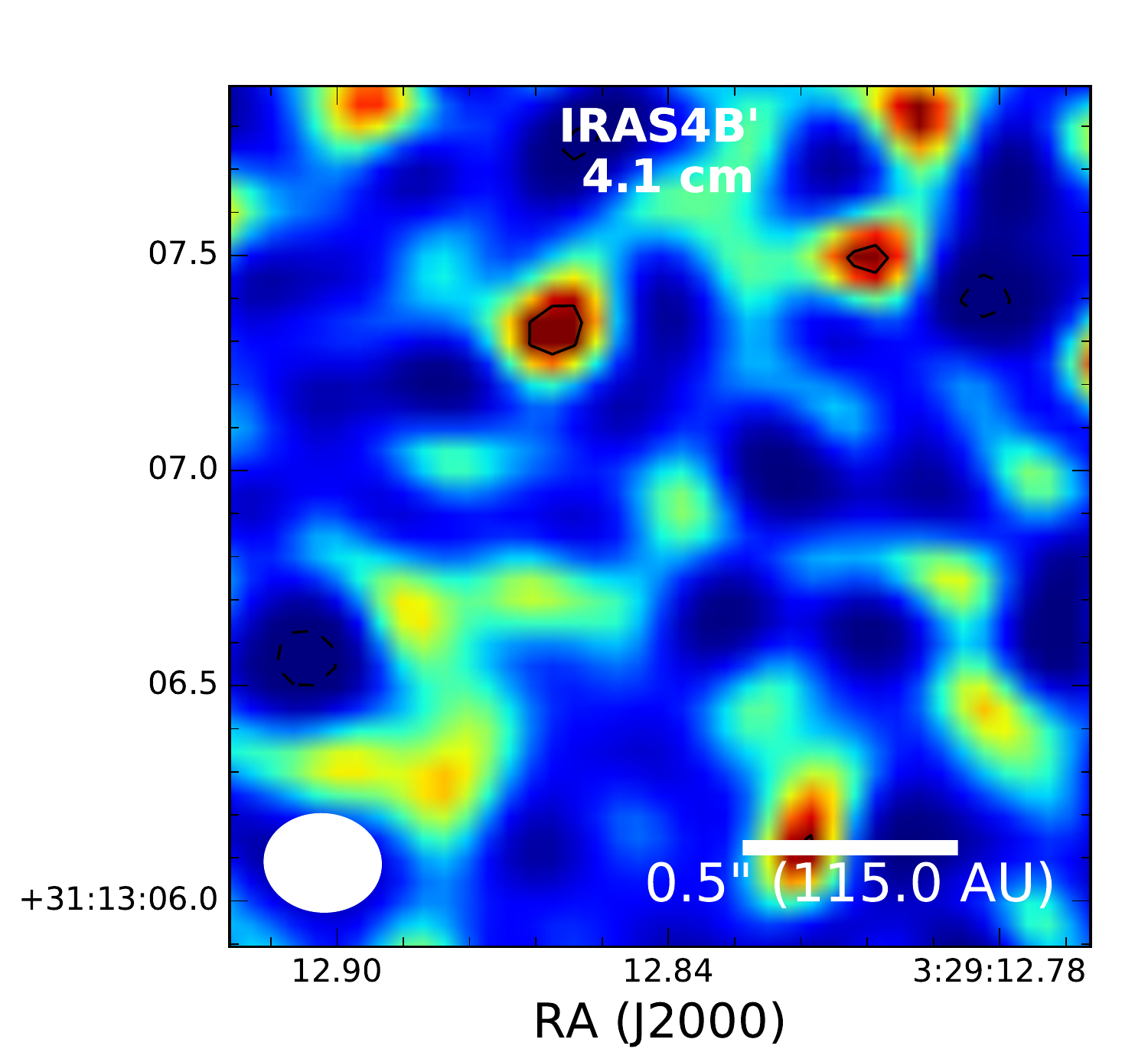}
  \includegraphics[width=0.24\linewidth]{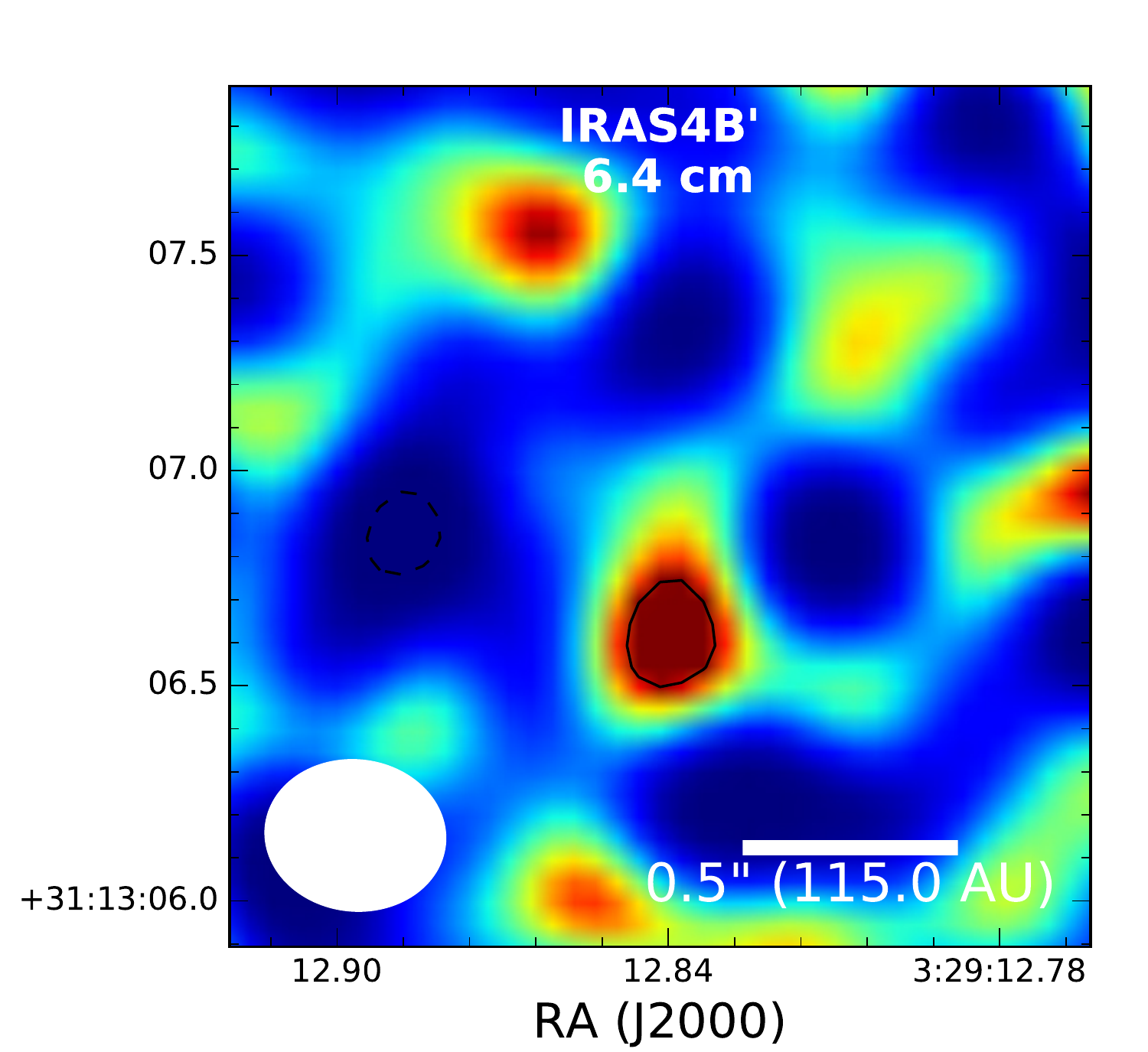}

  \includegraphics[width=0.24\linewidth]{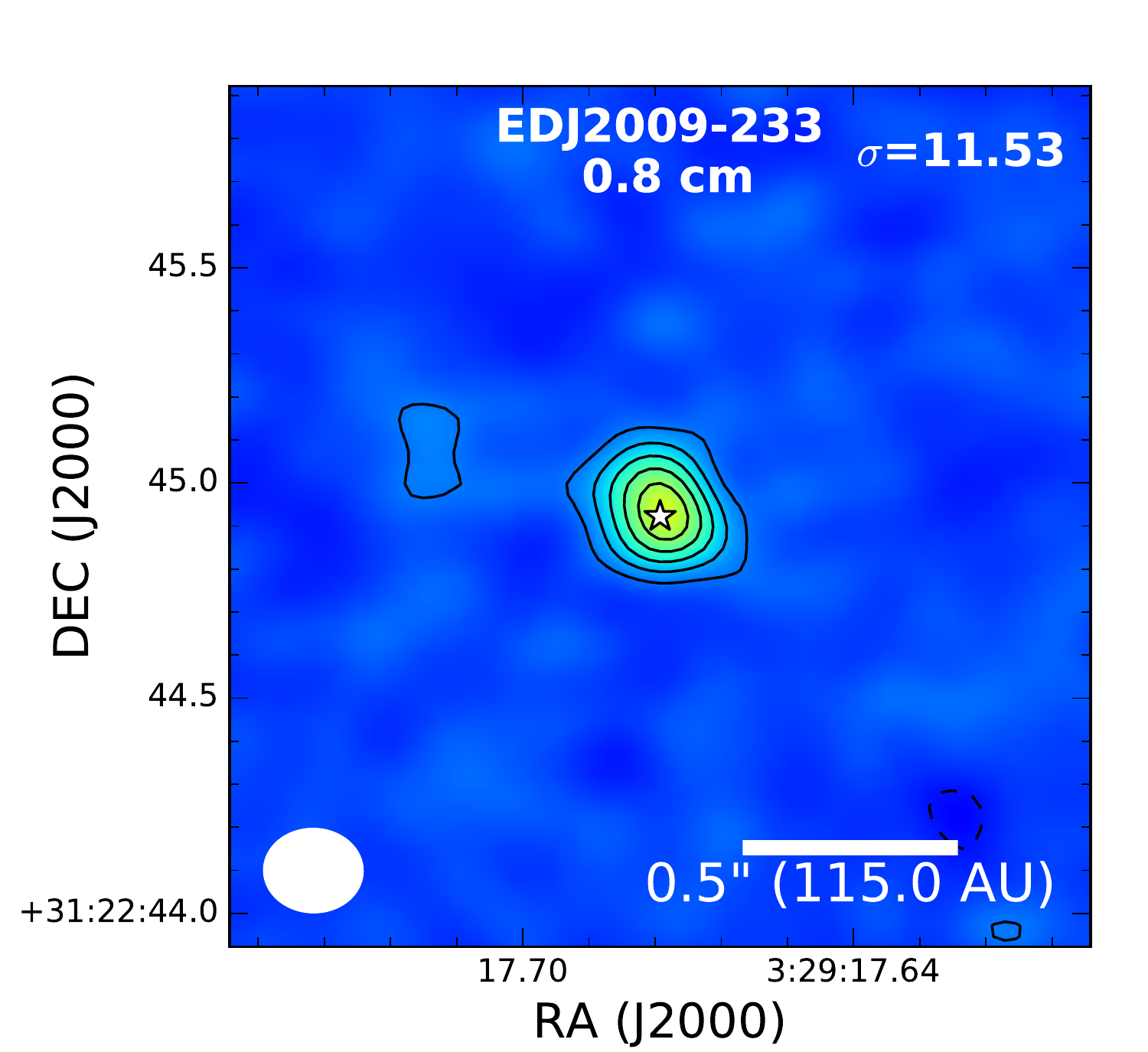}
  \includegraphics[width=0.24\linewidth]{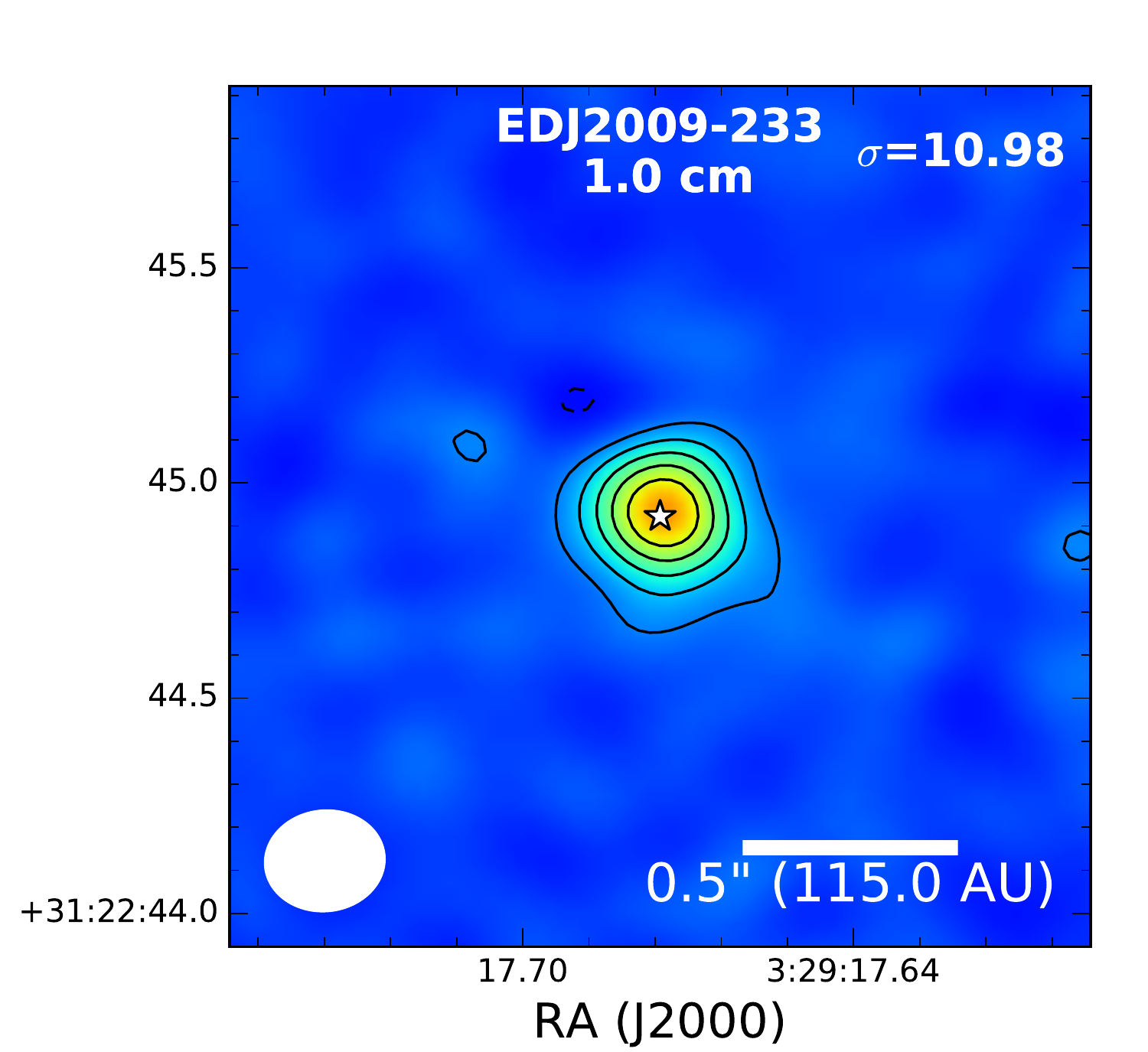}
  \includegraphics[width=0.24\linewidth]{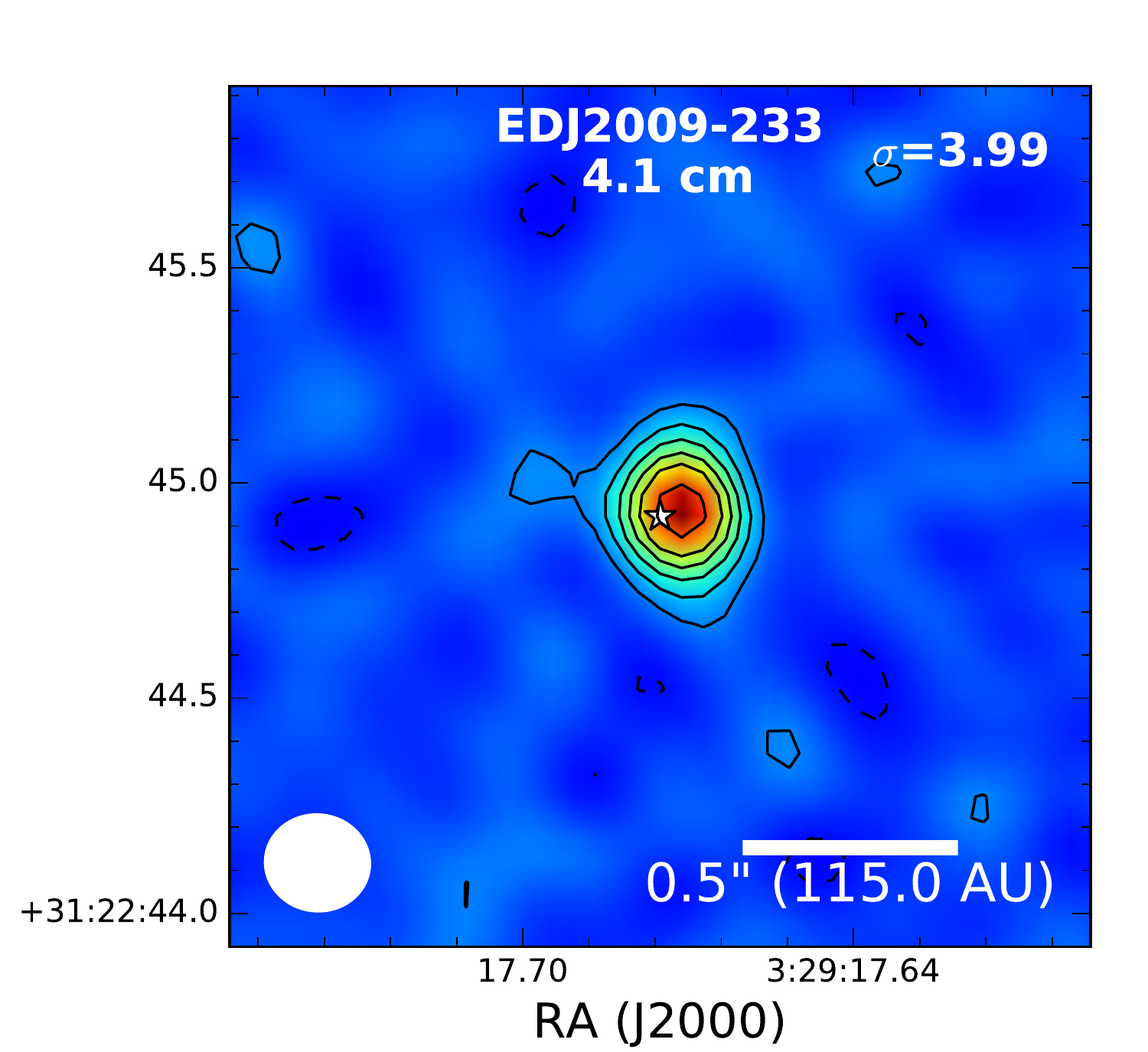}
  \includegraphics[width=0.24\linewidth]{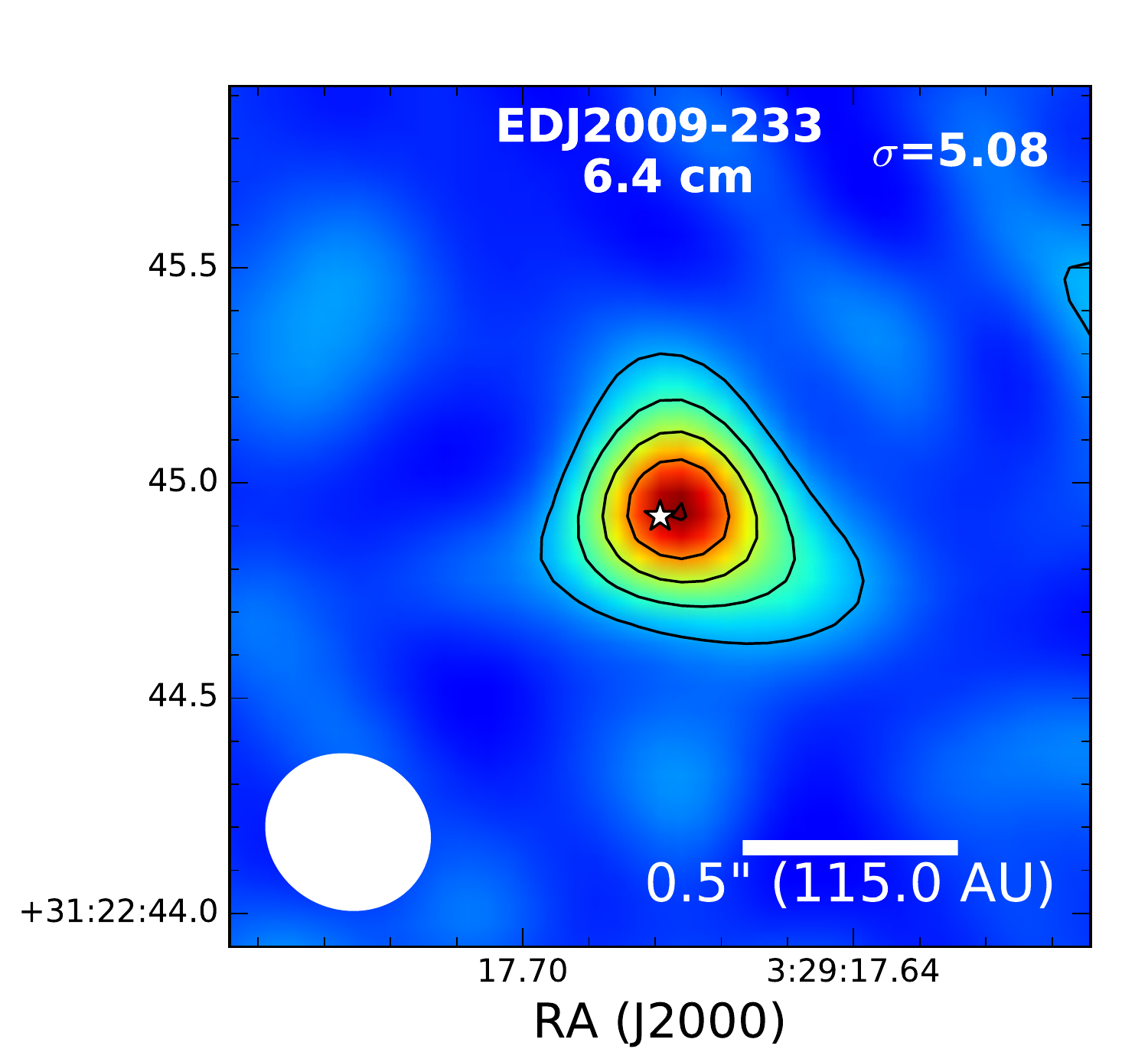}

\end{figure}
\begin{figure}

  \includegraphics[width=0.24\linewidth]{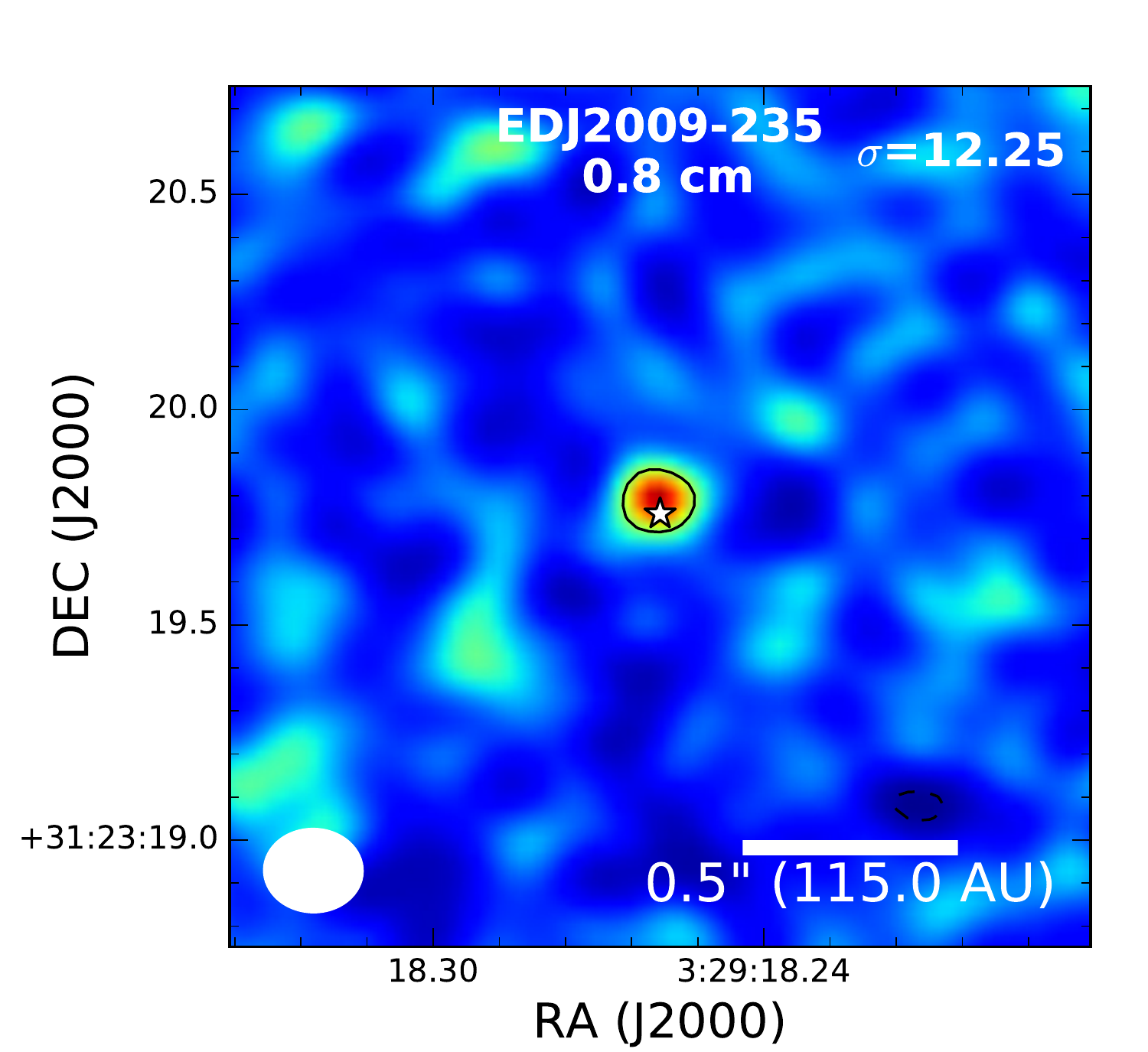}
  \includegraphics[width=0.24\linewidth]{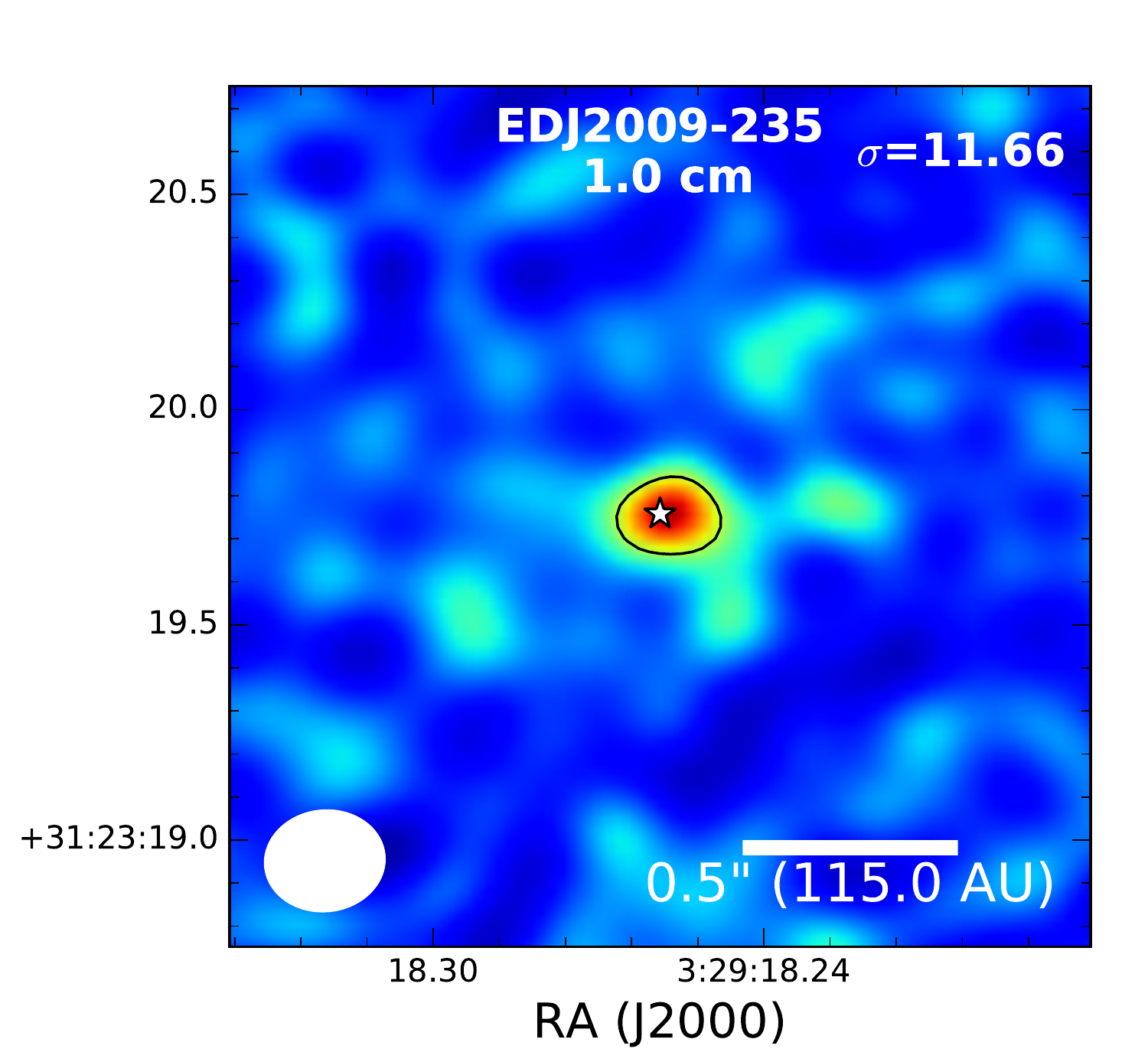}
  \includegraphics[width=0.24\linewidth]{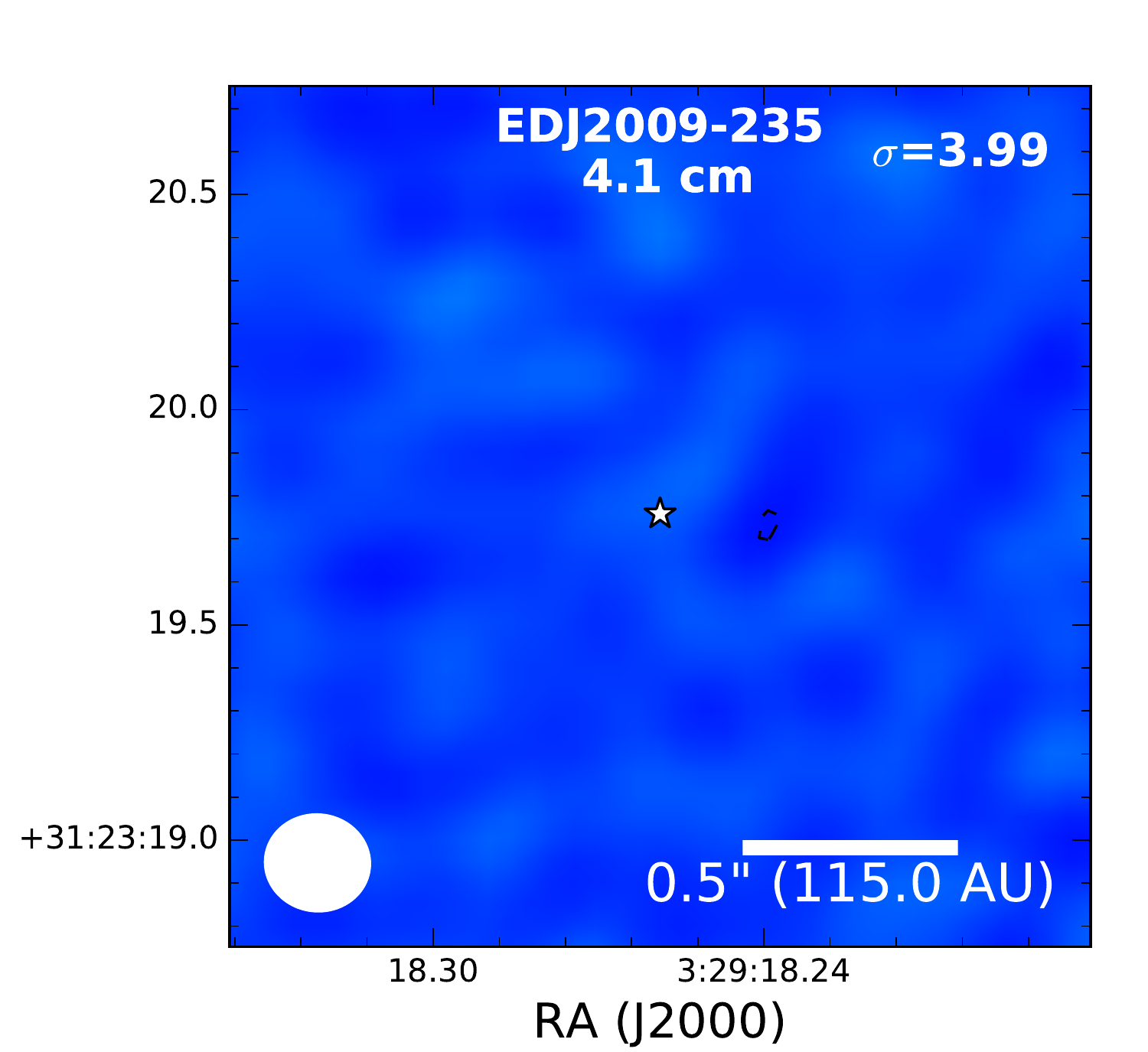}
  \includegraphics[width=0.24\linewidth]{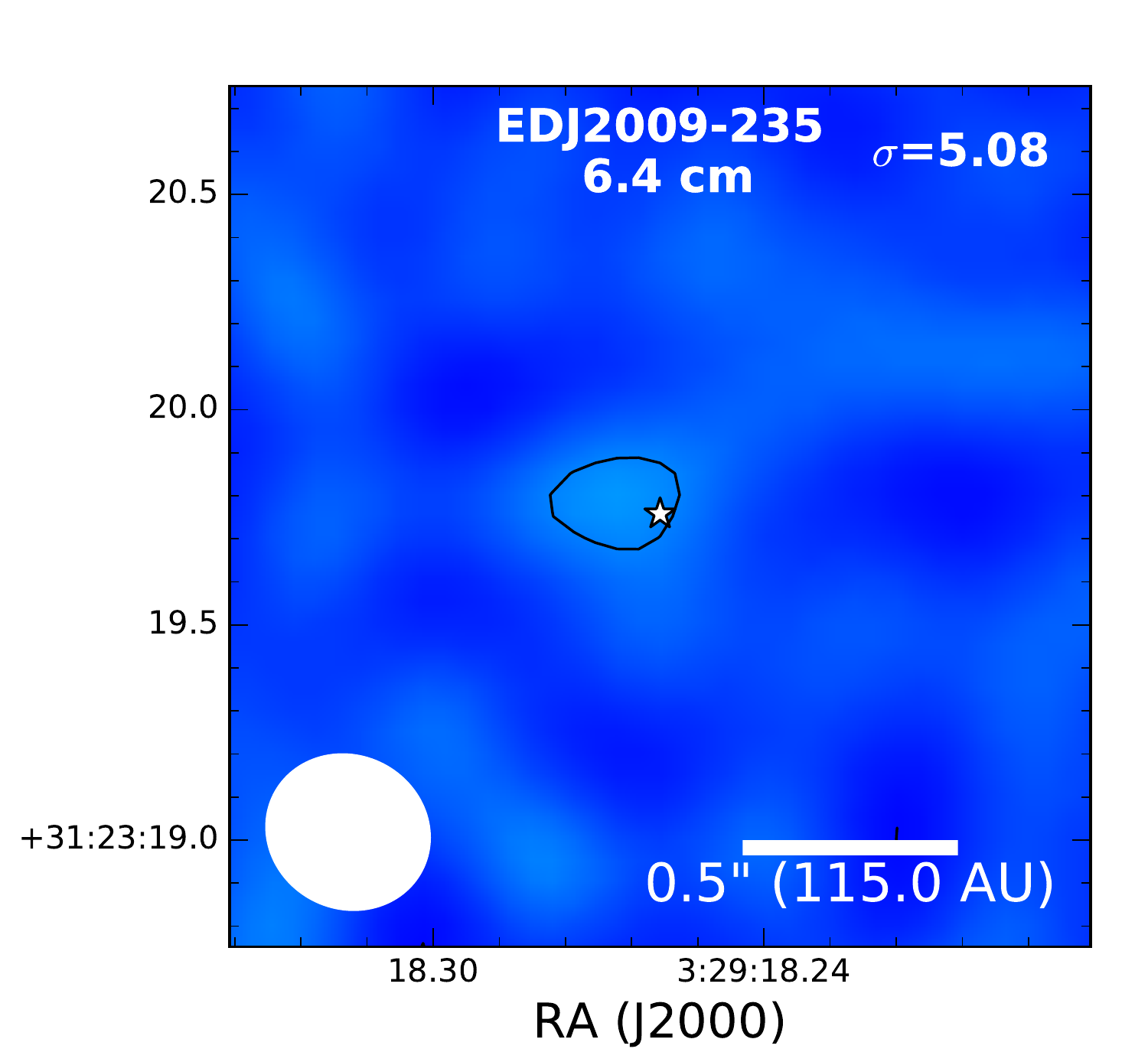}

  \includegraphics[width=0.24\linewidth]{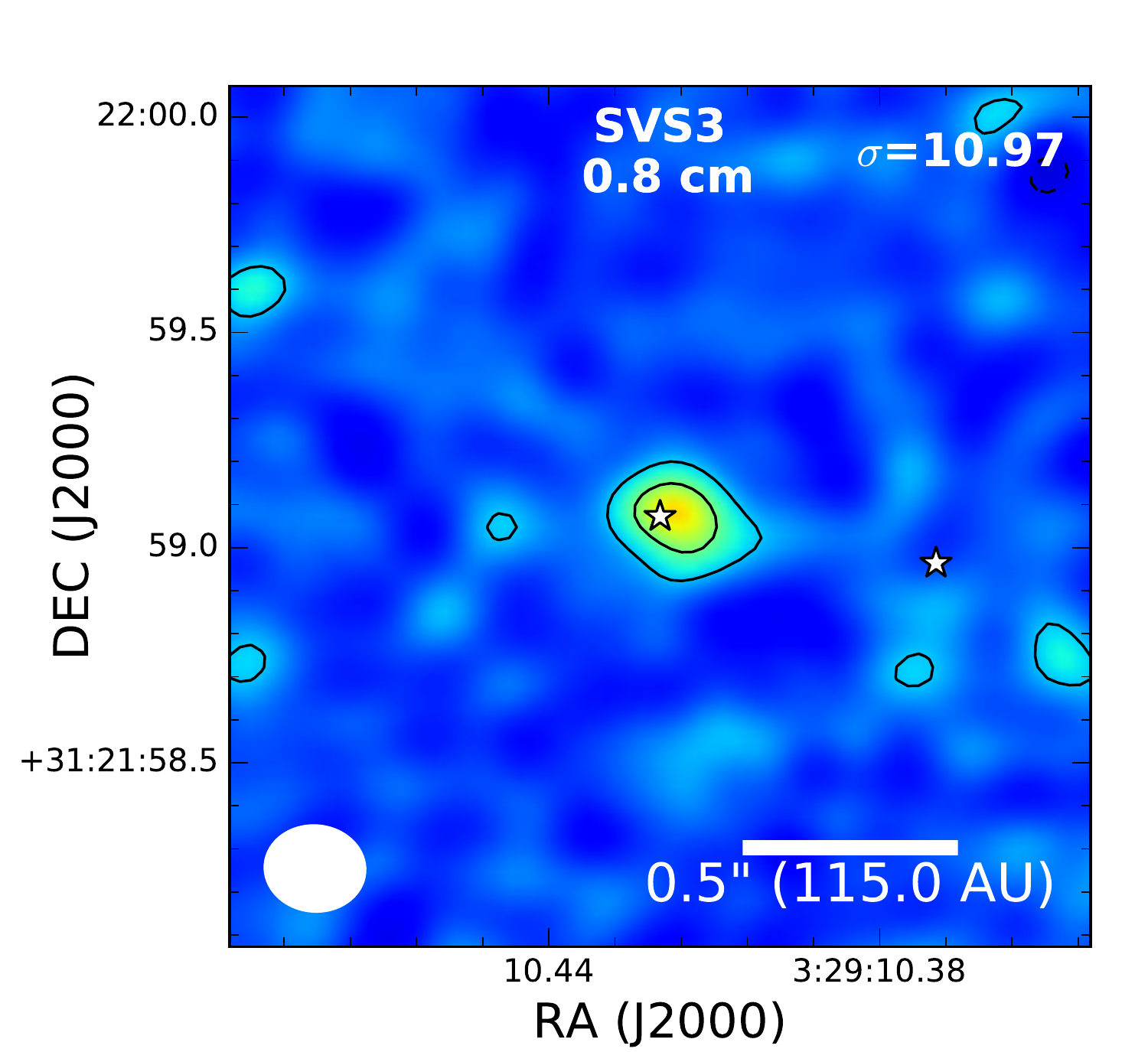}
  \includegraphics[width=0.24\linewidth]{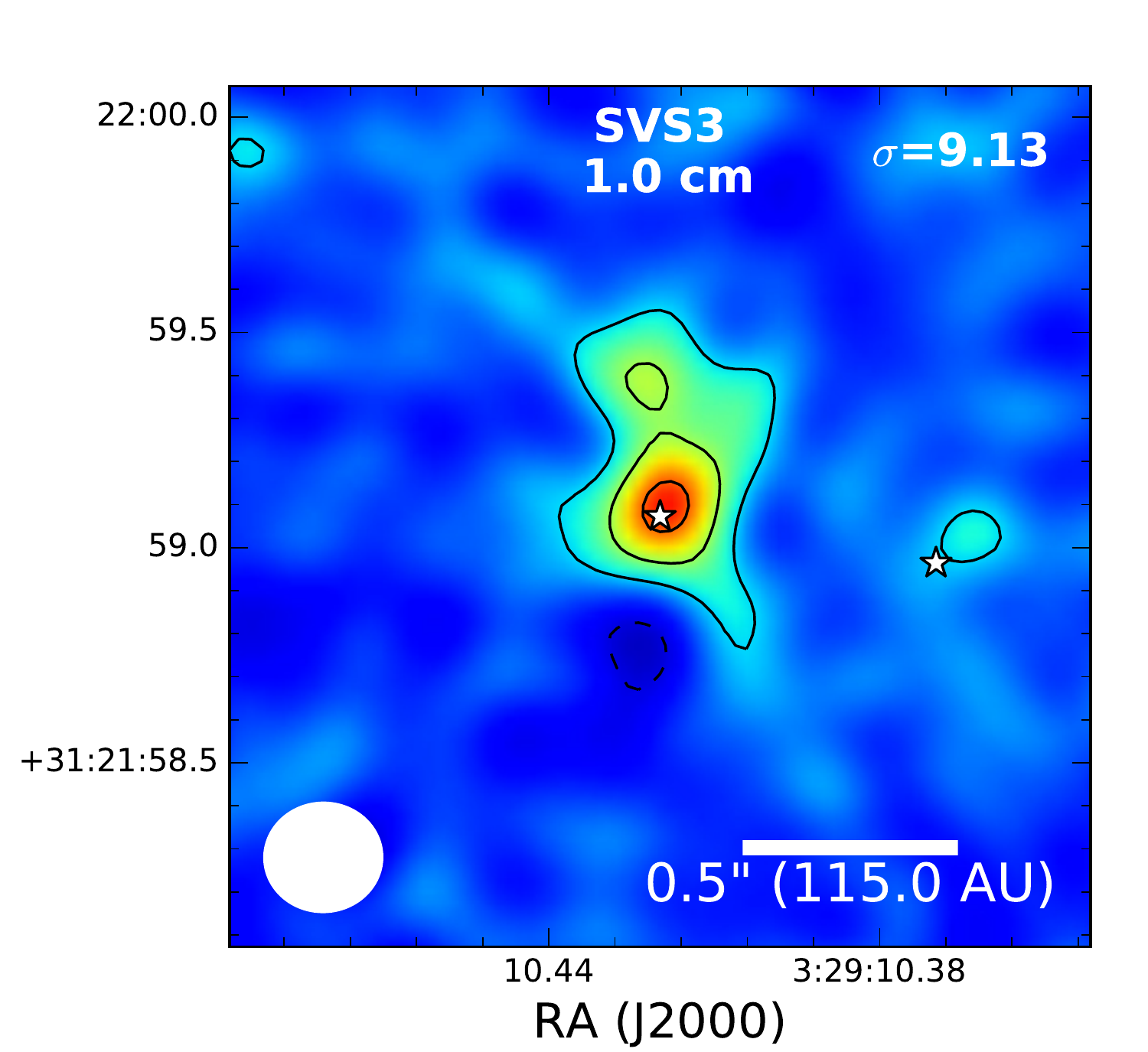}
  \includegraphics[width=0.24\linewidth]{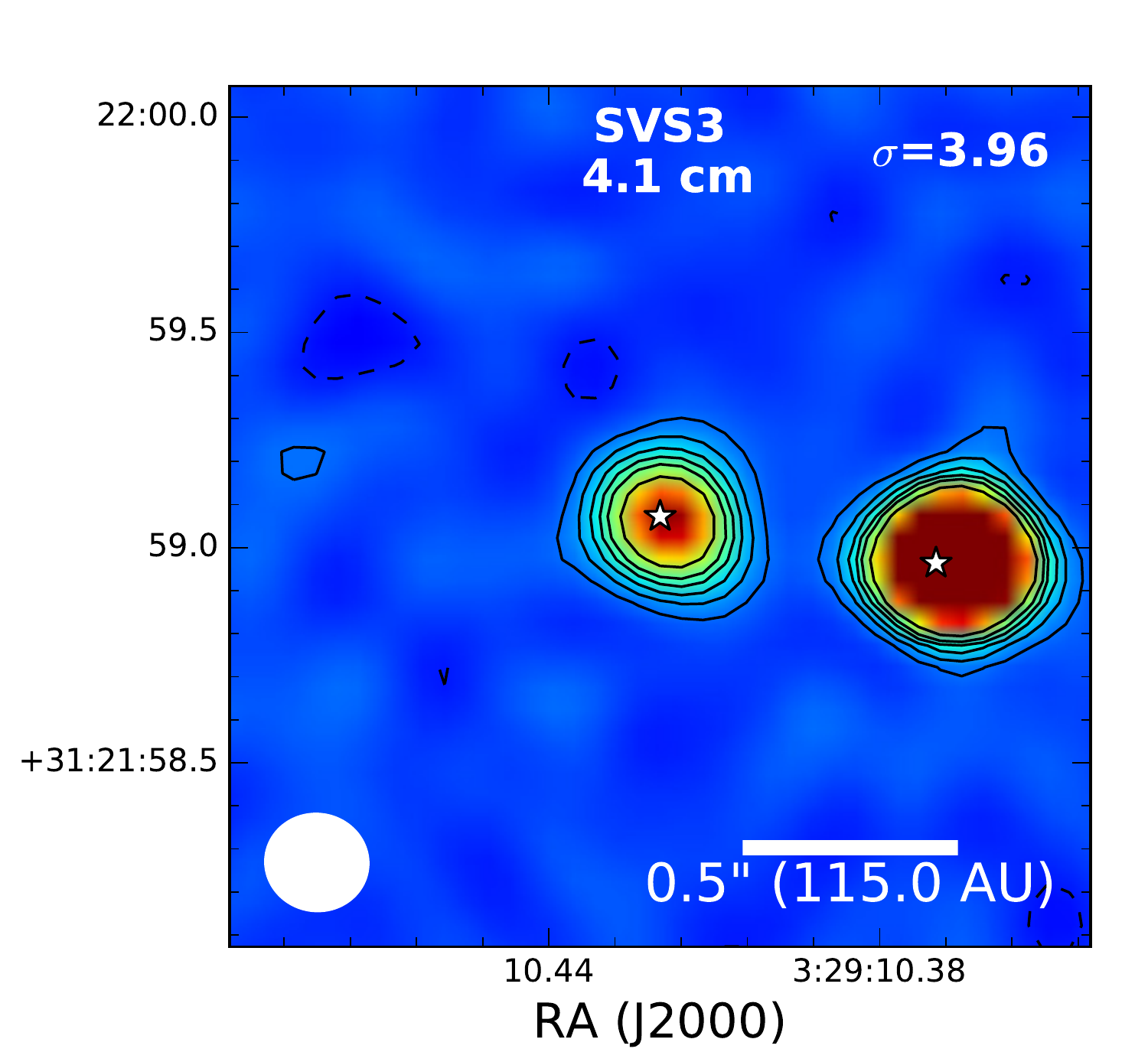}
  \includegraphics[width=0.24\linewidth]{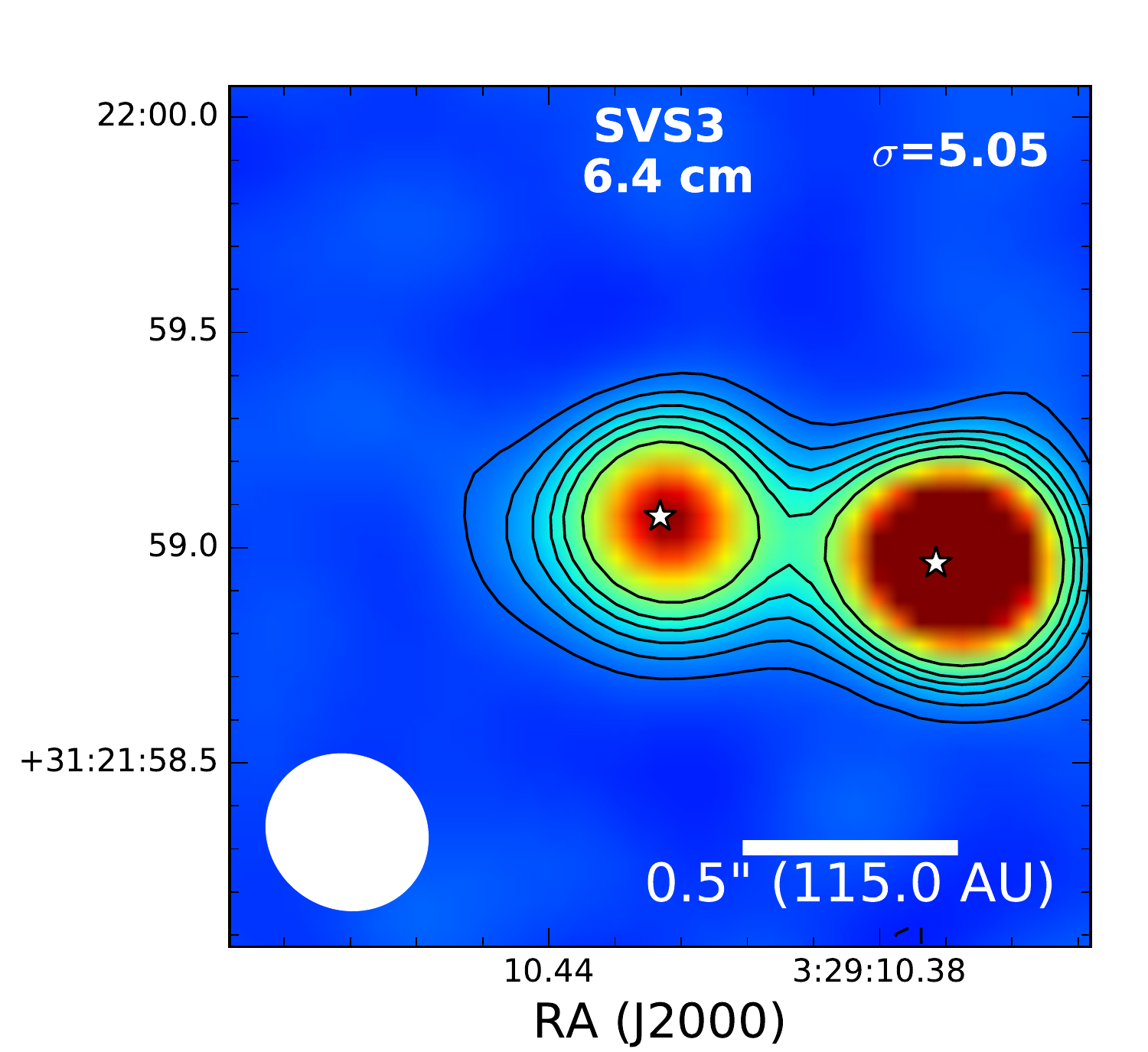}

\end{figure}

\clearpage

\startlongtable
\begin{deluxetable}{lcllllllll}

\tabletypesize{\scriptsize}
\tablecaption{Protostars of the VANDAM survey \label{tab:table1}}
\tablewidth{0pt}
\tablehead{
\colhead{Name$^{a}$} &
\colhead{RA$^{b}$} &
\colhead{Dec$^{b}$} &
\colhead{Field$^{c}$} &
\colhead{Detected} &
\colhead{Detected} &
\colhead{Class} &
\colhead{$L_\textrm{bol}$$^{d}$} &
\colhead{$T_\textrm{bol}$$^{d}$} &\\
\colhead{} &
\colhead{(J2000)} &
\colhead{(J2000)} &
\colhead{} &
\colhead{6.4 cm} &
\colhead{4.1 cm} &
\colhead{} &
\colhead{(L$_{\odot}$)} &
\colhead{(K)} &\\
}
\startdata
     Per-emb-1 & 03:43:56.805 &  +32:00:50.201 &     PerC31 &        Y &        Y &        0 & 1.8$\pm$0.1 & 27.0$\pm$1.0 &\\
     Per-emb-2 & 03:32:17.927 &  +30:49:47.825 &     PerC23 &        Y &        Y &        0 & 0.9$\pm$0.1 & 27.0$\pm$1.0 &\\
     Per-emb-3 & 03:29:00.574 &  +31:12:00.204 &     PerC37 &        Y &        Y &        0 & 0.5$\pm$0.1 & 32.0$\pm$2.0 &\\
     Per-emb-4 & 03:28:39.101 &  +31:06:01.800 &     PerC12 &        - &        - &        0 & 0.2$\pm$0.0 & 31.0$\pm$3.0 &\\
     Per-emb-5 & 03:31:20.938 &  +30:45:30.273 &     PerC24 &        - &        Y &        0 & 1.3$\pm$0.1 & 32.0$\pm$2.0 &\\
     Per-emb-6 & 03:33:14.404 &  +31:07:10.714 &     PerC38 &        Y &        Y &        0 & 0.3$\pm$0.0 & 52.0$\pm$3.0 &\\
     Per-emb-7 & 03:30:32.681 &  +30:26:26.480 &     PerC15 &        - &        - &        0 & 0.1$\pm$0.1 & 37.0$\pm$4.0 &\\
     Per-emb-8 & 03:44:43.981 &  +32:01:35.210 &     PerC25 &        Y &        Y &        0 & 2.6$\pm$0.5 & 43.0$\pm$6.0 &\\
\enddata
\tablenotetext{a}{ Names: Per-emb-XX \citep{Enoch2009}, EDJ2009-XXX \citep{Evans2009}, SVS-X \citep{Strom1976}}
\tablenotetext{b}{ Coordinates from VANDAM Ka-band observations \citep{Tobin2016}}
\tablenotetext{c}{ Name of the observational field where source was closest to the peak of the primary beam response.}
\tablenotetext{d}{Values from  \cite{Enoch2009}, \cite{Sadavoy2014} and \cite{Young2015}}
\end{deluxetable}

\begin{deluxetable}{lcccccccccccccc}
\tabletypesize{\scriptsize}
\tablecaption{Summary of the observed fields \label{tab:table2}}
\tablewidth{0pt}
\tablehead{
\colhead{Name} &
\colhead{Extragalactic} &
\colhead{Protostars} &
\colhead{YSOc} &
\colhead{RMS} &
\colhead{Beam size \& PA} &
\colhead{Extragalactic} &
\colhead{Protostars} &
\colhead{YSOc} &
\colhead{RMS} &
\colhead{Beam size \& PA} &\\
\colhead{} &
\colhead{(6.4 cm)} &
\colhead{(6.4 cm)} &
\colhead{(6.4 cm)} &
\colhead{(6.4 cm)} &
\colhead{(6.4 cm)} &
\colhead{(4 cm)} &
\colhead{(4 cm)} &
\colhead{(4 cm)} &
\colhead{(4 cm)} &
\colhead{(4 cm)} &\\
\colhead{} &
\colhead{} &
\colhead{} &
\colhead{} &
\colhead{(mJy/beam)} &
\colhead{} &
\colhead{} &
\colhead{} &
\colhead{(mJy/beam)} &\\
}
\startdata
         PerC1 &  7 &  5 &  0 &  0.0053 & 0.36" x 0.52"  -76.40 &  9 &  3 &  1 &  0.0041  & 0.23" x 0.34"  -76.03   \\
         PerC2 & 11 &  4 &  0 &  0.0050 & 0.34" x 0.48"  79.18 &  7 &  4 &  0 &  0.0040  & 0.22" x 0.31"  78.94   \\
         PerC3 & 25 &  1 &  0 &  0.0050 & 0.35" x 0.38"  -82.95 & 11 &  1 &  0 &  0.0038  & 0.23" x 0.25"  89.22   \\
         PerC4 & 12 &  2 &  0 &  0.0049 & 0.35" x 0.36"  -84.58 & 13 &  1 &  0 &  0.0042  & 0.23" x 0.24"  87.64   \\
         PerC5 &  9 &  7 &  1 &  0.0051 & 0.35" x 0.39"  62.69 &  8 &  5 &  1 &  0.0040  & 0.22" x 0.24"  83.94   \\
         PerC6 & 11 &  1 &  0 &  0.0050 & 0.35" x 0.50"  -76.67 &  7 &  1 &  0 &  0.0040  & 0.22" x 0.32"  -76.80   \\
         PerC7 & 15 &  3 &  0 &  0.0049 & 0.35" x 0.37"  -83.60 &  9 &  1 &  0 &  0.0041  & 0.23" x 0.24"  -87.92   \\
         PerC8 & 19 & 20 &  0 &  0.0048 & 0.35" x 0.41"  86.58 & 13 & 20 &  1 &  0.0039  & 0.22" x 0.26"  87.31   \\
\enddata
\end{deluxetable}

\clearpage
\startlongtable
\begin{longrotatetable}
\begin{deluxetable}{lllrrrrrrrrrrr}
\tabletypesize{\scriptsize}
\tablecaption{C-band observation results \label{tab:table3}}
\tablewidth{0pt}
\tablehead{
\colhead{Name} &
\colhead{RA} &
\colhead{Dec} &
\colhead{$F_{\nu int}$$^{a}$} &
\colhead{$F_{\nu peak}$$^{b}$} &
\colhead{RMS} &
\colhead{$F_{\nu int}$$^{a}$} &
\colhead{$F_{\nu peak}$$^{b}$} &
\colhead{RMS} &
\colhead{Sp. Index} &
\colhead{Sp. Index} &\\
\colhead{} &
\colhead{} &
\colhead{} &
\colhead{(6.4 cm)} &
\colhead{(6.4 cm)} &
\colhead{(6.4 cm)}&
\colhead{(4.1 cm)} &
\colhead{(4.1 cm)} &
\colhead{(4.1 cm)} &
\colhead{Int.$^{c}$} &
\colhead{Peak$^{d}$}&\\
\colhead{} &
\colhead{(J2000)} &
\colhead{(J2000)} &
\colhead{(mJy)} &
\colhead{(mJy/beam)} &
\colhead{(mJy/beam)} &
\colhead{(mJy)} &
\colhead{(mJy/beam)} &
\colhead{(mJy/beam)} &
\colhead{} &
\colhead{}&\\
}
\startdata
     Per-emb-1 & 03:43:56.805 &  +32:00:50.201 & 0.0508$\pm$0.0065 & 0.0476 & 0.0052  & 0.0651$\pm$0.0059 & 0.0539 & 0.0043  & 0.55$\pm$0.34 & 0.27$\pm$0.30 & \\
     Per-emb-2 & 03:32:17.927 &  +30:49:47.825 & 0.0365$\pm$0.0065 & 0.0278 & 0.0050  & 0.0508$\pm$0.0056 & 0.0426 & 0.0039  & 0.73$\pm$0.46 & 0.94$\pm$0.44 & \\
   Per-emb-2-A & 03:32:17.931 &  +30:49:47.705 & 0.0365$\pm$0.0065 & 0.0278 & 0.0050  & 0.0508$\pm$0.0056 & 0.0426 & 0.0039  & 0.73$\pm$0.46 & 0.94$\pm$0.44 & \\
   Per-emb-2-B & 03:32:17.926 &  +30:49:47.750 & -77.0$\pm$-77.0 & -77.0 & -77.0  & -77.0$\pm$-77.0 & -77.0 & -77.0  & \nodata & \nodata  & \\
     Per-emb-3 & 03:29:00.574 &  +31:12:00.204 & 0.1012$\pm$0.0215 & 0.0784 & 0.0068  & 0.0763$\pm$0.0073 & 0.0673 & 0.0055  & -0.62$\pm$0.51 & -0.33$\pm$0.26 & \\
     Per-emb-4 & 03:28:39.101 &  +31:06:01.800 & \(<\)0.0160$\pm$0.0053 & \(<\)0.0160 & 0.0053  & \(<\)0.0140$\pm$0.0047 & \(<\)0.0140 & 0.0047  & \nodata & \nodata  & \\
     Per-emb-5 & 03:31:20.938 &  +30:45:30.273 & \(<\)0.0197$\pm$0.0000 & \(<\)0.0000 & 0.0000  & 0.0283$\pm$0.0184 & 0.0252 & 0.0040  & \(>\)0.80$\pm$1.44 & \(>\)-99.00$\pm$-99.00 & \\
   Per-emb-5-A & 03:31:20.943 &  +30:45:30.271 & \(<\)0.0197$\pm$0.0066 & \(<\)0.0000 & 0.0066  & 0.0283$\pm$0.0184 & 0.0252 & 0.0040  & \(>\)0.80$\pm$1.61 & \(>\)-99.00$\pm$-99.00 & \\
\enddata
\tablecomments{Flag $-77$ is used when non-detection results from the source being unresolved.}
\tablenotetext{\rm a}{$F_{\nu,{\rm int}}$ - Integrated flux resulting from area under 2D Gaussian fit to the source.}
\tablenotetext{\rm b}{$F_{\nu,{\rm peak}}$ - Peak value of the flux density on the source.}
\tablenotetext{\rm c}Sp. Index (Int.) -  Spectral index calculated with the integrated flux.
\tablenotetext{\rm d}Sp. Index (Peak) - Spectral index calculated with the peak flux.

\end{deluxetable}
\end{longrotatetable}

\clearpage

\startlongtable
\begin{longrotatetable}
\begin{deluxetable}{llrrrrrrrrrrrr}
\tabletypesize{\scriptsize}
\tablecaption{Young Stellar Object candidates \label{tab:table4}}
\tablewidth{0pt}
\tablehead{
\colhead{Name$^a$} &
\colhead{RA} &
\colhead{Dec} &
\colhead{$F_{\nu,int}$} &
\colhead{$F_{\nu,peak}$} &
\colhead{RMS} &
\colhead{$F_{\nu,int}$} &
\colhead{$F_{\nu,peak}$} &
\colhead{RMS} &
\colhead{Sp. Index} &
\colhead{Sp. Index} &\\
\colhead{} &
\colhead{} &
\colhead{} &
\colhead{(6.4 cm)} &
\colhead{(6.4 cm)} &
\colhead{(6.4 cm)}&
\colhead{(4.1 cm)} &
\colhead{(4.1 cm)} &
\colhead{(4.1 cm)} &
\colhead{Int.} &
\colhead{Peak}&\\
\colhead{} &
\colhead{(J2000)} &
\colhead{(J2000)} &
\colhead{(mJy)} &
\colhead{(mJy/beam)} &
\colhead{(mJy/beam)} &
\colhead{(mJy)} &
\colhead{(mJy/beam)} &
\colhead{(mJy/beam)} &
\colhead{} &
\colhead{}&\\
}
\startdata
     FOW2011-6 & 03:44:20.41 &  +32:01:58.65 & 1.0652$\pm$0.0290 & 0.7647 & 0.0199  & \nodata & \nodata & \nodata  & \nodata & \nodata & \\
2MASSJ03283692 & 03:28:36.95 &  +31:17:35.28 & 0.3637$\pm$0.0296 & 0.2805 & 0.0250  & \nodata & \nodata & \nodata  & \nodata & \nodata & \\
   RAC97-VLA38 & 03:28:50.74 &  +31:22:25.27 & 1.5280$\pm$0.0247 & 1.1583 & 0.0176  & \nodata & \nodata & \nodata  & \nodata & \nodata & \\
2MASSJ03285097 & 03:28:50.98 &  +31:23:47.85 & 0.7210$\pm$0.0216 & 0.5409 & 0.0162  & \nodata & \nodata & \nodata  & \nodata & \nodata & \\
2MASSJ03290031 & 03:29:00.33 &  +31:13:38.49 & 0.0895$\pm$0.0133 & 0.0882 & 0.0128  & \nodata & \nodata & \nodata  & \nodata & \nodata & \\
2MASSJ03290462 & 03:29:04.62 &  +31:20:28.74 & \(<\)0.0203$\pm$0.0068 & \(<\)0.0203 & 0.0068  & 0.0347$\pm$0.0064 & 0.0391 & 0.0064  & \(>\)1.18$\pm$0.84 & \(>\)1.44$\pm$0.36 & \\
    OTS2008-68 & 03:29:15.88 &  +31:16:21.49 & 0.1270$\pm$0.0112 & 0.0844 & 0.0068  & 0.0943$\pm$0.0372 & 0.0410 & 0.0095  & -0.66$\pm$0.89 & -1.59$\pm$0.54 & \\
   RAC97-VLA31 & 03:29:16.62 &  +31:16:48.83 & 0.1006$\pm$0.0072 & 0.1026 & 0.0073  & 0.0656$\pm$0.0106 & 0.0796 & 0.0115  & -0.94$\pm$0.39 & -0.56$\pm$0.36 & \\
\enddata
\tablenotetext{a}{References: FOW2011 - \cite{Forbrich2011}, RAC97 - \cite{Rodriguez1999},
2MASS - 2-Micron All Sky Survey \citep{Cutri2003}, l*IC348LRL - \cite{Luhman1998}, EDJ2009 - \cite{Evans2009}, OTS2008 - \cite{Oasa2008}, GMM2008 - \cite{Gutermuth2008}
}
\end{deluxetable}
\end{longrotatetable}

\startlongtable
\begin{longrotatetable}
\begin{deluxetable}{llrrrrrrrrrrrr}
\tabletypesize{\scriptsize}
\rotate
\tablecaption{Known non-protostellar sources \label{tab:table5}}
\tablewidth{0pt}
\tablehead{
\colhead{Type$^a$} &
\colhead{Name$^b$} &
\colhead{RA} &
\colhead{Dec} &
\colhead{$F_{\nu,int}$} &
\colhead{$F_{\nu,peak}$} &
\colhead{RMS} &
\colhead{$F_{\nu,int}$} &
\colhead{$F_{\nu,peak}$} &
\colhead{RMS} &
\colhead{Sp. Index} &
\colhead{Sp. Index} &\\
\colhead{} &
\colhead{} &
\colhead{} &
\colhead{} &
\colhead{(6.4 cm)} &
\colhead{(6.4 cm)} &
\colhead{(6.4 cm)} &
\colhead{(4.1 cm)} &
\colhead{(4.1 cm)} &
\colhead{(4.1 cm)} &
\colhead{Int.} &
\colhead{Peak}&\\
\colhead{} &
\colhead{} &
\colhead{(J2000)} &
\colhead{(J2000)} &
\colhead{(mJy)} &
\colhead{(mJy/b)} &
\colhead{(mJy/b)} &
\colhead{(mJy)} &
\colhead{(mJy/b)} &%
\colhead{(mJy/b)} &
\colhead{} &
\colhead{}&\\
}
\startdata
           *iC & Cl*IC348LRL1888 & 03:44:19.78 &  +31:59:18.85 & 0.0535$\pm$0.0073 & 0.0454 & 0.0055  & 0.0507$\pm$0.0064 & 0.0426 & 0.0045  & -0.11$\pm$0.41 & -0.14$\pm$0.35 & \\
           Rad &      FOW2011-7 & 03:44:31.47 &  +32:00:39.43 & 0.3100$\pm$0.0455 & 0.0685 & 0.0067  & 0.1292$\pm$0.0608 & 0.0312 & 0.0077  & -1.93$\pm$1.08 & -1.37$\pm$0.59 & \\
             * & PSZ2003J034436.5 & 03:44:36.46 &  +32:03:13.60 & 0.1227$\pm$0.0147 & 0.1019 & 0.0108  & \(<\)0.0910$\pm$0.0303 & \(<\)0.0910 & 0.0303  & \(<\)-0.66$\pm$0.78 & \(<\)-0.25$\pm$0.77 & \\
             X &       CXOPZ134 & 03:44:36.90 &  +32:01:23.30 & 1.1031$\pm$0.0188 & 0.9293 & 0.0131  & \nodata & \nodata & \nodata  & \nodata & \nodata & \\
           Rad &    RAC97-VLA37 & 03:28:37.2 &  +31:21:25.45 & 0.5419$\pm$0.0148 & 0.4454 & 0.0121  & \nodata & \nodata & \nodata  & \nodata & \nodata & \\
             X &     WMW2010120 & 03:28:59.66 &  +31:25:42.88 & 0.1378$\pm$0.0125 & 0.1207 & 0.0109  & 0.1812$\pm$0.1133 & 0.1366 & 0.0409  & 0.60$\pm$1.39 & 0.27$\pm$0.69 & \\
           Rad &    RAC97-VLA32 & 03:29:20.70 &  +31:15:49.69 & 0.7378$\pm$0.0531 & 0.4459 & 0.0346  & \nodata & \nodata & \nodata  & \nodata & \nodata & \\
           Rad &      FOW201126 & 03:29:30.93 &  +31:22:11.85 & 1.2263$\pm$0.0273 & 0.8864 & 0.0180  & \nodata & \nodata & \nodata  & \nodata & \nodata & \\
\enddata
\tablenotetext{a}{Types from the SIMBAD catalog: Rad - radio soruce; *iC - star in cluster; * - star; X - X-ray source; 
IR - infrared source; PoC - Part of Cloud; TT* - T-Tauri star; Or* - variable Star of Orion Type; dS* - variable Star of delta Sct Type; ** - double or multiple star}
\tablenotetext{b}{References: Cl*IC348LRL - \cite{Luhman1998}, FOW2011 - \cite{Forbrich2011}, PSZ2003 - \cite{Preibisch2003}, CXOPZ - \cite{Preibisch2001}, RAC97 - \cite{Rodriguez1999},
WMW2010 - \cite{Winston2010}, XMMU - XMM-Newton data Unregistered, EYG2006 - \citep{Enoch2006}, AR2002 - \cite{Anglada2002}, NVSS - The NRAO VLA Sky Survey \citep{Condon1998}, 
SSTc2D - Spitzer c2d Legacy \citep{Young2004}, GFT2002 - \cite{Getman2002}, 2MASS - 2-Micron All Sky Survey \citep{Cutri2003}, RZP2014 - \cite{Rodriguez2014}, LMG94 - \cite{Ladd1994}, RR98 - 
\cite{Rodriguez1998}, KJT2007 - \cite{Kirk2007}, V*VXXX - General Catalog of Variable Stars \citep{Kukarkin1971}, ASR - \cite{Aspin1994}, MBO - \cite{Wilking2004}
}
\end{deluxetable}
\end{longrotatetable}

\startlongtable
\begin{longrotatetable}
\begin{deluxetable}{llrrrrrrrrrrrr}
\tabletypesize{\scriptsize}
\rotate
\tablecaption{New detections reported in \cite{Pech2016} \label{tab:table6}}
\tablewidth{0pt}
\tablehead{
\colhead{Type} &
\colhead{Name} &
\colhead{RA} &
\colhead{Dec} &
\colhead{$F_{\nu,int}$} &
\colhead{$F_{\nu,peak}$} &
\colhead{RMS} &
\colhead{$F_{\nu,int}$} &
\colhead{$F_{\nu,peak}$} &
\colhead{RMS} &
\colhead{Sp. Index} &
\colhead{Sp. Index} &\\
\colhead{} &
\colhead{} &
\colhead{} &
\colhead{} &
\colhead{(6.4 cm)} &
\colhead{(6.4 cm)} &
\colhead{(6.4 cm)} &
\colhead{(4.1 cm)} &
\colhead{(4.1 cm)} &
\colhead{(4.1 cm)} &
\colhead{Int.} &
\colhead{Peak}&\\
\colhead{} &
\colhead{} &
\colhead{(J2000)} &
\colhead{(J2000)} &
\colhead{(mJy)} &
\colhead{(mJy/b)} &
\colhead{(mJy/b)} &
\colhead{(mJy)} &
\colhead{(mJy/b)} &
\colhead{(mJy/b)} &
\colhead{} &
\colhead{}&\\
}
\startdata
J032903.14+312752.6 & 03:29:03.16 &  +31:27:52.58 & 0.0885$\pm$0.0132 & 0.0983 & 0.0151  & \nodata & \nodata & \nodata  & \nodata & \nodata & \\
J032819.46+311831.0 & 03:28:19.48 &  +31:18:31.20 & 0.7570$\pm$0.0345 & 0.4897 & 0.0223  & \nodata & \nodata & \nodata  & \nodata & \nodata & \\
J032825.98+311616.0 & 03:28:25.99 &  +31:16:16.07 & 0.3582$\pm$0.0106 & 0.2967 & 0.0088  & 0.1343$\pm$0.0193 & 0.1180 & 0.0167  & -2.16$\pm$0.32 & -2.03$\pm$0.32 & \\
J032832.41+311245.3 & 03:28:32.43 &  +31:12:45.52 & 0.1830$\pm$0.0213 & 0.1566 & 0.0204  & \nodata & \nodata & \nodata  & \nodata & \nodata & \\
J034327.28+320028.1 & 03:43:27.31 &  +32:00:28.34 & 0.1417$\pm$0.0172 & 0.1319 & 0.0148  & \nodata & \nodata & \nodata  & \nodata & \nodata & \\
J034330.40+320758.4 & 03:43:30.42 &  +32:07:58.84 & 0.2882$\pm$0.0204 & 0.2194 & 0.0172  & \nodata & \nodata & \nodata  & \nodata & \nodata & \\
J034342.10+320225.2 & 03:43:42.11 &  +32:02:25.34 & 0.1072$\pm$0.0094 & 0.0953 & 0.0077  & 0.0733$\pm$0.0356 & 0.0495 & 0.0120  & -0.84$\pm$1.09 & -1.44$\pm$0.56 & \\
J034448.89+320125.0 & 03:44:48.89 &  +32:01:25.07 & 0.1212$\pm$0.0064 & 0.1143 & 0.0051  & 0.0816$\pm$0.0055 & 0.0670 & 0.0041  & -0.87$\pm$0.19 & -1.18$\pm$0.17 & \\
\enddata
\end{deluxetable}
\end{longrotatetable}

\startlongtable
\begin{longrotatetable}
\begin{deluxetable}{llrrrrrrrrrrrr}
\tabletypesize{\scriptsize}
\rotate
\tablecaption{New detections from VANDAM survey \label{tab:table7}}
\tablewidth{0pt}
\tablehead{
\colhead{Name} &
\colhead{RA} &
\colhead{Dec} &
\colhead{$F_{\nu,int}$} &
\colhead{$F_{\nu,peak}$} &
\colhead{RMS} &
\colhead{$F_{\nu,int}$} &
\colhead{$F_{\nu,peak}$} &
\colhead{RMS} &
\colhead{Sp. Index} &
\colhead{Sp. Index} &\\
\colhead{} &
\colhead{} &
\colhead{} &
\colhead{(6.4 cm)} &
\colhead{(6.4 cm)} &
\colhead{(6.4 cm)} &
\colhead{(4.1 cm)} &
\colhead{(4.1 cm)} &
\colhead{(4.1 cm)} &
\colhead{Int.} &
\colhead{Peak}&\\
\colhead{} &
\colhead{(J2000)} &
\colhead{(J2000)} &
\colhead{(mJy)} &
\colhead{(mJy/b)} &
\colhead{(mJy/b)} &
\colhead{(mJy)} &
\colhead{(mJy/b)} &%
\colhead{(mJy/b)} &
\colhead{} &
\colhead{}&\\
}
\startdata
034357.31+315858.86 & 03:43:57.31 &  +31:58:58.86 & 0.0604$\pm$0.0101 & 0.0624 & 0.0100  & \(<\)0.0850$\pm$0.0283 & \(<\)0.0850 & 0.0283  & \(<\)0.75$\pm$0.82 & \(<\)0.68$\pm$0.82 & \\
034408.82+320020.63 & 03:44:08.82 &  +32:00:20.64 & 0.0439$\pm$0.0069 & 0.0406 & 0.0060  & 0.0441$\pm$0.0173 & 0.0286 & 0.0062  & 0.01$\pm$0.93 & -0.77$\pm$0.58 & \\
034415.23+315749.09 & 03:44:15.24 &  +31:57:49.09 & 0.0747$\pm$0.0075 & 0.0694 & 0.0065  & 0.1245$\pm$0.0076 & 0.1177 & 0.0070  & 1.13$\pm$0.26 & 1.16$\pm$0.24 & \\
034422.04+320322.79 & 03:44:22.05 &  +32:03:22.79 & \(<\)0.0208$\pm$0.0069 & \(<\)0.0208 & 0.0069  & 0.0446$\pm$0.0073 & 0.0561 & 0.0084  & \(>\)1.68$\pm$0.82 & \(>\)2.19$\pm$0.33 & \\
034425.40+320249.66 & 03:44:25.41 &  +32:02:49.66 & 0.0367$\pm$0.0066 & 0.0421 & 0.0067  & 0.0459$\pm$0.0181 & 0.0308 & 0.0076  & 0.49$\pm$0.96 & -0.69$\pm$0.65 & \\
032846.12+312050.13 & 03:28:46.12 &  +31:20:50.13 & 0.0428$\pm$0.0082 & 0.0442 & 0.0073  & 0.0516$\pm$0.0293 & 0.0379 & 0.0105  & 0.41$\pm$1.32 & -0.34$\pm$0.71 & \\
032848.98+312313.01 & 03:28:48.98 &  +31:23:13.02 & 0.0455$\pm$0.0079 & 0.0392 & 0.0065  & 0.0319$\pm$0.0232 & 0.0252 & 0.0076  & -0.78$\pm$1.64 & -0.97$\pm$0.76 & \\
032850.33+312101.60 & 03:28:50.33 &  +31:21:01.60 & 0.0551$\pm$0.0078 & 0.0505 & 0.0062  & 0.0498$\pm$0.0200 & 0.0334 & 0.0066  & -0.22$\pm$0.94 & -0.91$\pm$0.51 & \\
\enddata
\tablenotetext{a}{Types from the SIMBAD catalog: Rad - radio soruce; *iC - star in cluster; * - star; X - X-ray source; IR - infrared source; PoC - Part of Cloud; TT* - T-Tauri star; Or* - variable Star of Orion Type; dS* - variable Star of delta Sct Type; ** - double or multiple star}
\end{deluxetable}
\end{longrotatetable}

\clearpage

\begin{deluxetable}{lrrrrrrrrrr}
\tabletypesize{\scriptsize}
\tablecaption{Binary systems \label{tab:table8}}
\tablewidth{0pt}
\tablehead{
\colhead{System} &
\colhead{Separation} &
\colhead{Separation} &
\colhead{$log(F_1/F_2)$} &
\colhead{$log(F_1/F_2)$} &
\colhead{$\Delta \alpha$} &
\colhead{$log(M_\textrm{dust,1}/M_\textrm{dust,2})$} &\\
\colhead{} &
\colhead{(")} &
\colhead{(au)} &
\colhead{(6.4 cm)} &
\colhead{(4.1 cm)} &
\colhead{}&\\
}
\startdata
Per-emb-12 &   1.83 & 429.40 &  -0.28 &  -0.07 &   1.02 &   1.22 & \\
Per-emb-22 &   0.75 & 176.55 &   0.47 &   0.47 &   0.02 &   0.44 & \\
Per-emb-27 &   0.62 & 145.74 &  $>$0.59 &   0.55 & $>$0.18 & 0.68 & \\
Per-emb-33 (A+C) &   0.98 & 229.62 &  -0.73 &  $<$-1.04 & $>$1.57 &  0.66 & \\
 Per-emb-35 &   1.91 & 448.40 &   0.63 &   0.44 &   0.97 &  -0.17 & \\
Per-emb-40 &   0.39 &  91.92 &   0.70 &   0.95 &   1.27 &   0.64 & \\
Per-emb-44 &   0.30 &  70.53 &  -0.04 &  -0.07 &   0.15 &   0.51 & \\
Per-emb-49 &   0.31 &  73.52 &  -0.48 &   0.12 &   3.06 &   0.55 & \\
   L1448NW &   0.25 &  58.98 &   $<$0.03 &   0.43 & $<$1.22 & -0.01 & \\
EDJ2009-269 &   0.54 & 127.78 &          - &   0.26 &   - &  -0.10 & \\
      SVS3 &   0.65 & 152.77 &  -0.21 &  -0.41 &   1.04 &    - & \\
\enddata
\end{deluxetable}

\startlongtable
\begin{deluxetable}{lrrrrrrrl}
\tabletypesize{\scriptsize}
\tablecaption{Protostellar disk masses \label{tab:table9}}
\tablewidth{0pt}
\tablehead{
\colhead{Name} &
\colhead{$F_{\nu}$} &
\colhead{$F_{\nu}$} &
\colhead{$F_{\textrm{dust}}$/$F_{\textrm{9mm}}$} &
\colhead{$M_{\textrm{gas+dust}}$} &
\colhead{$M_{\textrm{gas+dust}}$} &
\colhead{Flag$^a$} &\\
\colhead{} &
\colhead{(9 mm)} &
\colhead{(corrected)} &
\colhead{} &
\colhead{(9 mm)} &
\colhead{(corrected)} &
\colhead{} &\\
\colhead{} &
\colhead{(mJy)} &
\colhead{(mJy)} &
\colhead{\%}&
\colhead{(M$_\odot$)} &
\colhead{(M$_\odot$)} & \\
}
\startdata
     Per-emb-1 &   0.72 &   0.57 &  79 &  0.152$\pm$ 0.004 &   0.121$\pm$ 0.004 &  a &\\
     Per-emb-2 &   2.04 &   1.89 &  92 &  0.432$\pm$ 0.019 &   0.400$\pm$ 0.019 &  a &\\
   Per-emb-2-A & - & - & - & - & - & - &\\
   Per-emb-2-B & - & - & - & - & - & - &\\
     Per-emb-3 &   0.35 &   0.32 &  91 &  0.074$\pm$ 0.004 &   0.068$\pm$ 0.004 &  a &\\
     Per-emb-4 & $<$  0.02 & $<$  0.02 & - & $<$ 0.005$\pm$ 0.002 & $<$ 0.005$\pm$ 0.002 &  e &\\
     Per-emb-5 &   1.05 &   1.02 &  97 &  0.223$\pm$ 0.007 &   0.217$\pm$ 0.007 &  b &\\
   Per-emb-5-A & - & - & - & - & - & - &\\
\enddata
\tablenotetext{a}{Flags: a - regular fit; b - one of the C-band channels with non-detection; c - excess of the free-free emission; d - non-detection in both C-band channels; e - non-detection in both Ka- and C-band;
 f - flat or negative spectral index in Ka-band;  Flags d, e, and f, were treated as upper limits in the analysis.}
\end{deluxetable}

\clearpage

\begin{deluxetable}{lllllll}

\tabletypesize{\footnotesize}
\tablecaption{Correlations between radio flux and disk masses with line luminosities \label{tab:table10}}
\tablewidth{0pt}
\tablehead{
\colhead{} &
\colhead{CO} &
\colhead{H\raisebox{-.4ex}{\scriptsize 2}O} &
\colhead{OH} &
\colhead{[O I]} &
}
\startdata
4.1 cm \\
Class 0   &  0.31 & 0.24  & 0.3                       & 0.27            &\\
Class I  & -0.19 & -0.10 & \textbf{0.64}             &  \textbf{0.52}  &\\
Class 0+I &  0.24 &  0.19 & \underline{\textbf{0.41}} & 0.34            &\\
6.4 cm  \\
Class 0  &  0.26 & 0.39  & \textbf{0.43}        & 0.28 &\\
Class I  & -0.18 & -0.13 & \textbf{0.52}        & {\bf 0.54} &\\
 Class 0+I &  0.16 & 0.28  & \underline{\bf 0.43} & 0.33 & \\
\enddata
\tablecomments {Spearman’s rank correlation coefficient ($\rho$) is listed. Bold font is for correlations $\rho>0.4$. Significant correlations ($P < 5\% $) are underlined.}
\end{deluxetable}

\end{document}